\documentclass[10pt,twocolumn,letterpaper]{pfbook} % book class customized for perfbook
\pdfoutput=1

\ifdefined\pdfsuppresswarningpagegroup \pdfsuppresswarningpagegroup=1 \fi

\usepackage[full]{textcomp} % use symbols in TS1 encoding
\usepackage{lmodern}
\usepackage[T1]{fontenc} % use postscript type 1 fonts
\usepackage[defaultsups,helvratio=0.9]{newtxtext} % use nice, standard fonts for roman

\usepackage{microtype}
\UseMicrotypeSet[protrusion]{basicmath} % disable protrusion for tt fonts
\usepackage{etoolbox}

\usepackage{float}
\floatstyle{ruled}
\newfloat{listing}{tbp}{lst}[chapter]
\floatname{listing}{Listing}
\usepackage{lscape}
\usepackage{epsfig}
\usepackage{subfig}
\newsubfloat{listing}
\usepackage{varwidth}
\usepackage{graphicx}
\usepackage{rotating}
\usepackage{setspace}
\usepackage[shortlabels,inline]{enumitem}
\setlist[description]{style=unboxed}
\newlist{sequence}{enumerate}{10}
\setlist[sequence]{label*=\arabic*}
\usepackage{ifthen}
\usepackage[table,svgnames]{xcolor}
\usepackage[shortcuts]{extdash}
\usepackage{changepage}
\usepackage{listings}
\usepackage{pifont} % symbols for qqz reference points and carriagereturn
\usepackage{gensymb} % symbols for both text and math modes such as \degree and \micro
\usepackage{verbatimbox}[2014/01/30] % for centering verbatim listing in figure environment
\usepackage{amsmath} % lineno v5.0 (loaded via fvextra) needs amsmath in front
\usepackage{fancyvrb}
\usepackage{fvextra}[2016/09/02]
\usepackage[bottom]{footmisc} % place footnotes under floating figures/tables
\usepackage{tabularx}
\usepackage[hyphens]{url}
\usepackage{threeparttable}
\usepackage{titlesec}[2016/03/21] % Suppress number in paragraph heading
\usepackage{fmtcount}
\usepackage{draftwatermark}[2015/02/19]
\SetWatermarkAngle{0.0}
\SetWatermarkFontSize{8pt}
\SetWatermarkScale{1.0}
\SetWatermarkLightness{.6}
\SetWatermarkHorCenter{.85\paperwidth}
\SetWatermarkVerCenter{.95\paperheight}
\SetWatermarkText{\texttt{\commitid}}
\usepackage[breakable,skins]{tcolorbox}
\usepackage[split,makeindex]{splitidx}
\usepackage[nottoc]{tocbibind}
\usepackage[columns=3,totoc,indentunit=12pt,justific=raggedright,font=small,columnsep=.15in]{idxlayout}
\usepackage{parnotes} % for footnotes in tabularx
\usepackage[bookmarks=true,bookmarksnumbered=true,pdfborder={0 0 0},linktoc=all]{hyperref}
\usepackage[all]{hypcap} % for going to the top of figure and table
\usepackage{mfirstuc}[=v2.07] % v2.08 or later is not compatible with our
\usepackage[toc,nopostdot,acronym]{glossaries}[=v4.49]
\usepackage{glossaries-extra}[=v1.48]
\usepackage[longragged]{glossaries-extra-stylemods}[=v1.48]

\usepackage{epigraph}[2020/01/02] % latest version prevents orphaned epigraph
\usepackage[xspace]{ellipsis}
\usepackage{braket} % for \ket{} macro in QC section
\usepackage{siunitx} % for \num{} macro
\usepackage{multirow}
\usepackage{noindentafter}
\NoIndentAfterCmd{\epigraph}
\usepackage[all]{nowidow}
\titleformat{\paragraph}[runin]{\normalfont\normalsize\bfseries}{}{0pt}{}

\newboolean{qqzbg}
\setboolean{qqzbg}{true} % overriden by target specific setting
\newcommand{\IfQqzBg}[2]{\ifthenelse{\boolean{qqzbg}}{#1}{#2}}
\newboolean{noqqz}
\setboolean{noqqz}{false}
\newcommand{\IfNoQqz}[2]{\ifthenelse{\boolean{noqqz}}{#1}{#2}}

\date{June 11, 2023 \\ Release v2023.06.11a }
\newcommand{\commityear}{2023}
\newcommand{\commitid}{v2023.06.11a}
\IfQqzBg{}{\setboolean{qqzbg}{true}}
\tcbsetforeverylayer{autoparskip}
\usepackage{qqz}
\usepackage{origpub}

\newboolean{inbook}
\setboolean{inbook}{true}
\newcommand{\IfInBook}[2]{\ifthenelse{\boolean{inbook}}{#1}{#2}}
\newboolean{twocolumn}
\setboolean{twocolumn}{true}
\newcommand{\IfTwoColumn}[2]{\ifthenelse{\boolean{twocolumn}}{#1}{#2}}
\newboolean{hardcover}
\setboolean{hardcover}{false}
\newcommand{\IfHardCover}[2]{\ifthenelse{\boolean{hardcover}}{#1}{#2}}
\newboolean{ebooksize}
\setboolean{ebooksize}{false}
\newcommand{\IfEbookSize}[2]{\ifthenelse{\boolean{ebooksize}}%
  {\ignorespaces#1\ignorespaces}{\ignorespaces#2\ignorespaces}}
\newboolean{afourpaper}
\setboolean{afourpaper}{false}
\newcommand{\IfAfourPaper}[2]{\ifthenelse{\boolean{afourpaper}}{#1}{#2}}
\newboolean{sansserif}
\setboolean{sansserif}{false}
\newcommand{\IfSansSerif}[2]{\ifthenelse{\boolean{sansserif}}%
  {\ignorespaces#1\ignorespaces}{\ignorespaces#2\ignorespaces}}
\newboolean{lmttforcode}
\setboolean{lmttforcode}{true}
\newcommand{\IfLmttForCode}[2]{\ifthenelse{\boolean{lmttforcode}}{#1}{#2}}
\newboolean{tblcptop}
\setboolean{tblcptop}{true}
\newcommand{\IfTblCpTop}[2]{\ifthenelse{\boolean{tblcptop}}{#1}{#2}}
\newboolean{nimbusavail}
\setboolean{nimbusavail}{false}
\newcommand{\IfNimbusAvail}[2]{\ifthenelse{\boolean{nimbusavail}}{#1}{#2}}
\newboolean{colorlinks}
\setboolean{colorlinks}{false}
\newcommand{\IfColorLinks}[2]{\ifthenelse{\boolean{colorlinks}}{#1}{#2}}
\newboolean{toarxiv}
\setboolean{toarxiv}{true}
\newcommand{\IfToArxiv}[2]{\ifthenelse{\boolean{toarxiv}}{#1}{#2}}
\newboolean{indexon}
\setboolean{indexon}{true}
\newcommand{\IfIndexOn}[2]{\ifthenelse{\boolean{indexon}}{#1}{#2}}
\newboolean{indexhl}
\setboolean{indexhl}{false}
\newcommand{\IfIndexHl}[2]{\ifthenelse{\boolean{indexhl}}{#1}{#2}}
\newboolean{indexhier}
\setboolean{indexhier}{true}
\newcommand{\IfIndexHier}[2]{\ifthenelse{\boolean{indexhier}}{#1}{#2}}

\makeatletter
\IfEbookSize{ % for ebook size build (more than 1000 pages)
\renewcommand*\@pnumwidth{2.2em}
}{}
\IfSansSerif{	% sans serif (Helvetica clone)
\renewcommand*\l@section{\@dottedtocline{1}{1.5em}{2.4em}}
\renewcommand*\l@subsection{\@dottedtocline{2}{3.9em}{3.7em}}
}{		% serif (Times Roman clone)
\renewcommand*\l@subsection{\@dottedtocline{2}{3.8em}{3.4em}}
}
\makeatother

\IfEbookSize{
\usepackage[section]{placeins}
}{
\usepackage{placeins}
}
\newindex[API Index]{api} % index for API
\newindex[People Name Index]{ppl} % index for People Name
\newcommand{\categapi}[1]{~{\scriptsize (#1)}}
\IfIndexHl{
\newcommand{\hlindex}[1]{\textcolor{DarkGreen}{#1}}
}{
\newcommand{\hlindex}[1]{#1}
}
\newcommand{\ucindex}[1]{%
  \lowercase{\def\temp{#1}}%
  \expandafter\index\expandafter{\temp@\makefirstuc{\temp}}%
}
\newcommand{\ucindexh}[2]{%
  \lowercase{\def\temp{#1}}%
  \lowercase{\def\tempb{#2}}%
  \expandafter\index\expandafter{\temp@\makefirstuc{\temp}!\tempb}%
}
\newcommand{\ucindexhm}[2]{%
  \lowercase{\def\temp{#1}}%
  \expandafter\index\expandafter{\temp@\makefirstuc{\temp}!#2}%
}
\newcommand{\indexhraw}[2]{%
  \expandafter\index\expandafter{#1!#2}%
}
\IfIndexHier{
\newcommand{\indexh}[3]{\ucindexh{#2}{#3}}

\newcommand{\indexhmr}[3]{\ucindexhm{#2}{#3}}
}{
\newcommand{\indexh}[3]{\ucindex{#1}}

\newcommand{\indexhmr}[3]{\index{#1}}
}

\newcommand{\IX}[1]{\ucindex{#1}\hlindex{#1}} % put with first letter capitalized into general index
\newcommand{\IXr}[1]{\index{#1}\hlindex{#1}} % put as is into general index
\newcommand{\IXpl}[1]{\ucindex{#1}\hlindex{#1s}} % put with first letter capitalized into general index for plural
\newcommand{\IXplr}[1]{\index{#1}\hlindex{#1s}} % put as is into general index for plural
\newcommand{\IXplx}[2]{\ucindex{#1}\hlindex{#1#2}} % put as is into general index for plural of exeptional form
\newcommand{\IXalt}[2]{\ucindex{#2}\hlindex{#1}} % put alternative with first letter capitalized into general index
\newcommand{\IXaltr}[2]{\index{#2}\hlindex{#1}} % put alternative as is into general index
\newcommand{\IXh}[2]{\indexh{#1 #2}{#2}{#1}\hlindex{#1 #2}}
\newcommand{\IXhpl}[2]{\indexh{#1 #2}{#2}{#1}\hlindex{#1 #2s}}

\newcommand{\IXhmrpl}[2]{\indexhmr{#1 #2}{#2}{#1}\hlindex{#1 #2s}}
\newcommand{\IXalth}[3]{\indexh{#1}{#3}{#2}\hlindex{#1}}

\newcommand{\BF}[1]{\textbf{#1}}
\newcommand{\IXB}[1]{\ucindex{#1|BF}\hlindex{#1}} % put with first letter capitalized into general index
\newcommand{\IXBpl}[1]{\ucindex{#1|BF}\hlindex{#1s}} % put with first letter capitalized into general index for plural
\newcommand{\IXBh}[2]{\indexh{#1 #2|BF}{#2|BF}{#1}\hlindex{#1 #2}}

\newcommand{\IXBhmr}[2]{\indexhmr{#1 #2|BF}{#2|BF}{#1}\hlindex{#1 #2}}

\newcommand{\IXBalth}[3]{\indexh{#1|BF}{#3|BF}{#2}\hlindex{#1}}

\newcommand{\GL}[1]{\underline{#1}}
\newcommand{\IXG}[1]{\ucindex{#1|GL}\hlindex{#1}} % put with first letter capitalized into general index
\newcommand{\IXGr}[1]{\index{#1|GL}\hlindex{#1}} % put as is into general index
\newcommand{\IXGalt}[2]{\ucindex{#2|GL}\hlindex{#1}} % put alternative with first letter capitalized into general index
\newcommand{\IXGaltr}[2]{\index{#2|GL}\hlindex{#1}} % put alternative as is into general index
\newcommand{\IXGh}[2]{\indexh{#1 #2|GL}{#2|GL}{#1}\hlindex{#1 #2}}

\newcommand{\IXGalth}[3]{\indexh{#1|GL}{#3|GL}{#2}\hlindex{#1}}

\newcommand{\IXGalthmr}[3]{\indexhmr{#1|GL}{#3|GL}{#2}\hlindex{#1}}
\newcommand{\apic}[1]{\hlindex{\co{#1}}\sindex[api]{#1@\co{#1}\categapi{c}}}
\newcommand{\apig}[1]{\hlindex{\co{#1}}\sindex[api]{#1@\co{#1}\categapi{g}}}
\newcommand{\apipx}[1]{\hlindex{\co{#1}}\sindex[api]{#1@\co{#1}\categapi{px}}}
\newcommand{\apik}[1]{\hlindex{\co{#1}}\sindex[api]{#1@\co{#1}\categapi{k}}}
\newcommand{\apikh}[1]{\hlindex{\co{#1}}\sindex[api]{#1@\co{#1}\categapi{kh}}}
\newcommand{\apipf}[1]{\hlindex{\co{#1}}\sindex[api]{#1@\co{#1}\categapi{pf}}}
\newcommand{\apiur}[1]{\hlindex{\co{#1}}\sindex[api]{#1@\co{#1}\categapi{ur}}}

\newcommand{\apialtc}[2]{\hlindex{\co{#1}}\sindex[api]{#2@\co{#2}\categapi{c}}}
\newcommand{\apialtg}[2]{\hlindex{\co{#1}}\sindex[api]{#2@\co{#2}\categapi{g}}}
\newcommand{\apialtk}[2]{\hlindex{\co{#1}}\sindex[api]{#2@\co{#2}\categapi{k}}}
\newcommand{\ppl}[2]{\hlindex{#1 #2}\index{#2, #1}} % forename surname in text, "surname, forename" into ppl index
\newcommand{\pplmdl}[2]{\hlindex{#1~#2}\index{#2, #1}} % for abbreviated middle name
\newcommand{\pplsur}[2]{\hlindex{#2}\index{#2, #1}} % surname in text, "surname, givenname" into ppl index
\IfTwoColumn{}{
  \setboolean{colorlinks}{true}
  \IfEbookSize{}{
    \renewcommand\footnotelayout{%
      \advance\leftskip 0.0in
      \advance\rightskip 0.7in
    }
}}

\IfColorLinks{
\hypersetup{colorlinks=true,allcolors=MediumBlue}
}{}

\IfToArxiv{
\hypersetup{
    colorlinks=true,
    linkcolor=black,
    citecolor=black,
    filecolor=black,
    urlcolor=black,
}
}{}

\IfNimbusAvail{
\usepackage{nimbusmononarrow}
}{}
\renewcommand*\ttdefault{lmtt}
\IfEbookSize{
  \newcommand{\OneColumnHSpace}[1]{}
}{
  \newcommand{\OneColumnHSpace}[1]{\IfTwoColumn{}{\hspace*{#1}}}
}

\IfSansSerif{
\renewcommand{\familydefault}{\sfdefault}
\normalfont
\usepackage[slantedGreek,scaled=.96]{newtxsf}
}{
\usepackage[slantedGreek]{newtxmath} % math package to be used with newtxtext
\AtBeginDocument{%
  \DeclareFontShape{\encodingdefault}{\rmdefault}{m}{sl}{<-> ptmro7t}{}%
  \DeclareFontShape{\encodingdefault}{\rmdefault}{b}{sl}{<-> ptmbo7t}{}%
  \DeclareFontShape{\encodingdefault}{\rmdefault}{bx}{sl}{<->ssub * ptm/b/sl}{}%
}
}
\usepackage{biolinum}
\renewcommand{\sfdefault}{qhv}

\newcommand{\LstLineNo}{\makebox[5ex][r]{\arabic{VerbboxLineNo}\hspace{2ex}}}

\usepackage{bm} % for bold math mode fonts --- should be after math mode font choice
\usepackage{booktabs}
\usepackage{arydshln}
\definecolor{lightgray}{gray}{0.9} % for coloring alternate rows in table

\fvset{fontsize=\scriptsize,obeytabs=true}
\IfTwoColumn{
\fvset{tabsize=2}
}{
\fvset{tabsize=8}
}
\DefineVerbatimEnvironment{VerbatimL}{Verbatim}%
{numbers=left,numbersep=5pt,xleftmargin=9pt}
\AfterEndEnvironment{VerbatimL}{\vspace*{-9pt}}
\DefineVerbatimEnvironment{VerbatimLL}{Verbatim}% for snippet inside list
{numbers=left,numbersep=5pt,xleftmargin=9pt}
\AfterEndEnvironment{VerbatimLL}{\vspace*{-5pt}}
\DefineVerbatimEnvironment{VerbatimN}{Verbatim}%
{numbers=left,numbersep=3pt,xleftmargin=5pt,xrightmargin=5pt,frame=single}
\DefineVerbatimEnvironment{VerbatimU}{Verbatim}%
{numbers=none,xleftmargin=5pt,xrightmargin=5pt,samepage=true,frame=single}

\IfLmttForCode{
\AtBeginEnvironment{verbatim}{\renewcommand{\ttdefault}{lmtt}}
\AtBeginEnvironment{verbbox}{\renewcommand{\ttdefault}{lmtt}}
\AtBeginEnvironment{table}{\renewcommand{\ttdefault}{lmtt}}
\AtBeginEnvironment{table*}{\renewcommand{\ttdefault}{lmtt}}
\AtBeginEnvironment{sidewaystable*}{\renewcommand{\ttdefault}{lmtt}}
\AtBeginEnvironment{minipage}{\renewcommand{\ttdefault}{lmtt}}
\AtBeginEnvironment{listing}{\renewcommand{\ttdefault}{lmtt}}
\AtBeginEnvironment{listing*}{\renewcommand{\ttdefault}{lmtt}}
\fvset{fontfamily=lmtt}
}{}

\IfTblCpTop{
\floatstyle{plaintop}
\restylefloat{table}
\addtolength{\abovecaptionskip}{-2.5pt}
\setlength{\abovetopsep}{-2pt}
}{}
\usepackage[capitalise,noabbrev,nosort]{cleveref}
\crefname{subsubsubappendix}{Appendix}{Appendices}
\crefname{sublisting}{Listing}{Listings}
\crefname{sequencei}{step}{steps}
\Crefname{sequencei}{Step}{Steps}
\crefname{enumi}{item}{items}
\Crefname{enumi}{Item}{Items}
\crefname{page}{page}{pages}
\Crefname{page}{Page}{Pages}
\Crefformat{equation}{Equation~#2#1#3}
\crefformat{equation}{Eq.~#2#1#3}

\newcommand{\crefthro}[2]{%
  \namecrefs{#1}~\ref{#1} through~\ref{#2}%
}

\newcounter{lblcount}
\newcommand{\clnrefp}[2]{%
  \setcounter{lblcount}{0}% Restart label count
  \renewcommand*{\do}[1]{\stepcounter{lblcount}}% Count label
  \docsvlist{#1}% Process list and count labels
  \def\nextitem{\def\nextitem{, }}% Separator
  \ifnum\value{lblcount}=1 #2~\lnref{#1}%
  \else\ifnum\value{lblcount}=2 {#2}s~%
  \renewcommand*{\do}[1]{%
    \addtocounter{lblcount}{-1}%
    \ifnum\value{lblcount}=0 { }and~\else\nextitem\fi\lnref{##1}}% How to process each label
  \else {#2}s~%
  \renewcommand*{\do}[1]{%
    \addtocounter{lblcount}{-1}%
    \ifnum\value{lblcount}=0 , and~\else\nextitem\fi\lnref{##1}}% How to process each label
  \fi%
  \docsvlist{#1}% Process list
  \fi%
}
\newcommand{\clnrefpraw}[2]{%
  \setcounter{lblcount}{0}% Restart label count
  \renewcommand*{\do}[1]{\stepcounter{lblcount}}% Count label
  \docsvlist{#1}% Process list and count labels
  \def\nextitem{\def\nextitem{, }}% Separator
  \ifnum\value{lblcount}=1 #2~\lnrefraw{#1}%
  \else\ifnum\value{lblcount}=2 {#2}s~%
  \renewcommand*{\do}[1]{%
    \addtocounter{lblcount}{-1}%
    \ifnum\value{lblcount}=0 { }and~\else\nextitem\fi\lnrefraw{##1}}% How to process each label
  \else {#2}s~%
  \renewcommand*{\do}[1]{%
    \addtocounter{lblcount}{-1}%
    \ifnum\value{lblcount}=0 , and~\else\nextitem\fi\lnrefraw{##1}}% How to process each label
  \fi%
  \docsvlist{#1}% Process list
  \fi%
}
\newcommand{\clnref}[1]{\clnrefp{#1}{line}}
\newcommand{\Clnref}[1]{\clnrefp{#1}{Line}}
\newcommand{\clnrefr}[1]{\clnrefpraw{#1}{line}}

\newcommand{\clnrefrange}[2]{lines~\lnref{#1}\==\lnref{#2}}
\newcommand{\Clnrefrange}[2]{Lines~\lnref{#1}\==\lnref{#2}}
\newcommand{\clnrefthro}[2]{lines~\lnref{#1} through~\lnref{#2}}

\newcommand{\pararef}[1]{Paragraph ``\nameref{#1}'' on Page~\pageref{#1}}

\newlength{\twocolumnwidth}
\newlength{\onecolumntextwidth}
\IfTwoColumn{
  \setlength{\twocolumnwidth}{\columnwidth}
  
  \IfHardCover{
    \usepackage[papersize={8.25in,10.75in},body={6.5in,8.25in},twocolumn,columnsep=0.25in]{geometry}
  }{
    \IfAfourPaper{
      \usepackage[a4paper,body={6.5in,8.25in},twocolumn,columnsep=0.25in]{geometry}
    }{
      \usepackage[letterpaper,body={6.5in,8.25in},twocolumn,columnsep=0.25in]{geometry}      
}}}{ % One Column
  \setlength{\twocolumnwidth}{3.125in}
  \IfEbookSize {
    \usepackage[papersize={4.5in,6.3in},margin=0.2in,footskip=0.2in,
      headsep=0.0335in,headheight=0.1665in,onecolumn,twoside=false]{geometry}
    \sloppy
    \setlength{\onecolumntextwidth}{4.1in}
    \usepackage{fancyhdr}
    \fancypagestyle{plain}{%
      \fancyhf{} % clear all header and footer fields
      
      \rhead{\textcolor{Grey}{\scriptsize\thepage}}
    }
    \pagestyle{plain}
    \titleformat{\chapter}[display]{\normalfont\bfseries}
                {\Large\chaptertitlename~\thechapter}{0pt}{\LARGE}
    \titlespacing*{\chapter}{0pt}{*1}{*2}
  }{
  \IfHardCover{
    \usepackage[papersize={8.25in,10.75in},body={4.75in,8.25in},onecolumn]{geometry}
  }{
    \IfAfourPaper{
    \usepackage[a4paper,body={4.75in,8.25in},onecolumn]{geometry}
    }{
    \usepackage[letterpaper,body={4.75in,8.25in},onecolumn]{geometry}
  }}}
  \geometry{hcentering=true} % horizontal centering for 1c layouts
}
\IfAfourPaper{
  \geometry{vcentering=true} % vertical centering for A4 paper
}{
  \geometry{vmarginratio=3:4}
}

\IfAfourPaper{
\SetWatermarkVerCenter{.92\paperheight}
}{}

\IfHardCover{
\SetWatermarkVerCenter{.95\paperheight}
}{}

\IfEbookSize{
\SetWatermarkHorCenter{.8\paperwidth}
\SetWatermarkVerCenter{.99\paperheight}
\newsavebox\ebbox
\newcommand{\ebresizewidth}[1]{\resizebox{\textwidth}{!}{#1}}
\newcommand{\ebresizewidthsw}[1]{\resizebox{.95\textheight}{!}{#1}}
\newcommand{\ebresizeverb}[2]{%
  \begin{lrbox}{\ebbox}%
    \begin{varwidth}{\textwidth}%
      {#2}%
    \end{varwidth}%
  \end{lrbox}%
  \resizebox{#1\textwidth}{!}{\usebox{\ebbox}}%
  \vspace*{-7pt}%
}
\newcommand\ebFloatBarrier{\FloatBarrier}
}{
\newcommand{\ebresizewidth}[1]{#1}
\newcommand{\ebresizewidthsw}[1]{#1}
\newcommand{\ebresizeverb}[2]{#2}
\newcommand\ebFloatBarrier{}
}

\IfTwoColumn{
\newcommand{\tcresizewidth}[1]{\resizebox{\columnwidth}{!}{#1}}
}{
\newcommand{\tcresizewidth}[1]{#1}
}

\newabbreviationstyle{pf-long-short}{
  \glsxtrAccSuppAbbrSetFirstLongAttrs\glscategorylabel
  \renewcommand*{\CustomAbbreviationFields}{%
    name={\glsxtrlongshortname},
    sort={\the\glsshorttok},
    first={\protect\glsfirstlongemfont{\the\glslongtok}%
     \protect\glsxtrfullsep{\the\glslabeltok}%
     \glsxtrparen{\protect\glsfirstabbrvfont{\the\glsshorttok}}},%
    firstplural={\protect\glsfirstlongemfont{\the\glslongpltok}%
     \protect\glsxtrfullsep{\the\glslabeltok}%
     \glsxtrparen{\protect\glsfirstabbrvfont{\the\glsshortpltok}}},%
    text={\protect\glsabbrvfont{\the\glsshorttok}},%
    plural={\protect\glsabbrvfont{\the\glsshortpltok}},%
    description={\protect\glslongfont{\the\glslongtok}},
    user1={\protect\index{\the\glslongtok\space(\the\glsshorttok)%
      @\makefirstuc{\the\glslongtok}\space(\the\glsshorttok)}},%
    user2={\protect\index{\the\glslongtok\space(\the\glsshorttok)%
      @\makefirstuc{\the\glslongtok}\space[\the\glsshorttok]}},%
    user3={\protect\index{\the\glslongtok\space(\the\glsshorttok)%
      @\makefirstuc{\the\glslongtok}\space<\the\glsshorttok>}}%
  }%
  \renewcommand*{\GlsXtrPostNewAbbreviation}{%
    \glshasattribute{\the\glslabeltok}{regular}%
    {%
      \glssetattribute{\the\glslabeltok}{regular}{false}%
    }%
    {}%
  }%
}%
{%
}

\newabbreviationstyle{pf-long-short-mod}{
  \glsxtrAccSuppAbbrSetFirstLongAttrs\glscategorylabel
  \renewcommand*{\CustomAbbreviationFields}{%
    name={\glsxtrlongshortname},
    sort={\the\glsshorttok},
    first={\protect\glsfirstlongemfont{\the\glslongtok}%
     \protect\glsxtrfullsep{\the\glslabeltok}%
     \glsxtrparen{\protect\glsfirstabbrvfont{\the\glsshorttok}}},%
    firstplural={\protect\glsfirstlongemfont{\the\glslongpltok}%
     \protect\glsxtrfullsep{\the\glslabeltok}%
     \glsxtrparen{\protect\glsfirstabbrvfont{\the\glsshortpltok}}},%
    text={\protect\glsabbrvfont{\the\glsshorttok}},%
    plural={\protect\glsabbrvfont{\the\glsshortpltok}},%
    description={\protect\glslongfont{\the\glslongtok}}
  }%
  \renewcommand*{\GlsXtrPostNewAbbreviation}{%
    \glshasattribute{\the\glslabeltok}{regular}%
    {%
      \glssetattribute{\the\glslabeltok}{regular}{false}%
    }%
    {}%
  }%
}%
{%
}

\setabbreviationstyle[acronym]{pf-long-short}
\setabbreviationstyle{pf-long-short-mod}

\IfIndexOn{
\newcommand{\acr}[1]{\gls{#1}} % print acronym via dictionary
\newcommand{\acrpl}[1]{\glspl{#1}} % print acronym via dictionary (plural)
\newcommand{\acrl}[1]{\glsxtrlong{#1}} % print acronym via dictionary (long form)
\newcommand{\acrlpl}[1]{\glsxtrlongpl{#1}} % print acronym via dictionary (long form, plural)
\newcommand{\acrf}[1]{\glsxtrlong{#1} (\glsxtrshort{#1})} % print acronym via dictionary (full form)
\newcommand{\acrfpl}[1]{\glsxtrlongpl{#1} (\glsxtrshortpl{#1})} % print acronym via dictionary (full form, plural)
\newcommand{\Acrfpl}[1]{\Glsxtrlongpl{#1} (\glsxtrshortpl{#1})} % print acronym via dictionary (full form, upper case, plural)
\newcommand{\acrfst}[1]{\glsreset{#1}\gls{#1}} % print acronym via dictionary (first form)
\newcommand{\Acrfst}[1]{\glsreset{#1}\Gls{#1}} % print acronym via dictionary (first form, upper case)
\newcommand{\acrml}[1]{\glsxtrlong*{#1:m}} % print acronym via dictionary (long form)
\newcommand{\acrmf}[1]{\glsxtrlong*{#1:m} (\glsxtrshort{#1})} % print acronym via dictionary (full form)
\newcommand{\acrmfst}[1]{\emph{\glsxtrlong*{#1:m}} (\glsxtrshort{#1})} % print acronym via dictionary (first form)
\newcommand{\Acrmfst}[1]{\emph{\Glsxtrlong*{#1:m}} (\glsxtrshort{#1})} % print acronym via dictionary (first form, upper case)
}{
\newcommand{\acr}[1]{\gls*{#1}} % print acronym via dictionary
\newcommand{\acrpl}[1]{\glspl*{#1}} % print acronym via dictionary (plural)
\newcommand{\acrl}[1]{\glsxtrlong*{#1}} % print acronym via dictionary (long form)
\newcommand{\acrlpl}[1]{\glsxtrlongpl*{#1}} % print acronym via dictionary (long form, plural)
\newcommand{\acrf}[1]{\glsxtrlong*{#1} (\glsxtrshort*{#1})} % print acronym via dictionary (full form)
\newcommand{\acrfpl}[1]{\glsxtrlongpl*{#1} (\glsxtrshortpl*{#1})} % print acronym via dictionary (full form, plural)
\newcommand{\Acrfpl}[1]{\Glsxtrlongpl*{#1} (\glsxtrshortpl*{#1})} % print acronym via dictionary (full form, upper case, plural)
\newcommand{\acrfst}[1]{\glsreset{#1}\gls*{#1}} % print acronym via dictionary (first form)
\newcommand{\Acrfst}[1]{\glsreset{#1}\Gls*{#1}} % print acronym via dictionary (first form, upper case)
\newcommand{\acrml}[1]{\acrl{#1:m}} % print acronym via dictionary (long form)
\newcommand{\acrmf}[1]{\acrf{#1:m}} % print acronym via dictionary (full form)
\newcommand{\acrmfst}[1]{\acrfst{#1:m}} % print acronym via dictionary (first form)
\newcommand{\Acrmfst}[1]{\Acrfst{#1:m}} % print acronym via dictionary (first form, upper case)
}

\newcommand{\IXacr}[1]{\glsuseri{#1}\acr{#1}} % put index via acronym dictionary
\newcommand{\IXacrpl}[1]{\glsuseri{#1}\acrpl{#1}} % put index via acronym dictionary (plural)
\newcommand{\IXacrl}[1]{\glsuseri{#1}\acrl{#1}} % put index via acronym dictionary (long form)
\newcommand{\IXacrlpl}[1]{\glsuseri{#1}\acrlpl{#1}} % put index via acronym dictionary (long form, plural, plural)
\newcommand{\IXacrf}[1]{\glsuseri{#1}\acrf{#1}} % put index via acronym dictionary (full form)
\newcommand{\IXacrfpl}[1]{\glsuseri{#1}\acrfpl{#1}} % put index via acronym dictionary (full form, plural)
\newcommand{\IXAcrfpl}[1]{\glsuseri{#1}\Acrfpl{#1}} % put index via acronym dictionary (full form, upper case, plural)
\newcommand{\IXacrfst}[1]{\glsuseri{#1}\acrfst{#1}} % put index via acronym dictionary (first form)
\newcommand{\IXBacrfst}[1]{\glsuserii{#1}\acrfst{#1}} % put index via acronym dictionary (first form)
\newcommand{\IXacrml}[1]{\glsuseri{#1}\acrml{#1}} % put index via acronym dictionary (long form)
\newcommand{\IXacrmf}[1]{\glsuseri{#1}\acrmf{#1}} % put index via acronym dictionary (full form)
\newcommand{\IXacrmfst}[1]{\glsuseri{#1}\acrmfst{#1}} % put index via acronym dictionary (first form)
\newcommand{\IXAcrmfst}[1]{\glsuseri{#1}\Acrmfst{#1}} % put index via acronym dictionary (first form, upper case)
\newcommand{\IXaltacr}[2]{\glsuseri{#2}\hlindex{#1}} % put index of acronym #2 to alternative string (#1)

\newacronym{cas}{CAS}{compare and swap}
\newabbreviation{cas:m}{CAS}{compare-and-swap}
\newacronym{cbmc}{CBMC}{C~bounded model checker}
\newacronym{dcas}{DCAS}{double compare-and-swap}
\newabbreviation{dcas:m}{DCAS}{double-compare-and-swap}
\newacronym{dsp}{DSP}{digital signal processor}
\newacronym{ebr}{EBR}{epoch-based reclamation}
\newacronym{fpga}{FPGA}{field\-/programmable gate array}
\newacronym{gpgpu}{GPGPU}{general\-/purpose graphical processing unit}
\newacronym{irq}{IRQ}{interrupt request}
\newacronym{ipi}{IPI}{inter\-/processor interrupt}
\newacronym{kcsan}{KCSAN}{kernel concurrency sanitizer}
\newacronym{lkmm}{LKMM}{Linux kernel memory consistency model}
\newacronym{mpi}{MPI}{Message Passing Interface}
\newacronym{nbs}{NBS}{non-blocking synchronization}
\newacronym{nmi}{NMI}{non-maskable interrupt}
\newacronym{nuca}{NUCA}{non-uniform cache architecture}
\newacronym{numa}{NUMA}{non-uniform memory architecture}
\newacronym{qsbr}{QSBR}{quiescent-state-based reclamation}
\newabbreviation{qsbr:m}{QSBR}{quiescent-state-based-reclamation}
\newacronym{raii}{RAII}{resource acquisition is initialization}
\newabbreviation{raii:m}{RAII}{resource-acquisition-is-initialization}
\newacronym{rcu}{RCU}{read-copy update}
\newacronym{smp}{SMP}{symmetric multiprocessing}
\newabbreviation{smp:m}{SMP}{symmetric-multiprocessing}
\newacronym{sql}{SQL}{Structured Query Language}
\newacronym{tle}{TLE}{transactional lock elision}
\newacronym{tm}{TM}{transactional memory}
\newacronym{htm}{HTM}{hardware transactional memory}
\newacronym{stm}{STM}{software transactional memory}
\newacronym{upc}{UPC}{Universal Parallel~C}
\newacronym{utm}{UTM}{unbounded transactional memory}
\newabbreviation{utm:m}{UTM}{unbounded-transactional-memory}

\glsunsetall

\makeglossaries

\begin{document}

%%HTMLSKIP
\lstset{
 literate={\_}{}{0\discretionary{\_}{}{\_}}%
  {\_\_}{}{0\discretionary{\_\_}{}{\_\_}}%
  {->}{}{0\discretionary{->}{}{->}}%
}
%%HTMLNOSKIP
\newcommand{\co}[1]{\lstinline[breaklines=true,breakatwhitespace=true]{#1}}
\newcommand{\nbco}[1]{\hbox{\lstinline[breaklines=false,breakatwhitespace=false]{#1}}} % no break lines for short snippet
\newcommand{\qco}[1]{``\nbco{#1}''} % \nbco with quotation marks
\newcommand{\tco}[1]{\texttt{\detokenize{#1}}} % for code in tabular environment
% \tco{} will break at spaces but not at underscores
\newcommand{\qtco}[1]{``\hbox{\tco{#1}}''} % \tco with quotation marks
\newcommand{\lopt}[1]{\tco{-}\tco{-}\tco{#1}} % to avoid "--" to endash conversion
\newcommand{\nf}[1]{\textnormal{#1}} % to return to normal font
\newcommand{\qop}[1]{{\sffamily #1}} % QC operator such as H, T, S, etc.

\DeclareRobustCommand{\euler}{\ensuremath{\mathrm{e}}}
\DeclareRobustCommand{\O}[1]{\ensuremath{\mathcal{O}\left(#1\right)}}
\DeclareRobustCommand{\Node}[1]{Node~{\ensuremath{#1}}}
\newcommand{\Power}[1]{POWER#1}
\newcommand{\ARM}[1]{Arm{#1}}
\newcommand{\ARMv}[1]{Armv{#1}}
\newcommand{\GNUC}{GNU~C}
\newcommand{\GCC}{GCC}
\newcommand{\IRQ}{IRQ}
\newcommand{\rt}{\mbox{-rt}} % to prevent line break behind "-"

\let\epigraphorig\epigraph
\renewcommand{\epigraph}[2]{\epigraphorig{\biolinum\emph{#1}}{\biolinum\scshape\footnotesize #2}}
\IfEbookSize{
  \newcommand{\Epigraph}[2]{\epigraph{#1}{#2}}
}{
  \newcommand{\Epigraph}[2]{\epigraphhead[65]{\epigraph{#1}{#2}}}
}

\input{ushyphex} % Hyphenation exceptions for US English from hyphenex package
\hyphenation{
  REQACK
  COPART
  GPGPU
  GPGPUs
  Schr\"o-din-ger
}
 % Hyphenation exceptions for perfbook

\title{
  Is Parallel Programming Hard, And, If So, \\
  What Can You Do About It?}
\author{
	Edited by: \\
	\\
	Paul E. McKenney \\
	Facebook \\
	\href{mailto:paulmck@kernel.org}{paulmck@kernel.org} \\
} % end author
% \date{\ }

\setcounter{secnumdepth}{4} % Enable counter for paragraph
%\fvset{fontsize=\scriptsize,numbers=left,numbersep=5pt,xleftmargin=9pt,obeytabs=true,tabsize=2}
\newcommand{\lnlblbase}{}
\newcommand{\lnlbl}[1]{\phantomsection\label{\lnlblbase:#1}}
\newlength{\lnlblraise}
\setlength{\lnlblraise}{6pt}
\AtBeginEnvironment{VerbatimN}{%
\renewcommand{\lnlbl}[1]{%
\raisebox{\lnlblraise}{\phantomsection\label{\lnlblbase:#1}}}%
}
\newcommand{\lnrefbase}{}
\newcommand{\lnref}[1]{\ref{\lnrefbase:#1}}
\newcommand{\lnrefraw}[1]{\ref{#1}}

\newenvironment{fcvlabel}[1][]{\renewcommand{\lnlblbase}{#1}%
\ignorespaces}{\ignorespacesafterend}
\newenvironment{fcvref}[1][]{\renewcommand{\lnrefbase}{#1}%
\ignorespaces}{\ignorespacesafterend}

\frontmatter

\maketitle
\IfTwoColumn{
  \onecolumn\begin{adjustwidth*}{.95in}{.8in}
  \addtolength{\parindent}{6pt}
}{}
% legal.tex
% mainfile: perfbook.tex
% SPDX-License-Identifier: CC-BY-SA-3.0

\section*{Legal Statement}

This work represents the views of the editor and the authors and does not
necessarily represent the view of their respective employers. \\
\\
Trademarks:
\begin{itemize}
\item	IBM, z~Systems, and PowerPC are trademarks or registered trademarks
	of International Business Machines Corporation in the United
	States, other countries, or both.
\item	Linux is a registered trademark of Linus Torvalds.
\item	Intel, Itanium, Intel Core, and Intel Xeon are trademarks
	of Intel Corporation or its subsidiaries in the United States,
	other countries, or both.
\item	Arm is a registered trademark of Arm Limited (or its subsidiaries)
	in the US and/or elsewhere.
\item	SPARC is a registered trademark of SPARC International, Inc.
	Products bearing SPARC trademarks are based on an architecture
	developed by Sun Microsystems, Inc.
\item	Other company, product, and service names may be trademarks or
	service marks of such companies.
\end{itemize}

The non-source-code text and images in this document are provided under
the terms of the Creative Commons Attribution-Share Alike 3.0 United
States license.\footnote{
	\url{https://creativecommons.org/licenses/by-sa/3.0/us/}}
In brief, you may use the contents of this document for any purpose,
personal, commercial, or otherwise, so long as attribution to the
authors is maintained.
Likewise, the document may be modified, and derivative works and
translations made available, so long as such modifications and
derivations are offered to the public on equal terms as the
non-source-code text and images in the original document.

Source code is covered by various versions of the GPL\@.\footnote{
	\url{https://www.gnu.org/licenses/gpl-2.0.html}}
Some of this code is GPLv2-only, as it derives from the Linux kernel,
while other code is GPLv2-or-later.
See the comment headers of the individual source files within the
CodeSamples directory in the git archive\footnote{
	\url{git://git.kernel.org/pub/scm/linux/kernel/git/paulmck/perfbook.git}}
for the exact licenses.
If you are unsure of the license for a given code fragment,
you should assume GPLv2-only.

Combined work {\textcopyright}~2005--\commityear\ by Paul E. McKenney.
Each individual contribution is copyright by its contributor at the time
of contribution, as recorded in the git archive.

\tableofcontents
\IfTwoColumn{
  \end{adjustwidth*}\twocolumn
}{}

\mainmatter

% howto/howto.tex
% mainfile: ../perfbook.tex
% SPDX-License-Identifier: CC-BY-SA-3.0

\QuickQuizChapter{chp:How To Use This Book}{How To Use This Book}{qqzhowto}
\Epigraph{If you would only recognize that life is hard, things would be so
	  much easier for you.}{Louis D. Brandeis}

The purpose of this book is to help you program
shared-memory parallel systems without risking your sanity.\footnote{
	Or, perhaps more accurately, without much greater risk to your
	sanity than that incurred by non-parallel programming.
	Which, come to think of it, might not be saying all that much.}
Nevertheless, you should think of the information in this book as a
foundation on which to build, rather than as a completed cathedral.
Your mission, if you choose to accept, is to help make further progress
in the exciting field of parallel programming---progress that will
in time render this book obsolete.

Parallel programming in the 21\textsuperscript{st} century is no longer
focused solely on science, research, and grand-challenge projects.
And this is all to the good, because it means that parallel programming
is becoming an engineering discipline.
Therefore, as befits an engineering discipline, this book examines
specific parallel-programming tasks and describes how to approach them.
In some surprisingly common cases, these tasks can be automated.

This book is written in the hope that presenting the engineering
discipline underlying successful
parallel-programming projects will free a new generation of parallel hackers
from the need to slowly and painstakingly reinvent old wheels, enabling
them to instead focus their energy and creativity on new frontiers.
However, what you get from this book will be determined by what you
put into it.
It is hoped that simply reading this book will be helpful,
and that working the Quick Quizzes will be even more helpful.
However, the best results come from applying the techniques taught
in this book to real-life problems.
As always, practice makes perfect.

But no matter how you approach it, we sincerely hope that parallel
programming brings you at least as much fun, excitement, and challenge
that it has brought to us!

\section{Roadmap}
\label{sec:howto:Roadmap}
\epigraph{Cat:
		Where are you going? \\
	  Alice:
		Which way should I go? \\
	  Cat:
		That depends on where you are going. \\
	  Alice:
		I don't know. \\
	  Cat:
		Then it doesn't matter which way you go.}
	 {Lewis Carroll, \emph{Alice in Wonderland}}

This book is a handbook of widely applicable and heavily
used design techniques, rather than
a collection of optimal algorithms with tiny areas of applicability.
You are currently reading \cref{chp:How To Use This Book}, but
you knew that already.
\Cref{chp:Introduction} gives a high-level overview of parallel
programming.

\Cref{chp:Hardware and its Habits} introduces shared-memory
parallel hardware.
After all, it is difficult to write good parallel code unless you
understand the underlying hardware.
Because hardware constantly evolves, this chapter will always be
out of date.
We will nevertheless do our best to keep up.
\Cref{chp:Tools of the Trade} then provides a very brief overview
of common shared-memory parallel-programming primitives.

\Cref{chp:Counting} takes an in-depth look at parallelizing
one of the simplest problems imaginable, namely counting.
Because almost everyone has an excellent grasp of counting, this chapter
is able to delve into many important parallel-programming issues without
the distractions of more-typical computer-science problems.
My impression is that this chapter has seen the greatest use in
parallel-programming coursework.

\Cref{chp:Partitioning and Synchronization Design}
introduces a number of design-level methods of addressing the issues
identified in \cref{chp:Counting}.
It turns out that it is important to address parallelism at
the design level when feasible:
To paraphrase \pplsur{Edsger W.}{Dijkstra}~\cite{Dijkstra:1968:LEG:362929.362947},
``retrofitted parallelism considered grossly
suboptimal''~\cite{PaulEMcKenney2012HOTPARsuboptimal}.

The next three chapters examine three important approaches to
synchronization.
\Cref{chp:Locking} covers locking, which is still not only the
workhorse of production-quality parallel programming, but is also widely
considered to be parallel programming's worst villain.
\Cref{chp:Data Ownership} gives a brief overview of data ownership,
an often overlooked but remarkably pervasive and powerful approach.
Finally, \cref{chp:Deferred Processing} introduces a number of
deferred-processing mechanisms, including reference counting,
hazard pointers, sequence locking, and RCU\@.

\Cref{chp:Data Structures} applies the lessons of previous
chapters to hash tables, which are heavily used due
to their excellent partitionability, which (usually) leads to excellent
performance and scalability.

As many have learned to their sorrow, parallel programming without
validation is a sure path to abject failure.
\Cref{chp:Validation} covers various forms of testing.
It is of course impossible to test reliability into your program
after the fact, so \cref{chp:Formal Verification}
follows up with a brief overview of a couple of practical approaches to
formal verification.

\Cref{chp:Putting It All Together}
contains a series of moderate-sized parallel programming problems.
The difficulty of these problems vary, but should be appropriate for
someone who has mastered the material in the previous chapters.

\Cref{sec:advsync:Advanced Synchronization}
looks at advanced synchronization methods, including
non-blocking synchronization and parallel real-time computing,
while \cref{chp:Advanced Synchronization: Memory Ordering}
covers the advanced topic of memory ordering.
\Cref{chp:Ease of Use} follows up with some ease-of-use advice.
\Cref{chp:Conflicting Visions of the Future}
looks at a few possible future directions, including
shared-memory parallel system design, software and hardware transactional
memory, and functional programming for parallelism.
Finally, \cref{chp:Looking Forward and Back} reviews the material in
this book and its origins.

This chapter is followed by a number of appendices.
The most popular of these appears to be
\cref{chp:app:whymb:Why Memory Barriers?},
which delves even further into memory ordering.
\Cref{chp:app:Answers to Quick Quizzes}
contains the answers to the infamous Quick Quizzes, which are discussed in
the next section.

\section{Quick Quizzes}
\label{sec:howto:Quick Quizzes}
\epigraph{Undertake something difficult, otherwise you will never grow.}
	 {Abbreviated from Ronald E.~Osburn}

``Quick quizzes'' appear throughout this book, and the answers may
be found in
\cref{chp:app:Answers to Quick Quizzes} starting on
\cpageref{chp:app:Answers to Quick Quizzes}.
Some of them are based on material in which that quick quiz
appears, but others require you to think beyond that section, and,
in some cases, beyond the realm of current knowledge.
As with most endeavors, what you get out of this book is largely
determined by what you are willing to put into it.
Therefore, readers who make a genuine effort to solve a quiz before
looking at the answer
find their effort repaid handsomely with increased understanding
of parallel programming.

\QuickQuizSeries{%
\QuickQuizB{
	Where are the answers to the Quick Quizzes found?
}\QuickQuizAnswerB{
	In \cref{chp:app:Answers to Quick Quizzes} starting on
	\cpageref{chp:app:Answers to Quick Quizzes}.
	Hey, I thought I owed you an easy one!
}\QuickQuizEndB
\QuickQuizM{
	Some of the Quick Quiz questions seem to be from the viewpoint
	of the reader rather than the author.
	Is that really the intent?
}\QuickQuizAnswerM{
	Indeed it is!
	Many are questions that Paul E.~McKenney would probably have
	asked if he was a novice student in a class covering this material.
	It is worth noting that Paul was taught most of this material by
	parallel hardware and software, not by professors.
	In Paul's experience, professors are much more likely to provide
	answers to verbal questions than are parallel systems, recent
	advances in voice-activated assistants notwithstanding.
	Of course, we could have a lengthy debate over which of professors
	or parallel systems provide the most useful answers to these sorts
	of questions,
	but for the time being let's just agree that usefulness of
	answers varies widely across the population both of professors
	and of parallel systems.

	Other quizzes are quite similar to actual questions that have been
	asked during conference presentations and lectures covering the
	material in this book.
	A few others are from the viewpoint of the author.
}\QuickQuizEndM
\QuickQuizE{
	These Quick Quizzes are just not my cup of tea.
	What can I do about it?
}\QuickQuizAnswerE{
Here are a few possible strategies:

\begin{enumerate}
\item	Just ignore the Quick Quizzes and read the rest of
	the book.
	You might miss out on the interesting material in
	some of the Quick Quizzes, but the rest of the book
	has lots of good material as well.
	This is an eminently reasonable approach if your main
	goal is to gain a general understanding of the material
	or if you are skimming through the book to find a
	solution to a specific problem.
\item	Look at the answer immediately rather than investing
	a large amount of time in coming up with your own
	answer.
	This approach is reasonable when a given Quick Quiz's
	answer holds the key to a specific problem you are
	trying to solve.
	This approach is also reasonable if you want a somewhat
	deeper understanding of the material, but when you do not
	expect to be called upon to generate parallel solutions given
	only a blank sheet of paper.
\item	If you find the Quick Quizzes distracting but impossible
	to ignore, you can always clone the \LaTeX{} source for
	this book from the git archive.
	You can then run the command \co{make nq}, which will
	produce a \co{perfbook-nq.pdf}.
	This PDF contains unobtrusive boxed tags where the Quick Quizzes
	would otherwise be, and gathers each chapter's Quick Quizzes
	at the end of that chapter in the classic textbook style.
\item	Learn to like (or at least tolerate) the Quick Quizzes.
	Experience indicates that quizzing yourself periodically
	while reading greatly increases comprehension and depth
	of understanding.
\end{enumerate}

Note that the quick quizzes are hyperlinked to the answers and vice versa.
Click either the ``Quick Quiz'' heading or the small black square
to move to the beginning of the answer.
From the answer, click on the heading or the small black square to
move to the beginning of the quiz, or, alternatively, click on the
small white square at the end of the answer to move to the end of the
corresponding quiz.
}\QuickQuizEndE
}

In short, if you need a deep
understanding of the material, then you should invest some time
into answering the Quick Quizzes.
Don't get me wrong, passively reading the material can be quite
valuable, but gaining full problem-solving capability really
does require that you practice solving problems.
Similarly, gaining full code-production capability really does
require that you practice producing code.

\QuickQuiz{
	If passively reading this book doesn't get me full problem-solving
	and code-production capabilities, what on earth is the point???
}\QuickQuizAnswer{
	For those preferring analogies, coding concurrent software is
	similar to playing music in that there are good uses for many
	different levels of talent and skill.
	Not everyone needs to devote their entire live to becoming a
	concert pianist.
	In fact, for every such virtuoso, there are a great many lesser
	pianists whose of music is welcomed by their friends and families.
	But these lesser pianists are probably doing something else to
	support themselves, and so it is with concurrent coding.

	One potential benefit of passively reading this book is the ability
	to read and understand modern concurrent code.
	This ability might in turn permit you to:

	\begin{enumerate}
	\item	See what the kernel does so that you can check to see
		if a proposed use case is valid.
	\item	Chase down a kernel bug.
	\item	Use information in the kernel to more easily chase down
		a userspace bug.
	\item	Produce a fix for a kernel bug.
	\item	Create a straightforward kernel feature, whether from
		scratch or using the modern copy-pasta development
		methodology.
	\end{enumerate}

	If you are proficient with straightforward uses of locks and
	atomic operations, passively reading this book should enable
	you to successfully apply modern concurrency techniques.

	And finally, if your job is to coordinate the activities of
	developers making use of modern concurrency techniques, passively
	reading this book might help you understand what on earth they
	are talking about.
}\QuickQuizEnd

I learned this the hard way during coursework for my late-in-life
Ph.D\@.
I was studying a familiar topic, and was surprised at how few of
the chapter's exercises I could answer off the top of my head.\footnote{
	So I suppose that it was just as well that my professors refused
	to let me waive that class!}
Forcing myself to answer the questions greatly increased my
retention of the material.
So with these Quick Quizzes I am not asking you to do anything
that I have not been doing myself.

Finally, the most common learning disability is thinking that
you already understand the material at hand.
The quick quizzes can be an extremely effective cure.

\section{Alternatives to This Book}
\label{sec:Alternatives to This Book}
\epigraph{Between two evils I always pick the one I never tried before.}
	 {Mae West}

As \pplsur{Donald}{Knuth} learned the hard way, if you want your book
to be finite, it must be focused.
This book focuses on shared-memory parallel programming, with an
emphasis on software that lives near the bottom of the software stack,
such as operating-system kernels, parallel data-management systems,
low-level libraries, and the like.
The programming language used by this book is C.

If you are interested in other aspects of parallelism, you might well
be better served by some other book.
Fortunately, there are many alternatives available to you:

\begin{enumerate}
\item	If you prefer a more academic and rigorous treatment of
	parallel programming,
	you might like \pplsur{Maurice P.}{Herlihy}'s and \pplsur{Nir}{Shavit}'s
	textbook~\cite{HerlihyShavit2008Textbook,HerlihyShavit2020Textbook}.
	This book starts with an interesting combination
	of low-level primitives at high levels of abstraction
	from the hardware, and works its way through locking
	and simple data structures including lists, queues,
	hash tables, and counters, culminating with transactional
	memory, all in Java.
	\ppl{Michael}{Scott}'s textbook~\cite{MichaelScott2013Textbook}
	approaches similar material with more of a
	software-engineering focus, and, as far as I know, is
	the first formally published academic textbook with
	section devoted to RCU\@.

	\pplsur{Maurice P.}{Herlihy}, \pplsur{Nir}{Shavit},
	\pplsur{Victor}{Luchangco}, and \pplsur{Michael}{Spear} did
	catch up in their second edition~\cite{HerlihyShavit2020Textbook}
	by adding short sections on hazard pointers and on \acr{rcu},
	with the latter in the guise of \acr{ebr}\@.\footnote{
		Albeit an implementation that contains a reader-preemption
		bug noted by \ppl{Richard}{Bornat}.}
	They also include a brief history of both, albeit with an
	abbreviated history of \acr{rcu} that picks up almost a year after
	it was accepted into the Linux kernel and more than 20~years
	after \pplsur{H. T.}{Kung}'s and \pplsur{Philip L.}{Lehman}'s
	landmark paper~\cite{Kung80}.
	Those wishing a deeper view of the history may find it in
	this book's \cref{sec:defer:RCU Related Work}.

	However, readers who might otherwise suspect a hostile attitude
	towards RCU on the part of this textbook's first author should
	refer to the last full sentence on the first page of one of his
	papers~\cite{Balmau:2016:FRM:2935764.2935790}.
	This sentence reads ``QSBR [a particular class of RCU
	implementations] is fast and can be applied to virtually any
	data structure.''
	These are clearly not the words of someone who is hostile
	towards RCU.
\item	If you would like an academic treatment of parallel
	programming from a programming\-/language\-/pragmatics viewpoint,
	you might be interested in the concurrency chapter from
	\pplsur{Michael}{Scott}'s
	textbook~\cite{MichaelScott2006Textbook,MichaelScott2015Textbook}
	on programming-language pragmatics.
\item	If you are interested in an object-oriented patternist
	treatment of parallel programming focussing on C++,
	you might try Volumes~2 and~4 of \pplsur{Douglas C.}{Schmidt}'s POSA
	series~\cite{SchmidtStalRohnertBuschmann2000v2Textbook,
	BuschmannHenneySchmidt2007v4Textbook}.
	Volume~4 in particular has some interesting chapters
	applying this work to a warehouse application.
	The realism of this example is attested to by
	the section entitled ``Partitioning the Big Ball of Mud'',
	in which the problems inherent in parallelism often take a back
	seat to getting one's head around a real-world application.
\item	If you want to work with Linux-kernel device drivers,
	then \pplsur{Jonathan}{Corbet}'s, \pplsur{Alessandro}{Rubini}'s,
	and \pplsur{Greg}{Kroah-Hartman}'s
	``Linux Device Drivers''~\cite{CorbetRubiniKroahHartman}
	is indispensable, as is the Linux Weekly News web site
	(\url{https://lwn.net/}).
	There is a large number of books and resources on
	the more general topic of Linux kernel internals.
\item	If your primary focus is scientific and technical computing,
	and you prefer a patternist approach,
	you might try \pplsur{Timothy G.}{Mattson} et al.'s
	textbook~\cite{Mattson2005Textbook}.
	It covers Java, C/C++, OpenMP, and MPI\@.
	Its patterns are admirably focused first on design,
	then on implementation.
\item	If your primary focus is scientific and technical computing,
	and you are interested in GPUs, CUDA, and MPI, you
	might check out \ppl{Norm}{Matloff}'s ``Programming on
	Parallel Machines''~\cite{NormMatloff2017ParProcBook}.
	Of course, the GPU vendors have quite a bit of additional
	information~\cite{AMD2020ROCm,CyrilZeller2011GPGPUbasics,NVidia2017GPGPU,NVidia2017GPGPU-university}.
\item	If you are interested in POSIX Threads, you might take
	a look at \pplmdl{David R.}{Butenhof}'s book~\cite{Butenhof1997pthreads}.
	In addition,
	\ppl{W.~Richard}{Stevens}'s book~\cite{WRichardStevens1992,WRichardStevens2013}
	covers UNIX and POSIX, and \ppl{Stewart}{Weiss}'s lecture
	notes~\cite{StewartWeiss2013UNIX} provide an
	thorough and accessible introduction with a good set of
	examples.
\item	If you are interested in C++11, you might like
	\ppl{Anthony}{Williams}'s ``C++ Concurrency in Action:
	Practical Multithreading''~\cite{AnthonyWilliams2012,AnthonyWilliams2019}.
\item	If you are interested in C++, but in a Windows environment,
	you might try \ppl{Herb}{Sutter}'s ``Effective Concurrency''
	series in
	Dr.~Dobbs Journal~\cite{HerbSutter2008EffectiveConcurrency}.
	This series does a reasonable job of presenting a
	commonsense approach to parallelism.
\item	If you want to try out Intel Threading Building Blocks,
	then perhaps \ppl{James}{Reinders}'s book~\cite{Reinders2007Textbook}
	is what you are looking for.
\item	Those interested in learning how various types of multi-processor
	hardware
	cache organizations affect the implementation of kernel
	internals should take a look at \ppl{Curt}{Schimmel}'s classic
	treatment of this subject~\cite{Schimmel:1994:USM:175689}.
\item	If you are looking for a hardware view, \pplsur{John L.}{Hennessy}'s and
	\pplsur{David A.}{Patterson}'s classic
	textbook~\cite{Hennessy2017,Hennessy2011} is well worth a read.
	A ``Readers Digest'' version of this tome geared for scientific
	and technical workloads (bashing big arrays) may be found in
	\ppl{Andrew}{Chien}'s
	textbook~\cite{AndrewChien2022ComputerArchitectureScientists}.
	If you are looking for an academic textbook on memory ordering from
	a more hardware-centric viewpoint,
	that of \ppl{Daniel}{Sorin} et al.~\cite{DanielJSorin2011MemModel,%
	VijayNagarajan2020MemModel}
	is highly recommended.
	For a memory-ordering tutorial from a Linux-kernel viewpoint,
	\ppl{Paolo}{Bonzini}'s LWN series is a good place to
	start~\cite{PaoloBonzini2021lockless1,PaoloBonzini2021lockless2,PaoloBonzini2021lockless3,PaoloBonzini2021lockless4,PaoloBonzini2021lockless5,PaoloBonzini2021lockless6}.
\item	Those wishing to learn about the Rust language's
	support for low-level concurrency should refer to \ppl{Mara}{Bos}'s
	book~\cite{MaraBos2023RustAtomicsAndLocks}.
\item	Finally, those using Java might be well-served by \ppl{Doug}{Lea}'s
	textbooks~\cite{DougLea1997Textbook,Goetz2007Textbook}.
\end{enumerate}

However, if you are interested in principles of parallel design
for low-level software, especially software written in C, read on!

\section{Sample Source Code}
\label{sec:howto:Sample Source Code}
\epigraph{Use the source, Luke!}{Unknown Star Wars fan}

This book discusses its fair share of source code, and in many cases
this source code may be found in the \path{CodeSamples} directory
of this book's git tree.
For example, on UNIX systems, you should be able to type the following:

\begin{VerbatimU}
find CodeSamples -name rcu_rcpls.c -print
\end{VerbatimU}

This command will locate the file \path{rcu_rcpls.c}, which is called out in
\cref{chp:app:``Toy'' RCU Implementations}\@.
Non-UNIX systems have their own well-known ways of locating files by filename.

\section{Whose Book Is This?}
\label{sec:howto:Whose Book Is This?}
\epigraph{If you become a teacher, by your pupils you'll be taught.}
	 {Oscar Hammerstein II}

As the cover says, the editor is one Paul E.~McKenney.
However, the editor does accept contributions via the
\href{mailto:perfbook@vger.kernel.org}
{\nolinkurl{perfbook@vger.kernel.org}} email list.
These contributions can be in pretty much any form, with popular
approaches including text emails,
patches against the book's \LaTeX{} source, and even \co{git pull} requests.
Use whatever form works best for you.

To create patches or \co{git pull} requests, you will need the
\LaTeX{} source to the book, which is at
\url{git://git.kernel.org/pub/scm/linux/kernel/git/paulmck/perfbook.git},
or, alternatively,
\url{https://git.kernel.org/pub/scm/linux/kernel/git/paulmck/perfbook.git}.
You will of course also need \co{git} and \LaTeX{}, which are
available as part of most mainstream Linux distributions.
Other packages may be required, depending on the distribution you use.
The required list of packages for a few popular distributions is listed
in the file \path{FAQ-BUILD.txt} in the \LaTeX{} source to the book.

\begin{listing}
\begin{VerbatimL}[breaklines=true,breakafter=/,
        breakaftersymbolpre=\raisebox{-.7ex}{\textcolor{darkgray}{\Pisymbol{psy}{191}}},
	breaksymbolleft=\textcolor{darkgray}{\tiny\ensuremath{\hookrightarrow}},
        numbers=none,xleftmargin=0pt]
git clone git://git.kernel.org/pub/scm/linux/kernel/git/paulmck/perfbook.git
cd perfbook
# You may need to install a font. See item 1 in FAQ.txt.
make                     # -jN for parallel build
evince perfbook.pdf &    # Two-column version
make perfbook-1c.pdf
evince perfbook-1c.pdf & # One-column version for e-readers
make help                # Display other build options
\end{VerbatimL}
\caption{Creating an Up-To-Date PDF}
\label{lst:howto:Creating a Up-To-Date PDF}
\end{listing}

To create and display a current \LaTeX{} source tree of this book,
use the list of Linux commands shown in
\cref{lst:howto:Creating a Up-To-Date PDF}.
In some environments, the \co{evince} command that displays \path{perfbook.pdf}
may need to be replaced, for example, with \co{acroread}.
The \co{git clone} command need only be used the first time you
create a PDF, subsequently, you can run the commands shown in
\cref{lst:howto:Generating an Updated PDF} to pull in any updates
and generate an updated PDF\@.
The commands in
\cref{lst:howto:Generating an Updated PDF}
must be run within the \path{perfbook} directory created by the commands
shown in
\cref{lst:howto:Creating a Up-To-Date PDF}.

\begin{listing}
\begin{VerbatimL}[numbers=none,xleftmargin=0pt]
git remote update
git checkout origin/master
make                     # -jN for parallel build
evince perfbook.pdf &    # Two-column version
make perfbook-1c.pdf
evince perfbook-1c.pdf & # One-column version for e-readers
\end{VerbatimL}
\caption{Generating an Updated PDF}
\label{lst:howto:Generating an Updated PDF}
\end{listing}

PDFs of this book are sporadically posted at
\url{https://kernel.org/pub/linux/kernel/people/paulmck/perfbook/perfbook.html}
and at
\url{http://www.rdrop.com/users/paulmck/perfbook/}.

The actual process of contributing patches and sending \co{git pull}
requests is similar to that of the Linux kernel, which is documented
here:
\url{https://www.kernel.org/doc/html/latest/process/submitting-patches.html}.
One important requirement is that each patch (or commit, in the case
of a \co{git pull} request) must contain a valid \co{Signed-off-by:} line,
which has the following format:

\begin{VerbatimU}
Signed-off-by: My Name <myname@example.org>
\end{VerbatimU}

Please see \url{https://lkml.org/lkml/2007/1/15/219} for an example
patch with a \co{Signed-off-by:} line.
Note well that the \co{Signed-off-by:} line has a very specific meaning,
namely that you are certifying that:

\begin{enumerate}[label={(\alph*)}]
\item	The contribution was created in whole or in part
	by me and I have the right to submit it under
	the open source license indicated in the file; or
\item	The contribution is based upon previous work
	that, to the best of my knowledge, is covered
	under an appropriate open source license and I
	have the right under that license to submit that
	work with modifications, whether created in whole
	or in part by me, under the same open source
	license (unless I am permitted to submit under
	a different license), as indicated in the file; or
\item	The contribution was provided directly to me by
	some other person who certified (a), (b) or (c)
	and I have not modified it.
\item	I understand and agree that this project and the
	contribution are public and that a record of the
	contribution (including all personal information
	I submit with it, including my sign-off) is
	maintained indefinitely and may be redistributed
	consistent with this project or the open source
	license(s) involved.
\end{enumerate}

This is quite similar to the Developer's Certificate of Origin (DCO)
1.1 used by the Linux kernel.
You must use your real name:
I unfortunately cannot accept pseudonymous or anonymous contributions.

The language of this book is American English, however, the open-source
nature of this book permits translations, and I personally encourage them.
The open-source licenses covering this book additionally allow you
to sell your translation, if you wish.
I do request that you send me a copy of the translation (hardcopy if
available), but this is a request made as a professional courtesy,
and is not in any way a prerequisite to the permission that you already
have under the Creative Commons and GPL licenses.
Please see the \co{FAQ.txt} file in the source tree for a list of
translations currently in progress.
I consider a translation effort to be ``in progress'' once at least one
chapter has been fully translated.

There are many styles under the ``American English'' rubric.
The style for this particular book is documented in
\cref{chp:app:styleguide:Style Guide}.

As noted at the beginning of this section, I am this book's editor.
However, if you choose to contribute, it will be your book as well.
In that spirit, I offer you \cref{chp:Introduction}, our introduction.

	\setlength{\parskip}{0.0pt plus 1ex}
	\input{qqzhowto}
	\setlength{\parskip}{0.0pt plus 1.0pt}% return to default

% intro/intro.tex
% mainfile: ../perfbook.tex
% SPDX-License-Identifier: CC-BY-SA-3.0

\QuickQuizChapter{chp:Introduction}{Introduction}{qqzintro}
\Epigraph{If parallel programming is so hard, why are there so many
	  parallel programs?}{Unknown}

Parallel programming has earned a reputation as one of the most
difficult areas a hacker can tackle.
Papers and textbooks warn of the perils of \IX{deadlock}, \IX{livelock},
\IXpl{race condition}, non-determinism,
\IXaltr{Amdahl's-Law}{Amdahl's Law} limits to scaling,
and excessive realtime latencies.
And these perils are quite real; we authors have accumulated uncounted
% 2020:
%	30 for Paul E. McKenney
years of experience along with the resulting emotional scars,
grey hairs, and hair loss.

However, new technologies that are difficult to use at introduction
invariably become easier over time.
For example, the once-rare ability to drive a car is now
commonplace in many countries.
This dramatic change came about for two basic reasons:
\begin{enumerate*}[(1)]
\item Cars became cheaper and more readily available, so that
more people had the opportunity to learn to drive, and
\item Cars became easier to operate due to automatic transmissions,
automatic chokes, automatic starters, greatly improved reliability,
and a host of other technological improvements.
\end{enumerate*}

The same is true for many other technologies, including computers.
It is no longer necessary to operate a keypunch in order to program.
Spreadsheets allow most non-programmers to get results from their computers
that would have required a team of specialists a few decades ago.
Perhaps the most compelling example is web-surfing and content creation,
which since the early 2000s has been easily done by
untrained, uneducated people using various now-commonplace
social-networking tools.
As recently as 1968, such content creation was a far-out research
project~\cite{DouglasEngelbart1968}, described at
the time as
``like a UFO landing on the White House lawn''~\cite{ScottGriffen2000}.
% http://www.ibiblio.org/pioneers/engelbart.html
% http://www.histech.rwth-aachen.de/www/quellen/engelbart/ahi62index.html

Therefore, if you wish to argue that parallel programming will remain
as difficult as it is currently perceived by many to be, it is you
who bears the burden of proof, keeping in mind the many centuries of
counter-examples in many fields of endeavor.

\section{Historic Parallel Programming Difficulties}
\label{sec:intro:Historic Parallel Programming Difficulties}
\epigraph{Not the power to remember, but its very opposite, the power to
	  forget, is a necessary condition for our existence.}
	 {Sholem Asch}

As indicated by its title, this book takes a different approach.
Rather than complain about the difficulty of parallel programming,
it instead examines the reasons why parallel programming is
difficult, and then works to help the reader to overcome these
difficulties.
As will be seen, these difficulties have historically fallen into several
categories, including:

\begin{enumerate}
\item	The historic high cost and relative rarity of parallel systems.
\item	The typical researcher's and practitioner's lack of experience
	with parallel systems.
\item	The paucity of publicly accessible parallel code.
\item	The lack of a widely understood engineering discipline of
	parallel programming.
\item	The high \IX{overhead} of communication relative to that of processing,
	even in tightly coupled shared-memory computers.
\end{enumerate}

Many of these historic difficulties are well on the way to being overcome.
First, over the past few decades, the cost of parallel systems
has decreased from many multiples of that of a house to that of a
modest meal, courtesy of \IXr{Moore's Law}~\cite{GordonMoore1965MooresLaw}.
Papers calling out the advantages of multicore CPUs were published
as early as 1996~\cite{Olukotun96}.
IBM introduced simultaneous multi-threading
into its high-end \Power{} family in 2000, and multicore in 2001.
Intel introduced hyperthreading into its commodity Pentium line in
November 2000, and both AMD and Intel introduced
dual-core CPUs in 2005.
Sun followed with the multicore/multi-threaded Niagara in late 2005.
In fact, by 2008, it was becoming difficult
to find a single-CPU desktop system, with single-core CPUs being
relegated to netbooks and embedded devices.
By 2012, even smartphones were starting to sport multiple CPUs.
By 2020, safety-critical software standards started addressing
concurrency.

Second, the advent of low-cost and readily available multicore systems
means that the once-rare experience of parallel programming is
now available to almost all researchers and practitioners.
In fact, parallel systems have long been within the budget of students
and hobbyists.
We can therefore expect greatly increased levels of invention and
innovation surrounding parallel systems, and that increased familiarity
will over time make the once prohibitively expensive field of parallel
programming much more friendly and commonplace.

Third, in the 20\textsuperscript{th} century, large systems of
highly parallel software were almost always closely guarded proprietary
secrets.
In happy contrast, the 21\textsuperscript{st} century has seen numerous
open-source (and thus publicly available) parallel software projects,
including the Linux kernel~\cite{Torvalds2.6kernel},
database systems~\cite{PostgreSQL2008,MySQL2008},
and message-passing systems~\cite{OpenMPI2008,BOINC2008}.
This book will draw primarily from the Linux kernel, but will
provide much material suitable for user-level applications.

Fourth, even though the large-scale parallel\-/programming projects of
the 1980s and 1990s were almost all proprietary projects, these
projects have seeded other communities with cadres of developers who
understand the engineering discipline required to develop production-quality
parallel code.
A major purpose of this book is to present this engineering discipline.

Unfortunately, the fifth difficulty, the high cost of communication
relative to that of processing, remains largely in force.
This difficulty has been receiving increasing attention during
the new millennium.
However, according to \ppl{Stephen}{Hawking},
the finite speed of light and the atomic
nature of matter will limit progress in this
area~\cite{BryanGardiner2007,GordonMoore03a}.
Fortunately, this difficulty has been in force since the late 1980s,
so that the aforementioned engineering discipline has evolved practical
and effective strategies for handling it.
In addition, hardware designers are increasingly aware of these issues,
so perhaps future hardware will be more friendly to parallel software,
as discussed in \cref{sec:cpu:Hardware Free Lunch?}.

\QuickQuiz{
	Come on now!!!
	Parallel programming has been known to be exceedingly
	hard for many decades.
	You seem to be hinting that it is not so hard.
	What sort of game are you playing?
}\QuickQuizAnswer{
	If you really believe that parallel programming is exceedingly
	hard, then you should have a ready answer to the question
	``Why is parallel programming hard?''
	One could list any number of reasons, ranging from deadlocks to
	race conditions to testing coverage, but the real answer is that
	{\em it is not really all that hard}.
	After all, if parallel programming was really so horribly difficult,
	how could a large number of open-source projects, ranging from Apache
	to MySQL to the Linux kernel, have managed to master it?

	A better question might be:
	``Why is parallel programming \emph{perceived} to be so difficult?''
	To see the answer, let's go back to the year 1991.
	Paul McKenney was walking across the parking lot to Sequent's
	benchmarking center carrying six dual-80486 Sequent Symmetry CPU
	boards, when he suddenly realized that he was carrying several
	times the price of the house he had just purchased.\footnote{
		Yes, this sudden realization {\em did} cause him to walk quite
		a bit more carefully.
		Why do you ask?}
	This high cost of parallel systems meant that
	parallel programming was restricted to a privileged few who
	worked for an employer who either manufactured or could afford to
	purchase machines costing upwards of \$100,000---in 1991 dollars US.

	In contrast, in 2020, Paul finds himself typing these words on a
	six-core x86 laptop.
	Unlike the dual-80486 CPU boards, this laptop also contains
	64\,GB of main memory, a 1\,TB solid-state disk, a display, Ethernet,
	USB ports, wireless, and Bluetooth.
	And the laptop is more than an order of magnitude cheaper than
	even one of those dual-80486 CPU boards, even before taking inflation
	into account.

	Parallel systems have truly arrived.
	They are no longer the sole domain of a privileged few, but something
	available to almost everyone.

	The earlier restricted availability of parallel hardware is
	the \emph{real} reason that parallel programming is considered
	so difficult.
	After all, it is quite difficult to learn to program even the simplest
	machine if you have no access to it.
	Since the age of rare and expensive parallel machines is for the most
	part behind us, the age during which
	parallel programming is perceived to be mind-crushingly difficult is
	coming to a close.\footnote{
		Parallel programming is in some ways more difficult than
		sequential programming, for example, parallel validation
		is more difficult.
		But no longer mind-crushingly difficult.}
}\QuickQuizEnd

However, even though parallel programming might not be as hard as
is commonly advertised, it is often more work than is sequential
programming.

\QuickQuiz{
	How could parallel programming \emph{ever} be as easy
	as sequential programming?
}\QuickQuizAnswer{
	It depends on the programming environment.
	SQL~\cite{DIS9075SQL92} is an underappreciated success
	story, as it permits programmers who know nothing about parallelism
	to keep a large parallel system productively busy.
	We can expect more variations on this theme as parallel
	computers continue to become cheaper and more readily available.
	For example, one possible contender in the scientific and
	technical computing arena is MATLAB*P,
	which is an attempt to automatically parallelize common
	matrix operations.

	Finally, on Linux and UNIX systems, consider the following
	shell command:

	\begin{VerbatimU}
	get_input | grep "interesting" | sort
	\end{VerbatimU}

	This shell pipeline runs the \co{get_input}, \co{grep},
	and \co{sort} processes in parallel.
	There, that wasn't so hard, now was it?

	In short, parallel programming is just as easy as sequential
	programming---at least in those environments that hide the parallelism
	from the user!
}\QuickQuizEnd

It therefore makes sense to consider alternatives to parallel programming.
However, it is not possible to reasonably consider parallel-programming
alternatives without understanding parallel-programming goals.
This topic is addressed in the next section.

\section{Parallel Programming Goals}
\label{sec:intro:Parallel Programming Goals}
\epigraph{If you don't know where you are going, you will end up somewhere
	  else.}{Yogi Berra}

The three major goals of parallel programming (over and above those
of sequential programming) are as follows:

\begin{enumerate}
\item	\IX{Performance}.
\item	\IX{Productivity}.
\item	\IX{Generality}.
\end{enumerate}

Unfortunately, given the current state of the art, it is possible to
achieve at best two of these three goals for any given parallel program.
These three goals therefore form the \emph{iron triangle of parallel
programming},
a triangle upon which overly optimistic hopes all too often come to
grief.\footnote{
	Kudos to Michael Wong for naming the iron triangle.}

\QuickQuizSeries{%
\QuickQuizB{
	Oh, really???
	What about correctness, maintainability, robustness, and so on?
}\QuickQuizAnswerB{
	These are important goals, but they are just as important for
	sequential programs as they are for parallel programs.
	Therefore, important though they are, they do not belong on
	a list specific to parallel programming.
}\QuickQuizEndB
\QuickQuizM{
	And if correctness, maintainability, and robustness don't
	make the list, why do productivity and generality?
}\QuickQuizAnswerM{
	Given that parallel programming is perceived to be much harder
	than sequential programming, productivity is tantamount and
	therefore must not be omitted.
	Furthermore, high-productivity parallel-programming environments
	such as SQL serve a specific purpose, hence generality must
	also be added to the list.
}\QuickQuizEndM
\QuickQuizM{
	Given that parallel programs are much harder to prove
	correct than are sequential programs, again, shouldn't
	correctness \emph{really} be on the list?
}\QuickQuizAnswerM{
	From an engineering standpoint, the difficulty in proving
	correctness, either formally or informally, would be important
	insofar as it impacts the primary goal of productivity.
	So, in cases where correctness proofs are important, they
	are subsumed under the ``productivity'' rubric.
}\QuickQuizEndM
\QuickQuizE{
	What about just having fun?
}\QuickQuizAnswerE{
	Having fun is important as well, but, unless you are a hobbyist,
	would not normally be a \emph{primary} goal.
	On the other hand, if you \emph{are} a hobbyist, go wild!
}\QuickQuizEndE
}

Each of these goals is elaborated upon in the following sections.

\subsection{Performance}
\label{sec:intro:Performance}

Performance is the primary goal behind most parallel-programming effort.
After all, if performance is not a concern, why not do yourself a favor:
Just write sequential code, and be happy?
It will very likely be easier
and you will probably get done much more quickly.

\QuickQuiz{
	Are there no cases where parallel programming is about something
	other than performance?
}\QuickQuizAnswer{
	There certainly are cases where the problem to be solved is
	inherently parallel, for example, Monte Carlo methods and
	some numerical computations.
	Even in these cases, however, there will be some amount of
	extra work managing the parallelism.

	Parallelism is also sometimes used for reliability.
	For but one example,
	triple-modulo redundancy has three systems run in parallel
	and vote on the result.
	In extreme cases, the three systems will be independently
	implemented using different algorithms and technologies.
}\QuickQuizEnd

Note that ``performance'' is interpreted broadly here,
including for example \IX{scalability} (performance per CPU) and \IX{efficiency}
(performance per watt).

\begin{figure}
\centering
\resizebox{3in}{!}{\includegraphics{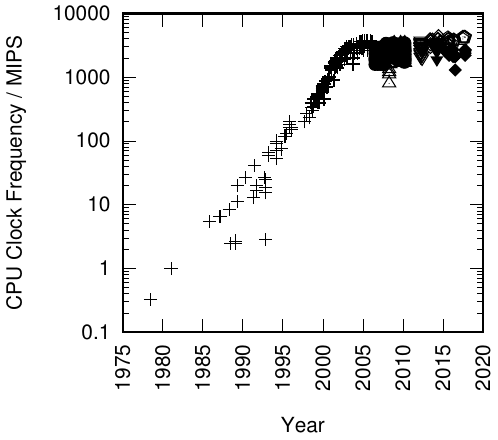}}
\caption{MIPS/Clock-Frequency Trend for Intel CPUs}
\label{fig:intro:Clock-Frequency Trend for Intel CPUs}
\end{figure}

That said, the focus of performance has shifted from hardware to
parallel software.
This change in focus is due to the fact that, although \IXr{Moore's Law}
continues to deliver increases in transistor density, it has ceased to
provide the traditional single-threaded performance increases.
This can be seen in
\cref{fig:intro:Clock-Frequency Trend for Intel CPUs},\footnote{
	This plot shows clock frequencies for newer CPUs theoretically
	capable of retiring one or more instructions per clock, and MIPS
	(millions of instructions per second, usually from the old
	Dhrystone benchmark)
	for older CPUs requiring multiple clocks to execute even the
	simplest instruction.
	The reason for shifting between these two measures is that the
	newer CPUs' ability to retire multiple instructions per clock is
	typically limited by memory-system performance.
	Furthermore, the benchmarks commonly used on the older CPUs
	are obsolete, and it is difficult to run the newer benchmarks
	on systems containing the old CPUs, in part because it is hard
	to find working instances of the old CPUs.}
which shows that writing single-threaded code and simply waiting
a year or two for the CPUs to catch up may no longer be an option.
Given the recent trends on the part of all major manufacturers towards
multicore/multithreaded systems, parallelism is the way to go for
those wanting to avail themselves of the full performance of their
systems.

\QuickQuiz{
	Why not instead rewrite programs from inefficient scripting
	languages to C or C++?
}\QuickQuizAnswer{
	If the developers, budget, and time is available for such a
	rewrite, and if the result will attain the required levels
	of performance on a single CPU, this can be a reasonable
	approach.
}\QuickQuizEnd

Even so, the first goal is performance rather than scalability,
especially given that the easiest way to attain linear scalability
is to reduce the performance of each CPU~\cite{LinusTorvalds2001a}.
Given a four-CPU system, which would you prefer?
A program that provides 100 transactions per second on a single CPU,
but does not scale at all?
Or a program that provides 10 transactions per second on a single CPU,
but scales perfectly?
The first program seems like a better bet, though the answer might
change if you happened to have a 32-CPU system.

That said, just because you have multiple CPUs is not necessarily
in and of itself a reason to use them all, especially given the
recent decreases in price of multi-CPU systems.
The key point to understand is that parallel programming is primarily
a performance optimization, and, as such, it is one potential optimization
of many.
If your program is fast enough as currently written, there is no
reason to optimize, either by parallelizing it or by applying any
of a number of potential sequential optimizations.\footnote{
	Of course, if you are a hobbyist whose primary interest is
	writing parallel software, that is more than enough reason to
	parallelize whatever software you are interested in.}
By the same token, if you are looking to apply parallelism as an
optimization to a sequential program, then you will need to compare
parallel algorithms to the best sequential algorithms.
This may require some care, as far too many publications ignore the
sequential case when analyzing the performance of parallel algorithms.

\subsection{Productivity}
\label{sec:intro:Productivity}

\EQuickQuiz{
	Why all this prattling on about non-technical issues???
	And not just \emph{any} non-technical issue, but \emph{productivity}
	of all things?
	Who cares?
}\EQuickQuizAnswer{
	If you are a pure hobbyist, perhaps you don't need to care.
	But even pure hobbyists will often care about how much they
	can get done, and how quickly.
	After all, the most popular hobbyist tools are usually those
	that are the best suited for the job, and an important part of
	the definition of ``best suited'' involves productivity.
	And if someone is paying you to write parallel code, they will
	very likely care deeply about your productivity.
	And if the person paying you cares about something, you would
	be most wise to pay at least some attention to it!

	Besides, if you \emph{really} didn't care about productivity,
	you would be doing it by hand rather than using a computer!
}\EQuickQuizEnd

\IX{Productivity} has been becoming increasingly important in recent decades.
To see this, consider that the price of early computers was tens
of millions of dollars at
a time when engineering salaries were but a few thousand dollars a year.
If dedicating a team of ten engineers to such a machine would improve
its performance, even by only 10\,\%, then their salaries
would be repaid many times over.

One such machine was the CSIRAC, the oldest still-intact stored-program
computer, which was put into operation in
1949~\cite{CSIRACMuseumVictoria,CSIRACUniversityMelbourne}.
Because this machine was built before the transistor era, it was constructed
of 2,000 vacuum tubes, ran with a clock frequency of 1\,kHz,
consumed 30\,kW of power, and weighed more than three metric tons.
Given that this machine had but 768 words of RAM, it is safe to say that
it did not suffer from the productivity issues that often plague
today's large-scale software projects.

Today, it would be quite difficult to purchase a machine with so
little computing power.
Perhaps the closest equivalents
are 8-bit embedded microprocessors exemplified by the venerable
Z80~\cite{z80Wikipedia}, but even the old Z80 had a CPU clock
frequency more than 1,000 times faster than the CSIRAC\@.
The Z80 CPU had 8,500 transistors, and could be purchased in 2008
for less than \$2 US per unit in 1,000-unit quantities.
In stark contrast to the CSIRAC, software-development costs are
anything but insignificant for the Z80.

\begin{figure}
\centering
\resizebox{3in}{!}{\includegraphics{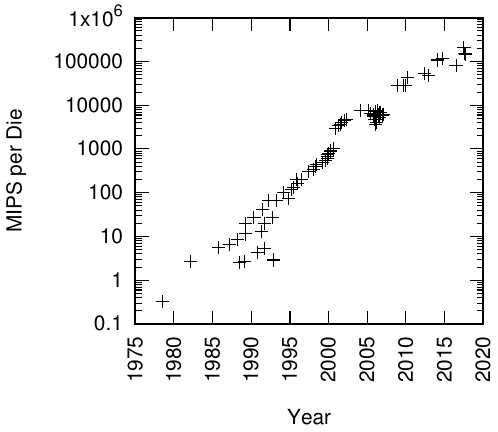}}
\caption{MIPS per Die for Intel CPUs}
\label{fig:intro:MIPS per Die for Intel CPUs}
\end{figure}

The CSIRAC and the Z80 are two points in a long-term trend, as can be
seen in
\cref{fig:intro:MIPS per Die for Intel CPUs}.
This figure plots an approximation to computational power per die
over the past four decades, showing an impressive six-order-of-magnitude
increase over a period of forty years.
Note that the advent of multicore CPUs has permitted this increase to
continue apace despite the clock-frequency wall encountered in 2003,
albeit courtesy of dies supporting more than 50 hardware threads each.

One of the inescapable consequences of the rapid decrease in
the cost of hardware is that software productivity becomes increasingly
important.
It is no longer sufficient merely to make efficient use of the hardware:
It is now necessary to make extremely efficient use of software
developers as well.
This has long been the case for sequential hardware, but
parallel hardware has become a low-cost commodity only recently.
Therefore, only recently has high productivity become critically important
when creating parallel software.

\QuickQuiz{
	Given how cheap parallel systems have become, how can anyone
	afford to pay people to program them?
}\QuickQuizAnswer{
	There are a number of answers to this question:
	\begin{enumerate}
	\item	Given a large computational cluster of parallel machines,
		the aggregate cost of the cluster can easily justify
		substantial developer effort, because the development
		cost can be spread over the large number of machines.
	\item	Popular software that is run by tens of millions of users
		can easily justify substantial developer effort,
		as the cost of this development can be spread over the tens
		of millions of users.
		Note that this includes things like kernels and system
		libraries.
	\item	If the low-cost parallel machine is controlling the operation
		of a valuable piece of equipment, then the cost of this
		piece of equipment might easily justify substantial
		developer effort.
	\item	If the software for the low-cost parallel machine produces an
		extremely valuable result (e.g., energy savings),
		then this valuable result might again justify substantial
		developer cost.
	\item	Safety-critical systems protect lives, which can clearly
		justify very large developer effort.
	\item	Hobbyists and researchers might instead seek knowledge,
		experience, fun, or glory.
	\end{enumerate}
	So it is not the case that the decreasing cost of hardware renders
	software worthless, but rather that it is no longer possible to
	``hide'' the cost of software development within the cost of
	the hardware, at least not unless there are extremely large
	quantities of hardware.
}\QuickQuizEnd

Perhaps at one time, the sole purpose of parallel software was performance.
Now, however, productivity is gaining the spotlight.

\subsection{Generality}
\label{sec:intro:Generality}

One way to justify the high cost of developing parallel software
is to strive for maximal \IX{generality}.
All else being equal, the cost of a more-general software artifact
can be spread over more users than that of a less-general one.
In fact, this economic force explains much of the maniacal focus
on portability, which can be seen as an important special case
of generality.\footnote{
	Kudos to Michael Wong for pointing this out.}

Unfortunately, generality often comes at the cost of performance,
productivity, or both.
For example, portability is often achieved via adaptation layers,
which inevitably exact a performance penalty.
To see this more generally, consider the following popular parallel
programming environments:

\begin{description}
\item[C/C++ ``Locking Plus Threads'':] This category, which includes
	POSIX Threads (pthreads)~\cite{OpenGroup1997pthreads},
	Windows Threads, and numerous
	operating-system kernel environments, offers excellent performance
	(at least within the confines of a single SMP system)
	and also offers good generality.
	Pity about the relatively low productivity.
\item[Java:] This general purpose and inherently multithreaded
	programming environment	is widely believed to offer much higher
	productivity than C or C++, courtesy of the automatic garbage collector
	and the rich set of class libraries.
	However, its performance, though greatly improved in the early
	2000s, lags that of C and C++.
\item[MPI:] This Message Passing Interface~\cite{MPIForum2008} powers
	the largest scientific and technical computing clusters in
	the world and offers unparalleled performance and scalability.
	In theory, it is general purpose, but it is mainly used
	for scientific and technical computing.
	Its productivity is believed by many to be even lower than that
	of C/C++ ``locking plus threads'' environments.
\item[OpenMP:] This set of compiler directives can be used
	to parallelize loops.
	It is thus quite specific to this
	task, and this specificity often limits its performance.
	It is, however, much easier to use than MPI or C/C++
	``locking plus threads.''
\item[SQL:] Structured Query Language~\cite{DIS9075SQL92} is
	specific to relational database queries.
	However, its performance is quite good as measured by the
	Transaction Processing Performance Council (TPC)
	benchmark results~\cite{TPC}.
	Productivity is excellent; in fact, this parallel programming
	environment enables people to make good use of a large parallel
	system despite having little or no knowledge of parallel
	programming concepts.
\end{description}

\begin{figure}
\centering
\resizebox{2.5in}{!}{\includegraphics{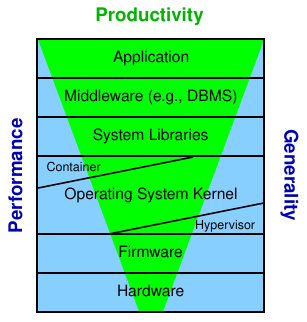}}
\caption{Software Layers and Performance, Productivity, and Generality}
\label{fig:intro:Software Layers and Performance; Productivity; and Generality}
\end{figure}

The nirvana of parallel programming environments, one that offers
world-class performance, productivity, and generality, simply does
not yet exist.
Until such a nirvana appears, it will be necessary to make engineering
tradeoffs among performance, productivity, and generality.
One such tradeoff is depicted by the green ``iron triangle''\footnote{
	Kudos to Michael Wong for coining ``iron triangle.''}
shown in
\cref{fig:intro:Software Layers and Performance; Productivity; and Generality},
which shows how productivity becomes increasingly important at the upper layers
of the system stack,
while performance and generality become increasingly important at the
lower layers of the system stack.
The huge development costs incurred at the lower layers
must be spread over equally huge numbers of users
(hence the importance of generality), and
performance lost in lower layers cannot easily be
recovered further up the stack.
In the upper layers of the stack, there might be very few users for a given
specific application, in which case productivity concerns are paramount.
This explains the tendency towards ``bloatware'' further up the stack:
Extra hardware is often cheaper than extra developers.
This book is intended for developers working near the bottom
of the stack, where performance and generality are of greatest concern.

\begin{figure}
\centering
\resizebox{3in}{!}{\includegraphics{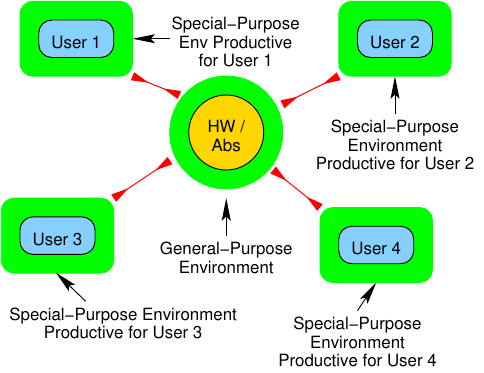}}
\caption{Tradeoff Between Productivity and Generality}
\label{fig:intro:Tradeoff Between Productivity and Generality}
\end{figure}

It is important to note that a tradeoff between productivity and
generality has existed for centuries in many fields.
For but one example, a nailgun is more productive than a hammer for
driving nails, but in contrast to the nailgun, a hammer can be used for
many things besides driving nails.
It should therefore be no surprise to see similar tradeoffs
appear in the field of parallel computing.
This tradeoff is shown schematically in
\cref{fig:intro:Tradeoff Between Productivity and Generality}.
Here, users~1, 2, 3, and~4 have specific jobs that they need the computer
to help them with.
The most productive possible language or environment for a given user is one
that simply does that user's job, without requiring any programming,
configuration, or other setup.

\QuickQuiz{
	This is a ridiculously unachievable ideal!
	Why not focus on something that is achievable in practice?
}\QuickQuizAnswer{
	This is eminently achievable.
	The cellphone is a computer that can be used to make phone
	calls and to send and receive text messages with little or
	no programming or configuration on the part of the end user.

	This might seem to be a trivial example at first glance,
	but if you consider it carefully you will see that it is
	both simple and profound.
	When we are willing to sacrifice generality, we can achieve
	truly astounding increases in productivity.
	Those who indulge in excessive generality will therefore fail to set
	the productivity bar high enough to succeed near the top of the
	software stack.
	This fact of life even has its own acronym:
	YAGNI, or ``You Ain't Gonna Need It.''
}\QuickQuizEnd

Unfortunately, a system that does the job required by user~1 is
unlikely to do user~2's job.
In other words, the most productive languages and environments are
domain-specific, and thus by definition lacking generality.

Another option is to tailor a given programming language or environment
to the hardware system (for example, low-level languages such as
assembly, C, C++, or Java) or to some abstraction (for example,
Haskell, Prolog, or Snobol), as is shown by the circular region near
the center of
\cref{fig:intro:Tradeoff Between Productivity and Generality}.
These languages can be considered to be general in the sense that they
are equally ill-suited to the jobs required by users~1, 2, 3, and~4.
In other words, their generality comes at the expense of
decreased productivity when compared to domain-specific languages
and environments.
Worse yet, a language that is tailored to a given abstraction
is likely to suffer from performance and scalability problems
unless and until it can be efficiently mapped to real hardware.

Is there no escape from iron triangle's three conflicting goals of
performance, productivity, and generality?

It turns out that there often is an escape, for example,
using the alternatives to parallel programming discussed in the next section.
After all, parallel programming can be a great deal of fun, but
it is not always the best tool for the job.

\section{Alternatives to Parallel Programming}
\label{sec:intro:Alternatives to Parallel Programming}
\epigraph{Experiment is folly when experience shows the way.}
	 {Roger M. Babson}

In order to properly consider alternatives to parallel programming,
you must first decide on what exactly you expect the parallelism
to do for you.
As seen in \cref{sec:intro:Parallel Programming Goals},
the primary goals of parallel programming are performance, productivity,
and generality.
Because this book is intended for developers working on
performance-critical code near the bottom of the software stack,
the remainder of this section focuses primarily on performance improvement.

It is important to keep in mind that parallelism is but one way to
improve performance.
Other well-known approaches include the following, in roughly increasing
order of difficulty:

\begin{enumerate}
\item	Run multiple instances of a sequential application.
\item	Make the application use existing parallel software.
\item	Optimize the serial application.
\end{enumerate}

These approaches are covered in the following sections.

\subsection{Multiple Instances of a Sequential Application}
\label{sec:intro:Multiple Instances of a Sequential Application}

Running multiple instances of a sequential application can allow you
to do parallel programming without actually doing parallel programming.
There are a large number of ways to approach this, depending on the
structure of the application.

If your program is analyzing a large number of different scenarios,
or is analyzing a large number of independent data sets, one easy
and effective approach is to create a single sequential program that
carries out a single analysis, then use any of a number of scripting
environments (for example the \co{bash} shell) to run a number of
instances of that sequential program in parallel.
In some cases, this approach can be easily extended to a cluster of
machines.

This approach may seem like cheating, and in fact some denigrate such
programs as ``\IX{embarrassingly parallel}''.
And in fact, this approach does have some potential disadvantages,
including increased memory consumption, waste of CPU cycles recomputing
common intermediate results, and increased copying of data.
However, it is often  extremely productive, garnering extreme performance
gains with little or no added effort.

\subsection{Use Existing Parallel Software}
\label{sec:intro:Use Existing Parallel Software}

There is no longer any shortage of parallel software environments that
can present a single-threaded programming environment,
including relational
databases~\cite{Date82},
web-application servers, and map-reduce environments.
For example, a common design provides a separate process for each
user, each of which generates SQL from user queries.
This per-user SQL is run against a common relational database, which
automatically runs the users' queries concurrently.
The per-user programs are responsible only for the user interface,
with the relational database taking full responsibility for the
difficult issues surrounding parallelism and persistence.

In addition, there are a growing number of parallel library functions,
particularly for numeric computation.
Even better, some libraries take advantage of special\-/purpose
hardware such as vector units and general\-/purpose graphical processing
units (GPGPUs).

Taking this approach often sacrifices some performance, at least when
compared to carefully hand-coding a fully parallel application.
However, such sacrifice is often well repaid by a huge reduction in
development effort.

\QuickQuiz{
	Wait a minute!
	Doesn't this approach simply shift the development effort from
	you to whoever wrote the existing parallel software you are using?
}\QuickQuizAnswer{
	Exactly!
	And that is the whole point of using existing software.
	One team's work can be used by many other teams, resulting in a
	large decrease in overall effort compared to all teams
	needlessly reinventing the wheel.
}\QuickQuizEnd

\subsection{Performance Optimization}
\label{sec:intro:Performance Optimization}

Up through the early 2000s, CPU clock frequencies doubled every 18 months.
It was therefore usually more important to create new functionality than to
carefully optimize performance.
Now that \IXr{Moore's Law} is ``only'' increasing transistor density instead
of increasing both transistor density and per-transistor performance,
it might be a good time to rethink the importance of performance
optimization.
After all, new hardware generations no longer bring significant
single-threaded performance improvements.
Furthermore, many performance optimizations can also conserve energy.

From this viewpoint, parallel programming is but another performance
optimization, albeit one that is becoming much more attractive
as parallel systems become cheaper and more readily available.
However, it is wise to keep in mind that the speedup available from
parallelism is limited to roughly the number of CPUs
(but see \cref{sec:SMPdesign:Beyond Partitioning}
for an interesting exception).
In contrast, the speedup available from traditional single-threaded
software optimizations can be much larger.
For example, replacing a long linked list with a hash table
or a search tree can improve performance by many orders of magnitude.
This highly optimized single-threaded program might run much
faster than its unoptimized parallel counterpart, making parallelization
unnecessary.
Of course, a highly optimized parallel program would be even better,
aside from the added development effort required.

Furthermore, different programs might have different performance
bottlenecks.
For example, if your program spends most of its time
waiting on data from your disk drive,
using multiple CPUs will probably just increase the time wasted waiting
for the disks.
In fact, if the program was reading from a single large file laid out
sequentially on a rotating disk, parallelizing your program might
well make it a lot slower due to the added seek overhead.
You should instead optimize the data layout so that
the file can be smaller (thus faster to read), split the file into chunks
which can be accessed in parallel from different drives,
cache frequently accessed data in main memory,
or, if possible,
reduce the amount of data that must be read.

\QuickQuiz{
	What other bottlenecks might prevent additional CPUs from
	providing additional performance?
}\QuickQuizAnswer{
	There are any number of potential bottlenecks:
	\begin{enumerate}
	\item	Main memory.
		If a single thread consumes all available
		memory, additional threads will simply page themselves
		silly.
	\item	Cache.
		If a single thread's cache footprint completely
		fills any shared CPU cache(s), then adding more threads
		will simply thrash those affected caches, as will be
		seen in \cref{chp:Data Structures}.
	\item	Memory bandwidth.
		If a single thread consumes all available
		memory bandwidth, additional threads will simply
		result in additional queuing on the system interconnect.
	\item	I/O bandwidth.
		If a single thread is I/O bound,
		adding more threads will simply result in them all
		waiting in line for the affected I/O resource.
	\end{enumerate}

	Specific hardware systems might have any number of additional
	bottlenecks.
	The fact is that every resource which is shared between
	multiple CPUs or threads is a potential bottleneck.
}\QuickQuizEnd

Parallelism can be a powerful optimization technique, but
it is not the only such technique, nor is it appropriate for all
situations.
Of course, the easier it is to parallelize your program, the
more attractive parallelization becomes as an optimization.
Parallelization has a reputation of being quite difficult,
which leads to the question ``exactly what makes parallel
programming so difficult?''

\section{What Makes Parallel Programming Hard?}
\label{sec:intro:What Makes Parallel Programming Hard?}
\epigraph{Real difficulties can be overcome; it is only the imaginary
	  ones that are unconquerable.}{Theodore N.~Vail}

\OriginallyPublished{Section}{sec:intro:What Makes Parallel Programming Hard?}{What Makes Parallel Programming Hard?}{a Portland State University Technical Report}{PaulEMcKenney2009ProgrammingHard}

It is important to note that the difficulty of parallel programming
is as much a human-factors issue as it is a set of technical properties of the
parallel programming problem.
We do need human beings to be able to tell parallel
systems what to do, otherwise known as programming.
But parallel programming involves two-way communication, with
a program's performance and scalability being the communication from
the machine to the human.
In short, the human writes a program telling the computer what to do,
and the computer critiques this program via the resulting performance and
scalability.
Therefore, appeals to abstractions or to mathematical analyses will
often be of severely limited utility.

In the Industrial Revolution, the interface between human and machine
was evaluated by human-factor studies, then called time-and-motion
studies.
Although there have been a few human-factor studies examining parallel
programming~\cite{RyanEccles2005HPCSNovice,RyanEccles2006HPCSNoviceNeeds,
LorinHochstein2005SC,DuaneSzafron1994PEMPDS}, these studies have
been extremely narrowly focused, and hence unable to demonstrate any
general results.
Furthermore, given that the normal range of programmer productivity
spans more than an order of magnitude, it is unrealistic to expect
an affordable study to be capable of detecting (say) a 10\,\% difference
in productivity.
Although the multiple-order-of-magnitude differences that such studies
\emph{can} reliably detect are extremely valuable, the most impressive
improvements tend to be based on a long series of 10\,\% improvements.

We must therefore take a different approach.

\begin{figure}
\centering
\resizebox{3in}{!}{\includegraphics{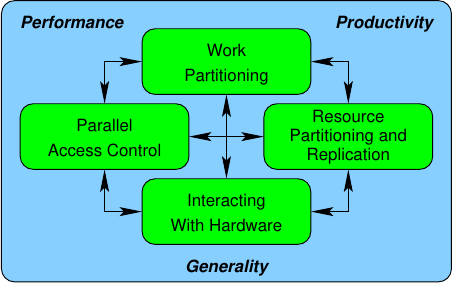}}
\caption{Categories of Tasks Required of Parallel Programmers}
\label{fig:intro:Categories of Tasks Required of Parallel Programmers}
\end{figure}

One such approach is to carefully consider the tasks that parallel
programmers must undertake that are not required of sequential programmers.
We can then evaluate how well a given programming language or environment
assists the developer with these tasks.
These tasks fall into the four categories shown in
\cref{fig:intro:Categories of Tasks Required of Parallel Programmers},
each of which is covered in the following sections.

\subsection{Work Partitioning}
\label{sec:intro:Work Partitioning}

Work partitioning is absolutely required for parallel execution:
If there is but one ``glob'' of work, then it can be executed by at
most one CPU at a time, which is by definition sequential execution.
However, partitioning the work requires great care.
For example, uneven partitioning can result in sequential execution
once the small partitions have completed~\cite{GeneAmdahl1967AmdahlsLaw}.
In less extreme cases, load balancing can be used to fully utilize
available hardware and restore performance and scalability.

Although partitioning can greatly improve performance and scalability,
it can also increase complexity.
For example, partitioning can complicate handling of global
errors and events:
A parallel program may need to carry out non-trivial synchronization
in order to safely process such global events.
More generally, each partition requires some sort of communication:
After all, if
a given thread did not communicate at all, it would have no effect and
would thus not need to be executed.
However, because communication incurs overhead, careless partitioning choices
can result in severe performance degradation.

Furthermore, the number of concurrent threads must often be controlled,
as each such thread occupies common resources, for example,
space in CPU caches.
If too many threads are permitted to execute concurrently, the
CPU caches will overflow, resulting in high cache miss rate, which in
turn degrades performance.
Conversely, large numbers of threads are often required to
overlap computation and I/O so as to fully utilize I/O devices.

\QuickQuiz{
	Other than CPU cache capacity, what might require limiting the
	number of concurrent threads?
}\QuickQuizAnswer{
	There are any number of potential limits on the number of
	threads:
	\begin{enumerate}
	\item	Main memory.
		Each thread consumes some memory
		(for its stack if nothing else), so that excessive
		numbers of threads can exhaust memory, resulting
		in excessive paging or memory-allocation failures.
	\item	I/O bandwidth.
		If each thread initiates a given
		amount of mass-storage I/O or networking traffic,
		excessive numbers of threads can result in excessive
		I/O queuing delays, again degrading performance.
		Some networking protocols may be subject to timeouts
		or other failures if there are so many threads that
		networking events cannot be responded to in a timely
		fashion.
	\item	Synchronization overhead.
		For many synchronization protocols, excessive numbers
		of threads can result in excessive spinning, blocking,
		or rollbacks, thus degrading performance.
	\end{enumerate}

	Specific applications and platforms may have any number of additional
	limiting factors.
}\QuickQuizEnd

Finally, permitting threads to execute concurrently greatly increases
the program's state space, which can make the program difficult to
understand and debug, degrading productivity.
All else being equal, smaller state spaces having more regular structure
are more easily understood, but this is a human-factors statement as much
as it is a technical or mathematical statement.
Good parallel designs might have extremely large state spaces, but
nevertheless be easy to understand due to their regular structure,
while poor designs can be impenetrable despite having a comparatively
small state space.
The best designs exploit embarrassing parallelism, or transform the
problem to one having an embarrassingly parallel solution.
In either case, ``embarrassingly parallel'' is in fact
an embarrassment of riches.
The current state of the art enumerates good designs; more work is
required to make more general judgments on
state-space size and structure.

\subsection{Parallel Access Control}
\label{sec:Parallel Access Control}

Given a single-threaded sequential program, that single
thread has full access to all of the program's resources.
These resources are most often in-memory data structures, but can be CPUs,
memory (including caches), I/O devices, computational accelerators, files,
and much else besides.

The first parallel-access-control issue is whether the form of access to
a given resource depends on that resource's location.
For example, in many message-passing environments, local-variable
access is via expressions and assignments,
while remote-variable access uses an entirely different
syntax, usually involving messaging.
The POSIX Threads environment~\cite{OpenGroup1997pthreads},
Structured Query Language (SQL)~\cite{DIS9075SQL92}, and
partitioned global address-space (PGAS) environments
such as Universal Parallel C (UPC)~\cite{ElGhazawi2003UPC,UPCConsortium2013}
offer implicit access,
while Message Passing Interface (MPI)~\cite{MPIForum2008} offers
explicit access because access to remote data requires explicit
messaging.

The other parallel-access-control issue is how threads coordinate
access to the resources.
This coordination is carried out by
the very large number of synchronization mechanisms
provided by various parallel languages and environments,
including message passing, locking, transactions,
reference counting, explicit timing, shared atomic variables, and data
ownership.
Many traditional parallel-programming concerns such as \IX{deadlock},
\IX{livelock}, and transaction rollback stem from this coordination.
This framework can be elaborated to include comparisons
of these synchronization mechanisms, for example locking vs.\@ transactional
memory~\cite{McKenney2007PLOSTM}, but such elaboration is beyond the
scope of this section.
(See
\cref{sec:future:Transactional Memory,%
sec:future:Hardware Transactional Memory}
for more information on transactional memory.)

\QuickQuiz{
	Just what is ``explicit timing''???
}\QuickQuizAnswer{
	Where each thread is given access to some set of resources during
	an agreed-to slot of time.
	For example, a parallel program with eight threads might be
	organized into eight-millisecond time intervals, so that the
	first thread is given access during the first millisecond of
	each interval, the second thread during the second millisecond,
	and so on.
	This approach clearly requires carefully synchronized clocks
	and careful control of execution times, and therefore should
	be used with considerable caution.

	In fact, outside of hard realtime environments, you almost
	certainly want to use something else instead.
	Explicit timing is nevertheless worth a mention, as it is
	always there when you need it.
}\QuickQuizEnd

\subsection{Resource Partitioning and Replication}
\label{sec:Resource Partitioning and Replication}

The most effective parallel algorithms and systems exploit resource
parallelism, so much so that it is
usually wise to begin parallelization by partitioning your write-intensive
resources and replicating frequently accessed read-mostly resources.
The resource in question is most frequently data, which might be
partitioned over computer systems, mass-storage devices, \IXplr{NUMA node},
CPU cores (or dies or hardware threads), pages, cache lines, instances
of synchronization primitives, or critical sections of code.
For example, partitioning over locking primitives is termed
``\IXh{data}{locking}''~\cite{Beck85}.

Resource partitioning is frequently application dependent.
For example, numerical applications frequently partition matrices
by row, column, or sub-matrix, while commercial applications frequently
partition write-intensive data structures and replicate
read-mostly data structures.
Thus, a commercial application might assign the data for a
given customer to a given few computers out of a large cluster.
An application might statically partition data, or dynamically
change the partitioning over time.

Resource partitioning is extremely effective, but
it can be quite challenging for complex multilinked data
structures.

\subsection{Interacting With Hardware}
\label{sec:Interacting With Hardware}

Hardware interaction is normally the domain of the operating system,
the compiler, libraries, or other software-environment infrastructure.
However, developers working with novel hardware features and components
will often need to work directly with such hardware.
In addition, direct access to the hardware can be required when squeezing
the last drop of performance out of a given system.
In this case, the developer may need to tailor or configure the application
to the cache geometry, system topology, or interconnect protocol of the
target hardware.

In some cases, hardware may be considered to be a resource which
is subject to partitioning or access control, as described in
the previous sections.

\subsection{Composite Capabilities}
\label{sec:Composite Capabilities}

\begin{figure}
\centering
\resizebox{3in}{!}{\includegraphics{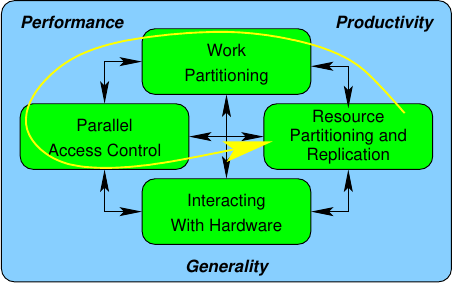}}
\caption{Ordering of Parallel-Programming Tasks}
\label{fig:intro:Ordering of Parallel-Programming Tasks}
\end{figure}

Although these four capabilities are fundamental,
good engineering practice uses composites of
these capabilities.
For example, the data-parallel approach first
partitions the data so as to minimize the need for
inter-partition communication, partitions the code accordingly,
and finally maps data partitions and threads so as to maximize
throughput while minimizing inter-thread communication,
as shown in
\cref{fig:intro:Ordering of Parallel-Programming Tasks}.
The developer can then
consider each partition separately, greatly reducing the size
of the relevant state space, in turn increasing productivity.
Even though some problems are non-partitionable,
clever transformations into forms permitting partitioning can
sometimes greatly enhance
both performance and scalability~\cite{PanagiotisMetaxas1999PDCS}.

\subsection{How Do Languages and Environments Assist With These Tasks?}
\label{sec:intro:How Do Languages and Environments Assist With These Tasks?}

Although many environments require the developer to deal manually
with these tasks, there are long-standing environments that bring
significant automation to bear.
The poster child for these environments is SQL, many implementations
of which automatically parallelize single large queries and also
automate concurrent execution of independent queries and updates.

These four categories of tasks must be carried out in all parallel
programs, but that of course does not necessarily mean that the developer
must manually carry out these tasks.
We can expect to see ever-increasing automation of these four tasks
as parallel systems continue to become cheaper and more readily available.

\QuickQuiz{
	Are there any other obstacles to parallel programming?
}\QuickQuizAnswer{
	There are a great many other potential obstacles to parallel
	programming.
	Here are a few of them:
	\begin{enumerate}
	\item	The only known algorithms for a given project might
		be inherently sequential in nature.
		In this case, either avoid parallel programming
		(there being no law saying that your project \emph{has}
		to run in parallel) or invent a new parallel algorithm.
	\item	The project allows binary-only plugins that share the same
		address space, such that no one developer has access to
		all of the source code for the project.
		Because many parallel bugs, including deadlocks, are
		global in nature, such binary-only plugins pose a severe
		challenge to current software development methodologies.
		This might well change, but for the time being, all
		developers of parallel code sharing a given address space
		need to be able to see \emph{all} of the code running in
		that address space.
	\item	The project contains heavily used APIs that were designed
		without regard to
		parallelism~\cite{HagitAttiya2011LawsOfOrder,Clements:2013:SCR:2517349.2522712}.
		Some of the more ornate features of the System V
		message-queue API form a case in point.
		Of course, if your project has been around for a few
		decades, and its developers did not have access to
		parallel hardware, it undoubtedly has at least
		its share of such APIs.
	\item	The project was implemented without regard to parallelism.
		Given that there are a great many techniques that work
		extremely well in a sequential environment, but that
		fail miserably in parallel environments, if your project
		ran only on sequential hardware for most of its lifetime,
		then your project undoubtably has at least its share of
		parallel-unfriendly code.
	\item	The project was implemented without regard to good
		software-development practice.
		The cruel truth is that shared-memory parallel
		environments are often much less forgiving of sloppy
		development practices than are sequential environments.
		You may be well-served to clean up the existing design
		and code prior to attempting parallelization.
	\item	The people who originally did the development on your
		project have since moved on, and the people remaining,
		while well able to maintain it or add small features,
		are unable to make ``big animal'' changes.
		In this case, unless you can work out a very simple
		way to parallelize your project, you will probably
		be best off leaving it sequential.
		That said, there are a number of simple approaches that
		you might use
		to parallelize your project, including running multiple
		instances of it, using a parallel implementation of
		some heavily used library function, or making use of
		some other parallel project, such as a database.
	\end{enumerate}

	One can argue that many of these obstacles are non-technical
	in nature, but that does not make them any less real.
	In short, parallelization of a large body of code
	can be a large and complex effort.
	As with any large and complex effort, it makes sense to
	do your homework beforehand.
}\QuickQuizEnd

\section{Discussion}
\label{sec:intro:Discussion}
\epigraph{Until you try, you don't know what you can't do.}
	 {Henry James}

This section has given an overview of the difficulties with, goals of,
and alternatives to parallel programming.
This overview was followed by a discussion of
what can make parallel programming hard, along with a high-level
approach for dealing with parallel programming's difficulties.
Those who still insist that parallel programming is impossibly difficult
should review some of the older guides to parallel
programmming~\cite{SQNTParallel,AndrewDBirrell1989Threads,Beck85,Inman85}.
The following quote from Andrew Birrell's
monograph~\cite{AndrewDBirrell1989Threads} is especially telling:

\begin{quote}
	Writing concurrent programs has a reputation for being exotic
	and difficult.
	I~believe it is neither.
	You need a system that provides you with good primitives
	and suitable libraries,
	you need a basic caution and carefulness, you need an armory of
	useful techniques, and you need to know of the common pitfalls.
	I~hope that this paper has helped you towards sharing my belief.
\end{quote}

The authors of these older guides were well up to the parallel programming
challenge back in the 1980s.
As such, there are simply no excuses for refusing to step up to the
parallel-programming challenge here in the 21\textsuperscript{st} century!

We are now ready to proceed to the next chapter, which dives into the
relevant properties of the parallel hardware underlying our parallel
software.

	\setlength{\parskip}{0.0pt plus 1ex}
	\input{qqzintro}
	\setlength{\parskip}{0.0pt plus 1.0pt}% return to default

% cpu/cpu.tex
% mainfile: ../perfbook.tex
% SPDX-License-Identifier: CC-BY-SA-3.0

\QuickQuizChapter{chp:Hardware and its Habits}{Hardware and its Habits}{qqzcpu}
\Epigraph{Premature abstraction is the root of all evil.}
	 {A cast of thousands}

Most people intuitively understand that passing messages between systems
is more expensive than performing simple calculations within the confines
of a single system.
But it is also the case that communicating among threads within the
confines of a single shared-memory system can also be quite expensive.
This chapter therefore looks at the cost of synchronization and communication
within a shared-memory system.
These few pages can do no more than scratch the surface of shared-memory
parallel hardware design; readers desiring more detail would do well
to start with a recent edition of \pplsur{John L.}{Hennessy}'s and
\pplsur{David A.}{Patterson}'s classic
text~\cite{Hennessy2017,Hennessy95a}.

\QuickQuiz{
	Why should parallel programmers bother learning low-level
	properties of the hardware?
	Wouldn't it be easier, better, and more elegant to remain at
	a higher level of abstraction?
}\QuickQuizAnswer{
	It might well be easier to ignore the detailed properties of
	the hardware, but in most cases it would be quite foolish
	to do so.
	If you accept that the only purpose of parallelism is to
	increase performance, and if you further accept that
	performance depends on detailed properties of the hardware,
	then it logically follows that parallel programmers are going
	to need to know at least a few hardware properties.

	This is the case in most engineering disciplines.
	Would \emph{you} want to use a bridge designed by an
	engineer who did not understand the properties of
	the concrete and steel making up that bridge?
	If not, why would you expect a parallel programmer to be
	able to develop competent parallel software without at least
	\emph{some} understanding of the underlying hardware?
}\QuickQuizEnd

% cpu/overview.tex
% mainfile: ../perfbook.tex
% SPDX-License-Identifier: CC-BY-SA-3.0

\section{Overview}
\label{sec:cpu:Overview}
\epigraph{Mechanical Sympathy:
	  Hardware and software working together in harmony.}
	 {Martin Thompson}

Careless reading of computer-system specification sheets might lead one
to believe that CPU performance is a footrace on a clear track, as
illustrated in \cref{fig:cpu:CPU Performance at its Best},
where the race always goes to the swiftest.

\begin{figure}
\centering
\resizebox{3in}{!}{\includegraphics{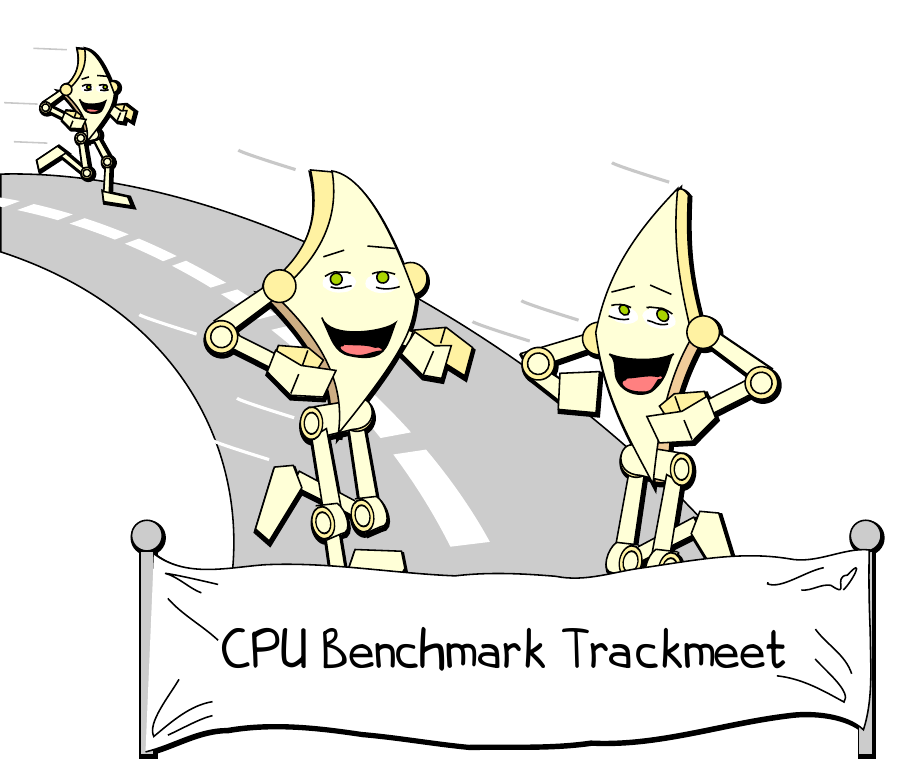}}
\caption{CPU Performance at its Best}
\ContributedBy{Figure}{fig:cpu:CPU Performance at its Best}{Melissa Broussard}
\end{figure}

Although there are a few CPU-bound benchmarks that approach the ideal case
shown in \cref{fig:cpu:CPU Performance at its Best},
the typical program more closely resembles an obstacle course than
a race track.
This is because the internal architecture of CPUs has changed dramatically
over the past few decades, courtesy of \IXr{Moore's Law}.
These changes are described in the following sections.

\subsection{Pipelined CPUs}
\label{sec:cpu:Pipelined CPUs}

\begin{figure}
\centering
\resizebox{3in}{!}{\includegraphics{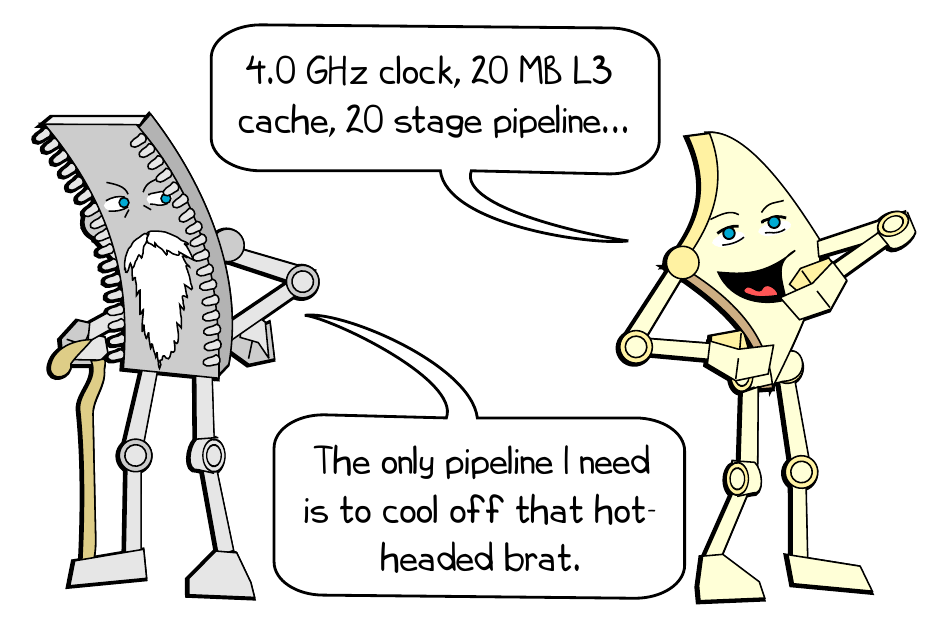}}
\caption{CPUs Old and New}
\ContributedBy{Figure}{fig:cpu:CPUs Old and New}{Melissa Broussard}
\end{figure}

In the 1980s, the typical microprocessor fetched an instruction, decoded
it, and executed it, typically taking \emph{at least} three clock cycles
to complete one instruction before even starting the next.
In contrast, the CPU of the late 1990s and of the 2000s execute many
instructions simultaneously, using \emph{pipelines}; \emph{superscalar}
techniques; \emph{out-of-order} instruction and data handling;
\emph{speculative execution}, and
more~\cite{Hennessy2017,Hennessy2011}
in order to optimize the flow of instructions and data through the CPU\@.
Some cores have more than one hardware thread, which is variously called
\emph{simultaneous multithreading} (SMT) or \emph{hyperthreading}
(HT)~\cite{JFennel1973SMT},
each of which appears as
an independent CPU to software, at least from a functional viewpoint.
These modern hardware features can greatly improve performance, as
illustrated by \cref{fig:cpu:CPUs Old and New}.

Achieving full performance with a CPU having a long pipeline requires
highly predictable control flow through the program.
Suitable control flow can be provided by a program that executes primarily
in tight loops, for example, arithmetic on large matrices or vectors.
The CPU can then correctly predict that the branch at the end of the loop
will be taken in almost all cases,
allowing the pipeline to be kept full and the CPU to execute at full speed.

\begin{figure}
\centering
\resizebox{3in}{!}{\includegraphics{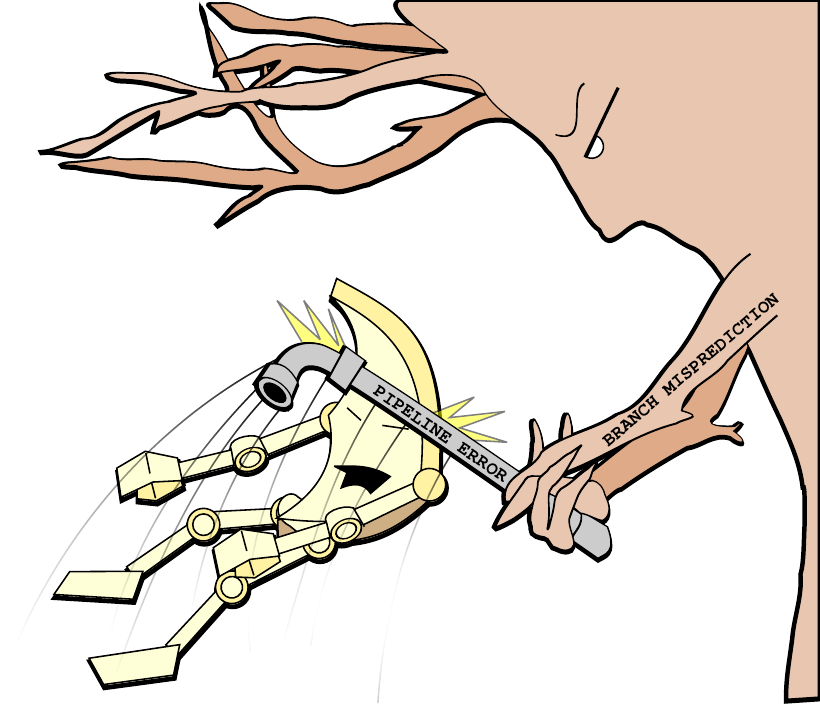}}
\caption{CPU Meets a Pipeline Flush}
\ContributedBy{Figure}{fig:cpu:CPU Meets a Pipeline Flush}{Melissa Broussard}
\end{figure}

However, branch prediction is not always so easy.
For example, consider a program with many loops, each of which iterates
a small but random number of times.
For another example, consider an old-school object-oriented program with
many virtual objects that can reference many different real objects, all
with different implementations for frequently invoked member functions,
resulting in many calls through pointers.
In these cases, it is difficult or even
impossible for the CPU to predict where the next branch might lead.
Then either the CPU must stall waiting for execution to proceed far
enough to be certain where that branch leads, or it must guess and
then proceed using speculative execution.
Although guessing works extremely well for programs with predictable
control flow, for unpredictable branches (such as those in binary search)
the guesses will frequently be wrong.
A wrong guess can be expensive because the CPU must discard any
speculatively executed instructions following the corresponding
branch, resulting in a pipeline flush.
If pipeline flushes appear too frequently, they drastically reduce
overall performance, as fancifully depicted in
\cref{fig:cpu:CPU Meets a Pipeline Flush}.

\begin{figure}
\centering
\resizebox{3in}{!}{\includegraphics{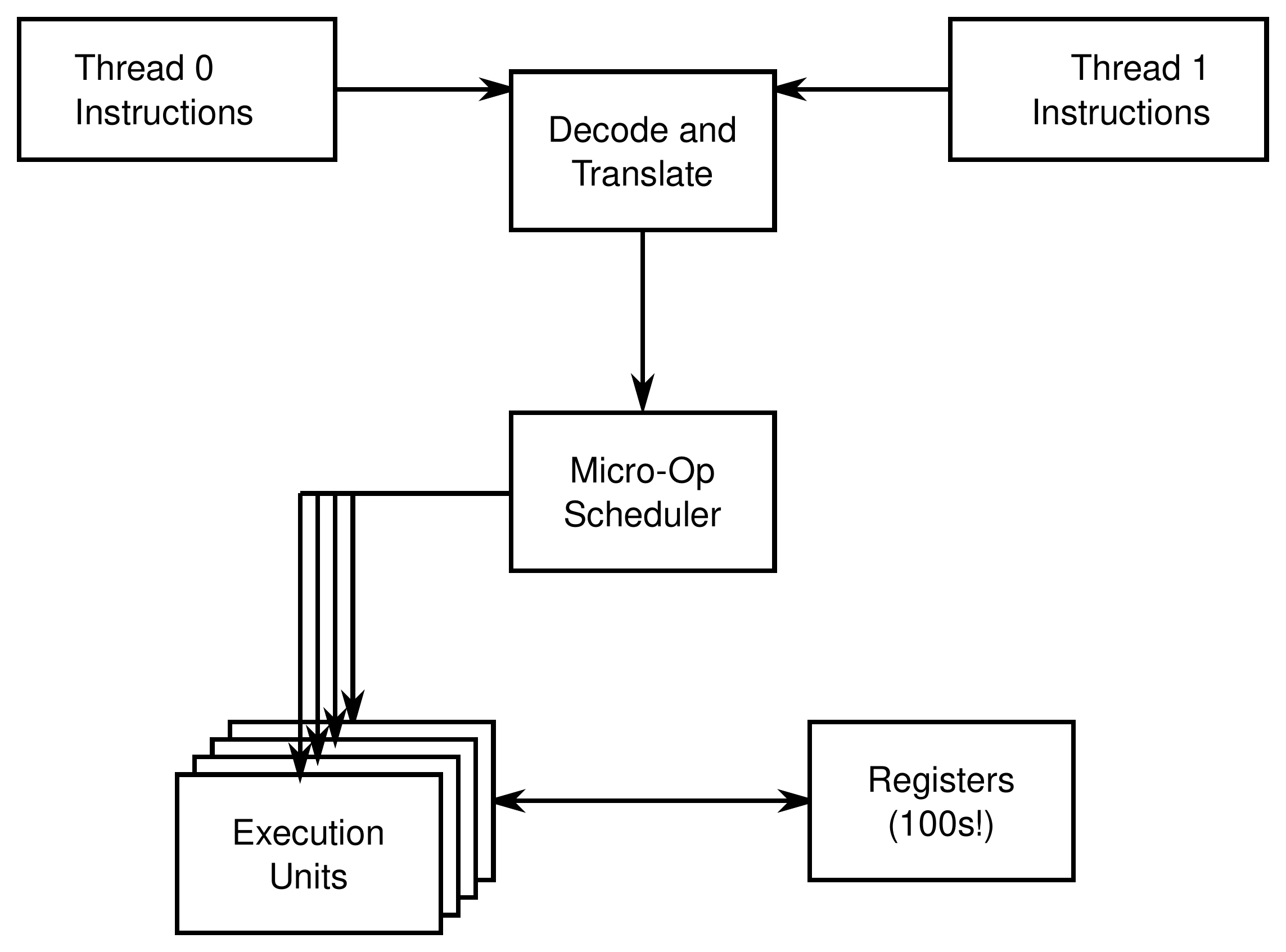}}
\caption{Rough View of Modern Micro-Architecture}
\label{fig:cpu:Rough View of Modern Micro-Architecture}
\end{figure}

This gets even worse in the increasingly common case of hyperthreading
(or SMT, if you prefer), especially on pipelined superscalar out-of-order
CPU featuring speculative execution.
In this increasingly common case, all the hardware threads sharing
a core also share that core's resources, including registers, cache,
execution units, and so on.
The instructions are often decoded into micro-operations, and use of the
shared execution units and the hundreds of hardware registers is often
coordinated by a micro-operation scheduler.
A rough diagram of such a two-threaded core is shown in
\cref{fig:cpu:Rough View of Modern Micro-Architecture},
and more accurate (and thus more complex) diagrams are available in
textbooks and scholarly papers.\footnote{
	Here is one example for a late-2010s Intel core:
	\url{https://en.wikichip.org/wiki/intel/microarchitectures/skylake_(server)}.}
Therefore, the execution of one hardware thread can and often is perturbed
by the actions of other hardware threads sharing that core.

Even if only one hardware thread is active (for example, in old-school
CPU designs where there is only one thread), counterintuitive results
are quite common.
Execution units often have overlapping capabilities, so that a CPU's
choice of execution unit can result in pipeline stalls due to contention
for that execution unit from later instructions.
In theory, this contention is avoidable, but in practice CPUs must choose
very quickly and without the benefit of clairvoyance.
In particular, adding an instruction to a tight loop can sometimes
actually cause execution to \emph{speed up}.

Unfortunately, pipeline flushes and shared-resource contention are not
the only hazards in the obstacle course that modern CPUs must run.
The next section covers the hazards of referencing memory.

\subsection{Memory References}
\label{sec:cpu:Memory References}

In the 1980s, it often took less time for a microprocessor to load a value
from memory than it did to execute an instruction.
More recently, microprocessors might execute hundreds or even thousands
of instructions in the time required to access memory.
This disparity is due to the fact that \IXr{Moore's Law} has increased CPU
performance at a much greater rate than it has decreased memory \IX{latency},
in part due to the rate at which memory sizes have grown.
For example, a typical 1970s minicomputer might have 4\,KB (yes, kilobytes,
not megabytes, let alone gigabytes or terabytes) of main memory, with
single-cycle access.\footnote{
	It is only fair to add that each of these single cycles
	lasted no less than 1.6 \emph{microseconds}.}
Present-day CPU designers still can construct a 4\,KB memory with single-cycle
access, even on systems with multi-GHz clock frequencies.
And in fact they frequently do construct such memories, but they now
call them ``level-0 caches'', plus they can be quite a bit bigger than 4\,KB.

\begin{figure}
\centering
\resizebox{3in}{!}{\includegraphics{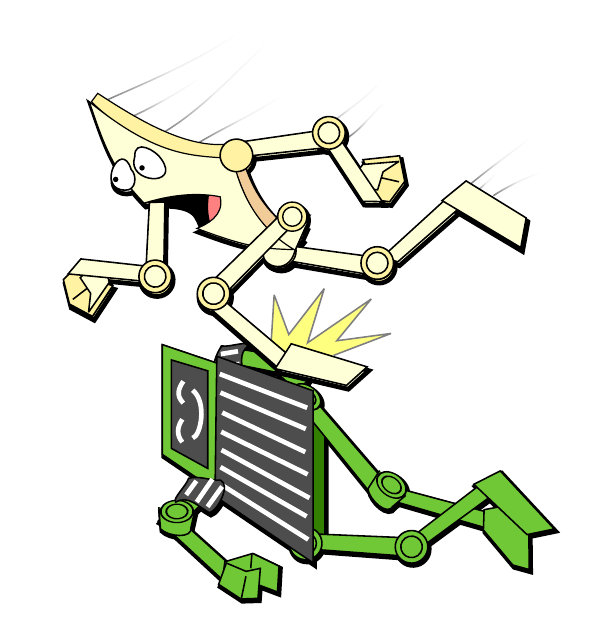}}
\caption{CPU Meets a Memory Reference}
\ContributedBy{Figure}{fig:cpu:CPU Meets a Memory Reference}{Melissa Broussard}
\end{figure}

Although the large caches found on modern microprocessors can do quite
a bit to help combat memory-access latencies,
these caches require highly predictable data-access patterns to
successfully hide those latencies.
Unfortunately, common operations such as traversing a linked list
have extremely unpredictable memory-access patterns---after all,
if the pattern was predictable, us software types would not bother
with the pointers, right?
Therefore, as shown in
\cref{fig:cpu:CPU Meets a Memory Reference},
memory references often pose severe obstacles to modern CPUs.

Thus far, we have only been considering obstacles that can arise during
a given CPU's execution of single-threaded code.
Multi-threading presents additional obstacles to the CPU, as
described in the following sections.

\subsection{Atomic Operations}
\label{sec:cpu:Atomic Operations}

One such obstacle is \IX{atomic} operations.
The problem here is that the whole idea of an atomic operation conflicts with
the piece-at-a-time assembly-line operation of a CPU pipeline.
To hardware designers' credit, modern CPUs use a number of extremely clever
tricks to make such operations \emph{look} atomic even though they
are in fact being executed piece-at-a-time,
with one common trick being to identify all the cachelines containing the
data to be atomically operated on,
ensure that these cachelines are owned by the CPU executing the
atomic operation, and only then proceed with the atomic operation
while ensuring that these cachelines remained owned by this CPU\@.
Because all the data is private to this CPU, other CPUs are unable to
interfere with the atomic operation despite the piece-at-a-time nature
of the CPU's pipeline.
Needless to say, this sort of trick can require that
the pipeline must be delayed or even flushed in order to
perform the setup operations that
permit a given atomic operation to complete correctly.

\begin{figure}
\centering
\resizebox{3in}{!}{\includegraphics{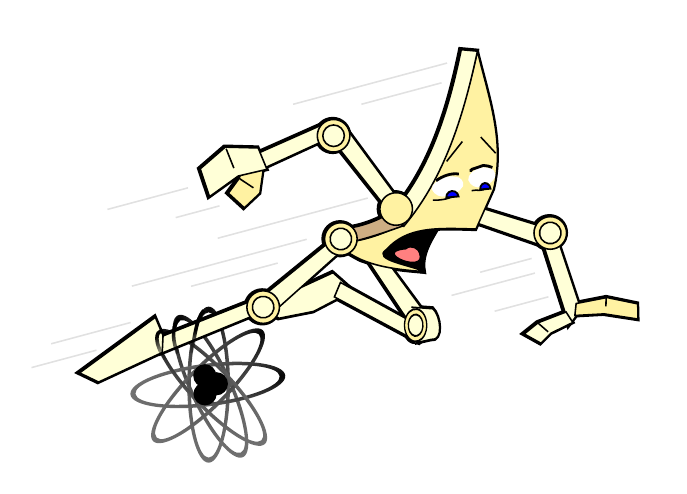}}
\caption{CPU Meets an Atomic Operation}
\ContributedBy{Figure}{fig:cpu:CPU Meets an Atomic Operation}{Melissa Broussard}
\end{figure}

In contrast, when executing a non-atomic operation, the CPU can load
values from cachelines as they appear and place the results in the
store buffer, without the need to wait for cacheline ownership.
Although there are a number of hardware optimizations that can sometimes
hide cache latencies, the resulting effect on performance is all too
often as depicted in
\cref{fig:cpu:CPU Meets an Atomic Operation}.

Unfortunately, atomic operations usually apply only to single elements
of data.
Because many parallel algorithms require that ordering constraints
be maintained between updates of multiple data elements, most CPUs
provide memory barriers.
These memory barriers also serve as performance-sapping obstacles,
as described in the next section.

\QuickQuiz{
	What types of machines would allow atomic operations on
	multiple data elements?
}\QuickQuizAnswer{
	One answer to this question is that it is often possible to
	pack multiple elements of data into a single machine word,
	which can then be manipulated atomically.

	A more trendy answer would be machines supporting transactional
	memory~\cite{DBLomet1977SIGSOFT,Knight:1986:AMF:319838.319854,Herlihy93a}.
	By early 2014, several mainstream systems provided limited
	hardware transactional memory implementations, which is covered
	in more detail in
	\cref{sec:future:Hardware Transactional Memory}.
	The jury is still out on the applicability of software transactional
	memory~\cite{McKenney2007PLOSTM,DonaldEPorter2007TRANSACT,
	ChistopherJRossbach2007a,CalinCascaval2008tmtoy,
	AleksandarDragovejic2011STMnotToy,AlexanderMatveev2012PessimisticTM},
	which is covered in \cref{sec:future:Transactional Memory}.
}\QuickQuizEnd

\subsection{Memory Barriers}
\label{sec:cpu:Memory Barriers}

\IXpl{Memory barrier} will be considered in more detail in
\cref{chp:Advanced Synchronization: Memory Ordering} and
\cref{chp:app:whymb:Why Memory Barriers?}\@.
In the meantime, consider the following simple lock-based \IX{critical
section}:

\begin{VerbatimN}
spin_lock(&mylock);
a = a + 1;
spin_unlock(&mylock);
\end{VerbatimN}

\begin{figure}
\centering
\resizebox{3in}{!}{\includegraphics{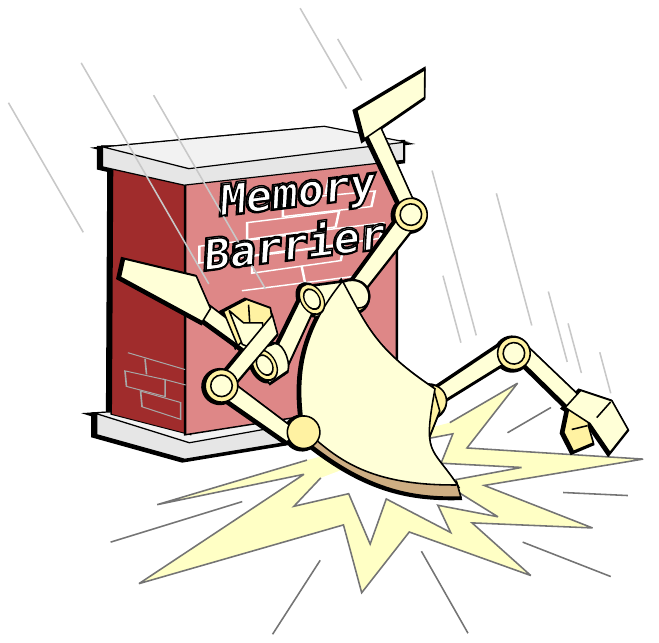}}
\caption{CPU Meets a Memory Barrier}
\ContributedBy{Figure}{fig:cpu:CPU Meets a Memory Barrier}{Melissa Broussard}
\end{figure}

If the CPU were not constrained to execute these statements in the
order shown, the effect would be that the variable ``a'' would be
incremented without the protection of ``mylock'', which would certainly
defeat the purpose of acquiring it.
To prevent such destructive reordering, locking primitives contain
either explicit or implicit memory barriers.
Because the whole purpose of these memory barriers is to prevent reorderings
that the CPU would otherwise undertake in order to increase performance,
memory barriers almost always reduce performance, as depicted in
\cref{fig:cpu:CPU Meets a Memory Barrier}.

As with atomic operations, CPU designers have been working hard to
reduce \IXh{memory-barrier}{overhead}, and have made substantial progress.

\subsection{Thermal Throttling}
\label{sec:cpu:Thermal Throttling}

\begin{figure}
\centering
\resizebox{3in}{!}{\includegraphics{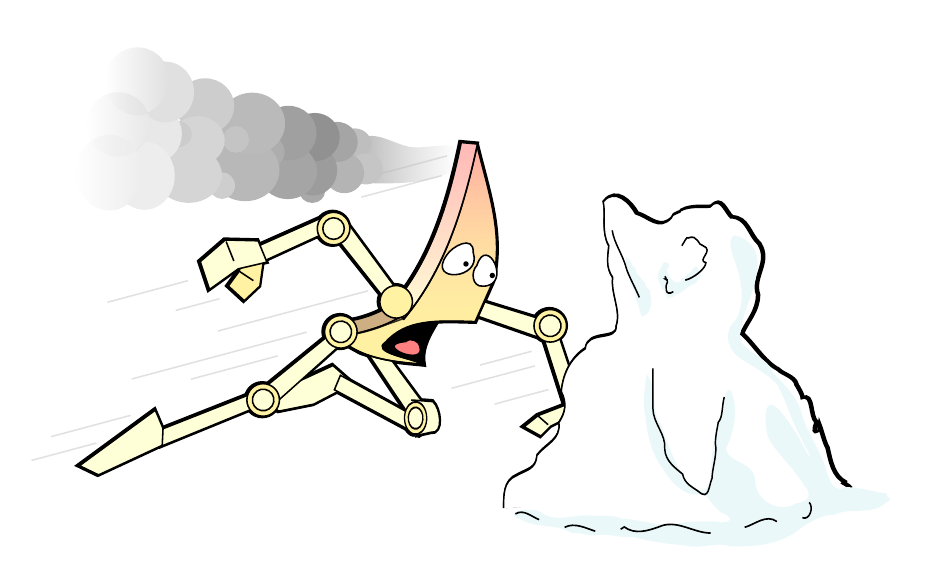}}
\caption{CPU Encounters Thermal Throttling}
\ContributedBy{Figure}{fig:cpu:Encounters Thermal Throttling}{Melissa Broussard, remixed}
\end{figure}

One increasingly common frustrating experience is to carefully
micro-optimize a critical code path, greatly reducing the number of
clock cycles consumed by that code path, only to find that the
wall-clock time consumed by that code has actually \emph{increased}.

Welcome to modern thermal throttling.

If you reduced the number of clock cycles by making more effective
use of the CPU's functional units, you will have increased the
power consumed by that CPU\@.
This will in turn increase the amount of heat dissipated by that CPU\@.
If this heat dissipation exceeds the cooling system's capacity, the
system will thermally throttle that CPU, for example, by reducing
its clock frequency, as fancifully depicted by the snow penguin in
\cref{fig:cpu:Encounters Thermal Throttling}.

If performance is of the essence, the proper fix is improved cooling,
an approach loved by serious gamers and by overclockers.\footnote{
	Some of whom make good use of liquid nitrogen.}
But if you cannot modify your computer's cooling system, perhaps because
you are renting it from a cloud provider, then you will need to take
some other optimization approach.
For example, you might need to apply algorithmic optimizations instead
of hardware-centric micro-optimizations.
Alternatively, perhaps you can parallelize your code, spreading the
work (and thus the heat) over multiple CPU cores.

\subsection{Cache Misses}
\label{sec:cpu:Cache Misses}

\begin{figure}
\centering
\resizebox{3in}{!}{\includegraphics{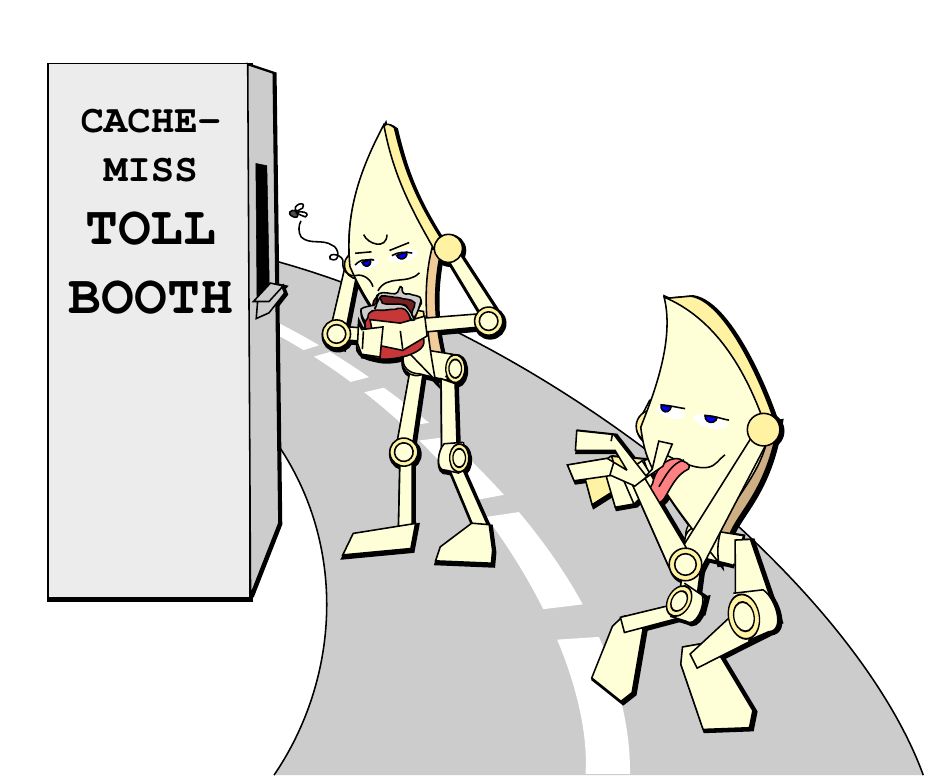}}
\caption{CPU Meets a Cache Miss}
\ContributedBy{Figure}{fig:cpu:CPU Meets a Cache Miss}{Melissa Broussard}
\end{figure}

An additional multi-threading obstacle to CPU performance is
the ``cache miss''.
As noted earlier, modern CPUs sport large caches in order to reduce the
performance penalty that would otherwise be incurred due to high memory
latencies.
However, these caches are actually counter-productive for variables that
are frequently shared among CPUs.
This is because when a given CPU wishes to modify the variable, it is
most likely the case that some other CPU has modified it recently.
In this case, the variable will be in that other CPU's cache, but not
in this CPU's cache, which will therefore incur an expensive cache miss
(see \cref{sec:app:whymb:Cache Structure} for more detail).
Such cache misses form a major obstacle to CPU performance, as shown
in \cref{fig:cpu:CPU Meets a Cache Miss}.

\QuickQuiz{
	So have CPU designers also greatly reduced the overhead of
	cache misses?
}\QuickQuizAnswer{
	Unfortunately, not so much.
	There has been some reduction given constant numbers of CPUs,
	but the finite speed of light and the atomic nature of
	matter limits their ability to reduce cache-miss overhead
	for larger systems.
	\Cref{sec:cpu:Hardware Free Lunch?}
	discusses some possible avenues for possible future progress.
}\QuickQuizEnd

\subsection{I/O Operations}
\label{sec:cpu:I/O Operations}

\begin{figure}
\centering
\resizebox{3in}{!}{\includegraphics{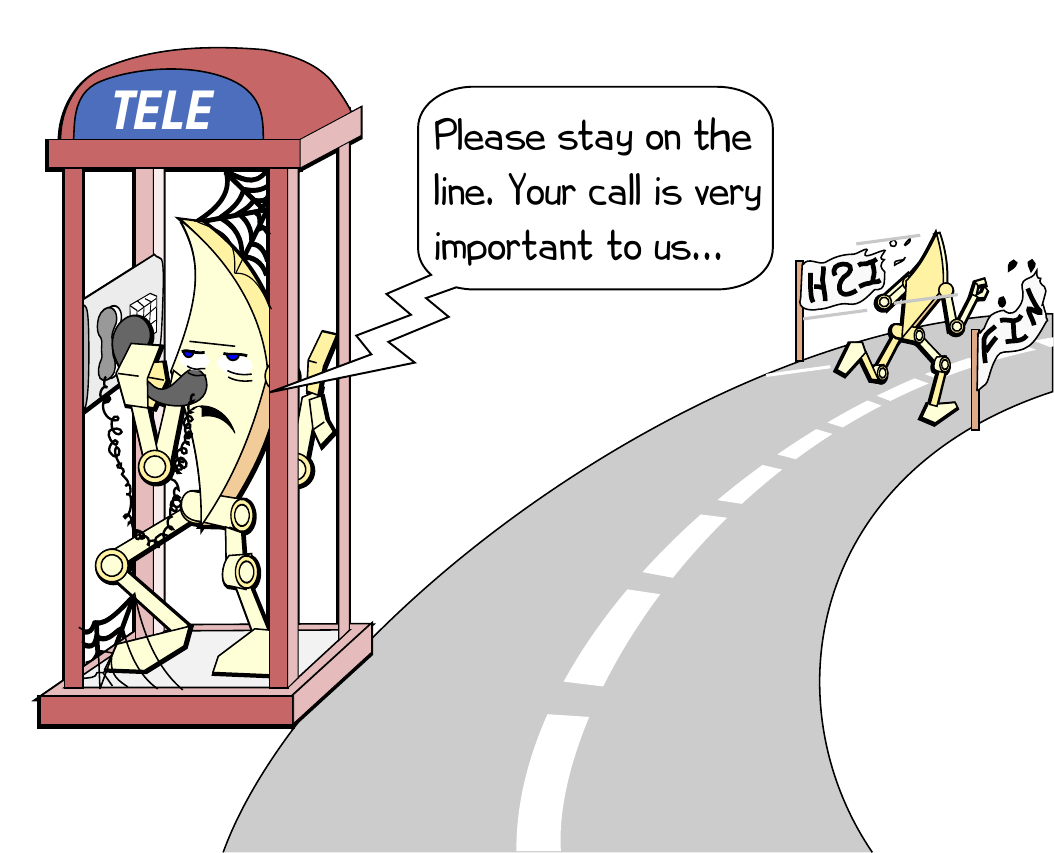}}
\caption{CPU Waits for I/O Completion}
\ContributedBy{Figure}{fig:cpu:CPU Waits for I/O Completion}{Melissa Broussard}
\end{figure}

A cache miss can be thought of as a CPU-to-CPU I/O operation, and as
such is one of the cheapest I/O operations available.
I/O operations involving networking, mass storage, or (worse yet) human
beings pose much greater obstacles than the internal obstacles called
out in the prior sections,
as illustrated by
\cref{fig:cpu:CPU Waits for I/O Completion}.

This is one of the differences between shared-memory and distributed-system
parallelism:
Shared-memory parallel programs must normally deal with no
obstacle worse than a cache miss, while a distributed parallel program
will typically incur the larger network communication latencies.
In both cases, the relevant latencies can be thought of as a cost of
communication---a cost that would be absent in a sequential program.
Therefore, the ratio between the overhead of the communication to
that of the actual work being performed is a key design parameter.
A major goal of parallel hardware design is to reduce this ratio as
needed to achieve the relevant performance and scalability goals.
In turn, as will be seen in
\cref{chp:Partitioning and Synchronization Design},
a major goal of parallel software design is to reduce the
frequency of expensive operations like communications cache misses.

Of course, it is one thing to say that a given operation is an obstacle,
and quite another to show that the operation is a \emph{significant}
obstacle.
This distinction is discussed in the following sections.

% cpu/overheads.tex
% mainfile: ../perfbook.tex
% SPDX-License-Identifier: CC-BY-SA-3.0

\section{Overheads}
\label{sec:cpu:Overheads}
\epigraph{Don't design bridges in ignorance of materials, and don't design
	  low-level software in ignorance of the underlying hardware.}
	 {Unknown}

This section presents actual \IXpl{overhead} of the obstacles to performance
listed out in the previous section.
However, it is first necessary to get a rough view of hardware system
architecture, which is the subject of the next section.

\subsection{Hardware System Architecture}
\label{sec:cpu:Hardware System Architecture}

\begin{figure}
\centering
\resizebox{3in}{!}{\includegraphics{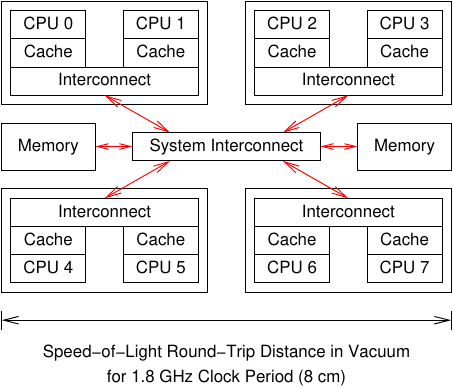}}
\caption{System Hardware Architecture}
\label{fig:cpu:System Hardware Architecture}
\end{figure}

\Cref{fig:cpu:System Hardware Architecture}
shows a rough schematic of an eight-core computer system.
Each die has a pair of CPU cores, each with its cache, as well as an
interconnect allowing the pair of CPUs to communicate with each other.
The system interconnect allows the four dies to communicate with each
other and with main memory.

Data moves through this system in units of ``\IXpl{cache line}'', which
are power-of-two fixed-size aligned blocks of memory, usually ranging
from 32 to 256 bytes in size.
When a CPU loads a variable from memory to one of its registers, it must
first load the cacheline containing that variable into its cache.
Similarly, when a CPU stores a value from one of its registers into
memory, it must also load the cacheline containing that variable into
its cache, but must also ensure that no other CPU has a copy of that
cacheline.

For example, if CPU~0 were to write to a variable whose cacheline
resided in CPU~7's cache, the following over-simplified sequence of
events might ensue:

\begin{enumerate}
\item	CPU~0 checks its local cache, and does not find the cacheline.
	It therefore records the write in its store buffer.
\item	A request for this cacheline is forwarded to CPU~0's and~1's
	interconnect, which checks CPU~1's local cache, and does not
	find the cacheline.
\item	This request is forwarded to the system interconnect, which
	checks with the other three dies, learning that the cacheline
	is held by the die containing CPU~6 and~7.
\item	This request is forwarded to CPU~6's and~7's interconnect, which
	checks both CPUs' caches, finding the value in CPU~7's cache.
\item	CPU~7 forwards the cacheline to its interconnect, and also
	flushes the cacheline from its cache.
\item	CPU~6's and~7's interconnect forwards the cacheline to the
	system interconnect.
\item	The system interconnect forwards the cacheline to CPU~0's and~1's
	interconnect.
\item	CPU~0's and~1's interconnect forwards the cacheline to CPU~0's
	cache.
\item	CPU~0 can now complete the write, updating the relevant portions
	of the newly arrived cacheline from the value previously
	recorded in the store buffer.
\end{enumerate}

\QuickQuizSeries{%
\QuickQuizB{
	This is a \emph{simplified} sequence of events?
	How could it \emph{possibly} be any more complex?
}\QuickQuizAnswerB{
	This sequence ignored a number of possible complications,
	including:

	\begin{enumerate}
	\item	Other CPUs might be concurrently attempting to perform
		memory-reference operations involving this same cacheline.
	\item	The cacheline might have been replicated read-only in
		several CPUs' caches, in which case, it would need to
		be flushed from their caches.
	\item	CPU~7 might have been operating on the cache line when
		the request for it arrived, in which case CPU~7 might
		need to hold off the request until its own operation
		completed.
	\item	CPU~7 might have ejected the cacheline from its cache
		(for example, in order to make room for other data),
		so that by the time that the request arrived, the
		cacheline was on its way to memory.
	\item	A correctable error might have occurred in the cacheline,
		which would then need to be corrected at some point before
		the data was used.
	\end{enumerate}

	Production-quality cache-coherence mechanisms are extremely
	complicated due to these sorts of
	considerations~\cite{Hennessy95a,DavidECuller1999,MiloMKMartin2012scale,DanielJSorin2011MemModel}.
}\QuickQuizEndB
\QuickQuizE{
	Why is it necessary to flush the cacheline from CPU~7's cache?
}\QuickQuizAnswerE{
	If the cacheline was not flushed from CPU~7's cache, then
	CPUs~0 and~7 might have different values for the same set
	of variables in the cacheline.
	This sort of incoherence greatly complicates parallel software,
	which is why wise hardware architects avoid it.
}\QuickQuizEndE
}

This simplified sequence is just the beginning of a discipline called
\emph{cache-coherency protocols}~\cite{Hennessy95a,DavidECuller1999,MiloMKMartin2012scale,DanielJSorin2011MemModel},
which is discussed in more detail in
\cref{chp:app:whymb:Why Memory Barriers?}\@.
As can be seen in the sequence of events triggered by a \IXacr{cas} operation,
a single instruction can cause considerable protocol traffic, which
can significantly degrade your parallel program's performance.

Fortunately, if a given variable is being frequently read during a time
interval during which it is never updated, that variable can be replicated
across all CPUs' caches.
This replication permits all CPUs to enjoy extremely fast access to
this \emph{read-mostly} variable.
\Cref{chp:Deferred Processing} presents synchronization
mechanisms that take full advantage of this important hardware read-mostly
optimization.

\subsection{Costs of Operations}
\label{sec:cpu:Costs of Operations}

\begin{table}
%\rowcolors{1}{}{lightgray}
\renewcommand*{\arraystretch}{1.1}
\centering\small
\tcresizewidth{
\begin{tabular}
  {
    ll
    S[table-format = 9.1]
    S[table-format = 9.1]
    r
  }
	\toprule
	\multicolumn{2}{l}{Operation}
			& \multicolumn{1}{r}{Cost (ns)}
				   & {\parbox[b]{.7in}{\raggedleft Ratio\\(cost/clock)}}
					    & CPUs \\
	\midrule
	\multicolumn{2}{l}{Clock period}
			   &   0.5 &    1.0 &		\\
	\midrule
	\multicolumn{2}{l}{Same-CPU}
			   &       &        & 0		\\
		& CAS      &   7.0 &   14.6 &		\\
		& lock     &  15.4 &   32.3 &		\\
	\midrule
	\multicolumn{2}{l}{On-Core}
			   &       &        & 224	\\
		& Blind CAS&   7.2 &   15.2 &		\\
		& CAS	   &  18.0 &   37.7 & 		\\
        \midrule
	\multicolumn{2}{l}{Off-Core}
			   &       &	    & 1--27	\\
		& Blind CAS&  47.5 &   99.8 & 225--251	\\
		& CAS	   & 101.9 &  214.0 &		\\
        \midrule
	\multicolumn{2}{l}{Off-Socket}
			   &       &        & 28--111	\\
		& Blind CAS& 148.8 &  312.5 & 252--335	\\
		& CAS	   & 442.9 &  930.1 &		\\
        \midrule
	\multicolumn{2}{l}{Cross-Interconnect}
			   &       &        & 112--223	\\
		& Blind CAS& 336.6 &  706.8 & 336--447	\\
		& CAS	   & 944.8 & 1984.2 &		\\
	\midrule
	\multicolumn{2}{l}{Off-System}
				&		&	      & \\
		& Comms Fabric  &         5 000 &      10 500 & \\
		& Global Comms  &   195 000 000 & 409 500 000 & \\
	\bottomrule
\end{tabular}
}
\caption{CPU 0 View of Synchronization Mechanisms on 8-Socket System With Intel Xeon Platinum 8176 CPUs @ 2.10\,GHz}
\label{tab:cpu:CPU 0 View of Synchronization Mechanisms on 8-Socket System With Intel Xeon Platinum 8176 CPUs at 2.10GHz}
\end{table}

The overheads of some common operations important to parallel programs are
displayed in
\cref{tab:cpu:CPU 0 View of Synchronization Mechanisms on 8-Socket System With Intel Xeon Platinum 8176 CPUs at 2.10GHz}.
This system's clock period rounds to 0.5\,ns.
Although it is not unusual for modern microprocessors to be able to
retire multiple instructions per clock period, the operations' costs are
nevertheless normalized to a clock period in the third column, labeled
``Ratio''.
The first thing to note about this table is the large values of many of
the ratios.

The same-CPU \acrmf{cas} operation consumes about seven
nanoseconds, a duration more than ten times that of the clock period.
CAS is an atomic operation in which the hardware compares the contents
of the specified memory location to a specified ``old'' value, and if
they compare equal, stores a specified ``new'' value, in which case the
CAS operation succeeds.
If they compare unequal, the memory location keeps its (unexpected) value,
and the CAS operation fails.
The operation is atomic in that the hardware guarantees that the memory
location will not be changed between the compare and the store.
CAS functionality is provided by the \co{lock;cmpxchg} instruction on x86.

The ``same-CPU'' prefix means that the CPU now performing the CAS operation
on a given variable was also the last CPU to access this variable, so
that the corresponding cacheline is already held in that CPU's cache.
Similarly, the same-CPU lock operation (a ``round trip'' pair consisting
of a lock acquisition and release) consumes more than fifteen nanoseconds,
or more than thirty clock cycles.
The lock operation is more expensive than CAS because it requires two
atomic operations on the lock data structure, one for acquisition and
the other for release.

On-core operations involving interactions between the hardware threads
sharing a single core are about the same cost as same-CPU operations.
This should not be too surprising, given that these two hardware threads
also share the full cache hierarchy.

In the case of the blind CAS, the software specifies the old value
without looking at the memory location.
This approach is appropriate when attempting to acquire a lock.
If the unlocked state is represented by zero and the locked state
is represented by the value one, then a CAS operation on the lock
that specifies zero for the old value and one for the new value
will acquire the lock if it is not already held.
The key point is that there is only one access to the memory
location, namely the CAS operation itself.

In contrast, a normal CAS operation's old value is derived from
some earlier load.
For example, to implement an atomic increment, the current value of
that location is loaded and that value is incremented to produce the
new value.
Then in the CAS operation, the value actually loaded would be specified
as the old value and the incremented value as the new value.
If the value had not been changed between the load and the CAS, this
would increment the memory location.
However, if the value had in fact changed, then the old value would
not match, causing a miscompare that would result in the CAS operation
failing.
The key point is that there are now two accesses to the memory location,
the load and the CAS\@.

Thus, it is not surprising that on-core blind CAS consumes only about
seven nanoseconds, while on-core CAS consumes about 18 nanoseconds.
The non-blind case's extra load does not come for free.
That said, the overhead of these operations are similar to same-CPU
CAS and lock, respectively.

\QuickQuiz{
	\Cref{tab:cpu:CPU 0 View of Synchronization Mechanisms on 8-Socket System With Intel Xeon Platinum 8176 CPUs at 2.10GHz}
	shows CPU~0 sharing a core with CPU~224.
	Shouldn't that instead be CPU~1???
}\QuickQuizAnswer{
	It is easy to be sympathetic to this view, but the file
	\path{/sys/devices/system/cpu/cpu0/cache/index0/shared_cpu_list}
	really does contain the string \co{0,224}.
	Therefore, CPU~0's hyperthread twin really is CPU~224.
	Some people speculate that this numbering allows naive applications
	and schedulers to perform better, citing the fact that on many
	workloads the second hyperthread does not provide a huge
	amount of additional performance.
	This speculation assumes that naive applications and schedulers
	would utilize CPUs in numerical order, leaving aside the weaker
	hyperthread twin CPUs until all cores are in use.
}\QuickQuizEnd

A blind CAS involving CPUs in different cores but on the same socket
consumes almost fifty nanoseconds, or almost one hundred clock cycles.
The code used for this cache-miss measurement passes the cache line
back and forth between a pair of CPUs, so this cache miss is satisfied
not from memory, but rather from the other CPU's cache.
A non-blind CAS operation, which as noted earlier must look at the old
value of the variable as well as store a new value, consumes over one
hundred nanoseconds, or more than two hundred clock cycles.
Think about this a bit.
In the time required to do \emph{one} CAS operation, the CPU could have
executed more than \emph{two hundred} normal instructions.
This should demonstrate the limitations not only of fine-grained locking,
but of any other synchronization mechanism relying on fine-grained
global agreement.

If the pair of CPUs are on different sockets, the operations are considerably
more expensive.
A blind CAS operation consumes almost 150~nanoseconds, or more than
three hundred clock cycles.
A normal CAS operation consumes more than 400~nanoseconds, or almost
\emph{one thousand} clock cycles.

Worse yet, not all pairs of sockets are created equal.
This particular system appears to be constructed as a pair of four-socket
components, with additional latency penalties when the CPUs reside
in different components.
In this case, a blind CAS operation consumes more than three hundred
nanoseconds, or more than seven hundred clock cycles.
A CAS operation consumes almost a full microsecond, or almost two
thousand clock cycles.

\QuickQuizLabelRel{\QspeedOfLightAtoms}{1} % cann't put label inside QQSeries

\QuickQuizSeries{%
\QuickQuizB{
	Surely the hardware designers could be persuaded to improve
	this situation!
	Why have they been content with such abysmal performance
	for these single-instruction operations?
}\QuickQuizAnswerB{
	The hardware designers \emph{have} been working on this
	problem, and have consulted with no less a luminary than
	the late physicist Stephen Hawking.
	Hawking's observation was that the hardware designers have
	two basic problems~\cite{BryanGardiner2007}:

	\begin{enumerate}
	\item	The finite speed of light, and
	\item	The atomic nature of matter.
	\end{enumerate}

\begin{table}
%\rowcolors{1}{}{lightgray}
\renewcommand*{\arraystretch}{1.1}
\centering\small
\begin{tabular}
  {
    ll
    S[table-format = 9.1]
    S[table-format = 9.1]
  }
	\toprule
	\multicolumn{2}{l}{Operation}
			& \multicolumn{1}{r}{Cost (ns)}
			& {\parbox[b]{.7in}{\raggedleft Ratio\\(cost/clock)}} \\
	\midrule
	\multicolumn{2}{l}{Clock period}
			&           0.4	&           1.0 \\
        \midrule
	\multicolumn{2}{l}{Same-CPU}
			&		&		\\
	& CAS		&          12.2	&          33.8 \\
	& lock		&          25.6	&          71.2 \\
        \midrule
        \multicolumn{2}{l}{On-Core}
			&		&		\\
	& Blind CAS	&          12.9	&          35.8 \\
	& CAS		&           7.0	&          19.4 \\
	\midrule
        \multicolumn{2}{l}{Off-Core}
			&		&		\\
	& Blind CAS	&          31.2	&          86.6 \\
	& CAS		&          31.2	&          86.5 \\
	\midrule
	\multicolumn{2}{l}{Off-Socket}
			&		&		\\
	& Blind CAS	&          92.4	&         256.7 \\
	& CAS		&          95.9	&         266.4 \\
	\midrule
	\multicolumn{2}{l}{Off-System}
			&		&		\\
	& Comms Fabric	&       2 600   &       7 220   \\
	& Global Comms	& 195 000 000	& 542 000 000   \\
	\bottomrule
\end{tabular}
\caption{Performance of Synchronization Mechanisms on 16-CPU 2.8\,GHz Intel X5550 (Nehalem) System}
\label{tab:cpu:Performance of Synchronization Mechanisms on 16-CPU 2.8GHz Intel X5550 (Nehalem) System}
\end{table}

	The first problem limits raw speed, and the second limits
	miniaturization, which in turn limits frequency.
	And even this sidesteps the power-consumption issue that
	is currently limiting production frequencies to well below
	10\,GHz.

	In addition,
	\cref{tab:cpu:CPU 0 View of Synchronization Mechanisms on 8-Socket System With Intel Xeon Platinum 8176 CPUs at 2.10GHz}
	on
	\cpageref{tab:cpu:CPU 0 View of Synchronization Mechanisms on 8-Socket System With Intel Xeon Platinum 8176 CPUs at 2.10GHz}
	represents a reasonably large system with no fewer than 448~hardware
	threads.
	Smaller systems often achieve better latency, as may be seen in
	\cref{tab:cpu:Performance of Synchronization Mechanisms on 16-CPU 2.8GHz Intel X5550 (Nehalem) System},
	which represents a much smaller system with only 16~hardware threads.
	A similar view is provided by the rows of
	\cref{tab:cpu:CPU 0 View of Synchronization Mechanisms on 8-Socket System With Intel Xeon Platinum 8176 CPUs at 2.10GHz}
	down to and including the two ``Off-Core'' rows.

\begin{table}
%\rowcolors{1}{}{lightgray}
\renewcommand*{\arraystretch}{1.1}
\centering\small
\tcresizewidth{
\begin{tabular}
  {
    ll
    S[table-format = 9.1]
    S[table-format = 9.1]
    r
  }
	\toprule
	\multicolumn{2}{l}{Operation}
		& \multicolumn{1}{r}{Cost (ns)}
			& {\parbox[b]{.7in}{\raggedleft Ratio\\(cost/clock)}}
			& CPUs \\
	\midrule
	\multicolumn{2}{l}{Clock period}
				     &   0.5 &    1.0 &			  \\
        \midrule
	\multicolumn{2}{l}{Same-CPU} &       &        &	0		  \\
	& CAS			     &   6.2 &   13.6 &			  \\
	& lock			     &  13.5 &   29.6 &			  \\
        \midrule
	\multicolumn{2}{l}{On-Core}  &       &        &	6		  \\
	& Blind CAS		     &   6.5 &   14.3 &			  \\
	& CAS			     &  16.2 &   35.6 &			  \\
        \midrule
	\multicolumn{2}{l}{Off-Core} &       &        &	1--5		  \\
	& Blind CAS		     &  22.2 &   48.8 & 7--11		  \\
	& CAS			     &  53.6 &  117.9 &			  \\
	\midrule
	\multicolumn{2}{l}{Off-System}&       &        &		  \\
	& Comms Fabric		      & 5 000 & 11 000 &		  \\
	& Global Comms		      & 195 000 000 & 429 000 000 &	  \\
	\bottomrule
\end{tabular}
}
\caption{CPU 0 View of Synchronization Mechanisms on 12-CPU Intel Core i7-8750H CPU @ 2.20\,GHz}
\label{tab:cpu:CPU 0 View of Synchronization Mechanisms on 12-CPU Intel Core i7-8750H CPU @ 2.20GHz}
\end{table}

	Furthermore, newer small-scale single-socket systems such
	as the laptop on which I am typing this also have more
	reasonable latencies, as can be seen in
	\cref{tab:cpu:CPU 0 View of Synchronization Mechanisms on 12-CPU Intel Core i7-8750H CPU @ 2.20GHz}.

	Alternatively, a 64-CPU system in the mid 1990s had
	cross-interconnect latencies in excess of five microseconds,
	so even the eight-socket 448-hardware-thread monster shown in
	\cref{tab:cpu:CPU 0 View of Synchronization Mechanisms on 8-Socket System With Intel Xeon Platinum 8176 CPUs at 2.10GHz}
	represents more than a five-fold improvement over its
	25-years-prior counterparts.

	Integration of hardware threads in a single core and multiple
	cores on a die have improved latencies greatly, at least within the
	confines of a single core or single die.
	There has been some improvement in overall system latency,
	but only by about a factor of two.
	Unfortunately, neither the speed of light nor the atomic nature
	of matter has changed much in the past few
	years~\cite{NoBugsHare2016CPUoperations}.
	Therefore, spatial and temporal locality are first-class concerns
	for concurrent software, even when running on relatively
	small systems.

	\Cref{sec:cpu:Hardware Free Lunch?}
	looks at what else hardware designers might be
	able to do to ease the plight of parallel programmers.
}\QuickQuizEndB
\QuickQuizE{
	\Cref{tab:cpu:Performance of Synchronization Mechanisms on 16-CPU 2.8GHz Intel X5550 (Nehalem) System}
	in the answer to \QuickQuizARef{\QspeedOfLightAtoms} on
	\cpageref{tab:cpu:Performance of Synchronization Mechanisms on 16-CPU 2.8GHz Intel X5550 (Nehalem) System}
	says that on-core CAS is faster than both of same-CPU CAS and
	on-core blind CAS\@.
	What is happening there?
}\QuickQuizAnswerE{
	I \emph{was} surprised by the data I obtained and did a rigorous
	check of their validity.
	I got the same result persistently.
	One theory that might explain the observation would be:
	The two threads in the core are able to overlap their accesses,
	while the single CPU must do everything sequentially.
	Unfortunately, there seems to be no public documentation explaining
	why the Intel X5550 (Nehalem) system behaved like that.
}\QuickQuizEndE
}                 % End of \QuickQuizSeries

\begin{table}
\rowcolors{1}{}{lightgray}
\renewcommand*{\arraystretch}{1.1}
\centering\small
\begin{tabular}{lrrrrr}
	\toprule
	Level &  Scope & Line Size &   Sets & Ways &    Size \\
	\midrule
	L0    &   Core &        64 &     64 &    8 &     32K \\
	L1    &   Core &        64 &     64 &    8 &     32K \\
	L2    &   Core &        64 &   1024 &   16 &   1024K \\
	L3    & Socket &        64 & 57,344 &   11 & 39,424K \\
	\bottomrule
\end{tabular}
\caption{Cache Geometry for 8-Socket System With Intel Xeon Platinum 8176 CPUs @ 2.10\,GHz}
\label{tab:cpu:Cache Geometry for 8-Socket System With Intel Xeon Platinum 8176 CPUs @ 2.10GHz}
\end{table}

Unfortunately, the high speed of within-core and within-socket communication
does not come for free.
First, there are only two CPUs within a given core and only 56 within
a given socket, compared to 448 across the system.
Second, as shown in
\cref{tab:cpu:Cache Geometry for 8-Socket System With Intel Xeon Platinum 8176 CPUs @ 2.10GHz},
the on-core caches are quite small compared to the on-socket caches, which
are in turn quite small compared to the 1.4\,TB of memory configured on
this system.
Third, again referring to the figure, the caches are organized as
a hardware hash table with a limited number of items per bucket.
For example, the raw size of the L3 cache (``Size'') is almost 40\,MB, but each
bucket (``Line'') can only hold 11 blocks of memory (``Ways''), each
of which can be at most 64 bytes (``Line Size'').
This means that only 12 bytes of memory (admittedly at carefully chosen
addresses) are required to overflow this 40\,MB cache.
On the other hand, equally careful choice of addresses might make good
use of the entire 40\,MB.

Spatial locality of reference is clearly extremely important, as is
spreading the data across memory.

I/O operations are even more expensive.
As shown in the ``Comms Fabric'' row,
high performance (and expensive!\@) communications fabric, such as
InfiniBand or any number of proprietary interconnects, has a latency
of roughly five microseconds for an end-to-end round trip, during which
time more than \emph{ten thousand} instructions might have been executed.
Standards-based communications networks often require some sort of
protocol processing, which further increases the \IX{latency}.
Of course, geographic distance also increases latency, with the
speed-of-light through optical fiber latency around the world coming to
roughly 195 \emph{milliseconds}, or more than 400 million clock
cycles, as shown in the ``Global Comms'' row.

% Reference for Infiniband latency:
% http://www.hpcadvisorycouncil.com/events/2014/swiss-workshop/presos/Day_1/1_Mellanox.pdf
%     page 6/76 'Leading Interconnect, Leading Performance'
% Needs updating...

\QuickQuiz{
	These numbers are insanely large!
	How can I possibly get my head around them?
}\QuickQuizAnswer{
	Get a roll of toilet paper.
	In the USA, each roll will normally have somewhere around
	350--500 sheets.
	Tear off one sheet to represent a single clock cycle, setting it aside.
	Now unroll the rest of the roll.

	The resulting pile of toilet paper will likely represent a single
	\IXacr{cas} cache miss.

	For the more-expensive inter-system communications latencies,
	use several rolls (or multiple cases) of toilet paper to represent
	the communications latency.

	Important safety tip:
	Make sure to account for the needs of those you live with when
	appropriating toilet paper, especially
	in 2020 or during a similar time when store shelves are free of
	toilet paper and much else besides.

	Furthermore, for those working on kernel code, a CPU disabling
	interrupts across a cache miss is analogous to you holding your
	breath while unrolling a roll of toilet paper.
	How many rolls of toilet paper can \emph{you} unroll while holding
	your breath?
	You might wish to avoid disabling interrupts across that many
	cache misses.\footnote{
		Kudos to Matthew Wilcox for this holding-breath analogy.}
}\QuickQuizEnd

\subsection{Hardware Optimizations}
\label{sec:cpu:Hardware Optimizations}

It is only natural to ask how the hardware is helping, and the answer
is ``Quite a bit!''

One hardware optimization is large cachelines.
This can provide a big performance boost, especially when software is
accessing memory sequentially.
For example, given a 64-byte cacheline and software accessing 64-bit
variables, the first access will still be slow due to speed-of-light
delays (if nothing else), but the remaining seven can be quite fast.
However, this optimization has a dark side, namely \IX{false sharing},
which happens when different variables in the same cacheline are
being updated by different CPUs, resulting in a high cache-miss rate.
Software can use the alignment directives available in many compilers
to avoid false sharing, and adding such directives is a common step
in tuning parallel software.

A second related hardware optimization is cache prefetching, in which
the hardware reacts to consecutive accesses by prefetching subsequent
cachelines, thereby evading speed-of-light delays for these
subsequent cachelines.
Of course, the hardware must use simple heuristics to determine when
to prefetch, and these heuristics can be fooled by the complex data-access
patterns in many applications.
Fortunately, some CPU families allow for this by providing special
prefetch instructions.
Unfortunately, the effectiveness of these instructions in the general
case is subject to some dispute.

A third hardware optimization is the store buffer, which allows a string
of store instructions to execute quickly even when the stores are to
non-consecutive addresses and when none of the needed cachelines are
present in the CPU's cache.
The dark side of this optimization is memory misordering, for which see
\cref{chp:Advanced Synchronization: Memory Ordering}.

A fourth hardware optimization is speculative execution, which can
allow the hardware to make good use of the store buffers without
resulting in memory misordering.
The dark side of this optimization can be energy inefficiency and
lowered performance if the speculative execution goes awry and must
be rolled back and retried.
Worse yet, the advent of
Spectre and Meltdown~\cite{JannHorn2018MeltdownSpectre}
made it apparent that hardware speculation can also enable side-channel
attacks that defeat memory-protection hardware so as to allow unprivileged
processes to read memory that they should not have access to.
It is clear that the combination of speculative execution and cloud
computing needs more than a bit of rework!

A fifth hardware optimization is large caches, allowing individual
CPUs to operate on larger datasets without incurring expensive cache
misses.
Although large caches can degrade \IXh{energy}{efficiency} and \IXh{cache-miss}
{latency}, the ever-growing cache sizes on production microprocessors
attests to the power of this optimization.

A final hardware optimization is read-mostly replication, in which
data that is frequently read but rarely updated is present in all
CPUs' caches.
This optimization allows the read-mostly data to be accessed
exceedingly efficiently, and is the subject of
\cref{chp:Deferred Processing}.

\begin{figure}
\centering
\resizebox{3in}{!}{\includegraphics{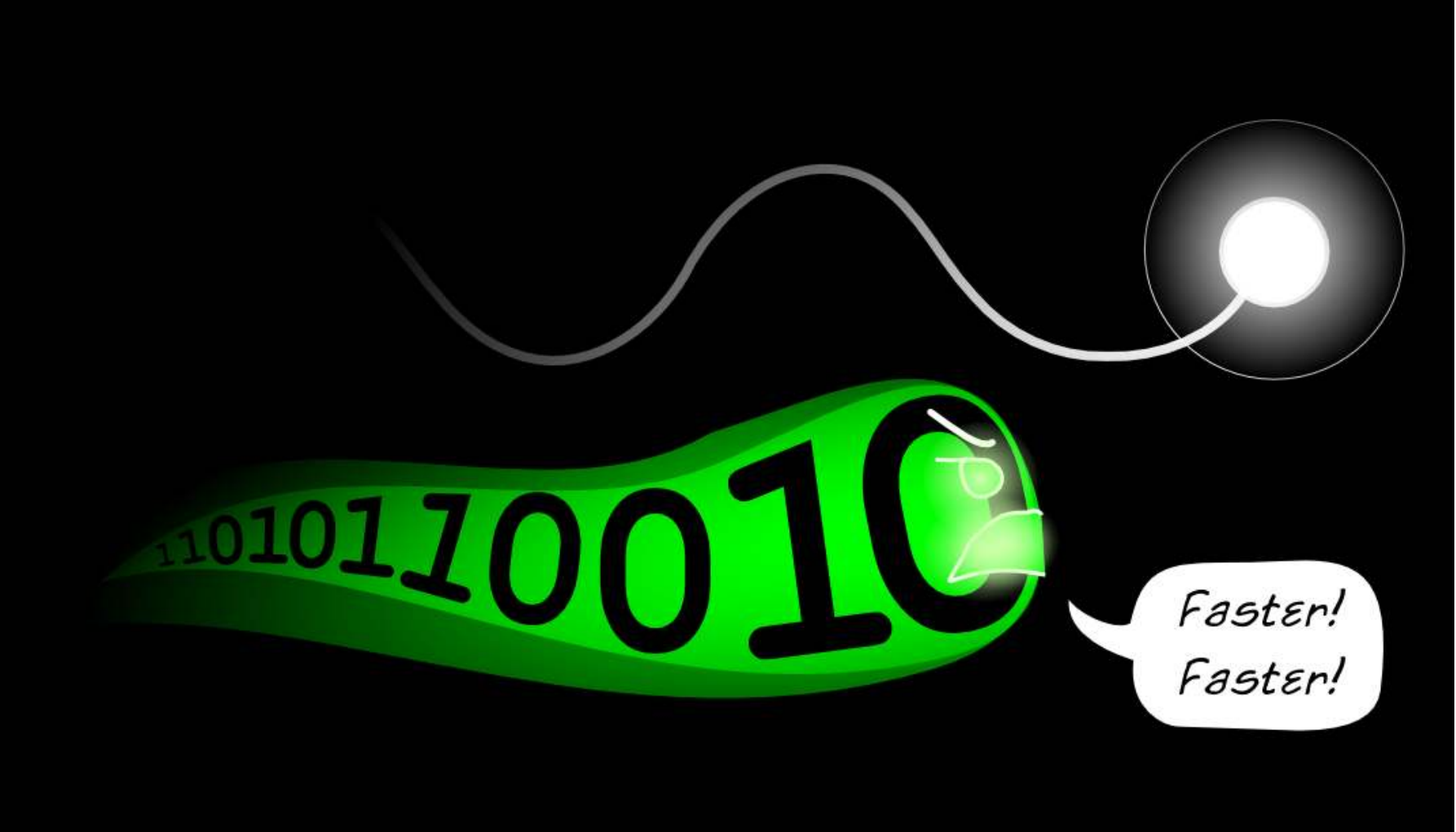}}
\caption{Hardware and Software:
				On Same Side}
\ContributedBy{Figure}{fig:cpu:Hardware and Software: On Same Side}{Melissa Broussard}
\end{figure}

In short, hardware and software engineers are really on the same side,
with both trying to make computers go fast despite the best efforts of
the laws of physics, as fancifully depicted in
\cref{fig:cpu:Hardware and Software: On Same Side}
where our data stream is trying its best to exceed the speed of light.
The next section discusses some additional things that the hardware engineers
might (or might not) be able to do, depending on how well recent
research translates to practice.
Software's contribution to this noble goal is outlined in the remaining
chapters of this book.

% cpu/hwfreelunch.tex
% mainfile: ../perfbook.tex
% SPDX-License-Identifier: CC-BY-SA-3.0

\section{Hardware Free Lunch?}
\label{sec:cpu:Hardware Free Lunch?}
\epigraph{The great trouble today is that there are too many people looking
	  for someone else to do something for them.
	  The solution to most of our troubles is to be found in everyone
	  doing something for themselves.}
	 {Henry Ford, updated}

The major reason that concurrency has been receiving so much focus over
the past few years is the end of \IXaltr{Moore's-Law}{Moore's Law}
induced single-threaded
performance increases
(or ``free lunch''~\cite{HerbSutter2008EffectiveConcurrency}),
as shown in
\cref{fig:intro:Clock-Frequency Trend for Intel CPUs} on
\cpageref{fig:intro:Clock-Frequency Trend for Intel CPUs}.
This section briefly surveys a few ways that hardware designers
might bring back the ``free lunch''.

However, the preceding section presented some substantial hardware
obstacles to exploiting concurrency.
One severe physical limitation that hardware designers face is the
finite speed of light.
As noted in
\cref{fig:cpu:System Hardware Architecture} on
\cpageref{fig:cpu:System Hardware Architecture},
light can manage only about an 8-centimeters round trip in a vacuum
during the duration of a 1.8\,GHz clock period.
This distance drops to about 3~centimeters for a 5\,GHz clock.
Both of these distances are relatively small compared to the size
of a modern computer system.

To make matters even worse, electric waves in silicon move from three to
thirty times more slowly than does light in a vacuum, and common
clocked logic constructs run still more slowly, for example, a
memory reference may need to wait for a local cache lookup to complete
before the request may be passed on to the rest of the system.
Furthermore, relatively low speed and high power drivers are required
to move electrical signals from one silicon die to another, for example,
to communicate between a CPU and main memory.

\QuickQuiz{
	But individual electrons don't move anywhere near that fast,
	even in conductors!!!
	The electron drift velocity in a conductor under semiconductor
	voltage levels is on the order of only one \emph{millimeter}
	per second.
	What gives???
}\QuickQuizAnswer{
	Electron drift velocity tracks the long-term movement of individual
	electrons.
	It turns out that individual electrons bounce around quite
	randomly, so that their instantaneous speed is very high, but
	over the long term, they don't move very far.
	In this, electrons resemble long-distance commuters, who
	might spend most of their time traveling at full highway
	speed, but over the long term go nowhere.
	These commuters' speed might be 70 miles per hour
	(113 kilometers per hour), but their long-term drift velocity
	relative to the planet's surface is zero.

	Therefore, we should pay attention not to the electrons'
	drift velocity, but to their instantaneous velocities.
	However, even their instantaneous velocities are nowhere near
	a significant fraction of the speed of light.
	Nevertheless, the measured velocity of electric waves
	in conductors \emph{is} a substantial fraction of the
	speed of light, so we still have a mystery on our hands.

	The other trick is that electrons interact with each other at
	significant distances (from an atomic perspective, anyway),
	courtesy of their negative charge.
	This interaction is carried out by photons, which \emph{do}
	move at the speed of light.
	So even with electricity's electrons, it is photons
	doing most of the fast footwork.

	Extending the commuter analogy, a driver might use a smartphone
	to inform other drivers of an accident or congestion, thus
	allowing a change in traffic flow to propagate much faster
	than the instantaneous velocity of the individual cars.
	Summarizing the analogy between electricity and traffic flow:

	\begin{enumerate}
	\item	The (very low) drift velocity of an electron is similar
		to the long-term velocity of a commuter, both being
		very nearly zero.
	\item	The (still rather low) instantaneous velocity of
		an electron is similar to the instantaneous velocity
		of a car in traffic.
		Both are much higher than the drift velocity, but
		quite small compared to the rate at which changes
		propagate.
	\item	The (much higher) propagation velocity of an electric
		wave is primarily due to photons transmitting
		electromagnetic force among the electrons.
		Similarly, traffic patterns can change quite quickly
		due to communication among drivers.
		Not that this is necessarily of much help to the
		drivers already stuck in traffic, any more than it
		is to the electrons already pooled in a given capacitor.
	\end{enumerate}

	Of course, to fully understand this topic, you should read
	up on electrodynamics.
}\QuickQuizEnd

There are nevertheless some technologies (both hardware and software)
that might help improve matters:

\begin{enumerate}
\item	3D integration,
\item	Novel materials and processes,
\item	Substituting light for electricity,
\item	Special-purpose accelerators, and
\item	Existing parallel software.
\end{enumerate}

Each of these is described in one of the following sections.

\subsection{3D Integration}
\label{sec:cpu:3D Integration}

3-dimensional integration (3DI) is the practice of bonding
very thin silicon dies to each other in a vertical stack.
This practice provides potential benefits, but also poses
significant fabrication challenges~\cite{JohnKnickerbocker2008:3DI}.

\begin{figure}
\centering
\resizebox{3in}{!}{\includegraphics{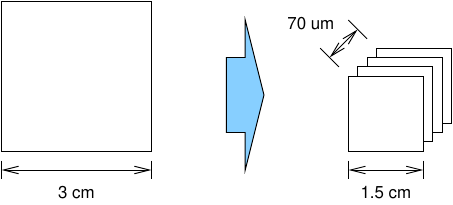}}
\caption{Latency Benefit of 3D Integration}
\label{fig:cpu:Latency Benefit of 3D Integration}
\end{figure}

Perhaps the most important benefit of 3DI is decreased path length through
the system, as shown in
\cref{fig:cpu:Latency Benefit of 3D Integration}.
A 3-centimeter silicon die is replaced with a stack of four 1.5-centimeter
dies, in theory decreasing the maximum path through the system by a factor
of two, keeping in mind that each layer is quite thin.
In addition, given proper attention to design and placement,
long horizontal electrical connections (which are both slow and
power hungry) can be replaced by short vertical electrical connections,
which are both faster and more power efficient.

However, delays due to levels of clocked logic will not be decreased
by 3D integration, and significant manufacturing, testing, power-supply,
and heat-dissipation problems must be solved for 3D integration to
reach production while still delivering on its promise.
The heat-dissipation problems might be solved using
semiconductors based on diamond, which is a good conductor
for heat, but an electrical insulator.
That said, it remains difficult to grow large single diamond crystals,
to say nothing of slicing them into wafers.
In addition, it seems unlikely that any of these technologies will be able to
deliver the exponential increases to which some people have become accustomed.
That said, they may be necessary steps on the path to the late Jim Gray's
``smoking hairy golf balls''~\cite{JimGray2002SmokingHairyGolfBalls}.

\subsection{Novel Materials and Processes}
\label{sec:cpu:Novel Materials and Processes}

Stephen Hawking is said to have claimed that semiconductor manufacturers
have but two fundamental problems:
\begin{enumerate*}[(1)]
\item The finite speed of light and
\item The atomic nature of matter~\cite{BryanGardiner2007}.
\end{enumerate*}
It is possible that semiconductor manufacturers are approaching these
limits, but there are nevertheless a few avenues of research and
development focused on working around these fundamental limits.

One workaround for the atomic nature of matter are so-called
``high-K dielectric'' materials, which allow larger devices to mimic the
electrical properties of infeasibly small devices.
These materials pose some severe fabrication challenges, but nevertheless
may help push the frontiers out a bit farther.
Another more-exotic workaround stores multiple bits in a single electron,
relying on the fact that a given electron can exist at a number of
energy levels.
It remains to be seen if this particular approach can be made to work
reliably in production semiconductor devices.

Another proposed workaround is the ``quantum dot'' approach that
allows much smaller device sizes, but which is still in the research
stage.

One challenge is that many recent hardware-device-level breakthroughs
require very tight control of which atoms are placed
where~\cite{MichaelJKelly2017DeviceLevel}.
It therefore seems likely that whoever finds a good way to hand-place
atoms on each of the billions of devices on a chip will have most
excellent bragging rights, if nothing else!

\subsection{Light, Not Electrons}
\label{sec:cpu:Light; Not Electrons}

Although the speed of light would be a hard limit, the fact is that
semiconductor devices are limited by the speed of electricity rather
than that of light, given that electric waves in semiconductor materials
move at between 3\,\% and 30\,\% of the speed of light in a vacuum.
The use of copper connections on silicon devices is one way to increase
the speed of electricity, and it is quite possible that additional
advances will push closer still to the actual speed of light.
In addition, there have been some experiments with tiny optical fibers
as interconnects within and between chips, based on the fact that
the speed of light in glass is more than 60\,\% of the speed of light
in a vacuum.
One obstacle to such optical fibers is the inefficiency conversion
between electricity and light and vice versa, resulting in both
power-consumption and heat-dissipation problems.

That said, absent some fundamental advances in the field of physics,
any exponential increases in the speed of data flow
will be sharply limited by the actual speed of light in a vacuum.

\subsection{Special-Purpose Accelerators}
\label{sec:cpu:Special-Purpose Accelerators}

A general-purpose CPU working on a specialized problem is often spending
significant time and energy doing work that is only tangentially related
to the problem at hand.
For example, when taking the dot product of a pair of vectors, a
general-purpose CPU will normally use a loop (possibly unrolled)
with a loop counter.
Decoding the instructions, incrementing the loop counter, testing this
counter, and branching back to the
top of the loop are in some sense wasted effort:
The real goal is instead to multiply corresponding elements of the
two vectors.
Therefore, a specialized piece of hardware designed specifically to
multiply vectors could get the job done more quickly and with less
energy consumed.

This is in fact the motivation for the vector instructions present in
many commodity microprocessors.
Because these instructions operate on multiple data items simultaneously,
they would permit a dot product to be computed with less instruction-decode
and loop overhead.

Similarly, specialized hardware can more efficiently encrypt and decrypt,
compress and decompress, encode and decode, and many other tasks besides.
Unfortunately, this efficiency does not come for free.
A computer system incorporating this specialized hardware will contain
more transistors, which will consume some power even when not in use.
Software must be modified to take advantage of this specialized hardware,
and this specialized hardware must be sufficiently generally useful
that the high up-front hardware-design costs can be spread over enough
users to make the specialized hardware affordable.
In part due to these sorts of economic considerations, specialized
hardware has thus far appeared only for a few application areas,
including graphics processing (GPUs), vector processors (MMX, SSE,
and VMX instructions), and, to a lesser extent, encryption.
And even in these areas, it is not always easy to realize the expected
performance gains, for example, due to thermal
throttling~\cite{VladKrasnov2017SIMDfreqscale,DanielLemire2018SIMDfreqscale,TravisDowns2020SIMDfreqscale}.

Unlike the server and PC arena, smartphones have long used a wide
variety of hardware accelerators.
These hardware accelerators are often used for media decoding,
so much so that a high-end MP3 player might be able to play audio
for several minutes---with its CPU fully powered off the entire time.
The purpose of these accelerators is to improve energy efficiency
and thus extend battery life:
Special purpose hardware can often compute more efficiently than
can a general-purpose CPU\@.
This is another example of the principle called out in
\cref{sec:intro:Generality}:
\IX{Generality} is almost never free.

Nevertheless, given the end of \IXaltr{Moore's-Law}{Moore's Law}-induced
single-threaded performance increases, it seems safe to assume that
increasing varieties of special-purpose hardware will appear.

\subsection{Existing Parallel Software}
\label{sec:cpu:Existing Parallel Software}

Although multicore CPUs seem to have taken the computing industry
by surprise, the fact remains that shared-memory parallel computer
systems have been commercially available for more than a quarter
century.
This is more than enough time for significant parallel software
to make its appearance, and it indeed has.
Parallel operating systems are quite commonplace, as are parallel
threading libraries, parallel relational database management systems, 
and parallel numerical software.
Use of existing parallel software can go a long ways towards solving any
parallel-software crisis we might encounter.

Perhaps the most common example is the parallel relational database
management system.
It is not unusual for single-threaded programs, often written in
high-level scripting languages, to access a central relational
database concurrently.
In the resulting highly parallel system, only the database need actually
deal directly with parallelism.
A very nice trick when it works!

% cpu/swdesign.tex
% mainfile: ../perfbook.tex
% SPDX-License-Identifier: CC-BY-SA-3.0

\section{Software Design Implications}
\label{sec:cpu:Software Design Implications}
\epigraph{One ship drives east and another west \\
	  While the self-same breezes blow; \\
	  'Tis the set of the sail and not the gail \\
	  That bids them where to go.}
	 {Ella Wheeler Wilcox}

The values of the ratios in
\cref{tab:cpu:CPU 0 View of Synchronization Mechanisms on 8-Socket System With Intel Xeon Platinum 8176 CPUs at 2.10GHz}
are critically important, as they limit the
efficiency of a given parallel application.
To see this, suppose that the parallel application uses \IXacr{cas}
operations to communicate among threads.
These CAS operations will typically involve a cache miss, that is, assuming
that the threads are communicating primarily with each other rather than
with themselves.
Suppose further that the unit of work corresponding to each CAS communication
operation takes 300\,ns, which is sufficient time to compute several
floating-point transcendental functions.
Then about half of the execution time will be consumed by the CAS
communication operations!
This in turn means that a two-CPU system running such a parallel program
would run no faster than a sequential implementation running on a
single CPU\@.

The situation is even worse in the distributed-system case, where the
latency of a single communications operation might take as long as
thousands or even millions of floating-point operations.
This illustrates how important it is for communications operations to
be extremely infrequent and to enable very large quantities of processing.

\QuickQuiz{
	Given that distributed-systems communication is so horribly
	expensive, why does anyone bother with such systems?
}\QuickQuizAnswer{
	There are a number of reasons:

	\begin{enumerate}
	\item	Shared-memory multiprocessor systems have strict size limits.
		If you need more than a few thousand CPUs, you have no
		choice but to use a distributed system.
	\item	Large shared-memory systems tend to be more expensive
		per unit computation than their smaller counterparts.
	\item	Large shared-memory systems tend to have much longer
		cache-miss latencies than do smaller system.
		To see this, compare
		\cref{tab:cpu:CPU 0 View of Synchronization Mechanisms on 8-Socket System With Intel Xeon Platinum 8176 CPUs at 2.10GHz}
		on \cpageref{tab:cpu:CPU 0 View of Synchronization Mechanisms on 8-Socket System With Intel Xeon Platinum 8176 CPUs at 2.10GHz}
		with
		\cref{tab:cpu:CPU 0 View of Synchronization Mechanisms on 12-CPU Intel Core i7-8750H CPU @ 2.20GHz}.
	\item	The distributed-systems communications operations do
		not necessarily use much CPU, so that computation can
		proceed in parallel with message transfer.
	\item	Many important problems are ``embarrassingly parallel'',
		so that extremely large quantities of processing may
		be enabled by a very small number of messages.
		SETI@HOME~\cite{SETIatHOME2008}
		was but one example of such an application.
		These sorts of applications can make good use of networks
		of computers despite extremely long communications
		latencies.
	\end{enumerate}

	Thus, large shared-memory systems tend to be used for applications
	that benefit from faster latencies than can be provided by
	distributed computing, and particularly for those applications
	that benefit from a large shared memory.

	It is likely that continued work on parallel applications will
	increase the number of embarrassingly parallel applications that
	can run well on machines and/or clusters having long communications
	latencies, reductions in cost being the driving force that it is.
	That said, greatly reduced hardware latencies would be an
	extremely welcome development, both for single-system and
	for distributed computing.
}\QuickQuizEnd

The lesson should be quite clear:
Parallel algorithms must be explicitly designed with these hardware
properties firmly in mind.
One approach is to run nearly independent threads.
The less frequently the threads communicate, whether by \IX{atomic} operations,
locks, or explicit messages, the better the application's performance
and scalability will be.
This approach will be touched on in \cref{chp:Counting},
explored in \cref{chp:Partitioning and Synchronization Design},
and taken to its logical extreme in \cref{chp:Data Ownership}.

Another approach is to make sure that any sharing be read-mostly, which
allows the CPUs' caches to replicate the read-mostly data, in turn
allowing all CPUs fast access.
This approach is touched on in
\cref{sec:count:Eventually Consistent Implementation},
and explored more deeply in \cref{chp:Deferred Processing}.

In short, achieving excellent parallel performance and scalability means
striving for embarrassingly parallel algorithms and implementations,
whether by careful choice of data structures and algorithms, use of
existing parallel applications and environments, or transforming the
problem into an embarrassingly parallel form.

\QuickQuiz{
	OK, if we are going to have to apply distributed-programming
	techniques to shared-memory parallel programs, why not just
	always use these distributed techniques and dispense with
	shared memory?
}\QuickQuizAnswer{
	Because it is often the case that only a small fraction of
	the program is performance-critical.
	Shared-memory parallelism allows us to focus distributed-programming
	techniques on that small fraction, allowing simpler shared-memory
	techniques to be used on the non-performance-critical bulk of
	the program.
}\QuickQuizEnd

So, to sum up:

\begin{enumerate}
\item	The good news is that multicore systems are inexpensive and
	readily available.
\item	More good news:
	The overhead of many synchronization operations is much lower
	than it was on parallel systems from the early 2000s.
\item	The bad news is that the overhead of cache misses is still high,
	especially on large systems.
\end{enumerate}

The remainder of this book describes ways of handling this bad news.

In particular,
\cref{chp:Tools of the Trade} will cover some of the low-level
tools used for parallel programming,
\cref{chp:Counting} will investigate problems and solutions to
parallel counting, and
\cref{chp:Partitioning and Synchronization Design}
will discuss design disciplines that promote performance and scalability.

	\setlength{\parskip}{0.0pt plus 1ex}
	\input{qqzcpu}
	\setlength{\parskip}{0.0pt plus 1.0pt}% return to default

% toolsoftrade/toolsoftrade.tex
% mainfile: ../perfbook.tex
% SPDX-License-Identifier: CC-BY-SA-3.0

\QuickQuizChapter{chp:Tools of the Trade}{Tools of the Trade}{qqztoolsoftrade}
\Epigraph{You are only as good as your tools, and your tools are only
	  as good as you are.}{Unknown}

This chapter provides a brief introduction to some basic tools of the
parallel-programming trade, focusing mainly on those available to
user applications running on operating systems similar to Linux.
\Cref{sec:toolsoftrade:Scripting Languages} begins with
scripting languages,
\cref{sec:toolsoftrade:POSIX Multiprocessing}
describes the multi-process parallelism supported by the POSIX API and
touches on POSIX threads,
\cref{sec:toolsoftrade:Alternatives to POSIX Operations}
presents analogous operations in other environments, and finally,
\cref{sec:toolsoftrade:The Right Tool for the Job: How to Choose?}
helps to choose the tool that will get the job done.

\QuickQuiz{
	You call these tools???
	They look more like low-level synchronization primitives to me!
}\QuickQuizAnswer{
	They look that way because they are in fact low-level synchronization
	primitives.
	And they are in fact the fundamental tools for building low-level
	concurrent software.
}\QuickQuizEnd

Please note that this chapter provides but a brief introduction.
More detail is available from the references (and from the Internet),
and more information will be provided in later chapters.

\section{Scripting Languages}
\label{sec:toolsoftrade:Scripting Languages}
\epigraph{The supreme excellence is simplicity.}
	 {Henry Wadsworth Longfellow, simplified}
% The original:
% "In character, in manner, in style, in all things, the supreme
% excellence is simplicity."

The Linux shell scripting languages provide simple but effective ways
of managing parallelism.
For example, suppose that you had a program \co{compute_it}
that you needed to run twice with two different sets of arguments.
This can be accomplished using UNIX shell scripting as follows:

\input{CodeSamples/toolsoftrade/parallel=compute_it.fcv}

\begin{figure}
\centering
\resizebox{3in}{!}{\includegraphics{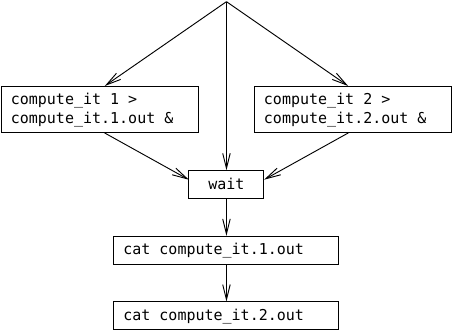}}
\caption{Execution Diagram for Parallel Shell Execution}
\label{fig:toolsoftrade:Execution Diagram for Parallel Shell Execution}
\end{figure}

\begin{fcvref}[ln:toolsoftrade:parallel:compute_it]
\Clnref{comp1,comp2} launch two instances of this
program, redirecting their
output to two separate files, with the \co{&} character directing the
shell to run the two instances of the program in the background.
\Clnref{wait} waits for both instances to complete, and
\clnref{cat1,cat2} display their output.
\end{fcvref}
The resulting execution is as shown in
\cref{fig:toolsoftrade:Execution Diagram for Parallel Shell Execution}:
The two instances of \co{compute_it} execute in parallel,
\co{wait} completes after both of them do, and then the two instances
of \co{cat} execute sequentially.
% @@@ Maui scheduler, load balancing, BOINC, and so on.

\QuickQuizSeries{%
\QuickQuizB{
	But this silly shell script isn't a \emph{real} parallel program!
	Why bother with such trivia???
}\QuickQuizAnswerB{
	Because you should \emph{never} forget the simple stuff!

	Please keep in mind that the title of this book is
	``Is Parallel Programming Hard, And, If So, What Can You Do About It?''.
	One of the most effective things you can do about it is to
	avoid forgetting the simple stuff!
	After all, if you choose to do parallel programming the hard
	way, you have no one but yourself to blame.
}\QuickQuizEndB
\QuickQuizE{
	Is there a simpler way to create a parallel shell script?
	If so, how?
	If not, why not?
}\QuickQuizAnswerE{
	One straightforward approach is the shell pipeline:

\begin{VerbatimU}
grep $pattern1 | sed -e 's/a/b/' | sort
\end{VerbatimU}

	For a sufficiently large input file,
	\co{grep} will pattern-match in parallel with \co{sed}
	editing and with the input processing of \co{sort}.
	See the file \path{parallel.sh} for a demonstration of
	shell-script parallelism and pipelining.
}\QuickQuizEndE
}

For another example, the \co{make} software-build scripting language
provides a \co{-j} option that specifies how much parallelism should be
introduced into the build process.
Thus, typing \co{make -j4} when building a Linux kernel specifies that
up to four build steps be executed concurrently.

It is hoped that these simple examples convince you that parallel
programming need not always be complex or difficult.

\QuickQuiz{
	But if script-based parallel programming is so easy, why
	bother with anything else?
}\QuickQuizAnswer{
	In fact, it is quite likely that a very large fraction of
	parallel programs in use today are script-based.
	However, script-based parallelism does have its limitations:
	\begin{enumerate}
	\item	Creation of new processes is usually quite heavyweight,
		involving the expensive \co{fork()} and \co{exec()}
		system calls.
	\item	Sharing of data, including pipelining, typically involves
		expensive file I/O.
	\item	The reliable synchronization primitives available to
		scripts also typically involve expensive file I/O.
	\item	Scripting languages are often too slow, but are often
		quite useful when coordinating execution of long-running
		programs written in lower-level programming languages.
	\end{enumerate}
	These limitations require that script-based parallelism use
	coarse-grained parallelism, with each unit of work having
	execution time of at least tens of milliseconds, and preferably
	much longer.

	Those requiring finer-grained parallelism are well advised to
	think hard about their problem to see if it can be expressed
	in a coarse-grained form.
	If not, they should consider using other parallel-programming
	environments, such as those discussed in
	\cref{sec:toolsoftrade:POSIX Multiprocessing}.
}\QuickQuizEnd

\section{POSIX Multiprocessing}
\label{sec:toolsoftrade:POSIX Multiprocessing}
\epigraph{A camel is a horse designed by committee.}{Unknown}

This section scratches the surface of the
POSIX environment, including pthreads~\cite{OpenGroup1997pthreads},
as this environment is readily available and widely implemented.
\Cref{sec:toolsoftrade:POSIX Process Creation and Destruction}
provides a glimpse of the POSIX \co{fork()} and related primitives,
\cref{sec:toolsoftrade:POSIX Thread Creation and Destruction}
touches on thread creation and destruction,
\cref{sec:toolsoftrade:POSIX Locking} gives a brief overview
of POSIX locking, and, finally,
\cref{sec:toolsoftrade:POSIX Reader-Writer Locking} describes a
specific lock which can be used for data that is read by many threads and only
occasionally updated.

\subsection{POSIX Process Creation and Destruction}
\label{sec:toolsoftrade:POSIX Process Creation and Destruction}

Processes are created using the \apipx{fork()} primitive, they may
be destroyed using the \apipx{kill()} primitive, they may destroy
themselves using the \apipx{exit()} primitive.
A~process executing a \co{fork()} primitive is said to be the ``parent''
of the newly created process.
A~parent may wait on its children using the \apipx{wait()} primitive.

Please note that the examples in this section are quite simple.
Real-world applications using these primitives might need to manipulate
signals, file descriptors, shared memory segments, and any number of
other resources.
In addition, some applications need to take specific actions if a given
child terminates, and might also need to be concerned with the reason
that the child terminated.
These issues can of course add substantial complexity to the code.
For more information, see any of a number of textbooks on the
subject~\cite{WRichardStevens1992,StewartWeiss2013UNIX}.

\begin{listing}
\begin{fcvlabel}[ln:toolsoftrade:forkjoin:main]
\begin{VerbatimL}[commandchars=\%\[\]]
pid = fork();%lnlbl[fork]
if (pid == 0) {%lnlbl[if]
	/* child */%lnlbl[child]
} else if (pid < 0) {%lnlbl[else]
	/* parent, upon error */%lnlbl[errora]
	perror("fork");
	exit(EXIT_FAILURE);%lnlbl[errorb]
} else {
	/* parent, pid == child ID */%lnlbl[parent]
}
\end{VerbatimL}
\end{fcvlabel}
\caption{Using the \tco{fork()} Primitive}
\label{lst:toolsoftrade:Using the fork() Primitive}
\end{listing}

\begin{fcvref}[ln:toolsoftrade:forkjoin:main]
If \co{fork()} succeeds, it returns twice, once for the parent
and again for the child.
The value returned from \co{fork()} allows the caller to tell
the difference, as shown in
\cref{lst:toolsoftrade:Using the fork() Primitive}
(\path{forkjoin.c}).
\Clnref{fork} executes the \co{fork()} primitive, and saves
its return value in local variable \co{pid}.
\Clnref{if} checks to see if \co{pid} is zero, in which case,
this is the child, which continues on to execute \clnref{child}.
As noted earlier, the child may terminate via the \co{exit()} primitive.
Otherwise, this is the parent, which checks for an error return from
the \co{fork()} primitive on \clnref{else}, and prints an error
and exits on \clnrefrange{errora}{errorb} if so.
Otherwise, the \co{fork()} has executed successfully, and the parent
therefore executes \clnref{parent} with the variable \co{pid}
containing the process ID of the child.
\end{fcvref}

\begin{listing}
\input{CodeSamples/api-pthreads/api-pthreads=waitall.fcv}
\caption{Using the \tco{wait()} Primitive}
\label{lst:toolsoftrade:Using the wait() Primitive}
\end{listing}

\begin{fcvref}[ln:api-pthreads:api-pthreads:waitall]
The parent process may use the \apipx{wait()} primitive to wait for its children
to complete.
However, use of this primitive is a bit more complicated than its shell-script
counterpart, as each invocation of \co{wait()} waits for but one child
process.
It is therefore customary to wrap \co{wait()} into a function similar
to the \apipx{waitall()} function shown in
\cref{lst:toolsoftrade:Using the wait() Primitive}
(\path{api-pthreads.h}),
with this \co{waitall()} function having semantics similar to the
shell-script \co{wait} command.
Each pass through the loop spanning \clnrefrange{loopa}{loopb}
waits on one child process.
\Clnref{wait} invokes the \co{wait()} primitive, which blocks
until a child process exits, and returns that child's process ID\@.
If the process ID is instead $-1$, this indicates that the \co{wait()}
primitive was unable to wait on a child.
If so, \clnref{ECHILD} checks for the \co{ECHILD} errno, which
indicates that there are no more child processes, so that
\clnref{break} exits the loop.
Otherwise, \clnref{perror,exit} print an error and exit.
\end{fcvref}

\QuickQuiz{
	Why does this \co{wait()} primitive need to be so complicated?
	Why not just make it work like the shell-script \co{wait} does?
}\QuickQuizAnswer{
	Some parallel applications need to take special action when
	specific children exit, and therefore need to wait for each
	child individually.
	In addition, some parallel applications need to detect the
	reason that the child died.
	As we saw in \cref{lst:toolsoftrade:Using the wait() Primitive},
	it is not hard to build a \co{waitall()} function out of
	the \co{wait()} function, but it would be impossible to
	do the reverse.
	Once the information about a specific child is lost, it is lost.
}\QuickQuizEnd

\begin{listing}
\input{CodeSamples/toolsoftrade/forkjoinvar=main.fcv}
\caption{Processes Created Via \tco{fork()} Do Not Share Memory}
\label{lst:toolsoftrade:Processes Created Via fork() Do Not Share Memory}
\end{listing}

\begin{fcvref}[ln:toolsoftrade:forkjoinvar:main]
It is critically important to note that the parent and child do \emph{not}
share memory.
This is illustrated by the program shown in
\cref{lst:toolsoftrade:Processes Created Via fork() Do Not Share Memory}
(\path{forkjoinvar.c}),
in which the child sets a global variable \co{x} to 1 on \clnref{setx},
prints a message on \clnref{print:c}, and exits on \clnref{exit:s}.
The parent continues at \clnref{waitall}, where it waits on the child,
and on \clnref{print:p} finds that its copy of the variable \co{x} is
still zero.
The output is thus as follows:
\end{fcvref}

\begin{VerbatimU}
Child process set x=1
Parent process sees x=0
\end{VerbatimU}

\QuickQuiz{
	Isn't there a lot more to \co{fork()} and \co{wait()}
	than discussed here?
}\QuickQuizAnswer{
	Indeed there is, and
	it is quite possible that this section will be expanded in
	future versions to include messaging features (such as UNIX
	pipes, TCP/IP, and shared file I/O) and memory mapping
	(such as \co{mmap()} and \co{shmget()}).
	In the meantime, there are any number of textbooks that cover
	these primitives in great detail,
	and the truly motivated can read manpages, existing parallel
	applications using these primitives, as well as the
	source code of the Linux-kernel implementations themselves.

	It is important to note that the parent process in
	\cref{lst:toolsoftrade:Processes Created Via fork() Do Not Share Memory}
	waits until after the child terminates to do its \co{printf()}.
	Using \co{printf()}'s buffered I/O concurrently to the same file
	from multiple processes is non-trivial, and is best avoided.
	If you really need to do concurrent buffered I/O,
	consult the documentation for your OS\@.
	For UNIX/Linux systems, Stewart Weiss's lecture notes provide
	a good introduction with informative
	examples~\cite{StewartWeiss2013UNIX}.
}\QuickQuizEnd

The finest-grained parallelism requires shared memory, and
this is covered in
\cref{sec:toolsoftrade:POSIX Thread Creation and Destruction}.
That said, shared-memory parallelism can be significantly more complex
than fork-join parallelism.

\subsection{POSIX Thread Creation and Destruction}
\label{sec:toolsoftrade:POSIX Thread Creation and Destruction}

\begin{fcvref}[ln:toolsoftrade:pcreate:mythread]
To create a thread within an existing process, invoke the
\apipx{pthread_create()} primitive, for example, as shown on
\clnref{create:a,create:b} of
\cref{lst:toolsoftrade:Threads Created Via pthread-create() Share Memory}
(\path{pcreate.c}).
The first argument is a pointer to a \co{pthread_t} in which to store the
ID of the thread to be created, the second \co{NULL} argument is a pointer
to an optional \co{pthread_attr_t}, the third argument is the function
(in this case, \co{mythread()})
that is to be invoked by the new thread, and the last \co{NULL} argument
is the argument that will be passed to \co{mythread()}.
\end{fcvref}

\begin{listing}
\input{CodeSamples/toolsoftrade/pcreate=mythread.fcv}
\caption{Threads Created Via \tco{pthread_create()} Share Memory}
\label{lst:toolsoftrade:Threads Created Via pthread-create() Share Memory}
\end{listing}

In this example, \co{mythread()} simply returns, but it could instead
call \apipx{pthread_exit()}.

\QuickQuiz{
	If the \co{mythread()} function in
	\cref{lst:toolsoftrade:Threads Created Via pthread-create() Share Memory}
	can simply return, why bother with \co{pthread_exit()}?
}\QuickQuizAnswer{
	In this simple example, there is no reason whatsoever.
	However, imagine a more complex example, where \co{mythread()}
	invokes other functions, possibly separately compiled.
	In such a case, \co{pthread_exit()} allows these other functions
	to end the thread's execution without having to pass some sort
	of error return all the way back up to \co{mythread()}.
}\QuickQuizEnd

\begin{fcvref}[ln:toolsoftrade:pcreate:mythread]
The \apipx{pthread_join()} primitive, shown on \clnref{join},
is analogous to
the fork-join \apipx{wait()} primitive.
It blocks until the thread specified by the \co{tid} variable completes
execution, either by invoking \apipx{pthread_exit()} or by returning from
the thread's top-level function.
The thread's exit value will be stored through the pointer passed as the
second argument to \co{pthread_join()}.
The thread's exit value is either the value passed to \co{pthread_exit()}
or the value returned by the thread's top-level function, depending on
how the thread in question exits.
\end{fcvref}

The program shown in
\cref{lst:toolsoftrade:Threads Created Via pthread-create() Share Memory}
produces output as follows, demonstrating that memory is in fact
shared between the two threads:

\begin{VerbatimU}
Child process set x=1
Parent process sees x=1
\end{VerbatimU}

Note that this program carefully makes sure that only one of the threads
stores a value to variable \co{x} at a time.
Any situation in which one thread might be storing a value to a given
variable while some other thread either loads from or stores to that
same variable is termed a \emph{\IX{data race}}.
Because the C language makes no guarantee that the results of a data race
will be in any way reasonable, we need some way of safely accessing
and modifying data concurrently, such as the locking primitives discussed
in the following section.

But your data races are benign, you say?
Well, maybe they are.
But please do everyone (yourself included) a big favor and read
\cref{sec:toolsoftrade:Shared-Variable Shenanigans}
very carefully.
As compilers optimize more and more aggressively, there are fewer and
fewer truly benign data races.

\QuickQuiz{
	If the C language makes no guarantees in presence of a data
	race, then why does the Linux kernel have so many data races?
	Are you trying to tell me that the Linux kernel is completely
	broken???
}\QuickQuizAnswer{
	Ah, but the Linux kernel is written in a carefully selected
	superset of the C language that includes special GNU
	extensions, such as asms, that permit safe execution even
	in presence of data races.
	In addition, the Linux kernel does not run on a number of
	platforms where data races would be especially problematic.
	For an example, consider embedded systems with 32-bit pointers
	and 16-bit busses.
	On such a system, a data race involving a store to and a load
	from a given pointer might well result in the load returning the
	low-order 16 bits of the old value of the pointer concatenated
	with the high-order 16 bits of the new value of the pointer.

	Nevertheless, even in the Linux kernel, data races can be
	quite dangerous and should be avoided where
	feasible~\cite{JonCorbet2012ACCESS:ONCE}.
}\QuickQuizEnd

\subsection{POSIX Locking}
\label{sec:toolsoftrade:POSIX Locking}

The POSIX standard allows the programmer to avoid data races via
``POSIX locking''.
POSIX locking features a number of primitives, the most fundamental
of which are \apipx{pthread_mutex_lock()} and \apipx{pthread_mutex_unlock()}.
These primitives operate on locks, which are of type \apipx{pthread_mutex_t}.
These locks may be declared statically and initialized with
\apipx{PTHREAD_MUTEX_INITIALIZER}, or they may be allocated dynamically
and initialized using the \apipx{pthread_mutex_init()} primitive.
The demonstration code in this section will take the former course.

The \co{pthread_mutex_lock()} primitive ``acquires'' the specified lock,
and the \co{pthread_mutex_unlock()} ``releases'' the specified lock.
Because these are ``exclusive'' locking primitives,
only one thread at a time may ``hold'' a given lock at a given time.
For example, if a pair of threads attempt to acquire the same lock
concurrently, one of the pair will be ``granted'' the lock first, and
the other will wait until the first thread releases the lock.
A simple and reasonably useful programming model permits a given data item
to be accessed only while holding the corresponding
lock~\cite{Hoare74}.

\QuickQuiz{
	What if I want several threads to hold the same lock at the
	same time?
}\QuickQuizAnswer{
	The first thing you should do is to ask yourself why you would
	want to do such a thing.
	If the answer is ``because I have a lot of data that is read
	by many threads, and only occasionally updated'', then
	POSIX reader-writer locks might be what you are looking for.
	These are introduced in
	\cref{sec:toolsoftrade:POSIX Reader-Writer Locking}.

	Another way to get the effect of multiple threads holding
	the same lock is for one thread to acquire the lock, and
	then use \co{pthread_create()} to create the other threads.
	The question of why this would ever be a good idea is left
	to the reader.
}\QuickQuizEnd

\begin{listing}
\ebresizeverb{.85}{
\input{CodeSamples/toolsoftrade/lock=reader_writer.fcv}
}
\caption{Demonstration of Exclusive Locks}
\label{lst:toolsoftrade:Demonstration of Exclusive Locks}
\end{listing}

\begin{fcvref}[ln:toolsoftrade:lock:reader_writer]
This exclusive-locking property is demonstrated using the code shown in
\cref{lst:toolsoftrade:Demonstration of Exclusive Locks}
(\path{lock.c}).
\Clnref{lock_a} defines and initializes a POSIX lock named \co{lock_a}, while
\clnref{lock_b} similarly defines and initializes a lock named \co{lock_b}.
\Clnref{x} defines and initializes a shared variable~\co{x}.
\end{fcvref}

\begin{fcvref}[ln:toolsoftrade:lock:reader_writer:reader]
\Clnrefrange{b}{e} define a function \co{lock_reader()} which repeatedly
reads the shared variable \co{x} while holding
the lock specified by \co{arg}.
\Clnref{cast} casts \co{arg} to a pointer to a \co{pthread_mutex_t}, as
required by the \co{pthread_mutex_lock()} and \co{pthread_mutex_unlock()}
primitives.
\end{fcvref}

\QuickQuizSeries{%
\QuickQuizB{
	Why not simply make the argument to \co{lock_reader()}
	on \clnrefr{ln:toolsoftrade:lock:reader_writer:reader:b} of
	\cref{lst:toolsoftrade:Demonstration of Exclusive Locks}
	be a pointer to a \co{pthread_mutex_t}?
}\QuickQuizAnswerB{
	Because we will need to pass \co{lock_reader()} to
	\co{pthread_create()}.
	Although we could cast the function when passing it to
	\co{pthread_create()}, function casts are quite a bit
	uglier and harder to get right than are simple pointer casts.
}\QuickQuizEndB
\QuickQuizE{
	\begin{fcvref}[ln:toolsoftrade:lock:reader_writer]
	What is the \apik{READ_ONCE()} on
	\clnref{reader:read_x,writer:inc} and the
	\apik{WRITE_ONCE()} on \clnref{writer:inc} of
	\cref{lst:toolsoftrade:Demonstration of Exclusive Locks}?
	\end{fcvref}
}\QuickQuizAnswerE{
	These macros constrain the compiler so as to prevent it from
	carrying out optimizations that would be problematic for concurrently
	accessed shared variables.
	They don't constrain the CPU at all, other than by preventing
	reordering of accesses to a given single variable.
	Note that this single-variable constraint does apply to the
	code shown in
	\cref{lst:toolsoftrade:Demonstration of Exclusive Locks}
	because only the variable \co{x} is accessed.

	For more information on \co{READ_ONCE()} and \co{WRITE_ONCE()},
	please see
	\cref{sec:toolsoftrade:Atomic Operations (gcc Classic)}.
	For more information on ordering accesses to multiple variables
	by multiple threads, please see
	\cref{chp:Advanced Synchronization: Memory Ordering}.
	In the meantime, \co{READ_ONCE(x)} has much in common with
	the \GCC\  intrinsic \co{__atomic_load_n(&x, __ATOMIC_RELAXED)}
	and \co{WRITE_ONCE(x, v)} has much in common with the \GCC\
	intrinsic \co{__atomic_store_n(&x, v, __ATOMIC_RELAXED)}.
}\QuickQuizEndE
}

\begin{fcvref}[ln:toolsoftrade:lock:reader_writer:reader]
\Clnrefrange{acq:b}{acq:e} acquire the specified
\co{pthread_mutex_t}, checking
for errors and exiting the program if any occur.
\Clnrefrange{loop:b}{loop:e} repeatedly check the value of \co{x},
printing the new value
each time that it changes.
\Clnref{sleep} sleeps for one millisecond, which allows this demonstration
to run nicely on a uniprocessor machine.
\Clnrefrange{rel:b}{rel:e} release the \co{pthread_mutex_t},
again checking for
errors and exiting the program if any occur.
Finally, \clnref{return} returns \co{NULL}, again to match the function type
required by \co{pthread_create()}.
\end{fcvref}

\QuickQuiz{
	Writing four lines of code for each acquisition and release
	of a \co{pthread_mutex_t} sure seems painful!
	Isn't there a better way?
}\QuickQuizAnswer{
	Indeed!
	And for that reason, the \co{pthread_mutex_lock()} and
	\co{pthread_mutex_unlock()} primitives are normally wrapped
	in functions that do this error checking.
	Later on, we will wrap them with the Linux kernel
	\co{spin_lock()} and \co{spin_unlock()} APIs.
}\QuickQuizEnd

\begin{fcvref}[ln:toolsoftrade:lock:reader_writer:writer]
\Clnrefrange{b}{e} of
\cref{lst:toolsoftrade:Demonstration of Exclusive Locks}
show \co{lock_writer()}, which
periodically updates the shared variable \co{x} while holding the
specified \co{pthread_mutex_t}.
As with \co{lock_reader()}, \clnref{cast} casts \co{arg} to a pointer
to \co{pthread_mutex_t},
\clnrefrange{acq:b}{acq:e} acquire the specified lock,
and \clnrefrange{rel:b}{rel:e} release it.
While holding the lock, \clnrefrange{loop:b}{loop:e}
increment the shared variable \co{x},
sleeping for five milliseconds between each increment.
Finally, \clnrefrange{rel:b}{rel:e} release the lock.
\end{fcvref}

\begin{listing}
\input{CodeSamples/toolsoftrade/lock=same_lock.fcv}
\caption{Demonstration of Same Exclusive Lock}
\label{lst:toolsoftrade:Demonstration of Same Exclusive Lock}
\end{listing}

\begin{fcvref}[ln:toolsoftrade:lock:same_lock]
\Cref{lst:toolsoftrade:Demonstration of Same Exclusive Lock}
shows a code fragment that runs \co{lock_reader()} and
\co{lock_writer()} as threads using the same lock, namely, \co{lock_a}.
\Clnrefrange{cr:reader:b}{cr:reader:e} create a thread
running \co{lock_reader()}, and then
\clnrefrange{cr:writer:b}{cr:writer:e} create a thread
running \co{lock_writer()}.
\Clnrefrange{wait:b}{wait:e} wait for both threads to complete.
The output of this code fragment is as follows:
\end{fcvref}

\begin{VerbatimU}
Creating two threads using same lock:
lock_reader(): x = 0
\end{VerbatimU}

Because both threads are using the same lock, the \co{lock_reader()}
thread cannot see any of the intermediate values of \co{x} produced
by \co{lock_writer()} while holding the lock.

\QuickQuiz{
	Is ``x = 0'' the only possible output from the code fragment
	shown in
	\cref{lst:toolsoftrade:Demonstration of Same Exclusive Lock}?
	If so, why?
	If not, what other output could appear, and why?
}\QuickQuizAnswer{
	No.
	The reason that ``x = 0'' was output was that \co{lock_reader()}
	acquired the lock first.
	Had \co{lock_writer()} instead acquired the lock first, then
	the output would have been \mbox{``x = 3''}.
	However, because the code fragment started \co{lock_reader()} first
	and because this run was performed on a multiprocessor,
	one would normally expect \co{lock_reader()} to acquire the
	lock first.
	Nevertheless, there are no guarantees, especially on a busy system.
}\QuickQuizEnd

\begin{listing}
\input{CodeSamples/toolsoftrade/lock=diff_lock.fcv}
\caption{Demonstration of Different Exclusive Locks}
\label{lst:toolsoftrade:Demonstration of Different Exclusive Locks}
\end{listing}

\Cref{lst:toolsoftrade:Demonstration of Different Exclusive Locks}
shows a similar code fragment, but this time using different locks:
\co{lock_a} for \co{lock_reader()} and \co{lock_b} for
\co{lock_writer()}.
The output of this code fragment is as follows:

\begin{VerbatimU}
Creating two threads w/different locks:
lock_reader(): x = 0
lock_reader(): x = 1
lock_reader(): x = 2
lock_reader(): x = 3
\end{VerbatimU}

Because the two threads are using different locks, they do not exclude
each other, and can run concurrently.
The \co{lock_reader()} function can therefore see the intermediate
values of \co{x} stored by \co{lock_writer()}.

\QuickQuizSeries{%
\QuickQuizB{
	Using different locks could cause quite a bit of confusion,
	what with threads seeing each others' intermediate states.
	So should well-written parallel programs restrict themselves
	to using a single lock in order to avoid this kind of confusion?
}\QuickQuizAnswerB{
	Although it is sometimes possible to write a program using a
	single global lock that both performs and scales well, such
	programs are exceptions to the rule.
	You will normally need to use multiple locks to attain good
	performance and scalability.

	One possible exception to this rule is ``transactional memory'',
	which is currently a research topic.
	Transactional\-/memory semantics can be loosely thought of as those
	of a single global lock with optimizations permitted and
	with the addition of rollback~\cite{HansJBoehm2009HOTPAR}.
}\QuickQuizEndB
\QuickQuizM{
	In the code shown in
	\cref{lst:toolsoftrade:Demonstration of Different Exclusive Locks},
	is \co{lock_reader()} guaranteed to see all the values produced
	by \co{lock_writer()}?
	Why or why not?
}\QuickQuizAnswerM{
	No.
	On a busy system, \co{lock_reader()} might be preempted
	for the entire duration of \co{lock_writer()}'s execution,
	in which case it would not see \emph{any} of \co{lock_writer()}'s
	intermediate states for \co{x}.
}\QuickQuizEndM
\QuickQuizE{
	Wait a minute here!!!
	\Cref{lst:toolsoftrade:Demonstration of Same Exclusive Lock}
	didn't initialize shared variable~\co{x},
	so why does it need to be initialized in
	\cref{lst:toolsoftrade:Demonstration of Different Exclusive Locks}?
}\QuickQuizAnswerE{
	See \clnrefr{ln:toolsoftrade:lock:reader_writer:x} of
	\cref{lst:toolsoftrade:Demonstration of Exclusive Locks}.
	Because the code in
	\cref{lst:toolsoftrade:Demonstration of Same Exclusive Lock}
	ran first, it could rely on the compile-time initialization of~\co{x}.
	The code in
	\cref{lst:toolsoftrade:Demonstration of Different Exclusive Locks}
	ran next, so it had to re-initialize~\co{x}.
}\QuickQuizEndE
}

Although there is quite a bit more to POSIX exclusive locking, these
primitives provide a good start and are in fact sufficient in a great
many situations.
The next section takes a brief look at POSIX reader-writer locking.

\subsection{POSIX Reader-Writer Locking}
\label{sec:toolsoftrade:POSIX Reader-Writer Locking}

The POSIX API provides a reader-writer lock, which is represented by
a \apipx{pthread_rwlock_t}.
As with \apipx{pthread_mutex_t}, \co{pthread_rwlock_t} may be statically
initialized via \apipx{PTHREAD_RWLOCK_INITIALIZER} or dynamically
initialized via the \apipx{pthread_rwlock_init()} primitive.
The \apipx{pthread_rwlock_rdlock()} primitive read-acquires the
specified \apipx{pthread_rwlock_t}, the \apipx{pthread_rwlock_wrlock()}
primitive write-acquires it, and the \apipx{pthread_rwlock_unlock()}
primitive releases it.
Only a single thread may write-hold a given \apipx{pthread_rwlock_t}
at any given time, but multiple threads may read-hold a given
\apipx{pthread_rwlock_t}, at least while there is no thread
currently write-holding it.

As you might expect, \IXhpl{reader-writer}{lock} are designed for read-mostly
situations.
In these situations, a reader-writer lock can provide greater scalability
than can an \IXh{exclusive}{lock} because the exclusive lock is by definition
limited to a single thread holding the lock at any given time, while
the reader-writer lock permits
an arbitrarily large number of readers to concurrently hold the lock.
However, in practice, we need to know how much additional scalability is
provided by reader-writer locks.

\begin{listing}
\input{CodeSamples/toolsoftrade/rwlockscale=reader.fcv}
\caption{Measuring Reader-Writer Lock Scalability}
\label{lst:toolsoftrade:Measuring Reader-Writer Lock Scalability}
\end{listing}

\begin{fcvref}[ln:toolsoftrade:rwlockscale:reader]
\Cref{lst:toolsoftrade:Measuring Reader-Writer Lock Scalability}
(\path{rwlockscale.c})
shows one way of measuring reader-writer lock scalability.
\Clnref{rwlock} shows the definition and initialization of the reader-writer
lock, \clnref{holdtm} shows the \co{holdtime} argument controlling the
time each thread holds the reader-writer lock,
\clnref{thinktm} shows the \co{thinktime} argument controlling the time between
the release of the reader-writer lock and the next acquisition,
\clnref{rdcnts} defines the \co{readcounts} array into which each reader thread
places the number of times it acquired the lock, and
\clnref{nrdrun} defines the \co{nreadersrunning} variable, which
determines when all reader threads have started running.

\Clnrefrange{goflag:b}{goflag:e} define \co{goflag},
which synchronizes the start and the
end of the test.
This variable is initially set to \co{GOFLAG_INIT}, then set to
\co{GOFLAG_RUN} after all the reader threads have started, and finally
set to \co{GOFLAG_STOP} to terminate the test run.
\end{fcvref}

\begin{fcvref}[ln:toolsoftrade:rwlockscale:reader:reader]
\Clnrefrange{b}{e} define \co{reader()}, which is the reader thread.
\Clnref{atmc_inc} atomically increments the \co{nreadersrunning} variable
to indicate that this thread is now running, and
\clnrefrange{wait:b}{wait:e} wait for the test to start.
The \apik{READ_ONCE()} primitive forces the compiler to fetch \co{goflag}
on each pass through the loop---the compiler would otherwise be within its
rights to assume that the value of \co{goflag} would never change.
\end{fcvref}

\QuickQuizSeries{%
\QuickQuizB{
	Instead of using \apik{READ_ONCE()} everywhere, why not just
	declare \co{goflag} as \co{volatile} on
	\clnrefr{ln:toolsoftrade:rwlockscale:reader:goflag:e} of
	\cref{lst:toolsoftrade:Measuring Reader-Writer Lock Scalability}?
}\QuickQuizAnswerB{
	A \co{volatile} declaration is in fact a reasonable alternative in
	this particular case.
	However, use of \apik{READ_ONCE()} has the benefit of clearly
	flagging to the reader that \co{goflag} is subject to concurrent
	reads and updates.
	Note that \apik{READ_ONCE()} is especially useful in cases where
	most of the accesses are protected by a lock (and thus \emph{not}
	subject to change), but where a few of the accesses are made outside
	of the lock.
	Using a \co{volatile} declaration in this case would make it harder
	for the reader to note the special accesses outside of the lock,
	and would also make it harder for the compiler to generate good
	code under the lock.
}\QuickQuizEndB
\QuickQuizM{
	\apik{READ_ONCE()} only affects the compiler, not the CPU\@.
	Don't we also need memory barriers to make sure
	that the change in \co{goflag}'s value propagates to the
	CPU in a timely fashion in
	\cref{lst:toolsoftrade:Measuring Reader-Writer Lock Scalability}?
}\QuickQuizAnswerM{
	No, memory barriers are not needed and won't help here.
	Memory barriers only enforce ordering among multiple memory
	references:
	They absolutely do not guarantee to expedite the propagation
	of data from one part of the system to another.\footnote{
		There have been persistent rumors of hardware in which
		memory barriers actually do expedite propagation of data,
		but no confirmed sightings.}
	This leads to a quick rule of thumb:
	You do not need memory barriers unless you are using more
	than one variable to communicate between multiple threads.

	But what about \co{nreadersrunning}?
	Isn't that a second variable used for communication?
	Indeed it is, and there really are the needed memory-barrier
	instructions buried in \apig{__sync_fetch_and_add()},
	which make sure that the thread proclaims its presence
	before checking to see if it should start.
}\QuickQuizEndM
\QuickQuizE{
	Would it ever be necessary to use \apik{READ_ONCE()} when accessing
	a per-thread variable, for example, a variable declared using
	\GCC's \apig{__thread} storage class?
}\QuickQuizAnswerE{
	It depends.
	If the per-thread variable was accessed only from its thread,
	and never from a signal handler, then no.
	Otherwise, it is quite possible that \apik{READ_ONCE()} is needed.
	We will see examples of both situations in
	\cref{sec:count:Signal-Theft Limit Counter Implementation}.

	This leads to the question of how one thread can gain access to
	another thread's \apig{__thread} variable, and the answer is that
	the second thread must store a pointer to its \apig{__thread}
	variable somewhere that the first thread has access to.
	One common approach is to maintain a linked list with one
	element per thread, and to store the address of each thread's
	\apig{__thread} variable in the corresponding element.
}\QuickQuizEndE
}

\begin{fcvref}[ln:toolsoftrade:rwlockscale:reader:reader]
The loop spanning \clnrefrange{loop:b}{loop:e} carries out the performance test.
\Clnrefrange{acq:b}{acq:e} acquire the lock,
\clnrefrange{hold:b}{hold:e} hold the lock for the specified
number of microseconds,
\clnrefrange{rel:b}{rel:e} release the lock,
and \clnrefrange{think:b}{think:e} wait for the specified
number of microseconds before re-acquiring the lock.
\Clnref{count} counts this lock acquisition.

\Clnref{mov_cnt} moves the lock-acquisition count to this thread's element of the
\co{readcounts[]} array, and \clnref{return} returns, terminating this thread.
\end{fcvref}

\begin{figure}
\centering
\resizebox{3in}{!}{\includegraphics{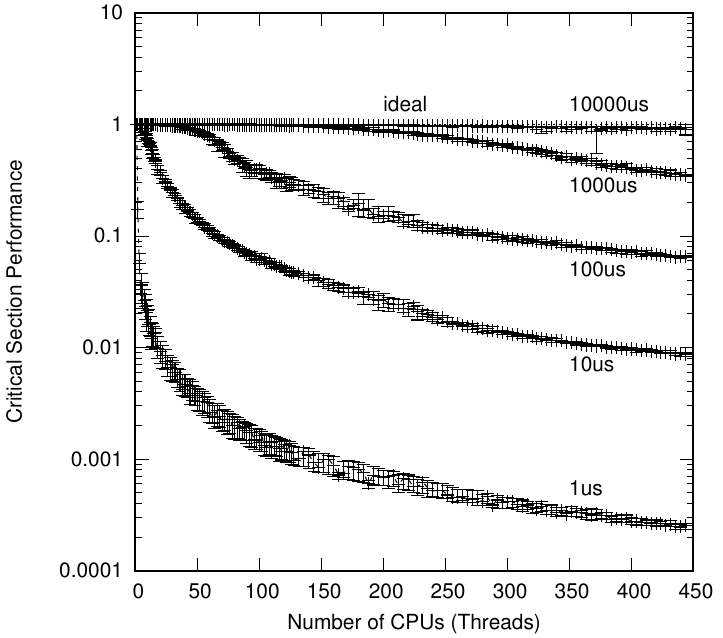}}
\caption{Reader-Writer Lock Scalability vs.\@ Microseconds in Critical Section on 8-Socket System With Intel Xeon Platinum 8176 CPUs @ 2.10GHz}
\label{fig:toolsoftrade:Reader-Writer Lock Scalability vs. Microseconds in Critical Section}
\end{figure}

\Cref{fig:toolsoftrade:Reader-Writer Lock Scalability vs. Microseconds in Critical Section}
shows the results of running this test on a 224-core Xeon system
with two hardware threads per core for a total of 448 software-visible
CPUs.
The \co{thinktime} parameter was zero for all these tests, and the
\co{holdtime} parameter set to values ranging from one microsecond (``1us''
on the graph) to 10,000\,microseconds (``10000us'' on the graph).
The actual value plotted is:
\begin{equation}
	\frac{L_N}{N L_1}
\end{equation}
where $N$ is the number of threads in the current run,
$L_N$ is the total number of lock acquisitions by all $N$ threads in the
current run, and
$L_1$ is the number of lock acquisitions in a single-threaded run.
Given ideal hardware and software scalability, this value will always
be 1.0.

As can be seen in the figure, reader-writer locking scalability is
decidedly non-ideal, especially for smaller sizes of \IXpl{critical
section}.
To see why read-acquisition can be so slow, consider
that all the acquiring threads must update the \co{pthread_rwlock_t}
data structure.
Therefore, if all 448 executing threads attempt to
read-acquire the reader-writer lock concurrently, they must update
this underlying \co{pthread_rwlock_t} one at a time.
One lucky thread might do so almost immediately, but the least-lucky
thread must wait for all the other 447 threads to do their updates.
This situation will only get worse as you add CPUs.
Note also the logscale y-axis.
Even though the 10,000\,microsecond trace appears quite ideal, it has
in fact degraded by about 10\,\% from ideal.

\QuickQuizSeries{%
\QuickQuizB{
	Isn't comparing against single-CPU throughput a bit harsh?
}\QuickQuizAnswerB{
	Not at all.
	In fact, this comparison was, if anything, overly lenient.
	A more balanced comparison would be against single-CPU
	throughput with the locking primitives commented out.
}\QuickQuizEndB
\QuickQuizM{
	But one microsecond is not a particularly small size for
	a critical section.
	What do I do if I need a much smaller critical section, for
	example, one containing only a few instructions?
}\QuickQuizAnswerM{
	If the data being read \emph{never} changes, then you do not
	need to hold any locks while accessing it.
	If the data changes sufficiently infrequently, you might be
	able to checkpoint execution, terminate all threads, change
	the data, then restart at the checkpoint.

	Another approach is to keep a single exclusive lock per
	thread, so that a thread read-acquires the larger aggregate
	reader-writer lock by acquiring its own lock, and write-acquires
	by acquiring all the per-thread locks~\cite{WilsonCHsieh92a}.
	This can work quite well for readers, but causes writers
	to incur increasingly large overheads as the number of threads
	increases.

	Some other ways of efficiently handling very small critical
	sections are described in \cref{chp:Deferred Processing}.
}\QuickQuizEndM
\QuickQuizE{
	The system used is a few years old, and new hardware should
	be faster.
	So why should anyone worry about reader-writer locks being slow?
}\QuickQuizAnswerE{
	In general, newer hardware is improving.
	However, it will need to improve several orders of magnitude
	to permit reader-writer lock to achieve ideal performance on
	448 CPUs.
	Worse yet, the greater the number of CPUs, the larger the
	required performance improvement.
	The performance problems of reader-writer locking are therefore
	very likely to be with us for quite some time to come.
}\QuickQuizEndE
}

Despite these limitations, reader-writer locking is quite useful in many
cases, for example when the readers must do high-latency file or network I/O\@.
There are alternatives, some of which will be presented in
\cref{chp:Counting,chp:Deferred Processing}.

\subsection{Atomic Operations (\GCC\ Classic)}
\label{sec:toolsoftrade:Atomic Operations (gcc Classic)}

\Cref{fig:toolsoftrade:Reader-Writer Lock Scalability vs. Microseconds in Critical Section}
shows that the overhead of reader-writer locking is most severe for the
smallest critical sections, so it would be nice to have some other way
of protecting tiny critical sections.
One such way uses \IX{atomic} operations.
\begin{fcvref}[ln:toolsoftrade:rwlockscale:reader:reader]
We have seen an atomic operation already, namely the
\apig{__sync_fetch_and_add()} primitive on \clnref{atmc_inc} of
\cref{lst:toolsoftrade:Measuring Reader-Writer Lock Scalability}.
This primitive atomically adds the value of its second argument to
the value referenced by its first argument, returning the old value
(which was ignored in this case).
If a pair of threads concurrently execute \co{__sync_fetch_and_add()} on
the same variable, the resulting value of the variable will include
the result of both additions.
\end{fcvref}

The \GNUC\ compiler offers a number of additional atomic operations,
including \apig{__sync_fetch_and_sub()},
\apig{__sync_fetch_and_or()},
\apig{__sync_fetch_and_and()},
\apig{__sync_fetch_and_xor()}, and
\apig{__sync_fetch_and_nand()}, all of which return the old value.
If you instead need the new value, you can instead use the
\apig{__sync_add_and_fetch()},
\apig{__sync_sub_and_fetch()},
\apig{__sync_or_and_fetch()},
\apig{__sync_and_and_fetch()},
\apig{__sync_xor_and_fetch()}, and
\apig{__sync_nand_and_fetch()} primitives.

\QuickQuiz{
	Is it really necessary to have both sets of primitives?
}\QuickQuizAnswer{
	Strictly speaking, no.
	One could implement any member of the second set using the
	corresponding member of the first set.
	For example, one could implement \apig{__sync_nand_and_fetch()}
	in terms of \apig{__sync_fetch_and_nand()} as follows:

\begin{VerbatimU}
tmp = v;
ret = __sync_fetch_and_nand(p, tmp);
ret = ~ret & tmp;
\end{VerbatimU}

	It is similarly possible to implement \apig{__sync_fetch_and_add()},
	\apig{__sync_fetch_and_sub()}, and \apig{__sync_fetch_and_xor()}
	in terms of their post-value counterparts.

	However, the alternative forms can be quite convenient, both
	for the programmer and for the compiler/library implementor.
}\QuickQuizEnd

The classic \IXacrml{cas} operation is provided by a pair of
primitives, \apig{__sync_bool_compare_and_swap()} and
\apig{__sync_val_compare_and_swap()}.
Both of these primitives atomically update a location to a new value,
but only if its prior value was equal to the specified old value.
The first variant returns 1 if the operation succeeded and 0 if it
failed, for example, if the prior value was not equal to the specified
old value.
The second variant returns the prior value of the location, which, if
equal to the specified old value, indicates that the operation succeeded.
Either of the \acrml{cas} operation is ``universal'' in the sense
that any atomic operation on a single location can be implemented in
terms of \acrml{cas}, though the earlier operations are often
more efficient where they apply.
The \acrml{cas} operation is also capable of serving as the basis
for a wider set of atomic operations, though the more elaborate of
these often suffer from complexity, scalability, and performance
problems~\cite{MauriceHerlihy90a}.

\QuickQuiz{
	Given that these atomic operations will often be able to
	generate single atomic instructions that are directly
	supported by the underlying instruction set, shouldn't
	they be the fastest possible way to get things done?
}\QuickQuizAnswer{
	Unfortunately, no.
	See \cref{chp:Counting} for some stark counterexamples.
}\QuickQuizEnd

The \apig{__sync_synchronize()} primitive issues a ``\IX{memory barrier}'',
which constrains both the compiler's and the CPU's ability to reorder
operations, as discussed in
\cref{chp:Advanced Synchronization: Memory Ordering}.
In some cases, it is sufficient to constrain the compiler's ability
to reorder operations, while allowing the CPU free rein, in which
case the \apik{barrier()} primitive may be used.
\begin{fcvref}[ln:toolsoftrade:lock:reader_writer:reader]
In some cases, it is only necessary to ensure that the compiler
avoids optimizing away a given memory read, in which case the
\apik{READ_ONCE()} primitive may be used, as it was on \clnref{read_x} of
\cref{lst:toolsoftrade:Demonstration of Exclusive Locks}.
\end{fcvref}
Similarly, the \apik{WRITE_ONCE()} primitive may be used to prevent the
compiler from optimizing away a given memory write.
These last three primitives are not provided directly by \GCC,
but may be implemented straightforwardly as shown in
\cref{lst:toolsoftrade:Compiler Barrier Primitive (for GCC)},
and all three are discussed at length in
\cref{sec:toolsoftrade:Accessing Shared Variables}.
Alternatively, \apialtk{READ_ONCE(x)}{READ_ONCE()} has much in common with
the \GCC\  intrinsic
\apialtg{__atomic_load_n(&x, __ATOMIC_RELAXED)}{__atomic_load_n()}
and \apik{WRITE_ONCE()} has much in common with the \GCC\ 
intrinsic \apialtg{__atomic_store_n(&x, v, __ATOMIC_RELAXED)}{__atomic_store_n()}.

\begin{listing}
\input{CodeSamples/api-pthreads/api-pthreads=compiler_barrier.fcv}
\caption{Compiler Barrier Primitive (for \GCC)}
\label{lst:toolsoftrade:Compiler Barrier Primitive (for GCC)}
\end{listing}

\QuickQuiz{
	What happened to \apikh{ACCESS_ONCE()}?
}\QuickQuizAnswer{
	In the 2018 v4.15 release, the Linux kernel's \apikh{ACCESS_ONCE()} was
	replaced by \apik{READ_ONCE()} and \apik{WRITE_ONCE()} for reads and
	writes, respectively~\cite{JonCorbet2012ACCESS:ONCE,
	JonathanCorbet2014ACCESS:ONCEcompilerBugs,
	MarkRutland2017ACCESS:ONCE:remove}.
	\apikh{ACCESS_ONCE()} was introduced as a helper in RCU code, but was
	promoted to core API soon afterward~\cite{
	PaulEMcKenney2007ACCESS:ONCE:rcu,
	LinusTorvalds2008ACCESS:ONCE:move}.
	Linux kernel's \apik{READ_ONCE()} and \apik{WRITE_ONCE()} have
	evolved into complex forms that look quite different than
	the original \apikh{ACCESS_ONCE()} implementation due to the
	need to support access-once semantics for large structures,
	but with the possibility of load/store tearing if the structure
	cannot be loaded and stored with a single machine instruction.
}\QuickQuizEnd

\subsection{Atomic Operations (C11)}
\label{sec:toolsoftrade:Atomic Operations (C11)}

The C11 standard added \IX{atomic} operations,
including loads (\apic{atomic_load()}),
stores (\apic{atomic_store()}),
memory barriers (\apic{atomic_thread_fence()} and
\apic{atomic_signal_fence()}), and read-modify-write atomics.
The read-modify-write atomics include
\apic{atomic_fetch_add()},
\apic{atomic_fetch_sub()},
\apic{atomic_fetch_and()},
\apic{atomic_fetch_xor()},
\apic{atomic_exchange()},
\apic{atomic_compare_exchange_strong()},
and
\apic{atomic_compare_exchange_weak()}.
These operate in a manner similar to those described in
\cref{sec:toolsoftrade:Atomic Operations (gcc Classic)},
but with the addition of memory-order arguments to \co{_explicit}
variants of all of the operations.
Without memory-order arguments, all the atomic operations are
fully ordered, and the arguments permit weaker orderings.
For example, ``\apialtc{atomic_load_explicit(&a, memory_order_relaxed)}
{atomic_load_explicit()}''
is vaguely similar to the Linux kernel's ``\apik{READ_ONCE()}''.\footnote{
	Memory ordering is described in more detail in
	\cref{chp:Advanced Synchronization: Memory Ordering} and
	\cref{chp:app:whymb:Why Memory Barriers?}.}

\subsection{Atomic Operations (Modern \GCC)}
\label{sec:toolsoftrade:Atomic Operations (Modern gcc)}

One restriction of the C11 atomics is that they apply only to special
atomic types, which can be problematic.
The \GNUC\ compiler therefore provides \IX{atomic} intrinsics, including
\apig{__atomic_load()},
\apig{__atomic_load_n()},
\apig{__atomic_store()},
\apig{__atomic_store_n()},
\apig{__atomic_thread_fence()}, etc.
These intrinsics offer the same semantics as their C11 counterparts,
but may be used on plain non-atomic objects.
Some of these intrinsics may be passed a memory-order argument from
this list:
\apig{__ATOMIC_RELAXED},
\apig{__ATOMIC_CONSUME},
\apig{__ATOMIC_ACQUIRE},
\apig{__ATOMIC_RELEASE},
\apig{__ATOMIC_ACQ_REL}, and
\apig{__ATOMIC_SEQ_CST}.

\subsection{Per-Thread Variables}
\label{sec:toolsoftrade:Per-Thread Variables}

Per-thread variables, also called thread-specific data, thread-local
storage, and other less-polite names, are used extremely
heavily in concurrent code, as will be explored in
\cref{chp:Counting,chp:Data Ownership}.
POSIX supplies the \apipx{pthread_key_create()} function to create a
per-thread variable (and return the corresponding key),
\apipx{pthread_key_delete()} to delete the per-thread variable corresponding
to key,
\apipx{pthread_setspecific()} to set the value of the current thread's
variable corresponding to the specified key,
and \apipx{pthread_getspecific()} to return that value.

A number of compilers (including \GCC) provide a \apig{__thread} specifier
that may be used in a variable definition to designate that variable
as being per-thread.
The name of the variable may then be used normally to access the
value of the current thread's instance of that variable.
Of course, \apig{__thread} is much easier to use than the POSIX
thead-specific data, and so \co{__thread} is usually preferred for
code that is to be built only with \GCC\ or other compilers supporting
\co{__thread}.

Fortunately, the C11 standard introduced a \apic{_Thread_local} keyword
that can be used in place of \apig{__thread}.
In the fullness of time, this new keyword should combine the ease of use
of \co{__thread} with the portability of POSIX thread-specific data.

\section{Alternatives to POSIX Operations}
\label{sec:toolsoftrade:Alternatives to POSIX Operations}
\epigraph{The strategic marketing paradigm of Open Source is a massively
	  parallel drunkard's walk filtered by a Darwinistic process.}
	 {Bruce Perens}

Unfortunately, threading operations, locking primitives, and atomic
operations were in reasonably wide use long before the various standards
committees got around to them.
As a result, there is considerable variation in how these operations
are supported.
It is still quite common to find these operations implemented in
assembly language, either for historical reasons or to obtain better
performance in specialized circumstances.
For example, \GCC's \co{__sync_} family of primitives all provide full
memory-ordering semantics, which in the past motivated many developers
to create their own implementations for situations where the full memory
ordering semantics are not required.
The following sections show some alternatives from the Linux kernel
and some historical primitives used by this book's sample code.

\subsection{Organization and Initialization}
\label{sec:toolsoftrade:Organization and Initialization}

Although many environments do not require any special initialization
code, the code samples in this book start with a call to \apipf{smp_init()},
which initializes a mapping from \apipx{pthread_t} to consecutive integers.
The userspace RCU library\footnote{
	See \cref{sec:defer:Read-Copy Update (RCU)} for more information
	on RCU\@.}
similarly requires a call to \apiur{rcu_init()}.
Although these calls can be hidden in environments (such as that of
\GCC) that support constructors,
most of the RCU flavors supported by the userspace RCU library
also require each thread invoke \apiur{rcu_register_thread()} upon thread
creation and \apiur{rcu_unregister_thread()} before thread exit.

In the case of the Linux kernel, it is a philosophical question as to
whether the kernel does not require calls to special initialization
code or whether the kernel's boot-time code is in fact the required
initialization code.

\subsection{Thread Creation, Destruction, and Control}
\label{sec:toolsoftrade:Thread Creation; Destruction; and Control}

The Linux kernel uses
\apik{struct task_struct} pointers to track kthreads,
\apik{kthread_create()} to create them,
\apik{kthread_should_stop()} to externally suggest that they stop
(which has no POSIX equivalent),\footnote{
	POSIX environments can work around the lack of
	\apik{kthread_should_stop()} by using a properly synchronized
	boolean flag in conjunction with \co{pthread_join()}.}
\apik{kthread_stop()} to wait for them to stop, and
\apik{schedule_timeout_interruptible()} for a timed wait.
There are quite a few additional kthread-management APIs, but this
provides a good start, as well as good search terms.

The CodeSamples API focuses on ``threads'', which are a locus of
control.\footnote{
	There are many other names for similar software constructs, including
	``process'', ``task'', ``fiber'', ``event'', ``execution agent'',
	and so on.
	Similar design principles apply to all of them.}
Each such thread has an identifier of type \apipf{thread_id_t},
and no two threads running at a given time will have the same
identifier.
Threads share everything except for per-thread local state,\footnote{
	How is that for a circular definition?}
which includes program counter and stack.

The thread API is shown in
\cref{lst:toolsoftrade:Thread API}, and members are described in the
following section.

\begin{listing}
\begin{VerbatimL}[numbers=none,xleftmargin=2pt]
int smp_thread_id(void)
thread_id_t create_thread(void *(*func)(void *), void *arg)
for_each_thread(t)
for_each_running_thread(t)
void *wait_thread(thread_id_t tid)
void wait_all_threads(void)
\end{VerbatimL}
\caption{Thread API}
\label{lst:toolsoftrade:Thread API}
\end{listing}

\subsubsection{API Members}

\begin{description}[style=nextline]
\item[\tco{create_thread()}]
The \apipf{create_thread()} primitive creates a new thread,
starting the new thread's execution
at the function \co{func} specified by \apipf{create_thread()}'s
first argument, and passing it the argument specified by
\apipf{create_thread()}'s second argument.
This newly created thread will terminate when it returns from the
starting function specified by \co{func}.
The \apipf{create_thread()} primitive returns the \apipf{thread_id_t}
corresponding to the newly created child thread.

This primitive will abort the program if more than \apipf{NR_THREADS}
threads are created, counting the one implicitly created by running
the program.
\apipf{NR_THREADS} is a compile-time constant that may be modified,
though some systems may have an upper bound for the allowable number
of threads.

\item[\tco{smp_thread_id()}]
Because the \apipf{thread_id_t} returned from \apipf{create_thread()} is
system-dependent, the \apipf{smp_thread_id()} primitive returns a thread
index corresponding to the thread making the request.
This index is guaranteed to be less than the maximum number of threads
that have been in existence since the program started,
and is therefore useful for bitmasks, array indices, and
the like.

\item[\tco{for_each_thread()}]
The \apipf{for_each_thread()} macro loops through all threads that exist,
including all threads that \emph{would} exist if created.
This macro is useful for handling the per-thread variables
introduced in \cref{sec:toolsoftrade:Per-Thread Variables}.

\item[\tco{for_each_running_thread()}]
The \apipf{for_each_running_thread()}
macro loops through only those threads that currently exist.
It is the caller's responsibility to synchronize with thread
creation and deletion if required.

\item[\tco{wait_thread()}]
The \apipf{wait_thread()} primitive waits for completion of the thread
specified by the \co{thread_id_t} passed to it.
This in no way interferes with the execution of the specified thread;
instead, it merely waits for it.
Note that \apipf{wait_thread()} returns the value that was returned by
the corresponding thread.

\item[\tco{wait_all_threads()}]
The \apipf{wait_all_threads()}
primitive waits for completion of all currently running threads.
It is the caller's responsibility to synchronize with thread creation
and deletion if required.
However, this primitive is normally used to clean up at the end of
a run, so such synchronization is normally not needed.

\end{description}

\subsubsection{Example Usage}

\Cref{lst:toolsoftrade:Example Child Thread} (\path{threadcreate.c})
shows an example hello-world-like child thread.
As noted earlier, each thread is allocated its own stack, so
each thread has its own private \co{arg} argument and \co{myarg} variable.
Each child simply prints its argument and its \apipf{smp_thread_id()}
before exiting.
Note that the \co{return} statement on
\clnrefr{ln:intro:threadcreate:thread_test:return} terminates the thread,
returning a \co{NULL} to whoever invokes \apipf{wait_thread()} on this
thread.

\begin{listing}
\input{CodeSamples/intro/threadcreate=thread_test.fcv}
\caption{Example Child Thread}
\label{lst:toolsoftrade:Example Child Thread}
\end{listing}

\begin{fcvref}[ln:intro:threadcreate:main]
The parent program is shown in
\cref{lst:toolsoftrade:Example Parent Thread}.
It invokes \co{smp_init()} to initialize the threading system on
\clnref{smp_init},
parses arguments on \clnrefrange{parse:b}{parse:e},
and announces its presence on \clnref{announce}.
It creates the specified number of child threads on
\clnrefrange{create:b}{create:e},
and waits for them to complete on \clnref{wait}.
Note that \apipf{wait_all_threads()} discards the threads return values,
as in this case they are all \co{NULL}, which is not very interesting.
\end{fcvref}

\begin{listing}
\input{CodeSamples/intro/threadcreate=main.fcv}
\caption{Example Parent Thread}
\label{lst:toolsoftrade:Example Parent Thread}
\end{listing}

\QuickQuiz{
	What happened to the Linux-kernel equivalents to \apipx{fork()}
	and \apipx{wait()}?
}\QuickQuizAnswer{
	They don't really exist.
	All tasks executing within the Linux kernel share memory,
	at least unless you want to do a huge amount of memory-mapping
	work by hand.
}\QuickQuizEnd

\subsection{Locking}
\label{sec:toolsoftrade:Locking}

A good starting subset of the Linux kernel's locking API is shown in
\cref{lst:toolsoftrade:Locking API},
each API element being described in the following section.
This book's CodeSamples locking API closely follows that of the Linux kernel.

\begin{listing}
\begin{VerbatimL}[numbers=none]
void spin_lock_init(spinlock_t *sp);
void spin_lock(spinlock_t *sp);
int spin_trylock(spinlock_t *sp);
void spin_unlock(spinlock_t *sp);
\end{VerbatimL}
\caption{Locking API}
\label{lst:toolsoftrade:Locking API}
\end{listing}

\subsubsection{API Members}

\begin{description}[style=nextline]

\item[\tco{spin_lock_init()}]
The \apik{spin_lock_init()} primitive initializes the specified
\apik{spinlock_t} variable, and must be invoked before
this variable is passed to any other spinlock primitive.

\item[\tco{spin_lock()}]
The \apik{spin_lock()} primitive acquires the specified spinlock,
if necessary, waiting until the spinlock becomes available.
In some environments, such as pthreads, this waiting will involve
blocking, while in others, such as the Linux kernel, it might involve
a CPU-bound spin loop.

The key point is that only one thread may hold a spinlock at any
given time.

\item[\tco{spin_trylock()}]
The \apik{spin_trylock()} primitive acquires the specified spinlock,
but only if it is immediately available.
It returns \co{true} if it was able to acquire the spinlock and
\co{false} otherwise.

\item[\tco{spin_unlock()}]
The \apik{spin_unlock()} primitive releases the specified spinlock,
allowing other threads to acquire it.

\end{description}

% \emph{@@@ likely need to add reader-writer locking.}

\subsubsection{Example Usage}

A spinlock named \co{mutex} may be used to protect a variable
\co{counter} as follows:

\begin{VerbatimU}
spin_lock(&mutex);
counter++;
spin_unlock(&mutex);
\end{VerbatimU}

\QuickQuiz{
	What problems could occur if the variable \co{counter} were
	incremented without the protection of \co{mutex}?
}\QuickQuizAnswer{
	On CPUs with load-store architectures, incrementing \co{counter}
	might compile into something like the following:

\begin{VerbatimU}
LOAD counter,r0
INC r0
STORE r0,counter
\end{VerbatimU}

	On such machines, two threads might simultaneously load the
	value of \co{counter}, each increment it, and each store the
	result.
	The new value of \co{counter} will then only be one greater
	than before, despite two threads each incrementing it.
}\QuickQuizEnd

However, the \apik{spin_lock()} and \apik{spin_unlock()} primitives
do have performance consequences, as will be seen in
\cref{chp:Data Structures}.

\subsection{Accessing Shared Variables}
\label{sec:toolsoftrade:Accessing Shared Variables}

It was not until 2011 that the C standard defined semantics for concurrent
read/write access to shared variables.
However, concurrent C code was being written at least a quarter century
earlier~\cite{Beck85,Inman85}.
This raises the question as to what today's greybeards did back
in long-past pre-C11 days.
A short answer to this question is ``they lived dangerously''.

\begin{listing}
\begin{fcvlabel}[ln:toolsoftrade:Living Dangerously Early 1990s Style]
\begin{VerbatimL}[commandchars=\\\{\}]
ptr = global_ptr;\lnlbl{temp}
if (ptr != NULL && ptr < high_address)
	do_low(ptr);
\end{VerbatimL}
\end{fcvlabel}
\caption{Living Dangerously Early 1990s Style}
\label{lst:toolsoftrade:Living Dangerously Early 1990s Style}
\end{listing}

\begin{listing}
\begin{fcvlabel}[ln:toolsoftrade:C Compilers Can Invent Loads]
\begin{VerbatimL}[commandchars=\\\{\}]
if (global_ptr != NULL &&\lnlbl{if:a}
    global_ptr < high_address)\lnlbl{if:b}
	do_low(global_ptr);\lnlbl{do_low}
\end{VerbatimL}
\end{fcvlabel}
\caption{C Compilers Can Invent Loads}
\label{lst:toolsoftrade:C Compilers Can Invent Loads}
\end{listing}

At least they would have been living dangerously had they been using
2021 compilers.
In (say) the early 1990s, compilers did fewer optimizations, in part
because there were fewer compiler writers and in part due to the
relatively small memories of that era.
Nevertheless, problems did arise, as shown in
\cref{lst:toolsoftrade:Living Dangerously Early 1990s Style},
which the compiler is within its rights to transform into
\cref{lst:toolsoftrade:C Compilers Can Invent Loads}.
As you can see, the temporary on
\clnrefr{ln:toolsoftrade:Living Dangerously Early 1990s Style:temp} of
\cref{lst:toolsoftrade:Living Dangerously Early 1990s Style}
has been optimized away, so that \co{global_ptr} will be loaded
up to three times.

\QuickQuiz{
	What is wrong with loading
	\cref{lst:toolsoftrade:Living Dangerously Early 1990s Style}'s
	\co{global_ptr} up to three times?
}\QuickQuizAnswer{
	Suppose that \co{global_ptr} is initially non-\co{NULL},
	but that some other thread sets \co{global_ptr} to \co{NULL}.
	\begin{fcvref}[ln:toolsoftrade:C Compilers Can Invent Loads]
	Suppose further that \clnref{if:a} of the transformed code
	(\cref{lst:toolsoftrade:C Compilers Can Invent Loads})
	executes just before \co{global_ptr} is set to \co{NULL} and
	\clnref{if:b} just after.
	Then \clnref{if:a} will conclude that
	\co{global_ptr} is non-\co{NULL},
	\clnref{if:b} will conclude that it is less than
	\co{high_address},
	so that \clnref{do_low} passes \co{do_low()} a \co{NULL} pointer,
	which \co{do_low()} just might not be prepared to deal with.
	\end{fcvref}

	Your editor made exactly this mistake in the DYNIX/ptx
	kernel's memory allocator in the early 1990s.
	Tracking down the bug consumed a holiday weekend not just
	for your editor, but also for several of his colleagues.
	In short, this is not a new problem, nor is it likely to
	go away on its own.
}\QuickQuizEnd

\Cref{sec:toolsoftrade:Shared-Variable Shenanigans}
describes additional problems caused by plain accesses,
\cref{sec:toolsoftrade:A Volatile Solution,%
sec:toolsoftrade:Assembling the Rest of a Solution}
describe some pre-C11 solutions.
Of course, where practical, direct C-language memory references
should be replaced by the primitives described in
\cref{sec:toolsoftrade:Atomic Operations (gcc Classic)}
or (especially)
\cref{sec:toolsoftrade:Atomic Operations (C11)}.
Use these primitives to avoid \IXpl{data race}, that is, ensure that if
there are multiple concurrent C-language accesses to a given variable,
all of those accesses are loads.

\subsubsection{Shared-Variable Shenanigans}
\label{sec:toolsoftrade:Shared-Variable Shenanigans}
\OriginallyPublished{Section}{sec:toolsoftrade:Shared-Variable Shenanigans}{Shared-Variable Shenanigans}{Linux Weekly News}{JadeAlglave2019WhoAfraidCompiler}
Given code that does \IXalt{plain loads and stores}{plain access},\footnote{
	That is, normal loads and stores instead of C11 atomics, inline
	assembly, or volatile accesses.}
the compiler is within
its rights to assume that the affected variables are neither accessed
nor modified by any other thread.
This assumption allows the compiler to carry out a large number of
transformations, including load tearing, store tearing,
load fusing, store fusing, code reordering, invented loads,
invented stores, store-to-load transformations, and dead-code elimination,
all of which work just fine in single-threaded code.
But concurrent code can be broken by each of these transformations,
or shared-variable shenanigans, as described below.

\begin{description}[labelsep=.4em]
\item[Load tearing] occurs when the compiler uses multiple load
instructions for a single access.
For example, the compiler could in theory compile the load from
\co{global_ptr} (see
\clnrefr{ln:toolsoftrade:Living Dangerously Early 1990s Style:temp} of
\cref{lst:toolsoftrade:Living Dangerously Early 1990s Style})
as a series of one-byte loads.
If some other thread was concurrently setting \co{global_ptr} to
\co{NULL}, the result might have all but one byte of the pointer
set to zero, thus forming a ``wild pointer''.
Stores using such a wild pointer could corrupt arbitrary
regions of memory, resulting in rare and difficult-to-debug crashes.

Worse yet, on (say) an 8-bit system with 16-bit pointers, the compiler
might have no choice but to use a pair of 8-bit instructions to access
a given pointer.
Because the C standard must support all manner of systems, the standard
cannot rule out load tearing in the general case.

\item[Store tearing] occurs when the compiler uses multiple store
instructions for a single access.
For example, one thread might store \co{0x12345678} to a four-byte integer
variable at the same time another thread stored \co{0xabcdef00}.
If the compiler used 16-bit stores for either access, the result
might well be \co{0x1234ef00}, which could come as quite a surprise to
code loading from this integer.
Nor is this a strictly theoretical issue.
For example, there are CPUs that feature small immediate instruction
fields, and on such CPUs, the compiler might split a 64-bit store into
two 32-bit stores in order to reduce the overhead of explicitly forming
the 64-bit constant in a register, even on a 64-bit CPU\@.
There are historical reports of this actually happening in
the wild (e.g.~\cite{KonstantinKhlebnikov2013gccstoretearing}),
but there is also a recent
report~\cite{WillDeacon2019StoreTearingReport}.\footnote{
	Note that this tearing can happen even on properly aligned
	and machine-word-sized accesses, and in this particular case,
	even for volatile stores.
	Some might argue that this behavior constitutes a bug in the
	compiler, but either way it illustrates the perceived value of
	store tearing from a compiler-writer viewpoint.
}

Of course, the compiler simply has no choice but to tear some stores
in the general case, given
the possibility of code using 64-bit integers running on a 32-bit system.
But for properly aligned machine-sized stores, \apik{WRITE_ONCE()} will
prevent store tearing.

\begin{listing}
\begin{fcvlabel}[ln:toolsoftrade:Preventing Load Fusing]
\begin{VerbatimLL}[commandchars=\\\{\}]
while (!need_to_stop)
	do_something_quickly();
\end{VerbatimLL}
\end{fcvlabel}
\caption{Inviting Load Fusing}
\label{lst:toolsoftrade:Inviting Load Fusing}
\end{listing}

\begin{listing}
\begin{fcvlabel}[ln:toolsoftrade:C Compilers Can Fuse Loads]
\begin{VerbatimLL}[commandchars=\\\[\]]
if (!need_to_stop)
	for (;;) {\lnlbl[loop:b]
		do_something_quickly();
		do_something_quickly();
		do_something_quickly();
		do_something_quickly();
		do_something_quickly();
		do_something_quickly();
		do_something_quickly();
		do_something_quickly();
		do_something_quickly();
		do_something_quickly();
		do_something_quickly();
		do_something_quickly();
		do_something_quickly();
		do_something_quickly();
		do_something_quickly();
		do_something_quickly();
	}\lnlbl[loop:e]
\end{VerbatimLL}
\end{fcvlabel}
\caption{C Compilers Can Fuse Loads}
\label{lst:toolsoftrade:C Compilers Can Fuse Loads}
\end{listing}

\item[Load fusing] occurs when the compiler uses the result of a
prior load from a given variable instead of repeating the load.
Not only is this sort of optimization just fine in single-threaded
code, it is often just fine in multithreaded code.
Unfortunately, the word ``often'' hides some truly annoying exceptions.

For example, suppose that a real-time system needs to invoke a
function named \co{do_something_quickly()} repeatedly until the
variable \co{need_to_stop} was set, and that the compiler can see
that \co{do_something_quickly()} does not store to \co{need_to_stop}.
One (unsafe) way to code this is shown in
\cref{lst:toolsoftrade:Inviting Load Fusing}.
The compiler might reasonably unroll this loop sixteen times in order
to reduce the per-invocation of the backwards branch at the end of the
loop.
Worse yet, because the compiler knows that \co{do_something_quickly()}
does not store to \co{need_to_stop}, the compiler could quite reasonably
decide to check this variable only once, resulting in the code shown in
\cref{lst:toolsoftrade:C Compilers Can Fuse Loads}.
\begin{fcvref}[ln:toolsoftrade:C Compilers Can Fuse Loads]
Once entered, the loop on
\clnrefrange{loop:b}{loop:e} will never exit, regardless of how
many times some other thread stores a non-zero value to \co{need_to_stop}.
\end{fcvref}
The result will at best be consternation, and might well also
include severe physical damage.

\begin{listing}
\begin{fcvlabel}[ln:toolsoftrade:C Compilers Can Fuse Non-Adjacent Loads]
\begin{VerbatimLL}[commandchars=\\\[\]]
int *gp; \lnlbl[gp]

void t0(void)
{
	WRITE_ONCE(gp, &myvar); \lnlbl[wgp]
}

void t1(void)
{
	p1 = gp; \lnlbl[p1]
	do_something(p1);
	p2 = READ_ONCE(gp); \lnlbl[p2]
	if (p2) { \lnlbl[if]
		do_something_else();
		p3 = *gp; \lnlbl[p3]
	}
}
\end{VerbatimLL}
\end{fcvlabel}
\caption{C Compilers Can Fuse Non-Adjacent Loads}
\label{lst:toolsoftrade:C Compilers Can Fuse Non-Adjacent Loads}
\end{listing}

\begin{fcvref}[ln:toolsoftrade:C Compilers Can Fuse Non-Adjacent Loads]
The compiler can fuse loads across surprisingly large spans of code.
For example, in
\cref{lst:toolsoftrade:C Compilers Can Fuse Non-Adjacent Loads},
\co{t0()} and \co{t1()} run concurrently, and \co{do_something()} and
\co{do_something_else()} are inline functions.
\Clnref{gp} declares pointer \co{gp}, which C initializes to \co{NULL}
by default.
At some point, \clnref{wgp} of \co{t0()} stores a non-\co{NULL}
pointer to \co{gp}.
Meanwhile, \co{t1()} loads from \co{gp} three times on
\clnref{p1,p2,p3}.
Given that \clnref{if} finds that \co{gp} is non-\co{NULL}, one might
hope that the dereference on \clnref{p3} would be guaranteed never
to fault.
Unfortunately, the compiler is within its rights to fuse the read on
\clnref{p1,p3}, which means that if \clnref{p1}
loads \co{NULL} and \clnref{p2} loads \co{&myvar}, \clnref{p3}
could load \co{NULL}, resulting in a fault.\footnote{
	\ppl{Will}{Deacon} reports that this happened in the Linux kernel.}
Note that the intervening \apik{READ_ONCE()} does not prevent the other
two loads from being fused, despite the fact that all three are loading
from the same variable.
\end{fcvref}

\QuickQuiz{
	Why does it matter whether \co{do_something()} and
	\co{do_something_else()} in
	\cref{lst:toolsoftrade:C Compilers Can Fuse Non-Adjacent Loads}
	are inline functions?
}\QuickQuizAnswer{
	\begin{fcvref}[ln:toolsoftrade:C Compilers Can Fuse Non-Adjacent Loads]
	Because \co{gp} is not a static variable, if either
	\co{do_something()} or \co{do_something_else()} were separately
	compiled, the compiler would have to assume that either or both
	of these two functions might change the value of \co{gp}.
	This possibility would force the compiler to reload \co{gp}
	on \clnref{p3}, thus avoiding the \co{NULL}-pointer dereference.
	\end{fcvref}
}\QuickQuizEnd

\item[Store fusing] can occur when the compiler notices a pair of successive
stores to a given variable with no intervening loads from that variable.
In this case, the compiler is within its rights to omit the first store.
This is never a problem in single-threaded code, and in fact it is
usually not a problem in correctly written concurrent code.
After all, if the two stores are executed in quick succession, there is
very little chance that some other thread could load the value from the
first store.

\begin{listing}
\begin{fcvlabel}[ln:toolsoftrade:C Compilers Can Fuse Stores]
\begin{VerbatimLL}[commandchars=\\\[\]]
void shut_it_down(void)
{
	status = SHUTTING_DOWN; /* BUGGY!!! */\lnlbl[store:a]
	start_shutdown();
	while (!other_task_ready) /* BUGGY!!! */\lnlbl[loop:b]
		continue;\lnlbl[loop:e]
	finish_shutdown();\lnlbl[finish]
	status = SHUT_DOWN; /* BUGGY!!! */\lnlbl[store:b]
	do_something_else();
}

void work_until_shut_down(void)
{
	while (status != SHUTTING_DOWN) /* BUGGY!!! */\lnlbl[until:loop:b]
		do_more_work();\lnlbl[until:loop:e]
	other_task_ready = 1; /* BUGGY!!! */\lnlbl[other:store]
}
\end{VerbatimLL}
\end{fcvlabel}
\caption{C Compilers Can Fuse Stores}
\label{lst:toolsoftrade:C Compilers Can Fuse Stores}
\end{listing}

However, there are exceptions, for example as shown in
\cref{lst:toolsoftrade:C Compilers Can Fuse Stores}.
\begin{fcvref}[ln:toolsoftrade:C Compilers Can Fuse Stores]
The function \co{shut_it_down()} stores to the shared
variable \co{status} on \clnref{store:a,store:b},
and so assuming that neither
\co{start_shutdown()} nor \co{finish_shutdown()} access \co{status},
the compiler could reasonably remove the store to \co{status} on
\clnref{store:a}.
Unfortunately, this would mean that \co{work_until_shut_down()} would
never exit its loop spanning
\clnref{until:loop:b,until:loop:e}, and thus would never set
\co{other_task_ready}, which would in turn mean that \co{shut_it_down()}
would never exit its loop spanning
\clnref{loop:b,loop:e}, even if
the compiler chooses not to fuse the successive loads from
\co{other_task_ready} on \clnref{loop:b}.

And there are more problems with the code in
\cref{lst:toolsoftrade:C Compilers Can Fuse Stores},
including code reordering.

\item[Code reordering] is a common compilation technique used to
combine common subexpressions, reduce register pressure, and
improve utilization of the many functional units available on
modern superscalar microprocessors.
It is also another reason why the code in
\cref{lst:toolsoftrade:C Compilers Can Fuse Stores}
is buggy.
For example, suppose that the \co{do_more_work()} function on
\clnref{until:loop:e}
does not access \co{other_task_ready}.
Then the compiler would be within its rights to move the assignment
to \co{other_task_ready} on
\clnref{other:store} to precede \clnref{until:loop:b}, which might
be a great disappointment for anyone hoping that the last call to
\co{do_more_work()} on \clnref{until:loop:e} happens before the call to
\co{finish_shutdown()} on \clnref{finish}.
\end{fcvref}

It might seem futile to prevent the compiler from changing the order of
accesses in cases where the underlying hardware is free to reorder them.
However, modern machines have \emph{exact exceptions} and
\emph{exact interrupts}, meaning that any interrupt or exception will
appear to have happened at a specific place in the instruction
stream.
This means that the handler will see the effect of all prior
instructions, but won't see the effect of any subsequent instructions.
\apik{READ_ONCE()} and \apik{WRITE_ONCE()} can therefore be used to
control communication between interrupted code and interrupt handlers,
independent of the ordering provided by the underlying hardware.\footnote{
	That said, the various standards committees would prefer that
	you use atomics or variables of type \apic{sig_atomic_t}, instead
	of \apik{READ_ONCE()} and \apik{WRITE_ONCE()}.}

\item[Invented loads] were illustrated by the code in
\cref{lst:toolsoftrade:Living Dangerously Early 1990s Style,%
lst:toolsoftrade:C Compilers Can Invent Loads},
in which the compiler optimized away a temporary variable,
thus loading from a shared variable more often than intended.

Invented loads can be a performance hazard.
These hazards can occur when a load of variable in a ``hot''
cacheline is hoisted out of an \co{if} statement.
These hoisting optimizations are not uncommon, and can cause significant
increases in cache misses, and thus significant degradation of
both performance and scalability.

\item[Invented stores] can occur in a number of situations.
\begin{fcvref}[ln:toolsoftrade:C Compilers Can Fuse Stores]
For example, a compiler emitting code for \co{work_until_shut_down()} in
\cref{lst:toolsoftrade:C Compilers Can Fuse Stores}
might notice that \co{other_task_ready} is not accessed by
\co{do_more_work()}, and stored to on \clnref{other:store}.
If \co{do_more_work()} was a complex inline function, it might be
necessary to do a register spill, in which case one attractive
place to use for temporary storage is \co{other_task_ready}.
After all, there are no accesses to it, so what is the harm?

Of course, a non-zero store to this variable at just the wrong time
would result in the \co{while} loop on
\clnref{loop:b} terminating
prematurely, again allowing \co{finish_shutdown()} to run
concurrently with \co{do_more_work()}.
Given that the entire point of this \co{while} appears to be to
prevent such concurrency, this is not a good thing.
\end{fcvref}

\begin{listing}
\begin{fcvlabel}[ln:toolsoftrade:Inviting an Invented Store]
\begin{VerbatimLL}[commandchars=\\\{\}]
if (condition)
	a = 1;
else
	do_a_bunch_of_stuff(&a);
\end{VerbatimLL}
\end{fcvlabel}
\caption{Inviting an Invented Store}
\label{lst:toolsoftrade:Inviting an Invented Store}
\end{listing}

\begin{listing}
\begin{fcvlabel}[ln:toolsoftrade:Compiler Invents an Invited Store]
\begin{VerbatimLL}[commandchars=\\\[\]]
a = 1;\lnlbl[store:uncond]
if (!condition) {
	a = 0;\lnlbl[store:cond]
	do_a_bunch_of_stuff(&a);
}
\end{VerbatimLL}
\end{fcvlabel}
\caption{Compiler Invents an Invited Store}
\label{lst:toolsoftrade:Compiler Invents an Invited Store}
\end{listing}

Using a stored-to variable as a temporary might seem outlandish,
but it is permitted by the standard.
Nevertheless, readers might be justified in wanting a less
outlandish example, which is provided by
\cref{lst:toolsoftrade:Inviting an Invented Store,%
lst:toolsoftrade:Compiler Invents an Invited Store}.

A compiler emitting code for
\cref{lst:toolsoftrade:Inviting an Invented Store}
might know that the value of \co{a} is initially zero,
which might be a strong temptation to optimize away one branch
by transforming this code to that in
\cref{lst:toolsoftrade:Compiler Invents an Invited Store}.
\begin{fcvref}[ln:toolsoftrade:Compiler Invents an Invited Store]
Here, \clnref{store:uncond} unconditionally stores \co{1} to \co{a}, then
resets the value back to zero on
\clnref{store:cond} if \co{condition} was not set.
This transforms the if-then-else into an if-then, saving one branch.
\end{fcvref}

\QuickQuiz{
	Ouch!
	So can't the compiler invent a store to a normal variable pretty
	much any time it likes?
}\QuickQuizAnswer{
	Thankfully, the answer is no.
	This is because the compiler is forbidden from introducing data races.
	The case of inventing a store just before a normal store is
	quite special:
	It is not possible for some other entity, be it CPU, thread,
	signal handler, or interrupt handler, to be able to see the
	invented store unless the code already has a data race, even
	without the invented store.
	And if the code already has a data race, it already invokes
	the dreaded spectre of undefined behavior, which allows the
	compiler to generate pretty much whatever code it wants,
	regardless of the wishes of the developer.

	But if the original store is volatile, as in \apik{WRITE_ONCE()},
	for all the compiler knows, there might be a side effect
	associated with the store that could signal some other thread,
	allowing data-race-free access to the variable.
	By inventing the store, the compiler might be introducing a
	data race, which it is not permitted to do.

	Furthermore, in
	\cref{lst:toolsoftrade:Compiler Invents an Invited Store},
	the address of that variable is passed to
	\co{do_a_bunch_of_stuff()}.
	If the compiler can see this function's definition, and
	notes that \co{a} is unconditionally stored to without
	any synchronization operations, then the compiler can be
	quite sure that it is not introducing a data race in this
	case.

	In the case of \co{volatile} and atomic variables, the compiler
	is specifically forbidden from inventing writes.
}\QuickQuizEnd

Finally, pre-C11 compilers could invent writes to unrelated
variables that happened to be adjacent to written-to
variables~\cite[Section 4.2]{Boehm:2005:TCI:1064978.1065042}.
This variant of invented stores has been outlawed by the prohibition
against compiler optimizations that invent data races.

\begin{listing}
\begin{fcvlabel}[ln:toolsoftrade:Inviting a Store-to-Load Conversion]
\begin{VerbatimLL}[commandchars=\\\[\]]
r1 = p;\lnlbl[load:p]
if (unlikely(r1))\lnlbl[if]
	do_something_with(r1);\lnlbl[dsw]
barrier();\lnlbl[barrier]
p = NULL;\lnlbl[null]
\end{VerbatimLL}
\end{fcvlabel}
\caption{Inviting a Store-to-Load Conversion}
\label{lst:toolsoftrade:Inviting a Store-to-Load Conversion}
\end{listing}

\item[Store-to-load transformations] can occur when the compiler notices
that a plain store might not actually change the value in memory.
\begin{fcvref}[ln:toolsoftrade:Inviting a Store-to-Load Conversion]
For example, consider
\cref{lst:toolsoftrade:Inviting a Store-to-Load Conversion}.
\Clnref{load:p} fetches \co{p}, but the \qco{if} statement on
\clnref{if} also tells the compiler that the developer thinks that
\co{p} is usually zero.\footnote{
	The \apik{unlikely()} function provides this hint to the compiler,
	and different compilers provide different ways of implementing
	\co{unlikely()}.}
The \apik{barrier()} statement on \clnref{barrier} forces the compiler
to forget the value of \co{p}, but one could imagine a compiler
choosing to remember the hint---or getting an additional hint via
feedback-directed optimization.
Doing so would cause the compiler to realize that \clnref{null}
is often an expensive no-op.
\end{fcvref}

\begin{listing}
\begin{fcvlabel}[ln:toolsoftrade:Compiler Converts a Store to a Load]
\begin{VerbatimLL}[commandchars=\\\[\]]
r1 = p;\lnlbl[load:p]
if (unlikely(r1))\lnlbl[if]
	do_something_with(r1);\lnlbl[dsw]
barrier();\lnlbl[barrier]
if (p != NULL)\lnlbl[if1]
	p = NULL;\lnlbl[null]
\end{VerbatimLL}
\end{fcvlabel}
\caption{Compiler Converts a Store to a Load}
\label{lst:toolsoftrade:Compiler Converts a Store to a Load}
\end{listing}

\begin{fcvref}[ln:toolsoftrade:Compiler Converts a Store to a Load]
Such a compiler might therefore guard the store of \co{NULL}
with a check, as shown on \clnref{if1,null} of
\cref{lst:toolsoftrade:Compiler Converts a Store to a Load}.
Although this transformation is often desirable, it could be problematic
if the actual store was required for ordering.
For example, a write memory barrier (Linux kernel \apik{smp_wmb()}) would
order the store, but not the load.
This situation might suggest use of \apik{smp_store_release()} over
\apik{smp_wmb()}.
\end{fcvref}

\item[Dead-code elimination] can occur when the compiler notices that
the value from a load is never used, or when a variable is stored to,
but never loaded from.
This can of course eliminate an access to a shared variable, which
can in turn defeat a memory-ordering primitive, which could cause your
concurrent code to act in surprising ways.
Experience thus far indicates that relatively few such surprises will
be at all pleasant.
Elimination of store-only variables is especially dangerous in cases
where external code locates the variable via symbol tables:
The compiler is necessarily ignorant of such external-code accesses,
and might thus eliminate a variable that the external code relies upon.

\end{description}

Reliable concurrent code clearly needs a way to cause the compiler to
preserve the number, order, and type of important accesses to shared
memory, a topic taken up by
\cref{sec:toolsoftrade:A Volatile Solution,%
sec:toolsoftrade:Assembling the Rest of a Solution},
which are up next.

\subsubsection{A Volatile Solution}
\label{sec:toolsoftrade:A Volatile Solution}

Although it is now much maligned, before the advent of C11 and
C++11~\cite{PeteBecker2011N3242}, the \apic{volatile} keyword was an
indispensable tool in the parallel programmer's toolbox.
This raises the question of exactly what \co{volatile} means,
a question that is not answered with excessive precision even
by more recent versions of this standard~\cite{RichardSmith2019N4800}.\footnote{
	JF Bastien thoroughly documented the history and use cases
	for the \co{volatile} keyword in
	C++~\cite{JFBastien2018DeprecatingVolatile}.}
This version guarantees that ``Accesses through \co{volatile}
\co{glvalues} are evaluated strictly according to the rules of the
abstract machine'',
that \co{volatile} accesses are side effects,
that they are one of the four forward-progress indicators,
and that their exact semantics are implementation-defined.
Perhaps the clearest guidance is provided by this non-normative note:

\begin{quote}
	\apic{volatile} is a hint to the implementation to avoid
	aggressive optimization involving the object because the value
	of the object might be changed by means undetectable by an
	implementation.
	Furthermore, for some implementations, \apic{volatile} might indicate
	that special hardware instructions are required to access
	the object.
	See 6.8.1 for detailed semantics.
	In general, the semantics of \apic{volatile} are intended to be the
	same in C++ as they are in C.
\end{quote}

This wording might be reassuring to those writing low-level code, except
for the fact that compiler writers are free to completely ignore
non-normative notes.
Parallel programmers might instead reassure themselves that compiler
writers would like to avoid breaking device drivers (though perhaps
only after a few ``frank and open'' discussions with device-driver
developers), and device drivers impose at least the following
constraints~\cite{PaulEMcKenney2016P0124R6-LKMM}:

\begin{enumerate}
\item	Implementations are forbidden from tearing an aligned volatile
	access when machine instructions of that access's size and type
	are available.\footnote{
		Note that this leaves unspecified what to do with 128-bit
		loads and stores on CPUs having 128-bit CAS but not
		128-bit loads and stores.}
	Concurrent code relies on this constraint to avoid unnecessary
	load and store tearing.
\item	Implementations must not assume anything about the semantics of
	a volatile access, nor, for any volatile access that returns a
	value, about the possible set of values that might be
	returned.\footnote{
		This is strongly implied by the implementation-defined
		semantics called out above.}
	Concurrent code relies on this constraint to avoid optimizations
	that are inapplicable given that other processors might be
	concurrently accessing the location in question.
\item	Aligned machine-sized non-mixed-size volatile accesses interact
	naturally with volatile assembly-code sequences before and after.
	This is necessary because some devices must be accessed using
	a combination of volatile MMIO accesses and special-purpose
	assembly-language instructions.
	Concurrent code relies on this constraint in order to achieve
	the desired ordering properties from combinations of volatile
	accesses and other means discussed in
	\cref{sec:toolsoftrade:Assembling the Rest of a Solution}.
\end{enumerate}

Concurrent code also relies on the first two constraints to avoid
undefined behavior that could result due to data races if any of the
accesses to a given object was either non-atomic or non-volatile,
assuming that all accesses are aligned and machine-sized.
The semantics of mixed-size accesses to the same locations are more
complex, and are left aside for the time being.

So how does \apic{volatile} stack up against the earlier examples?

\begin{listing}
\begin{fcvlabel}[ln:toolsoftrade:Avoiding Danger, 2018 Style]
\begin{VerbatimL}[commandchars=\\\{\}]
ptr = READ_ONCE(global_ptr);\lnlbl{temp}
if (ptr != NULL && ptr < high_address)
	do_low(ptr);
\end{VerbatimL}
\end{fcvlabel}
\caption{Avoiding Danger, 2018 Style}
\label{lst:toolsoftrade:Avoiding Danger; 2018 Style}
\end{listing}

Using \apik{READ_ONCE()} on
\clnrefr{ln:toolsoftrade:Living Dangerously Early 1990s Style:temp} of
\cref{lst:toolsoftrade:Living Dangerously Early 1990s Style}
avoids invented loads,
resulting in the code shown in
\cref{lst:toolsoftrade:Avoiding Danger; 2018 Style}.

\begin{listing}
\begin{fcvlabel}[ln:toolsoftrade:Preventing Load Fusing]
\begin{VerbatimL}[commandchars=\\\{\}]
while (!READ_ONCE(need_to_stop))
	do_something_quickly();
\end{VerbatimL}
\end{fcvlabel}
\caption{Preventing Load Fusing}
\label{lst:toolsoftrade:Preventing Load Fusing}
\end{listing}

As shown in
\cref{lst:toolsoftrade:Preventing Load Fusing},
\apik{READ_ONCE()} can also prevent the loop unrolling in
\cref{lst:toolsoftrade:C Compilers Can Fuse Loads}.

\begin{listing}
\begin{fcvlabel}[ln:toolsoftrade:Preventing Store Fusing and Invented Stores]
\begin{VerbatimL}[commandchars=\\\[\]]
void shut_it_down(void)
{
	WRITE_ONCE(status, SHUTTING_DOWN); /* BUGGY!!! */\lnlbl[store:a]
	start_shutdown();
	while (!READ_ONCE(other_task_ready)) /* BUGGY!!! */\lnlbl[loop:b]
		continue;\lnlbl[loop:e]
	finish_shutdown();\lnlbl[finish]
	WRITE_ONCE(status, SHUT_DOWN); /* BUGGY!!! */\lnlbl[store:b]
	do_something_else();
}

void work_until_shut_down(void)
{
	while (READ_ONCE(status) != SHUTTING_DOWN) /* BUGGY!!! */\lnlbl[until:loop:b]
		do_more_work();\lnlbl[until:loop:e]
	WRITE_ONCE(other_task_ready, 1); /* BUGGY!!! */\lnlbl[other:store]
}
\end{VerbatimL}
\end{fcvlabel}
\caption{Preventing Store Fusing and Invented Stores}
\label{lst:toolsoftrade:Preventing Store Fusing and Invented Stores}
\end{listing}

\apik{READ_ONCE()} and \apik{WRITE_ONCE()} can also be used to prevent the
store fusing and invented stores that were shown in
\cref{lst:toolsoftrade:C Compilers Can Fuse Stores},
with the result shown in
\cref{lst:toolsoftrade:Preventing Store Fusing and Invented Stores}.
However, this does nothing to prevent code reordering, which requires
some additional tricks taught in
\cref{sec:toolsoftrade:Assembling the Rest of a Solution}.

\begin{listing}
\begin{fcvlabel}[ln:toolsoftrade:Disinviting an Invented Store]
\begin{VerbatimL}[commandchars=\\\{\}]
if (condition)
	WRITE_ONCE(a, 1);
else
	do_a_bunch_of_stuff();
\end{VerbatimL}
\end{fcvlabel}
\caption{Disinviting an Invented Store}
\label{lst:toolsoftrade:Disinviting an Invented Store}
\end{listing}

Finally, \apik{WRITE_ONCE()} can be used to prevent the store invention
shown in
\cref{lst:toolsoftrade:Inviting an Invented Store},
with the resulting code shown in
\cref{lst:toolsoftrade:Disinviting an Invented Store}.

To summarize, the \apic{volatile} keyword can prevent load
tearing and store tearing in cases where the loads and stores are
machine-sized and properly aligned.
It can also prevent load fusing, store fusing, invented loads, and
invented stores.
However, although it does prevent the compiler from reordering \apic{volatile}
accesses with each other, it does nothing to prevent the
CPU from reordering these accesses.
Furthermore, it does nothing to prevent either compiler or CPU from
reordering non-\co{volatile} accesses with each other or with
\co{volatile} accesses.
Preventing these types of reordering requires the techniques described
in the next section.

\subsubsection{Assembling the Rest of a Solution}
\label{sec:toolsoftrade:Assembling the Rest of a Solution}

Additional ordering has traditionally been provided by recourse to
assembly language, for example, \GCC\ asm directives.
Oddly enough, these directives need not actually contain assembly language,
as exemplified by the \apik{barrier()} macro shown in
\cref{lst:toolsoftrade:Compiler Barrier Primitive (for GCC)}.

\begin{listing}
\begin{fcvlabel}[ln:toolsoftrade:Preventing C Compilers From Fusing Loads]
\begin{VerbatimL}[commandchars=\\\[\]]
while (!need_to_stop) {
	barrier(); \lnlbl[b1]
	do_something_quickly();
	barrier(); \lnlbl[b2]
}
\end{VerbatimL}
\end{fcvlabel}
\caption{Preventing C Compilers From Fusing Loads}
\label{lst:toolsoftrade:Preventing C Compilers From Fusing Loads}
\end{listing}

In the \apik{barrier()} macro, the \co{__asm__} introduces the asm
directive, the \co{__volatile__} prevents the compiler from optimizing
the asm away, the empty string specifies that no actual instructions
are to be emitted, and the final \co{"memory"} tells the compiler that
this do-nothing asm can arbitrarily change memory.
In response, the compiler will avoid moving any memory references across
the \apik{barrier()} macro.
This means that the real-time-destroying loop unrolling shown in
\cref{lst:toolsoftrade:C Compilers Can Fuse Loads}
can be prevented by adding \apik{barrier()} calls as shown on
\clnrefr{ln:toolsoftrade:Preventing C Compilers From Fusing Loads:b1,%
ln:toolsoftrade:Preventing C Compilers From Fusing Loads:b2}
of
\cref{lst:toolsoftrade:Preventing C Compilers From Fusing Loads}.
These two lines of code prevent the compiler from pushing the load from
\co{need_to_stop} into or past \co{do_something_quickly()} from either
direction.

\begin{listing}
\begin{fcvlabel}[ln:toolsoftrade:Preventing Reordering]
\begin{VerbatimL}[commandchars=\\\[\]]
void shut_it_down(void)
{
	WRITE_ONCE(status, SHUTTING_DOWN);
	smp_mb(); \lnlbl[mb1]
	start_shutdown();
	while (!READ_ONCE(other_task_ready))\lnlbl[loop:b]
		continue;
	smp_mb(); \lnlbl[mb2]
	finish_shutdown();
	smp_mb(); \lnlbl[mb3]
	WRITE_ONCE(status, SHUT_DOWN);
	do_something_else();
}

void work_until_shut_down(void)
{
	while (READ_ONCE(status) != SHUTTING_DOWN) {
		smp_mb(); \lnlbl[mb4]
		do_more_work();
	}
	smp_mb(); \lnlbl[mb5]
	WRITE_ONCE(other_task_ready, 1);\lnlbl[other:store]
}
\end{VerbatimL}
\end{fcvlabel}
\caption{Preventing Reordering}
\label{lst:toolsoftrade:Preventing Reordering}
\end{listing}

However, this does nothing to prevent the CPU from reordering the
references.
In many cases, this is not a problem because the hardware can only do
a certain amount of reordering.
However, there are cases such as
\cref{lst:toolsoftrade:C Compilers Can Fuse Stores} where the
hardware must be constrained.
\Cref{lst:toolsoftrade:Preventing Store Fusing and Invented Stores}
prevented store fusing and invention, and
\cref{lst:toolsoftrade:Preventing Reordering}
further prevents the remaining reordering by addition of
\apik{smp_mb()} on
\begin{fcvref}[ln:toolsoftrade:Preventing Reordering]
\clnref{mb1,mb2,mb3,mb4,mb5}.
\end{fcvref}
The \apik{smp_mb()} macro is similar to \apik{barrier()} shown in
\cref{lst:toolsoftrade:Compiler Barrier Primitive (for GCC)},
but with the empty string replaced by a string containing the
instruction for a full memory barrier, for example, \co{"mfence"}
on x86 or \co{"sync"} on PowerPC.

\QuickQuiz{
	But aren't full memory barriers very heavyweight?
	Isn't there a cheaper way to enforce the ordering needed in
	\cref{lst:toolsoftrade:Preventing Reordering}?
}\QuickQuizAnswer{
	As is often the case, the answer is ``it depends''.
	However, if only two threads are accessing the \co{status}
	and \co{other_task_ready} variables, then the
	\apik{smp_store_release()} and \apik{smp_load_acquire()}
	functions discussed in
	\cref{sec:toolsoftrade:Atomic Operations}
	will suffice.
}\QuickQuizEnd

Ordering is also provided by some read-modify-write atomic
operations, some of which are presented in
\cref{sec:toolsoftrade:Atomic Operations}.
In the general case, memory ordering can be quite subtle, as
discussed in
\cref{chp:Advanced Synchronization: Memory Ordering}.
The next section covers an alternative to memory ordering, namely
limiting or even entirely avoiding data races.

\subsubsection{Avoiding Data Races}
\label{sec:toolsoftrade:Avoiding Data Races}

\begin{quote}
``Doctor, it hurts my head when I think about concurrently accessing
shared variables!''

``Then stop concurrently accessing shared variables!!!''
\end{quote}

The doctor's advice might seem unhelpful, but
one time-tested way to avoid concurrently accessing shared variables
is access those variables only when holding a particular lock, as will
be discussed in \cref{chp:Locking}.
Another way is to access a given ``shared'' variable only from a given
CPU or thread, as will be discussed in
\cref{chp:Data Ownership}.
It is possible to combine these two approaches, for example, a given
variable might be modified only by a given CPU or thread while holding a
particular lock, and might be read either from that same CPU or thread
on the one hand, or from some other CPU or thread while holding that
same lock on the other.
In all of these situations, all accesses to the shared variables may
be plain C-language accesses.

Here is a list of situations
allowing plain loads and stores for some accesses to a given variable,
while requiring markings (such as \apik{READ_ONCE()} and \apik{WRITE_ONCE()})
for other accesses to that same variable:

\begin{enumerate}
\item	A shared variable is only modified by a given owning CPU or
	thread, but is read by other CPUs or threads.
	All stores must use \apik{WRITE_ONCE()}.
	The owning CPU or thread may use plain loads.
	Everything else must use \apik{READ_ONCE()} for loads.
\item	A shared variable is only modified while holding a given
	lock, but is read by code not holding that lock.
	All stores must use \apik{WRITE_ONCE()}.
	CPUs or threads holding the lock may use plain loads.
	Everything else must use \apik{READ_ONCE()} for loads.
\item	A shared variable is only modified while holding a given
	lock by a given owning CPU or thread, but is read by other
	CPUs or threads or by code not holding that lock.
	All stores must use \apik{WRITE_ONCE()}.
	The owning CPU or thread may use plain loads, as may any
	CPU or thread holding the lock.
	Everything else must use \apik{READ_ONCE()} for loads.
\item	A shared variable is only accessed by a given CPU or thread
	and by a signal or interrupt handler running in that CPU's
	or thread's context.
	The handler can use plain loads and stores, as can any code
	that has prevented the handler from being invoked, that is,
	code that has blocked signals and/or interrupts.
	All other code must use \apik{READ_ONCE()} and \apik{WRITE_ONCE()}.
\item	A shared variable is only accessed by a given CPU or thread
	and by a signal or interrupt handler running in that CPU's
	or thread's context, and the handler always restores the values of any
	variables that it has written before return.
	The handler can use plain loads and stores, as can any code
	that has prevented the handler from being invoked, that is,
	code that has blocked signals and/or interrupts.
	All other code can use plain loads, but must use \apik{WRITE_ONCE()}
	to prevent store tearing, store fusing, and invented stores.
\end{enumerate}

\QuickQuiz{
	What needs to happen if an interrupt or signal handler
	might itself be interrupted?
}\QuickQuizAnswer{
	Then that interrupt handler must follow the same rules that
	are followed by other interrupted code.
	Only those handlers that cannot be themselves interrupted
	or that access no variables shared with an interrupting handler
	may safely use plain accesses, and even then only if those
	variables cannot be concurrently accessed by some other CPU or
	thread.
}\QuickQuizEnd

In most other cases, loads from and stores to a shared variable must
use \apik{READ_ONCE()} and \apik{WRITE_ONCE()} or stronger, respectively.
But it bears repeating that neither \apik{READ_ONCE()} nor \apik{WRITE_ONCE()}
provide any ordering guarantees other than within the compiler.
See the above
\cref{sec:toolsoftrade:Assembling the Rest of a Solution} or
\cref{chp:Advanced Synchronization: Memory Ordering}
for information on such guarantees.

Examples of many of these data-race-avoidance patterns are presented in
\cref{chp:Counting}.

\subsection{Atomic Operations}
\label{sec:toolsoftrade:Atomic Operations}

The Linux kernel provides a wide variety of \IX{atomic} operations, but
those defined on type \apik{atomic_t} provide a good start.
Normal non-tearing reads and stores are provided by
\apik{atomic_read()} and \apik{atomic_set()}, respectively.
\IX{Acquire load} is provided by \apik{smp_load_acquire()} and
\IX{release store} by \apik{smp_store_release()}.

Non-value-returning fetch-and-add operations are provided by
\apik{atomic_add()}, \apik{atomic_sub()}, \apik{atomic_inc()}, and
\apik{atomic_dec()}, among others.
An atomic decrement that returns a reached-zero indication is provided
by both \apik{atomic_dec_and_test()} and \apik{atomic_sub_and_test()}.
An atomic add that returns the new value is provided by
\apik{atomic_add_return()}.
Both \apik{atomic_add_unless()} and \apik{atomic_inc_not_zero()} provide
conditional atomic operations, where nothing happens unless the
original value of the atomic variable is different than the value
specified (these are very handy for managing
\IXalt{reference counters}{reference count}, for example).

An atomic exchange operation is provided by \apik{atomic_xchg()}, and
the celebrated \acrmf{cas} operation is provided by
\apik{atomic_cmpxchg()}.
Both of these return the old value.
Many additional atomic RMW primitives are available in the Linux kernel,
see the \path{Documentation/atomic_t.txt} file in the Linux-kernel
source tree.\footnote{
	As of Linux kernel v5.11.
}

This book's CodeSamples API closely follows that of the Linux kernel.

\subsection{Per-CPU Variables}
\label{sec:toolsoftrade:Per-CPU Variables}

The Linux kernel uses \apik{DEFINE_PER_CPU()} to define a per-CPU variable,
\apik{this_cpu_ptr()} to form a reference to this CPU's instance of a
given per-CPU variable, \apik{per_cpu()} to access a specified CPU's
instance of a given per-CPU variable, along with many other special-purpose
per-CPU operations.

\Cref{lst:toolsoftrade:Per-Thread-Variable API}
shows this book's per-thread-variable API, which is patterned
after the Linux kernel's per-CPU-variable API\@.
This API provides the per-thread equivalent of global variables.
Although this API is, strictly speaking, not necessary,\footnote{
	You could instead use \apig{__thread} or \apic{_Thread_local}.}
it can provide a good userspace analogy to Linux kernel code.

\begin{listing}
\begin{VerbatimL}[numbers=none]
DEFINE_PER_THREAD(type, name)
DECLARE_PER_THREAD(type, name)
per_thread(name, thread)
__get_thread_var(name)
init_per_thread(name, v)
\end{VerbatimL}
\caption{Per-Thread-Variable API}
\label{lst:toolsoftrade:Per-Thread-Variable API}
\end{listing}

\QuickQuiz{
	How could you work around the lack of a per-thread-variable
	API on systems that do not provide it?
}\QuickQuizAnswer{
	One approach would be to create an array indexed by
	\apipf{smp_thread_id()}, and another would be to use a hash
	table to map from \apipf{smp_thread_id()} to an array
	index---which is in fact what this
	set of APIs does in pthread environments.

	Another approach would be for the parent to allocate a structure
	containing fields for each desired per-thread variable, then
	pass this to the child during thread creation.
	However, this approach can impose large software-engineering
	costs in large systems.
	To see this, imagine if all global variables in a large system
	had to be declared in a single file, regardless of whether or
	not they were C static variables!
}\QuickQuizEnd

\subsubsection{API Members}

\begin{description}[style=nextline]
\item[\tco{DEFINE_PER_THREAD()}]

The \apipf{DEFINE_PER_THREAD()} primitive defines a per-thread variable.
Unfortunately, it is not possible to provide an initializer in the way
permitted by the Linux kernel's \apik{DEFINE_PER_CPU()} primitive,
but there is an \apipf{init_per_thread()} primitive that permits easy
runtime initialization.

\item[\tco{DECLARE_PER_THREAD()}]
The \apipf{DECLARE_PER_THREAD()} primitive is a declaration in the C sense,
as opposed to a definition.
Thus, a \apipf{DECLARE_PER_THREAD()} primitive may be used to access
a per-thread variable defined in some other file.

\item[\tco{per_thread()}]
The \apipf{per_thread()} primitive accesses the specified thread's variable.

\item[\tco{__get_thread_var()}]
The \apipf{__get_thread_var()} primitive accesses the current thread's variable.

\item[\tco{init_per_thread()}]
The \apipf{init_per_thread()} primitive sets all threads' instances of
the specified variable to the specified value.
The Linux kernel accomplishes this via normal C initialization,
relying in clever use of linker scripts and code executed during
the CPU-online process.

\end{description}

\subsubsection{Usage Example}

Suppose that we have a counter that is incremented very frequently
but read out quite rarely.
As will become clear in
\cref{sec:count:Statistical Counters},
it is helpful to implement such a counter using a per-thread variable.
Such a variable can be defined as follows:

\begin{VerbatimU}
DEFINE_PER_THREAD(int, counter);
\end{VerbatimU}

The counter must be initialized as follows:

\begin{VerbatimU}
init_per_thread(counter, 0);
\end{VerbatimU}

A thread can increment its instance of this counter as follows:

\begin{VerbatimU}
p_counter = &__get_thread_var(counter);
WRITE_ONCE(*p_counter, *p_counter + 1);
\end{VerbatimU}

The value of the counter is then the sum of its instances.
A snapshot of the value of the counter can thus be collected
as follows:

\begin{VerbatimU}
for_each_thread(t)
  sum += READ_ONCE(per_thread(counter, t));
\end{VerbatimU}

Again, it is possible to gain a similar effect using other mechanisms,
but per-thread variables combine convenience and high performance,
as will be shown in more detail in
\cref{sec:count:Statistical Counters}.

\QuickQuiz{
	What do you do if you need a per-thread (not per-CPU!) variable
	in the Linux kernel?
}\QuickQuizAnswer{
	First, needing a per-thread variable is less likely than
	you might think.
	Per-CPU variables can often do a per-thread variable's job.
	For example, if you only need to do addition, bitwise AND,
	bitwise OR, exchange, or compare-and-exchange, then the
	\co{this_cpu_add()},
	\co{this_cpu_add_return()},
	\co{this_cpu_and()},
	\co{this_cpu_or()},
	\co{this_cpu_xchg()},
	\co{this_cpu_cmpxchg()}, and
	\co{this_cpu_cmpxchg_double()}
	operations, respectively, will do the job cheaply and atomically
	with respect to context switches, interrupt handlers, and softirq
	handlers, but \emph{not} non-maskable interrupts.

	Second, within a preemption-disabled region of code, for
	example, one surrounded by the \co{preempt_disable()} and
	\co{preempt_enable()} macros, the current task is guaranteed to
	remain executing on the current CPU\@.
	Therefore, while within one such region, any series of accesses
	to per-CPU variables is atomic with respect to context switches,
	though not with respect to interrupt handlers, softirq handlers,
	and non-maskable interrupts.
	But please be aware that a preemption-disabled region of code
	that runs for more than a few microseconds will not be looked upon
	with favor by people attempting to construct real-time systems.

	Third, a field added to the \co{task_struct} structure acts
	as set of per-task variables.
	However, there are those who keep a close eye on the size of
	this structure, and these people are likely to ask hard
	questions about the need for any added fields.
	Therefore, if your field is being added for some facility
	that is only built into some kernels, you should definitely
	place your new \co{task_struct} fields under an appropriate
	\co{#ifdef}.

	Fourth and finally, your per-task variable might instead
	be located in some other structure and protected by some
	synchronization mechanism that is already in use.
	For example, if your code must hold a given lock, can accesses
	to this storage instead be protected by that lock?
	The fact that this is at the end of the list notwithstanding,
	you should look into this possibility first, not last!
}\QuickQuizEnd

\section{The Right Tool for the Job:
					How to Choose?}
\label{sec:toolsoftrade:The Right Tool for the Job: How to Choose?}
\epigraph{If you get stuck, change your tools; it may free your thinking.}
	 {Paul Arden, abbreviated}

As a rough rule of thumb, use the simplest tool that will get the job done.
If you can, simply program sequentially.
If that is insufficient, try using a shell script to mediate parallelism.
If the resulting shell-script \co{fork()}/\co{exec()} overhead
(about 480 microseconds for a minimal C program on an Intel Core Duo
laptop) is too
large, try using the C-language \apipx{fork()} and \apipx{wait()} primitives.
If the overhead of these primitives (about 80 microseconds for a minimal
child process) is still too large, then you
might need to use the POSIX threading primitives, choosing the appropriate
locking and/or atomic-operation primitives.
If the overhead of the POSIX threading primitives (typically sub-microsecond)
is too great, then the primitives introduced in
\cref{chp:Deferred Processing} may be required.
Of course, the actual overheads will depend not only on your hardware,
but most critically on the manner in which you use the primitives.
Furthermore, always remember that inter-process communication and
message-passing can be good alternatives to shared-memory multithreaded
execution, especially when your code makes good use of the design
principles called out in
\cref{chp:Partitioning and Synchronization Design}.

\QuickQuiz{
	Wouldn't the shell normally use \apipx{vfork()} rather than
	\apipx{fork()}?
}\QuickQuizAnswer{
	It might well do that, however, checking is left as an exercise
	for the reader.
	But in the meantime, I hope that we can agree that \apipx{vfork()}
	is a variant of \apipx{fork()}, so that we can use \apipx{fork()}
	as a generic term covering both.
}\QuickQuizEnd

Because concurrency was added to the C standard several decades after
the C language was first used to build concurrent systems, there are
a number of ways of concurrently accessing shared variables.
All else being equal, the C11 standard operations described in
\cref{sec:toolsoftrade:Atomic Operations (C11)}
should be your first stop.
If you need to access a given shared variable both with \IXplx{plain access}{es}
and atomically, then the modern \GCC\ atomics described in
\cref{sec:toolsoftrade:Atomic Operations (Modern gcc)}
might work well for you.
If you are working on an old codebase that uses the classic \GCC\ \co{__sync}
API, then you should review
\cref{sec:toolsoftrade:Atomic Operations (gcc Classic)}
as well as the relevant \GCC\ documentation.
If you are working on the Linux kernel or similar codebase that
combines use of the \apic{volatile} keyword with inline assembly,
or if you need dependencies to provide ordering, look at the material
presented in \cref{sec:toolsoftrade:Accessing Shared Variables}
as well as that in
\cref{chp:Advanced Synchronization: Memory Ordering}.

Whatever approach you take, please keep in mind that randomly hacking
multi-threaded code is a spectacularly bad idea, especially given that
shared-memory parallel systems use your own intelligence against you:
The smarter you are, the deeper a hole you will dig for yourself before
you realize that you are in trouble~\cite{DeadlockEmpire2016}.
Therefore, it is necessary to make the right design choices as well as
the correct choice of individual primitives,
as will be discussed at length in subsequent chapters.

	\setlength{\parskip}{0.0pt plus 1ex}
	\input{qqztoolsoftrade}
	\setlength{\parskip}{0.0pt plus 1.0pt}% return to default

% count/count.tex
% mainfile: ../perfbook.tex
% SPDX-License-Identifier: CC-BY-SA-3.0

\QuickQuizChapter{chp:Counting}{Counting}{qqzcount}
\Epigraph{As easy as 1, 2, 3!}{Unknown}

Counting is perhaps the simplest and most natural thing a computer can do.
However, counting efficiently and scalably on a large
shared-memory multiprocessor can be quite challenging.
Furthermore, the simplicity of the underlying concept of counting
allows us to explore the fundamental issues of concurrency without
the distractions
of elaborate data structures or complex synchronization primitives.
Counting therefore provides an excellent introduction to
parallel programming.

This chapter covers a number of special cases for which there are simple,
fast, and scalable counting algorithms.
But first, let us find out how much you already know about concurrent
counting.

\EQuickQuiz{
	Why should efficient and scalable counting be hard???
	After all, computers have special hardware for the sole purpose
	of doing counting!!!
}\EQuickQuizAnswer{
	Because the straightforward counting algorithms, for example,
	atomic operations on a shared counter, either are slow and scale
	badly, or are inaccurate, as will be seen in
	\cref{sec:count:Why Isn't Concurrent Counting Trivial?}.
}\EQuickQuizEnd

\EQuickQuiz{
	{\bfseries Network-packet counting problem.}
	Suppose that you need to collect statistics on the number
	of networking packets transmitted and received.
	Packets might be transmitted or received by any CPU on the system.
	Suppose further that your system is capable of
	handling millions of packets per second per CPU, and that
	a systems-monitoring package reads the count every five seconds.
	How would you implement this counter?
}\EQuickQuizAnswer{
	Hint:
	The act of updating the counter must be blazingly fast, but
	because the counter is read out only about once in five million
	updates, the act of reading out the counter can be quite slow.
	In addition, the value read out normally need not be all that
	accurate---after all, since the counter is updated a thousand
	times per millisecond, we should be able to work with a value
	that is within a few thousand counts of the ``true value'',
	whatever ``true value'' might mean in this context.
	However, the value read out should maintain roughly the same
	absolute error over time.
	For example, a 1\,\% error might be just fine when the count
	is on the order of a million or so, but might be absolutely
	unacceptable once the count reaches a trillion.
	See \cref{sec:count:Statistical Counters}.
}\EQuickQuizEnd

\QuickQuizLabel{\QcountQstatcnt}

\EQuickQuiz{
	{\bfseries Approximate structure-allocation limit problem.}
	Suppose that you need to maintain a count of the number of
	structures allocated in order to fail any allocations
	once the number of structures in use exceeds a limit
	(say, 10,000).
	Suppose further that the structures are short-lived, the
	limit is rarely exceeded, and a ``sloppy'' approximate limit
	is acceptable.
}\EQuickQuizAnswer{
	Hint:
	The act of updating the counter must again be blazingly
	fast, but the counter is read out each time that the
	counter is increased.
	However, the value read out need not be accurate
	\emph{except} that it must distinguish approximately
	between values below the limit and values greater than or
	equal to the limit.
	See \cref{sec:count:Approximate Limit Counters}.
}\EQuickQuizEnd

\QuickQuizLabel{\QcountQapproxcnt}

\EQuickQuiz{
	{\bfseries Exact structure-allocation limit problem.}
	Suppose that you need to maintain a count of the number of
	structures allocated in order to fail any allocations
	once the number of structures in use exceeds an exact limit
	(again, say 10,000).
	Suppose further that these structures are short-lived,
	and that the limit is rarely exceeded, that there is almost
	always at least one structure in use, and suppose further
	still that it is necessary to know exactly when this counter reaches
	zero, for example, in order to free up some memory
	that is not required unless there is at least one structure
	in use.
}\EQuickQuizAnswer{
	Hint:
	The act of updating the counter must once again be blazingly
	fast, but the counter is read out each time that the
	counter is increased.
	However, the value read out need not be accurate
	\emph{except} that it absolutely must distinguish perfectly
	between values between the limit and zero on the one hand,
	and values that either are less than or equal to zero or
	are greater than or equal to the limit on the other hand.
	See \cref{sec:count:Exact Limit Counters}.
}\EQuickQuizEnd

\QuickQuizLabel{\QcountQexactcnt}

\EQuickQuiz{
	{\bfseries Removable I/O device access-count problem.}
	Suppose that you need to maintain a \IX{reference count} on a
	heavily used removable mass-storage device, so that you
	can tell the user when it is safe to remove the device.
	As usual, the user indicates a desire to remove the device, and
	the system tells the user when it is safe to do so.
}\EQuickQuizAnswer{
	Hint:
	Yet again, the act of updating the counter must be blazingly
	fast and scalable in order to avoid slowing down I/O operations,
	but because the counter is read out only when the
	user wishes to remove the device, the counter read-out
	operation can be extremely slow.
	Furthermore, there is no need to be able to read out
	the counter at all unless the user has already indicated
	a desire to remove the device.
	In addition, the value read out need not be accurate
	\emph{except} that it absolutely must distinguish perfectly
	between non-zero and zero values, and even then only when
	the device is in the process of being removed.
	However, once it has read out a zero value, it must act
	to keep the value at zero until it has taken some action
	to prevent subsequent threads from gaining access to the
	device being removed.
	See \cref{sec:count:Applying Exact Limit Counters}.
}\EQuickQuizEnd

\QuickQuizLabel{\QcountQIOcnt}

\Cref{sec:count:Why Isn't Concurrent Counting Trivial?}
shows why counting is non-trivial.
\Cref{sec:count:Statistical Counters,sec:count:Approximate Limit Counters}
investigate network-packet counting and approximate structure-allocation
limits, respectively.
\Cref{sec:count:Exact Limit Counters}
takes on exact structure-allocation limits.
Finally, \cref{sec:count:Parallel Counting Discussion}
presents performance measurements and discussion.

\Cref{sec:count:Why Isn't Concurrent Counting Trivial?,%
sec:count:Statistical Counters}
contain introductory material, while the remaining sections
are more advanced.

\section{Why Isn't Concurrent Counting Trivial?}
\label{sec:count:Why Isn't Concurrent Counting Trivial?}
\epigraph{Seek simplicity, and distrust it.}{Alfred North Whitehead}

Let's start with something simple, for example, the straightforward
use of arithmetic shown in
\cref{lst:count:Just Count!} (\path{count_nonatomic.c}).
\begin{fcvref}[ln:count:count_nonatomic:inc-read]
Here, we have a counter on \clnref{counter}, we increment it on
\clnref{inc}, and we read out its value on \clnref{read}.
What could be simpler?
\end{fcvref}

\QuickQuiz{
	One thing that could be simpler is \co{++} instead of that
	concatenation of \co{READ_ONCE()} and \co{WRITE_ONCE()}.
	Why all that extra typing???
}\QuickQuizAnswer{
	See \cref{sec:toolsoftrade:Shared-Variable Shenanigans}
	on \cpageref{sec:toolsoftrade:Shared-Variable Shenanigans}
	for more information on how the compiler can cause trouble,
	as well as how \co{READ_ONCE()} and \co{WRITE_ONCE()} can avoid
	this trouble.
}\QuickQuizEnd

\begin{listing}
\input{CodeSamples/count/count_nonatomic=inc-read.fcv}
\caption{Just Count!}
\label{lst:count:Just Count!}
\end{listing}

This approach has the additional advantage of being blazingly fast if
you are doing lots of reading and almost no incrementing, and on small
systems, the performance is excellent.

There is just one large fly in the ointment:
This approach can lose counts.
On my six-core x86 laptop, a short run invoked \co{inc_count()}
285,824,000 times, but the final value of the counter was only
35,385,525.
Although approximation does have a large place in computing, loss of
87\,\% of the counts is a bit excessive.

\QuickQuizSeries{%
\QuickQuizB{
	But can't a smart compiler prove that
	\clnrefr{ln:count:count_nonatomic:inc-read:inc}
	of
	\cref{lst:count:Just Count!}
	is equivalent to the \co{++} operator and produce an x86
	add-to-memory instruction?
	And won't the CPU cache cause this to be atomic?
}\QuickQuizAnswerB{
	Although the \co{++} operator \emph{could} be atomic, there
	is no requirement that it be so unless it is applied to a
	C11 \co{_Atomic} variable.
	And indeed, in the absence of \co{_Atomic}, \GCC\ often
	chooses to load the value to a register, increment
	the register, then store the value to memory, which is
	decidedly non-atomic.

	Furthermore, note the volatile casts in
	\co{READ_ONCE()} and \co{WRITE_ONCE()}, which tell
	the compiler that the location might well be an MMIO
	device register.
	Because MMIO registers are not cached, it would be unwise for
	the compiler to assume that the increment operation is atomic.
}\QuickQuizEndB
\QuickQuizE{
	The 8-figure accuracy on the number of failures indicates
	that you really did test this.
	Why would it be necessary to test such a trivial program,
	especially when the bug is easily seen by inspection?
}\QuickQuizAnswerE{
	Not only are there very few
	trivial parallel programs, and most days I am
	not so sure that there are many trivial sequential programs, either.

	No matter how small or simple the program, if you haven't tested
	it, it does not work.
	And even if you have tested it, Murphy's Law says that there will
	be at least a few bugs still lurking.

	Furthermore, while proofs of correctness certainly do have their
	place, they never will replace testing, including the
	\path{counttorture.h} test setup used here.
	After all, proofs are only as good as the assumptions that they
	are based on.
	Finally, proofs can be every bit as buggy as are programs!
}\QuickQuizEndE
}

\begin{listing}
\input{CodeSamples/count/count_atomic=inc-read.fcv}
\caption{Just Count Atomically!}
\label{lst:count:Just Count Atomically!}
\end{listing}

\begin{figure}
\centering
\resizebox{2.5in}{!}{\includegraphics{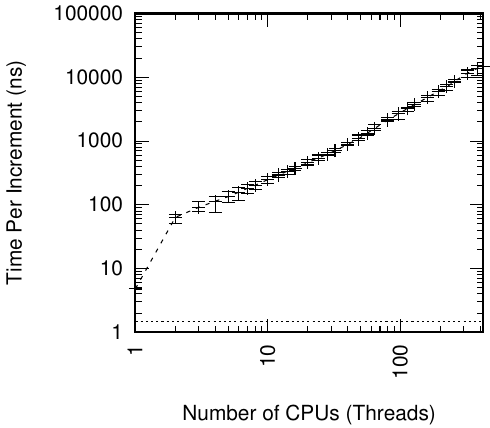}}
\caption{Atomic Increment Scalability on x86}
\label{fig:count:Atomic Increment Scalability on x86}
\end{figure}

The straightforward way to count accurately is to use \IX{atomic} operations,
as shown in
\cref{lst:count:Just Count Atomically!} (\path{count_atomic.c}).
\begin{fcvref}[ln:count:count_atomic:inc-read]
\Clnref{counter} defines an atomic variable,
\clnref{inc} atomically increments it, and
\clnref{read} reads it out.
\end{fcvref}
Because this is atomic, it keeps perfect count.
However, it is slower:
On my six-core x86 laptop, it is more than
twenty times slower than non-atomic increment, even
when only a single thread is incrementing.\footnote{
	Interestingly enough, non-atomically incrementing a counter will
	advance the counter more quickly than atomically incrementing
	the counter.
	Of course, if your only goal is to make the counter increase
	quickly, an easier approach is to simply assign a large value
	to the counter.
	Nevertheless, there is likely to be a role for algorithms that
	use carefully relaxed notions of correctness in order to gain
	greater performance and
	scalability~\cite{Andrews91textbook,Arcangeli03,10.5555/3241639.3241645,DavidUngar2011unsync}.}

This poor performance should not be a surprise, given the discussion in
\cref{chp:Hardware and its Habits},
nor should it be a surprise that the performance of atomic increment
gets slower as the number of CPUs and threads increase, as shown in
\cref{fig:count:Atomic Increment Scalability on x86}.
In this figure, the horizontal dashed line resting on the x~axis
is the ideal performance that would be achieved
by a perfectly scalable algorithm:
With such an algorithm, a given increment would incur the same
overhead that it would in a single-threaded program.
Atomic increment of a single global variable is clearly
decidedly non-ideal, and gets multiple orders of magnitude worse with
additional CPUs.

\QuickQuizSeries{%
\QuickQuizB{
	Why doesn't the horizontal dashed line on the x~axis meet the
	diagonal line at $x=1$?
}\QuickQuizAnswerB{
	Because of the overhead of the atomic operation.
	The dashed line on the x~axis represents the overhead of
	a single \emph{non-atomic} increment.
	After all, an \emph{ideal} algorithm would not only scale
	linearly, it would also incur no performance penalty compared
	to single-threaded code.

	This level of idealism may seem severe, but if it is good
	enough for \ppl{Linus}{Torvalds}, it is good enough for you.
}\QuickQuizEndB
\QuickQuizE{
	But atomic increment is still pretty fast.
	And incrementing a single variable in a tight loop sounds
	pretty unrealistic to me, after all, most of the program's
	execution should be devoted to actually doing work, not accounting
	for the work it has done!
	Why should I care about making this go faster?
}\QuickQuizAnswerE{
	In many cases, atomic increment will in fact be fast enough
	for you.
	In those cases, you should by all means use atomic increment.
	That said, there are many real-world situations where
	more elaborate counting algorithms are required.
	The canonical example of such a situation is counting packets
	and bytes in highly optimized networking stacks, where it is
	all too easy to find much of the execution time going into
	these sorts of accounting tasks, especially on large
	multiprocessors.

	In addition, as noted at the beginning of this chapter,
	counting provides an excellent view of the
	issues encountered in shared-memory parallel programs.
}\QuickQuizEndE
}

\begin{figure}
\centering
\resizebox{3in}{!}{\includegraphics{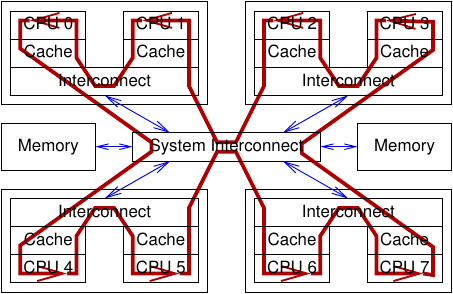}}
\caption{Data Flow For Global Atomic Increment}
\label{fig:count:Data Flow For Global Atomic Increment}
\end{figure}

\begin{figure}
\centering
\resizebox{3.2in}{!}{\includegraphics{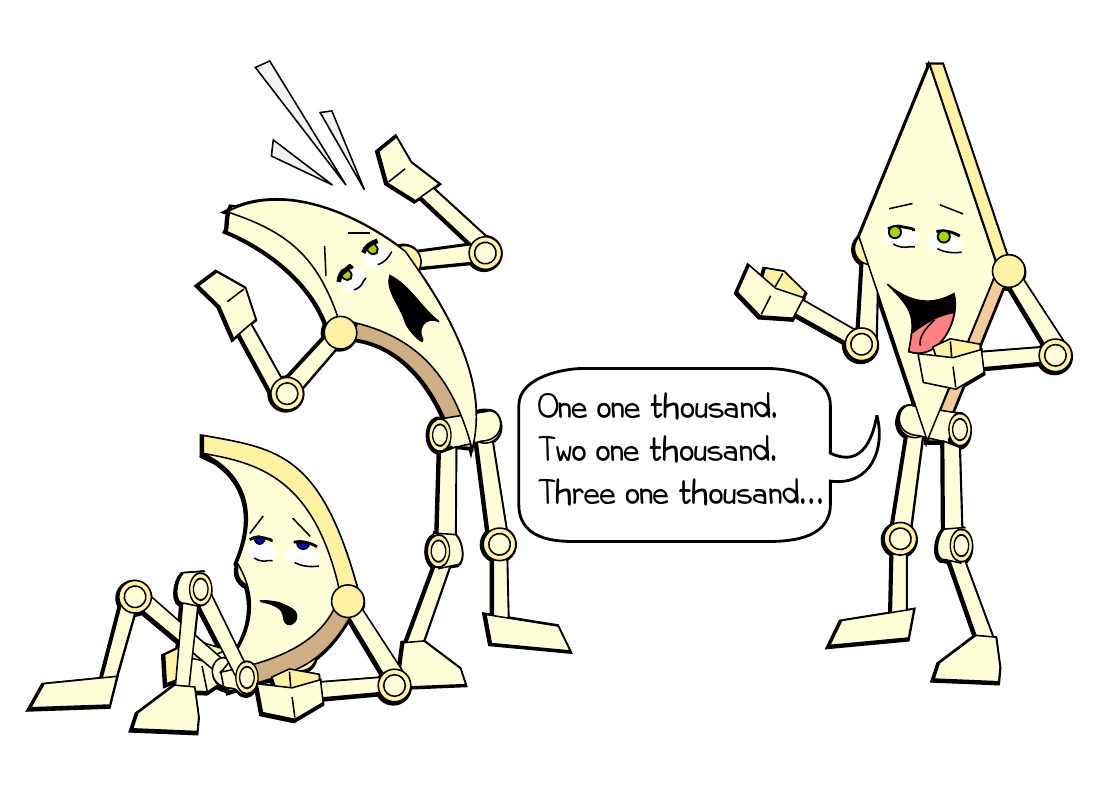}}
\caption{Waiting to Count}
\ContributedBy{Figure}{fig:count:Waiting to Count}{Melissa Broussard}
\end{figure}

For another perspective on global atomic increment, consider
\cref{fig:count:Data Flow For Global Atomic Increment}.
In order for each CPU to get a chance to increment a given
global variable, the \IX{cache line} containing that variable must
circulate among all the CPUs, as shown by the red arrows.
Such circulation will take significant time, resulting in
the poor performance seen in
\cref{fig:count:Atomic Increment Scalability on x86},
which might be thought of as shown in
\cref{fig:count:Waiting to Count}.
The following sections discuss high-performance counting, which
avoids the delays inherent in such circulation.

\QuickQuiz{
	But why can't CPU designers simply ship the addition operation to the
	data, avoiding the need to circulate the cache line containing
	the global variable being incremented?
}\QuickQuizAnswer{
	It might well be possible to do this in some cases.
	However, there are a few complications:
	\begin{enumerate}
	\item	If the value of the variable is required, then the
		thread will be forced to wait for the operation
		to be shipped to the data, and then for the result
		to be shipped back.
	\item	If the atomic increment must be ordered with respect
		to prior and/or subsequent operations, then the thread
		will be forced to wait for the operation to be shipped
		to the data, and for an indication that the operation
		completed to be shipped back.
	\item	Shipping operations among CPUs will likely require
		more lines in the system interconnect, which will consume
		more die area and more electrical power.
	\end{enumerate}
	But what if neither of the first two conditions holds?
	Then you should think carefully about the algorithms discussed
	in \cref{sec:count:Statistical Counters}, which achieve
	near-ideal performance on commodity hardware.

\begin{figure}
\centering
\resizebox{3in}{!}{\includegraphics{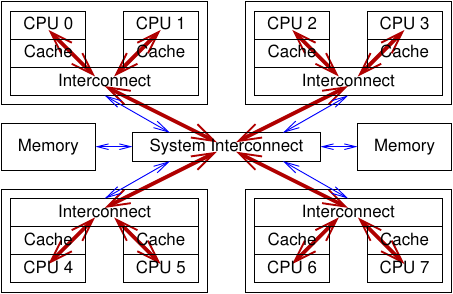}}
\caption{Data Flow For Global Combining-Tree Atomic Increment}
\label{fig:count:Data Flow For Global Combining-Tree Atomic Increment}
\end{figure}

	If either or both of the first two conditions hold, there is
	\emph{some} hope for improved hardware.
	One could imagine the hardware implementing a combining tree,
	so that the increment requests from multiple CPUs are combined
	by the hardware into a single addition when the combined request
	reaches the hardware.
	The hardware could also apply an order to the requests, thus
	returning to each CPU the return value corresponding to its
	particular atomic increment.
	This results in instruction latency that varies as $\O{\log N}$,
	where $N$ is the number of CPUs, as shown in
	\cref{fig:count:Data Flow For Global Combining-Tree Atomic Increment}.
	And CPUs with this sort of hardware optimization started to
	appear in 2011.

	This is a great improvement over the $\O{N}$ performance
	of current hardware shown in
	\cref{fig:count:Data Flow For Global Atomic Increment},
	and it is possible that hardware latencies might decrease
	further if innovations such as three-dimensional fabrication prove
	practical.
	Nevertheless, we will see that in some important special cases,
	software can do \emph{much} better.
}\QuickQuizEnd

\section{Statistical Counters}
\label{sec:count:Statistical Counters}
\epigraph{Facts are stubborn things, but statistics are pliable.}
	 {Mark Twain}

This section covers the common special case of statistical counters, where
the count is updated extremely frequently and the value is read out
rarely, if ever.
These will be used to solve the network-packet counting problem
posed in \QuickQuizRef{\QcountQstatcnt}.

\subsection{Design}

Statistical counting is typically handled by providing a counter per
thread (or CPU, when running in the kernel), so that each thread
updates its own counter, as was foreshadowed in
\cref{sec:toolsoftrade:Per-CPU Variables}
on \cpageref{sec:toolsoftrade:Per-CPU Variables}.
The aggregate value of the counters is read out by simply summing up
all of the threads' counters,
relying on the commutative and associative properties of addition.
This is an example of the Data Ownership pattern that will be introduced in
\cref{sec:SMPdesign:Data Ownership}
on \cpageref{sec:SMPdesign:Data Ownership}.

\QuickQuiz{
	But doesn't the fact that C's ``integers'' are limited in size
	complicate things?
}\QuickQuizAnswer{
	No, because modulo addition is still commutative and associative.
	At least as long as you use unsigned integers.
	Recall that in the C standard, overflow of signed integers results
	in undefined behavior, never mind the fact that machines that
	do anything other than wrap on overflow are quite rare these days.
	Unfortunately, compilers frequently carry out optimizations that
	assume that signed integers will not overflow, so if your code
	allows signed integers to overflow, you can run into trouble
	even on modern twos-complement hardware.

	That said, one potential source of additional complexity arises
	when attempting to gather (say) a 64-bit sum from 32-bit
	per-thread counters.
	Dealing with this added complexity is left as
	an exercise for the reader, for whom some of the techniques
	introduced later in this chapter could be quite helpful.
}\QuickQuizEnd

\subsection{Array-Based Implementation}
\label{sec:count:Array-Based Implementation}

One way to provide per-thread variables is to allocate an array with
one element per
thread (presumably cache aligned and padded to avoid false sharing).

\QuickQuiz{
	An array???
	But doesn't that limit the number of threads?
}\QuickQuizAnswer{
	It can, and in this toy implementation, it does.
	But it is not that hard to come up with an alternative
	implementation that permits an arbitrary number of threads,
	for example, using C11's \co{_Thread_local} facility,
	as shown in
	\cref{sec:count:Per-Thread-Variable-Based Implementation}.
}\QuickQuizEnd

\begin{listing}
\input{CodeSamples/count/count_stat=inc-read.fcv}
\caption{Array-Based Per-Thread Statistical Counters}
\label{lst:count:Array-Based Per-Thread Statistical Counters}
\end{listing}

Such an array can be wrapped into per-thread primitives, as shown in
\cref{lst:count:Array-Based Per-Thread Statistical Counters}
(\path{count_stat.c}).
\begin{fcvref}[ln:count:count_stat:inc-read]
\Clnref{define} defines an array containing a set of per-thread counters of
type \co{unsigned long} named, creatively enough, \co{counter}.

\Clnrefrange{inc:b}{inc:e}
show a function that increments the counters, using the
\apipf{__get_thread_var()} primitive to locate the currently running
thread's element of the \co{counter} array.
Because this element is modified only by the corresponding thread,
non-atomic increment suffices.
However, this code uses \co{WRITE_ONCE()} to prevent destructive compiler
optimizations.
For but one example, the compiler is within its rights to use a
to-be-stored-to location as temporary storage, thus writing what
would be for all intents and purposes garbage to that location
just before doing the desired store.
This could of course be rather confusing to anything attempting to
read out the count.
The use of \co{WRITE_ONCE()} prevents this optimization and others besides.

\QuickQuiz{
	What other nasty optimizations could \GCC\ apply?
}\QuickQuizAnswer{
	See \cref{sec:toolsoftrade:Shared-Variable Shenanigans,%
	sec:memorder:Compile-Time Consternation}
	for more information.
	One nasty optimization would be to apply common subexpression
	elimination to successive calls to the \co{read_count()} function,
	which might come as a surprise to code expecting changes in the
	values returned from successive calls to that function.
}\QuickQuizEnd

\Clnrefrange{read:b}{read:e}
show a function that reads out the aggregate value of the counter,
using the \apipf{for_each_thread()} primitive to iterate over the list of
currently running threads, and using the \apipf{per_thread()} primitive
to fetch the specified thread's counter.
This code also uses \co{READ_ONCE()} to ensure that the compiler doesn't
optimize these loads into oblivion.
For but one example, a pair of consecutive calls to \co{read_count()}
might be inlined, and an intrepid optimizer might notice that the same
locations were being summed and thus incorrectly conclude that it would
be simply wonderful to sum them once and use the resulting value twice.
This sort of optimization might be rather frustrating to people expecting
later \co{read_count()} calls to account for the activities of other
threads.
The use of \co{READ_ONCE()} prevents this optimization and others besides.
\end{fcvref}

\QuickQuizSeries{%
\QuickQuizB{
	How does the per-thread \co{counter} variable in
	\cref{lst:count:Array-Based Per-Thread Statistical Counters}
	get initialized?
}\QuickQuizAnswerB{
	The C standard specifies that the initial value of
	global variables is zero, unless they are explicitly initialized,
	thus implicitly initializing all the instances of \co{counter}
	to zero.
	Besides, in the common case where the user is interested only in
	differences between consecutive reads from statistical counters,
	the initial value is irrelevant.
}\QuickQuizEndB
\QuickQuizE{
	How is the code in
	\cref{lst:count:Array-Based Per-Thread Statistical Counters}
	supposed to permit more than one counter?
}\QuickQuizAnswerE{
	Indeed, this toy example does not support more than one counter.
	Modifying it so that it can provide multiple counters is left
	as an exercise to the reader.
}\QuickQuizEndE
}

\begin{figure}
\centering
\resizebox{3in}{!}{\includegraphics{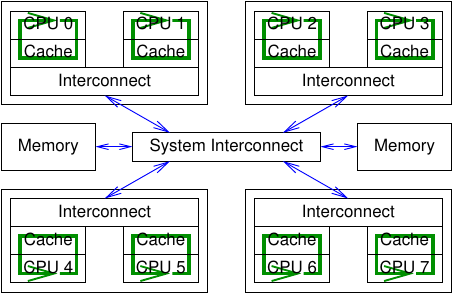}}
\caption{Data Flow For Per-Thread Increment}
\label{fig:count:Data Flow For Per-Thread Increment}
\end{figure}

This approach scales linearly with increasing number of updater threads
invoking \co{inc_count()}.
As is shown by the green arrows on each CPU in
\cref{fig:count:Data Flow For Per-Thread Increment},
the reason for this is that each CPU can make rapid progress incrementing
its thread's variable, without any expensive cross-system communication.
As such, this section solves the network-packet counting problem presented
at the beginning of this chapter.

\QuickQuiz{
	The read operation takes time to sum up the per-thread values,
	and during that time, the counter could well be changing.
	This means that the value returned by
	\co{read_count()} in
	\cref{lst:count:Array-Based Per-Thread Statistical Counters}
	will not necessarily be exact.
	Assume that the counter is being incremented at rate
	$r$ counts per unit time, and that \co{read_count()}'s
	execution consumes $\Delta$ units of time.
	What is the expected error in the return value?
}\QuickQuizAnswer{
	Let's do worst-case analysis first, followed by a less
	conservative analysis.

	In the worst case, the read operation completes immediately,
	but is then delayed for $\Delta$ time units before returning,
	in which case the worst-case error is simply $r \Delta$.

	This worst-case behavior is rather unlikely, so let us instead
	consider the case where the reads from each of the $N$
	counters is spaced equally over the time period $\Delta$.
	There will be $N+1$ intervals of duration $\frac{\Delta}{N+1}$
	between the $N$ reads.
	The rate $r$ of increments is expected to be spread evenly over
	the $N$ counters, for $\frac{r}{N}$ increments per unit time
	for each individual counter.
	The error due to the delay after the read from the last thread's
	counter will be given by $\frac{r \Delta}{N \left( N + 1 \right)}$,
	the second-to-last thread's counter by
	$\frac{2 r \Delta}{N \left( N + 1 \right)}$,
	the third-to-last by
	$\frac{3 r \Delta}{N \left( N + 1 \right)}$,
	and so on.
	The total error is given by the sum of the errors due to the
	reads from each thread's counter, which is:

	\begin{equation}
		\frac{r \Delta}{N \left( N + 1 \right)}
			\sum_{i = 1}^N i
	\end{equation}

	Expressing the summation in closed form yields:

	\begin{equation}
		\frac{r \Delta}{N \left( N + 1 \right)}
			\frac{N \left( N + 1 \right)}{2}
	\end{equation}

	Canceling yields the intuitively expected result:

	\begin{equation}
		\frac{r \Delta}{2}
	\label{eq:count:CounterErrorAverage}
	\end{equation}

	It is important to remember that error continues accumulating
	as the caller executes code making use of the count returned
	by the read operation.
	For example, if the caller spends time $t$ executing some
	computation based on the result of the returned count, the
	worst-case error will have increased to $r \left(\Delta + t\right)$.

	The expected error will have similarly increased to:

	\begin{equation}
		r \left( \frac{\Delta}{2} + t \right)
	\end{equation}

	Of course, it is sometimes unacceptable for the counter to
	continue incrementing during the read operation.
	\Cref{sec:count:Applying Exact Limit Counters}
	discusses a way to handle this situation.

	Thus far, we have been considering a counter that is only
	increased, never decreased.
	If the counter value is being changed by $r$ counts per unit
	time, but in either direction, we should expect the error
	to reduce.
	However, the worst case is unchanged because although the
	counter \emph{could} move in either direction, the worst
	case is when the read operation completes immediately,
	but then is delayed for $\Delta$ time units, during which
	time all the changes in the counter's value move it in
	the same direction, again giving us an absolute error
	of $r \Delta$.

	There are a number of ways to compute the average error,
	based on a variety of assumptions about the patterns of
	increments and decrements.
	For simplicity, let's assume that the $f$ fraction of
	the operations are decrements, and that the error of interest
	is the deviation from the counter's long-term trend line.
	Under this assumption, if $f$ is less than or equal to 0.5,
	each decrement will be canceled by an increment, so that
	$2f$ of the operations will cancel each other, leaving
	$1-2f$ of the operations being uncanceled increments.
	On the other hand, if $f$ is greater than 0.5, $1-f$ of
	the decrements are canceled by increments, so that the
	counter moves in the negative direction by $-1+2\left(1-f\right)$,
	which simplifies to $1-2f$, so that the counter moves an average
	of $1-2f$ per operation in either case.
	Therefore, that the long-term
	movement of the counter is given by $\left( 1-2f \right) r$.
	Plugging this into
	\cref{eq:count:CounterErrorAverage} yields:

	\begin{equation}
		\frac{\left( 1 - 2 f \right) r \Delta}{2}
	\end{equation}

	All that aside, in most uses of statistical counters, the
	error in the value returned by \co{read_count()} is
	irrelevant.
	This irrelevance is due to the fact that the time required
	for \co{read_count()} to execute is normally extremely
	small compared to the time interval between successive
	calls to \co{read_count()}.
}\QuickQuizEnd

\QuickQuizLabel{\StatisticalCounterAccuracy}

However, many implementations provide cheaper mechanisms for
per-thread data that are free from arbitrary array-size limits.
This is the topic of the next section.

\subsection{Per-Thread-Variable-Based Implementation}
\label{sec:count:Per-Thread-Variable-Based Implementation}

The C language, since C11, features a \apic{_Thread_local} storage class that
provides per-thread storage.\footnote{
	\GCC\ provides its own \apig{__thread} storage class, which was used
	in previous versions of this book.
	The two methods for specifying a thread-local variable are
	interchangeable when using \GCC\@.}
This can be used as shown in
\cref{lst:count:Per-Thread Statistical Counters} (\path{count_end.c})
to implement
a statistical counter that not only scales well and avoids arbitrary
thread-number limits, but that also incurs little or no performance
penalty to incrementers compared to simple non-atomic increment.

\begin{listing}
\input{CodeSamples/count/count_end=whole.fcv}
\caption{Per-Thread Statistical Counters}
\label{lst:count:Per-Thread Statistical Counters}
\end{listing}

\begin{fcvref}[ln:count:count_end:whole]
\Clnrefrange{var:b}{var:e} define needed variables:
\co{counter} is the per-thread counter
variable, the \co{counterp[]} array allows threads to access each others'
counters, \co{finalcount} accumulates the total as individual threads exit,
and \co{final_mutex} coordinates between threads accumulating the total
value of the counter and exiting threads.
\end{fcvref}

\QuickQuiz{
	Doesn't that explicit \co{counterp} array in
	\cref{lst:count:Per-Thread Statistical Counters}
	reimpose an arbitrary limit on the number of threads?
	Why doesn't the C language provide a \co{per_thread()} interface, similar
	to the Linux kernel's \co{per_cpu()} primitive, to allow
	threads to more easily access each others' per-thread variables?
}\QuickQuizAnswer{
	Why indeed?

	To be fair, user-mode thread-local storage faces some challenges
	that the Linux kernel gets to ignore.
	When a user-level thread exits, its per-thread variables all
	disappear, which complicates the problem of per-thread-variable
	access, particularly before the advent of user-level RCU
	(see \cref{sec:defer:Read-Copy Update (RCU)}).
	In contrast, in the Linux kernel, when a CPU goes offline,
	that CPU's per-CPU variables remain mapped and accessible.

	Similarly, when a new user-level thread is created, its
	per-thread variables suddenly come into existence.
	In contrast, in the Linux kernel, all per-CPU variables are
	mapped and initialized at boot time, regardless of whether
	the corresponding CPU exists yet, or indeed, whether the
	corresponding CPU will ever exist.

	A key limitation that the Linux kernel imposes is a compile-time
	maximum bound on the number of CPUs, namely, \co{CONFIG_NR_CPUS},
	along with a typically tighter boot-time bound of \co{nr_cpu_ids}.
	In contrast, in user space, there is not necessarily a hard-coded
	upper limit on the number of threads.

	Of course, both environments must handle dynamically loaded
	code (dynamic libraries in user space, kernel modules in the
	Linux kernel), which increases the complexity of per-thread
	variables.

	These complications make it significantly harder for user-space
	environments to provide access to other threads' per-thread
	variables.
	Nevertheless, such access is highly useful, and it is hoped
	that it will someday appear.

	In the meantime, textbook examples such as this one can use
	arrays whose limits can be easily adjusted by the user.
	Alternatively, such arrays can be dynamically allocated and
	expanded as needed at runtime.
	Finally, variable-length data structures such as linked
	lists can be used, as is done in the userspace RCU
	library~\cite{MathieuDesnoyers2009URCU,MathieuDesnoyers2012URCU}.
	This last approach can also reduce \IX{false sharing} in some cases.
}\QuickQuizEnd

\begin{fcvref}[ln:count:count_end:whole:inc]
The \co{inc_count()} function used by updaters is quite simple, as can
be seen on \clnrefrange{b}{e}.
\end{fcvref}

\begin{fcvref}[ln:count:count_end:whole:read]
The \co{read_count()} function used by readers is a bit more complex.
\Clnref{acquire} acquires a lock to exclude exiting threads, and
\clnref{release} releases it.
\Clnref{sum:init} initializes the sum to the count accumulated by those threads that
have already exited, and
\clnrefrange{loop:b}{loop:e} sum the counts being accumulated
by threads currently running.
Finally, \clnref{return} returns the sum.
\end{fcvref}

\QuickQuizSeries{%
\QuickQuizB{
	\begin{fcvref}[ln:count:count_end:whole:read]
	Doesn't the check for \co{NULL} on \clnref{check} of
	\cref{lst:count:Per-Thread Statistical Counters}
	add extra branch mispredictions?
	Why not have a variable set permanently to zero, and point
	unused counter-pointers to that variable rather than setting
	them to \co{NULL}?
	\end{fcvref}
}\QuickQuizAnswerB{
	This is a reasonable strategy.
	Checking for the performance difference is left as an exercise
	for the reader.
	However, please keep in mind that the fastpath is not
	\co{read_count()}, but rather \co{inc_count()}.
}\QuickQuizEndB
\QuickQuizE{
	Why on earth do we need something as heavyweight as a \emph{lock}
	guarding the summation in the function \co{read_count()} in
	\cref{lst:count:Per-Thread Statistical Counters}?
}\QuickQuizAnswerE{
	Remember, when a thread exits, its per-thread variables disappear.
	Therefore, if we attempt to access a given thread's per-thread
	variables after that thread exits, we will get a segmentation
	fault.
	The lock coordinates summation and thread exit, preventing this
	scenario.

	Of course, we could instead read-acquire a reader-writer lock,
	but \cref{chp:Deferred Processing} will introduce even
	lighter-weight mechanisms for implementing the required coordination.

	Another approach would be to use an array instead of a per-thread
	variable, which, as Alexey Roytman notes, would eliminate
	the tests against \co{NULL}.
	However, array accesses are often slower than accesses to
	per-thread variables, and use of an array would imply a
	fixed upper bound on the number of threads.
	Also, note that neither tests nor locks are needed on the
	\co{inc_count()} fastpath.
}\QuickQuizEndE
}

\begin{fcvref}[ln:count:count_end:whole:reg]
\Clnrefrange{b}{e} show the \co{count_register_thread()}
function, which
must be called by each thread before its first use of this counter.
This function simply sets up this thread's element of the \co{counterp[]}
array to point to its per-thread \co{counter} variable.
\end{fcvref}

\QuickQuiz{
	Why on earth do we need to acquire the lock in
	\co{count_register_thread()} in
	\cref{lst:count:Per-Thread Statistical Counters}?
	It is a single properly aligned machine-word store to a location
	that no other thread is modifying, so it should be atomic anyway,
	right?
}\QuickQuizAnswer{
	This lock could in fact be omitted, but better safe than
	sorry, especially given that this function is executed only at
	thread startup, and is therefore not on any critical path.
	Now, if we were testing on machines with thousands of CPUs,
	we might need to omit the lock, but on machines with ``only''
	a hundred or so CPUs, there is no need to get fancy.
}\QuickQuizEnd

\begin{fcvref}[ln:count:count_end:whole:unreg]
\Clnrefrange{b}{e} show the \co{count_unregister_thread()}
function, which
must be called prior to exit by each thread that previously called
\co{count_register_thread()}.
\Clnref{acquire} acquires the lock, and
\clnref{release} releases it, thus excluding any
calls to \co{read_count()} as well as other calls to
\co{count_unregister_thread()}.
\Clnref{add} adds this thread's \co{counter} to the global
\co{finalcount},
and then \clnref{NULL} \co{NULL}s out its \co{counterp[]} array entry.
A subsequent call to \co{read_count()} will see the exiting thread's
count in the global \co{finalcount}, and will skip the exiting thread
when sequencing through the \co{counterp[]} array, thus obtaining
the correct total.
\end{fcvref}

This approach gives updaters almost exactly the same performance as
a non-atomic add, and also scales linearly.
On the other hand, concurrent reads contend for a single global lock,
and therefore perform poorly and scale abysmally.
However, this is not a problem for statistical counters, where incrementing
happens often and readout happens almost never.
Of course, this approach is considerably more complex than the
array-based scheme, due to the fact that a given thread's per-thread
variables vanish when that thread exits.

\QuickQuiz{
	Fine, but the Linux kernel doesn't have to acquire a lock
	when reading out the aggregate value of per-CPU counters.
	So why should user-space code need to do this???
}\QuickQuizAnswer{
	Remember, the Linux kernel's per-CPU variables are always
	accessible, even if the corresponding CPU is offline---even
	if the corresponding CPU never existed and never will exist.

\begin{listing}
\input{CodeSamples/count/count_tstat=whole.fcv}
\caption{Per-Thread Statistical Counters With Lockless Summation}
\label{lst:count:Per-Thread Statistical Counters With Lockless Summation}
\end{listing}

	One workaround is to ensure that each thread continues to exist
	until all threads are finished, as shown in
	\cref{lst:count:Per-Thread Statistical Counters With Lockless Summation}
	(\path{count_tstat.c}).
	Analysis of this code is left as an exercise to the reader,
	however, please note that it requires tweaks in the
	\path{counttorture.h} counter-evaluation scheme.
	(Hint:
		See \co{#ifndef KEEP_GCC_THREAD_LOCAL}.)
	\Cref{chp:Deferred Processing} will introduce
	synchronization mechanisms that handle this situation in a much
	more graceful manner.
}\QuickQuizEnd

Both the array-based and \co{_Thread_local}-based approaches offer excellent
update-side performance and scalability.
However, these benefits result in large read-side expense for large
numbers of threads.
The next section shows one way to reduce read-side expense while
still retaining the update-side scalability.

\subsection{Eventually Consistent Implementation}
\label{sec:count:Eventually Consistent Implementation}

One way to retain update-side scalability while greatly improving
read-side performance is to weaken consistency requirements.
The counting algorithm in the previous section is guaranteed to
return a value between the value that an ideal counter would have
taken on near the beginning of \co{read_count()}'s execution and
that near the end of \co{read_count()}'s execution.
\emph{Eventual consistency}~\cite{WernerVogels:2009:EventuallyConsistent}
provides a weaker guarantee:
In absence of calls to \co{inc_count()}, calls to
\co{read_count()} will eventually return an accurate count.

We exploit eventual consistency by maintaining a global counter.
However, updaters only manipulate their per-thread counters.
A separate thread is provided to transfer counts from the per-thread
counters to the global counter.
Readers simply access the value of the global counter.
If updaters are active, the value used by the readers will be out of
date, however, once updates cease, the global counter will eventually
converge on the true value---hence this approach qualifies as
eventually consistent.

\begin{listing}
\ebresizeverb{.9}{\input{CodeSamples/count/count_stat_eventual=whole.fcv}}
\caption{Array-Based Per-Thread Eventually Consistent Counters}
\label{lst:count:Array-Based Per-Thread Eventually Consistent Counters}
\end{listing}

\begin{fcvref}[ln:count:count_stat_eventual:whole]
The implementation is shown in
\cref{lst:count:Array-Based Per-Thread Eventually Consistent Counters}
(\path{count_stat_eventual.c}).
\Clnrefrange{per_thr_cnt}{glb_cnt}
show the per-thread variable and the global variable that
track the counter's value, and \clnref{stopflag} shows \co{stopflag}
which is used to coordinate termination (for the case where we want
to terminate the program with an accurate counter value).
The \co{inc_count()} function shown on
\clnrefrange{inc:b}{inc:e} is similar to its
counterpart in
\cref{lst:count:Array-Based Per-Thread Statistical Counters}.
The \co{read_count()} function shown on
\clnrefrange{read:b}{read:e} simply returns the
value of the \co{global_count} variable.

However, the \co{count_init()} function on
\clnrefrange{init:b}{init:e}
creates the \co{eventual()} thread shown on
\clnrefrange{eventual:b}{eventual:e}, which
cycles through all the threads,
summing the per-thread local \co{counter} and storing the
sum to the \co{global_count} variable.
The \co{eventual()} thread waits an arbitrarily chosen one millisecond
between passes.

The \co{count_cleanup()} function on
\clnrefrange{cleanup:b}{cleanup:e} coordinates termination.
The call to \co{smp_load_acquire()} here and the call to \co{smp_store_release()}
in \co{eventual()} ensure that all updates to \co{global_count} are visible
to code following the call to \co{count_cleanup()}.

This approach gives extremely fast counter read-out while still
supporting linear counter-update scalability.
However, this excellent read-side performance and update-side scalability
comes at the cost of the additional thread running \co{eventual()}.
\end{fcvref}

\QuickQuizSeries{%
\QuickQuizB{
	Why doesn't \co{inc_count()} in
	\cref{lst:count:Array-Based Per-Thread Eventually Consistent Counters}
	need to use atomic instructions?
	After all, we now have multiple threads accessing the per-thread
	counters!
}\QuickQuizAnswerB{
	Because one of the two threads only reads, and because the
	variable is aligned and machine-sized, non-atomic instructions
	suffice.
	That said, the \co{READ_ONCE()} macro is used to prevent
	compiler optimizations that might otherwise prevent the
	counter updates from becoming visible to
	\co{eventual()}.\footnote{
		A simple definition of \co{READ_ONCE()} is shown in
		\cref{lst:toolsoftrade:Compiler Barrier Primitive (for GCC)}.}

	An older version of this algorithm did in fact use atomic
	instructions, kudos to Ersoy Bayramoglu for pointing out that
	they are in fact unnecessary.
	However, note that on a 32-bit system,
	the per-thread \co{counter} variables
	might need to be limited to 32 bits in order to sum them accurately,
	but with a 64-bit \co{global_count} variable to avoid overflow.
	In this case, it is necessary to zero the per-thread
	\co{counter} variables periodically in order to avoid overflow,
	which does require atomic instructions.
	It is extremely important to note that this zeroing cannot
	be delayed too long or overflow of the smaller per-thread
	variables will result.
	This approach therefore imposes real-time requirements on the
	underlying system, and in turn must be used with extreme care.

	In contrast, if all variables are the same size, overflow
	of any variable is harmless because the eventual sum
	will be modulo the word size.
}\QuickQuizEndB
\QuickQuizM{
	Won't the single global thread in the function \co{eventual()} of
	\cref{lst:count:Array-Based Per-Thread Eventually Consistent Counters}
	be just as severe a bottleneck as a global lock would be?
}\QuickQuizAnswerM{
	In this case, no.
	What will happen instead is that as the number of threads increases,
	the estimate of the counter
	value returned by \co{read_count()} will become more inaccurate.
}\QuickQuizEndM
\QuickQuizM{
	Won't the estimate returned by \co{read_count()} in
	\cref{lst:count:Array-Based Per-Thread Eventually Consistent Counters}
	become increasingly
	inaccurate as the number of threads rises?
}\QuickQuizAnswerM{
	Yes.
	If this proves problematic, one fix is to provide multiple
	\co{eventual()} threads, each covering its own subset of
	the other threads.
	In more extreme cases, a tree-like hierarchy of
	\co{eventual()} threads might be required.
}\QuickQuizEndM
\QuickQuizM{
	Given that in the eventually\-/consistent algorithm shown in
	\cref{lst:count:Array-Based Per-Thread Eventually Consistent Counters}
	both reads and updates have extremely low overhead
	and are extremely scalable, why would anyone bother with the
	implementation described in
	\cref{sec:count:Array-Based Implementation},
	given its costly read-side code?
}\QuickQuizAnswerM{
	The thread executing \co{eventual()} consumes CPU time.
	As more of these eventually\-/consistent counters are added,
	the resulting \co{eventual()} threads will eventually
	consume all available CPUs.
	This implementation therefore suffers a different sort of
	scalability limitation, with the scalability limit being in
	terms of the number of eventually consistent counters rather
	than in terms of the number of threads or CPUs.

	Of course, it is possible to make other tradeoffs.
	For example, a single thread could be created to handle all
	eventually\-/consistent counters, which would limit the
	overhead to a single CPU, but would result in increasing
	update-to-read latencies as the number of counters increased.
	Alternatively, that single thread could track the update rates
	of the counters, visiting the frequently\-/updated counters
	more frequently.
	In addition, the number of threads handling the counters could
	be set to some fraction of the total number of CPUs, and
	perhaps also adjusted at runtime.
	Finally, each counter could specify its latency, and
	deadline\-/scheduling techniques could be used to provide
	the required latencies to each counter.

	There are no doubt many other tradeoffs that could be made.
}\QuickQuizEndM
\QuickQuizE{
	What is the accuracy of the estimate returned by \co{read_count()} in
	\cref{lst:count:Array-Based Per-Thread Eventually Consistent Counters}?
}\QuickQuizAnswerE{
	A straightforward way to evaluate this estimate is to use the
	analysis derived in \QuickQuizARef{\StatisticalCounterAccuracy},
	but set $\Delta$ to the interval between the beginnings of
	successive runs of the \co{eventual()} thread.
	Handling the case where a given counter has multiple \co{eventual()}
	threads is left as an exercise for the reader.
}\QuickQuizEndE
}

\subsection{Discussion}

These three implementations show that it is possible to obtain
near-uniprocessor performance for statistical counters, despite running
on a parallel machine.

\QuickQuizSeries{%
\QuickQuizB{
	What fundamental difference is there between counting packets
	and counting the total number of bytes in the packets, given
	that the packets vary in size?
}\QuickQuizAnswerB{
	When counting packets, the counter is only incremented by the
	value one.
	On the other hand, when counting bytes, the counter might
	be incremented by largish numbers.

	Why does this matter?
	Because in the increment-by-one case, the value returned will
	be exact in the sense that the counter must necessarily have
	taken on that value at some point in time, even if it is impossible
	to say precisely when that point occurred.
	In contrast, when counting bytes, two different threads might
	return values that are inconsistent with any global ordering
	of operations.

	To see this, suppose that thread~0 adds the value three to its
	counter, thread~1 adds the value five to its counter, and
	threads~2 and~3 sum the counters.
	If the system is ``weakly ordered'' or if the compiler
	uses aggressive optimizations, thread~2 might find the
	sum to be three and thread~3 might find the sum to be five.
	The only possible global orders of the sequence of values
	of the counter are 0,3,8 and 0,5,8, and neither order is
	consistent with the results obtained.

	If you missed this one, you are not alone.
	Michael Scott used this question to stump Paul E.~McKenney
	during Paul's Ph.D. defense.
}\QuickQuizEndB
\QuickQuizE{
	Given that the reader must sum all the threads' counters, this
	counter-read operation could take a long time given large numbers
	of threads.
	Is there any way that the increment operation can remain
	fast and scalable while allowing readers to also enjoy
	not only reasonable performance and scalability, but also
	good accuracy?
}\QuickQuizAnswerE{
	One approach would be to maintain a global approximation
	to the value, similar to the approach described in
	\cref{sec:count:Eventually Consistent Implementation}.
	Updaters would increment their per-thread variable, but when it
	reached some predefined limit, atomically add it to a global
	variable, then zero their per-thread variable.
	This would permit a tradeoff between average increment overhead
	and accuracy of the value read out.
	In particular, it would allow sharp bounds on the read-side
	inaccuracy.

	Another approach makes use of the fact that readers often care
	only about certain transitions in value, not in the exact value.
	This approach is examined in
	\cref{sec:count:Approximate Limit Counters}.

	The reader is encouraged to think up and try out other approaches,
	for example, using a combining tree.
}\QuickQuizEndE
}

Given what has been presented in this section, you should now be able
to answer the Quick Quiz about statistical counters for networking
near the beginning of this chapter.

\section{Approximate Limit Counters}
\label{sec:count:Approximate Limit Counters}
\epigraph{An approximate answer to the right problem is worth a good deal
	  more than an exact answer to an approximate problem.}
	 {John Tukey}

Another special case of counting involves limit-checking.
For example, as noted in the approximate structure-allocation limit
problem in \QuickQuizRef{\QcountQapproxcnt},
suppose that you need to maintain a count of the number of
structures allocated in order to fail any allocations once the number
of structures in use exceeds a limit, in this case, 10,000.
Suppose further that these structures are short-lived, that this
limit is rarely exceeded, and that this limit is approximate in
that it is OK either to exceed it sometimes by some bounded amount
or to fail to reach it sometimes, again by some bounded amount.
See \cref{sec:count:Exact Limit Counters}
if you instead need the limit to be exact.

\subsection{Design}

One possible design for limit counters is to divide the limit of 10,000
by the number of threads, and give each thread a fixed pool of structures.
For example, given 100 threads, each thread would manage its own pool
of 100 structures.
This approach is simple, and in some cases works well, but it does not
handle the common case where a given structure is allocated by one
thread and freed by another~\cite{McKenney93}.
On the one hand, if a given thread takes credit for any structures it
frees, then the thread doing most of the allocating runs out
of structures, while the threads doing most of the freeing have lots
of credits that they cannot use.
On the other hand, if freed structures are credited to the CPU that
allocated them, it will be necessary for CPUs to manipulate each
others' counters, which will require expensive \IX{atomic} instructions
or other means of communicating between threads.\footnote{
	That said, if each structure will always be freed
	by the same CPU (or thread) that allocated it, then
	this simple partitioning approach works extremely well.}

In short, for many important workloads, we cannot fully partition the counter.
Given that partitioning the counters was what brought the excellent
update-side performance for the three schemes discussed in
\cref{sec:count:Statistical Counters}, this might be grounds
for some pessimism.
However, the eventually consistent algorithm presented in
\cref{sec:count:Eventually Consistent Implementation}
provides an interesting hint.
Recall that this algorithm kept two sets of books, a
per-thread \co{counter} variable for updaters and a \co{global_count}
variable for readers, with an \co{eventual()} thread that periodically
updated \co{global_count} to be eventually consistent with the values
of the per-thread \co{counter}.
The per-thread \co{counter} perfectly partitioned the counter value, while
\co{global_count} kept the full value.

For limit counters, we can use a variation on this theme where we
\emph{partially partition} the counter.
For example, consider four threads with each having not only a
per-thread \co{counter}, but also a per-thread maximum value (call
it \co{countermax}).

But then what happens if a given thread needs to increment its
\co{counter}, but \co{counter} is equal to its \co{countermax}?
The trick here is to move half of that thread's \co{counter} value
to a \co{globalcount}, then increment \co{counter}.
For example, if a given thread's \co{counter} and \co{countermax}
variables were both equal to 10, we do the following:

\begin{enumerate}
\item	Acquire a global lock.
\item	Add five to \co{globalcount}.
\item	To balance out the addition, subtract five from this
	thread's \co{counter}.
\item	Release the global lock.
\item	Increment this thread's \co{counter}, resulting in a value of six.
\end{enumerate}

Although this procedure still requires a global lock, that lock need only be
acquired once for every five increment operations, greatly reducing
that lock's level of \IXalt{contention}{lock contention}.
We can reduce this contention as low as we wish by increasing the
value of \co{countermax}.
However, the corresponding penalty for increasing the value of
\co{countermax} is reduced accuracy of \co{globalcount}.
To see this, note that on a four-CPU system, if \co{countermax}
is equal to ten, \co{globalcount} will be in error by at most
40 counts.
In contrast, if \co{countermax} is increased to 100, \co{globalcount}
might be in error by as much as 400 counts.

This raises the question of just how much we care about \co{globalcount}'s
deviation from the aggregate value of the counter, where this aggregate value
is the sum of \co{globalcount} and each thread's \co{counter} variable.
The answer to this question depends on how far the aggregate value is
from the counter's limit (call it \co{globalcountmax}).
The larger the difference between these two values, the larger \co{countermax}
can be without risk of exceeding the \co{globalcountmax} limit.
This means that the
value of a given thread's \co{countermax} variable can be set
based on this difference.
When far from the limit, the \co{countermax} per-thread variables are
set to large values to optimize for performance and scalability, while
when close to the limit, these same variables are set to small values
to minimize the error in the checks against the \co{globalcountmax} limit.

This design is an example of \emph{parallel fastpath}, which is an important
design pattern in which the common case executes with no expensive
instructions and no interactions between threads, but where occasional
use is also made of a more conservatively designed
(and higher overhead) global algorithm.
This design pattern is covered in more detail in
\cref{sec:SMPdesign:Parallel Fastpath}.

\subsection{Simple Limit Counter Implementation}
\label{sec:count:Simple Limit Counter Implementation}

\begin{figure}
\centering
\resizebox{2.5in}{!}{\includegraphics{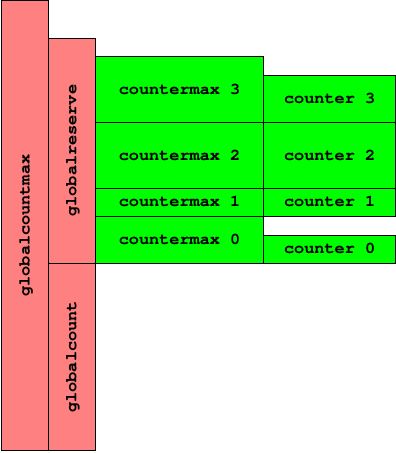}}
\caption{Simple Limit Counter Variable Relationships}
\label{fig:count:Simple Limit Counter Variable Relationships}
\end{figure}

\begin{fcvref}[ln:count:count_lim:variable]
\Cref{lst:count:Simple Limit Counter Variables}
shows both the per-thread and global variables used by this
implementation.
The per-thread \co{counter} and \co{countermax} variables are the
corresponding thread's local counter and the upper bound on that
counter, respectively.
The \co{globalcountmax} variable on
\clnref{globalcountmax} contains the upper
bound for the aggregate counter, and the \co{globalcount} variable
on \clnref{globalcount} is the global counter.
The sum of \co{globalcount} and each thread's \co{counter} gives
the aggregate value of the overall counter.
The \co{globalreserve} variable on
\clnref{globalreserve} is at least the sum of all of the
per-thread \co{countermax} variables.
\end{fcvref}
The relationship among these variables is shown by
\cref{fig:count:Simple Limit Counter Variable Relationships}:
\begin{enumerate}
\item	The sum of \co{globalcount} and \co{globalreserve} must
	be less than or equal to \co{globalcountmax}.
\item	The sum of all threads' \co{countermax} values must be
	less than or equal to \co{globalreserve}.
\item	Each thread's \co{counter} must be less than or equal to
	that thread's \co{countermax}.
\end{enumerate}

\begin{listing}
\input{CodeSamples/count/count_lim=variable.fcv}
\caption{Simple Limit Counter Variables}
\label{lst:count:Simple Limit Counter Variables}
\end{listing}

Each element of the \co{counterp[]} array references the corresponding
thread's \co{counter} variable, and, finally, the \co{gblcnt_mutex}
spinlock guards all of the global variables, in other words, no thread
is permitted to access or modify any of the global variables unless it
has acquired \co{gblcnt_mutex}.

\begin{listing}
\input{CodeSamples/count/count_lim=add_sub_read.fcv}
\caption{Simple Limit Counter Add, Subtract, and Read}
\label{lst:count:Simple Limit Counter Add; Subtract; and Read}
\end{listing}

\Cref{lst:count:Simple Limit Counter Add; Subtract; and Read}
shows the \co{add_count()}, \co{sub_count()}, and \co{read_count()}
functions (\path{count_lim.c}).

\QuickQuiz{
	Why does
	\cref{lst:count:Simple Limit Counter Add; Subtract; and Read}
	provide \co{add_count()} and \co{sub_count()} instead of the
	\co{inc_count()} and \co{dec_count()} interfaces show in
	\cref{sec:count:Statistical Counters}?
}\QuickQuizAnswer{
	Because structures come in different sizes.
	Of course, a limit counter corresponding to a specific size
	of structure might still be able to use
	\co{inc_count()} and \co{dec_count()}.
}\QuickQuizEnd

\begin{fcvref}[ln:count:count_lim:add_sub_read:add]
\Clnrefrange{b}{e} show \co{add_count()},
which adds the specified value \co{delta}
to the counter.
\Clnref{checklocal} checks to see if there is room for
\co{delta} on this thread's
\co{counter}, and, if so,
\clnref{add} adds it and \clnref{return:ls} returns success.
This is the \co{add_counter()} fastpath, and it does no atomic operations,
references only per-thread variables, and should not incur any cache misses.
\end{fcvref}

\begin{listing}
\begin{VerbatimL}[firstnumber=3]
	if (counter + delta <= countermax) {
		WRITE_ONCE(counter, counter + delta);
		return 1;
	}
\end{VerbatimL}
\caption{Intuitive Fastpath}
\label{lst:count:Intuitive Fastpath}
\end{listing}

\QuickQuiz{
	What is with the strange form of the condition on
	\clnrefr{ln:count:count_lim:add_sub_read:add:checklocal} of
	\cref{lst:count:Simple Limit Counter Add; Subtract; and Read}?
	Why not the more intuitive form of the fastpath shown in
	\cref{lst:count:Intuitive Fastpath}?
}\QuickQuizAnswer{
	Two words.
	``Integer overflow.''

	Try the formulation in \cref{lst:count:Intuitive Fastpath}
	with \co{counter} equal to~10 and
	\co{delta} equal to \co{ULONG_MAX}.
	Then try it again with the code shown in
	\cref{lst:count:Simple Limit Counter Add; Subtract; and Read}.

	A good understanding of integer overflow will be required for
	the rest of this example, so if you have never dealt with
	integer overflow before, please try several examples to get
	the hang of it.
	Integer overflow can sometimes be more difficult to get right
	than parallel algorithms!
}\QuickQuizEnd

\begin{fcvref}[ln:count:count_lim:add_sub_read:add]
If the test on
\clnref{checklocal} fails, we must access global variables, and thus
must acquire \co{gblcnt_mutex} on
\clnref{acquire}, which we release on \clnref{release:f}
in the failure case or on \clnref{release:s} in the success case.
\Clnref{globalize} invokes \co{globalize_count()}, shown in
\cref{lst:count:Simple Limit Counter Utility Functions},
which clears the thread-local variables, adjusting the global variables
as needed, thus simplifying global processing.
(But don't take \emph{my} word for it, try coding it yourself!)
\Clnref{checkglb:b,checkglb:e} check to see
if addition of \co{delta} can be accommodated,
with the meaning of the expression preceding the less-than sign shown in
\cref{fig:count:Simple Limit Counter Variable Relationships}
as the difference in height of the two red (leftmost) bars.
If the addition of \co{delta} cannot be accommodated, then
\clnref{release:f} (as noted earlier) releases \co{gblcnt_mutex} and
\clnref{return:gf}
returns indicating failure.

Otherwise, we take the slowpath.
\Clnref{addglb} adds \co{delta} to \co{globalcount}, and then
\clnref{balance} invokes \co{balance_count()} (shown in
\cref{lst:count:Simple Limit Counter Utility Functions})
in order to update both the global and the per-thread variables.
This call to \co{balance_count()}
will usually set this thread's \co{countermax} to re-enable the fastpath.
\Clnref{release:s} then releases
\co{gblcnt_mutex} (again, as noted earlier), and, finally,
\clnref{return:gs} returns indicating success.
\end{fcvref}

\QuickQuiz{
	Why does \co{globalize_count()} zero the per-thread variables,
	only to later call \co{balance_count()} to refill them in
	\cref{lst:count:Simple Limit Counter Add; Subtract; and Read}?
	Why not just leave the per-thread variables non-zero?
}\QuickQuizAnswer{
	That is in fact what an earlier version of this code did.
	But addition and subtraction are extremely cheap, and handling
	all of the special cases that arise is quite complex.
	Again, feel free to try it yourself, but beware of integer
	overflow!
}\QuickQuizEnd

\begin{fcvref}[ln:count:count_lim:add_sub_read:sub]
\Clnrefrange{b}{e} show \co{sub_count()},
which subtracts the specified
\co{delta} from the counter.
\Clnref{checklocal} checks to see if the per-thread counter can accommodate
this subtraction, and, if so, \clnref{sub} does the subtraction and
\clnref{return:ls} returns success.
These lines form \co{sub_count()}'s fastpath, and, as with
\co{add_count()}, this fastpath executes no costly operations.

If the fastpath cannot accommodate subtraction of \co{delta},
execution proceeds to the slowpath on
\clnrefrange{acquire}{return:gs}.
Because the slowpath must access global state, \clnref{acquire}
acquires \co{gblcnt_mutex}, which is released either by \clnref{release:f}
(in case of failure) or by \clnref{release:s} (in case of success).
\Clnref{globalize} invokes \co{globalize_count()}, shown in
\cref{lst:count:Simple Limit Counter Utility Functions},
which again clears the thread-local variables, adjusting the global variables
as needed.
\Clnref{checkglb} checks to see if the counter can accommodate subtracting
\co{delta}, and, if not, \clnref{release:f} releases \co{gblcnt_mutex}
(as noted earlier) and \clnref{return:gf} returns failure.
\end{fcvref}

\QuickQuizSeries{%
\QuickQuizB{
	Given that \co{globalreserve} counted against us in \co{add_count()},
	why doesn't it count for us in \co{sub_count()} in
	\cref{lst:count:Simple Limit Counter Add; Subtract; and Read}?
}\QuickQuizAnswerB{
	The \co{globalreserve} variable tracks the sum of all threads'
	\co{countermax} variables.
	The sum of these threads' \co{counter} variables might be anywhere
	from zero to \co{globalreserve}.
	We must therefore take a conservative approach, assuming that
	all threads' \co{counter} variables are full in \co{add_count()}
	and that they are all empty in \co{sub_count()}.

	But remember this question, as we will come back to it later.
}\QuickQuizEndB
\QuickQuizE{
	Suppose that one thread invokes \co{add_count()} shown in
	\cref{lst:count:Simple Limit Counter Add; Subtract; and Read},
	and then another thread invokes \co{sub_count()}.
	Won't \co{sub_count()} return failure even though the value of
	the counter is non-zero?
}\QuickQuizAnswerE{
	Indeed it will!
	In many cases, this will be a problem, as discussed in
	\cref{sec:count:Simple Limit Counter Discussion}, and
	in those cases the algorithms from
	\cref{sec:count:Exact Limit Counters}
	will likely be preferable.
}\QuickQuizEndE
}

\begin{fcvref}[ln:count:count_lim:add_sub_read:sub]
If, on the other hand, \clnref{checkglb} finds that the counter \emph{can}
accommodate subtracting \co{delta}, we complete the slowpath.
\Clnref{subglb} does the subtraction and then
\clnref{balance} invokes \co{balance_count()} (shown in
\cref{lst:count:Simple Limit Counter Utility Functions})
in order to update both global and per-thread variables
(hopefully re-enabling the fastpath).
Then \clnref{release:s} releases \co{gblcnt_mutex}, and
\clnref{return:gs} returns success.
\end{fcvref}

\QuickQuiz{
	Why have both \co{add_count()} and \co{sub_count()} in
	\cref{lst:count:Simple Limit Counter Add; Subtract; and Read}?
	Why not simply pass a negative number to \co{add_count()}?
}\QuickQuizAnswer{
	Given that \co{add_count()} takes an \co{unsigned} \co{long}
	as its argument, it is going to be a bit tough to pass it a
	negative number.
	And unless you have some anti-matter memory, there is little
	point in allowing negative numbers when counting the number
	of structures in use!

	All kidding aside, it would of course be possible to combine
	\co{add_count()} and \co{sub_count()}, however, the \co{if}
	conditions on the combined function would be more complex
	than in the current pair of functions, which would in turn
	mean slower execution of these fast paths.
}\QuickQuizEnd

\begin{fcvref}[ln:count:count_lim:add_sub_read:read]
\Clnrefrange{b}{e} show \co{read_count()},
which returns the aggregate value
of the counter.
It acquires \co{gblcnt_mutex} on \clnref{acquire}
and releases it on \clnref{release},
excluding global operations from \co{add_count()} and \co{sub_count()},
and, as we will see, also excluding thread creation and exit.
\Clnref{initsum} initializes local variable \co{sum} to the value of
\co{globalcount}, and then the loop spanning
\clnrefrange{loop:b}{loop:e} sums the
per-thread \co{counter} variables.
\Clnref{return} then returns the sum.
\end{fcvref}

\begin{listing}
\input{CodeSamples/count/count_lim=utility.fcv}
\caption{Simple Limit Counter Utility Functions}
\label{lst:count:Simple Limit Counter Utility Functions}
\end{listing}

\Cref{lst:count:Simple Limit Counter Utility Functions}
shows a number of utility functions used by the \co{add_count()},
\co{sub_count()}, and \co{read_count()} primitives shown in
\cref{lst:count:Simple Limit Counter Add; Subtract; and Read}.

\begin{fcvref}[ln:count:count_lim:utility:globalize]
\Clnrefrange{b}{e} show \co{globalize_count()},
which zeros the current thread's
per-thread counters, adjusting the global variables appropriately.
It is important to note that this function does not change the aggregate
value of the counter, but instead changes how the counter's current value
is represented.
\Clnref{add} adds the thread's \co{counter} variable to \co{globalcount},
and \clnref{zero} zeroes \co{counter}.
Similarly, \clnref{sub} subtracts the per-thread \co{countermax} from
\co{globalreserve}, and \clnref{zeromax} zeroes \co{countermax}.
It is helpful to refer to
\cref{fig:count:Simple Limit Counter Variable Relationships}
when reading both this function and \co{balance_count()}, which is next.
\end{fcvref}

\begin{fcvref}[ln:count:count_lim:utility:balance]
\Clnrefrange{b}{e} show \co{balance_count()},
which is roughly speaking
the inverse of \co{globalize_count()}.
This function's job is to set the current thread's
\co{countermax} variable to the largest value that avoids the risk
of the counter exceeding the \co{globalcountmax} limit.
Changing the current thread's \co{countermax} variable of course
requires corresponding adjustments to \co{counter}, \co{globalcount}
and \co{globalreserve}, as can be seen by referring back to
\cref{fig:count:Simple Limit Counter Variable Relationships}.
By doing this, \co{balance_count()} maximizes use of
\co{add_count()}'s and \co{sub_count()}'s low-overhead fastpaths.
As with \co{globalize_count()}, \co{balance_count()} is not permitted
to change the aggregate value of the counter.

\Clnrefrange{share:b}{share:e} compute this thread's share of
that portion of
\co{globalcountmax} that is not already covered by either
\co{globalcount} or \co{globalreserve}, and assign the
computed quantity to this thread's \co{countermax}.
\Clnref{adjreserve} makes the corresponding adjustment to \co{globalreserve}.
\Clnref{middle} sets this thread's \co{counter} to the middle of the range
from zero to \co{countermax}.
\Clnref{check} checks to see whether \co{globalcount} can in fact accommodate
this value of \co{counter}, and, if not,
\clnref{adjcounter} decreases \co{counter}
accordingly.
Finally, in either case,
\clnref{adjglobal} makes the corresponding adjustment to
\co{globalcount}.
\end{fcvref}

\QuickQuiz{
	\begin{fcvref}[ln:count:count_lim:utility:balance]
	Why set \co{counter} to \co{countermax / 2} in \clnref{middle} of
	\cref{lst:count:Simple Limit Counter Utility Functions}?
	Wouldn't it be simpler to just take \co{countermax} counts?
	\end{fcvref}
}\QuickQuizAnswer{
	\begin{fcvref}[ln:count:count_lim:utility:balance]
	First, it really is reserving \co{countermax} counts
	(see \clnref{adjreserve}), however,
	it adjusts so that only half of these are actually in use
	by the thread at the moment.
	This allows the thread to carry out at least \co{countermax / 2}
	increments or decrements before having to refer back to
	\co{globalcount} again.

	Note that the accounting in \co{globalcount} remains accurate,
	thanks to the adjustment in \clnref{adjglobal}.
	\end{fcvref}
}\QuickQuizEnd

\begin{figure*}
\centering
\IfEbookSize{
\resizebox{\textwidth}{!}{\includegraphics{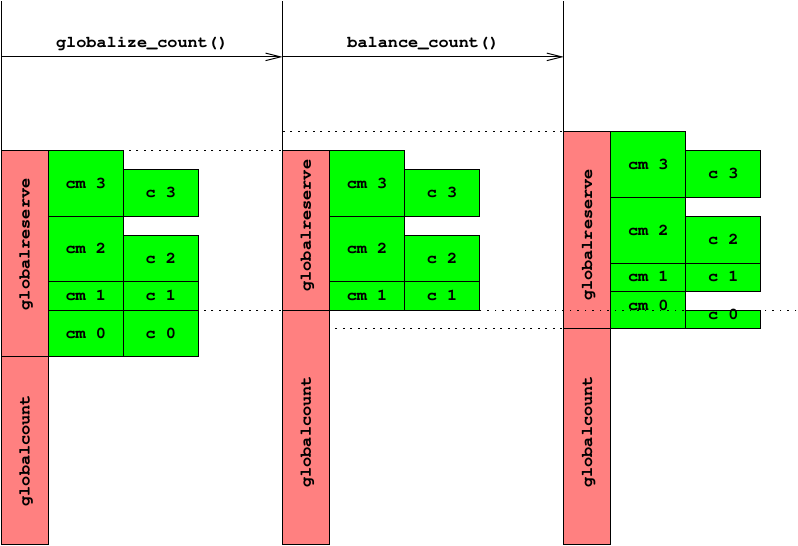}}
}{
\resizebox{5in}{!}{\includegraphics{count/globbal}}
}
\caption{Schematic of Globalization and Balancing}
\label{fig:count:Schematic of Globalization and Balancing}
\end{figure*}

It is helpful to look at a schematic depicting how the relationship
of the counters changes with the execution of first
\co{globalize_count()} and then \co{balance_count()}, as shown in
\cref{fig:count:Schematic of Globalization and Balancing}.
Time advances from left to right, with the leftmost configuration
roughly that of
\cref{fig:count:Simple Limit Counter Variable Relationships}.
The center configuration shows the relationship of these same counters
after \co{globalize_count()} is executed by thread~0.
As can be seen from the figure, thread~0's \co{counter} (``c~0'' in
the figure) is added to \co{globalcount}, while the value of
\co{globalreserve} is reduced by this same amount.
Both thread~0's \co{counter} and its \co{countermax}
(``cm~0'' in the figure) are reduced to zero.
The other three threads' counters are unchanged.
Note that this change did not affect the overall value of the counter,
as indicated by the bottommost dotted line connecting the leftmost
and center configurations.
In other words, the sum of \co{globalcount} and the four threads'
\co{counter} variables is the same in both configurations.
Similarly, this change did not affect the sum of \co{globalcount} and
\co{globalreserve}, as indicated by the upper dotted line.

The rightmost configuration shows the relationship of these counters
after \co{balance_count()} is executed, again by thread~0.
One-quarter of the remaining count, denoted by the vertical line extending
up from all three configurations, is added to thread~0's
\co{countermax} and half of that to thread~0's \co{counter}.
The amount added to thread~0's \co{counter} is also subtracted from
\co{globalcount} in order to avoid changing the overall value of the
counter (which is again the sum of \co{globalcount} and the three
threads' \co{counter} variables), again as indicated by the lowermost
of the two dotted lines connecting the center and rightmost configurations.
The \co{globalreserve} variable is also adjusted so that this variable
remains equal to the sum of the four threads' \co{countermax}
variables.
Because thread~0's \co{counter} is less than its \co{countermax},
thread~0 can once again increment the counter locally.

\QuickQuiz{
	In \cref{fig:count:Schematic of Globalization and Balancing},
	even though a quarter of the remaining count up to the limit is
	assigned to thread~0, only an eighth of the remaining count is
	consumed, as indicated by the uppermost dotted line connecting
	the center and the rightmost configurations.
	Why is that?
}\QuickQuizAnswer{
	The reason this happened is that thread~0's \co{counter} was
	set to half of its \co{countermax}.
	Thus, of the quarter assigned to thread~0, half of that quarter
	(one eighth) came from \co{globalcount}, leaving the other half
	(again, one eighth) to come from the remaining count.

	There are two purposes for taking this approach:
	\begin{enumerate*}[(1)]
	\item To allow thread~0 to use the fastpath for decrements as
	well as increments and
	\item To reduce the inaccuracies if all threads are monotonically
	incrementing up towards the limit.
	\end{enumerate*}
	To see this last point, step through the algorithm and watch
	what it does.
}\QuickQuizEnd

\begin{fcvref}[ln:count:count_lim:utility:register]
\Clnrefrange{b}{e} show \co{count_register_thread()},
which sets up state for
newly created threads.
This function simply installs
a pointer to the newly created thread's \co{counter} variable into
the corresponding entry of the \co{counterp[]} array under the protection
of \co{gblcnt_mutex}.
\end{fcvref}

\begin{fcvref}[ln:count:count_lim:utility:unregister]
Finally, \clnrefrange{b}{e} show \co{count_unregister_thread()},
which tears down
state for a soon-to-be-exiting thread.
\Clnref{acquire} acquires \co{gblcnt_mutex} and
\clnref{release} releases it.
\Clnref{globalize} invokes \co{globalize_count()}
to clear out this thread's
counter state, and \clnref{clear} clears this thread's entry in the
\co{counterp[]} array.
\end{fcvref}

\subsection{Simple Limit Counter Discussion}
\label{sec:count:Simple Limit Counter Discussion}

This type of counter is quite fast when aggregate values are near zero,
with some overhead due to the comparison and branch in both
\co{add_count()}'s and \co{sub_count()}'s fastpaths.
However, the use of a per-thread \co{countermax} reserve means that
\co{add_count()} can fail even when
the aggregate value of the counter is nowhere near \co{globalcountmax}.
Similarly, \co{sub_count()} can fail
even when the aggregate value of the counter is nowhere near zero.

In many cases, this is unacceptable.
Even if the \co{globalcountmax} is intended to be an approximate limit,
there is usually a limit to exactly how much approximation can be tolerated.
One way to limit the degree of approximation is to impose an upper limit
on the value of the per-thread \co{countermax} instances.
This task is undertaken in the next section.

\subsection{Approximate Limit Counter Implementation}
\label{sec:count:Approximate Limit Counter Implementation}

Because this implementation (\path{count_lim_app.c}) is quite similar to
that in the previous section
(\cref{lst:count:Simple Limit Counter Variables,%
lst:count:Simple Limit Counter Add; Subtract; and Read,%
lst:count:Simple Limit Counter Utility Functions}),
only the changes are shown here.
\Cref{lst:count:Approximate Limit Counter Variables}
is identical to
\cref{lst:count:Simple Limit Counter Variables},
with the addition of \co{MAX_COUNTERMAX}, which sets the maximum
permissible value of the per-thread \co{countermax} variable.

\begin{listing}
\input{CodeSamples/count/count_lim_app=variable.fcv}
\caption{Approximate Limit Counter Variables}
\label{lst:count:Approximate Limit Counter Variables}
\end{listing}

\begin{listing}
\input{CodeSamples/count/count_lim_app=balance.fcv}
\caption{Approximate Limit Counter Balancing}
\label{lst:count:Approximate Limit Counter Balancing}
\end{listing}

\begin{fcvref}[ln:count:count_lim_app:balance]
Similarly,
\cref{lst:count:Approximate Limit Counter Balancing}
is identical to the \co{balance_count()} function in
\cref{lst:count:Simple Limit Counter Utility Functions},
with the addition of
\clnref{enforce:b,enforce:e}, which enforce the
\co{MAX_COUNTERMAX} limit on the per-thread \co{countermax} variable.
\end{fcvref}

\subsection{Approximate Limit Counter Discussion}

These changes greatly reduce the limit inaccuracy seen in the previous version,
but present another problem:
Any given value of \co{MAX_COUNTERMAX} will cause a workload-dependent
fraction of accesses to fall off the fastpath.
As the number of threads increase, non-fastpath execution will become both
a performance and a scalability problem.
However, we will defer this problem and turn instead to counters
with exact limits.

\section{Exact Limit Counters}
\label{sec:count:Exact Limit Counters}
\epigraph{Exactitude can be expensive.
	  Spend wisely.}{Unknown}

To solve the exact structure-allocation limit problem noted in
\QuickQuizRef{\QcountQexactcnt},
we need a limit counter that can tell exactly when its limits are
exceeded.
One way of implementing such a limit counter is to
cause threads that have reserved counts to give them up.
One way to do this is to use \IX{atomic} instructions.
Of course, atomic instructions will slow down the fastpath, but on the
other hand, it would be silly not to at least give them a try.

\subsection{Atomic Limit Counter Implementation}
\label{sec:count:Atomic Limit Counter Implementation}

Unfortunately,
if one thread is to safely remove counts from another thread,
both threads will need to atomically manipulate that thread's
\co{counter} and \co{countermax} variables.
The usual way to do this is to combine these two variables into a
single variable,
for example, given a 32-bit variable, using the high-order 16 bits to
represent \co{counter} and the low-order 16 bits to represent
\co{countermax}.

\QuickQuiz{
	Why is it necessary to atomically manipulate the thread's
	\co{counter} and \co{countermax} variables as a unit?
	Wouldn't it be good enough to atomically manipulate them
	individually?
}\QuickQuizAnswer{
	This might well be possible, but great care is required.
	Note that removing \co{counter} without first zeroing
	\co{countermax} could result in the corresponding thread
	increasing \co{counter} immediately after it was zeroed,
	completely negating the effect of zeroing the counter.

	The opposite ordering, namely zeroing \co{countermax} and then
	removing \co{counter}, can also result in a non-zero
	\co{counter}.
	To see this, consider the following sequence of events:

	\begin{enumerate}
	\item	Thread~A fetches its \co{countermax}, and finds that
		it is non-zero.
	\item	Thread~B zeroes Thread~A's \co{countermax}.
	\item	Thread~B removes Thread~A's \co{counter}.
	\item	Thread~A, having found that its \co{countermax}
		is non-zero, proceeds to add to its \co{counter},
		resulting in a non-zero value for \co{counter}.
	\end{enumerate}

	Again, it might well be possible to atomically manipulate
	\co{countermax} and \co{counter} as separate variables,
	but it is clear that great care is required.
	It is also quite likely that doing so will slow down the
	fastpath.

	Exploring these possibilities are left as exercises for
	the reader.
}\QuickQuizEnd

\begin{listing}
\input{CodeSamples/count/count_lim_atomic=var_access.fcv}
\caption{Atomic Limit Counter Variables and Access Functions}
\label{lst:count:Atomic Limit Counter Variables and Access Functions}
\end{listing}

\begin{fcvref}[ln:count:count_lim_atomic:var_access:var]
The variables and access functions for a simple atomic limit counter
are shown in
\cref{lst:count:Atomic Limit Counter Variables and Access Functions}
(\path{count_lim_atomic.c}).
The \co{counter} and \co{countermax} variables in earlier algorithms
are combined into the single variable \co{counterandmax} shown on
\clnref{candmax}, with \co{counter} in the upper half and \co{countermax} in
the lower half.
This variable is of type \apik{atomic_t}, which has an underlying
representation of \co{int}.

\Clnrefrange{def:b}{def:e} show the definitions for \co{globalcountmax}, \co{globalcount},
\co{globalreserve}, \co{counterp}, and \co{gblcnt_mutex}, all of which
take on roles similar to their counterparts in
\cref{lst:count:Approximate Limit Counter Variables}.
\Clnref{CM_BITS} defines \co{CM_BITS}, which gives the number of bits in each half
of \co{counterandmax}, and \clnref{MAX_CMAX} defines \co{MAX_COUNTERMAX}, which
gives the maximum value that may be held in either half of
\co{counterandmax}.
\end{fcvref}

\QuickQuiz{
	In what way does
	\clnrefr{ln:count:count_lim_atomic:var_access:var:CM_BITS} of
	\cref{lst:count:Atomic Limit Counter Variables and Access Functions}
	violate the C standard?
}\QuickQuizAnswer{
	It assumes eight bits per byte.
	This assumption does hold for all current commodity microprocessors
	that can be easily assembled into shared-memory multiprocessors,
	but certainly does not hold for all computer systems that have
	ever run C code.
	(What could you do instead in order to comply with the C standard?
	What drawbacks would it have?)
}\QuickQuizEnd

\begin{fcvref}[ln:count:count_lim_atomic:var_access:split_int]
\Clnrefrange{b}{e} show the \co{split_counterandmax_int()}
function, which,
when given the underlying \co{int} from the
\co{atomic_t counterandmax} variable, splits it into its
\co{counter} (\co{c})
and \co{countermax} (\co{cm}) components.
\Clnref{msh} isolates the most-significant half of this \co{int},
placing the result as specified by argument \co{c},
and \clnref{lsh} isolates the least-significant half of this \co{int},
placing the result as specified by argument \co{cm}.
\end{fcvref}

\begin{fcvref}[ln:count:count_lim_atomic:var_access:split]
\Clnrefrange{b}{e} show the \co{split_counterandmax()} function, which
picks up the underlying \co{int} from the specified variable
on \clnref{int}, stores it as specified by the \co{old} argument on
\clnref{old}, and then invokes \co{split_counterandmax_int()} to split
it on \clnref{split_int}.
\end{fcvref}

\QuickQuiz{
	Given that there is only one \co{counterandmax} variable,
	why bother passing in a pointer to it on
	\clnrefr{ln:count:count_lim_atomic:var_access:split:func} of
	\cref{lst:count:Atomic Limit Counter Variables and Access Functions}?
}\QuickQuizAnswer{
	There is only one \co{counterandmax} variable \emph{per thread}.
	Later, we will see code that needs to pass other threads'
	\co{counterandmax} variables to \co{split_counterandmax()}.
}\QuickQuizEnd

\begin{fcvref}[ln:count:count_lim_atomic:var_access:merge]
\Clnrefrange{b}{e} show the \co{merge_counterandmax()} function, which
can be thought of as the inverse of \co{split_counterandmax()}.
\Clnref{merge} merges the \co{counter} and \co{countermax}
values passed in \co{c} and \co{cm}, respectively, and returns
the result.
\end{fcvref}

\QuickQuiz{
	Why does \co{merge_counterandmax()} in
	\cref{lst:count:Atomic Limit Counter Variables and Access Functions}
	return an \co{int} rather than storing directly into an
	\co{atomic_t}?
}\QuickQuizAnswer{
	Later, we will see that we need the \co{int} return to pass
	to the \co{atomic_cmpxchg()} primitive.
}\QuickQuizEnd

\begin{listing}
\ebresizeverb{.81}{\input{CodeSamples/count/count_lim_atomic=add_sub.fcv}}
\caption{Atomic Limit Counter Add and Subtract}
\label{lst:count:Atomic Limit Counter Add and Subtract}
\end{listing}

\Cref{lst:count:Atomic Limit Counter Add and Subtract}
shows the \co{add_count()} and \co{sub_count()} functions.

\begin{fcvref}[ln:count:count_lim_atomic:add_sub:add]
\Clnrefrange{b}{e} show \co{add_count()}, whose fastpath spans
\clnrefrange{fast:b}{return:fs},
with the remainder of the function being the slowpath.
\Clnrefrange{fast:b}{loop:e} of the fastpath form a compare-and-swap
(CAS) loop, with
the \apik{atomic_cmpxchg()} primitive on
\clnrefrange{atmcmpex}{loop:e} performing the
actual CAS\@.
\Clnref{split} splits the current thread's \co{counterandmax} variable into its
\co{counter} (in \co{c}) and \co{countermax} (in \co{cm}) components,
while placing the underlying \co{int} into \co{old}.
\Clnref{check} checks whether the amount \co{delta} can be accommodated
locally (taking care to avoid integer overflow), and if not,
\clnref{goto} transfers to the slowpath.
Otherwise, \clnref{merge} combines an updated \co{counter} value with the
original \co{countermax} value into \co{new}.
The \co{atomic_cmpxchg()} primitive on
\clnrefrange{atmcmpex}{loop:e} then atomically
compares this thread's \co{counterandmax} variable to \co{old},
updating its value to \co{new} if the comparison succeeds.
If the comparison succeeds, \clnref{return:fs} returns success, otherwise,
execution continues in the loop at \clnref{fast:b}.
\end{fcvref}

\QuickQuizSeries{%
\QuickQuizB{
	Yecch!
	Why the ugly \co{goto} on
	\clnrefr{ln:count:count_lim_atomic:add_sub:add:goto} of
	\cref{lst:count:Atomic Limit Counter Add and Subtract}?
	Haven't you heard of the \co{break} statement???
}\QuickQuizAnswerB{
	Replacing the \co{goto} with a \co{break} would require keeping
	a flag to determine whether or not
	\clnrefr{ln:count:count_lim_atomic:add_sub:add:return:fs}
	should return, which
	is not the sort of thing you want on a fastpath.
	If you really hate the \co{goto} that much, your best bet would
	be to pull the fastpath into a separate function that returned
	success or failure, with ``failure'' indicating a need for the
	slowpath.
	This is left as an exercise for goto-hating readers.
}\QuickQuizEndB
\QuickQuizE{
	\begin{fcvref}[ln:count:count_lim_atomic:add_sub:add]
	Why would the \co{atomic_cmpxchg()} primitive at
	\clnrefrange{atmcmpex}{loop:e} of
	\cref{lst:count:Atomic Limit Counter Add and Subtract}
	ever fail?
	After all, we picked up its old value on \clnref{split} and have not
	changed it!
	\end{fcvref}
}\QuickQuizAnswerE{
	\begin{fcvref}[ln:count:count_lim_atomic:add_sub:add]
	Later, we will see how the \co{flush_local_count()} function in
	\cref{lst:count:Atomic Limit Counter Utility Functions 1}
	might update this thread's \co{counterandmax} variable concurrently
	with the execution of the fastpath on
	\clnrefrange{fast:b}{loop:e} of
	\cref{lst:count:Atomic Limit Counter Add and Subtract}.
	\end{fcvref}
}\QuickQuizEndE
}

\begin{fcvref}[ln:count:count_lim_atomic:add_sub:add]
\Clnrefrange{slow:b}{return:ss} of
\cref{lst:count:Atomic Limit Counter Add and Subtract}
show \co{add_count()}'s slowpath, which is protected by \co{gblcnt_mutex},
which is acquired on \clnref{acquire} and released on
\clnref{release:f,release:s}.
\Clnref{globalize} invokes \co{globalize_count()},
which moves this thread's
state to the global counters.
\Clnrefrange{checkglb:b}{checkglb:e} check whether
the \co{delta} value can be accommodated by
the current global state, and, if not, \clnref{flush} invokes
\co{flush_local_count()} to flush all threads' local state to the
global counters, and then
\clnrefrange{checkglb:nb}{checkglb:ne} recheck whether \co{delta} can
be accommodated.
If, after all that, the addition of \co{delta} still cannot be accommodated,
then \clnref{release:f} releases \co{gblcnt_mutex} (as noted earlier), and
then \clnref{return:sf} returns failure.

Otherwise, \clnref{addglb} adds \co{delta} to the global counter,
\clnref{balance}
spreads counts to the local state if appropriate, \clnref{release:s} releases
\co{gblcnt_mutex} (again, as noted earlier), and finally,
\clnref{return:ss}
returns success.
\end{fcvref}

\begin{fcvref}[ln:count:count_lim_atomic:add_sub:sub]
\Clnrefrange{b}{e} of
\cref{lst:count:Atomic Limit Counter Add and Subtract}
show \co{sub_count()}, which is structured similarly to
\co{add_count()}, having a fastpath on
\clnrefrange{fast:b}{fast:e} and a slowpath on
\clnrefrange{slow:b}{slow:e}.
A line-by-line analysis of this function is left as an exercise to
the reader.
\end{fcvref}

\begin{listing}
\input{CodeSamples/count/count_lim_atomic=read.fcv}
\caption{Atomic Limit Counter Read}
\label{lst:count:Atomic Limit Counter Read}
\end{listing}

\begin{fcvref}[ln:count:count_lim_atomic:read]
\Cref{lst:count:Atomic Limit Counter Read} shows \co{read_count()}.
\Clnref{acquire} acquires \co{gblcnt_mutex} and
\clnref{release} releases it.
\Clnref{initsum} initializes local variable \co{sum} to the value of
\co{globalcount}, and the loop spanning
\clnrefrange{loop:b}{loop:e} adds the
per-thread counters to this sum, isolating each per-thread counter
using \co{split_counterandmax} on \clnref{split}.
Finally, \clnref{return} returns the sum.
\end{fcvref}

\begin{listing}
\input{CodeSamples/count/count_lim_atomic=utility1.fcv}
\caption{Atomic Limit Counter Utility Functions 1}
\label{lst:count:Atomic Limit Counter Utility Functions 1}
\end{listing}

\begin{listing}
\input{CodeSamples/count/count_lim_atomic=utility2.fcv}
\caption{Atomic Limit Counter Utility Functions 2}
\label{lst:count:Atomic Limit Counter Utility Functions 2}
\end{listing}

\Cref{lst:count:Atomic Limit Counter Utility Functions 1,%
lst:count:Atomic Limit Counter Utility Functions 2}
show the utility functions
\co{globalize_count()},
\co{flush_local_count()},
\co{balance_count()},
\co{count_register_thread()}, and
\co{count_unregister_thread()}.
\begin{fcvref}[ln:count:count_lim_atomic:utility1:globalize]
The code for \co{globalize_count()} is shown on
\clnrefrange{b}{e}
of \cref{lst:count:Atomic Limit Counter Utility Functions 1}, and
is similar to that of previous algorithms, with the addition of
\clnref{split}, which is now required to split out \co{counter} and
\co{countermax} from \co{counterandmax}.
\end{fcvref}

\begin{fcvref}[ln:count:count_lim_atomic:utility1:flush]
The code for \co{flush_local_count()}, which moves all threads' local
counter state to the global counter, is shown on
\clnrefrange{b}{e}.
\Clnref{checkrsv} checks to see if the value of
\co{globalreserve} permits
any per-thread counts, and, if not, \clnref{return:n} returns.
Otherwise, \clnref{initzero} initializes local variable \co{zero} to a combined
zeroed \co{counter} and \co{countermax}.
The loop spanning \clnrefrange{loop:b}{loop:e} sequences
through each thread.
\Clnref{checkp} checks to see if the current thread has counter state,
and, if so, \clnrefrange{atmxchg}{glbrsv} move that state
to the global counters.
\Clnref{atmxchg} atomically fetches the current thread's state
while replacing it with zero.
\Clnref{split} splits this state into its \co{counter}
(in local variable \co{c})
and \co{countermax} (in local variable \co{cm}) components.
\Clnref{glbcnt} adds this thread's \co{counter} to \co{globalcount}, while
\clnref{glbrsv} subtracts this thread's \co{countermax} from \co{globalreserve}.
\end{fcvref}

\QuickQuizSeries{%
\QuickQuizB{
	What stops a thread from simply refilling its
	\co{counterandmax} variable immediately after
	\co{flush_local_count()} on
	\clnrefr{ln:count:count_lim_atomic:utility1:flush:b} of
	\cref{lst:count:Atomic Limit Counter Utility Functions 1}
	empties it?
}\QuickQuizAnswerB{
	This other thread cannot refill its \co{counterandmax}
	until the caller of \co{flush_local_count()} releases the
	\co{gblcnt_mutex}.
	By that time, the caller of \co{flush_local_count()} will have
	finished making use of the counts, so there will be no problem
	with this other thread refilling---assuming that the value
	of \co{globalcount} is large enough to permit a refill.
}\QuickQuizEndB
\QuickQuizE{
	What prevents concurrent execution of the fastpath of either
	\co{add_count()} or \co{sub_count()} from interfering with
	the \co{counterandmax} variable while
	\co{flush_local_count()} is accessing it on
	\clnrefr{ln:count:count_lim_atomic:utility1:flush:atmxchg} of
	\cref{lst:count:Atomic Limit Counter Utility Functions 1}?
}\QuickQuizAnswerE{
	Nothing.
	Consider the following three cases:
	\begin{enumerate}
	\item	If \co{flush_local_count()}'s \co{atomic_xchg()} executes
		before the \co{split_counterandmax()} of either fastpath,
		then the fastpath will see a zero \co{counter} and
		\co{countermax}, and will thus transfer to the slowpath
		(unless of course \co{delta} is zero).
	\item	If \co{flush_local_count()}'s \co{atomic_xchg()} executes
		after the \co{split_counterandmax()} of either fastpath,
		but before that fastpath's \co{atomic_cmpxchg()},
		then the \co{atomic_cmpxchg()} will fail, causing the
		fastpath to restart, which reduces to case~1 above.
	\item	If \co{flush_local_count()}'s \co{atomic_xchg()} executes
		after the \co{atomic_cmpxchg()} of either fastpath,
		then the fastpath will (most likely) complete successfully
		before \co{flush_local_count()} zeroes the thread's
		\co{counterandmax} variable.
	\end{enumerate}
	Either way, the race is resolved correctly.
}\QuickQuizEndE
}

\begin{fcvref}[ln:count:count_lim_atomic:utility2]
\Clnrefrange{balance:b}{balance:e} on
\cref{lst:count:Atomic Limit Counter Utility Functions 2}
show the code for \co{balance_count()}, which refills
the calling thread's local \co{counterandmax} variable.
This function is quite similar to that of the preceding algorithms,
with changes required to handle the merged \co{counterandmax} variable.
Detailed analysis of the code is left as an exercise for the reader,
as it is with the \co{count_register_thread()} function starting on
\clnref{register:b} and the \co{count_unregister_thread()} function starting on
\clnref{unregister:b}.
\end{fcvref}

\QuickQuiz{
	Given that the \co{atomic_set()} primitive does a simple
	store to the specified \co{atomic_t}, how can
	\clnrefr{ln:count:count_lim_atomic:utility2:balance:atmcset} of
	\co{balance_count()} in
	\cref{lst:count:Atomic Limit Counter Utility Functions 2}
	work correctly in face of concurrent \co{flush_local_count()}
	updates to this variable?
}\QuickQuizAnswer{
	The caller of both \co{balance_count()} and
	\co{flush_local_count()} hold \co{gblcnt_mutex}, so
	only one may be executing at a given time.
}\QuickQuizEnd

The next section qualitatively evaluates this design.
\ebFloatBarrier

\subsection{Atomic Limit Counter Discussion}

This is the first implementation that actually allows the counter to
be run all the way to either of its limits, but it does so at the
expense of adding atomic operations to the fastpaths, which slow down
the fastpaths significantly on some systems.
Although some workloads might tolerate this slowdown, it is worthwhile
looking for algorithms with better write-side performance.
One such algorithm uses a signal handler to steal counts from other
threads.
Because signal handlers run in the context of the signaled thread,
atomic operations are not necessary, as shown in the next section.

\QuickQuiz{
	But signal handlers can be migrated to some other
	CPU while running.
	Doesn't this possibility require that atomic instructions
	and memory barriers are required to reliably communicate
	between a thread and a signal handler that interrupts that
	thread?
}\QuickQuizAnswer{
	No.
	If the signal handler is migrated to another CPU, then the
	interrupted thread is also migrated along with it.
}\QuickQuizEnd

\subsection{Signal-Theft Limit Counter Design}
\label{sec:count:Signal-Theft Limit Counter Design}

\begin{figure}
\centering
\resizebox{2in}{!}{\includegraphics{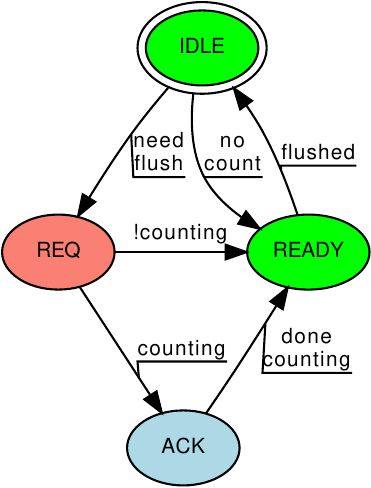}}
\caption{Signal-Theft State Machine}
\label{fig:count:Signal-Theft State Machine}
\end{figure}

Even though per-thread state will now be manipulated only by the
corresponding thread, there will still need to be synchronization
with the signal handlers.
This synchronization is provided by the state machine shown in
\cref{fig:count:Signal-Theft State Machine}.

The state machine starts out in the IDLE state, and when \co{add_count()}
or \co{sub_count()} find that the combination of the local thread's count
and the global count cannot accommodate the request, the corresponding
slowpath sets each thread's \co{theft} state to REQ (unless that thread
has no count, in which case it transitions directly to READY)\@.
Only the slowpath, which holds the \co{gblcnt_mutex} lock, is permitted to
transition from the IDLE state, as indicated by the green color.\footnote{
	For those with black-and-white versions of this book,
	IDLE and READY are green, REQ is red, and ACK is blue.}
The slowpath then sends a signal to each thread, and the corresponding
signal handler checks the corresponding thread's \co{theft} and
\co{counting} variables.
If the \co{theft} state is not REQ, then the signal handler is not
permitted to change the state, and therefore simply returns.
Otherwise, if the \co{counting} variable is set, indicating that
the current thread's fastpath is in progress, the signal handler
sets the \co{theft} state to ACK, otherwise to READY\@.

If the \co{theft} state is ACK,
only the fastpath is permitted to change
the \co{theft} state, as indicated by the blue color.
When the fastpath completes, it sets the \co{theft} state to READY\@.

Once the slowpath sees a thread's \co{theft} state is READY, the
slowpath is permitted to steal that thread's count.
The slowpath then sets that thread's \co{theft} state to IDLE\@.

\QuickQuizSeries{%
\QuickQuizB{
	In \cref{fig:count:Signal-Theft State Machine}, why is
	the REQ \co{theft} state colored red?
}\QuickQuizAnswerB{
	To indicate that only the fastpath is permitted to change the
	\co{theft} state, and that if the thread remains in this
	state for too long, the thread running the slowpath will
	resend the POSIX signal.
}\QuickQuizEndB
\QuickQuizE{
	In \cref{fig:count:Signal-Theft State Machine}, what is
	the point of having separate REQ and ACK \co{theft} states?
	Why not simplify the state machine by collapsing
	them into a single REQACK state?
	Then whichever of the signal handler or the fastpath gets there
	first could set the state to READY\@.
}\QuickQuizAnswerE{
	Reasons why collapsing the REQ and ACK states would be a very
	bad idea include:
	\begin{enumerate}
	\item	The slowpath uses the REQ and ACK states to determine
		whether the signal should be retransmitted.
		If the states were collapsed, the slowpath would have
		no choice but to send redundant signals, which would
		have the unhelpful effect of needlessly slowing down
		the fastpath.
	\item	The following race would result:
		\begin{enumerate}
		\item	The slowpath sets a given thread's state to REQACK.
		\item	That thread has just finished its fastpath, and
			notes the REQACK state.
		\item	The thread receives the signal, which also notes
			the REQACK state, and, because there is no fastpath
			in effect, sets the state to READY\@.
		\item	The slowpath notes the READY state, steals the
			count, and sets the state to IDLE, and completes.
		\item	The fastpath sets the state to READY, disabling
			further fastpath execution for this thread.
		\end{enumerate}
		The basic problem here is that the combined REQACK state
		can be referenced by both the signal handler and the
		fastpath.
		The clear separation maintained by the four-state
		setup ensures orderly state transitions.
	\end{enumerate}
	That said, you might well be able to make a three-state setup
	work correctly.
	If you do succeed, compare carefully to the four-state setup.
	Is the three-state solution really preferable, and why or why not?
}\QuickQuizEndE
}

\subsection{Signal-Theft Limit Counter Implementation}
\label{sec:count:Signal-Theft Limit Counter Implementation}

\begin{fcvref}[ln:count:count_lim_sig:data]
\Cref{lst:count:Signal-Theft Limit Counter Data}
(\path{count_lim_sig.c})
shows the data structures used by the signal-theft based counter
implementation.
\Clnrefrange{value:b}{value:e} define the states and values
for the per-thread theft state machine
described in the preceding section.
\Clnrefrange{var:b}{var:e} are similar to earlier implementations,
with the addition of
\clnref{maxp,theftp} to allow remote access to a
thread's \co{countermax}
and \co{theft} variables, respectively.
\end{fcvref}

\begin{listing}
\ebresizeverb{.9}{\input{CodeSamples/count/count_lim_sig=data.fcv}}
\caption{Signal-Theft Limit Counter Data}
\label{lst:count:Signal-Theft Limit Counter Data}
\end{listing}

\begin{fcvref}[ln:count:count_lim_sig:migration:globalize]
\Cref{lst:count:Signal-Theft Limit Counter Value-Migration Functions}
shows the functions responsible for migrating counts between per-thread
variables and the global variables.
\Clnrefrange{b}{e} show \co{globalize_count()},
which is identical to earlier
implementations.
\end{fcvref}
\begin{fcvref}[ln:count:count_lim_sig:migration:flush_sig]
\Clnrefrange{b}{e} show \co{flush_local_count_sig()},
which is the signal
handler used in the theft process.
\Clnref{check:REQ,return:n} check to see if
the \co{theft} state is REQ, and, if not
returns without change.
\Clnref{set:ACK} sets the \co{theft} state to ACK, and, if
\clnref{check:fast} sees that
this thread's fastpaths are not running, \clnref{set:READY} uses
\co{smp_store_release()} to set the \co{theft}
state to READY, further ensuring that any change to \co{counter} in
the fastpath happens before this change of \co{theft} to READY\@.
\end{fcvref}

\begin{listing}
\ebresizeverb{.839}{\input{CodeSamples/count/count_lim_sig=migration.fcv}}
\caption{Signal-Theft Limit Counter Value-Migration Functions}
\label{lst:count:Signal-Theft Limit Counter Value-Migration Functions}
\end{listing}

\QuickQuiz{
	In \cref{lst:count:Signal-Theft Limit Counter Value-Migration Functions},
	doesn't \co{flush_local_count_sig()} need stronger memory barriers?
}\QuickQuizAnswer{
	No, that \co{smp_store_release()} suffices because this code
	communicates only with \co{flush_local_count()}, and there is
	no need for store-to-load ordering.
}\QuickQuizEnd

\begin{fcvref}[ln:count:count_lim_sig:migration:flush]
\Clnrefrange{b}{e} show \co{flush_local_count()}, which is called from the
slowpath to flush all threads' local counts.
The loop spanning
\clnrefrange{loop:b}{loop:e} advances the \co{theft} state for each
thread that has local count, and also sends that thread a signal.
\Clnref{skip} skips any non-existent threads.
Otherwise, \clnref{checkmax} checks to see if the current thread holds any local
count, and, if not, \clnref{READY} sets the thread's \co{theft} state to READY
and \clnref{next} skips to the next thread.
Otherwise, \clnref{REQ} sets the thread's \co{theft} state to REQ and
\clnref{signal} sends the thread a signal.
\end{fcvref}

\QuickQuizSeries{%
\QuickQuizB{
	In \cref{lst:count:Signal-Theft Limit Counter Value-Migration Functions},
	why is it safe for
	\clnrefr{ln:count:count_lim_sig:migration:flush:checkmax}
	to directly access the other thread's
	\co{countermax} variable?
}\QuickQuizAnswerB{
	Because the other thread is not permitted to change the value
	of its \co{countermax} variable unless it holds the
	\co{gblcnt_mutex} lock.
	But the caller has acquired this lock, so it is not possible
	for the other thread to hold it, and therefore the other thread
	is not permitted to change its \co{countermax} variable.
	We can therefore safely access it---but not change it.
}\QuickQuizEndB
\QuickQuizM{
	In \cref{lst:count:Signal-Theft Limit Counter Value-Migration Functions},
	why doesn't
	\clnrefr{ln:count:count_lim_sig:migration:flush:signal}
	check for the current thread sending itself
	a signal?
}\QuickQuizAnswerM{
	There is no need for an additional check.
	The caller of \co{flush_local_count()} has already invoked
	\co{globalize_count()}, so the check on
	\clnrefr{ln:count:count_lim_sig:migration:flush:checkmax}
	will have succeeded, skipping the later \co{pthread_kill()}.
}\QuickQuizEndM
\QuickQuizE{
	The code shown in
	\cref{lst:count:Signal-Theft Limit Counter Data,%
	lst:count:Signal-Theft Limit Counter Value-Migration Functions}
	works with \GCC\ and POSIX\@.
	What would be required to make it also conform to the ISO C standard?
}\QuickQuizAnswerE{
	The \co{theft} variable must be of type \co{sig_atomic_t}
	to guarantee that it can be safely shared between the signal
	handler and the code interrupted by the signal.
}\QuickQuizEndE
}

\begin{fcvref}[ln:count:count_lim_sig:migration:flush]
The loop spanning \clnrefrange{loop2:b}{loop2:e} waits until each
thread reaches READY state,
then steals that thread's count.
\Clnrefrange{skip:nonexist}{next2} skip any non-existent threads,
and the loop spanning
\clnrefrange{loop3:b}{loop3:e} waits until the current
thread's \co{theft} state becomes READY\@.
\Clnref{block} blocks for a millisecond to avoid priority-inversion problems,
and if \clnref{check:REQ} determines that the thread's signal has not yet arrived,
\clnref{signal2} resends the signal.
Execution reaches \clnref{thiev:b} when the thread's \co{theft} state becomes
READY, so \clnrefrange{thiev:b}{thiev:e} do the thieving.
\Clnref{IDLE} then sets the thread's \co{theft} state back to IDLE\@.
\end{fcvref}

\QuickQuiz{
	In \cref{lst:count:Signal-Theft Limit Counter Value-Migration Functions},
	why does \clnrefr{ln:count:count_lim_sig:migration:flush:signal2}
	resend the signal?
}\QuickQuizAnswer{
	Because many operating systems over several decades have
	had the property of losing the occasional signal.
	Whether this is a feature or a bug is debatable, but
	irrelevant.
	The obvious symptom from the user's viewpoint will not be
	a kernel bug, but rather a user application hanging.

	\emph{Your} user application hanging!
}\QuickQuizEnd

\begin{fcvref}[ln:count:count_lim_sig:migration:balance]
\Clnrefrange{b}{e} show \co{balance_count()}, which is similar to that of
earlier examples.
\end{fcvref}

\begin{listing}
\input{CodeSamples/count/count_lim_sig=add.fcv}
\caption{Signal-Theft Limit Counter Add Function}
\label{lst:count:Signal-Theft Limit Counter Add Function}
\end{listing}

\begin{listing}
\input{CodeSamples/count/count_lim_sig=sub.fcv}
\caption{Signal-Theft Limit Counter Subtract Function}
\label{lst:count:Signal-Theft Limit Counter Subtract Function}
\end{listing}

\begin{fcvref}[ln:count:count_lim_sig:add]
\Cref{lst:count:Signal-Theft Limit Counter Add Function}
shows the \co{add_count()} function.
The fastpath spans \clnrefrange{fast:b}{return:fs}, and the slowpath
\clnrefrange{acquire}{return:ss}.
\Clnref{fast:b} sets the per-thread \co{counting} variable to 1 so that
any subsequent signal handlers interrupting this thread will
set the \co{theft} state to ACK rather than READY, allowing this
fastpath to complete properly.
\Clnref{barrier:1} prevents the compiler from reordering any of the fastpath body
to precede the setting of \co{counting}.
\Clnref{check:b,check:e} check to see
if the per-thread data can accommodate
the \co{add_count()} and if there is no ongoing theft in progress,
and if so \clnref{add:f} does the fastpath addition and
\clnref{fasttaken} notes that
the fastpath was taken.

In either case, \clnref{barrier:2} prevents the compiler from reordering the
fastpath body to follow \clnref{clearcnt}, which permits any subsequent signal
handlers to undertake theft.
\Clnref{barrier:3} again disables compiler reordering, and then
\clnref{check:ACK}
checks to see if the signal handler deferred the \co{theft}
state-change to READY, and, if so, \clnref{READY} uses
\co{smp_store_release()} to set the \co{theft} state to
READY, further ensuring that
any CPU that sees the READY state also sees the effects
of \clnref{add:f}.
If the fastpath addition at \clnref{add:f} was executed, then
\clnref{return:fs} returns
success.
\end{fcvref}

\begin{listing}
\input{CodeSamples/count/count_lim_sig=read.fcv}
\caption{Signal-Theft Limit Counter Read Function}
\label{lst:count:Signal-Theft Limit Counter Read Function}
\end{listing}

\begin{fcvref}[ln:count:count_lim_sig:add]
Otherwise, we fall through to the slowpath starting at \clnref{acquire}.
The structure of the slowpath is similar to those of earlier examples,
so its analysis is left as an exercise to the reader.
\end{fcvref}
Similarly, the structure of \co{sub_count()} on
\cref{lst:count:Signal-Theft Limit Counter Subtract Function}
is the same
as that of \co{add_count()}, so the analysis of \co{sub_count()} is also
left as an exercise for the reader, as is the analysis of
\co{read_count()} in
\cref{lst:count:Signal-Theft Limit Counter Read Function}.

\begin{listing}
\input{CodeSamples/count/count_lim_sig=initialization.fcv}
\caption{Signal-Theft Limit Counter Initialization Functions}
\label{lst:count:Signal-Theft Limit Counter Initialization Functions}
\end{listing}

\begin{fcvref}[ln:count:count_lim_sig:initialization:init]
\Clnrefrange{b}{e} of
\cref{lst:count:Signal-Theft Limit Counter Initialization Functions}
show \co{count_init()}, which set up \co{flush_local_count_sig()}
as the signal handler for \co{SIGUSR1},
enabling the \apipx{pthread_kill()} calls in \co{flush_local_count()}
to invoke \co{flush_local_count_sig()}.
The code for thread registry and unregistry is similar to that of
earlier examples, so its analysis is left as an exercise for the
reader.
\end{fcvref}

\subsection{Signal-Theft Limit Counter Discussion}

The signal-theft implementation runs more than eight times as fast as the
atomic implementation on my six-core x86 laptop.
Is it always preferable?

The signal-theft implementation would be vastly preferable on Pentium-4
systems, given their slow atomic instructions, but the old 80386-based
Sequent Symmetry systems would do much better with the shorter path
length of the atomic implementation.
However, this increased update-side performance comes at the
prices of higher read-side overhead:
Those POSIX signals are not free.
If ultimate performance is of the essence, you will need to measure
them both on the system that your application is to be deployed on.

\QuickQuiz{
	Not only are POSIX signals slow, sending one to each thread
	simply does not scale.
	What would you do if you had (say) 10,000 threads and needed
	the read side to be fast?
}\QuickQuizAnswer{
	One approach is to use the techniques shown in
	\cref{sec:count:Eventually Consistent Implementation},
	summarizing an approximation to the overall counter value in
	a single variable.
	Another approach would be to use multiple threads to carry
	out the reads, with each such thread interacting with a
	specific subset of the updating threads.
}\QuickQuizEnd

This is but one reason why high-quality APIs are so important:
They permit implementations to be changed as required by ever-changing
hardware performance characteristics.

\QuickQuiz{
	What if you want an exact limit counter to be exact only for
	its lower limit, but to allow the upper limit to be inexact?
}\QuickQuizAnswer{
	One simple solution is to overstate the upper limit by the
	desired amount.
	The limiting case of such overstatement results in the
	upper limit being set to the largest value that the counter is
	capable of representing.
}\QuickQuizEnd

\subsection{Applying Exact Limit Counters}
\label{sec:count:Applying Exact Limit Counters}

Although the exact limit counter implementations presented in this
section can be very useful, they are not much help if the counter's value
remains near zero at all times, as it might when counting the number
of outstanding accesses to an I/O device.
The high overhead of such near-zero counting is especially painful
given that we normally don't care how many references there are.
As noted in the removable I/O device access-count problem posed by
\QuickQuizRef{\QcountQIOcnt},
the number of accesses is irrelevant except in those rare cases when
someone is actually trying to remove the device.

One simple solution to this problem is to add a large ``bias''
(for example, one billion) to the
counter in order to ensure that the value is far enough from zero that
the counter can operate efficiently.
When someone wants to remove the device, this bias is subtracted from
the counter value.
Counting the last few accesses will be quite inefficient,
but the important point is that the many prior accesses will have been
counted at full speed.

\QuickQuiz{
	What else had you better have done when using a biased counter?
}\QuickQuizAnswer{
	You had better have set the upper limit to be large enough
	accommodate the bias, the expected maximum number of accesses,
	and enough ``slop'' to allow the counter to work efficiently
	even when the number of accesses is at its maximum.
}\QuickQuizEnd

Although a biased counter can be quite helpful and useful, it is only a
partial solution to the removable I/O device access-count problem
called out on
\cpageref{chp:Counting}.
When attempting to remove a device, we must not only know the precise
number of current I/O accesses, we also need to prevent any future
accesses from starting.
One way to accomplish this is to read-acquire a reader-writer lock
when updating the counter, and to write-acquire that same reader-writer
lock when checking the counter.
Code for doing I/O might be as follows:

\begin{fcvlabel}[ln:count:inline:I/O]
\begin{VerbatimN}[commandchars=\\\[\]]
read_lock(&mylock);		\lnlbl[acq]
if (removing) {			\lnlbl[check]
	read_unlock(&mylock);	\lnlbl[rel1]
	cancel_io();		\lnlbl[cancel]
} else {
	add_count(1);		\lnlbl[inc]
	read_unlock(&mylock);	\lnlbl[rel2]
	do_io();		\lnlbl[do]
	sub_count(1);		\lnlbl[dec]
}
\end{VerbatimN}
\end{fcvlabel}

\begin{fcvref}[ln:count:inline:I/O]
\Clnref{acq} read-acquires the lock, and either
\clnref{rel1} or~\lnref{rel2} releases it.
\Clnref{check} checks to see if the device is being removed, and, if so,
\clnref{rel1} releases the lock and
\clnref{cancel} cancels the I/O, or takes whatever
action is appropriate given that the device is to be removed.
Otherwise, \clnref{inc} increments the access count,
\clnref{rel2} releases the
lock, \clnref{do} performs the I/O, and
\clnref{dec} decrements the access count.
\end{fcvref}

\QuickQuiz{
	This is ridiculous!
	We are \emph{read}-acquiring a reader-writer lock to
	\emph{update} the counter?
	What are you playing at???
}\QuickQuizAnswer{
	Strange, perhaps, but true!
	Almost enough to make you think that the name
	``reader-writer lock'' was poorly chosen, isn't it?
}\QuickQuizEnd

The code to remove the device might be as follows:

\begin{fcvlabel}[ln:count:inline:remove]
\begin{VerbatimN}[commandchars=\\\[\]]
write_lock(&mylock);		\lnlbl[acq]
removing = 1;			\lnlbl[note]
sub_count(mybias);
write_unlock(&mylock);		\lnlbl[rel]
while (read_count() != 0)	\lnlbl[loop:b]
	poll(NULL, 0, 1);	\lnlbl[loop:e]
remove_device();		\lnlbl[remove]
\end{VerbatimN}
\end{fcvlabel}

\begin{fcvref}[ln:count:inline:remove]
\Clnref{acq} write-acquires the lock and
\clnref{rel} releases it.
\Clnref{note} notes that the device is being removed, and the loop spanning
\clnrefrange{loop:b}{loop:e} waits for any I/O operations to complete.
Finally, \clnref{remove} does any additional processing needed to prepare for
device removal.
\end{fcvref}

\QuickQuiz{
	What other issues would need to be accounted for in a real system?
}\QuickQuizAnswer{
	A huge number!

	Here are a few to start with:

	\begin{enumerate}
	\item	There could be any number of devices, so that the
		global variables are inappropriate, as are the
		lack of arguments to functions like \co{do_io()}.
	\item	Polling loops can be problematic in real systems,
		wasting CPU time and energy.
		In many cases, an event-driven design is far better,
		for example, where the last completing I/O wakes up the
		device-removal thread.
	\item	The I/O might fail, and so \co{do_io()} will likely
		need a return value.
	\item	If the device fails, the last I/O might never complete.
		In such cases, there might need to be some sort of
		timeout to allow error recovery.
	\item	Both \co{add_count()} and \co{sub_count()} can
		fail, but their return values are not checked.
	\item	Reader-writer locks do not scale well.
		One way of avoiding the high read-acquisition costs
		of reader-writer locks is presented in
		\cref{chp:Locking,chp:Deferred Processing}.
	\end{enumerate}
}\QuickQuizEnd

\section{Parallel Counting Discussion}
\label{sec:count:Parallel Counting Discussion}
\epigraph{This idea that there is generality in the specific is of
	  far-reaching importance.}
	 {Douglas R. Hofstadter}

This chapter has presented the reliability, performance, and
scalability problems with traditional counting primitives.
The C-language \co{++} operator is not guaranteed to function reliably in
multithreaded code, and atomic operations to a single variable neither
perform nor scale well.
This chapter therefore presented a number of counting algorithms that
perform and scale extremely well in certain special cases.

It is well worth reviewing the lessons from these counting algorithms.
To that end,
\cref{sec:count:Parallel Counting Validation}
overviews requisite validation,
\cref{sec:count:Parallel Counting Performance}
summarizes performance and scalability,
\cref{sec:count:Parallel Counting Specializations}
discusses the need for specialization,
and finally,
\cref{sec:count:Parallel Counting Lessons}
enumerates lessons learned and calls attention to later chapters that
will expand on these lessons.

\begin{table*}
\rowcolors{4}{}{lightgray}
\renewcommand*{\arraystretch}{1.1}
\small
\centering
\newcommand{\NA}{\cellcolor{white}}
\ebresizewidth{
\begin{tabular}{lrcS[table-format=2.1]S[table-format=3.0]S[table-format=4.0]
		  S[table-format=6.0]S[table-format=6.0]}
	\toprule
	\multirow{2}{*}{\begin{picture}(60,15)(0,-3)\put(0,0){Algorithm}
			\put(14,-10){(\path{count_*.c})}\end{picture}} &
	    & \multirow{2}{*}{\begin{picture}(6,50)(0,-24)\rotatebox{90}{Exact?}\end{picture}} &
		\multicolumn{1}{c}{\multirow{2}{*}{\begin{picture}(30,15)(0,-3)
			\put(0,0){Updates}\put(15,-10){(ns)}\end{picture}}} &
			\multicolumn{4}{c}{Reads (ns)} \\
	\cmidrule{5-8}
	    & Section & & &
				   \multicolumn{1}{r}{1 CPU} &
				      \multicolumn{1}{r}{8 CPUs} &
					 \multicolumn{1}{r}{64 CPUs} &
					    \multicolumn{1}{r}{420 CPUs} \\
		\midrule
		\path{stat} & \ref{sec:count:Array-Based Implementation} & \NA &
		 6.3 & 294 & 303   & 315     &    612 \\
	\path{stat_eventual} & \ref{sec:count:Eventually Consistent Implementation} & \NA &
		 6.4 &   1 &   1   &   1     &      1 \\
	\path{end} & \ref{sec:count:Per-Thread-Variable-Based Implementation} & \NA &
		 2.9 & 301 & 6 309 & 147 594 & 239 683 \\
	\path{end_rcu} & \ref{sec:together:RCU and Per-Thread-Variable-Based Statistical Counters} & \NA &
		 2.9 & 454 &   481 &     508 &   2 317 \\
	\midrule
	\path{lim} & \ref{sec:count:Simple Limit Counter Implementation} &
		N &  3.2 & 435 & 6 678 & 156 175 & 239 422 \\
	\path{lim_app} & \ref{sec:count:Approximate Limit Counter Implementation} &
		N &  2.4 & 485 & 7 041 & 173 108 & 239 682 \\
	\path{lim_atomic} & \ref{sec:count:Atomic Limit Counter Implementation} &
		Y & 19.7 & 513 & 7 085 & 199 957 & 239 450 \\
	\path{lim_sig} & \ref{sec:count:Signal-Theft Limit Counter Implementation} &
		Y &  4.7 & 519 & 6 805 & 120 000 & 238 811 \\
	\bottomrule
\end{tabular}
}
\caption{Statistical/Limit Counter Performance on x86}
\label{tab:count:Statistical/Limit Counter Performance on x86}
\end{table*}

\subsection{Parallel Counting Validation}
\label{sec:count:Parallel Counting Validation}

Many of the algorithms in this section are quite simple, so much so
that it is tempting to declare them to be correct by construction or
by inspection.
Unfortunately, it is all too easy for those carrying out the construction
or the inspection to become overconfident, tired, confused, or just plain
sloppy, all of which can result in bugs.
And early implementations of these limit counters have in fact contained
bugs, in some cases aided and abetted by the complexities inherent in
maintaining a 64-bit count on a 32-bit system.
Therefore, validation is not optional, even for the simple algorithms
presented in this chapter.

The statistical counters are tested for acting like counters
(``\path{counttorture.h}''), that is, that the aggregate sum in
the counter changes by the sum of the amounts added by the various
update-side threads.

The limit counters are also tested for acting like counters
(``\path{limtorture.h}''), and additionally checked for their ability
to accommodate the specified limit.

Both of these test suites produce performance data that is used
in \cref{sec:count:Parallel Counting Performance}.

Although this level of validation is good and sufficient for textbook
implementations such as these, it would be wise to apply additional
validation before putting similar algorithms into production.
\Cref{chp:Validation} describes additional approaches to testing, and
given the simplicity of most of these counting algorithms, most of
the techniques described in \cref{chp:Formal Verification} can also
be quite helpful.

\subsection{Parallel Counting Performance}
\label{sec:count:Parallel Counting Performance}

The top half of \cref{tab:count:Statistical/Limit Counter Performance on x86}
shows the performance of the four parallel statistical counting
algorithms.
All four algorithms provide near-perfect linear scalability for updates.
The per-thread-variable implementation (\path{count_end.c})
is significantly faster on
updates than the array-based implementation
(\path{count_stat.c}), but is slower at reads on large numbers of core,
and suffers severe \IX{lock contention} when there are many parallel readers.
This contention can be addressed using the deferred-processing
techniques introduced in
\cref{chp:Deferred Processing},
as shown on the \path{count_end_rcu.c} row of
\cref{tab:count:Statistical/Limit Counter Performance on x86}.
Deferred processing also shines on the \path{count_stat_eventual.c} row,
courtesy of eventual consistency.

\QuickQuizSeries{%
\QuickQuizB{
	On the \path{count_stat.c} row of
	\cref{tab:count:Statistical/Limit Counter Performance on x86},
	we see that the read-side scales linearly with the number of
	threads.
	How is that possible given that the more threads there are,
	the more per-thread counters must be summed up?
}\QuickQuizAnswerB{
	The read-side code must scan the entire fixed-size array, regardless
	of the number of threads, so there is no difference in performance.
	In contrast, in the last two algorithms, readers must do more
	work when there are more threads.
	In addition, the last two algorithms interpose an additional
	level of indirection because they map from integer thread ID
	to the corresponding \co{_Thread_local} variable.
}\QuickQuizEndB
\QuickQuizE{
	Even on the fourth row of
	\cref{tab:count:Statistical/Limit Counter Performance on x86},
	the read-side performance of these statistical counter
	implementations is pretty horrible.
	So why bother with them?
}\QuickQuizAnswerE{
	``Use the right tool for the job.''

	As can be seen from
	\cref{fig:count:Atomic Increment Scalability on x86},
	single-variable atomic increment need not apply for any job
	involving heavy use of parallel updates.
	In contrast, the algorithms shown in the top half of
	\cref{tab:count:Statistical/Limit Counter Performance on x86}
	do an excellent job of handling update-heavy situations.
	Of course, if you have a read-mostly situation, you should
	use something else, for example, an eventually consistent design
	featuring a single atomically incremented
	variable that can be read out using a single load,
	similar to the approach used in
	\cref{sec:count:Eventually Consistent Implementation}.
}\QuickQuizEndE
}

The bottom half of \cref{tab:count:Statistical/Limit Counter Performance on x86}
shows the performance of the parallel limit-counting algorithms.
Exact enforcement of the limits incurs a substantial update-side
performance penalty, although on this x86 system that penalty can be
reduced by substituting signals for atomic operations.
All of these implementations suffer from read-side lock contention
in the face of concurrent readers.

\QuickQuizSeries{%
\QuickQuizB{
	Given the performance data shown in the bottom half of
	\cref{tab:count:Statistical/Limit Counter Performance on x86},
	we should always prefer signals over atomic operations, right?
}\QuickQuizAnswerB{
	That depends on the workload.
	Note that on a 64-core system, you need more than
	one hundred non-atomic operations (with roughly
	a 40-nanosecond performance gain) to make up for even one
	signal (with almost a 5-\emph{microsecond} performance loss).
	Although there are no shortage of workloads with far greater
	read intensity, you will need to consider your particular
	workload.

	In addition, although memory barriers have historically been
	expensive compared to ordinary instructions, you should
	check this on the specific hardware you will be running.
	The properties of computer hardware do change over time,
	and algorithms must change accordingly.
}\QuickQuizEndB
\QuickQuizE{
	Can advanced techniques be applied to address the lock
	contention for readers seen in the bottom half of
	\cref{tab:count:Statistical/Limit Counter Performance on x86}?
}\QuickQuizAnswerE{
	One approach is to give up some update-side performance, as is
	done with scalable non-zero indicators
	(SNZI)~\cite{FaithEllen:2007:SNZI}.
	There are a number of other ways one might go about this, and these
	are left as exercises for the reader.
	Any number of approaches that apply hierarchy, which replace
	frequent global-lock acquisitions with local lock acquisitions
	corresponding to lower levels of the hierarchy, should work quite well.
}\QuickQuizEndE
}

In short, this chapter has demonstrated a number of counting algorithms
that perform and scale extremely well in a number of special cases.
But must our parallel counting be confined to special cases?
Wouldn't it be better to have a general algorithm that operated
efficiently in all cases?
The next section looks at these questions.

\subsection{Parallel Counting Specializations}
\label{sec:count:Parallel Counting Specializations}

The fact that these algorithms only work well in their respective special
cases might be considered a major problem with parallel programming in
general.
After all, the C-language \co{++} operator works just fine in single-threaded
code, and not just for special cases, but in general, right?

This line of reasoning does contain a grain of truth, but is in essence
misguided.
The problem is not parallelism as such, but rather scalability.
To understand this, first consider the C-language \co{++} operator.
The fact is that it does \emph{not} work in general, only for a restricted
range of numbers.
If you need to deal with 1,000-digit decimal numbers, the C-language \co{++}
operator will not work for you.

\QuickQuiz{
	The \co{++} operator works just fine for 1,000-digit numbers!
	Haven't you heard of operator overloading???
}\QuickQuizAnswer{
	In the C++ language, you might well be able to use \co{++}
	on a 1,000-digit number, assuming that you had access to a
	class implementing such numbers.
	But as of 2021, the C language does not permit operator overloading.
}\QuickQuizEnd

This problem is not specific to arithmetic.
Suppose you need to store and query data.
Should you use an ASCII file?
XML\@?
A relational database?
A linked list?
A dense array?
A B-tree?
A radix tree?
Or one of the plethora of other data
structures and environments that permit data to be stored and queried?
It depends on what you need to do, how fast you need it done, and how
large your data set is---even on sequential systems.

Similarly, if you need to count, your solution will depend on how large
of numbers you need to work with, how many CPUs need to be manipulating
a given number concurrently, how the number is to be used, and what
level of performance and scalability you will need.

Nor is this problem specific to software.
The design for a bridge meant to allow people to walk across a small brook
might be a simple as a single wooden plank.
But you would probably not use a plank to span the kilometers-wide mouth of
the Columbia River, nor would such a design be advisable for bridges
carrying concrete trucks.
In short, just as bridge design must change with increasing span and load,
so must software design change as the number of CPUs increases.
That said, it would be good to automate this process, so that the
software adapts to changes in hardware configuration and in workload.
There has in fact been some research into this sort of
automation~\cite{Appavoo03a,Soules03a}, and the Linux kernel does some
boot-time reconfiguration, including limited binary rewriting.
This sort of adaptation will become increasingly important as the
number of CPUs on mainstream systems continues to increase.

In short, as discussed in
\cref{chp:Hardware and its Habits},
the laws of physics constrain parallel software just as surely as they
constrain mechanical artifacts such as bridges.
These constraints force specialization, though in the case of software
it might be possible to automate the choice of specialization to
fit the hardware and workload in question.

Of course, even generalized counting is quite specialized.
We need to do a great number of other things with computers.
The next section relates what we have learned from counters to
topics taken up later in this book.

\subsection{Parallel Counting Lessons}
\label{sec:count:Parallel Counting Lessons}

The opening paragraph of this chapter promised that our study of counting
would provide an excellent introduction to parallel programming.
This section makes explicit connections between the lessons from
this chapter and the material presented in a number of later chapters.

The examples in this chapter have shown that an important scalability
and performance tool is \emph{partitioning}.
The counters might be fully partitioned, as in the statistical counters
discussed in \cref{sec:count:Statistical Counters},
or partially partitioned as in the limit counters discussed in
\cref{sec:count:Approximate Limit Counters,%
sec:count:Exact Limit Counters}.
Partitioning will be considered in far greater depth in
\cref{chp:Partitioning and Synchronization Design},
and partial parallelization in particular in
\cref{sec:SMPdesign:Parallel Fastpath}, where it is called
\emph{parallel fastpath}.

\QuickQuiz{
	But if we are going to have to partition everything, why bother
	with shared-memory multithreading?
	Why not just partition the problem completely and run as
	multiple processes, each in its own address space?
}\QuickQuizAnswer{
	Indeed, multiple processes with separate address spaces can be
	an excellent way to exploit parallelism, as the proponents of
	the fork-join methodology and the Erlang language would be very
	quick to tell you.
	However, there are also some advantages to shared-memory parallelism:
	\begin{enumerate}
	\item	Only the most performance-critical portions of the
		application must be partitioned, and such portions
		are usually a small fraction of the application.
	\item	Although cache misses are quite slow compared to
		individual register-to-register instructions,
		they are typically considerably faster than
		inter-process-communication primitives, which in
		turn are considerably faster than things like
		TCP/IP networking.
	\item	Shared-memory multiprocessors are readily available
		and quite inexpensive, so, in stark contrast to the
		1990s, there is little cost penalty for use of
		shared-memory parallelism.
	\end{enumerate}
	As always, use the right tool for the job!
}\QuickQuizEnd

The partially partitioned counting algorithms used locking to
guard the global data, and locking is the subject of
\cref{chp:Locking}.
In contrast, the partitioned data tended to be fully under the control of
the corresponding thread, so that no synchronization whatsoever was required.
This \emph{data ownership} will be introduced in
\cref{sec:SMPdesign:Data Ownership}
and discussed in more detail in
\cref{chp:Data Ownership}.

Because integer addition and subtraction are extremely cheap
compared to typical synchronization operations, achieving reasonable
scalability requires synchronization operations be used sparingly.
One way of achieving this is to batch the addition and subtraction
operations, so that a great many of these cheap operations are handled
by a single synchronization operation.
Batching optimizations of one sort or another are used by each of
the counting algorithms listed in
\cref{tab:count:Statistical/Limit Counter Performance on x86}.

Finally, the eventually consistent statistical counter discussed in
\cref{sec:count:Eventually Consistent Implementation}
showed how deferring activity (in that case, updating the global
counter) can provide substantial performance and scalability benefits.
This approach allows common case code to use much cheaper synchronization
operations than would otherwise be possible.
\Cref{chp:Deferred Processing} will examine a number of additional
ways that deferral can improve performance, scalability, and even
real-time response.

Summarizing the summary:

\begin{enumerate}
\item	Partitioning promotes performance and scalability.
\item	Partial partitioning, that is, partitioning applied only to
	common code paths, works almost as well.
\item	Partial partitioning can be applied to code (as in
	\cref{sec:count:Statistical Counters}'s statistical
	counters' partitioned updates and non-partitioned reads), but also
	across time (as in
	\cref{sec:count:Approximate Limit Counters}'s and
	\cref{sec:count:Exact Limit Counters}'s
	limit counters running fast when far from
	the limit, but slowly when close to the limit).
\item	Partitioning across time often batches updates locally
	in order to reduce the number of expensive global operations,
	thereby decreasing synchronization overhead, in turn
	improving performance and scalability.
	All the algorithms shown in
	\cref{tab:count:Statistical/Limit Counter Performance on x86}
	make heavy use of batching.
\item	Read-only code paths should remain read-only:
	Spurious synchronization writes to shared memory kill performance
	and scalability, as seen in the \path{count_end.c} row of
	\cref{tab:count:Statistical/Limit Counter Performance on x86}.
\item	Judicious use of delay promotes performance and scalability, as
	seen in \cref{sec:count:Eventually Consistent Implementation}.
\item	Parallel performance and scalability is usually a balancing act:
	Beyond a certain point, optimizing some code paths will degrade
	others.
	The \path{count_stat.c} and \path{count_end_rcu.c} rows of
	\cref{tab:count:Statistical/Limit Counter Performance on x86}
	illustrate this point.
\item	Different levels of performance and scalability will affect
	algorithm and data-structure design, as do a large number of
	other factors.
	\Cref{fig:count:Atomic Increment Scalability on x86}
	illustrates this point:
	Atomic increment might be completely acceptable for a two-CPU
	system, but nevertheless be completely inadequate for an eight-CPU system.
	% @@@ Rework the above sentence.
\end{enumerate}

\begin{figure}
\centering
\resizebox{3in}{!}{\includegraphics{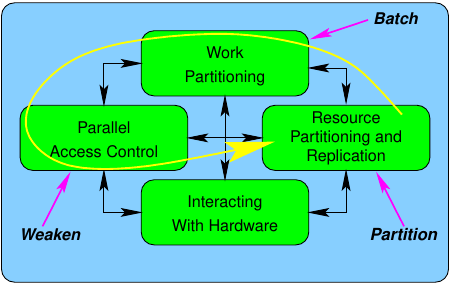}}
\caption{Optimization and the Four Parallel-Programming Tasks}
\label{fig:count:Optimization and the Four Parallel-Programming Tasks}
\end{figure}

Summarizing still further, we have the ``big three'' methods of
increasing performance and scalability, namely
(1)~\emph{partitioning} over CPUs or threads,
(2)~\emph{batching} so that more work can be done by each expensive
synchronization operation, and
(3)~\emph{weakening} synchronization operations where feasible.
As a rough rule of thumb, you should apply these methods in this order,
as was noted earlier in the discussion of
\cref{fig:intro:Ordering of Parallel-Programming Tasks}
on
\cpageref{fig:intro:Ordering of Parallel-Programming Tasks}.
The partitioning optimization applies to the
``Resource Partitioning and Replication'' bubble,
the batching optimization to the ``Work Partitioning'' bubble,
and the weakening optimization to the ``Parallel Access Control'' bubble,
as shown in
\cref{fig:count:Optimization and the Four Parallel-Programming Tasks}.
Of course, if you are using special-purpose hardware such as
digital signal processors (DSPs), field-programmable gate arrays (FPGAs),
or general-purpose graphical processing units (GPGPUs), you may need
to pay close attention to the ``Interacting With Hardware'' bubble
throughout the design process.
For example, the structure of a GPGPU's hardware threads and memory
connectivity might richly reward very careful partitioning
and batching design decisions.

In short, as noted at the beginning of this chapter, the simplicity
of counting have allowed us to explore many
fundamental concurrency issues without the distraction of
complex synchronization primitives or elaborate data structures.
Such synchronization primitives and data structures are covered
in later chapters.

	\setlength{\parskip}{0.0pt plus 1ex}
	\input{qqzcount}
	\setlength{\parskip}{0.0pt plus 1.0pt}% return to default

% SMPdesign/SMPdesign.tex
% mainfile: ../perfbook.tex
% SPDX-License-Identifier: CC-BY-SA-3.0

\QuickQuizChapter{chp:Partitioning and Synchronization Design}{Partitioning and Synchronization Design}{qqzSMPdesign}
\Epigraph{Divide and rule.}{Philip II of Macedon}

This chapter describes how to design software to take advantage of
modern commodity multicore systems by using idioms, or
``design patterns''~\cite{Alexander79,GOF95,SchmidtStalRohnertBuschmann2000v2Textbook},
to balance performance, scalability, and response time.
Correctly partitioned problems lead to simple, scalable, and
high-performance solutions, while poorly partitioned problems result
in slow and complex solutions.
This chapter will help you design partitioning into your code, with
some discussion of batching and weakening as well.
The word ``design'' is very important:
You should partition first, batch second, weaken third, and code fourth.
Changing this order often leads to poor performance and scalability
along with great frustration.\footnote{
	That other great dodge around the Laws of Physics, read-only
	replication, is covered in \cref{chp:Deferred Processing}.}

This chapter will also look at some specific problems, including:

\begin{enumerate}
\item	Constraints on the classic Dining Philosophers problem requiring
	that all the philophers be able to dine concurrently.
\label{sec:SMPdesign:Problems Dining Philosophers}
\item	Lock-based double-ended queue implementations that provide
	concurrency between operations on both ends of a given queue
	when there are many elements in the queue, but still work
	correctly when the queue contains only a few elements.
	(Or, for that matter, no elements.)
\label{sec:SMPdesign:Problems Double-Ended Queue}
\item	Summarizing the rough quality of a concurrent algorithm with only
	a few numbers.
\label{sec:SMPdesign:Problems Quality Assessment}
\item	Selecting the right granularity of partitioning.
\label{sec:SMPdesign:Problems Granularity}
\item	Comcurrent designs for applications that do not fully partition.
\label{sec:SMPdesign:Problems Parallel Fastpath}
\item	Obtaining more than 2x speedup from two CPUs.
\label{sec:SMPdesign:Problems Maze}
\end{enumerate}

To this end, \cref{sec:SMPdesign:Partitioning Exercises}
presents partitioning exercises,
\cref{sec:SMPdesign:Design Criteria} reviews partitionability
design criteria,
\cref{sec:SMPdesign:Synchronization Granularity}
discusses synchronization granularity selection,
\cref{sec:SMPdesign:Parallel Fastpath}
overviews important parallel-fastpath design patterns
that provide speed and scalability on common-case fastpaths while using
simpler less-scalable ``slow path'' fallbacks for unusual situations,
and finally
\cref{sec:SMPdesign:Beyond Partitioning}
takes a brief look beyond partitioning.

% SMPdesign/partexercises.tex
% mainfile: ../perfbook.tex
% SPDX-License-Identifier: CC-BY-SA-3.0

\section{Partitioning Exercises}
\label{sec:SMPdesign:Partitioning Exercises}
\epigraph{Whenever a theory appears to you as the only possible one,
	  take this as a sign that you have neither understood the theory
	  nor the problem which it was intended to solve.}
	  {Karl Popper}

Although partitioning is more widely understood than it was in the early
2000s, its value is still underappreciated.
\Cref{sec:SMPdesign:Dining Philosophers Problem}
therefore takes more highly parallel look at the classic Dining
Philosophers problem and
\cref{sec:SMPdesign:Double-Ended Queue}
revisits the double-ended queue.

\subsection{Dining Philosophers Problem}
\label{sec:SMPdesign:Dining Philosophers Problem}

\begin{figure}
\centering
\includegraphics[scale=.7]{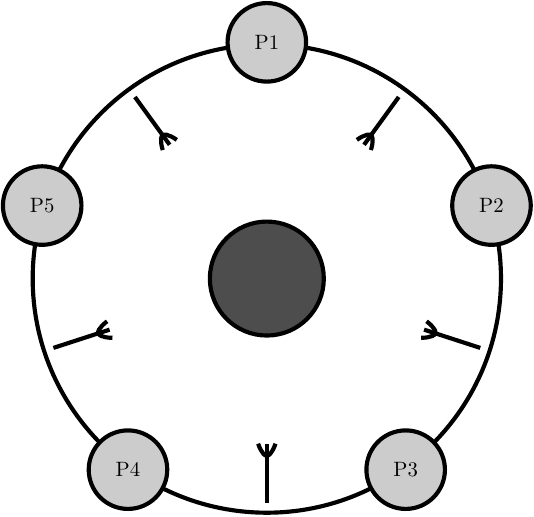}
\caption{Dining Philosophers Problem}
\ContributedBy{Figure}{fig:SMPdesign:Dining Philosophers Problem}{Kornilios Kourtis}
\end{figure}

\Cref{fig:SMPdesign:Dining Philosophers Problem} shows a diagram
of the classic \IX{Dining Philosophers problem}~\cite{Dijkstra1971HOoSP}.
This problem features five philosophers who do nothing but think and
eat a ``very difficult kind of spaghetti'' which requires two forks
to eat.\footnote{
	But feel free to instead think in terms of chopsticks.}
A given philosopher is permitted to use only the forks to his or her
immediate right and left, but will not put a given fork down until sated.

\begin{figure*}
\centering
\IfTwoColumn{
\resizebox{5in}{!}{\includegraphics{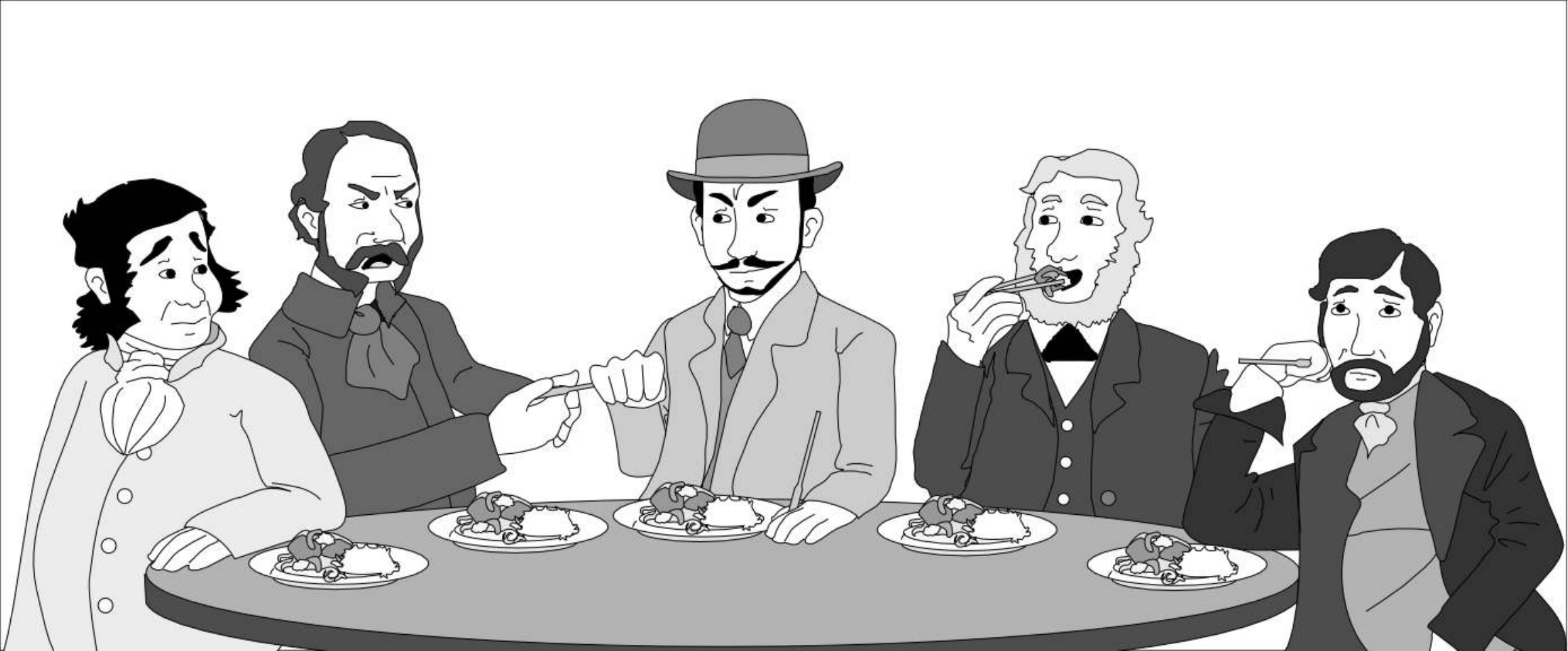}}
}{
\resizebox{\onecolumntextwidth}{!}{\includegraphics{cartoons/Dining-philosophers}}
}
\caption{Partial Starvation Is Also Bad}
\ContributedBy{Figure}{fig:cpu:Partial Starvation Is Also Bad}{Melissa Broussard}
\end{figure*}

The object is to construct an algorithm that, quite literally,
prevents \IX{starvation}.
One starvation scenario would be if all of the philosophers picked up
their leftmost forks simultaneously.
Because none of them will put down their fork until after they finished
eating, and because none of them may pick up their second fork until at
least one of them has finished eating, they all starve.
Please note that it is not sufficient to allow at least one philosopher
to eat.
As \cref{fig:cpu:Partial Starvation Is Also Bad}
shows, starvation of even a few of the philosophers is to be avoided.

\begin{figure}
\centering
\includegraphics[scale=.7]{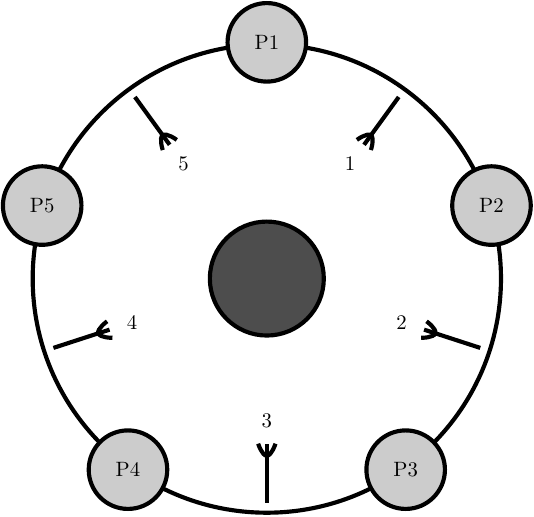}
\caption{Dining Philosophers Problem, Textbook Solution}
\ContributedBy{Figure}{fig:SMPdesign:Dining Philosophers Problem; Textbook Solution}{Kornilios Kourtis}
\end{figure}

\pplsur{Edsger W.}{Dijkstra}'s solution used a global semaphore,
which works fine assuming
negligible communications delays, an assumption that became invalid
in the late 1980s or early 1990s.\footnote{
	It is all too easy to denigrate Dijkstra from the viewpoint
	of the year 2021, more than 50 years after the fact.
	If you still feel the need to denigrate Dijkstra, my advice
	is to publish something, wait 50 years, and then see
	how well \emph{your} ideas stood the test of time.}
More recent solutions number the forks as shown in
\cref{fig:SMPdesign:Dining Philosophers Problem; Textbook Solution}.
Each philosopher picks up the lowest-numbered fork next to his or her
plate, then picks up the other fork.
The philosopher sitting in the uppermost position in the diagram thus
picks up the leftmost fork first, then the rightmost fork, while the
rest of the philosophers instead pick up their rightmost fork first.
Because two of the philosophers will attempt to pick up fork~1 first,
and because only one of those two philosophers will succeed,
there will be five forks available to four philosophers.
At least one of these four will have two forks, and will thus be able
to eat.

This general technique of numbering resources and acquiring them in
numerical order is heavily used as a deadlock-prevention technique.
However, it is easy to imagine a sequence of events that will result
in only one philosopher eating at a time even though all are hungry:

\begin{enumerate}
    \item P2 picks up fork~1, preventing P1 from taking a fork.
    \item P3 picks up fork~2.
    \item P4 picks up fork~3.
    \item P5 picks up fork~4.
    \item P5 picks up fork~5 and eats.
    \item P5 puts down forks~4 and 5.
    \item P4 picks up fork~4 and eats.
\end{enumerate}

In short, this algorithm can result in only one philosopher eating at
a given time, even when all five philosophers are hungry,
despite the fact that there are more than enough forks for two
philosophers to eat concurrently.
It should be possible to do better than this!

\begin{figure}
\centering
\includegraphics[scale=.7]{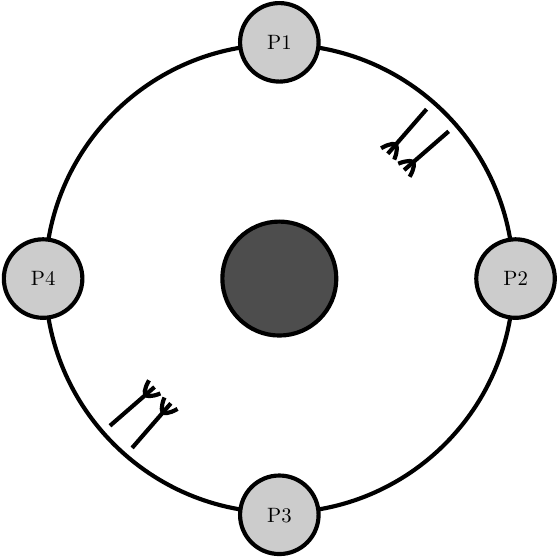}
\caption{Dining Philosophers Problem, Partitioned}
\ContributedBy{Figure}{fig:SMPdesign:Dining Philosophers Problem; Partitioned}{Kornilios Kourtis}
\end{figure}

One approach is shown in
\cref{fig:SMPdesign:Dining Philosophers Problem; Partitioned},
which includes four philosophers rather than five to better illustrate the
partition technique.
Here the upper and rightmost philosophers share a pair of forks,
while the lower and leftmost philosophers share another pair of forks.
If all philosophers are simultaneously hungry, at least two will
always be able to eat concurrently.
In addition, as shown in the figure, the forks can now be bundled
so that the pair are picked up and put down simultaneously, simplifying
the acquisition and release algorithms.

\QuickQuizSeries{%
\QuickQuizB{
	Is there a better solution to the Dining
	Philosophers Problem?
}\QuickQuizAnswerB{
	One such improved solution is shown in
	\cref{fig:SMPdesign:Dining Philosophers Problem; Fully Partitioned},
	where the philosophers are simply provided with an additional
	five forks.
	All five philosophers may now eat simultaneously, and there
	is never any need for philosophers to wait on one another.
	In addition, this approach offers greatly improved disease control.

\begin{figure}
\centering
\includegraphics[scale=.7]{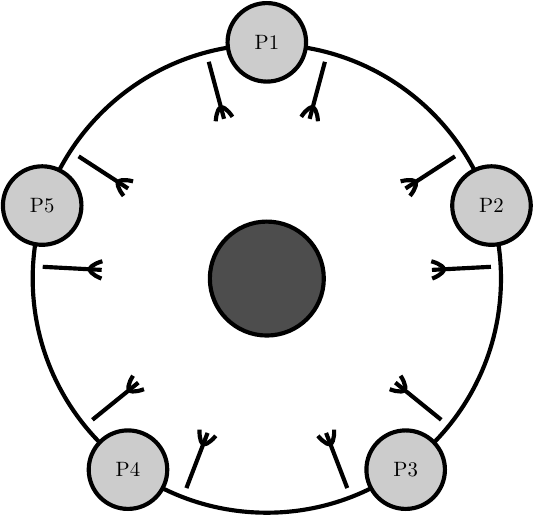}
\caption{Dining Philosophers Problem, Fully Partitioned}
\QContributedBy{Figure}{fig:SMPdesign:Dining Philosophers Problem; Fully Partitioned}{Kornilios Kourtis}
\end{figure}

	This solution might seem like cheating to some, but such
	``cheating'' is key to finding good solutions to many
	concurrency problems, as any hungry philosopher would agree.

	And this is one solution to the Dining Philosophers
	concurrent-consumption problem called out on
	\cpageref{sec:SMPdesign:Problems Dining Philosophers}.
}\QuickQuizEndB
\QuickQuizE{
	How would you valididate an algorithm alleged to solve the Dining
	Philosophers Problem?
}\QuickQuizAnswerE{
	Much depends on the details of the algorithm, but here are a
	couple of places to start.

	First, for algorithms in which picking up left-hand and right-hand
	forks are separate operations, start with all forks on the table.
	Then have all philosophers attempt to pick up their first fork.
	Once all philosophers either have their first fork or are waiting
	for someone to put down their first fork, have each non-waiting
	philosopher pick up their second fork.
	At this point in any starvation-free solution, at least one
	philosopher will be eating.
	If there were any waiting philosophers, repeat this test,
	preferably imposing random variations in timing.

	Second, create a stress test in which philosphers start and
	stop eating at random times.
	Generate starvation and fairness conditions and verify that
	these conditions are met.
	Here are a couple of example starvation and fairness conditions:

	\begin{enumerate}
	\item	If all other philosophers have stopped eating $N$ times
		since a given philosopher attempted to pick up a given
		fork, that philosopher should have succeeded in picking
		up that fork.
		For high-quality solutions using high-quality locking
		primitives (or high-quality atomic operations), $N=1$
		is doable.
	\item	Given an upper bound $T$ on the time any philosopher holds
		onto both forks before putting them down, the maximum
		waiting time for any philosopher should be bounded by
		$NT$ for some $N$ that is not hugely larger than the
		number of philosophers.
	\item	Generate some statistic representing the time from
		when philosophers attempt to pick up their first fork
		to the time when they start eating.
		The smaller this statistic, the better the solution.
		Mean, median, and maximum are all useful statistics,
		but examining the full distribution can also be
		enlightening.
	\end{enumerate}

	Readers are encouraged to actually try testing any of the
	solutions presented in this book, and especially testing solutions
	of their own devising.
}\QuickQuizEndE
}

This is an example of ``horizontal parallelism''~\cite{Inman85}
or ``data parallelism'',
so named because there is no dependency among the pairs of philosophers.
In a horizontally parallel data-processing system, a given item of data
would be processed by only one of a replicated set of software
components.

\QuickQuiz{
	And in just what sense can this ``horizontal parallelism'' be
	said to be ``horizontal''?
}\QuickQuizAnswer{
	Inman was working with protocol stacks, which are normally
	depicted vertically, with the application on top and the
	hardware interconnect on the bottom.
	Data flows up and down this stack.
	``Horizontal parallelism'' processes packets from different network
	connections in parallel, while ``vertical parallelism''
	handles different protocol-processing steps for a given
	packet in parallel.

	``Vertical parallelism'' is also called ``pipelining''.
}\QuickQuizEnd

\subsection{Double-Ended Queue}
\label{sec:SMPdesign:Double-Ended Queue}

A double-ended queue is a data structure containing a list of elements
that may be inserted or removed from either end~\cite{Knuth73}.
It has been claimed that a lock-based implementation permitting
concurrent operations on both ends of the double-ended queue is
difficult~\cite{DanGrossman2007TMGCAnalogy}.
This section shows how a partitioning design strategy can result
in a reasonably simple implementation, looking at three
general approaches in the following sections.
But first, how should we validate a concurrent double-ended queue?

\subsubsection{Double-Ended Queue Validation}
\label{sec:SMPdesign:Double-Ended Queue Validation}

A good place to start is with invariants.
For example, if elements are pushed onto one end of a double-ended
queue and popped off of the other, the order of those elements must
be preserved.
Similarly, if elements are pushed onto one end of the queue and popped
off of that same end, the order of those elements must be reversed.
Any element popped from the queue must have been most recently pushed
onto that queue, and if the queue is emptied, all elements pushed onto
it must have already been popped from it.

The beginnings of a test suite for concurrent double-ended queues
(``\path{deqtorture.h}'') provides the following checks:

\begin{enumerate}
\item	Element-ordering checks provided by \co{CHECK_SEQUENCE_PAIR()}.
\item	Checks that elements popped were most recently pushed, provided
	by \co{melee()}.
\item	Checks that elements pushed are popped before the queue is
	emptied, also provided by \co{melee()}.
\end{enumerate}

This suite includes both sequential and concurrent tests.
Although this suite is good and sufficient for textbook code, you
should test considerably more thoroughly for code intended for
production use.
\Cref{chp:Validation,chp:Formal Verification} cover a large array
of validation tools and techniques.

But with a prototype test suite in place, we are ready to look at the
double-ended-queue algorithms in the next sections.

\subsubsection{Left- and Right-Hand Locks}
\label{sec:SMPdesign:Left- and Right-Hand Locks}

\begin{figure}
\centering
\resizebox{3in}{!}{\includegraphics{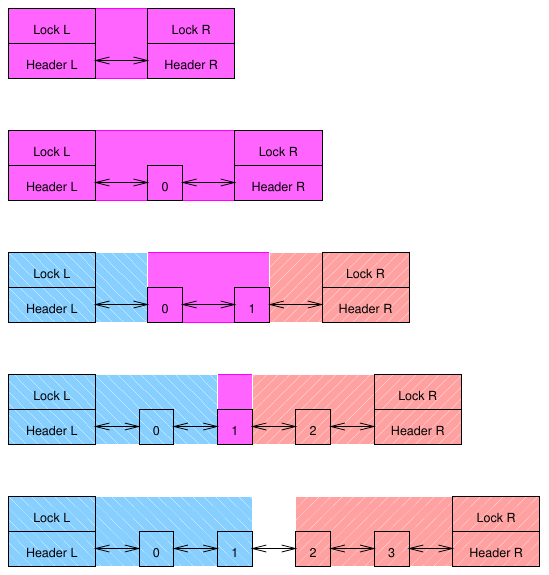}}
\caption{Double-Ended Queue With Left- and Right-Hand Locks}
\label{fig:SMPdesign:Double-Ended Queue With Left- and Right-Hand Locks}
\end{figure}

One seemingly straightforward approach would be to use a doubly
linked list with a left-hand lock
for left-hand-end enqueue and dequeue operations along with a right-hand
lock for right-hand-end operations, as shown in
\cref{fig:SMPdesign:Double-Ended Queue With Left- and Right-Hand Locks}.
However, the problem with this approach is that the two locks'
domains must overlap when there are fewer than four elements on the
list.
This overlap is due to the fact that removing any given element affects
not only that element, but also its left- and right-hand neighbors.
These domains are indicated by color in the figure, with blue
with downward stripes indicating
the domain of the left-hand lock, red with upward stripes
indicating the domain of the right-hand
lock, and purple (with no stripes) indicating overlapping domains.
Although it is possible to create an algorithm that works this way,
the fact that it has no fewer than five special cases should raise
a big red flag, especially given that concurrent activity at the other
end of the list can shift the queue from one special case to another
at any time.
It is far better to consider other designs.

\subsubsection{Compound Double-Ended Queue}
\label{sec:SMPdesign:Compound Double-Ended Queue}

\begin{figure}
\centering
\resizebox{3in}{!}{\includegraphics{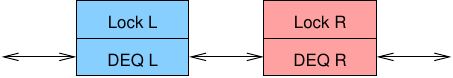}}
\caption{Compound Double-Ended Queue}
\label{fig:SMPdesign:Compound Double-Ended Queue}
\end{figure}

One way of forcing non-overlapping lock domains is shown in
\cref{fig:SMPdesign:Compound Double-Ended Queue}.
Two separate double-ended queues are run in tandem, each protected by
its own lock.
This means that elements must occasionally be shuttled from one of
the double-ended queues to the other, in which case both locks must
be held.
A simple lock hierarchy may be used to avoid \IX{deadlock}, for example,
always acquiring the left-hand lock before acquiring the right-hand lock.
This will be much simpler than applying two locks to the same
double-ended queue, as we can unconditionally left-enqueue elements
to the left-hand queue and right-enqueue elements to the right-hand
queue.
The main complication arises when dequeuing from an empty queue, in
which case it is necessary to:

\begin{enumerate}
\item	If holding the right-hand lock, release it and acquire the
	left-hand lock.
\item	Acquire the right-hand lock.
\item	Rebalance the elements across the two queues.
\item	Remove the required element if there is one.
\item	Release both locks.
\end{enumerate}

\QuickQuiz{
	In this compound double-ended queue implementation, what should
	be done if the queue has become non-empty while releasing
	and reacquiring the lock?
}\QuickQuizAnswer{
	In this case, simply dequeue an item from the non-empty
	queue, release both locks, and return.
}\QuickQuizEnd

The resulting code (\path{locktdeq.c}) is quite straightforward.
The rebalancing operation might well shuttle a given element back
and forth between the two queues, wasting time and possibly requiring
workload-dependent heuristics to obtain optimal performance.
Although this might well be the best approach in some cases, it is
interesting to try for an algorithm with greater determinism.

\subsubsection{Hashed Double-Ended Queue}
\label{sec:SMPdesign:Hashed Double-Ended Queue}

One of the simplest and most effective ways to deterministically
partition a data structure is to hash it.
It is possible to trivially hash a double-ended queue by assigning
each element a sequence number based on its position in the list,
so that the first element left-enqueued into an empty queue is numbered
zero and the first element right-enqueued into an empty queue is numbered
one.
A series of elements left-enqueued into an otherwise-idle queue would
be assigned decreasing numbers ($-1$, $-2$, $-3$, \ldots), while a series of
elements right-enqueued into an otherwise-idle queue would be assigned
increasing numbers (2, 3, 4, \ldots).
A key point is that it is not necessary to actually represent a given
element's number, as this number will be implied by its position in
the queue.

\begin{figure}
\centering
\resizebox{3in}{!}{\includegraphics{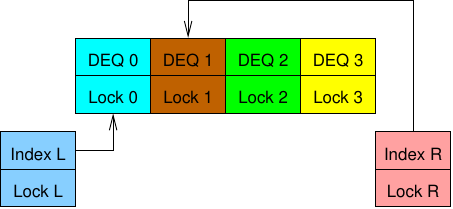}}
\caption{Hashed Double-Ended Queue}
\label{fig:SMPdesign:Hashed Double-Ended Queue}
\end{figure}

Given this approach, we assign one lock to guard the left-hand index,
one to guard the right-hand index, and one lock for each hash chain.
\Cref{fig:SMPdesign:Hashed Double-Ended Queue} shows the resulting
data structure given four hash chains.
Note that the lock domains do not overlap, and that deadlock is avoided
by acquiring the index locks before the chain locks, and by never
acquiring more than one lock of a given type (index or chain) at a time.

\begin{figure}
\centering
\resizebox{3in}{!}{\includegraphics{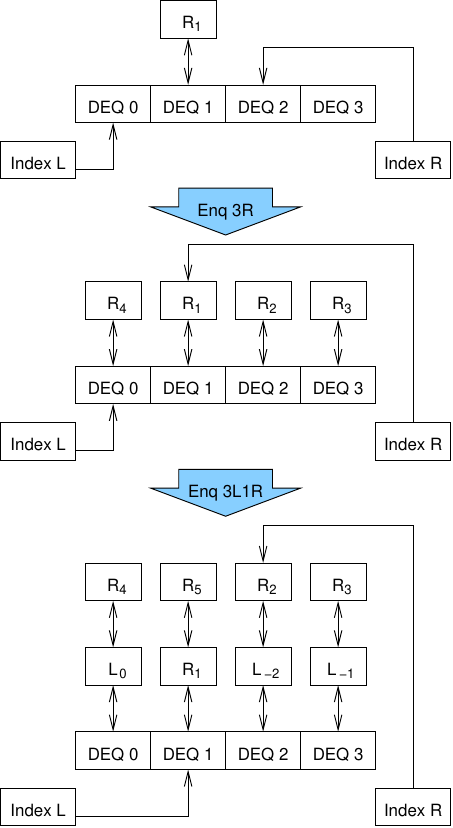}}
\caption{Hashed Double-Ended Queue After Insertions}
\label{fig:SMPdesign:Hashed Double-Ended Queue After Insertions}
\end{figure}

Each hash chain is itself a double-ended queue, and in this example,
each holds every fourth element.
The uppermost portion of
\cref{fig:SMPdesign:Hashed Double-Ended Queue After Insertions}
shows the state after a single element (``R$_1$'') has been
right-enqueued, with the right-hand index having been incremented to
reference hash chain~2.
The middle portion of this same figure shows the state after
three more elements have been right-enqueued.
As you can see, the indexes are back to their initial states
(see \cref{fig:SMPdesign:Hashed Double-Ended Queue}), however,
each hash chain is now non-empty.
The lower portion of this figure shows the state after three additional
elements have been left-enqueued and an additional element has been
right-enqueued.

From the last state shown in
\cref{fig:SMPdesign:Hashed Double-Ended Queue After Insertions},
a left-dequeue operation would return element ``L$_{-2}$'' and leave
the left-hand index referencing hash chain~2, which would then
contain only a single element (``R$_2$'').
In this state, a left-enqueue running concurrently with a right-enqueue
would result in \IX{lock contention}, but the probability of such contention
can be reduced to arbitrarily low levels by using a larger hash table.

\begin{figure}
\centering
\resizebox{1.5in}{!}{\includegraphics{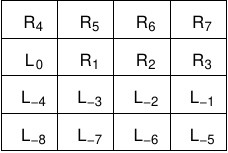}}
\caption{Hashed Double-Ended Queue With 16 Elements}
\label{fig:SMPdesign:Hashed Double-Ended Queue With 16 Elements}
\end{figure}

\Cref{fig:SMPdesign:Hashed Double-Ended Queue With 16 Elements}
shows how 16 elements would be organized in a four-hash-bucket
parallel double-ended queue.
Each underlying single-lock double-ended queue holds a one-quarter
slice of the full parallel double-ended queue.

\begin{listing}
\input{CodeSamples/SMPdesign/lockhdeq=struct_pdeq.fcv}
\caption{Lock-Based Parallel Double-Ended Queue Data Structure}
\label{lst:SMPdesign:Lock-Based Parallel Double-Ended Queue Data Structure}
\end{listing}

\Cref{lst:SMPdesign:Lock-Based Parallel Double-Ended Queue Data Structure}
shows the corresponding C-language data structure, assuming an
existing \co{struct deq} that provides a trivially locked
double-ended-queue implementation.
\begin{fcvref}[ln:SMPdesign:lockhdeq:struct_pdeq]
This data structure contains the left-hand lock on \clnref{llock},
the left-hand index on \clnref{lidx}, the right-hand lock on \clnref{rlock}
(which is cache-aligned in the actual implementation),
the right-hand index on \clnref{ridx}, and, finally, the hashed array
of simple lock-based double-ended queues on \clnref{bkt}.
A high-performance implementation would of course use padding or special
alignment directives to avoid \IX{false sharing}.
\end{fcvref}

\begin{listing}
\input{CodeSamples/SMPdesign/lockhdeq=pop_push.fcv}
\caption{Lock-Based Parallel Double-Ended Queue Implementation}
\label{lst:SMPdesign:Lock-Based Parallel Double-Ended Queue Implementation}
\end{listing}

\Cref{lst:SMPdesign:Lock-Based Parallel Double-Ended Queue Implementation}
(\path{lockhdeq.c})
shows the implementation of the enqueue and dequeue functions.\footnote{
	One could easily create a polymorphic implementation in any
	number of languages, but doing so is left as an exercise for
	the reader.}
Discussion will focus on the left-hand operations, as the right-hand
operations are trivially derived from them.

\begin{fcvref}[ln:SMPdesign:lockhdeq:pop_push:popl]
\Clnrefrange{b}{e} show \co{pdeq_pop_l()},
which left\-/dequeues and returns
an element if possible, returning \co{NULL} otherwise.
\Clnref{acq} acquires the left-hand spinlock,
and \clnref{idx} computes the
index to be dequeued from.
\Clnref{deque} dequeues the element, and,
if \clnref{check} finds the result to be
non-\co{NULL}, \clnref{record} records the new left-hand index.
Either way, \clnref{rel} releases the lock, and,
finally, \clnref{return} returns
the element if there was one, or \co{NULL} otherwise.
\end{fcvref}

\begin{fcvref}[ln:SMPdesign:lockhdeq:pop_push:pushl]
\Clnrefrange{b}{e} show \co{pdeq_push_l()},
which left-enqueues the specified
element.
\Clnref{acq} acquires the left-hand lock,
and \clnref{idx} picks up the left-hand
index.
\Clnref{enque} left-enqueues the specified element
onto the double-ended queue
indexed by the left-hand index.
\Clnref{update} then updates the left-hand index
and \clnref{rel} releases the lock.
\end{fcvref}

As noted earlier, the right-hand operations are completely analogous
to their left-handed counterparts, so their analysis is left as an
exercise for the reader.

\QuickQuiz{
	Is the hashed double-ended queue a good solution?
	Why or why not?
}\QuickQuizAnswer{
	The best way to answer this is to run \path{lockhdeq.c} on
	a number of different multiprocessor systems, and you are
	encouraged to do so in the strongest possible terms.
	One reason for concern is that each operation on this
	implementation must acquire not one but two locks.
	% Getting about 500 nanoseconds per element when used as
	% a queue on a 4.2GHz Power system.  This is roughly the same as
	% the version covered by a single lock.  Sequential (unlocked
	% variant is more than an order of magnitude faster!

	The first well-designed performance study will be cited.\footnote{
		The studies by Dalessandro
		et al.~\cite{LukeDalessandro:2011:ASPLOS:HybridNOrecSTM:deque}
		and Dice et al.~\cite{DavidDice:2010:SCA:HTM:deque} are
		excellent starting points.}
	Do not forget to compare to a sequential implementation!
}\QuickQuizEnd

\subsubsection{Compound Double-Ended Queue Revisited}
\label{sec:SMPdesign:Compound Double-Ended Queue Revisited}

This section revisits the compound double-ended queue, using a trivial
rebalancing scheme that moves all the elements from the non-empty
queue to the now-empty queue.

\QuickQuiz{
	Move \emph{all} the elements to the queue that became empty?
	In what possible universe is this brain-dead solution in any
	way optimal???
}\QuickQuizAnswer{
	It is optimal in the case where data flow switches direction only
	rarely.
	It would of course be an extremely poor choice if the double-ended
	queue was being emptied from both ends concurrently.
	This of course raises another question, namely, in what possible
	universe emptying from both ends concurrently would be a reasonable
	thing to do.
	Work-stealing queues are one possible answer to this question.
}\QuickQuizEnd

In contrast to the hashed implementation presented in
the previous section, the compound implementation will build on
a sequential implementation of a double-ended queue that uses
neither locks nor atomic operations.

\begin{listing}
\ebresizeverb{.97}{% Do not edit!
% Generated by utilities/fcvextract.pl.
\begin{fcvlabel}[ln:SMPdesign:locktdeq:pop_push]
\begin{VerbatimL}[commandchars=\\\@\$]
struct cds_list_head *pdeq_pop_l(struct pdeq *d)\lnlbl@popl:b$
{
	struct cds_list_head *e;

	spin_lock(&d->llock);\lnlbl@popl:acq:l$
	e = deq_pop_l(&d->ldeq);\lnlbl@popl:deq:ll$
	if (e == NULL) {
		spin_lock(&d->rlock);\lnlbl@popl:acq:r$
		e = deq_pop_l(&d->rdeq);\lnlbl@popl:deq:lr$
		cds_list_splice(&d->rdeq.chain, &d->ldeq.chain);\lnlbl@popl:move$
		CDS_INIT_LIST_HEAD(&d->rdeq.chain);\lnlbl@popl:init:r$
		spin_unlock(&d->rlock);\lnlbl@popl:rel:r$
	}\lnlbl@popl:skip$
	spin_unlock(&d->llock);\lnlbl@popl:rel:l$
	return e;
}\lnlbl@popl:e$

struct cds_list_head *pdeq_pop_r(struct pdeq *d)\lnlbl@popr:b$
{
	struct cds_list_head *e;

	spin_lock(&d->rlock);\lnlbl@popr:acq:r1$
	e = deq_pop_r(&d->rdeq);\lnlbl@popr:deq:rr1$
	if (e == NULL) {\lnlbl@popr:check1$
		spin_unlock(&d->rlock);\lnlbl@popr:rel:r1$
		spin_lock(&d->llock);\lnlbl@popr:acq:l$
		spin_lock(&d->rlock);\lnlbl@popr:acq:r2$
		e = deq_pop_r(&d->rdeq);\lnlbl@popr:deq:rr2$
		if (e == NULL) {\lnlbl@popr:check2$
			e = deq_pop_r(&d->ldeq);\lnlbl@popr:deq:rl$
			cds_list_splice(&d->ldeq.chain, &d->rdeq.chain);\lnlbl@popr:move$
			CDS_INIT_LIST_HEAD(&d->ldeq.chain);\lnlbl@popr:init:l$
		}
		spin_unlock(&d->llock);\lnlbl@popr:rel:l$
	}\lnlbl@popr:skip2$
	spin_unlock(&d->rlock);\lnlbl@popr:rel:r2$
	return e;
}\lnlbl@popr:e$

void pdeq_push_l(struct cds_list_head *e, struct pdeq *d)\lnlbl@pushl:b$
{
	spin_lock(&d->llock);\lnlbl@pushl:acq:l$
	deq_push_l(e, &d->ldeq);\lnlbl@pushl:que:l$
	spin_unlock(&d->llock);\lnlbl@pushl:rel:l$
}\lnlbl@pushl:e$

void pdeq_push_r(struct cds_list_head *e, struct pdeq *d)\lnlbl@pushr:b$
{
	spin_lock(&d->rlock);
	deq_push_r(e, &d->rdeq);
	spin_unlock(&d->rlock);
}\lnlbl@pushr:e$
\end{VerbatimL}
\end{fcvlabel}
}
\caption{Compound Parallel Double-Ended Queue Implementation}
\label{lst:SMPdesign:Compound Parallel Double-Ended Queue Implementation}
\end{listing}

\Cref{lst:SMPdesign:Compound Parallel Double-Ended Queue Implementation}
shows the implementation.
Unlike the hashed implementation, this compound implementation is
asymmetric, so that we must consider the \co{pdeq_pop_l()}
and \co{pdeq_pop_r()} implementations separately.

\QuickQuiz{
	Why can't the compound parallel double-ended queue
	implementation be symmetric?
}\QuickQuizAnswer{
	The need to avoid deadlock by imposing a lock hierarchy
	forces the asymmetry, just as it does in the fork-numbering
	solution to the Dining Philosophers Problem
	(see \cref{sec:SMPdesign:Dining Philosophers Problem}).
}\QuickQuizEnd

\begin{fcvref}[ln:SMPdesign:locktdeq:pop_push:popl]
The \co{pdeq_pop_l()} implementation is shown on
\clnrefrange{b}{e}
of the figure.
\Clnref{acq:l} acquires the left-hand lock,
which \clnref{rel:l} releases.
\Clnref{deq:ll} attempts to left-dequeue an element
from the left-hand underlying
double-ended queue, and, if successful,
skips \clnrefrange{acq:r}{skip} to simply
return this element.
Otherwise, \clnref{acq:r} acquires the right-hand lock, \clnref{deq:lr}
left-dequeues an element from the right-hand queue,
and \clnref{move} moves any remaining elements on the right-hand
queue to the left-hand queue, \clnref{init:r} initializes
the right-hand queue,
and \clnref{rel:r} releases the right-hand lock.
The element, if any, that was dequeued on \clnref{deq:lr} will be returned.
\end{fcvref}

\begin{fcvref}[ln:SMPdesign:locktdeq:pop_push:popr]
The \co{pdeq_pop_r()} implementation is shown on \clnrefrange{b}{e}
of the figure.
As before, \clnref{acq:r1} acquires the right-hand lock
(and \clnref{rel:r2}
releases it), and \clnref{deq:rr1} attempts to right-dequeue an element
from the right-hand queue, and, if successful,
skips \clnrefrange{rel:r1}{skip2}
to simply return this element.
However, if \clnref{check1} determines that there was no element to dequeue,
\clnref{rel:r1} releases the right-hand lock and
\clnrefrange{acq:l}{acq:r2} acquire both
locks in the proper order.
\Clnref{deq:rr2} then attempts to right-dequeue an element
from the right-hand
list again, and if \clnref{check2} determines that this second attempt has
failed, \clnref{deq:rl} right-dequeues an element from the left-hand queue
(if there is one available), \clnref{move} moves any remaining elements
from the left-hand queue to the right-hand queue, and \clnref{init:l}
initializes the left-hand queue.
Either way, \clnref{rel:l} releases the left-hand lock.
\end{fcvref}

\QuickQuizSeries{%
\QuickQuizB{
	Why is it necessary to retry the right-dequeue operation
	on \clnrefr{ln:SMPdesign:locktdeq:pop_push:popr:deq:rr2} of
	\cref{lst:SMPdesign:Compound Parallel Double-Ended Queue Implementation}?
}\QuickQuizAnswerB{
	\begin{fcvref}[ln:SMPdesign:locktdeq:pop_push:popr]
	This retry is necessary because some other thread might have
	enqueued an element between the time that this thread dropped
	\co{d->rlock} on \clnref{rel:r1} and the time that it reacquired this
	same lock on \clnref{acq:r2}.
	\end{fcvref}
}\QuickQuizEndB
\QuickQuizE{
	Surely the left-hand lock must \emph{sometimes} be available!!!
	So why is it necessary that
	\clnrefr{ln:SMPdesign:locktdeq:pop_push:popr:rel:r1} of
	\cref{lst:SMPdesign:Compound Parallel Double-Ended Queue Implementation}
	unconditionally release the right-hand lock?
}\QuickQuizAnswerE{
	It would be possible to use \co{spin_trylock()} to attempt
	to acquire the left-hand lock when it was available.
	However, the failure case would still need to drop the
	right-hand lock and then re-acquire the two locks in order.
	Making this transformation (and determining whether or not
	it is worthwhile) is left as an exercise for the reader.
}\QuickQuizEndE
}

\begin{fcvref}[ln:SMPdesign:locktdeq:pop_push:pushl]
The \co{pdeq_push_l()} implementation is shown on
\clnrefrange{b}{e} of
\cref{lst:SMPdesign:Compound Parallel Double-Ended Queue Implementation}.
\Clnref{acq:l} acquires the left-hand spinlock,
\clnref{que:l} left-enqueues the
element onto the left-hand queue, and finally \clnref{rel:l} releases
the lock.
\end{fcvref}
\begin{fcvref}[ln:SMPdesign:locktdeq:pop_push:pushr]
The \co{pdeq_push_r()} implementation (shown on \clnrefrange{b}{e})
is quite similar.
\end{fcvref}

\QuickQuiz{
	But in the case where data is flowing in only one direction,
	the algorithm shown in
	\cref{lst:SMPdesign:Compound Parallel Double-Ended Queue Implementation}
	will have both ends attempting to acquire the same lock
	whenever the consuming end empties its underlying
	double-ended queue.
	Doesn't that mean that sometimes this algorithm fails to
	provide concurrent access to both ends of the queue even
	when the queue contains an arbitrarily large number of elements?
}\QuickQuizAnswer{
	Indeed it does!

	But the same is true of other algorithms claiming this property.
	For example, in solutions using software transactional memory
	mechanisms based on hashed arrays of locks,
	the leftmost and rightmost elements' addresses will sometimes
	happen to hash to the same lock.
	These hash collisions will also prevent concurrent access.
	For another example, solutions using hardware transactional
	memory mechanisms with software
	fallbacks~\cite{Yoo:2013:PEI:2503210.2503232,RickMerrit2011PowerTM,ChristianJacobi2012MainframeTM}
	often use locking within those software fallbacks, and thus
	suffer (albeit hopefully rarely) from whatever concurrency
	limitations that these locking solutions suffer from.

	Therefore, as of 2021, all practical solutions to the
	concurrent double-ended queue problem fail to provide full
	concurrency in at least some circumstances, including the
	compound double-ended queue.
}\QuickQuizEnd

\subsubsection{Double-Ended Queue Discussion}
\label{sec:SMPdesign:Double-Ended Queue Discussion}

The compound implementation is somewhat more complex than the
hashed variant presented in
\cref{sec:SMPdesign:Hashed Double-Ended Queue},
but is still reasonably simple.
Of course, a more intelligent rebalancing scheme could be arbitrarily
complex, but the simple scheme shown here has been shown to
perform well compared to software
alternatives~\cite{LukeDalessandro:2011:ASPLOS:HybridNOrecSTM:deque}
and even compared to algorithms using hardware
assist~\cite{DavidDice:2010:SCA:HTM:deque}.
Nevertheless, the best we can hope for from such a scheme
is 2x scalability, as at most two threads can be holding the
dequeue's locks concurrently.
This limitation also applies to algorithms based on \IXacrl{nbs},
such as the compare-and-swap-based dequeue algorithm of
Michael~\cite{DBLP:conf/europar/Michael03}.\footnote{
	This paper is interesting in that it showed that special
	double-compare-and-swap (DCAS) instructions are not needed
	for lock-free implementations of double-ended queues.
	Instead, the common compare-and-swap (e.g., x86 cmpxchg)
	suffices.}

\QuickQuiz{
	Why are there not one but two solutions to the double-ended queue
	problem?
}\QuickQuizAnswer{
	There are actually at least three.
	The third, by Dominik Dingel, makes interesting use of
	reader-writer locking, and may be found in \path{lockrwdeq.c}.

	And so there is not one, but rather three solutions to the
	lock-based double-ended queue problem on
	\cpageref{sec:SMPdesign:Problems Double-Ended Queue}!
}\QuickQuizEnd

In fact, as noted by Dice et al.~\cite{DavidDice:2010:SCA:HTM:deque},
an unsynchronized single-threaded double-ended queue significantly
outperforms any of the parallel implementations they studied.
Therefore, the key point is that there can be significant overhead enqueuing to
or dequeuing from a shared queue, regardless of implementation.
This should come as no surprise in light of the material in
\cref{chp:Hardware and its Habits}, given the strict
first-in-first-out (FIFO) nature of these queues.

Furthermore, these strict FIFO queues are strictly FIFO only with
respect to
\emph{linearization points}~\cite{Herlihy:1990:LCC:78969.78972}\footnote{
	In short, a linearization point is a single point within a given
	function where that function can be said to have taken effect.
	In this lock-based implementation, the linearization points
	can be said to be anywhere within the critical section that
	does the work.}
that are not visible to the caller, in fact, in these examples,
the linearization points are buried in the lock-based critical
sections.
These queues are not strictly FIFO with respect to (say) the times at which
the individual operations started~\cite{AndreasHaas2012FIFOisnt}.
This indicates that the strict FIFO property is not all that valuable
in concurrent programs, and in fact, Kirsch et al.\ present less-strict
queues that provide improved performance and
scalability~\cite{ChristophMKirsch2012FIFOisntTR}.\footnote{
	Nir Shavit produced relaxed stacks for roughly the same
	reasons~\cite{Shavit:2011:DSM:1897852.1897873}.
	This situation leads some to believe that the linearization
	points are useful to theorists rather than developers, and
	leads others to wonder to what extent the designers of such
	data structures and algorithms were considering the needs
	of their users.}
All that said, if you are pushing all the data used by your concurrent
program through a single queue, you really need to rethink your
overall design.

\subsection{Partitioning Example Discussion}
\label{sec:SMPdesign:Partitioning Example Discussion}

The optimal solution to the dining philosophers problem given in
the answer to the Quick Quiz in
\cref{sec:SMPdesign:Dining Philosophers Problem}
is an excellent example of ``horizontal parallelism'' or
``data parallelism''.
The synchronization overhead in this case is nearly (or even exactly)
zero.
In contrast, the double-ended
queue implementations are examples of ``vertical parallelism'' or
``pipelining'', given that data moves from one thread to another.
The tighter coordination required for pipelining in turn requires
larger units of work to obtain a given level of \IX{efficiency}.

\QuickQuizSeries{%
\QuickQuizB{
	The tandem double-ended queue runs about twice as fast as
	the hashed double-ended queue, even when I increase the
	size of the hash table to an insanely large number.
	Why is that?
}\QuickQuizAnswerB{
	The hashed double-ended queue's locking design only permits
	one thread at a time at each end, and further requires
	two lock acquisitions for each operation.
	The tandem double-ended queue also permits one thread at a time
	at each end, and in the common case requires only one lock
	acquisition per operation.
	Therefore, the tandem double-ended queue should be expected to
	outperform the hashed double-ended queue.

	Can you create a double-ended queue that allows multiple
	concurrent operations at each end?
	If so, how?
	If not, why not?
}\QuickQuizEndB
\QuickQuizE{
	Is there a significantly better way of handling concurrency
	for double-ended queues?
}\QuickQuizAnswerE{
	One approach is to transform the problem to be solved
	so that multiple double-ended queues can be used in parallel,
	allowing the simpler single-lock double-ended queue to be used,
	and perhaps also replace each double-ended queue with a pair of
	conventional single-ended queues.
	Without such ``horizontal scaling'', the speedup is limited
	to 2.0.
	In contrast, horizontal-scaling designs can achieve very large
	speedups, and are especially attractive if there are multiple threads
	working either end of the queue, because in this
	multiple-thread case the dequeue
	simply cannot provide strong ordering guarantees.
	After all, the fact that a given thread removed an item first
	in no way implies that it will process that item
	first~\cite{AndreasHaas2012FIFOisnt}.
	And if there are no guarantees, we may as well obtain the
	performance benefits that come with refusing to provide these
	guarantees.
	% about twice as fast as hashed version on 4.2GHz Power.

	Regardless of whether or not the problem can be transformed
	to use multiple queues, it is worth asking whether work can
	be batched so that each enqueue and dequeue operation corresponds
	to larger units of work.
	This batching approach decreases contention on the queue data
	structures, which increases both performance and scalability,
	as will be seen in
	\cref{sec:SMPdesign:Synchronization Granularity}.
	After all, if you must incur high synchronization overheads,
	be sure you are getting your money's worth.

	Other researchers are working on other ways to take advantage
	of limited ordering guarantees in
	queues~\cite{ChristophMKirsch2012FIFOisntTR}.
}\QuickQuizEndE
}

These two examples show just how powerful partitioning can be in
devising parallel algorithms.
\Cref{sec:SMPdesign:Locking Granularity and Performance}
looks briefly at a third example, matrix multiply.
However, all three of these examples beg for more and better design
criteria for parallel programs, a topic taken up in the next section.

% SMPdesign/criteria.tex
% mainfile: ../perfbook.tex
% SPDX-License-Identifier: CC-BY-SA-3.0

\section{Design Criteria}
\label{sec:SMPdesign:Design Criteria}
\epigraph{One pound of learning requires ten pounds of commonsense to apply it.}
	 {Persian proverb}

One way to obtain the best performance and scalability is to simply
hack away until you converge on the best possible parallel program.
Unfortunately, if your program is other than microscopically tiny,
the space of possible parallel programs is so huge
that convergence is not guaranteed in the lifetime of the universe.
Besides, what exactly is the ``best possible parallel program''?
After all, \cref{sec:intro:Parallel Programming Goals}
called out no fewer than three parallel-programming goals of
\IX{performance}, \IX{productivity}, and \IX{generality},
and the best possible performance will likely come at a cost in
terms of productivity and generality.
We clearly need to be able to make higher-level choices at design
time in order to arrive at an acceptably good parallel program
before that program becomes obsolete.

However, more detailed design criteria are required to
actually produce a real-world design, a task taken up in this section.
This being the real world, these criteria often conflict to a
greater or lesser degree, requiring that the designer carefully
balance the resulting tradeoffs.

As such, these criteria may be thought of as the ``forces''
acting on the design, with particularly good tradeoffs between
these forces being called ``design patterns''~\cite{Alexander79,GOF95}.

The design criteria for attaining the three parallel-programming goals
are speedup,
contention, overhead, read-to-write ratio, and complexity:
\begin{description}
\item[Speedup:]  As noted in
	\cref{sec:intro:Parallel Programming Goals},
	increased performance is the major reason
	to go to all of the time and trouble
	required to parallelize it.
	Speedup is defined to be the ratio of the time required
	to run a sequential version of the program to the time
	required to run a parallel version.
\item[Contention:]  If more CPUs are applied to a parallel
	program than can be kept busy by that program,
	the excess CPUs are prevented from doing
	useful work by contention.
	This may be \IX{lock contention}, memory contention, or a host
	of other performance killers.
\item[Work-to-Synchronization Ratio:]  A uniprocessor,
	single\-/threaded, non-preemptible, and non\-/interruptible\footnote{
		Either by masking interrupts or by being oblivious to them.}
	version of a given parallel
	program would not need any synchronization primitives.
	Therefore, any time consumed by these primitives
	(including communication cache misses as well as
	\IXh{message}{latency}, locking primitives, atomic instructions,
	and \IXpl{memory barrier})
	is overhead that does not contribute directly to the useful
	work that the program is intended to accomplish.
	Note that the important measure is the
	relationship between the synchronization overhead
	and the overhead of the code in the \IX{critical section}, with larger
	critical sections able to tolerate greater synchronization overhead.
	The work-to-synchronization ratio is related to
	the notion of synchronization efficiency.
\item[Read-to-Write Ratio:]  A data structure that is
	rarely updated may often be replicated rather than partitioned,
	and furthermore may be protected with asymmetric
	synchronization primitives that reduce readers' synchronization
	overhead at the expense of that of writers, thereby
	reducing overall synchronization overhead.
	Corresponding optimizations are possible for frequently
	updated data structures, as discussed in
	\cref{chp:Counting}.
\item[Complexity:]  A parallel program is more complex than
	an equivalent sequential program because the parallel program
	has a much larger state space than does the sequential program,
	although large state spaces having regular structures can in
	some cases be easily understood.
	A parallel programmer must
	consider synchronization primitives, messaging, locking design,
	critical-section identification,
	and deadlock in the context of this larger state space.

	This greater complexity often translates
	to higher development and maintenance costs.
	Therefore, budgetary constraints can
	limit the number and types of modifications made to
	an existing program, since a given degree of speedup is
	worth only so much time and trouble.
	Worse yet, added complexity can actually \emph{reduce}
	performance and scalability.

	Therefore, beyond a certain point,
	there may be potential sequential optimizations
	that are cheaper and more effective than parallelization.
	As noted in
	\cref{sec:intro:Performance},
	parallelization is but one performance optimization of
	many, and is furthermore an optimization that applies
	most readily to CPU-based bottlenecks.
\end{description}
These criteria will act together to enforce a maximum speedup.
The first three criteria are deeply interrelated, so
the remainder of this section analyzes these
interrelationships.\footnote{
	A real-world parallel system will be subject to many additional
	design criteria, such as data-structure layout,
	memory size, memory-hierarchy latencies, bandwidth limitations,
	and I/O issues.}

Note that these criteria may also appear as part of the requirements
specification, and further that they are one solution to the problem
of summarizing the quality of a concurrent algorithm from
\cpageref{sec:SMPdesign:Problems Quality Assessment}.
For example, speedup may act as a relative desideratum
(``the faster, the better'')
or as an absolute requirement of the workload (``the system
must support at least 1,000,000 web hits per second'').
Classic design pattern languages describe relative desiderata as forces
and absolute requirements as context.

An understanding of the relationships between these design criteria can
be very helpful when identifying appropriate design tradeoffs for a
parallel program.
\begin{enumerate}
\item	The less time a program spends in exclusive-lock critical sections,
	the greater the potential speedup.
	This is a consequence of \IXr{Amdahl's Law}~\cite{GeneAmdahl1967AmdahlsLaw}
	because only one CPU may execute within a given
	exclusive-lock critical section at a given time.

	More specifically, for unbounded linear scalability, the fraction
	of time that the program spends in a given exclusive critical
	section must decrease as the number of CPUs increases.
	For example, a program will not scale to 10~CPUs
	unless it spends much less than one tenth of its time in the
	most-restrictive exclusive-lock critical section.
\item	Contention effects consume the excess CPU and/or
	wallclock time when the actual speedup is less than
	the number of available CPUs.
	The larger the gap between the number of CPUs
	and the actual speedup, the less efficiently the
	CPUs will be used.
	Similarly, the greater the desired efficiency, the smaller
	the achievable speedup.
\item	If the available synchronization primitives have
	high overhead compared to the critical sections
	that they guard, the best way to improve speedup
	is to reduce the number of times that the primitives
	are invoked.
	This can be accomplished by batching critical sections,
	using data ownership (see \cref{chp:Data Ownership}),
	using asymmetric primitives
	(see \cref{chp:Deferred Processing}),
	or by using a coarse-grained design such as \IXh{code}{locking}.
\item	If the critical sections have high overhead compared
	to the primitives guarding them, the best way
	to improve speedup is to increase parallelism
	by moving to reader/writer locking, \IXh{data}{locking}, asymmetric,
	or data ownership.
\item	If the critical sections have high overhead compared
	to the primitives guarding them and the data structure
	being guarded is read much more often than modified,
	the best way to increase parallelism is to move
	to reader/writer locking or asymmetric primitives.
\item	Many changes that improve SMP performance, for example,
	reducing lock contention, also improve real-time
	latencies~\cite{PaulMcKenney2005h}.
\end{enumerate}

\QuickQuiz{
	Don't all these problems with critical sections mean that
	we should just always use
	non-blocking synchronization~\cite{MauriceHerlihy90a},
	which don't have critical sections?
}\QuickQuizAnswer{
	Although non-blocking synchronization can be very useful
	in some situations, it is no panacea, as discussed in
	\cref{sec:advsync:Non-Blocking Synchronization}.
	Also, non-blocking synchronization really does have
	critical sections, as noted by Josh Triplett.
	For example, in a non-blocking algorithm based on
	compare-and-swap operations, the code starting at the
	initial load and continuing to the compare-and-swap
	is analogous to a lock-based critical section.
}\QuickQuizEnd

It is worth reiterating that contention has many guises, including
lock contention, memory contention, cache overflow, thermal throttling,
and much else besides.
This chapter looks primarily at lock and memory contention.

\section{Synchronization Granularity}
\label{sec:SMPdesign:Synchronization Granularity}
\epigraph{Doing little things well is a step toward doing big things better.}
	 {Harry F.~Banks}

\begin{figure}
\centering
\resizebox{1.2in}{!}{\includegraphics{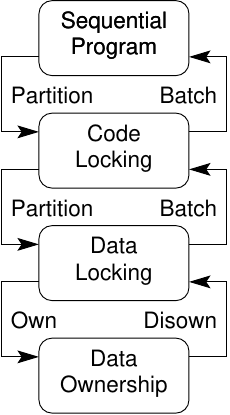}}
\caption{Design Patterns and Lock Granularity}
\label{fig:SMPdesign:Design Patterns and Lock Granularity}
\end{figure}

\Cref{fig:SMPdesign:Design Patterns and Lock Granularity}
gives a pictorial view of different levels of synchronization granularity,
each of which is described in one of the following sections.
These sections focus primarily on locking, but similar granularity
issues arise with all forms of synchronization.

\subsection{Sequential Program}
\label{sec:SMPdesign:Sequential Program}

\begin{figure}
\centering
\resizebox{3in}{!}{\includegraphics{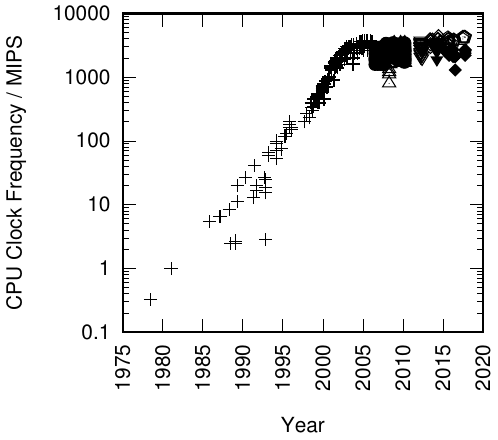}}
\caption{MIPS/Clock-Frequency Trend for Intel CPUs}
\label{fig:SMPdesign:Clock-Frequency Trend for Intel CPUs}
\end{figure}

\begin{figure}
\centering
\resizebox{3in}{!}{\includegraphics{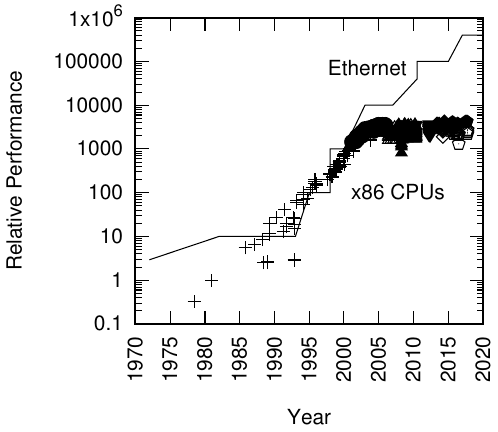}}
\caption{Ethernet Bandwidth vs.\@ Intel x86 CPU Performance}
\label{fig:SMPdesign:Ethernet Bandwidth vs. Intel x86 CPU Performance}
\end{figure}

If the program runs fast enough on a single processor, and
has no interactions with other processes, threads, or interrupt
handlers, you should
remove the synchronization primitives and spare yourself their
overhead and complexity.
Some years back, there were those who would argue that \IXr{Moore's Law}
would eventually force all programs into this category.
However, as can be seen in
\cref{fig:SMPdesign:Clock-Frequency Trend for Intel CPUs},
the exponential increase in single-threaded performance halted in
about 2003.
Therefore,
increasing performance will increasingly require parallelism.\footnote{
	This plot shows clock frequencies for newer CPUs theoretically
	capable of retiring one or more instructions per clock, and MIPS for
	older CPUs requiring multiple clocks to execute even the
	simplest instruction.
	The reason for taking this approach is that the newer CPUs'
	ability to retire multiple instructions per clock is typically
	limited by memory-system performance.}
Given that back in 2006 Paul typed the first version of this sentence
on a dual-core laptop, and further given that many of the graphs added
in 2020 were generated on a system with 56~hardware threads per socket,
parallelism is well and truly here.
It is also important to note that Ethernet bandwidth is continuing to
grow, as shown in
\cref{fig:SMPdesign:Ethernet Bandwidth vs. Intel x86 CPU Performance}.
This growth will continue to motivate multithreaded servers in order to
handle the communications load.

Please note that this does \emph{not} mean that you should code each
and every program in a multi-threaded manner.
Again, if a program runs quickly enough on a single processor,
spare yourself the overhead and complexity of SMP synchronization
primitives.
The simplicity of the hash-table lookup code in
\cref{lst:SMPdesign:Sequential-Program Hash Table Search}
underscores this point.\footnote{
	The examples in this section are taken from Hart et
	al.~\cite{ThomasEHart2006a}, adapted for clarity
	by gathering related code from multiple files.}
A key point is that speedups due to parallelism are normally
limited to the number of CPUs.
In contrast, speedups due to sequential optimizations, for example,
careful choice of data structure, can be arbitrarily large.

\QuickQuiz{
	What should you do to validate a hash table?
}\QuickQuizAnswer{
	Quite a bit, actually.

	See \cref{sec:datastruct:RCU-Protected Hash Table Validation}
	for a good starting point.
}\QuickQuizEnd

On the other hand, if you are not in this happy situation, read on!

\begin{listing}
\begin{VerbatimL}[commandchars=\\\@\$]
struct hash_table
{
	long nbuckets;
	struct node **buckets;
};

typedef struct node {
	unsigned long key;
	struct node *next;
} node_t;

int hash_search(struct hash_table *h, long key)
{
	struct node *cur;

	cur = h->buckets[key % h->nbuckets];
	while (cur != NULL) {
		if (cur->key >= key) {
			return (cur->key == key);
		}
		cur = cur->next;
	}
	return 0;
}
\end{VerbatimL}
\caption{Sequential-Program Hash Table Search}
\label{lst:SMPdesign:Sequential-Program Hash Table Search}
\end{listing}

% ./test_hash_null.exe 1000 0/100 1 1024 1
% ./test_hash_null.exe: nmilli: 1000 update/total: 0/100 nelements: 1 nbuckets: 1024 nthreads: 1
% ./test_hash_null.exe: avg = 96.2913  max = 98.2337  min = 90.4095  std = 2.95314
% ./test_hash_null.exe: nmilli: 1000 update/total: 0/100 nelements: 1 nbuckets: 1024 nthreads: 1
% ./test_hash_null.exe: avg = 91.5592  max = 97.3315  min = 89.9885  std = 2.88925
% ./test_hash_null.exe: nmilli: 1000 update/total: 0/100 nelements: 1 nbuckets: 1024 nthreads: 1
% ./test_hash_null.exe: avg = 93.3568  max = 106.162  min = 89.8828  std = 6.40418

\subsection{Code Locking}
\label{sec:SMPdesign:Code Locking}

\IXh{Code}{locking} is quite simple due to the fact that it uses only
global locks.\footnote{
	If your program instead has locks in data structures,
	or, in the case of Java, uses classes with synchronized
	instances, you are instead using ``data locking'', described
	in \cref{sec:SMPdesign:Data Locking}.}
It is especially
easy to retrofit an existing program to use code locking in
order to run it on a multiprocessor.
If the program has only a single shared resource, code locking
will even give optimal performance.
However, many of the larger and more complex programs
require much of the execution to
occur in \IXpl{critical section}, which in turn causes code locking
to sharply limits their scalability.

Therefore, you should use code locking on programs that spend
only a small fraction of their execution time in critical sections or
from which only modest scaling is required.
In addition, programs that primarily use the more scalable approaches
described in later sections often use code locking to handle rare error
cases or significant state transitions.
In these cases, code locking will provide a relatively simple
program that is very similar to its sequential counterpart,
as can be seen in
\cref{lst:SMPdesign:Code-Locking Hash Table Search}.
However, note that the simple return of the comparison in
\co{hash_search()} in
\cref{lst:SMPdesign:Sequential-Program Hash Table Search}
has now become three statements due to the need to release the
lock before returning.

\begin{listing}
\begin{fcvlabel}[ln:SMPdesign:Code-Locking Hash Table Search]
\begin{VerbatimL}[commandchars=\\\@\$]
spinlock_t hash_lock;

struct hash_table
{
	long nbuckets;
	struct node **buckets;
};

typedef struct node {
	unsigned long key;
	struct node *next;
} node_t;

int hash_search(struct hash_table *h, long key)
{
	struct node *cur;
	int retval;

	spin_lock(&hash_lock);				\lnlbl@acq$
	cur = h->buckets[key % h->nbuckets];
	while (cur != NULL) {
		if (cur->key >= key) {
			retval = (cur->key == key);
			spin_unlock(&hash_lock);	\lnlbl@rel1$
			return retval;
		}
		cur = cur->next;
	}
	spin_unlock(&hash_lock);			\lnlbl@rel2$
	return 0;
}
\end{VerbatimL}
\end{fcvlabel}
\caption{Code-Locking Hash Table Search}
\label{lst:SMPdesign:Code-Locking Hash Table Search}
\end{listing}

\begin{fcvref}[ln:SMPdesign:Code-Locking Hash Table Search]
Note that the \co{hash_lock} acquisition and release statements on
\clnref{acq,rel1,rel2} are mediating ownership of the hash table among
the CPUs wishing to concurrently access that hash table.
Another way of looking at this is that \co{hash_lock} is partitioning
time, thus giving each requesting CPU its own partition of time during
which it owns this hash table.
In addition, in a well-designed algorithm, there should be ample partitions
of time during which no CPU owns this hash table.
\end{fcvref}

\QuickQuiz{
	``Partitioning time''?
	Isn't that an odd turn of phrase?
}\QuickQuizAnswer{
	Perhaps so.

	But in the next section we will be partitioning space (that is,
	address space) as well as time.
	This nomenclature will permit us to partition spacetime, as
	opposed to (say) partitioning space but segmenting time.
}\QuickQuizEnd

Unfortunately, code locking is particularly prone to ``\IX{lock contention}'',
where multiple CPUs need to acquire the lock concurrently.
SMP programmers who have taken care of groups of small children
(or groups of older people who are acting like children) will immediately
recognize the danger of having only one of something,
as illustrated in \cref{fig:SMPdesign:Lock Contention}.

% ./test_hash_codelock.exe 1000 0/100 1 1024 1
% ./test_hash_codelock.exe: nmilli: 1000 update/total: 0/100 nelements: 1 nbuckets: 1024 nthreads: 1
% ./test_hash_codelock.exe: avg = 164.115  max = 170.388  min = 161.659  std = 3.21857
% ./test_hash_codelock.exe: nmilli: 1000 update/total: 0/100 nelements: 1 nbuckets: 1024 nthreads: 1
% ./test_hash_codelock.exe: avg = 181.17  max = 198.4  min = 162.459  std = 15.8585
% ./test_hash_codelock.exe: nmilli: 1000 update/total: 0/100 nelements: 1 nbuckets: 1024 nthreads: 1
% ./test_hash_codelock.exe: avg = 167.651  max = 189.014  min = 162.144  std = 10.6819

% ./test_hash_codelock.exe 1000 0/100 1 1024 2
% ./test_hash_codelock.exe: nmilli: 1000 update/total: 0/100 nelements: 1 nbuckets: 1024 nthreads: 2
% ./test_hash_codelock.exe: avg = 378.481  max = 385.971  min = 374.235  std = 4.05934
% ./test_hash_codelock.exe: nmilli: 1000 update/total: 0/100 nelements: 1 nbuckets: 1024 nthreads: 2
% ./test_hash_codelock.exe: avg = 753.414  max = 1015.28  min = 377.734  std = 294.942
% ./test_hash_codelock.exe: nmilli: 1000 update/total: 0/100 nelements: 1 nbuckets: 1024 nthreads: 2
% ./test_hash_codelock.exe: avg = 502.737  max = 980.924  min = 374.406  std = 239.383

One solution to this problem, named ``data locking'', is described
in the next section.

\begin{figure}
\centering
\resizebox{2.5in}{!}{\includegraphics{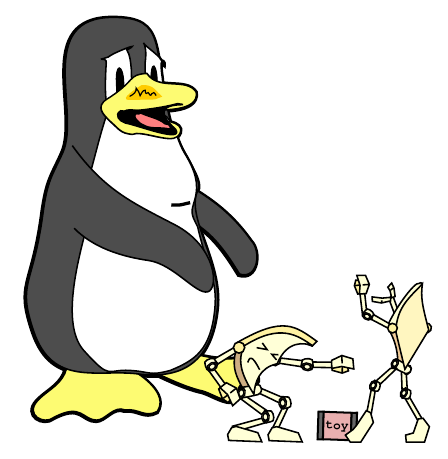}}
\caption{Lock Contention}
\ContributedBy{Figure}{fig:SMPdesign:Lock Contention}{Melissa Broussard}
\end{figure}

\subsection{Data Locking}
\label{sec:SMPdesign:Data Locking}

Many data structures may be partitioned,
with each partition of the data structure having its own lock.
Then the \IXpl{critical section} for each part of the data structure
can execute in parallel,
although only one instance of the critical section for a given
part could be executing at a given time.
You should use data locking when contention must
be reduced, and where synchronization overhead is not
limiting speedups.
Data locking reduces contention by distributing the instances
of the overly-large critical section across multiple data structures,
for example, maintaining per-hash-bucket critical sections in a
hash table, as shown in
\cref{lst:SMPdesign:Data-Locking Hash Table Search}.
The increased scalability again results in a slight increase in complexity
in the form of an additional data structure, the \co{struct bucket}.

\begin{listing}
\begin{VerbatimL}
struct hash_table
{
	long nbuckets;
	struct bucket **buckets;
};

struct bucket {
	spinlock_t bucket_lock;
	node_t *list_head;
};

typedef struct node {
	unsigned long key;
	struct node *next;
} node_t;

int hash_search(struct hash_table *h, long key)
{
	struct bucket *bp;
	struct node *cur;
	int retval;

	bp = h->buckets[key % h->nbuckets];
	spin_lock(&bp->bucket_lock);
	cur = bp->list_head;
	while (cur != NULL) {
		if (cur->key >= key) {
			retval = (cur->key == key);
			spin_unlock(&bp->bucket_lock);
			return retval;
		}
		cur = cur->next;
	}
	spin_unlock(&bp->bucket_lock);
	return 0;
}
\end{VerbatimL}
\caption{Data-Locking Hash Table Search}
\label{lst:SMPdesign:Data-Locking Hash Table Search}
\end{listing}

In contrast with the contentious situation
shown in \cref{fig:SMPdesign:Lock Contention},
data locking helps promote harmony, as illustrated by
\cref{fig:SMPdesign:Data Locking}---and in parallel programs,
this \emph{almost} always translates into increased performance and
scalability.
For this reason, data locking was heavily used by Sequent in its
kernels~\cite{Beck85,Inman85,Garg90,Dove90,McKenney92b,McKenney92a,McKenney93}.

Another way of looking at this is to think of each \co{->bucket_lock}
as mediating ownership not of the entire hash table as was done for code
locking, but only for the bucket corresponding to that \co{->bucket_lock}.
Each lock still partitions time, but the per-bucket-locking technique
also partitions the address space, so that the overall technique can be
said to partition spacetime.
If the number of buckets is large enough, this partitioning of space
should with high probability permit a given CPU immediate access to
a given hash bucket.

\begin{figure}
\centering
\resizebox{2.4in}{!}{\includegraphics{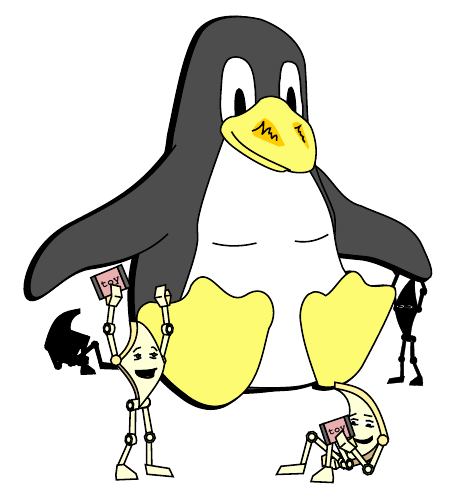}}
\caption{Data Locking}
\ContributedBy{Figure}{fig:SMPdesign:Data Locking}{Melissa Broussard}
\end{figure}

% ./test_hash_spinlock.exe 1000 0/100 1 1024 1
% ./test_hash_spinlock.exe: nmilli: 1000 update/total: 0/100 nelements: 1 nbuckets: 1024 nthreads: 1
% ./test_hash_spinlock.exe: avg = 158.118  max = 162.404  min = 156.199  std = 2.19391
% ./test_hash_spinlock.exe: nmilli: 1000 update/total: 0/100 nelements: 1 nbuckets: 1024 nthreads: 1
% ./test_hash_spinlock.exe: avg = 157.717  max = 162.446  min = 156.415  std = 2.36662
% ./test_hash_spinlock.exe: nmilli: 1000 update/total: 0/100 nelements: 1 nbuckets: 1024 nthreads: 1
% ./test_hash_spinlock.exe: avg = 158.369  max = 164.75  min = 156.501  std = 3.19454

% ./test_hash_spinlock.exe 1000 0/100 1 1024 2
% ./test_hash_spinlock.exe: nmilli: 1000 update/total: 0/100 nelements: 1 nbuckets: 1024 nthreads: 2
% ./test_hash_spinlock.exe: avg = 223.426  max = 422.948  min = 167.858  std = 100.136
% ./test_hash_spinlock.exe: nmilli: 1000 update/total: 0/100 nelements: 1 nbuckets: 1024 nthreads: 2
% ./test_hash_spinlock.exe: avg = 235.462  max = 507.134  min = 167.466  std = 135.836
% ./test_hash_spinlock.exe: nmilli: 1000 update/total: 0/100 nelements: 1 nbuckets: 1024 nthreads: 2
% ./test_hash_spinlock.exe: avg = 305.807  max = 481.685  min = 167.939  std = 132.589

However, as those who have taken care of small children can again attest,
even providing enough to go around is no guarantee of tranquillity.
The analogous situation can arise in SMP programs.
For example, the Linux kernel maintains a cache of files and directories
(called ``dcache'').
Each entry in this cache has its own lock, but the entries corresponding
to the root directory and its direct descendants are much more likely to
be traversed than are more obscure entries.
This can result in many CPUs contending for the locks of these popular
entries, resulting in a situation not unlike that
shown in \cref{fig:SMPdesign:Data and Skew}.

\begin{figure}
\centering
\resizebox{\twocolumnwidth}{!}{\includegraphics{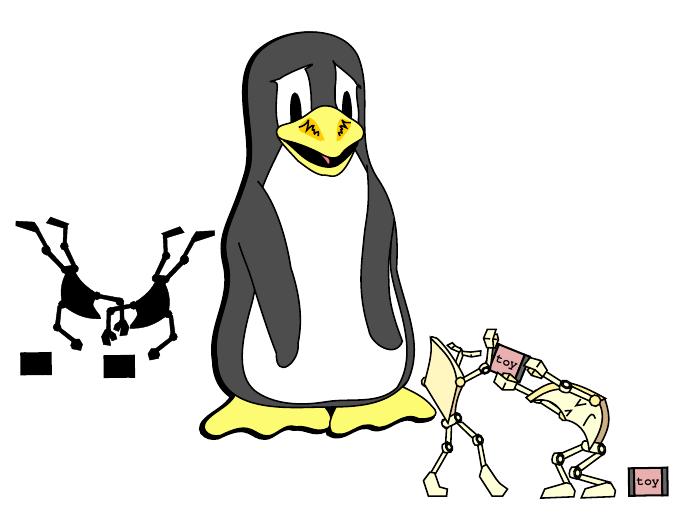}}
\caption{Data Locking and Skew}
\ContributedBy{Figure}{fig:SMPdesign:Data and Skew}{Melissa Broussard}
\end{figure}

In many cases, algorithms can be designed to reduce the instance of
data skew, and in some cases eliminate it entirely
(for example, in the Linux kernel's
dcache~\cite{McKenney04a,JonathanCorbet2010dcacheRCU,NeilBrown2015PathnameLookup,NeilBrown2015RCUwalk,NeilBrown2015PathnameSymlinks}).
Data locking is often used for partitionable data structures such as
hash tables, as well as in situations where multiple entities are each
represented by an instance of a given data structure.
The Linux-kernel task list is an example of the latter, each task
structure having its own \co{alloc_lock} and \co{pi_lock}.

A key challenge with data locking on dynamically allocated structures
is ensuring that the structure remains in existence while the lock is
being acquired~\cite{Gamsa99}.
The code in
\cref{lst:SMPdesign:Data-Locking Hash Table Search}
finesses this challenge by placing the locks in the statically allocated
hash buckets, which are never freed.
However, this trick would not work if the hash table were resizeable,
so that the locks were now dynamically allocated.
In this case, there would need to be some means to prevent the hash
bucket from being freed during the time that its lock was being acquired.

\QuickQuiz{
	What are some ways of preventing a structure from being freed while
	its lock is being acquired?
}\QuickQuizAnswer{
	Here are a few possible solutions to this
	\emph{\IX{existence guarantee}} problem:

	\begin{enumerate}
	\item	Provide a statically allocated lock that is held while
		the per-structure lock is being acquired, which is an
		example of hierarchical locking (see
		\cref{sec:SMPdesign:Hierarchical Locking}).
		Of course, using a single global lock for this purpose
		can result in unacceptably high levels of lock contention,
		dramatically reducing performance and scalability.
	\item	Provide an array of statically allocated locks, hashing
		the structure's address to select the lock to be acquired,
		as described in \cref{chp:Locking}.
		Given a hash function of sufficiently high quality, this
		avoids the scalability limitations of the single global
		lock, but in read-mostly situations, the lock-acquisition
		overhead can result in unacceptably degraded performance.
	\item	Use a garbage collector, in software environments providing
		them, so that a structure cannot be deallocated while being
		referenced.
		This works very well, removing the existence-guarantee
		burden (and much else besides) from the developer's
		shoulders, but imposes the overhead of garbage collection
		on the program.
		Although garbage-collection technology has advanced
		considerably in the past few decades, its overhead
		may be unacceptably high for some applications.
		In addition, some applications require that the developer
		exercise more control over the layout and placement of
		data structures than is permitted by most garbage collected
		environments.
	\item	As a special case of a garbage collector, use a global
		reference counter, or a global array of reference counters.
		These have strengths and limitations similar to those
		called out above for locks.
	\item	Use \emph{\IXpl{hazard pointer}}~\cite{MagedMichael04a}, which
		can be thought of as an inside-out reference count.
		Hazard-pointer-based algorithms maintain a per-thread list of
		pointers, so that the appearance of a given pointer on
		any of these lists acts as a reference to the corresponding
		structure.
		Hazard pointers are starting to see significant production use
		(see \cref{sec:defer:Production Uses of Hazard Pointers}).
	\item	Use transactional memory
		(TM)~\cite{Herlihy93a,DBLomet1977SIGSOFT,Shavit95},
		so that each reference and
		modification to the data structure in question is
		performed atomically.
		Although TM has engendered much excitement in recent years,
		and seems likely to be of some use in production software,
		developers should exercise some
		caution~\cite{Blundell2005DebunkTM,Blundell2006TMdeadlock,McKenney2007PLOSTM},
		particularly in performance-critical code.
		In particular, existence guarantees require that the
		transaction covers the full path from a global reference
		to the data elements being updated.
		For more on TM, including ways to overcome some of its
		weaknesses by combining it with other synchronization
		mechanisms, see
		\cref{sec:future:Transactional Memory,sec:future:Hardware Transactional Memory}.
	\item	Use RCU, which can be thought of as an extremely lightweight
		approximation to a garbage collector.
		Updaters are not permitted to free RCU-protected
		data structures that RCU readers might still be referencing.
		RCU is most heavily used for read-mostly data structures,
		and is discussed at length in
		\cref{sec:defer:Read-Copy Update (RCU)}.
	\end{enumerate}

	For more on providing existence guarantees, see
	\cref{chp:Locking,chp:Deferred Processing}.
}\QuickQuizEnd

\subsection{Data Ownership}
\label{sec:SMPdesign:Data Ownership}

Data ownership partitions a given data structure over the threads
or CPUs, so that each thread/CPU accesses its subset of the data
structure without any synchronization overhead whatsoever.
However, if one thread wishes to access some other thread's data,
the first thread is unable to do so directly.
Instead, the first thread must communicate with the second thread,
so that the second thread performs the operation on behalf of the
first, or, alternatively, migrates the data to the first thread.

Data ownership might seem arcane, but it is used very frequently:
\begin{enumerate}
\item	Any variables accessible by only one CPU or thread
	(such as {\tt auto} variables in C
	and C++) are owned by that CPU or process.
\item	An instance of a user interface owns the corresponding
	user's context.
	It is very common for applications interacting with parallel
	database engines to be written as if they were entirely
	sequential programs.
	Such applications own the user interface and his current
	action.
	Explicit parallelism is thus confined to the
	database engine itself.
\item	Parametric simulations are often trivially parallelized
	by granting each thread ownership of a particular region
	of the parameter space.
	There are also computing frameworks designed for this
	type of problem~\cite{BOINC2008}.
\end{enumerate}

If there is significant sharing, communication between the threads
or CPUs can result in significant complexity and overhead.
Furthermore, if the most-heavily used data happens to be that owned
by a single CPU, that CPU will be a ``\IX{hot spot}'', sometimes with
results resembling that shown in \cref{fig:SMPdesign:Data and Skew}.
However, in situations where no sharing is required, data ownership
achieves ideal performance, and with code that can be as simple
as the sequential-program case shown in
\cref{lst:SMPdesign:Sequential-Program Hash Table Search}.
Such situations are often referred to as ``\IX{embarrassingly
parallel}'', and, in the best case, resemble the situation
previously shown in \cref{fig:SMPdesign:Data Locking}.

% ./test_hash_null.exe 1000 0/100 1 1024 1
% ./test_hash_null.exe: nmilli: 1000 update/total: 0/100 nelements: 1 nbuckets: 1024 nthreads: 1
% ./test_hash_null.exe: avg = 96.2913  max = 98.2337  min = 90.4095  std = 2.95314
% ./test_hash_null.exe: nmilli: 1000 update/total: 0/100 nelements: 1 nbuckets: 1024 nthreads: 1
% ./test_hash_null.exe: avg = 91.5592  max = 97.3315  min = 89.9885  std = 2.88925
% ./test_hash_null.exe: nmilli: 1000 update/total: 0/100 nelements: 1 nbuckets: 1024 nthreads: 1
% ./test_hash_null.exe: avg = 93.3568  max = 106.162  min = 89.8828  std = 6.40418

% ./test_hash_null.exe 1000 0/100 1 1024 2
% ./test_hash_null.exe: nmilli: 1000 update/total: 0/100 nelements: 1 nbuckets: 1024 nthreads: 2
% ./test_hash_null.exe: avg = 45.4526  max = 46.4281  min = 45.1954  std = 0.487791
% ./test_hash_null.exe: nmilli: 1000 update/total: 0/100 nelements: 1 nbuckets: 1024 nthreads: 2
% ./test_hash_null.exe: avg = 46.0238  max = 49.2861  min = 45.1852  std = 1.63127
% ./test_hash_null.exe: nmilli: 1000 update/total: 0/100 nelements: 1 nbuckets: 1024 nthreads: 2
% ./test_hash_null.exe: avg = 46.6858  max = 52.6278  min = 45.1761  std = 2.97102

Another important instance of data ownership occurs when the data
is read-only, in which case,
all threads can ``own'' it via replication.

Where data locking partitions both the address space (with one
hash buckets per partition) and time (using per-bucket locks), data
ownership partitions only the address space.
The reason that data ownership need not partition time is because a
given thread or CPU is assigned permanent ownership of a given
address-space partition.

\QuickQuiz{
	But won't system boot and shutdown (or application startup
	and shutdown) be partitioning time, even for data ownership?
}\QuickQuizAnswer{
	You can indeed think in these terms.

	And if you are working on a persistent data store where state
	survives shutdown, thinking in these terms might even be useful.
}\QuickQuizEnd

Data ownership will be presented in more detail in
\cref{chp:Data Ownership}.

\subsection{Locking Granularity and Performance}
\label{sec:SMPdesign:Locking Granularity and Performance}

This section looks at locking granularity and performance from
a mathematical synchronization\-/efficiency viewpoint.
Readers who are uninspired by mathematics might choose to skip
this section.

The approach is to use a crude queueing model for the \IX{efficiency} of
synchronization mechanism that operate on a single shared global
variable, based on an M/M/1 queue.
M/M/1 queuing models are based on an exponentially distributed
``inter-arrival rate'' $\lambda$ and an exponentially distributed
``service rate'' $\mu$.
The inter-arrival rate $\lambda$ can be thought of as the average
number of synchronization operations per second that the system
would process if the synchronization were free, in other words,
$\lambda$ is an inverse measure of the overhead of each non-synchronization
unit of work.
For example, if each unit of work was a transaction, and if each transaction
took one millisecond to process, excluding synchronization overhead,
then $\lambda$ would be 1,000~transactions per second.

The service rate $\mu$ is defined similarly, but for the average
number of synchronization operations per second that the system
would process if the overhead of each transaction was zero, and
ignoring the fact that CPUs must wait on each other to complete
their synchronization operations, in other words, $\mu$ can be roughly
thought of as the synchronization overhead in absence of contention.
For example, suppose that each transaction's synchronization operation
involves an atomic increment instruction, and that a computer system is
able to do a private-variable atomic increment every 5~nanoseconds on
each CPU
(see \cref{fig:count:Atomic Increment Scalability on x86}).\footnote{
	Of course, if there are 8 CPUs all incrementing the same
	shared variable, then each CPU must wait at least 35~nanoseconds
	for each of the other CPUs to do its increment before consuming
	an additional 5~nanoseconds doing its own increment.
	In fact, the wait will be longer due to the need
	to move the variable from one CPU to another.}
The value of $\mu$ is therefore about 200,000,000 atomic increments
per second.

Of course, the value of $\lambda$ increases as increasing numbers of CPUs
increment a shared variable because each CPU is capable of processing
transactions independently (again, ignoring synchronization):

\begin{equation}
	\lambda = n \lambda_0
\end{equation}

Here, $n$ is the number of CPUs and $\lambda_0$ is the transaction-processing
capability of a single CPU\@.
Note that the expected time for a single CPU to execute a single transaction
in the absence of contention is $1 / \lambda_0$.

Because the CPUs have to ``wait in line'' behind each other to get their
chance to increment the single shared variable, we can use the M/M/1
queueing-model expression for the expected total waiting time:

\begin{equation}
	T = \frac{1}{\mu - \lambda}
\end{equation}

Substituting the above value of $\lambda$:

\begin{equation}
	T = \frac{1}{\mu - n \lambda_0}
\end{equation}

Now, the efficiency is just the ratio of the time required to process
a transaction in absence of synchronization ($1 / \lambda_0$)
to the time required including synchronization ($T + 1 / \lambda_0$):

\begin{equation}
	e = \frac{1 / \lambda_0}{T + 1 / \lambda_0}
\end{equation}

Substituting the above value for $T$ and simplifying:

\begin{equation}
	e = \frac{\frac{\mu}{\lambda_0} - n}{\frac{\mu}{\lambda_0} - (n - 1)}
\end{equation}

But the value of $\mu / \lambda_0$ is just the ratio of the time required
to process the transaction (absent synchronization overhead) to that of
the synchronization overhead itself (absent contention).
If we call this ratio $f$, we have:

\begin{equation}
	e = \frac{f - n}{f - (n - 1)}
\end{equation}

\begin{figure}
\centering
\resizebox{2.5in}{!}{\includegraphics{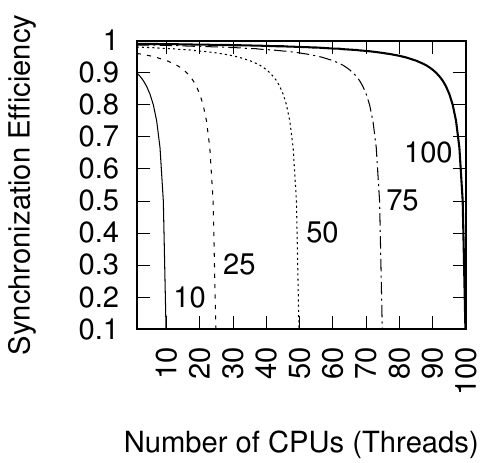}}
\caption{Synchronization Efficiency}
\label{fig:SMPdesign:Synchronization Efficiency}
\end{figure}

\Cref{fig:SMPdesign:Synchronization Efficiency} plots the synchronization
efficiency $e$ as a function of the number of CPUs/threads $n$ for
a few values of the overhead ratio $f$.
For example, again using the 5-nanosecond atomic increment, the $f=10$
line corresponds to each CPU attempting an atomic increment every
50~nanoseconds, and the $f=100$ line corresponds to each CPU attempting
an atomic increment every 500~nanoseconds, which in turn corresponds to
some hundreds (perhaps thousands) of instructions.
Given that each trace drops off sharply with increasing numbers of
CPUs or threads, we can conclude that
synchronization mechanisms based on
atomic manipulation of a single global shared variable will not
scale well if used heavily on current commodity hardware.
This is an abstract mathematical depiction of the forces leading
to the parallel counting algorithms that were discussed in
\cref{chp:Counting}.
Your real-world mileage may differ.

Nevertheless, the concept of efficiency is useful, and even in cases
having little or no formal synchronization.
Consider for example a matrix multiply, in which the columns of one
matrix are multiplied (via ``dot product'') by the rows of another,
resulting in an entry in a third matrix.
Because none of these operations conflict, it is possible to partition
the columns of the first matrix among a group of threads, with each thread
computing the corresponding columns of the result matrix.
The threads can therefore operate entirely independently, with no
synchronization overhead whatsoever, as is done in
\path{matmul.c}.
One might therefore expect a perfect efficiency of 1.0.

\begin{figure}
\centering
\resizebox{2.5in}{!}{\includegraphics{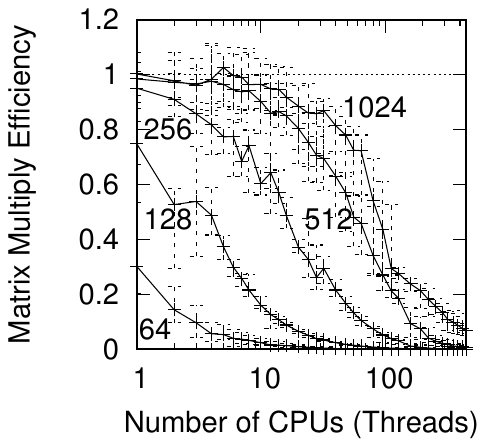}}
\caption{Matrix Multiply Efficiency}
\label{fig:SMPdesign:Matrix Multiply Efficiency}
\end{figure}

However,
\cref{fig:SMPdesign:Matrix Multiply Efficiency}
tells a different story, especially for a 64-by-64 matrix multiply,
which never gets above an efficiency of about 0.3, even when running
single-threaded, and drops sharply as more threads are added.\footnote{
	In contrast to the smooth traces of
	\cref{fig:SMPdesign:Synchronization Efficiency},
	the wide error bars and jagged traces of
	\cref{fig:SMPdesign:Matrix Multiply Efficiency}
	gives evidence of its real-world nature.}
The 128-by-128 matrix does better, but still fails to demonstrate
much performance increase with added threads.
The 256-by-256 matrix does scale reasonably well, but only up to a handful
of CPUs.
The 512-by-512 matrix multiply's efficiency is measurably less
than 1.0 on as few as 10 threads, and even the 1024-by-1024 matrix
multiply deviates noticeably from perfection at a few tens of threads.
Nevertheless, this figure clearly demonstrates the performance and
scalability benefits of batching:
If you must incur synchronization overhead, you may as well get your
money's worth, which is the solution to the problem of deciding on
granularity of synchronization put forth on
\cpageref{sec:SMPdesign:Problems Granularity}.

\QuickQuiz{
	How can a single-threaded 64-by-64 matrix multiple possibly
	have an efficiency of less than 1.0?
	Shouldn't all of the traces in
	\cref{fig:SMPdesign:Matrix Multiply Efficiency}
	have efficiency of exactly 1.0 when running on one thread?
}\QuickQuizAnswer{
	The \path{matmul.c} program creates the specified number of
	worker threads, so even the single-worker-thread case incurs
	thread-creation overhead.
	Making the changes required to optimize away thread-creation
	overhead in the single-worker-thread case is left as an
	exercise to the reader.
}\QuickQuizEnd

Given these inefficiencies,
it is worthwhile to look into more-scalable approaches
such as the data locking described in
\cref{sec:SMPdesign:Data Locking}
or the parallel-fastpath approach discussed in the next section.

\QuickQuizSeries{%
\QuickQuizB{
	How are data-parallel techniques going to help with matrix
	multiply?
	It is \emph{already} data parallel!!!
}\QuickQuizAnswerB{
	I am glad that you are paying attention!
	This example serves to show that although data parallelism can
	be a very good thing, it is not some magic wand that automatically
	wards off any and all sources of inefficiency.
	Linear scaling at full performance, even to ``only'' 64 threads,
	requires care at all phases of design and implementation.

	In particular, you need to pay careful attention to the
	size of the partitions.
	For example, if you split a 64-by-64 matrix multiply across
	64 threads, each thread gets only 64 floating-point multiplies.
	The cost of a floating-point multiply is minuscule compared to
	the overhead of thread creation, and cache-miss overhead
	also plays a role in spoiling the theoretically perfect scalability
	(and also in making the traces so jagged).
	The full 448~hardware threads would require a matrix with
	hundreds of thousands of rows and columns to attain good
	scalability, but by that point GPGPUs become quite attractive,
	especially from a price/performance viewpoint.

	Moral:
	If you have a parallel program with variable input,
	always include a check for the input size being too small to
	be worth parallelizing.
	And when it is not helpful to parallelize, it is not helpful
	to incur the overhead required to spawn a thread, now is it?
}\QuickQuizEndB
\QuickQuizE{
	What did you do to validate this matrix multiply algorithm?
}\QuickQuizAnswerE{
	For this simple approach, very little.

	However, the validation of production-quality matrix multiply
	requires great care and attention.
	Some cases require careful handling of floating-point rounding
	errors, others involve complex sparse-matrix data structures,
	and still others make use of special-purpose arithmetic hardware
	such as vector units or GPGPUs.
	Adequate tests for handling of floating-point rounding errors
	can be especially challenging.
}\QuickQuizEndE
}

\section{Parallel Fastpath}
\label{sec:SMPdesign:Parallel Fastpath}
\epigraph{There are two ways of meeting difficulties:
	  You alter the difficulties, or you alter yourself to meet them.}
	 {Phyllis Bottome}

Fine-grained (and therefore \emph{usually} higher-performance)
designs are typically more complex than are coarser-grained designs.
In many cases, most of the overhead is incurred by a small fraction
of the code~\cite{Knuth73}.
So why not focus effort on that small fraction?

This is the idea behind the parallel-fastpath design pattern, to aggressively
parallelize the common-case code path without incurring the complexity
that would be required to aggressively parallelize the entire algorithm.
You must understand not only the specific algorithm you wish
to parallelize, but also the workload that the algorithm will
be subjected to.
Great creativity and design effort is often required to construct
a parallel fastpath.

Parallel fastpath combines different patterns (one for the
fastpath, one elsewhere) and is therefore a template pattern.
The following instances of parallel
fastpath occur often enough to warrant their own patterns,
as depicted in \cref{fig:SMPdesign:Parallel-Fastpath Design Patterns}:

\begin{figure}
\centering
\resizebox{2.3in}{!}{\includegraphics{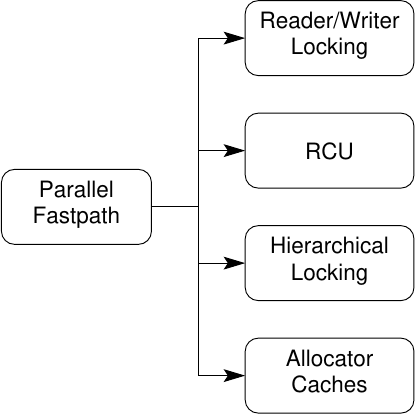}}
\caption{Parallel-Fastpath Design Patterns}
\label{fig:SMPdesign:Parallel-Fastpath Design Patterns}
\end{figure}

\begin{enumerate}
\item	Reader/Writer Locking
	(described below in \cref{sec:SMPdesign:Reader/Writer Locking}).
\item	Read-copy update (RCU), which may be used as a high-performance
	replacement for reader/writer locking, is introduced in
	\cref{sec:defer:Read-Copy Update (RCU)}.
	Other alternatives include hazard pointers
	(\cref{sec:defer:Hazard Pointers})
	and sequence locking (\cref{sec:defer:Sequence Locks}).
	These alternatives will not be discussed further in this chapter.
\item   Hierarchical Locking~(\cite{McKenney95b}), which is touched upon
	in \cref{sec:SMPdesign:Hierarchical Locking}.
\item	Resource Allocator Caches~(\cite{McKenney95b,McKenney93}).
	See \cref{sec:SMPdesign:Resource Allocator Caches}
	for more detail.
\end{enumerate}

\subsection{Reader/Writer Locking}
\label{sec:SMPdesign:Reader/Writer Locking}

If synchronization overhead is negligible (for example, if the program
uses coarse-grained parallelism with large \IXpl{critical section}), and if
only a small fraction of the critical sections modify data, then allowing
multiple readers to proceed in parallel can greatly increase scalability.
Writers exclude both readers and each other.
There are many implementations of reader-writer locking, including
the POSIX implementation described in
\cref{sec:toolsoftrade:POSIX Reader-Writer Locking}.
\Cref{lst:SMPdesign:Reader-Writer-Locking Hash Table Search}
shows how the hash search might be implemented using reader-writer locking.

\begin{listing}
\begin{VerbatimL}[commandchars=\\\@\$]
rwlock_t hash_lock;

struct hash_table
{
	long nbuckets;
	struct node **buckets;
};

typedef struct node {
	unsigned long key;
	struct node *next;
} node_t;

int hash_search(struct hash_table *h, long key)
{
	struct node *cur;
	int retval;

	read_lock(&hash_lock);
	cur = h->buckets[key % h->nbuckets];
	while (cur != NULL) {
		if (cur->key >= key) {
			retval = (cur->key == key);
			read_unlock(&hash_lock);
			return retval;
		}
		cur = cur->next;
	}
	read_unlock(&hash_lock);
	return 0;
}
\end{VerbatimL}
\caption{Reader-Writer-Locking Hash Table Search}
\label{lst:SMPdesign:Reader-Writer-Locking Hash Table Search}
\end{listing}

Reader/writer locking is a simple instance of asymmetric locking.
Snaman~\cite{Snaman87} describes a more ornate six-mode
asymmetric locking design used in several clustered systems.
Locking in general and reader-writer locking in particular is described
extensively in
\cref{chp:Locking}.

\subsection{Hierarchical Locking}
\label{sec:SMPdesign:Hierarchical Locking}

The idea behind hierarchical locking is to have a coarse-grained lock
that is held only long enough to work out which fine-grained lock
to acquire.
\Cref{lst:SMPdesign:Hierarchical-Locking Hash Table Search}
shows how our hash-table search might be adapted to do hierarchical
locking, but also shows the great weakness of this approach:
We have paid the overhead of acquiring a second lock, but we only
hold it for a short time.
In this case, the data-locking approach would be simpler
and likely perform better.

\begin{listing}
\begin{fcvlabel}[ln:SMPdesign:Hierarchical-Locking Hash Table Search]
\begin{VerbatimL}[commandchars=\\\@\$]
struct hash_table
{
	long nbuckets;
	struct bucket **buckets;
};

struct bucket {
	spinlock_t bucket_lock;
	node_t *list_head;
};

typedef struct node {
	spinlock_t node_lock;
	unsigned long key;
	struct node *next;
} node_t;

int hash_search(struct hash_table *h, long key)
{
	struct bucket *bp;
	struct node *cur;
	int retval;

	bp = h->buckets[key % h->nbuckets];
	spin_lock(&bp->bucket_lock);
	cur = bp->list_head;
	while (cur != NULL) {
		if (cur->key >= key) {
			spin_lock(&cur->node_lock);
			spin_unlock(&bp->bucket_lock);
			retval = (cur->key == key);\lnlbl@retval$
			spin_unlock(&cur->node_lock);
			return retval;
		}
		cur = cur->next;
	}
	spin_unlock(&bp->bucket_lock);
	return 0;
}
\end{VerbatimL}
\end{fcvlabel}
\caption{Hierarchical-Locking Hash Table Search}
\label{lst:SMPdesign:Hierarchical-Locking Hash Table Search}
\end{listing}

\QuickQuiz{
	In what situation would hierarchical locking work well?
}\QuickQuizAnswer{
	If the comparison on
	\clnrefr{ln:SMPdesign:Hierarchical-Locking Hash Table Search:retval} of
	\cref{lst:SMPdesign:Hierarchical-Locking Hash Table Search}
	were replaced by a much heavier-weight operation,
	then releasing \co{bp->bucket_lock} \emph{might} reduce lock
	contention enough to outweigh the overhead of the extra
	acquisition and release of \co{cur->node_lock}.
}\QuickQuizEnd

\subsection{Resource Allocator Caches}
\label{sec:SMPdesign:Resource Allocator Caches}

This section presents a simplified schematic of a parallel fixed-block-size
memory allocator.
More detailed descriptions may be found in the
literature~\cite{McKenney92a,McKenney93,Bonwick01slab,McKenney01e,JasonEvans2011jemalloc,ChrisKennelly2020tcmalloc}
or in the Linux kernel~\cite{Torvalds2.6kernel}.

\subsubsection{Parallel Resource Allocation Problem}

The basic problem facing a parallel memory allocator is the tension
between the need to provide extremely fast memory allocation and
freeing in the common case and the need to efficiently distribute
memory in face of unfavorable allocation and freeing patterns.

To see this tension, consider a straightforward application of
data ownership to this problem---simply carve up memory so that
each CPU owns its share.
For example, suppose that a system with 12~CPUs has 64~gigabytes
of memory, for example, the laptop I am using right now.
We could simply assign each CPU a five-gigabyte region of memory,
and allow each CPU to allocate from its own region, without the need
for locking and its complexities and overheads.
Unfortunately, this scheme fails when CPU~0 only allocates memory and
CPU~1 only frees it, as happens in simple producer-consumer workloads.

The other extreme, \IXh{code}{locking}, suffers from excessive
\IX{lock contention}
and overhead~\cite{McKenney93}.

\subsubsection{Parallel Fastpath for Resource Allocation}
\label{sec:SMPdesign:Parallel Fastpath for Resource Allocation}

\begin{figure}
\centering
\resizebox{3in}{!}{\includegraphics{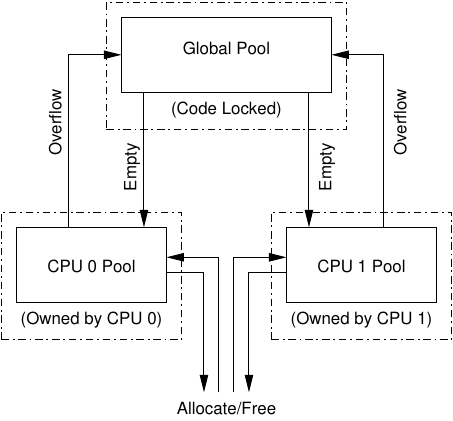}}
\caption{Allocator Cache Schematic}
\label{fig:SMPdesign:Allocator Cache Schematic}
\end{figure}

The commonly used solution uses parallel fastpath with each CPU
owning a modest cache of blocks, and with a large code-locked
shared pool for additional blocks.
To prevent any given CPU from monopolizing the memory blocks,
we place a limit on the number of blocks that can be in each CPU's
cache.
In a two-CPU system, the flow of memory blocks will be as shown
in \cref{fig:SMPdesign:Allocator Cache Schematic}:
When a given CPU is trying to free a block when its pool is full,
it sends blocks to the global pool, and, similarly, when that CPU
is trying to allocate a block when its pool is empty, it retrieves
blocks from the global pool.

\subsubsection{Data Structures}

The actual data structures for a ``toy'' implementation of allocator
caches are shown in
\cref{lst:SMPdesign:Allocator-Cache Data Structures}
(``\path{smpalloc.c}'').
The ``Global Pool'' of \cref{fig:SMPdesign:Allocator Cache Schematic}
is implemented by \co{globalmem} of type \co{struct globalmempool},
and the two CPU pools by the per-thread variable \co{perthreadmem} of
type \co{struct perthreadmempool}.
Both of these data structures have arrays of pointers to blocks
in their \co{pool} fields, which are filled from index zero upwards.
Thus, if \co{globalmem.pool[3]} is \co{NULL}, then the remainder of
the array from index 4 up must also be \co{NULL}.
The \co{cur} fields contain the index of the highest-numbered full
element of the \co{pool} array, or $-1$ if all elements are empty.
All elements from \co{globalmem.pool[0]} through
\co{globalmem.pool[globalmem.cur]} must be full, and all the rest
must be empty.\footnote{
	Both pool sizes (\co{TARGET_POOL_SIZE} and
	\co{GLOBAL_POOL_SIZE}) are unrealistically small, but this small
	size makes it easier to single-step the program in order to get
	a feel for its operation.}

\begin{listing}
\input{CodeSamples/SMPdesign/smpalloc=data_struct.fcv}
\caption{Allocator-Cache Data Structures}
\label{lst:SMPdesign:Allocator-Cache Data Structures}
\end{listing}

\begin{figure}
\centering
\resizebox{2.6in}{!}{\includegraphics{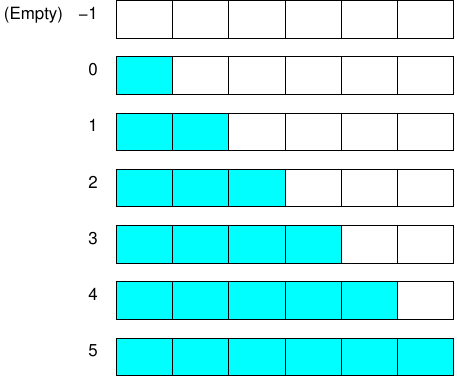}}
\caption{Allocator Pool Schematic}
\label{fig:SMPdesign:Allocator Pool Schematic}
\end{figure}

The operation of the pool data structures is illustrated by
\cref{fig:SMPdesign:Allocator Pool Schematic},
with the six boxes representing the array of pointers making up
the \co{pool} field, and the number preceding them representing
the \co{cur} field.
The shaded boxes represent non-\co{NULL} pointers, while the empty
boxes represent \co{NULL} pointers.
An important, though potentially confusing, invariant of this
data structure is that the \co{cur} field is always one
smaller than the number of non-\co{NULL} pointers.

\subsubsection{Allocation Function}

\begin{fcvref}[ln:SMPdesign:smpalloc:alloc]
The allocation function \co{memblock_alloc()} may be seen in
\cref{lst:SMPdesign:Allocator-Cache Allocator Function}.
\Clnref{pick} picks up the current thread's per-thread pool,
and \clnref{chk:empty} checks to see if it is empty.

If so, \clnrefrange{ack}{rel} attempt to refill it
from the global pool
under the spinlock acquired on \clnref{ack} and released on \clnref{rel}.
\Clnrefrange{loop:b}{loop:e} move blocks from the global
to the per-thread pool until
either the local pool reaches its target size (half full) or
the global pool is exhausted, and \clnref{set} sets the per-thread pool's
count to the proper value.

In either case, \clnref{chk:notempty} checks for the per-thread
pool still being
empty, and if not, \clnrefrange{rem:b}{rem:e} remove a block and return it.
Otherwise, \clnref{ret:NULL} tells the sad tale of memory exhaustion.
\end{fcvref}

\begin{listing}
\input{CodeSamples/SMPdesign/smpalloc=alloc.fcv}
\caption{Allocator-Cache Allocator Function}
\label{lst:SMPdesign:Allocator-Cache Allocator Function}
\end{listing}

\subsubsection{Free Function}

\begin{fcvref}[ln:SMPdesign:smpalloc:free]
\Cref{lst:SMPdesign:Allocator-Cache Free Function} shows
the memory-block free function.
\Clnref{get} gets a pointer to this thread's pool, and
\clnref{chk:full} checks to see if this per-thread pool is full.

If so, \clnrefrange{acq}{empty:e} empty half of the per-thread pool
into the global pool,
with \clnref{acq,rel} acquiring and releasing the spinlock.
\Clnrefrange{loop:b}{loop:e} implement the loop moving blocks
from the local to the
global pool, and \clnref{set} sets the per-thread pool's count to the proper
value.

In either case, \clnref{place} then places the newly freed block into the
per-thread pool.
\end{fcvref}

\begin{listing}
\input{CodeSamples/SMPdesign/smpalloc=free.fcv}
\caption{Allocator-Cache Free Function}
\label{lst:SMPdesign:Allocator-Cache Free Function}
\end{listing}

\QuickQuiz{
	Doesn't this resource-allocator design resemble that of
	the approximate limit counters covered in
	\cref{sec:count:Approximate Limit Counters}?
}\QuickQuizAnswer{
	Indeed it does!
	We are used to thinking of allocating and freeing memory,
	but the algorithms in
	\cref{sec:count:Approximate Limit Counters}
	are taking very similar actions to allocate and free
	``count''.
}\QuickQuizEnd

\subsubsection{Performance}

\begin{figure}
\centering
\resizebox{2.5in}{!}{\includegraphics{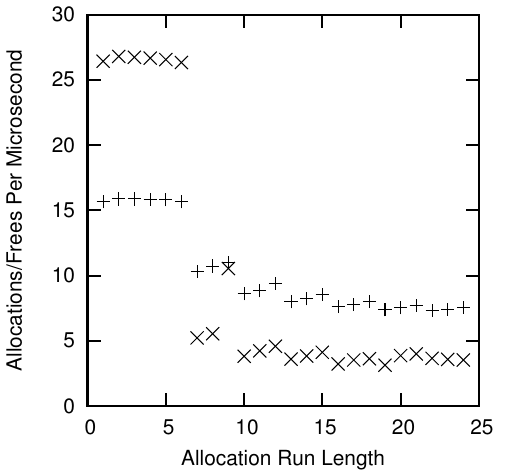}}
\caption{Allocator Cache Performance}
\label{fig:SMPdesign:Allocator Cache Performance}
\end{figure}

Rough performance results\footnote{
	This data was not collected in a statistically meaningful way,
	and therefore should be viewed with great skepticism and suspicion.
	Good data-collection and -reduction practice is discussed
	in \cref{chp:Validation}.
	That said, repeated runs gave similar results, and these results
	match more careful evaluations of similar algorithms.}
are shown in
\cref{fig:SMPdesign:Allocator Cache Performance},
running on a dual-core Intel x86 running at 1\,GHz (4300 bogomips per CPU)
with at most six blocks allowed in each CPU's cache.
In this micro-benchmark,
each thread repeatedly allocates a group of blocks and then frees all
the blocks in that group, with
the number of blocks in the group being the ``allocation run length''
displayed on the x-axis.
The y-axis shows the number of successful allocation/free pairs per
microsecond---failed allocations are not counted.
The ``X''s are from a two-thread run, while the ``+''s are from a
single-threaded run.

Note that run lengths up to six scale linearly and give excellent performance,
while run lengths greater than six show poor performance and almost always
also show \emph{negative} scaling.
It is therefore quite important to size \co{TARGET_POOL_SIZE}
sufficiently large,
which fortunately is usually quite easy to do in actual
practice~\cite{McKenney01e}, especially given today's large memories.
For example, in most systems, it is quite reasonable to set
\co{TARGET_POOL_SIZE} to 100, in which case allocations and frees
are guaranteed to be confined to per-thread pools at least 99\,\% of
the time.

As can be seen from the figure, the situations where the common-case
data-ownership applies (run lengths up to six) provide greatly improved
performance compared to the cases where locks must be acquired.
Avoiding synchronization in the common case will be a recurring theme through
this book.

\QuickQuizSeries{%
\QuickQuizB{
	In \cref{fig:SMPdesign:Allocator Cache Performance},
	there is a pattern of performance rising with increasing run
	length in groups of three samples, for example, for run lengths
	10, 11, and 12.
	Why?
}\QuickQuizAnswerB{
	This is due to the per-CPU target value being three.
	A run length of 12 must acquire the global-pool lock twice,
	while a run length of 13 must acquire the global-pool lock
	three times.
}\QuickQuizEndB
\QuickQuizE{
	Allocation failures were observed in the two-thread
	tests at run lengths of 19 and greater.
	Given the global-pool size of 40 and the per-thread target
	pool size $s$ of three, number of threads $n$ equal to two,
	and assuming that the per-thread pools are initially
	empty with none of the memory in use, what is the smallest allocation
	run length $m$ at which failures can occur?
	(Recall that each thread repeatedly allocates $m$ block of memory,
	and then frees the $m$ blocks of memory.)
	Alternatively, given $n$ threads each with pool size $s$, and
	where each thread repeatedly first allocates $m$ blocks of memory
	and then frees those $m$ blocks, how large must the global pool
	size be?
	\emph{Note:}
	Obtaining the correct answer will require you to
	examine the \path{smpalloc.c} source code, and very likely
	single-step it as well.
	You have been warned!
}\QuickQuizAnswerE{
	This solution is adapted from one put forward by Alexey Roytman.
	It is based on the following definitions:

	\begin{description}
	\item[$g$]	Number of blocks globally available.
	\item[$i$]	Number of blocks left in the initializing thread's
			per-thread pool.
			(This is one reason you needed to look at the code!)
	\item[$m$]	Allocation/free run length.
	\item[$n$]	Number of threads, excluding the initialization thread.
	\item[$p$]	Per-thread maximum block consumption, including
			both the blocks actually allocated and the blocks
			remaining in the per-thread pool.
	\end{description}

	The values $g$, $m$, and $n$ are given.
	The value for $p$ is $m$ rounded up to the next multiple of $s$,
	as follows:

	\begin{equation}
		p = s \left \lceil \frac{m}{s} \right \rceil
	\label{sec:SMPdesign:p}
	\end{equation}

	The value for $i$ is as follows:

	\begin{equation}
		i = \left \{
			\begin{array}{l}
				g \pmod{2 s} = 0: 2 s \\
				g \pmod{2 s} \ne 0: g \pmod{2 s}
			\end{array}
		    \right .
	\label{sec:SMPdesign:i}
	\end{equation}

	\begin{figure}
	\centering
	\resizebox{3in}{!}{\includegraphics{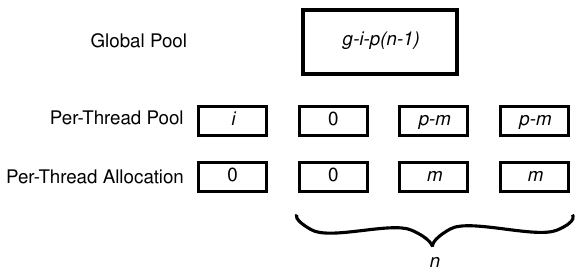}}
	\caption{Allocator Cache Run-Length Analysis}
	\label{fig:SMPdesign:Allocator Cache Run-Length Analysis}
	\end{figure}

	The relationships between these quantities are shown in
	\cref{fig:SMPdesign:Allocator Cache Run-Length Analysis}.
	The global pool is shown on the top of this figure, and
	the ``extra'' initializer thread's per-thread pool and
	per-thread allocations are the left-most pair of boxes.
	The initializer thread has no blocks allocated, but has
	$i$ blocks stranded in its per-thread pool.
	The rightmost two pairs of boxes are the per-thread pools and
	per-thread allocations of threads holding the maximum possible
	number of blocks, while the second-from-left pair of boxes
	represents the thread currently trying to allocate.

	The total number of blocks is $g$, and adding up the per-thread
	allocations and per-thread pools, we see that the global pool
	contains $g-i-p(n-1)$ blocks.
	If the allocating thread is to be successful, it needs at least
	$m$ blocks in the global pool, in other words:

	\begin{equation}
		g - i - p(n - 1) \ge m
	\label{sec:SMPdesign:g-vs-m}
	\end{equation}

	The question has $g=40$, $s=3$, and $n=2$.
	\Cref{sec:SMPdesign:i} gives $i=4$, and
	\cref{sec:SMPdesign:p} gives $p=18$ for $m=18$
	and $p=21$ for $m=19$.
	Plugging these into \cref{sec:SMPdesign:g-vs-m}
	shows that $m=18$ will not overflow, but that $m=19$ might
	well do so.

	The presence of $i$ could be considered to be a bug.
	After all, why allocate memory only to have it stranded in
	the initialization thread's cache?
	One way of fixing this would be to provide a \co{memblock_flush()}
	function that flushed the current thread's pool into the
	global pool.
	The initialization thread could then invoke this function
	after freeing all of the blocks.
}\QuickQuizEndE
}

\subsubsection{Validation}

Validation of this simple allocator spawns a specified number of threads,
with each thread repeatedly allocating a specified number of memory
blocks and then deallocating them.
This simple regimen suffices to exercise both the per-thread caches
and the global pool, as can be seen in
\cref{fig:SMPdesign:Allocator Cache Performance}.

Much more aggressive validation is required for memory allocators that
are to be used in production.
The test suites for tcmalloc~\cite{ChrisKennelly2020tcmalloc} and
jemalloc~\cite{JasonEvans2011jemalloc} are instructive, as are the
tests for the Linux kernel's memory allocator.

\subsubsection{Real-World Design}

The toy parallel resource allocator was quite simple, but real-world
designs expand on this approach in a number of ways.

First, real-world allocators are required to handle a wide range
of allocation sizes, as opposed to the single size shown in this
toy example.
One popular way to do this is to offer a fixed set of sizes, spaced
so as to balance external and internal \IX{fragmentation}, such as in
the late-1980s BSD memory allocator~\cite{McKusick88}.
Doing this would mean that the ``globalmem'' variable would need
to be replicated on a per-size basis, and that the associated
lock would similarly be replicated, resulting in \IXh{data}{locking}
rather than the toy program's code locking.

Second, production-quality systems must be able to repurpose memory,
meaning that they must be able to coalesce blocks into larger structures,
such as pages~\cite{McKenney93}.
This coalescing will also need to be protected by a lock, which again
could be replicated on a per-size basis.

Third, coalesced memory must be returned to the underlying memory
system, and pages of memory must also be allocated from the underlying
memory system.
The locking required at this level will depend on that of the underlying
memory system, but could well be code locking.
Code locking can often be tolerated at this level, because this
level is so infrequently reached in well-designed systems~\cite{McKenney01e}.

Concurrent userspace allocators face similar
challenges~\cite{ChrisKennelly2020tcmalloc,JasonEvans2011jemalloc}.

Despite this real-world design's greater complexity, the underlying
idea is the same---repeated application of parallel fastpath,
as shown in
\cref{fig:app:questions:Schematic of Real-World Parallel Allocator}.

And ``parallel fastpath'' is one of the solutions to the non-partitionable
application problem put forth on
\cpageref{sec:SMPdesign:Problems Parallel Fastpath}.

\begin{table}
\rowcolors{1}{}{lightgray}
\small
\centering
\renewcommand*{\arraystretch}{1.25}
\begin{tabularx}{\twocolumnwidth}{ll>{\raggedright\arraybackslash}X}
\toprule
Level	& Locking & Purpose \\
\midrule
Per-thread pool	  & Data ownership & High-speed allocation \\
Global block pool & Data locking   & Distributing blocks among threads \\
Coalescing	  & Data locking   & Combining blocks into pages \\
System memory	  & Code locking   & Memory from/to system \\
\bottomrule
\end{tabularx}
\caption{Schematic of Real-World Parallel Allocator}
\label{fig:app:questions:Schematic of Real-World Parallel Allocator}
\end{table}

% SMPdesign/beyond.tex
% mainfile: ../perfbook.tex
% SPDX-License-Identifier: CC-BY-SA-3.0

\section{Beyond Partitioning}
\label{sec:SMPdesign:Beyond Partitioning}
\OriginallyPublished{Section}{sec:SMPdesign:Beyond Partitioning}{Retrofitted Parallelism Considered Grossly Sub-Optimal}{4\textsuperscript{th} USENIX Workshop on Hot Topics on Parallelism}{PaulEMcKenney2012HOTPARsuboptimal}
\epigraph{It is all right to aim high if you have plenty of ammunition.}
	 {Hawley R. Everhart}

This chapter has discussed how data partitioning can be used to design
simple linearly scalable parallel programs.
\Cref{sec:SMPdesign:Data Ownership} hinted at the possibilities
of data replication, which will be used to great effect in
\cref{sec:defer:Read-Copy Update (RCU)}.

The main goal of applying partitioning and replication is to achieve
linear speedups, in other words, to ensure that the total amount of
work required does not increase significantly as the number of CPUs
or threads increases.
A problem that can be solved via partitioning and/or replication,
resulting in linear speedups, is \emph{\IX{embarrassingly parallel}}.
But can we do better?

To answer this question, let us examine the solution of
labyrinths and mazes.
Of course, labyrinths and mazes have been objects of fascination for
millennia~\cite{WikipediaLabyrinth},
so it should come as no surprise that they are generated and solved
using computers, including biological
computers~\cite{AndrewAdamatzky2011SlimeMold},
GPGPUs~\cite{ChristerEricson2008GPUMaze}, and even
discrete hardware~\cite{MIT:TRMag:MemristorMazes}.
Parallel solution of mazes is sometimes used as a class project in
universities~\cite{ETHZurich:FS2011maze,UMD:CMSC433maze} and
as a vehicle to demonstrate the benefits of parallel-programming
frameworks~\cite{RonFosner2010maze}.

Common advice is to use a parallel work-queue algorithm
(PWQ)~\cite{ETHZurich:FS2011maze,RonFosner2010maze}.
This section evaluates this advice by comparing PWQ
against a sequential algorithm (SEQ) and also against
an alternative parallel algorithm, in all cases solving randomly generated
square mazes.
\Cref{sec:SMPdesign:Work-Queue Parallel Maze Solver} discusses PWQ,
\cref{sec:SMPdesign:Alternative Parallel Maze Solver} discusses an alternative
parallel algorithm,
\cref{sec:SMPdesign:Performance Comparison I} analyzes its anomalous performance,
\cref{sec:SMPdesign:Alternative Sequential Maze Solver} derives an improved
sequential algorithm from the alternative parallel algorithm,
\cref{sec:SMPdesign:Performance Comparison II} makes further performance
comparisons,
and finally
\cref{sec:SMPdesign:Future Directions and Conclusions}
presents future directions and concluding remarks.

\subsection{Work-Queue Parallel Maze Solver}
\label{sec:SMPdesign:Work-Queue Parallel Maze Solver}

PWQ is based on SEQ, which is shown in
\cref{lst:SMPdesign:SEQ Pseudocode}
(pseudocode for \path{maze_seq.c}).
The maze is represented by a 2D array of cells and
a linear-array-based work queue named \co{->visited}.

\begin{listing}
\begin{fcvlabel}[ln:SMPdesign:SEQ Pseudocode]
\begin{VerbatimL}[commandchars=\\\@\$]
int maze_solve(maze *mp, cell sc, cell ec)
{
	cell c = sc;
	cell n;
	int vi = 0;

	maze_try_visit_cell(mp, c, c, &n, 1);		\lnlbl@initcell$
	for (;;) {					\lnlbl@loop:b$
		while (!maze_find_any_next_cell(mp, c, &n)) {\lnlbl@loop2:b$
			if (++vi >= mp->vi)		\lnlbl@ifge$
				return 0;
			c = mp->visited[vi].c;
		}					\lnlbl@loop2:e$
		do {					\lnlbl@loop3:b$
			if (n == ec) {
				return 1;
			}
			c = n;
		} while (maze_find_any_next_cell(mp, c, &n));\lnlbl@loop3:e$
		c = mp->visited[vi].c;			\lnlbl@finalize$
	}						\lnlbl@loop:e$
}
\end{VerbatimL}
\end{fcvlabel}
\caption{SEQ Pseudocode}
\label{lst:SMPdesign:SEQ Pseudocode}
\end{listing}

\begin{fcvref}[ln:SMPdesign:SEQ Pseudocode]
\Clnref{initcell} visits the initial cell, and each iteration of the loop spanning
\clnrefrange{loop:b}{loop:e} traverses passages headed by one cell.
The loop spanning
\clnrefrange{loop2:b}{loop2:e} scans the \co{->visited[]} array for a
visited cell with an unvisited neighbor, and the loop spanning
\clnrefrange{loop3:b}{loop3:e} traverses one fork of the submaze
headed by that neighbor.
\Clnref{finalize} initializes for the next pass through the outer loop.
\end{fcvref}

\begin{listing}
\begin{fcvlabel}[ln:SMPdesign:SEQ Helper Pseudocode]
\begin{VerbatimL}[commandchars=\\\@\$]
int maze_try_visit_cell(struct maze *mp, cell c, cell t, \lnlbl@try:b$
                        cell *n, int d)
{
	if (!maze_cells_connected(mp, c, t) ||	\lnlbl@try:chk:adj$
	    (*celladdr(mp, t) & VISITED))	\lnlbl@try:chk:not:visited$
		return 0;			\lnlbl@try:ret:failure$
	*n = t;					\lnlbl@try:nextcell$
	mp->visited[mp->vi] = t;		\lnlbl@try:recordnext$
	mp->vi++;				\lnlbl@try:next:visited$
	*celladdr(mp, t) |= VISITED | d;	\lnlbl@try:mark:visited$
	return 1;				\lnlbl@try:ret:success$
}						\lnlbl@try:e$

int maze_find_any_next_cell(struct maze *mp, cell c, \lnlbl@find:b$
                            cell *n)
{
	int d = (*celladdr(mp, c) & DISTANCE) + 1;	\lnlbl@find:curplus1$

	if (maze_try_visit_cell(mp, c, prevcol(c), n, d))\lnlbl@find:chk:prevcol$
		return 1;				\lnlbl@find:ret:prevcol$
	if (maze_try_visit_cell(mp, c, nextcol(c), n, d))\lnlbl@find:chk:nextcol$
		return 1;				\lnlbl@find:ret:nextcol$
	if (maze_try_visit_cell(mp, c, prevrow(c), n, d))\lnlbl@find:chk:prevrow$
		return 1;				\lnlbl@find:ret:prevrow$
	if (maze_try_visit_cell(mp, c, nextrow(c), n, d))\lnlbl@find:chk:nextrow$
		return 1;				\lnlbl@find:ret:nextrow$
	return 0;					\lnlbl@find:ret:false$
}							\lnlbl@find:e$
\end{VerbatimL}
\end{fcvlabel}
\caption{SEQ Helper Pseudocode}
\label{lst:SMPdesign:SEQ Helper Pseudocode}
\end{listing}

\begin{fcvref}[ln:SMPdesign:SEQ Helper Pseudocode:try]
The pseudocode for \co{maze_try_visit_cell()} is shown on
\clnrefrange{b}{e}
of \cref{lst:SMPdesign:SEQ Helper Pseudocode}
(\path{maze.c}).
\Clnref{chk:adj} checks to see if cells \co{c} and \co{t} are
adjacent and connected,
while \clnref{chk:not:visited} checks to see if cell \co{t} has
not yet been visited.
The \co{celladdr()} function returns the address of the specified cell.
If either check fails, \clnref{ret:failure} returns failure.
\Clnref{nextcell} indicates the next cell,
\clnref{recordnext} records this cell in the next
slot of the \co{->visited[]} array,
\clnref{next:visited} indicates that this slot
is now full, and \clnref{mark:visited} marks this cell as visited and also records
the distance from the maze start.
\Clnref{ret:success} then returns success.
\end{fcvref}

\begin{fcvref}[ln:SMPdesign:SEQ Helper Pseudocode:find]
The pseudocode for \co{maze_find_any_next_cell()} is shown on
\clnrefrange{b}{e}
of \cref{lst:SMPdesign:SEQ Helper Pseudocode}
(\path{maze.c}).
\Clnref{curplus1} picks up the current cell's distance plus 1,
while \clnref{chk:prevcol,chk:nextcol,chk:prevrow,chk:nextrow}
check the cell in each direction, and
\clnref{ret:prevcol,ret:nextcol,ret:prevrow,ret:nextrow}
return true if the corresponding cell is a candidate next cell.
The \co{prevcol()}, \co{nextcol()}, \co{prevrow()}, and \co{nextrow()}
each do the specified array-index-conversion operation.
If none of the cells is a candidate, \clnref{ret:false} returns false.
\end{fcvref}

\begin{figure}
\centering
\resizebox{1.2in}{!}{\includegraphics{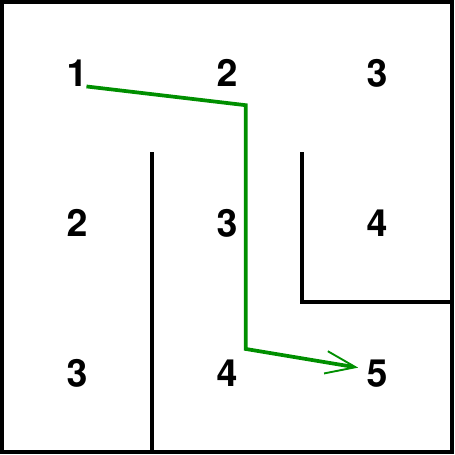}}
\caption{Cell-Number Solution Tracking}
\label{fig:SMPdesign:Cell-Number Solution Tracking}
\end{figure}

The path is recorded in the maze by counting the number of cells from
the starting point, as shown in
\cref{fig:SMPdesign:Cell-Number Solution Tracking},
where the starting cell is in the upper left and the ending cell is
in the lower right.
Starting at the ending cell and following
consecutively decreasing cell numbers traverses the solution.

\begin{figure}
\centering
\resizebox{2.2in}{!}{\includegraphics{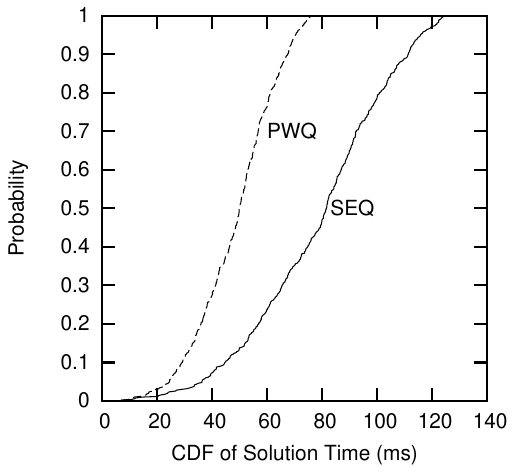}}
\caption{CDF of Solution Times For SEQ and PWQ}
\label{fig:SMPdesign:CDF of Solution Times For SEQ and PWQ}
\end{figure}

The parallel work-queue solver is a straightforward parallelization
of the algorithm shown in
\cref{lst:SMPdesign:SEQ Pseudocode,lst:SMPdesign:SEQ Helper Pseudocode}.
\begin{fcvref}[ln:SMPdesign:SEQ Pseudocode]
\Clnref{ifge} of \cref{lst:SMPdesign:SEQ Pseudocode} must use fetch-and-add,
and the local variable \co{vi} must be shared among the various threads.
\end{fcvref}
\begin{fcvref}[ln:SMPdesign:SEQ Helper Pseudocode:try]
\Clnref{chk:not:visited,mark:visited} of
\cref{lst:SMPdesign:SEQ Helper Pseudocode} must be
combined into a CAS loop, with CAS failure indicating a loop in the
maze.
\Clnrefrange{recordnext}{next:visited} of this listing must use
fetch-and-add to arbitrate concurrent
attempts to record cells in the \co{->visited[]} array.
\end{fcvref}

This approach does provide significant speedups on a dual-CPU
Lenovo W500 running at 2.53\,GHz, as shown in
\cref{fig:SMPdesign:CDF of Solution Times For SEQ and PWQ},
which shows the cumulative distribution functions (CDFs) for the solution
times of the two algorithms, based on the solution of 500 different square
500-by-500 randomly generated mazes.
The substantial overlap
of the projection of the CDFs onto the x-axis will be addressed in
\cref{sec:SMPdesign:Performance Comparison I}.

Interestingly enough, the sequential solution-path tracking works unchanged
for the parallel algorithm.
However, this uncovers a significant weakness in the parallel algorithm:
At most one thread may be making progress along the solution path at
any given time.
This weakness is addressed in the next section.

\subsection{Alternative Parallel Maze Solver}
\label{sec:SMPdesign:Alternative Parallel Maze Solver}

Youthful maze solvers are often urged to start at both ends, and
this advice has been repeated more recently in the context of automated
maze solving~\cite{UMD:CMSC433maze}.
This advice amounts to partitioning, which has been a powerful
parallelization strategy
in the context of parallel programming for both operating-system
kernels~\cite{Beck85,Inman85} and
applications~\cite{DavidAPatterson2010TroubleMulticore}.
This section applies this strategy, using two child threads that start
at opposite ends of the solution path, and takes a brief look at the
performance and scalability consequences.

\begin{listing}
\begin{fcvlabel}[ln:SMPdesign:Partitioned Parallel Solver Pseudocode]
\begin{VerbatimL}[commandchars=\\\@\$]
int maze_solve_child(maze *mp, cell *visited, cell sc)	\lnlbl@b$
{
	cell c;
	cell n;
	int vi = 0;

	myvisited = visited; myvi = &vi;		\lnlbl@store:ptr$
	c = visited[vi];				\lnlbl@retrieve$
	do {
		while (!maze_find_any_next_cell(mp, c, &n)) {
			if (visited[++vi].row < 0)
				return 0;
			if (READ_ONCE(mp->done))	\lnlbl@chk:done1$
				return 1;
			c = visited[vi];
		}
		do {
			if (READ_ONCE(mp->done))	\lnlbl@chk:done2$
				return 1;
			c = n;
		} while (maze_find_any_next_cell(mp, c, &n));
		c = visited[vi];
	} while (!READ_ONCE(mp->done));		\lnlbl@chk:done3$
	return 1;
}
\end{VerbatimL}
\end{fcvlabel}
\caption{Partitioned Parallel Solver Pseudocode}
\label{lst:SMPdesign:Partitioned Parallel Solver Pseudocode}
\end{listing}

\begin{fcvref}[ln:SMPdesign:Partitioned Parallel Solver Pseudocode]
The partitioned parallel algorithm (PART), shown in
\cref{lst:SMPdesign:Partitioned Parallel Solver Pseudocode}
(\path{maze_part.c}),
is similar to SEQ, but has a few important differences.
First, each child thread has its own \co{visited} array, passed in by
the parent as shown on \clnref{b},
which must be initialized to all [$-1$, $-1$].
\Clnref{store:ptr} stores a pointer to this array into the per-thread variable
\co{myvisited} to allow access by helper functions, and similarly stores
a pointer to the local visit index.
Second, the parent visits the first cell on each child's behalf,
which the child retrieves on \clnref{retrieve}.
Third, the maze is solved as soon as one child locates a cell that has
been visited by the other child.
When \co{maze_try_visit_cell()} detects this,
it sets a \co{->done} field in the maze structure.
Fourth, each child must therefore periodically check the \co{->done}
field, as shown on \clnref{chk:done1,chk:done2,chk:done3}.
The \co{READ_ONCE()} primitive must disable any compiler
optimizations that might combine consecutive loads or that
might reload the value.
A C++1x volatile relaxed load suffices~\cite{RichardSmith2019N4800}.
Finally, the \co{maze_find_any_next_cell()} function must use
compare-and-swap to mark a cell as visited, however
no constraints on ordering are required beyond those provided by
thread creation and join.
\end{fcvref}

\begin{listing}
\begin{fcvlabel}[ln:SMPdesign:Partitioned Parallel Helper Pseudocode]
\begin{VerbatimL}[commandchars=\\\@\$]
int maze_try_visit_cell(struct maze *mp, int c, int t,
                        int *n, int d)
{
	cell_t t;
	cell_t *tp;
	int vi;

	if (!maze_cells_connected(mp, c, t))		\lnlbl@chk:conn:b$
		return 0;				\lnlbl@chk:conn:e$
	tp = celladdr(mp, t);
	do {						\lnlbl@loop:b$
		t = READ_ONCE(*tp);
		if (t & VISITED) {			\lnlbl@chk:visited$
			if ((t & TID) != mytid)		\lnlbl@chk:other$
				mp->done = 1;		\lnlbl@located$
			return 0;			\lnlbl@ret:fail$
		}
	} while (!CAS(tp, t, t | VISITED | myid | d));	\lnlbl@loop:e$
	*n = t;						\lnlbl@update:new$
	vi = (*myvi)++;					\lnlbl@update:visited:b$
	myvisited[vi] = t;				\lnlbl@update:visited:e$
	return 1;					\lnlbl@ret:success$
}
\end{VerbatimL}
\end{fcvlabel}
\caption{Partitioned Parallel Helper Pseudocode}
\label{lst:SMPdesign:Partitioned Parallel Helper Pseudocode}
\end{listing}

\begin{fcvref}[ln:SMPdesign:Partitioned Parallel Helper Pseudocode]
The pseudocode for \co{maze_find_any_next_cell()} is identical to that shown in
\cref{lst:SMPdesign:SEQ Helper Pseudocode},
but the pseudocode for \co{maze_try_visit_cell()} differs, and
is shown in
\cref{lst:SMPdesign:Partitioned Parallel Helper Pseudocode}.
\Clnrefrange{chk:conn:b}{chk:conn:e}
check to see if the cells are connected, returning failure
if not.
The loop spanning \clnrefrange{loop:b}{loop:e} attempts to mark
the new cell visited.
\Clnref{chk:visited} checks to see if it has already been visited, in which case
\clnref{ret:fail} returns failure, but only after \clnref{chk:other}
checks to see if
we have encountered the other thread, in which case \clnref{located} indicates
that the solution has been located.
\Clnref{update:new} updates to the new cell,
\clnref{update:visited:b,update:visited:e} update this thread's visited
array, and \clnref{ret:success} returns success.
\end{fcvref}

\begin{figure}
\centering
\resizebox{2.2in}{!}{\includegraphics{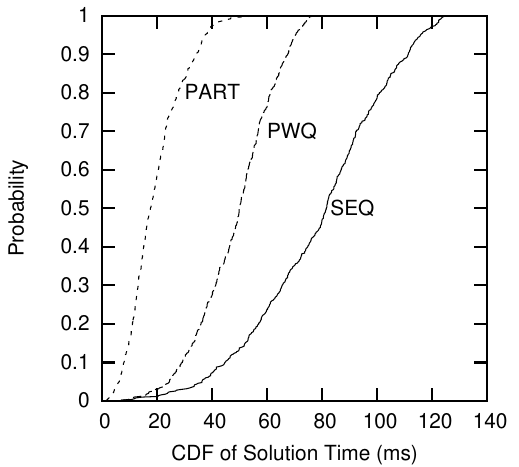}}
\caption{CDF of Solution Times For SEQ, PWQ, and PART}
\label{fig:SMPdesign:CDF of Solution Times For SEQ; PWQ; and PART}
\end{figure}

Performance testing revealed a surprising anomaly, shown in
\cref{fig:SMPdesign:CDF of Solution Times For SEQ; PWQ; and PART}.
The median solution time for PART (17 milliseconds)
is more than four times faster than that of SEQ (79 milliseconds),
despite running on only two threads.

The first reaction to such a dramatic performance anomaly is to check
for bugs, which suggests stringent validation be applied.
This is the topic of the next section.

\subsection{Maze Validation}
\label{sec:SMPdesign:Maze Validation}

Much of the validation effort comprised consistency checks, which
can be located by searching for \co{ABORT()} in
\path{CodeSamples/SMPdesign/maze/*.c}.
Examples checks include:

\begin{enumerate}
\item	Maze solution steps that end up outside of the maze.
\item	Mazes that suddenly have zero or fewer rows or columns.
\item	Newly created mazes with unreachable cells.
\item	Mazes that have no solution.
\item	Discontinuous maze solutions.
\item	Attempts to start the maze solver outside of the maze.
\item	Mazes whose solution path is longer than the number of cells
	in the maze.
\item	Subsolutions by different threads cross each other.
\item	Memory-allocation failure.
\item	System-call failure.
\end{enumerate}

Additional manual validation was applied by Paul's wife, who greatly
enjoys solving puzzles.

However, if this maze software was to be used in production, whatever
that might mean, it would be wise to construct an independent maze
\co{fsck} program.
Nevertheless, the mazes and solutions all proved to be quite valid.
The next section therefore more deeply analyzes the scalability anomaly
called out in \cref{sec:SMPdesign:Alternative Parallel Maze Solver}.

\subsection{Performance Comparison I}
\label{sec:SMPdesign:Performance Comparison I}

\begin{figure}
\centering
\resizebox{2.2in}{!}{\includegraphics{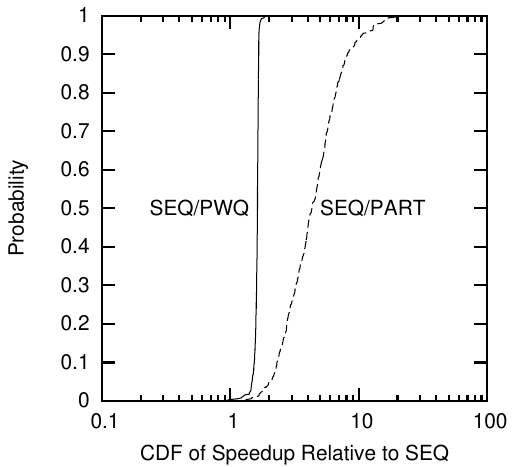}}
\caption{CDF of SEQ/PWQ and SEQ/PART Solution-Time Ratios}
\label{fig:SMPdesign:CDF of SEQ/PWQ and SEQ/PART Solution-Time Ratios}
\end{figure}

Although the algorithms were in fact finding valid solutions to valid
mazes, the plot of CDFs in
\cref{fig:SMPdesign:CDF of Solution Times For SEQ; PWQ; and PART}
assumes independent data points.
This is not the case:
The performance tests randomly generate a maze,
and then run all solvers on that maze.
It therefore makes sense to plot the CDF of the ratios of
solution times for each generated maze,
as shown in
\cref{fig:SMPdesign:CDF of SEQ/PWQ and SEQ/PART Solution-Time Ratios},
greatly reducing the CDFs' overlap.
This plot reveals that for some mazes, PART
is more than \emph{forty} times faster than SEQ\@.
In contrast, PWQ is never more than about
two times faster than SEQ\@.
A forty-times speedup on two threads demands explanation.
After all, this is not merely embarrassingly parallel, where partitionability
means that adding threads does not increase the overall computational cost.
It is instead \emph{\IX{humiliatingly parallel}}:
Adding threads significantly reduces the overall computational cost,
resulting in large algorithmic superlinear speedups.

\begin{figure}
\centering
\resizebox{1.4in}{!}{\includegraphics{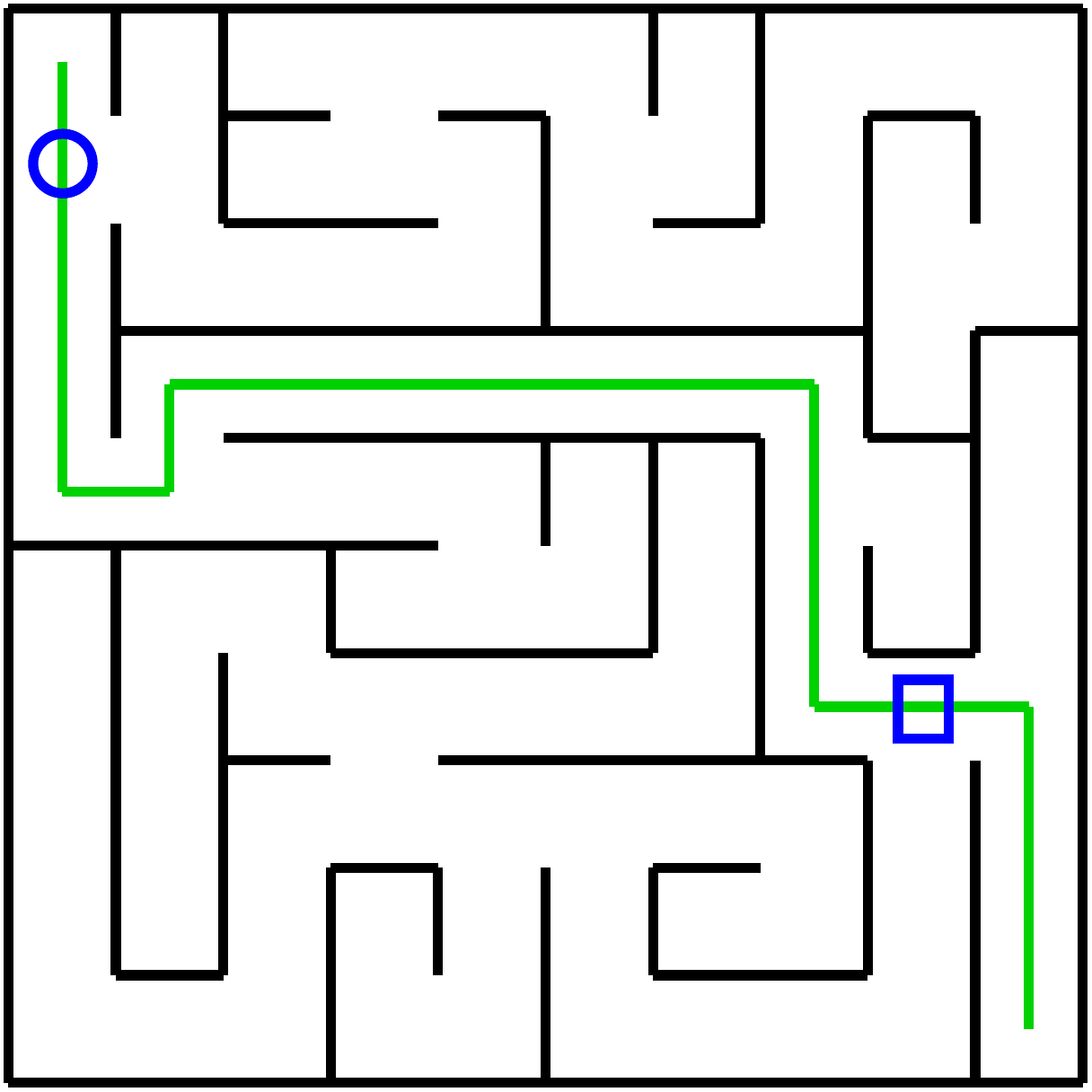}}
\caption{Reason for Small Visit Percentages}
\label{fig:SMPdesign:Reason for Small Visit Percentages}
\end{figure}

\begin{figure}
\centering
\resizebox{2.2in}{!}{\includegraphics{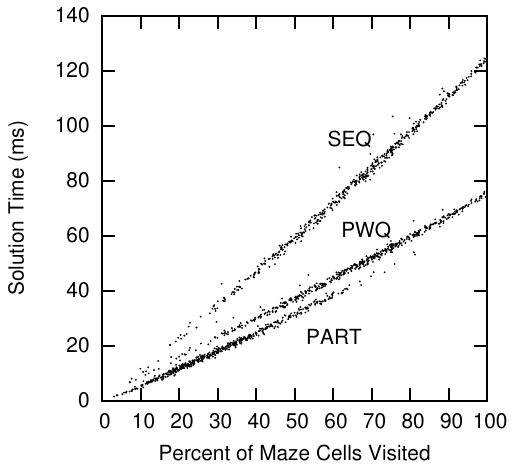}}
\caption{Correlation Between Visit Percentage and Solution Time}
\label{fig:SMPdesign:Correlation Between Visit Percentage and Solution Time}
\end{figure}

Further investigation showed that
PART sometimes visited fewer than 2\,\% of the maze's cells,
while SEQ and PWQ never visited fewer than about 9\,\%.
The reason for this difference is shown by
\cref{fig:SMPdesign:Reason for Small Visit Percentages}.
If the thread traversing the solution from the upper left reaches
the circle, the other thread cannot reach
the upper-right portion of the maze.
Similarly, if the other thread reaches the square,
the first thread cannot reach the lower-left
portion of the maze.
Therefore, PART will likely visit a small fraction
of the non-solution-path cells.
In short, the superlinear speedups are due to threads getting in each
others' way.
This is a sharp contrast with decades of experience with
parallel programming, where workers have struggled
to keep threads \emph{out} of each others' way.

\Cref{fig:SMPdesign:Correlation Between Visit Percentage and Solution Time}
confirms a strong correlation between cells visited and solution time
for all three methods.
The slope of PART's scatterplot is smaller than that of SEQ,
indicating that PART's pair of threads visits a given fraction
of the maze faster than can SEQ's single thread.
PART's scatterplot is also weighted toward small visit
percentages, confirming that PART does less total work, hence
the observed humiliating parallelism.
This humiliating parallelism also provides more than 2x speedup on two
CPUs, as put forth in
\cpageref{sec:SMPdesign:Problems Maze}.

\begin{figure}
\centering
\resizebox{1.4in}{!}{\includegraphics{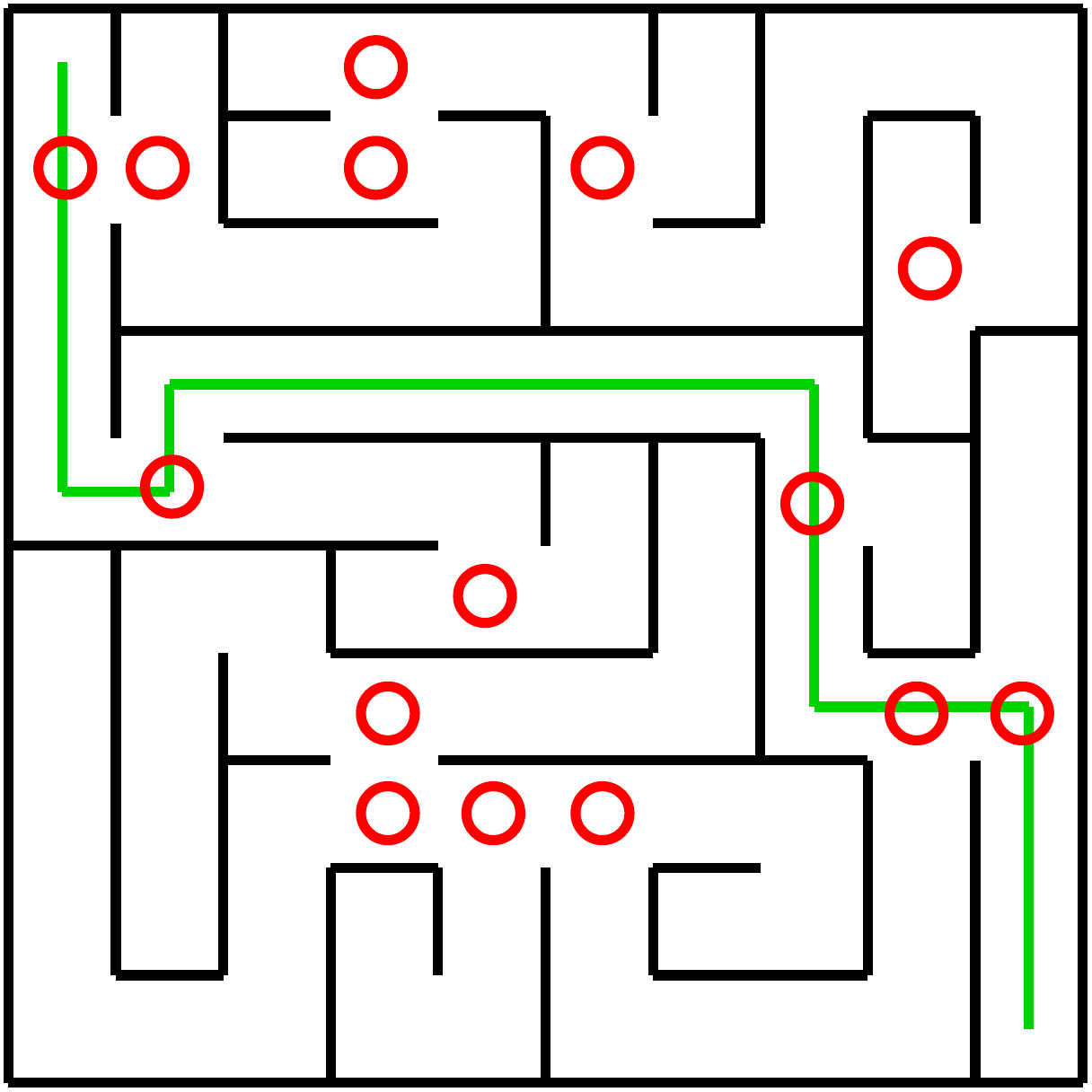}}
\caption{PWQ Potential Contention Points}
\label{fig:SMPdesign:PWQ Potential Contention Points}
\end{figure}

The fraction of cells visited by PWQ is similar to that of SEQ\@.
In addition, PWQ's solution time is greater than that of PART,
even for equal visit fractions.
The reason for this is shown in
\cref{fig:SMPdesign:PWQ Potential Contention Points}, which has a red
circle on each cell with more than two neighbors.
Each such cell can result in contention in PWQ, because
one thread can enter but two threads can exit, which hurts
performance, as noted earlier in this chapter.
In contrast, PART can incur such contention but once, namely
when the solution is located.
Of course, SEQ never contends.

\QuickQuiz{
	Given that a 2D maze achieved 4x speedup on two CPUs, would
	a 3D maze achieve an 8x speedup on two CPUs?
}\QuickQuizAnswer{
	This is an excellent question that is left to a suitably
	interested and industrious reader.
}\QuickQuizEnd

\begin{figure}
\centering
\resizebox{2.2in}{!}{\includegraphics{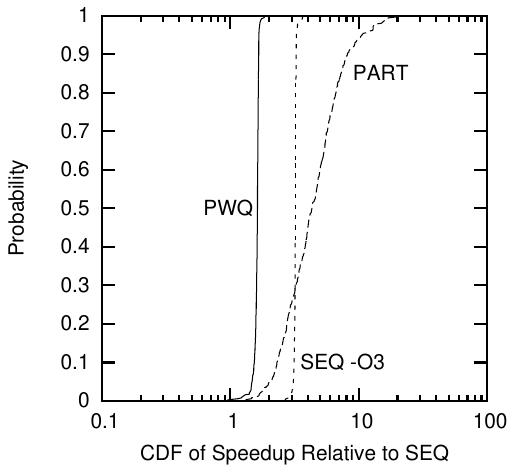}}
\caption{Effect of Compiler Optimization (-O3)}
\label{fig:SMPdesign:Effect of Compiler Optimization (-O3)}
\end{figure}

Although PART's speedup is impressive, we should not neglect sequential
optimizations.
\Cref{fig:SMPdesign:Effect of Compiler Optimization (-O3)} shows that
SEQ, when compiled with -O3, is about twice as fast
as unoptimized PWQ, approaching the performance of unoptimized PART\@.
Compiling all three algorithms with -O3 gives results similar to
(albeit faster than) those shown in
\cref{fig:SMPdesign:CDF of SEQ/PWQ and SEQ/PART Solution-Time Ratios},
except that PWQ provides almost no speedup compared
to SEQ, in keeping with \IXr{Amdahl's Law}~\cite{GeneAmdahl1967AmdahlsLaw}.
However, if the goal is to double performance compared to unoptimized
SEQ, as opposed to achieving optimality, compiler
optimizations are quite attractive.

Cache alignment and padding often improves performance by reducing
\IX{false sharing}.
However, for these maze-solution algorithms, aligning and padding the
maze-cell array \emph{degrades} performance by up to 42\,\% for 1000x1000 mazes.
Cache locality is more important than avoiding
false sharing, especially for large mazes.
For smaller 20-by-20 or 50-by-50 mazes, aligning and padding can produce
up to a 40\,\% performance improvement for PART,
but for these small sizes, SEQ performs better anyway because there
is insufficient time for PART to make up for the overhead of
thread creation and destruction.

In short, the partitioned parallel maze solver is an interesting example
of an algorithmic superlinear speedup.
If ``algorithmic superlinear speedup'' causes cognitive dissonance,
please proceed to the next section.

\subsection{Alternative Sequential Maze Solver}
\label{sec:SMPdesign:Alternative Sequential Maze Solver}

\begin{figure}
\centering
\resizebox{2.2in}{!}{\includegraphics{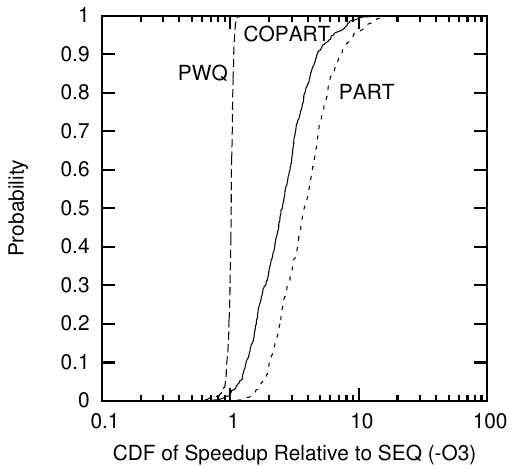}}
\caption{Partitioned Coroutines}
\label{fig:SMPdesign:Partitioned Coroutines}
\end{figure}

The presence of algorithmic superlinear speedups suggests simulating
parallelism via co-routines, for example, manually switching context
between threads on each pass through the main do-while loop in
\cref{lst:SMPdesign:Partitioned Parallel Solver Pseudocode}.
This context switching is straightforward because the context
consists only of the variables \co{c} and \co{vi}:
Of the numerous ways to achieve the effect, this is a good tradeoff
between context-switch overhead and visit percentage.
As can be seen in
\cref{fig:SMPdesign:Partitioned Coroutines},
this coroutine algorithm (COPART) is quite effective, with the performance
on one thread being within about 30\,\% of PART on two threads
(\path{maze_2seq.c}).

\subsection{Performance Comparison II}
\label{sec:SMPdesign:Performance Comparison II}

\begin{figure}
\centering
\resizebox{2.2in}{!}{\includegraphics{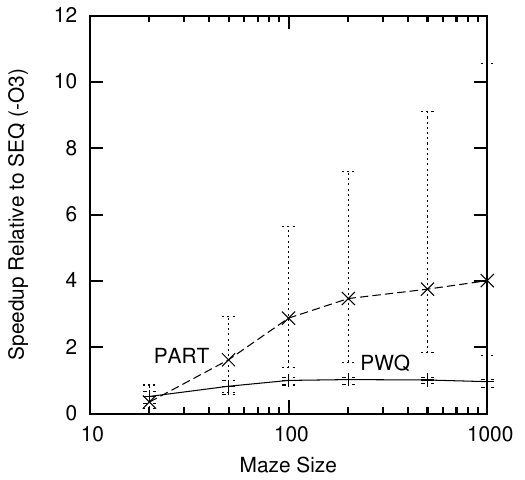}}
\caption{Varying Maze Size vs.\@ SEQ}
\label{fig:SMPdesign:Varying Maze Size vs. SEQ}
\end{figure}

\begin{figure}
\centering
\resizebox{2.2in}{!}{\includegraphics{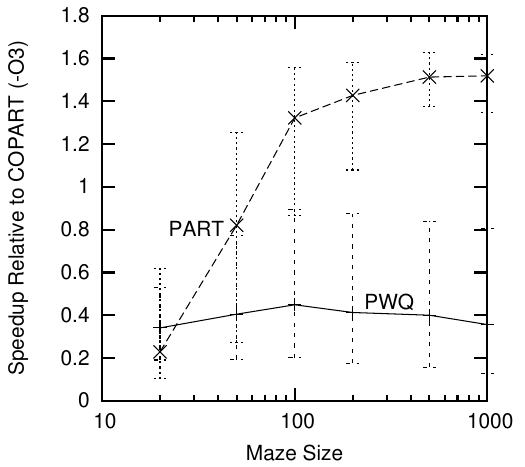}}
\caption{Varying Maze Size vs.\@ COPART}
\label{fig:SMPdesign:Varying Maze Size vs. COPART}
\end{figure}

\Cref{fig:SMPdesign:Varying Maze Size vs. SEQ,%
fig:SMPdesign:Varying Maze Size vs. COPART}
show the effects of varying maze size, comparing both PWQ and PART
running on two threads
against either SEQ or COPART, respectively, with 90\=/percent\-/confidence
error bars.
PART shows superlinear scalability against SEQ and modest scalability
against COPART for 100-by-100 and larger mazes.
PART exceeds theoretical energy-efficiency breakeven against COPART at roughly
the 200-by-200 maze size, given that power consumption rises as roughly
the square of the frequency for high frequencies~\cite{TrevorMudge2000Power},
so that 1.4x scaling on two threads consumes the same energy
as a single thread at equal solution speeds.
In contrast, PWQ shows poor scalability against both SEQ and COPART
unless unoptimized:
\Cref{fig:SMPdesign:Varying Maze Size vs. SEQ,%
fig:SMPdesign:Varying Maze Size vs. COPART}
were generated using -O3.

\begin{figure}
\centering
\resizebox{2.2in}{!}{\includegraphics{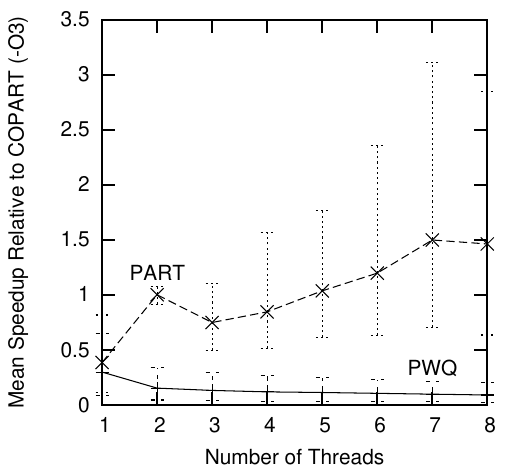}}
\caption{Mean Speedup vs.\@ Number of Threads, 1000x1000 Maze}
\label{fig:SMPdesign:Mean Speedup vs. Number of Threads; 1000x1000 Maze}
\end{figure}

\Cref{fig:SMPdesign:Mean Speedup vs. Number of Threads; 1000x1000 Maze}
shows the performance of PWQ and PART relative to COPART\@.
For PART runs with more than two threads, the additional threads were
started evenly spaced along the diagonal connecting the starting and
ending cells.
Simplified link-state routing~\cite{BERT-87} was used to
detect early termination on PART runs with more than two threads
(the solution is flagged when
a thread is connected to both beginning and end).
PWQ performs quite poorly, but
PART hits breakeven at two threads and again at five threads, achieving
modest speedups beyond five threads.
Theoretical energy efficiency breakeven is within the 90\=/percent\-/confidence
interval for seven and eight threads.
The reasons for the peak at two threads are (1) the lower complexity
of termination detection in the two-thread case and (2) the fact that
there is a lower probability of the third and subsequent threads making
useful forward progress:
Only the first two threads are guaranteed to start on the solution line.
This disappointing performance compared to results in
\cref{fig:SMPdesign:Varying Maze Size vs. COPART}
is due to the less-tightly integrated hardware available in the
larger and older Xeon system running at 2.66\,GHz.

\QuickQuiz{
	Why place the third, fourth, and so on threads on the diagonal?
	Why not instead distribute them evenly around the maze?
}\QuickQuizAnswer{
	There are indeed a great many ways to distribute the extra
	threads.
	Evaluation of distribution strategies is left to a suitably
	interested and industrious reader.
}\QuickQuizEnd

\subsection{Future Directions and Conclusions}
\label{sec:SMPdesign:Future Directions and Conclusions}

Much future work remains.
First, this section applied only one technique used by human maze solvers.
Others include following walls to exclude portions of the maze
and choosing internal starting points based on the
locations of previously traversed paths.
Second, different choices of
starting and ending points might favor different algorithms.
Third, although placement of the PART algorithm's
first two threads is straightforward, there are any number of
placement schemes for the remaining threads.
Optimal placement might well depend on the starting and ending points.
Fourth, study of unsolvable mazes and cyclic mazes
is likely to produce interesting results.
Fifth, the lightweight C++11 atomic operations might improve performance.
Sixth, it would be interesting to compare the speedups for
three-dimensional mazes (or of even higher-order mazes).
Finally, for mazes, humiliating parallelism indicated a
more-efficient sequential implementation using coroutines.
Do humiliatingly parallel algorithms always lead to more-efficient
sequential implementations, or are there inherently humiliatingly parallel
algorithms for which coroutine context-switch overhead overwhelms the
speedups?

This section demonstrated and analyzed parallelization of maze-solution
algorithms.
A conventional work-queue-based algorithm did well only when compiler
optimizations were disabled, suggesting that some prior results obtained
using high-level/overhead languages will be invalidated
by advances in optimization.

This section gave a clear example where approaching parallelism
as a first-class optimization technique rather than as a derivative of a
sequential algorithm paves the way for an improved sequential algorithm.
High-level design-time application of parallelism is likely to be a
fruitful field of study.
This section took the problem of solving mazes from mildly scalable
to humiliatingly parallel and back again.
It is hoped that this experience will motivate work on parallelism
as a first-class design-time whole-application optimization technique,
rather than as a grossly suboptimal after-the-fact micro-optimization
to be retrofitted into existing programs.

\section{Partitioning, Parallelism, and Optimization}
\label{sec:SMPdesign:Partitioning; Parallelism; and Optimization}
\epigraph{Knowledge is of no value unless you put it into practice.}
	 {Anton Chekhov}

Most important, although this chapter has demonstrated that applying
parallelism at the design level gives excellent results, this final
section shows that this is not enough.
For search problems such as maze solution, this section has shown that
search strategy is even more important than parallel design.
Yes, for this particular type of maze, intelligently applying parallelism
identified a superior search strategy, but this sort of luck is no
substitute for a clear focus on search strategy itself.

As noted back in \cref{sec:intro:Parallel Programming Goals},
parallelism is but one potential optimization of many.
A successful design needs to focus on the most important optimization.
Much though I might wish to claim otherwise, that optimization might
or might not be parallelism.

However, for the many cases where parallelism is the right optimization,
the next section covers that synchronization workhorse, locking.

	\setlength{\parskip}{0.0pt plus 1ex}
	\input{qqzSMPdesign}
	\setlength{\parskip}{0.0pt plus 1.0pt}% return to default

% locking/locking.tex
% mainfile: ../perfbook.tex
% SPDX-License-Identifier: CC-BY-SA-3.0

\QuickQuizChapter{chp:Locking}{Locking}{qqzlocking}
\Epigraph{Locking is the worst general-purpose synchronization mechanism
	  except for all those other mechanisms that
	  have been tried from time to time.}{With apologies
	  to the memory of Winston Churchill and to whoever he was
	  quoting}

In recent concurrency research, locking often plays the role of villain.
\IX{Locking} stands accused of inciting deadlocks, convoying, \IX{starvation},
\IX{unfairness}, \IXpl{data race}, and all manner of other concurrency sins.
Interestingly enough, the role of workhorse in production-quality
shared-memory parallel software is also played by locking.
This chapter will look into this dichotomy between villain and
hero, as fancifully depicted in
\cref{fig:locking:Locking: Villain or Slob?,%
fig:locking:Locking: Workhorse or Hero?}.

There are a number of reasons behind this Jekyll-and-Hyde dichotomy:

\begin{enumerate}
\item	Many of locking's sins have pragmatic design solutions that
	work well in most cases, for example:
	\begin{enumerate}
	\item	Use of lock hierarchies to avoid deadlock.
	\item	Deadlock-detection tools, for example, the Linux kernel's
		lockdep facility~\cite{JonathanCorbet2006lockdep}.
	\item	Locking-friendly data structures, such as
		arrays, hash tables, and radix trees, which will
		be covered in \cref{chp:Data Structures}.
	\end{enumerate}
\item	Some of locking's sins are problems only at high levels of
	contention, levels reached only by poorly designed programs.
\item	Some of locking's sins are avoided by using other synchronization
	mechanisms in concert with locking.
	These other mechanisms include
	statistical counters
	(see \cref{chp:Counting}),
	reference counters
	(see \cref{sec:defer:Reference Counting}),
	hazard pointers
	(see \cref{sec:defer:Hazard Pointers}),
	sequence-locking readers
	(see \cref{sec:defer:Sequence Locks}),
	RCU
	(see \cref{sec:defer:Read-Copy Update (RCU)}),
	and simple non-blocking data structures
	(see \cref{sec:advsync:Non-Blocking Synchronization}).
\item	Until quite recently, almost all large shared-memory parallel
	programs were developed in secret, so that it was not easy
	to learn of these pragmatic solutions.
\item	Locking works extremely well for some software artifacts
	and extremely poorly for others.
	Developers who have worked on artifacts for which locking
	works well can be expected to have a much more positive
	opinion of locking than those who have worked on artifacts
	for which locking works poorly, as will be discussed in
	\cref{sec:locking:Locking: Hero or Villain?}.
\item	All good stories need a villain, and locking has a long and
	honorable history serving as a research-paper whipping boy.
\end{enumerate}

\QuickQuiz{
	Just how can serving as a whipping boy be considered to be
	in any way honorable???
}\QuickQuizAnswer{
	The reason locking serves as a research-paper whipping boy is
	because it is heavily used in practice.
	In contrast, if no one used or cared about locking, most research
	papers would not bother even mentioning it.
}\QuickQuizEnd

This chapter will give an overview of a number of ways to avoid locking's
more serious sins.

\begin{figure}
\centering
\resizebox{2in}{!}{\includegraphics{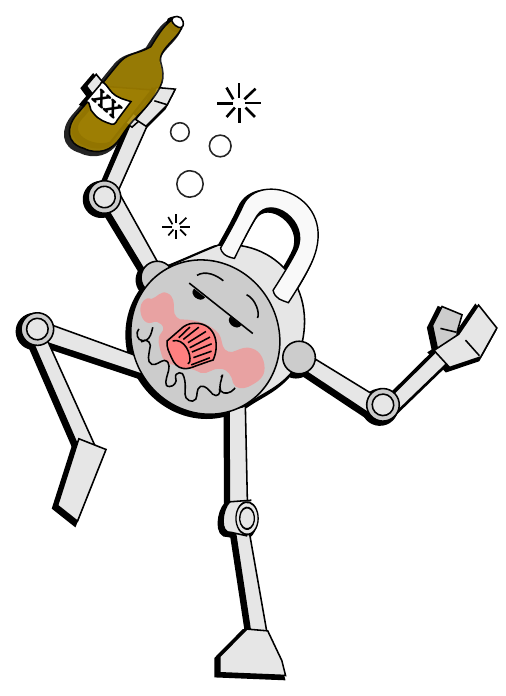}}
\caption{Locking:
		  Villain or Slob?}
\ContributedBy{Figure}{fig:locking:Locking: Villain or Slob?}{Melissa Broussard}
\end{figure}

\begin{figure}
\centering
\resizebox{2in}{!}{\includegraphics{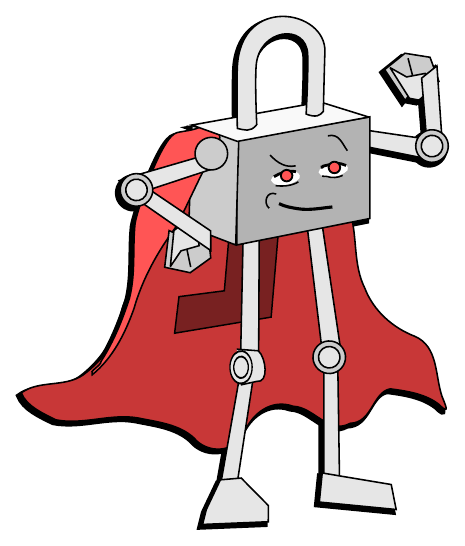}}
\caption{Locking:
		  Workhorse or Hero?}
\ContributedBy{Figure}{fig:locking:Locking: Workhorse or Hero?}{Melissa Broussard}
\end{figure}

\section{Staying Alive}
\label{sec:locking:Staying Alive}
\epigraph{I work to stay alive.}{Bette Davis}

Given that locking stands accused of deadlock and starvation,
one important concern for shared-memory parallel developers is
simply staying alive.
The following sections therefore cover deadlock, \IX{livelock}, starvation,
unfairness, and inefficiency.

\subsection{Deadlock}
\label{sec:locking:Deadlock}

\IXB{Deadlock} occurs when each member of a group of threads is holding at
least one lock while at the same time waiting on a lock held by a member
of that same group.
This happens even in groups containing a single thread when that thread
attempts to acquire a non-recursive lock that it already holds.
Deadlock can therefore occur even given but one thread and one lock!

Without some sort of external intervention, deadlock is forever.
No thread can acquire the lock it is waiting on until that
lock is released by the thread holding it, but the thread holding
it cannot release it until the holding thread acquires the lock that
it is in turn waiting on.

\begin{figure}
\centering
\resizebox{1.5in}{!}{\includegraphics{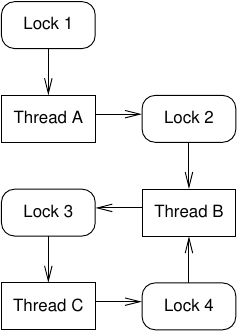}}
\caption{Deadlock Cycle}
\label{fig:locking:Deadlock Cycle}
\end{figure}

We can create a directed-graph representation of a deadlock scenario
with nodes for threads and locks, as shown in
\cref{fig:locking:Deadlock Cycle}.
An arrow from a lock to a thread indicates that the thread holds
the lock, for example, Thread~B holds Locks~2 and~4.
An arrow from a thread to a lock indicates that the thread is waiting
on the lock, for example, Thread~B is waiting on Lock~3.

A deadlock scenario will always contain at least one deadlock cycle.
In \cref{fig:locking:Deadlock Cycle}, this cycle is
Thread~B, Lock~3, Thread~C, Lock~4, and back to Thread~B.

\QuickQuiz{
	But the definition of lock-based deadlock only said that each
	thread was holding at least one lock and waiting on another lock
	that was held by some thread.
	How do you know that there is a cycle?
}\QuickQuizAnswer{
	Suppose that there is no cycle in the graph.
	We would then have a directed acyclic graph (DAG), which would
	have at least one leaf node.

	If this leaf node was a lock, then we would have a thread
	that was waiting on a lock that wasn't held by any thread,
	counter to the definition.
	In this case the thread would immediately acquire the lock.

	On the other hand, if this leaf node was a thread, then
	we would have a thread that was not waiting on any lock,
	again counter to the definition.
	And in this case, the thread would either be running or
	be blocked on something that is not a lock.
	In the first case, in the absence of infinite-loop bugs,
	the thread will eventually release the lock.
	In the second case, in the absence of a failure-to-wake bug,
	the thread will eventually wake up and release the lock.\footnote{
		Of course, one type of failure-to-wake bug is a
		deadlock that involves not only locks, but also non-lock
		resources.
		But the question really did say ``lock-based deadlock''!}

	Therefore, given this definition of lock-based deadlock, there
	must be a cycle in the corresponding graph.
}\QuickQuizEnd

Although there are some software environments such as database systems
that can recover from an existing deadlock, this approach requires either
that one of the threads be killed or that a lock be forcibly stolen from
one of the threads.
This killing and forcible stealing works well for transactions,
but is often problematic for kernel and application-level use of locking:
Dealing with the resulting partially updated structures can be extremely
complex, hazardous, and error-prone.

Therefore, kernels and applications should instead avoid deadlocks.
Deadlock-avoidance strategies include locking hierarchies
(\cref{sec:locking:Locking Hierarchies}),
local locking hierarchies
(\cref{sec:locking:Local Locking Hierarchies}),
layered locking hierarchies
(\cref{sec:locking:Layered Locking Hierarchies}),
temporal locking hierarchies
(\cref{sec:locking:Temporal Locking Hierarchies}),
strategies for dealing with APIs containing pointers to locks
(\cref{sec:locking:Locking Hierarchies and Pointers to Locks}),
conditional locking
(\cref{sec:locking:Conditional Locking}),
acquiring all needed locks first
(\cref{sec:locking:Acquire Needed Locks First}),
single-lock-at-a-time designs
(\cref{sec:locking:Single-Lock-at-a-Time Designs}),
and strategies for signal/interrupt handlers
(\cref{sec:locking:Signal/Interrupt Handlers}).
Although there is no deadlock-avoidance strategy that works perfectly
for all situations, there is a good selection of tools to choose from.

\subsubsection{Locking Hierarchies}
\label{sec:locking:Locking Hierarchies}

Locking hierarchies order the locks and prohibit acquiring locks out
of order.
In \cref{fig:locking:Deadlock Cycle},
we might order the locks numerically, thus forbidding a thread
from acquiring a given lock if it already holds a lock
with the same or a higher number.
Thread~B has violated this hierarchy because it is attempting to
acquire Lock~3 while holding Lock~4.
This violation permitted the deadlock to occur.

Again, to apply a locking hierarchy, order the locks and prohibit
out-of-order lock acquisition.
For different types of locks, it is helpful to have a
carefully considered hierarchy from one type to the next.
For many instances of the same type of lock, for example, a per-node
lock in a search tree, the traditional approach is to carry out lock
acquisition in order of the addresses of the locks to be acquired.
Either way, in large program, it is wise to use tools such as the Linux-kernel
\co{lockdep}~\cite{JonathanCorbet2006lockdep}
to enforce your locking hierarchy.

\subsubsection{Local Locking Hierarchies}
\label{sec:locking:Local Locking Hierarchies}

However, the global nature of locking hierarchies makes them difficult to
apply to library functions.
After all, when a program using a given library function has not yet
been written, how can the poor library-function implementor possibly
follow the yet-to-be-defined locking hierarchy?

One special (but common) case is when the library function does not
invoke any of the caller's code.
In this case, the caller's locks will never be acquired while holding
any of the library's locks, so that there cannot be a deadlock cycle
containing locks from both the library and the caller.

\QuickQuiz{
	Are there any exceptions to this rule, so that there really could be
	a deadlock cycle containing locks from both the library and
	the caller, even given that the library code never invokes
	any of the caller's functions?
}\QuickQuizAnswer{
	Indeed there are!
	Here are a few of them:
	\begin{enumerate}
	\item	If one of the library function's arguments is a pointer
		to a lock that this library function acquires, and if
		the library function holds one of its locks while
		acquiring the caller's lock, then we could have a
		deadlock cycle involving both caller and library locks.
	\item	If one of the library functions returns a pointer to
		a lock that is acquired by the caller, and if the
		caller acquires one of its locks while holding the
		library's lock, we could again have a deadlock
		cycle involving both caller and library locks.
	\item	If one of the library functions acquires a lock and
		then returns while still holding it, and if the caller
		acquires one of its locks, we have yet another way
		to create a deadlock cycle involving both caller
		and library locks.
	\item	If the caller has a signal handler that acquires
		locks, then the deadlock cycle can involve both
		caller and library locks.
		In this case, however, the library's locks are
		innocent bystanders in the deadlock cycle.
		That said, please note that acquiring a lock from
		within a signal handler is a no-no in many
		environments---it is not just a bad idea, it
		is unsupported.
		But if you absolutely must acquire a lock in a signal
		handler, be sure to block that signal while holding that
		same lock in thread context, and also while holding any
		other locks acquired while that same lock is held.
	\end{enumerate}
}\QuickQuizEnd

\begin{figure}
\centering
\resizebox{3in}{!}{\includegraphics{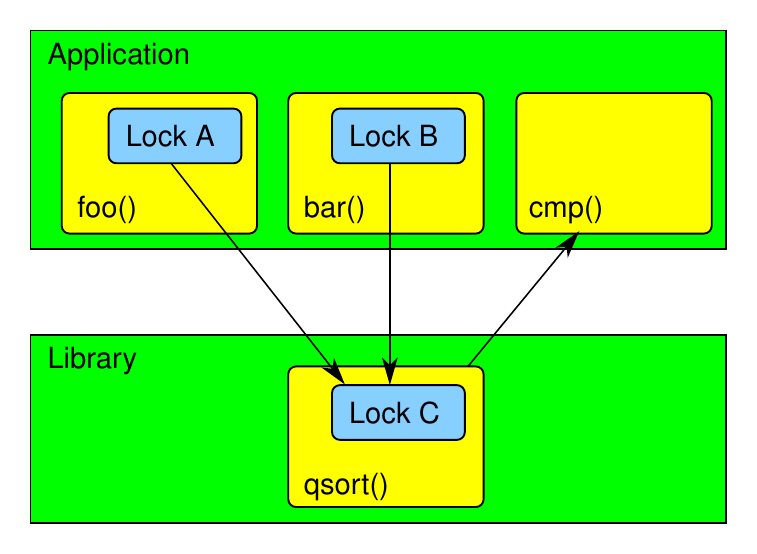}}
\caption{No \tco{qsort()} Compare-Function Locking}
\label{fig:locking:No qsort() Compare-Function Locking}
\end{figure}

\begin{figure}
\centering
\resizebox{3in}{!}{\includegraphics{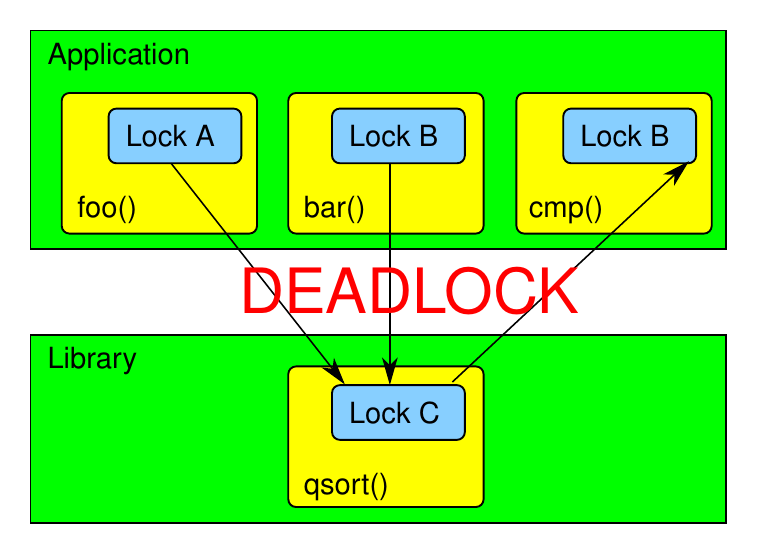}}
\caption{Without \tco{qsort()} Local Locking Hierarchy}
\label{fig:locking:Without qsort() Local Locking Hierarchy}
\end{figure}

But suppose that a library function does invoke the caller's code.
For example, \co{qsort()} invokes a caller-provided comparison function.
Now, normally this comparison function will operate on unchanging local
data, so that it need not acquire locks, as shown in
\cref{fig:locking:No qsort() Compare-Function Locking}.
But maybe someone is crazy enough to sort a collection whose keys
are changing, thus requiring that the comparison function acquire locks,
which might result in deadlock, as shown in
\cref{fig:locking:Without qsort() Local Locking Hierarchy}.
How can the library function avoid this deadlock?

The golden rule in this case is ``Release all locks before invoking
unknown code.''
To follow this rule, the \co{qsort()} function must release all of its
locks before invoking the comparison function.
Thus \co{qsort()} will not be holding any of its locks while the comparison
function acquires any of the caller's locks, thus avoiding deadlock.

\QuickQuiz{
	But if \co{qsort()} releases all its locks before invoking
	the comparison function, how can it protect against races
	with other \co{qsort()} threads?
}\QuickQuizAnswer{
	By privatizing the data elements being compared
	(as discussed in \cref{chp:Data Ownership})
	or through use of deferral mechanisms such as
	reference counting (as discussed in
	\cref{chp:Deferred Processing}).
	Or through use of layered locking hierarchies, as described
	in \cref{sec:locking:Layered Locking Hierarchies}.

	On the other hand, changing a key in a list that is
	currently being sorted is at best rather brave.
}\QuickQuizEnd

\begin{figure}
\centering
\resizebox{3in}{!}{\includegraphics{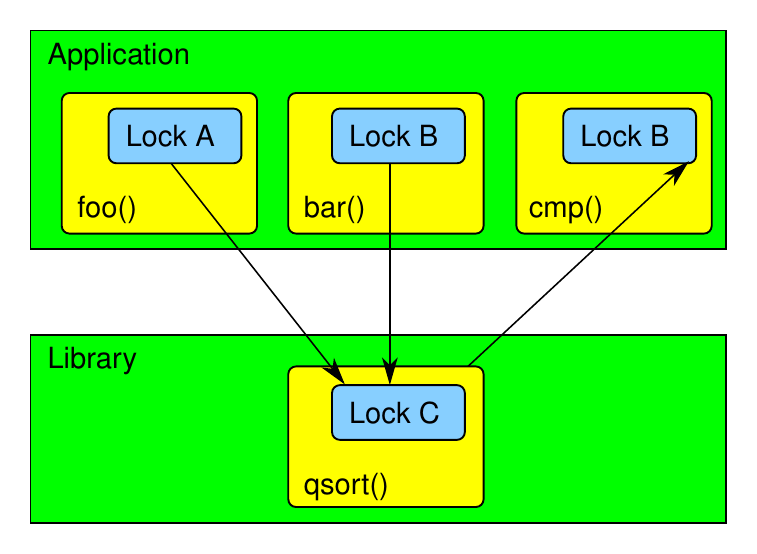}}
\caption{Local Locking Hierarchy for \tco{qsort()}}
\label{fig:locking:Local Locking Hierarchy for qsort()}
\end{figure}

To see the benefits of local locking hierarchies, compare
\cref{fig:locking:Without qsort() Local Locking Hierarchy,%
fig:locking:Local Locking Hierarchy for qsort()}.
In both figures, application functions \co{foo()} and \co{bar()}
invoke \co{qsort()} while holding Locks~A and~B, respectively.
Because this is a parallel implementation of \co{qsort()}, it acquires
Lock~C\@.
Function \co{foo()} passes function \co{cmp()} to \co{qsort()},
and \co{cmp()} acquires Lock~B\@.
Function \co{bar()} passes a simple integer-comparison function (not
shown) to \co{qsort()}, and this simple function does not acquire any
locks.

Now, if \co{qsort()} holds Lock~C while calling \co{cmp()} in violation
of the golden release-all-locks rule above, as shown in
\cref{fig:locking:Without qsort() Local Locking Hierarchy},
deadlock can occur.
To see this, suppose that one thread invokes \co{foo()} while a second
thread concurrently invokes \co{bar()}.
The first thread will acquire Lock~A and the second thread will acquire
Lock~B\@.
If the first thread's call to \co{qsort()} acquires Lock~C, then it
will be unable to acquire Lock~B when it calls \co{cmp()}.
But the first thread holds Lock~C, so the second thread's call to
\co{qsort()} will be unable to acquire it, and thus unable to release
Lock~B, resulting in deadlock.

In contrast, if \co{qsort()} releases Lock~C before invoking the
comparison function, which is unknown code from \co{qsort()}'s perspective,
then deadlock is avoided as shown in
\cref{fig:locking:Local Locking Hierarchy for qsort()}.

If each module releases all locks before invoking unknown code, then
deadlock is avoided if each module separately avoids deadlock.
This rule therefore greatly simplifies deadlock analysis and greatly
improves modularity.

Nevertheless, this golden rule comes with a warning.
When you release those locks, any state that they protect is subject
to arbitrary changes, changes that are all too easy for the function's
caller to forget, resulting in subtle and difficult-to-reproduce bugs.
Because the \co{qsort()} comparison function rarely acquires locks,
let's switch to a different example.

Consider the recursive tree iterator in
\cref{lst:locking:Recursive Tree Iterator} (\path{rec_tree_itr.c}).
The iterator visits every node in the tree, invoking a user's callback
function.
The tree lock is released before the invocation and re-acquired after return.
This code makes dangerous assumptions:
\begin{enumerate*}[(1)]
\item	The number of children of the current node has not changed,
\item	The ancestors stored on the stack by the recursion are still
	there, and
\item	The visited node itself has not been removed and freed.
\end{enumerate*}
A few of these hazards can be encountered if one thread calls
\co{tree_add()} while another thread releases the tree's lock to run a
callback function.

\QuickQuiz{
	So the iterating thread may or may not observe the added child.
	What is the big deal?
}\QuickQuizAnswer{
	There are at least two hazards in this situation.

	One is indeed that the number of children may or may not be
	observed to have changed.
	While that would be consistent with \co{tree_add()} being called
	either before or after the iterator started, it is better not
	left to the vagaries of the compiler.
	A more serious problem is that \co{realloc()} may not be able
	to extend the array in place, causing the heap to free the
	one used by the iterator and replace it with another block of
	memory.
	If the \co{children} pointer is not re-read then the iterating
	thread will access invalid memory (either free or reclaimed).
}\QuickQuizEnd

One strategy is to ensure that state is preserved despite the lock being
released, for example, by acquiring a reference on a node to prevent it
from being freed.
Alternatively, the state can be re-initialized once the lock is
re-acquired after the callback function returns.

\begin{listing}
\input{CodeSamples/locking/rec_tree_itr=tree_for_each.fcv}
\caption{Recursive Tree Iterator}
\label{lst:locking:Recursive Tree Iterator}
\end{listing}

\subsubsection{Layered Locking Hierarchies}
\label{sec:locking:Layered Locking Hierarchies}

\begin{figure}
\centering
\resizebox{3in}{!}{\includegraphics{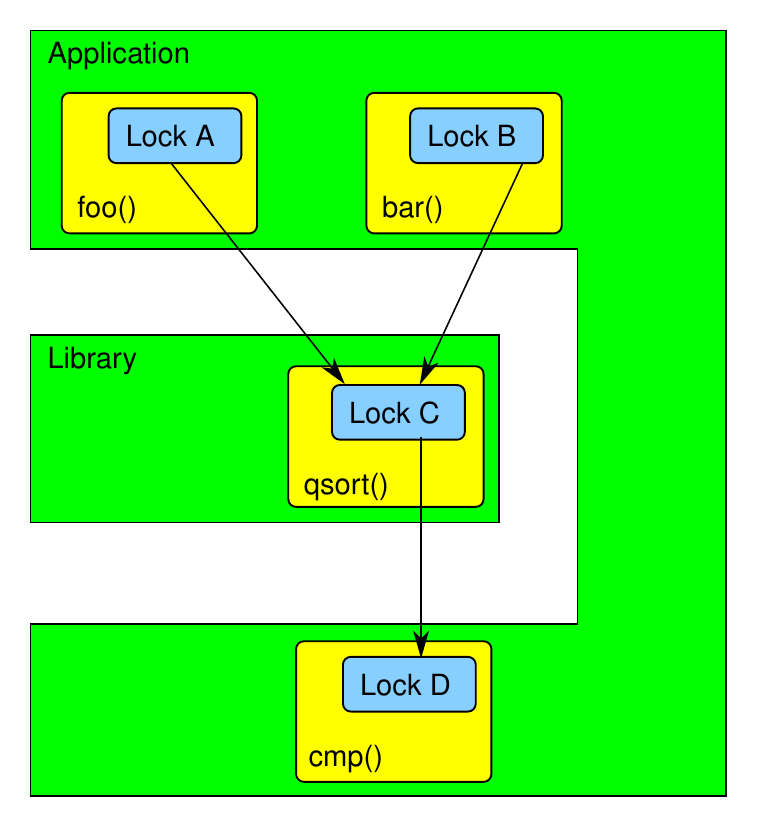}}
\caption{Layered Locking Hierarchy for \tco{qsort()}}
\label{fig:locking:Layered Locking Hierarchy for qsort()}
\end{figure}

Unfortunately, it might be infeasible to preserve state on the one hand
or to re-initialize it on the other, thus ruling out a local locking
hierarchy where all locks are released before invoking unknown code.
However, we can instead construct a layered locking hierarchy, as shown in
\cref{fig:locking:Layered Locking Hierarchy for qsort()}.
Here, the \co{cmp()} function uses a new Lock~D that is acquired after
all of Locks~A, B, and~C, avoiding deadlock.
We therefore have three layers to the global deadlock hierarchy, the
first containing Locks~A and~B, the second containing Lock~C, and
the third containing Lock~D\@.

\begin{listing}
\input{CodeSamples/locking/locked_list=start_next.fcv}
\caption{Concurrent List Iterator}
\label{lst:locking:Concurrent List Iterator}
\end{listing}

Please note that it is not typically possible to mechanically
change \co{cmp()} to use the new Lock~D\@.
Quite the opposite:
It is often necessary to make profound design-level modifications.
Nevertheless, the effort required for such modifications is normally
a small price to pay in order to avoid deadlock.
More to the point, this potential deadlock should preferably be detected
at design time, before any code has been generated!

For another example where releasing all locks before invoking unknown
code is impractical, imagine an iterator over a linked list, as shown in
\cref{lst:locking:Concurrent List Iterator} (\path{locked_list.c}).
The \co{list_start()} function acquires a lock on the list and returns
the first element (if there is one), and
\co{list_next()} either returns a pointer to the next element in the list
or releases the lock and returns \co{NULL} if the end of the list has
been reached.

\begin{listing}
\input{CodeSamples/locking/locked_list=list_print.fcv}
\caption{Concurrent List Iterator Usage}
\label{lst:locking:Concurrent List Iterator Usage}
\end{listing}

\begin{fcvref}[ln:locking:locked_list:list_print:ints]
\Cref{lst:locking:Concurrent List Iterator Usage} shows how
this list iterator may be used.
\Clnrefrange{b}{e} define the \co{list_ints} element
containing a single integer,
\end{fcvref}
\begin{fcvref}[ln:locking:locked_list:list_print:print]
and \clnrefrange{b}{e} show how to iterate over the list.
\Clnref{start} locks the list and fetches a pointer to the first element,
\clnref{entry} provides a pointer to our enclosing \co{list_ints} structure,
\clnref{print} prints the corresponding integer, and
\clnref{next} moves to the next element.
This is quite simple, and hides all of the locking.
\end{fcvref}

That is, the locking remains hidden as long as the code processing each
list element does not itself acquire a lock that is held across some
other call to \co{list_start()} or \co{list_next()}, which results in
deadlock.
We can avoid the deadlock by layering the locking hierarchy
to take the list-iterator locking into account.

This layered approach can be extended to an arbitrarily large number of layers,
but each added layer increases the complexity of the locking design.
Such increases in complexity are particularly inconvenient for some
types of object-oriented designs, in which control passes back and forth
among a large group of objects in an undisciplined manner.\footnote{
	One name for this is ``object-oriented spaghetti code.''}
This mismatch between the habits of object-oriented design and the
need to avoid deadlock is an important reason why parallel programming
is perceived by some to be so difficult.

Some alternatives to highly layered locking hierarchies are covered in
\cref{chp:Deferred Processing}.

\subsubsection{Temporal Locking Hierarchies}
\label{sec:locking:Temporal Locking Hierarchies}

One way to avoid deadlock is to defer acquisition of one of the
conflicting locks.
This approach is used in Linux-kernel RCU, whose \apik{call_rcu()}
function is invoked by the Linux-kernel scheduler while holding
its locks.
This means that \co{call_rcu()} cannot always safely invoke the scheduler
to do a wakeup, for example, in order to wake up an RCU kthread in order
to start the new grace period that is required by the callback queued
by \co{call_rcu()}.

\QuickQuiz{
	What do you mean ``cannot always safely invoke the scheduler''?
	Either \co{call_rcu()} can or cannot safely invoke the scheduler,
	right?
}\QuickQuizAnswer{
	It really does depend.

	The scheduler locks are always held with interrupts disabled.
	Therefore, if \co{call_rcu()} is invoked with interrupts
	enabled, no scheduler locks are held, and \co{call_rcu()}
	can safely call into the scheduler.
	Otherwise, if interrupts are disabled, one of the scheduler
	locks \emph{might} be held, so \co{call_rcu()} must play it
	safe and refrain from calling into the scheduler.
}\QuickQuizEnd

However, grace periods last for many milliseconds, so waiting another
millisecond before starting a new grace period is not normally a problem.
Therefore, if \co{call_rcu()} detects a possible deadlock with the
scheduler, it arranges to start the new grace period later, either
within a timer handler or within the scheduler-clock interrupt handler,
depending on configuration.
Because no scheduler locks are held across either handler, deadlock
is successfully avoided.

The overall approach is thus to adhere to a locking hierarchy by deferring
lock acquisition to an environment in which no locks are held.

\subsubsection{Locking Hierarchies and Pointers to Locks}
\label{sec:locking:Locking Hierarchies and Pointers to Locks}

Although there are some exceptions, an external API containing a pointer
to a lock is very often a misdesigned API\@.
Handing an internal lock to some other software component is after all
the antithesis of information hiding, which is in turn a key design
principle.

\QuickQuiz{
	Name one common situation where a pointer to a lock is passed
	into a function.
}\QuickQuizAnswer{
	Locking primitives, of course!
}\QuickQuizEnd

One exception is functions that hand off some entity,
where the caller's lock must be held until the handoff is complete,
but where the lock must be released before the function returns.
One example of such a function is the POSIX \apipx{pthread_cond_wait()}
function, where passing a pointer to a \apipx{pthread_mutex_t}
prevents hangs due to lost wakeups.

\QuickQuiz{
	Doesn't the fact that \co{pthread_cond_wait()} first releases the
	mutex and then re-acquires it eliminate the possibility of deadlock?
}\QuickQuizAnswer{
	Absolutely not!

	Consider a program that acquires \co{mutex_a}, and then
	\co{mutex_b}, in that order, and then passes \co{mutex_a}
	to \co{pthread_cond_wait()}.
	Now, \co{pthread_cond_wait()} will release \co{mutex_a}, but
	will re-acquire it before returning.
	If some other thread acquires \co{mutex_a} in the meantime
	and then blocks on \co{mutex_b}, the program will deadlock.
}\QuickQuizEnd

In short, if you find yourself exporting an API with a pointer to a
lock as an argument or as the return value, do yourself a favor and
carefully reconsider your API design.
It might well be the right thing to do, but experience indicates that
this is unlikely.

\subsubsection{Conditional Locking}
\label{sec:locking:Conditional Locking}

But suppose that there is no reasonable locking hierarchy.
This can happen in real life, for example, in some types of layered
network protocol stacks where packets flow in both directions, for
example, in implementations of distributed lock managers.
In the networking case, it might be necessary to hold the locks from
both layers when passing a packet from one layer to another.
Given that packets travel both up and down the protocol stack, this
is an excellent recipe for deadlock, as illustrated in
\cref{lst:locking:Protocol Layering and Deadlock}.
\begin{fcvref}[ln:locking:Protocol Layering and Deadlock]
Here, a packet moving down the stack towards the wire must acquire
the next layer's lock out of order.
Given that packets moving up the stack away from the wire are acquiring
the locks in order, the lock acquisition in \clnref{acq} of the listing
can result in deadlock.
\end{fcvref}

\begin{listing}
\begin{fcvlabel}[ln:locking:Protocol Layering and Deadlock]
\begin{VerbatimL}[commandchars=\\\{\}]
spin_lock(&lock2);
layer_2_processing(pkt);
nextlayer = layer_1(pkt);
spin_lock(&nextlayer->lock1);	\lnlbl{acq}
spin_unlock(&lock2);
layer_1_processing(pkt);
spin_unlock(&nextlayer->lock1);
\end{VerbatimL}
\end{fcvlabel}
\caption{Protocol Layering and Deadlock}
\label{lst:locking:Protocol Layering and Deadlock}
\end{listing}

One way to avoid deadlocks in this case is to impose a locking hierarchy,
but when it is necessary to acquire a lock out of order, acquire it
conditionally, as shown in
\cref{lst:locking:Avoiding Deadlock Via Conditional Locking}.
\begin{fcvref}[ln:locking:Avoiding Deadlock Via Conditional Locking]
Instead of unconditionally acquiring the layer-1 lock, \clnref{trylock}
conditionally acquires the lock using the \apik{spin_trylock()} primitive.
This primitive acquires the lock immediately if the lock is available
(returning non-zero), and otherwise returns zero without acquiring the lock.

\begin{listing}
\begin{fcvlabel}[ln:locking:Avoiding Deadlock Via Conditional Locking]
\begin{VerbatimL}[commandchars=\\\[\]]
retry:
	spin_lock(&lock2);
	layer_2_processing(pkt);
	nextlayer = layer_1(pkt);
	if (!spin_trylock(&nextlayer->lock1)) {	\lnlbl[trylock]
		spin_unlock(&lock2);		\lnlbl[rel2]
		spin_lock(&nextlayer->lock1);	\lnlbl[acq1]
		spin_lock(&lock2);		\lnlbl[acq2]
		if (layer_1(pkt) != nextlayer) {\lnlbl[recheck]
			spin_unlock(&nextlayer->lock1);
			spin_unlock(&lock2);
			goto retry;
		}
	}
	spin_unlock(&lock2);
	layer_1_processing(pkt);		\lnlbl[l1_proc]
	spin_unlock(&nextlayer->lock1);
\end{VerbatimL}
\end{fcvlabel}
\caption{Avoiding Deadlock Via Conditional Locking}
\label{lst:locking:Avoiding Deadlock Via Conditional Locking}
\end{listing}

If \co{spin_trylock()} was successful, \clnref{l1_proc} does the needed
layer-1 processing.
Otherwise, \clnref{rel2} releases the lock, and
\clnref{acq1,acq2} acquire them in
the correct order.
Unfortunately, there might be multiple networking devices on
the system (e.g., Ethernet and WiFi), so that the \co{layer_1()}
function must make a routing decision.
This decision might change at any time, especially if the system
is mobile.\footnote{
	And, in contrast to the 1900s, mobility is the common case.}
Therefore, \clnref{recheck} must recheck the decision, and if it has changed,
must release the locks and start over.
\end{fcvref}

\QuickQuizSeries{%
\QuickQuizB{
	Can the transformation from
	\cref{lst:locking:Protocol Layering and Deadlock} to
	\cref{lst:locking:Avoiding Deadlock Via Conditional Locking}
	be applied universally?
}\QuickQuizAnswerB{
	Absolutely not!

	This transformation assumes that the
	\co{layer_2_processing()} function is idempotent, given that
	it might be executed multiple times on the same packet when
	the \co{layer_1()} routing decision changes.
	Therefore, in real life, this transformation can become
	arbitrarily complex.
}\QuickQuizEndB
\QuickQuizE{
	But the complexity in
	\cref{lst:locking:Avoiding Deadlock Via Conditional Locking}
	is well worthwhile given that it avoids deadlock, right?
}\QuickQuizAnswerE{
	Maybe.

	If the routing decision in \co{layer_1()} changes often enough,
	the code will always retry, never making forward progress.
	This is termed ``\IX{livelock}'' if no thread makes any
	forward progress or ``\IX{starvation}''
	if some threads make forward progress but others do not
	(see \cref{sec:locking:Livelock and Starvation}).
}\QuickQuizEndE
}

\subsubsection{Acquire Needed Locks First}
\label{sec:locking:Acquire Needed Locks First}

In an important special case of conditional locking, all needed
locks are acquired before any processing is carried out, where
the needed locks might be identified by hashing the addresses
of the data structures involved.
In this case, processing need not be idempotent:
If it turns out to be impossible to acquire a given lock without
first releasing one that was already acquired, just release all
the locks and try again.
Only once all needed locks are held will any processing be carried out.

However, this procedure can result in \emph{livelock}, which will
be discussed in
\cref{sec:locking:Livelock and Starvation}.

\QuickQuiz{
	When using the ``acquire needed locks first'' approach described in
	\cref{sec:locking:Acquire Needed Locks First},
	how can livelock be avoided?
}\QuickQuizAnswer{
	Provide an additional global lock.
	If a given thread has repeatedly tried and failed to acquire the needed
	locks, then have that thread unconditionally acquire the new
	global lock, and then unconditionally acquire any needed locks.
	(Suggested by Doug Lea.)
}\QuickQuizEnd

A related approach, two-phase locking~\cite{PhilipABernstein1987},
has seen long production use in transactional database systems.
In the first phase of a two-phase locking transaction, locks are
acquired but not released.
Once all needed locks have been acquired, the transaction enters the
second phase, where locks are released, but not acquired.
This locking approach allows databases to provide serializability
guarantees for their transactions, in other words, to guarantee
that all values seen and produced by the transactions are consistent
with some global ordering of all the transactions.
Many such systems rely on the ability to abort transactions, although
this can be simplified by avoiding making any changes to shared data
until all needed locks are acquired.
Livelock and deadlock are issues in such systems, but practical
solutions may be found in any of a number of database textbooks.

\subsubsection{Single-Lock-at-a-Time Designs}
\label{sec:locking:Single-Lock-at-a-Time Designs}

In some cases, it is possible to avoid nesting locks, thus avoiding
deadlock.
For example, if a problem is perfectly partitionable, a single
lock may be assigned to each partition.
Then a thread working on a given partition need only acquire the one
corresponding lock.
Because no thread ever holds more than one lock at a time,
deadlock is impossible.

However, there must be some mechanism to ensure that the needed data
structures remain in existence during the time that neither lock is
held.
One such mechanism is discussed in
\cref{sec:locking:Lock-Based Existence Guarantees}
and several others are presented in
\cref{chp:Deferred Processing}.

\subsubsection{Signal/Interrupt Handlers}
\label{sec:locking:Signal/Interrupt Handlers}

Deadlocks involving signal handlers are often quickly dismissed by
noting that it is not legal to invoke \apipx{pthread_mutex_lock()} from
within a signal handler~\cite{OpenGroup1997pthreads}.
However, it is possible (though often unwise) to hand-craft
locking primitives that \emph{can} be invoked from signal handlers.
Besides which, almost all operating-system kernels permit locks to
be acquired from within interrupt handlers, which are analogous
to signal handlers.

The trick is to block signals (or disable interrupts, as the case may be)
when acquiring any lock that might be acquired within a signal
(or an interrupt) handler.
Furthermore, if holding such a lock, it is illegal to attempt to
acquire any lock that is ever acquired
outside of a signal handler without blocking signals.

\QuickQuiz{
	Suppose Lock~A is never acquired within a signal handler,
	but Lock~B is acquired both from thread context and by signal
	handlers.
	Suppose further that Lock~A is sometimes acquired with signals
	unblocked.
	Why is it illegal to acquire Lock~A holding Lock~B?
}\QuickQuizAnswer{
	Because this would lead to deadlock.
	Given that Lock~A is sometimes held outside of a signal
	handler without blocking signals, a signal might be handled while
	holding this lock.
	The corresponding signal handler might then acquire
	Lock~B, so that Lock~B is acquired while holding Lock~A\@.
	Therefore, if we also acquire Lock~A while holding Lock~B,
	we will have a deadlock cycle.
	Note that this problem exists even if signals are blocked while
	holding Lock~B.

	This is another reason to be very careful with locks that are
	acquired within interrupt or signal handlers.
	But the Linux kernel's lock dependency checker knows about this
	situation and many others as well, so please do make full use
	of it!
}\QuickQuizEnd

If a lock is acquired by the handlers for several signals, then each
and every one of these signals must be blocked whenever that lock is
acquired, even when that
lock is acquired within a signal handler.

\QuickQuiz{
	How can you legally block signals within a signal handler?
}\QuickQuizAnswer{
	One of the simplest and fastest ways to do so is to use
	the \co{sa_mask} field of the \co{struct sigaction} that
	you pass to \co{sigaction()} when setting up the signal.
}\QuickQuizEnd

Unfortunately, blocking and unblocking signals can be expensive in
some operating systems, notably including Linux, so performance
concerns often mean that locks acquired in signal handlers are only
acquired in signal handlers, and that lockless synchronization
mechanisms are used to communicate between application code and
signal handlers.

Or that signal handlers are avoided completely except for handling
fatal errors.

\QuickQuiz{
	If acquiring locks in signal handlers is such a bad idea, why
	even discuss ways of making it safe?
}\QuickQuizAnswer{
	Because these same rules apply to the interrupt handlers used in
	operating-system kernels and in some embedded applications.

	In many application environments, acquiring locks in signal
	handlers is frowned upon~\cite{OpenGroup1997pthreads}.
	However, that does not stop clever developers from (perhaps
	unwisely) fashioning home-brew locks out of atomic operations.
	And atomic operations are in many cases perfectly legal in
	signal handlers.
}\QuickQuizEnd

\subsubsection{Discussion}
\label{sec:locking:Locking Hierarchy Discussion}

There are a large number of deadlock-avoidance strategies available to
the shared-memory parallel programmer, but there are sequential
programs for which none of them is a good fit.
This is one of the reasons that expert programmers have more than
one tool in their toolbox:
Locking is a powerful concurrency tool, but there are jobs better
addressed with other tools.

\QuickQuiz{
	Given an object-oriented application that passes control freely
	among a group of objects such that there is no straightforward
	locking hierarchy,\footnote{
		Also known as ``object-oriented spaghetti code.''}
	layered or otherwise, how can this
	application be parallelized?
}\QuickQuizAnswer{
	There are a number of approaches:
	\begin{enumerate}
	\item	In the case of parametric search via simulation,
		where a large number of simulations will be run
		in order to converge on (for example) a good design
		for a mechanical or electrical device, leave the
		simulation single-threaded, but run many instances
		of the simulation in parallel.
		This retains the object-oriented design, and gains
		parallelism at a higher level, and likely also avoids
		both deadlocks and synchronization overhead.
	\item	Partition the objects into groups such that there
		is no need to operate on objects in
		more than one group at a given time.
		Then associate a lock with each group.
		This is an example of a single-lock-at-a-time
		design, which discussed in
		\cref{sec:locking:Single-Lock-at-a-Time Designs}.
	\item	Partition the objects into groups such that threads
		can all operate on objects in the groups in some
		groupwise ordering.
		Then associate a lock with each group, and impose a
		locking hierarchy over the groups.
	\item	Impose an arbitrarily selected hierarchy on the locks,
		and then use conditional locking if it is necessary
		to acquire a lock out of order, as was discussed in
		\cref{sec:locking:Conditional Locking}.
	\item	Before carrying out a given group of operations, predict
		which locks will be acquired, and attempt to acquire them
		before actually carrying out any updates.
		If the prediction turns out to be incorrect, drop
		all the locks and retry with an updated prediction
		that includes the benefit of experience.
		This approach was discussed in
		\cref{sec:locking:Acquire Needed Locks First}.
	\item	Use transactional memory.
		This approach has a number of advantages and disadvantages
		which will be discussed in
		\crefrange{sec:future:Transactional Memory}{sec:future:Hardware Transactional Memory}.
	\item	Refactor the application to be more concurrency-friendly.
		This would likely also have the side effect of making
		the application run faster even when single-threaded, but might
		also make it more difficult to modify the application.
	\item	Use techniques from later chapters in addition to locking.
	\end{enumerate}
}\QuickQuizEnd

Nevertheless, the strategies described in this section have proven
quite useful in many settings.

\subsection{Livelock and Starvation}\ucindex{Livelock|BF}\ucindex{Starvation|BF}
\label{sec:locking:Livelock and Starvation}

Although conditional locking can be an effective deadlock-avoidance
mechanism, it can be abused.
Consider for example the beautifully symmetric example shown in
\cref{lst:locking:Abusing Conditional Locking}.
This example's beauty hides an ugly livelock.
To see this, consider the following sequence of events:

\begin{listing}
\begin{fcvlabel}[ln:locking:Abusing Conditional Locking]
\begin{VerbatimL}[commandchars=\\\[\]]
void thread1(void)
{
retry:					\lnlbl[thr1:retry]
	spin_lock(&lock1);		\lnlbl[thr1:acq1]
	do_one_thing();
	if (!spin_trylock(&lock2)) {	\lnlbl[thr1:try2]
		spin_unlock(&lock1);    \lnlbl[thr1:rel1]
		goto retry;
	}
	do_another_thing();
	spin_unlock(&lock2);
	spin_unlock(&lock1);
}

void thread2(void)
{
retry:					\lnlbl[thr2:retry]
	spin_lock(&lock2);		\lnlbl[thr2:acq2]
	do_a_third_thing();
	if (!spin_trylock(&lock1)) {	\lnlbl[thr2:try1]
		spin_unlock(&lock2);	\lnlbl[thr2:rel2]
		goto retry;
	}
	do_a_fourth_thing();
	spin_unlock(&lock1);
	spin_unlock(&lock2);
}
\end{VerbatimL}
\end{fcvlabel}
\caption{Abusing Conditional Locking}
\label{lst:locking:Abusing Conditional Locking}
\end{listing}

\begin{fcvref}[ln:locking:Abusing Conditional Locking]
\begin{enumerate}
\item	Thread~1 acquires \co{lock1} on \clnref{thr1:acq1}, then invokes
	\co{do_one_thing()}.
\item	Thread~2 acquires \co{lock2} on \clnref{thr2:acq2}, then invokes
	\co{do_a_third_thing()}.
\item	Thread~1 attempts to acquire \co{lock2} on \clnref{thr1:try2},
	but fails because Thread~2 holds it.
\item	Thread~2 attempts to acquire \co{lock1} on \clnref{thr2:try1},
	but fails because Thread~1 holds it.
\item	Thread~1 releases \co{lock1} on \clnref{thr1:rel1},
	then jumps to \co{retry} at \clnref{thr1:retry}.
\item	Thread~2 releases \co{lock2} on \clnref{thr2:rel2},
	and jumps to \co{retry} at \clnref{thr2:retry}.
\item	The livelock dance repeats from the beginning.
\end{enumerate}
\end{fcvref}

\QuickQuiz{
	How can the livelock shown in
	\cref{lst:locking:Abusing Conditional Locking}
	be avoided?
}\QuickQuizAnswer{
	\Cref{lst:locking:Avoiding Deadlock Via Conditional Locking}
	provides some good hints.
	In many cases, livelocks are a hint that you should revisit your
	locking design.
	Or visit it in the first place if your locking design
	``just grew''.

	That said, one good-and-sufficient approach due to Doug Lea
	is to use conditional locking as described in
	\cref{sec:locking:Conditional Locking}, but combine this
	with acquiring all needed locks first, before modifying shared
	data, as described in
	\cref{sec:locking:Acquire Needed Locks First}.
	If a given critical section retries too many times,
	unconditionally acquire
	a global lock, then unconditionally acquire all the needed locks.
	This avoids both deadlock and livelock, and scales reasonably
	assuming that the global lock need not be acquired too often.
}\QuickQuizEnd

Livelock can be thought of as an extreme form of starvation where
a group of threads starves, rather than just one of them.\footnote{
	Try not to get too hung up on the exact definitions of terms
	like livelock, starvation, and unfairness.
	Anything that causes a group of threads to fail to make adequate
	forward progress is a bug that needs to be fixed, and debating
	names doesn't fix bugs.}

\begin{listing}
\begin{fcvlabel}[ln:locking:Conditional Locking and Exponential Backoff]
\begin{VerbatimL}[commandchars=\\\[\]]
void thread1(void)
{
	unsigned int wait = 1;
retry:
	spin_lock(&lock1);
	do_one_thing();
	if (!spin_trylock(&lock2)) {
		spin_unlock(&lock1);
		sleep(wait);
		wait = wait << 1;
		goto retry;
	}
	do_another_thing();
	spin_unlock(&lock2);
	spin_unlock(&lock1);
}

void thread2(void)
{
	unsigned int wait = 1;
retry:
	spin_lock(&lock2);
	do_a_third_thing();
	if (!spin_trylock(&lock1)) {
		spin_unlock(&lock2);
		sleep(wait);
		wait = wait << 1;
		goto retry;
	}
	do_a_fourth_thing();
	spin_unlock(&lock1);
	spin_unlock(&lock2);
}
\end{VerbatimL}
\end{fcvlabel}
\caption{Conditional Locking and Exponential Backoff}
\label{lst:locking:Conditional Locking and Exponential Backoff}
\end{listing}

Livelock and starvation are serious issues in software transactional
memory implementations, and so the concept of \emph{contention
manager} has been introduced to encapsulate these issues.
In the case of locking, simple exponential backoff can often address
livelock and starvation.
The idea is to introduce exponentially increasing delays before each
retry, as shown in
\cref{lst:locking:Conditional Locking and Exponential Backoff}.

\QuickQuiz{
	What problems can you spot in the code in
	\cref{lst:locking:Conditional Locking and Exponential Backoff}?
}\QuickQuizAnswer{
	Here are a couple:
	\begin{enumerate}
	\item	A one-second wait is way too long for most uses.
		Wait intervals should begin with roughly the time
		required to execute the critical section, which will
		normally be in the microsecond or millisecond range.
	\item	The code does not check for overflow.
		On the other hand, this bug is nullified
		by the previous bug:
		32 bits worth of seconds is more than 50 years.
	\end{enumerate}
}\QuickQuizEnd

For better results, backoffs should be bounded, and
even better high-contention results are obtained via queued
locking~\cite{Anderson90}, which is discussed more in
\cref{sec:locking:Other Exclusive-Locking Implementations}.
Of course, best of all is to use a good parallel design that avoids
these problems by maintaining low \IX{lock contention}.

\subsection{Unfairness}
\label{sec:locking:Unfairness}

\begin{figure}
\centering
\resizebox{3in}{!}{\includegraphics{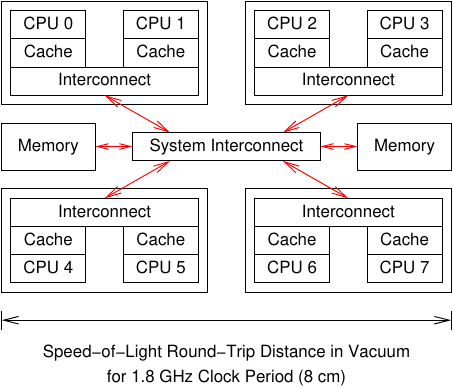}}
\caption{System Architecture and Lock Unfairness}
\label{fig:locking:System Architecture and Lock Unfairness}
\end{figure}

\IXB{Unfairness} can be thought of as a less-severe form of starvation,
where a subset of threads contending for a given lock are granted
the lion's share of the acquisitions.
This can happen on machines with shared caches or \IXacr{numa} characteristics,
for example, as shown in
\cref{fig:locking:System Architecture and Lock Unfairness}.
If CPU~0 releases a lock that all the other CPUs are attempting
to acquire, the interconnect shared between CPUs~0 and~1 means that
CPU~1 will have an advantage over CPUs~2--7.
Therefore CPU~1 will likely acquire the lock.
If CPU~1 holds the lock long enough for CPU~0 to be requesting the
lock by the time CPU~1 releases it and vice versa, the lock can
shuttle between CPUs~0 and~1, bypassing CPUs~2--7.

\QuickQuiz{
	Wouldn't it be better just to use a good parallel design
	so that lock contention was low enough to avoid unfairness?
}\QuickQuizAnswer{
	It would be better in some sense, but there are situations
	where it can be appropriate to use
	designs that sometimes result in high lock contentions.

	For example, imagine a system that is subject to a rare error
	condition.
	It might well be best to have a simple error-handling design
	that has poor performance and scalability for the duration of
	the rare error condition, as opposed to a complex and
	difficult-to-debug design that is helpful only when one of
	those rare error conditions is in effect.

	That said, it is usually worth putting some effort into
	attempting to produce a design that both simple as well as
	efficient during error conditions, for example by partitioning
	the problem.
}\QuickQuizEnd

\subsection{Inefficiency}
\label{sec:locking:Inefficiency}

Locks are implemented using atomic instructions and \IXpl{memory barrier},
and often involve cache misses.
As we saw in \cref{chp:Hardware and its Habits},
these instructions are quite expensive, roughly two
orders of magnitude greater overhead than simple instructions.
This can be a serious problem for locking:
If you protect a single instruction with a lock, you will increase
the overhead by a factor of one hundred.
Even assuming perfect scalability, \emph{one hundred} CPUs would
be required to keep up with a single CPU executing the same code
without locking.

\begin{figure}
\centering
\resizebox{2.0in}{!}{\includegraphics{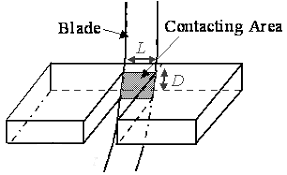}}
\caption{Saw Kerf}
\label{fig:locking:Saw Kerf}
\end{figure}

This situation is not confined to locking.
\Cref{fig:locking:Saw Kerf}
shows how this same principle applies to the age-old activity of
sawing wood.
As can be seen in the figure, sawing a board converts a small piece of
that board (the width of the saw blade) into sawdust.
Of course, locks partition time instead of sawing wood,\footnote{
	That is, locking is temporal synchronization.
	Mechanisms that synchronize both temporally and spatially
	are described in \cref{chp:Deferred Processing}.}
but just like sawing wood, using locks to partition time wastes some of
that time due to lock overhead and (worse yet) lock contention.
One important difference is that if someone saws a board into too-small
pieces, the resulting conversion of most of that board into sawdust will
be immediately obvious.
In contrast, it is not always obvious that a given lock acquisition
is wasting excessive amounts of time.

And this situation underscores the importance of the
synchronization\-/granularity tradeoff discussed in
\cref{sec:SMPdesign:Synchronization Granularity},
especially \cref{fig:SMPdesign:Synchronization Efficiency}:
Too coarse a granularity will limit scalability, while too fine a
granularity will result in excessive synchronization overhead.

Acquiring a lock might be expensive, but once held, the CPU's caches
are an effective performance booster, at least for large \IXpl{critical section}.
In addition, once a lock is held, the data protected by that lock can
be accessed by the lock holder without interference from other threads.

\QuickQuiz{
	How might the lock holder be interfered with?
}\QuickQuizAnswer{
	If the data protected by the lock is in the same cache line
	as the lock itself, then attempts by other CPUs to acquire
	the lock will result in expensive cache misses on the part
	of the CPU holding the lock.
	This is a special case of \IX{false sharing}, which can also occur
	if a pair of variables protected by different locks happen
	to share a cache line.
	In contrast, if the lock is in a different cache line than
	the data that it protects, the CPU holding the lock will
	usually suffer a cache miss only on first access to a given
	variable.

	Of course, the downside of placing the lock and data into separate
	cache lines is that the code will incur two cache misses rather
	than only one in the uncontended case.
	As always, choose wisely!
}\QuickQuizEnd

The Rust programming language takes lock/data association a step further
by allowing the developer to make a compiler-visible association between
a lock and the data that it protects~\cite{RalfJung2021RustSafeSysProg}.
When such an association has been made, attempts to access the data
without the benefit of the corresponding lock will result in a
compile-time diagnostic.
The hope is that this will greatly reduce the frequency of this class
of bugs.
Of course, this approach does not apply straightforwardly to cases
where the data to be locked is distributed throughout the nodes of
some data structure or when that which is locked is purely abstract,
for example, when a small subset of state-machine transitions is to
be protected by a given lock.
For this reason, Rust allows locks to be associated with types rather
than data items or even to be associated with nothing at all.
This last option permits Rust to emulate traditional locking use cases,
but is not popular among Rust developers.
Perhaps the Rust community will come up with other mechanisms tailored
to other locking use cases.

\section{Types of Locks}
\label{sec:locking:Types of Locks}
\epigraph{Only locks in life are what you think you know, but don't.
	  Accept your ignorance and try something new.}
	 {Dennis Vickers}

There are a surprising number of types of locks, more than this
short chapter can possibly do justice to.
The following sections discuss
exclusive locks (\cref{sec:locking:Exclusive Locks}),
reader-writer locks (\cref{sec:locking:Reader-Writer Locks}),
multi-role locks (\cref{sec:locking:Beyond Reader-Writer Locks}),
and scoped locking (\cref{sec:locking:Scoped Locking}).

\subsection{Exclusive Locks}
\label{sec:locking:Exclusive Locks}

\IXhpl{Exclusive}{lock} are what they say they are:
Only one thread may hold the lock at a time.
The holder of such a lock thus has exclusive access to all data protected
by that lock, hence the name.

Of course, this all assumes that this lock is held across all accesses
to data purportedly protected by the lock.
Although there are some tools that can help (see for example
\cref{sec:formal:Axiomatic Approaches and Locking}),
the ultimate responsibility for ensuring that the lock is always acquired
when needed rests with the developer.

\QuickQuiz{
	Does it ever make sense to have an exclusive lock acquisition
	immediately followed by a release of that same lock, that is,
	an empty critical section?
}\QuickQuizAnswer{
	Empty lock-based critical sections are rarely used, but they
	do have their uses.
	The point is that the semantics of exclusive locks have two
	components:
	\begin{enumerate*}[(1)]
	\item The familiar data-protection semantic and
	\item A messaging semantic, where releasing a given lock notifies
	a waiting acquisition of that same lock.
	\end{enumerate*}
	An empty critical section uses the messaging component without
	the data-protection component.

	The rest of this answer provides some example uses of empty
	critical sections, however, these examples should be considered
	``gray magic.''\footnote{
		Thanks to Alexey Roytman for this description.}
	As such, empty critical sections are almost never used in practice.
	Nevertheless, pressing on into this gray area \ldots

	One historical use of empty critical sections appeared in the
	networking stack of the 2.4 Linux kernel through use of a
	read-side-scalable reader-writer lock called \co{brlock}
	for ``big reader lock''.
	This use case is a way of approximating the semantics of read-copy
	update (RCU), which is discussed in
	\cref{sec:defer:Read-Copy Update (RCU)}.
	And in fact this Linux-kernel use case has been replaced
	with RCU\@.

	The empty-lock-critical-section idiom can also be used to
	reduce lock contention in some situations.
	For example, consider a multithreaded user-space application where
	each thread processes units of work maintained in a per-thread
	list, where threads are prohibited from touching each others'
	lists~\cite{PaulEMcKenney2012EmptyLocks}.
	There could also be updates that require that all previously
	scheduled units of work have completed before the update can
	progress.
	One way to handle this is to schedule a unit of work on each
	thread, so that when all of these units of work complete, the
	update may proceed.

	In some applications, threads can come and go.
	For example, each thread might correspond to one user of the
	application, and thus be removed when that user logs out or
	otherwise disconnects.
	In many applications, threads cannot depart atomically:
	They must instead explicitly unravel themselves from various
	portions of the application using a specific sequence of actions.
	One specific action will be refusing to accept further requests
	from other threads, and another specific action will be disposing
	of any remaining units of work on its list, for example, by
	placing these units of work in a global work-item-disposal list
	to be taken by one of the remaining threads.
	(Why not just drain the thread's work-item list by executing
	each item?
	Because a given work item might generate more work items, so
	that the list could not be drained in a timely fashion.)

	If the application is to perform and scale well, a good locking
	design is required.
	One common solution is to have a global lock (call it \co{G})
	protecting the entire
	process of departing (and perhaps other things as well),
	with finer-grained locks protecting the
	individual unraveling operations.

	Now, a departing thread must clearly refuse to accept further
	requests before disposing of the work on its list, because
	otherwise additional work might arrive after the disposal action,
	which would render that disposal action ineffective.
	So simplified pseudocode for a departing thread might be as follows:

	\begin{enumerate}
	\item	Acquire lock \co{G}.
	\item	Acquire the lock guarding communications.
	\item	Refuse further communications from other threads.
	\item	Release the lock guarding communications.
	\item	Acquire the lock guarding the global work-item-disposal list.
	\item	Move all pending work items to the global
		work-item-disposal list.
	\item	Release the lock guarding the global work-item-disposal list.
	\item	Release lock \co{G}.
	\end{enumerate}

	Of course, a thread that needs to wait for all pre-existing work
	items will need to take departing threads into account.
	To see this, suppose that this thread starts waiting for all
	pre-existing work items just after a departing thread has refused
	further communications from other threads.
	How can this thread wait for the departing thread's work items
	to complete, keeping in mind that threads are not allowed to
	access each others' lists of work items?

	One straightforward approach is for this thread to acquire \co{G}
	and then the lock guarding the global work-item-disposal list, then
	move the work items to its own list.
	The thread then release both locks,
	places a work item on the end of its own list,
	and then wait for all of the work items that it placed on each thread's
	list (including its own) to complete.

	This approach does work well in many cases, but if special
	processing is required for each work item as it is pulled in
	from the global work-item-disposal list, the result could be
	excessive contention on \co{G}.
	One way to avoid that contention is to acquire \co{G} and then
	immediately release it.
	Then the process of waiting for all prior work items look
	something like the following:

	\begin{enumerate}
	\item	Set a global counter to one and initialize a condition
		variable to zero.
	\item	Send a message to all threads to cause them to atomically
		increment the global counter, and then to enqueue a
		work item.
		The work item will atomically decrement the global
		counter, and if the result is zero, it will set a
		condition variable to one.
	\item	Acquire \co{G}, which will wait on any currently departing
		thread to finish departing.
		Because only one thread may depart at a time, all the
		remaining threads will have already received the message
		sent in the preceding step.
	\item	Release \co{G}.
	\item	Acquire the lock guarding the global work-item-disposal list.
	\item	Move all work items from the global work-item-disposal list
		to this thread's list, processing them as needed along the way.
	\item	Release the lock guarding the global work-item-disposal list.
	\item	Enqueue an additional work item onto this thread's list.
		(As before, this work item will atomically decrement
		the global counter, and if the result is zero, it will
		set a condition variable to one.)
	\item	Wait for the condition variable to take on the value one.
	\end{enumerate}

	Once this procedure completes, all pre-existing work items are
	guaranteed to have completed.
	The empty critical sections are using locking for messaging as
	well as for protection of data.
}\QuickQuizEnd

\QuickQuizLabel{\QlockingQemptycriticalsection}

It is important to note that unconditionally acquiring an exclusive lock
has two effects:
\begin{enumerate*}[(1)]
\item Waiting for all prior holders of that lock to release it and
\item Blocking any other acquisition attempts until the lock is released.
\end{enumerate*}
As a result, at lock acquisition time, any concurrent acquisitions of
that lock must be partitioned into prior holders and subsequent
holders.
Different types of exclusive locks use different partitioning
strategies~\cite{BjoernBrandenburgPhD,Guerraoui:2019:LPA:3319851.3301501},
for example:

\begin{enumerate}
\item	Strict FIFO, with acquisitions starting earlier acquiring
	the lock earlier.
\item	Approximate FIFO, with acquisitions starting sufficiently
	earlier acquiring the lock earlier.
\item	FIFO within priority level, with higher-priority threads
	acquiring the lock earlier than any lower-priority threads
	attempting to acquire the lock at about the same time, but so
	that some FIFO ordering applies for threads of the same priority.
\item	Random, so that the new lock holder is chosen randomly from
	all threads attempting acquisition, regardless of timing.
\item
	Unfair, so that a given acquisition might never acquire the lock
	(see \cref{sec:locking:Unfairness}).
\end{enumerate}

Unfortunately, locking implementations with stronger guarantees
typically incur higher overhead, motivating the wide variety of locking
implementations in production use.
For example, real-time systems often require some degree of FIFO
ordering within priority level, and much else besides
(see \cref{sec:advsync:Event-Driven Real-Time Support}),
while non-realtime systems subject to high contention might require
only enough ordering to avoid starvation, and finally, non-realtime
systems designed to avoid contention might not need fairness at all.

\subsection{Reader-Writer Locks}
\label{sec:locking:Reader-Writer Locks}

\IXhpl{Reader-writer}{lock}~\cite{Courtois71}
permit any number of readers to hold the lock
concurrently on the one hand or a single writer to hold the lock
on the other.
In theory, then, reader-writer locks should allow excellent scalability
for data that is read often and written rarely.
In practice, the scalability will depend on the reader-writer lock
implementation.

The classic reader-writer lock implementation involves a set of
counters and flags that are manipulated atomically.
This type of implementation suffers from the same problem as does
exclusive locking for short critical sections:
The overhead of acquiring and releasing the lock
is about two orders of magnitude greater than the overhead
of a simple instruction.
Of course, if the critical section is long enough, the overhead of
acquiring and releasing the lock becomes negligible.
However, because only
one thread at a time can be manipulating the lock, the required
critical-section size increases with the number of CPUs.

It is possible to design a reader-writer lock that is much more
favorable to readers through use of per-thread exclusive
locks~\cite{WilsonCHsieh92a}.
To read, a thread acquires only its own lock.
To write, a thread acquires all locks.
In the absence of writers, each reader incurs only atomic-instruction
and memory-barrier overhead, with no cache misses, which is quite
good for a locking primitive.
Unfortunately, writers must incur cache misses as well as atomic-instruction
and memory-barrier overhead---multiplied by the number of threads.

In short, reader-writer locks can be quite useful in a number of
situations, but each type of implementation does have its drawbacks.
The canonical use case for reader-writer locking involves very long
\IXhpl{read-side}{critical section}, preferably measured in hundreds of microseconds
or even milliseconds.

As with exclusive locks, a reader-writer lock acquisition cannot complete
until all prior conflicting holders of that lock have released it.
If a lock is read-held, then read acquisitions can complete immediately,
but write acquisitions must wait until there are no longer any readers
holding the lock.
If a lock is write-held, then all acquisitions must wait until the writer
releases the lock.
Again as with exclusive locks, different reader-writer lock implementations
provide different degrees of FIFO ordering to readers on the one hand
and to writers on the other.

But suppose a large number of readers hold the lock and a writer is waiting
to acquire the lock.
Should readers be allowed to continue to acquire the lock, possibly
starving the writer?
Similarly, suppose that a writer holds the lock and that a large number
of both readers and writers are waiting to acquire the lock.
When the current writer releases the lock, should it be given to a reader
or to another writer?
If it is given to a reader, how many readers should be allowed to acquire
the lock before the next writer is permitted to do so?

There are many possible answers to these questions, with different
levels of complexity, overhead, and fairness.
Different implementations might have different costs, for example,
some types of reader-writer locks incur extremely large latencies
when switching from read-holder to write-holder mode.
Here are a few possible approaches:

\begin{enumerate}
\item
	Reader-preference implementations unconditionally favor readers
	over writers, possibly allowing write acquisitions to be
	indefinitely blocked.
\item	Batch-fair implementations ensure that when both readers and writers
	are acquiring the lock, both have reasonable access via batching.
	For example, the lock might admit five readers per CPU, then two
	writers, then five more readers per CPU, and so on.
\item	Writer-preference implementations unconditionally favor
	writers over readers, possibly allowing read acquisitions to be
	indefinitely blocked.
\end{enumerate}

Of course, these distinctions matter only under conditions of high
lock contention.

Please keep the waiting/blocking dual nature of locks firmly in mind.
This will be revisited in \cref{chp:Deferred Processing}'s discussion
of scalable high-performance special-purpose alternatives to locking.

\subsection{Beyond Reader-Writer Locks}
\label{sec:locking:Beyond Reader-Writer Locks}

\begin{table}
\renewcommand*{\arraystretch}{1.2}
\newcommand{\x}{\textcolor{gray!20}{\rule{7pt}{7pt}}}
\newcommand{\rothead}[1]{\begin{picture}(6,65)(0,0)\rotatebox{90}{#1}\end{picture}}
\small
\centering
\begin{tabular}{lcccccc}
	\toprule
	& \rothead{Null (Not Held)}
	& \rothead{Concurrent Read}
	& \rothead{Concurrent Write}
	& \rothead{Protected Read}
	& \rothead{Protected Write}
	& \rothead{Exclusive}
	\\
%				  NL   CR   CW     PR   PW   EX
	\cmidrule(r){1-1} \cmidrule{2-7}
	Null (Not Held)		& \x & \x & \x   & \x & \x & \x \\
	Concurrent Read		& \x & \x & \x   & \x & \x &  X \\
	Concurrent Write	& \x & \x & \x   &  X &  X &  X \\
	Protected Read		& \x & \x &  X   & \x &  X &  X \\
	Protected Write		& \x & \x &  X   &  X &  X &  X \\
	Exclusive		& \x &  X &  X   &  X &  X &  X \\
	\bottomrule
\end{tabular}
\caption{VAX/VMS Distributed Lock Manager Policy}
\label{tab:locking:VAX/VMS Distributed Lock Manager Policy}
\end{table}

Reader-writer locks and exclusive locks differ in their admission policy:
Exclusive locks allow at most one holder, while reader-writer locks
permit an arbitrary number of read-holders (but only one write-holder).
There is a very large number of possible admission policies, one of
which is that of the VAX/VMS distributed lock
manager (DLM)~\cite{Snaman87}, which is shown in
\cref{tab:locking:VAX/VMS Distributed Lock Manager Policy}.
Blank cells indicate compatible modes, while cells containing ``X''
indicate incompatible modes.

The VAX/VMS DLM uses six modes.
For purposes of comparison, exclusive
locks use two modes (not held and held), while reader-writer locks
use three modes (not held, read held, and write held).

The first mode is null, or not held.
This mode is compatible with all other modes, which is to be expected:
If a thread is not holding a lock, it should not prevent any
other thread from acquiring that lock.

The second mode is concurrent read, which is compatible with every other
mode except for exclusive.
The concurrent-read mode might be used to accumulate approximate
statistics on a data structure, while permitting updates to proceed
concurrently.

The third mode is concurrent write, which is compatible with null,
concurrent read, and concurrent write.
The concurrent-write mode might be used to update approximate statistics,
while still permitting reads and concurrent updates to proceed
concurrently.

The fourth mode is protected read, which is compatible with null,
concurrent read, and protected read.
The protected-read mode might be used to obtain a consistent snapshot
of the data structure, while permitting reads but not updates to
proceed concurrently.

The fifth mode is protected write, which is compatible with null and
concurrent read.
The protected-write mode might be used to carry out updates to a data
structure that could interfere with protected readers but which could
be tolerated by concurrent readers.

The sixth and final mode is exclusive, which is compatible only with null.
The exclusive mode is used when it is necessary to exclude all other accesses.

It is interesting to note that exclusive locks and reader-writer locks
can be emulated by the VAX/VMS DLM\@.
Exclusive locks would use only the null and exclusive modes, while
reader-writer locks might use the null, protected-read, and
protected-write modes.

\QuickQuiz{
	Is there any other way for the VAX/VMS DLM to emulate
	a reader-writer lock?
}\QuickQuizAnswer{
	There are in fact several.
	One way would be to use the null, protected-read, and exclusive
	modes.
	Another way would be to use the null, protected-read, and
	concurrent-write modes.
	A third way would be to use the null, concurrent-read, and
	exclusive modes.
}\QuickQuizEnd

Although the VAX/VMS DLM policy has seen widespread production use
for distributed databases, it does not appear to be used much in
shared-memory applications.
One possible reason for this is that the greater communication overheads
of distributed databases can hide the greater overhead of the
VAX/VMS DLM's more-complex admission policy.

Nevertheless, the VAX/VMS DLM is an interesting illustration of just
how flexible the concepts behind locking can be.
It also serves as a very simple introduction to the locking schemes
used by modern DBMSes, which can have more than thirty locking modes,
compared to VAX/VMS's six.

\subsection{Scoped Locking}
\label{sec:locking:Scoped Locking}

The locking primitives discussed thus far require explicit acquisition and
release primitives, for example, \co{spin_lock()} and \co{spin_unlock()},
respectively.
Another approach is to use the object-oriented \IXacrmfst{raii}
pattern~\cite{MargaretAEllis1990Cplusplus}.\footnote{
	Though more clearly expressed at
	\url{https://www.stroustrup.com/bs_faq2.html\#finally}.}
This pattern is often applied to auto variables in languages like C++,
where the corresponding \emph{constructor} is invoked upon entry to
the object's scope, and the corresponding \emph{destructor} is invoked
upon exit from that scope.
This can be applied to locking by having the constructor acquire the
lock and the destructor free it.

This approach can be quite useful, in fact in 1990 I was convinced that it
was the only type of locking that was needed.\footnote{
	My later work with parallelism at Sequent Computer Systems very
	quickly disabused me of this misguided notion.}
One very nice property of RAII locking is that you don't need to carefully
release the lock on each and every code path that exits that scope,
a property that can eliminate a troublesome set of bugs.

However, RAII locking also has a dark side.
RAII makes it quite difficult to encapsulate lock acquisition and release,
for example, in iterators.
In many iterator implementations, you would like to acquire the lock in
the iterator's ``start'' function and release it in the iterator's ``stop''
function.
RAII locking instead requires that the lock acquisition and release take
place in the same level of scoping, making such encapsulation difficult or
even impossible.

Strict RAII locking also prohibits overlapping critical sections, due
to the fact that scopes must nest.
This prohibition makes it difficult or impossible to express a number of
useful constructs, for example, locking trees
that mediate between multiple concurrent attempts to assert an event.
Of an arbitrarily large group of concurrent attempts, only one need succeed,
and the best strategy for the remaining attempts is for them to fail as
quickly and painlessly as possible.
Otherwise, lock contention becomes pathological on large systems
(where ``large'' is many hundreds of CPUs).
Therefore, C++17~\cite{RichardSmith2019N4800} has escapes from strict RAII
in its \co{unique_lock} class, which allows the scope of the critical
section to be controlled to roughly the same extent as can be achieved
with explicit lock acquisition and release primitives.

\begin{figure}
\centering
\resizebox{3in}{!}{\includegraphics{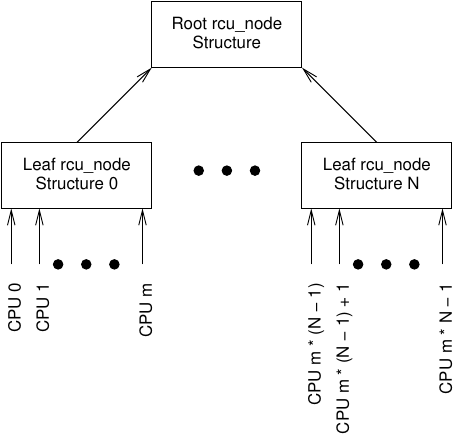}}
\caption{Locking Hierarchy}
\label{fig:locking:Locking Hierarchy}
\end{figure}

Example strict-RAII-unfriendly data structures from Linux-kernel RCU
are shown in
\cref{fig:locking:Locking Hierarchy}.
Here, each CPU is assigned a leaf \co{rcu_node} structure, and each
\co{rcu_node} structure has a pointer to its parent (named, oddly
enough, \co{->parent}), up to the root \co{rcu_node} structure,
which has a \co{NULL} \co{->parent} pointer.
The number of child \co{rcu_node} structures per parent can vary,
but is typically 32 or 64.
Each \co{rcu_node} structure also contains a lock named \co{->fqslock}.

The general approach is a \emph{tournament}, where
a given CPU conditionally acquires its
leaf \co{rcu_node} structure's \co{->fqslock}, and, if successful,
attempt to acquire that of the parent, then release that of the child.
In addition, at each level, the CPU checks a global \co{gp_flags}
variable, and if this variable indicates that some other CPU has
asserted the event, the first CPU drops out of the competition.
This acquire-then-release sequence continues until either the
\co{gp_flags} variable indicates that someone else won the tournament,
one of the attempts to acquire an \co{->fqslock} fails, or
the root \co{rcu_node} structure's \co{->fqslock} has been acquired.
If the root \co{rcu_node} structure's \co{->fqslock} is acquired,
a function named \co{do_force_quiescent_state()} is invoked.

\begin{listing}
\begin{fcvlabel}[ln:locking:Conditional Locking to Reduce Contention]
\begin{VerbatimL}[commandchars=\\\[\]]
void force_quiescent_state(struct rcu_node *rnp_leaf)
{
	int ret;
	struct rcu_node *rnp = rnp_leaf;
	struct rcu_node *rnp_old = NULL;

	for (; rnp != NULL; rnp = rnp->parent) {	\lnlbl[loop:b]
		ret = (READ_ONCE(gp_flags)) ||		\lnlbl[flag_set]
		       !raw_spin_trylock(&rnp->fqslock);\lnlbl[trylock]
		if (rnp_old != NULL)			\lnlbl[non_NULL]
			raw_spin_unlock(&rnp_old->fqslock);\lnlbl[rel1]
		if (ret)				\lnlbl[giveup]
			return;				\lnlbl[return]
		rnp_old = rnp;				\lnlbl[save]
	}						\lnlbl[loop:e]
	if (!READ_ONCE(gp_flags)) {			\lnlbl[flag_not_set]
		WRITE_ONCE(gp_flags, 1);		\lnlbl[set_flag]
		do_force_quiescent_state();		\lnlbl[invoke]
		WRITE_ONCE(gp_flags, 0);		\lnlbl[clr_flag]
	}
	raw_spin_unlock(&rnp_old->fqslock);		\lnlbl[rel2]
}
\end{VerbatimL}
\end{fcvlabel}
\caption{Conditional Locking to Reduce Contention}
\label{lst:locking:Conditional Locking to Reduce Contention}
\end{listing}

Simplified code to implement this is shown in
\cref{lst:locking:Conditional Locking to Reduce Contention}.
The purpose of this function is to mediate between CPUs who have concurrently
detected a need to invoke the \co{do_force_quiescent_state()} function.
At any given time, it only makes sense for one instance of
\co{do_force_quiescent_state()} to be active, so if there are multiple
concurrent callers, we need at most one of them to actually invoke
\co{do_force_quiescent_state()}, and we need the rest to (as quickly and
painlessly as possible) give up and leave.

\begin{fcvref}[ln:locking:Conditional Locking to Reduce Contention]
To this end, each pass through the loop spanning \clnrefrange{loop:b}{loop:e} attempts
to advance up one level in the \co{rcu_node} hierarchy.
If the \co{gp_flags} variable is already set (\clnref{flag_set}) or if the attempt
to acquire the current \co{rcu_node} structure's \co{->fqslock} is
unsuccessful (\clnref{trylock}), then local variable \co{ret} is set to 1.
If \clnref{non_NULL} sees that local variable \co{rnp_old} is non-\co{NULL},
meaning that we hold \co{rnp_old}'s \co{->fqs_lock},
\clnref{rel1} releases this lock (but only after the attempt has been made
to acquire the parent \co{rcu_node} structure's \co{->fqslock}).
If \clnref{giveup} sees that either \clnref{flag_set} or~\lnref{trylock}
saw a reason to give up,
\clnref{return} returns to the caller.
Otherwise, we must have acquired the current \co{rcu_node} structure's
\co{->fqslock}, so \clnref{save} saves a pointer to this structure in local
variable \co{rnp_old} in preparation for the next pass through the loop.

If control reaches \clnref{flag_not_set}, we won the tournament, and now holds the
root \co{rcu_node} structure's \co{->fqslock}.
If \clnref{flag_not_set} still sees that the global variable \co{gp_flags} is zero,
\clnref{set_flag} sets \co{gp_flags} to one, \clnref{invoke} invokes
\co{do_force_quiescent_state()},
and \clnref{clr_flag} resets \co{gp_flags} back to zero.
Either way, \clnref{rel2} releases the root \co{rcu_node} structure's
\co{->fqslock}.
\end{fcvref}

\QuickQuizSeries{%
\QuickQuizB{
	The code in
	\cref{lst:locking:Conditional Locking to Reduce Contention}
	is ridiculously complicated!
	Why not conditionally acquire a single global lock?
}\QuickQuizAnswerB{
	Conditionally acquiring a single global lock does work very well,
	but only for relatively small numbers of CPUs.
	To see why it is problematic in systems with many hundreds of
	CPUs, look at
	\cref{fig:count:Atomic Increment Scalability on x86}.
}\QuickQuizEndB
\QuickQuizE{
	\begin{fcvref}[ln:locking:Conditional Locking to Reduce Contention]
	Wait a minute!
	If we ``win'' the tournament on \clnref{flag_not_set} of
	\cref{lst:locking:Conditional Locking to Reduce Contention},
	we get to do all the work of \co{do_force_quiescent_state()}.
	Exactly how is that a win, really?
	\end{fcvref}
}\QuickQuizAnswerE{
	How indeed?
	This just shows that in concurrency, just as in life, one
	should take care to learn exactly what winning entails before
	playing the game.
}\QuickQuizEndE
}

This function illustrates the not-uncommon pattern of hierarchical
locking.
This pattern is difficult to implement using strict RAII locking,\footnote{
	Which is why many RAII locking implementations provide a way
	to leak the lock out of the scope that it was acquired and into
	the scope in which it is to be released.
	However, some object must mediate the scope leaking, which can
	add complexity compared to non-RAII explicit locking primitives.}
just like the iterator encapsulation noted earlier, and so explicit
lock/unlock primitives (or C++17-style \co{unique_lock} escapes) will
be required for the foreseeable future.

\section{Locking Implementation Issues}
\label{sec:locking:Locking Implementation Issues}
\epigraph{When you translate a dream into reality, it's never a full
	  implementation.
	  It is easier to dream than to do.}
	 {Shai Agassi}

Developers are almost always best-served by using whatever locking
primitives are provided by the system, for example, the POSIX
pthread mutex locks~\cite{OpenGroup1997pthreads,Butenhof1997pthreads}.
Nevertheless, studying sample implementations can be helpful,
as can considering the challenges posed by extreme workloads and
environments.

\subsection{Sample Exclusive-Locking Implementation Based on Atomic Exchange}
\label{sec:locking:Sample Exclusive-Locking Implementation Based on Atomic Exchange}

\begin{fcvref}[ln:locking:xchglock:lock_unlock]
This section reviews the implementation shown in
\cref{lst:locking:Sample Lock Based on Atomic Exchange}.
The data structure for this lock is just an \co{int}, as shown on
\clnref{typedef}, but could be any integral type.
The initial value of this lock is zero, meaning ``unlocked'',
as shown on \clnref{initval}.
\end{fcvref}

\begin{listing}
\input{CodeSamples/locking/xchglock=lock_unlock.fcv}
\caption{Sample Lock Based on Atomic Exchange}
\label{lst:locking:Sample Lock Based on Atomic Exchange}
\end{listing}

\QuickQuiz{
	\begin{fcvref}[ln:locking:xchglock:lock_unlock]
	Why not rely on the C language's default initialization of
	zero instead of using the explicit initializer shown on
	\clnref{initval} of
	\cref{lst:locking:Sample Lock Based on Atomic Exchange}?
	\end{fcvref}
}\QuickQuizAnswer{
	Because this default initialization does not apply to locks
	allocated as auto variables within the scope of a function.
}\QuickQuizEnd

\begin{fcvref}[ln:locking:xchglock:lock_unlock:lock]
Lock acquisition is carried out by the \co{xchg_lock()} function
shown on \clnrefrange{b}{e}.
This function uses a nested loop, with the outer loop repeatedly
atomically exchanging the value of the lock with the value one
(meaning ``locked'').
If the old value was already the value one (in other words, someone
else already holds the lock), then the inner loop (\clnrefrange{inner:b}{inner:e})
spins until the lock is available, at which point the outer loop
makes another attempt to acquire the lock.
\end{fcvref}

\QuickQuiz{
	\begin{fcvref}[ln:locking:xchglock:lock_unlock:lock]
	Why bother with the inner loop on \clnrefrange{inner:b}{inner:e} of
	\cref{lst:locking:Sample Lock Based on Atomic Exchange}?
	Why not simply repeatedly do the atomic exchange operation
	on \clnref{atmxchg}?
	\end{fcvref}
}\QuickQuizAnswer{
	\begin{fcvref}[ln:locking:xchglock:lock_unlock:lock]
	Suppose that the lock is held and that several threads
	are attempting to acquire the lock.
	In this situation, if these threads all loop on the atomic
	exchange operation, they will ping-pong the cache line
	containing the lock among themselves, imposing load
	on the interconnect.
	In contrast, if these threads are spinning in the inner
	loop on \clnrefrange{inner:b}{inner:e},
	they will each spin within their own
	caches, placing negligible load on the interconnect.
	\end{fcvref}
}\QuickQuizEnd

\begin{fcvref}[ln:locking:xchglock:lock_unlock:unlock]
Lock release is carried out by the \co{xchg_unlock()} function
shown on \clnrefrange{b}{e}.
\Clnref{atmxchg} atomically exchanges the value zero (``unlocked'') into
the lock, thus marking it as having been released.
\end{fcvref}

\QuickQuiz{
	\begin{fcvref}[ln:locking:xchglock:lock_unlock:unlock]
	Why not simply store zero into the lock word on \clnref{atmxchg} of
	\cref{lst:locking:Sample Lock Based on Atomic Exchange}?
	\end{fcvref}
}\QuickQuizAnswer{
	This can be a legitimate implementation, but only if
	this store is preceded by a memory barrier and makes use
	of \co{WRITE_ONCE()}.
	The memory barrier is not required when the \co{xchg()}
	operation is used because this operation implies a
	full memory barrier due to the fact that it returns
	a value.
}\QuickQuizEnd

This lock is a simple example of a test-and-set lock~\cite{Segall84},
but very similar
mechanisms have been used extensively as pure spinlocks in production.

\subsection{Other Exclusive-Locking Implementations}
\label{sec:locking:Other Exclusive-Locking Implementations}

There are a great many other possible implementations of locking based
on atomic instructions, many of which are reviewed in the classic paper
by Mellor-Crummey and Scott~\cite{MellorCrummey91a}.
These implementations represent different points in a multi-dimensional
design tradeoff~\cite{Guerraoui:2019:LPA:3319851.3301501,HugoGuirouxPhD,McKenney96a}.
For example,
the atomic-exchange-based test-and-set lock presented in the previous
section works well when contention is low and has the advantage
of small memory footprint.
It avoids giving the lock to threads that cannot use it, but as
a result can suffer from \IX{unfairness} or even \IX{starvation} at high
contention levels.

In contrast, ticket lock~\cite{MellorCrummey91a}, which was once used
in the Linux kernel, avoids unfairness at high contention levels.
However, as a consequence of its strict FIFO discipline, it can grant
the lock to a thread that is currently unable to use it, perhaps due
to that thread being preempted or interrupted.
On the other hand, it is important to avoid getting too worried about the
possibility of preemption and interruption.
After all, in many cases, this preemption and interruption could just
as well happen just after the lock was
acquired.\footnote{
	Besides, the best way of handling high lock contention is to avoid
	it in the first place!
	There are nevertheless some situations where high lock contention
	is the lesser of the available evils, and in any case, studying
	schemes that deal with high levels of contention is a good mental
	exercise.}

All locking implementations where waiters spin on a single memory
location, including both test-and-set locks and ticket locks,
suffer from performance problems at high contention levels.
The problem is that the thread releasing the lock must update the
value of the corresponding memory location.
At low contention, this is not a problem:
The corresponding cache line is very likely still local to and
writeable by the thread holding the lock.
In contrast, at high levels of contention, each thread attempting to
acquire the lock will have a read-only copy of the \IX{cache line}, and
the lock holder will need to invalidate all such copies before it
can carry out the update that releases the lock.
In general, the more CPUs and threads there are, the greater the
overhead incurred when releasing the lock under conditions of
high contention.

This negative scalability has motivated a number of different
queued-lock
implementations~\cite{Anderson90,Graunke90,MellorCrummey91a,Wisniewski94,Craig93,Magnusson94,Takada93},
some of which are used in recent versions of the Linux
kernel~\cite{JonathanCorbet2014qspinlocks}.
Queued locks avoid high cache-invalidation overhead by assigning each
thread a queue element.
These queue elements are linked together into a queue that governs the
order that the lock will be granted to the waiting threads.
The key point is that each thread spins on its own queue element,
so that the lock holder need only invalidate the first element from
the next thread's CPU's cache.
This arrangement greatly reduces the overhead of lock handoff at high
levels of contention.

More recent queued-lock implementations also take the system's architecture
into account, preferentially granting locks locally, while also taking
steps to avoid
starvation~\cite{McKenney02e,radovic03hierarchical,radovic02efficient,BenJackson02,McKenney02d}.
Many of these can be thought of as analogous to the elevator algorithms
traditionally used in scheduling disk I/O.

Unfortunately, the same scheduling logic that improves the \IX{efficiency}
of queued locks at high contention also increases their overhead at
low contention.
Beng-Hong Lim and Anant Agarwal therefore combined a simple test-and-set
lock with a queued lock, using the test-and-set lock at low levels of
contention and switching to the queued lock at high levels of
contention~\cite{BengHongLim94}, thus getting low overhead at low levels
of contention and getting fairness and high throughput at high levels
of contention.
Browning et al.\ took a similar approach, but avoided the use of a separate
flag, so that the test-and-set fast path uses the same sequence of
instructions that would be used in a simple test-and-set
lock~\cite{LukeBrowning2005SimpleLockNUMAAware}.
This approach has been used in production.

Another issue that arises at high levels of contention is when the
lock holder is delayed, especially when the delay is due to
preemption, which can result in \emph{priority inversion},
where a low-priority thread holds a lock, but is preempted
by a medium priority CPU-bound thread, which results in
a high-priority process blocking while attempting to acquire the
lock.
The result is that the CPU-bound medium-priority process is preventing the
high-priority process from running.
One solution is \emph{priority inheritance}~\cite{Lampson1980Mesa},
which has been widely used for real-time
computing~\cite{LuiSha1990PriorityInheritance,JonathanCorbet2006PriorityInheritance},
despite some lingering controversy over this
practice~\cite{Yodaiken2004FSM,DougLocke2002a}.

Another way to avoid priority inversion is to prevent preemption
while a lock is held.
Because preventing preemption while locks are held also improves throughput,
most proprietary UNIX kernels offer some form of scheduler-conscious
synchronization mechanism~\cite{Kontothanassis97a},
largely due to the efforts of a certain sizable database vendor.
These mechanisms usually take the form of a hint that preemption
should be avoided in a given region of code, with this hint typically
being placed in a machine register.
These hints frequently take the form of a bit set in a particular
machine register, which enables extremely low per-lock-acquisition overhead
for these mechanisms.
In contrast, Linux avoids these hints.
Instead, the Linux kernel community's response to requests for
scheduler-conscious synchronization was a mechanism called
\emph{futexes}~\cite{HubertusFrancke2002Futex,IngoMolnar2006RobustFutexes,StevenRostedt2006piFutexes,UlrichDrepper2011Futexes}.

Interestingly enough, atomic instructions are not strictly needed to
implement locks~\cite{Dijkstra65a,Lamport74a}.
An excellent exposition of the issues surrounding locking implementations
based on simple loads and stores may be found in Herlihy's and
Shavit's textbook~\cite{HerlihyShavit2008Textbook,HerlihyShavit2020Textbook}.
The main point echoed here is that such implementations currently
have little practical application, although a careful study of
them can be both entertaining and enlightening.
Nevertheless, with one exception described below, such study is left
as an exercise for the reader.

Gamsa et al.~\cite[Section 5.3]{Gamsa99} describe a token-based
mechanism in which a token circulates among the CPUs.
When the token reaches a given CPU, it has exclusive
access to anything protected by that token.
There are any number of schemes that may be used to implement
the token-based mechanism, for example:

\begin{enumerate}
\item	Maintain a per-CPU flag, which is initially
	zero for all but one CPU\@.
	When a CPU's flag is non-zero, it holds the token.
	When it finishes with the token, it zeroes its flag and
	sets the flag of the next CPU to one (or to any other
	non-zero value).
\item	Maintain a per-CPU counter, which is initially set
	to the corresponding CPU's number, which we assume
	to range from zero to $N-1$, where $N$ is the number
	of CPUs in the system.
	When a CPU's counter is greater than that of the next
	CPU (taking counter wrap into account), the first CPU holds the token.
	When it is finished with the token, it sets the next
	CPU's counter to a value one greater than its own counter.
\end{enumerate}

\QuickQuizSeries{%
\QuickQuizB{
	How can you tell if one counter is greater than another,
	while accounting for counter wrap?
}\QuickQuizAnswerB{
	In the C language, the following macro correctly handles this:

\begin{VerbatimU}
#define ULONG_CMP_LT(a, b) \
        (ULONG_MAX / 2 < (a) - (b))
\end{VerbatimU}

	Although it is tempting to simply subtract two signed integers,
	this should be avoided because signed overflow is undefined
	in the C language.
	For example, if the compiler knows that one of the values is
	positive and the other negative, it is within its rights to
	simply assume that the positive number is greater than the
	negative number, even though subtracting the negative number
	from the positive number might well result in overflow and
	thus a negative number.

	How could the compiler know the signs of the two numbers?
	It might be able to deduce it based on prior assignments
	and comparisons.
	In this case, if the per-CPU counters were signed, the compiler
	could deduce that they were always increasing in value, and
	then might assume that they would never go negative.
	This assumption could well lead the compiler to generate
	unfortunate code~\cite{PaulEMcKenney2012SignedOverflow,JohnRegehr2010UndefinedBehavior}.
}\QuickQuizEndB
\QuickQuizE{
	Which is better, the counter approach or the flag approach?
}\QuickQuizAnswerE{
	The flag approach will normally suffer fewer cache misses,
	but a better answer is to try both and see which works best
	for your particular workload.
}\QuickQuizEndE
}

This lock is unusual in that a given CPU cannot necessarily acquire it
immediately, even if no other CPU is using it at the moment.
Instead, the CPU must wait until the token comes around to it.
This is useful in cases where CPUs need periodic access to the \IX{critical
section}, but can tolerate variances in token-circulation rate.
Gamsa et al.~\cite{Gamsa99} used it to implement a variant of
read-copy update (see \cref{sec:defer:Read-Copy Update (RCU)}),
but it could also be used to protect periodic per-CPU operations such
as flushing per-CPU caches used by memory allocators~\cite{McKenney93},
garbage-collecting per-CPU data structures, or flushing per-CPU
data to shared storage (or to mass storage, for that matter).

The Linux kernel now uses queued spinlocks~\cite{JonathanCorbet2014qspinlocks},
but because of the complexity of implementations that provide good
performance across the range of contention levels, the path has not
always been smooth~\cite{CatalinMarinas2018qspinlockTLA,WillDeacon2018qspinlock}.
As increasing numbers of people gain familiarity with parallel hardware
and parallelize increasing amounts of code, we can continue to expect more
special-purpose locking primitives to appear, see for example Guerraoui et
al.~\cite{Guerraoui:2019:LPA:3319851.3301501,HugoGuirouxPhD}.
Nevertheless, you should carefully consider this important safety tip:
Use the standard synchronization primitives whenever humanly possible.
The big advantage of the standard synchronization primitives over
roll-your-own efforts is that the standard primitives are typically
\emph{much} less bug-prone.\footnote{
	And yes, I have done at least my share of roll-your-own
	synchronization primitives.
	However, you will notice that my hair is much greyer than
	it was before I started doing that sort of work.
	Coincidence?
	Maybe.
	But are you \emph{really} willing to risk your own hair turning
	prematurely grey?}

% locking/locking-existence.tex
% mainfile: ../perfbook.tex
% SPDX-License-Identifier: CC-BY-SA-3.0

\section{Lock-Based Existence Guarantees}
\label{sec:locking:Lock-Based Existence Guarantees}
\epigraph{Existence precedes and rules essence.}{Jean-Paul Sartre}

A key challenge in parallel programming is to provide
\emph{\IXpl{existence guarantee}}~\cite{Gamsa99},
so that attempts to access a given object can rely on that object
being in existence throughout a given access attempt.

\begin{listing}
\begin{fcvlabel}[ln:locking:Per-Element Locking Without Existence Guarantees]
\begin{VerbatimL}[commandchars=\\\@\$]
int delete(int key)
{
	int b;
	struct element *p;

	b = hashfunction(key);
	p = hashtable[b];
	if (p == NULL || p->key != key)		\lnlbl@chk_first$
		return 0;
	spin_lock(&p->lock);			\lnlbl@acq$
	hashtable[b] = NULL;			\lnlbl@NULL$
	spin_unlock(&p->lock);
	kfree(p);
	return 1;				\lnlbl@return1$
}
\end{VerbatimL}
\end{fcvlabel}
\caption{Per-Element Locking Without Existence Guarantees (Buggy!)}
\label{lst:locking:Per-Element Locking Without Existence Guarantees (Buggy!)}
\end{listing}

In some cases, existence guarantees are implicit:

\begin{enumerate}
\item	Global variables and static local variables in the base module
	will exist as long as the application is running.
\item	Global variables and static local variables in a loaded module
	will exist as long as that module remains loaded.
\item	A module will remain loaded as long as at least one of its functions
	has an active instance.
\item	A given function instance's on-stack variables will exist until
	that instance returns.
\item	If you are executing within a given function or have been
	called (directly or indirectly) from that function,
	then the given function has an active instance.
\end{enumerate}

These implicit existence guarantees are straightforward, though
bugs involving implicit existence guarantees really can happen.

\QuickQuiz{
	How can relying on implicit existence guarantees result in
	a bug?
}\QuickQuizAnswer{
	Here are some bugs resulting from improper use of implicit
	existence guarantees:
	\begin{enumerate}
	\item	A program writes the address of a global variable to
		a file, then a later instance of that same program
		reads that address and attempts to dereference it.
		This can fail due to address-space randomization,
		to say nothing of recompilation of the program.
	\item	A module can record the address of one of its variables
		in a pointer located in some other module, then attempt
		to dereference that pointer after the module has
		been unloaded.
	\item	A function can record the address of one of its on-stack
		variables into a global pointer, which some other
		function might attempt to dereference after that function
		has returned.
	\end{enumerate}
	I am sure that you can come up with additional possibilities.
}\QuickQuizEnd

But the more interesting---and troublesome---guarantee involves
heap memory:
A dynamically allocated data structure will exist until it is freed.
The problem to be solved is to synchronize the freeing of the structure
with concurrent accesses to that same structure.
One way to do this is with \emph{explicit guarantees}, such as locking.
If a given structure may only be freed while holding a given lock, then holding
that lock guarantees that structure's existence.

But this guarantee depends on the existence of the lock itself.
One straightforward way to guarantee the lock's existence is to
place the lock in a global variable, but global locking has the disadvantage
of limiting scalability.
One way of providing scalability that improves as the size of the
data structure increases is to place a lock in each element of the
structure.
Unfortunately, putting the lock that is to protect a data element
in the data element itself is subject to subtle \IXpl{race condition},
as shown in
\cref{lst:locking:Per-Element Locking Without Existence Guarantees (Buggy!)}.

\QuickQuiz{
	\begin{fcvref}[ln:locking:Per-Element Locking Without Existence Guarantees]
	What if the element we need to delete is not the first element
	of the list on \clnref{chk_first} of
	\cref{lst:locking:Per-Element Locking Without Existence Guarantees (Buggy!)}?
	\end{fcvref}
}\QuickQuizAnswer{
	This is a very simple hash table with no chaining, so the only
	element in a given bucket is the first element.
	The reader is invited to adapt this example to a hash table with
	full chaining.
}\QuickQuizEnd

\begin{fcvref}[ln:locking:Per-Element Locking Without Existence Guarantees]
To see one of these race conditions, consider the following sequence
of events:
\begin{enumerate}
\item	Thread~0 invokes \co{delete(0)}, and reaches \clnref{acq} of
	the listing, acquiring the lock.
\item	Thread~1 concurrently invokes \co{delete(0)}, reaching
	\clnref{acq}, but spins on the lock because Thread~0 holds it.
\item	Thread~0 executes \clnrefrange{NULL}{return1}, removing the element from
	the hashtable, releasing the lock, and then freeing the
	element.
\item	Thread~0 continues execution, and allocates memory, getting
	the exact block of memory that it just freed.
\item	Thread~0 then initializes this block of memory as some
	other type of structure.
\item	Thread~1's \co{spin_lock()} operation fails due to the
	fact that what it believes to be \co{p->lock} is no longer
	a spinlock.
\end{enumerate}
Because there is no existence guarantee, the identity of the
data element can change while a thread is attempting to acquire
that element's lock on \clnref{acq}!
\end{fcvref}

\begin{listing}
\begin{fcvlabel}[ln:locking:Per-Element Locking With Lock-Based Existence Guarantees]
\begin{VerbatimL}[commandchars=\\\@\$]
int delete(int key)
{
	int b;
	struct element *p;
	spinlock_t *sp;

	b = hashfunction(key);
	sp = &locktable[b];
	spin_lock(sp);				\lnlbl@acq$
	p = hashtable[b];			\lnlbl@getp$
	if (p == NULL || p->key != key) {
		spin_unlock(sp);
		return 0;
	}
	hashtable[b] = NULL;
	spin_unlock(sp);
	kfree(p);
	return 1;
}
\end{VerbatimL}
\end{fcvlabel}
\caption{Per-Element Locking With Lock-Based Existence Guarantees}
\label{lst:locking:Per-Element Locking With Lock-Based Existence Guarantees}
\end{listing}

\begin{fcvref}[ln:locking:Per-Element Locking With Lock-Based Existence Guarantees]
One way to fix this example is to use a hashed set of global locks, so
that each hash bucket has its own lock, as shown in
\cref{lst:locking:Per-Element Locking With Lock-Based Existence Guarantees}.
This approach allows acquiring the proper lock (on \clnref{acq}) before
gaining a pointer to the data element (on \clnref{getp}).
Although this approach works quite well for elements contained in a
single partitionable data structure such as the hash table shown in the
listing, it can be problematic if a given data element can be a member
of multiple hash tables or given more-complex data structures such
as trees or graphs.
Not only can these problems be solved, but the solutions also form
the basis of lock-based software transactional memory
implementations~\cite{Shavit95,DaveDice2006DISC}.
However,
\cref{chp:Deferred Processing}
describes simpler---and faster---ways of providing existence guarantees.
\end{fcvref}

% @@@ Optimized sharded locking from P1726R5, and pointer-zap implications.

\section{Locking:
		  Hero or Villain?}
\label{sec:locking:Locking: Hero or Villain?}
\epigraph{You either die a hero or you live long enough to see yourself
	  become the villain.}
	 {Aaron Eckhart \emph{as} Harvey Dent}

As is often the case in real life, locking can be either hero or villain,
depending on how it is used and on the problem at hand.
In my experience, those writing whole applications are happy with
locking, those writing parallel libraries are less happy, and those
parallelizing existing sequential libraries are extremely unhappy.
The following sections discuss some reasons for these differences in
viewpoints.

\subsection{Locking For Applications:
				      Hero!}
\label{sec:locking:Locking For Applications: Hero!}

When writing an entire application (or entire kernel), developers have
full control of the design, including the synchronization design.
Assuming that the design makes good use of partitioning, as discussed in
\cref{chp:Partitioning and Synchronization Design}, locking
can be an extremely effective synchronization mechanism, as demonstrated
by the heavy use of locking in production-quality parallel software.

Nevertheless, although such software usually bases most of its
synchronization design on locking, such software also almost always
makes use of other synchronization mechanisms, including
special counting algorithms (\cref{chp:Counting}),
data ownership (\cref{chp:Data Ownership}),
reference counting (\cref{sec:defer:Reference Counting}),
hazard pointers (\cref{sec:defer:Hazard Pointers}),
sequence locking (\cref{sec:defer:Sequence Locks}), and
read-copy update (\cref{sec:defer:Read-Copy Update (RCU)}).
In addition, practitioners use tools for deadlock
detection~\cite{JonathanCorbet2006lockdep},
lock acquisition/release balancing~\cite{JonathanCorbet2004sparse},
cache-miss analysis~\cite{ValgrindHomePage},
hardware-counter-based profiling~\cite{LinuxKernelPerfWiki,OProfileHomePage},
and many more besides.

Given careful design, use of a good combination of synchronization
mechanisms, and good tooling, locking works quite well for applications
and kernels.

\subsection{Locking For Parallel Libraries:
					    Just Another Tool}
\label{sec:locking:Locking For Parallel Libraries: Just Another Tool}

Unlike applications and kernels, the designer of a library cannot
know the locking design of the code that the library will be interacting
with.
In fact, that code might not be written for years to come.
Library designers therefore have less control and must exercise more
care when laying out their synchronization design.

Deadlock is of course of particular concern, and the techniques discussed
in \cref{sec:locking:Deadlock} need to be applied.
One popular deadlock-avoidance strategy is therefore to ensure that
the library's locks are independent subtrees of the enclosing program's
locking hierarchy.
However, this can be harder than it looks.

One complication was discussed in
\cref{sec:locking:Local Locking Hierarchies}, namely
when library functions call into application code, with \co{qsort()}'s
comparison-function argument being a case in point.
Another complication is the interaction with signal handlers.
If an application signal handler is invoked from a signal received within
the library function, deadlock can ensue just as surely as
if the library function had called the signal handler directly.
A final complication occurs for those library functions that can be used
between a \co{fork()}/\co{exec()} pair, for example, due to use of
the \co{system()} function.
In this case, if your library function was holding a lock at the time of
the \co{fork()}, then the child process will begin life with that lock held.
Because the thread that will release the lock is running in the parent
but not the child, if the child calls your library function, deadlock
will ensue.

The following strategies may be used to avoid deadlock problems in these cases:

\begin{enumerate}
\item	Don't use either callbacks or signals.
\item	Don't acquire locks from within callbacks or signal handlers.
\item	Let the caller control synchronization.
\item	Parameterize the library API to delegate locking to caller.
\item	Explicitly avoid callback deadlocks.
\item	Explicitly avoid signal-handler deadlocks.
\item	Avoid invoking \co{fork()}.
\end{enumerate}

Each of these strategies is discussed in one of the following sections.

\subsubsection{Use Neither Callbacks Nor Signals}
\label{sec:locking:Use Neither Callbacks Nor Signals}

If a library function avoids callbacks and the application as a whole
avoids signals, then any locks acquired by that library function will
be leaves of the locking-hierarchy tree.
This arrangement avoids deadlock, as discussed in
\cref{sec:locking:Locking Hierarchies}.
Although this strategy works extremely well where it applies,
there are some applications that must use signal handlers,
and there are some library functions (such as the \co{qsort()} function
discussed in
\cref{sec:locking:Local Locking Hierarchies})
that require callbacks.

The strategy described in the next section can often be used in these cases.

\subsubsection{Avoid Locking in Callbacks and Signal Handlers}
\label{sec:locking:Avoid Locking in Callbacks and Signal Handlers}

If neither callbacks nor signal handlers acquire locks, then they
cannot be involved in deadlock cycles, which allows straightforward
locking hierarchies to once again consider library functions to
be leaves on the locking-hierarchy tree.
This strategy works very well for most uses of \co{qsort}, whose
callbacks usually simply compare the two values passed in to them.
This strategy also works wonderfully for many signal handlers,
especially given that acquiring locks from within signal handlers
is generally frowned upon~\cite{TheOpenGroup1997SUS},\footnote{
	But the standard's words do not stop clever coders from creating
	their own home-brew locking primitives from atomic operations.}
but can fail if the application needs to manipulate complex data structures
from a signal handler.

Here are some ways to avoid acquiring locks in signal handlers even
if complex data structures must be manipulated:

\begin{enumerate}
\item	Use simple data structures based on \IXacrl{nbs},
	as will be discussed in
	\cref{sec:advsync:Simple NBS}.
\item	If the data structures are too complex for reasonable use of
	non-blocking synchronization, create a queue that allows
	non-blocking enqueue operations.
	In the signal handler, instead of manipulating the complex
	data structure, add an element to the queue describing the
	required change.
	A separate thread can then remove elements from the queue and
	carry out the required changes using normal locking.
	There are a number of readily available implementations of
	concurrent
	queues~\cite{ChristophMKirsch2012FIFOisntTR,MathieuDesnoyers2009URCU,MichaelScott96}.
\end{enumerate}

This strategy should be enforced with occasional manual or (preferably)
automated inspections of callbacks and signal handlers.
When carrying out these inspections, be wary of clever coders who
might have (unwisely) created home-brew locks from atomic operations.

\subsubsection{Caller Controls Synchronization}
\label{sec:locking:Caller Controls Synchronization}

Letting the caller control synchronization
works extremely well when the library functions are operating
on independent caller-visible instances of a data structure, each
of which may be synchronized separately.
For example, if the library functions operate on a search tree,
and if the application needs a large number of independent search
trees, then the application can associate a lock with each tree.
The application then acquires and releases locks as needed, so
that the library need not be aware of parallelism at all.
Instead, the application controls the parallelism, so that locking
can work very well, as was discussed in
\cref{sec:locking:Locking For Applications: Hero!}.

However, this strategy fails if the
library implements a data structure that requires internal
concurrency, for example, a hash table or a parallel sort.
In this case, the library absolutely must control its own
synchronization.

\subsubsection{Parameterize Library Synchronization}
\label{sec:locking:Parameterize Library Synchronization}

The idea here is to add arguments to the library's API to specify
which locks to acquire, how to acquire and release them, or both.
This strategy allows the application to take on the global task of
avoiding deadlock by specifying which locks to acquire (by passing in
pointers to the locks in question) and how to
acquire them (by passing in pointers to lock acquisition and release
functions),
but also allows a given library function to control its own concurrency
by deciding where the locks should be acquired and released.

In particular, this strategy allows the lock acquisition and release
functions to block signals as needed without the library code needing to
be concerned with which signals need to be blocked by which locks.
The separation of concerns used by this strategy can be quite effective,
but in some cases the strategies laid out in the following sections
can work better.

That said, passing explicit pointers to locks to external APIs must
be very carefully considered, as discussed in
\cref{sec:locking:Locking Hierarchies and Pointers to Locks}.
Although this practice is sometimes the right thing to do, you should do
yourself a favor by looking into alternative designs first.

\subsubsection{Explicitly Avoid Callback Deadlocks}
\label{sec:locking:Explicitly Avoid Callback Deadlocks}

The basic rule behind this strategy was discussed in
\cref{sec:locking:Local Locking Hierarchies}:
``Release all locks before invoking unknown code.''
This is usually the best approach because it allows the application to
ignore the library's locking hierarchy:
The library remains a leaf or isolated subtree of the application's
overall locking hierarchy.

In cases where it is not possible to release all locks before invoking
unknown code, the layered locking hierarchies described in
\cref{sec:locking:Layered Locking Hierarchies} can work well.
For example, if the unknown code is a signal handler, this implies that
the library function block signals across all lock acquisitions, which
can be complex and slow.
Therefore, in cases where signal handlers (probably unwisely) acquire
locks, the strategies in the next section may prove helpful.

\subsubsection{Explicitly Avoid Signal-Handler Deadlocks}
\label{sec:locking:Explicitly Avoid Signal-Handler Deadlocks}

Suppose that a given library function is known to acquire locks,
but does not block signals.
Suppose further that it is necessary to invoke that function both from
within and outside of a signal handler, and that it is not permissible
to modify this library function.
Of course, if no special action is taken, then if a signal arrives
while that library function is holding its lock, deadlock can occur
when the signal handler invokes that same library function,
which in turn attempts to re-acquire that same lock.

Such deadlocks can be avoided as follows:

\begin{enumerate}
\item	If the application invokes the library function from
	within a signal handler, then that signal must be blocked
	every time that the library function is invoked from outside
	of a signal handler.
\item	If the application invokes the library function
	while holding a lock acquired within a given signal
	handler, then that signal must be blocked every time that the
	library function is called outside of a signal handler.
\end{enumerate}

These rules can be enforced by using tools similar to
the Linux kernel's lockdep lock dependency
checker~\cite{JonathanCorbet2006lockdep}.
One of the great strengths of lockdep is that it is not fooled by
human intuition~\cite{StevenRostedt2011locdepCryptic}.

\subsubsection{Library Functions Used Between \tco{fork()} and \tco{exec()}}
\label{sec:locking:Library Functions Used Between fork() and exec()}

As noted earlier, if a thread executing a library function is holding
a lock at the time that some other thread invokes \apipx{fork()}, the
fact that the parent's memory is copied to create the child means that
this lock will be born held in the child's context.
The thread that will release this lock is running in the parent, but
not in the child, which means that although the parent's copy of this
lock will be released, the child's copy never will be.
Therefore, any attempt on the part of the child to invoke that same
library function (thus acquiring that same lock) will result in deadlock.

A pragmatic and straightforward way of solving this problem is
to \co{fork()} a child process while the process is still single-threaded,
and have this child process remain single-threaded.
Requests to create further child processes can then be communicated
to this initial child process, which can safely carry out any
needed \co{fork()} and \apipx{exec()} system calls on behalf of its
multi-threaded parent process.

Another rather less pragmatic and straightforward solution to this problem
is to have the library function check to see if the owner of the lock
is still running, and if not, ``breaking'' the lock by re-initializing
and then acquiring it.
However, this approach has a couple of vulnerabilities:

\begin{enumerate}
\item	The data structures protected by that lock are likely to
	be in some intermediate state, so that naively breaking the lock
	might result in arbitrary memory corruption.
\item	If the child creates additional threads, two threads might
	break the lock concurrently, with the result that both
	threads believe they own the lock.
	This could again result in arbitrary memory corruption.
\end{enumerate}

The \apipx{pthread_atfork()} function is provided to help deal with these situations.
The idea is to register a triplet of functions, one to be called by the
parent before the \co{fork()}, one to be called by the parent after the
\co{fork()}, and one to be called by the child after the \co{fork()}.
Appropriate cleanups can then be carried out at these three points.

Be warned, however, that coding of \co{pthread_atfork()} handlers is
quite subtle in general.
The cases where \co{pthread_atfork()} works best are cases where the
data structure in question can simply be re-initialized by the child.
Which might be one reason why the POSIX standard forbids use of any
non-async-signal-safe functions between the \co{fork()} and the
\co{exec()}, which rules out acquisition of locks during that time.

Other alternatives to \co{fork()}/\co{exec()} include \co{posix_spawn()}
and \co{io_uring_spawn()}~\cite{JoshTriplett2022io_uring_spawn,JakeEdge2022io_uring_spawn}.

\subsubsection{Parallel Libraries:
				   Discussion}
\label{sec:locking:Parallel Libraries: Discussion}

Regardless of the strategy used, the description of the library's API
must include a clear description of that strategy and how the caller
should interact with that strategy.
In short, constructing parallel libraries using locking is possible,
but not as easy as constructing a parallel application.

\subsection{Locking For Parallelizing Sequential Libraries:
							    Villain!}
\label{sec:locking:Locking For Parallelizing Sequential Libraries: Villain!}

With the advent of readily available low-cost multicore systems,
a common task is parallelizing an existing library that was designed
with only single-threaded use in mind.
This all-too-common disregard for parallelism can result in a library
API that is severely flawed from a parallel-programming viewpoint.
Candidate flaws include:

\begin{enumerate}
\item	Implicit prohibition of partitioning.
\item	Callback functions requiring locking.
\item	Object-oriented spaghetti code.
\end{enumerate}

These flaws and the consequences for locking are discussed in the following
sections.

\subsubsection{Partitioning Prohibited}
\label{sec:locking:Partitioning Prohibited}

Suppose that you were writing a single-threaded hash-table implementation.
It is easy and fast to maintain an exact count of the total number of items
in the hash table, and also easy and fast to return this exact count on each
addition and deletion operation.
So why not?

One reason is that exact counters do not perform or scale well on
multicore systems, as was
seen in \cref{chp:Counting}.
As a result, the parallelized implementation of the hash table will not
perform or scale well.

So what can be done about this?
One approach is to return an approximate count, using one of the algorithms
from \cref{chp:Counting}.
Another approach is to drop the element count altogether.

Either way, it will be necessary to inspect uses of the hash table to see
why the addition and deletion operations need the exact count.
Here are a few possibilities:

\begin{enumerate}
\item	Determining when to resize the hash table.
	In this case, an approximate count should work quite well.
	It might also be useful to trigger the resizing operation from
	the length of the longest chain, which can be computed and
	maintained in a nicely partitioned per-chain manner.
\item	Producing an estimate of the time required to traverse the
	entire hash table.
	An approximate count works well in this case, also.
\item	For diagnostic purposes, for example, to check for items being
	lost when transferring them to and from the hash table.
	This clearly requires an exact count.
	However, given that this usage is diagnostic in nature, it might
	suffice to maintain the lengths of the hash chains, then to
	infrequently sum them
	up while locking out addition and deletion operations.
\end{enumerate}

It turns out that there is now a strong theoretical basis for some of the
constraints that performance and scalability place on a parallel library's
APIs~\cite{HagitAttiya2011LawsOfOrder,Attiya:2011:LOE:1925844.1926442,PaulEMcKenney2011SNC}.
Anyone designing a parallel library needs to pay close attention to
those constraints.

Although it is all too easy to blame locking for what are really problems
due to a concurrency-unfriendly API, doing so is not helpful.
On the other hand, one has little choice but to sympathize with the
hapless developer who made this choice in (say) 1985.
It would have been a rare and courageous developer to anticipate the
need for parallelism at that time, and it would have required an
even more rare combination of brilliance and luck to actually arrive
at a good parallel-friendly API\@.

Times change, and code must change with them.
That said, there might be a huge number of users of a popular library,
in which case an incompatible change to the API would be quite foolish.
Adding a parallel-friendly API to complement the existing heavily used
sequential-only API is usually the best course of action.

Nevertheless, human nature being what it is, we can expect our hapless
developer to be more likely to complain about locking than about his
or her own poor (though understandable) API design choices.

\subsubsection{Deadlock-Prone Callbacks}
\label{sec:locking:Deadlock-Prone Callbacks}

\Cref{sec:locking:Local Locking Hierarchies,%
sec:locking:Layered Locking Hierarchies,%
sec:locking:Locking For Parallel Libraries: Just Another Tool}
described how undisciplined use of callbacks can result in locking
woes.
These sections also described how to design your library function to
avoid these problems, but it is unrealistic to expect a 1990s programmer
with no experience in parallel programming to have followed such a design.
Therefore, someone attempting to parallelize an existing callback-heavy
single-threaded library will likely have many opportunities to curse
locking's villainy.

If there are a very large number of uses of a callback-heavy library,
it may be wise to again add a parallel-friendly API to the library in
order to allow existing users to convert their code incrementally.
Alternatively, some advocate use of transactional memory in these cases.
While the jury is still out on transactional memory,
\cref{sec:future:Transactional Memory} discusses its strengths and
weaknesses.
It is important to note that hardware transactional memory
(discussed in
\cref{sec:future:Hardware Transactional Memory})
cannot help here unless the hardware transactional memory implementation
provides \IXpl{forward-progress guarantee}, which few do.
Other alternatives that appear to be quite practical (if less heavily
hyped) include the methods discussed in
\cref{sec:locking:Conditional Locking,sec:locking:Acquire Needed Locks First},
as well as those that will be discussed in
\cref{chp:Data Ownership,chp:Deferred Processing}.

\subsubsection{Object-Oriented Spaghetti Code}
\label{sec:locking:Object-Oriented Spaghetti Code}

Object-oriented programming went mainstream sometime in the 1980s or
1990s, and as a result there is a huge amount of single-threaded
object-oriented code in production.
Although object orientation can be a valuable software technique,
undisciplined use of objects can easily result in object-oriented
spaghetti code.
In object-oriented spaghetti code, control flits from object to object
in an essentially random manner, making the code hard to understand
and even harder, and perhaps impossible, to accommodate a locking hierarchy.

Although many might argue that such code should be cleaned up in any
case, such things are much easier to say than to do.
If you are tasked with parallelizing such a beast, you can reduce the
number of opportunities to curse locking by using the techniques
described in
\cref{sec:locking:Conditional Locking,sec:locking:Acquire Needed Locks First},
as well as those that will be discussed in
\cref{chp:Data Ownership,chp:Deferred Processing}.
This situation appears to be the use case that inspired transactional
memory, so it might be worth a try as well.
That said, the choice of synchronization mechanism should be made in
light of the hardware habits discussed in
\cref{chp:Hardware and its Habits}.
After all, if the overhead of the synchronization mechanism is orders of
magnitude more than that of the operations being protected, the results
are not going to be pretty.

And that leads to a question well worth asking in these situations:
Should the code remain sequential?
For example, perhaps parallelism should be introduced at the process level
rather than the thread level.
In general, if a task is proving extremely hard, it is worth some time
spent thinking about not only alternative ways to accomplish that
particular task, but also alternative tasks that might better solve
the problem at hand.

\section{Summary}
\label{sec:locking:Summary}
\epigraph{Achievement unlocked.}{Unknown}

Locking is perhaps the most widely used and most generally useful
synchronization tool.
However, it works best when designed into an application
or library from the beginning.
Given the large quantity of pre-existing single-threaded code that might
need to one day run in parallel, locking should therefore not be the
only tool in your parallel-programming toolbox.
The next few chapters will discuss other tools, and how they can best
be used in concert with locking and with each other.

	\setlength{\parskip}{0.0pt plus 1ex}
	\input{qqzlocking}
	\setlength{\parskip}{0.0pt plus 1.0pt}% return to default

% owned/owned.tex
% mainfile: ../perfbook.tex
% SPDX-License-Identifier: CC-BY-SA-3.0

\QuickQuizChapter{chp:Data Ownership}{Data Ownership}{qqzowned}
\Epigraph{It is mine, I tell you.
	  My own.
	  My precious.
	  Yes, my precious.}
	 {Gollum in \emph{The Fellowship of the Ring}, J.R.R.~Tolkien}

One of the simplest ways to avoid the synchronization overhead that
comes with locking is to parcel the data out among the threads (or,
in the case of kernels, CPUs)
so that a given piece of data is accessed and modified by only one
of the threads.
Interestingly enough, data ownership covers each of the ``big three''
parallel design techniques:
It partitions over threads (or CPUs, as the case may be),
it batches all local operations,
and its elimination of synchronization operations is weakening
carried to its logical extreme.
It should therefore be no surprise that data ownership is heavily used:
Even novices use it almost instinctively.
In fact, it is so heavily used that this chapter will not introduce any
new examples, but will instead refer back to those of previous chapters.

\QuickQuiz{
	What form of data ownership is extremely difficult
	to avoid when creating shared-memory parallel programs
	(for example, using pthreads) in C or C++?
}\QuickQuizAnswer{
	Use of auto variables in functions.
	By default, these are private to the thread executing the
	current function.
}\QuickQuizEnd

There are a number of approaches to data ownership.
\Cref{sec:owned:Multiple Processes} presents the logical extreme
in data ownership, where each thread has its own private address space.
\Cref{sec:owned:Partial Data Ownership and pthreads} looks at
the opposite extreme, where the data is shared, but different threads
own different access rights to the data.
\Cref{sec:owned:Function Shipping} describes function shipping,
which is a way of allowing other threads to have indirect access to
data owned by a particular thread.
\Cref{sec:owned:Designated Thread} describes how designated
threads can be assigned ownership of a specified function and the
related data.
\Cref{sec:owned:Privatization} discusses improving performance
by transforming algorithms with shared data to instead use data ownership.
Finally, \cref{sec:owned:Other Uses of Data Ownership} lists
a few software environments that feature data ownership as a
first-class citizen.

\section{Multiple Processes}
\label{sec:owned:Multiple Processes}
\epigraph{A man's home is his castle}{Ancient Laws of England}

\Cref{sec:toolsoftrade:Scripting Languages}
introduced the following example:

\begin{VerbatimN}[samepage=true]
compute_it 1 > compute_it.1.out &
compute_it 2 > compute_it.2.out &
wait
cat compute_it.1.out
cat compute_it.2.out
\end{VerbatimN}

This example runs two instances of the \co{compute_it} program in
parallel, as separate processes that do not share memory.
Therefore, all data in a given process is owned by that process,
so that almost the entirety of data in the above example is owned.
This approach almost entirely eliminates synchronization overhead.
The resulting combination of extreme simplicity and optimal performance
is obviously quite attractive.

\QuickQuizSeries{%
\QuickQuizB{
	What synchronization remains in the example shown in
	\cref{sec:owned:Multiple Processes}?
}\QuickQuizAnswerB{
	The creation of the threads via the \co{sh} \co{&} operator
	and the joining of thread via the \co{sh} \co{wait}
	command.

	Of course, if the processes explicitly share memory, for
	example, using the \co{shmget()} or \co{mmap()} system
	calls, explicit synchronization might well be needed when
	acccessing or updating the shared memory.
	The processes might also synchronize using any of the following
	interprocess communications mechanisms:
	\begin{enumerate}
	\item	System V semaphores.
	\item	System V message queues.
	\item	UNIX-domain sockets.
	\item	Networking protocols, including TCP/IP, UDP, and a whole
		host of others.
	\item	File locking.
	\item	Use of the \co{open()} system call with the
		\co{O_CREAT} and \co{O_EXCL} flags.
	\item	Use of the \co{rename()} system call.
	\end{enumerate}
	A complete list of possible synchronization mechanisms is left
	as an exercise to the reader, who is warned that it will be
	an extremely long list.
	A surprising number of unassuming system calls can be pressed
	into service as synchronization mechanisms.
}\QuickQuizEndB
\QuickQuizE{
	Is there any shared data in the example shown in
	\cref{sec:owned:Multiple Processes}?
}\QuickQuizAnswerE{
	That is a philosophical question.

	Those wishing the answer ``no'' might argue that processes by
	definition do not share memory.

	Those wishing to answer ``yes'' might list a large number of
	synchronization mechanisms that do not require shared memory,
	note that the kernel will have some shared state, and perhaps
	even argue that the assignment of process IDs (PIDs) constitute
	shared data.

	Such arguments are excellent intellectual exercise, and are
	also a wonderful way of feeling intelligent and scoring points
	against hapless classmates or colleagues, but are mostly a way
	of avoiding getting anything useful done.
}\QuickQuizEndE
}

This same pattern can be written in C as well as in \co{sh}, as illustrated by
\cref{lst:toolsoftrade:Using the fork() Primitive,%
lst:toolsoftrade:Using the wait() Primitive}.

It bears repeating that these trivial forms of parallelism are not in
any way cheating or ducking responsibility, but are rather simple and
elegant ways to make your code run faster.
It is fast, scales well, is easy to program, easy to maintain, and
gets the job done.
In addition, taking this approach (where applicable) allows the developer
more time to focus on other things whether these things might involve
applying sophisticated single-threaded optimizations to \co{compute_it}
on the one hand, or applying sophisticated parallel-programming patterns
to portions of the code where this approach is inapplicable.
What is not to like?

The next section discusses the use of data ownership in shared-memory
parallel programs.

\section{Partial Data Ownership and pthreads}
\label{sec:owned:Partial Data Ownership and pthreads}
\epigraph{Give thy mind more to what thou hast than to what thou hast not.}
	 {Marcus Aurelius Antoninus}

Concurrent counting (see \cref{chp:Counting}) uses data ownership heavily,
but adds a twist.
Threads are not allowed to modify data owned by other threads,
but they are permitted to read it.
In short, the use of shared memory allows more nuanced notions
of ownership and access rights.

For example, consider the per-thread statistical counter implementation
shown in
\cref{lst:count:Per-Thread Statistical Counters} on
\cpageref{lst:count:Per-Thread Statistical Counters}.
Here, \co{inc_count()} updates only the corresponding thread's
instance of \co{counter},
while \co{read_count()} accesses, but does not modify, all
threads' instances of \co{counter}.

\QuickQuiz{
	Does it ever make sense to have partial data ownership where
	each thread reads only its own instance of a per-thread variable,
	but writes to other threads' instances?
}\QuickQuizAnswer{
	Amazingly enough, yes.
	One example is a simple message-passing system where threads post
	messages to other threads' mailboxes, and where each thread
	is responsible for removing any message it sent once that message
	has been acted on.
	Implementation of such an algorithm is left as an exercise for
	the reader, as is identifying other algorithms with similar
	ownership patterns.
}\QuickQuizEnd

Partial data ownership is also common within the Linux kernel.
For example, a given CPU might be permitted to read a given set of its
own per-CPU variables only with interrupts disabled, another CPU might
be permitted to read that same set of the first CPU's per-CPU variables
only when holding the corresponding per-CPU lock.
Then that given CPU would be permitted to update this set of its own
per-CPU variables if it both has interrupts disabled and holds its
per-CPU lock.
This arrangement can be thought of as a reader-writer lock that allows
each CPU very low-overhead access to its own set of per-CPU variables.
There are a great many variations on this theme.

For its own part, pure data ownership is also both common and useful,
for example, the per-thread memory-allocator caches discussed in
\cref{sec:SMPdesign:Resource Allocator Caches}
starting on
\cpageref{sec:SMPdesign:Resource Allocator Caches}.
In this algorithm, each thread's cache is completely private to that
thread.

\section{Function Shipping}
\label{sec:owned:Function Shipping}
\epigraph{If the mountain will not come to Muhammad, then Muhammad must
	  go to the mountain.}
	 {\emph{Essays}, Francis Bacon}

The previous section described a weak form of data ownership where
threads reached out to other threads' data.
This can be thought of as bringing the data to the functions that
need it.
An alternative approach is to send the functions to the data.

Such an approach is illustrated in
\cref{sec:count:Signal-Theft Limit Counter Design}
beginning on
\cpageref{sec:count:Signal-Theft Limit Counter Design},
in particular the \co{flush_local_count_sig()} and
\co{flush_local_count()} functions in
\cref{lst:count:Signal-Theft Limit Counter Value-Migration Functions}
on
\cpageref{lst:count:Signal-Theft Limit Counter Value-Migration Functions}.

The \co{flush_local_count_sig()} function is a signal handler that
acts as the shipped function.
The \co{pthread_kill()} function in \co{flush_local_count()}
sends the signal---shipping the function---and then waits until
the shipped function executes.
This shipped function has the not-unusual added complication of
needing to interact with any concurrently executing \co{add_count()}
or \co{sub_count()} functions (see
\cref{lst:count:Signal-Theft Limit Counter Add Function}
on
\cpageref{lst:count:Signal-Theft Limit Counter Add Function} and
\cref{lst:count:Signal-Theft Limit Counter Subtract Function}
on
\cpageref{lst:count:Signal-Theft Limit Counter Subtract Function}).

\QuickQuiz{
	What mechanisms other than POSIX signals may be used for function
	shipping?
}\QuickQuizAnswer{
	There is a very large number of such mechanisms, including:
	\begin{enumerate}
	\item	System V message queues.
	\item	Shared-memory dequeue (see
		\cref{sec:SMPdesign:Double-Ended Queue}).
	\item	Shared-memory mailboxes.
	\item	UNIX-domain sockets.
	\item	TCP/IP or UDP, possibly augmented by any number of
		higher-level protocols, including RPC, HTTP,
		XML, SOAP, and so on.
	\end{enumerate}
	Compilation of a complete list is left as an exercise to
	sufficiently single-minded readers, who are warned that the
	list will be extremely long.
}\QuickQuizEnd

\section{Designated Thread}
\label{sec:owned:Designated Thread}
\epigraph{Let a man practice the profession which he best knows.}
	 {Cicero}

The earlier sections describe ways of allowing each thread to keep its
own copy or its own portion of the data.
In contrast, this section describes a functional-decomposition approach,
where a special designated thread owns the rights to the data
that is required to do its job.
\begin{fcvref}[ln:count:count_stat_eventual:whole:eventual]
The eventually consistent counter implementation described in
\cref{sec:count:Eventually Consistent Implementation} provides an example.
This implementation has a designated thread that runs the
\co{eventual()} function shown on \clnrefrange{b}{e} of
\cref{lst:count:Array-Based Per-Thread Eventually Consistent Counters}.
This \co{eventual()} thread periodically pulls the per-thread counts
into the global counter, so that accesses to the global counter will,
as the name says, eventually converge on the actual value.
\end{fcvref}

\QuickQuiz{
	\begin{fcvref}[ln:count:count_stat_eventual:whole:eventual]
	But none of the data in the \co{eventual()} function shown on
	\clnrefrange{b}{e} of
	\cref{lst:count:Array-Based Per-Thread Eventually Consistent Counters}
	is actually owned by the \co{eventual()} thread!
	In just what way is this data ownership???
	\end{fcvref}
}\QuickQuizAnswer{
	\begin{fcvref}[ln:count:count_stat_eventual:whole]
	The key phrase is ``owns the rights to the data''.
	In this case, the rights in question are the rights to access
	the per-thread \co{counter} variable defined on \clnref{per_thr_cnt}
	of the listing.
	This situation is similar to that described in
	\cref{sec:owned:Partial Data Ownership and pthreads}.

	However, there really is data that is owned by the \co{eventual()}
	thread, namely the \co{t} and \co{sum} variables defined on
	\clnref{t,sum} of the listing.

	For other examples of designated threads, look at the kernel
	threads in the Linux kernel, for example, those created by
	\co{kthread_create()} and \co{kthread_run()}.
	\end{fcvref}
}\QuickQuizEnd

\section{Privatization}
\label{sec:owned:Privatization}
\epigraph{There is, of course, a difference between what a man seizes
	  and what he really possesses.}
	 {Pearl S.~Buck}

One way of improving the performance and scalability of a shared-memory
parallel program is to transform it so as to convert shared data to
private data that is owned by a particular thread.

An excellent example of this is shown in the answer to one of the
Quick Quizzes in
\cref{sec:SMPdesign:Dining Philosophers Problem},
which uses privatization to produce a solution to the
Dining Philosophers problem with much better performance and scalability
than that of the standard textbook solution.
The original problem has five philosophers sitting around the table
with one fork between each adjacent pair of philosophers, which permits
at most two philosophers to eat concurrently.

We can trivially privatize this problem by providing an additional five
forks, so that each philosopher has his or her own private pair of forks.
This allows all five philosophers to eat concurrently, and also offers
a considerable reduction in the spread of certain types of disease.

In other cases, privatization imposes costs.
For example, consider the simple limit counter shown in
\cref{lst:count:Simple Limit Counter Add; Subtract; and Read} on
\cpageref{lst:count:Simple Limit Counter Add; Subtract; and Read}.
This is an example of an algorithm where threads can read each others'
data, but are only permitted to update their own data.
A quick review of the algorithm shows that the only cross-thread
accesses are in the summation loop in \co{read_count()}.
If this loop is eliminated, we move to the more-efficient pure
data ownership, but at the cost of a less-accurate result
from \co{read_count()}.

\QuickQuiz{
	Is it possible to obtain greater accuracy while still
	maintaining full privacy of the per-thread data?
}\QuickQuizAnswer{
	Yes.
	One approach is for \co{read_count()} to add the value
	of its own per-thread variable.
	This maintains full ownership and performance, but only
	a slight improvement in accuracy, particularly on systems
	with very large numbers of threads.

	Another approach is for \co{read_count()} to use function
	shipping, for example, in the form of per-thread signals.
	This greatly improves accuracy, but at a significant performance
	cost for \co{read_count()}.

	However, both of these methods have the advantage of eliminating
	cache thrashing for the common case of updating counters.
}\QuickQuizEnd

Partial privatization is also possible, with some synchronization
requirements, but less than in the fully shared case.
Some partial-privatization possibilities were explored in
\cref{sec:toolsoftrade:Avoiding Data Races}.
\Cref{chp:Deferred Processing} will introduce a temporal component
to data ownership by providing ways of safely taking public data
structures private.

In short, privatization is a powerful tool in the parallel programmer's
toolbox, but it must nevertheless be used with care.
Just like every other synchronization primitive, it has the potential
to increase complexity while decreasing performance and scalability.

\section{Other Uses of Data Ownership}
\label{sec:owned:Other Uses of Data Ownership}
\epigraph{Everything comes to us that belongs to us if we create the
	  capacity to receive it.}
	 {Rabindranath Tagore}

Data ownership works best when the data can be partitioned so that there
is little or no need for cross thread access or update.
Fortunately, this situation is reasonably common, and in a wide variety
of parallel-programming environments.

Examples of data ownership include:

\begin{enumerate}
\item	All message-passing environments, such as MPI~\cite{MPIForum2008}
	and BOINC~\cite{BOINC2008}.
\item	Map-reduce~\cite{MapReduce2008MIT}.
\item	Client-server systems, including RPC, web services, and
	pretty much any system with a back-end database server.
\item	Shared-nothing database systems.
\item	Fork-join systems with separate per-process address spaces.
\item	Process-based parallelism, such as the Erlang language.
\item	Private variables, for example, C-language on-stack auto variables,
	in threaded environments.
\item	Many parallel linear-algebra algorithms, especially those
	well-suited for GPGPUs.\footnote{
		But note that a great many other classes of applications
		have also been ported to
		GPGPUs~\cite{NormMatloff2017ParProcBook,AMD2020ROCm,NVidia2017GPGPU,NVidia2017GPGPU-university}.}
\item	Operating-system kernels adapted for networking, where each connection
	(also called \emph{flow}~\cite{Shenker89,ZhangPhD,McKenney90})
	is assigned to a specific thread.
	One recent example of this approach is the IX operating
	system~\cite{Belay:2016:IOS:3014162.2997641}.
	IX does have some shared data structures, which use synchronization
	mechanisms to be described in
	\cref{sec:defer:Read-Copy Update (RCU)}.
\end{enumerate}

Data ownership is perhaps the most underappreciated synchronization
mechanism in existence.
When used properly, it delivers unrivaled simplicity, performance,
and scalability.
Perhaps its simplicity costs it the respect that it deserves.
Hopefully a greater appreciation for the subtlety and power of data ownership
will lead to greater level of respect, to say nothing of leading to
greater performance and scalability coupled with reduced complexity.

% populate with problems showing benefits of coupling data ownership with
% other approaches. For example, work-stealing schedulers. Perhaps also
% move memory allocation here, though its current location is quite good.

	\setlength{\parskip}{0.0pt plus 1ex}
	\input{qqzowned}
	\setlength{\parskip}{0.0pt plus 1.0pt}% return to default

% defer/defer.tex
% mainfile: ../perfbook.tex
% SPDX-License-Identifier: CC-BY-SA-3.0

\QuickQuizChapter{chp:Deferred Processing}{Deferred Processing}{qqzdefer}
\Epigraph{All things come to those who wait.}{Violet Fane}

The strategy of deferring work goes back before the dawn of recorded
history.
It has occasionally been derided as procrastination or even as sheer laziness.
However, in the last few decades workers have recognized this strategy's value
in simplifying and streamlining parallel algorithms~\cite{Kung80,HMassalinPhD}.
Believe it or not, ``laziness'' in parallel programming often outperforms and
out-scales industriousness!
These performance and scalability benefits stem from the fact that
deferring work can enable weakening of synchronization primitives,
thereby reducing synchronization overhead.

Those who are willing and able to read and understand this chapter
will uncover many mysteries, including:

\begin{enumerate}
\item	The reference-counting trap that awaits unwary developers of
	concurrent code.
\label{sec:defer:Mysteries reference-counting trap}
\item	A concurrent reference counter that avoids not only this trap,
	but also avoids expensive atomic read-modify-write accesses,
	and in addition avoids as well as writes of any kind to the data
	structure being traversed.
\label{sec:defer:Mysteries hazard pointers}
\item	The under-appreciated restricted form of software transactional
	memory that is used heavily within the Linux kernel.
\label{sec:defer:Mysteries sequence locking}
\item	A synchronization primitive that allows a concurrently updated
	linked data structure to be traversed using exactly the same
	sequence of machine instructions that might be used to traverse
	a sequential implementation of that same data structure.
\label{sec:defer:Mysteries RCU}
\item	A synchronization primitive whose use cases are far more
	conceptually more complex than is the primitive itself.
\label{sec:defer:Mysteries RCU Use Cases}
\item	How to choose among the various deferred-processing primitives.
\label{sec:defer:Mysteries Which to choose}
\end{enumerate}

General approaches of work deferral include
reference counting (\cref{sec:defer:Reference Counting}),
hazard pointers (\cref{sec:defer:Hazard Pointers}),
sequence locking (\cref{sec:defer:Sequence Locks}),
and RCU (\cref{sec:defer:Read-Copy Update (RCU)}).
Finally, \cref{sec:defer:Which to Choose?}
describes how to choose among the work-deferral schemes covered in
this chapter and \cref{sec:defer:What About Updates?}
discusses updates.
But first, \cref{sec:defer:Running Example} will introduce an example
algorithm that will be used to compare and contrast these approaches.

\section{Running Example}
\label{sec:defer:Running Example}
\epigraph{An ounce of application is worth a ton of abstraction.}
	 {Booker T.~Washington}

This chapter will use a simplified packet-routing algorithm to demonstrate
the value of these approaches and to allow them to be compared.
Routing algorithms are used in operating-system kernels to
deliver each outgoing TCP/IP packet to the appropriate network interface.
This particular algorithm is a simplified version of the classic 1980s
packet-train-optimized algorithm used in BSD UNIX~\cite{VanJacobson88},
consisting of a simple linked list.\footnote{
	In other words, this is not OpenBSD, NetBSD, or even
	FreeBSD, but none other than Pre-BSD\@.}
Modern routing algorithms use more complex data structures, however a
simple algorithm will help highlight issues specific to parallelism in
a straightforward setting.

\begin{figure}
\centering
\resizebox{3in}{!}{\includegraphics{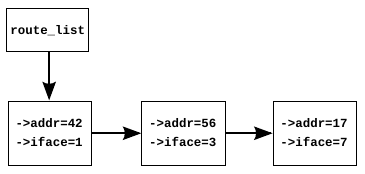}}
\caption{Pre-BSD Packet Routing List}
\label{fig:defer:Pre-BSD Packet Routing List}
\end{figure}

We further simplify the algorithm by reducing the search key from
a quadruple consisting of source and destination IP addresses and
ports all the way down to a simple integer.
The value looked up and returned will also be a simple integer,
so that the data structure is as shown in
\cref{fig:defer:Pre-BSD Packet Routing List}, which
directs packets with address~42 to interface~1, address~56 to
interface~3, and address~17 to interface~7.
This list will normally be searched frequently and updated rarely.
In \cref{chp:Hardware and its Habits}
we learned that the best ways to evade inconvenient laws of physics, such as
the finite speed of light and the atomic nature of matter, is to
either partition the data or to rely on read-mostly sharing.
This chapter applies read-mostly sharing techniques to Pre-BSD packet
routing.

\begin{listing}
\input{CodeSamples/defer/route_seq=lookup_add_del.fcv}
\caption{Sequential Pre-BSD Routing Table}
\label{lst:defer:Sequential Pre-BSD Routing Table}
\end{listing}

\Cref{lst:defer:Sequential Pre-BSD Routing Table} (\path{route_seq.c})
shows a simple single-threaded implementation corresponding to
\cref{fig:defer:Pre-BSD Packet Routing List}.
\begin{fcvref}[ln:defer:route_seq:lookup_add_del:entry]
\Clnrefrange{b}{e} define a \co{route_entry} structure and
\clnref{header} defines
the \co{route_list} header.
\end{fcvref}
\begin{fcvref}[ln:defer:route_seq:lookup_add_del:lookup]
\Clnrefrange{b}{e} define \co{route_lookup()}, which sequentially searches
\co{route_list}, returning the corresponding \co{->iface}, or
\co{ULONG_MAX} if there is no such route entry.
\end{fcvref}
\begin{fcvref}[ln:defer:route_seq:lookup_add_del:add]
\Clnrefrange{b}{e} define \co{route_add()}, which allocates a
\co{route_entry} structure, initializes it, and adds it to the
list, returning \co{-ENOMEM} in case of memory-allocation failure.
\end{fcvref}
\begin{fcvref}[ln:defer:route_seq:lookup_add_del:del]
Finally, \clnrefrange{b}{e} define \co{route_del()}, which removes and
frees the specified \co{route_entry} structure if it exists,
or returns \co{-ENOENT} otherwise.
\end{fcvref}

This single-threaded implementation serves as a prototype for the various
concurrent implementations in this chapter, and also as an estimate of
ideal scalability and performance.

% defer/refcnt.tex
% mainfile: ../perfbook.tex
% SPDX-License-Identifier: CC-BY-SA-3.0

\section{Reference Counting}
\label{sec:defer:Reference Counting}
\epigraph{I am never letting you go!}{Unknown}

\IXalt{Reference counting}{reference count}
tracks the number of references to a given object in
order to prevent that object from being prematurely freed.
As such, it has a long and honorable history of use dating back to
at least an early 1960s Weizenbaum
paper~\cite{Weizenbaum:1963:SLP:367593.367617}.
Weizenbaum discusses reference counting as if it was already well-known,
so it likely dates back to the 1950s or even to the 1940s.
And perhaps even further, given that people repairing large dangerous
machines have long used a mechanical reference-counting technique
implemented via padlocks.
Before entering the machine, each worker locks a padlock onto the
machine's on/off switch, thus preventing the machine from being powered
on while that worker is inside.
Reference counting is thus an excellent time-honored candidate for a
concurrent implementation of Pre-BSD routing.

\begin{listing}
\input{CodeSamples/defer/route_refcnt=lookup.fcv}
\caption{Reference-Counted Pre-BSD Routing Table Lookup (BUGGY!!!)}
\label{lst:defer:Reference-Counted Pre-BSD Routing Table Lookup}
\end{listing}

\begin{listing}
\input{CodeSamples/defer/route_refcnt=add_del.fcv}
\caption{Reference-Counted Pre-BSD Routing Table Add\slash Delete (BUGGY!!!)}
\label{lst:defer:Reference-Counted Pre-BSD Routing Table Add/Delete}
\end{listing}

To that end,
\cref{lst:defer:Reference-Counted Pre-BSD Routing Table Lookup}
shows data structures and the \co{route_lookup()} function and
\cref{lst:defer:Reference-Counted Pre-BSD Routing Table Add/Delete}
shows the \co{route_add()} and \co{route_del()} functions
(all at \path{route_refcnt.c}).
Since these algorithms are quite similar to the sequential algorithm
shown in
\cref{lst:defer:Sequential Pre-BSD Routing Table},
only the differences will be discussed.

\begin{fcvref}[ln:defer:route_refcnt:lookup:entry]
Starting with
\cref{lst:defer:Reference-Counted Pre-BSD Routing Table Lookup},
\clnref{refcnt} adds the actual reference counter,
\clnref{freed} adds a \co{->re_freed}
use-after-free check field,
\clnref{routelock} adds the \co{routelock} that will
be used to synchronize concurrent updates,
\end{fcvref}
\begin{fcvref}[ln:defer:route_refcnt:lookup:re_free]
and \clnrefrange{b}{e} add \co{re_free()}, which sets
\co{->re_freed}, enabling \co{route_lookup()} to check for
use-after-free bugs.
\end{fcvref}
\begin{fcvref}[ln:defer:route_refcnt:lookup:lookup]
In \co{route_lookup()} itself,
\clnrefrange{relprev:b}{relprev:e} release the reference
count of the prior element and free it if the count becomes zero,
and \clnrefrange{acq:b}{acq:e} acquire a reference on the new element,
with \clnref{check_uaf,abort} performing the use-after-free check.
\end{fcvref}

\QuickQuiz{
	Why bother with a use-after-free check?
}\QuickQuizAnswer{
	To greatly increase the probability of finding bugs.
	A small torture-test program
	(\path{routetorture.h}) that allocates and frees only
	one type of structure can tolerate a surprisingly
	large amount of use-after-free misbehavior.
	See \cref{fig:debugging:Number of Tests Required for 99 Percent Confidence Given Failure Rate}
	on \cpageref{fig:debugging:Number of Tests Required for 99 Percent Confidence Given Failure Rate}
	and the related discussion in
	\cref{sec:debugging:Hunting Heisenbugs}
	starting on
	\cpageref{sec:debugging:Hunting Heisenbugs}
	for more on the importance
	of increasing the probability of finding bugs.
}\QuickQuizEnd

\begin{fcvref}[ln:defer:route_refcnt:add_del]
In \cref{lst:defer:Reference-Counted Pre-BSD Routing Table Add/Delete},
\clnref{acq1,rel1,acq2,rel2,rel3} introduce locking to synchronize
concurrent updates.
\Clnref{init:freed} initializes the \co{->re_freed} use-after-free-check field,
and finally \clnrefrange{re_free:b}{re_free:e} invoke
\co{re_free()} if the new value of
the reference count is zero.
\end{fcvref}

\QuickQuiz{
	Why doesn't \co{route_del()} in
	\cref{lst:defer:Reference-Counted Pre-BSD Routing Table Add/Delete}
	use reference counts to
	protect the traversal to the element to be freed?
}\QuickQuizAnswer{
	Because the traversal is already protected by the lock, so
	no additional protection is required.
}\QuickQuizEnd

\begin{figure}
\centering
\resizebox{2.5in}{!}{\includegraphics{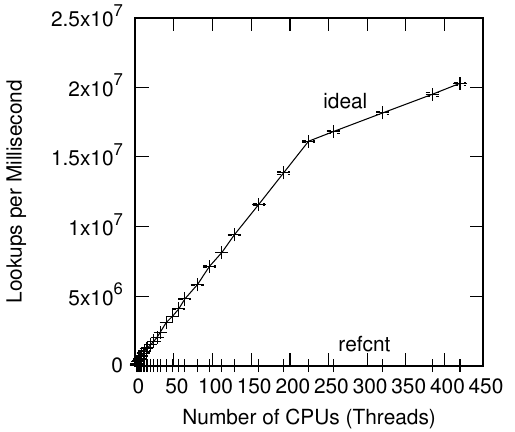}}
\caption{Pre-BSD Routing Table Protected by Reference Counting}
\label{fig:defer:Pre-BSD Routing Table Protected by Reference Counting}
\end{figure}

\Cref{fig:defer:Pre-BSD Routing Table Protected by Reference Counting}
shows the performance and scalability of reference counting on a
read-only workload with a ten-element list running on an eight-socket
28-core-per-socket hyperthreaded 2.1\,GHz x86 system with a total of
448 hardware threads (\path{hps.2019.12.02a/lscpu.hps}).
The ``ideal'' trace was generated by running the sequential code shown in
\cref{lst:defer:Sequential Pre-BSD Routing Table},
which works only because this is a read-only workload.
The reference-counting performance is abysmal and its scalability even
more so, with the ``refcnt'' trace indistinguishable from the x-axis.
This should be no surprise in view of
\cref{chp:Hardware and its Habits}:
The reference-count acquisitions and releases have added frequent
shared-memory writes to an otherwise read-only workload, thus
incurring severe retribution from the laws of physics.
As well it should, given that all the wishful thinking in the world
is not going to increase the speed of light or decrease the size of
the atoms used in modern digital electronics.

\QuickQuizSeries{%
\QuickQuizB{
	Why the break in the ``ideal'' line  at 224 CPUs in
	\cref{fig:defer:Pre-BSD Routing Table Protected by Reference Counting}?
	Shouldn't it be a straight line?
}\QuickQuizAnswerB{
	The break is due to hyperthreading.
	On this particular system, the first hardware thread in each
	core within a socket have consecutive CPU numbers,
	followed by the first hardware threads in each core for the
	other sockets,
	and finally followed by the second hardware thread in each core
	on all the sockets.
	On this particular system, CPU numbers 0--27 are the first
	hardware threads in each of the 28 cores in the first socket,
	numbers 28--55 are the first hardware threads in each of the
	28 cores in the second socket, and so on, so that numbers 196--223
	are the first hardware threads in each of the 28 cores in
	the eighth socket.
	Then CPU numbers 224--251 are the second hardware threads in each 
	of the 28 cores of the first socket, numbers 252--279 are the
	second hardware threads in each of the 28 cores of the second
	socket, and so on until numbers 420--447 are the second hardware
	threads in each of the 28 cores of the eighth socket.

	Why does this matter?

	Because the two hardware threads of a given core share resources,
	and this workload seems to allow a single hardware thread to
	consume more than half of the relevant resources within its core.
	Therefore, adding the second hardware thread of that core adds
	less than one might hope.
	Other workloads might gain greater benefit from each core's
	second hardware thread, but much depends on the details of both
	the hardware and the workload.
}\QuickQuizEndB
\QuickQuizE{
	Shouldn't the refcnt trace in
	\cref{fig:defer:Pre-BSD Routing Table Protected by Reference Counting}
	be at least a little bit off of the x-axis???
}\QuickQuizAnswerE{
	Define ``a little bit.''

\begin{figure}
\centering
\resizebox{2.5in}{!}{\includegraphics{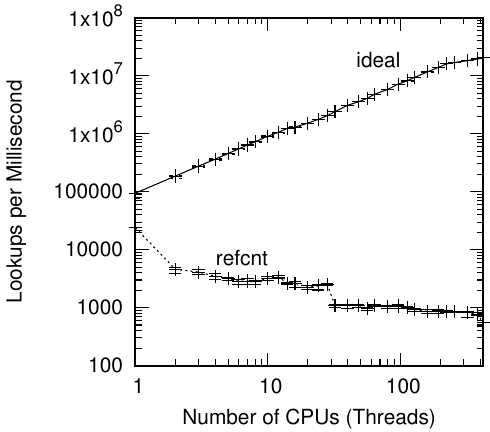}}
\caption{Pre-BSD Routing Table Protected by Reference Counting, Log Scale}
\label{fig:defer:Pre-BSD Routing Table Protected by Reference Counting; Log Scale}
\end{figure}

	\Cref{fig:defer:Pre-BSD Routing Table Protected by Reference Counting; Log Scale}
	shows the same data, but on a log-log plot.
	As you can see, the refcnt line drops below 5,000 at two CPUs.
	This means that the refcnt performance at two CPUs is more than
	one thousand times smaller than the first y-axis tick of
	$5 \times 10^6$ in
	\cref{fig:defer:Pre-BSD Routing Table Protected by Reference Counting}.
	Therefore, the depiction of the performance of reference counting
	shown in
	\cref{fig:defer:Pre-BSD Routing Table Protected by Reference Counting}
	is all too accurate.
}\QuickQuizEndE
}

But it gets worse.

Running multiple updater threads repeatedly invoking
\co{route_add()} and \co{route_del()} will quickly encounter the
\co{abort()} statement on
\clnrefr{ln:defer:route_refcnt:lookup:lookup:abort} of
\cref{lst:defer:Reference-Counted Pre-BSD Routing Table Lookup},
which indicates a use-after-free bug.
This in turn means that the reference counts are not only profoundly
degrading scalability and performance, but also failing to provide
the needed protection.

One sequence of events leading to the use-after-free bug is as follows,
given the list shown in
\cref{fig:defer:Pre-BSD Packet Routing List}:

\begin{fcvref}[ln:defer:route_refcnt:lookup]
\begin{enumerate}
\item	Thread~A looks up address~42, reaching
	\clnref{lookup:check_NULL} of
	\co{route_lookup()} in
	\cref{lst:defer:Reference-Counted Pre-BSD Routing Table Lookup}.
	In other words, Thread~A has a pointer to the first element,
	but has not yet acquired a reference to it.
\item	Thread~B invokes \co{route_del()} in
	\cref{lst:defer:Reference-Counted Pre-BSD Routing Table Add/Delete}
	to delete the route entry for address~42.
	It completes successfully, and because this entry's \co{->re_refcnt}
	field was equal to the value one, it invokes
	\co{re_free()} to set the \co{->re_freed} field and to free the entry.
\item	Thread~A continues execution of \co{route_lookup()}.
	Its \co{rep} pointer is non-\co{NULL}, but
	\clnref{lookup:check_uaf} sees that
	its \co{->re_freed} field is non-zero,
	so \clnref{lookup:abort} invokes
	\co{abort()}.
\end{enumerate}
\end{fcvref}

The problem is that the reference count is located in the object
to be protected, but that means that there is no protection during
the instant in time when the reference count itself is being acquired!
This is the reference-counting counterpart of a locking issue noted
by Gamsa et al.~\cite{Gamsa99}.
One could imagine using a global lock or reference count to protect
the per-route-entry reference-count acquisition, but this would
result in severe contention issues.
Although algorithms exist that allow safe reference-count acquisition
in a concurrent environment~\cite{Valois95a}, they are not only extremely
complex and error-prone~\cite{MagedMichael95a}, but also provide
terrible performance and scalability~\cite{ThomasEHart2007a}.

In short, concurrency has most definitely reduced the usefulness
of reference counting!
Of course, as with other synchronization primitives, reference counts
also have well-known ease-of-use shortcomings.
These can result in memory leaks on the one hand or premature
freeing on the other.

And this is the reference-counting trap that awaits unwary developers of
concurrent code, noted back on
\cpageref{sec:defer:Mysteries reference-counting trap}.

\QuickQuiz{
	If concurrency has ``most definitely reduced the usefulness
	of reference counting'', why are there so many reference
	counters in the Linux kernel?
}\QuickQuizAnswer{
	That sentence did say ``reduced the usefulness'', not
	``eliminated the usefulness'', now didn't it?

	Please see
	\cref{sec:together:Refurbish Reference Counting},
	which discusses some of the techniques that the Linux kernel
	uses to take advantage of reference counting in a highly
	concurrent environment.
}\QuickQuizEnd

It is sometimes helpful to look at a problem in an entirely different
way in order to successfully solve it.
To this end, the next section describes what could be thought of as
an inside-out reference count that provides decent performance and
scalability.

% defer/hazptr.tex
% mainfile: ../perfbook.tex
% From an C++ Standards Committee meeting:  "Can I hazptr cheezeberger?"

\section{Hazard Pointers}
\label{sec:defer:Hazard Pointers}
\epigraph{If in doubt, turn it inside out.}{Zara Carpenter}

One way of avoiding problems with concurrent reference counting
is to implement the reference counters
inside out, that is, rather than incrementing an integer stored in the
data element, instead store a pointer to that data element in
per-CPU (or per-thread) lists.
Each element of these lists is called a
\emph{\IX{hazard pointer}}~\cite{MagedMichael04a}.\footnote{
	Also independently invented by others~\cite{HerlihyLM02}.}
The value of a given data element's ``virtual reference counter'' can
then be obtained by counting the number of hazard pointers referencing
that element.
Therefore, if that element has been rendered inaccessible to readers,
and there are no longer any hazard pointers referencing it, that element
may safely be freed.

\begin{listing}
\input{CodeSamples/defer/hazptr=record_clear.fcv}
\caption{Hazard-Pointer Recording and Clearing}
\label{lst:defer:Hazard-Pointer Recording and Clearing}
\end{listing}

Of course, this means that hazard-pointer acquisition must be carried
out quite carefully in order to avoid destructive races with concurrent
deletion.
\begin{fcvref}[ln:defer:hazptr:record_clear]
One implementation is shown in
\cref{lst:defer:Hazard-Pointer Recording and Clearing},
which shows \co{hp_try_record()} on \clnrefrange{htr:b}{htr:e},
\co{hp_record()} on \clnrefrange{hr:b}{hr:e}, and
\co{hp_clear()} on
\clnrefrange{hc:b}{hc:e} (\path{hazptr.h}).

The \co{hp_try_record()} macro on \clnref{htr:e} is simply a casting
wrapper for the \co{_h_t_r_impl()} function, which attempts to store
the pointer referenced by \co{p} into the hazard pointer referenced
by \co{hp}.
If successful, it returns the value of the stored pointer.
If it fails due to that pointer being \co{NULL}, it returns \co{NULL}.
Finally, if it fails due to racing with an update, it returns a special
\co{HAZPTR_POISON} token.

\QuickQuiz{
	Given that papers on hazard pointers use the bottom bits
	of each pointer to mark deleted elements, what is up with
	\co{HAZPTR_POISON}?
}\QuickQuizAnswer{
	The published implementations of hazard pointers used
	non-blocking synchronization techniques for insertion
	and deletion.
	These techniques require that readers traversing the
	data structure ``help'' updaters complete their updates,
	which in turn means that readers need to look at the successor
	of a deleted element.

	In contrast, we will be using locking to synchronize updates,
	which does away with the need for readers to help updaters
	complete their updates, which in turn allows us to leave
	pointers' bottom bits alone.
	This approach allows read-side code to be simpler and faster.
}\QuickQuizEnd

\Clnref{htr:ro1} reads the pointer to the object to be protected.
If \clnref{htr:race1} finds that this pointer was either \co{NULL} or
the special \co{HAZPTR_POISON} deleted-object token, it returns
the pointer's value to inform the caller of the failure.
Otherwise, \clnref{htr:store} stores the pointer into the specified
hazard pointer, and \clnref{htr:mb} forces full ordering of that
store with the reload of the original pointer on \clnref{htr:ro2}.
(See \cref{chp:Advanced Synchronization: Memory Ordering}
for more information on memory ordering.)
If the value of the original pointer has not changed, then the hazard
pointer protects the pointed-to object, and in that case,
\clnref{htr:success} returns a pointer to that object, which also
indicates success to the caller.
Otherwise, if the pointer changed between the two \co{READ_ONCE()}
invocations, \clnref{htr:race2} indicates failure.

\QuickQuiz{
	Why does \co{hp_try_record()} in
	\cref{lst:defer:Hazard-Pointer Recording and Clearing}
	take a double indirection to the data element?
	Why not \co{void *} instead of \co{void **}?
}\QuickQuizAnswer{
	Because \co{hp_try_record()} must check for concurrent modifications.
	To do that job, it needs a pointer to a pointer to the element,
	so that it can check for a modification to the pointer to the
	element.
}\QuickQuizEnd

The \co{hp_record()} function is quite straightforward:
It repeatedly invokes \co{hp_try_record()} until the return value
is something other than \co{HAZPTR_POISON}.

\QuickQuiz{
	Why bother with \co{hp_try_record()}?
	Wouldn't it be easier to just use the failure-immune
	\co{hp_record()} function?
}\QuickQuizAnswer{
	It might be easier in some sense, but as will be seen in the
	Pre-BSD routing example, there are situations for which
	\co{hp_record()} simply does not work.
}\QuickQuizEnd

The \co{hp_clear()} function is even more straightforward, with
an \co{smp_mb()} to force full ordering between the caller's uses
of the object protected by the hazard pointer and the setting of
the hazard pointer to \co{NULL}.
\end{fcvref}

\begin{listing}
\ebresizeverb{.91}{\input{CodeSamples/defer/hazptr=scan_free.fcv}}
\caption{Hazard-Pointer Scanning and Freeing}
\label{lst:defer:Hazard-Pointer Scanning and Freeing}
\end{listing}

\begin{fcvref}[ln:defer:hazptr:scan_free:free]
Once a hazard-pointer-protected object has been removed from its
linked data structure, so that it is now inaccessible to future
hazard-pointer readers, it is passed to \co{hazptr_free_later()},
which is shown on \clnrefrange{b}{e} of
\cref{lst:defer:Hazard-Pointer Scanning and Freeing}
(\path{hazptr.c}).
\Clnref{enq:b,enq:e}
enqueue the object on a per-thread list \co{rlist}
and \clnref{count} counts the object in \co{rcount}.
If \clnref{check} sees that a sufficiently large number of objects are now
queued, \clnref{scan} invokes \co{hazptr_scan()} to attempt to
free some of them.
\end{fcvref}

\begin{fcvref}[ln:defer:hazptr:scan_free:scan]
The \co{hazptr_scan()} function is shown on \clnrefrange{b}{e}
of the listing.
This function relies on a fixed maximum number of threads (\co{NR_THREADS})
and a fixed maximum number of hazard pointers per thread (\co{K}),
which allows a fixed-size array of hazard pointers to be used.
Because any thread might need to scan the hazard pointers, each thread
maintains its own array, which is referenced by the per-thread variable
\co{gplist}.
If \clnref{check} determines that this thread has not yet allocated its
\co{gplist}, \clnrefrange{alloc:b}{alloc:e} carry out the allocation.
The \IX{memory barrier} on \clnref{mb1} ensures that all threads see the
removal of all objects by this thread before
\clnrefrange{loop:b}{loop:e} scan
all of the hazard pointers, accumulating non-NULL pointers into
the \co{plist} array and counting them in \co{psize}.
The memory barrier on \clnref{mb2} ensures that the reads of
the hazard pointers
happen before any objects are freed.
\Clnref{sort} then sorts this array to enable use of binary search below.

\Clnref{rem:b,rem:e}
remove all elements from this thread's list of
to-be-freed objects, placing them on the local \co{tmplist}
and \clnref{zero} zeroes the count.
Each pass through the loop spanning
\clnrefrange{loop2:b}{loop2:e} processes each
of the to-be-freed objects.
\Clnref{rem1st:b,rem1st:e}
remove the first object from \co{tmplist},
and if \clnref{chkhazp:b,chkhazp:e}
determine that there is a hazard pointer
protecting this object, \clnrefrange{back:b}{back:e}
place it back onto \co{rlist}.
Otherwise, \clnref{free} frees the object.
\end{fcvref}

\begin{listing}
\input{CodeSamples/defer/route_hazptr=lookup.fcv}
\caption{Hazard-Pointer Pre-BSD Routing Table Lookup}
\label{lst:defer:Hazard-Pointer Pre-BSD Routing Table Lookup}
\end{listing}

The Pre-BSD routing example can use hazard pointers as shown in
\cref{lst:defer:Hazard-Pointer Pre-BSD Routing Table Lookup}
for data structures and \co{route_lookup()}, and in
\cref{lst:defer:Hazard-Pointer Pre-BSD Routing Table Add/Delete}
for \co{route_add()} and \co{route_del()}
(\path{route_hazptr.c}).
As with reference counting, the hazard-pointers implementation
is quite similar to the sequential algorithm shown in
\cref{lst:defer:Sequential Pre-BSD Routing Table}
on
\cpageref{lst:defer:Sequential Pre-BSD Routing Table},
so only differences will be discussed.

\begin{fcvref}[ln:defer:route_hazptr:lookup]
Starting with
\cref{lst:defer:Hazard-Pointer Pre-BSD Routing Table Lookup},
\clnref{hh} shows the \co{->hh} field used to queue objects pending
hazard-pointer free,
\clnref{re_freed} shows the \co{->re_freed} field used to detect
use-after-free bugs, and \clnref{tryrecord} invokes
\co{hp_try_record()} to attempt to acquire a hazard pointer.
If the return value is \co{NULL}, \clnref{NULL} returns a not-found
indication to the caller.
If the call to \co{hp_try_record()} raced with deletion, \clnref{deleted}
branches back to \clnref{retry}'s \co{retry} to re-traverse the list
from the beginning.
The \co{do}--\co{while} loop falls through when the desired element is
located, but if this element has already been freed, \clnref{abort}
terminates the program.
Otherwise, the element's \co{->iface} field is returned to the caller.

Note that \clnref{tryrecord} invokes \co{hp_try_record()} rather
than the easier-to-use \co{hp_record()}, restarting the full search
upon \co{hp_try_record()} failure.
And such restarting is absolutely required for correctness.
To see this, consider a hazard-pointer-protected linked list
containing elements~A, B, and~C that is subjected to the following
sequence of events:
\end{fcvref}

\begin{enumerate}
\item	Thread~0 stores a hazard pointer to element~B
	(having presumably traversed to element~B from element~A).
\item	Thread~1 removes element~B from the list, which sets
	the pointer from element~B to element~C to the special
	\co{HAZPTR_POISON} value in order to mark the deletion.
	Because Thread~0 has a hazard pointer to element~B,
	it cannot yet be freed.
\item	Thread~1 removes element~C from the list.
	Because there are no hazard pointers referencing element~C,
	it is immediately freed.
\item	Thread~0 attempts to acquire a hazard pointer to now-removed
	element~B's successor, but \co{hp_try_record()} returns the
	\co{HAZPTR_POISON} value, forcing the caller to restart its
	traversal from the beginning of the list.
\end{enumerate}

Which is a very good thing, because B's successor is the now-freed
element~C, which means that Thread~0's subsequent accesses might have
resulted in arbitrarily horrible memory corruption, especially if the
memory for element~C had since been re-allocated for some other purpose.
Therefore, hazard-pointer readers must typically restart the full
traversal in the face of a concurrent deletion.
Often the restart must go back to some global (and thus immortal) pointer,
but it is sometimes possible to restart at some intermediate location
if that location is guaranteed to still be live, for example, due to
the current thread holding a lock, a reference count, etc.

\QuickQuiz{
	Readers must ``typically'' restart?
	What are some exceptions?
}\QuickQuizAnswer{
	If the pointer emanates from a global variable or is otherwise
	not subject to being freed, then \co{hp_record()} may be
	used to repeatedly attempt to record the hazard pointer,
	even in the face of concurrent deletions.

	In certain cases, restart can be avoided by using link counting
	as exemplified by the UnboundedQueue and ConcurrentHashMap data
	structures implemented in Folly open-source library.\footnote{
		\url{https://github.com/facebook/folly}}
}\QuickQuizEnd

Because algorithms using hazard pointers might be restarted at any
step of their traversal through the linked data structure, such algorithms
must typically take care to avoid making any changes to the data
structure until after they have acquired all the hazard pointers that
are required for the update in question.

\QuickQuiz{
	But don't these restrictions on hazard pointers also apply
	to other forms of reference counting?
}\QuickQuizAnswer{
	Yes and no.
	These restrictions apply only to reference-counting mechanisms whose
	reference acquisition can fail.
}\QuickQuizEnd

These hazard-pointer restrictions result in great benefits to readers,
courtesy of the fact that the hazard pointers are stored local to each
CPU or thread, which in turn allows traversals to be carried out without
any writes to the data structures being traversed.
Referring back to
\cref{fig:count:Optimization and the Four Parallel-Programming Tasks}
on
\cpageref{fig:count:Optimization and the Four Parallel-Programming Tasks},
hazard pointers enable the CPU caches to do resource replication, which
in turn allows weakening of the parallel-access-control mechanism,
thus boosting performance and scalability.

Another advantage of restarting hazard pointers traversals is a reduction in
minimal memory footprint:
Any object not currently referenced by some hazard pointer may be
immediately freed.
In contrast,
\cref{sec:defer:Read-Copy Update (RCU)}
will discuss a mechanism that avoids read-side retries (and minimizes
read-side overhead), but which can result in a much larger memory
footprint.

\begin{listing}
\input{CodeSamples/defer/route_hazptr=add_del.fcv}
\caption{Hazard-Pointer Pre-BSD Routing Table Add\slash Delete}
\label{lst:defer:Hazard-Pointer Pre-BSD Routing Table Add/Delete}
\end{listing}

\begin{fcvref}[ln:defer:route_hazptr:add_del]
The \co{route_add()} and \co{route_del()} functions are shown in
\cref{lst:defer:Hazard-Pointer Pre-BSD Routing Table Add/Delete}.
\Clnref{init_freed} initializes \co{->re_freed},
\clnref{poison} poisons the \co{->re_next} field of the newly removed
object, and
\clnref{free_later} passes that object to the
\co{hazptr_free_later()} function, which will free that object once it
is safe to do so.
The spinlocks work the same as in
\cref{lst:defer:Reference-Counted Pre-BSD Routing Table Add/Delete}.
\end{fcvref}

\begin{figure}
\centering
\resizebox{2.5in}{!}{\includegraphics{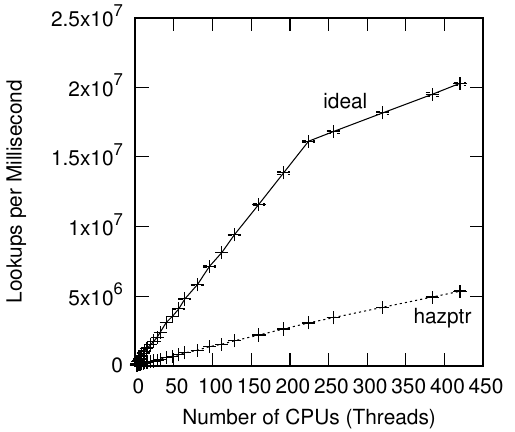}}
\caption{Pre-BSD Routing Table Protected by Hazard Pointers}
\label{fig:defer:Pre-BSD Routing Table Protected by Hazard Pointers}
\end{figure}

\Cref{fig:defer:Pre-BSD Routing Table Protected by Hazard Pointers}
shows the hazard-pointers-protected Pre-BSD routing algorithm's
performance on the same read-only workload as for
\cref{fig:defer:Pre-BSD Routing Table Protected by Reference Counting}.
Although hazard pointers scale far better than does reference counting,
hazard pointers still require readers to do writes to shared
memory (albeit with much improved locality of reference),
and also require a full memory barrier and retry check for each
object traversed.
Therefore, hazard-pointers performance is still far short of ideal.
On the other hand, unlike naive approaches to concurrent
reference-counting, hazard pointers not only operate correctly for
workloads involving concurrent updates, but also exhibit excellent
scalability.
Additional performance comparisons with other mechanisms may be found in
\cref{chp:Data Structures}
and in other publications~\cite{ThomasEHart2007a,McKenney:2013:SDS:2483852.2483867,MagedMichael04a}.

\QuickQuizSeries{%
\QuickQuizB{
	\Cref{fig:defer:Pre-BSD Routing Table Protected by Hazard Pointers}
	shows no sign of hyperthread-induced flattening at 224 threads.
	Why is that?
}\QuickQuizAnswerB{
	Modern microprocessors are complicated beasts, so significant
	skepticism is appropriate for any simple answer.
	That aside, the most likely reason is the full memory barriers
	required by hazard-pointers readers.
	Any delays resulting from those memory barriers would make time
	available to the other hardware thread sharing the core, resulting
	in greater scalability at the expense of per-hardware-thread
	performance.
}\QuickQuizEndB
\QuickQuizE{
	The paper ``Structured Deferral:
	Synchronization via
	Procrastination''~\cite{McKenney:2013:SDS:2483852.2483867}
	shows that hazard pointers have near-ideal performance.
	Whatever happened in
	\cref{fig:defer:Pre-BSD Routing Table Protected by Hazard Pointers}???
}\QuickQuizAnswerE{
	First,
	\cref{fig:defer:Pre-BSD Routing Table Protected by Hazard Pointers}
	has a linear y-axis, while most of the graphs in the
	``Structured Deferral'' paper have logscale y-axes.
	Next, that paper uses lightly-loaded hash tables, while
	\cref{fig:defer:Pre-BSD Routing Table Protected by Hazard Pointers}'s
	uses a 10-element simple linked list, which means that hazard pointers
	face a larger memory-barrier penalty in this workload than in
	that of the ``Structured Deferral'' paper.
	Finally, that paper used an older modest-sized x86 system, while
	a much newer and larger system was used to generate the data
	shown in
	\cref{fig:defer:Pre-BSD Routing Table Protected by Hazard Pointers}.

	In addition, use of pairwise asymmetric
	barriers~\cite{Windows2008FlushProcessWriteBuffers,JonathanCorbet2010sys-membarrier,Linuxmanpage2018sys-membarrier}
	has been proposed to eliminate the read-side hazard-pointer
	memory barriers on systems supporting this notion~\cite{DavidGoldblatt2018asymmetricFences},
	which might improve the performance of hazard pointers beyond
	what is shown in the figure.

	As always, your mileage may vary.
	Given the difference in performance, it is clear that hazard
	pointers give you the best performance either for
	very large data structures (where the memory-barrier overhead
	will at least partially overlap cache-miss penalties) and
	for data structures such as hash tables where a lookup
	operation needs a minimal number of hazard pointers.
}\QuickQuizEndE
}

And hazard pointers are the concurrent reference counter mentioned
on \cpageref{sec:defer:Mysteries hazard pointers}.
The next section attempts to improve on hazard pointers by using
sequence locks, which avoid both read-side writes and per-object memory
barriers.

\IfTwoColumn{}{\FloatBarrier}
% defer/seqlock.tex
% mainfile: ../perfbook.tex
% SPDX-License-Identifier: CC-BY-SA-3.0

\section{Sequence Locks}
\label{sec:defer:Sequence Locks}
\epigraph{It'll be just like starting over.}{John Lennon}

\begin{figure}
\centering
\resizebox{3in}{!}{\includegraphics{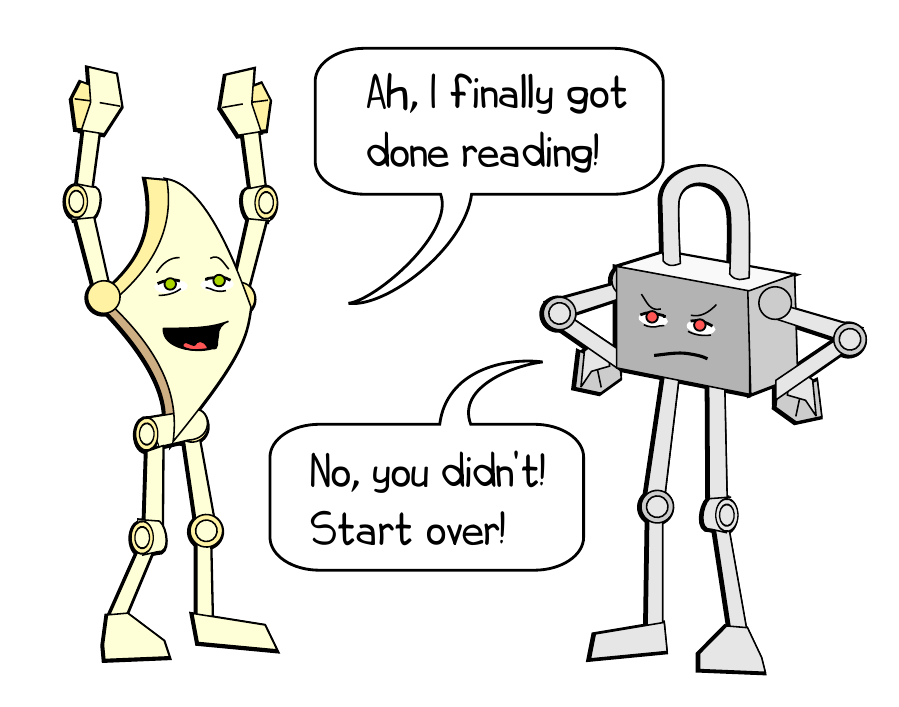}}
\caption{Reader And Uncooperative Sequence Lock}
\label{fig:defer:Reader And Uncooperative Sequence Lock}
\end{figure}

The published sequence-lock
record~\cite{10.1145/800212.806505,10.1145/359863.359878}
extends back as far as that of reader-writer locking, but sequence locks
nevertheless remain in relative obscurity.
Sequence locks are used in the Linux kernel for read-mostly data that
must be seen in a consistent state by readers.
However, unlike reader-writer locking, readers do not exclude writers.
Instead, like hazard pointers, sequence locks force readers to
\emph{retry} an operation if they detect activity from a concurrent writer.
As can be seen from
\cref{fig:defer:Reader And Uncooperative Sequence Lock},
it is important to design code using sequence locks so that readers
very rarely need to retry.

\QuickQuiz{
	Why isn't this sequence-lock discussion in \cref{chp:Locking},
	you know, the one on \emph{locking}?
}\QuickQuizAnswer{
	The sequence-lock mechanism is really a combination of two
	separate synchronization mechanisms, sequence counts and
	locking.
	In fact, the sequence-count mechanism is available separately
	in the Linux kernel via the
	\co{write_seqcount_begin()} and \co{write_seqcount_end()}
	primitives.

	However, the combined \co{write_seqlock()} and
	\co{write_sequnlock()} primitives are used much more heavily
	in the Linux kernel.
	More importantly, many more people will understand what you
	mean if you say ``sequence lock'' than if you say
	``sequence count''.

	So this section is entitled ``Sequence Locks'' so that people
	will understand what it is about just from the title, and
	it appears in the ``Deferred Processing'' because (1) of the
	emphasis on the ``sequence count'' aspect of ``sequence locks''
	and (2) because a ``sequence lock'' is much more than merely
	a lock.
}\QuickQuizEnd

\begin{listing}
\begin{VerbatimL}
do {
	seq = read_seqbegin(&test_seqlock);
	/* read-side access. */
} while (read_seqretry(&test_seqlock, seq));
\end{VerbatimL}
\caption{Sequence-Locking Reader}
\label{lst:defer:Sequence-Locking Reader}
\end{listing}

\begin{listing}
\begin{VerbatimL}
write_seqlock(&test_seqlock);
/* Update */
write_sequnlock(&test_seqlock);
\end{VerbatimL}
\caption{Sequence-Locking Writer}
\label{lst:defer:Sequence-Locking Writer}
\end{listing}

The key component of sequence locking is the sequence number, which has
an even value in the absence of updaters and an odd value if there
is an update in progress.
Readers can then snapshot the value before and after each access.
If either snapshot has an odd value, or if the two snapshots differ,
there has been a concurrent update, and the reader must discard
the results of the access and then retry it.
Readers therefore use the \co{read_seqbegin()} and \co{read_seqretry()}
functions shown in \cref{lst:defer:Sequence-Locking Reader}
when accessing data protected by a sequence lock.
Writers must increment the value before and after each update,
and only one writer is permitted at a given time.
Writers therefore use the \co{write_seqlock()} and \co{write_sequnlock()}
functions shown in \cref{lst:defer:Sequence-Locking Writer}
when updating data protected by a sequence lock.

As a result, sequence-lock-protected data can have an arbitrarily
large number of concurrent readers, but only one writer at a time.
Sequence locking is used in the Linux kernel to protect calibration
quantities used for timekeeping.
It is also used in pathname traversal to detect concurrent rename operations.

\begin{listing}
\input{CodeSamples/defer/seqlock=impl.fcv}
\caption{Sequence-Locking Implementation}
\label{lst:defer:Sequence-Locking Implementation}
\end{listing}

A simple implementation of sequence locks is shown in
\cref{lst:defer:Sequence-Locking Implementation}
(\path{seqlock.h}).
\begin{fcvref}[ln:defer:seqlock:impl:typedef]
The \co{seqlock_t} data structure is shown on
\clnrefrange{b}{e}, and contains
the sequence number along with a lock to serialize writers.
\end{fcvref}
\begin{fcvref}[ln:defer:seqlock:impl:init]
\Clnrefrange{b}{e} show \co{seqlock_init()}, which, as the name indicates,
initializes a \co{seqlock_t}.
\end{fcvref}

\begin{fcvref}[ln:defer:seqlock:impl:read_seqbegin]
\Clnrefrange{b}{e} show \co{read_seqbegin()}, which begins a sequence-lock
\IXh{read-side}{critical section}.
\Clnref{fetch} takes a snapshot of the sequence counter, and
\clnref{mb} orders
this snapshot operation before the caller's critical section.
Finally, \clnref{ret} returns the value of the snapshot (with the least-significant
bit cleared), which the caller
will pass to a later call to \co{read_seqretry()}.
\end{fcvref}

\QuickQuiz{
	Why not have \co{read_seqbegin()} in
	\cref{lst:defer:Sequence-Locking Implementation}
	check for the low-order bit being set, and retry
	internally, rather than allowing a doomed read to start?
}\QuickQuizAnswer{
	That would be a legitimate implementation.
	However, if the workload is read-mostly, it would likely
	increase the overhead of the common-case successful read,
	which could be counter-productive.
	However, given a sufficiently large fraction of updates
	and sufficiently high-overhead readers, having the
	check internal to \co{read_seqbegin()} might be preferable.
}\QuickQuizEnd

\begin{fcvref}[ln:defer:seqlock:impl:read_seqretry]
\Clnrefrange{b}{e} show \co{read_seqretry()}, which returns \co{true} if there
was at least one writer since the time of the corresponding
call to \co{read_seqbegin()}.
\Clnref{mb} orders the caller's prior critical section before \clnref{fetch}'s
fetch of the new snapshot of the sequence counter.
\Clnref{ret} checks whether the sequence counter has changed,
in other words, whether there has been at least one writer, and returns
\co{true} if so.
\end{fcvref}

\QuickQuizSeries{%
\QuickQuizB{
	Why is the \co{smp_mb()} on
	\clnrefr{ln:defer:seqlock:impl:read_seqretry:mb} of
	\cref{lst:defer:Sequence-Locking Implementation}
	needed?
}\QuickQuizAnswerB{
	If it was omitted, both the compiler and the CPU would be
	within their rights to move the critical section preceding
	the call to \co{read_seqretry()} down below this function.
	This would prevent the sequence lock from protecting the
	critical section.
	The \co{smp_mb()} primitive prevents such reordering.
}\QuickQuizEndB
\QuickQuizM{
	Can't weaker memory barriers be used in the code in
	\cref{lst:defer:Sequence-Locking Implementation}?
}\QuickQuizAnswerM{
	In older versions of the Linux kernel, no.

	\begin{fcvref}[ln:defer:seqlock:impl]
	In very new versions of the Linux kernel,
	\clnref{read_seqbegin:fetch} could use
	\co{smp_load_acquire()} instead of \co{READ_ONCE()}, which
	in turn would allow the \co{smp_mb()} on
	\clnref{read_seqbegin:mb} to be dropped.
	Similarly, \clnref{write_sequnlock:inc} could use an
	\co{smp_store_release()}, for
	example, as follows:

\begin{VerbatimU}
smp_store_release(&slp->seq, READ_ONCE(slp->seq) + 1);
\end{VerbatimU}

	This would allow the \co{smp_mb()} on
	\clnref{write_sequnlock:mb} to be dropped.
	\end{fcvref}
}\QuickQuizEndM
\QuickQuizE{
	What prevents sequence-locking updaters from starving readers?
}\QuickQuizAnswerE{
	Nothing.
	This is one of the weaknesses of sequence locking, and as a
	result, you should use sequence locking only in read-mostly
	situations.
	Unless of course read-side starvation is acceptable in your
	situation, in which case, go wild with the sequence-locking updates!
}\QuickQuizEndE
}

\begin{fcvref}[ln:defer:seqlock:impl:write_seqlock]
\Clnrefrange{b}{e} show \co{write_seqlock()}, which simply acquires the lock,
increments the sequence number, and executes a memory barrier to ensure
that this increment is ordered before the caller's critical section.
\end{fcvref}
\begin{fcvref}[ln:defer:seqlock:impl:write_sequnlock]
\Clnrefrange{b}{e} show \co{write_sequnlock()}, which executes a memory barrier
to ensure that the caller's critical section is ordered before the
increment of the sequence number on \clnref{inc}, then releases the lock.
\end{fcvref}

\QuickQuizSeries{%
\QuickQuizB{
	What if something else serializes writers, so that the lock
	is not needed?
}\QuickQuizAnswerB{
	In this case, the \co{->lock} field could be omitted, as it
	is in \co{seqcount_t} in the Linux kernel.
}\QuickQuizEndB
\QuickQuizE{
	Why isn't \co{seq} on
	\clnrefr{ln:defer:seqlock:impl:typedef:seq} of
	\cref{lst:defer:Sequence-Locking Implementation}
	\co{unsigned} rather than \co{unsigned long}?
	After all, if \co{unsigned} is good enough for the Linux
	kernel, shouldn't it be good enough for everyone?
}\QuickQuizAnswerE{
	Not at all.
	The Linux kernel has a number of special attributes that allow
	it to ignore the following sequence of events:
	\begin{enumerate}
	\item	Thread~0 executes \co{read_seqbegin()}, picking up
		\co{->seq} in
		\clnrefr{ln:defer:seqlock:impl:read_seqbegin:fetch},
		noting that the value is even,
		and thus returning to the caller.
	\item	Thread~0 starts executing its read-side critical section,
		but is then preempted for a long time.
	\item	Other threads repeatedly invoke \co{write_seqlock()} and
		\co{write_sequnlock()}, until the value of \co{->seq}
		overflows back to the value that Thread~0 fetched.
	\item	Thread~0 resumes execution, completing its read-side
		critical section with inconsistent data.
	\item	Thread~0 invokes \co{read_seqretry()}, which incorrectly
		concludes that Thread~0 has seen a consistent view of
		the data protected by the sequence lock.
	\end{enumerate}

	The Linux kernel uses sequence locking for things that are
	updated rarely, with time-of-day information being a case
	in point.
	This information is updated at most once per millisecond,
	so that seven weeks would be required to overflow the counter.
	If a kernel thread was preempted for seven weeks, the Linux
	kernel's soft-lockup code would be emitting warnings every two
	minutes for that entire time.

	In contrast, with a 64-bit counter, more than five centuries
	would be required to overflow, even given an update every
	\emph{nano}second.
	Therefore, this implementation uses a type for \co{->seq}
	that is 64 bits on 64-bit systems.
}\QuickQuizEndE
}

\begin{listing}
\input{CodeSamples/defer/route_seqlock=lookup.fcv}
\caption{Sequence-Locked Pre-BSD Routing Table Lookup (BUGGY!!!)}
\label{lst:defer:Sequence-Locked Pre-BSD Routing Table Lookup}
\end{listing}

\begin{listing}
\input{CodeSamples/defer/route_seqlock=add_del.fcv}
\caption{Sequence-Locked Pre-BSD Routing Table Add\slash Delete (BUGGY!!!)}
\label{lst:defer:Sequence-Locked Pre-BSD Routing Table Add/Delete}
\end{listing}

So what happens when sequence locking is applied to the Pre-BSD
routing table?
\Cref{lst:defer:Sequence-Locked Pre-BSD Routing Table Lookup}
shows the data structures and \co{route_lookup()}, and
\cref{lst:defer:Sequence-Locked Pre-BSD Routing Table Add/Delete}
shows \co{route_add()} and \co{route_del()} (\path{route_seqlock.c}).
This implementation is once again similar to its counterparts in earlier
sections, so only the differences will be highlighted.

\begin{fcvref}[ln:defer:route_seqlock:lookup]
In
\cref{lst:defer:Sequence-Locked Pre-BSD Routing Table Lookup},
\clnref{struct:re_freed} adds \co{->re_freed}, which is checked on
\clnref{lookup:chk_freed,lookup:abort}.
\Clnref{struct:sl} adds a sequence lock, which is used by \co{route_lookup()}
\end{fcvref}
\begin{fcvref}[ln:defer:route_seqlock:lookup:lookup]
on \clnref{r_sqbegin,r_sqretry1,r_sqretry2},
with \clnref{goto_retry1,goto_retry2} branching back to
the \co{retry} label on \clnref{retry}.
The effect is to retry any lookup that runs concurrently with an update.
\end{fcvref}

\begin{fcvref}[ln:defer:route_seqlock:add_del]
In
\cref{lst:defer:Sequence-Locked Pre-BSD Routing Table Add/Delete},
\clnref{add:w_sqlock,add:w_squnlock,del:w_sqlock,%
del:w_squnlock1,del:w_squnlock2}
acquire and release the sequence lock,
while \clnref{add:clr_freed,del:set_freed} handle \co{->re_freed}.
This implementation is therefore quite straightforward.
\end{fcvref}

\begin{figure}
\centering
\resizebox{2.5in}{!}{\includegraphics{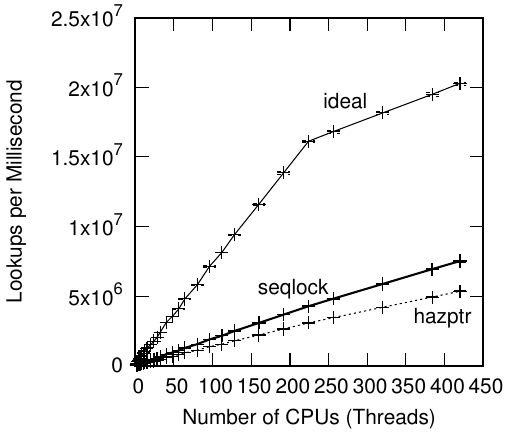}}
\caption{Pre-BSD Routing Table Protected by Sequence Locking}
\label{fig:defer:Pre-BSD Routing Table Protected by Sequence Locking}
\end{figure}

It also performs better on the read-only workload, as can be seen in
\cref{fig:defer:Pre-BSD Routing Table Protected by Sequence Locking},
though its performance is still far from ideal.
Worse yet, it suffers use-after-free failures.
The problem is that the reader might encounter a segmentation violation
due to accessing an already-freed structure before \co{read_seqretry()}
has a chance to warn of the concurrent update.

\QuickQuiz{
	Can this bug be fixed?
	In other words, can you use sequence locks as the \emph{only}
	synchronization mechanism protecting a linked list supporting
	concurrent addition, deletion, and lookup?
}\QuickQuizAnswer{
	One trivial way of accomplishing this is to surround all
	accesses, including the read-only accesses, with
	\co{write_seqlock()} and \co{write_sequnlock()}.
	Of course, this solution also prohibits all read-side
	parallelism, resulting in massive lock contention,
	and furthermore could just as easily be implemented
	using simple locking.

	If you do come up with a solution that uses \co{read_seqbegin()}
	and \co{read_seqretry()} to protect read-side accesses, make
	sure that you correctly handle the following sequence of events:

	\begin{enumerate}
	\item	CPU~0 is traversing the linked list, and picks up a pointer
		to list element~A.
	\item	CPU~1 removes element~A from the list and frees it.
	\item	CPU~2 allocates an unrelated data structure, and gets
		the memory formerly occupied by element~A\@.
		In this unrelated data structure, the memory previously
		used for element~A's \co{->next} pointer is now occupied
		by a floating-point number.
	\item	CPU~0 picks up what used to be element~A's \co{->next}
		pointer, gets random bits, and therefore gets a
		segmentation fault.
	\end{enumerate}

	One way to protect against this sort of problem requires use
	of ``type-safe memory'', which will be discussed in
	\cref{sec:defer:Type-Safe Memory}.
	Roughly similar solutions are possible using the hazard pointers
	discussed in
	\cref{sec:defer:Hazard Pointers}.
	But in either case, you would be using some other synchronization
	mechanism in addition to sequence locks!
}\QuickQuizEnd

As hinted on
\cpageref{sec:defer:Mysteries sequence locking},
both the read-side and write-side critical sections of a sequence lock
can be thought of as transactions, and sequence locking therefore can
be thought of as a limited form of transactional memory, which will be
discussed in \cref{sec:future:Transactional Memory}.
The limitations of sequence locking are:
\begin{enumerate*}[(1)]
\item Sequence locking restricts updates and
\item Sequence locking does not permit traversal of pointers
to objects that might be freed by updaters.
\end{enumerate*}
These limitations are of course overcome by transactional memory, but
can also be overcome by combining other synchronization primitives
with sequence locking.

Sequence locks allow writers to defer readers, but not vice versa.
This can result in \IX{unfairness} and even \IX{starvation}
in writer-heavy workloads.\footnote{
	Dmitry Vyukov describes one way to reduce (but, sadly, not eliminate)
	reader starvation:
	\url{http://www.1024cores.net/home/lock-free-algorithms/reader-writer-problem/improved-lock-free-seqlock}.}
On the other hand, in the absence of writers, sequence-lock readers are
reasonably fast and scale linearly.
It is only human to want the best of both worlds:
Fast readers without the possibility of read-side failure,
let alone starvation.
In addition, it would also be nice to overcome sequence locking's limitations
with pointers.
The following section presents a synchronization mechanism with exactly
these properties.

\IfTwoColumn{}{\FloatBarrier}
% defer/rcu.tex
% mainfile: ../perfbook.tex
% SPDX-License-Identifier: CC-BY-SA-3.0

\section{Read-Copy Update (RCU)}
\label{sec:defer:Read-Copy Update (RCU)}
\epigraph{``Free'' is a \emph{very} good price!}{Tom Peterson}

All of the mechanisms discussed in the preceding sections
used one of a number of approaches to defer specific actions
until they may be carried out safely.
The reference counters discussed in
\cref{sec:defer:Reference Counting}
use explicit counters to defer actions that could disturb readers,
which results in read-side contention and thus poor scalability.
The hazard pointers covered by
\cref{sec:defer:Hazard Pointers}
uses implicit counters in the guise of per-thread lists of pointer.
This avoids read-side contention, but requires readers to do stores and
conditional branches, as well as either \IXhpl{full}{memory barrier}
in read-side primitives or real-time-unfriendly \IXacrlpl{ipi} in
update-side primitives.\footnote{
	In some important special cases, this extra work can be avoided
	by using link counting as exemplified by the UnboundedQueue
	and ConcurrentHashMap data structures implemented in Folly
	open-source library
	(\url{https://github.com/facebook/folly}).}
The sequence lock presented in
\cref{sec:defer:Sequence Locks}
also avoids read-side contention, but does not protect pointer
traversals and, like hazard pointers, requires either full memory barriers
in read-side primitives, or inter-processor interrupts in update-side
primitives.
These schemes' shortcomings raise the question of
whether it is possible to do better.

This section introduces \IXBacrfst{rcu}, which provides
an API that allows readers to be associated with regions in the source code,
rather than with expensive updates to frequently updated shared data.
The remainder of this
section examines RCU from a number of different perspectives.
\Cref{sec:defer:Introduction to RCU} provides the classic
introduction to RCU,
\cref{sec:defer:RCU Fundamentals} covers fundamental RCU
concepts,
\cref{sec:defer:RCU Linux-Kernel API} presents the Linux-kernel
API,
\cref{sec:defer:RCU Usage} introduces some common RCU use cases,
and finally
\cref{sec:defer:RCU Related Work} covers recent work related
to RCU.

Although RCU has gained a reputation for being subtle and difficult,
when used properly, it is quite straightforward.
In fact, no less an authority than Butler Lampson classifies it as easy
concurrency~\cite[Chapter 3]{Apt-Hoare2022Dijkstra}.

% defer/rcuintro.tex
% mainfile: ../perfbook.tex
% SPDX-License-Identifier: CC-BY-SA-3.0

\subsection{Introduction to RCU}
\label{sec:defer:Introduction to RCU}

The approaches discussed in the preceding sections have provided
good scalability but decidedly non-ideal performance for the
Pre-BSD routing table.
Therefore, in the spirit of ``only those who have gone too far
know how far you can go'',\footnote{
	With apologies to T.~S.~Eliot.}
we will go all the way, looking into algorithms in which concurrent
readers might well execute exactly the same sequence of assembly language
instructions as would a single-threaded lookup, despite the presence of
concurrent updates.
Of course, this laudable goal might raise serious implementability
questions, but we cannot possibly succeed if we don't even try!

And should we succeed, we will have uncovered yet another of the
mysteries set forth on
\cpageref{sec:defer:Mysteries RCU}.

\subsubsection{Minimal Insertion and Deletion}
\label{sec:defer:Minimal Insertion and Deletion}

\begin{figure}
\centering
\resizebox{3in}{!}{\includegraphics{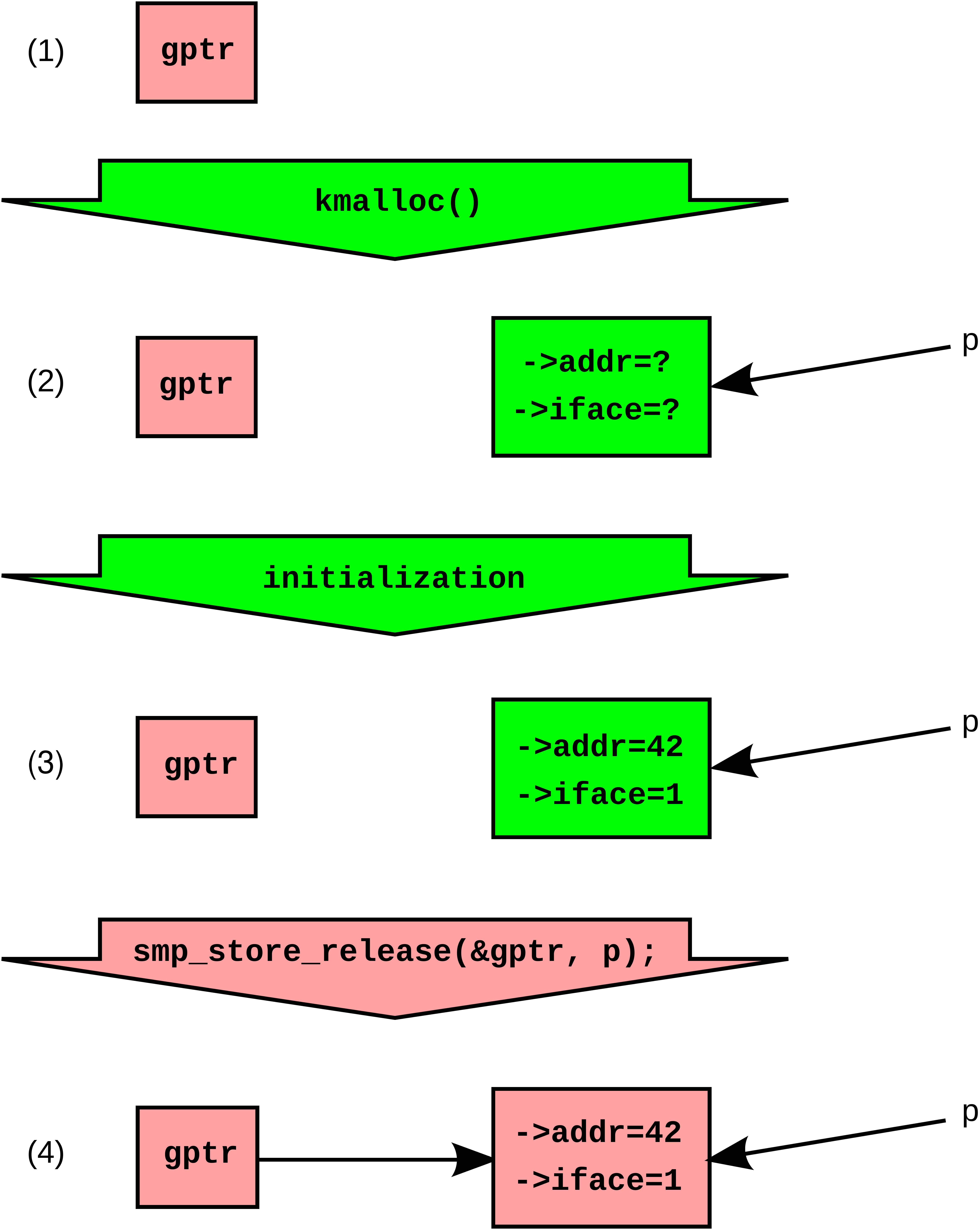}}
\caption{Insertion With Concurrent Readers}
\label{fig:defer:Insertion With Concurrent Readers}
\end{figure}

To minimize implementability concerns, we focus on a minimal
data structure, which consists of a single global pointer that is either
\co{NULL} or references a single structure.
Minimal though it might be, this data structure is heavily used in
production~\cite{GeoffRomer2018C++DeferredReclamationP0561R4}.
A classic approach for insertion is shown in
\cref{fig:defer:Insertion With Concurrent Readers},
which shows four states with time advancing from top to bottom.
The first row shows the initial state, with \co{gptr} equal to \co{NULL}.
In the second row, we have allocated a structure which is uninitialized,
as indicated by the question marks.
In the third row, we have initialized the structure.
Finally, in the fourth and final row, we have updated \co{gptr} to
reference the newly allocated and initialized element.

We might hope that this assignment to \co{gptr} could use a simple
C-language assignment statement.
Unfortunately,
\cref{sec:toolsoftrade:Shared-Variable Shenanigans}
dashes these hopes.
Therefore, the updater cannot use a simple C-language assignment, but
must instead use \co{smp_store_release()} as shown in the figure,
or, as will be seen, \co{rcu_assign_pointer()}.

Similarly, one might hope that readers could use a single C-language
assignment to fetch the value of \co{gptr}, and be guaranteed to either
get the old value of \co{NULL} or to get the newly installed pointer,
but either way see a valid result.
Unfortunately, \cref{sec:toolsoftrade:Shared-Variable Shenanigans}
dashes these hopes as well.
To obtain this guarantee, readers must instead use \co{READ_ONCE()},
or, as will be seen, \co{rcu_dereference()}.
However, on most modern computer systems, each of these read-side primitives
can be implemented with a single load instruction, exactly the instruction
that would normally be used in single-threaded code.

Reviewing \cref{fig:defer:Insertion With Concurrent Readers}
from the viewpoint of readers, in the first three states all readers
see \co{gptr} having the value \co{NULL}.
Upon entering the fourth state, some readers might see \co{gptr} still
having the value \co{NULL} while others might see it referencing the
newly inserted element, but after some time, all readers will see this
new element.
At all times, all readers will see \co{gptr} as containing a valid pointer.
Therefore, it really is possible to add new data to linked data structures
while allowing concurrent readers to execute the same sequence of machine
instructions that is normally used in single-threaded code.
This no-cost approach to concurrent reading provides excellent performance
and scalability, and also is eminently suitable for real-time use.

\begin{figure}
\centering
\resizebox{3in}{!}{\includegraphics{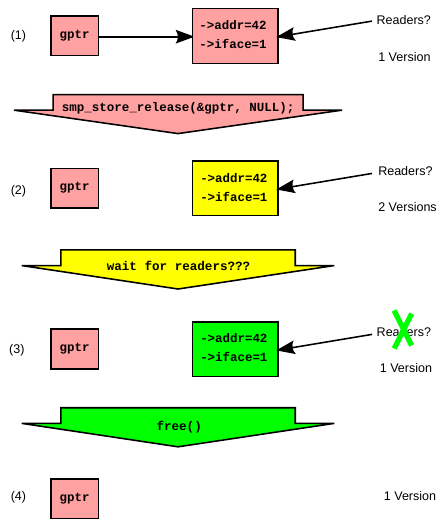}}
\caption{Deletion With Concurrent Readers}
\label{fig:defer:Deletion With Concurrent Readers}
\end{figure}

Insertion is of course quite useful, but sooner or later, it will also
be necessary to delete data.
As can be seen in
\cref{fig:defer:Deletion With Concurrent Readers},
the first step is easy.
Again taking the lessons from
\cref{sec:toolsoftrade:Shared-Variable Shenanigans}
to heart, \co{smp_store_release()} is used to \co{NULL} the pointer,
thus moving from the first row to the second in the figure.
At this point, pre-existing readers see the old structure with
\co{->addr} of~42 and \co{->iface} of~1, but new readers will see
a \co{NULL} pointer, that is, concurrent readers can disagree on
the state, as indicated by the ``2~Versions'' in the figure.

\QuickQuizSeries{%
\QuickQuizB{
	Why does
	\cref{fig:defer:Deletion With Concurrent Readers}
	use \co{smp_store_release()} given that it is storing
	a \co{NULL} pointer?
	Wouldn't \co{WRITE_ONCE()} work just as well in this case,
	given that there is no structure initialization to order
	against the store of the \co{NULL} pointer?
}\QuickQuizAnswerB{
	Yes, it would.

	Because a \co{NULL} pointer is being assigned, there is nothing
	to order against, so there is no need for \co{smp_store_release()}.
	In contrast, when assigning a non-\co{NULL} pointer, it is
	necessary to use \co{smp_store_release()} in order to ensure
	that initialization of the pointed-to structure is carried
	out before assignment of the pointer.

	In short, \co{WRITE_ONCE()} would work, and would
	save a little bit of CPU time on some architectures.
	However, as we will see, software-engineering concerns
	will motivate use of a special \co{rcu_assign_pointer()}
	that is quite similar to \co{smp_store_release()}.
}\QuickQuizEndB
\QuickQuizE{
	Readers running concurrently with each other and with the procedure
	outlined in
	\cref{fig:defer:Deletion With Concurrent Readers}
	can disagree on the value of \co{gptr}.
	Isn't that just a wee bit problematic???
}\QuickQuizAnswerE{
	Not necessarily.

	As hinted at in
	\cref{sec:cpu:Hardware Optimizations,sec:cpu:Hardware Free Lunch?},
	speed-of-light delays mean that a computer's data is always
	stale compared to whatever external reality that data is intended
	to model.

	Real-world algorithms therefore absolutely must tolerate
	inconsistancies between external reality and the in-computer
	data reflecting that reality.
	Many of those algorithms are also able to tolerate some degree
	of inconsistency within the in-computer data.
	\Cref{sec:datastruct:RCU-Protected Hash Table Discussion}
	discusses this point in more detail.

	Please note that this need to tolerate inconsistent and stale
	data is not limited to RCU\@.
	It also applies to reference counting, hazard pointers, sequence
	locks, and even to some locking use cases.
	For example, if you compute some quantity while holding a lock,
	but use that quantity after releasing that lock,
	you might well be using stale data.
	After all, the data that quantity is based on might change
	arbitrarily as soon as the lock is released.

	So yes, RCU readers can see stale and inconsistent data, but no,
	this is not necessarily problematic.
	And, when needed, there are RCU usage patterns that avoid both
	staleness and inconsistency~\cite{Arcangeli03}.
}\QuickQuizEndE
}

We get back to a single version simply by waiting for all the
pre-existing readers to complete, as shown in row~3.
At that point, all the pre-existing readers are done, and no later
reader has a path to the old data item, so there can no longer be
any readers referencing it.
It may therefore be safely freed, as shown on row~4.

Thus, given a way to wait for pre-existing readers to complete,
it is possible to both add data to and remove data from a linked
data structure, despite the readers executing the same sequence
of machine instructions that would be appropriate for single-threaded
execution.
So perhaps going all the way was not too far after all!

But how can we tell when all of the pre-existing readers have in
fact completed?
This question is the topic of \cref{sec:defer:Waiting for Readers}.
But first, the next section defines RCU's core API.

\subsubsection{Core RCU API}
\label{sec:defer:Core RCU API}

The full Linux-kernel API is quite extensive, with more than one
hundred API members.
However, this section will confine itself to six core RCU API members,
which suffices for the upcoming sections introducing RCU and covering
its fundamentals.
The full API is covered in \cref{sec:defer:RCU Linux-Kernel API}.

Three members of the core APIs are used by readers.
The \apik{rcu_read_lock()} and \apik{rcu_read_unlock()} functions delimit
\IXhmrpl{RCU read-side}{critical section}.
These may be nested, so that one \co{rcu_read_lock()}--\co{rcu_read_unlock()}
pair can be enclosed within another.
In this case, the nested set of RCU read-side critical sections act as
one large critical section covering the full extent of the nested set.
The third read-side API member, \apik{rcu_dereference()}, fetches an
\IXr{RCU-protected pointer}.
Conceptually, \co{rcu_dereference()} simply loads from memory, but we
will see in
\cref{sec:defer:Publish-Subscribe Mechanism}
that \co{rcu_dereference()} must prevent the compiler and (in
one case) the CPU from reordering its load with later memory operations
that dereference this pointer.

\QuickQuiz{
	What is an RCU-protected pointer?
}\QuickQuizAnswer{
	A pointer to \IXr{RCU-protected data}.
	RCU-protected data is in turn a block of dynamically allocated
	memory whose freeing will be deferred such that an RCU grace
	period will elapse between the time that there were no longer
	any RCU-reader-accessible pointers to that block and the time
	that that block is freed.
	This ensures that no RCU readers will have access to that block at
	the time that it is freed.

	RCU-protected pointers must be handled carefully.
	For example, any reader that intends to dereference an
	RCU-protected pointer must use \co{rcu_dereference()} (or
	stronger) to load that pointer.
	In addition, any updater must use \co{rcu_assign_pointer()}
	(or stronger) to store to that pointer.
}\QuickQuizEnd

The other three members of the core APIs are used by updaters.
The \apik{synchronize_rcu()} function implements the ``wait for readers''
operation from \cref{fig:defer:Deletion With Concurrent Readers}.
The \apik{call_rcu()} function is the asynchronous counterpart of
\co{synchronize_rcu()} by invoking the specified function after
all pre-existing RCU readers have completed.
Finally, the \apik{rcu_assign_pointer()} macro is used to update an
RCU-protected pointer.
Conceptually, this is simply an assignment statement, but we will
see in
\cref{sec:defer:Publish-Subscribe Mechanism}
that \co{rcu_assign_pointer()} must prevent the compiler and the CPU
from reordering this assignment to precede any prior assignments used
to initialize the pointed-to structure.

\begin{table*}
\renewcommand*{\arraystretch}{1.25}
\rowcolors{2}{}{lightgray}
\small
\centering
\ebresizewidth{
\begin{tabular}{llp{2.4in}}
\toprule
&
	Primitive &
		Purpose \\
\midrule
\cellcolor{white}\emph{Readers} &
	\tco{rcu_read_lock()} &
		Start an RCU read-side critical section. \\
&
	\tco{rcu_read_unlock()} &
		End an RCU read-side critical section. \\
\cellcolor{white}&
	\tco{rcu_dereference()} &
		Safely load an RCU-protected pointer. \\
\midrule
\emph{Updaters} &
	\tco{synchronize_rcu()} &
		Wait for all pre-existing RCU read-side critical sections to
		complete. \\
\cellcolor{white}&
	\tco{call_rcu()} &
		Invoke the specified function after all pre-existing RCU read-side
		critical sections complete. \\
&
	\tco{rcu_assign_pointer()} &
		Safely update an RCU-protected pointer. \\
\bottomrule
\end{tabular}
}
\caption{Core RCU API}
\label{tab:defer:Core RCU API}
\end{table*}

\QuickQuiz{
	What does \co{synchronize_rcu()} do if it starts at about
	the same time as an \co{rcu_read_lock()}?
}\QuickQuizAnswer{
	If a \co{synchronize_rcu()} cannot prove that it started
	before a given \co{rcu_read_lock()}, then it must wait
	for the corresponding \co{rcu_read_unlock()}.
}\QuickQuizEnd

The core RCU API is summarized in \cref{tab:defer:Core RCU API} for
easy reference.
With that, we are ready to continue this introduction to RCU with
the key RCU operation, waiting for readers.

\subsubsection{Waiting for Readers}
\label{sec:defer:Waiting for Readers}

It is tempting to base the reader-waiting functionality of
\co{synchronize_rcu()} and \co{call_rcu()} on a reference counter
updated by \co{rcu_read_lock()} and \co{rcu_read_unlock()}, but
\cref{fig:count:Atomic Increment Scalability on x86}
in
\cref{chp:Counting}
shows that concurrent reference counting results in extreme overhead.
This extreme overhead was confirmed in the specific case of
reference counters in
\cref{fig:defer:Pre-BSD Routing Table Protected by Reference Counting}
on
\cpageref{fig:defer:Pre-BSD Routing Table Protected by Reference Counting}.
Hazard pointers profoundly reduce this overhead, but, as we saw in
\cref{fig:defer:Pre-BSD Routing Table Protected by Hazard Pointers}
on
\cpageref{fig:defer:Pre-BSD Routing Table Protected by Hazard Pointers},
not to zero.
Nevertheless, many RCU implementations use counters with carefully
controlled cache locality.

A second approach observes that memory synchronization is expensive,
and therefore uses registers instead, namely each CPU's or thread's
program counter (PC), thus imposing no overhead on readers, at least
in the absence of concurrent updates.
The updater polls each relevant PC, and if that PC is not within read-side
code, then the corresponding CPU or thread is within a quiescent state,
in turn signaling the completion of any reader that might have access
to the newly removed data element.
Once all CPU's or thread's PCs have been observed to be outside of any
reader, the \IX{grace period} has completed.
Please note that this approach poses some serious challenges, including
memory ordering, functions that are \emph{sometimes} invoked from readers,
and ever-exciting code-motion optimizations.
Nevertheless, this approach is said to be used in
production~\cite{MikeAsh2015Apple}.

A third approach is to simply wait for a fixed period of time that is
long enough to comfortably exceed the lifetime of any reasonable
reader~\cite{Jacobson93,AjuJohn95}.
This can work quite well in hard real-time systems~\cite{YuxinRen2018RTRCU},
but in less exotic
settings, Murphy says that it is critically important to be prepared
even for unreasonably long-lived readers.
To see this, consider the consequences of failing do so:
A data item will be freed while the unreasonable reader is still
referencing it, and that item might well be immediately reallocated,
possibly even as a data item of some other type.
The unreasonable reader and the unwitting reallocator would then
be attempting to use the same memory for two very different purposes.
The ensuing mess will be exceedingly difficult to debug.

A fourth approach is to wait forever, secure in the knowledge that
doing so will accommodate even the most unreasonable reader.
This approach is also called ``leaking memory'', and has a bad reputation
due to the fact that memory leaks often require untimely and
inconvenient reboots.
Nevertheless, this is a viable strategy when the update rate and the
uptime are both sharply bounded.
For example, this approach could work well in a high-availability
cluster where systems were periodically crashed in order to ensure
that cluster really remained highly available.\footnote{
	The program that forces the periodic crashing is sometimes
	known as a ``chaos monkey'':
	\url{https://netflix.github.io/chaosmonkey/}.
	However, it might also be a mistake to neglect chaos caused
	by systems running for too long.}
Leaking the memory is also a viable strategy in environments having
garbage collectors, in which case the garbage collector can be thought
of as plugging the leak~\cite{Kung80}.
However, if your environment lacks a garbage collector, read on!

A fifth approach avoids the period crashes in favor of periodically
``stopping the world'', as exemplified by the traditional stop-the-world
garbage collector.
This approach was also heavily used during the decades before
ubiquitous connectivity, when it was common practice to power systems
off at the end of each working day.
However, in today's always-connected always-on world, stopping the world
can gravely degrade response times, which has been one motivation for the
development of concurrent garbage collectors~\cite{DavidFBacon2003RTGC}.
Furthermore, although we need all pre-existing readers to complete, we do
not need them all to complete at the same time.

This observation leads to the sixth approach, which is stopping
one CPU or thread at a time.
This approach has the advantage of not degrading reader response times
at all, let alone gravely.
Furthermore, numerous applications already have states (termed
\emph{\IXpl{quiescent state}}) that can be
reached only after all pre-existing readers are done.
In transaction-processing systems, the time between a pair of
successive transactions might be a quiescent state.
In reactive systems, the state between a pair of successive events
might be a quiescent state.
Within non-preemptive operating-systems kernels, a context switch can be
a quiescent state~\cite{McKenney98}.
Either way, once all CPUs and/or threads have passed through a quiescent
state, the system is said to have completed a \emph{grace period},
at which point all readers in existence at the start of that grace period
are guaranteed to have completed.
As a result, it is also guaranteed to be safe to free any removed data
items that were removed prior to the start of that grace period.\footnote{
	It is possible to do much more with RCU than simply defer
	reclamation of memory, but deferred reclamation is RCU's most
	common use case, and is therefore an excellent place to start.
	For an example of the more general case of deferred execution,
	please see phased state change in \cref{sec:defer:Phased State
	Change}.}

Within a non-preemptive operating-system kernel, for context switch to be
a valid quiescent state, readers must be prohibited from blocking while
referencing a given instance data structure obtained via the \co{gptr}
pointer shown in
\cref{fig:defer:Insertion With Concurrent Readers,%
fig:defer:Deletion With Concurrent Readers}.
This no-blocking constraint is consistent with similar constraints
on pure spinlocks, where a CPU is forbidden from blocking while
holding a spinlock.
Without this constraint, all CPUs might be consumed by threads
spinning attempting to acquire a spinlock held by a blocked thread.
The spinning threads will not relinquish their CPUs until they acquire
the lock, but the thread holding the lock cannot possibly release it
until one of the spinning threads relinquishes a CPU\@.
This is a classic deadlock situation, and this \IX{deadlock} is avoided
by forbidding blocking while holding a spinlock.

Again, this same constraint is imposed on reader threads dereferencing
\co{gptr}:
Such threads are not allowed to block until after
they are done using the pointed-to data item.
Returning to the second row of
\cref{fig:defer:Deletion With Concurrent Readers},
where the updater has just completed executing the \co{smp_store_release()},
imagine that CPU~0 executes a context switch.
Because readers are not permitted to block while traversing the linked
list, we are guaranteed that all prior readers that might have been running on
CPU~0 will have completed.
Extending this line of reasoning to the other CPUs, once each CPU has
been observed executing a context switch, we are guaranteed that all
prior readers have completed, and that there are no longer any reader
threads referencing the newly removed data element.
The updater can then safely free that data element, resulting in the
state shown at the bottom of
\cref{fig:defer:Deletion With Concurrent Readers}.

\begin{figure}
\centering
\resizebox{3in}{!}{\includegraphics{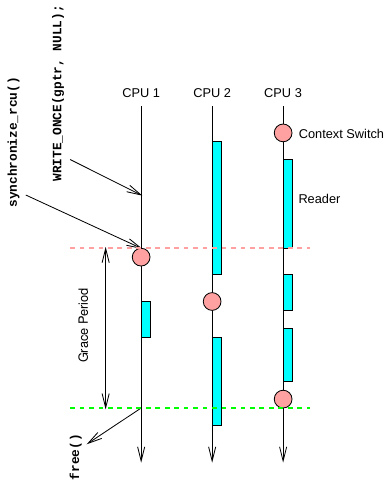}}
\caption{QSBR\@:
		 Waiting for Pre-Existing Readers}
\label{fig:defer:QSBR: Waiting for Pre-Existing Readers}
\end{figure}

This approach is termed \IXacrfst{qsbr}~\cite{ThomasEHart2006a}.
A QSBR schematic is shown in
\cref{fig:defer:QSBR: Waiting for Pre-Existing Readers},
with time advancing from the top of the figure to the bottom.
The cyan-colored boxes depict RCU read-side critical sections,
each of which begins with \co{rcu_read_lock()} and ends with
\co{rcu_read_unlock()}.
CPU~1 does the \co{WRITE_ONCE()} that removes the current data
item (presumably having previously read the pointer value and
availed itself of appropriate synchronization), then waits
for readers.
This wait operation results in an immediate context switch, which is a
quiescent state (denoted by the pink circle), which in turn means that
all prior reads on CPU~1 have completed.
Next, CPU~2 does a context switch, so that all readers on CPUs~1 and~2
are now known to have completed.
Finally, CPU~3 does a context switch.
At this point, all readers throughout the entire system are known to have
completed, so the grace period ends, permitting \co{synchronize_rcu()} to
return to its caller, in turn permitting CPU~1 to free the old data item.

\QuickQuiz{
	In \cref{fig:defer:QSBR: Waiting for Pre-Existing Readers},
	the last of CPU~3's readers that could possibly have
	access to the old data item ended before the grace period
	even started!
	So why would anyone bother waiting until CPU~3's later context
	switch???
}\QuickQuizAnswer{
	Because that waiting is exactly what enables readers to use
	the same sequence of instructions that is appropriate for
	single-theaded situations.
	In other words, this additional ``redundant'' waiting enables
	excellent read-side performance, scalability, and real-time
	response.
}\QuickQuizEnd

\subsubsection{Toy Implementation}
\label{sec:defer:Toy Implementation}

Although production-quality QSBR implementations can be quite complex,
a toy non-preemptive Linux-kernel implementation is quite simple:

\begin{VerbatimN}[samepage=true]
void synchronize_rcu(void)
{
	int cpu;

	for_each_online_cpu(cpu)
		sched_setaffinity(current->pid, cpumask_of(cpu));
}
\end{VerbatimN}

The \co{for_each_online_cpu()} primitive iterates over all CPUs, and
the \co{sched_setaffinity()} function causes the current thread to
execute on the specified CPU, which forces the destination CPU to execute
a context switch.
Therefore, once the \co{for_each_online_cpu()} has completed, each CPU
has executed a context switch, which in turn guarantees that
all pre-existing reader threads have completed.

\begin{listing}
\begin{fcvlabel}[ln:defer:Insertion and Deletion With Concurrent Readers]
\begin{VerbatimL}[commandchars=\\\[\]]
struct route *gptr;

int access_route(int (*f)(struct route *rp))
{
	int ret = -1;
	struct route *rp;

	rcu_read_lock();
	rp = rcu_dereference(gptr);
	if (rp)
		ret = f(rp);		\lnlbl[access_rp]
	rcu_read_unlock();
	return ret;
}

struct route *ins_route(struct route *rp)
{
	struct route *old_rp;

	spin_lock(&route_lock);
	old_rp = gptr;
	rcu_assign_pointer(gptr, rp);
	spin_unlock(&route_lock);
	return old_rp;
}

int del_route(void)
{
	struct route *old_rp;

	spin_lock(&route_lock);
	old_rp = gptr;
	RCU_INIT_POINTER(gptr, NULL);
	spin_unlock(&route_lock);
	synchronize_rcu();
	free(old_rp);
	return !!old_rp;
}
\end{VerbatimL}
\end{fcvlabel}
\caption{Insertion and Deletion With Concurrent Readers}
\label{lst:defer:Insertion and Deletion With Concurrent Readers}
\end{listing}

Please note that this approach is \emph{not} production quality.
Correct handling of a number of corner cases and the need for a number
of powerful optimizations mean that production-quality implementations
are quite complex.
In addition, RCU implementations for preemptible environments
require that readers actually do something, which in non-real-time
Linux-kernel environments can be as simple as defining
\co{rcu_read_lock()} and \co{rcu_read_unlock()} as \co{preempt_disable()}
and \co{preempt_enable()}, respectively.\footnote{
	Some toy RCU implementations that handle preempted
	read-side critical sections are shown in
	\cref{chp:app:``Toy'' RCU Implementations}\@.}
However, this simple non-preemptible approach is conceptually complete,
and demonstrates that it really is possible to provide read-side
synchronization at zero cost, even in the face of concurrent updates.
In fact,
\cref{lst:defer:Insertion and Deletion With Concurrent Readers}
shows how reading (\co{access_route()}),
\cref{fig:defer:Insertion With Concurrent Readers}'s
insertion (\co{ins_route()}) and
\cref{fig:defer:Deletion With Concurrent Readers}'s
deletion (\co{del_route()}) can
be implemented.
(A slightly more capable routing table is shown in
\cref{sec:defer:RCU for Pre-BSD Routing}.)

\QuickQuizSeries{%
\QuickQuizB{
	What is the point of \co{rcu_read_lock()} and \co{rcu_read_unlock()} in
	\cref{lst:defer:Insertion and Deletion With Concurrent Readers}?
	Why not just let the quiescent states speak for themselves?
}\QuickQuizAnswerB{
	Recall that readers are not permitted to pass through a quiescent
	state.
	For example, within the Linux kernel, RCU readers are not permitted
	to execute a context switch.
	Use of \co{rcu_read_lock()} and \co{rcu_read_unlock()} enables
	debug checks for improperly placed quiescent states, making it
	easy to find bugs that would otherwise be difficult to find,
	intermittent, and quite destructive.
}\QuickQuizEndB
\QuickQuizE{
	What is the point of \co{rcu_dereference()}, \co{rcu_assign_pointer()}
	and \co{RCU_INIT_POINTER()} in
	\cref{lst:defer:Insertion and Deletion With Concurrent Readers}?
	Why not just use \co{READ_ONCE()}, \co{smp_store_release()}, and
	\co{WRITE_ONCE()}, respectively?
}\QuickQuizAnswerE{
	The RCU-specific APIs do have similar semantics to the suggested
	replacements, but also enable static-analysis debugging checks
	that complain if an RCU-specific API is invoked on a non-RCU
	pointer and vice versa.
}\QuickQuizEndE
}

Referring back to
\cref{lst:defer:Insertion and Deletion With Concurrent Readers},
note that \co{route_lock} is used to synchronize between concurrent updaters
invoking \co{ins_route()} and \co{del_route()}.
However, this lock is not acquired by readers invoking \co{access_route()}:
Readers are instead protected by the QSBR techniques described in
\cref{sec:defer:Waiting for Readers}.

Note that \co{ins_route()} simply returns the old value of \co{gptr}, which
\cref{fig:defer:Insertion With Concurrent Readers} assumed would
always be \co{NULL}.
This means that it is the caller's responsibility to figure out what to
do with a non-\co{NULL} value, a task complicated by the fact that
readers might still be referencing it for an indeterminate period of time.
Callers might use one of the following approaches:

\begin{enumerate}
\item	Use \co{synchronize_rcu()} to safely free the pointed-to structure.
	Although this approach is correct from an RCU perspective, it
	arguably has software-engineering leaky-API problems.
\item	Trip an assertion if the returned pointer is non-\co{NULL}.
\item	Pass the returned pointer to a later invocation of
	\co{ins_route()} to restore the earlier value.
\end{enumerate}

In contrast, \co{del_route()} uses \co{synchronize_rcu()} and
\co{free()} to safely free the newly deleted data item.

\QuickQuiz{
	But what if the old structure needs to be freed, but the caller
	of \co{ins_route()} cannot block, perhaps due to performance
	considerations or perhaps because the caller is executing within
	an RCU read-side critical section?
}\QuickQuizAnswer{
	A \co{call_rcu()} function, which is described in
	\cref{sec:defer:Wait For Pre-Existing RCU Readers},
	permits asynchronous grace-period waits.
}\QuickQuizEnd

This example shows one general approach to reading and updating
RCU-protected data structures, however, there is quite a variety
of use cases, several of which are covered in
\cref{sec:defer:RCU Usage}.

In summary, it is in fact possible to create concurrent linked data
structures that can be traversed by readers executing the same sequence
of machine instructions that would be executed by single-threaded readers.
The next section summarizes RCU's high-level properties.

\subsubsection{RCU Properties}
\label{sec:defer:RCU Properties}

A key RCU property is that reads need not wait for updates.
This property enables RCU implementations to provide low-cost or even
no-cost readers, resulting in low overhead and excellent scalability.
This property also allows RCU readers and updaters to make useful
concurrent forward progress.
In contrast, conventional synchronization primitives must enforce strict
mutual exclusion using expensive instructions, thus increasing overhead
and degrading scalability, but also typically prohibiting readers and
updaters from making useful concurrent forward progress.

\QuickQuiz{
	Doesn't \cref{sec:defer:Sequence Locks}'s seqlock
	also permit readers and updaters to make useful concurrent
	forward progress?
}\QuickQuizAnswer{
	Yes and no.
	Although seqlock readers can run concurrently with
	seqlock writers, whenever this happens, the \co{read_seqretry()}
	primitive will force the reader to retry.
	This means that any work done by a seqlock reader running concurrently
	with a seqlock updater will be discarded and then redone upon retry.
	So seqlock readers can \emph{run} concurrently with updaters,
	but they cannot actually get any work done in this case.

	In contrast, RCU readers can perform useful work even in presence
	of concurrent RCU updaters.

	However, both reference counters
	(\cref{sec:defer:Reference Counting})
	and hazard pointers
	(\cref{sec:defer:Hazard Pointers})
	really do permit useful concurrent forward progress for both
	updaters and readers, just at somewhat greater cost.
	Please see
	\cref{sec:defer:Which to Choose?}
	for a comparison of these different solutions to the
	deferred-reclamation problem.
}\QuickQuizEnd

As noted earlier, RCU delimits readers with \co{rcu_read_lock()} and
\co{rcu_read_unlock()}, and ensures that each reader has a coherent view
of each object (see \cref{fig:defer:Deletion With Concurrent Readers}) by
maintaining multiple versions of objects and using update-side primitives
such as \co{synchronize_rcu()} to ensure that objects are not
freed until after the completion of all readers that might be using them.
RCU uses \co{rcu_assign_pointer()} and \co{rcu_dereference()} to provide
efficient and scalable mechanisms for publishing and reading new versions
of an object, respectively.
These mechanisms distribute the work among read and
update paths in such a way as to make read paths extremely fast, using
replication and weakening optimizations in a manner similar to
\IXpl{hazard pointer}, but without the need for read-side retries.
In some cases, including \co{CONFIG_PREEMPT=n} Linux kernels,
RCU's read-side primitives have zero overhead.

But are these properties actually useful in practice?
This question is taken up by the next section.

\subsubsection{Practical Applicability}
\label{sec:defer:Practical Applicability}

\begin{figure}
\centering
\resizebox{3in}{!}{\includegraphics{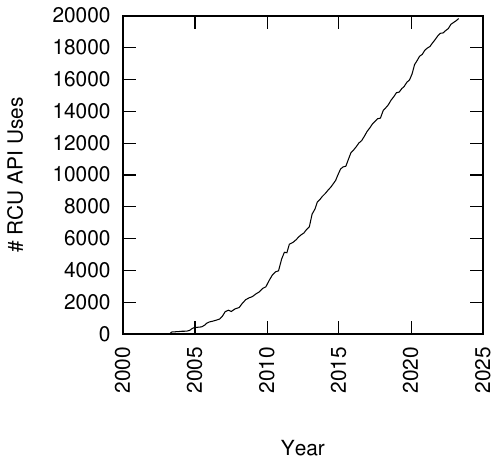}}
\caption{RCU Usage in the Linux Kernel}
\label{fig:defer:RCU Usage in the Linux Kernel}
\end{figure}

RCU has been used in the Linux kernel since
October 2002~\cite{Torvalds2.5.43}.
Use of the RCU API has increased substantially since that time,
as can be seen in
\cref{fig:defer:RCU Usage in the Linux Kernel}.
RCU has enjoyed heavy use both prior to and since its acceptance
in the Linux kernel, as discussed in
\cref{sec:defer:RCU Related Work}.
In short, RCU enjoys wide practical applicability.

The minimal example discussed in this section is a good introduction to RCU\@.
However, effective use of RCU often requires that you think differently
about your problem.
It is therefore useful to examine RCU's fundamentals, a task taken up
by the following section.

% defer/rcufundamental.tex
% mainfile: ../perfbook.tex
% SPDX-License-Identifier: CC-BY-SA-3.0

\subsection{RCU Fundamentals}
\label{sec:defer:RCU Fundamentals}
\OriginallyPublished{Section}{sec:defer:RCU Fundamentals}{RCU Fundamentals}{Linux Weekly News}{PaulEMcKenney2007WhatIsRCUFundamentally}

This section re-examines the ground covered in the previous section, but
independent of any particular example or use case.
People who prefer to live their lives very close to the actual code may
wish to skip the underlying fundamentals presented in this section.

RCU is made up of three fundamental mechanisms, the first being
used for insertion, the second being used for deletion, and the third
being used to allow readers to tolerate concurrent insertions and deletions.
\Cref{sec:defer:Publish-Subscribe Mechanism}
describes the publish-subscribe mechanism used for insertion,
\cref{sec:defer:Wait For Pre-Existing RCU Readers}
describes how waiting for pre-existing RCU readers enables deletion,
and
\cref{sec:defer:Maintain Multiple Versions of Recently Updated Objects}
discusses how maintaining multiple versions of recently updated objects
permits concurrent insertions and deletions.
Finally,
\cref{sec:defer:Summary of RCU Fundamentals}
summarizes RCU fundamentals.

\subsubsection{Publish-Subscribe Mechanism}
\label{sec:defer:Publish-Subscribe Mechanism}

Because RCU readers are not excluded by RCU updaters, an RCU-protected
data structure might change while a reader accesses it.
The accessed data item might be moved, removed, or replaced.
Because the data structure does not ``hold still'' for the reader,
each reader's access can be thought of as subscribing to the current
version of the RCU-protected data item.
For their part, updaters can be thought of as publishing new versions.

% @@@ Merge usage section into "Which to Choose?" and suggest choices.

\begin{figure}
\centering
\resizebox{3in}{!}{\includegraphics{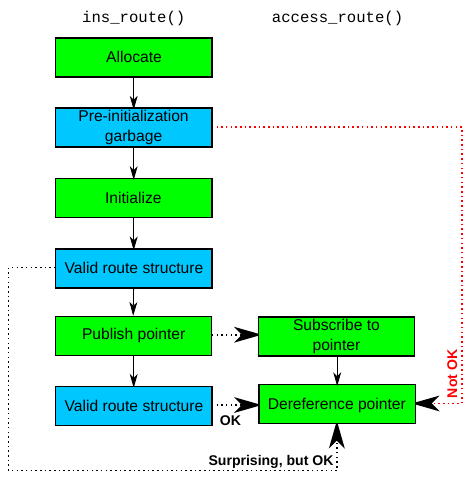}}
\caption{Publication/Subscription Constraints}
\label{fig:defer:Publication/Subscription Constraints}
\end{figure}

Unfortunately, as laid out in
\cref{sec:toolsoftrade:Shared-Variable Shenanigans}
and reiterated in
\cref{sec:defer:Minimal Insertion and Deletion},
it is unwise to use \IXplx{plain access}{es} for these publication and subscription
operations.
It is instead necessary to inform both the compiler and the CPU
of the need for care, as can be seen from
\cref{fig:defer:Publication/Subscription Constraints},
which illustrates interactions between concurrent executions of
\co{ins_route()} (and its caller) and \co{access_route()} from
\cref{lst:defer:Insertion and Deletion With Concurrent Readers}.

The \co{ins_route()} column from
\cref{fig:defer:Publication/Subscription Constraints}
shows \co{ins_route()}'s caller allocating a new \co{route} structure,
which then contains pre-initialization garbage.
The caller then initializes the newly allocated structure, and then
invokes \co{ins_route()} to publish a pointer to the new \co{route}
structure.
Publication does not affect the contents of the structure, which
therefore remain valid after publication.

The \co{access_route()} column from this same figure shows
the pointer being subscribed to and dereferenced.
This dereference operation absolutely must see a valid \co{route}
structure rather than pre-initialization garbage because referencing
garbage could result in memory corruption, crashes, and hangs.
As noted earlier, avoiding such garbage means that the publish and
subscribe operations must inform both the compiler and the CPU of the
need to maintain the needed ordering.

Publication is carried out by \co{rcu_assign_pointer()}, which ensures that
\co{ins_route()}'s caller's initialization is ordered before the actual
publication operation's store of the pointer.
In addition, \co{rcu_assign_pointer()} must be atomic in the sense that 
concurrent readers see either the old value of the pointer or the new
value of the pointer, but not some mash-up of these two values.
These requirements are met by the C11 store-release operation, and in
fact in the Linux kernel, \co{rcu_assign_pointer()} is defined in terms
of \co{smp_store_release()}, which is similar to C11 store-release.

Note that if concurrent updates are required, some sort of synchronization
mechanism will be required to mediate among multiple concurrent
\co{rcu_assign_pointer()} calls on the same pointer.
In the Linux kernel, locking is the mechanism of choice, but pretty
much any synchronization mechanism may be used.
An example of a particularly lightweight synchronization mechanism is
\cref{chp:Data Ownership}'s data ownership:
If each pointer is owned by a particular thread, then that thread may
execute \co{rcu_assign_pointer()} on that pointer with no additional
synchronization overhead.

\QuickQuiz{
	Wouldn't use of data ownership for RCU updaters mean that
	the updates could use exactly the same sequence of instructions
	as would the corresponding single-threaded code?
}\QuickQuizAnswer{
	Sometimes, for example, on TSO systems such as x86 or the IBM
	mainframe where a store-release operation emits a single store
	instruction.
	However, weakly ordered systems must also emit a memory barrier
	or perhaps a store-release instruction.
	In addition, removing data requires quite a bit of additional
	work because it is necessary to wait for pre-existing readers
	before freeing the removed data.
}\QuickQuizEnd

Subscription is carried out by \co{rcu_dereference()}, which orders
the subscription operation's load from the pointer is before the
dereference.
Similar to \co{rcu_assign_pointer()}, \co{rcu_dereference()} must be
atomic in the sense that the value loaded must be that from a single
store, for example, the compiler must not tear the load.\footnote{
	That is, the compiler must not break the load into multiple
	smaller loads, as described under ``load tearing'' in
	\cref{sec:toolsoftrade:Shared-Variable Shenanigans}.}
Unfortunately, compiler support for \co{rcu_dereference()} is at best
a work in progress~\cite{PaulEMcKennneyConsumeP0190R4,PaulEMcKenney2017markconsumeP0462R1,JFBastien2018P0750R1consume}.
In the meantime, the Linux kernel relies on volatile loads, the details of
the various CPU architectures, coding
restrictions~\cite{PaulEMcKenney2014rcu-dereference},
and, on DEC Alpha~\cite{ALPHA2002}, a memory-barrier instruction.
However, on other architectures, \co{rcu_dereference()} typically
emits a single load instruction, just as would the equivalent single-threaded
code.
The coding restrictions are described in more detail in
\cref{sec:memorder:Address- and Data-Dependency Difficulties},
however, the common case of field selection (\qtco{->}) works quite well.
Software that does not require the ultimate in read-side performance
can instead use C11 \IXpl{acquire load}, which provide the needed ordering and
more, albeit at a cost.
It is hoped that lighter-weight compiler support for \co{rcu_dereference()}
will appear in due course.

In short, use of \co{rcu_assign_pointer()} for publishing pointers and
use of \co{rcu_dereference()} for subscribing to them successfully avoids the
``Not OK'' garbage loads depicted in
\cref{fig:defer:Publication/Subscription Constraints}.
These two primitives can therefore be used to add new data to linked
structures without disrupting concurrent readers.

\QuickQuiz{
	But suppose that updaters are adding and removing multiple data
	items from a linked list while a reader is iterating over that
	same list.
	Specifically, suppose that a list initially contains elements
	A, B, and~C, and that an updater removes element A and then
	adds a new element D at the end of the list.
	The reader might well see \{A, B, C, D\}, when that sequence of
	elements never actually ever existed!
	In what alternate universe would that qualify as ``not disrupting
	concurrent readers''???
}\QuickQuizAnswer{
	In the universe where an iterating reader is only required to
	traverse elements that were present throughout the full duration
	of the iteration.
	In the example, that would be elements B and~C\@.
	Because elements A and~D were each present for only part of the
	iteration, the reader is permitted to iterate over them, but not
	obliged to.
	Note that this supports the common case where the reader is simply
	looking up a single item, and does not know or care about the
	presence or absence of other items.

	If stronger consistency is required, then higher-cost
	synchronization mechanisms are required, for example, sequence
	locking or reader-writer locking.
	But if stronger consistency is \emph{not} required (and it very often
	is not), then why pay the higher cost?
}\QuickQuizEnd

Adding data to a linked structure without disrupting readers is a good thing,
as are the cases where this can be done with no added read-side cost compared
to single-threaded readers.
However, in most cases it is also necessary to remove data, and this is the
subject of the next section.

\subsubsection{Wait For Pre-Existing RCU Readers}
\label{sec:defer:Wait For Pre-Existing RCU Readers}

In its most basic form, RCU is a way of waiting for things to finish.
Of course, there are a great many other ways of waiting for things to
finish, including reference counts, reader-writer locks, events, and so on.
The great advantage of RCU is that it can wait for each of
(say) 20,000 different things without having to explicitly
track each and every one of them, and without having to worry about
the performance degradation, scalability limitations, complex deadlock
scenarios, and memory-leak hazards that are inherent in schemes
using explicit tracking.

In RCU's case, each of the things waited on is called an
\emph{\IXBhmr{RCU read-side}{critical section}}.
As noted in
\cref{tab:defer:Core RCU API}, an RCU read-side critical
section starts with an \co{rcu_read_lock()} primitive, and ends with a
corresponding \co{rcu_read_unlock()} primitive.
RCU read-side critical sections can be nested, and may contain pretty
much any code, as long as that code does not contain a quiescent state.
For example, within the Linux kernel, it is illegal to sleep within
an RCU read-side critical section because a context switch is a quiescent
state.\footnote{
	However, a special form of RCU called SRCU~\cite{PaulEMcKenney2006c}
	does permit general sleeping in SRCU read-side critical sections.}
If you abide by these conventions, you can use RCU to wait for \emph{any}
pre-existing RCU read-side critical section to complete, and
\co{synchronize_rcu()} uses indirect means to do the actual
waiting~\cite{MathieuDesnoyers2012URCU,McKenney:2013:SDS:2483852.2483867}.

\begin{figure}
\centering
\resizebox{3in}{!}{\includegraphics{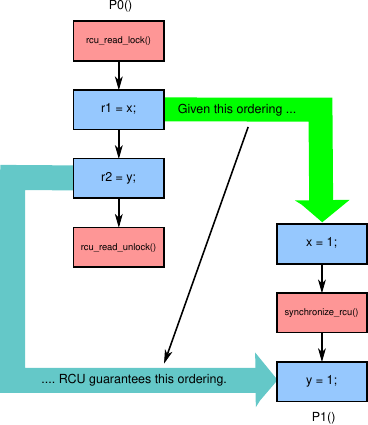}}
\caption{RCU Reader and Later Grace Period}
\label{fig:defer:RCU Reader and Later Grace Period}
\end{figure}

The relationship between an RCU read-side critical section and a later
RCU grace period is an if-then relationship, as illustrated by
\cref{fig:defer:RCU Reader and Later Grace Period}.
If any portion of a given critical section precedes the beginning of
a given grace period, then RCU guarantees that all of that critical
section will precede the end of that grace period.
In the figure, \co{P0()}'s access to \co{x} precedes \co{P1()}'s access
to this same variable, and thus also precedes the grace period generated
by \co{P1()}'s call to \co{synchronize_rcu()}.
It is therefore guaranteed that \co{P0()}'s access to \co{y} will precede
\co{P1()}'s access.
In this case, if \co{r1}'s final value is 0, then \co{r2}'s final value
is guaranteed to also be 0.

\QuickQuiz{
	What other final values of \co{r1} and \co{r2} are possible in
	\cref{fig:defer:RCU Reader and Later Grace Period}?
}\QuickQuizAnswer{
	The \co{r1 == 0 && r2 == 0} possibility was called out in the text.
	Given that \co{r1 == 0} implies \co{r2 == 0}, we know that
	\co{r1 == 0 && r2 == 1} is forbidden.
	The following discussion will show that both
	\co{r1 == 1 && r2 == 1} and \co{r1 == 1 && r2 == 0} are possible.
}\QuickQuizEnd

\begin{figure}
\centering
\resizebox{3in}{!}{\includegraphics{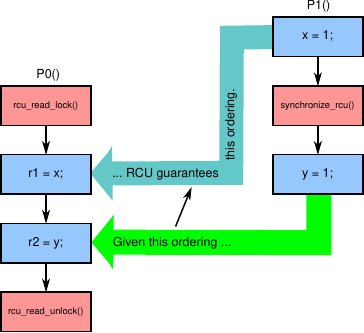}}
\caption{RCU Reader and Earlier Grace Period}
\label{fig:defer:RCU Reader and Earlier Grace Period}
\end{figure}

The relationship between an RCU read-side critical section and an earlier
RCU grace period is also an if-then relationship, as illustrated by
\cref{fig:defer:RCU Reader and Earlier Grace Period}.
If any portion of a given critical section follows the end of
a given grace period, then RCU guarantees that all of that critical
section will follow the beginning of that grace period.
In the figure, \co{P0()}'s access to \co{y} follows \co{P1()}'s access
to this same variable, and thus follows the grace period generated by
\co{P1()}'s call to \co{synchronize_rcu()}.
It is therefore guaranteed that \co{P0()}'s access to \co{x} will follow
\co{P1()}'s access.
In this case, if \co{r2}'s final value is 1, then \co{r1}'s final value
is guaranteed to also be 1.

\QuickQuiz{
	What would happen if the order of \co{P0()}'s two accesses was
	reversed in
	\cref{fig:defer:RCU Reader and Earlier Grace Period}?
}\QuickQuizAnswer{
	Absolutely nothing would change.
	The fact that \co{P0()}'s loads from \co{x} and \co{y} are
	in the same RCU read-side critical section suffices;
	their order is irrelevant.
}\QuickQuizEnd

\begin{figure}
\centering
\resizebox{3in}{!}{\includegraphics{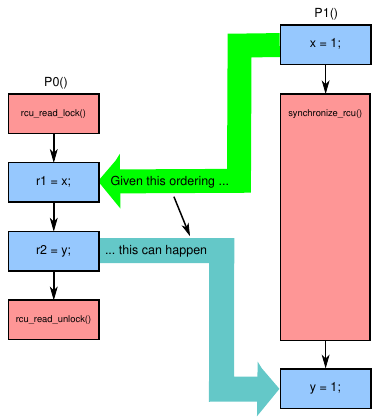}}
\caption{RCU Reader Within Grace Period}
\label{fig:defer:RCU Reader Within Grace Period}
\end{figure}

Finally, as shown in
\cref{fig:defer:RCU Reader Within Grace Period},
an RCU read-side critical section can be completely overlapped by
an RCU grace period.
In this case, \co{r1}'s final value is 1 and \co{r2}'s final value is 0.

However, it cannot be the case that \co{r1}'s final value is 0 and \co{r2}'s
final value is 1.
This would mean that an RCU read-side critical section had completely
overlapped a grace period, which is forbidden (or at the very least
constitutes a bug in RCU)\@.
RCU's wait-for-readers guarantee therefore has two parts:
\begin{enumerate*}[(1)]
\item If any part of a given RCU read-side critical section precedes
the beginning of a given grace period, then the entirety of that
critical section precedes the end of that grace period.
\item If any part of a given RCU read-side critical section follows
the end of a given grace period, then the entirety of that
critical section follows the beginning of that grace period.
\end{enumerate*}
This definition is sufficient for almost all RCU-based algorithms, but
for those wanting more,
simple executable formal models of RCU are available
as part of Linux kernel v4.17 and later, as discussed in
\cref{sec:formal:Axiomatic Approaches and RCU}.
In addition, RCU's ordering properties are examined in much
greater detail in \cref{sec:memorder:RCU}.

\QuickQuiz{
	What would happen if \co{P0()}'s accesses in
	\crefrange{fig:defer:RCU Reader and Later Grace Period}{fig:defer:RCU Reader Within Grace Period}
	were stores?
}\QuickQuizAnswer{
	The exact same ordering rules would apply, that is,
	(1)~If any part of \co{P0()}'s RCU read-side critical section
	preceded the beginning of \co{P1()}'s grace period, all of
	\co{P0()}'s RCU read-side critical section would precede the
	end of \co{P1()}'s grace period, and
	(2)~If any part of \co{P0()}'s RCU read-side critical section
	followed the end of \co{P1()}'s grace period, all of \co{P0()}'s
	RCU read-side critical section would follow the beginning of
	\co{P1()}'s grace period.

	It might seem strange to have RCU read-side critical sections
	containing writes, but this capability is not only permitted,
	but also highly useful.
	For example, the Linux kernel frequently carries out an
	RCU-protected traversal of a linked data structure and then
	acquires a reference to the destination data element.
	Because this data element must not be freed in the meantime,
	that element's reference counter must necessarily be incremented
	within the traversal's RCU read-side critical section.
	However, that increment entails a write to memory.
	Therefore, it is a very good thing that memory writes are
	permitted within RCU read-side critical sections.

	If having writes in RCU read-side critical sections still seems
	strange, please review
	\cref{sec:count:Applying Exact Limit Counters},
	which presented a use case for writes in reader-writer locking
	read-side critical sections.
}\QuickQuizEnd

Although RCU's wait-for-readers capability really is sometimes used to
order the assignment of values to variables as shown in
\crefrange{fig:defer:RCU Reader and Later Grace Period}
{fig:defer:RCU Reader Within Grace Period},
it is more frequently used to safely free data elements removed from
a linked structure, as was done in
\cref{sec:defer:Introduction to RCU}.
The general process is illustrated by the following pseudocode:

\begin{enumerate}
\item	Make a change, for example, remove an element from a linked list.
\item	Wait for all pre-existing RCU read-side critical sections to
	completely finish (for example, by using
	\co{synchronize_rcu()}).
\item	Clean up, for example, free the element that was replaced above.
\end{enumerate}

\begin{figure}
\centering
\IfEbookSize{
\resizebox{2in}{!}{\includegraphics{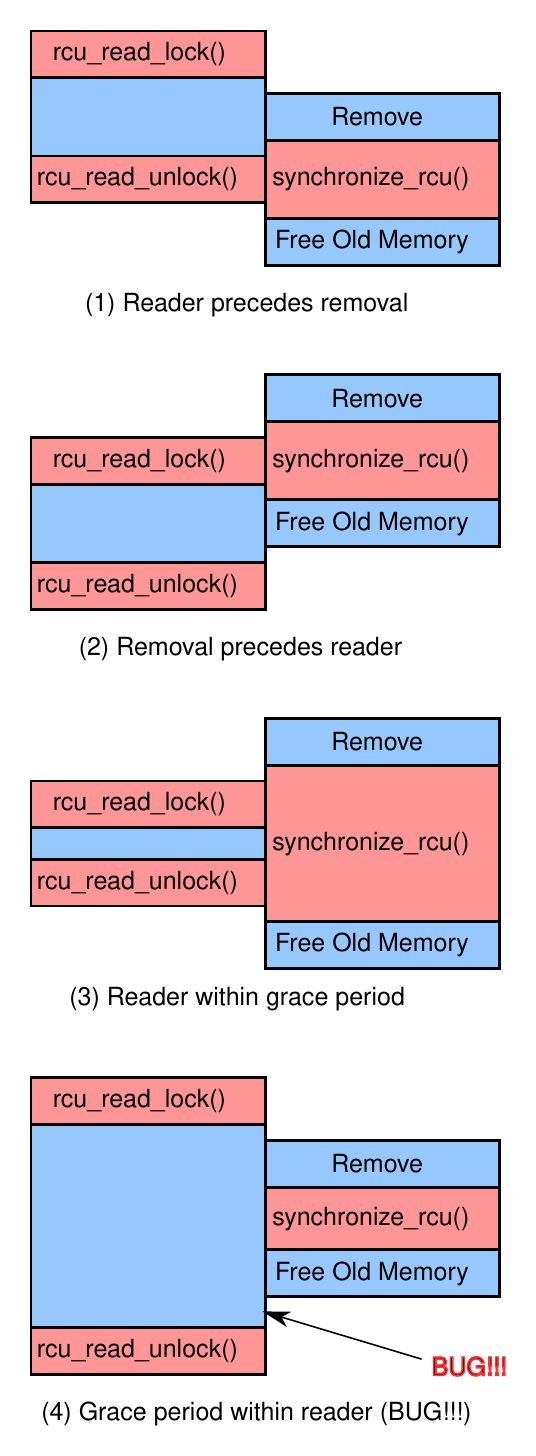}}
}{
\resizebox{2.5in}{!}{\includegraphics{defer/RCUGPorderingSummary}}
}
\caption{Summary of RCU Grace-Period Ordering Guarantees}
\label{fig:defer:Summary of RCU Grace-Period Ordering Guarantees}
\end{figure}

This more abstract procedure requires a more abstract diagram than
\crefrange{fig:defer:RCU Reader and Later Grace Period}
{fig:defer:RCU Reader Within Grace Period},
which are specific to a particular litmus test.
After all, an RCU implementation must work correctly regardless of
the form of the RCU updates and the RCU read-side critical sections.
\Cref{fig:defer:Summary of RCU Grace-Period Ordering Guarantees}
fills this need, showing the four possible scenarios, with time
advancing from top to bottom within each scenario.
Within each scenario, an RCU reader is represented by the left-hand
stack of boxes and RCU updater by the right-hand stack.

In the first scenario, the reader starts execution before the
updater starts the removal, so it is possible that this reader
has a reference to the removed data element.
Therefore, the updater must not free this element until after the
reader completes.
In the second scenario, the reader does not start execution until
after the removal has completed.
The reader cannot possibly obtain a reference to the already-removed
data element, so this element may be freed before the reader completes.
The third scenario is like the second, but illustrates that even when the
reader cannot possibly obtain a reference to an element, it is still
permissible to defer the freeing of that element until after the
reader completes.
In the fourth and final scenario, the reader starts execution before
the updater starts removing the data element, but this element
is (incorrectly) freed before the reader completed.
A correct RCU implementation will not allow this fourth scenario to
occur.
This diagram thus illustrates RCU's wait-for-readers functionality:
Given a grace period, each reader ends before the end of that grace
period, starts after the beginning of that grace period, or both, in
which case it is wholly contained within that grace period.

Because RCU readers can make forward progress while updates
are in progress, different readers might disagree about the state
of the data structure, a topic taken up by the next section.

\subsubsection{Maintain Multiple Versions of Recently Updated Objects}
\label{sec:defer:Maintain Multiple Versions of Recently Updated Objects}

This section discusses how RCU accommodates synchronization-free readers
by maintaining multiple versions of data.
Because these synchronization-free readers provide very weak temporal
synchronization, RCU users compensate via spatial synchronization.
Spatial synchronization was discussed in
\cref{chp:Partitioning and Synchronization Design}, and is heavily used
in practice to obtain good performance and scalability.
In this section, spatial synchronization will be used to attain a weak
(but useful) form of correctness as well as excellent performance and
scalability.

\Cref{fig:defer:Deletion With Concurrent Readers}
in
\cref{sec:defer:Minimal Insertion and Deletion}
showed a simple variant of spatial synchronization, in which different
readers running concurrently with \co{del_route()}
(see \cref{lst:defer:Insertion and Deletion With Concurrent Readers})
might see the old \co{route} structure or an empty list, but either
way get a valid result.
Of course, a closer look at
\cref{fig:defer:Insertion With Concurrent Readers}
shows that calls to \co{ins_route()} can also result in concurrent
readers seeing different versions:
Either the initial empty list or the newly inserted \co{route} structure.
Note that both reference counting
(\cref{sec:defer:Reference Counting})
and hazard pointers
(\cref{sec:defer:Hazard Pointers})
can also cause concurrent readers to see different versions, but
RCU's lightweight readers make this more likely.

\begin{figure}
\centering
\resizebox{3in}{!}{\includegraphics{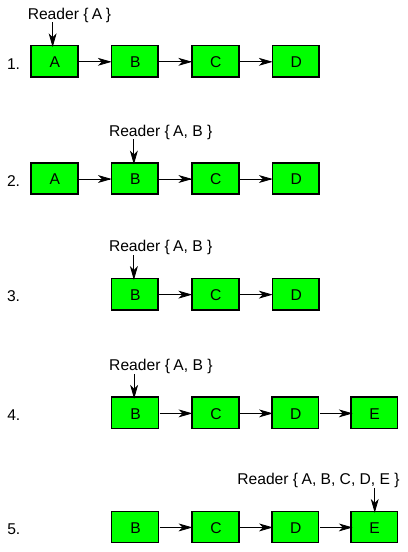}}
\caption{Multiple RCU Data-Structure Versions}
\label{fig:defer:Multiple RCU Data-Structure Versions}
\end{figure}

However, maintaining multiple weakly consistent versions can provide
some surprises.
For example, consider
\cref{fig:defer:Multiple RCU Data-Structure Versions},
in which a reader is traversing a linked list that is concurrently
updated.\footnote{
	RCU linked-list APIs may be found in
	\cref{sec:defer:RCU Linux-Kernel API}.}
In the first row of the figure, the reader is referencing data item~A,
and in the second row, it advances to~B, having thus far seen A followed by~B\@.
In the third row, an updater removes element~A and in the fourth row
an updater adds element~E to the end of the list.
In the fifth and final row, the reader completes its traversal, having
seeing elements~A through~E\@.

Except that there was no time at which such a list existed.
This situation might be even more surprising than that shown in
\cref{fig:defer:Deletion With Concurrent Readers},
in which different concurrent readers see different versions.
In contrast, in
\cref{fig:defer:Multiple RCU Data-Structure Versions}
the reader sees a version that never actually existed!

One way to resolve this strange situation is via weaker semanitics.
A reader traversal must encounter any data item that was present
during the full traversal (B, C, and~D), and might or might not
encounter data items that were present for only part of the
traversal (A and~E)\@.
Therefore, in this particular case, it is perfectly legitimate for
the reader traversal to encounter all five elements.
If this outcome is problematic, another way to resolve this situation is
through use of stronger synchronization mechanisms, such as reader-writer
locking, or clever use of timestamps and versioning,
as discussed in \cref{sec:defer:Quasi Multi-Version Concurrency Control}.
Of course, stronger mechanisms will be more expensive, but then again
the engineering life is all about choices and tradeoffs.

Strange though this situation might seem, it is entirely consistent with
the real world.
As we saw in
\cref{sec:cpu:Overheads},
the finite speed of light cannot be ignored within a computer system,
and it most certainly cannot be ignored outside of this system.
This in turn means that any data within the system representing state
in the real world outside of the system is always and forever outdated,
and thus inconsistent with the real world.
Therefore, it is quite possible that the sequence \{A, B, C, D, E\}
occurred in the real world, but due to speed-of-light delays was
never represented in the computer system's memory.
In this case, the reader's surprising traversal would correctly reflect
reality.

As a result, algorithms operating on real-world data must account for
inconsistent data, either by tolerating inconsistencies or by taking
steps to exclude or reject them.
In many cases, these algorithms are also perfectly capable of dealing
with inconsistencies within the system.

The pre-BSD packet routing example laid out in
\cref{sec:defer:Running Example}
is a case in point.
The contents of a routing list is set by routing protocols, and these
protocols feature significant delays (seconds or even minutes) to avoid
routing instabilities.
Therefore, once a routing update reaches a given system,
it might well have been sending packets the wrong way for quite some time.
Sending a few more packets the wrong way for the few microseconds during
which the update is in flight is clearly not a problem because the same
higher-level protocol actions that deal with delayed routing updates
will also deal with internal inconsistencies.

Nor is Internet routing the only situation tolerating inconsistencies.
To repeat, any algorithm in which data within a system tracks
outside-of-system state must tolerate inconsistencies, which includes
security policies (often set by committees of humans), storage configuration,
and WiFi access points, to say nothing of removable hardware such as
microphones, headsets, cameras, mice, printers, and much else besides.
Furthermore, the large number of Linux-kernel RCU API uses shown in
\cref{fig:defer:RCU Usage in the Linux Kernel},
combined with the Linux kernel's heavy use of reference counting
and with increasing use of \IXpl{hazard pointer} in other projects, demonstrates
that tolerance for such inconsistencies is more common than one might
imagine.

One root cause of this common-case tolerance of inconsistencies is
that single-item lookups are much more common in practice than are
full-data-structure traversals.
After all, full-data-structure traversals are much more expensive than
single-item lookups, so developers are motivated to avoid such traversals.
Not only are concurrent updates less likely to affect a single-item
lookup than they are a full traversal, but it is also the case that an
isolated single-item lookup has no way of detecting such inconsistencies.
As a result, in the common case, such inconsistencies are not just
tolerable, they are in fact invisible.

In such cases, RCU readers can be considered to be fully ordered with
updaters, despite the fact that these readers might be executing the
exact same sequence of machine instructions that would be executed by
a single-threaded program, as hinted on
\cpageref{sec:defer:Mysteries RCU}.
For example, referring back to
\cref{lst:defer:Insertion and Deletion With Concurrent Readers}
on \cpageref{lst:defer:Insertion and Deletion With Concurrent Readers},
suppose that each reader thread invokes \co{access_route()} exactly
once during its lifetime, and that there is no other communication among
reader and updater threads.
\begin{fcvref}[ln:defer:Insertion and Deletion With Concurrent Readers]
Then each invocation of \co{access_route()} can be ordered after the
\co{ins_route()} invocation that produced the \co{route} structure
accessed by \clnref{access_rp} of the listing in \co{access_route()}
and ordered before any subsequent
\co{ins_route()} or \co{del_route()} invocation.
\end{fcvref}

In summary, maintaining multiple versions is exactly what enables the
extremely low overheads of RCU readers, and as noted earlier, many
algorithms are unfazed by multiple versions.
However, there are algorithms that absolutely cannot handle multiple versions.
There are techniques for adapting such algorithms to
RCU~\cite{PaulEdwardMcKenneyPhD}, for example, the use of sequence locking
described in \cref{sec:together:Correlated Data Elements}.

\paragraph{Exercises}
\label{sec:defer:Exercises}

These examples assumed that a mutex was held across the entire
update operation, which would mean that there could be at most two
versions of the list active at a given time.

\QuickQuizSeries{%
\QuickQuizB{
	How would you modify the deletion example to permit more than two
	versions of the list to be active?
}\QuickQuizAnswerB{
	One way of accomplishing this is as shown in
	\cref{lst:defer:Concurrent RCU Deletion}.

\begin{listing}
\begin{VerbatimL}
spin_lock(&mylock);
p = search(head, key);
if (p == NULL)
	spin_unlock(&mylock);
else {
	list_del_rcu(&p->list);
	spin_unlock(&mylock);
	synchronize_rcu();
	kfree(p);
}
\end{VerbatimL}
\caption{Concurrent RCU Deletion}
\label{lst:defer:Concurrent RCU Deletion}
\end{listing}

	Note that this means that multiple concurrent deletions might be
	waiting in \co{synchronize_rcu()}.
}\QuickQuizEndB
\QuickQuizM{
	How many RCU versions of a given list can be
	active at any given time?
}\QuickQuizAnswerM{
	That depends on the synchronization design.
	If a semaphore protecting the update is held across the grace period,
	then there can be at most two versions, the old and the new.

	However, suppose that only the search, the update, and the
	\co{list_replace_rcu()} were protected by a lock, so that
	the \co{synchronize_rcu()} was outside of that lock, similar
	to the code shown in
	\cref{lst:defer:Concurrent RCU Deletion}.
	Suppose further that a large number of threads undertook an
	RCU replacement at about the same time, and that readers
	are also constantly traversing the data structure.

	Then the following sequence of events could occur, starting from
	the end state of
	\cref{fig:defer:Multiple RCU Data-Structure Versions}:

	\begin{enumerate}
	\item	Thread~A traverses the list, obtaining a reference to
		Element~C.
	\item	Thread~B replaces Element~C with a new
		Element~F, then waits for its \co{synchronize_rcu()}
		call to return.
	\item	Thread~C traverses the list, obtaining a reference to
		Element~F.
	\item	Thread~D replaces Element~F with a new
		Element~G, then waits for its \co{synchronize_rcu()}
		call to return.
	\item	Thread~E traverses the list, obtaining a reference to
		Element~G.
	\item	Thread~F replaces Element~G with a new
		Element~H, then waits for its \co{synchronize_rcu()}
		call to return.
	\item	Thread~G traverses the list, obtaining a reference to
		Element~H.
	\item	And the previous two steps repeat quickly with additional
		new elements, so that all of them happen before any of
		the \co{synchronize_rcu()} calls return.
	\end{enumerate}

	Thus, there can be an arbitrary number of versions active,
	limited only by memory and by how many updates could be completed
	within a grace period.
	But please note that data structures that are updated so frequently
	are not likely to be good candidates for RCU\@.
	Nevertheless, RCU can handle high update rates when necessary.
}\QuickQuizEndM
\QuickQuizE{
	How can the per-update overhead of RCU be reduced?
}\QuickQuizAnswerE{
	The most effective way to reduce the per-update overhead
	of RCU is to increase the number of updates served by
	a given grace period.
	This works because the per-grace period overhead is nearly
	independent of the number of updates served by that
	grace period.

	One way to do this is to delay the start of a given grace
	period in the hope that more updates requiring that grace
	period appear in the meantime.
	Another way is to slow down execution of the grace period
	in the hope that more updates requiring an additional
	grace period will accumulate in the meantime.

	There are many other possible optimizations, and fanatically
	devoted readers are referred to the Linux-kernel RCU
	implementation.
}\QuickQuizEndE
}

\subsubsection{Summary of RCU Fundamentals}
\label{sec:defer:Summary of RCU Fundamentals}

This section has described the three fundamental components of RCU-based
algorithms:

\begin{enumerate}
\item	A publish-subscribe mechanism for adding new data featuring
	\co{rcu_assign_pointer()} for update-side publication and
	\co{rcu_dereference()} for read-side subscription,

\item	A way of waiting for pre-existing RCU readers to finish
	based on readers being delimited by \co{rcu_read_lock()}
	and \co{rcu_read_unlock()} on the one hand and
	updaters waiting via \co{synchronize_rcu()} or
	\co{call_rcu()} on the other
	(see \cref{sec:memorder:RCU} for a formal description),
	and

\item	A discipline of maintaining multiple versions to permit
	change without harming or unduly delaying concurrent RCU readers.
\end{enumerate}

\QuickQuiz{
	How can RCU updaters possibly delay RCU readers, given that
	neither \co{rcu_read_lock()} nor \co{rcu_read_unlock()}
	spin or block?
}\QuickQuizAnswer{
	The modifications undertaken by a given RCU updater will cause the
	corresponding CPU to invalidate cache lines containing the data,
	forcing the CPUs running concurrent RCU readers to incur expensive
	cache misses.
	(Can you design an algorithm that changes a data structure
	\emph{without}
	inflicting expensive cache misses on concurrent readers?
	On subsequent readers?)
}\QuickQuizEnd

These three RCU components allow data to be updated in the face of concurrent
readers that might be executing the same sequence of machine instructions
that would be used by a reader in a single-threaded implementation.
These RCU components can be combined in different ways to implement a
surprising variety of different types of RCU-based algorithms, a number
of which are presented in
\cref{sec:defer:RCU Usage}.
However, it is usually better to work at higher levels of abstraction.
To this end, the next section describes the Linux-kernel API, which
includes simple data structures such as lists.

% defer/rcuAPI.tex
% mainfile: ../perfbook.tex
% SPDX-License-Identifier: CC-BY-SA-3.0

\subsection{RCU Linux-Kernel API}
\label{sec:defer:RCU Linux-Kernel API}
\OriginallyPublished{Section}{sec:defer:RCU Linux-Kernel API}{RCU Linux-Kernel API}{Linux Weekly News}{PaulEMcKenney2008WhatIsRCUAPI}

This section looks at RCU from the viewpoint of its Linux-kernel API\@.\footnote{
	Userspace RCU's API is documented
	elsewhere~\cite{PaulMcKenney2013LWNURCU}.}
\Cref{sec:defer:RCU has a Family of Wait-to-Finish APIs}
presents RCU's wait-to-finish APIs,
\cref{sec:defer:RCU has Publish-Subscribe and Version-Maintenance APIs}
presents RCU's publish-subscribe and version-maintenance APIs,
\cref{sec:defer:RCU has List-Processing APIs}
presents RCU's list-processing APIs,
\cref{sec:defer:RCU Has Diagnostic APIs}
presents RCU's diagnostic APIs, and
\cref{sec:defer:Where Can RCU's APIs Be Used?}
describes in which contexts RCU's various APIs may be used.
Finally,
\cref{sec:defer:So; What is RCU Really?}
presents concluding remarks.

Readers who are not excited about kernel internals may wish to skip
ahead to \cref{sec:defer:RCU Usage}
on \cpageref{sec:defer:RCU Usage},
but preferably after reviewing the next section covering software-engineering
considerations.

\subsubsection{RCU API and Software Engineering}
\label{sec:defer:RCU API and Software Engineering}

Readers who have looked ahead to
\cref{tab:defer:RCU Wait-to-Finish APIs,%
tab:defer:RCU Publish-Subscribe and Version Maintenance APIs,%
tab:defer:RCU-Protected List APIs,%
tab:defer:RCU Diagnostic APIs}
might have noted that the full list of Linux-kernel APIs sports more
than 100 members.
This is in sharp (and perhaps dismaying) contrast to the mere six
API members shown in
\cref{tab:defer:Core RCU API}.
This situation clearly raises the question ``Why so many???''

This question is answered more thoroughly in the following sections,
but in the meantime the rest of this section summarizes the motivations.

There is a wise old saying to the effect of ``To err is human.''
This means that purpose of a significant fraction of the RCU API is to
provide diagnostics, most notably in \cref{tab:defer:RCU Diagnostic APIs},
but elsewhere as well.

Important causes of human error are the limits of the human brain,
for example, the limited capacity of short-term memory.
The toy examples shown in this book do not stress these limits.
This is out of necessity:
Many readers push their cognitive limits while learning new material,
so the examples need to be kept simple.

These examples therefore keep \co{rcu_dereference()} invocations
in the same function as the enclosing \co{rcu_read_lock()} and
\co{rcu_read_unlock()} calls.
In contrast, real-world software must frequently invoke these API members
from different functions, and even from different translation units.
The Linux kernel RCU API has therefore expanded to accommodate lockdep,
which allows \co{rcu_dereference()} and friends to complain if it is
not protected by \co{rcu_read_lock()}.
Linux-kernel RCU also checks for some double-free errors, infinite
loops in RCU read-side critical sections, and attempts to invoke
quiescent states within RCU read-side critical sections.

Another way that real-world software accommodates the limits of human
cognition is through abstraction.
The Linux-kernel API therefore includes members that operate on lists
in addition to the pointer-oriented core API of
\cref{tab:defer:Core RCU API}.
The Linux kernel itself also provides RCU-protected hash tables and
search trees.

Operating-systems kernels such as Linux operate near the bottom of the
``iron triangle'' of the software stack shown in
\cref{fig:intro:Software Layers and Performance; Productivity; and Generality},
where performance is critically important.
There are thus specialized variants of a number of RCU APIs for use on
fastpaths, for example, as discussed in
\cref{sec:defer:RCU has Publish-Subscribe and Version-Maintenance APIs},
\co{RCU_INIT_POINTER()} may be used in
place of \co{rcu_assign_pointer()} in cases where the RCU-protected pointer
is being assigned to \co{NULL} or when that pointer is not yet accessible
by readers.
Use of \co{RCU_INIT_POINTER()} allows the compiler more leeway in
selecting instructions and carrying out optimizations, thus increasing
performance.

On the other hand, when used incorrectly \co{RCU_INIT_POINTER()} can
result in silent memory corruption, so please be careful!
Yes, in some cases, the kernel can check for inappropriate use of
RCU API members from a given kernel context, but the constraints of
\co{RCU_INIT_POINTER()} use are not yet checkable.

Finally, within the Linux kernel, the aforementioned limits of human
cognition are compounded by the variety and severity of workloads running
on Linux.
As of v5.16, this has given rise to no fewer than five flavors of
RCU, each designed to provide different performance, scalability,
response-time, and energy efficiency tradeoffs to RCU readers and writers.
These RCU flavors are the subject of the next section.

\subsubsection{RCU has a Family of Wait-to-Finish APIs}
\label{sec:defer:RCU has a Family of Wait-to-Finish APIs}

The most straightforward answer to ``what is RCU'' is that RCU is
an API\@.
For example, the RCU implementation used in the Linux kernel is
summarized by
\cref{tab:defer:RCU Wait-to-Finish APIs},
which shows the wait-for-readers portions of the RCU, ``sleepable'' RCU
(SRCU), Tasks RCU, and generic APIs, respectively,
and by
\cref{tab:defer:RCU Publish-Subscribe and Version Maintenance APIs},
which shows the publish-subscribe portions of the
API~\cite{PaulEMcKenney2019RCUAPI}.\footnote{
	This citation covers v4.20 and later.
	Documetation for earlier versions of the Linux-kernel RCU API may
	be found elsewhere~\cite{PaulEMcKenney2008WhatIsRCUAPI,PaulEMcKenney2014RCUAPI}.}

\begin{sidewaystable*}[tbp]
\rowcolors{1}{}{lightgray}
\renewcommand*{\arraystretch}{1.3}
\centering
\caption{RCU Wait-to-Finish APIs}
\label{tab:defer:RCU Wait-to-Finish APIs}
\scriptsize\IfEbookSize{\hspace*{-1.8in}}{\hspace*{-.125in}}
\ebresizeverb{0.7}{
\begin{tabularx}{8.5in}{>{\raggedright\arraybackslash}p{0.94in}
    >{\raggedright\arraybackslash}X
    >{\raggedright\arraybackslash}X
    >{\raggedright\arraybackslash}p{1.1in}
    >{\raggedright\arraybackslash}p{1.35in}
    >{\raggedright\arraybackslash}p{1.45in}}
\toprule
&
    {\bf RCU}: Original &
	{\bf SRCU}: Sleeping readers &
	    {\bf Tasks RCU}: Free tracing trampolines &
		{\bf Tasks RCU Rude}: Free idle-task tracing trampolines &
		    {\bf Tasks RCU Trace}: Protect sleepable BPF programs \\
\midrule
{\bf Initialization and Cleanup} &
    &
	\tco{DEFINE_SRCU()} \tco{DEFINE_STATIC_SRCU()}
	\tco{init_srcu_struct()} \tco{cleanup_srcu_struct()} &
	    &
		&
		    \\
{\bf Read-side critical-section markers} &
    \tco{rcu_read_lock()}~! \tco{rcu_read_unlock()}~!
    \tco{rcu_read_lock_bh()} \tco{rcu_read_unlock_bh()}
    \tco{rcu_read_lock_sched()} \tco{rcu_read_unlock_sched()}
    (Plus anything disabing bottom halves, preemption, or interrupts.) &
	\tco{srcu_read_lock()} \tco{srcu_read_unlock()} &
	    Voluntary context switch &
		Voluntary context switch and preempt-enable regions of code &
		    \tco{rcu_read_lock_trace()} \tco{rcu_read_unlock_trace()} \\
{\bf Update-side primitives (synchronous) } &
    { \tco{synchronize_rcu()}
      \tco{synchronize_net()}
      \tco{synchronize_rcu_expedited()} } &
	\tco{synchronize_srcu()} \tco{synchronize_srcu_expedited()} &
	    \tco{synchronize_rcu_tasks()} &
		\tco{synchronize_rcu_tasks_rude()} &
		    \tco{synchronize_rcu_tasks_trace()} \\
{\bf Update-side primitives (asynchronous / callback) } &
    \tco{call_rcu()} ! &
	\tco{call_srcu()} &
	    \tco{call_rcu_tasks()} &
		\tco{call_rcu_tasks_rude()} &
		    \tco{call_rcu_tasks_trace()} \\
{\bf Update-side primitives (wait for callbacks) } &
    \tco{rcu_barrier()} &
	\tco{srcu_barrier()} &
	    \tco{rcu_barrier_tasks()} &
		\tco{rcu_barrier_tasks_rude()} &
		    \tco{rcu_barrier_tasks_trace()} \\
{\bf Update-side primitives (initiate / wait)} &
    \tco{get_state_synchronize_rcu()}
    \tco{cond_synchronize_rcu()} &
	&
	    &
		&
		    \\
{\bf Update-side primitives (free memory) } &
    \tco{kfree_rcu()} &
	&
	    &
		&
		    \\
{\bf Type-safe memory } &
    \tco{SLAB_TYPESAFE_BY_RCU} &
	&
	    &
		&
		    \\
{\bf Read side constraints } &
    No blocking (only preemption) &
	No \tco{synchronize_srcu()} with same \tco{srcu_struct} &
	    No voluntary context switch &
		Neither blocking nor preemption &
			No RCU tasks trace grace period \\
{\bf Read side overhead } &
    CPU-local accesses (\tco{barrier()} on \tco{PREEMPT=n}) &
	Simple instructions, memory barriers &
	    Free &
		CPU-local accesses (free on \tco{PREEMPT=n}) &
		    CPU-local accesses \\
{\bf Asynchronous update-side overhead } &
    sub-microsecond &
	sub-microsecond &
	    sub-microsecond &
		sub-microsecond &
		    sub-microsecond \\
{\bf Grace-period latency } &
    10s of milliseconds &
        Milliseconds &
	    Seconds &
		Milliseconds &
		    10s of milliseconds \\
{\bf Expedited grace-period latency } &
    10s of microseconds &
        Microseconds &
	    N/A &
		N/A &
		    N/A \\
\bottomrule
\end{tabularx}
}
\end{sidewaystable*}

If you are new to RCU, you might consider focusing on just one
of the columns in
\cref{tab:defer:RCU Wait-to-Finish APIs},
each of which summarizes one member of the Linux kernel's RCU API family.
For example, if you are primarily interested in understanding how RCU
is used in the Linux kernel, ``RCU'' would be the place to start,
as it is used most frequently.
On the other hand, if you want to understand RCU for its own sake,
``Tasks RCU'' has the simplest API\@.
You can always come back for the other columns later.

If you are already familiar with RCU, these tables can
serve as a useful reference.

\QuickQuiz{
	Why do some of the cells in
	\cref{tab:defer:RCU Wait-to-Finish APIs}
	have exclamation marks (``!'')?
}\QuickQuizAnswer{
	The API members with exclamation marks (\co{rcu_read_lock()},
	\co{rcu_read_unlock()}, and \co{call_rcu()}) were the
	only members of the Linux RCU API that Paul E. McKenney was aware
	of back in the mid-90s.
	During this timeframe, he was under the mistaken impression that
	he knew all that there is to know about RCU\@.
}\QuickQuizEnd

The ``RCU'' column corresponds to the consolidation of the
three Linux-kernel RCU
implementations~\cite{PaulEMcKenney2019RCUCVE,McKenney:2019:CRS:3319647.3325836},
in which RCU read-side critical sections start with
\apik{rcu_read_lock()}, \apik{rcu_read_lock_bh()}, or \apik{rcu_read_lock_sched()}
and end with \apik{rcu_read_unlock()}, \apik{rcu_read_unlock_bh()},
or \apik{rcu_read_unlock_sched()}, respectively.
Any region of code that disables bottom halves, interrupts, or preemption
also acts as an RCU read-side critical section.
RCU read-side critical sections may be nested.
The corresponding synchronous update-side primitives,
\apik{synchronize_rcu()} and \apik{synchronize_rcu_expedited()}, along with
their synonym \apik{synchronize_net()}, wait for any type of currently
executing RCU read-side critical sections to complete.
The length of this wait is known as a ``\IX{grace period}'', and
\apik{synchronize_rcu_expedited()} is designed to reduce \IXh{grace-period}
{latency} at the expense of increased CPU overhead and IPIs.
The asynchronous update-side primitive, \apik{call_rcu()},
invokes a specified function with a specified argument after a
subsequent grace period.
For example, \co{call_rcu(p,f);} will result in
the ``RCU callback'' \co{f(p)}
being invoked after a subsequent grace period.
There are situations,
such as when unloading a Linux-kernel module that uses \co{call_rcu()},
when it is necessary to wait for all
outstanding RCU callbacks to complete~\cite{PaulEMcKenney2007rcubarrier}.
The \apik{rcu_barrier()} primitive does this job.

\QuickQuizSeries{%
\QuickQuizB{
	How do you prevent a huge number of RCU read-side critical
	sections from indefinitely blocking a \co{synchronize_rcu()}
	invocation?
}\QuickQuizAnswerB{
	There is no need to do anything to prevent RCU read-side
	critical sections from indefinitely blocking a
	\co{synchronize_rcu()} invocation, because the
	\co{synchronize_rcu()} invocation need wait only for
	\emph{pre-existing} RCU read-side critical sections.
	So as long as each RCU read-side critical section is
	of finite duration, RCU grace periods will also remain
	finite.
}\QuickQuizEndB
\QuickQuizM{
	The \co{synchronize_rcu()} API waits for all pre-existing
	interrupt handlers to complete, right?
}\QuickQuizAnswerM{
	In v4.20 and later Linux kernels,
	yes~\cite{PaulEMcKenney2019RCUCVE,McKenney:2019:CRS:3319647.3325836}.

	But not in earlier kernels, and especially not when using
	preemptible RCU\@!
	You instead want \apik{synchronize_irq()}.
	Alternatively, you can place calls to \co{rcu_read_lock()}
	and \co{rcu_read_unlock()} in the specific interrupt handlers that
	you want \co{synchronize_rcu()} to wait for.
	But even then, be careful, as preemptible RCU will not be guaranteed
	to wait for that portion of the interrupt handler preceding the
	\co{rcu_read_lock()} or following the \co{rcu_read_unlock()}.
}\QuickQuizEndM
\QuickQuizE{
	What is the difference between \co{synchronize_rcu()} and
	\co{rcu_barrier()}?
}\QuickQuizAnswerE{
	They wait on different things.
	While \co{synchronize_rcu()} waits for pre-existing RCU read-side
	critical sections to complete, \co{rcu_barrier()} instead waits
	for callbacks from prior calls to \co{call_rcu()} to be invoked.

\begin{listing}
\begin{fcvlabel}[ln:defer:synchonize-rcu vs rcu-barrier]
\begin{VerbatimL}[commandchars=\\\@\$]
	do_something_1();			\lnlbl@ds1$
	rcu_read_lock();			\lnlbl@rrl$
	do_something_2();			\lnlbl@ds2$
	call_rcu(&p->rh, f);			\lnlbl@cr$
	do_something_3();			\lnlbl@ds3$
	rcu_read_unlock();			\lnlbl@rrul$
	do_something_4();			\lnlbl@ds4$
	// f(&p->rh) invoked			\lnlbl@cb$
	do_something_5();			\lnlbl@ds5$
\end{VerbatimL}
\end{fcvlabel}
\caption{\tco{synchronize_rcu()} vs. \tco{rcu_barrier()}}
\label{lst:defer:synchronize-rcu vs. rcu-barrier}
\end{listing}

	This distinction is illustrated by
	\cref{lst:defer:synchronize-rcu vs. rcu-barrier}, which
	shows code being executed by a given CPU\@.
	For simplicity, assume that no other CPU is executing
	\co{rcu_read_lock()}, \co{rcu_read_unlock()}, or
	\co{call_rcu()}.

\begin{table*}
\renewcommand*{\arraystretch}{1.2}
\centering
\small
\begin{fcvref}[ln:defer:synchonize-rcu vs rcu-barrier]
\begin{tabular}{lll}
\toprule
            & \multicolumn{2}{c}{Must Wait Until (in \cref{lst:defer:synchronize-rcu vs. rcu-barrier}):} \\
\cmidrule{2-3}
\multicolumn{1}{c}{Invoked at:} & \multicolumn{1}{c}{\tco{synchronize_rcu()}}
					& \multicolumn{1}{c}{\tco{rcu_barrier()}} \\
\cmidrule(r){1-1} \cmidrule{2-3}
\tco{do_something_1()} & 			 & \\
\tco{do_something_2()} & \tco{rcu_read_unlock()} (\clnref{rrul}) & \\
\tco{do_something_3()} & \tco{rcu_read_unlock()} (\clnref{rrul})
						 & \tco{f(&p->rh)} (\clnref{cb}) \\
\tco{do_something_4()} &			 & \tco{f(&p->rh)} (\clnref{cb}) \\
\tco{do_something_5()} & 			 & \\
\bottomrule
\end{tabular}
\end{fcvref}
\caption{\tco{synchronize_rcu()} vs. \tco{rcu_barrier()}}
\label{tab:defer:synchonize-rcu vs rcu-barrier}
\end{table*}

	\Cref{tab:defer:synchonize-rcu vs rcu-barrier}
	shows how long each primitive must wait if invoked
	concurrently with each of the \co{do_something_*()}
	functions, with empty cells indicating that no
	waiting is necessary.
	As you can see, \co{synchronize_rcu()} need not wait unless
	it is in an RCU read-side critical section, in which case
	it must wait for the \co{rcu_read_unlock()} that ends that
	critical section.
	In contrast, RCU read-side critical sections have no effect
	on \co{rcu_barrier()}.
	However, when \co{rcu_barrier()} executes after a
	\co{call_rcu()} invocation, it must wait until the
	corresponding RCU callback is invoked.

	All that said, there is a special case where each call to
	\co{rcu_barrier()} can be replaced by a direct call to
	\co{synchronize_rcu()}, and that is where \co{synchronize_rcu()}
	is implemented in terms of \co{call_rcu()} and where there is
	a single global list of callbacks.
	But please do not do this in portable code!!!
}\QuickQuizEndE
}

Finally, RCU may be used to provide
\IX{type-safe memory}~\cite{Cheriton96a}, as described in
\cref{sec:defer:Type-Safe Memory}.
In the context of RCU, type-safe memory guarantees that a given
data element will not change type during any RCU read-side critical section
that accesses it.
To make use of RCU-based type-safe memory, pass
\apik{SLAB_TYPESAFE_BY_RCU} to \apik{kmem_cache_create()}.

The ``SRCU'' column in
\cref{tab:defer:RCU Wait-to-Finish APIs}
displays a specialized RCU API that permits general sleeping in SRCU
read-side critical
sections~\cite{PaulEMcKenney2006c}
delimited by \apik{srcu_read_lock()} and \apik{srcu_read_unlock()}.
However, unlike RCU, SRCU's \apik{srcu_read_lock()} returns a value that
must be passed into the corresponding \apik{srcu_read_unlock()}.
This difference is due to the fact that the SRCU user allocates an
\apik{srcu_struct} for each distinct SRCU usage, so that there is no
convenient place to store a per-task reader-nesting count.
(Keep in mind that although the Linux kernel provides dynamically
allocated per-CPU storage, there is not yet dynamically allocated
per-task storage.)

A given \co{srcu_struct} structure may be defined as a global
variable with \co{DEFINE_SRCU()} if the structure must be used in
multiple translation units, or with \co{DEFINE_STATIC_SRCU()} otherwise.
For example, \co{DEFINE_SRCU(my_srcu)} would create a global variable
named \co{my_srcu} that could be used by any file in the program.
Alternatively, an \co{srcu_struct} structure may be either an on-stack
variable or a dynamically allocated region of memory.
In both of these non-global-variable cases, the memory must be initialized
using \co{init_srcu_struct()} prior to its first use and cleaned up
using \co{cleanup_srcu_struct()} after its last use (but before the
underlying storage disappears).

However they are created, these distinct \co{srcu_struct} structures
prevent SRCU read-side critical sections from blocking unrelated
\apik{synchronize_srcu()} and \apik{synchronize_srcu_expedited()}
invocations.
Of course, use of either \apik{synchronize_srcu()} or
\apik{synchronize_srcu_expedited()} within an SRCU read-side critical
section can result in self-deadlock, so should be avoided.
As with RCU, SRCU's \co{synchronize_srcu_expedited()} decreases
grace-period latency compared to \co{synchronize_srcu()}, but at
the expense of increased CPU overhead.

\QuickQuiz{
	Under what conditions can \co{synchronize_srcu()} be safely
	used within an SRCU read-side critical section?
}\QuickQuizAnswer{
	In principle, you can use either \co{synchronize_srcu()} or
	\co{synchronize_srcu_expedited()} with a given \co{srcu_struct}
	within an SRCU read-side critical section that uses some other
	\co{srcu_struct}.
	In practice, however, doing this is almost certainly a bad idea.
	In particular, the code shown in
	\cref{lst:defer:Multistage SRCU Deadlocks}
	could still result in deadlock.

\begin{listing}
\begin{VerbatimL}
idx = srcu_read_lock(&ssa);
synchronize_srcu(&ssb);
srcu_read_unlock(&ssa, idx);

/* . . . */

idx = srcu_read_lock(&ssb);
synchronize_srcu(&ssa);
srcu_read_unlock(&ssb, idx);
\end{VerbatimL}
\caption{Multistage SRCU Deadlocks}
\label{lst:defer:Multistage SRCU Deadlocks}
\end{listing}
}\QuickQuizEnd

Similar to normal RCU, self-deadlock can be avoided using the
asynchronous \apik{call_srcu()} function.
However, special care must be taken when using \co{call_srcu()} because
a single task could register SRCU callbacks very quickly.
Given that SRCU allows readers to block for arbitrary periods of
time, this could consume an arbitrarily large quantity of memory.
In contrast, given the synchronous \co{synchronize_srcu()}
interface, a given task must finish waiting for a given grace period
before it can start waiting for the next one.

Also similar to RCU, there is an \apik{srcu_barrier()} function that waits
for all prior \co{call_srcu()} callbacks to be invoked.

In other words, SRCU compensates for its extremely weak
forward-progress guarantees by permitting the developer to restrict
its scope.

The ``Tasks RCU'' column in
\cref{tab:defer:RCU Wait-to-Finish APIs} displays a specialized
RCU API that mediates freeing of the trampolines used in Linux-kernel
tracing.
These trampolines are used to transfer control from a point in the
code being traced to the code doing the actual tracing.
It is of course necessary to ensure that all code executing within
a given trampoline has finished before freeing that trampoline.

Changes to the code being traced are typically limited to a single jump
or call instruction, and thus cannot accommodate the sequence of code
required to implement \co{rcu_read_lock()} and \co{rcu_read_unlock()}.
Nor can the trampoline contain these calls to \co{rcu_read_lock()} and
\co{rcu_read_unlock()}.
To see this, consider a CPU that is just about to start executing a
given trampoline.
Because it has not yet executed the \co{rcu_read_lock()}, that
trampoline could be freed at any time, which would come as a fatal
surprise to this CPU\@.
Therefore, trampolines cannot be protected by synchronization primitives
executed in either the traced code or in the trampoline itself.
Which does raise the question of exactly how the trampoline is to be
protected.

The key to answering this question is to note that trampoline code
never contains code that either directly or indirectly does a
voluntary context switch.
This code might be preempted, but it will never directly or indirectly
invoke \apik{schedule()}.
This suggests a variant of RCU having voluntary context switches and
idle execution as its only quiescent states.
This variant is Tasks RCU\@.

Tasks RCU is unusual in having no read-side marking functions, which
is good given that its main use case has nowhere to put such markings.
Instead, calls to \co{schedule()} serve directly as quiescent states.
Updates can use \apik{synchronize_rcu_tasks()} to wait for all pre-existing
trampoline execution to complete, or they can use its asynchronous
counterpart, \apik{call_rcu_tasks()}.
There is also an \apik{rcu_barrier_tasks()} that waits for completion
of callbacks corresponding to all prior invocations of \co{call_rcu_tasks()}.
There is no \co{synchronize_rcu_tasks_expedited()} because there has
not yet been a request for it, though implementing a useful variant of
it would not be free of challenges.

\QuickQuiz{
	In a kernel built with \co{CONFIG_PREEMPT_NONE=y}, won't
	\co{synchronize_rcu()} wait for all trampolines, given
	that preemption is disabled and that trampolines never
	directly or indirectly invoke \co{schedule()}?
}\QuickQuizAnswer{
	You are quite right!

	In fact, in nonpreemptible kernels, \co{synchronize_rcu_tasks()}
	is a wrapper around \co{synchronize_rcu()}.
}\QuickQuizEnd

The ``Tasks RCU Rude'' column provides a more effective variant
of the toy implementation presented in
\cref{sec:defer:Toy Implementation}.
This variant causes each CPU to execute a context switch,
so that any voluntary context switch or any preemptible region of
code can serve as a quiescent state.
The Tasks RCU Rude variant uses the Linux-kernel workqueues facility to
force concurrent context switches, in contrast to the serial
CPU-by-CPU approach taken by the toy implementation.
The API mirrors that of Tasks RCU, including the lack of explicit
read-side markers.

Finally, the ``Tasks RCU Trace'' column provides an RCU implementation
with functionality similar to that of SRCU, except with much faster
read-side markers.\footnote{
	And thus is unusual for the Tasks RCU family for having
	explicit read-side markers!}
However, this speed is a consequence of the fact that these markers
do not execute memory-barrier instructions, which means that Tasks RCU
Trace grace periods must often send IPIs to all CPUs and must always
scan the entire task list, thus degrading real-time response and
consuming considerable CPU time.
Nevertheless, in the absence of readers, the resulting grace-period
latency is reasonably short, rivaling that of RCU\@.

\subsubsection{RCU has Publish-Subscribe and Version-Maintenance APIs}
\label{sec:defer:RCU has Publish-Subscribe and Version-Maintenance APIs}

Fortunately, the RCU publish-subscribe and version-maintenance
primitives shown in
\cref{tab:defer:RCU Publish-Subscribe and Version Maintenance APIs}
apply to all of the variants of RCU discussed above.
This commonality can allow more code to be shared, and reduces API
proliferation.
The original purpose of the RCU publish-subscribe APIs was to
bury memory barriers into these APIs, so that Linux kernel
programmers could use RCU without needing to become expert on
the memory-ordering models of each of the 20+ CPU families
that Linux supports~\cite{Spraul01}.

\begin{table*}
\renewcommand*{\arraystretch}{1.15}
\footnotesize
\centering\OneColumnHSpace{-.4in}
\ebresizewidth{
\begin{tabular}{llp{2.2in}}
\toprule
Category &
	Primitives &
		Overhead \\
\midrule
Pointer publish &
	\tco{rcu_assign_pointer()} &
		Memory barrier \\
&
	\tco{rcu_replace_pointer()} &
		Memory barrier (two of them on Alpha) \\
&
	\tco{rcu_pointer_handoff()} &
		Simple instructions \\
&
	\tco{RCU_INIT_POINTER()} &
		Simple instructions \\
&
	\tco{RCU_POINTER_INITIALIZER()} &
		Compile-time constant \\
\midrule
Pointer subscribe (traversal) &
	\tco{rcu_access_pointer()} &
		Simple instructions \\
&
	\tco{rcu_dereference()} &
		Simple instructions (memory barrier on Alpha) \\
&
	\tco{rcu_dereference_check()} &
		Simple instructions (memory barrier on Alpha) \\
&
	\tco{rcu_dereference_protected()} &
		Simple instructions \\
&
	\tco{rcu_dereference_raw()} &
		Simple instructions (memory barrier on Alpha) \\
&
	\tco{rcu_dereference_raw_notrace()} &
		Simple instructions (memory barrier on Alpha) \\
\bottomrule
\end{tabular}
}
\caption{RCU Publish-Subscribe and Version Maintenance APIs}
\label{tab:defer:RCU Publish-Subscribe and Version Maintenance APIs}
\end{table*}

These primitives operate directly on pointers, and are useful for
creating RCU-protected linked data structures, such as RCU-protected
arrays and trees.
The special case of linked lists is handled by a separate set of
APIs described in
\cref{sec:defer:RCU has List-Processing APIs}.

The first category publishes pointers to new data items.
The \apik{rcu_assign_pointer()} primitive ensures that any
prior initialization remains ordered before the assignment to the
pointer on weakly ordered machines.
The \apik{rcu_replace_pointer()} primitive updates the pointer just like
\co{rcu_assign_pointer()} does, but also returns the previous value,
just like \co{rcu_dereference_protected()} (see below) would, including
the lockdep expression.
This replacement is convenient when the updater must both publish a new
pointer and free the structure referenced by the old pointer.

\QuickQuizSeries{%
\QuickQuizB{
	Normally, any pointer subject to \co{rcu_dereference()} \emph{must}
	always be updated using one of the pointer-publish functions in
	\cref{tab:defer:RCU Publish-Subscribe and Version Maintenance APIs},
	for example, \co{rcu_assign_pointer()}.

	What is an exception to this rule?
}\QuickQuizAnswerB{
	One such exception is when a multi-element linked
	data structure is initialized as a unit while inaccessible to other
	CPUs, and then a single \co{rcu_assign_pointer()} is used
	to plant a global pointer to this data structure.
	The initialization-time pointer assignments need not use
	\co{rcu_assign_pointer()}, though any such assignments that
	happen after the structure is globally visible \emph{must} use
	\co{rcu_assign_pointer()}.

	However, unless this initialization code is on an impressively hot
	code-path, it is probably wise to use \co{rcu_assign_pointer()}
	anyway, even though it is in theory unnecessary.
	It is all too easy for a ``minor'' change to invalidate your cherished
	assumptions about the initialization happening privately.
}\QuickQuizEndB
\QuickQuizE{
	Are there any downsides to the fact that these traversal and update
	primitives can be used with any of the RCU API family members?
}\QuickQuizAnswerE{
	It can sometimes be difficult for automated
	code checkers such as ``sparse'' (or indeed for human beings) to
	work out which type of RCU read-side critical section a given
	RCU traversal primitive corresponds to.
	For example, consider the code shown in
	\cref{lst:defer:Diverse RCU Read-Side Nesting}.

\begin{listing}
\begin{VerbatimL}
rcu_read_lock();
preempt_disable();
p = rcu_dereference(global_pointer);

/* . . . */

preempt_enable();
rcu_read_unlock();
\end{VerbatimL}
\caption{Diverse RCU Read-Side Nesting}
\label{lst:defer:Diverse RCU Read-Side Nesting}
\end{listing}

	Is the \co{rcu_dereference()} primitive in a vanilla RCU critical
	section or an RCU Sched critical section?
	What would you have to do to figure this out?

	But perhaps after the consolidation of the RCU flavors in
	the v4.20 Linux kernel we no longer need to care!
}\QuickQuizEndE
}

The \apik{rcu_pointer_handoff()} primitive simply returns its sole argument,
but is useful to tooling checking for pointers being leaked from
RCU read-side critical sections.
Use of \co{rcu_pointer_handoff()} indicates to such tooling that protection
of the structure in question has been handed off from RCU to some other
mechanism, such as locking or reference counting.

The \apik{RCU_INIT_POINTER()} macro can be used to initialize RCU-protected
pointers that have not yet been exposed to readers, or alternatively,
to set RCU-protected pointers to \co{NULL}.
In these restricted cases, the memory-barrier instructions provided by
\co{rcu_assign_pointer()} are not needed.
Similarly, \apik{RCU_POINTER_INITIALIZER()} provides a \GCC-style
structure initializer to allow easy initialization of RCU-protected
pointers in structures.

The second category subscribes to pointers to data items, or,
alternatively, safely traverses RCU-protected pointers.
Again, simply loading these pointers using C-language accesses
could result in seeing pre-initialization garbage in the pointed-to data.
Similarly, loading these pointer by any means outside of an RCU
read-side critical section could result in the pointed-to object being
freed at any time.
However, if the pointer is merely to be tested and not dereferenced,
the freeing of the pointed-to object is not necessarily a problem.
In this case, \apik{rcu_access_pointer()} may be used.
Normally, however, RCU read-side protection is required, and so the
\apik{rcu_dereference()} primitive uses the Linux kernel's \co{lockdep}
facility~\cite{JonathanCorbet2006lockdep} to verify that this
\co{rcu_dereference()} invocation is under the protection of
\co{rcu_read_lock()}, \co{srcu_read_lock()}, or some other RCU read-side
marker.
In contrast, the \co{rcu_access_pointer()} primitive does not involve
\co{lockdep}, and thus will not provoke \co{lockdep} complaints when
used outside of an RCU read-side critical section.

Another situation where protection is not required is when update-side code
accesses the RCU-protected pointer while holding the update-side lock.
The \apik{rcu_dereference_protected()} API member is provided for this
situation.
Its first parameter is the RCU-protected pointer, and the second
parameter takes a lockdep expression describing which locks must be
held in order for the access to be safe.
Code invoked both from readers and updaters can use
\apik{rcu_dereference_check()}, which also takes a lockdep expression, but
which may also be invoked from read-side code not holding the locks.
In some cases, the lockdep expressions can be very complex, for example,
when using fine-grained locking, any of a very large number of locks
might be held, and it might be quite difficult to work out which applies.
In these (hopefully rare) cases, \apik{rcu_dereference_raw()} provides
protection but does not check for being invoked within a reader or with
any particular lock being held.
The \apik{rcu_dereference_raw_notrace()} API member acts similarly, but
cannot be traced, and may therefore be safely used by tracing code.

Although pretty much any linked structure can be accessed by manipulating
pointers, higher-level structures can be quite helpful.
The next section therefore looks at various sorts of RCU-protected
linked lists used by the Linux kernel.

\subsubsection{RCU has List-Processing APIs}
\label{sec:defer:RCU has List-Processing APIs}

\begin{figure}
\centering
\resizebox{3in}{!}{\includegraphics{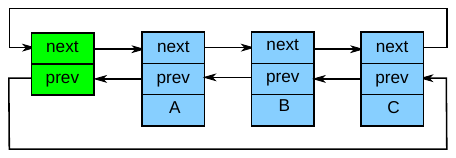}}
\caption{Linux Circular Linked List (\tco{list})}
\label{fig:defer:Linux Circular Linked List (list)}
\end{figure}

\begin{figure}
\centering
\resizebox{3in}{!}{\includegraphics{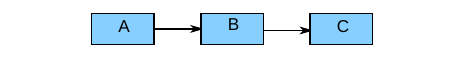}}
\caption{Linux Linked List Abbreviated}
\label{fig:defer:Linux Linked List Abbreviated}
\end{figure}

Although \co{rcu_assign_pointer()} and
\co{rcu_dereference()} can in theory be used to construct any
conceivable RCU-protected data structure, in practice it is often better
to use higher-level constructs.
Therefore, the \co{rcu_assign_pointer()} and
\co{rcu_dereference()}
primitives have been embedded in special RCU variants of Linux's
list-manipulation API\@.
Linux has four variants of doubly linked list, the circular
\co{struct list_head} and the linear
\co{struct hlist_head}/\co{struct hlist_node},
\co{struct hlist_nulls_head}/\co{struct hlist_nulls_node}, and
\co{struct hlist_bl_head}/\co{struct hlist_bl_node}
pairs.
The former is laid out as shown in
\cref{fig:defer:Linux Circular Linked List (list)},
where the green (leftmost) boxes represent the list header and the blue
(rightmost three) boxes represent the elements in the list.
This notation is cumbersome, and will therefore be abbreviated as shown in
\cref{fig:defer:Linux Linked List Abbreviated},
which shows only the non-header (blue) elements.

\begin{figure}
\centering
\resizebox{3in}{!}{\includegraphics{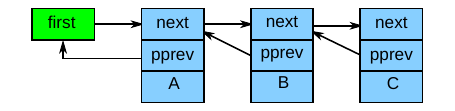}}
\caption{Linux Linear Linked List (\tco{hlist})}
\label{fig:defer:Linux Linear Linked List (hlist)}
\end{figure}

Linux's \co{hlist}\footnote{
	The ``h'' stands for hashtable, in which it reduces memory
	use by half compared to Linux's double-pointer circular
	linked list.}
is a linear list, which means that
it needs only one pointer for the header rather than the two
required for the circular list, as shown in
\cref{fig:defer:Linux Linear Linked List (hlist)}.
Thus, use of \co{hlist} can halve the memory consumption for the hash-bucket
arrays of large hash tables.
As before, this notation is cumbersome, so \co{hlist} structures will
be abbreviated in the same way \co{list_head}-style lists are, as shown in
\cref{fig:defer:Linux Linked List Abbreviated}.

A variant of Linux's \co{hlist}, named \co{hlist_nulls}, provides multiple
distinct \co{NULL} pointers, but otherwise uses the same layout as shown in
\cref{fig:defer:Linux Linear Linked List (hlist)}.
In this variant, a \co{->next} pointer having a zero low-order bit is
considered to be a pointer.
However, if the low-order bit is set to one, the upper bits identify
the type of \co{NULL} pointer.
This type of list is used to allow lockless readers to detect when a
node has been moved from one list to another.
For example, each bucket of a hash table might use its index to mark
its \co{NULL} pointer.
Should a reader encounter a \co{NULL} pointer not matching the index of
the bucket it started from, that reader knows that an element it was
traversing was moved to some other bucket during the traversal, taking
that reader with it.
The reader can use the \apik{is_a_nulls()} function (which returns \co{true}
if passed an \co{hlist_nulls} \co{NULL} pointer) to determine when
it reaches the end of a list, and the \apik{get_nulls_value()} function
(which returns its argument's \co{NULL}-pointer identifier) to fetch
the type of \co{NULL} pointer.
When \co{get_nulls_value()} returns an unexpected value, the reader
can take corrective action, for example, restarting its traversal from
the beginning.

\QuickQuiz{
	But what if an \co{hlist_nulls} reader gets moved to some other
	bucket and then back again?
}\QuickQuizAnswer{
	One way to handle this is to always move nodes to the beginning
	of the destination bucket, ensuring that when the reader reaches
	the end of the list having a matching \co{NULL} pointer, it will
	have searched the entire list.

	Of course, if there are too many move operations in a hash table
	with many elements per bucket, the reader might never reach the
	end of a list.
	One way of avoiding this in the common case is to keep hash
	tables well-tuned, thus with short lists.
	One way of detecting the problem and handling it is for the
	reader to terminate the search after traversing some large
	number of nodes, acquire the update-side lock, and redo the
	search, but this might introduce deadlocks.
	Another way of avoiding the problem entirely is for readers to
	search within RCU read-side critical sections, and to wait for
	an RCU grace period between successive updates.
	An intermediate position might wait for an RCU grace period
	every $N$ updates, for some suitable value of $N$.
}\QuickQuizEnd

More information on \co{hlist_nulls} is available in the Linux-kernel
source tree, with helpful example code provided in the
\path{rculist_nulls.rst} file (\path{rculist_nulls.txt} in older kernels).

Another variant of Linux's \co{hlist} incorporates bit-locking,
and is named \co{hlist_bl}.
This variant uses the same layout as shown in
\cref{fig:defer:Linux Linear Linked List (hlist)},
but reserves the low-order bit of the head pointer (``first'' in the
figure) to lock the list.
This approach also reduces memory usage, as it allows what would otherwise
be a separate spinlock to be stored with the pointer itself.

\begin{sidewaystable*}[htbp]
\rowcolors{1}{}{lightgray}
\renewcommand*{\arraystretch}{1.3}
\centering
\caption{RCU-Protected List APIs}
\label{tab:defer:RCU-Protected List APIs}
\footnotesize
\newlength{\cwa}\newlength{\cwb}\newlength{\cwc}\newlength{\cwd}
\IfNimbusAvail{
  \renewcommand{\ttdefault}{NimbusMonoN}
  \setlength{\cwa}{1.9in}\setlength{\cwb}{2.1in}
  \setlength{\cwc}{1.8in}\setlength{\cwd}{1.6in}
}{
  \setlength{\cwa}{1.95in}\setlength{\cwb}{2.15in}
  \setlength{\cwc}{1.9in}\setlength{\cwd}{1.7in}
}
\ebresizewidthsw{
\begin{tabular}{>{\raggedright\arraybackslash}p{\cwa}
    >{\raggedright\arraybackslash}p{\cwb}
    >{\raggedright\arraybackslash}p{\cwc}
    >{\raggedright\arraybackslash}p{\cwd}}
\toprule
\pmb{\tco{list}}: Circular doubly linked list &
    \pmb{\tco{hlist}}: Linear doubly linked list &
	\pmb{\tco{hlist_nulls}}: Linear doubly linked list with marked
	NULL pointer, with up to 31~bits of marking &
	    \pmb{\tco{hlist_bl}}: Linear doubly linked list with bit locking \\
\midrule
\multicolumn{4}{l}{{\bf Structures}} \\
\tco{struct list_head} &
    \tco{struct}{\tt ~}\tco{hlist_head} ~~~~~~~~~~~~~~
    \tco{struct}{\tt ~}\tco{hlist_node} &
	\tco{struct}{\tt ~}\tco{hlist_nulls_head}
	\tco{struct}{\tt ~}\tco{hlist_nulls_node} &
	    \tco{struct}{\tt ~}\tco{hlist_bl_head}
	    \tco{struct}{\tt ~}\tco{hlist_bl_node} \\
\multicolumn{4}{l}{{\bf Initialization}} \\
&
    \tco{INIT_LIST_HEAD_RCU()} &
	&
	    \\
\multicolumn{4}{l}{{\bf Full traversal}} \\
\tco{list_for_each_entry_rcu()}
\tco{list_for_each_entry_lockless()} &
    \tco{hlist_for_each_entry_rcu()}
    \tco{hlist_for_each_entry_rcu_bh()}
    \tco{hlist_for_each_entry_rcu_notrace()} &
	\tco{hlist_nulls_for_each_entry_rcu()}
	\tco{hlist_nulls_for_each_entry_safe()} &
	    \tco{hlist_bl_for_each_entry_rcu()} \\
\multicolumn{4}{l}{{\bf Resume traversal}} \\
\tco{list_for_each_entry_continue_rcu()}
\tco{list_for_each_entry_from_rcu()} &
    \tco{hlist_for_each_entry_continue_rcu()}
    \tco{hlist_for_each_entry_continue_rcu_bh()}
    \tco{hlist_for_each_entry_from_rcu()} &
	&
	    \\
\multicolumn{4}{l}{{\bf Stepwise traversal}} \\
\tco{list_entry_rcu()}
\tco{list_entry_lockless()}
\tco{list_first_or_null_rcu()}
\tco{list_next_rcu()}
\tco{list_next_or_null_rcu()} &
    \multicolumn{1}{p{1.2in}}{\tco{hlist_first_rcu()}
			      \tco{hlist_next_rcu()}
			      \tco{hlist_pprev_rcu()}} &
	\tco{hlist_nulls_first_rcu()}
	\tco{hlist_nulls_next_rcu()} &
	    \tco{hlist_bl_first_rcu()} \\
\multicolumn{4}{l}{{\bf Add}} \\
\multicolumn{1}{p{1.2in}}{\tco{list_add_rcu()}
			  \tco{list_add_tail_rcu()}} &
    \tco{hlist_add_before_rcu()}
    \tco{hlist_add_behind_rcu()}
    \tco{hlist_add_head_rcu()}
    \tco{hlist_add_tail_rcu()} &
	\tco{hlist_nulls_add_head_rcu()} &
	    \tco{hlist_bl_add_head_rcu()}
	    \tco{hlist_bl_set_first_rcu()} \\
\multicolumn{4}{l}{{\bf Delete}} \\
\tco{list_del_rcu()} &
    \multicolumn{1}{p{1.2in}}{\tco{hlist_del_rcu()}
			      \tco{hlist_del_init_rcu()}} &
	\tco{hlist_nulls_del_rcu()}
	\tco{hlist_nulls_del_init_rcu()} &
	    \tco{hlist_bl_del_rcu()}
	    \tco{hlist_bl_del_init_rcu()} \\
\multicolumn{4}{l}{{\bf Replace}} \\
\tco{list_replace_rcu()} &
    \tco{hlist_replace_rcu()} &
	&
	    \\
\multicolumn{4}{l}{{\bf Splice}} \\
\tco{list_splice_init_rcu()} &
    \tco{list_splice_tail_init_rcu()} &
	&
	    \\
\bottomrule
\end{tabular}
}
\end{sidewaystable*}

The API members for these linked-list variants are summarized in
\cref{tab:defer:RCU-Protected List APIs}.
More information is available in the \path{Documentation/RCU}
directory of the Linux-kernel source tree and at
Linux Weekly News~\cite{PaulEMcKenney2019RCUAPI}.

However, the remainder of this section expands on the use of
\apik{list_replace_rcu()}, given that this API member gave RCU its name.
This API member is used to carry out more complex updates in which an
element in the middle of the list having multiple fields is atomically
updated, so that a given reader sees either the old set of values or
the new set of values, but not a mixture of the two sets.
For example, each node of a linked list might have integer fields
\co{->a}, \co{->b}, and~\co{->c}, and it might be necessary to update
a given node's fields from 5, 6, and~7 to 5, 2, and~3, respectively.

The code implementing this atomic update is straightforward:

\begin{fcvlabel}[ln:defer:Canonical RCU Replacement Example (2nd)]
\begin{VerbatimN}[samepage=true,commandchars=\\\[\],firstnumber=15]
q = kmalloc(sizeof(*p), GFP_KERNEL);	\lnlbl[kmalloc]
*q = *p;				\lnlbl[copy]
q->b = 2;				\lnlbl[update1]
q->c = 3;				\lnlbl[update2]
list_replace_rcu(&p->list, &q->list);	\lnlbl[replace]
synchronize_rcu();			\lnlbl[sync_rcu]
kfree(p);				\lnlbl[kfree]
\end{VerbatimN}
\end{fcvlabel}

\begin{figure}
\centering
\IfEbookSize{
\resizebox{2in}{!}{\includegraphics{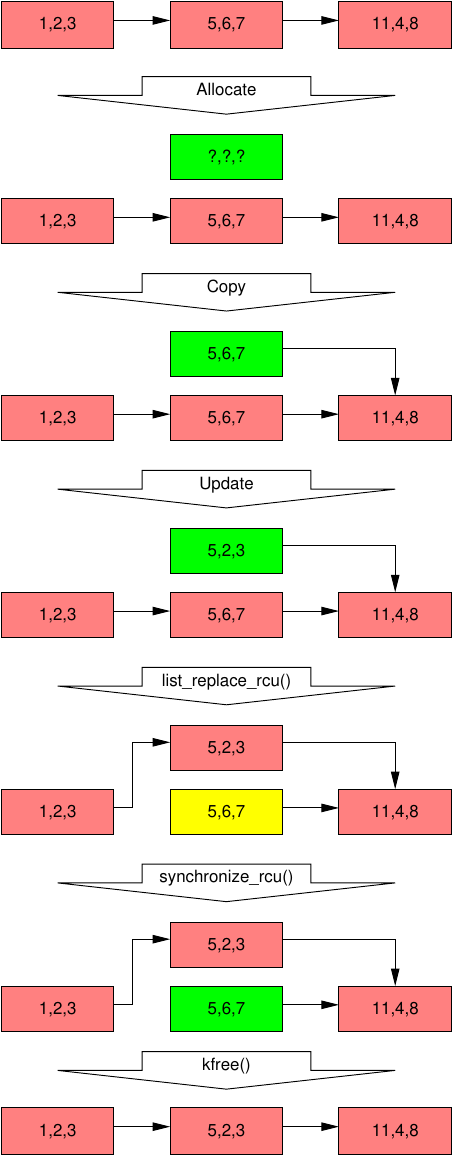}}
}{
\resizebox{2.7in}{!}{\includegraphics{defer/RCUReplacement}}
}
\caption{RCU Replacement in Linked List}
\label{fig:defer:RCU Replacement in Linked List}
\end{figure}

The following discussion walks through this code, using
\cref{fig:defer:RCU Replacement in Linked List} to illustrate
the state changes.
The triples in each element represent the values of fields \co{->a},
\co{->b}, and~\co{->c}, respectively.
The red-shaded elements might be referenced by readers,
and because readers do not synchronize directly with updaters,
readers might run concurrently with this entire replacement process.
Please note that backwards pointers and the link from the tail to the
head are omitted for clarity.

The initial state of the list, including the pointer \co{p},
is the same as for the deletion example, as shown on the
first row of the figure.

The following text describes how to replace the \co{5,6,7} element
with \co{5,2,3} in such a way that any given reader sees one of these
two values.

\begin{fcvref}[ln:defer:Canonical RCU Replacement Example (2nd)]
\Clnref{kmalloc} allocates a replacement element,
resulting in the state as shown in the second row of
\cref{fig:defer:RCU Replacement in Linked List}.
At this point, no reader can hold a reference to the newly allocated
element (as indicated by its green shading), and it is uninitialized
(as indicated by the question marks).

\Clnref{copy} copies the old element to the new one, resulting in the
state as shown in the third row of
\cref{fig:defer:RCU Replacement in Linked List}.
The newly allocated element still cannot be referenced by readers, but
it is now initialized.

\Clnref{update1} updates \co{q->b} to the value ``2'', and
\clnref{update2} updates \co{q->c} to the value ``3'',
as shown on the fourth row of
\cref{fig:defer:RCU Replacement in Linked List}.
Note that the newly allocated structure is still inaccessible to readers.

Now, \clnref{replace} does the replacement, so that the new element is
finally visible to readers, and hence is shaded red, as shown on
the fifth row of
\cref{fig:defer:RCU Replacement in Linked List}.
At this point, as shown below, we have two versions of the list.
Pre-existing readers might see the \co{5,6,7} element (which is
therefore now shaded yellow), but
new readers will instead see the \co{5,2,3} element.
But any given reader is guaranteed to see one set of values or the
other, not a mixture of the two.

After the \co{synchronize_rcu()} on \clnref{sync_rcu} returns,
a grace period will have elapsed, and so all reads that started before the
\apik{list_replace_rcu()} will have completed.
In particular, any readers that might have been holding references
to the \co{5,6,7} element are guaranteed to have exited
their RCU read-side critical sections, and are thus prohibited from
continuing to hold a reference.
Therefore, there can no longer be any readers holding references
to the old element, as indicated its green shading in the sixth row of
\cref{fig:defer:RCU Replacement in Linked List}.
As far as the readers are concerned, we are back to having a single version
of the list, but with the new element in place of the old.

After the \apik{kfree()} on \clnref{kfree} completes, the list will
appear as shown on the final row of
\cref{fig:defer:RCU Replacement in Linked List}.
\end{fcvref}

Despite the fact that RCU was named after the replacement case,
the vast majority of RCU usage within the Linux kernel relies on
the simple independent insertion and deletion, as was shown in
\cref{fig:defer:Multiple RCU Data-Structure Versions} in
\cref{sec:defer:Maintain Multiple Versions of Recently Updated Objects}.

The next section looks at APIs that assist developers in debugging
their code that makes use of RCU\@.

\subsubsection{RCU Has Diagnostic APIs}
\label{sec:defer:RCU Has Diagnostic APIs}

\begin{table}
\renewcommand*{\arraystretch}{1.15}
\footnotesize
\centering
\begin{tabular}{ll}
\toprule
Category &
	Primitives \\
\midrule
Mark RCU pointer &
	\tco{__rcu} \\
\midrule
Debug-object support &
	\tco{init_rcu_head()} \\
&	\tco{destroy_rcu_head()} \\
&	\tco{init_rcu_head_on_stack()} \\
&	\tco{destroy_rcu_head_on_stack()} \\
\midrule
Stall-warning control &
	\tco{rcu_cpu_stall_reset()} \\
\midrule
Callback checking &
	\tco{rcu_head_init()} \\
&	\tco{rcu_head_after_call_rcu()} \\
\midrule
lockdep support &
	\tco{rcu_read_lock_held()} \\
&	\tco{rcu_read_lock_bh_held()} \\
&	\tco{rcu_read_lock_sched_held()} \\
&	\tco{srcu_read_lock_held()} \\
&	\tco{rcu_is_watching()} \\
&	\tco{RCU_LOCKDEP_WARN()} \\
&	\tco{RCU_NONIDLE()} \\
&	\tco{rcu_sleep_check()} \\
\bottomrule
\end{tabular}
\caption{RCU Diagnostic APIs}
\label{tab:defer:RCU Diagnostic APIs}
\end{table}

\Cref{tab:defer:RCU Diagnostic APIs}
shows RCU's diagnostic APIs.

The \co{__rcu} tag marks an RCU-protected pointer, for example,
\qtco{struct foo __rcu *p;}.
Pointers that might be passed to \co{rcu_dereference()} can be marked,
but pointers holding values returned from \co{rcu_dereference()}
should not be.
Providing these markings on variables, structure fields, function
parameters, and return values allows the Linux kernel's \co{sparse}
tool to detect situations where RCU-protected pointers are
incorrectly accessed using plain C-language loads and stores.

Debug-object support is automatic for any \apik{rcu_head} structures
that are part of a structure obtained from the Linux kernel's
memory allocators, but those building their own special-purpose
memory allocators can use \apik{init_rcu_head()} and \apik{destroy_rcu_head()}
at allocation and free time, respectively.
Those using \co{rcu_head} structures allocated on the function-call
stack (it happens!\@) may use \apik{init_rcu_head_on_stack()}
before first use and \apik{destroy_rcu_head_on_stack()} after last use,
but before returning from the function.
Debug-object support allows detection of bugs involving passing the
same \co{rcu_head} structure to \co{call_rcu()} and friends in
quick succession, which is the \co{call_rcu()} counterpart to the
infamous double-free class of memory-allocation bugs.

Stall-warning control is provided by \apik{rcu_cpu_stall_reset()}, which
allows the caller to suppress RCU CPU stall warnings for the remainder
of the current grace period.
RCU CPU stall warnings help pinpoint situations where an RCU read-side
critical section runs for an excessive length of time, and it is useful
for things like kernel debuggers to be able to suppress them, for example,
when encountering a breakpoint.

Callback checking is provided by \apik{rcu_head_init()} and
\apik{rcu_head_after_call_rcu()}.
The former is invoked on an \co{rcu_head} structure before it is passed
to \co{call_rcu()}, and then \co{rcu_head_after_call_rcu()} will
check to see if the callback has been invoked with the specified
function.

Support for lockdep~\cite{JonathanCorbet2006lockdep} includes
\apik{rcu_read_lock_held()},
\apik{rcu_read_lock_bh_held()},
\apik{rcu_read_lock_sched_held()}, and
\apik{srcu_read_lock_held()},
each of which returns \co{true} if invoked within the corresponding
type of RCU read-side critical section.

\QuickQuiz{
	Why isn't there a \co{rcu_read_lock_tasks_held()} for Tasks RCU?
}\QuickQuizAnswer{
	Because Tasks RCU does not have read-side markers.
	Instead, Tasks RCU read-side critical sections are
	bounded by voluntary context switches.
}\QuickQuizEnd

Because \co{rcu_read_lock()} cannot be used from the idle loop,
and because energy-efficiency concerns have caused the idle loop
to become quite ornate, \apik{rcu_is_watching()} returns \co{true} if
invoked in a context where use of \co{rcu_read_lock()} is legal.
Note again that \co{srcu_read_lock()} may be used from idle and
even offline CPUs, which means that \co{rcu_is_watching()} does not
apply to SRCU\@.

\apik{RCU_LOCKDEP_WARN()} emits a warning if lockdep is enabled and if
its argument evaluates to \co{true}.
For example, \co{RCU_LOCKDEP_WARN(!rcu_read_lock_held())} would emit a
warning if invoked outside of an RCU read-side critical section.

\apik{RCU_NONIDLE()} may be used to force RCU to watch when executing
the statement that is passed in as the sole argument.
For example, \co{RCU_NONIDLE(WARN_ON(!rcu_is_watching()))}
would never emit a warning.
However, changes in the 2020--2021 timeframe extend RCU's reach deeper
into the idle loop, which should greatly reduce or even eliminate the
need for \co{RCU_NONIDLE()}.

Finally,  \apik{rcu_sleep_check()} emits a warning if invoked within
an RCU, RCU-bh, or RCU-sched read-side critical section.

\subsubsection{Where Can RCU's APIs Be Used?}
\label{sec:defer:Where Can RCU's APIs Be Used?}

\begin{figure}
\centering
\resizebox{3in}{!}{\includegraphics{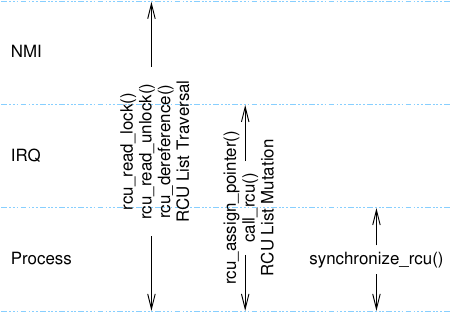}}
\caption{RCU API Usage Constraints}
\label{fig:defer:RCU API Usage Constraints}
\end{figure}

\Cref{fig:defer:RCU API Usage Constraints}
shows which APIs may be used in which in-kernel environments.
The RCU read-side primitives may be used in any environment, including NMI,
the RCU mutation and asynchronous grace-period primitives may be used in any
environment other than NMI, and, finally, the RCU synchronous grace-period
primitives may be used only in process context.
The RCU list-traversal primitives include \apik{list_for_each_entry_rcu()},
\apik{hlist_for_each_entry_rcu()}, etc.
Similarly, the RCU list-mutation primitives include
\apik{list_add_rcu()}, \apik{hlist_del_rcu()}, etc.

Note that primitives from other families of RCU may be substituted,
for example, \co{srcu_read_lock()} may be used in any context
in which \co{rcu_read_lock()} may be used.

\subsubsection{So, What \emph{is} RCU Really?}
\label{sec:defer:So; What is RCU Really?}

At its core, RCU is nothing more nor less than an API that supports
publication and subscription for insertions, waiting for all RCU readers
to complete, and maintenance of multiple versions.
That said, it is possible to build higher-level constructs
on top of RCU, including the reader-writer-locking, reference-counting,
and existence-guarantee constructs listed in
\cref{sec:defer:RCU Usage}.
Furthermore, I have no doubt that the Linux community will continue to
find interesting new uses for RCU,
just as they do for any of a number of synchronization
primitives throughout the kernel.

Of course, a more-complete view of RCU would also include
all of the things you can do with these APIs.

However, for many people, a complete view of RCU must include sample
RCU implementations.
\Cref{chp:app:``Toy'' RCU Implementations} therefore presents a series
of ``toy'' RCU implementations of increasing complexity and capability,
though others might prefer the classic
``User-Level Implementations of Read-Copy
Update''~\cite{MathieuDesnoyers2012URCU}.
For everyone else, the next section gives an overview of some RCU use cases.

% defer/rcuusage.tex
% mainfile: ../perfbook.tex
% SPDX-License-Identifier: CC-BY-SA-3.0

\subsection{RCU Usage}
\label{sec:defer:RCU Usage}
\OriginallyPublished{Section}{sec:defer:RCU Usage}{RCU Usage}{Linux Weekly News}{PaulEMcKenney2008WhatIsRCUUsage}

This section answers the question ``What is RCU?'' from the viewpoint
of the uses to which RCU can be put.
Because RCU is most frequently used to replace some existing mechanism,
we look at it primarily in terms of its relationship to such mechanisms,
as listed in \cref{tab:defer:RCU Usage}
and as displayed in \cref{fig:defer:Relationships Between RCU Use Cases}.
Following the sections listed in this table,
\cref{sec:defer:RCU Usage Summary} provides a summary.

\begin{table}
\renewcommand*{\arraystretch}{1.2}
\centering
\small
\begin{tabular}{ll}
\toprule
Mechanism RCU Replaces & Page \\
\midrule
RCU for pre-BSD routing &
	\pageref{sec:defer:RCU for Pre-BSD Routing} \\
Wait for pre-existing things to finish &
	\pageref{sec:defer:Wait for Pre-Existing Things to Finish} \\
Phased state change &
	\pageref{sec:defer:Phased State Change} \\
Add-only list (publish/subscribe) &
	\pageref{sec:defer:Add-Only List} \\
Type-safe memory &
	\pageref{sec:defer:Type-Safe Memory} \\
Existence Guarantee &
	\pageref{sec:defer:Existence Guarantee} \\
Light-weight garbage collector &
	\pageref{sec:defer:Light-Weight Garbage Collector} \\
Delete-only list &
	\pageref{sec:defer:Delete-Only List} \\
Quasi reader-writer lock &
	\pageref{sec:defer:Quasi Reader-Writer Lock} \\
Quasi reference count &
	\pageref{sec:defer:Quasi Reference Count} \\
Quasi multi-version concurrency control (MVCC) &
	\pageref{sec:defer:Quasi Multi-Version Concurrency Control} \\
\bottomrule
\end{tabular}
\caption{RCU Usage}
\label{tab:defer:RCU Usage}
\end{table}

\subsubsection{RCU for Pre-BSD Routing}
\label{sec:defer:RCU for Pre-BSD Routing}

In contrast to the later sections, this section focuses on a very
specific use case for the purpose of comparison with other mechanisms.

\Cref{lst:defer:RCU Pre-BSD Routing Table Lookup,%
lst:defer:RCU Pre-BSD Routing Table Add/Delete}
show code for an RCU-protected Pre-BSD routing table
(\path{route_rcu.c}).
The former shows data structures and \co{route_lookup()},
and the latter shows \co{route_add()} and \co{route_del()}.

\begin{listing}
\input{CodeSamples/defer/route_rcu=lookup.fcv}
\caption{RCU Pre-BSD Routing Table Lookup}
\label{lst:defer:RCU Pre-BSD Routing Table Lookup}
\end{listing}

\begin{listing}
\input{CodeSamples/defer/route_rcu=add_del.fcv}
\caption{RCU Pre-BSD Routing Table Add/Delete}
\label{lst:defer:RCU Pre-BSD Routing Table Add/Delete}
\end{listing}

\begin{fcvref}[ln:defer:route_rcu:lookup]
In \cref{lst:defer:RCU Pre-BSD Routing Table Lookup},
\clnref{rh} adds the \co{->rh} field used by RCU reclamation,
\clnref{re_freed} adds the \co{->re_freed} use-after-free-check field,
\clnref{lock,unlock1,unlock2}
add RCU read-side protection,
and \clnref{chk_freed,abort} add the use-after-free check.
\end{fcvref}
\begin{fcvref}[ln:defer:route_rcu:add_del]
In \cref{lst:defer:RCU Pre-BSD Routing Table Add/Delete},
\clnref{add:lock,add:unlock,del:lock,%
del:unlock1,del:unlock2} add update-side locking,
\clnref{add:add_rcu,del:del_rcu} add RCU update-side protection,
\clnref{del:call_rcu} causes \co{route_cb()} to be invoked after
a grace period elapses,
and \clnrefrange{cb:b}{cb:e} define \co{route_cb()}.
This is minimal added code for a working concurrent implementation.
\end{fcvref}

\begin{figure}
\centering
\resizebox{2.5in}{!}{\includegraphics{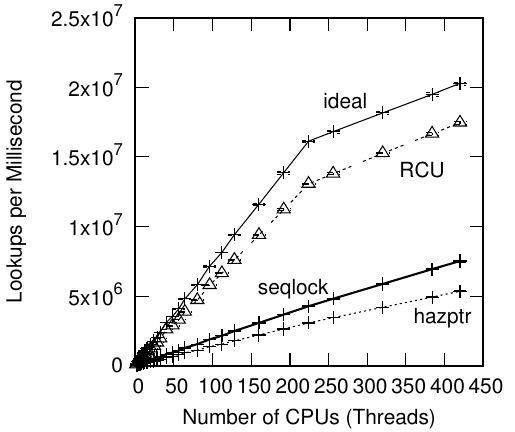}}
\caption{Pre-BSD Routing Table Protected by RCU}
\label{fig:defer:Pre-BSD Routing Table Protected by RCU}
\end{figure}

\Cref{fig:defer:Pre-BSD Routing Table Protected by RCU}
shows the performance on the read-only workload.
RCU scales quite well, and offers nearly ideal performance.
However, this data was generated using the \co{RCU_SIGNAL}
flavor of userspace
RCU~\cite{MathieuDesnoyers2009URCU,PaulMcKenney2013LWNURCU},
for which \co{rcu_read_lock()} and \co{rcu_read_unlock()}
generate a small amount of code.
What happens for the \IXacr{qsbr} flavor of RCU, which generates no code at all
for \co{rcu_read_lock()} and \co{rcu_read_unlock()}?
(See \cref{sec:defer:Introduction to RCU},
and especially
\cref{fig:defer:QSBR: Waiting for Pre-Existing Readers},
for a discussion of RCU QSBR\@.)

The answer to this is shown in
\cref{fig:defer:Pre-BSD Routing Table Protected by RCU QSBR},
which shows that RCU QSBR's performance and scalability actually exceeds
that of the ideal synchronization-free workload.

\QuickQuizSeries{%
\QuickQuizB{
	Wait, what???
	How can RCU QSBR possibly be better than ideal?
	Just what rubbish definition of ideal would fail to be the best
	of all possible results???
}\QuickQuizAnswerB{
	This is an excellent question, and the answer is that modern
	CPUs and compilers are extremely complex.
	But before getting into that, it is well worth noting that
	RCU QSBR's performance advantage appears only in the
	one-hardware-thread-per-core regime.
	Once the system is fully loaded, RCU QSBR's performance drops
	back to ideal.

	The RCU variant of the \co{route_lookup()} search loop actually
	has one more x86 instruction than does the sequential version,
	namely the \co{lea} in the sequence
	\co{cmp}, \co{je}, \co{mov}, \co{cmp}, \co{lea}, and \co{jne}.
	This extra instruction is due to the \co{rcu_head} structure
	at the beginning of the RCU variant's \co{route_entry} structure,
	so that, unlike the sequential variant, the RCU variant's
	\co{->re_next.next} pointer has a non-zero offset.
	Back in the 1980s, this additional \co{lea} instruction might
	have reliably resulted in the RCU variant being slower, but we
	are now in the 21\textsuperscript{st} century, and the 1980s
	are long gone.

	But those of you who read
	\cref{sec:cpu:Pipelined CPUs}
	carefully already knew all of this!

	These counter-intuitive results of course means that any
	performance result on modern microprocessors must be subject to
	some skepticism.
	In theory, it really does not make sense to obtain performance
	results that are better than ideal, but it really can happen
	on modern microprocessors.
	Such results can be thought of as similar to the celebrated
	super-linear speedups (see
	\cref{sec:SMPdesign:Beyond Partitioning}
	for one such example), that is, of interest but also of limited
	practical importance.
	Nevertheless, one of the strengths of RCU is that its read-side
	overhead is so low that tiny effects such as this one are visible
	in real performance measurements.

\begin{figure}
\centering
% Run with initial rcu_head structure in route_entry moved down.
\resizebox{2.5in}{!}{\includegraphics{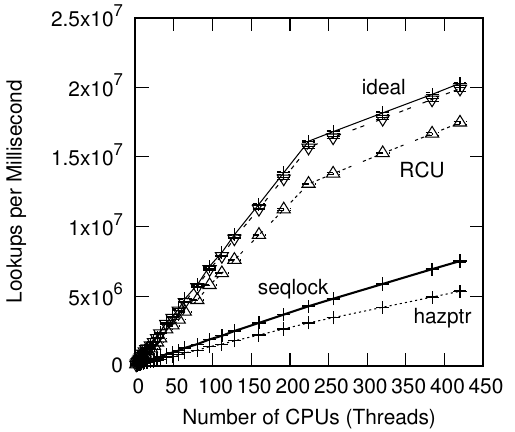}}
\caption{Pre-BSD Routing Table Protected by RCU QSBR With Non-Initial \tco{rcu_head}}
\label{fig:defer:Pre-BSD Routing Table Protected by RCU QSBR With Non-Initial rcu-head}
\end{figure}

	This raises the question as to what would happen if the
	\co{rcu_head} structure were to be moved so that RCU's
	\co{->re_next.next} pointer also had zero offset, just the
	same as the sequential variant.
	And the answer, as can be seen in
	\cref{fig:defer:Pre-BSD Routing Table Protected by RCU QSBR With Non-Initial rcu-head},
	is that this causes RCU QSBR's performance to decrease to where
	it is still very nearly ideal, but no longer super-ideal.
}\QuickQuizEndB
\QuickQuizE{
	Given RCU QSBR's read-side performance, why bother with any
	other flavor of userspace RCU?
}\QuickQuizAnswerE{
	Because RCU QSBR places constraints on the overall application
	that might not be tolerable,
	for example, requiring that each and every thread in the
	application regularly pass through a quiescent state.
	Among other things, this means that RCU QSBR is not helpful
	to library writers, who might be better served by other
	flavors of userspace RCU~\cite{PaulMcKenney2013LWNURCU}.
}\QuickQuizEndE
}

\begin{figure}
\centering
% Run with initial rcu_head structure at beginning of route_entry structure.
% This need to run twice is the reason for the oddball directory.
% @@@ Make this generate both files with a single run?
\resizebox{2.5in}{!}{\includegraphics{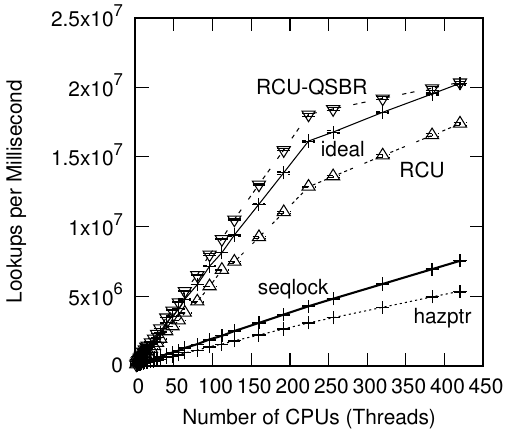}}
\caption{Pre-BSD Routing Table Protected by RCU QSBR}
\label{fig:defer:Pre-BSD Routing Table Protected by RCU QSBR}
\end{figure}

\begin{figure*}
\centering
\IfTwoColumn{
  \resizebox{5.5in}{!}{\includegraphics{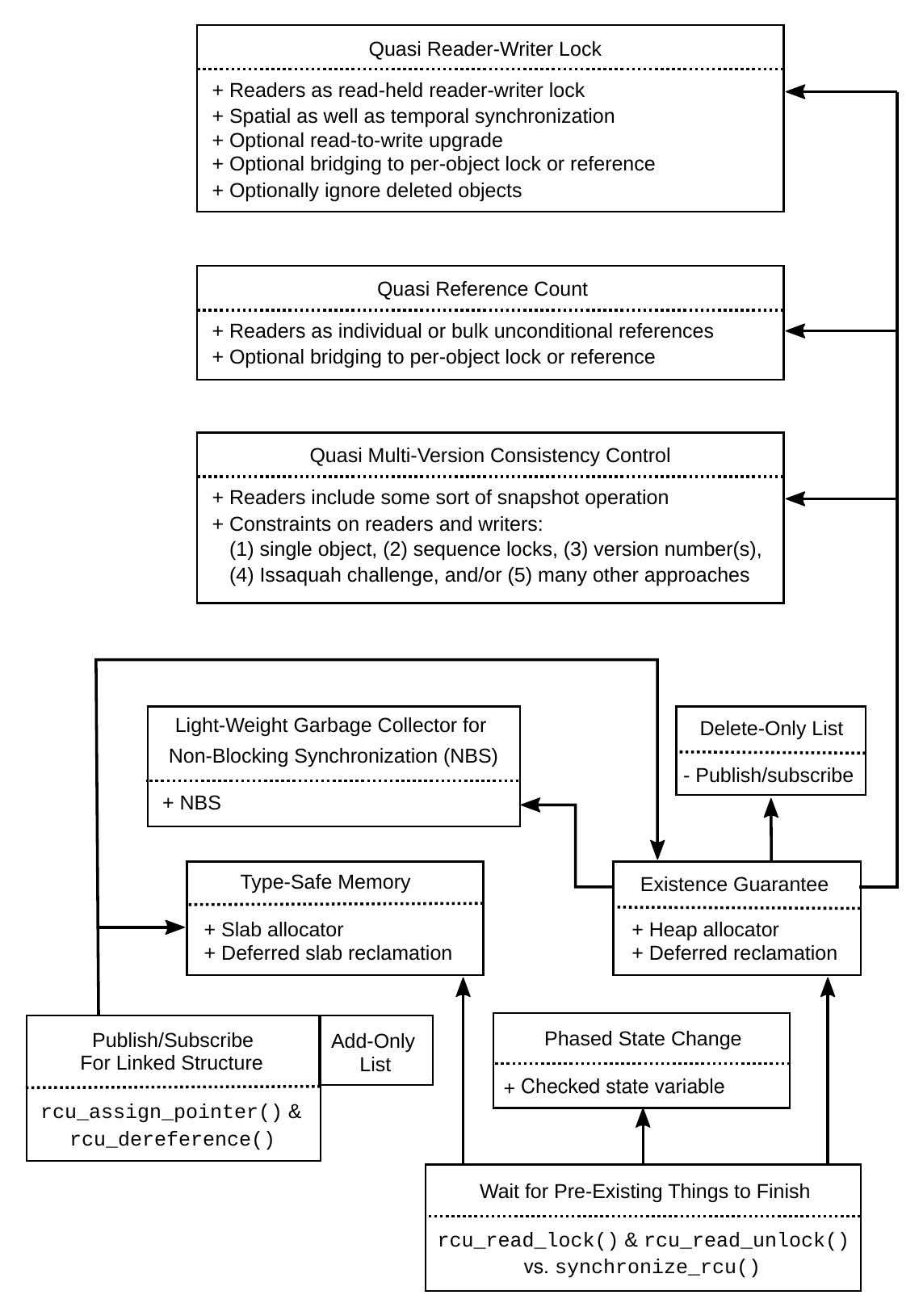}}
}{
  \resizebox{.96\textwidth}{!}{\includegraphics{defer/RCUusecases}}
  % eb builds require .96
}
\caption{Relationships Between RCU Use Cases}
\label{fig:defer:Relationships Between RCU Use Cases}
\end{figure*}

Although Pre-BSD routing is an excellent RCU use case, it is worthwhile
looking at the relationships betweeen the wider spectrum of use cases
shown in
\cref{fig:defer:Relationships Between RCU Use Cases}.
This task is taken up by the following sections.

While reading these sections, please ask yourself which of these
use cases best describes Pre-BSD routing.

\subsubsection{Wait for Pre-Existing Things to Finish}
\label{sec:defer:Wait for Pre-Existing Things to Finish}

As noted in \cref{sec:defer:RCU Fundamentals}
an important component
of RCU is a way of waiting for RCU readers to finish.
One of
RCU's great strength is that it allows you to wait for each of
thousands of different things to finish without having to explicitly
track each and every one of them, and without incurring
the performance degradation, scalability limitations, complex deadlock
scenarios, and memory-leak hazards that are inherent in schemes that
use explicit tracking.

In this section, we will show how \co{synchronize_sched()}'s
read-side counterparts (which include anything that disables preemption,
along with hardware operations and
primitives that disable interrupts) permit you to interaction with
non-maskable interrupt
(NMI) handlers, which is quite difficult using locking.
This approach has been called ``Pure RCU''~\cite{PaulEdwardMcKenneyPhD},
and it is used in a few places in the Linux kernel.

The basic form of such ``Pure RCU'' designs is as follows:

\begin{enumerate}
\item	Make a change, for example, to the way that the OS reacts to an NMI\@.
\item	Wait for all pre-existing read-side critical sections to
	completely finish (for example, by using the
	\co{synchronize_sched()} primitive).\footnote{
		In Linux kernel v5.1 and later, \co{synchronize_sched()} has
		been subsumed into \co{synchronize_rcu()}.}
	The key observation here is that subsequent RCU read-side critical
	sections are guaranteed to see whatever change was made.
\item	Clean up, for example, return status indicating that the
	change was successfully made.
\end{enumerate}

The remainder of this section presents example code adapted from
the Linux kernel.
In this example, the \co{nmi_stop()} function in the now-defunct
oprofile facility uses \co{synchronize_sched()} to ensure that all
in-flight NMI notifications have completed before freeing the associated
resources.
A simplified version of this code is shown in
\cref{lst:defer:Using RCU to Wait for NMIs to Finish}.

\begin{listing}
\begin{fcvlabel}[ln:defer:Using RCU to Wait for NMIs to Finish]
\begin{VerbatimL}[commandchars=\\\@\$]
struct profile_buffer {				\lnlbl@struct:b$
	long size;
	atomic_t entry[0];
};						\lnlbl@struct:e$
static struct profile_buffer *buf = NULL;	\lnlbl@struct:buf$

void nmi_profile(unsigned long pcvalue)		\lnlbl@nmi_profile:b$
{
	struct profile_buffer *p = rcu_dereference(buf);\lnlbl@nmi_profile:rcu_deref$

	if (p == NULL)				\lnlbl@nmi_profile:if_NULL$
		return;				\lnlbl@nmi_profile:ret:a$
	if (pcvalue >= p->size)			\lnlbl@nmi_profile:if_oor$
		return;				\lnlbl@nmi_profile:ret:b$
	atomic_inc(&p->entry[pcvalue]);		\lnlbl@nmi_profile:inc$
}						\lnlbl@nmi_profile:e$

void nmi_stop(void)				\lnlbl@nmi_stop:b$
{
	struct profile_buffer *p = buf;		\lnlbl@nmi_stop:fetch$

	if (p == NULL)				\lnlbl@nmi_stop:if_NULL$
		return;				\lnlbl@nmi_stop:ret$
	rcu_assign_pointer(buf, NULL);		\lnlbl@nmi_stop:NULL$
	synchronize_sched();			\lnlbl@nmi_stop:sync_sched$
	kfree(p);				\lnlbl@nmi_stop:kfree$
}						\lnlbl@nmi_stop:e$
\end{VerbatimL}
\end{fcvlabel}
\caption{Using RCU to Wait for NMIs to Finish}
\label{lst:defer:Using RCU to Wait for NMIs to Finish}
\end{listing}

\begin{fcvref}[ln:defer:Using RCU to Wait for NMIs to Finish:struct]
\Clnrefrange{b}{e} define a \co{profile_buffer} structure, containing a
size and an indefinite array of entries.
\Clnref{buf} defines a pointer to a profile buffer, which is
presumably initialized elsewhere to point to a dynamically allocated
region of memory.
\end{fcvref}

\begin{fcvref}[ln:defer:Using RCU to Wait for NMIs to Finish:nmi_profile]
\Clnrefrange{b}{e} define the \co{nmi_profile()} function,
which is called from within an NMI handler.
As such, it cannot be preempted, nor can it be interrupted by a normal
interrupt handler, however, it is still subject to delays due to cache misses,
ECC errors, and cycle stealing by other hardware threads within the same
core.
\Clnref{rcu_deref} gets a local pointer to the profile buffer using the
\co{rcu_dereference()} primitive to ensure memory ordering on
DEC Alpha, and
\clnref{if_NULL,ret:a} exit from this function if there is no
profile buffer currently allocated, while \clnref{if_oor,ret:b}
exit from this function if the \co{pcvalue} argument
is out of range.
Otherwise, \clnref{inc} increments the profile-buffer entry indexed
by the \co{pcvalue} argument.
Note that storing the size with the buffer guarantees that the
range check matches the buffer, even if a large buffer is suddenly
replaced by a smaller one.
\end{fcvref}

\begin{fcvref}[ln:defer:Using RCU to Wait for NMIs to Finish:nmi_stop]
\Clnrefrange{b}{e} define the \co{nmi_stop()} function,
where the caller is responsible for mutual exclusion (for example,
holding the correct lock).
\Clnref{fetch} fetches a pointer to the profile buffer, and
\clnref{if_NULL,ret} exit the function if there is no buffer.
Otherwise, \clnref{NULL} \co{NULL}s out the profile-buffer pointer
(using the \co{rcu_assign_pointer()} primitive to maintain
memory ordering on weakly ordered machines),
and \clnref{sync_sched} waits for an RCU Sched grace period to elapse,
in particular, waiting for all non-preemptible regions of code,
including NMI handlers, to complete.
Once execution continues at \clnref{kfree}, we are guaranteed that
any instance of \co{nmi_profile()} that obtained a
pointer to the old buffer has returned.
It is therefore safe to free the buffer, in this case using the
\co{kfree()} primitive.
\end{fcvref}

\QuickQuiz{
	Suppose that the \co{nmi_profile()} function was preemptible.
	What would need to change to make this example work correctly?
}\QuickQuizAnswer{
	One approach would be to use
	\co{rcu_read_lock()} and \co{rcu_read_unlock()}
	in \co{nmi_profile()}, and to replace the
	\co{synchronize_sched()} with \co{synchronize_rcu()},
	perhaps as shown in
	\cref{lst:defer:Using RCU to Wait for Mythical Preemptible NMIs to Finish}.
\begin{listing}
\begin{VerbatimL}
struct profile_buffer {
	long size;
	atomic_t entry[0];
};
static struct profile_buffer *buf = NULL;

void nmi_profile(unsigned long pcvalue)
{
	struct profile_buffer *p;

	rcu_read_lock();
	p = rcu_dereference(buf);
	if (p == NULL) {
		rcu_read_unlock();
		return;
	}
	if (pcvalue >= p->size) {
		rcu_read_unlock();
		return;
	}
	atomic_inc(&p->entry[pcvalue]);
	rcu_read_unlock();
}

void nmi_stop(void)
{
	struct profile_buffer *p = buf;

	if (p == NULL)
		return;
	rcu_assign_pointer(buf, NULL);
	synchronize_rcu();
	kfree(p);
}
\end{VerbatimL}
\caption{Using RCU to Wait for Mythical Preemptible NMIs to Finish}
\label{lst:defer:Using RCU to Wait for Mythical Preemptible NMIs to Finish}
\end{listing}

	But why on earth would an NMI handler be preemptible???
}\QuickQuizEnd

In short, RCU makes it easy to dynamically switch among profile
buffers (you just \emph{try} doing this efficiently with atomic
operations, or at all with locking!).
This is a rare use of RCU in its pure form.
RCU is normally used at higher levels of abstraction, as
will be shown in the following sections.

\subsubsection{Phased State Change}
\label{sec:defer:Phased State Change}

\begin{figure}
\centering
\resizebox{\twocolumnwidth}{!}{\includegraphics{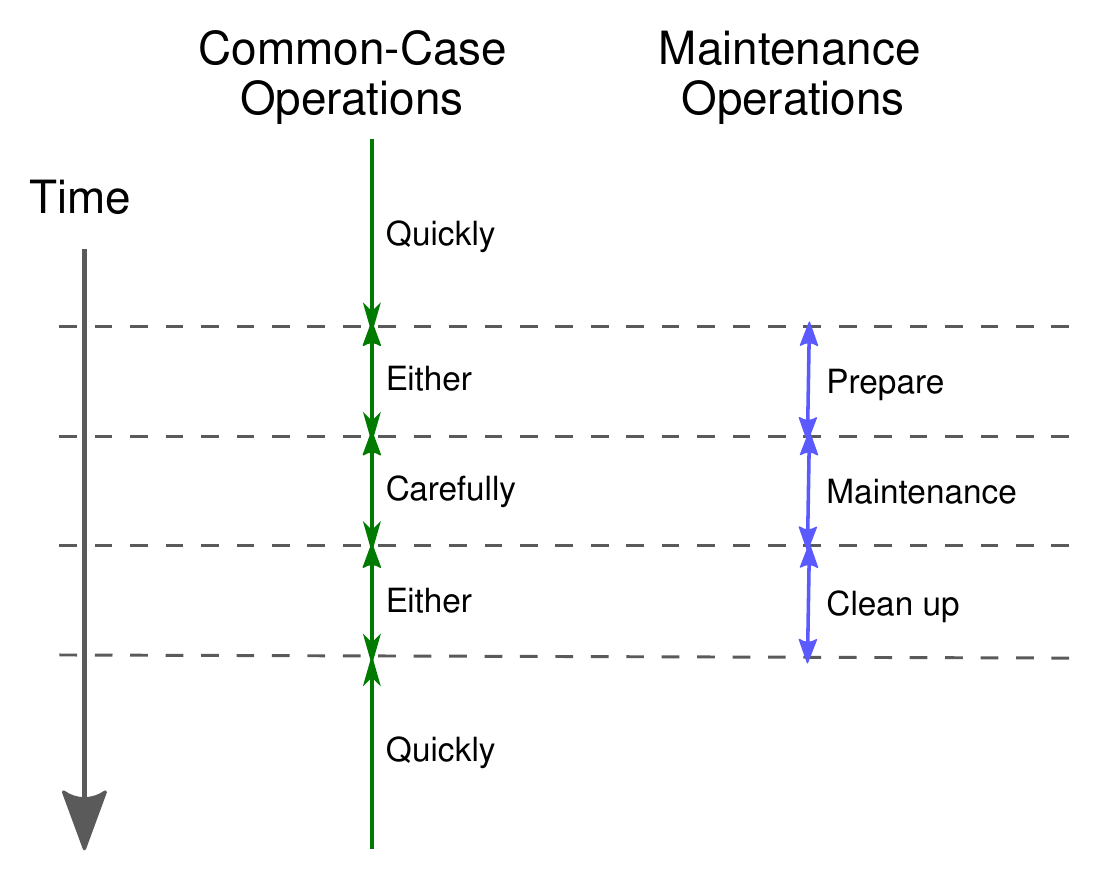}}
\caption{Phased State Change for Maintenance Operation}
\label{fig:defer:Phased State Change for Maintenance Operation}
\end{figure}

\Cref{fig:defer:Phased State Change for Maintenance Operation}
shows a timeline for an example phased state change to efficiently
handle maintenance operations.
If there is no maintenance operation in progress, common-case operations
must proceed quickly, for example, without acquiring a reader-writer lock.
However, if there is a maintenance operation in progress, the common-case
operations must be undertaken carefully, taking into account added
complexities due to their running concurrently with that maintenance
operation.
This means that common-case operations will incur higher overhead during
maintenance operations, which is one reason that maintenance operations
are normally scheduled to take place during times of low load.

In the figure, these apparently conflicting requirements are resolved by
having a prepare phase prior to the maintenance operation and a cleanup
phase after it, during which the common-case operations can proceed
either quickly or carefully.

\begin{listing}
\begin{fcvlabel}[ln:defer:Phased State Change for Maintenance Operations]
\begin{VerbatimL}[commandchars=\\\@\$]
bool be_careful;

void cco(void)
{
	rcu_read_lock();			\lnlbl@rrl$
	if (READ_ONCE(be_careful))		\lnlbl@if$
		cco_carefully();		\lnlbl@careful$
	else
		cco_quickly();			\lnlbl@quick$
	rcu_read_unlock();			\lnlbl@rul$
}

void maint(void)
{
	WRITE_ONCE(be_careful, true);		\lnlbl@tocareful$
	synchronize_rcu();			\lnlbl@sr1$
	do_maint();				\lnlbl@maint$
	synchronize_rcu();			\lnlbl@sr2$
	WRITE_ONCE(be_careful, false);		\lnlbl@toquick$
}
\end{VerbatimL}
\end{fcvlabel}
\caption{Phased State Change for Maintenance Operations}
\label{lst:defer:Phased State Change for Maintenance Operations}
\end{listing}

\begin{fcvref}[ln:defer:Phased State Change for Maintenance Operations]
Example pseudo-code for this phased state change is shown in
\cref{lst:defer:Phased State Change for Maintenance Operations}.
The common-case operations are carried out by \co{cco()} within an RCU
read-side critical section extending from \clnref{rrl} to \clnref{rul}.
Here, \clnref{if} checks a global \co{be_careful} flag, invoking
\co{cco_carefully()} or \co{cco_quickly()}, as indicated.
\end{fcvref}

\begin{fcvref}[ln:defer:Phased State Change for Maintenance Operations]
This allows the \co{maint()} function to set the \co{be_careful} flag
on \clnref{tocareful} and wait for an RCU grace period on \clnref{sr1}.
When control reaches \clnref{maint}, all \co{cco()} functions that saw a
\co{false} value of \co{be_careful} (and thus which might invoke
the \co{cco_quickly()} function) will have completed their operations,
so that all currently executing \co{cco()} functions will be invoking
\co{cco_carefully()}.
This means that it is safe for the \co{do_maint()} function to be
invoked.
\Clnref{sr2} then waits for all \co{cco()} functions that might have
run concurrently with \co{do_maint()} to complete, and finally
\clnref{toquick} sets the \co{be_careful} flag back to \co{false}.
\end{fcvref}

\QuickQuizSeries{%
\QuickQuizB{
	What is the point of the second call to \co{synchronize_rcu()}
	in function
	\co{maint()} in \cref{lst:defer:Phased State Change for Maintenance Operations}?
	Isn't it OK for any \co{cco()} invocations in the clean-up
	phase to invoke either \co{cco_carefully()} or \co{cco_quickly()}?
}\QuickQuizAnswerB{
	The problem is that there is no ordering between the \co{cco()}
	function's load from \co{be_careful} and any memory loads
	executed by the \co{cco_quickly()} function.
	Because there is no ordering, without that second call to
	\co{syncrhonize_rcu()}, memory ordering could cause loads
	in \co{cco_quickly()} to overlap with stores by \co{do_maint()}.

	Another alternative would be to compensate for the removal of
	that second call to \co{synchronize_rcu()} by changing the
	\co{READ_ONCE()} to \co{smp_load_acquire()} and the
	\co{WRITE_ONCE()} to \co{smp_store_release()}, thus restoring
	the needed ordering.
}\QuickQuizEndB

\QuickQuizE{
	How can you be sure that the code shown in
	\co{maint()} in \cref{lst:defer:Phased State Change for Maintenance Operations}
	really works?\@
}\QuickQuizAnswerE{
	By one popular school of thought, you cannot.

	But in this case, those willing to jump ahead to
	\cref{chp:Formal Verification}
	and
	\cref{chp:Advanced Synchronization: Memory Ordering}
	might find a couple of LKMM litmus tests to be interesting
	(\path{C-RCU-phased-state-change-1.litmus} and
	\path{C-RCU-phased-state-change-2.litmus}).
	These tests could be argued to demonstrate that this code
	and a variant of it really do work.
}\QuickQuizEndE
}% End of \QuickQuizSeries

Phased state change allows frequent operations to use light-weight
checks, without the need for expensive lock acquisitions or atomic
read-modify-write operations, and is used in the Linux kernel in the
guise of \co{rcu_sync}~\cite{OlegNesterov2013rcusync} to implement a
variant of reader-writer semaphores with lightweight readers.
Phased state change adds only a checked state variable
to the wait-to-finish use case
(\cref{sec:defer:Wait for Pre-Existing Things to Finish}),
thus also residing at a rather low level of abstraction.

\subsubsection{Add-Only List}
\label{sec:defer:Add-Only List}

Add-only data structures, exemplified by the add-only list, can be used
for a surprisingly common set of use cases, perhaps most commonly the
logging of changes.
Add-only data structures are a pure use of RCU's underlying
publish/subscribe mechanism.

An add-only variant of a pre-BSD routing table can be derived from
\cref{lst:defer:RCU Pre-BSD Routing Table Lookup,lst:defer:RCU Pre-BSD Routing Table Add/Delete}.
Because there is no deletion, the \co{route_del()} and \co{route_cb()}
functions may be dispensed with, along with the \co{->rh}
and \co{->re_freed} fields of the \co{route_entry} structure, the
\co{rcu_read_lock()}, the \co{rcu_read_unlock()} invocations in the
\co{route_lookup()} function, and all uses of the \co{->re_freed} field
in all remaining functions.

Of course, if there are many concurrent invocations of the \co{route_add()}
function, there will be heavy contention on \co{routelock}, and if lockless
techniques are used, heavy memory contention on \co{routelist}.
The usual way to avoid this contention is to use a concurrency-friendly
data structure such as a hash table (see \cref{chp:Data Structures}).
Alternatively, per-CPU data structures might be periodically merged
into a single global data structure.

On the other hand, if there is never any deletion, extended time periods
featuring many concurrent invocations of \co{route_add()} will eventually
consume all available memory.
Therefore, most RCU-protected data structures also implement deletion.

\subsubsection{Type-Safe Memory}
\label{sec:defer:Type-Safe Memory}

A number of lockless algorithms do not require that a given data
element keep the same identity through a given RCU read-side critical
section referencing it---but only if that data element retains the
same type.
In other words, these lockless algorithms can tolerate a given data
element being freed and reallocated as the same type of structure
while they are referencing it, but must prohibit a change in type.
This guarantee, called ``\IX{type-safe memory}'' in
academic literature~\cite{Cheriton96a},
is weaker than the \IXpl{existence guarantee} discussed
in \cref{sec:defer:Existence Guarantee},
and is therefore quite a bit harder to work with.
Type-safe memory algorithms in the Linux kernel make use of slab caches,
specially marking these caches with \co{SLAB_TYPESAFE_BY_RCU}
so that RCU is used when returning a freed-up
slab to system memory.
This use of RCU guarantees that any in-use element of
such a slab will remain in that slab, thus retaining its type,
for the duration of any pre-existing RCU read-side critical sections.

\QuickQuiz{
	But what if there is an arbitrarily long series of RCU
	read-side critical sections in multiple threads, so that at
	any point in time there is at least one thread in the system
	executing in an RCU read-side critical section?
	Wouldn't that prevent any data from a \co{SLAB_TYPESAFE_BY_RCU}
	slab ever being returned to the system, possibly resulting
	in OOM events?
}\QuickQuizAnswer{
	There could certainly be an arbitrarily long period of time
	during which at least one thread is always in an RCU read-side
	critical section.
	However, the key words in the description in
	\cref{sec:defer:Type-Safe Memory}
	are ``in-use'' and ``pre-existing''.
	Keep in mind that a given RCU read-side critical section is
	conceptually only permitted to gain references to data elements
	that were visible to readers during that critical section.
	Furthermore, remember that a slab cannot be returned to the
	system until all of its data elements have been freed, in fact,
	the RCU grace period cannot start until after they have all been
	freed.

	Therefore, the slab cache need only wait for those RCU read-side
	critical sections that started before the freeing of the last element
	of the slab.
	This in turn means that any RCU grace period that begins after
	the freeing of the last element will do---the slab may be returned
	to the system after that grace period ends.
}\QuickQuizEnd

It is important to note that \co{SLAB_TYPESAFE_BY_RCU} will
\emph{in no way}
prevent \co{kmem_cache_alloc()} from immediately reallocating
memory that was just now freed via \co{kmem_cache_free()}!
In fact, the \co{SLAB_TYPESAFE_BY_RCU}-protected data structure
just returned by \co{rcu_dereference()} might be freed and reallocated
an arbitrarily large number of times, even when under the protection
of \co{rcu_read_lock()}.
Instead, \co{SLAB_TYPESAFE_BY_RCU} operates by preventing
\co{kmem_cache_free()}
from returning a completely freed-up slab of data structures
to the system until after an RCU grace period elapses.
In short, although a given RCU read-side critical section might see a
given \co{SLAB_TYPESAFE_BY_RCU} data element being freed and reallocated
arbitrarily often, the element's type is guaranteed not to change until
that critical section has completed.

These algorithms therefore typically use a validation step that checks
to make sure that the newly referenced data structure really is the one
that was requested~\cite[Section~2.5]{LaninShasha1986TSM}.
These validation checks require that portions of the data structure
remain untouched by the free-reallocate process.
Such validation checks are usually very hard to get right, and can
hide subtle and difficult bugs.

Therefore, although type-safety-based lockless algorithms can be extremely
helpful in a very few difficult situations, you should instead use existence
guarantees where possible.
Simpler is after all almost always better!
On the other hand, type-safety-based lockless algorithms can
provide improved cache locality, and thus improved performance.
This improved cache locality is provided by the fact that such
algorithms can immediately reallocate a newly freed block of memory.
In contrast, algorithms based on existence guarantees must wait for
all pre-existing readers before reallocating memory, by which time
that memory may have been ejected from CPU caches.

As can be seen in \cref{fig:defer:Relationships Between RCU Use Cases},
RCU's type-safe-memory use case combines both the wait-to-finish
and publish-subscribe components, but in the Linux kernel also includes
the slab allocator's deferred reclamation specified by the
\co{SLAB_TYPESAFE_BY_RCU} flag.

\subsubsection{Existence Guarantee}
\label{sec:defer:Existence Guarantee}

Gamsa et al.~\cite{Gamsa99}
discuss \IXpl{existence guarantee} and describe how a mechanism
resembling RCU can be used to provide these existence guarantees
(see Section~5 on page~7 of the PDF), and
\cref{sec:locking:Lock-Based Existence Guarantees}
discusses how to guarantee existence via locking, along with the
ensuing disadvantages of doing so.
The effect is that if any RCU-protected data element is accessed
within an RCU read-side critical section, that data element is
guaranteed to remain in existence for the duration of that RCU
read-side critical section.

\begin{listing}
\begin{fcvlabel}[ln:defer:Existence Guarantees Enable Per-Element Locking]
\begin{VerbatimL}[commandchars=\\\@\$]
int delete(int key)
{
	struct element *p;
	int b;

	b = hashfunction(key);			\lnlbl@hash$
	rcu_read_lock();			\lnlbl@rdlock$
	p = rcu_dereference(hashtable[b]);
	if (p == NULL || p->key != key) {	\lnlbl@chkkey$
		rcu_read_unlock();		\lnlbl@rdunlock1$
		return 0;			\lnlbl@ret_0:a$
	}
	spin_lock(&p->lock);			\lnlbl@acq$
	if (hashtable[b] == p && p->key == key) {\lnlbl@chkkey2$
		rcu_read_unlock();		\lnlbl@rdunlock2$
		rcu_assign_pointer(hashtable[b], NULL);\lnlbl@remove$
		spin_unlock(&p->lock);		\lnlbl@rel1$
		synchronize_rcu();		\lnlbl@sync_rcu$
		kfree(p);			\lnlbl@kfree$
		return 1;			\lnlbl@ret_1$
	}
	spin_unlock(&p->lock);			\lnlbl@rel2$
	rcu_read_unlock();			\lnlbl@rdunlock3$
	return 0;				\lnlbl@ret_0:b$
}
\end{VerbatimL}
\end{fcvlabel}
\caption{Existence Guarantees Enable Per-Element Locking}
\label{lst:defer:Existence Guarantees Enable Per-Element Locking}
\end{listing}

\begin{fcvref}[ln:defer:Existence Guarantees Enable Per-Element Locking]
\Cref{lst:defer:Existence Guarantees Enable Per-Element Locking}
demonstrates how RCU-based existence guarantees can enable
per-element locking via a function that deletes an element from
a hash table.
\Clnref{hash} computes a hash function, and \clnref{rdlock} enters an RCU
read-side critical section.
If \clnref{chkkey} finds that the corresponding bucket of the hash table is
empty or that the element present is not the one we wish to delete,
then \clnref{rdunlock1} exits the RCU read-side critical section and
\clnref{ret_0:a}
indicates failure.
\end{fcvref}

\QuickQuiz{
	What if the element we need to delete is not the first element
	of the list on
	\clnrefr{ln:defer:Existence Guarantees Enable Per-Element Locking:chkkey} of
	\cref{lst:defer:Existence Guarantees Enable Per-Element Locking}?
}\QuickQuizAnswer{
	As with the (bug-ridden)
	\cref{lst:locking:Per-Element Locking Without Existence Guarantees (Buggy!)},
	this is a very simple hash table with no chaining, so the only
	element in a given bucket is the first element.
	The reader is again invited to adapt this example to a hash table with
	full chaining.
	Less energetic readers might wish to refer to
	\cref{chp:Data Structures}.
}\QuickQuizEnd

\begin{fcvref}[ln:defer:Existence Guarantees Enable Per-Element Locking]
Otherwise, \clnref{acq} acquires the update-side spinlock, and
\clnref{chkkey2} then checks that the element is still the one that we want.
If so, \clnref{rdunlock2} leaves the RCU read-side critical section,
\clnref{remove} removes it from the table, \clnref{rel1} releases
the lock, \clnref{sync_rcu} waits for all pre-existing RCU read-side critical
sections to complete, \clnref{kfree} frees the newly removed element,
and \clnref{ret_1} indicates success.
If the element is no longer the one we want, \clnref{rel2} releases
the lock, \clnref{rdunlock3} leaves the RCU read-side critical section,
and \clnref{ret_0:b} indicates failure to delete the specified key.
\end{fcvref}

\QuickQuizSeries{%
\QuickQuizB{
	\begin{fcvref}[ln:defer:Existence Guarantees Enable Per-Element Locking]
	Why is it OK to exit the RCU read-side critical section on
	\clnref{rdunlock2} of
	\cref{lst:defer:Existence Guarantees Enable Per-Element Locking}
	before releasing the lock on \clnref{rel1}?
	\end{fcvref}
}\QuickQuizAnswerB{
	\begin{fcvref}[ln:defer:Existence Guarantees Enable Per-Element Locking]
	First, please note that the second check on \clnref{chkkey2} is
	necessary because some other
	CPU might have removed this element while we were waiting
	to acquire the lock.
	However, the fact that we were in an RCU read-side critical section
	while acquiring the lock guarantees that this element could not
	possibly have been re-allocated and re-inserted into this
	hash table.
	Furthermore, once we acquire the lock, the lock itself guarantees
	the element's existence, so we no longer need to be in an
	RCU read-side critical section.

	The question as to whether it is necessary to re-check the
	element's key is left as an exercise to the reader.
	% A re-check is necessary if the key can mutate or if it is
	% necessary to reject deleted entries (in cases where deletion
	% is recorded by mutating the key.
	\end{fcvref}
}\QuickQuizEndB
\QuickQuizM{
	\begin{fcvref}[ln:defer:Existence Guarantees Enable Per-Element Locking]
	Why not exit the RCU read-side critical section on
	\clnref{rdunlock3} of
	\cref{lst:defer:Existence Guarantees Enable Per-Element Locking}
	before releasing the lock on \clnref{rel2}?
	\end{fcvref}
}\QuickQuizAnswerM{
	Suppose we reverse the order of these two lines.
	Then this code is vulnerable to the following sequence of
	events:
	\begin{enumerate}
	\begin{fcvref}[ln:defer:Existence Guarantees Enable Per-Element Locking]
	\item	CPU~0 invokes \co{delete()}, and finds the element
		to be deleted, executing through \clnref{rdunlock2}.
		It has not yet actually deleted the element, but
		is about to do so.
	\item	CPU~1 concurrently invokes \co{delete()}, attempting
		to delete this same element.
		However, CPU~0 still holds the lock, so CPU~1 waits
		for it at \clnref{acq}.
	\item	CPU~0 executes \clnref{remove,rel1},
		and blocks at \clnref{sync_rcu} waiting for CPU~1
		to exit its RCU read-side critical section.
	\item	CPU~1 now acquires the lock, but the test on \clnref{chkkey2}
		fails because CPU~0 has already removed the element.
		CPU~1 now executes \clnref{rel2}
		(which we switched with \clnref{rdunlock3}
		for the purposes of this Quick Quiz)
		and exits its RCU read-side critical section.
	\item	CPU~0 can now return from \co{synchronize_rcu()},
		and thus executes \clnref{kfree}, sending the element to
		the freelist.
	\item	CPU~1 now attempts to release a lock for an element
		that has been freed, and, worse yet, possibly
		reallocated as some other type of data structure.
		This is a fatal memory-corruption error.
	\end{fcvref}
	\end{enumerate}
}\QuickQuizEndM
\QuickQuizE{
	The RCU-based algorithm shown in
	\cref{lst:defer:Existence Guarantees Enable Per-Element Locking}
	locks very similar to that in
	\cref{lst:locking:Per-Element Locking With Lock-Based Existence Guarantees},
	so why should the RCU-based approach be any better?
}\QuickQuizAnswerE{
	\Cref{lst:defer:Existence Guarantees Enable Per-Element Locking}
	replaces the per-element \co{spin_lock()} and \co{spin_unlock()}
	shown in
	\cref{lst:locking:Per-Element Locking With Lock-Based Existence Guarantees}
	with a much cheaper \co{rcu_read_lock()} and \co{rcu_read_unlock()},
	thus greatly improving both performance and scalability.
	For more detail, please see
	\cref{sec:datastruct:RCU-Protected Hash Table Performance}.
}\QuickQuizEndE
}

Alert readers will recognize this as only a slight variation on
the original wait-to-finish theme
(\cref{sec:defer:Wait for Pre-Existing Things to Finish}),
adding publish/subscribe, linked structures, a heap allocator
(typically), and deferred reclamation, as shown in
\cref{fig:defer:Relationships Between RCU Use Cases}.
They might also note the deadlock-immunity advantages over the lock-based
existence guarantees discussed in
\cref{sec:locking:Lock-Based Existence Guarantees}.

\subsubsection{Light-Weight Garbage Collector}
\label{sec:defer:Light-Weight Garbage Collector}

A not-uncommon exclamation made by people first learning about
RCU is ``RCU is sort of like a garbage collector!''
This exclamation has a large grain of truth, but it can also be
misleading.

Perhaps the best way to think of the relationship between RCU
and automatic garbage collectors (GCs) is that RCU resembles
a GC in that the \emph{timing} of collection is automatically
determined, but that RCU differs from a GC in that:
\begin{enumerate*}[(1)]
\item The programmer must manually indicate when a given data
structure is eligible to be collected and
\item The programmer must manually mark the RCU read-side critical
sections where references might be held.
\end{enumerate*}

Despite these differences, the resemblance does go quite deep.
In fact, the first RCU-like mechanism I am aware of used a
reference-count-based garbage collector to handle the grace
periods~\cite{Kung80}, and the connection between RCU and
garbage collection has been noted more
recently~\cite{HarshalSheth2016goRCU}.

The light-weight garbage collector use case is very similar to
the existence-guarantee use case, adding only the desired non-blocking
algorithm to the mix.
This light-weight garbage collector use case can also be used in
conjunction with the existence guarantees described in the next section.

\subsubsection{Delete-Only List}
\label{sec:defer:Delete-Only List}

The delete-only list is the less-popular counterpart to the add-only list
covered in \cref{sec:defer:Add-Only List}, and can be thought of as the
existence-guarantee use case, but without the publish/subscribe component,
as shown in \cref{fig:defer:Relationships Between RCU Use Cases}.
A delete-only list can be used when the universe of possible members of
the list is known at initialization, and where members can be removed.
For example, elements of the list might represent hardware elements of
the system that are subject to failure, but cannot be repaired or
replaced without a reboot.

An delete-only variant of a pre-BSD routing table can be derived from
\cref{lst:defer:RCU Pre-BSD Routing Table Lookup,lst:defer:RCU Pre-BSD Routing Table Add/Delete}.
Because there is no addition, the \co{route_add()} function may be
dispensed with, or, alternatively, its use might be restricted to
initialization time.
In theory, the \co{route_lookup()} function can use a non-RCU iterator,
though in the Linux kernel this will result in complaints from debug code.
In addition, the incremental cost of an RCU iterator is usually
negligible.

As a result, delete-only situations typically use algorithms and data
structures that are designed for addition as well as deletion.

\subsubsection{Quasi Reader-Writer Lock}
\label{sec:defer:Quasi Reader-Writer Lock}

Perhaps the most common use of RCU within the Linux kernel is as
a replacement for reader-writer locking in read-intensive situations.
Nevertheless, this use of RCU was not immediately apparent to me
at the outset.
In fact, I chose to implement a lightweight reader-writer
lock~\cite{WilsonCHsieh92a}\footnote{
	Similar to \co{brlock} in the 2.4 Linux kernel and to
	\co{lglock} in more recent Linux kernels.}
before implementing a general-purpose RCU implementation
back in the early 1990s.
Each and every one of the uses I envisioned for the lightweight reader-writer
lock was instead implemented using RCU\@.
In fact, it was more than
three years before the lightweight reader-writer lock saw its first use.
Boy, did I feel foolish!

The key similarity between RCU and reader-writer locking is that
both have read-side critical sections that can execute concurrently.
In fact, in some cases, it is possible to mechanically substitute RCU API
members for the corresponding reader-writer lock API members.
But first, why bother?

Advantages of RCU include performance,
deadlock immunity, and realtime latency.
There are, of course, limitations to RCU, including the fact that
readers and updaters run concurrently, that low-priority RCU readers
can block high-priority threads waiting for a grace period to elapse,
and that grace-period latencies can extend for many milliseconds.
These advantages and limitations are discussed in the following paragraphs.

\paragraph{Performance}

\begin{figure}
\centering
\resizebox{2.5in}{!}{\includegraphics{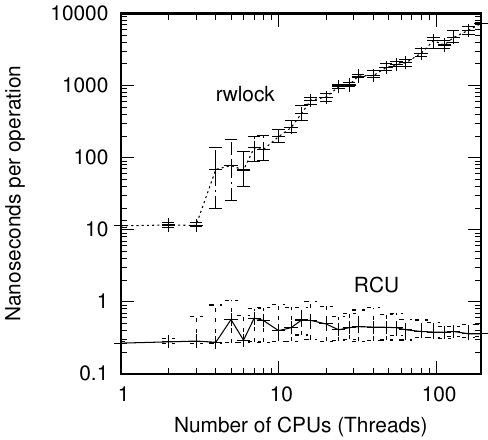}}
\caption{Performance Advantage of RCU Over Reader-Writer Locking}
\label{fig:defer:Performance Advantage of RCU Over Reader-Writer Locking}
\end{figure}

The read-side performance advantages of Linux-kernel RCU over
reader-writer locking are shown in
\cref{fig:defer:Performance Advantage of RCU Over Reader-Writer Locking},
which was generated on a 448-CPU 2.10\,GHz Intel x86 system.

\QuickQuizSeries{%
\QuickQuizB{
	WTF\@?
	How the heck do you expect me to believe that RCU can have less
	than a 300-picosecond overhead when the clock period at 2.10\,GHz
	is almost 500\,picoseconds?
}\QuickQuizAnswerB{
	First, consider that the inner loop used to
	take this measurement is as follows:

\begin{VerbatimN}
	for (i = nloops; i >= 0; i--) {
		rcu_read_lock();
		rcu_read_unlock();
	}
\end{VerbatimN}

	Next, consider the effective definitions of \co{rcu_read_lock()}
	and \co{rcu_read_unlock()}:

\begin{VerbatimN}
#define rcu_read_lock()   barrier()
#define rcu_read_unlock() barrier()
\end{VerbatimN}

	These definitions constrain compiler code-movement optimizations
	involving memory references, but emit no instructions in and
	of themselves.
	However, if the loop variable is maintained in a register,
	the accesses to \co{i} will not count as memory references.
	Furthermore, the compiler can do loop unrolling,
	allowing the resulting code to ``execute'' multiple passes
	through the loop body simply by incrementing \co{i} by
	some value larger than the value 1.

	So the ``measurement'' of 267 picoseconds is simply the fixed
	overhead of the timing measurements divided by the number of
	passes through the inner loop containing the calls
	to \co{rcu_read_lock()} and \co{rcu_read_unlock()}, plus
	the code to manipulate \co{i} divided by the loop-unrolling
	factor.
	And therefore, this measurement really is in error, in fact,
	it exaggerates the overhead by an arbitrary number of orders
	of magnitude.
	After all, in terms of machine instructions emitted, the actual
	overheads of \co{rcu_read_lock()} and of \co{rcu_read_unlock()}
	are each precisely zero.

	It is not just every day that a timing measurement of 267
	picoseconds turns out to be an overestimate!
}\QuickQuizEndB

\QuickQuizM{
	Didn't an earlier edition of this book show RCU read-side
	overhead way down in the sub-picosecond range?
	What happened???
}\QuickQuizAnswerM{
	Excellent memory!!!
	The overhead in some early releases was in fact roughly
	100~femtoseconds.

	What happened was that RCU usage spread more broadly through the
	Linux kernel, including into code that takes page faults.
	Back at that time, \co{rcu_read_lock()} and \co{rcu_read_unlock()}
	were complete no-ops in \co{CONFIG_PREEMPT=n} kernels.
	Unfortunately, that situation allowed the compiler to reorder
	page-faulting memory accesses into RCU read-side critical
	sections.
	Of course, page faults can block, which destroys those critical
	sections.

	Nor was this a theoretical problem:
	A failure actually manifested in 2019.
	\ppl{Herbert}{Xu} tracked down this failure down and
	\ppl{Linus}{Torvalds}
	therefore queued a commit to upgrade \co{rcu_read_lock()} and
	\co{rcu_read_unlock()} to unconditionally include a call to
	\co{barrier()}~\cite{LinusTorvalds2019:RCUreader.barrier}.
	And although \co{barrier()} emits no code, it does constrain
	compiler optimizations.
	And so the price of widespread RCU usage is slightly higher
	\co{rcu_read_lock()} and \co{rcu_read_unlock()} overhead.
	As such, Linux-kernel RCU has proven to be a victim of its
	own success.

	Of course, it is also the case that the older results were obtained
	on a different system than were those shown in
	\cref{fig:defer:Performance Advantage of RCU Over Reader-Writer Locking}.
	So which change had the most effect, Linus's commit or the change in
	the system?
	This question is left as an exercise to the reader.
}\QuickQuizEndM

\QuickQuizE{
	Why is there such large variation for the \co{RCU} trace in
	\cref{fig:defer:Performance Advantage of RCU Over Reader-Writer Locking}?
}\QuickQuizAnswerE{
	Keep in mind that this is a log-log plot, so those large-seeming
	\co{RCU} variances in reality span only a few hundred picoseconds.
	And that is such a short time that anything could cause it.
	However, given that the variance decreases with both small and
	large numbers of CPUs, one hypothesis is that the variation is
	due to migrations from one CPU to another.

	Yes, these measurements were taken with interrupts disabled, but
	they were also taken within a guest OS, so that preemption was
	still possible at the hypervisor level.
	In addition, the system featured hyperthreading and a single
	hardware thread running this RCU workload is able to consume
	more than half of the core's resources.
	Therefore, the overall throughput varies depending on how many
	of a given guest OS's CPUs share cores.
	Attempting to reduce these variations by running the guest OSes
	at real-time priority (as suggested by Joel Fernandes) is left
	as an exercise for the reader.
}\QuickQuizEndE
}                 % End of \QuickQuizSeries

Note that reader-writer locking is more than an order of magnitude slower
than RCU on a single CPU, and is more than \emph{four} orders of magnitude
slower on 192~CPUs.
In contrast, RCU scales quite well.
In both cases, the error bars cover the full range of the measurements
from 30~runs, with the line being the median.

\begin{figure}
\centering
\resizebox{2.5in}{!}{\includegraphics{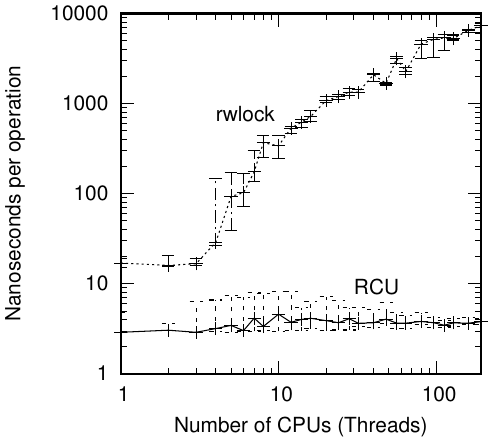}}
\caption{Performance Advantage of Preemptible RCU Over Reader-Writer Locking}
\label{fig:defer:Performance Advantage of Preemptible RCU Over Reader-Writer Locking}
\end{figure}

A more moderate view may be obtained from a \co{CONFIG_PREEMPT} kernel,
though RCU still beats reader-writer locking by between a factor of seven
on a single CPU and by three orders of magnitude on 192~CPUs, as shown in
\cref{fig:defer:Performance Advantage of Preemptible RCU Over Reader-Writer Locking},
which was generated on the same 448-CPU 2.10\,GHz x86 system.
Note the high variability of reader-writer locking at larger numbers of CPUs.
The error bars span the full range of data.

\QuickQuiz{
	Given that the system had no fewer than 448~hardware threads,
	why only 192~CPUs?
}\QuickQuizAnswer{
	Because the script (\path{rcuscale.sh}) that generates this data
	spawns a guest operating system for each set of points gathered,
	and on this particular system, both \co{qemu} and KVM limit the
	number of CPUs that may be configured into a given guest OS\@.
	Yes, it would have been possible to run a few more CPUs, but
	192 is a nice round number from a binary perspective, given
	that 256 is infeasible.
}\QuickQuizEnd

\begin{figure}
\centering
\resizebox{2.5in}{!}{\includegraphics{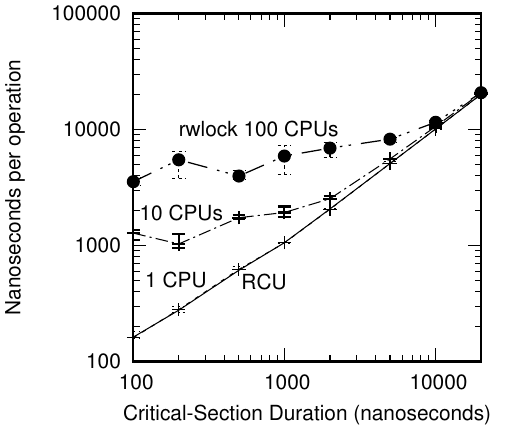}}
\caption{Comparison of RCU to Reader-Writer Locking as Function of Critical-Section Duration, 192 CPUs}
\label{fig:defer:Comparison of RCU to Reader-Writer Locking as Function of Critical-Section Duration}
\end{figure}

Of course, the low performance of reader-writer locking in
\cref{fig:defer:Performance Advantage of RCU Over Reader-Writer Locking,%
fig:defer:Performance Advantage of Preemptible RCU Over Reader-Writer Locking}
is exaggerated by the unrealistic zero-length critical sections.
The performance advantages of RCU decrease as the overhead of the critical
sections increase, as shown in
\cref{fig:defer:Comparison of RCU to Reader-Writer Locking as Function of Critical-Section Duration},
which was run on the same system as the previous plots.
Here, the y-axis represents the sum of the overhead of the read-side
primitives and that of the critical section and the x-axis represents
the critical-section overhead in nanoseconds.
But please note the logscale y~axis, which means that the small
separations between the traces still represent significant differences.
This figure shows non-preemptible RCU, but given that preemptible RCU's
read-side overhead is only about three nanoseconds, its plot would be
nearly identical to
\cref{fig:defer:Comparison of RCU to Reader-Writer Locking as Function of Critical-Section Duration}.

\QuickQuiz{
	Why the larger error ranges for the submicrosecond durations in
	\cref{fig:defer:Comparison of RCU to Reader-Writer Locking as Function of Critical-Section Duration}?
}\QuickQuizAnswer{
	Because smaller disturbances result in greater relative errors
	for smaller measurements.
	Also, the Linux kernel's \co{ndelay()} nanosecond-scale primitive
	is (as of 2020) less accurate than is the \co{udelay()} primitive
	used for the data for durations of a microsecond or more.
	It is instructive to compare to the zero-length case shown in
	\cref{fig:defer:Performance Advantage of RCU Over Reader-Writer Locking}.
}\QuickQuizEnd

There are three traces for reader-writer locking, with the upper trace
being for 100~CPUs, the next for 10~CPUs, and the lowest for 1~CPU\@.
The greater the number of CPUs and the shorter the critical sections,
the greater is RCU's performance advantage.
These performance advantages are underscored by the fact that 100-CPU
systems are no longer uncommon and that a number of system calls (and
thus any RCU read-side critical sections that they contain) complete
within microseconds.

In addition, as is discussed in the next paragraph,
RCU read-side primitives are almost entirely deadlock-immune.

\paragraph{Deadlock Immunity}

Although RCU offers significant performance advantages for
read-mostly workloads, one of the primary reasons for creating
RCU in the first place was in fact its immunity to read-side
deadlocks.
This immunity stems from the fact that
RCU read-side primitives do not block, spin, or even
do backwards branches, so that their execution time is deterministic.
It is therefore impossible for them to participate in a deadlock
cycle.

\QuickQuiz{
	Is there an exception to this deadlock immunity, and if so,
	what sequence of events could lead to deadlock?
}\QuickQuizAnswer{
	One way to cause a deadlock cycle involving
	RCU read-side primitives is via the following (illegal) sequence
	of statements:

\begin{VerbatimU}
rcu_read_lock();
synchronize_rcu();
rcu_read_unlock();
\end{VerbatimU}

	The \co{synchronize_rcu()} cannot return until all
	pre-existing RCU read-side critical sections complete, but
	is enclosed in an RCU read-side critical section that cannot
	complete until the \co{synchronize_rcu()} returns.
	The result is a classic self-deadlock---you get the same
	effect when attempting to write-acquire a reader-writer lock
	while read-holding it.

	Note that this self-deadlock scenario does not apply to
	RCU QSBR, because the context switch performed by the
	\co{synchronize_rcu()} would act as a quiescent state
	for this CPU, allowing a grace period to complete.
	However, this is if anything even worse, because data used
	by the RCU read-side critical section might be freed as a
	result of the grace period completing.
	Plus Linux kernel's lockdep facility will yell at you.

	In short, do not invoke synchronous RCU update-side primitives, which
	are listed in
	\cref{tab:defer:RCU Wait-to-Finish APIs},
	from within an RCU read-side critical section.

	In addition, within the Linux kernel, RCU uses the scheduler
	and the scheduler uses RCU\@.
	In some cases, both RCU and the scheduler must take care to
	avoid deadlock.
}\QuickQuizEnd

An interesting consequence of RCU's read-side deadlock immunity is
that it is possible to unconditionally upgrade an RCU
reader to an RCU updater.
Attempting to do such an upgrade with reader-writer locking results
in deadlock.
A sample code fragment that does an RCU read-to-update upgrade follows:

\begin{VerbatimN}[samepage=true]
rcu_read_lock();
list_for_each_entry_rcu(p, &head, list_field) {
	do_something_with(p);
	if (need_update(p)) {
		spin_lock(my_lock);
		do_update(p);
		spin_unlock(&my_lock);
	}
}
rcu_read_unlock();
\end{VerbatimN}

Note that \co{do_update()} is executed under
the protection of the lock \emph{and} under RCU read-side protection.

Another interesting consequence of RCU's deadlock immunity is its
immunity to a large class of priority inversion problems.
For example, low-priority RCU readers cannot prevent a high-priority
RCU updater from acquiring the update-side lock.
Similarly, a low-priority RCU updater cannot prevent high-priority
RCU readers from entering an RCU read-side critical section.

\QuickQuiz{
	Immunity to both deadlock and priority inversion???
	Sounds too good to be true.
	Why should I believe that this is even possible?
}\QuickQuizAnswer{
	It really does work.
	After all, if it didn't work, the Linux kernel would not run.
}\QuickQuizEnd

\paragraph{Realtime Latency}

Because RCU read-side primitives neither spin nor block, they offer
excellent realtime latencies.
In addition, as noted earlier, this means that they are
immune to priority inversion
involving the RCU read-side primitives and locks.

However, RCU is susceptible to more subtle priority-inversion scenarios,
for example, a high-priority process blocked waiting for an RCU
grace period to elapse can be blocked by low-priority RCU readers
in \rt\ kernels.
This can be solved by using RCU priority
boosting~\cite{PaulEMcKenney2007BoostRCU,DinakarGuniguntala2008IBMSysJ}.

However, use of RCU priority boosting requires that \co{rcu_read_unlock()}
do deboosting, which entails acquiring scheduler locks.
Some care is therefore required within the scheduler and RCU to avoid
deadlocks, which as of the v5.15 Linux kernel requires RCU to avoid
invoking the scheduler while holding any of RCU's locks.

This in turn means that \co{rcu_read_unlock()} is not always lockless
when RCU priority boosting is enabled.
However, \co{rcu_read_unlock()} will still be lockless if its
critical section was not priority-boosted.
Furthermore, critical sections will not be priority boosted unless they
are preempted, or, in -rt kernels, they acquire non-raw spinlocks.
This means that \co{rcu_read_unlock()} will normally be lockless from the
perspective of the highest priority task running on any given CPU.

\paragraph{RCU Readers and Updaters Run Concurrently}

Because RCU readers never spin nor block, and because updaters are not
subject to any sort of rollback or abort semantics, RCU readers and
updaters really can run concurrently.
This means that RCU readers might access stale data, and might even
see inconsistencies, either of which can render conversion from reader-writer
locking to RCU non-trivial.

\begin{figure}
\centering
\resizebox{3in}{!}{\includegraphics{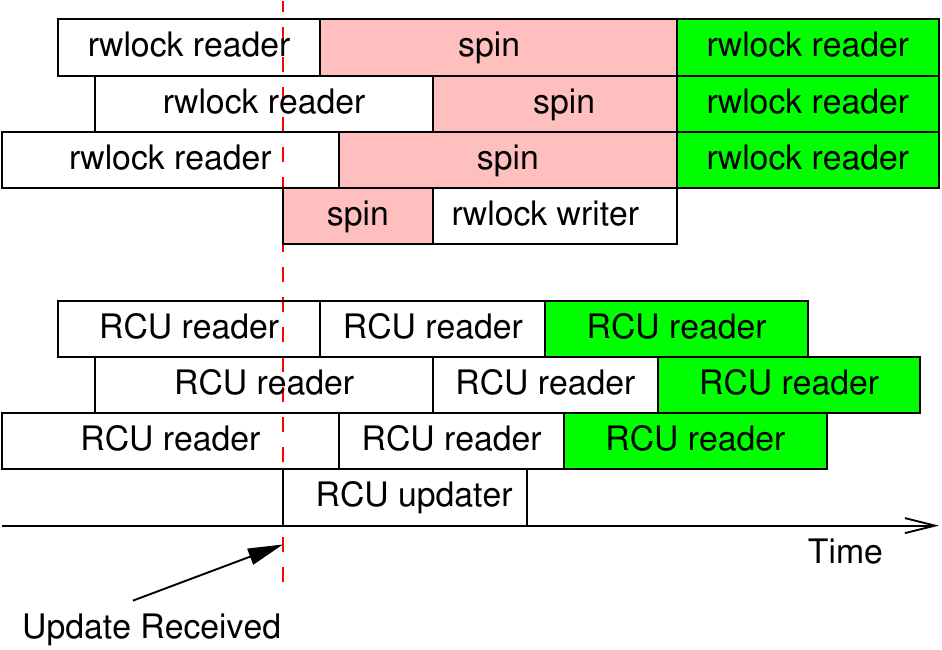}}
\caption{Response Time of RCU vs.\@ Reader-Writer Locking}
\label{fig:defer:Response Time of RCU vs. Reader-Writer Locking}
\end{figure}

However, in a surprisingly large number of situations, inconsistencies and
stale data are not problems.
The classic example is the networking routing table.
Because routing updates can take considerable time to reach a given
system (seconds or even minutes), the system will have been sending
packets the wrong way for quite some time when the update arrives.
It is usually not a problem to continue sending updates the wrong
way for a few additional milliseconds.
Furthermore, because RCU updaters can make changes without waiting for
RCU readers to finish,
the RCU readers might well see the change more quickly than would
batch-fair
reader-writer-locking readers, as shown in
\cref{fig:defer:Response Time of RCU vs. Reader-Writer Locking}.

\QuickQuiz{
	But how many other algorithms really tolerate stale and
	inconsistent data?
}\QuickQuizAnswer{
	Quite a few!

	Please keep in mind that the finite speed of light means that
	data reaching a given computer system is at least slightly stale
	at the time that it arrives, and extremely stale in the case
	of astronomical data.
	The finite speed of light also places a sharp limit on the
	consistency of data arriving from different sources of via
	different paths.

	You might as well face the fact that the laws of physics
	are incompatible with naive notions of perfect freshness and
	consistency.
}\QuickQuizEnd

Once the update is received, the rwlock writer cannot proceed until the
last reader completes, and subsequent readers cannot proceed until the
writer completes.
However, these subsequent readers are guaranteed to see the new value,
as indicated by the green shading of the rightmost boxes.
In contrast, RCU readers and updaters do not block each other, which permits
the RCU readers to see the updated values sooner.
Of course, because their execution overlaps that of the RCU updater,
\emph{all} of the RCU readers might well see updated values, including
the three readers that started before the update.
Nevertheless only the green-shaded rightmost RCU readers
are \emph{guaranteed} to see the updated values.

Reader-writer locking and RCU simply provide different guarantees.
With reader-writer locking, any reader that begins after the writer begins
is guaranteed to see new values, and any reader that attempts to
begin while the writer is spinning might or might not see new values,
depending on the reader/writer preference of the rwlock implementation in
question.
In contrast, with RCU, any reader that begins after the updater completes
is guaranteed to see new values, and any reader that completes after the
updater begins might or might not see new values, depending on timing.

The key point here is that, although reader-writer locking does
indeed guarantee consistency within the confines of the computer system,
there are situations where this consistency comes at the price of
increased \emph{inconsistency} with the outside world, courtesy of
the finite speed of light and the non-zero size of atoms.
In other words, reader-writer locking obtains internal consistency at the
price of silently stale data with respect to the outside world.

Note that if a value is computed while read-holding a reader-writer
lock, and then that value is used after that lock is released, then
this reader-writer-locking use case is using stale data.
After all, the quantities that this value is based on could change
at any time after that lock is released.
This sort of reader-writer-locking use case is often easy to
convert to RCU, as will be shown in
\cref{lst:defer:Converting Reader-Writer Locking to RCU: Data,%
lst:defer:Converting Reader-Writer Locking to RCU: Search,%
lst:defer:Converting Reader-Writer Locking to RCU: Deletion}
and the accompanying text.

\paragraph{Low-Priority RCU Readers Can Block High-Priority Reclaimers}

In Realtime RCU~\cite{DinakarGuniguntala2008IBMSysJ} or
SRCU~\cite{PaulEMcKenney2006c},
a preempted reader will prevent a grace period from completing, even if
a high-priority task is blocked waiting for that grace period to complete.
Realtime RCU can avoid this problem by substituting \co{call_rcu()}
for \co{synchronize_rcu()} or by using RCU priority
boosting~\cite{PaulEMcKenney2007BoostRCU,DinakarGuniguntala2008IBMSysJ}.
It might someday be necessary to augment SRCU and RCU Tasks Trace with
priority boosting, but not before a clear real-world need is demonstrated.

\QuickQuiz{
	If Tasks RCU Trace might someday be priority boosted, why
	not also Tasks RCU and Tasks RCU Rude?
}\QuickQuizAnswer{
	Maybe, but these are less likely.

	In the case of Tasks RCU, recall that the quiescent state is
	a voluntary context switch.
	Thus, all tasks not blocked after a voluntary context switch
	might need to be boosted, and the mechanics of deboosting would
	not likely be at all pretty.

	In the case of Tasks RCU Rude, as was the case with the old
	RCU Sched, any preemptible region of code is a quiescent state.
	Thus, the only tasks that might need boosting are those currently
	running with preemption disabled.
	But boosting the priority of a preemption-disabled task has no
	effect.
	It therefore seems doubly unlikely that priority boosting will
	ever be introduced to Tasks RCU Rude, at least in its current
	form.
}\QuickQuizEnd

\paragraph{RCU Grace Periods Extend for Many Milliseconds}

With the exception of userspace
RCU~\cite{MathieuDesnoyers2009URCU,PaulMcKenney2013LWNURCU},
expedited grace periods, and several of the ``toy''
RCU implementations described in
\cref{chp:app:``Toy'' RCU Implementations},
RCU grace periods extend milliseconds.
Although there are a number of techniques to render such long
delays harmless, including use of the asynchronous interfaces
(\co{call_rcu()} and \co{call_rcu_bh()}) or of the polling interfaces
(\co{get_state_synchronize_rcu()}, \co{start_poll_synchronize_rcu()},
and \co{poll_state_synchronize_rcu()}), this situation is a major reason
for the rule of thumb that RCU be used in read-mostly situations.

As noted in \cref{sec:defer:RCU Linux-Kernel API}, within the Linux
kernel, shorter grace periods may be obtained via expedited grace
periods, for example, by invoking \co{synchronize_rcu_expedited()}
instead of \co{synchronize_rcu()}.
Expedited grace periods can reduce delays to as little as a few tens of
microseconds, albeit at the expense of higher CPU utilization and IPIs.
The added IPIs can be especially unwelcome in some real-time workloads.

\paragraph{Code:
		 Reader-Writer Locking vs.\@ RCU}

In the best case, the conversion from reader-writer locking to RCU
is quite simple, as shown in
\cref{lst:defer:Converting Reader-Writer Locking to RCU: Data,%
lst:defer:Converting Reader-Writer Locking to RCU: Search,%
lst:defer:Converting Reader-Writer Locking to RCU: Deletion},
all taken from
Wikipedia~\cite{WikipediaRCU}.

\begin{listing*}
{ \scriptsize
\begin{verbbox}
 1 struct el {                           1 struct el {
 2   struct list_head lp;                2   struct list_head lp;
 3   long key;                           3   long key;
 4   spinlock_t mutex;                   4   spinlock_t mutex;
 5   int data;                           5   int data;
 6   /* Other data fields */             6   /* Other data fields */
 7 };                                    7 };
 8 DEFINE_RWLOCK(listmutex);             8 DEFINE_SPINLOCK(listmutex);
 9 LIST_HEAD(head);                      9 LIST_HEAD(head);
\end{verbbox}
}
\hspace*{0.9in}\OneColumnHSpace{-0.5in}
\IfEbookSize{\hspace*{-1.05in}}{}\theverbbox
\caption{Converting Reader-Writer Locking to RCU\@:
						    Data}
\label{lst:defer:Converting Reader-Writer Locking to RCU: Data}
\end{listing*}

\begin{listing*}
{ \scriptsize
\begin{verbbox}
 1 int search(long key, int *result)     1 int search(long key, int *result)
 2 {                                     2 {
 3   struct el *p;                       3   struct el *p;
 4                                       4
 5   read_lock(&listmutex);              5   rcu_read_lock();
 6   list_for_each_entry(p, &head, lp) { 6   list_for_each_entry_rcu(p, &head, lp) {
 7     if (p->key == key) {              7     if (p->key == key) {
 8       *result = p->data;              8       *result = p->data;
 9       read_unlock(&listmutex);        9       rcu_read_unlock();
10       return 1;                      10       return 1;
11     }                                11     }
12   }                                  12   }
13   read_unlock(&listmutex);           13   rcu_read_unlock();
14   return 0;                          14   return 0;
15 }                                    15 }
\end{verbbox}
}
\hspace*{0.9in}\OneColumnHSpace{-0.5in}
\IfEbookSize{\hspace*{-1.05in}}{}\theverbbox
\caption{Converting Reader-Writer Locking to RCU\@:
						    Search}
\label{lst:defer:Converting Reader-Writer Locking to RCU: Search}
\end{listing*}

\begin{listing*}
{ \scriptsize
\begin{verbbox}
 1 int delete(long key)                  1 int delete(long key)
 2 {                                     2 {
 3   struct el *p;                       3   struct el *p;
 4                                       4
 5   write_lock(&listmutex);             5   spin_lock(&listmutex);
 6   list_for_each_entry(p, &head, lp) { 6   list_for_each_entry(p, &head, lp) {
 7     if (p->key == key) {              7     if (p->key == key) {
 8       list_del(&p->lp);               8       list_del_rcu(&p->lp);
 9       write_unlock(&listmutex);       9       spin_unlock(&listmutex);
                                        10       synchronize_rcu();
10       kfree(p);                      11       kfree(p);
11       return 1;                      12       return 1;
12     }                                13     }
13   }                                  14   }
14   write_unlock(&listmutex);          15   spin_unlock(&listmutex);
15   return 0;                          16   return 0;
16 }                                    17 }
\end{verbbox}
}
\hspace*{0.9in}\OneColumnHSpace{-0.5in}
\IfEbookSize{\hspace*{-1.05in}}{}\theverbbox
\caption{Converting Reader-Writer Locking to RCU\@:
						    Deletion}
\label{lst:defer:Converting Reader-Writer Locking to RCU: Deletion}
\end{listing*}

However, the transformation is not always this straightforward.
This is because neither the \co{spin_lock()} nor the
\co{synchronize_rcu()} in
\cref{lst:defer:Converting Reader-Writer Locking to RCU: Deletion}
exclude the readers in
\cref{lst:defer:Converting Reader-Writer Locking to RCU: Search}.
First, the \co{spin_lock()} does not interact in any way with
\co{rcu_read_lock()} and \co{rcu_read_unlock()}, thus not excluding them.
Second, although both \co{write_lock()} and \co{synchronize_rcu()}
wait for pre-existing readers, only \co{write_lock()} prevents
subsequent readers from commencing.\footnote{
	Kudos to whoever pointed this out to Paul.}
Thus, \co{synchronize_rcu()} cannot exclude readers.
Nevertheless, a great many situations using reader-writer locking can
be converted to RCU\@.

More-elaborate cases of replacing reader-writer locking with RCU
may be found
elsewhere~\cite{NeilBrown2015PathnameLookup,NeilBrown2015RCUwalk}.

\paragraph{Semantics:
		      Reader-Writer Locking vs.\@ RCU}

Expanding on the previous section, reader-writer locking semantics can
be roughly and informally summarized by the following three temporal
constraints:

\begin{enumerate}
\item	Write-side acquisitions wait for any read-holders to release
	the lock.
\item	Writer-side acquisitions wait for any write-holder to release
	the lock.
\item	Read-side acquisitions wait for any write-holder to release
	the lock.
\end{enumerate}

RCU dispenses entirely with constraint~\#3 and weakens the other two
as follows:

\begin{enumerate}
\item	Writers wait for any pre-existing read-holders before progressing
	to the destructive phase of their update (usually the freeing of
	memory).
\item	Writers synchronize with each other as needed.
\end{enumerate}

It is of course this weakening that permits RCU implementations to attain
excellent performance and scalability.
It also allows RCU to implement the aforementioned unconditional
read-to-write upgrade that is so attractive and so deadlock-prone in
reader-writer locking.
Code using RCU can compensate for this weakening in a surprisingly large
number of ways, but most commonly by imposing spatial constraints:

\begin{enumerate}
\item	New data is placed in newly allocated memory.
\item	Old data is freed, but only after:
	\begin{enumerate}
	\item	That data has been unlinked so as to be inaccessible
		to later readers, and
	\item	A subsequent RCU grace period has elapsed.
	\end{enumerate}
\end{enumerate}

Of course, there are some reader-writer-locking use cases for which
RCU's weakened semantics are inappropriate, but experience in the Linux
kernel indicates that more than 80\% of reader-writer locks can in fact
be replaced by RCU\@.
For example, a common reader-writer-locking use case computes some value
while holding the lock and then uses that value after releasing that lock.
This use case results in stale data, and therefore often accommodates
RCU's weaker semantics.

\begin{listing}
\input{CodeSamples/defer/singleton=get.fcv}
\caption{RCU Singleton Get}
\label{lst:defer:Singleton Get}
\end{listing}

\begin{listing}
\input{CodeSamples/defer/singleton=set.fcv}
\caption{RCU Singleton Set}
\label{lst:defer:Singleton Set}
\end{listing}

\begin{fcvref}[ln:defer:singleton:get]
This interaction of temporal and spatial constraints is illustrated
by the RCU singleton data structure illustrated in
\cref{fig:defer:Insertion With Concurrent Readers,fig:defer:Deletion With Concurrent Readers}.
This structure is defined on \clnrefrange{myconfig.b}{myconfig.e} of
\cref{lst:defer:Singleton Get}, and contains two integer fields,
\co{->a} and \co{->b} (\path{singleton.c}).
The current instance of this structure is referenced by the \co{curconfig}
pointer defined on \clnref{myconfig.e}.
\end{fcvref}

\begin{fcvref}[ln:defer:singleton:get]
The fields of the current structure are passed back through the
\co{cur_a} and \co{cur_b} parameters to the \co{get_config()} function
defined on \clnrefrange{get_config.b}{get_config.e}.
These two fields can be slightly out of date, but they absolutely must
be consistent with each other.
The \co{get_config()} function provides this consistency
within the RCU read-side critical section starting on
\clnref{rrl} and ending on either \clnref{rrul1} or \clnref{rrul2},
which provides the needed temporal synchronization.
\Clnref{rderef} fetches the pointer to the current \co{myconfig} structure.
This structure will be used regardless of any concurrent changes due
to calls to the \co{set_config()} function, thus providing the needed
spatial synchronization.
If \clnref{nullchk} determines that the \co{curconfig} pointer was
\co{NULL}, \clnref{retfail} returns failure.
Otherwise, \clnref{copya,copyb} copy out the \co{->a} and \co{->b} fields
and \clnref{retsuccess} returns success.
These \co{->a} and \co{->b} fields are from the same \co{myconfig}
structure, and the RCU read-side critical section prevents this structure
from being freed, thus guaranteeing that these two fields are consistent
with each other.
\end{fcvref}

\begin{fcvref}[ln:defer:singleton:set]
The structure is updated by the \co{set_config()} function shown in
\cref{lst:defer:Singleton Set}.
\Clnrefrange{allocinit.b}{allocinit.e} allocate and initialize a
new \co{myconfig} structure.
\Clnref{xchg} atomically exchanges a pointer to this new structure with
the pointer to the old structure in \co{curconfig}, while also providing
full memory ordering both before and after the \co{xchg()} operation,
thus providing the needed updater/reader spatial synchronization on
the one hand and the needed updater/updater synchronization on the other.
If \clnref{if} determines that the pointer to the old structure was
in fact non-\co{NULL}, \clnref{sr} waits for a grace period (thus providing
the needed reader/updater temporal synchronization) and \clnref{free}
frees the old structure, safe in the knowledge that there are no
longer any readers still referencing it.
\end{fcvref}

\begin{figure*}
\centering
\resizebox{\onecolumntextwidth}{!}{\includegraphics{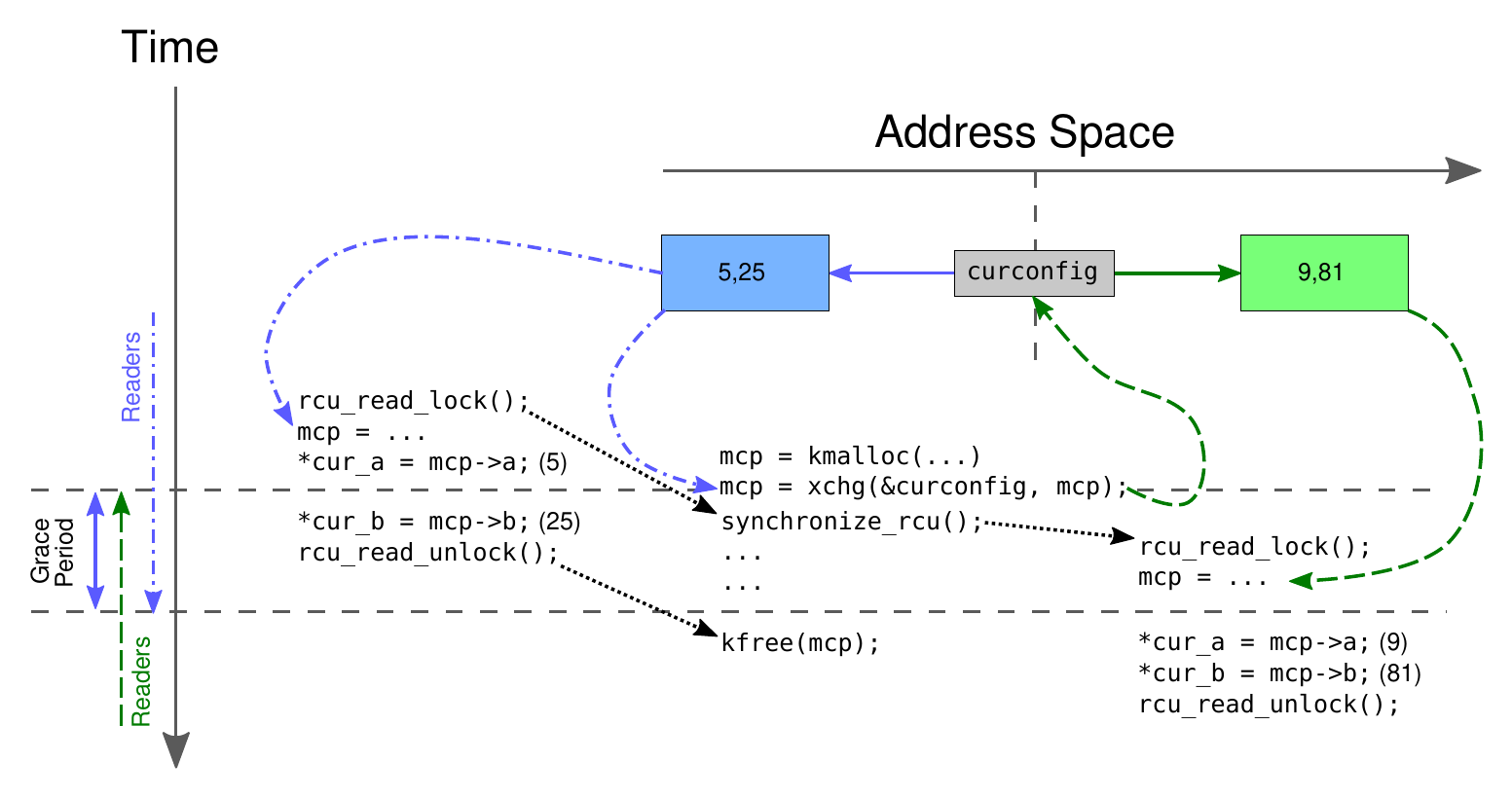}}
\caption{RCU Spatial/Temporal Synchronization}
\label{fig:defer:RCU Spatial/Temporal Synchronization}
\end{figure*}

\Cref{fig:defer:RCU Spatial/Temporal Synchronization} shows an abbreviated
representation of \co{get_config()} on the left and right and a similarly
abbreviated representation of \co{set_config()} in the middle.
Time advances from top to bottom, and the address space of the objects
referenced by \co{curconfig} advances from left to right.
The boxes with comma-separated numbers each represent a \co{myconfig}
structure, with the constraint that \co{->b} is the square of \co{->a}.
Each blue dash-dotted arrow represents an interaction with the old structure
(on the left, containing ``5,25'') and each green dashed arrow represents
an interaction with the new structure (on the right, containing ``9,81'').

The black dotted arrows represent temporal relationships between RCU readers
on the left and right and the RCU grace period at center, with each
arrow pointing from an older event to a newer event.
The call to \co{synchronize_rcu()} followed the leftmost \co{rcu_read_lock()},
and therefore that \co{synchronize_rcu()} invocation must not return
until after the corresponding \co{rcu_read_unlock()}.
In contrast, the call to \co{synchronize_rcu()} precedes the
rightmost \co{rcu_read_lock()}, which allows the return from that same
\co{synchronize_rcu()} to ignore the corresponding \co{rcu_read_unlock()}.
These temporal relationships prevent the \co{myconfig} structures from
being freed while RCU readers are still accessing them.

The two horizontal grey dashed lines represent the period of time during
which different readers get different results, however, each reader
will see one and only one of the two objects.
All readers that end before the first horizontal line will see the
leftmost \co{myconfig} structure, and all readers that start after the
second horizontal line will see the rightmost structure.
Between the two lines, that is, during the grace period, different
readers might see different objects, but as long as each reader
loads the \co{curconfig} pointer only once, each reader will see
a consistent view of its \co{myconfig} structure.

\QuickQuiz{
	But doesn't the RCU grace period start sometime after the
	call to \co{synchronize_rcu()} rather than in the middle
	of that \co{xchg()} statement?
}\QuickQuizAnswer{
	Which grace period, exactly?

	The updater is required to wait for at least one grace
	period that starts at or some time after the removal,
	in this case, the \co{xchg()}.
	So in
	\cref{fig:defer:RCU Spatial/Temporal Synchronization},
	the indicated grace period starts as early as theoretically
	possible and extends to the return from \co{synchronize_rcu()}.
	This is a perfectly legal grace period corresponding to the
	change carried out by that \co{xchg()} statement.
}\QuickQuizEnd

In short, when operating on a suitable linked data structure, RCU
combines temporal and spatial synchronization in order to approximate
reader-writer locking, with RCU read-side critical sections acting as
the reader-writer-locking reader, as shown in
\cref{fig:defer:Relationships Between RCU Use Cases,fig:defer:RCU Spatial/Temporal Synchronization}.
RCU's temporal synchronization is provided by the read-side markers,
for example, \co{rcu_read_lock()} and \co{rcu_read_unlock()}, as well as
the update-side grace-period primitives, for example, \co{synchronize_rcu()}
or \co{call_rcu()}.
The spatial synchronization is provided by the read-side
\co{rcu_dereference()} family of primitives, each of which subscribes
to a version published by \co{rcu_assign_pointer()}.\footnote{
	Preferably with both \co{rcu_dereference()} and
	\co{rcu_assign_pointer()} being embedded in higher-level APIs.}
RCU's combining of temporal and spatial synchronization contrasts to
the schemes presented in
\cref{sec:SMPdesign:Code Locking,sec:SMPdesign:Data Locking,sec:locking:Inefficiency},
in which temporal and spatial synchronization are provided separately by
locking and by static data-structure layout, respectively.

\QuickQuiz{
	Is RCU the only synchronization mechanism that combines temporal
	and spatial synchronization in this way?
}\QuickQuizAnswer{
	Not at all.

	Hazard pointers can be considered to combine temporal and spatial
	synchronization in a similar manner.
	Referring to
	\cref{lst:defer:Hazard-Pointer Recording and Clearing},
	the \co{hp_record()} function's acquisition of a reference
	provides both spatial and temporal synchronization, subscribing
	to a version and marking the start of a reference, respectively.
	This function therefore combines the effects of RCU's
	\co{rcu_read_lock()} and \co{rcu_dereference()}.
	Referring now to
	\cref{lst:defer:Hazard-Pointer Scanning and Freeing},
	the \co{hp_clear()} function's release of a reference provides
	temporal synchronization marking the end of a reference, and is
	thus similar to RCU's \co{rcu_read_unlock()}.
	The \co{hazptr_free_later()} function's retiring of a
	hazard-pointer-protected object provides temporal synchronization,
	similar to RCU's \co{call_rcu()}.
	The primitives used to mutate a hazard-pointer-protected
	structure provide spatial synchronization, similar to RCU's
	\co{rcu_assign_pointer()}.

	Alternatively, one could instead come at hazard pointers by
	analogy with reference counting.
}\QuickQuizEnd

\subsubsection{Quasi Reference Count}
\label{sec:defer:Quasi Reference Count}

Because grace periods are not allowed to complete while
there is an RCU read-side critical section in progress,
the RCU read-side primitives may be used as a restricted
reference-counting mechanism.
For example, consider the following code fragment:

\begin{VerbatimN}
rcu_read_lock();  /* acquire reference. */
p = rcu_dereference(head);
/* do something with p. */
rcu_read_unlock();  /* release reference. */
\end{VerbatimN}

The combination of the \co{rcu_read_lock()} and \co{rcu_dereference()}
primitives can be thought of as acquiring a reference to \co{p},
because a grace period starting after the \co{rcu_dereference()}
assignment to \co{p} cannot possibly end until after we reach the matching
\co{rcu_read_unlock()}.
This reference-counting scheme is restricted in that it is forbidden
to wait for RCU grace periods within RCU read-side critical sections,
and also forbidden to hand off an RCU read-side critical section's
references from one task to another.

Regardless of these restrictions,
the following code can safely delete \co{p}:

\begin{VerbatimN}
spin_lock(&mylock);
p = head;
rcu_assign_pointer(head, NULL);
spin_unlock(&mylock);
/* Wait for all references to be released. */
synchronize_rcu();
kfree(p);
\end{VerbatimN}

The assignment to \co{head} prevents any future references
to \co{p} from being acquired, and the \co{synchronize_rcu()}
waits for any previously acquired references to be released.

\QuickQuiz{
	But wait!
	This is exactly the same code that might be used when thinking
	of RCU as a replacement for reader-writer locking!
	What gives?
}\QuickQuizAnswer{
	This is an effect of the Law of Toy Examples:
	Beyond a certain point, the code fragments look the same.
	The only difference is in how we think about the code.
	For example, what does an \co{atomic_inc()} operation do?
	It might be acquiring another explicit reference to an object
	to which we already have a reference, it might be incrementing
	an often-read/seldom-updated statistical counter, it might
	be checking into an HPC-style barrier, or any of a number of
	other things.

	However, these differences can be extremely important.
	For but one example of the importance, consider that if we think
	of RCU as a restricted reference counting scheme, we would never
	be fooled into thinking that the updates would exclude the RCU
	read-side critical sections.

	It nevertheless is often useful to think of RCU as a replacement
	for reader-writer locking, for example, when you are replacing
	reader-writer locking with RCU\@.
}\QuickQuizEnd

Of course, RCU can also be combined with traditional reference counting,
as discussed in
\cref{sec:together:Refurbish Reference Counting}.

\begin{figure}
\centering
\resizebox{2.5in}{!}{\includegraphics{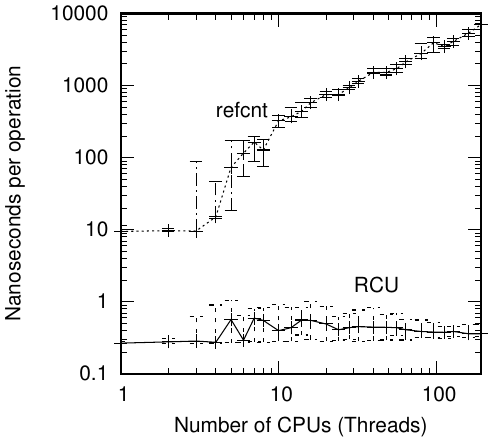}}
\caption{Performance of RCU vs.\@ Reference Counting}
\label{fig:defer:Performance of RCU vs. Reference Counting}
\end{figure}

\begin{figure}
\centering
\resizebox{2.5in}{!}{\includegraphics{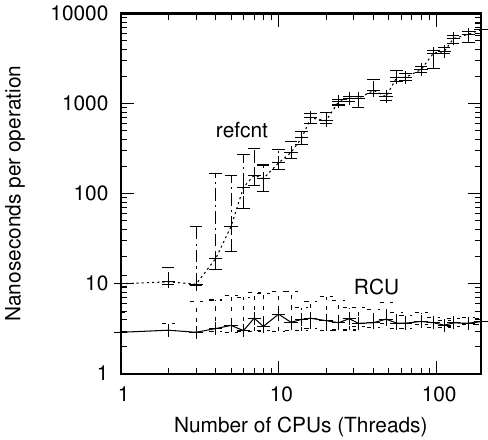}}
\caption{Performance of Preemptible RCU vs.\@ Reference Counting}
\label{fig:defer:Performance of Preemptible RCU vs. Reference Counting}
\end{figure}

But why bother?
Again, part of the answer is performance, as shown in
\cref{fig:defer:Performance of RCU vs. Reference Counting,%
fig:defer:Performance of Preemptible RCU vs. Reference Counting},
again showing data taken on a 448-CPU 2.1\,GHz Intel x86 system
for non-preemptible and preemptible Linux-kernel RCU, respectively.
Non-preemptible RCU's advantage over
\IXalt{reference counting}{reference count} ranges from
more than an order of magnitude at one CPU up to about four orders of
magnitude at 192~CPUs.
Preemptible RCU's advantage ranges from about a factor of three at
one CPU up to about three orders of magnitude at 192~CPUs.

\begin{figure}
\centering
\resizebox{2.5in}{!}{\includegraphics{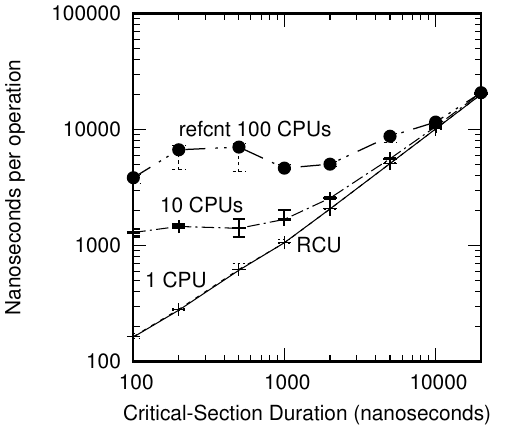}}
\caption{Response Time of RCU vs.\@ Reference Counting, 192 CPUs}
\label{fig:defer:Response Time of RCU vs. Reference Counting}
\end{figure}

However, as with reader-writer locking, the performance advantages of
RCU are most pronounced for short-duration critical sections and for
large numbers of CPUs, as shown in
\cref{fig:defer:Response Time of RCU vs. Reference Counting}
for the same system.
In addition, as with reader-writer locking, many system calls (and thus
any RCU read-side critical sections that they contain) complete in
a few microseconds.

Although traditional reference counters are usually associated with a
specific data structure, or perhaps a specific group of data structures,
this approach does have some disadvantages.
For example, maintaining a single global reference counter for a large
variety of data structures typically results in bouncing the cache line
containing the reference count.
As we saw in
\crefrange{fig:defer:Performance of RCU vs. Reference Counting}{fig:defer:Response Time of RCU vs. Reference Counting},
such cache-line bouncing can severely degrade performance.

In contrast, RCU's lightweight \co{rcu_read_lock()},
\co{rcu_dereference()}, and \co{rcu_read_unlock()} read-side primitives
permit extremely frequent read-side usage with negligible performance
degradation.
Except that the calls to \co{rcu_dereference()} are not doing anything
specific to acquire a reference to the pointed-to object.
The heavy lifting is instead done by the \co{rcu_read_lock()} and
\co{rcu_read_unlock()} primitives and their interactions with RCU
grace periods.

And ignoring those calls to \co{rcu_dereference()} permits RCU to be
thought of as a ``bulk reference-counting'' mechanism, where each call
to \co{rcu_read_lock()} obtains a reference on each and every RCU-protected
object, and with little or no overhead.
However, the restrictions that go with RCU can be quite onerous.
For example, in many cases, the Linux-kernel prohibition against
sleeping while in an RCU read-side critical section would defeat the
entire purpose.
Such cases might be better served by the hazard pointers mechanism
described in \cref{sec:defer:Hazard Pointers}.
Cases where code rarely sleeps have been handled by using RCU as a
reference count in the common non-sleeping case and by bridging
to an explicit reference counter when sleeping is necessary.

Alternatively, situations where a reference must be held by a single task
across a section of code that sleeps may be accommodated with Sleepable
RCU (SRCU)~\cite{PaulEMcKenney2006c}.
This fails to cover the not-uncommon situation where a reference is ``passed''
from one task to another, for example, when a reference is acquired
when starting an I/O and released in the corresponding completion
interrupt handler.
Again, such cases might be better handled by explicit reference counters
or by hazard pointers.

Of course, SRCU brings restrictions of its own, namely that the
return value from \co{srcu_read_lock()} be passed into the
corresponding \co{srcu_read_unlock()}, and that no SRCU primitives
be invoked from hardware interrupt handlers or from \IXacrf{nmi}
handlers.
The jury is still out as to how much of a problem is presented by
this restriction, and as to how it can best be handled.

However, in the common case where references are held within the confines
of a single CPU or task, RCU can be used as high-performance and highly
scalable reference-counting mechanism.

As shown in \cref{fig:defer:Relationships Between RCU Use Cases},
quasi reference counts add RCU readers as individual or bulk
reference counts, possibly also bridging to reference counters
in corner cases.

\subsubsection{Quasi Multi-Version Concurrency Control}
\label{sec:defer:Quasi Multi-Version Concurrency Control}

RCU can also be thought of as a simplified multi-version concurrency
control (MVCC) mechanism with weak consistency criteria.
The multi-version aspects were touched upon in
\cref{sec:defer:Maintain Multiple Versions of Recently Updated Objects}.
However, in its native form, RCU provides version consistency only
within a given RCU-protected data element.

Nevertheless, there are situations where consistency and fresh data are
required across multiple data elements.
Fortunately, there are a number of approaches that avoid inconsistency
and stale data, including the following:

\begin{enumerate}
\item	Enclose RCU readers within sequence-locking readers, forcing
	the RCU readers to be retried should an update occur,
	as described in
	\cref{sec:together:Correlated Data Elements}
	and
	\cref{sec:together:Atomic Move}.
\item	Place the data that must be consistent into a single element
	of a linked data structure, and refrain from updating those
	fields within any element visible to RCU readers.
	RCU readers gaining a reference to any such element are then
	guaranteed to see consistent values.
	See \cref{sec:together:Correlated Fields} for additional details.
\item	Use a per-element lock that guards a ``deleted'' flag to allow
	RCU readers to reject stale
	data~\cite{PaulEdwardMcKenneyPhD,Arcangeli03}.
\item	Provide an existence flag that is referenced by all data elements
	whose update is to appear atomic to RCU
	readers~\cite{PaulEMcKennneyAtomicTreeN4037,PaulEMcKennneyAtomicTreeCPPCON2014,PaulEMcKenneyIssaquahUpdate2015,PaulEMcKenney2016IssaquahACMApp,PaulEMcKenney2016IssaquahCPPCON}.
\item	Use one of a wide range of counter-based
	methods~\cite{PaulEMcKenney2008cyclicRCU,PaulEMcKenney2010cyclicRCU,PaulEMcKenney2011cyclicparallelRCU,PaulEMcKenney2014cyclicRCU,Matveev:2015:RLS:2815400.2815406,Kim:2019:MSR:3297858.3304040}.
	In these approaches, updaters maintain a version number and maintain links
	to old versions of a given piece of data.
	Readers take a snapshot of the current version number, and, if necessary,
	traverse the links to find a version consistent with that snapshot.
\end{enumerate}

In short, when using RCU to approximate multi-version concurrency control,
you only pay for the level of consistency that you actually need.

As shown in \cref{fig:defer:Relationships Between RCU Use Cases},
quasi multi-version concurrency control is based on existence guarantees,
adding read-side snapshot operations and constraints on readers and
writers, the exact form of the constraint being dictated by the
consistency requirements, as summarized above.

\subsubsection{RCU Usage Summary}
\label{sec:defer:RCU Usage Summary}

At its core, RCU is nothing more nor less than an API that provides:

\begin{enumerate}
\item	A publish-subscribe mechanism for adding new data,
\item	A way of waiting for pre-existing RCU readers to finish, and
\item	A discipline of maintaining multiple versions to permit change
	without harming or unduly delaying concurrent RCU readers.
\end{enumerate}

That said, it is possible to build higher-level constructs on top of RCU,
including the various use cases described in the earlier sections.
Furthermore, I have no doubt that new use cases will continue to be
found for RCU, as well as for any of a number of other synchronization
primitives.
And so it is that RCU's use cases are conceptually more complex than
is RCU itself, as hinted on
\cpageref{sec:defer:Mysteries RCU Use Cases}.

\QuickQuiz{
	Which of these use cases best describes the Pre-BSD routing
	example in
	\cref{sec:defer:RCU for Pre-BSD Routing}?
}\QuickQuizAnswer{
	Pre-BSD routing could be argued to fit into either
	quasi reader-writer lock, quasi reference count, or
	quasi multi-version concurrency control.
	The code is the same either way.
	This is similar to things like \co{atomic_inc()}, another tool
	that can be put to a great many uses.
}\QuickQuizEnd

\begin{figure}
\centering
\resizebox{3in}{!}{\includegraphics{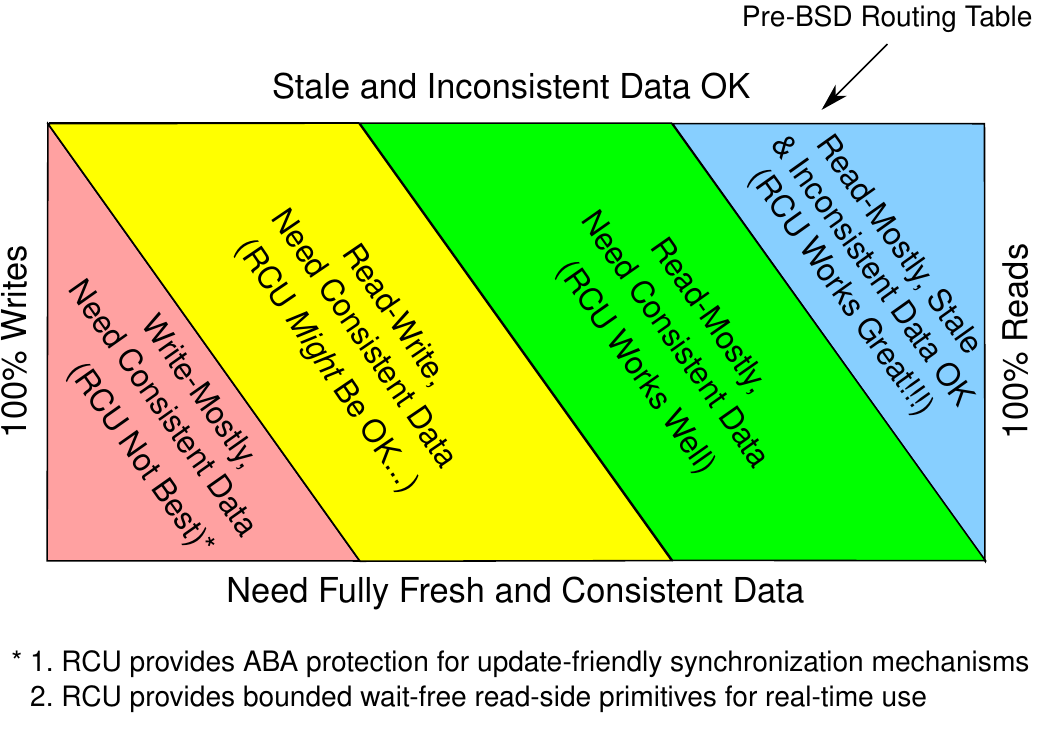}}
\caption{RCU Areas of Applicability}
\label{fig:defer:RCU Areas of Applicability}
\end{figure}

In the meantime,
\cref{fig:defer:RCU Areas of Applicability}
shows some rough rules of thumb on where RCU is most helpful.

As shown in the blue box in the upper-right corner of the figure, RCU
works best if you have read-mostly data where stale and inconsistent
data is permissible (but see below for more information on stale and
inconsistent data).
The canonical example of this case in the Linux kernel is routing tables.
Because it may have taken many seconds or even minutes for the
routing updates to propagate across the Internet, the system
has been sending packets the wrong way for quite some time.
Having some small probability of continuing to send some of them the wrong
way for a few more milliseconds is almost never a problem.

If you have a read-mostly workload where consistent data is required,
RCU works well, as shown by the green ``read-mostly, need consistent data''
box.
One example of this case is the Linux kernel's mapping from user-level
System-V semaphore IDs to the corresponding in-kernel data structures.
Semaphores tend to be used far more frequently than they are created
and destroyed, so this mapping is read-mostly.
However, it would be erroneous to perform a semaphore operation on
a semaphore that has already been deleted.
This need for consistency is handled by using the lock in the
in-kernel semaphore data structure, along with a ``deleted''
flag that is set when deleting a semaphore.
If a user ID maps to an in-kernel data structure with the
``deleted'' flag set, the data structure is ignored, so that
the user ID is flagged as invalid.

Although this requires that the readers acquire a lock for the
data structure representing the semaphore itself,
it allows them to dispense with locking for the
mapping data structure.
The readers therefore locklessly
traverse the tree used to map from ID to data structure,
which in turn greatly improves performance, scalability, and
real-time response.

As indicated by the yellow ``read-write'' box, RCU can also be useful
for read-write
workloads where consistent data is required, although usually in
conjunction with a number of other synchronization primitives.
For example, the directory-entry cache in recent Linux kernels uses RCU in
conjunction with sequence locks, per-CPU locks, and per-data-structure
locks to allow lockless traversal of pathnames in the common case.
Although RCU can be very beneficial in this read-write case, the
corresponding code is often more complex than that of the read-mostly
cases.

Finally, as indicated by the red box in the lower-left corner
of the figure, update-mostly workloads requiring consistent
data are rarely good places to use RCU, though there are some
exceptions~\cite{MathieuDesnoyers2012URCU}.
For example, as noted in
\cref{sec:defer:Type-Safe Memory},
within the Linux kernel, the \co{SLAB_TYPESAFE_BY_RCU}
slab-allocator flag provides type-safe memory to RCU readers, which can
greatly simplify \IXacrl{nbs} and other lockless
algorithms.
In addition, if the rare readers are on critical code paths on real-time
systems, use of RCU for those readers might provide real-time response
benefits that more than make up for the increased update-side overhead,
as discussed in \cref{sec:advsync:The Role of RCU}.

In short, RCU is an API that includes a publish-subscribe mechanism for
adding new data, a way of waiting for pre-existing RCU readers to finish,
and a discipline of maintaining multiple versions to allow updates to
avoid harming or unduly delaying concurrent RCU readers.
This RCU API is best suited for read-mostly situations, especially if
stale and inconsistent data can be tolerated by the application.

% defer/rcurelated.tex
% mainfile: ../perfbook.tex
% SPDX-License-Identifier: CC-BY-SA-3.0

\subsection{RCU Related Work}
\label{sec:defer:RCU Related Work}
\OriginallyPublished{Section}{sec:defer:RCU Related Work}{RCU Related Work}{Linux Weekly News}{PaulEMcKenney2014ReadMostly}
\OriginallyPublished{Section}{sec:defer:RCU Related Work}{RCU Related Work}{Linux Weekly News}{PaulEMcKenney2015ReadMostly}

The first known mention of anything resembling RCU took the form of a bug
report from
\ppl{Donald}{Knuth}~\cite[page 413 of Fundamental Algorithms]{Knuth73}
against \ppl{Joseph}{Weizenbaum}'s SLIP list-processing facility for
FORTRAN~\cite{Weizenbaum:1963:SLP:367593.367617}.
Knuth was justified in reporting the bug, as SLIP had no notion of
any sort of grace-period guarantee.

The first known non-bug-report mention of anything resembling RCU appeared
in \pplsur{H. T.}{Kung}'s and \pplsur{Philip L.}{Lehman}'s landmark
paper~\cite{Kung80}.
There was some additional use of this technique in
academia~\cite{Manber82,Manber84,BarbaraLiskov1988ArgusCACM,Pugh90,Andrews91textbook,Pu95a,Cowan96a,Rastogi:1997:LPV:645923.671017,Gamsa99},
but much of the work in this area was instead carried out by
practitioners~\cite{RichardRashid87a,Hennessy89,Jacobson93,AjuJohn95,Slingwine95,Slingwine97,Slingwine98,McKenney98}.

\QuickQuiz{
	Garbage collectors?
	Passive serialization?
	System reference points?
	Quiescent states?
	Aging?
	Generations?
	Why on earth couldn't the knuckleheads working on these early
	papers bring themselves to agree on a common terminology???
}\QuickQuizAnswer{
	There were multiple independent inventions of mechanisms
	vaguely resembling RCU\@.
	Each group of inventors was unaware of the others, so each
	made up its own terminology as a matter of course.
	And the different terminology made it quite difficult for
	any one group to find any of the others.

	Sorry, but life is like that sometimes!
}\QuickQuizEnd

By the year 2000, the initiative had passed to open-source projects,
most notably the Linux kernel
community~\cite{RustyRussell2000a,RustyRussell2000b,McKenney01b,McKenney01a,McKenney02a,Arcangeli03}.\footnote{
	A list of citations with well over 200 entries may be found in
	\co{bib/RCU.bib} in the {\LaTeX} source for this book.}
RCU was accepted into the Linux kernel in late 2002, with many subsequent
improvements for scalability, robustness, real-time response, energy
efficiency, and specialized use cases.
As of 2023, Linux-kernel RCU is still under active development.

\QuickQuiz{
	Why didn't Kung's and Lehman's paper result in immediate use
	of RCU?
}\QuickQuizAnswer{
	One reason is that Kung and Lehman were simply ahead of their
	time.
	Another reason was that their approach, ground-breaking though
	it was, did not take a number of software-engineering and
	performance issues into account.

	To see that they were ahead of their time, consider that three
	years after their paper was published, Paul was working on a
	PDP-11 system running BSD 2.8.
	This system lacked any sort of automatic configuration, which
	meant that any hardware modification, including adding a new
	disk drive, required hand-editing and rebuilding the kernel.
	Furthermore, this was a single-CPU system, which meant that
	full-system synchronization was a simple matter of disabling
	interrupts.

	Fast-forward a number of years, and multicore systems permitting
	runtime changes in hardware configuration were commonplace.
	This meant that the hardware configuration data that was implicitly
	represented in 1980s kernel source code was now a mutable
	data structure that was accessed on every I/O\@.
	Such data structures rarely change, but could change at any time.
	And this read-mostly property applies to many other new-age
	data structures, including those concerning networking (rare in
	the 1980s), security policies (physical locks in the 1980s),
	software configuration (immutable at runtime in the 1980s),
	and much else besides.
	There was thus much more opportunity for RCU to demonstrate its
	benefits in the 1990s and 2000s than there was in the 1980s.

	Kung's and Lehman's software-engineering sins included failing
	to mark readers (thus presenting debugging difficulties),
	failing to provide a clean RCU API (thus tying their mechanism
	to a specific data structure), and failing to allow for any
	post-grace-period operation other than freeing memory (thus
	disallowing a number of RCU use cases).

	Kung and Lehman presented two garbage-collection strategies.
	The first waited for all processes running at a given time
	to terminate, which represented another software-engineering
	sin that ruled out their mechanism's use in software that
	runs indefinitely.
	The second used per-object reference counting, which greatly
	complicates their read-side code (thus representing yet
	another software-engineering sin), and, on modern hardware,
	results in severe cache-miss overhead (thus representing a
	performance sin, see for example
	\cref{fig:defer:Performance of RCU vs. Reference Counting,fig:defer:Performance of Preemptible RCU vs. Reference Counting}).

	Despite this long list of software-engineering and performance
	sins, Kung's and Lehman's paper remains a truly impressive piece
	of work, especially considering that much of the later work
	(both independent and not) committed these same sins, plus others
	as well.
}\QuickQuizEnd

However, in the mid 2010s, there was a welcome upsurge in RCU research
and development across a number of communities and
institutions~\cite{FransKaashoek2015ParallelOSHistory}.
\Cref{sec:defer:RCU Uses} describes uses of RCU,
\cref{sec:defer:RCU Implementations} describes RCU implementations
(as well as work that both creates and uses an implementation),
and finally,
\cref{sec:defer:RCU Validation} describes verification and validation
of RCU and its uses.

\subsubsection{RCU Uses}
\label{sec:defer:RCU Uses}

\ppl{Phil}{Howard} and \ppl{Jon}{Walpole} of Portland State University
(PSU) have
applied RCU to red-black
trees~\cite{PhilHowardPhD,PhilHoward2011RCUTMRBTree} combined with updates
synchronized using software transactional memory.
\ppl{Josh}{Triplett} and \ppl{Jon}{Walpole} (again of PSU)
applied RCU to resizable
hash tables~\cite{JoshTriplettPhD,Triplett:2011:RPHash,JonCorbet2014RCUhash1,JonCorbet2014RCUhash2}.
Other RCU-protected resizable hash tables have been created by
\ppl{Herbert}{Xu}~\cite{HerbertXu2010RCUResizeHash} and by
\ppl{Mathieu}{Desnoyers}~\cite{PaulMcKenney2013LWNURCUhash}.

\ppl{Austin}{Clements}, \ppl{Frans}{Kaashoek}, and \ppl{Nickolai}{Zeldovich}
of MIT created an RCU-optimized balanced binary tree
(Bonsai)~\cite{AustinClements2012RCULinux:mmapsem}, and applied this
tree to the Linux kernel's VM subsystem in order to reduce read-side
contention on the Linux kernel's \co{mmap_sem}.
This work resulted in order-of-magnitude speedups and scalability up to
at least 80 CPUs for a microbenchmark featuring large numbers of minor
page faults.
This is similar to a patch developed earlier by
\ppl{Peter}{Zijlstra}~\cite{PeterZijlstra2014SpeculativePageFault}, and both
were limited by the fact that, at the time, filesystem data structures
were not safe for RCU readers.
\pplsur{Austin}{Clements} et al.\ avoided this limitation by
optimizing the page-fault
path for anonymous pages only.
More recently, filesystem data structures have been made safe for RCU
readers~\cite{JonathanCorbet2010dcacheRCU,JonathanCorbet2011dcacheRCUbug},
so perhaps this work can be implemented for all page types, not just
anonymous pages---\ppl{Peter}{Zijlstra} has, in fact, recently prototyped
exactly this, and \ppl{Laurent}{Dufour} \ppl{Michel}{Lespinasse} have
continued work along these lines.
For their part, \ppl{Matthew}{Wilcox} and \ppl{Liam}{Howlett} are working
towards use of RCU to enable fine-grained locking of and lockless access
to other memory-management data structures.

\ppl{Yandong}{Mao} and \ppl{Robert}{Morris} of MIT and \ppl{Eddie}{Kohler} of
Harvard University created another RCU-protected tree named
Masstree~\cite{Mao:2012:CCF:2168836.2168855} that combines ideas from B+
trees and tries.
Although this tree is about 2.5x slower than an RCU-protected hash table,
it supports operations on key ranges, unlike hash tables.
In addition, Masstree supports efficient storage of objects with long
shared key prefixes and, furthermore, provides persistence via logging
to mass storage.

The paper notes that Masstree's performance rivals that of memcached, even
given that Masstree is persistently storing updates and memcached is not.
The paper also compares Masstree's performance to the persistent
datastores MongoDB, VoltDB, and Redis, reporting significant performance
advantages for Masstree, in some cases exceeding two orders of magnitude.
Another paper~\cite{Tu:2013:STM:2517349.2522713}, by \ppl{Stephen}{Tu},
\ppl{Wenting}{Zheng}, \ppl{Barbara}{Liskov}, and \ppl{Samuel}{Madden}
of MIT and \pplsur{Eddie}{Kohler},
applies Masstree to an in-memory database named Silo, achieving 700K
transactions per second (42M transactions per minute) on a well-known
transaction-processing benchmark.
Interestingly enough, Silo guarantees linearizability without incurring
the overhead of grace periods while holding locks.

\ppl{Maya}{Arbel} and \ppl{Hagit}{Attiya} of Technion took a more rigorous
approach~\cite{MayaArbel2014RCUtree} to an RCU-protected search tree that,
like Masstree, allows concurrent updates.
This paper includes a proof of correctness, including proof that all
operations on this tree are \IX{linearizable}.
Unfortunately, this implementation achieves linearizability by incurring
the full latency of grace-period waits while holding locks, which degrades
scalability of update-only workloads.
One way around this problem is to abandon
linearizability~\cite{AndreasHaas2012FIFOisnt,PaulEMcKennneyAtomicTreeN4037},
however, Arbel and Attiya instead created an RCU variant that reduces
low-end grace-period latency.
Of course, nothing comes for free, and this RCU variant appears to hit
a scalability limit at about 32 CPUs.
Although there is much to be said for dropping linearizability, thus
gaining both performance and scalability, it is very good to see academics
experimenting with alternative RCU implementations.

\subsubsection{RCU Implementations}
\label{sec:defer:RCU Implementations}

\ppl{Timothy}{Harris} created a time-based user-space
RCU~\cite{TimothyLHarris2001} that improves on those created previously
by Jacobson~\cite{Jacobson93} and John~\cite{AjuJohn95}.
These prior two time-based approaches each assume a sharp upper bound on
reader duration, which can work correctly in hard real-time systems.
In non-real-time systems, this type of approach is subject to failure
when readers are interrupted, preempted, or otherwise delayed.
However, the fact that such a failure-prone implementation would be
independently invented twice shows the depth of the need for RCU-like
mechanisms.
\ppl{Timothy}{Harris} improves upon these two earlier efforts by
requiring each reader to take a snapshot of a global timebase before
starting its read-side traversal.
Freeing a reader-visible object is then deferred until all processes'
reader snapshots indicate a time following that of the removal of that object.
However, global timebases can be expensive and inaccurate on some systems.

\ppl{Keir}{Fraser} created a user-space RCU named \IXacr{ebr} for use in
non-blocking synchronization and software transactional
memory~\cite{KeirAnthonyFraserPhD,UCAM-CL-TR-579,KeirFraser2007withoutLocks}.
This work improves on that of \ppl{Timothy}{Harris} by replacing the
global clock with a software counter, thus eliminating much of the
expense and all of the inaccuracy associated with commodity-system
global clocks of that time.
Interestingly enough, this work cites Linux-kernel RCU on the one hand,
but also inspired the name \acr{qsbr} for the original non-preemptible
Linux-kernel RCU implementation.

\ppl{Mathieu}{Desnoyers} created a user-space RCU for use in
tracing~\cite{MathieuDesnoyers2009URCU,MathieuDesnoyersPhD,MathieuDesnoyers2012URCU,PaulMcKenney2013LWNURCU,PaulMcKenney2013LWNURCUhash,PaulMcKenney2013LWNURCUhashAPI,PaulMcKenney2013LWNURCUqueuestack,PaulMcKenney2013LWNURCUqueuestackAPI,PaulMcKenney2013LWNURCUatomicop,PaulMcKenney2013LWNURCUmenagerie,PaulMcKenney2013LWNURCUAPI,PaulMcKenney2013LWNURCUlist,PaulMcKenney2013LWNURCUmenagerieRCU},
which has seen use in a number of projects~\cite{MikeDay2013RCUqemu}.

Researchers at Charles University in Prague have also been
working on RCU implementations, including dissertations by
\ppl{Andrej}{Podzimek}~\cite{AndrejPodzimek2010masters} and
\ppl{Adam}{Hraska}~\cite{AdamHraska2013RCUHelenOS}.

\ppl{Yujie}{Liu} (Lehigh University), \ppl{Victor}{Luchangco} (Oracle Labs), and
\ppl{Michael}{Spear} (also Lehigh)~\cite{Liu:2013:MSA:2549695.2549732}
pressed scalable non-zero indicators
(SNZI)~\cite{FaithEllen:2007:SNZI} into service as a grace-period
mechanism.
The intended use is to implement software transactional memory
(see \cref{sec:future:Transactional Memory}), which
imposes linearizability requirements, which in turn seems to
limit scalability.

RCU-like mechanisms are also finding their way into Java.
\pplsur{KC}{Sivaramakrishnan} et al.~\cite{Sivaramakrishnan:2012:ERB:2258996.2259005}
use an RCU-like mechanism to eliminate the read barriers that are
otherwise required when interacting with Java's garbage collector,
resulting in significant performance improvements.

\ppl{Ran}{Liu}, \ppl{Heng}{Zhang}, and \ppl{Haibo}{Chen} of
Shanghai Jiao Tong University
created a specialized variant of RCU that they used for an optimized
``passive reader-writer lock''~\cite{RanLiu2014PassiveRWLock}, similar to
those created by \ppl{Gautham}{Shenoy}~\cite{GauthamShenoy2006RCUrwlock} and
\ppl{Srivatsa}{Bhat}~\cite{SrivatsaSBhat2014RCUrwlock}.
The Liu et al.\ paper is interesting from a number of
perspectives~\cite{PaulEMcKenney2014ReadMostly}.

\ppl{Mike}{Ash} posted~\cite{MikeAsh2015Apple} a description of an RCU-like
primitive in Apple's Objective-C runtime.
This approach identifies read-side critical sections via designated
code ranges, thus qualifying as another method of achieving
zero read-side overhead, albeit one that poses some interesting
practical challenges for large read-side critical sections that
span multiple functions.

\ppl{Pedro}{Ramalhete} and \ppl{Andreia}{Correia}~\cite{PedroRmalhete2015PoorMansRCU}
produced ``Poor Man's RCU'', which, despite using a pair of
\IXhpl{reader-writer}{lock}, manages to provide \IXalt{lock-free}{lock free}
\IXpl{forward-progress guarantee} to
readers~\cite{PaulEMcKenney2015ReadMostly}.

\ppl{Maya}{Arbel} and \ppl{Adam}{Morrison}~\cite{Arbel:2015:PRR:2858788.2688518}
produced ``Predicate RCU'', which works hard to reduce grace-period
duration in order to efficiently support algorithms that hold
update-side locks across grace periods.
This results in reduced batching of updates into grace periods
and reduced scalability, but does succeed in providing short
grace periods.

\QuickQuiz{
	Why not just drop the lock before waiting for the grace
	period, or using something like \co{call_rcu()}
	instead of waiting for a grace period?
}\QuickQuizAnswer{
	The authors wished to support \IX{linearizable} tree
	operations, so that concurrent additions to, deletions
	from, and searches of the tree would appear to execute
	in some globally agreed-upon order.
	In their search trees, this requires holding locks
	across grace periods.
	(It is probably better to drop linearizability as a
	requirement in most cases, but linearizability is a
	surprisingly popular (and costly!\@) requirement.)
}\QuickQuizEnd

\ppl{Alexander}{Matveev} (MIT), \ppl{Nir}{Shavit} (MIT and Tel-Aviv University),
\ppl{Pascal}{Felber} (University of Neuch\^{a}tel), and \ppl{Patrick}{Marlier} (also
University of Neuch\^{a}tel)~\cite{Matveev:2015:RLS:2815400.2815406}
produced an RCU-like mechanism that can be thought of as
software transactional memory that explicitly marks
read-only transactions.
Their use cases require holding locks across grace periods, which limits
scalability~\cite{PaulEMcKenney2015ReadMostly,PaulEMcKenney2015ReadMostlySidebar}.
This appears to be the first academic RCU-related work to
make good use of the \co{rcutorture} test suite, and also the
first to have submitted a performance improvement to Linux-kernel
RCU, which was accepted into v4.4.

\ppl{Alexander}{Matveev}'s RLU was followed up by MV-RLU from
\ppl{Jaeho}{Kim} et al.~\cite{Kim:2019:MSR:3297858.3304040}.
This work improves scalability over RLU by permitting multiple concurrent
updates, by avoiding holding locks across grace periods, and by using
asynchronous grace periods, for example, \co{call_rcu()} instead of
\co{synchronize_rcu()}.
This paper also made some interesting performance-evaluation choices that
are discussed further in
\cref{sec:future:Deferred Reclamation}
on
\cpageref{sec:future:Deferred Reclamation}.

\ppl{Adam}{Belay} et al.~created an RCU implementation that guards the
data structures used by TCP/IP's address-resolution protocol (ARP)
in their IX operating system~\cite{Belay:2016:IOS:3014162.2997641}.

\ppl{Geoff}{Romer} and \ppl{Andrew}{Hunter} (both at Google) proposed
a cell-based API for RCU
protection of singleton data structures for inclusion in the
C++ standard~\cite{GeoffRomer2018C++DeferredReclamationP0561R4}.

\ppl{Dimitrios}{Siakavaras} et al.~have applied
HTM and RCU to search trees~\cite{Siakavaras2017CombiningHA,DimitriosSiakavaras2020RCU-HTM-B+Trees},
\ppl{Christina}{Giannoula} et al.~have used HTM and RCU to color
graphs~\cite{ChristinaGiannoula2018HTM-RCU-graphcoloring},
and
\ppl{SeongJae}{Park} et al.~have used HTM and RCU to optimize high-contention
locking on \IXacr{numa} systems.

\ppl{Alex}{Kogan} et al.~applied RCU to the construction of range locking
for scalable address spaces~\cite{AlexKogan2020RCUrangelocks}.

Production uses of RCU are listed in
\cref{sec:defer:Production Uses of RCU}.

\subsubsection{RCU Validation}
\label{sec:defer:RCU Validation}

In early 2017, it is commonly recognized that almost any bug is a potential
security exploit, so validation and verification are first-class concerns.

Researchers at Stony Brook University have produced an RCU-aware data-race
detector~\cite{AbhinavDuggal2010Masters,JustinSeyster2012PhD,Seyster:2011:RFA:2075416.2075425}.
\ppl{Alexey}{Gotsman} of IMDEA, \ppl{Noam}{Rinetzky} of Tel Aviv University,
and \ppl{Hongseok}{Yang} of the University of Oxford have published a
paper~\cite{AlexeyGotsman2012VerifyGraceExtended} expressing the formal
semantics of RCU in terms of separation logic, and have continued with
other aspects of concurrency.

\ppl{Joseph}{Tassarotti} (Carnegie-Mellon University), \ppl{Derek}{Dreyer} (Max
Planck Institute for Software Systems), and \ppl{Viktor}{Vafeiadis}
(also MPI-SWS)~\cite{JosephTassarotti2015RCUproof}
produced a manual formal proof of correctness of the \IXacrf{qsbr}
variant of userspace
RCU~\cite{MathieuDesnoyers2009URCU,MathieuDesnoyers2012URCU}.
\ppl{Lihao}{Liang} (University of Oxford), \pplmdl{Paul E.}{McKenney} (IBM),
\ppl{Daniel}{Kroening}, and \ppl{Tom}{Melham}
(both also Oxford)~\cite{LihaoLiang2016VerifyTreeRCU}
used the \IXacrf{cbmc}~\cite{EdmundClarke2004CBMC}
to produce a mechanical proof of correctness of a significant portion
of Linux-kernel Tree RCU\@.
\ppl{Lance}{Roy}~\cite{LanceRoy2017CBMC-SRCU} used CBMC to produce a similar
proof of correctness for a significant portion of Linux-kernel
sleepable RCU (SRCU)~\cite{PaulEMcKenney2006c}.
Finally, \ppl{Michalis}{Kokologiannakis} and \ppl{Konstantinos}{Sagonas}
(National Technical University of
Athens)~\cite{MichalisKokologiannakis2017NidhuggRCU,MichalisKokologiannakis2019RCUstatelessModelCheck}
used the Nighugg tool~\cite{CarlLeonardsson2014Nidhugg}
to produce a mechanical proof of correctness of a somewhat larger
portion of Linux-kernel Tree RCU\@.

None of these efforts located any bugs other than bugs injected into
RCU specifically to test the verification tools.
In contrast,
\ppl{Alex}{Groce} (Oregon State University), \ppl{Iftekhar}{Ahmed},
\ppl{Carlos}{Jensen} (both also OSU), and \pplmdl{Paul E.}{McKenney}
(IBM)~\cite{Groce:2015:VMC:2916135.2916190}
automatically mutated Linux-kernel RCU's source code to test the
coverage of the \co{rcutorture} test suite.
The effort found several holes in this suite's coverage, one of which
was hiding a real bug (since fixed) in Tiny RCU\@.

With some luck, all of this validation work will eventually result in
more and better tools for validating concurrent code.

% defer/whichtochoose.tex
% mainfile: ../perfbook.tex
% SPDX-License-Identifier: CC-BY-SA-3.0

\section{Which to Choose?}
\label{sec:defer:Which to Choose?}
\epigraph{Choose always the way that seems the best, however rough it
	  may be; custom will soon render it easy and agreeable.}
	  {Pythagoras}

\Cref{sec:defer:Which to Choose? (Overview)}
provides a high-level overview and then
\cref{sec:defer:Which to Choose? (Details)}
provides a more detailed view
of the differences between the deferred-processing techniques presented
in this chapter.
This discussion assumes a linked data structure that is large enough
that readers do not hold references from one traversal to another,
and where elements might be added to and removed from the structure
at any location and at any time.
\Cref{sec:defer:Which to Choose? (Production Use)}
then points out a few publicly visible production uses of
\IXpl{hazard pointer}, sequence locking, and RCU\@.
This discussion should help you to make an informed choice between
these techniques.

\subsection{Which to Choose?
			      (Overview)}
\label{sec:defer:Which to Choose? (Overview)}

\begin{table*}
\rowcolors{1}{}{lightgray}
\renewcommand*{\arraystretch}{1.25}
\footnotesize
\centering\OneColumnHSpace{-.3in}
\ebresizewidth{
\begin{tabularx}{5.3in}{>{\raggedright\arraybackslash}p{1.1in}
    >{\raggedright\arraybackslash}p{1.0in}
    >{\raggedright\arraybackslash}X
    >{\raggedright\arraybackslash}X
    >{\raggedright\arraybackslash}p{.9in}}
	\toprule
	Property
		& Reference Counting
			& Hazard Pointers
				& Sequence Locks
					& RCU \\
%		  RC	  HP	  SL	  RCU \\
	\midrule
	Readers
		& Slow and unscalable
			& Fast and scalable
				& Fast and scalable
					& Fast and scalable \\
	Memory Overhead
		& Counter per object
			& Pointer per reader per object
				& No protection
					& None \\
	Duration of Protection
		& Can be long
			& Can be long
				& No protection
					& User must bound duration \\
	Need for Traversal Retries
		& If object deleted
			& If object deleted
				& If any update
					& Never \\
	\bottomrule
\end{tabularx}
}
\caption{Which Deferred Technique to Choose?
					     (Overview)}
\label{tab:defer:Which Deferred Technique to Choose? (Overview)}
\end{table*}

\Cref{tab:defer:Which Deferred Technique to Choose? (Overview)}
shows a few high-level properties that distinguish the deferred-reclamation
techniques from one another.

The ``Readers'' row summarizes the results presented in
\cref{fig:defer:Pre-BSD Routing Table Protected by RCU QSBR},
which shows that all but \IXalt{reference counting}{reference count}
enjoy reasonably fast and scalable readers.

The ``Memory Overhead'' row evaluates each technique's need
for external storage with which to record reader protection.
RCU relies on quiescent states, and thus needs no storage to represent
readers, whether within or outside of the object.
Reference counting can use a single integer within each object in the
structure, and no additional storage is required.
Hazard pointers require external-to-object pointers be provisioned,
and that there be sufficient pointers for each CPU or thread to
track all the objects being referenced at any given time.
Given that most hazard-pointer-based traversals require only a few
hazard pointers, this is not normally a problem in practice.
Of course, sequence locks provides no pointer-traversal protection,
which is why it is normally used on static data.

\QuickQuiz{
	Why can't users dynamically allocate the hazard pointers as they
	are needed?
}\QuickQuizAnswer{
	They can, but at the expense of additional reader-traversal
	overhead and, in some environments, the need to handle
	memory-allocation failure.
}\QuickQuizEnd

The ``Duration of Protection'' describes constraints (if any) on how
long a period of time a user may protect a given object.
Reference counting and hazard pointers can both protect objects for
extended time periods with no untoward side effects, but
maintaining an RCU reference to even one object prevents all other RCU
from being freed.
RCU readers must therefore be relatively short in order to avoid running
the system out of memory, with special-purpose implementations such
as SRCU, Tasks RCU, and Tasks Trace RCU being exceptions to this rule.
Again, sequence locks provide no pointer-traversal protection,
which is why it is normally used on static data.

The ``Need for Traversal Retries'' row tells whether a new reference to
a given object may be acquired unconditionally, as it can with RCU, or
whether the reference acquisition can fail, resulting in a retry
operation, which is the case for reference counting, hazard pointers,
and sequence locks.
In the case of reference counting and hazard pointers, retries are only
required if an attempt to acquire a reference to a given object while
that object is in the process of being deleted, a topic covered in more
detail in the next section.
Sequence locking must of course retry its critical section should it
run concurrently with any update.

\QuickQuiz{
	But don't Linux-kernel \co{kref} reference counters allow
	guaranteed unconditional reference acquisition?
}\QuickQuizAnswer{
	Yes they do, but the guarantee only applies unconditionally
	in cases where a reference is already held.
	With this in mind, please review the paragraph at the beginning of
	\cref{sec:defer:Which to Choose?}, especially the part
	saying ``large enough that readers do not hold references from
	one traversal to another''.
}\QuickQuizEnd

Of course, different rows will have different levels of importance in
different situations.
For example, if your current code is having read-side scalability problems
with hazard pointers, then it does not matter that hazard pointers can require
retrying reference acquisition because your current code already handles
this.
Similarly, if response-time considerations already limit the duration
of reader traversals, as is often the case in kernels and low-level
applications, then it does not matter that RCU has duration-limit
requirements because your code already meets them.
In the same vein, if readers must already write to the objects that they
are traversing, the read-side overhead of reference counters might
not be so important.
Of course, if the data to be protected is in statically allocated variables,
then sequence locking's inability to protect pointers is irrelevant.

Finally, there is some work on dynamically switching between hazard
pointers and RCU based on dynamic sampling of
delays~\cite{Balmau:2016:FRM:2935764.2935790}.
This defers the choice between hazard pointers and RCU to runtime,
and delegates responsibility for the decision to the software.

Nevertheless, this table should be of great help when choosing between
these techniques.
But those wishing more detail should continue on to the next section.

\subsection{Which to Choose?
			     (Details)}
\label{sec:defer:Which to Choose? (Details)}

\begin{table*}
\rowcolors{1}{}{lightgray}
\renewcommand*{\arraystretch}{1.25}
\footnotesize
\centering\OneColumnHSpace{-.3in}
\ebresizewidth{
\begin{tabularx}{5.3in}{>{\raggedright\arraybackslash}p{1.1in}
    >{\raggedright\arraybackslash}p{1.2in}
    >{\raggedright\arraybackslash}X
    >{\raggedright\arraybackslash}X
    >{\raggedright\arraybackslash}p{.9in}}
	\toprule
	Property
		& Reference Counting
			& Hazard Pointers
				& Sequence Locks
					& RCU \\
%		  RC	  HP	  SL	  RCU \\
	\midrule
	Existence Guarantees
		& Complex
			& Yes
				& No
					& Yes \\
	Updates and Readers Progress Concurrently
		& Yes
			& Yes
				& No
					& Yes \\
	Contention Among Readers
		& High
			& None
				& None
					& None \\
	Reader Per\-/Critical\-/Section Overhead
		& N/A
			& N/A
				& Two \tco{smp_mb()}
					& Ranges from none to two
					  \tco{smp_mb()} \\
	Reader Per-Object Traversal Overhead
		& Read-modify-write atomic operations, memory\-/barrier
		  instructions, and cache misses
			& \tco{smp_mb()}\parnote[*]{This \co{smp_mb()} can be
				downgraded to a compiler \co{barrier()} by using
				the Linux-kernel \co{membarrier()} system call.}
				& None, but unsafe
					& None (volatile accesses) \\
	Reader Forward Progress Guarantee
		& Lock free
			& Lock free
				& Blocking
					& Bounded wait free \\
	Reader Reference Acquisition
		& Can fail (conditional)
			& Can fail (conditional)
				& Unsafe
					& Cannot fail (unconditional) \\
	Memory Footprint
		& Bounded
			& Bounded
				& Bounded
					& Unbounded \\
	Reclamation Forward Progress
		& Lock free
			& Lock free
				& N/A
					& Blocking \\
	Automatic Reclamation
		& Yes
			& Use Case
				& N/A
					& Use Case \\
	Lines of Code
		& 94
			& 79
				& 79
					& 73 \\
	\bottomrule
\end{tabularx}
}
\begin{minipage}{\onecolumntextwidth}
	\vspace*{1ex}
	\parnotes
\end{minipage}
\caption{Which Deferred Technique to Choose?
					     (Details)}
\label{tab:defer:Which Deferred Technique to Choose?  (Details)}
\end{table*}

\Cref{tab:defer:Which Deferred Technique to Choose? (Details)}
provides more-detailed rules of thumb that can help you choose among the
four deferred-processing techniques presented in this chapter.

As shown in the ``Existence Guarantee'' row,
if you need \IXpl{existence guarantee} for linked
data elements, you must use reference counting, hazard pointers, or RCU\@.
Sequence locks do not provide existence guarantees, instead providing
detection of updates, retrying any read-side critical sections
that do encounter an update.

Of course, as shown in the ``Updates and Readers Progress Concurrently''
row, this detection of updates implies
that sequence locking does not permit updaters and readers to make forward
progress concurrently.
After all, preventing such forward progress is the whole point of using
sequence locking in the first place!
This situation points the way to using sequence locking in conjunction
with reference counting, hazard pointers, or RCU in order to provide
both existence guarantees and update detection.
In fact, the Linux kernel combines RCU and sequence locking in
this manner during pathname lookup.

The ``Contention Among Readers'', ``Reader Per-Critical-Section Overhead'',
and ``Reader Per-Object Traversal Overhead'' rows give a rough sense of
the read-side overhead of these techniques.
The overhead of reference counting can be quite large, with
contention among readers along with a fully ordered read-modify-write
atomic operation required for each and every object traversed.
Hazard pointers incur the overhead of a \IX{memory barrier}
for each data element
traversed, and sequence locks incur the overhead of a pair of memory barriers
for each attempt to execute the critical section.
The overhead of RCU implementations vary from nothing to that of a pair of
memory barriers for each read-side critical section, thus providing RCU
with the best performance, particularly for read-side critical sections
that traverse many data elements.
Of course, the read-side overhead of all deferred-processing variants can
be reduced by batching, so that each read-side operation covers more data.

\QuickQuiz{
	But didn't the answer to one of the quick quizzes in
	\cref{sec:defer:Hazard Pointers}
	say that pairwise asymmetric barriers could eliminate the
	read-side \co{smp_mb()} from hazard pointers?
}\QuickQuizAnswer{
	Yes, it did.
	However, doing this could be argued to change hazard-pointers
	``Reclamation Forward Progress'' row (discussed later) from
	lock-free to blocking because a CPU spinning with interrupts
	disabled in the kernel would prevent the update-side portion of
	the asymmetric barrier from completing.
	In the Linux kernel, such blocking could in theory be prevented
	by building the kernel with \co{CONFIG_NO_HZ_FULL}, designating
	the relevant CPUs as \co{nohz_full} at boot time, ensuring that
	only one thread was ever runnable on a given CPU at a given
	time, and avoiding ever calling into the kernel.
	Alternatively, you could ensure that the kernel was free of any
	bugs that might cause CPUs to spin with interrupts disabled.

	Given that CPUs spinning in the Linux kernel with interrupts
	disabled seems to be rather rare, one might counter-argue that
	asymmetric-barrier hazard-pointer updates are non-blocking
	in practice, if not in theory.
}\QuickQuizEnd

The ``Reader Forward Progress Guarantee'' row shows that only RCU
has a \IXalth{bounded wait-free}{bounded}{wait free}
\IX{forward-progress guarantee}, which means that
it can carry out a finite traversal by executing a bounded number of
instructions.

The ``Reader Reference Acquisition'' row indicates that only RCU is
capable of unconditionally acquiring references.
The entry for sequence locks is ``Unsafe'' because, again, sequence locks
detect updates rather than acquiring references.
Reference counting and hazard pointers both require that traversals be
restarted from the beginning if a given acquisition fails.
To see this, consider a linked list containing objects~A, B, C, and~D,
in that order, and the following series of events:

\begin{enumerate}
\item	A reader acquires a reference to object~B.
\item	An updater removes object~B, but refrains from freeing it because
	the reader holds a reference.
	The list now contains objects~A, C, and~D, and
	object~B's \co{->next} pointer is set to \co{HAZPTR_POISON}.
\item	The updater removes object~C, so that the list now contains
	objects~A and~D\@.
	Because there is no reference to object~C, it is immediately freed.
\item	The reader tries to advance to the successor of the object
	following the now-removed object~B, but the poisoned
	\co{->next} pointer prevents this.
	Which is a good thing, because object~B's \co{->next} pointer
	would otherwise point to the freelist.
\item	The reader must therefore restart its traversal from the head
	of the list.
\end{enumerate}

Thus, when failing to acquire a reference, a hazard-pointer or
reference-counter traversal must restart that traversal from the
beginning.
In the case of nested linked data structures, for example, a
tree containing linked lists, the traversal must be restarted from
the outermost data structure.
This situation gives RCU a significant ease-of-use advantage.

However, RCU's ease-of-use advantage does not come
for free, as can be seen in the ``Memory Footprint'' row.
RCU's support of unconditional reference acquisition means that
it must avoid freeing any object reachable by a given
RCU reader until that reader completes.
RCU therefore has an unbounded memory footprint, at least unless updates
are throttled.
In contrast, reference counting and hazard pointers need to retain only
those data elements actually referenced by concurrent readers.

This tension between memory footprint and acquisition
failures is sometimes resolved within the Linux kernel by combining use
of RCU and reference counters.
RCU is used for short-lived references, which means that RCU read-side
critical sections can be short.
These short RCU read-side critical sections in turn mean that the corresponding
RCU \IXpl{grace period} can also be short, which limits the memory footprint.
For the few data elements that need longer-lived references, reference
counting is used.
This means that the complexity of reference-acquisition failure only
needs to be dealt with for those few data elements:
The bulk of the reference acquisitions are unconditional, courtesy of RCU\@.
See \cref{sec:together:Refurbish Reference Counting}
for more information on combining reference counting with other
synchronization mechanisms.

The ``Reclamation Forward Progress'' row shows that hazard pointers
can provide non-blocking updates~\cite{MagedMichael04a,HerlihyLM02}.
Reference counting might or might not, depending on the implementation.
However, sequence locking cannot provide non-blocking updates, courtesy
of its update-side lock.
RCU updaters must wait on readers, which also rules out fully non-blocking
updates.
However, there are situations in which the only blocking operation is
a wait to free memory, which results in a situation that, for many
purposes, is as good as non-blocking~\cite{MathieuDesnoyers2012URCU}.

As shown in the ``Automatic Reclamation'' row, only reference
counting can automate freeing of memory, and even then only
for non-cyclic data structures.
Certain use cases for hazard pointers and RCU can provide automatic
reclamation using \emph{link counts}, which can be thought of as
reference counts, but applying only to incoming links from other
parts of the data structure~\cite{MagedMichael2018FollyHazptr}.

Finally, the ``Lines of Code'' row shows the size of the Pre-BSD
Routing Table implementations, giving a rough idea of relative ease of use.
That said, it is important to note that the reference-counting and
sequence-locking implementations are buggy, and that a correct
reference-counting implementation is considerably
more complex~\cite{Valois95a,MagedMichael95a}.
For its part, a correct sequence-locking implementation requires
the addition of some other synchronization mechanism, for example,
hazard pointers or RCU, so that sequence locking detects concurrent
updates and the other mechanism provides safe reference acquisition.

As more experience is gained using these techniques, both separately
and in combination, the rules of thumb laid out in this section will
need to be refined.
However, this section does reflect the current state of the art.

\subsection{Which to Choose?
			     (Production Use)}
\label{sec:defer:Which to Choose? (Production Use)}

This section points out a few publicly visible production uses of
hazard pointers, sequence locking, and RCU\@.
Reference counting is omitted, not because it is unimportant, but rather
because it is not only used pervasively, but heavily documented in textbooks
going back a half century.
One of the hoped-for benefits of listing production uses of these other
techniques is to provide examples to study---or to find bugs in, as
the case may be.\footnote{
	Kudos to Mathias Stearn, Matt Wilson, David Goldblatt, LiveJournal
	user fanf, Nadav Har'El, Avi Kivity, Dmitry Vyukov, Raul Guitterez
	S., Twitter user @peo3, Paolo Bonzini, and Thomas Monjalon for
	locating a great many of these use cases.}

\subsubsection{Production Uses of Hazard Pointers}
\label{sec:defer:Production Uses of Hazard Pointers}

In 2010, Keith Bostic added hazard pointers to
WiredTiger~\cite{KeithBostic2010WiredTigerhazptr}.
MongoDB 3.0, released in 2015, included WiredTiger and thus hazard pointers.

In 2011, Samy Al Bahra added hazard pointers to the Concurrency Kit
library~\cite{SamyAlBahra2011ckhp}.

In 2014, Maxim Khizhinsky added hazard pointers to
libcds~\cite{MaximKhizhinsky2014libcdsHazptr}.

In 2015, David Gwynne introduced shared reference pointers, a form
of hazard pointers, to OpenBSD~\cite{DavidGwynne2015srp}.

In 2017--2018, the Rust-language
\co{arc-swap}~\cite{MichalVaner2018arc-swapHazptr} and
\co{conc}~\cite{crates.io.user.ticki2017concHazptr}
crates rolled their own implementations of hazard pointers.

In 2018, Maged Michael added hazard pointers to Facebook's Folly
library~\cite{MagedMichael2018FollyHazptr}, where it is used heavily.

\subsubsection{Production Uses of Sequence Locking}
\label{sec:defer:Production Uses of Sequence Locking}

The Linux kernel added sequence locking to v2.5.60 in
2003~\cite{JonathanCorbet2003seqlock}, having been generalized from
an ad-hoc technique used in x86's implementation of the
\co{gettimeofday()} system call.

In 2011, Samy Al Bahra added sequence locking to the Concurrency Kit
library~\cite{SamyAlBahra2011ckseqlock}.

Paolo Bonzini added a simple sequence-lock to the QEMU emulator in
2013~\cite{PaoloBonzini2013QEMUseqlock}.

Alexis Menard abstracted a sequence-lock implementation in Chromium
in 2016~\cite{AlexisMenard2016ChromiumSeqLock}.

A simple sequence locking implementation was added to \co{jemalloc()}
in 2018~\cite{DavidGoldblatt2018seqlock}.
The eigen library also has a special-purpose queue that is managed by
a mechanism resembling sequence locking.

\subsubsection{Production Uses of RCU}
\label{sec:defer:Production Uses of RCU}

IBM's VM/XA is adopted passive serialization, a mechanism similar to
RCU, some time in the 1980s~\cite{Hennessy89}.

DYNIX/ptx adopted RCU in 1993~\cite{McKenney98,Slingwine95}.

The Linux kernel adopted Dipankar Sarma's implementation of RCU in
2002~\cite{Torvalds2.5.43}.

The userspace RCU project started in 2009~\cite{MathieuDesnoyers2009URCU}.

The Knot DNS project started using the userspace RCU
library in 2010~\cite{LubosSlovak2010KnotDNSRCU}.
That same year, the OSv kernel added an RCU
implementation~\cite{AviKivity2013OSvRCU},
later adding an RCU-protected linked list~\cite{AviKivity2013OSvRCUlist}
and an RCU-protected hash table~\cite{AviKivity2013OSvRCUhash}.

In 2011, Samy Al Bahra added epochs
(a form of RCU~\cite{UCAM-CL-TR-579,KeirFraser2007withoutLocks})
to the Concurrency Kit
library~\cite{SamyAlBahra2011ckEpoch}.

NetBSD began using the aforementioned passive serialization with v6.0 in
2012~\cite{NetBSD2012pserialize}.
Among other things, passive serialization is used in
NetBSD packet filter (NPF)~\cite{MindaugasRasiukevicius2014NPFRCU}.

Paolo Bonzini added RCU support to the QEMU emulator in 2015 via a
friendly fork of the userspace RCU
library~\cite{MikeDay2013RCUqemu,PaoloBonzini2013QEMURCU}.

In 2015, Maxim Khizhinsky added RCU to
libcds~\cite{MaxKhiszinsky2015C++RCU}.

Mindaugas Rasiukevicius implemented libqsbr in 2016, which features
\IXacr{qsbr} and \IXacrf{ebr}~\cite{MindaugasRasiukevicius2016libqsbr},
both of which are types of implementations of RCU\@.

Sheth et al.~\cite{HarshalSheth2016goRCU}
demonstrated the value of leveraging Go's garbage
collector to provide RCU-like functionality, and
the Go programming language provides a \co{Value} type that can
provide this functionality.\footnote{
	See \url{https://golang.org/pkg/sync/atomic/\#Value}, particularly
	the ``Example (ReadMostly)''.}

Matt Klein describes an RCU-like mechanism that is used in the Envoy
Proxy~\cite{MattKlein2017EnvoyRCU}.

Honnappa Nagarahalli added an RCU library to the Data Plane Development
Kit (DPDK) in 2018~\cite{HonnappaNagarahalli2018dpdkRCU}.

Stjepan Glavina merged an epoch-based RCU implementation into the
\co{crossbeam} set of concurrency-support ``crates'' for the Rust
language~\cite{StjepanGlavina2018RustRCU}.

Jason Donenfeld produced an RCU implementations as part of his port of
WireGuard to Windows~NT kernel~\cite{JasonDonenfeld2021:WindowsNTwireguardRCU}.

Finally, any garbage-collected concurrent language (not just Go!\@) gets
the update side of an RCU implementation at zero incremental cost.

\subsubsection{Summary of Production Uses}
\label{sec:defer:Summary of Production Uses}

Perhaps the time will come when sequence locking, hazard pointers, and
RCU are all as heavily used and as well known as are reference counters.
Until that time comes, the current production uses of these mechanisms
should help guide the choice of mechanism as well as showing how best
to apply each of them.
And with that, we have uncovered the last of the mysteries put forth on
\cpageref{sec:defer:Mysteries Which to choose}.

The next section discusses updates, a ticklish issue for many of the
read-mostly mechanisms described in this chapter.

% defer/updates.tex
% mainfile: ../perfbook.tex
% SPDX-License-Identifier: CC-BY-SA-3.0

\section{What About Updates?}
\label{sec:defer:What About Updates?}
\epigraph{The only thing constant in life is change.}
	 {Fran\c{c}ois de la Rochefoucauld}

The deferred-processing techniques called out in this chapter are most
directly applicable to read-mostly situations, which begs the question
``But what about updates?''
After all, increasing the performance and scalability of readers is all
well and good, but it is only natural to also want great performance and
scalability for writers.

We have already seen one situation featuring high performance and
scalability for writers, namely the counting algorithms surveyed in
\cref{chp:Counting}.
These algorithms featured partially partitioned data structures so
that updates can operate locally, while the more-expensive reads
must sum across the entire data structure.
Silas Boyd-Wickhizer has generalized this notion to produce
OpLog, which he has applied to
Linux-kernel pathname lookup, VM reverse mappings, and the \co{stat()} system
call~\cite{SilasBoydWickizerPhD}.

Another approach, called ``Disruptor'', is designed for applications
that process high-volume streams of input data.
The approach is to rely on single-producer-single-consumer FIFO queues,
minimizing the need for synchronization~\cite{AdrianSutton2013LCA:Disruptor}.
For Java applications, Disruptor also has the virtue of minimizing use
of the garbage collector.

And of course, where feasible, fully partitioned or ``sharded'' systems
provide excellent performance and scalability, as noted in
\cref{chp:Partitioning and Synchronization Design}.

The next chapter will look at updates in the context of several types
of data structures.

	\setlength{\parskip}{0.0pt plus 1ex}
	\input{qqzdefer}
	\setlength{\parskip}{0.0pt plus 1.0pt}% return to default

% datastruct/datastruct.tex
% mainfile: ../perfbook.tex
% SPDX-License-Identifier: CC-BY-SA-3.0

\QuickQuizChapter{chp:Data Structures}{Data Structures}{qqzdatastruct}
\Epigraph{Bad programmers worry about the code.
	  Good programmers worry about data structures and their
	  relationships.}
	 {Linus Torvalds}

Serious discussions of algorithms include time complexity of their
data structures~\cite{ThomasHCorman2001Algorithms}.
However, for parallel programs, the time complexity includes concurrency
effects because these effects can be overwhelmingly large, as shown in
\cref{chp:Hardware and its Habits}.
In other words, a good programmer's data-structure relationships
include those aspects related to concurrency.

This chapter will expose a number of complications:

\begin{enumerate}
\item	Data structures designed in full accordance with the
	good advice given in
	\cref{chp:Partitioning and Synchronization Design}
	can nonetheless abjectly fail to scale on some types of systems.
\label{sec:datastruct:NUMA-induced failure to scale}
\item	Data structures designed in full accordance with the
	good advice given in both
	\cref{chp:Partitioning and Synchronization Design}
	and
	\cref{chp:Deferred Processing}
	can \emph{still} abjectly fail to scale on some types of systems.
\label{sec:datastruct:Cache-size-induced failure to scale}
\item	Even \emph{read-only synchronization-free} data-structure traversal 
	can fail to scale on some types of systems.
\label{sec:datastruct:Cache-size-induced failure to scale redux}
\item	Data-structure traverals avoiding the aforementioned complications
	can still be impeded by concurrent updates.
\label{sec:datastruct:Update-activity failure to scale}
\end{enumerate}

\Cref{sec:datastruct:Motivating Application}
presents the motivating application for this chapter's data structures.
\Cref{chp:Partitioning and Synchronization Design} showed how
partitioning improves scalability, so
\cref{sec:datastruct:Partitionable Data Structures}
discusses partitionable data structures.
\Cref{chp:Deferred Processing} described how deferring some
actions can greatly improve both performance and scalability,
a topic taken up by
\cref{sec:datastruct:Read-Mostly Data Structures}.
\Cref{sec:datastruct:Non-Partitionable Data Structures}
looks at a non-partitionable data structure, splitting
it into read-mostly and partitionable portions,
which improves both performance and scalability.
Because this chapter cannot delve into the details of every concurrent
data structure,
\cref{sec:datastruct:Other Data Structures}
surveys a few of the important ones.
Although the best performance and scalability results from design rather
than after-the-fact micro-optimization, micro-optimization is nevertheless
necessary for the absolute best possible performance and scalability,
as described in
\cref{sec:datastruct:Micro-Optimization}.
Finally, \cref{sec:datastruct:Summary}
presents a summary of this chapter.

\section{Motivating Application}
\label{sec:datastruct:Motivating Application}
\epigraph{The art of doing mathematics consists in finding that special
	  case which contains all the germs of generality.}
	 {David Hilbert}

We will use the Schr\"odinger's Zoo application to evaluate
performance~\cite{McKenney:2013:SDS:2483852.2483867}.
Schr\"odinger has a zoo containing a large number of animals, and
he would like to track them using an in-memory database with
each animal in the zoo represented by a data item in this database.
Each animal has a unique name that is used as a key, with a variety
of data tracked for each animal.

Births, captures, and purchases result in insertions, while deaths,
releases, and sales result in deletions.
Because Schr\"odinger's zoo contains a large quantity of short-lived
animals, including mice and insects, the database must handle
high update rates.
Those interested in Schr\"odinger's animals can query them, and
Schr\"odinger has noted suspiciously query rates for his cat, so much
so that he suspects that his mice might be checking up on their nemesis.
Whatever their source, Schr\"odinger's application must handle high
query rates to a single data element.

As we will see, this simple application can be a challenge to concurrent
data structures.

\section{Partitionable Data Structures}
\label{sec:datastruct:Partitionable Data Structures}
\epigraph{Finding a way to live the simple life today is the most
	  complicated task.}
	 {Henry A. Courtney, updated}

There are a huge number of data structures in use today, so much so
that there are multiple textbooks covering them.
This section focuses on a single data structure, namely the hash table.
This focused approach allows a much deeper investigation of how concurrency
interacts with data structures, and also focuses on a data structure
that is heavily used in practice.
\Cref{sec:datastruct:Hash-Table Design}
overviews the design, and
\cref{sec:datastruct:Hash-Table Implementation}
presents the implementation.
Finally,
\cref{sec:datastruct:Hash-Table Performance}
discusses the resulting performance and scalability.

\subsection{Hash-Table Design}
\label{sec:datastruct:Hash-Table Design}

\Cref{chp:Partitioning and Synchronization Design}
emphasized the need to apply partitioning in order to attain
respectable performance and scalability, so partitionability
must be a first-class criterion when selecting data structures.
This criterion is well satisfied by that workhorse of parallelism,
the hash table.
Hash tables are conceptually simple, consisting of an array of
\emph{hash buckets}.
A \emph{hash function} maps from a given element's \emph{key}
to the hash bucket that this element will be stored in.
Each hash bucket therefore heads up a linked list of elements,
called a \emph{hash chain}.
When properly configured, these hash chains will be quite short,
permitting a hash table to access its elements extremely efficiently.

\QuickQuiz{
	But chained hash tables are but one type of many.
	Why the focus on chained hash tables?
}\QuickQuizAnswer{
	Chained hash tables are completely partitionable, and thus
	well-suited to concurrent use.
	There are other completely-partitionable hash tables, for
	example, split-ordered list~\cite{OriShalev2006SplitOrderListHash},
	but they are considerably more complex.
	We therefore start with chained hash tables.
}\QuickQuizEnd

In addition, each bucket has its own lock, so that elements in different
buckets of the hash table may be added, deleted, and looked up completely
independently.
A large hash table with a large number of buckets (and thus locks), with
each bucket containing a small number of elements should therefore provide
excellent scalability.

\subsection{Hash-Table Implementation}
\label{sec:datastruct:Hash-Table Implementation}

\begin{figure}
\centering
\resizebox{3in}{!}{\includegraphics{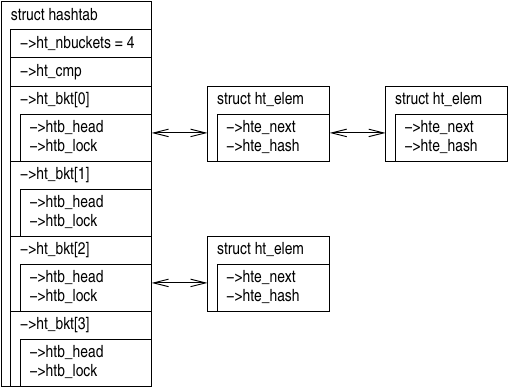}}
\caption{Hash-Table Data-Structure Diagram}
\label{fig:datastruct:Hash-Table Data-Structure Diagram}
\end{figure}

\begin{fcvref}[ln:datastruct:hash_bkt:struct]
\Cref{lst:datastruct:Hash-Table Data Structures}
(\path{hash_bkt.c})
shows a set of data structures used in a simple fixed-sized hash
table using chaining and per-hash-bucket locking, and
\cref{fig:datastruct:Hash-Table Data-Structure Diagram}
diagrams how they fit together.
The \co{hashtab} structure (\clnrefrange{tab:b}{tab:e} in
\cref{lst:datastruct:Hash-Table Data Structures})
contains four \co{ht_bucket} structures
(\clnrefrange{bucket:b}{bucket:e} in
\cref{lst:datastruct:Hash-Table Data Structures}),
with the \co{->ht_nbuckets} field controlling the number of buckets
and the \co{->ht_cmp} field holding the pointer to key-comparison
function.
Each such bucket contains a list header \co{->htb_head} and
a lock \co{->htb_lock}.
The list headers chain \co{ht_elem} structures
(\clnrefrange{elem:b}{elem:e} in
\cref{lst:datastruct:Hash-Table Data Structures})
through their
\co{->hte_next} fields, and each \co{ht_elem} structure also caches
the corresponding element's hash value in the \co{->hte_hash} field.
The \co{ht_elem} structure is included in a larger structure
which might contain a complex key.
\end{fcvref}

\begin{listing}
\input{CodeSamples/datastruct/hash/hash_bkt=struct.fcv}
\caption{Hash-Table Data Structures}
\label{lst:datastruct:Hash-Table Data Structures}
\end{listing}

\Cref{fig:datastruct:Hash-Table Data-Structure Diagram}
shows bucket~0 containing two elements and bucket~2 containing one.

\begin{fcvref}[ln:datastruct:hash_bkt:map_lock:map]
\Cref{lst:datastruct:Hash-Table Mapping and Locking}
shows mapping and locking functions.
\Clnref{b,e}
show the macro \co{HASH2BKT()}, which maps from a hash value
to the corresponding \co{ht_bucket} structure.
This macro uses a simple modulus:
If more aggressive hashing is required,
the caller needs to implement it when mapping from key to hash value.
The remaining two functions acquire and release the \co{->htb_lock}
corresponding to the specified hash value.
\end{fcvref}

\begin{listing}
\input{CodeSamples/datastruct/hash/hash_bkt=map_lock.fcv}
\caption{Hash-Table Mapping and Locking}
\label{lst:datastruct:Hash-Table Mapping and Locking}
\end{listing}

\begin{fcvref}[ln:datastruct:hash_bkt:lookup]
\Cref{lst:datastruct:Hash-Table Lookup}
shows \co{hashtab_lookup()},
which returns a pointer to the element with the specified hash and key if it
exists, or \co{NULL} otherwise.
This function takes both a hash value and a pointer to the key because
this allows users of this function to use arbitrary keys and
arbitrary hash functions.
\Clnref{map} maps from the hash value to a pointer to the corresponding
hash bucket.
Each pass through the loop spanning
\clnrefrange{loop:b}{loop:e} examines one element
of the bucket's hash chain.
\Clnref{hashmatch} checks to see if the hash values match, and if not,
\clnref{next}
proceeds to the next element.
\Clnref{keymatch} checks to see if the actual key matches, and if so,
\clnref{return} returns a pointer to the matching element.
If no element matches, \clnref{ret_NULL} returns \co{NULL}.
\end{fcvref}

\begin{listing}
\input{CodeSamples/datastruct/hash/hash_bkt=lookup.fcv}
\caption{Hash-Table Lookup}
\label{lst:datastruct:Hash-Table Lookup}
\end{listing}

\QuickQuiz{
	\begin{fcvref}[ln:datastruct:hash_bkt:lookup]
	But isn't the double comparison on
	\clnrefrange{hashmatch}{return} in
	\cref{lst:datastruct:Hash-Table Lookup} inefficient
	in the case where the key fits into an unsigned long?
	\end{fcvref}
}\QuickQuizAnswer{
	Indeed it is!
	However, hash tables quite frequently store information with
	keys such as character strings that do not necessarily fit
	into an unsigned long.
	Simplifying the hash-table implementation for the case where
	keys always fit into unsigned longs is left as an exercise
	for the reader.
}\QuickQuizEnd

\begin{listing}
\input{CodeSamples/datastruct/hash/hash_bkt=add_del.fcv}
\caption{Hash-Table Modification}
\label{lst:datastruct:Hash-Table Modification}
\end{listing}

\Cref{lst:datastruct:Hash-Table Modification}
shows the \co{hashtab_add()} and \co{hashtab_del()} functions
that add and delete elements from the hash table, respectively.

\begin{fcvref}[ln:datastruct:hash_bkt:add_del:add]
The \co{hashtab_add()} function simply sets the element's hash
value on \clnref{set}, then adds it to the corresponding bucket on
\clnref{add:b,add:e}.
\end{fcvref}
The \co{hashtab_del()} function simply removes the specified element
from whatever hash chain it is on, courtesy of the doubly linked
nature of the hash-chain lists.
Before calling either of these two functions, the caller is required to
ensure that no other thread is accessing
or modifying this same bucket, for example, by invoking
\co{hashtab_lock()} beforehand.

\begin{listing}
\input{CodeSamples/datastruct/hash/hash_bkt=alloc_free.fcv}
\caption{Hash-Table Allocation and Free}
\label{lst:datastruct:Hash-Table Allocation and Free}
\end{listing}

\Cref{lst:datastruct:Hash-Table Allocation and Free}
shows \co{hashtab_alloc()} and \co{hashtab_free()},
which do hash-table allocation and freeing, respectively.
\begin{fcvref}[ln:datastruct:hash_bkt:alloc_free:alloc]
Allocation begins on
\clnrefrange{alloc:b}{alloc:e} with allocation of the underlying memory.
If \clnref{chk_NULL} detects that memory has been exhausted,
\clnref{ret_NULL} returns
\co{NULL} to the caller.
Otherwise, \clnref{set_nbck,set_cmp} initialize
the number of buckets and the pointer to key-comparison function,
and the loop
spanning \clnrefrange{loop:b}{loop:e} initializes the buckets themselves,
including the chain list header on
\clnref{init_head} and the lock on \clnref{init_lock}.
Finally, \clnref{return} returns a pointer to the newly allocated hash table.
\end{fcvref}
\begin{fcvref}[ln:datastruct:hash_bkt:alloc_free:free]
The \co{hashtab_free()} function on
\clnrefrange{b}{e} is straightforward.
\end{fcvref}

\subsection{Hash-Table Performance}
\label{sec:datastruct:Hash-Table Performance}

\begin{figure}
\centering
\resizebox{2.5in}{!}{\includegraphics{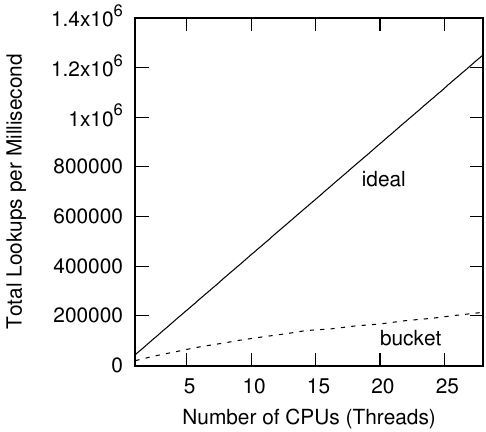}}
\caption{Read-Only Hash-Table Performance For Schr\"odinger's Zoo}
\label{fig:datastruct:Read-Only Hash-Table Performance For Schroedinger's Zoo}
\end{figure}

The performance results for a single 28-core socket of a 2.1\,GHz
Intel Xeon system using a bucket-locked hash table
with 262,144 buckets are shown in
\cref{fig:datastruct:Read-Only Hash-Table Performance For Schroedinger's Zoo}.
The performance does scale nearly linearly, but it falls a far short
of the ideal performance level, even at only 28~CPUs.
Part of this shortfall is due to the fact that the lock acquisitions and
releases incur no cache misses on a single CPU, but do incur misses
on two or more CPUs.

\begin{figure}
\centering
\resizebox{2.5in}{!}{\includegraphics{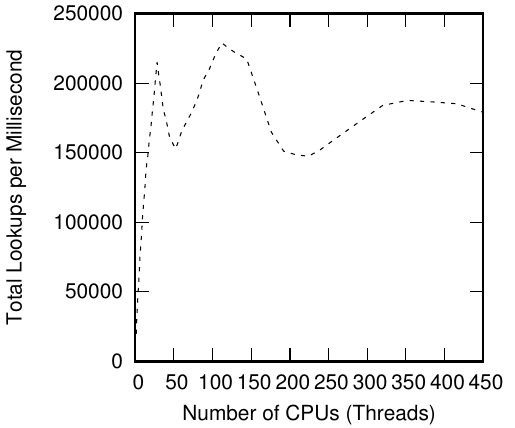}}
\caption{Read-Only Hash-Table Performance For Schr\"odinger's Zoo, 448 CPUs}
\label{fig:datastruct:Read-Only Hash-Table Performance For Schroedinger's Zoo; 448 CPUs}
\end{figure}

And things only get worse with more CPUs, as can be seen in
\cref{fig:datastruct:Read-Only Hash-Table Performance For Schroedinger's Zoo; 448 CPUs}.
We do not need to show ideal performance:
The performance for 29~CPUs and beyond is all too clearly worse than abysmal.
This clearly underscores the dangers of extrapolating performance from a
modest number of CPUs.

Of course, one possible reason for the collapse in performance might be
that more hash buckets are needed.
We can test this by increasing the number of hash buckets.

\QuickQuiz{
	Instead of simply increasing the number of hash buckets,
	wouldn't it be better to cache-align the existing hash buckets?
}\QuickQuizAnswer{
	The answer depends on a great many things.
	If the hash table has a large number of elements per bucket, it
	would clearly be better to increase the number of hash buckets.
	On the other hand, if the hash table is lightly loaded,
	the answer depends on the hardware, the effectiveness of the
	hash function, and the workload.
	Interested readers are encouraged to experiment.
}\QuickQuizEnd

\begin{figure}
\centering
\resizebox{2.5in}{!}{\includegraphics{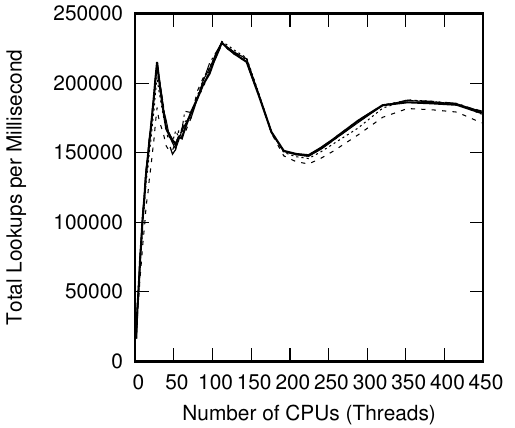}}
\caption{Read-Only Hash-Table Performance For Schr\"odinger's Zoo, Varying Buckets}
\label{fig:datastruct:Read-Only Hash-Table Performance For Schroedinger's Zoo; Varying Buckets}
\end{figure}

However, as can be seen in
\cref{fig:datastruct:Read-Only Hash-Table Performance For Schroedinger's Zoo; Varying Buckets},
changing the number of buckets has almost no effect:
Scalability is still abysmal.
In particular, we still see a sharp dropoff at 29~CPUs and beyond,
clearly demonstrating the complication put forward on
\cpageref{sec:datastruct:NUMA-induced failure to scale}.
And just as clearly, something else is going on.

The problem is that this is a multi-socket system, with CPUs~0--27
and~225--251 mapped to the first socket as shown in
\cref{fig:datastruct:NUMA Topology of System Under Test}.
Test runs confined to the first 28~CPUs therefore perform quite
well, but tests that involve socket~0's CPUs~0--27 as well as
socket~1's CPU~28 incur the overhead of passing data across
socket boundaries.
This can severely degrade performance, as was discussed in
\cref{sec:cpu:Hardware System Architecture}.
In short, large multi-socket systems require good locality of reference
in addition to full partitioning.
The remainder of this chapter will discuss ways of providing good
locality of reference within the hash table itself, but in the
meantime please note that one other way to provide good locality
of reference would be to place large data elements in the hash
table.
For example, Schr\"odinger might attain excellent cache locality by
placing photographs or even videos of his animals in each element of
the hash table.
But for those needing hash tables containing small data elements,
please read on!

\QuickQuiz{
	Given the negative scalability of the Schr\"odinger's
	Zoo application across sockets, why not just run multiple
	copies of the application, with each copy having a subset
	of the animals and confined to run on a single socket?
}\QuickQuizAnswer{
	You can do just that!
	In fact, you can extend this idea to large clustered systems,
	running one copy of the application on each node of the cluster.
	This practice is called ``sharding'', and is heavily used in
	practice by large web-based
	retailers~\cite{DeCandia:2007:DAH:1323293.1294281}.

	However, if you are going to shard on a per-socket basis within
	a multisocket system, why not buy separate smaller and cheaper
	single-socket systems, and then run one shard of the database
	on each of those systems?
}\QuickQuizEnd

One key property of the Schr\"odinger's-zoo runs discussed thus far is that
they are all read-only.
This makes the performance degradation due to lock-acquisition-induced
cache misses all the more painful.
Even though we are not updating the underlying hash table itself, we are
still paying the price for writing to memory.
Of course, if the hash table was never going to be updated, we could dispense
entirely with mutual exclusion.
This approach is quite straightforward and is left as an exercise for the
reader.
But even with the occasional update, avoiding writes avoids cache
misses, and allows the read-mostly data to be replicated across all
the caches, which in turn promotes locality of reference.

The next section therefore examines optimizations that can be carried out in
read-mostly cases where updates are rare, but could happen at any time.

\setlength\dashlinedash{1pt}
\setlength\dashlinegap{2pt}

\begin{figure}
\renewcommand*{\arraystretch}{1.2}
\centering
\begin{tabular}{r||c:c}
	& \multicolumn{2}{c}{Hyperthread} \\
	\cline{2-3}
	Socket & 0 &  1 \\
	\hline
	\hline
	0 &    0--27 & 224--251 \\
	\cdashline{2-3}
	1 &   28--55 & 252--279 \\
	\cdashline{2-3}
	2 &   56--83 & 280--307 \\
	\cdashline{2-3}
	3 &  84--111 & 308--335 \\
	\cdashline{2-3}
	4 & 112--139 & 336--363 \\
	\cdashline{2-3}
	5 & 140--167 & 364--391 \\
	\cdashline{2-3}
	6 & 168--195 & 392--419 \\
	\cdashline{2-3}
	7 & 196--223 & 420--447 \\
\end{tabular}
\caption{NUMA Topology of System Under Test}
\label{fig:datastruct:NUMA Topology of System Under Test}
\end{figure}

\section{Read-Mostly Data Structures}
\label{sec:datastruct:Read-Mostly Data Structures}
\epigraph{Adapt the remedy to the disease.}{Chinese proverb}

Although partitioned data structures can offer excellent scalability,
\IXacr{numa} effects can result in severe degradations of both performance and
scalability.
In addition,
the need for read-side synchronization can degrade performance in
read-mostly situations.
However, we can achieve both performance and scalability by using
RCU, which was introduced in
\cref{sec:defer:Read-Copy Update (RCU)}.
Similar results can be achieved using \IXpl{hazard pointer}
(\path{hazptr.c})~\cite{MagedMichael04a}, which will be included in
the performance results shown in this
section~\cite{McKenney:2013:SDS:2483852.2483867}.

\subsection{RCU-Protected Hash Table Implementation}
\label{sec:datastruct:RCU-Protected Hash Table Implementation}

For an RCU-protected hash table with per-bucket locking,
updaters use locking as shown in
\cref{sec:datastruct:Partitionable Data Structures},
but readers use RCU\@.
The data structures remain as shown in
\cref{lst:datastruct:Hash-Table Data Structures},
and the \co{HASH2BKT()}, \co{hashtab_lock()}, and \co{hashtab_unlock()}
functions remain as shown in
\cref{lst:datastruct:Hash-Table Mapping and Locking}.
However, readers use the lighter-weight concurrency-control embodied
by \co{hashtab_lock_lookup()} and \co{hashtab_unlock_lookup()}
shown in
\cref{lst:datastruct:RCU-Protected Hash-Table Read-Side Concurrency Control}.

\begin{listing}
\input{CodeSamples/datastruct/hash/hash_bkt_rcu=lock_unlock.fcv}
\caption{RCU-Protected Hash-Table Read-Side Concurrency Control}
\label{lst:datastruct:RCU-Protected Hash-Table Read-Side Concurrency Control}
\end{listing}

\Cref{lst:datastruct:RCU-Protected Hash-Table Lookup}
shows \co{hashtab_lookup()} for the RCU-protected per-bucket-locked
hash table.
This is identical to that in
\cref{lst:datastruct:Hash-Table Lookup}
except that \apiur{cds_list_for_each_entry()} is replaced
by \apiur{cds_list_for_each_entry_rcu()}.
Both of these primitives traverse the hash chain referenced
by \co{htb->htb_head} but \co{cds_list_for_each_entry_rcu()} also
correctly enforces memory ordering in case of concurrent insertion.
This is an important difference between these two hash-table implementations:
Unlike the pure per-bucket-locked implementation, the RCU protected
implementation allows lookups to run concurrently with insertions
and deletions, and RCU-aware primitives like
\co{cds_list_for_each_entry_rcu()} are required to correctly handle
this added concurrency.
Note also that \co{hashtab_lookup()}'s caller must be within an
RCU read-side critical section, for example, the caller must invoke
\co{hashtab_lock_lookup()} before invoking \co{hashtab_lookup()}
(and of course invoke \co{hashtab_unlock_lookup()} some time afterwards).

\begin{listing}
\input{CodeSamples/datastruct/hash/hash_bkt_rcu=lookup.fcv}
\caption{RCU-Protected Hash-Table Lookup}
\label{lst:datastruct:RCU-Protected Hash-Table Lookup}
\end{listing}

\QuickQuiz{
	But if elements in a hash table can be removed concurrently
	with lookups, doesn't that mean that a lookup could return
	a reference to a data element that was removed immediately
	after it was looked up?
}\QuickQuizAnswer{
	Yes it can!
	This is why \co{hashtab_lookup()} must be invoked within an
	RCU read-side critical section, and it is why
	\co{hashtab_add()} and \co{hashtab_del()} must also use
	RCU-aware list-manipulation primitives.
	Finally, this is why the caller of \co{hashtab_del()} must
	wait for a grace period (e.g., by calling \co{synchronize_rcu()})
	before freeing the removed element.
	This will ensure that all RCU readers that might reference
	the newly removed element have completed before that element
	is freed.
}\QuickQuizEnd

\begin{listing}
\input{CodeSamples/datastruct/hash/hash_bkt_rcu=add_del.fcv}
\caption{RCU-Protected Hash-Table Modification}
\label{lst:datastruct:RCU-Protected Hash-Table Modification}
\end{listing}

\Cref{lst:datastruct:RCU-Protected Hash-Table Modification}
shows \co{hashtab_add()} and \co{hashtab_del()}, both of which
are quite similar to their counterparts in the non-RCU hash table
shown in
\cref{lst:datastruct:Hash-Table Modification}.
The \co{hashtab_add()} function uses \apiur{cds_list_add_rcu()} instead
of \apiur{cds_list_add()} in order to ensure proper ordering when
an element is added to the hash table at the same time that it is
being looked up.
The \co{hashtab_del()} function uses \apiur{cds_list_del_rcu()} instead
of \apiur{cds_list_del_init()} to allow for the case where an element is
looked up just before it is deleted.
Unlike \co{cds_list_del_init()}, \co{cds_list_del_rcu()} leaves the
forward pointer intact, so that \co{hashtab_lookup()} can traverse
to the newly deleted element's successor.

Of course, after invoking \co{hashtab_del()}, the caller must wait for
an RCU \IX{grace period} (e.g., by invoking \co{synchronize_rcu()}) before
freeing or otherwise reusing the memory for the newly deleted element.

\subsection{RCU-Protected Hash Table Validation}
\label{sec:datastruct:RCU-Protected Hash Table Validation}

Although the topic of validation is covered in detail in
\cref{chp:Validation}, the fact is that a hash table with lockless
RCU-protected lookups needs special attention to validation sooner rather
than later.

The test suite (``\path{hashtorture.h}'') contains a \co{smoketest()}
function that verifies that a specific series of single-threaded
additions, deletions, and lookups give the expected results.

Concurrent test runs put each updater thread in control of its portion of
the elements, which allows assertions checking for the following issues:

\begin{enumerate}
\item	A just-now-to-be-added element already being in the table
	according to \co{hastab_lookup()}.
\item	A just-now-to-be-added element being marked as being in the
	table by its \co{->in_table} flag.
\item	A just-now-to-be-deleted element not being in the table according
	to \co{hastab_lookup()}.
\item	A just-now-to-be-deleted element being marked as not being in
	the table by its \co{->in_table} flag.
\end{enumerate}

In addition, concurrent test runs run lookups concurrently with updates
in order to catch all manner of data-structure corruption problems.
Some runs also continually resize the hash table concurrently with both
lookups and updates to verify correct behavior, and also to verify that
resizes do not unduly delay either readers or updaters.

Finally, the concurrent tests output statistics that can be used to
track down performance and scalabilty issues, which provides the raw
data used by
\cref{sec:datastruct:RCU-Protected Hash Table Performance}.

\QuickQuiz{
	The \path{hashtorture.h} file contains more than 1,000 lines!
	Is that a comprehensive test or what???
}\QuickQuizAnswer{
	What.

	The \path{hashtorture.h} tests are a good start and suffice
	for a textbook algorithm.
	If this code was to be used in production, much more testing
	would be required:

	\begin{enumerate}
	\item	Have some subset of elements that always reside in the
		table, and verify that lookups always find these elements
		regardless of the number and type of concurrent updates
		in flight.
	\item	Pair an updater with one or more readers, verifying that
		after an element is added, once a reader successfully
		looks up that element, all later lookups succeed.
		The definition of ``later'' will depend on the table's
		consistency requirements.
	\item	Pair an updater with one or more readers, verifying that
		after an element is deleted, once a reader's lookup
		of that element fails, all later lookups also fail.
	\end{enumerate}

	There are many more tests where those came from, the exact
	nature of which depend on the details of the requirements
	on your particular hash table.
}\QuickQuizEnd

All code requires significant validation effort, and high-performance
concurrent code requires more validation than most.

\subsection{RCU-Protected Hash Table Performance}
\label{sec:datastruct:RCU-Protected Hash Table Performance}

\begin{figure}
\centering
\resizebox{2.5in}{!}{\includegraphics{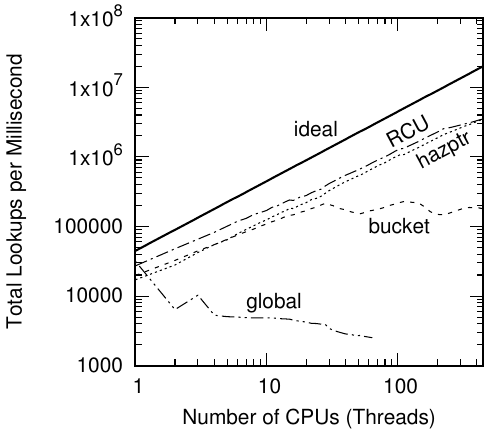}}
\caption{Read-Only RCU-Protected Hash-Table Performance For Schr\"odinger's Zoo}
\label{fig:datastruct:Read-Only RCU-Protected Hash-Table Performance For Schroedinger's Zoo}
\end{figure}

\Cref{fig:datastruct:Read-Only RCU-Protected Hash-Table Performance For Schroedinger's Zoo}
shows the read-only performance of RCU-protected and hazard-pointer-protected
hash tables against the previous section's per-bucket-locked implementation.
As you can see, both RCU and hazard pointers perform and scale
much better than per-bucket locking because read-only
replication avoids NUMA effects.
The difference increases with larger numbers of threads.
Results from a globally locked implementation are also shown, and as expected
the results are even worse than those of the per-bucket-locked implementation.
RCU does slightly better than hazard pointers.

\begin{figure}
\centering
\resizebox{2.5in}{!}{\includegraphics{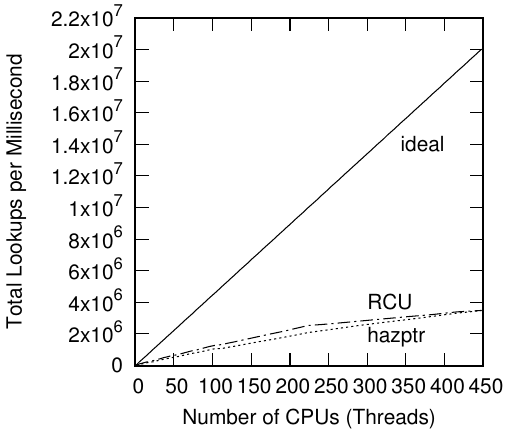}}
\caption{Read-Only RCU-Protected Hash-Table Performance For Schr\"odinger's Zoo, Linear Scale}
\label{fig:datastruct:Read-Only RCU-Protected Hash-Table Performance For Schroedinger's Zoo; Linear Scale}
\end{figure}

\Cref{fig:datastruct:Read-Only RCU-Protected Hash-Table Performance For Schroedinger's Zoo; Linear Scale}
shows the same data on a linear scale.
This drops the global-locking trace into the x-axis, but allows the
non-ideal performance of RCU and hazard pointers to be more readily
discerned.
Both show a change in slope at 224~CPUs, and this is due to hardware
multithreading.
At 224 and fewer CPUs, each thread has a core to itself.
In this regime, RCU does better than does hazard pointers because the
latter's read-side \IXpl{memory barrier} result in dead time within the core.
In short, RCU is better able to utilize a core from a single hardware
thread than is hazard pointers.

This situation changes above 224~CPUs.
Because RCU is using more than half of each core's resources from a
single hardware thread, RCU gains relatively little benefit from the
second hardware thread in each core.
The slope of the hazard-pointers trace also decreases at 224~CPUs, but
less dramatically,
because the second hardware thread is able to fill in the time
that the first hardware thread is stalled due to \IXh{memory-barrier}{latency}.
As we will see in later sections, this second-hardware-thread
advantage depends on the workload.

\begin{figure}
\centering
\resizebox{2.5in}{!}{\includegraphics{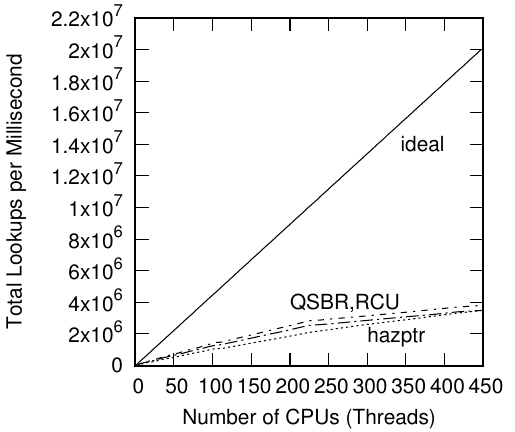}}
\caption{Read-Only RCU-Protected Hash-Table Performance For Schr\"odinger's Zoo including QSBR, Linear Scale}
\label{fig:datastruct:Read-Only RCU-Protected Hash-Table Performance For Schroedinger's Zoo including QSBR; Linear Scale}
\end{figure}

But why is RCU's performance a factor of five less than ideal?
One possibility is that the per-thread counters manipulated by
\co{rcu_read_lock()} and \co{rcu_read_unlock()} are slowing things down.
\Cref{fig:datastruct:Read-Only RCU-Protected Hash-Table Performance For Schroedinger's Zoo including QSBR; Linear Scale}
therefore adds the results for the \IXacr{qsbr} variant of RCU, whose read-side
primitives do nothing.
And although QSBR does perform slightly better than does RCU, it is still
about a factor of five short of ideal.

\begin{figure}
\centering
\resizebox{2.5in}{!}{\includegraphics{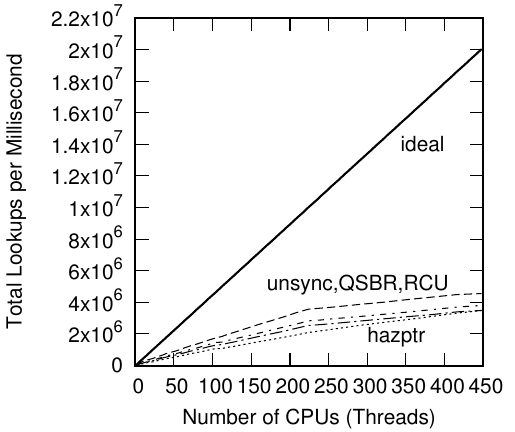}}
\caption{Read-Only RCU-Protected Hash-Table Performance For Schr\"odinger's Zoo including QSBR and Unsynchronized, Linear Scale}
\label{fig:datastruct:Read-Only RCU-Protected Hash-Table Performance For Schroedinger's Zoo including QSBR and Unsynchronized; Linear Scale}
\end{figure}

\Cref{fig:datastruct:Read-Only RCU-Protected Hash-Table Performance For Schroedinger's Zoo including QSBR and Unsynchronized; Linear Scale}
adds completely unsynchronized results, which works because this
is a read-only benchmark with nothing to synchronize.
Even with no synchronization whatsoever, performance still falls far
short of ideal, thus demonstrating two more complications on
\cpageref{sec:datastruct:Cache-size-induced failure to scale redux}.

The problem is that this system has sockets with 28 cores, which have
the modest cache sizes shown in
\cref{tab:cpu:Cache Geometry for 8-Socket System With Intel Xeon Platinum 8176 CPUs @ 2.10GHz}
on \cpageref{tab:cpu:Cache Geometry for 8-Socket System With Intel Xeon Platinum 8176 CPUs @ 2.10GHz}.
Each hash bucket (\co{struct ht_bucket}) occupies 56~bytes and each
element (\co{struct zoo_he}) occupies 72~bytes for the RCU and QSBR runs.
The benchmark generating
\cref{fig:datastruct:Read-Only RCU-Protected Hash-Table Performance For Schroedinger's Zoo including QSBR and Unsynchronized; Linear Scale}
used 262,144 buckets and up to 262,144 elements, for a total of
33,554,448~bytes, which not only overflows the 1,048,576-byte L2 caches
by more than a factor of thirty, but is also uncomfortably close to the
L3 cache size of 40,370,176~bytes, especially given that this cache has
only 11~ways.
This means that L2 cache collisions will be the rule and also that L3
cache collisions will not be uncommon, so that the resulting cache misses
will degrade performance.
In this case, the bottleneck is not in the CPU, but rather in the hardware
memory system.
% This data was generated wtih 262,144 buckets and elements.
% RCU hash ht_elem structure is 40 bytes plus 32 more from zoo_he for 72.
% struct ht_bucket is 56 bytes.
% For "# B" RCU runs, hash buckets weigh in at 14,680,080 bytes and the hash
%    elements up to 18,874,368 bytes.  The total is 33,554,448 bytes, which
%    in theory fits into the 40,370,176-byte L3 cache, but definitely not
%    the 1,048,576-byte L2 cache, much less the 32,768-byte L0 and L1 caches.
%    The L3 is 11-way set associative.
%    So we have 8192 4K pages of data (buckets and elements) that must
%    fit into 896 11-way set-associated cache sets.

Additional evidence for this memory-system bottleneck may be found by
examining the unsynchronized code.
This code does not need locks, so each hash bucket occupies only 16~bytes
compared to the 56~bytes for RCU and QSBR\@.
Similarly, each hash-table element occupies only 56~bytes compared to the
72~bytes for RCU and QSBR\@.
So it is unsurprising that the single-CPU unsynchronized run performs up
to about half again faster than that of either QSBR or RCU\@.

\QuickQuiz{
	How can we be so sure that the hash-table size is at fault here,
	especially given that
	\cref{fig:datastruct:Read-Only Hash-Table Performance For Schroedinger's Zoo; Varying Buckets}
	on \cpageref{fig:datastruct:Read-Only Hash-Table Performance For Schroedinger's Zoo; Varying Buckets}
	shows that varying hash-table size has almost
	no effect?
	Might the problem instead be something like \IX{false sharing}?
}\QuickQuizAnswer{
	Excellent question!

	False sharing requires writes, which are not featured in the
	unsynchronized and RCU runs of this lookup-only
	benchmark.
	The problem is therefore not false sharing.

\begin{figure}
\centering
\resizebox{3in}{!}{\includegraphics{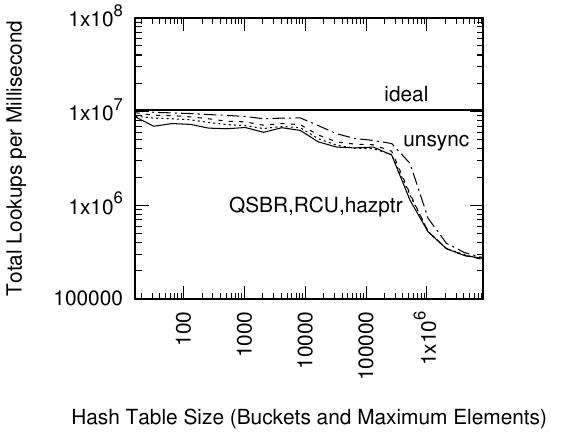}}
\caption{Read-Only RCU-Protected Hash-Table Performance For Schr\"odinger's Zoo at 448 CPUs, Varying Table Size}
\label{fig:datastruct:Read-Only RCU-Protected Hash-Table Performance For Schr\"odinger's Zoo at 448 CPUs; Varying Table Size}
\end{figure}

	Still unconvinced?
	Then look at the log-log plot in
	\cref{fig:datastruct:Read-Only RCU-Protected Hash-Table Performance For Schr\"odinger's Zoo at 448 CPUs; Varying Table Size},
	which shows performance for 448~CPUs as a function of the
	hash-table size, that is, number of buckets and maximum number
	of elements.
	A hash-table of size 1,024 has 1,024~buckets and contains
	at most 1,024~elements, with the average occupancy being
	512~elements.
	Because this is a read-only benchmark, the actual occupancy is
	always equal to the average occupancy.

	This figure shows near-ideal performance below about 8,000~elements,
	that is, when the hash table comprises less than 1\,MB of data.
	This near-ideal performance is consistent with that for the
	pre-BSD routing table shown in
	\cref{fig:defer:Pre-BSD Routing Table Protected by RCU}
	on \cpageref{fig:defer:Pre-BSD Routing Table Protected by RCU},
	even at 448~CPUs.
	However, the performance drops significantly (this is a log-log
	plot) at about 8,000~elements, which is where the 1,048,576-byte
	L2 cache overflows.
	Performance falls off a cliff (even on this log-log plot) at about
	300,000~elements, where the 40,370,176-byte L3 cache overflows.
	This demonstrates that the memory-system bottleneck is profound,
	degrading performance by well in excess of an order of magnitude
	for the large hash tables.
	This should not be a surprise, as the size-8,388,608 hash table
	occupies about 1\,GB of memory, overflowing the L3 caches by
	a factor of 25.

	The reason that
	\cref{fig:datastruct:Read-Only Hash-Table Performance For Schroedinger's Zoo; Varying Buckets}
	on \cpageref{fig:datastruct:Read-Only Hash-Table Performance For Schroedinger's Zoo; Varying Buckets}
	shows little effect is that its data was gathered from
	bucket-locked hash tables, where locking overhead and contention
	drowned out cache-capacity effects.
	In contrast, both RCU and hazard-pointers readers avoid stores
	to shared data, which means that the cache-capacity effects come
	to the fore.

	Still not satisfied?
	Find a multi-socket system and run this code, making use of
	whatever performance-counter hardware is available.
	This hardware should allow you to track down the precise cause
	of any slowdowns exhibited on your particular system.
	The experience gained by doing this exercise will be extremely
	valuable, giving you a significant advantage over those whose
	understanding of this issue is strictly theoretical.\footnote{
		Of course, a theoretical understanding beats no
		understanding.}
}\QuickQuizEnd

What if the memory footprint is reduced still further?
\Cref{fig:defer:Pre-BSD Routing Table Protected by RCU QSBR With Non-Initial rcu-head}
on \cpageref{fig:defer:Pre-BSD Routing Table Protected by RCU QSBR With Non-Initial rcu-head}
shows that RCU attains very nearly ideal performance on the much smaller
data structure represented by the pre-BSD routing table.

\QuickQuiz{
	The memory system is a serious bottleneck on this big system.
	Why bother putting 448~CPUs on a system without giving them
	enough memory bandwidth to do something useful???
}\QuickQuizAnswer{
	It would indeed be a bad idea to use this large and expensive
	system for a workload consisting solely of simple hash-table
	lookups of small data elements.
	However, this system is extremely useful for a great many
	workloads that feature more processing and less memory accessing.
	For example, some in-memory databases run extremely well on
	this class of system, albeit when running much more complex
	sets of queries than performed by the benchmarks in this chapter.
	For example, such systems might be processing images or video
	streams stored in each element, providing further performance
	benefits due to the fact that the resulting sequential memory
	accesses will make better use of the available memory bandwidth
	than will a pure pointer-following workload.

	But let this be a lesson to you.
	Modern computer systems come in a great many shapes and sizes,
	and great care is frequently required to select one that suits
	your application.
	And perhaps even more frequently, significant care and work is
	required to adjust your application to the specific computer
	systems at hand.
}\QuickQuizEnd

\begin{figure}
\centering
\resizebox{2.5in}{!}{\includegraphics{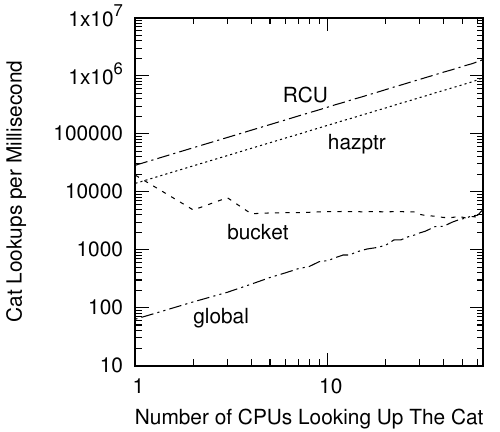}}
\caption{Read-Side Cat-Only RCU-Protected Hash-Table Performance For Schr\"odinger's Zoo at 64 CPUs}
\label{fig:datastruct:Read-Side Cat-Only RCU-Protected Hash-Table Performance For Schroedinger's Zoo at 64 CPUs}
\end{figure}

As noted earlier, Schr\"odinger is surprised by the popularity of his
cat~\cite{ErwinSchroedinger1935Cat}, but recognizes the need to reflect
this popularity in his design.
\Cref{fig:datastruct:Read-Side Cat-Only RCU-Protected Hash-Table Performance For Schroedinger's Zoo at 64 CPUs}
shows the results of 64-CPU runs, varying the number of CPUs that are
doing nothing but looking up the cat.
Both RCU and hazard pointers respond well to this challenge, but bucket
locking scales negatively, eventually performing as badly as global
locking.
This should not be a surprise because if all CPUs are doing nothing
but looking up the cat, the lock corresponding to the cat's bucket
is for all intents and purposes a global lock.

This cat-only benchmark illustrates one potential problem with
fully partitioned sharding approaches.
Only the CPUs associated with the cat's
partition is able to access the cat, limiting the cat-only
throughput.
Of course, a great many applications have good load-spreading
properties, and for these applications sharding works
quite well.
However, sharding does not handle ``\IXpl{hot spot}'' very well, with
the hot spot exemplified by Schr\"odinger's cat being but one case
in point.

\begin{figure}
\centering
\resizebox{2.5in}{!}{\includegraphics{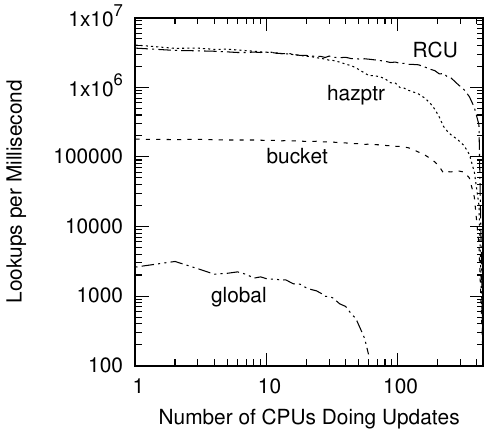}}
\caption{Read-Side RCU-Protected Hash-Table Performance For Schr\"odinger's Zoo in the Presence of Updates}
\label{fig:datastruct:Read-Side RCU-Protected Hash-Table Performance For Schroedinger's Zoo in the Presence of Updates}
\end{figure}

If we were only ever going to read the data, we would not need any
concurrency control to begin with.
\Cref{fig:datastruct:Read-Side RCU-Protected Hash-Table Performance For Schroedinger's Zoo in the Presence of Updates}
therefore shows the effect of updates on readers.
At the extreme left-hand side of this graph, all but one of the CPUs
are doing lookups, while to the right all 448~CPUs are doing updates.
For all four implementations, the number of lookups per millisecond
decreases as the number of updating CPUs increases, of course reaching
zero lookups per millisecond when all 448~CPUs are updating.
Both hazard pointers and RCU do well compared to per-bucket locking
because their readers do not increase update-side lock contention.
RCU does well relative to hazard pointers as the number of updaters
increases due to the latter's read-side memory barriers, which incur
greater overhead, especially in the presence of updates, and particularly
when execution involves more than one socket.
It therefore seems likely that modern hardware heavily optimizes memory-barrier
execution, greatly reducing memory-barrier overhead in the read-only case.
% @@@ When there is a section covering trees, reference sharding with
% @@@ CPUs being permitted to directly look up popular entries.

\begin{figure}
\centering
\resizebox{2.5in}{!}{\includegraphics{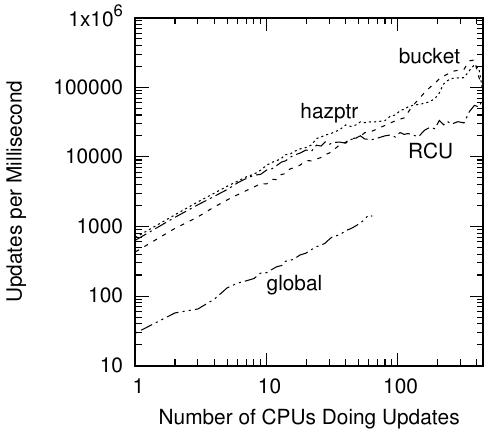}}
\caption{Update-Side RCU-Protected Hash-Table Performance For Schr\"odinger's Zoo}
\label{fig:datastruct:Update-Side RCU-Protected Hash-Table Performance For Schroedinger's Zoo}
\end{figure}

Where
\cref{fig:datastruct:Read-Side RCU-Protected Hash-Table Performance For Schroedinger's Zoo in the Presence of Updates}
showed the effect of increasing update rates on lookups,
\cref{fig:datastruct:Update-Side RCU-Protected Hash-Table Performance For Schroedinger's Zoo}
shows the effect of increasing update rates on the updates themselves.
Again, at the left-hand side of the figure all but one of the CPUs are
doing lookups and at the right-hand side of the figure all 448~CPUs are
doing updates.
Hazard pointers and RCU start off with a significant advantage because,
unlike bucket locking, readers do not exclude updaters.
However, as the number of updating CPUs increases, update-side overhead
starts to make its presence known, first for RCU and then for hazard
pointers.
Of course, all three of these implementations beat global locking.

It is quite possible that the differences in lookup performance observed in
\cref{fig:datastruct:Read-Side RCU-Protected Hash-Table Performance For Schroedinger's Zoo in the Presence of Updates}
are affected by the differences in update rates.
One way to check this is to artificially throttle the update rates of
per-bucket locking and hazard pointers to match that of RCU\@.
Doing so does not significantly improve the lookup performance of
per-bucket locking, nor does it close the gap between hazard pointers
and RCU\@.
However, removing the read-side memory barriers from hazard pointers
(thus resulting in an unsafe implementation) does nearly close the gap
between hazard pointers and RCU\@.
Although this unsafe hazard-pointer implementation will
usually be reliable enough for benchmarking purposes, it is absolutely
not recommended for production use.

\QuickQuiz{
	The dangers of extrapolating from 28~CPUs to 448~CPUs was
	made quite clear in
	\cref{sec:datastruct:Hash-Table Performance}.
	Would extrapolating up from 448~CPUs be any safer?
}\QuickQuizAnswer{
	In theory, no, it isn't any safer, and a useful exercise would be
	to run these programs on larger systems.
	In practice, there are only a very few systems with more than
	448~CPUs, in contrast to the huge number having more than 28~CPUs.
	This means that although it is dangerous to extrapolate beyond
	448~CPUs, there is very little need to do so.

	In addition, other testing has shown that RCU read-side primitives
	offer consistent performance and scalability up to at least 1024~CPUs.
	However, it is useful to review
	\cref{fig:datastruct:Read-Only RCU-Protected Hash-Table Performance For Schr\"odinger's Zoo at 448 CPUs; Varying Table Size}
	and its associated commentary.
	You see, unlike the 448-CPU system that provided this data,
	the system enjoying linear scalability up to 1024~CPUs boasted
	excellent memory bandwidth.
}\QuickQuizEnd

And this situation exposes yet another of the complications listed on
\cpageref{sec:datastruct:Update-activity failure to scale}.

% @@@ Testing strategy.  Summarize hashtorture, add QQ for additional
% @@@ things that could be done.  Note difference between textbook
% @@@ and production algorithms, also the need for clear understanding
% @@@ of what level of consistency is required.

\subsection{RCU-Protected Hash Table Discussion}
\label{sec:datastruct:RCU-Protected Hash Table Discussion}

One consequence of the RCU and hazard-pointer implementations is
that a pair of concurrent readers might disagree on the state of
the cat.
For example, one of the readers might have fetched the pointer to
the cat's data structure just before it was removed, while another
reader might have fetched this same pointer just afterwards.
The first reader would then believe that the cat was alive, while
the second reader would believe that the cat was dead.

This situation is completely fitting for Schr\"odinger's
cat, but it turns out that it is quite reasonable for normal
non-quantum cats as well.
After all, it is impossible to determine exactly when an animal is born
or dies.

To see this, let's suppose that we detect a cat's death by heartbeat.
This raise the question of exactly how long we should wait after the
last heartbeat before declaring death.
It is clearly ridiculous to wait only one millisecond, because then
a healthy living cat would have to be declared dead---and then
resurrected---more than once per second.
It is equally ridiculous to wait a full month, because by that time
the poor cat's death would have made itself very clearly known
via olfactory means.

\begin{figure}
\centering
\resizebox{3in}{!}{\includegraphics{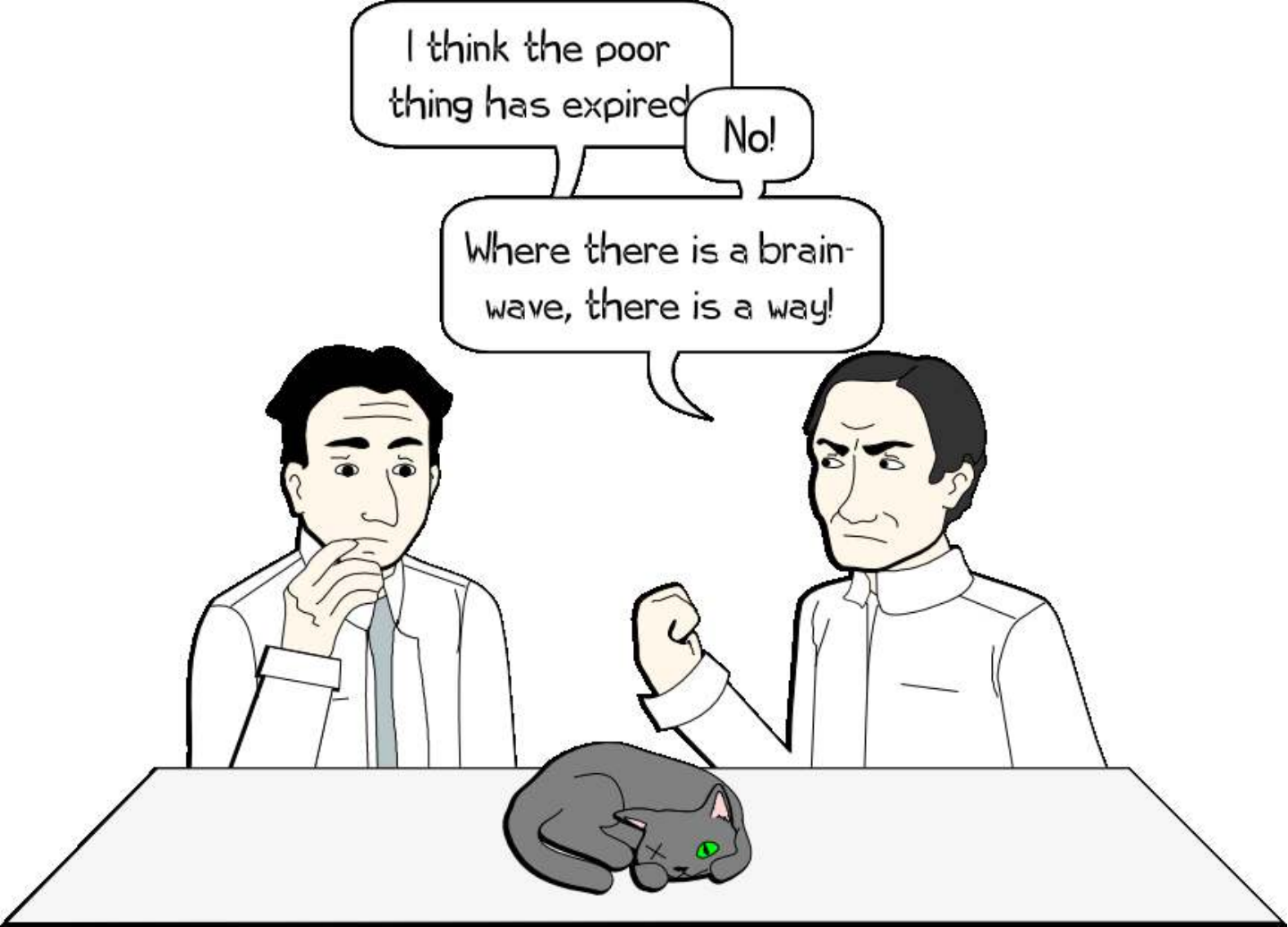}}
\caption{Even Veterinarians Disagree!}
\ContributedBy{Figure}{fig:datastruct:Even Veterinarians Disagree}{Melissa Broussard}
\end{figure}

Because an animal's heart can stop for some seconds and then start up
again, there is a tradeoff between timely recognition of death and
probability of false alarms.
It is quite possible that a pair of veterinarians might disagree on
the time to wait between the last heartbeat and the declaration of
death.
For example, one veterinarian might declare death thirty seconds after
the last heartbeat, while another might insist on waiting a full
minute.
In this case, the two veterinarians would disagree on the state of the
cat for the second period of thirty seconds following the last heartbeat,
as fancifully depicted in
\cref{fig:datastruct:Even Veterinarians Disagree}.

\pplsur{Weiner}{Heisenberg} taught us to live with this sort of
uncertainty~\cite{WeinerHeisenberg1927Uncertain}, which is a good
thing because computing hardware and software acts similarly.
For example, how do you know that a piece of computing hardware
has failed?
Often because it does not respond in a timely fashion.
Just like the cat's heartbeat, this results in a window of
uncertainty as to whether or not the hardware has really failed,
as opposed to just being slow.

Furthermore, most computing systems are intended to interact with
the outside world.
Consistency with the outside world is therefore of paramount importance.
However, as we saw in
\cref{fig:defer:Response Time of RCU vs. Reader-Writer Locking}
on
\cpageref{fig:defer:Response Time of RCU vs. Reader-Writer Locking},
increased internal consistency can come at the expense of degraded
external consistency.
Techniques such as RCU and hazard pointers give up some degree of
internal consistency to attain improved external consistency.

In short, internal consistency is not necessarily a natural part of all
problem domains, and often incurs great expense in terms of performance,
scalability, consistency with the outside
world~\cite{AndreasHaas2012FIFOisnt,AndreasHaas2013CFRelaxedQueues,10.5555/3241639.3241645},
or all of the above.

\section{Non-Partitionable Data Structures}
\label{sec:datastruct:Non-Partitionable Data Structures}
\epigraph{Don't be afraid to take a big step if one is indicated.
	  You can't cross a chasm in two small steps.}
	 {David Lloyd George}

Fixed-size hash tables are perfectly partitionable, but resizable hash
tables pose partitioning challenges when growing or shrinking, as
fancifully depicted in
\cref{fig:datastruct:Partitioning Problems}.
However, it turns out that it is possible to construct high-performance
scalable RCU-protected hash tables, as described in the following sections.

\begin{figure}
\centering
\resizebox{3in}{!}{\includegraphics{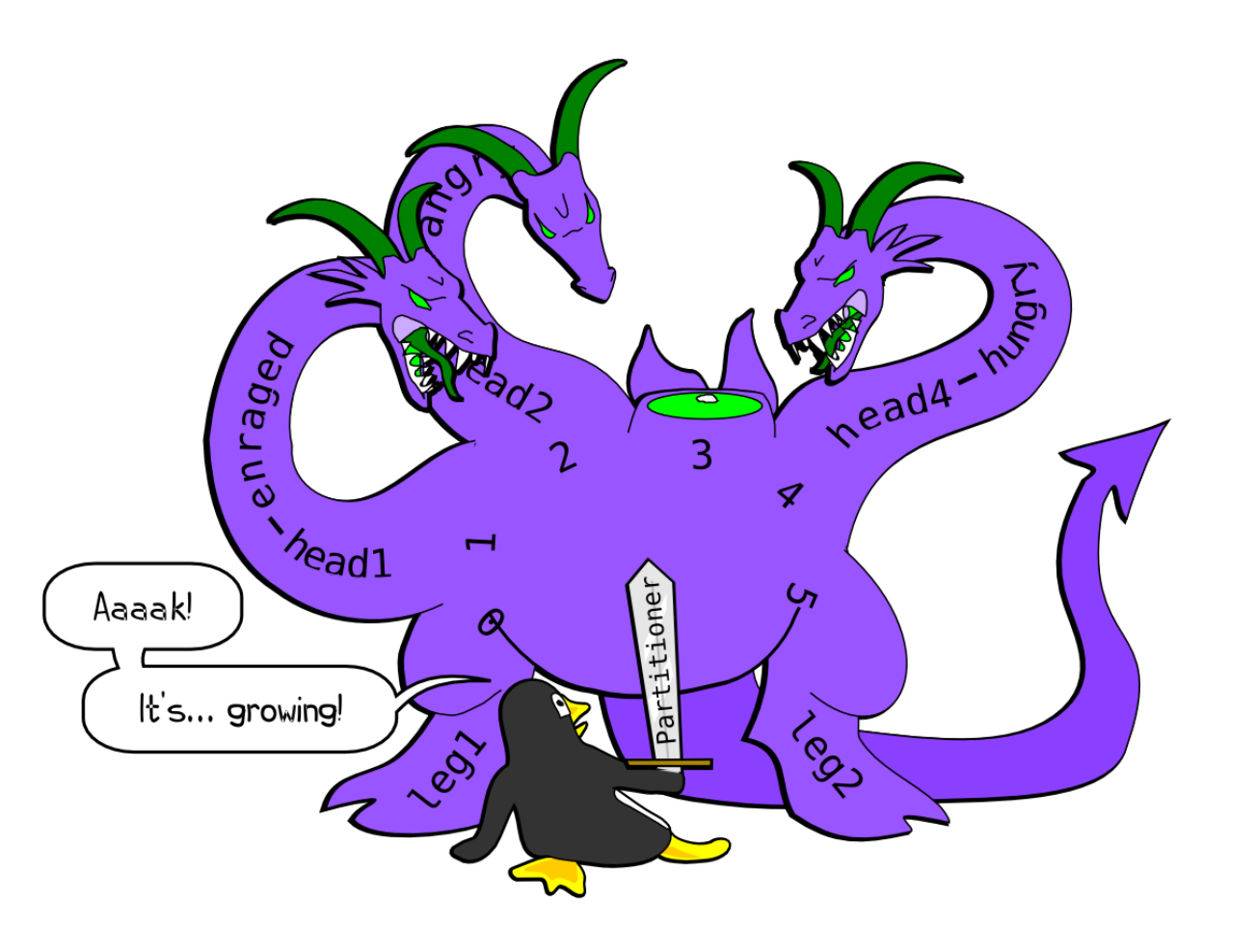}}
\caption{Partitioning Problems}
\ContributedBy{Figure}{fig:datastruct:Partitioning Problems}{Melissa Broussard}
\end{figure}

\subsection{Resizable Hash Table Design}
\label{sec:datastruct:Resizable Hash Table Design}

In happy contrast to the situation in the early 2000s, there are now
no fewer than three different types of scalable RCU-protected hash
tables.
The first (and simplest) was developed for the Linux kernel by
\ppl{Herbert}{Xu}~\cite{HerbertXu2010RCUResizeHash}, and is described in the
following sections.
The other two are covered briefly in
\cref{sec:datastruct:Other Resizable Hash Tables}.

The key insight behind the first hash-table implementation is that
each data element can have two sets of list pointers, with one set
currently being used by RCU readers (as well as by non-RCU updaters)
and the other being used to construct a new resized hash table.
This approach allows lookups, insertions, and deletions to all run
concurrently with a resize operation (as well as with each other).

\begin{figure}
\centering
\resizebox{3in}{!}{\includegraphics{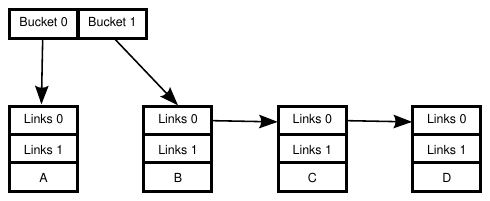}}
\caption{Growing a Two-List Hash Table, State~(a)}
\label{fig:datastruct:Growing a Two-List Hash Table; State (a)}
\end{figure}

\begin{figure}
\centering
\resizebox{3in}{!}{\includegraphics{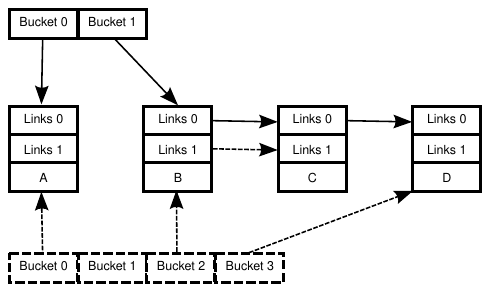}}
\caption{Growing a Two-List Hash Table, State~(b)}
\label{fig:datastruct:Growing a Two-List Hash Table; State (b)}
\end{figure}

\begin{figure}
\centering
\resizebox{3in}{!}{\includegraphics{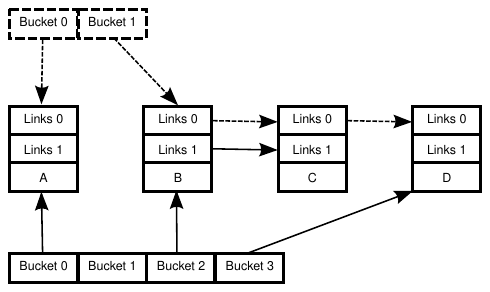}}
\caption{Growing a Two-List Hash Table, State~(c)}
\label{fig:datastruct:Growing a Two-List Hash Table; State (c)}
\end{figure}

\begin{figure}
\centering
\resizebox{3in}{!}{\includegraphics{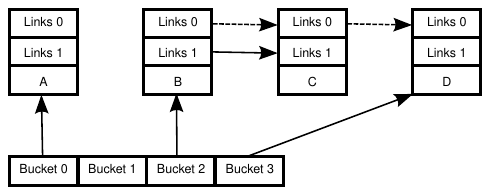}}
\caption{Growing a Two-List Hash Table, State~(d)}
\label{fig:datastruct:Growing a Two-List Hash Table; State (d)}
\end{figure}

The resize operation proceeds as shown in
\crefrange{fig:datastruct:Growing a Two-List Hash Table; State (a)}
{fig:datastruct:Growing a Two-List Hash Table; State (d)},
with the initial two-bucket state shown in
\cref{fig:datastruct:Growing a Two-List Hash Table; State (a)}
and with time advancing from figure to figure.
The initial state uses the zero-index links to chain the elements into
hash buckets.
A four-bucket array is allocated, and the one-index links are used to
chain the elements into these four new hash buckets.
This results in state~(b) shown in
\cref{fig:datastruct:Growing a Two-List Hash Table; State (b)},
with readers still using the original two-bucket array.

The new four-bucket array is exposed to readers and then a grace-period
operation waits for all readers, resulting in state~(c), shown in
\cref{fig:datastruct:Growing a Two-List Hash Table; State (c)}.
In this state, all readers are using the new four-bucket array,
which means that the old two-bucket array may now be freed, resulting
in state~(d), shown in
\cref{fig:datastruct:Growing a Two-List Hash Table; State (d)}.

This design leads to a relatively straightforward implementation,
which is the subject of the next section.

\subsection{Resizable Hash Table Implementation}
\label{sec:datastruct:Resizable Hash Table Implementation}

\begin{fcvref}[ln:datastruct:hash_resize:data]
Resizing is accomplished by the classic approach of inserting a level
of indirection, in this case, the \co{ht} structure shown on
\clnrefrange{ht:b}{ht:e} of
\cref{lst:datastruct:Resizable Hash-Table Data Structures}
(\path{hash_resize.c}).
The \co{hashtab} structure shown on
\clnrefrange{hashtab:b}{hashtab:e} contains only a
pointer to the current \co{ht} structure along with a spinlock that
is used to serialize concurrent attempts to resize the hash table.
If we were to use a traditional lock- or atomic-operation-based
implementation, this \co{hashtab} structure could become a severe bottleneck
from both performance and scalability viewpoints.
However, because resize operations should be relatively infrequent,
we should be able to make good use of RCU\@.

\begin{listing}
\input{CodeSamples/datastruct/hash/hash_resize=data.fcv}
\caption{Resizable Hash-Table Data Structures}
\label{lst:datastruct:Resizable Hash-Table Data Structures}
\end{listing}

The \co{ht} structure represents a specific size of the hash table,
as specified by the \co{->ht_nbuckets} field on \clnref{ht:nbuckets}.
The size is stored in the same structure containing the array of
buckets (\co{->ht_bkt[]} on
\clnref{ht:bkt}) in order to avoid mismatches between
the size and the array.
The \co{->ht_resize_cur} field on
\clnref{ht:resize_cur} is equal to $-1$ unless a resize
operation
is in progress, in which case it indicates the index of the bucket whose
elements are being inserted into the new hash table, which is referenced
by the \co{->ht_new} field on \clnref{ht:new}.
If there is no resize operation in progress, \co{->ht_new} is \co{NULL}.
Thus, a resize operation proceeds by allocating a new \co{ht} structure
and referencing it via the \co{->ht_new} pointer, then advancing
\co{->ht_resize_cur} through the old table's buckets.
When all the elements have been added to the new table, the new
table is linked into the \co{hashtab} structure's \co{->ht_cur} field.
Once all old readers have completed, the old hash table's \co{ht} structure
may be freed.

The \co{->ht_idx} field on
\clnref{ht:idx} indicates which of the two sets of
list pointers are being used by this instantiation of the hash table,
and is used to index the \co{->hte_next[]} array in the \co{ht_elem}
structure on \clnref{ht_elem:next}.

The \co{->ht_cmp()}, \co{->ht_gethash()}, and \co{->ht_getkey()} fields on
\clnrefrange{ht:cmp}{ht:getkey}
collectively define the per-element key and the hash function.
The \co{->ht_cmp()} function compares a specified key with that of
the specified element,
the \co{->ht_gethash()} calculates the specified key's hash,
and \co{->ht_getkey()} extracts the key from the enclosing data
element.

The \co{ht_lock_state} shown on \clnrefrange{hls:b}{hls:e}
is used to communicate lock state from a new \co{hashtab_lock_mod()}
to \co{hashtab_add()}, \co{hashtab_del()}, and \co{hashtab_unlock_mod()}.
This state prevents the algorithm from being redirected to the wrong
bucket during concurrent resize operations.

The \co{ht_bucket} structure is the same as before, and the
\co{ht_elem} structure differs from that of previous implementations
only in providing a two-element array of list pointer sets in place of
the prior single set of list pointers.

In a fixed-sized hash table, bucket selection is quite straightforward:
Simply transform the hash value to the corresponding bucket index.
In contrast, when resizing, it is also necessary to determine which
of the old and new sets of buckets to select from.
If the bucket that would be selected from the old table has already
been distributed into the new table, then the bucket should be selected
from the new table as well as from the old table.
Conversely, if the bucket that would be selected from the old table
has not yet been distributed, then the bucket should be selected from
the old table.
\end{fcvref}

\begin{listing}
\input{CodeSamples/datastruct/hash/hash_resize=get_bucket.fcv}
\caption{Resizable Hash-Table Bucket Selection}
\label{lst:datastruct:Resizable Hash-Table Bucket Selection}
\end{listing}

\begin{fcvref}[ln:datastruct:hash_resize:get_bucket]
Bucket selection is shown in
\cref{lst:datastruct:Resizable Hash-Table Bucket Selection},
which shows \co{ht_get_bucket()} on
\clnrefrange{single:b}{single:e} and \co{ht_search_bucket()} on
\clnrefrange{hsb:b}{hsb:e}.
The \co{ht_get_bucket()} function returns a reference to the bucket
corresponding to the specified key in the specified hash table, without
making any allowances for resizing.
It also stores the bucket index corresponding to the key into the location
referenced by parameter~\co{b} on
\clnref{single:gethash}, and the corresponding
hash value corresponding to the key into the location
referenced by parameter~\co{h} (if non-\co{NULL}) on \clnref{single:h}.
\Clnref{single:return} then returns a reference to the corresponding bucket.

The \co{ht_search_bucket()} function searches for the specified key
within the specified hash-table version.
\Clnref{hsb:get_curbkt} obtains a reference to the bucket corresponding
to the specified key.
The loop spanning \clnrefrange{hsb:loop:b}{hsb:loop:e} searches
that bucket, so that if \clnref{hsb:match} detects a match,
\clnref{hsb:ret_match} returns a pointer to the enclosing data element.
Otherwise, if there is no match,
\clnref{hsb:ret_NULL} returns \co{NULL} to indicate
failure.
\end{fcvref}

\QuickQuiz{
	How does the code in
	\cref{lst:datastruct:Resizable Hash-Table Bucket Selection}
	protect against the resizing process progressing past the
	selected bucket?
}\QuickQuizAnswer{
	It does not provide any such protection.
	That is instead the job of the update-side concurrency-control
	functions described next.
}\QuickQuizEnd

This implementation of \co{ht_get_bucket()} and \co{ht_search_bucket()}
permits lookups and modifications to run concurrently with a resize
operation.

\begin{listing}
\input{CodeSamples/datastruct/hash/hash_resize=lock_unlock_mod.fcv}
\caption{Resizable Hash-Table Update-Side Concurrency Control}
\label{lst:datastruct:Resizable Hash-Table Update-Side Concurrency Control}
\end{listing}

Read-side concurrency control is provided by RCU as was shown in
\cref{lst:datastruct:RCU-Protected Hash-Table Read-Side Concurrency Control},
but the update-side concurrency\-/control functions
\co{hashtab_lock_mod()} and \co{hashtab_unlock_mod()}
must now deal with the possibility of a
concurrent resize operation as shown in
\cref{lst:datastruct:Resizable Hash-Table Update-Side Concurrency Control}.

\begin{fcvref}[ln:datastruct:hash_resize:lock_unlock_mod:l]
The \co{hashtab_lock_mod()} spans
\clnrefrange{b}{e} in the listing.
\Clnref{rcu_lock} enters an RCU read-side critical section to prevent
the data structures from being freed during the traversal,
\clnref{refhashtbl} acquires a reference to the current hash table, and then
\clnref{refbucket} obtains a reference to the bucket in this hash table
corresponding to the key.
\Clnref{acq_bucket} acquires that bucket's lock, which will prevent any concurrent
resizing operation from distributing that bucket, though of course it
will have no effect if that bucket has already been distributed.
\Clnrefrange{lsp0b}{lsp0e} store the bucket pointer and
pointer-set index into their respective fields in the
\co{ht_lock_state} structure, which communicates the information to
\co{hashtab_add()}, \co{hashtab_del()}, and \co{hashtab_unlock_mod()}.
\Clnref{ifresized} then checks to see if a concurrent resize
operation has already distributed this bucket across the new hash table,
and if not, \clnref{lsp1_1} indicates that there is no
already-resized hash bucket and
\clnref{fastret1} returns with the selected hash bucket's
lock held (thus preventing a concurrent resize operation from distributing
this bucket) and also within an RCU read-side critical section.
\IX{Deadlock} is avoided because the old table's locks are always acquired
before those of the new table, and because the use of RCU prevents more
than two versions from existing at a given time, thus preventing a
deadlock cycle.

Otherwise, a concurrent resize operation has already distributed this
bucket, so \clnref{new_hashtbl} proceeds to the new hash table,
\clnref{get_newbkt} selects the bucket corresponding to the key,
and \clnref{acq_newbkt} acquires the bucket's lock.
\Clnrefrange{lsp1b}{lsp1e} store the bucket pointer and
pointer-set index into their respective fields in the
\co{ht_lock_state} structure, which again communicates this information to
\co{hashtab_add()}, \co{hashtab_del()}, and \co{hashtab_unlock_mod()}.
Because this bucket has already been resized and because
\co{hashtab_add()} and \co{hashtab_del()} affect both the old and the
new \co{ht_bucket} structures, two locks are held, one on each of the
two buckets.
Additionally, both elements of each array in \co{ht_lock_state} structure
are used, with the \co{[0]} element pertaining to the old \co{ht_bucket}
structure and the \co{[1]} element pertaining to the new structure.
Once again, \co{hashtab_lock_mod()} exits within an RCU read-side critical
section.
\end{fcvref}

\begin{fcvref}[ln:datastruct:hash_resize:lock_unlock_mod:ul]
The \co{hashtab_unlock_mod()} function releases the lock(s) acquired by
\co{hashtab_lock_mod()}.
\Clnref{relbkt0} releases the lock on the old \co{ht_bucket} structure.
In the unlikely event that \clnref{ifbkt1} determines that a resize
operation is in progress, \clnref{relbkt1} releases the lock on the
new \co{ht_bucket} structure.
Either way, \clnref{rcu_unlock} exits the RCU read-side critical
section.
\end{fcvref}

\QuickQuiz{
	Suppose that one thread is inserting an element into the
	hash table during a resize operation.
	What prevents this insertion from being lost due to a subsequent
	resize operation completing before the insertion does?
}\QuickQuizAnswer{
	The second resize operation will not be able to move beyond
	the bucket into which the insertion is taking place due to
	the insertion holding the lock(s) on one or both of the hash
	buckets in the hash tables.
	Furthermore, the insertion operation takes place within an
	RCU read-side critical section.
	As we will see when we examine the \co{hashtab_resize()}
	function, this means that each resize operation uses
	\co{synchronize_rcu()} invocations to wait for the insertion's
	read-side critical section to complete.
}\QuickQuizEnd

\begin{listing}
\input{CodeSamples/datastruct/hash/hash_resize=access.fcv}
\caption{Resizable Hash-Table Access Functions}
\label{lst:datastruct:Resizable Hash-Table Access Functions}
\end{listing}

Now that we have bucket selection and concurrency control in place,
we are ready to search and update our resizable hash table.
The \co{hashtab_lookup()}, \co{hashtab_add()}, and \co{hashtab_del()}
functions are shown in
\cref{lst:datastruct:Resizable Hash-Table Access Functions}.

\begin{fcvref}[ln:datastruct:hash_resize:access:lkp]
The \co{hashtab_lookup()} function on
\clnrefrange{b}{e} of the listing does
hash lookups.
\Clnref{get_curtbl} fetches the current hash table and
\clnref{get_curbkt} searches the bucket corresponding to the
specified key.
\Clnref{ret} returns a pointer to the searched-for element
or \co{NULL} when the search fails.
The caller must be within an RCU read-side critical section.
\end{fcvref}

\QuickQuiz{
	The \co{hashtab_lookup()} function in
	\cref{lst:datastruct:Resizable Hash-Table Access Functions}
	ignores concurrent resize operations.
	Doesn't this mean that readers might miss an element that was
	previously added during a resize operation?
}\QuickQuizAnswer{
	No.
	As we will see soon,
	the \co{hashtab_add()} and \co{hashtab_del()} functions
	keep the old hash table up-to-date while a resize operation
	is in progress.
}\QuickQuizEnd

\begin{fcvref}[ln:datastruct:hash_resize:access:add]
The \co{hashtab_add()} function on \clnrefrange{b}{e} of the listing adds
new data elements to the hash table.
\Clnref{htbp} picks up the current \co{ht_bucket} structure into which the
new element is to be added, and \clnref{i} picks up the index of
the pointer pair.
\Clnref{add} adds the new element to the current hash bucket.
If \clnref{ifnew} determines that this bucket has been distributed
to a new version of the hash table, then \clnref{addnew} also adds the
new element to the corresponding new bucket.
The caller is required to handle concurrency, for example, by invoking
\co{hashtab_lock_mod()} before the call to \co{hashtab_add()} and invoking
\co{hashtab_unlock_mod()} afterwards.
\end{fcvref}

\begin{fcvref}[ln:datastruct:hash_resize:access:del]
The \co{hashtab_del()} function on
\clnrefrange{b}{e} of the listing removes
an existing element from the hash table.
\Clnref{i} picks up the index of the pointer pair
and \clnref{del} removes the specified element from the current table.
If \clnref{ifnew} determines that this bucket has been distributed
to a new version of the hash table, then \clnref{delnew} also removes
the specified element from the corresponding new bucket.
As with \co{hashtab_add()}, the caller is responsible for concurrency
control and this concurrency control suffices for synchronizing with
a concurrent resize operation.
\end{fcvref}

\QuickQuiz{
	The \co{hashtab_add()} and \co{hashtab_del()} functions in
	\cref{lst:datastruct:Resizable Hash-Table Access Functions}
	can update two hash buckets while a resize operation is progressing.
	This might cause poor performance if the frequency of resize operation
	is not negligible.
	Isn't it possible to reduce the cost of updates in such cases?
}\QuickQuizAnswer{
	Yes, at least assuming that a slight increase in the cost of
	\co{hashtab_lookup()} is acceptable.
	One approach is shown in
	\cref{lst:datastruct:Resizable Hash-Table Access Functions (Fewer Updates),%
	lst:datastruct:Resizable Hash-Table Update-Side Locking Function (Fewer Updates)}
	(\path{hash_resize_s.c}).

\begin{listing}
\input{CodeSamples/datastruct/hash/hash_resize_s=access.fcv}
\caption{Resizable Hash-Table Access Functions (Fewer Updates)}
\label{lst:datastruct:Resizable Hash-Table Access Functions (Fewer Updates)}
\end{listing}

\begin{listing}
\input{CodeSamples/datastruct/hash/hash_resize_s=lock_mod.fcv}
\caption{Resizable Hash-Table Update-Side Locking Function (Fewer Updates)}
\label{lst:datastruct:Resizable Hash-Table Update-Side Locking Function (Fewer Updates)}
\end{listing}

	This version of \co{hashtab_add()} adds an element to
	either the old bucket if it is not resized yet, or to the new
	bucket if it has been resized, and \co{hashtab_del()} removes
	the specified element from any buckets into which it has been inserted.
	The \co{hashtab_lookup()} function searches the new bucket
	if the search of the old bucket fails, which has the disadvantage
	of adding overhead to the lookup fastpath.
	The alternative \co{hashtab_lock_mod()} returns the locking
	state of the new bucket in \co{->hbp[0]} and \co{->hls_idx[0]}
	if resize operation is in progress, instead of the perhaps
	more natural choice of \co{->hbp[1]} and \co{->hls_idx[1]}.
	However, this less-natural choice has the advantage of simplifying
	\co{hashtab_add()}.

	Further analysis of the code is left as an exercise for the reader.
}\QuickQuizEnd

\begin{listing*}
\input{CodeSamples/datastruct/hash/hash_resize=resize.fcv}
\caption{Resizable Hash-Table Resizing}
\label{lst:datastruct:Resizable Hash-Table Resizing}
\end{listing*}

\begin{fcvref}[ln:datastruct:hash_resize:resize]
The actual resizing itself is carried out by \co{hashtab_resize}, shown in
\cref{lst:datastruct:Resizable Hash-Table Resizing} on
\cpageref{lst:datastruct:Resizable Hash-Table Resizing}.
\Clnref{trylock} conditionally acquires the top-level \co{->ht_lock}, and if
this acquisition fails, \clnref{ret_busy} returns \co{-EBUSY} to indicate that
a resize is already in progress.
Otherwise, \clnref{get_curtbl} picks up a reference to the current hash table,
and \clnrefrange{alloc:b}{alloc:e} allocate a new hash table of the desired size.
If a new set of hash/key functions have been specified, these are
used for the new table, otherwise those of the old table are preserved.
If \clnref{chk_nomem} detects memory-allocation failure,
\clnref{rel_nomem} releases \co{->ht_lock}
and \clnref{ret_nomem} returns a failure indication.

\Clnref{get_curidx} picks up the current table's index and
\clnref{put_curidx} stores its inverse to
the new hash table, thus ensuring that the two hash tables avoid overwriting
each other's linked lists.
\Clnref{set_newtbl} then starts the bucket-distribution process by
installing a reference to the new table into the \co{->ht_new} field of
the old table.
\Clnref{sync_rcu} ensures that all readers who are not aware of the
new table complete before the resize operation continues.

Each pass through the loop spanning \clnrefrange{loop:b}{loop:e} distributes the contents
of one of the old hash table's buckets into the new hash table.
\Clnref{get_oldcur} picks up a reference to the old table's current bucket
and \clnref{acq_oldcur} acquires that bucket's spinlock.
\end{fcvref}

\QuickQuiz{
	\begin{fcvref}[ln:datastruct:hash_resize:resize]
	In the \co{hashtab_resize()} function in
	\cref{lst:datastruct:Resizable Hash-Table Resizing},
	what guarantees that the update to \co{->ht_new} on \clnref{set_newtbl}
	will be seen as happening before the update to \co{->ht_resize_cur}
	on \clnref{update_resize} from the perspective of
	\co{hashtab_add()} and \co{hashtab_del()}?
	In other words, what prevents \co{hashtab_add()}
	and \co{hashtab_del()} from dereferencing
	a \co{NULL} pointer loaded from \co{->ht_new}?
	\end{fcvref}
}\QuickQuizAnswer{
	\begin{fcvref}[ln:datastruct:hash_resize:resize]
	The \co{synchronize_rcu()} on \clnref{sync_rcu} of
	\cref{lst:datastruct:Resizable Hash-Table Resizing}
	ensures that all pre-existing RCU readers have completed between
	the time that we install the new hash-table reference on
	\clnref{set_newtbl} and the time that we update \co{->ht_resize_cur} on
	\clnref{update_resize}.
	This means that any reader that sees a non-negative value
	of \co{->ht_resize_cur} cannot have started before the
	assignment to \co{->ht_new}, and thus must be able to see
	the reference to the new hash table.

	And this is why the update-side \co{hashtab_add()} and
	\co{hashtab_del()} functions must be enclosed
	in RCU read-side critical sections, courtesy of
	\co{hashtab_lock_mod()} and \co{hashtab_unlock_mod()} in
	\cref{lst:datastruct:Resizable Hash-Table Update-Side Concurrency Control}.
	\end{fcvref}
}\QuickQuizEnd

\begin{fcvref}[ln:datastruct:hash_resize:resize]
Each pass through the loop spanning
\clnrefrange{loop_list:b}{loop_list:e} adds one data element
from the current old-table bucket to the corresponding new-table bucket,
holding the new-table bucket's lock during the add operation.
\Clnref{update_resize} updates
\co{->ht_resize_cur} to indicate that this bucket has been distributed.
Finally, \clnref{rel_oldcur} releases the old-table bucket lock.

Execution reaches \clnref{rcu_assign} once all old-table buckets have been distributed
across the new table.
\Clnref{rcu_assign} installs the newly created table as the current one, and
\clnref{sync_rcu_2} waits for all old readers (who might still be referencing
the old table) to complete.
Then \clnref{rel_master} releases the resize-serialization lock,
\clnref{free} frees
the old hash table, and finally \clnref{ret_success} returns success.
\end{fcvref}

\QuickQuiz{
	\begin{fcvref}[ln:datastruct:hash_resize:resize]
	Why is there a \co{WRITE_ONCE()} on \clnref{update_resize}
	in \cref{lst:datastruct:Resizable Hash-Table Resizing}?
	\end{fcvref}
}\QuickQuizAnswer{
	\begin{fcvref}[ln:datastruct:hash_resize:lock_unlock_mod]
	Together with the \co{READ_ONCE()}
	on \clnref{l:ifresized} in \co{hashtab_lock_mod()}
	of \cref{lst:datastruct:Resizable Hash-Table Update-Side Concurrency Control},
	it tells the compiler that the non-initialization accesses
	to \co{->ht_resize_cur} must remain because reads
	from \co{->ht_resize_cur} really can race with writes,
	just not in a way to change the ``if'' conditions.
	\end{fcvref}
}\QuickQuizEnd

\subsection{Resizable Hash Table Discussion}
\label{sec:datastruct:Resizable Hash Table Discussion}

\begin{figure}
\centering
\resizebox{2.7in}{!}{\includegraphics{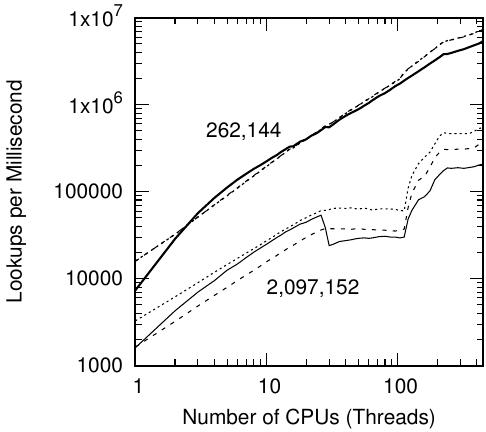}}
\caption{Overhead of Resizing Hash Tables Between 262,144 and 524,288 Buckets vs.\@ Total Number of Elements}
\label{fig:datastruct:Overhead of Resizing Hash Tables Between 262;144 and 524;288 Buckets vs. Total Number of Elements}
\end{figure}

\Cref{fig:datastruct:Overhead of Resizing Hash Tables Between 262;144 and 524;288 Buckets vs. Total Number of Elements}
compares resizing hash tables to their fixed-sized counterparts
for 262,144 and 2,097,152 elements in the hash table.
The figure shows three traces for each element count, one
for a fixed-size 262,144-bucket hash table, another for a
fixed-size 524,288-bucket hash table, and a third for a resizable
hash table that shifts back and forth between 262,144 and 524,288
buckets, with a one-millisecond pause between each resize operation.

The uppermost three traces are for the 262,144-element hash table.\footnote{
	You see only two traces?
	The dashed one is composed of two traces that differ
	only slightly, hence the irregular-looking dash pattern.}
The dashed trace corresponds to the two fixed-size hash tables,
and the solid trace to the resizable hash table.
In this case, the short hash chains cause normal lookup overhead
to be so low that the overhead of resizing dominates over most
of the range.
In particular, the entire hash table fits into L3 cache.
% Nevertheless, the larger fixed-size hash table has a significant
% performance advantage, so that resizing can be quite beneficial,
% at least given sufficient time between resizing operations: One
% millisecond is clearly too short a time.

The lower three traces are for the 2,097,152-element hash table.
The upper dashed trace corresponds to the 262,144-bucket fixed-size
hash table, the solid trace in the middle for low CPU counts and at
the bottom for high CPU counts to the resizable hash table,
and the other trace to the 524,288-bucket fixed-size hash table.
The fact that there are now an average of eight elements per bucket
can only be expected to produce a sharp decrease in performance,
as in fact is shown in the graph.
But worse yet, the hash-table elements occupy 128\,MB, which overflows
each socket's 39\,MB L3 cache, with performance consequences analogous
to those described in \cref{sec:cpu:Costs of Operations}.
The resulting cache overflow means that the memory system is involved
even for a read-only benchmark, and as you can see from the sublinear
portions of the lower three traces, the memory system can be a serious
bottleneck.

\QuickQuiz{
	How much of the difference in performance between the large and
	small hash tables shown in
	\cref{fig:datastruct:Overhead of Resizing Hash Tables Between 262;144 and 524;288 Buckets vs. Total Number of Elements}
	was due to long hash chains and how much was due to
	memory-system bottlenecks?
}\QuickQuizAnswer{
	The easy way to answer this question is to do another run with
	2,097,152 elements, but this time also with 2,097,152 buckets,
	thus bringing the average number of elements per bucket back down
	to unity.

\begin{figure}
\centering
\resizebox{2.7in}{!}{\includegraphics{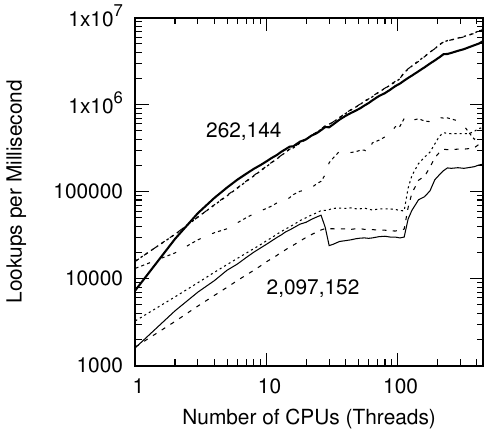}}
\caption{Effect of Memory-System Bottlenecks on Hash Tables}
\label{fig:datastruct:Effect of Memory-System Bottlenecks on Hash Tables}
\end{figure}

	The results are shown by the triple-dashed new trace in
	the middle of
	\cref{fig:datastruct:Effect of Memory-System Bottlenecks on Hash Tables}.
	The other six traces are identical to their counterparts in
	\cref{fig:datastruct:Overhead of Resizing Hash Tables Between 262;144 and 524;288 Buckets vs. Total Number of Elements}
	on \cpageref{fig:datastruct:Overhead of Resizing Hash Tables Between 262;144 and 524;288 Buckets vs. Total Number of Elements}.
	The gap between this new trace and the lower set of three
	traces is a rough measure of how much of the difference in
	performance was due to hash-chain length, and the gap between
	the new trace and the upper set of three traces is a rough measure
	of how much of that difference was due to memory-system bottlenecks.
	The new trace starts out slightly below its 262,144-element
	counterpart at a single CPU, showing that cache capacity is
	degrading performance slightly even on that single CPU\@.\footnote{
		Yes, as far as hardware architects are concerned,
		caches are part of the memory system.}
	This is to be expected, given that unlike its smaller counterpart,
	the 2,097,152-bucket hash table does not fit into the L3 cache.
	This new trace rises just past 28~CPUs, which is also to be
	expected.
	This rise is due to the fact that the 29\textsuperscript{th}
	CPU is on another socket, which brings with it an additional
	39\,MB of cache as well as additional memory bandwidth.

	But the large hash table's advantage over that of the hash table
	with 524,288~buckets (but still 2,097,152 elements) decreases
	with additional CPUs, which is consistent with the bottleneck
	residing in the memory system.
	Above about 400~CPUs, the 2,097,152-bucket hash table is
	actually outperformed slightly by the 524,288-bucket hash
	table.
	This should not be a surprise because the memory system is
	the bottleneck and the larger number of buckets increases this
	workload's memory footprint.

	The alert reader will have noted the word ``rough'' above
	and might be interested in a more detailed analysis.
	Such readers are invited to run similar benchmarks, using
	whatever performance counters or hardware-analysis tools
	they might have available.
	This can be a long and complex journey, but those brave enough
	to embark on it will be rewarded with detailed knowledge of
	hardware performance and its effect on software.
}\QuickQuizEnd

Referring to the last column of
\cref{tab:cpu:CPU 0 View of Synchronization Mechanisms on 8-Socket System With Intel Xeon Platinum 8176 CPUs at 2.10GHz},
we recall that the first 28~CPUs are in the first socket, on a
one-CPU-per-core basis, which explains the sharp decrease in performance
of the resizable hash table beyond 28~CPUs.
Sharp though this decrease is, please recall that it is due to constant
resizing back and forth.
It would clearly be better to resize once to 524,288 buckets,
or, even better, do a single eight-fold resize to 2,097,152 elements,
thus dropping the average number of elements per bucket down to the
level enjoyed by the runs producing the upper three traces.

The key point from this data is that the RCU-protected resizable hash
table performs and scales almost as well as does its fixed-size counterpart.
The performance during an actual resize operation of course suffers
somewhat due to the cache misses causes by the updates to each element's
pointers, and this effect is most pronounced when the memory system
becomes a bottleneck.
This indicates that hash tables should be resized by substantial amounts,
and that hysteresis should be applied to prevent performance degradation
due to too-frequent resize operations.
In memory-rich environments, hash-table sizes should furthermore
be increased much more aggressively than they are decreased.

Another key point is that although the \co{hashtab} structure is
non-partitionable, it is also read-mostly, which suggests the use
of RCU\@.
Given that the performance and scalability of this resizable hash table is
very nearly that of RCU-protected fixed-sized hash tables, we must
conclude that this approach was quite successful.

Finally, it is important to note that insertions, deletions, and
lookups can proceed concurrently with a resize operation.
This concurrency is
critically important when resizing large hash tables, especially
for applications that must meet severe response-time constraints.

Of course, the \co{ht_elem} structure's
pair of pointer sets does impose some memory overhead,
which is taken up in the next section.

\subsection{Other Resizable Hash Tables}
\label{sec:datastruct:Other Resizable Hash Tables}

One shortcoming of the resizable hash table described earlier in this
section is memory consumption.
Each data element has two pairs of linked-list pointers rather than just
one.
Is it possible to create an RCU-protected resizable hash table that
makes do with just one pair?

It turns out that the answer is ``yes''.
Josh Triplett et al.~\cite{Triplett:2011:RPHash}
produced a \emph{relativistic hash table} that incrementally
splits and combines corresponding hash chains so that readers always
see valid hash chains at all points during the resizing operation.
This incremental splitting and combining relies on the fact that it
is harmless for a reader to see a data element that should be in some
other hash chain:
When this happens, the reader will simply ignore the extraneous data
element due to key mismatches.

\begin{figure}
\centering
\resizebox{3in}{!}{\includegraphics{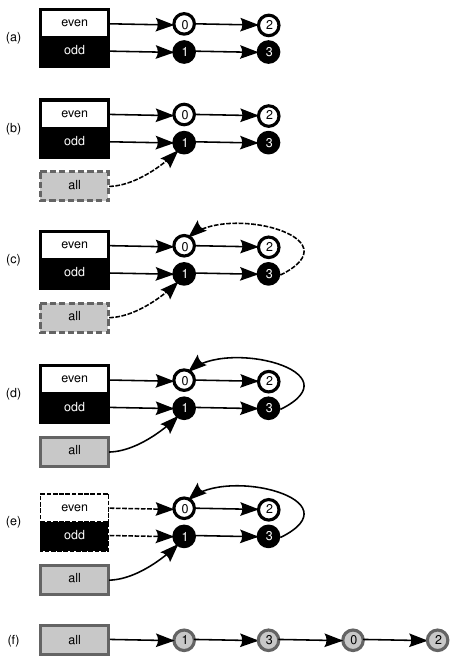}}
\caption{Shrinking a Relativistic Hash Table}
\label{fig:datastruct:Shrinking a Relativistic Hash Table}
\end{figure}

The process of shrinking a relativistic hash table by a factor of two
is shown in
\cref{fig:datastruct:Shrinking a Relativistic Hash Table},
in this case shrinking a two-bucket hash table into a one-bucket
hash table, otherwise known as a linear list.
This process works by coalescing pairs of buckets in the old larger hash
table into single buckets in the new smaller hash table.
For this process to work correctly, we clearly need to constrain the hash
functions for the two tables.
One such constraint is to use the same underlying hash function for
both tables, but to throw out the low-order bit when shrinking from
large to small.
For example, the old two-bucket hash table would
use the two top bits of the value, while the new one-bucket hash table
could use the top bit of the value.
In this way, a given pair of adjacent even and odd buckets in the old
large hash table can be coalesced into a single bucket in the new small
hash table, while still having a single hash value cover all of the
elements in that single bucket.

The initial state is shown at the top of the figure, with time advancing
from top to bottom, starting with initial state~(a).
The shrinking process begins by allocating the new smaller array of
buckets, and having each bucket of this new smaller array reference
the first element of one of the buckets of the corresponding pair in
the old large hash table, resulting in state~(b).

Then the two hash chains are linked together, resulting in state~(c).
In this state, readers looking up an even-numbered element see no change,
and readers looking up elements~1 and~3 likewise see no change.
However, readers looking up some other odd number will also traverse
elements~0 and~2.
This is harmless because any odd number will compare not-equal to these
two elements.
There is some performance loss, but on the other hand, this is exactly
the same performance loss that will be experienced once the new small
hash table is fully in place.

Next, the new small hash table is made accessible to readers, resulting
in state~(d).
Note that older readers might still be traversing the old large hash
table, so in this state both hash tables are in use.

The next step is to wait for all pre-existing readers to complete,
resulting in state~(e).
In this state, all readers are using the new small hash table, so that
the old large hash table's buckets may be freed, resulting in the final
state~(f).

\begin{figure}
\centering
\resizebox{3in}{!}{\includegraphics{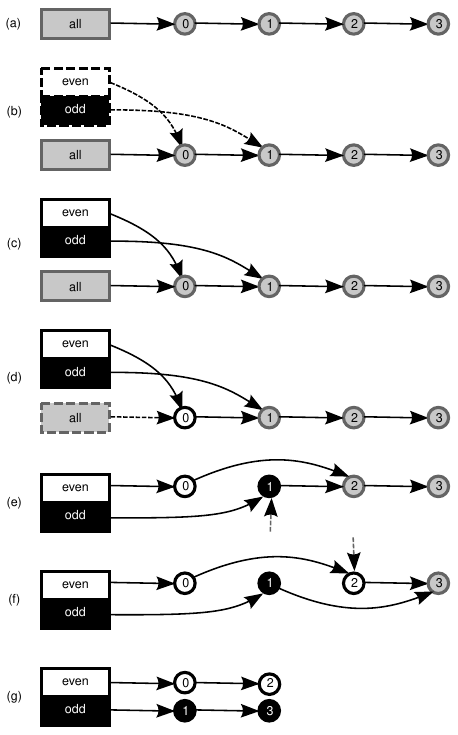}}
\caption{Growing a Relativistic Hash Table}
\label{fig:datastruct:Growing a Relativistic Hash Table}
\end{figure}

Growing a relativistic hash table reverses the shrinking process,
but requires more grace-period steps, as shown in
\cref{fig:datastruct:Growing a Relativistic Hash Table}.
The initial state~(a) is at the top of this figure, with time advancing
from top to bottom.

We start by allocating the new large two-bucket hash table, resulting
in state~(b).
Note that each of these new buckets references the first element destined
for that bucket.
These new buckets are published to readers, resulting in state~(c).
After a grace-period operation, all readers are using the new large
hash table, resulting in state~(d).
In this state, only those readers traversing the even-values hash bucket
traverse element~0, which is therefore now colored white.

At this point, the old small hash buckets may be freed, although many
implementations use these old buckets to track progress ``unzipping''
the list of items into their respective new buckets.
The last even-numbered element in the first consecutive run of such
elements now has its pointer-to-next updated to reference the following
even-numbered element.
After a subsequent grace-period operation, the result is state~(e).
The vertical arrow indicates the next element to be unzipped, and
element~1 is now colored black to indicate that only those readers traversing
the odd-values hash bucket may reach it.

Next, the last odd-numbered element in the first consecutive run of such
elements now has its pointer-to-next updated to reference the following
odd-numbered element.
After a subsequent grace-period operation, the result is state~(f).
A final unzipping operation (including a grace-period operation)
results in the final state~(g).

In short, the relativistic hash table reduces the number of per-element
list pointers at the expense of additional grace periods incurred during
resizing.
These additional grace periods are usually not a problem because
insertions, deletions, and lookups may proceed concurrently with
a resize operation.

It turns out that it is possible to reduce the per-element memory overhead
from a pair of pointers to a single pointer, while still retaining
$\O{1}$ deletions.
This is accomplished by augmenting split-order
list~\cite{OriShalev2006SplitOrderListHash}
with RCU
protection~\cite{MathieuDesnoyers2009URCU,PaulMcKenney2013LWNURCUhash}.
The data elements in the hash table are arranged into a single
sorted linked list, with each hash bucket referencing the first element
in that bucket.
Elements are deleted by setting low-order bits in their pointer-to-next
fields, and these elements are removed from the list by later traversals
that encounter them.

This RCU-protected split-order list is complex, but offers lock-free
progress guarantees for all insertion, deletion, and lookup operations.
Such guarantees can be important in real-time applications.
An implementation is available from recent versions of the userspace RCU
library~\cite{MathieuDesnoyers2009URCU}.

\section{Other Data Structures}
\label{sec:datastruct:Other Data Structures}
\epigraph{All life is an experiment.
	  The more experiments you make the better.}
	 {Ralph Waldo Emerson}

The preceding sections have focused on data structures that enhance
concurrency due to partitionability
(\cref{sec:datastruct:Partitionable Data Structures}),
efficient handling of read-mostly access patterns
(\cref{sec:datastruct:Read-Mostly Data Structures}),
or application of read-mostly techniques to avoid
non-partitionability
(\cref{sec:datastruct:Non-Partitionable Data Structures}).
This section gives a brief review of other data structures.

One of the hash table's greatest advantages for parallel use is that it
is fully partitionable, at least while not being resized.
One way of preserving the partitionability and the size independence is
to use a radix tree, which is also called a trie.
Tries partition the search key, using each successive key partition
to traverse the next level of the trie.
As such, a trie can be thought of as a set of nested hash tables,
thus providing the required partitionability.
One disadvantage of tries is that a sparse key space can result in
inefficient use of memory.
There are a number of compression techniques that may be used to
work around this disadvantage, including hashing the key value to
a smaller keyspace before the
traversal~\cite{RobertOlsson2007Trash}.
Radix trees are heavily used in practice, including in the Linux
kernel~\cite{NickPiggin2006radixtree}.

One important special case of both a hash table and a trie is what is
perhaps the oldest of data structures, the array and its multi-dimensional
counterpart, the matrix.
The fully partitionable nature of matrices is exploited heavily in
concurrent numerical algorithms.

Self-balancing trees are heavily used in sequential code, with
AVL trees and red-black trees being perhaps the most well-known
examples~\cite{ThomasHCorman2001Algorithms}.
Early attempts to parallelize AVL trees were complex and not necessarily
all that efficient~\cite{Ellis80},
however, more recent work on red-black trees provides better
performance and scalability by using RCU for readers and hashed arrays
of locks\footnote{
	In the guise of swissTM~\cite{AleksandarDragovejic2011STMnotToy},
	which is a variant of software transactional memory in which
	the developer flags non-shared accesses.}
to protect reads and updates,
respectively~\cite{PhilHoward2011RCUTMRBTree,PhilipWHoward2013RCUrbtree}.
It turns out that red-black trees rebalance aggressively, which works
well for sequential programs, but not necessarily so well for parallel
use.
Recent work has therefore made use of RCU-protected ``bonsai trees''
that rebalance less aggressively~\cite{AustinClements2012RCULinux:mmapsem},
trading off optimal tree depth to gain more efficient concurrent updates.

Concurrent skip lists lend themselves well to RCU readers, and in fact
represents an early academic use of a technique resembling
RCU~\cite{Pugh90}.

Concurrent double-ended queues were discussed in
\cref{sec:SMPdesign:Double-Ended Queue},
and concurrent stacks and queues have a long history~\cite{Treiber86},
though not normally the most impressive performance or scalability.
They are nevertheless a common feature of concurrent
libraries~\cite{PaulMcKenney2013LWNURCUqueuestack}.
Researchers have recently proposed relaxing the ordering constraints
of stacks and queues~\cite{Shavit:2011:DSM:1897852.1897873},
with some work indicating that relaxed-ordered queues actually have
better ordering properties than do strict FIFO
queues~\cite{AndreasHaas2012FIFOisnt,ChristophMKirsch2012FIFOisntTR,AndreasHaas2013CFRelaxedQueues}.

It seems likely that continued work with concurrent data structures will
produce novel algorithms with surprising properties.

\section{Micro-Optimization}
\label{sec:datastruct:Micro-Optimization}
\epigraph{The devil is in the details.}{Unknown}

The data structures shown in this chapter were coded straightforwardly,
with no adaptation to the underlying system's cache hierarchy.
In addition, many of the implementations used pointers to functions
for key-to-hash conversions and other frequent operations.
Although this approach provides simplicity and portability, in many
cases it does give up some performance.

The following sections touch on specialization, memory conservation,
and hardware considerations.
Please do not mistake these short sections for a definitive treatise
on this subject.
Whole books have been written on optimizing to a specific CPU, let
alone to the set of CPU families in common use today.

\subsection{Specialization}
\label{sec:datastruct:Specialization}

The resizable hash table presented in
\cref{sec:datastruct:Non-Partitionable Data Structures}
used an opaque type for the key.
This allows great flexibility, permitting any sort of key to be
used, but it also incurs significant overhead due to the calls via
of pointers to functions.
Now, modern hardware uses sophisticated branch-prediction techniques
to minimize this overhead, but on the other hand, real-world software
is often larger than can be accommodated even by today's large
hardware branch-prediction tables.
This is especially the case for calls via pointers, in which case
the branch prediction hardware must record a pointer in addition
to branch-taken/branch-not-taken information.

This overhead can be eliminated by specializing a hash-table implementation
to a given key type and hash function, for example, by using C++ templates.
Doing so eliminates the \co{->ht_cmp()}, \co{->ht_gethash()}, and
\co{->ht_getkey()} function pointers in the \co{ht} structure shown in
\cref{lst:datastruct:Resizable Hash-Table Data Structures} on
\cpageref{lst:datastruct:Resizable Hash-Table Data Structures}.
It also eliminates the corresponding calls through these pointers,
which could allow the compiler to inline the resulting fixed functions,
eliminating not only the overhead of the call instruction, but the
argument marshalling as well.

\QuickQuiz{
	How much do these specializations really save?
	Are they really worth it?
}\QuickQuizAnswer{
	The answer to the first question is left as an exercise to
	the reader.
	Try specializing the resizable hash table and see how much
	performance improvement results.
	The second question cannot be answered in general, but must
	instead be answered with respect to a specific use case.
	Some use cases are extremely sensitive to performance and
	scalability, while others are less so.
}\QuickQuizEnd

All that aside, one of the great benefits of modern hardware compared
to that available when I first started learning to program back in
the early 1970s is that much less specialization is required.
This allows much greater productivity than was possible back in the
days of four-kilobyte address spaces.

\subsection{Bits and Bytes}
\label{sec:datastruct:Bits and Bytes}

The hash tables discussed in this chapter made almost no attempt to conserve
memory.
For example, the \co{->ht_idx} field in the \co{ht} structure in
\cref{lst:datastruct:Resizable Hash-Table Data Structures} on
\cpageref{lst:datastruct:Resizable Hash-Table Data Structures}
always has a value of either zero or one, yet takes up a full 32 bits
of memory.
It could be eliminated, for example, by stealing a bit from the
\co{->ht_resize_key} field.
This works because the \co{->ht_resize_key} field is large enough to
address every byte of memory and the \co{ht_bucket} structure
is more than one byte long, so that
the \co{->ht_resize_key} field must have several bits to spare.

This sort of bit-packing trick is frequently used in data structures
that are highly replicated, as is the \co{page} structure in the Linux
kernel.
However, the resizable hash table's \co{ht} structure is not all that
highly replicated.
It is instead the \co{ht_bucket} structures we should focus on.
There are two major opportunities for shrinking the \co{ht_bucket} structure:
\begin{enumerate*}[(1)]
\item Placing the \tco{->htb_lock} field in a low-order bit of one of the
\tco{->htb_head} pointers and
\item Reducing the number of pointers required.
\end{enumerate*}

The first opportunity might make use of bit-spinlocks in the Linux
kernel, which are provided by the \path{include/linux/bit_spinlock.h}
header file.
These are used in space-critical data structures in the Linux kernel,
but are not without their disadvantages:

\begin{enumerate}
\item	They are significantly slower than the traditional spinlock
	primitives.
\item	They cannot participate in the lockdep deadlock detection
	tooling in the Linux kernel~\cite{JonathanCorbet2006lockdep}.
\item	They do not record lock ownership, further complicating
	debugging.
\item	They do not participate in priority boosting in \rt\ kernels,
	which means that preemption must be disabled when holding
	bit spinlocks, which can degrade real-time latency.
\end{enumerate}

Despite these disadvantages, bit-spinlocks are extremely useful when
memory is at a premium.

One aspect of the second opportunity was covered in
\cref{sec:datastruct:Other Resizable Hash Tables},
which presented resizable hash tables that require only one
set of bucket-list pointers in place of the pair of sets required
by the resizable hash table presented in
\cref{sec:datastruct:Non-Partitionable Data Structures}.
Another approach would be to use singly linked bucket lists in
place of the doubly linked lists used in this chapter.
One downside of this approach is that deletion would then require
additional overhead, either by marking the outgoing pointer
for later removal
or by searching the bucket list for the element being deleted.

In short, there is a tradeoff between minimal memory overhead on
the one hand, and performance and simplicity on the other.
Fortunately, the relatively large memories available on modern
systems have allowed us to prioritize performance and simplicity
over memory overhead.
However, even though the year~2022's pocket-sized smartphones sport
many gigabytes of memory and its mid-range servers sport terabytes, it
is sometimes necessary to take extreme measures to reduce memory overhead.

\subsection{Hardware Considerations}
\label{sec:datastruct:Hardware Considerations}

Modern computers typically move data between CPUs and main memory in
fixed-sized blocks that range in size from 32 bytes to 256 bytes.
These blocks are called \emph{\IXpl{cache line}}, and are extremely important
to high performance and scalability, as was discussed in
\cref{sec:cpu:Overheads}.
One timeworn way to kill both performance and scalability is to
place incompatible variables into the same cacheline.
For example, suppose that a resizable hash table data element had
the \co{ht_elem} structure in the same cacheline as a frequently incremented
counter.
The frequent incrementing would cause the cacheline to be present at
the CPU doing the incrementing, but nowhere else.
If other CPUs attempted to traverse the hash bucket list containing
that element, they would incur expensive cache misses, degrading both
performance and scalability.

\begin{listing}
\begin{VerbatimL}
struct hash_elem {
	struct ht_elem e;
	long __attribute__ ((aligned(64))) counter;
};
\end{VerbatimL}
\caption{Alignment for 64-Byte Cache Lines}
\label{lst:datastruct:Alignment for 64-Byte Cache Lines}
\end{listing}

One way to solve this problem on systems with 64-byte cache line is shown in
\cref{lst:datastruct:Alignment for 64-Byte Cache Lines}.
Here \GCC's \co{aligned} attribute is used to force the \co{->counter}
and the \co{ht_elem} structure into separate cache lines.
This would allow CPUs to traverse the hash bucket list at full speed
despite the frequent incrementing.

Of course, this raises the question ``How did we know that cache lines
are 64 bytes in size?''
On a Linux system, this information may be obtained from the
\path{/sys/devices/system/cpu/cpu*/cache/} directories, and it is even
possible to make the installation process rebuild the application to
accommodate the system's hardware structure.
However, this would be more difficult if you wanted your application to
also run on non-Linux systems.
Furthermore, even if you were content to run only on Linux, such a
self-modifying installation poses validation challenges.
For example, systems with 32-byte cachelines might work well, but
performance might suffer on systems with 64-byte cachelines due
to \IX{false sharing}.

Fortunately, there are some rules of thumb that work reasonably well in
practice, which were gathered into a 1995
paper~\cite{BenjaminGamsa95a}.\footnote{
	A number of these rules are paraphrased and expanded on here
	with permission from Orran Krieger.}
The first group of rules involve rearranging structures to accommodate
cache geometry:

\begin{enumerate}
\item	Place read-mostly data far from frequently updated data.
	For example, place read-mostly data at the beginning of the
	structure and frequently updated data at the end.
	Place data that is rarely accessed in between.
\item	If the structure has groups of fields such that each group is
	updated by an independent code path, separate these groups
	from each other.
	Again, it can be helpful to place rarely accessed data
	between the groups.
	In some cases, it might also make sense to place each such group
	into a separate structure referenced by the original structure.
\item	Where possible, associate update-mostly data with a CPU, thread,
	or task.
	We saw several very effective examples of this rule of thumb
	in the counter implementations in
	\cref{chp:Counting}.
\item	Going one step further, partition your data on a per-CPU,
	per-thread, or per-task basis, as was discussed in
	\cref{chp:Data Ownership}.
\end{enumerate}

There has been some work towards automated trace-based rearrangement
of structure
fields~\cite{Golovanevsky:2010:TDL:2174824.2174835}.
This work might well ease one of the more painstaking tasks
required to get excellent performance and scalability from
multithreaded software.

An additional set of rules of thumb deal with locks:

\begin{enumerate}
\item	Given a heavily contended lock protecting data that is
	frequently modified, take one of the following approaches:
	\begin{enumerate}
	\item	Place the lock in a different cacheline than the data
		that it protects.
	\item	Use a lock that is adapted for high contention, such
		as a queued lock.
	\item	Redesign to reduce lock contention.
		(This approach is best, but is not always trivial.)
	\end{enumerate}
\item	Place uncontended locks into the same cache line as the data
	that they protect.
	This approach means that the cache miss that brings the
	lock to the current CPU also brings its data.
\item	Protect read-mostly data with \IXpl{hazard pointer}, RCU, or, for
	long-duration critical sections, reader-writer locks.
\end{enumerate}

Of course, these are rules of thumb rather than absolute rules.
Some experimentation is required to work out which are most applicable
to a given situation.

\section{Summary}
\label{sec:datastruct:Summary}
\epigraph{There's only one thing more painful than learning from
	  experience, and that is not learning from experience.}
	 {Archibald MacLeish}

This chapter has focused primarily on hash tables, including resizable
hash tables, which are not fully partitionable.
\Cref{sec:datastruct:Other Data Structures} gave a quick
overview of a few non-hash-table data structures.
Nevertheless, this exposition of hash tables is an excellent introduction
to the many issues surrounding high-performance scalable data access,
including:

\begin{enumerate}
\item	Fully partitioned data structures work well on small systems,
	for example, single-socket systems.
\item	Larger systems require locality of reference as well as
	full partitioning.
\item	Read-mostly techniques, such as hazard pointers and RCU,
	provide good locality of reference for read-mostly workloads,
	and thus provide excellent performance and scalability even
	on larger systems.
\item	Read-mostly techniques also work well on some types of
	non-partitionable data structures, such as resizable hash tables.
\item	Large data structures can overflow CPU caches, reducing performance
	and scalability.
\item	Additional performance and scalability can be obtained by
	specializing the data structure to a specific workload,
	for example, by replacing a general key with a 32-bit integer.
\item	Although requirements for portability and for extreme performance
	often conflict, there are some data-structure-layout techniques
	that can strike a good balance between these two sets of
	requirements.
\end{enumerate}

That said, performance and scalability are of little use without reliability,
so the next chapter covers validation.

	\setlength{\parskip}{0.0pt plus 1ex}
	\input{qqzdatastruct}
	\setlength{\parskip}{0.0pt plus 1.0pt}% return to default

% debugging/debugging.tex
% mainfile: ../perfbook.tex
% SPDX-License-Identifier: CC-BY-SA-3.0

\QuickQuizChapter{chp:Validation}{Validation}{qqzdebugging}
\Epigraph{If it is not tested, it doesn't work.}{Unknown}

I have had a few parallel programs work the first time, but that is only
because I have written an extremely large number parallel programs over
the past few decades.
And I have had far more parallel programs that fooled me into thinking
that they were working correctly the first time than actually were working
the first time.

I thus need to validate my parallel programs.
The basic trick behind validation, is to realize that the computer knows
what is wrong.
It is therefore your job to force it to tell you.
This chapter can therefore be thought of as a short course in
machine interrogation.
But you can leave the good-cop/bad-cop routine at home.
This chapter covers much more sophisticated and effective methods,
especially given that most computers couldn't tell a good cop from a
bad cop, at least as far as we know.

A longer course may be found in many recent books on validation, as
well as at least one older but valuable
one~\cite{GlenfordJMyers1979}.
Validation is an extremely important topic that cuts across all forms
of software, and is worth intensive study in its own right.
However, this book is primarily about concurrency, so this chapter will do
little more than scratch the surface of this critically important topic.

\Cref{sec:debugging:Introduction}
introduces the philosophy of debugging.
\Cref{sec:debugging:Tracing}
discusses tracing,
\cref{sec:debugging:Assertions}
discusses assertions, and
\cref{sec:debugging:Static Analysis}
discusses static analysis.
\Cref{sec:debugging:Code Review}
describes some unconventional approaches to code review that can
be helpful when the fabled 10,000 eyes happen not to be looking at your code.
\Cref{sec:debugging:Probability and Heisenbugs}
overviews the use of probability for validating parallel software.
Because performance and scalability are first-class requirements
for parallel programming,
\cref{sec:debugging:Performance Estimation} covers these
topics.
Finally,
\cref{sec:debugging:Summary}
gives a fanciful summary and a short list of statistical traps to avoid.

But never forget that the three best debugging tools are a thorough
understanding of the requirements, a solid design, and a good night's
sleep!

\section{Introduction}
\label{sec:debugging:Introduction}
%
% \epigraph{If debugging is the process of removing software bugs, then
% 	  programming must be the process of putting them in.}
% 	 {Edsger W.~Dijkstra}
\epigraph{Debugging is like being the detective in a crime movie where
	  you are also the murderer.}
	 {Filipe Fortes}
% https://twitter.com/fortes/status/399339918213652480?lang=en Nov 9, 2013

\Cref{sec:debugging:Where Do Bugs Come From?}
discusses the sources of bugs, and
\cref{sec:debugging:Required Mindset}
overviews the mindset required when validating software.
\Cref{sec:debugging:When Should Validation Start?}
discusses when you should start validation, and
\cref{sec:debugging:The Open Source Way} describes the
surprisingly effective open-source regimen of code review and
community testing.

\subsection{Where Do Bugs Come From?}
\label{sec:debugging:Where Do Bugs Come From?}

Bugs come from developers.
The basic problem is that the human brain did not evolve with computer
software in mind.
Instead, the human brain evolved in concert with other human brains and
with animal brains.
Because of this history, the following three characteristics of computers
often come as a shock to human intuition:

\begin{enumerate}
\item	Computers lack common sense, despite huge sacrifices at the
	altar of artificial intelligence.
\item	Computers fail to understand user intent, or more formally,
	computers generally lack a theory of mind.
\item	Computers cannot do anything useful with a fragmentary plan,
	instead requiring that every detail of all possible scenarios
	be spelled out in full.
\end{enumerate}

The first two points should be uncontroversial, as they are illustrated
by any number of failed products, perhaps most famously Clippy and
Microsoft Bob.
By attempting to relate to users as people, these two products raised
common-sense and theory-of-mind expectations that they proved incapable
of meeting.
Perhaps the set of software assistants are now available on smartphones
will fare better, but as of 2021 reviews are mixed.
That said, the developers working on them by all accounts still develop
the old way:
The assistants might well benefit end users, but not so much their own
developers.

This human love of fragmentary plans deserves more explanation,
especially given that it is a classic two-edged sword.
This love of fragmentary plans is apparently due to the assumption that
the person carrying out the plan will have (1)~common sense and (2)~a good
understanding of the intent and requirements driving the plan.
This latter assumption is especially likely to hold in the common case
where the person doing the planning and the person carrying out the plan
are one and the same:
In this case, the plan will be revised almost
subconsciously as obstacles arise, especially when that person has the
a good understanding of the problem at hand.
In fact, the love of fragmentary plans has served human beings well,
in part because it is better to take random actions that have a some
chance of locating food than to starve to death while attempting to plan
the unplannable.
However, the usefulness of fragmentary plans in the
everyday life of which we are all experts is no guarantee of their future
usefulness in stored-program computers.

Furthermore, the need to follow fragmentary plans has had important effects
on the human psyche, due to the fact
that throughout much of human history, life was often difficult and dangerous.
It should come as no surprise that executing a fragmentary plan that has
a high probability of a violent encounter with sharp teeth and claws
requires almost insane levels of optimism---a level of optimism that actually
is present in most human beings.
These insane levels of optimism extend to self-assessments of programming
ability, as evidenced by the effectiveness of (and the controversy over)
code-interviewing techniques~\cite{RegBraithwaite2007FizzBuzz}.
In fact, the clinical term for a human being with less-than-insane
levels of optimism is ``clinically depressed''.
Such people usually have extreme difficulty functioning in their daily
lives, underscoring the perhaps counter-intuitive importance of insane
levels of optimism to a normal, healthy life.
Furtheremore, if you are not insanely optimistic, you are less likely
to start a difficult but worthwhile project.\footnote{
	There are some famous exceptions to this rule of thumb.
	Some people take on difficult or risky projects in order to at
	least a temporarily escape from their depression.
	Others have nothing to lose:
	The project is literally a matter of life or death.}

\QuickQuiz{
	When in computing is it necessary to follow a
	fragmentary plan?
}\QuickQuizAnswer{
	There are any number of situations, but perhaps the most important
	situation is when no one has ever created anything resembling
	the program to be developed.
	In this case, the only way to create a credible plan is to
	implement the program, create the plan, and implement it a
	second time.
	But whoever implements the program for the first time has no
	choice but to follow a fragmentary plan because any detailed
	plan created in ignorance cannot survive first contact with
	the real world.

	And perhaps this is one reason why evolution has favored insanely
	optimistic human beings who are happy to follow fragmentary plans!
}\QuickQuizEnd

An important special case is the project that, while valuable, is not
valuable enough to justify the time required to implement it.
This special case is quite common, and one early symptom is the
unwillingness of the decision-makers to invest enough to actually
implement the project.
A natural reaction is for the developers to produce an unrealistically
optimistic estimate in order to be permitted to start the project.
If the organization is strong enough and its decision-makers ineffective
enough, the project might succeed despite the resulting schedule slips
and budget overruns.
However, if the organization is not strong enough and if the decision-makers
fail to cancel the project as soon as it becomes clear that the estimates
are garbage, then the project might well kill the organization.
This might result in another organization picking up the project and
either completing it, canceling it, or being killed by it.
A given project might well succeed only after killing several
organizations.
One can only hope that the organization that eventually makes a success
of a serial-organization-killer project maintains a suitable
level of humility, lest it be killed by its next such project.

\QuickQuiz{
	Who cares about the organization?
	After all, it is the project that is important!
}\QuickQuizAnswer{
	Yes, projects are important, but if you like being paid for your
	work, you need organizations as well as projects.
}\QuickQuizEnd

Important though insane levels of optimism might be, they are a key source
of bugs (and perhaps failure of organizations).
The question is therefore ``How to maintain the optimism required to start
a large project while at the same time injecting enough reality to keep
the bugs down to a dull roar?''
The next section examines this conundrum.

\subsection{Required Mindset}
\label{sec:debugging:Required Mindset}

When carrying out any validation effort, keep the following
definitions firmly in mind:

\begin{enumerate}
\item	The only bug-free programs are trivial programs.
\item	A reliable program has no known bugs.
\end{enumerate}

From these definitions, it logically follows that any reliable
non-trivial program contains at least one bug that you do not
know about.
Therefore, any validation effort undertaken on a non-trivial program
that fails to find any bugs is itself a failure.
A good validation is therefore an exercise in destruction.
This means that if you are the type of person who enjoys breaking things,
validation is just job for you.

\begin{SaveVerbatim}{VerbDebuggingQQZ}
        real    0m0.132s
        user    0m0.040s
        sys     0m0.008s
\end{SaveVerbatim}

\QuickQuiz{
	Suppose that you are writing a script that processes the
	output of the \co{time} command, which looks as follows:

	\begin{center}
	\fbox{\BUseVerbatim[boxwidth=2in,baseline=c]{VerbDebuggingQQZ}}
	\end{center}

	The script is required to check its input for errors, and to
	give appropriate diagnostics if fed erroneous \co{time} output.
	What test inputs should you provide to this program to test it
	for use with \co{time} output generated by single-threaded programs?
}\QuickQuizAnswer{
	Can you say ``Yes'' to all the following questions?

	\begin{enumerate}
	\item	Do you have a test case in which all the time is
		consumed in user mode by a CPU-bound program?
	\item	Do you have a test case in which all the time is
		consumed in system mode by a CPU-bound program?
	\item	Do you have a test case in which all three times
		are zero?
	\item	Do you have a test case in which the \qco{user} and \qco{sys}
		times sum to more than the \qco{real} time?
		(This would of course be completely legitimate in
		a multithreaded program.)
	\item	Do you have a set of tests cases in which one of the
		times uses more than one second?
	\item	Do you have a set of tests cases in which one of the
		times uses more than ten seconds?
	\item	Do you have a set of test cases in which one of the
		times has non-zero minutes?
		(For example, \qco{15m36.342s}.)
	\item	Do you have a set of test cases in which one of the
		times has a seconds value of greater than 60?
	\item	Do you have a set of test cases in which one of the
		times overflows 32 bits of milliseconds?
		64 bits of milliseconds?
	\item	Do you have a set of test cases in which one of the
		times is negative?
	\item	Do you have a set of test cases in which one of the
		times has a positive minutes value but a negative
		seconds value?
	\item	Do you have a set of test cases in which one of the
		times omits the \qco{m} or the \qco{s}?
	\item	Do you have a set of test cases in which one of the
		times is non-numeric?
		(For example, \qco{Go Fish}.)
	\item	Do you have a set of test cases in which one of the
		lines is omitted?
		(For example, where there is a \qco{real} value and
		a \qco{sys} value, but no \qco{user} value.)
	\item	Do you have a set of test cases where one of the
		lines is duplicated?
		Or duplicated, but with a different time value for
		the duplicate?
	\item	Do you have a set of test cases where a given line
		has more than one time value?
		(For example, \qco{real 0m0.132s 0m0.008s}.)
	\item	Do you have a set of test cases containing random
		characters?
	\item	In all test cases involving invalid input, did you
		generate all permutations?
	\item	For each test case, do you have an expected outcome
		for that test?
	\end{enumerate}

	If you did not generate test data for a substantial number of
	the above cases, you will need to cultivate a more destructive
	attitude in order to have a chance of generating high-quality
	tests.

	Of course, one way to economize on destructiveness is to
	generate the tests with the to-be-tested source code at hand,
	which is called white-box testing (as opposed to black-box testing).
	However, this is no panacea:
	You will find that it is all too easy to find your thinking
	limited by what the program can handle, thus failing to generate
	truly destructive inputs.
}\QuickQuizEnd

But perhaps you are a super-programmer whose code is always perfect
the first time every time.
If so, congratulations!
Feel free to skip this chapter, but
I do hope that you will forgive my skepticism.
You see, I have too many people who claimed to be able to write perfect
code the first time, which is not too surprising given the previous
discussion of optimism and over-confidence.
And even if you really are a super-programmer, you just might
find yourself debugging lesser mortals' work.

One approach for the rest of us is to alternate between our normal
state of insane optimism
(Sure, I can program that!\@) and severe pessimism
(It seems to work, but I just know that there have to be more bugs hiding
in there somewhere!).
It helps if you enjoy breaking things.
If you don't, or if your joy in breaking things is limited to breaking
\emph{other} people's things, find someone who does love breaking your
code and have them help you break it.

Another helpful frame of mind is to hate it when other people find bugs in
your code.
This hatred can help motivate you to torture your code beyond all reason
in order to increase the probability that you will be the one to find
the bugs.
Just make sure to suspend this hatred long enough to sincerely thank
anyone who does find a bug in your code!
After all, by so doing, they saved you the trouble of tracking it down,
and possibly at great personal expense dredging through your code.

Yet another helpful frame of mind is studied skepticism.
You see, believing that you understand the code means you can learn
absolutely nothing about it.
Ah, but you know that you completely understand the code because you
wrote or reviewed it?
Sorry, but the presence of bugs suggests that your understanding is at
least partially fallacious.
One cure is to write down what you know to be true and double-check this
knowledge, as discussed in
\crefrange{sec:debugging:Tracing}{sec:debugging:Code Review}.
Objective reality \emph{always} overrides whatever you
might think you know.

One final frame of mind is to consider the possibility that someone's
life depends on your code being correct.
One way of looking at this is that consistently making good things happen
requires a lot of focus on a lot of bad things that might happen, with
an eye towards preventing or otherwise handling those bad things.\footnote{
	For more on this philosophy, see the chapter entitled
	``The Power of Negative Thinking''
	from Chris Hadfield's excellent book entitled
	``An Astronaut's Guide to Life on Earth.''}
The prospect of these bad things might also motivate you to torture your
code into revealing the whereabouts of its bugs.

This wide variety of frames of mind opens the door to
the possibility of multiple people with different frames of
mind contributing to the project, with varying levels of optimism.
This can work well, if properly organized.

\begin{figure}
\centering
\resizebox{2in}{!}{\includegraphics{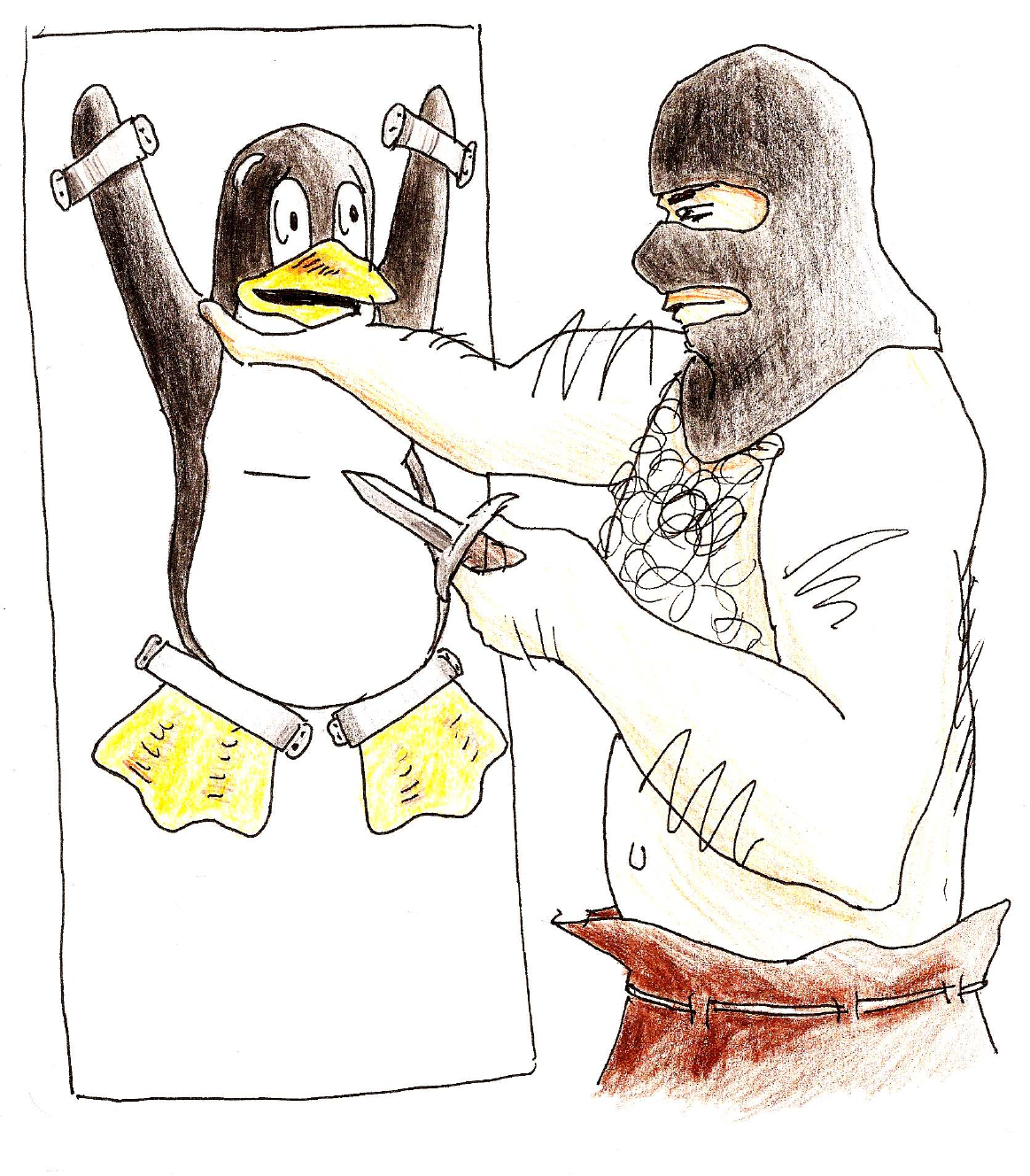}}
\caption{Validation and the Geneva Convention}
\ContributedBy{Figure}{fig:debugging:Validation and the Geneva Convention}{Melissa Broussard}
\end{figure}

\begin{figure}
\centering
\resizebox{2in}{!}{\includegraphics{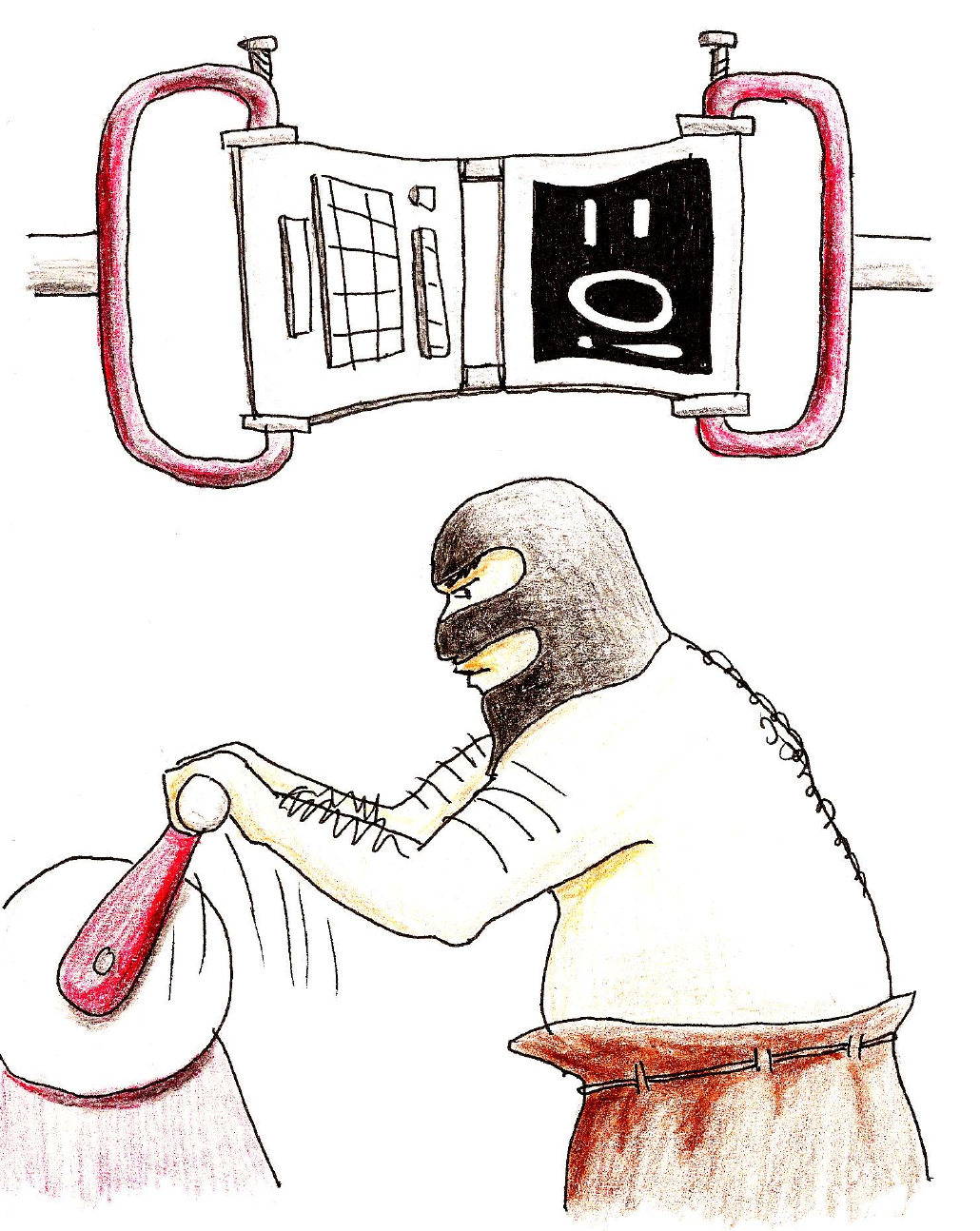}}
\caption{Rationalizing Validation}
\ContributedBy{Figure}{fig:debugging:Rationalizing Validation}{Melissa Broussard}
\end{figure}

Some people might see vigorous validation as a form of torture, as
depicted in
\cref{fig:debugging:Validation and the Geneva Convention}.\footnote{
	The cynics among us might question whether these people are
	afraid that validation will find bugs that they will then be
	required to fix.}
Such people might do well to remind themselves that, Tux cartoons aside,
they are really torturing an inanimate object, as shown in
\cref{fig:debugging:Rationalizing Validation}.
Rest assured that those who fail to torture their code are doomed to be
tortured by it!

However, this leaves open the question of exactly when during the project
lifetime validation should start, a topic taken up by the next section.

\subsection{When Should Validation Start?}
\label{sec:debugging:When Should Validation Start?}

Validation should start exactly when the project starts.

To see this, consider that tracking down a bug is much harder in a large
program than in a small one.
Therefore, to minimize the time and effort required to track down bugs,
you should test small units of code.
Although you won't find all the bugs this way, you will find a substantial
fraction, and it will be much easier to find and fix the ones you do find.
Testing at this level can also alert you to larger flaws in your overall
design, minimizing the time you waste writing code that is broken
by design.

But why wait until you have code before validating your design?\footnote{
	The old saying ``First we must code, then we have incentive to
	think'' notwithstanding.}
Hopefully reading \cref{chp:Hardware and its Habits,%
chp:Tools of the Trade} provided you with the information
required to avoid some regrettably common design flaws,
but discussing your design with a colleague or even simply writing it
down can help flush out additional flaws.

However, it is all too often the case that waiting to start validation
until you have a design is waiting too long.
Mightn't your natural level of optimism caused you to start the design
before you fully understood the requirements?
The answer to this question will almost always be ``yes''.
One good way to avoid flawed requirements is to get to know your users.
To really serve them well, you will have to live among them.

\QuickQuiz{
	You are asking me to do all this validation BS before
	I even start coding???
	That sounds like a great way to never get started!!!
}\QuickQuizAnswer{
	If it is your project, for example, a hobby, do what you like.
	Any time you waste will be your own, and you have no one else
	to answer to for it.
	And there is a good chance that the time will not be completely
	wasted.
	For example, if you are embarking on a first-of-a-kind project,
	the requirements are in some sense unknowable anyway.
	In this case, the best approach might be to quickly prototype
	a number of rough solutions, try them out, and see what works
	best.

	On the other hand, if you are being paid to produce a system that
	is broadly similar to existing systems, you owe it to your users,
	your employer, and your future self to validate early and often.
}\QuickQuizEnd

First-of-a-kind projects often use different methodologies such as
rapid prototyping or agile.
Here, the main goal of early prototypes are not to create correct
implementations, but rather to learn the project's requirements.
But this does not mean that you omit validation; it instead means that
you approach it differently.

One such approach takes a Darwinian view, with the validation suite
eliminating code that is not fit to solve the problem at hand.
From this viewpoint, a vigorous validation suite is essential to the
fitness of your software.
However, taking this approach to its logical conclusion is quite humbling,
as it requires us developers to admit that our carefully crafted changes
to the codebase are, from a Darwinian standpoint, random mutations.
On the other hand, this conclusion is supported by long experience
indicating that seven percent of fixes introduce at least one
bug~\cite{RexBlack2012SQA}.

How vigorous should your validation suite be?
If the bugs it finds aren't threatening the very foundations of your
software design, then it is not yet vigorous enough.
After all, your design is just as prone to bugs as is your code, and
the earlier you find and fix the bugs in your design, the less time
you will waste coding those design bugs.

\QuickQuiz{
	Are you actually suggesting that it is possible to test
	correctness into software???
	Everyone knows that is impossible!!!
}\QuickQuizAnswer{
	Please note that the text used the word ``validation'' rather
	than the word ``testing''.
	The word ``validation'' includes formal methods as well as
	testing, for more on which please see
	\cref{chp:Formal Verification}.

	But as long as we are bringing up things that everyone should
	know, let's remind ourselves that Darwinian evolution is
	not about correctness, but rather about survival.
	As is software.
	My goal as a developer is not that my software be attractive
	from a theoretical viewpoint, but rather that it survive
	whatever its users throw at it.

	Although the notion of correctness does have its uses, its
	fundamental limitation is that the specification against which
	correctness is judged will also have bugs.
	This means nothing more nor less than that traditional correctness
	proofs prove that the code in question contains the intended
	set of bugs!

	Alternative definitions of correctness instead focus on the
	lack of problematic properties, for example, proving that the
	software has no use-after-free bugs, no \co{NULL} pointer
	dereferences, no array-out-of-bounds references, and so on.
	Make no mistake, finding and eliminating such classes of bugs
	can be highly useful.
	But the fact remains that the lack of certain classes of bugs
	does nothing to demonstrate fitness for any specific purpose.

	Therefore, usage-driven validation remains critically important.

	Besides, it is also impossible to verify correctness into your
	software, especially given the problematic need to verify both
	the verifier and the specification.
}\QuickQuizEnd

It is worth reiterating that this advice applies to first-of-a-kind
projects.
If you are instead doing a project in a well-explored area, you would
be quite foolish to refuse to learn from previous experience.
But you should still start validating right at the beginning of the
project, but hopefully guided by others' hard-won knowledge of both
requirements and pitfalls.

An equally important question is ``When should validation stop?''
The best answer is ``Some time after the last change.''
Every change has the potential to create a bug, and thus every
change must be validated.
Furthermore, validation development should continue through the
full lifetime of the project.
After all, the Darwinian perspective above implies that bugs are
adapting to your validation suite.
Therefore, unless you continually improve your validation suite, your
project will naturally accumulate hordes of validation-suite-immune bugs.

But life is a tradeoff, and every bit of time invested in validation
suites as a bit of time that cannot be invested in directly improving
the project itself.
These sorts of choices are never easy, and it can be just as damaging
to overinvest in validation as it can be to underinvest.
But this is just one more indication that life is not easy.

Now that we have established that you should start validation when
you start the project (if not earlier!), and that both validation and
validation development should continue throughout the lifetime of that
project, the following sections cover a number of validation techniques
and methods that have proven their worth.

\subsection{The Open Source Way}
\label{sec:debugging:The Open Source Way}

The open-source programming methodology has proven quite effective, and
includes a regimen of intense code review and testing.

I can personally attest to the effectiveness of the open-source community's
intense code review.
One of my first patches to the Linux kernel involved a distributed
filesystem where one node might write to a given file that another node
has mapped into memory.
In this case, it is necessary to invalidate the affected pages from
the mapping in order to allow the filesystem to maintain coherence
during the write operation.
I coded up a first attempt at a patch, and, in keeping with the open-source
maxim ``post early, post often'', I posted the patch.
I then considered how I was going to test it.

But before I could even decide on an overall test strategy, I got a
reply to my posting pointing out a few bugs.
I fixed the bugs and reposted the patch, and returned to thinking
out my test strategy.
However, before I had a chance to write any test code, I received
a reply to my reposted patch, pointing out more bugs.
This process repeated itself many times, and I am not sure that I
ever got a chance to actually test the patch.

This experience brought home the truth of the open-source saying:
Given enough eyeballs, all bugs are shallow~\cite{EricSRaymond99b}.

However, when you post some code or a given patch, it is worth
asking a few questions:

\begin{enumerate}
\item	How many of those eyeballs are actually going to look at your code?
\item	How many will be experienced and clever enough to actually find
	your bugs?
\item	Exactly when are they going to look?
\end{enumerate}

I was lucky:
There was someone out there who wanted the functionality
provided by my patch, who had long experience with distributed filesystems,
and who looked at my patch almost immediately.
If no one had looked at my patch, there would have been no review, and
therefore none of those bugs would have been located.
If the people looking at my patch had lacked experience with distributed
filesystems, it is unlikely that they would have found all the bugs.
Had they waited months or even years to look, I likely would have forgotten
how the patch was supposed to work, making it much more difficult to
fix them.

However, we must not forget the second tenet of the open-source development,
namely intensive testing.
For example, a great many people test the Linux kernel.
Some test patches as they are submitted, perhaps even yours.
Others test the -next tree, which is helpful, but there is likely to be
several weeks or even months delay between the time that you write the
patch and the time that it appears in the -next tree, by which time the
patch will not be quite as fresh in your mind.
Still others test maintainer trees, which often have a similar time delay.

Quite a few people don't test code until it is committed to mainline,
or the master source tree (Linus's tree in the case of the Linux kernel).
If your maintainer won't accept your patch until it has been tested,
this presents you with a deadlock situation:
Your patch won't be accepted until it is tested, but it won't be tested
until it is accepted.
Nevertheless, people who test mainline code are still relatively
aggressive, given that many people and organizations do not test code
until it has been pulled into a Linux distro.

And even if someone does test your patch, there is no guarantee that they
will be running the hardware and software configuration and workload
required to locate your bugs.

Therefore, even when writing code for an open-source project, you need to
be prepared to develop and run your own test suite.
Test development is an underappreciated and very valuable skill, so be
sure to take full advantage of any existing test suites available to
you.
Important as test development is, we must leave further discussion of it
to books dedicated to that topic.
The following sections therefore discuss locating bugs in your code given that
you already have a good test suite.

\section{Tracing}
\label{sec:debugging:Tracing}
\epigraph{The machine knows what is wrong.
	  Make it tell you.}{Unknown}

When all else fails, add a \co{printk()}!
Or a \co{printf()}, if you are working with user-mode C-language applications.

The rationale is simple:
If you cannot figure out how execution reached a given point in the code,
sprinkle print statements earlier in the code to work out what happened.
You can get a similar effect, and with more convenience and flexibility,
by using a debugger such as gdb (for user applications) or kgdb
(for debugging Linux kernels).
Much more sophisticated tools exist, with some of the more recent
offering the ability to rewind backwards in time from the point
of failure.

These brute-force testing tools are all valuable, especially now
that typical systems have more than 64K of memory and CPUs running
faster than 4\,MHz.
Much has been
written about these tools, so this chapter will add only a little more.

However, these tools all have a serious shortcoming when you need a
fastpath to tell you what is going wrong, namely, these tools often have
excessive overheads.
There are special tracing technologies for this purpose, which typically
leverage data ownership techniques
(see \cref{chp:Data Ownership})
to minimize the overhead of runtime data collection.
One example within the Linux kernel is
``trace events''~\cite{StevenRostedt2010perfTraceEventP1,StevenRostedt2010perfTraceEventP2,StevenRostedt2010perfTraceEventP3,StevenRostedt2010perfHP+DeathlyMacros},
which uses per-CPU buffers to allow data to be collected with
extremely low overhead.
Even so, enabling tracing can sometimes change timing enough to
hide bugs, resulting in \emph{heisenbugs}, which are discussed in
\cref{sec:debugging:Probability and Heisenbugs}
and especially \cref{sec:debugging:Hunting Heisenbugs}.
In the kernel, BPF can do data reduction in the kernel, reducing
the overhead of transmitting the needed information from the kernel
to userspace~\cite{BrendanGregg2019BPFperftools}.
In userspace code, there is a huge number of tools that can help you.
One good starting point is Brendan Gregg's blog.\footnote{
	\url{http://www.brendangregg.com/blog/}}

Even if you avoid heisenbugs, other pitfalls await you.
For example, although the machine really does know all,
what it knows is almost always way more than your head can hold.
For this reason, high-quality test suites normally come with sophisticated
scripts to analyze the voluminous output.
But beware---scripts will only notice what you tell them to.
My \co{rcutorture} scripts are a case in point:
Early versions of those scripts were quite satisfied with a test run
in which RCU \IXpl{grace period} stalled indefinitely.
This of course resulted in the scripts being modified to detect RCU
grace-period stalls, but this does not change the fact that the scripts
will only detect problems that I make them detect.
But note well that unless you have a solid design, you won't know what
your script should check for!

Another problem with tracing and especially with \co{printk()} calls
is that their overhead can rule out production use.
In such cases, assertions can be helpful.

\section{Assertions}
\label{sec:debugging:Assertions}
\epigraph{No man really becomes a fool until he stops asking questions.}
	 {Charles P. Steinmetz}

Assertions are usually implemented in the following manner:

\begin{VerbatimN}
if (something_bad_is_happening())
	complain();
\end{VerbatimN}

This pattern is often encapsulated into C-preprocessor macros or
language intrinsics, for example, in the Linux kernel, this might
be represented as \co{WARN_ON(something_bad_is_happening())}.
Of course, if \co{something_bad_is_happening()} quite frequently,
the resulting output might obscure reports of other problems,
in which case
\co{WARN_ON_ONCE(something_bad_is_happening())} might be more appropriate.

\QuickQuiz{
	How can you implement \co{WARN_ON_ONCE()}?
}\QuickQuizAnswer{
	If you don't mind \co{WARN_ON_ONCE()} sometimes warning more
	than once, simply maintain a static variable that is initialized
	to zero.
	If the condition triggers, check the variable, and
	if it is non-zero, return.
	Otherwise, set it to one, print the message, and return.

	If you really need the message to never appear more than once,
	you can use an atomic exchange operation in place of ``set it
	to one'' above.
	Print the message only if the atomic exchange operation returns
	zero.
}\QuickQuizEnd

In parallel code, one bad something that might happen is that
a function expecting to be called under a particular lock might be called
without that lock being held.
Such functions sometimes have header comments stating something like
``The caller must hold \co{foo_lock} when calling this function'', but
such a comment does no good unless someone actually reads it.
An executable statement carries far more weight.
The Linux kernel's lockdep
facility~\cite{JonathanCorbet2006lockdep,StevenRostedt2011locdepCryptic}
therefore provides a \co{lockdep_assert_held()} function that checks
whether the specified lock is held.
Of course, lockdep incurs significant overhead, and thus might not be
helpful in production.

An especially bad parallel-code something is unexpected concurrent
access to data.
The \IXBacrfst{kcsan}~\cite{JonathanCorbet2019KCSAN}
uses existing markings such as \co{READ_ONCE()} and \co{WRITE_ONCE()}
to determine which concurrent accesses deserve warning messages.
KCSAN has a significant false-positive rate, especially from the
viewpoint of developers thinking in terms of C as assembly language
with additional syntax.
KCSAN therefore provides a \co{data_race()} construct to forgive
known-benign \IXpl{data race}, and also the \co{ASSERT_EXCLUSIVE_ACCESS()}
and \co{ASSERT_EXCLUSIVE_WRITER()} assertions to explicitly check for data
races~\cite{MarcoElver2020FearDataRaceDetector1,MarcoElver2020FearDataRaceDetector2}.

So what can be done in cases where checking is necessary, but where the
overhead of runtime checking cannot be tolerated?
One approach is static analysis, which is discussed in the next section.

\section{Static Analysis}
\label{sec:debugging:Static Analysis}
\epigraph{A lot of automation isn't a replacement of
	  humans but of mind-numbing behavior.}
	 {Summarized from Stewart Butterfield}

Static analysis is a validation technique where one program takes a second
program as input, reporting errors and vulnerabilities located in this
second program.
Interestingly enough, almost all programs are statically analyzed
by their compilers or interpreters.
These tools are far from perfect, but their ability to locate
errors has improved immensely over the past few decades, in part because
they now have much more than 64K bytes of memory in which to carry out their
analyses.

The original UNIX \co{lint} tool~\cite{StephenJohnson1977lint} was
quite useful, though much of its functionality has since been incorporated
into C compilers.
There are nevertheless lint-like tools in use to this day.
The sparse static analyzer~\cite{JonathanCorbet2004sparse}
finds higher-level issues in the Linux kernel, including:

\begin{enumerate}
\item	Misuse of pointers to user-space structures.
\item	Assignments from too-long constants.
\item	Empty \co{switch} statements.
\item	Mismatched lock acquisition and release primitives.
\item	Misuse of per-CPU primitives.
\item	Use of RCU primitives on non-RCU pointers and vice versa.
\end{enumerate}

Although it is likely that compilers will continue to increase their
static-analysis capabilities, the sparse static analyzer demonstrates
the benefits of static analysis outside of the compiler, particularly
for finding application-specific bugs.
\Crefrange{sec:formal:SAT Solvers}{sec:formal:Stateless Model Checkers}
describe more sophisticated forms of static analysis.

\section{Code Review}
\label{sec:debugging:Code Review}
\epigraph{If a man speaks of my virtues, he steals from me;
	  if he speaks of my vices, then he is my teacher.}
	 {Chinese proverb}

Code review is a special case of static analysis with human beings doing
the analysis.
This section covers inspection, walkthroughs, and self-inspection.

\subsection{Inspection}
\label{sec:debugging:Inspection}

Traditionally, formal code inspections take place in face-to-face meetings
with formally defined roles:
Moderator, developer, and one or two other participants.
The developer reads through the code, explaining what it is doing and
why it works.
The one or two other participants ask questions and raise issues,
hopefully exposing the author's invalid assumptions, while the moderator's
job is to resolve any resulting conflicts and take notes.
This process can be extremely effective at locating bugs, particularly
if all of the participants are familiar with the code at hand.

However, this face-to-face formal procedure does not necessarily
work well in the global Linux kernel community.
Instead, individuals review code separately and provide comments via
email or IRC\@.
The note-taking is provided by email archives or IRC logs, and moderators
volunteer their services as required by the occasional flamewar.
This process also works reasonably well, particularly if all of the
participants are familiar with the code at hand.
In fact, one advantage of the Linux kernel community approach over
traditional formal inspections is the greater probability of contributions
from people \emph{not} familiar with the code, who might not be blinded
by the author's invalid assumptions, and who might also test the code.

\QuickQuiz{
	Just what invalid assumptions are you accusing Linux kernel
	hackers of harboring???
}\QuickQuizAnswer{
	Those wishing a complete answer to this question are encouraged
	to search the Linux kernel \co{git} repository for commits
	containing the string \qco{Fixes:}.
	There were many thousands of them just in the year 2020, including
	fixes for the following invalid assumptions:

	\begin{enumerate}
	\item	Testing for a non-zero denominator will prevent
		divide-by-zero errors.
		(Hint:
		Suppose that the test uses 64-bit arithmetic
		but that the division uses 32-bit arithmetic.)
	\item	Userspace can be trusted to zero out versioned data
		structures used to communicate with the kernel.
		(Hint:
		Sometimes userspace has no idea how large the
		data structure is.)
	\item	Outdated TCP duplicate selective acknowledgement (D-SACK)
		packets can be completely ignored.
		(Hint:
		These packets might also contain other information.)
	\item	All CPUs are little-endian.
	\item	Once a data structure is no longer needed, all of its
		memory may be immediately freed.
	\item	All devices can be initialized while in standby mode.
	\item	Developers can be trusted to consistently do correct
		hexidecimal arithmetic.
	\end{enumerate}

	Those who look at these commits in greater detail will conclude
	that invalid assumptions are the rule, not the exception.
}\QuickQuizEnd

It is quite likely that the Linux kernel community's review process
is ripe for improvement:

\begin{enumerate}
\item	There is sometimes a shortage of people with the time and
	expertise required to carry out an effective review.
\item	Even though all review discussions are archived, they are
	often ``lost'' in the sense that insights are forgotten and
	people fail to look up the discussions.
	This can result in re-insertion of the same old bugs.
\item	It is sometimes difficult to resolve flamewars when they do
	break out, especially when the combatants have disjoint
	goals, experience, and vocabulary.
\end{enumerate}

Perhaps some of the needed improvements will be provided by
continuous-integration-style testing, but there are many bugs more
easily found by review than by testing.
When reviewing, therefore, it is worthwhile to look at relevant documentation
in commit logs, bug reports, and LWN articles.
This documentation can help you quickly build up the required expertise.

\subsection{Walkthroughs}
\label{sec:debugging:Walkthroughs}

A traditional code walkthrough is similar to a formal inspection,
except that the group
``plays computer'' with the code, driven by specific test cases.
A typical walkthrough team has a moderator, a secretary (who records
bugs found), a testing expert (who generates the test cases) and
perhaps one to two others.
These can be extremely effective, albeit also extremely time-consuming.

It has been some decades since I have participated in a formal
walkthrough, and I suspect that a present-day walkthrough would
use single-stepping debuggers.
One could imagine a particularly sadistic procedure as follows:

\begin{enumerate}
\item	The tester presents the test case.
\item	The moderator starts the code under a debugger, using the
	specified test case as input.
\item	Before each statement is executed, the developer is required
	to predict the outcome of the statement and explain why
	this outcome is correct.
\item	If the outcome differs from that predicted by the developer,
	this is taken as a potential bug.
\item	In parallel code, a ``concurrency shark'' asks what code
	might execute concurrently with this code, and why such
	concurrency is harmless.
\end{enumerate}

Sadistic, certainly.
Effective?
Maybe.
If the participants have a good understanding of the requirements,
software tools, data structures, and algorithms, then walkthroughs
can be extremely effective.
If not, walkthroughs are often a waste of time.

\subsection{Self-Inspection}
\label{sec:debugging:Self-Inspection}

Although developers are usually not all that effective at inspecting
their own code, there are a number of situations where there is no
reasonable alternative.
For example, the developer might be the only person authorized to look
at the code, other qualified developers might all be too busy, or
the code in question might be sufficiently bizarre that the developer
is unable to convince anyone else to take it seriously until after
demonstrating a prototype.
In these cases, the following procedure can be quite helpful,
especially for complex parallel code:

\begin{enumerate}
\item	Write design document with requirements, diagrams for data structures,
	and rationale for design choices.
\item	Consult with experts, updating the design document as needed.
\item	Write the code in pen on paper, correcting errors as you go.
	Resist the temptation to refer to pre-existing nearly identical code
	sequences, instead, copy them.
\item	At each step, articulate and question your assumptions,
	inserting assertions or constructing tests to check them.
\item	If there were errors, copy the code in pen on fresh paper, correcting
	errors as you go.
	Repeat until the last two copies are identical.
\item	Produce proofs of correctness for any non-obvious code.
\item	Use a source-code control system.
	Commit early; commit often.
\item	Test the code fragments from the bottom up.
\item	When all the code is integrated (but preferably before),
	do full-up functional and stress testing.
\item	Once the code passes all tests, write code-level documentation,
	perhaps as an extension to the design document discussed above.
	Fix both the code and the test code as needed.
\end{enumerate}

When I follow this procedure for new RCU code, there are normally only
a few bugs left at the end.
With a few prominent (and embarrassing)
exceptions~\cite{PaulEMcKenney2011RCU3.0trainwreck},
I usually manage to locate these bugs before others do.
That said, this is getting more difficult over time as the number and
variety of Linux-kernel users increases.

\QuickQuizSeries{%
\QuickQuizB{
	Why would anyone bother copying existing code in pen on paper???
	Doesn't that just increase the probability of transcription errors?
}\QuickQuizAnswerB{
	If you are worried about transcription errors, please allow me
	to be the first to introduce you to a really cool tool named
	\co{diff}.
	In addition, carrying out the copying can be quite valuable:
	\begin{enumerate}
	\item	If you are copying a lot of code, you are probably failing
		to take advantage of an opportunity for abstraction.
		The act of copying code can provide great motivation
		for abstraction.
	\item	Copying the code gives you an opportunity to think about
		whether the code really works in its new setting.
		Is there some non-obvious constraint, such as the need
		to disable interrupts or to hold some lock?
	\item	Copying the code also gives you time to consider whether
		there is some better way to get the job done.
	\end{enumerate}
	So, yes, copy the code!
}\QuickQuizEndB
\QuickQuizM{
	This procedure is ridiculously over-engineered!
	How can you expect to get a reasonable amount of software
	written doing it this way???
}\QuickQuizAnswerM{
	Indeed, repeatedly copying code by hand is laborious and slow.
	However, when combined with heavy-duty stress testing and
	proofs of correctness, this approach is also extremely effective
	for complex parallel code where ultimate performance and
	reliability are required and where debugging is difficult.
	The Linux-kernel RCU implementation is a case in point.

	On the other hand, if you are writing a simple single-threaded
	shell script, then you would be best-served by a different
	methodology.
	For example, enter each command one at a time into an interactive
	shell with a test data set to make sure that it does what you
	want, then copy-and-paste the successful commands into your
	script.
	Finally, test the script as a whole.

	If you have a friend or colleague who is willing to help out,
	pair programming can work very well, as can any number of
	formal design- and code-review processes.

	And if you are writing code as a hobby, then do whatever you like.

	In short, different types of software need different development
	methodologies.
}\QuickQuizEndM
\QuickQuizE{
	What do you do if, after all the pen-on-paper copying, you find
	a bug while typing in the resulting code?
}\QuickQuizAnswerE{
	The answer, as is often the case, is ``it depends''.
	If the bug is a simple typo, fix that typo and continue typing.
	However, if the bug indicates a design flaw, go back to pen
	and paper.
}\QuickQuizEndE
}

The above procedure works well for new code, but what if you need to
inspect code that you have already written?
You can of course apply the above procedure for old code in the special
case where you wrote one to throw away~\cite{Brooks79},
but the following approach can also be helpful in less desperate
circumstances:

\begin{enumerate}
\item	Using your favorite documentation tool (\LaTeX{}, HTML,
	OpenOffice, or straight ASCII), describe the high-level
	design of the code in question.
	Use lots of diagrams to illustrate the data structures
	and how these structures are updated.
\item	Make a copy of the code, stripping away all comments.
\item	Document what the code does statement by statement.
\item	Fix bugs as you find them.
\end{enumerate}

This works because describing the code in detail is an excellent way to spot
bugs~\cite{GlenfordJMyers1979}.
This second procedure is also a good way to get your head around
someone else's code, although the first step often suffices.

Although review and inspection by others is probably more efficient and
effective, the above procedures can be quite helpful in cases where
for whatever reason it is not feasible to involve others.

At this point, you might be wondering how to write parallel code without
having to do all this boring paperwork.
Here are some time-tested ways of accomplishing this:

\begin{enumerate}
\item	Write a sequential program that scales through use of
	available parallel library functions.
\item	Write sequential plug-ins for a parallel framework,
	such as map-reduce, BOINC, or a web-application server.
\item	Fully partition your problems, then implement sequential
	program(s) that run in parallel without communication.
\item	Stick to one of the application areas (such as linear algebra)
	where tools can automatically decompose and parallelize
	the problem.
\item	Make extremely disciplined use of parallel-programming
	primitives, so that the resulting code is easily seen to be correct.
	But beware:
	It is always tempting to break the rules ``just a little bit''
	to gain better performance or scalability.
	Breaking the rules often results in general breakage.
	That is, unless you carefully do the paperwork described in this
	section.
\end{enumerate}

But the sad fact is that even if you do the paperwork or use one of
the above ways to more-or-less safely avoid paperwork,
there will be bugs.
If nothing else, more users and a greater variety of users will expose
more bugs more quickly, especially if those users are doing things
that the original developers did not consider.
The next section describes how to handle the probabilistic bugs that
occur all too commonly when validating parallel software.

\QuickQuiz{
	Wait!
	Why on earth would an abstract piece of software fail only
	sometimes???
}\QuickQuizAnswer{
	Because complexity and concurrency can produce results that
	are indistinguishable from
	randomness~\cite{PeterOkech2009InherentRandomness}.
	For example, a bug in Linux-kernel RCU required the following
	to hold before that bug would manifest:
	\begin{enumerate}
	\item	The kernel was built for HPC or real-time use, so that
		a given CPU's RCU work could be offloaded to some other
		CPU\@.
	\item	An offloaded CPU went offline just after generating a
		large quantity of RCU work.
	\item	A special \co{rcu_barrier()} API was invoked just at
		this time.
	\item	The RCU work from the newly offlined CPU was still being
		processed after \co{rcu_barrier()} returned.
	\item	One of these remaining RCU work items was related to
		the code invoking the \co{rcu_barrier()}.
	\end{enumerate}
	Making this bug manifest therefore required considerable luck
	or great testing skill.
	But the testing skill could be effective only if the bug was
	known, which of course it was not.
	Therefore, the manifesting of this bug was very well modeled
	as a probabilistic process.
}\QuickQuizEnd

\section{Probability and Heisenbugs}
\label{sec:debugging:Probability and Heisenbugs}
\epigraph{With both heisenbugs and impressionist art, the closer you
	  get, the less you see.}
	 {Unknown}

So your parallel program fails sometimes.
But you used techniques from the earlier sections to locate
the problem and now have a fix in place!
Congratulations!!!

\begin{figure}
\centering
\resizebox{3in}{!}{\includegraphics{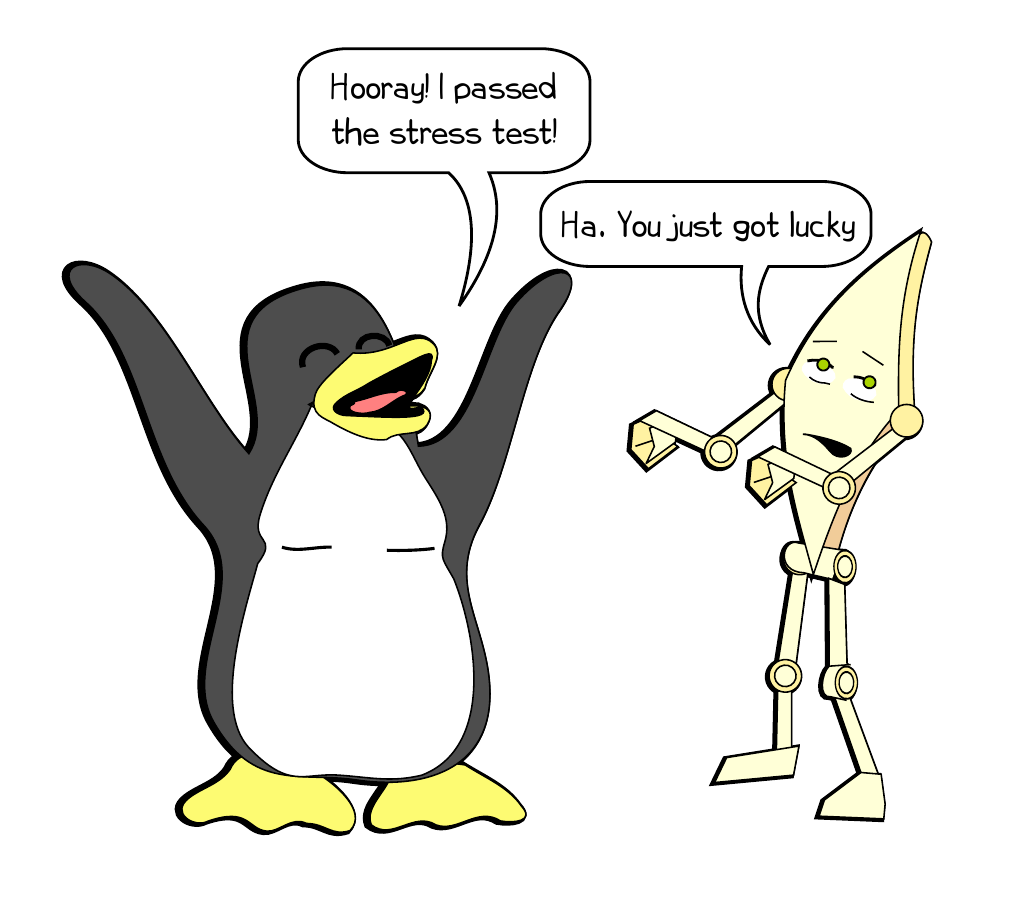}}
\caption{Passed on Merits?
			   Or Dumb Luck?}
\ContributedBy{Figure}{fig:cpu:Passed-the-stress-test}{Melissa Broussard}
\end{figure}

Now the question is just how much testing is required in order to be
certain that
you actually fixed the bug, as opposed to just reducing the probability
of it occurring on the one hand, having fixed only one of several
related bugs on the other hand, or made some ineffectual unrelated
change on yet a third hand.
In short, what is the answer to the eternal question posed by
\cref{fig:cpu:Passed-the-stress-test}?

Unfortunately, the honest answer is that an infinite amount of testing
is required to attain absolute certainty.

\QuickQuiz{
	Suppose that you had a very large number of systems at your
	disposal.
	For example, at current cloud prices, you can purchase a
	huge amount of CPU time at low cost.
	Why not use this approach to get close enough to certainty
	for all practical purposes?
}\QuickQuizAnswer{
	This approach might well be a valuable addition to your
	validation arsenal.
	But it does have limitations that rule out ``for all practical
	purposes'':
	\begin{enumerate}
	\item	Some bugs have extremely low probabilities of occurrence,
		but nevertheless need to be fixed.
		For example, suppose that the Linux kernel's RCU
		implementation had a bug that is triggered only once
		per million years of machine time on average.
		A million years of CPU time is hugely expensive even on
		the cheapest cloud platforms, but we could expect
		this bug to result in more than 50 failures per day
		on the more than 20~billion Linux instances in the
		world as of 2017.
	\item	The bug might well have zero probability of occurrence
		on your particular cloud-computing test setup, which
		means that you won't see it no matter how much machine
		time you burn testing it.
		For but one example, there are RCU bugs that appear
		only in preemptible kernels, and also other RCU bugs
		that appear only in non-preemptible kernels.
	\end{enumerate}
	Of course, if your code is small enough, formal validation
	may be helpful, as discussed in
	\cref{chp:Formal Verification}.
	But beware:
	Formal validation of your code will not find errors in your
	assumptions, misunderstanding of the requirements,
	misunderstanding of the software or hardware primitives you use,
	or errors that you did not think to construct a proof for.
}\QuickQuizEnd

But suppose that we are willing to give up absolute certainty in favor
of high probability.
Then we can bring powerful statistical tools to bear on this problem.
However, this section will focus on simple statistical tools.
These tools are extremely helpful, but please note that reading this
section is not a substitute for statistics classes.\footnote{
	Which I most highly recommend.
	The few statistics courses I have taken have provided value
	far beyond that of the time I spent on them.}

For our start with simple statistical tools, we need to decide whether
we are doing discrete or continuous testing.
Discrete testing features well-defined individual test runs.
For example, a boot-up test of a Linux kernel patch is an example
of a discrete test:
The kernel either comes up or it does not.
Although you might spend an hour boot-testing your kernel, the number of
times you attempted to boot the kernel and the number of times the
boot-up succeeded would often be of more interest than the length
of time you spent testing.
Functional tests tend to be discrete.

On the other hand, if my patch involved RCU, I would probably run
\co{rcutorture}, which is a kernel module that, strangely enough, tests RCU\@.
Unlike booting the kernel, where the appearance of a login prompt
signals the successful end of a discrete test, \co{rcutorture} will happily
continue torturing RCU until either the kernel crashes or until you
tell it to stop.
The duration of the \co{rcutorture} test is usually of more
interest than the number of times you started and stopped it.
Therefore, \co{rcutorture} is an example of a continuous test, a category
that includes many stress tests.

Statistics for discrete tests are simpler and more familiar than those
for continuous tests, and furthermore the statistics for discrete tests
can often be pressed into service for continuous tests, though with some
loss of accuracy.
We therefore start with discrete tests.

\subsection{Statistics for Discrete Testing}
\label{sec:debugging:Statistics for Discrete Testing}

Suppose a bug has a 10\,\% chance of occurring in a given run and that
we do five runs.
How do we compute the probability of at least one run failing?
Here is one way:

\begin{enumerate}
\item	Compute the probability of a given run succeeding, which is 90\,\%.
\item	Compute the probability of all five runs succeeding, which
	is 0.9 raised to the fifth power, or about 59\,\%.
\item	Because either all five runs succeed, or at least one fails,
	subtract the 59\,\% expected success rate from 100\,\%, yielding
	a 41\,\% expected failure rate.
\end{enumerate}

For those preferring formulas, call the probability of a single failure $f$.
The probability of a single success is then $1-f$ and the probability
that all of $n$ tests will succeed is $S_n$:

\begin{equation}
	S_n = \left(1-f\right)^n
\end{equation}

The probability of failure is $1-S_n$, or:

\begin{equation}
	F_n = 1-\left(1-f\right)^n
\label{eq:debugging:Binomial Failure Rate}
\end{equation}

\QuickQuiz{
	Say what???
	When I plug the earlier five-test 10\,\%-failure-rate example into
	the formula, I get 59,050\,\% and that just doesn't make sense!!!
}\QuickQuizAnswer{
	You are right, that makes no sense at all.

	Remember that a probability is a number between zero and one,
	so that you need to divide a percentage by 100 to get a
	probability.
	So 10\,\% is a probability of 0.1, which gets a probability
	of 0.4095, which rounds to 41\,\%, which quite sensibly
	matches the earlier result.
}\QuickQuizEnd

So suppose that a given test has been failing 10\,\% of the time.
How many times do you have to run the test to be 99\,\% sure that
your supposed fix actually helped?

Another way to ask this question is ``How many times would we need
to run the test to cause the probability of failure to rise above 99\,\%?''
After all, if we were to run the test enough times that the probability
of seeing at least one failure becomes 99\,\%, if there are no failures,
there is only 1\,\% probability of this ``success'' being due to dumb luck.
And if we plug $f=0.1$ into
\cref{eq:debugging:Binomial Failure Rate} and vary $n$,
we find that 43 runs gives us a 98.92\,\% chance of at least one test failing
given the original 10\,\% per-test failure rate,
while 44 runs gives us a 99.03\,\% chance of at least one test failing.
So if we run the test on our fix 44 times and see no failures, there
is a 99\,\% probability that our fix really did help.

But repeatedly plugging numbers into
\cref{eq:debugging:Binomial Failure Rate}
can get tedious, so let's solve for $n$:

\begin{eqnarray}
	F_n = 1-\left(1-f\right)^n \\
	1 - F_n = \left(1-f\right)^n \\
	\log \left(1 - F_n\right) = n \; \log \left(1 - f\right)
\end{eqnarray}

Finally the number of tests required is given by:

\begin{equation}
	n = \frac{\log\left(1 - F_n\right)}{\log\left(1 - f\right)}
\label{eq:debugging:Binomial Number of Tests Required}
\end{equation}

Plugging $f=0.1$ and $F_n=0.99$ into
\cref{eq:debugging:Binomial Number of Tests Required}
gives 43.7, meaning that we need 44 consecutive successful test
runs to be 99\,\% certain that our fix was a real improvement.
This matches the number obtained by the previous method, which
is reassuring.

\QuickQuiz{
	In \cref{eq:debugging:Binomial Number of Tests Required},
	are the logarithms base-10, base-2, or base-$\euler$?
}\QuickQuizAnswer{
	It does not matter.
	You will get the same answer no matter what base of logarithms
	you use because the result is a pure ratio of logarithms.
	The only constraint is that you use the same base for both
	the numerator and the denominator.
}\QuickQuizEnd

\begin{figure}
\centering
\resizebox{2.5in}{!}{\includegraphics{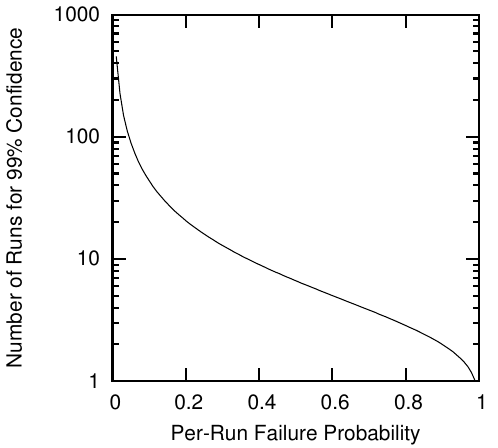}}
\caption{Number of Tests Required for 99 Percent Confidence Given Failure Rate}
\label{fig:debugging:Number of Tests Required for 99 Percent Confidence Given Failure Rate}
\end{figure}

\Cref{fig:debugging:Number of Tests Required for 99 Percent Confidence Given Failure Rate}
shows a plot of this function.
Not surprisingly, the less frequently each test run fails, the more
test runs are required to be 99\,\% confident that the bug has been
fixed.
If the bug caused the test to fail only 1\,\% of the time, then a
mind-boggling 458 test runs are required.
As the failure probability decreases, the number of test runs required
increases, going to infinity as the failure probability goes to zero.

The moral of this story is that when you have found a rarely occurring
bug, your testing job will be much easier if you can come up with
a carefully targeted test with a much higher failure rate.
For example, if your targeted test raised the failure rate from 1\,\%
to 30\,\%, then the number of runs required for 99\,\% confidence
would drop from 458 to a more tractable 13.

But these thirteen test runs would only give you 99\,\% confidence that
your fix had produced ``some improvement''.
Suppose you instead want to have 99\,\% confidence that your fix reduced
the failure rate by an order of magnitude.
How many failure-free test runs are required?

An order of magnitude improvement from a 30\,\% failure rate would be
a 3\,\% failure rate.
Plugging these numbers into
\cref{eq:debugging:Binomial Number of Tests Required} yields:

\begin{equation}
	n = \frac{\log\left(1 - 0.99\right)}{\log\left(1 - 0.03\right)} = 151.2
\end{equation}

So our order of magnitude improvement requires roughly an order of
magnitude more testing.
Certainty is impossible, and high probabilities are quite expensive.
This is why making tests run more quickly and making failures more
probable are essential skills in the development of highly reliable
software.
These skills will be covered in
\cref{sec:debugging:Hunting Heisenbugs}.

\subsection{Statistics Abuse for Discrete Testing}
\label{sec:debugging:Statistics Abuse for Discrete Testing}

But suppose that you have a continuous test that fails about three
times every ten hours, and that you fix the bug that you believe was
causing the failure.
How long do you have to run this test without failure to be 99\,\% certain
that you reduced the probability of failure?

Without doing excessive violence to statistics, we could simply
redefine a one-hour run to be a discrete test that has a 30\,\%
probability of failure.
Then the results of in the previous section tell us that if the test
runs for 13 hours without failure, there is a 99\,\% probability that
our fix actually improved the program's reliability.

A dogmatic statistician might not approve of this approach, but the sad
fact is that the errors introduced by this sort of statistical abuse are
usually quite small compared to the errors in your failure-rate estimates.
Nevertheless, the next section takes a more rigorous approach.

\subsection{Statistics for Continuous Testing}
\label{sec:debuggingStatistics for Continuous Testing}

The fundamental formula for failure probabilities is the Poisson
distribution:

\begin{equation}
	F_m = \frac{\lambda^m}{m!} \euler^{-\lambda}
\label{eq:debugging:Poisson Probability}
\end{equation}

Here $F_m$ is the probability of $m$ failures in the test and
$\lambda$ is the expected failure rate per unit time.
A rigorous derivation may be found in any advanced probability
textbook, for example, Feller's classic ``An Introduction to Probability
Theory and Its Applications''~\cite{Feller58}, while a more
intuitive derivation may be found in the first edition of
this book~\cite[Equations 11.8--11.26]{McKenney2014ParallelProgramming-e1}.

Let's try reworking the example from
\cref{sec:debugging:Statistics Abuse for Discrete Testing}
using the Poisson distribution.
Recall that this example involved a test with a 30\,\% failure rate per
hour, and that the question was how long the test would need to run
error-free
on a alleged fix to be 99\,\% certain that the fix actually reduced the
failure rate.
In this case, $m$ is zero, so that
\cref{eq:debugging:Poisson Probability} reduces to:

\begin{equation}
	F_0 =  \euler^{-\lambda}
\end{equation}

Solving this requires setting $F_0$
to 0.01 and solving for $\lambda$, resulting in:

\begin{equation}
	\lambda = - \ln 0.01 = 4.6
\end{equation}

Because we get $0.3$ failures per hour, the number of hours required
is $4.6/0.3 = 14.3$, which is within 10\,\% of the 13 hours
calculated using the method in
\cref{sec:debugging:Statistics Abuse for Discrete Testing}.
Given that you normally won't know your failure rate to anywhere near
10\,\%, the simpler method described in
\cref{sec:debugging:Statistics Abuse for Discrete Testing}
is almost always good and sufficient.

More generally, if we have $n$ failures per unit time, and we want to
be $P$\,\% certain that a fix reduced the failure rate, we can use the
following formula:

\begin{equation}
	T = - \frac{1}{n} \ln \frac{100 - P}{100}
\label{eq:debugging:Error-Free Test Duration}
\end{equation}

\QuickQuiz{
	Suppose that a bug causes a test failure three times per hour
	on average.
	How long must the test run error-free to provide 99.9\,\%
	confidence that the fix significantly reduced the probability
	of failure?
}\QuickQuizAnswer{
	We set $n$ to $3$ and $P$ to $99.9$ in
	\cref{eq:debugging:Error-Free Test Duration}, resulting in:

	\begin{equation}
		T = - \frac{1}{3} \ln \frac{100 - 99.9}{100} = 2.3
	\end{equation}

	If the test runs without failure for 2.3 hours, we can be 99.9\,\%
	certain that the fix reduced the probability of failure.
}\QuickQuizEnd

As before, the less frequently the bug occurs and the greater the
required level of confidence, the longer the required error-free test run.

Suppose that a given test fails about once every hour, but after a bug
fix, a 24-hour test run fails only twice.
Assuming that the failure leading to the bug is a random occurrence,
what is the probability that the small number of
failures in the second run was due to random chance?
In other words, how confident should we be that the fix actually
had some effect on the bug?
This probability may be calculated by summing
\cref{eq:debugging:Poisson Probability} as follows:

\begin{equation}
	F_0 + F_1 + \dots + F_{m - 1} + F_m =
		\sum_{i=0}^m \frac{\lambda^i}{i!} \euler^{-\lambda}
\end{equation}

This is the Poisson cumulative distribution function, which can be
written more compactly as:

\begin{equation}
	F_{i \le m} = \sum_{i=0}^m \frac{\lambda^i}{i!} \euler^{-\lambda}
\label{eq:debugging:Possion CDF}
\end{equation}

Here $m$ is the actual number of errors in the long test run
(in this case, two) and $\lambda$ is expected number of errors
in the long test run (in this case, 24).
Plugging $m=2$ and $\lambda=24$ into this expression gives the probability
of two or fewer failures as about
$1.2 \times 10^{-8}$, in other words, we have a high level of confidence
that the fix actually had some relationship to the bug.\footnote{
	Of course, this result in no way excuses you from finding and
	fixing the bug(s) resulting in the remaining two failures!}

\QuickQuizSeries{%
\QuickQuizB{
	Doing the summation of all the factorials and exponentials
	is a real pain.
	Isn't there an easier way?
}\QuickQuizAnswerB{
	One approach is to use the open-source symbolic manipulation
	program named ``maxima''.
	Once you have installed this program, which is a part of many
	Linux distributions, you can run it and give the
	\co{load(distrib);} command followed by any number of
	\co{bfloat(cdf_poisson(m,l));} commands, where the \co{m} is
	replaced by the desired value of $m$ (the actual number of failures in
	actual test) and the \co{l} is replaced by the desired value of
	$\lambda$ (the expected number of failures in the actual test).

	In particular, the \co{bfloat(cdf_poisson(2,24));} command
	results in \co{1.181617112359357b-8}, which matches the value
	given by \cref{eq:debugging:Possion CDF}.

\begin{table}
\renewcommand*{\arraystretch}{1.25}
\rowcolors{3}{}{lightgray}
\small
\centering
\begin{tabular}{rrrr}
	\toprule
		& \multicolumn{3}{c}{Improvement} \\
		\cmidrule(l){2-4}
	Certainty (\%)
		& Any
			& 10x
				& 100x \\
	\cmidrule{1-1} \cmidrule(l){2-4}
	90.0	& 2.3	& 23.0	& 230.0  \\
	95.0	& 3.0	& 30.0	& 300.0  \\
	99.0	& 4.6	& 46.1	& 460.5  \\
	99.9	& 6.9	& 69.1	& 690.7  \\
	\bottomrule
\end{tabular}
\caption{Human-Friendly Poisson-Function Display}
\label{tab:debugging:Human-Friendly Poisson-Function Display}
\end{table}

	Another approach is to recognize that in this real world,
	it is not all that useful to compute (say) the duration of a test
	having two or fewer errors that would give a 76.8\,\% confidence
	of a 349.2x improvement in reliability.
	Instead, human beings tend to focus on specific values, for
	example, a 95\,\% confidence of a 10x improvement.
	People also greatly prefer error-free test runs, and so should
	you because doing so reduces your required test durations.
	Therefore, it is quite possible that the values in
	\cref{tab:debugging:Human-Friendly Poisson-Function Display}
	will suffice.
	Simply look up the desired confidence and degree of improvement,
	and the resulting number will give you the required
	error-free test duration in terms of the expected time for
	a single error to appear.
	So if your pre-fix testing suffered one failure per hour, and the
	powers that be require a 95\,\% confidence of a 10x improvement,
	you need a 30-hour error-free run.

	Alternatively, you can use the rough-and-ready method described in
	\cref{sec:debugging:Statistics Abuse for Discrete Testing}.
}\QuickQuizEndB
\QuickQuizE{
	But wait!!!
	Given that there has to be \emph{some} number of failures
	(including the possibility of zero failures), shouldn't
	\cref{eq:debugging:Possion CDF}
	approach the value $1$ as $m$ goes to infinity?
}\QuickQuizAnswerE{
	Indeed it should.
	And it does.

	To see this, note that $\euler^{-\lambda}$ does not depend on $i$,
	which means that it can be pulled out of the summation as follows:

	\begin{equation}
		\euler^{-\lambda} \sum_{i=0}^\infty \frac{\lambda^i}{i!}
	\end{equation}

	The remaining summation is exactly the Taylor series for
	$\euler^\lambda$, yielding:

	\begin{equation}
		\euler^{-\lambda} \euler^\lambda
	\end{equation}

	The two exponentials are reciprocals, and therefore cancel,
	resulting in exactly $1$, as required.
}\QuickQuizEndE
}

The Poisson distribution is a powerful tool for analyzing test results,
but the fact is that in this last example there were still two remaining
test failures in a 24-hour test run.
Such a low failure rate results in very long test runs.
The next section discusses counter-intuitive ways of improving this situation.

\subsection{Hunting Heisenbugs}
\label{sec:debugging:Hunting Heisenbugs}

This line of thought also helps explain heisenbugs:
Adding tracing and assertions can easily reduce the probability
of a bug appearing, which
is why extremely lightweight tracing and assertion mechanism are
so critically important.

The term ``\IXB{heisenbug}'' was inspired by the \pplsur{Weiner}{Heisenberg}
\IX{Uncertainty Principle} from quantum physics, which states that
it is impossible to
exactly quantify a given particle's position and velocity at any given
point in time~\cite{WeinerHeisenberg1927Uncertain}.
Any attempt to more accurately measure that particle's position will
result in increased uncertainty of its velocity and vice versa.
Similarly, attempts to track down
the heisenbug causes its symptoms to radically change or even disappear
completely.\footnote{
	The term ``heisenbug'' is a misnomer, as most heisenbugs are
	fully explained by the \emph{observer effect} from classical
	physics.
	Nevertheless, the name has stuck.}
Of course, adding debugging overhead can and sometimes does make bugs
more probable.
But developers are more likely to remember the frustration of a
disappearing heisenbug than the joy inspired by the bug becoming
more easily reproduced!

If the field of physics inspired the name of this problem, it is only
fair that the field of physics should inspire the solution.
Fortunately, particle physics is up to the task:
Why not create an \IXBalth{anti-heisenbug}{anti-}{heisenbug}
to annihilate the heisenbug?
Or, perhaps more accurately, to annihilate the heisen-ness of
the heisenbug?
Although producing an anti-heisenbug for a given heisenbug is more an
art than a science, the following sections describe a number of ways to
do just that:

\begin{enumerate}
\item	Add delay to race-prone regions (\cref{sec:debugging:Add Delay}).
\item	Increase workload intensity
	(\cref{sec:debugging:Increase Workload Intensity}).
\item	Isolate suspicious subsystems
	(\cref{sec:debugging:Isolate Suspicious Subsystems}).
\item	Simulate unusual events (\cref{sec:debugging:Simulate Unusual Events}).
\item	Count near misses (\cref{sec:debugging:Count Near Misses}).
\end{enumerate}

These are followed by discussion in
\cref{sec:debugging:Heisenbug Discussion}.

\subsubsection{Add Delay}
\label{sec:debugging:Add Delay}

Consider the count-lossy code in
\cref{sec:count:Why Isn't Concurrent Counting Trivial?}.
Adding \co{printf()} statements will likely greatly reduce or even
eliminate the lost counts.
However, converting the load-add-store sequence to a load-add-delay-store
sequence will greatly increase the incidence of lost counts (try it!).
Once you spot a bug involving a \IX{race condition}, it is frequently possible
to create an anti-heisenbug by adding delay in this manner.

Of course, this begs the question of how to find the race condition in
the first place.
Although very lucky developers might accidentally create delay-based
anti-heisenbugs when adding debug code, this is in general a dark art.
Nevertheless, there are a number of things you can do to find your race
conditions.

One approach is to recognize that race conditions often end up corrupting
some of the data involved in the race.
It is therefore good practice to double-check the synchronization of
any corrupted data.
Even if you cannot immediately recognize the race condition, adding
delay before and after accesses to the corrupted data might change
the failure rate.
By adding and removing the delays in an organized fashion (e.g., binary
search), you might learn more about the workings of the race condition.

\QuickQuiz{
	How is this approach supposed to help if the corruption affected some
	unrelated pointer, which then caused the corruption???
}\QuickQuizAnswer{
	Indeed, that can happen.
	Many CPUs have hardware-debugging facilities that can help you
	locate that unrelated pointer.
	Furthermore, if you have a core dump, you can search the core
	dump for pointers referencing the corrupted region of memory.
	You can also look at the data layout of the corruption, and
	check pointers whose type matches that layout.

	You can also step back and test the modules making up your
	program more intensively, which will likely confine the corruption
	to the module responsible for it.
	If this makes the corruption vanish, consider adding additional
	argument checking to the functions exported from each module.

	Nevertheless, this is a hard problem, which is why I used the
	words ``a bit of a dark art''.
}\QuickQuizEnd

Another important approach is to
vary the software and hardware configuration and look for statistically
significant differences in failure rate.
For example, back in the 1990s, it was common practice to test on systems
having CPUs running at different clock rates, which tended to make some
types of race conditions more probable.
One way of getting a similar effect today is to test on multi-socket
systems, thus incurring the large delays described in
\cref{sec:cpu:Overheads}.

However you choose to add delays, you can then look more intensively at
the code implicated by those delays that make the greatest difference
in failure rate.
It might be helpful to test that code in isolation, for example.

One important aspect of software configuration is the history of
changes, which is why \co{git bisect} is so useful.
Bisection of the change history can provide very valuable clues as
to the nature of the heisenbug, in this case presumably by locating
a commit that shows a change in the software's response to the addition
or removal of a given delay.

\QuickQuiz{
	But I did the bisection, and ended up with a huge commit.
	What do I do now?
}\QuickQuizAnswer{
	A huge commit?
	Shame on you!
	This is but one reason why you are supposed to keep the commits small.

	And that is your answer:
	Break up the commit into bite-sized pieces and bisect the pieces.
	In my experience, the act of breaking up the commit is often
	sufficient to make the bug painfully obvious.
}\QuickQuizEnd

Once you locate the suspicious section of code, you can then introduce
delays to attempt to increase the probability of failure.
As we have seen, increasing the probability of failure makes it much
easier to gain high confidence in the corresponding fix.

However, it is sometimes quite difficult to track down the problem using
normal debugging techniques.
The following sections present some other alternatives.

\subsubsection{Increase Workload Intensity}
\label{sec:debugging:Increase Workload Intensity}

It is often the case that a given test suite places relatively
low stress on a given subsystem, so that a small change in timing
can cause a heisenbug to disappear.
One way to create an anti-heisenbug for this case is to increase
the workload intensity, which has a good chance of increasing the
bug's probability.
If the probability is increased sufficiently, it may be possible to
add lightweight diagnostics such as tracing without causing the
bug to vanish.

How can you increase the workload intensity?
This depends on the program, but here are some things to try:

\begin{enumerate}
\item	Add more CPUs.
\item	If the program uses networking, add more network adapters
	and more or faster remote systems.
\item	If the program is doing heavy I/O when the problem occurs,
	either (1) add more storage devices, (2) use faster storage
	devices, for example, substitute SSDs for disks,
	or (3) use a RAM-based filesystem to substitute main
	memory for mass storage.
\item	Change the size of the problem, for example, if doing a parallel
	matrix multiply, change the size of the matrix.
	Larger problems may introduce more complexity, but smaller
	problems often increase the level of contention.
	If you aren't sure whether you should go large or go small,
	just try both.
\end{enumerate}

However, it is often the case that the bug is in a specific subsystem,
and the structure of the program limits the amount of stress that can
be applied to that subsystem.
The next section addresses this situation.

\subsubsection{Isolate Suspicious Subsystems}
\label{sec:debugging:Isolate Suspicious Subsystems}

If the program is structured such that it is difficult or impossible
to apply much stress to a subsystem that is under suspicion,
a useful anti-heisenbug is a stress test that tests that subsystem in
isolation.
The Linux kernel's \co{rcutorture} module takes exactly this approach with RCU\@:
Applying more stress to RCU than is feasible in a production environment
increases the probability that RCU bugs will be found during testing
rather than in production.\footnote{
	Though sadly not increased to probability one.}

In fact, when creating a parallel program, it is wise to stress-test
the components separately.
Creating such component-level stress tests can seem like a waste of time,
but a little bit of component-level testing can save a huge amount
of system-level debugging.

\subsubsection{Simulate Unusual Events}
\label{sec:debugging:Simulate Unusual Events}

Heisenbugs are sometimes due to unusual events, such as
memory-allocation failure, conditional-lock-acquisition failure,
CPU-hotplug operations, timeouts, packet losses, and so on.
One way to construct an anti-heisenbug for this class of heisenbug
is to introduce spurious failures.

For example, instead of invoking \co{malloc()} directly, invoke
a wrapper function that uses a random number to decide whether
to return \co{NULL} unconditionally on the one hand, or to actually
invoke \co{malloc()} and return the resulting pointer on the other.
Inducing spurious failures is an excellent way to bake robustness into
sequential programs as well as parallel programs.

\QuickQuiz{
	Why don't conditional-locking primitives provide this
	spurious-failure functionality?
}\QuickQuizAnswer{
	There are locking algorithms that depend on conditional-locking
	primitives telling them the truth.
	For example, if conditional-lock failure signals that
	some other thread is already working on a given job,
	spurious failure might cause that job to never get done,
	possibly resulting in a hang.
}\QuickQuizEnd

\subsubsection{Count Near Misses}
\label{sec:debugging:Count Near Misses}

Bugs are often all-or-nothing things, so that a bug either happens
or not, with nothing in between.
However, it is sometimes possible to define a \emph{near miss} where
the bug does not result in a failure, but has likely manifested.
For example, suppose your code is making a robot walk.
The robot's falling down constitutes a bug in your program, but
stumbling and recovering might constitute a near miss.
If the robot falls over only once per hour, but stumbles every few
minutes, you might be able to speed up your debugging progress by
counting the number of stumbles in addition to the number of falls.

In concurrent programs, timestamping can sometimes be used to detect
near misses.
For example, locking primitives incur significant delays, so if there is
a too-short delay between a pair of operations that are supposed
to be protected by different acquisitions of the same lock, this too-short
delay might be counted as a near miss.\footnote{
	Of course, in this case, you might be better off using
	whatever \co{lock_held()} primitive is available
	in your environment.
	If there isn't a \co{lock_held()} primitive, create one!}

\begin{figure}
\centering
\resizebox{3in}{!}{\includegraphics{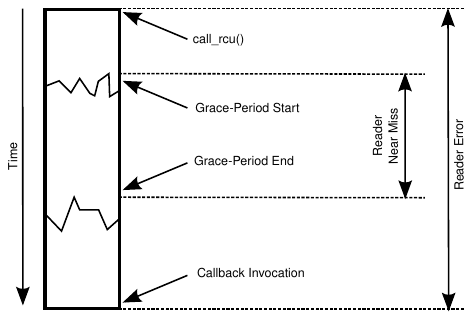}}
\caption{RCU Errors and Near Misses}
\label{fig:debugging:RCU Errors and Near Misses}
\end{figure}

For example, a low-probability bug in RCU priority boosting occurred
roughly once every hundred hours of focused \co{rcutorture} testing.
Because it would take almost 500 hours of failure-free testing to be
99\,\% certain that the bug's probability had been significantly reduced,
the \co{git bisect} process
to find the failure would be painfully slow---or would require an extremely
large test farm.
Fortunately, the RCU operation being tested included not only a wait for
an RCU grace period, but also a previous wait for the grace period to start
and a subsequent wait for an RCU callback to be
invoked after completion of the RCU grace period.
This distinction between an \co{rcutorture} error and near miss is
shown in
\cref{fig:debugging:RCU Errors and Near Misses}.
To qualify as a full-fledged error, an RCU read-side critical section
must extend from the \co{call_rcu()} that initiated a grace period,
through the remainder of the previous grace period, through the
entirety of the grace period initiated by the \co{call_rcu()}
(denoted by the region between the jagged lines), and
through the delay from the end of that grace period to the callback
invocation, as indicated by the ``Error'' arrow.
However, the formal definition of RCU prohibits RCU read-side critical
sections from extending across a single grace period, as indicated by
the ``Near Miss'' arrow.
This suggests using near misses as the error condition, however, this
can be problematic because different CPUs can have different opinions
as to exactly where a given
grace period starts and ends, as indicated by the jagged lines.\footnote{
	In real life, these lines can be much more jagged because idle
	CPUs can be completely unaware of a great many recent grace
	periods.}
Using the near misses as the error condition could therefore result
in false positives, which need to be avoided in the automated
\co{rcutorture} testing.

By sheer dumb luck, \co{rcutorture} happens to include some statistics that
are sensitive to the near-miss version of the grace period.
As noted above, these statistics are subject to false positives due to
their unsynchronized access to RCU's state variables,
but these false positives turn out to be extremely rare on strongly
ordered systems such as the IBM mainframe and x86, occurring less than
once per thousand hours of testing.

These near misses occurred roughly once per hour, about two orders of
magnitude more frequently than the actual errors.
Use of these near misses allowed the bug's root cause to be identified
in less than a week and a high degree of confidence in the fix to be
built in less than a day.
In contrast, excluding the near misses in favor of the real errors would
have required months of debug and validation time.

To sum up near-miss counting, the general approach is to replace counting
of infrequent failures with more-frequent near misses that are believed
to be correlated with those failures.
These near-misses can be considered an anti-heisenbug to the real failure's
heisenbug because the near-misses, being more frequent, are likely to
be more robust in the face of changes to your code, for example, the
changes you make to add debugging code.

\subsubsection{Heisenbug Discussion}
\label{sec:debugging:Heisenbug Discussion}

The alert reader might have noticed that this section was fuzzy and
qualitative, in stark contrast to the precise mathematics of
\cref{sec:debugging:Statistics for Discrete Testing,%
sec:debugging:Statistics Abuse for Discrete Testing,%
sec:debuggingStatistics for Continuous Testing}.
If you love precision and mathematics, you may be disappointed to
learn that the situations to which this section applies are far
more common than those to which the preceding sections apply.

In fact, the common case is that although you might have reason to believe
that your code has bugs, you have no idea what those bugs are, what
causes them, how likely they are to appear, or what conditions affect
their probability of appearance.
In this all-too-common case, statistics cannot help you.\footnote{
	Although if you know what your program is supposed to do and
	if your program is small enough (both less likely that you
	might think), then the formal-verification tools described in
	\cref{chp:Formal Verification}
	can be helpful.}
That is to say, statistics cannot help you \emph{directly}.
But statistics can be of great indirect help---\emph{if} you have the
humility required to admit that you make mistakes, that you can reduce the
probability of these mistakes (for example, by getting enough sleep), and
that the number and type of mistakes you made in the past is indicative of
the number and type of mistakes that you are likely to make in the future.
For example, I have a deplorable tendency to forget to write a small
but critical portion of the initialization code, and frequently get most
or even all of a parallel program correct---except for a stupid
omission in initialization.
Once I was willing to admit to myself that I am prone to this type of
mistake, it was easier (but not easy!\@) to force myself to double-check
my initialization code.
Doing this allowed me to find numerous bugs ahead of time.

When your quick bug hunt morphs into a long-term quest, it is important
to log everything you have tried and what happened.
In the common case where the software is changing during the course of
your quest, make sure to record the exact version of the software to
which each log entry applies.
From time to time, reread the entire log in order to make connections
between clues encountered at different times.
Such rereading is especially important upon encountering a surprising
test result, for example, I reread my log upon realizing that what I
thought was a failure of the hypervisor to schedule a vCPU was instead
an interrupt storm preventing that vCPU from making forward progress
on the interrupted code.
If the code you are debugging is new to you, this log is also an
excellent place to document the relationships between code and data
structures.
Keeping a log when you are furiously chasing a difficult bug might seem
like needless paperwork, but it has on many occasions saved me from
debugging around and around in circles, which can waste far more time
than keeping a log ever could.

Using Taleb's nomenclature~\cite{NassimTaleb2007BlackSwan},
a white swan is a bug that we can reproduce.
We can run a large number of tests, use ordinary statistics to
estimate the bug's probability, and use ordinary statistics again
to estimate our confidence in a proposed fix.
An unsuspected bug is a black swan.
We know nothing about it, we have no tests that have yet caused it
to happen, and statistics is of no help.
Studying our own behavior, especially the number and types of mistakes
we make, can turn black swans into grey swans.
We might not know exactly what the bugs are, but we have some idea of
their number and maybe also of their type.
Ordinary statistics is still of no help (at least not until we are
able to reproduce one of the bugs), but robust\footnote{
	That is to say brutal.}
testing methods can be of great help.
The goal, therefore, is to use experience and good validation practices
to turn the black swans grey, focused testing and analysis to turn the
grey swans white, and ordinary methods to fix the white swans.

That said, thus far, we have focused solely on bugs in the parallel program's
functionality.
However, performance is a first-class requirement for a parallel program.
Otherwise, why not write a sequential program?
To repurpose Kipling, our goal when writing parallel code is to fill
the unforgiving second with sixty minutes worth of distance run.
The next section therefore discusses a number of performance bugs that
would be happy to thwart this Kiplingesque goal.

\section{Performance Estimation}
\label{sec:debugging:Performance Estimation}
\epigraph{There are lies, damn lies, statistics, and benchmarks.}
	 {Unknown}

Parallel programs usually have performance and scalability requirements,
after all, if performance is not an issue, why not use a sequential
program?
Ultimate performance and linear scalability might not be necessary, but
there is little use for a parallel program that runs slower than its
optimal sequential counterpart.
And there really are cases where every microsecond matters and every
nanosecond is needed.
Therefore, for parallel programs, insufficient performance is just as
much a bug as is incorrectness.

\QuickQuizSeries{%
\QuickQuizB{
	That is ridiculous!!!
	After all, isn't getting the correct answer later than one would like
	better than getting an incorrect answer???
}\QuickQuizAnswerB{
	This question fails to consider the option of choosing not to
	compute the answer at all, and in doing so, also fails to consider
	the costs of computing the answer.
	For example, consider short-term weather forecasting, for which
	accurate models exist, but which require large (and expensive)
	clustered supercomputers, at least if you want to actually run
	the model faster than the weather.

	And in this case, any performance bug that prevents the model from
	running faster than the actual weather prevents any forecasting.
	Given that the whole purpose of purchasing the large clustered
	supercomputers was to forecast weather, if you cannot run the
	model faster than the weather, you would be better off not running
	the model at all.

	More severe examples may be found in the area of safety-critical
	real-time computing.
}\QuickQuizEndB
\QuickQuizE{
	But if you are going to put in all the hard work of parallelizing
	an application, why not do it right?
	Why settle for anything less than optimal performance and
	linear scalability?
}\QuickQuizAnswerE{
	Although I do heartily salute your spirit and aspirations,
	you are forgetting that there may be high costs due to delays
	in the program's completion.
	For an extreme example, suppose that a 40\,\% performance shortfall
	from a single-threaded application is causing one person to die
	each day.
	Suppose further that in a day you could hack together a
	quick and dirty
	parallel program that ran 50\,\% faster on an eight-CPU system
	than the sequential version, but that an optimal parallel
	program would require four months of painstaking design, coding,
	debugging, and tuning.

	It is safe to say that more than 100 people would prefer the
	quick and dirty version.
}\QuickQuizEndE
}

Validating a parallel program must therfore include validating its
performance.
But validating performance means having a workload to run and performance
criteria with which to evaluate the program at hand.
These needs are often met by \emph{performance benchmarks}, which
are discussed in the next section.

\subsection{Benchmarking}
\label{sec:debugging:Benchmarking}

Frequent abuse aside, benchmarks are both useful and heavily used,
so it is not helpful to be too dismissive of them.
Benchmarks span the range from ad hoc test jigs to international
standards, but regardless of their level of formality, benchmarks
serve four major purposes:

\begin{enumerate}
\item	Providing a fair framework for comparing competing implementations.
\item	Focusing competitive energy on improving implementations in ways
	that matter to users.
\item	Serving as example uses of the implementations being benchmarked.
\item	Serving as a marketing tool to highlight your software
	against your competitors' offerings.
\end{enumerate}

Of course, the only completely fair framework is the intended
application itself.
So why would anyone who cared about fairness in benchmarking
bother creating imperfect benchmarks rather than simply
using the application itself as the benchmark?

Running the actual application is in fact the best approach where it is practical.
Unfortunately, it is often impractical for the following reasons:

\begin{enumerate}
\item	The application might be proprietary, and you
	might not have the right to run the intended application.
\item	The application might require more hardware
	than you have access to.
\item	The application might use data that you cannot
	access, for example, due to privacy regulations.
\item	The application might take longer than is convenient to
	reproduce a performance or scalability problem.\footnote{
		Microbenchmarks can help, but
		please see \cref{sec:debugging:Microbenchmarking}.}
\end{enumerate}

Creating a benchmark that approximates
the application can help overcome these obstacles.
A carefully constructed benchmark can help promote performance,
scalability, \IXh{energy}{efficiency}, and much else besides.
However, be careful to avoid investing too much into the benchmarking
effort.
It is after all important to invest at least a little into the
application itself~\cite{Gray91}.

\subsection{Profiling}
\label{sec:debugging:Profiling}

In many cases, a fairly small portion of your software is responsible
for the majority of the performance and scalability shortfall.
However, developers are notoriously unable to identify the actual
bottlenecks by inspection.
For example, in the case of a kernel buffer allocator, all attention focused
on a search of a dense array which turned out to represent only
a few percent of the allocator's execution time.
An execution profile collected via a logic analyzer focused attention
on the cache misses that were actually responsible for the majority
of the problem~\cite{McKenney93}.

An old-school but quite effective method of tracking down performance
and scalability bugs is to run your program under a debugger,
then periodically interrupt it, recording the stacks of all threads
at each interruption.
The theory here is that if something is slowing down your program,
it has to be visible in your threads' executions.

That said, there are a number of tools
that will usually do a much better job of helping you to focus your
attention where it will do the most good.
Two popular choices are \co{gprof} and \co{perf}.
To use \co{perf} on a single-process program, prefix your command
with \co{perf record}, then after the command completes, type
\co{perf report}.
There is a lot of work on tools for performance debugging of multi-threaded
programs, which should make this important job easier.
Again, one good starting point is Brendan Gregg's blog.\footnote{
	\url{http://www.brendangregg.com/blog/}}

\subsection{Differential Profiling}
\label{sec:debugging:Differential Profiling}

Scalability problems will not necessarily be apparent unless you are running
on very large systems.
However, it is sometimes possible to detect impending scalability problems
even when running on much smaller systems.
One technique for doing this is called \emph{differential profiling}.

The idea is to run your workload under two different sets of conditions.
For example, you might run it on two CPUs, then run it again on four
CPUs.
You might instead vary the load placed on the system, the number of
network adapters, the number of mass-storage devices, and so on.
You then collect profiles of the two runs, and mathematically combine
corresponding profile measurements.
For example, if your main concern is scalability, you might take the
ratio of corresponding measurements, and then sort the ratios into
descending numerical order.
The prime scalability suspects will then be sorted to the top of the
list~\cite{McKenney95a,McKenney99b}.

Some tools such as \co{perf} have built-in differential-profiling
support.

\subsection{Microbenchmarking}
\label{sec:debugging:Microbenchmarking}

Microbenchmarking can be useful when deciding which algorithms or
data structures are worth incorporating into a larger body of software
for deeper evaluation.

One common approach to microbenchmarking is to measure the time,
run some number of iterations of the code
under test, then measure the time again.
The difference between the two times divided by the number of iterations
gives the measured time required to execute the code under test.

Unfortunately, this approach to measurement allows any number of errors
to creep in, including:

\begin{enumerate}
\item	The measurement will include some of the overhead of
	the time measurement.
	This source of error can be reduced to an arbitrarily small
	value by increasing the number of iterations.
\item	The first few iterations of the test might incur cache misses
	or (worse yet) page faults that might inflate the measured
	value.
	This source of error can also be reduced by increasing the
	number of iterations, or it can often be eliminated entirely
	by running a few warm-up iterations before starting the
	measurement period.
	Most systems have ways of detecting whether a given process
	incurred a page fault, and you should make use of this to
	reject runs whose performance has been thus impeded.
\item	Some types of interference, for example, random memory errors,
	are so rare that they can be dealt with by running a number
	of sets of iterations of the test.
	If the level of interference was statistically significant,
	any performance outliers could be rejected statistically.
\item	Any iteration of the test might be interfered with by other
	activity on the system.
	Sources of interference include other applications, system
	utilities and daemons, device interrupts, firmware interrupts
	(including system management interrupts, or SMIs),
	virtualization, memory errors, and much else besides.
	Assuming that these sources of interference occur randomly,
	their effect can be minimized by reducing the number of
	iterations.
\item	Thermal throttling can understate scalability because increasing
	CPU activity increases heat generation, and on systems without
	adequate cooling (most of them!), this can result in the CPU
	frequency decreasing as the number of CPUs increases.\footnote{
		Systems with adequate cooling tend to look like gaming systems.}
	Of course, if you are testing an application to evaluate its
	expected behavior when run in production, such thermal throttling
	is simply a fact of life.
	Otherwise, if you are interested in theoretical scalability,
	use a system with adequate cooling or reduce the CPU clock rate
	to a level that the cooling system can handle.
\end{enumerate}

The first and fourth sources of interference provide conflicting advice,
which is one sign that we are living in the real world.

\QuickQuiz{
	But what about other sources of error, for example, due to
	interactions between caches and memory layout?
}\QuickQuizAnswer{
	Changes in memory layout can indeed result in unrealistic
	decreases in execution time.
	For example, suppose that a given microbenchmark almost
	always overflows the L0 cache's associativity, but with just the right
	memory layout, it all fits.
	If this is a real concern, consider running your microbenchmark
	using huge pages (or within the kernel or on bare metal) in
	order to completely control the memory layout.

	But note that there are many different possible memory-layout
	bottlenecks.
	Benchmarks sensitive to memory bandwidth (such as those involving
	matrix arithmetic) should spread the running threads across the
	available cores and sockets to maximize memory parallelism.
	They should also spread the data across \IXplr{NUMA node}, memory
	controllers, and DRAM chips to the extent possible.
	In contrast, benchmarks sensitive to memory latency (including
	most poorly scaling applications) should instead maximize
	locality, filling each core and socket in turn before adding
	another one.
}\QuickQuizEnd

The following sections discuss ways of dealing with these measurement
errors, with
\cref{sec:debugging:Isolation}
covering isolation techniques that may be used to prevent some forms of
interference,
and with
\cref{sec:debugging:Detecting Interference}
covering methods for detecting interference so as to reject measurement
data that might have been corrupted by that interference.

\subsection{Isolation}
\label{sec:debugging:Isolation}

The Linux kernel provides a number of ways to isolate a group of
CPUs from outside interference.

First, let's look at interference by other processes, threads, and tasks.
The POSIX \co{sched_setaffinity()} system call may be used to move
most tasks off of a given set of CPUs and to confine your tests to
that same group.
The Linux-specific user-level \co{taskset} command may be used for
the same purpose, though both \co{sched_setaffinity()} and
\co{taskset} require elevated permissions.
Linux-specific control groups (cgroups) may be used for this same purpose.
This approach can be quite effective at reducing interference, and
is sufficient in many cases.
However, it does have limitations, for example, it cannot do anything
about the per-CPU kernel threads that are often used for housekeeping
tasks.

One way to avoid interference from per-CPU kernel threads is to run
your test at a high real-time priority, for example, by using
the POSIX \co{sched_setscheduler()} system call.
However, note that if you do this, you are implicitly taking on
responsibility for avoiding infinite loops, because otherwise
your test can prevent part of the kernel from functioning.
This is an example of the Spiderman Principle:
``With great power comes great responsibility.''
And although the default real-time throttling settings often address
such problems, they might do so by causing your real-time threads
to miss their deadlines.

These approaches can greatly reduce, and perhaps even eliminate,
interference from processes, threads, and tasks.
However, it does nothing to prevent interference from device
interrupts, at least in the absence of threaded interrupts.
Linux allows some control of threaded interrupts via the
\path{/proc/irq} directory, which contains numerical directories, one
per interrupt vector.
Each numerical directory contains \co{smp_affinity} and
\co{smp_affinity_list}.
Given sufficient permissions, you can write a value to these files
to restrict interrupts to the specified set of CPUs.
For example, either
``\co{echo 3 > /proc/irq/23/smp_affinity}''
or
``\co{echo 0-1 > /proc/irq/23/smp_affinity_list}''
would confine interrupts on vector~23 to CPUs~0 and~1,
at least given sufficient privileges.
You can use ``\co{cat /proc/interrupts}'' to obtain a list of the interrupt
vectors on your system, how many are handled by each CPU, and what
devices use each interrupt vector.

Running a similar command for all interrupt vectors on your system
would confine interrupts to CPUs~0 and~1, leaving the remaining CPUs
free of interference.
Or mostly free of interference, anyway.
It turns out that the scheduling-clock interrupt fires on each CPU
that is running in user mode.\footnote{
	Frederic Weisbecker leads up a \co{NO_HZ_FULL}
	adaptive-ticks project
	that allows scheduling-clock interrupts to be disabled
	on CPUs that have only one runnable task.
	As of 2021, this is largely complete.}
In addition you must take care to ensure that the set of CPUs that you
confine the interrupts to is capable of handling the load.

But this only handles processes and interrupts running in the same
operating-system instance as the test.
Suppose that you are running the test in a guest OS that is itself
running on a hypervisor, for example, Linux running KVM\@?
Although you can in theory apply the same techniques at the hypervisor
level that you can at the guest-OS level, it is quite common for
hypervisor-level operations to be restricted to authorized personnel.
In addition, none of these techniques work against firmware-level
interference.

\QuickQuiz{
	Wouldn't the techniques suggested to isolate the code under
	test also affect that code's performance, particularly if
	it is running within a larger application?
}\QuickQuizAnswer{
	Indeed it might, although in most microbenchmarking efforts
	you would extract the code under test from the enclosing
	application.
	Nevertheless, if for some reason you must keep the code under
	test within the application, you will very likely need to use
	the techniques discussed in
	\cref{sec:debugging:Detecting Interference}.
}\QuickQuizEnd

Of course, if it is in fact the interference that is producing the
behavior of interest, you will instead need to promote interference,
in which case being unable to prevent it is not a problem.
But if you really do need interference-free measurements, then instead
of preventing the interference, you might need to detect the interference
as described in the next section.

\subsection{Detecting Interference}
\label{sec:debugging:Detecting Interference}

If you cannot prevent interference, perhaps you can detect it
and reject results from any affected test runs.
\Cref{sec:debugging:Detecting Interference Via Measurement}
describes methods of rejection involving additional measurements,
while \cref{sec:debugging:Detecting Interference Via Statistics}
describes statistics-based rejection.

\subsubsection{Detecting Interference Via Measurement}
\label{sec:debugging:Detecting Interference Via Measurement}

%	Sources of interference include other applications, system
%	utilities and daemons, device interrupts, firmware interrupts
%	(including system management interrupts, or SMIs),
%	virtualization, memory errors, and much else besides.

Many systems, including Linux, provide means for determining after the
fact whether some forms of interference have occurred.
For example, process-based interference results in context switches,
which, on Linux-based systems, are visible in
\path{/proc/<PID>/sched} via the \co{nr_switches} field.
Similarly, interrupt-based interference can be detected via the
\path{/proc/interrupts} file.

\begin{listing}
\begin{fcvlabel}[ln:debugging:Using getrusage() to Detect Context Switches]
\begin{VerbatimL}
#include <sys/time.h>
#include <sys/resource.h>

/* Return 0 if test results should be rejected. */
int runtest(void)
{
	struct rusage ru1;
	struct rusage ru2;

	if (getrusage(RUSAGE_SELF, &ru1) != 0) {
		perror("getrusage");
		abort();
	}
	/* run test here. */
	if (getrusage(RUSAGE_SELF, &ru2 != 0) {
		perror("getrusage");
		abort();
	}
	return (ru1.ru_nvcsw == ru2.ru_nvcsw &&
	        ru1.runivcsw == ru2.runivcsw);
}
\end{VerbatimL}
\end{fcvlabel}
\caption{Using \tco{getrusage()} to Detect Context Switches}
\label{lst:debugging:Using getrusage() to Detect Context Switches}
\end{listing}

Opening and reading files is not the way to low overhead, and it is
possible to get the count of context switches for a given thread
by using the \co{getrusage()} system call, as shown in
\cref{lst:debugging:Using getrusage() to Detect Context Switches}.
This same system call can be used to detect minor page faults (\co{ru_minflt})
and major page faults (\co{ru_majflt}).

Unfortunately, detecting memory errors and firmware interference is quite
system-specific, as is the detection of interference due to virtualization.
Although avoidance is better than detection, and detection is better than
statistics, there are times when one must avail oneself of statistics,
a topic addressed in the next section.

\subsubsection{Detecting Interference Via Statistics}
\label{sec:debugging:Detecting Interference Via Statistics}

Any statistical analysis will be based on assumptions about the data,
and performance microbenchmarks often support the following assumptions:

\begin{enumerate}
\item	Smaller measurements are more likely to be accurate than
	larger measurements.
\item	The measurement uncertainty of good data is known.
\item	A reasonable fraction of the test runs will result in good data.
\end{enumerate}

The fact that smaller measurements are more likely to be accurate than
larger measurements suggests that sorting the measurements in increasing
order is likely to be productive.\footnote{
	To paraphrase the old saying, ``Sort first and ask questions later.''}
The fact that the measurement uncertainty is known allows us to accept
measurements within this uncertainty of each other:
If the effects of interference are large compared to this uncertainty,
this will ease rejection of bad data.
Finally, the fact that some fraction (for example, one third) can be
assumed to be good allows us to blindly accept the first portion of the
sorted list, and this data can then be used to gain an estimate of the
natural variation of the measured data, over and above the assumed
measurement error.

The approach is to take the specified number of leading elements from the
beginning of the sorted list, and use these to estimate a typical
inter-element delta, which in turn may be multiplied by the number of
elements in the list to obtain an upper bound on permissible values.
The algorithm then repeatedly considers the next element of the list.
If it falls below the upper bound, and if the distance between
the next element and the previous element is not too much greater than
the average inter-element distance for the portion of the list accepted
thus far, then the next element is accepted and the process repeats.
Otherwise, the remainder of the list is rejected.

\begin{listing}
\input{CodeSamples/debugging/datablows=whole.fcv}
\caption{Statistical Elimination of Interference}
\label{lst:debugging:Statistical Elimination of Interference}
\end{listing}

\Cref{lst:debugging:Statistical Elimination of Interference}
shows a simple \co{sh}/\co{awk} script implementing this notion.
Input consists of an x-value followed by an arbitrarily long list of y-values,
and output consists of one line for each input line, with fields as follows:

\begin{enumerate}
\item	The x-value.
\item	The average of the selected data.
\item	The minimum of the selected data.
\item	The maximum of the selected data.
\item	The number of selected data items.
\item	The number of input data items.
\end{enumerate}

This script takes three optional arguments as follows:

\begin{description}
\item	[\lopt{divisor}\nf{:}] Number of segments to divide the list
	into, for example, a divisor of four means that the first quarter of
	the data elements will be assumed to be good.
	This defaults to three.
\item	[\lopt{relerr}\nf{:}] Relative measurement error.
	The script assumes that values that differ by less than this
	error are for all intents and purposes equal.
	This defaults to 0.01, which is equivalent to 1\,\%.
\item	[\lopt{trendbreak}\nf{:}] Ratio of inter-element spacing
	constituting a break in the trend of the data.
	For example, if the average spacing in the data accepted so far
	is 1.5, then if the trend-break ratio is 2.0, then if the next
	data value differs from the last one by more than 3.0, this
	constitutes a break in the trend.
	(Unless of course, the relative error is greater than 3.0, in
	which case the ``break'' will be ignored.)
\end{description}

\begin{fcvref}[ln:debugging:datablows:whole]
\Clnrefrange{param:b}{param:e} of
\cref{lst:debugging:Statistical Elimination of Interference}
set the default values for the parameters, and
\clnrefrange{parse:b}{parse:e} parse
any command-line overriding of these parameters.
\end{fcvref}
\begin{fcvref}[ln:debugging:datablows:whole:awk]
The \co{awk} invocation on \clnref{invoke} sets the values of the
\co{divisor}, \co{relerr}, and \co{trendbreak} variables to their
\co{sh} counterparts.
In the usual \co{awk} manner,
\clnrefrange{copy:b}{end} are executed on each input
line.
The loop spanning \clnref{copy:b,copy:e} copies
the input y-values to the
\co{d} array, which \clnref{asort} sorts into increasing order.
\Clnref{comp_i} computes the number of trustworthy y-values
by applying \co{divisor} and rounding up.

\Clnrefrange{delta}{comp_max:e} compute the \co{maxdelta}
lower bound on the upper bound of y-values.
To this end, \clnref{maxdelta} multiplies the difference in values over
the trusted region of data by the \co{divisor}, which projects the
difference in values across the trusted region across the entire
set of y-values.
However, this value might well be much smaller than the relative error,
so \clnref{maxdelta1} computes the absolute error (\co{d[i] * relerr})
and adds
that to the difference \co{delta} across the trusted portion of the data.
\Clnref{comp_max:b,comp_max:e} then compute the maximum of
these two values.

Each pass through the loop spanning \clnrefrange{add:b}{add:e}
attempts to add another
data value to the set of good data.
\Clnrefrange{chk_engh}{break} compute the trend-break delta,
with \clnref{chk_engh} disabling this
limit if we don't yet have enough values to compute a trend,
and with \clnref{mul_avr} multiplying \co{trendbreak} by the average
difference between pairs of data values in the good set.
If \clnref{chk_max} determines that the candidate data value would exceed the
lower bound on the upper bound (\co{maxdelta}) \emph{and}
that the difference between the candidate data value
and its predecessor exceeds the trend-break difference (\co{maxdiff}),
then \clnref{break} exits the loop:
We have the full good set of data.

\Clnrefrange{comp_stat:b}{comp_stat:e} then compute and print
statistics.
\end{fcvref}

\QuickQuizSeries{%
\QuickQuizB{
	This approach is just plain weird!
	Why not use means and standard deviations, like we were taught
	in our statistics classes?
}\QuickQuizAnswerB{
	Because mean and standard deviation were not designed to do this job.
	To see this, try applying mean and standard deviation to the
	following data set, given a 1\,\% relative error in measurement:

	\begin{quote}
		49,548.4 49,549.4 49,550.2 49,550.9 49,550.9 49,551.0
		49,551.5 49,552.1 49,899.0 49,899.3 49,899.7 49,899.8
		49,900.1 49,900.4 52,244.9 53,333.3 53,333.3 53,706.3
		53,706.3 54,084.5
	\end{quote}

	The problem is that mean and standard deviation do not rest on
	any sort of measurement-error assumption, and they will therefore
	see the difference between the values near 49,500 and those near
	49,900 as being statistically significant, when in fact they are
	well within the bounds of estimated measurement error.

	Of course, it is possible to create a script similar to
	that in
	\cref{lst:debugging:Statistical Elimination of Interference}
	that uses standard deviation rather than absolute difference
	to get a similar effect,
	and this is left as an exercise for the interested reader.
	Be careful to avoid divide-by-zero errors arising from strings
	of identical data values!
}\QuickQuizEndB
\QuickQuizE{
	But what if all the y-values in the trusted group of data
	are exactly zero?
	Won't that cause the script to reject any non-zero value?
}\QuickQuizAnswerE{
	Indeed it will!
	But if your performance measurements often produce a value of
	exactly zero, perhaps you need to take a closer look at your
	performance-measurement code.

	Note that many approaches based on mean and standard deviation
	will have similar problems with this sort of dataset.
}\QuickQuizEndE
}

Although statistical interference detection can be quite useful, it should
be used only as a last resort.
It is far better to avoid interference in the first place
(\cref{sec:debugging:Isolation}), or, failing that,
detecting interference via measurement
(\cref{sec:debugging:Detecting Interference Via Measurement}).

\section{Summary}
\label{sec:debugging:Summary}
%
% \epigraph{To err is human---but it feels devine.}{\emph{Mae West}}
\epigraph{To err is human!
	  Stop being human!!!}{Ed Nofziger}

Although validation never will be an exact science, much can be gained
by taking an organized approach to it, as an organized approach will
help you choose the right validation tools for your job, avoiding
situations like the one fancifully depicted in
\cref{fig:debugging:Choose Validation Methods Wisely}.

\begin{figure}
\centering
\resizebox{3in}{!}{\includegraphics{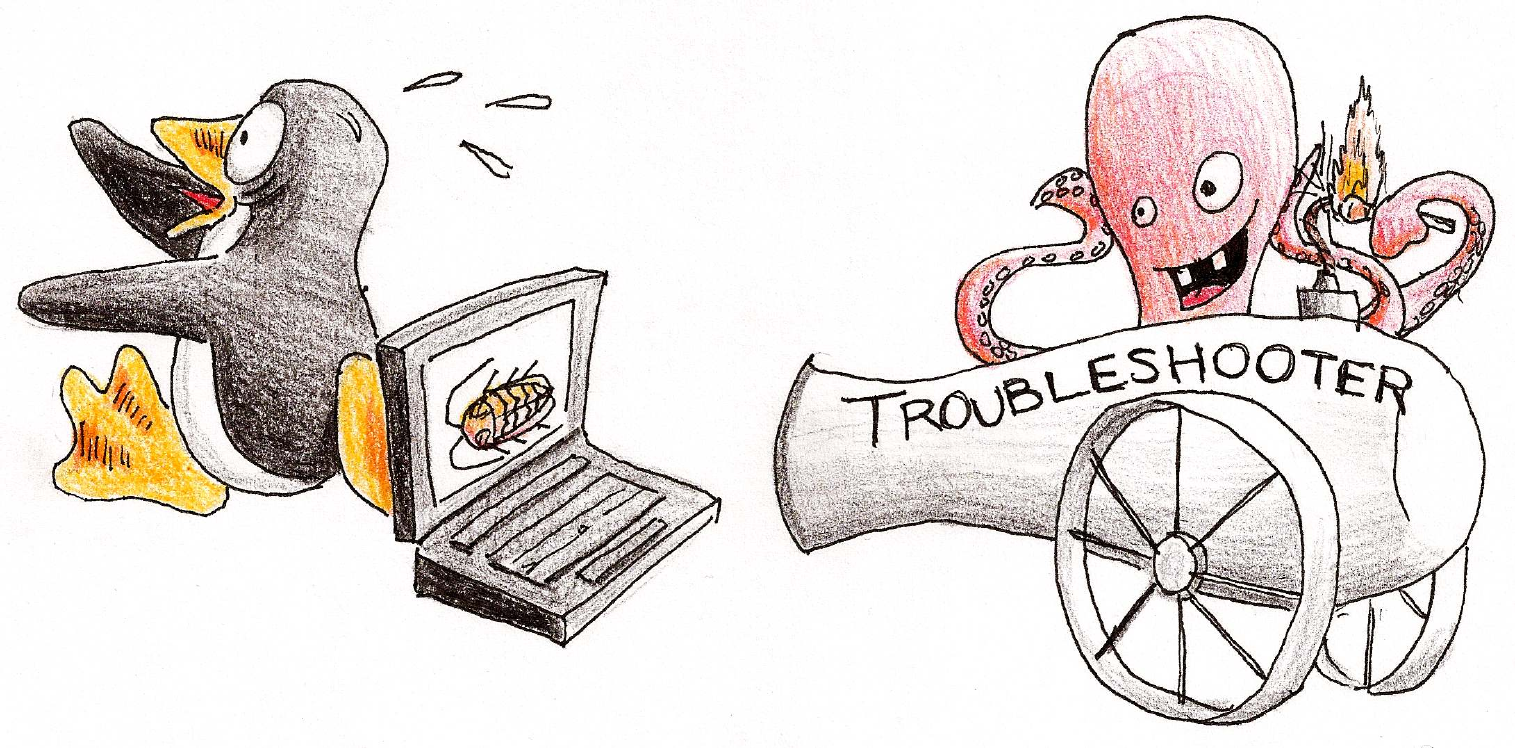}}
\caption{Choose Validation Methods Wisely}
\ContributedBy{Figure}{fig:debugging:Choose Validation Methods Wisely}{Melissa Broussard}
\end{figure}

A key choice is that of statistics.
Although the methods described in this chapter work very well most of
the time, they do have their limitations, courtesy of the Halting
Problem~\cite{AlanMTuring1937HaltingProblem,GeoffreyKPullum2000HaltingProblem}.
Fortunately for us, there is a huge number of special cases in which
we can not only work out whether a program will halt, but also
estimate how long it will run before halting, as discussed in
\cref{sec:debugging:Performance Estimation}.
Furthermore, in cases where a given program might or might not work
correctly, we can often establish estimates for what fraction of the
time it will work correctly, as discussed in
\cref{sec:debugging:Probability and Heisenbugs}.

Nevertheless, unthinking reliance on these estimates is brave to the
point of foolhardiness.
After all, we are summarizing a huge mass of complexity in code and
data structures down to a single solitary number.
Even though we can get away with such bravery a surprisingly large
fraction of the time, abstracting all that code and data away will
occasionally cause severe problems.

One possible problem is variability, where repeated runs give wildly
different results.
This problem is often addressed using standard deviation, however, using
two numbers to summarize the behavior of a large and complex program is
about as brave as using only one number.
In computer programming, the surprising thing is that use of the
mean or the mean and standard deviation are often sufficient.
Nevertheless, there are no guarantees.

One cause of variation is confounding factors.
For example, the CPU time consumed by a linked-list search will depend
on the length of the list.
Averaging together runs with wildly different list lengths will
probably not be useful, and adding a standard deviation to the mean
will not be much better.
The right thing to do would be control for list length, either by
holding the length constant or to measure CPU time as a function of
list length.

Of course, this advice assumes that you are aware of the confounding
factors, and Murphy says that you will not be.
I have been involved in projects that had confounding factors as diverse
as air conditioners (which drew considerable power at startup, thus
causing the voltage supplied to the computer to momentarily drop too
low, sometimes resulting in failure), cache state (resulting in odd
variations in performance), I/O errors (including disk errors, packet
loss, and duplicate Ethernet MAC addresses), and even porpoises (which
could not resist playing with an array of transponders, which could be
otherwise used for high-precision acoustic positioning and navigation).
And this is but one reason why a good night's sleep is such an effective
debugging tool.

In short, validation always will require some measure of the behavior of
the system.
To be at all useful, this measure must be a severe summarization of the
system, which in turn means that it can be misleading.
So as the saying goes, ``Be careful.
It is a real world out there.''

But what if you are working on the Linux kernel, which as of 2017 was
estimated to have more than 20 billion instances running throughout
the world?
In that case, a bug that occurs once every million years on a single system
will be encountered more than 50 times per day across the installed base.
A test with a 50\,\% chance of encountering this bug in a one-hour run
would need to increase that bug's probability of occurrence by more than
ten orders of magnitude, which poses a severe challenge to
today's testing methodologies.
One important tool that can sometimes be applied with good effect to
such situations is formal verification, the subject of the next chapter,
and, more speculatively, \cref{sec:future:Formal Regression Testing?}.

The topic of choosing a validation plan, be it testing, formal
verification, or both, is taken up by
\cref{sec:formal:Choosing a Validation Plan}.

	\setlength{\parskip}{0.0pt plus 1ex}
	\input{qqzdebugging}
	\setlength{\parskip}{0.0pt plus 1.0pt}% return to default

% formal/formal.tex
% mainfile: ../perfbook.tex
% SPDX-License-Identifier: CC-BY-SA-3.0

\QuickQuizChapter{chp:Formal Verification}{Formal Verification}{qqzformal}
\Epigraph{Beware of bugs in the above code; I have only proved it correct,
	  not tried it.}{Donald Knuth}

\OriginallyPublished{Chapter}{chp:Formal Verification}{Formal Verification}{Linux Weekly News}{PaulEMcKenney2007QRCUspin,PaulEMcKenney2008dynticksRCU,PaulEMcKenney2011ppcmem}

Parallel algorithms can be hard to write, and even harder to debug.
Testing, though essential, is insufficient, as fatal \IXpl{race condition}
can have extremely low probabilities of occurrence.
Proofs of correctness can be valuable, but in the end are just as
prone to human error as is the original algorithm.
In addition, a proof of correctness cannot be expected to find errors
in your assumptions, shortcomings in the requirements,
misunderstandings of the underlying software or hardware primitives,
or errors that you did not think to construct a proof for.
This means that formal methods can never replace testing.
Nevertheless, formal methods can be a valuable addition to your validation
toolbox.

It would be very helpful to have a tool that could somehow locate
all race conditions.
A number of such tools exist, for example,
\cref{sec:formal:State-Space Search} provides an
introduction to the general-purpose state-space search tools Promela and Spin,
\cref{sec:formal:Special-Purpose State-Space Search}
similarly introduces the special-purpose ppcmem tool,
\cref{sec:formal:Axiomatic Approaches}
looks at an example axiomatic approach,
\cref{sec:formal:SAT Solvers}
briefly overviews SAT solvers,
\cref{sec:formal:Stateless Model Checkers}
briefly overviews stateless model checkers,
\cref{sec:formal:Summary}
sums up use of formal-verification tools for verifying parallel algorithms,
and finally
\cref{sec:formal:Choosing a Validation Plan}
discusses how to decide how much and what type of validation to apply
to a given software project.

% formal/spinhint.html
% mainfile: ../perfbook.tex
% SPDX-License-Identifier: CC-BY-SA-3.0

\section{State-Space Search}
\label{sec:formal:State-Space Search}
\epigraph{Follow every byway / Every path you know.}
	 {\emph{Climb Every Mountain}, Rodgers \& Hammerstein}

This section features the general-purpose Promela and Spin tools,
which may be used to carry out a full
state-space search of many types of multi-threaded code.
They are used to verifying data communication protocols.
\Cref{sec:formal:Promela and Spin}
introduces Promela and Spin, including a couple of warm-up exercises
verifying both non-atomic and atomic increment.
\Cref{sec:formal:How to Use Promela}
describes use of Promela, including example command lines and a
comparison of Promela syntax to that of C\@.
\Cref{sec:formal:Promela Example: Locking}
shows how Promela may be used to verify locking,
\cref{sec:formal:Promela Example: QRCU}
uses Promela to verify an unusual implementation of RCU named ``QRCU'',
and finally
\cref{sec:formal:Promela Parable: dynticks and Preemptible RCU}
applies Promela to early versions of RCU's dyntick-idle implementation.

\subsection{Promela and Spin}
\label{sec:formal:Promela and Spin}

\IX{Promela} is a language designed to help verify protocols, but which
can also be used to verify small parallel algorithms.
You recode your algorithm and correctness constraints in the C-like
language Promela, and then use \IX{Spin} to translate it into a C program
that you can compile and run.
The resulting program carries out a full state-space search of your
algorithm, either verifying or finding counter-examples for
assertions that you can associate with in your Promela program.

This full-state search can be extremely powerful, but can also be a two-edged
sword.
If your algorithm is too complex or your Promela implementation is
careless, there might be more states than fit in memory.
Furthermore, even given sufficient memory, the state-space search might
well run for longer than the expected lifetime of the universe.
Therefore, use this tool for compact but complex parallel algorithms.
Attempts to naively apply it to even moderate-scale algorithms (let alone
the full Linux kernel) will end badly.

Promela and Spin may be downloaded from
\url{https://spinroot.com/spin/whatispin.html}.

The above site also gives links to Gerard Holzmann's excellent
book~\cite{Holzmann03a} on Promela and Spin,
as well as searchable online references starting at:
\url{https://www.spinroot.com/spin/Man/index.html}.

The remainder of this section describes how to use Promela to debug
parallel algorithms, starting with simple examples and progressing to
more complex uses.

\subsubsection{Warm-Up:
			Non-Atomic Increment}
\label{sec:formal:Warm-Up: Non-Atomic Increment}

\begin{fcvref}[ln:formal:promela:increment:whole]
\Cref{lst:formal:Promela Code for Non-Atomic Increment}
demonstrates the textbook \IX{race condition}
resulting from non-atomic increment.
\Clnref{nprocs} defines the number of processes to run (we will vary this
to see the effect on state space), \clnref{count} defines the counter,
and \clnref{prog} is used to implement the assertion that appears on
\clnrefrange{assert:b}{assert:e}.

\begin{listing}
\input{CodeSamples/formal/promela/increment=whole.fcv}
\caption{Promela Code for Non-Atomic Increment}
\label{lst:formal:Promela Code for Non-Atomic Increment}
\end{listing}

\Clnrefrange{proc:b}{proc:e} define a process that increments
the counter non-atomically.
The argument \co{me} is the process number, set by the initialization
block later in the code.
Because simple Promela statements are each assumed atomic, we must
break the increment into the two statements on
\clnrefrange{incr:b}{incr:e}.
The assignment on \clnref{setprog} marks the process's completion.
Because the Spin system will fully search the state space, including
all possible sequences of states, there is no need for the loop
that would be used for conventional stress testing.

\Clnrefrange{init:b}{init:e} are the initialization block,
which is executed first.
\Clnrefrange{doinit:b}{doinit:e} actually do the initialization,
while \clnrefrange{assert:b}{assert:e}
perform the assertion.
Both are atomic blocks in order to avoid unnecessarily increasing
the state space:
Because they are not part of the algorithm proper,
we lose no verification coverage by making them atomic.

The \co{do-od} construct on \clnrefrange{dood1:b}{dood1:e}
implements a Promela loop,
which can be thought of as a C \co{for (;;)} loop containing a
\co{switch} statement that allows expressions in case labels.
The condition blocks (prefixed by \co{::})
are scanned non-deterministically,
though in this case only one of the conditions can possibly hold at a given
time.
The first block of the \co{do-od} from
\clnrefrange{block1:b}{block1:e}
initializes the i-th
incrementer's progress cell, runs the i-th incrementer's process, and
then increments the variable \co{i}.
The second block of the \co{do-od} on
\clnref{block2} exits the loop once
these processes have been started.

The atomic block on \clnrefrange{assert:b}{assert:e} also contains
a similar \co{do-od}
loop that sums up the progress counters.
The \co{assert()} statement on \clnref{assert} verifies that
if all processes
have been completed, then all counts have been correctly recorded.
\end{fcvref}

You can build and run this program as follows:

\begin{VerbatimU}
spin -a increment.spin      # Translate the model to C
cc -DSAFETY -o pan pan.c    # Compile the model
./pan                       # Run the model
\end{VerbatimU}

\begin{listing}
\begin{Verbatim}
pan:1: assertion violated
        ((sum<2)||(counter==2)) (at depth 22)
pan: wrote increment.spin.trail

(Spin Version 6.4.8 -- 2 March 2018)
Warning: Search not completed
        + Partial Order Reduction

Full statespace search for:
        never claim             - (none specified)
        assertion violations    +
        cycle checks            - (disabled by -DSAFETY)
        invalid end states      +

State-vector 48 byte, depth reached 24, errors: 1
       45 states, stored
       13 states, matched
       58 transitions (= stored+matched)
       53 atomic steps
hash conflicts:         0 (resolved)

Stats on memory usage (in Megabytes):
    0.003       equivalent memory usage for states
                  (stored*(State-vector + overhead))
    0.290       actual memory usage for states
  128.000       memory used for hash table (-w24)
    0.534       memory used for DFS stack (-m10000)
  128.730       total actual memory usage
\end{Verbatim}
\vspace*{-9pt}
\caption{Non-Atomic Increment Spin Output}
\label{lst:formal:Non-Atomic Increment Spin Output}
\end{listing}

This will produce output as shown in
\cref{lst:formal:Non-Atomic Increment Spin Output}.
The first line tells us that our assertion was violated (as expected
given the non-atomic increment!).
The second line that a \co{trail} file was written describing how the
assertion was violated.
The ``Warning'' line reiterates that all was not well with our model.
The second paragraph describes the type of state-search being carried out,
in this case for assertion violations and invalid end states.
The third paragraph gives state-size statistics:
This small model had only 45 states.
The final line shows memory usage.

The \co{trail} file may be rendered human-readable as follows:

\begin{VerbatimU}
spin -t -p increment.spin
\end{VerbatimU}

\begin{listing*}
\ebresizeverb{.9}{
\begin{Verbatim}
using statement merging
  1:    proc  0 (:init::1) increment.spin:21 (state 1)  [i = 0]
  2:    proc  0 (:init::1) increment.spin:23 (state 2)  [((i<2))]
  2:    proc  0 (:init::1) increment.spin:24 (state 3)  [progress[i] = 0]
Starting incrementer with pid 1
  3:    proc  0 (:init::1) increment.spin:25 (state 4)  [(run incrementer(i))]
  4:    proc  0 (:init::1) increment.spin:26 (state 5)  [i = (i+1)]
  5:    proc  0 (:init::1) increment.spin:23 (state 2)  [((i<2))]
  5:    proc  0 (:init::1) increment.spin:24 (state 3)  [progress[i] = 0]
Starting incrementer with pid 2
  6:    proc  0 (:init::1) increment.spin:25 (state 4)  [(run incrementer(i))]
  7:    proc  0 (:init::1) increment.spin:26 (state 5)  [i = (i+1)]
  8:    proc  0 (:init::1) increment.spin:27 (state 6)  [((i>=2))]
  9:    proc  0 (:init::1) increment.spin:22 (state 10) [break]
 10:    proc  2 (incrementer:1) increment.spin:11 (state 1)     [temp = counter]
 11:    proc  1 (incrementer:1) increment.spin:11 (state 1)     [temp = counter]
 12:    proc  2 (incrementer:1) increment.spin:12 (state 2)     [counter = (temp+1)]
 13:    proc  2 (incrementer:1) increment.spin:13 (state 3)     [progress[me] = 1]
 14: proc 2 terminates
 15:    proc  1 (incrementer:1) increment.spin:12 (state 2)     [counter = (temp+1)]
 16:    proc  1 (incrementer:1) increment.spin:13 (state 3)     [progress[me] = 1]
 17: proc 1 terminates
 18:    proc  0 (:init::1) increment.spin:31 (state 12) [i = 0]
 18:    proc  0 (:init::1) increment.spin:32 (state 13) [sum = 0]
 19:    proc  0 (:init::1) increment.spin:34 (state 14) [((i<2))]
 19:    proc  0 (:init::1) increment.spin:35 (state 15) [sum = (sum+progress[i])]
 19:    proc  0 (:init::1) increment.spin:36 (state 16) [i = (i+1)]
 20:    proc  0 (:init::1) increment.spin:34 (state 14) [((i<2))]
 20:    proc  0 (:init::1) increment.spin:35 (state 15) [sum = (sum+progress[i])]
 20:    proc  0 (:init::1) increment.spin:36 (state 16) [i = (i+1)]
 21:    proc  0 (:init::1) increment.spin:37 (state 17) [((i>=2))]
 22:    proc  0 (:init::1) increment.spin:33 (state 21) [break]
spin: increment.spin:39, Error: assertion violated
spin: text of failed assertion: assert(((sum<2)||(counter==2)))
 23:    proc  0 (:init::1) increment.spin:39 (state 22) [assert(((sum<2)||(counter==2)))]
spin: trail ends after 23 steps
#processes: 1
                counter = 1
                progress[0] = 1
                progress[1] = 1
 23:    proc  0 (:init::1) increment.spin:41 (state 24) <valid end state>
3 processes created
\end{Verbatim}}
\IfEbookSize{\vspace*{7pt}}{\vspace*{-9pt}}
\caption{Non-Atomic Increment Error Trail}
\label{lst:formal:Non-Atomic Increment Error Trail}
\end{listing*}

This gives the output shown in
\cref{lst:formal:Non-Atomic Increment Error Trail}.
As can be seen, the first portion of the init block created both
incrementer processes, both of which first fetched the counter,
then both incremented and stored it, losing a count.
The assertion then triggered, after which the global state is displayed.

\subsubsection{Warm-Up:
			Atomic Increment}
\label{sec:formal:Warm-Up: Atomic Increment}

It is easy to fix this example by placing the body of the incrementer
processes in an atomic block as shown in
\cref{lst:formal:Promela Code for Atomic Increment}.
One could also have simply replaced the pair of statements with
\co{counter = counter + 1}, because Promela statements are
atomic.
Either way, running this modified model gives us an error-free traversal
of the state space, as shown in
\cref{lst:formal:Atomic Increment Spin Output}.

\begin{listing}
\input{CodeSamples/formal/promela/atomicincrement=incrementer.fcv}
\caption{Promela Code for Atomic Increment}
\label{lst:formal:Promela Code for Atomic Increment}
\end{listing}

\begin{listing}
\begin{Verbatim}
(Spin Version 6.4.8 -- 2 March 2018)
        + Partial Order Reduction

Full statespace search for:
        never claim             - (none specified)
        assertion violations    +
        cycle checks            - (disabled by -DSAFETY)
        invalid end states      +

State-vector 48 byte, depth reached 22, errors: 0
       52 states, stored
       21 states, matched
       73 transitions (= stored+matched)
       68 atomic steps
hash conflicts:         0 (resolved)

Stats on memory usage (in Megabytes):
    0.004       equivalent memory usage for states
                  (stored*(State-vector + overhead))
    0.290       actual memory usage for states
  128.000       memory used for hash table (-w24)
    0.534       memory used for DFS stack (-m10000)
  128.730       total actual memory usage

unreached in proctype incrementer
        (0 of 5 states)
unreached in init
        (0 of 24 states)
\end{Verbatim}
\vspace*{-9pt}
\caption{Atomic Increment Spin Output}
\label{lst:formal:Atomic Increment Spin Output}
\end{listing}

\Cref{tab:advsync:Memory Usage of Increment Model}
shows the number of states and memory consumed
as a function of number of incrementers modeled
(by redefining \co{NUMPROCS}):

\begin{table}
\rowcolors{1}{}{lightgray}
\small
\renewcommand*{\arraystretch}{1.2}
\centering
\begin{tabular}{S[table-format = 1.0]S[table-format = 7.0]S[table-format = 3.1]}
	\toprule
	\multicolumn{1}{l}{\# incrementers} &
		\multicolumn{1}{r}{\# states} &
			\multicolumn{1}{r}{total memory usage (MB)} \\
	\midrule
	1 &		        11 &        128.7 \\
	2 &		        52 &        128.7 \\
	3 &		       372 &        128.7 \\
	4 &		     3 496 &        128.9 \\
	5 &		    40 221 &        131.7 \\
	6 &		   545 720 &        174.0 \\
	7 &		 8 521 446 &        881.9 \\
	\bottomrule
\end{tabular}
\caption{Memory Usage of Increment Model}
\label{tab:advsync:Memory Usage of Increment Model}
\end{table}

Running unnecessarily large models is thus subtly discouraged, although
882\,MB is well within the limits of modern desktop and laptop machines.

With this example under our belt, let's take a closer look at the
commands used to analyze Promela models and then look at more
elaborate examples.

\subsection{How to Use Promela}
\label{sec:formal:How to Use Promela}

Given a source file \path{qrcu.spin}, one can use the following commands:

\begin{description}[style=nextline]
\item	[\tco{spin -a qrcu.spin}]
	Create a file \path{pan.c} that fully searches the state machine.
\item	[\tco{cc -DSAFETY [-DCOLLAPSE] [-DMA=N] -o pan pan.c}]
	Compile the generated state-machine search.
	The \co{-DSAFETY} generates optimizations that are appropriate
	if you have only assertions (and perhaps \co{never} statements).
	If you have liveness, fairness, or forward-progress checks,
	you may need to compile without \co{-DSAFETY}.
	If you leave off \co{-DSAFETY} when you could have used it,
	the program will let you know.

	The optimizations produced by \co{-DSAFETY} greatly speed things
	up, so you should use it when you can.
	An example situation where you cannot use \co{-DSAFETY} is
	when checking for \IXpl{livelock} (AKA ``non-progress cycles'')
	via \co{-DNP}.

	The optional \co{-DCOLLAPSE} generates code for a state vector
	compression mode.

	Another optional flag \co{-DMA=N} generates code for a slow
	but aggressive state-space memory compression mode.
\item	[\tco{./pan [-mN] [-wN]}]
	This actually searches the state space.
	The number of states can reach into the tens of millions with
	very small state machines, so you will need a machine with
	large memory.
	For example, \path{qrcu.spin} with 3~updaters and 2~readers required
	10.5\,GB of memory even with the \co{-DCOLLAPSE} flag.

	If you see a message from \co{./pan} saying:
	``\co{error: max search depth too small}'', you need to increase
	the maximum depth by a \co{-mN} option for a complete search.
	The default is \co{-m10000}.

	The \co{-wN} option specifies the hashtable size.
	The default for full state-space search is \co{-w24}.\footnote{
		As of Spin Version 6.4.6 and 6.4.8.
		In the online manual of Spin dated 10 July 2011, the
		default for exhaustive search mode is said to be \co{-w19},
		which does not meet the actual behavior.}

	If you aren't sure whether your machine has enough memory,
	run \co{top} in one window and \co{./pan} in another.
	Keep the focus on the \co{./pan} window so that you can quickly
	kill execution if need be.
	As soon as CPU time drops much below 100\,\%, kill \co{./pan}.
	If you have removed focus from the window running \co{./pan},
	you may wait a long time for the windowing system to grab
	enough memory to do anything for you.

	Another option to avoid memory exhaustion is the
	\co{-DMEMLIM=N} compiler flag.
	\co{-DMEMLIM=2000} would set the maximum of 2\,GB.

	Don't forget to capture the output, especially
	if you are working on a remote machine.

	If your model includes forward-progress checks, you will likely
	need to enable ``weak fairness'' via the \co{-f} command-line
	argument to \co{./pan}.
	If your forward-progress checks involve \co{accept} labels,
	you will also need the \co{-a} argument.
	% forward reference to model: formal.2009.02.19a in
	% /home/linux/git/userspace-rcu/formal-model.
\item	[\tco{spin -t -p qrcu.spin}]
	Given \co{trail} file output by a run that encountered an
	error, output the sequence of steps leading to that error.
	The \co{-g} flag will also include the values of changed
	global variables, and the  \co{-l} flag will also include
	the values of changed local variables.
\end{description}

\subsubsection{Promela Peculiarities}
\label{sec:formal:Promela Peculiarities}

Although all computer languages have underlying similarities,
Promela will provide some surprises to people used to coding in C,
C++, or Java.

\begin{enumerate}
\item	In C, \qco{;} terminates statements.
	In Promela it separates them.
	Fortunately, more recent versions of Spin have become
	much more forgiving of ``extra'' semicolons.
\item	Promela's looping construct, the \co{do} statement, takes
	conditions.
	This \co{do} statement closely resembles a looping if-then-else
	statement.
\item	In C's \co{switch} statement, if there is no matching case, the whole
	statement is skipped.
	In Promela's equivalent, confusingly called \co{if}, if there is
	no matching guard expression, you get an error without a
	recognizable corresponding error message.
	So, if the error output indicates an innocent line of code,
	check to see if you left out a condition from an \co{if} or \co{do}
	statement.
\item	When creating stress tests in C, one usually races suspect operations
	against each other repeatedly.
	In Promela, one instead sets up a single race, because Promela
	will search out all the possible outcomes from that single race.
	Sometimes you do need to loop in Promela, for example,
	if multiple operations overlap, but
	doing so greatly increases the size of your state space.
\item	In C, the easiest thing to do is to maintain a loop counter to track
	progress and terminate the loop.
	In Promela, loop counters must be avoided like the plague
	because they cause the state space to explode.
	On the other hand, there is no penalty for infinite loops in
	Promela as long as none of the variables monotonically increase
	or decrease---Promela will figure out how many passes through
	the loop really matter, and automatically prune execution beyond
	that point.
\item	In C torture-test code, it is often wise to keep per-task control
	variables.
	They are cheap to read, and greatly aid in debugging the test code.
	In Promela, per-task control variables should be used only when
	there is no other alternative.
	To see this, consider a 5-task verification with one bit each
	to indicate completion.
	This gives 32 states.
	In contrast, a simple counter would have only six states,
	more than a five-fold reduction.
	That factor of five might not seem like a problem, at least
	not until you are struggling with a verification program
	possessing more than 150 million states consuming more
	than 10\,GB of memory!
\item	One of the most challenging things both in C torture-test code and
	in Promela is formulating good assertions.
	Promela also allows \co{never} claims that act like an assertion
	replicated between every line of code.
\item	Dividing and conquering is extremely helpful in Promela in keeping
	the state space under control.
	Splitting a large model into two roughly equal halves will result
	in the state space of each half being roughly the square root of
	the whole.
	For example, a million-state combined model might reduce to a
	pair of thousand-state models.
	Not only will Promela handle the two smaller models much more
	quickly with much less memory, but the two smaller algorithms
	are easier for people to understand.
\end{enumerate}

\subsubsection{Promela Coding Tricks}
\label{sec:formal:Promela Coding Tricks}

Promela was designed to analyze protocols, so using it on parallel programs
is a bit abusive.
The following tricks can help you to abuse Promela safely:

\begin{enumerate}
\item	Memory reordering.
	Suppose you have a pair of statements copying globals \co{x} and \co{y}
	to locals \co{r1} and \co{r2}, where ordering matters (e.g., unprotected
	by locks), but where you have no \IXpl{memory barrier}.
	This can be modeled in Promela as follows:

\begin{VerbatimN}[samepage=true]
if
:: 1 -> r1 = x;
        r2 = y
:: 1 -> r2 = y;
        r1 = x
fi
\end{VerbatimN}

	The two branches of the \co{if} statement will be selected
	nondeterministically, since they both are available.
	Because the full state space is searched, \emph{both} choices
	will eventually be made in all cases.

	Of course, this trick will cause your state space to explode
	if used too heavily.
	In addition, it requires you to anticipate possible reorderings.

\item	State reduction.
	If you have complex assertions, evaluate them under \co{atomic}.
	After all, they are not part of the algorithm.
	One example of a complex assertion (to be discussed in more
	detail later) is as shown in
	\cref{lst:formal:Complex Promela Assertion}.

	There is no reason to evaluate this assertion
	non-atomically, since it is not actually part of the algorithm.
	Because each statement contributes to state, we can reduce
	the number of useless states by enclosing it in an \co{atomic}
	block as shown in
	\cref{lst:formal:Atomic Block for Complex Promela Assertion}.

\item	Promela does not provide functions.
	You must instead use C preprocessor macros.
	However, you must use them carefully in order to avoid
	combinatorial explosion.
\end{enumerate}

\begin{listing}
\begin{VerbatimL}
i = 0;
sum = 0;
do
:: i < N_QRCU_READERS ->
	sum = sum + (readerstart[i] == 1 &&
	             readerprogress[i] == 1);
	i++
:: i >= N_QRCU_READERS ->
	assert(sum == 0);
	break
od
\end{VerbatimL}
\caption{Complex Promela Assertion}
\label{lst:formal:Complex Promela Assertion}
\end{listing}

\begin{listing}
\begin{VerbatimL}
atomic {
	i = 0;
	sum = 0;
	do
	:: i < N_QRCU_READERS ->
		sum = sum + (readerstart[i] == 1 &&
		             readerprogress[i] == 1);
		i++
	:: i >= N_QRCU_READERS ->
		assert(sum == 0);
		break
	od
}
\end{VerbatimL}
\caption{Atomic Block for Complex Promela Assertion}
\label{lst:formal:Atomic Block for Complex Promela Assertion}
\end{listing}

Now we are ready for further examples.

\subsection{Promela Example:
			     Locking}
\label{sec:formal:Promela Example: Locking}

\begin{fcvref}[ln:formal:promela:lock:whole]
Since locks are generally useful, \co{spin_lock()} and
\co{spin_unlock()}
macros are provided in \path{lock.h}, which may be included from
multiple Promela models, as shown in
\cref{lst:formal:Promela Code for Spinlock}.
The \co{spin_lock()} macro contains an infinite \co{do-od} loop
spanning \clnrefrange{dood:b}{dood:e},
courtesy of the single guard expression of ``1'' on \clnref{one}.
The body of this loop is a single atomic block that contains
an \co{if-fi} statement.
The \co{if-fi} construct is similar to the \co{do-od} construct, except
that it takes a single pass rather than looping.
If the lock is not held on \clnref{notheld}, then
\clnref{acq} acquires it and
\clnref{break} breaks out of the enclosing \co{do-od} loop (and also exits
the atomic block).
On the other hand, if the lock is already held on \clnref{held},
we do nothing (\co{skip}), and fall out of the \co{if-fi} and the
atomic block so as to take another pass through the outer
loop, repeating until the lock is available.
\end{fcvref}

\begin{listing}
\input{CodeSamples/formal/promela/lock=whole.fcv}
\caption{Promela Code for Spinlock}
\label{lst:formal:Promela Code for Spinlock}
\end{listing}

The \co{spin_unlock()} macro simply marks the lock as no
longer held.

Note that memory barriers are not needed because Promela assumes
full ordering.
In any given Promela state, all processes agree on both the current
state and the order of state changes that caused us to arrive at
the current state.
This is analogous to the ``sequentially consistent'' memory model
used by a few computer systems (such as 1990s MIPS and PA-RISC\@).
As noted earlier, and as will be seen in a later example,
weak memory ordering must be explicitly coded.

\begin{listing}
\input{CodeSamples/formal/promela/lock=spin.fcv}
\caption{Promela Code to Test Spinlocks}
\label{lst:formal:Promela Code to Test Spinlocks}
\end{listing}

\begin{fcvref}[ln:formal:promela:lock:spin]
These macros are tested by the Promela code shown in
\cref{lst:formal:Promela Code to Test Spinlocks}.
This code is similar to that used to test the increments,
with the number of locking processes defined by the \co{N_LOCKERS}
macro definition on \clnref{nlockers}.
The mutex itself is defined on \clnref{mutex},
an array to track the lock owner
on \clnref{array}, and \clnref{sum} is used by assertion
code to verify that only one process holds the lock.
\end{fcvref}

\begin{fcvref}[ln:formal:promela:lock:spin:locker]
The locker process is on \clnrefrange{b}{e}, and simply loops forever
acquiring the lock on \clnref{lock}, claiming it on \clnref{claim},
unclaiming it on \clnref{unclaim}, and releasing it on \clnref{unlock}.
\end{fcvref}

\begin{fcvref}[ln:formal:promela:lock:spin:init]
The init block on \clnrefrange{b}{e} initializes the current locker's
\co{havelock} array entry on \clnref{array}, starts the current locker on
\clnref{start}, and advances to the next locker on \clnref{next}.
Once all locker processes are spawned, the \co{do-od} loop
moves to \clnref{chkassert}, which checks the assertion.
\Clnref{sum,j} initialize the control variables,
\clnrefrange{atm:b}{atm:e} atomically sum the \co{havelock} array entries,
\clnref{assert} is the assertion, and \clnref{break} exits the loop.
\end{fcvref}

We can run this model by placing the two code fragments of
\cref{lst:formal:Promela Code for Spinlock,%
lst:formal:Promela Code to Test Spinlocks} into
files named \path{lock.h} and \path{lock.spin}, respectively, and then running
the following commands:

\begin{VerbatimU}
spin -a lock.spin
cc -DSAFETY -o pan pan.c
./pan
\end{VerbatimU}

\begin{listing}
\begin{Verbatim}
(Spin Version 6.4.8 -- 2 March 2018)
        + Partial Order Reduction

Full statespace search for:
        never claim             - (none specified)
        assertion violations    +
        cycle checks            - (disabled by -DSAFETY)
        invalid end states      +

State-vector 52 byte, depth reached 360, errors: 0
      576 states, stored
      929 states, matched
     1505 transitions (= stored+matched)
      368 atomic steps
hash conflicts:         0 (resolved)

Stats on memory usage (in Megabytes):
    0.044       equivalent memory usage for states
                  (stored*(State-vector + overhead))
    0.288       actual memory usage for states
  128.000       memory used for hash table (-w24)
    0.534       memory used for DFS stack (-m10000)
  128.730       total actual memory usage

unreached in proctype locker
        lock.spin:19, state 20, "-end-"
        (1 of 20 states)
unreached in init
        (0 of 22 states)
\end{Verbatim}
\vspace*{-9pt}
\caption{Output for Spinlock Test}
\label{lst:formal:Output for Spinlock Test}
\end{listing}

The output will look something like that shown in
\cref{lst:formal:Output for Spinlock Test}.
As expected, this run has no assertion failures (\qco{errors: 0}).

\QuickQuizSeries{%
\QuickQuizB{
	Why is there an unreached statement in locker?
	After all, isn't this a \emph{full} state-space
	search?
}\QuickQuizAnswerB{
	The locker process is an infinite loop, so control
	never reaches the end of this process.
	However, since there are no monotonically increasing variables,
	Promela is able to model this infinite loop with a small
	number of states.
}\QuickQuizEndB
\QuickQuizE{
	What are some Promela code-style issues with this example?
}\QuickQuizAnswerE{
	There are several:
	\begin{enumerate}
	\item	The declaration of \co{sum} should be moved to within
		the init block, since it is not used anywhere else.
	\item	The assertion code should be moved outside of the
		initialization loop.
		The initialization loop can then be placed in an atomic
		block, greatly reducing the state space (by how much?).
	\item	The atomic block covering the assertion code should
		be extended to include the initialization of \co{sum}
		and \co{j}, and also to cover the assertion.
		This also reduces the state space (again, by how
		much?).
	\end{enumerate}
}\QuickQuizEndE
}

\subsection{Promela Example:
			     QRCU}
\label{sec:formal:Promela Example: QRCU}

This final example demonstrates a real-world use of Promela on Oleg
Nesterov's
QRCU~\cite{OlegNesterov2006QRCU,OlegNesterov2006aQRCU},
but modified to speed up the \co{synchronize_qrcu()}
fastpath.

But first, what is QRCU?

QRCU is a variant of SRCU~\cite{PaulEMcKenney2006c}
that trades somewhat higher read overhead
(atomic increment and decrement on a global variable) for extremely
low grace-period latencies.
If there are no readers, the grace period will be detected in less
than a microsecond, compared to the multi-millisecond grace-period
latencies of most other RCU implementations.

\begin{enumerate}
\item	There is a \co{qrcu_struct} that defines a QRCU domain.
	Like SRCU (and unlike other variants of RCU) QRCU's action
	is not global, but instead focused on the specified
	\co{qrcu_struct}.
\item	There are \co{qrcu_read_lock()} and \co{qrcu_read_unlock()}
	primitives that delimit QRCU read-side critical sections.
	The corresponding \co{qrcu_struct} must be passed into
	these primitives, and the return value from \co{qrcu_read_lock()}
	must be passed to \co{qrcu_read_unlock()}.

	For example:

\begin{VerbatimU}
idx = qrcu_read_lock(&my_qrcu_struct);
/* read-side critical section. */
qrcu_read_unlock(&my_qrcu_struct, idx);
\end{VerbatimU}

\item	There is a \co{synchronize_qrcu()} primitive that blocks until
	all pre-existing QRCU read-side critical sections complete,
	but, like SRCU's \co{synchronize_srcu()}, QRCU's
	\co{synchronize_qrcu()} need wait only for those read-side
	critical sections that are using the same \co{qrcu_struct}.

	For example, \co{synchronize_qrcu(&your_qrcu_struct)}
	would \emph{not} need to wait on the earlier QRCU read-side
	critical section.
	In contrast, \co{synchronize_qrcu(&my_qrcu_struct)}
	\emph{would} need to wait, since it shares the same
	\co{qrcu_struct}.
\end{enumerate}

A Linux-kernel patch for QRCU has been
produced~\cite{PaulMcKenney2007QRCUpatch},
but is unlikely to ever be included in the Linux kernel.

\begin{listing}
\input{CodeSamples/formal/promela/qrcu=gvar.fcv}
\caption{QRCU Global Variables}
\label{lst:formal:QRCU Global Variables}
\end{listing}

Returning to the Promela code for QRCU, the global variables are as shown in
\cref{lst:formal:QRCU Global Variables}.
This example uses locking and includes \path{lock.h}.
Both the number of readers and writers can be varied using the
two \co{#define} statements, giving us not one but two ways to create
combinatorial explosion.
The \co{idx} variable controls which of the two elements of the \co{ctr}
array will be used by readers, and the \co{readerprogress} variable
allows an assertion to determine when all the readers are finished
(since a QRCU update cannot be permitted to complete until all
pre-existing readers have completed their QRCU read-side critical
sections).
The \co{readerprogress} array elements have values as follows,
indicating the state of the corresponding reader:

\begin{enumerate}[label={\arabic*}:,start=0,itemsep=0pt]
\item	Not yet started.
\item	Within QRCU read-side critical section.
\item	Finished with QRCU read-side critical section.
\end{enumerate}

Finally, the \co{mutex} variable is used to serialize updaters' slowpaths.

\begin{listing}
\input{CodeSamples/formal/promela/qrcu=reader.fcv}
\caption{QRCU Reader Process}
\label{lst:formal:QRCU Reader Process}
\end{listing}

\begin{fcvref}[ln:formal:promela:qrcu:reader]
QRCU readers are modeled by the \co{qrcu_reader()} process shown in
\cref{lst:formal:QRCU Reader Process}.
A \co{do-od} loop spans \clnrefrange{do}{od},
with a single guard of ``1''
on \clnref{one} that makes it an infinite loop.
\Clnref{curidx} captures the current value of the global index,
and \clnrefrange{atm:b}{atm:e}
atomically increment it (and break from the infinite loop)
if its value was non-zero (\co{atomic_inc_not_zero()}).
\Clnref{cs:entry} marks entry into the RCU read-side critical section, and
\clnref{cs:exit} marks exit from this critical section,
both lines for the benefit of
the \co{assert()} statement that we shall encounter later.
\Clnref{atm:dec} atomically decrements the same counter that we incremented,
thereby exiting the RCU read-side critical section.
\end{fcvref}

\begin{listing}
\input{CodeSamples/formal/promela/qrcu=sum_unordered.fcv}
\caption{QRCU Unordered Summation}
\label{lst:formal:QRCU Unordered Summation}
\end{listing}

\begin{fcvref}[ln:formal:promela:qrcu:sum_unordered]
The C-preprocessor macro shown in
\cref{lst:formal:QRCU Unordered Summation}
sums the pair of counters so as to emulate weak memory ordering.
\Clnrefrange{fetch:b}{fetch:e} fetch one of the counters,
and \clnref{sum_other} fetches the other
of the pair and sums them.
The atomic block consists of a single \co{do-od} statement.
This \co{do-od} statement (spanning \clnrefrange{do}{od}) is unusual in that
it contains two unconditional
branches with guards on \clnref{g1,g2}, which causes Promela to
non-deterministically choose one of the two (but again, the full
state-space search causes Promela to eventually make all possible
choices in each applicable situation).
The first branch fetches the zero-th counter and sets \co{i} to 1 (so
that \clnref{sum_other} will fetch the first counter), while the second
branch does the opposite, fetching the first counter and setting \co{i}
to 0 (so that \clnref{sum_other} will fetch the second counter).
\end{fcvref}

\QuickQuiz{
	Is there a more straightforward way to code the \co{do-od} statement?
}\QuickQuizAnswer{
	Yes.
	Replace it with \co{if-fi} and remove the two \co{break} statements.
}\QuickQuizEnd

\begin{listing}
\ebresizeverb{.88}{\input{CodeSamples/formal/promela/qrcu=updater.fcv}}
\caption{QRCU Updater Process}
\label{lst:formal:QRCU Updater Process}
\end{listing}

\begin{fcvref}[ln:formal:promela:qrcu:updater]
With the \co{sum_unordered} macro in place, we can now proceed
to the update-side process shown in
\cref{lst:formal:QRCU Updater Process}.
The update-side process repeats indefinitely, with the corresponding
\co{do-od} loop ranging over \clnrefrange{do}{od}.
Each pass through the loop first snapshots the global \co{readerprogress}
array into the local \co{readerstart} array on
\clnrefrange{atm1:b}{atm1:e}.
This snapshot will be used for the assertion on \clnref{assert}.
\Clnref{sum_unord} invokes \co{sum_unordered}, and then
\clnrefrange{reinvoke:b}{reinvoke:e}
re-invoke \co{sum_unordered} if the fastpath is potentially
usable.

\Clnrefrange{slow:b}{slow:e} execute the slowpath code if need be, with
\clnref{acq,rel} acquiring and releasing the update-side lock,
\clnrefrange{flip_idx:b}{flip_idx:e} flipping the index, and
\clnrefrange{wait:b}{wait:e} waiting for
all pre-existing readers to complete.

\Clnrefrange{atm2:b}{atm2:e} then compare the current values
in the \co{readerprogress}
array to those collected in the \co{readerstart} array,
forcing an assertion failure should any readers that started before
this update still be in progress.
\end{fcvref}

\QuickQuizSeries{%
\QuickQuizB{
	\begin{fcvref}[ln:formal:promela:qrcu:updater]
	Why are there atomic blocks at \clnrefrange{atm1:b}{atm1:e}
	and \clnrefrange{atm2:b}{atm2:e}, when the operations
	within those atomic
	blocks have no atomic implementation on any current
	production microprocessor?
	\end{fcvref}
}\QuickQuizAnswerB{
	Because those operations are for the benefit of the
	assertion only.
	They are not part of the algorithm itself.
	There is therefore no harm in marking them atomic, and
	so marking them greatly reduces the state space that must
	be searched by the Promela model.
}\QuickQuizEndB
\QuickQuizE{
	\begin{fcvref}[ln:formal:promela:qrcu:updater]
	Is the re-summing of the counters on
	\clnrefrange{reinvoke:b}{reinvoke:e}
	\emph{really} necessary?
	\end{fcvref}
}\QuickQuizAnswerE{
	Yes.
	To see this, delete these lines and run the model.

	Alternatively, consider the following sequence of steps:

	\begin{enumerate}
	\item	One process is within its RCU read-side critical
		section, so that the value of \co{ctr[0]} is zero and
		the value of \co{ctr[1]} is two.
	\item	An updater starts executing, and sees that the sum of
		the counters is two so that the fastpath cannot be
		executed.
		It therefore acquires the lock.
	\item	A second updater starts executing, and fetches the value
		of \co{ctr[0]}, which is zero.
	\item	The first updater adds one to \co{ctr[0]}, flips
		the index (which now becomes zero), then subtracts
		one from \co{ctr[1]} (which now becomes one).
	\item	The second updater fetches the value of \co{ctr[1]},
		which is now one.
	\item	The second updater now incorrectly concludes that it
		is safe to proceed on the fastpath, despite the fact
		that the original reader has not yet completed.
	\end{enumerate}
}\QuickQuizEndE
}

\begin{listing}
\input{CodeSamples/formal/promela/qrcu=init.fcv}
\caption{QRCU Initialization Process}
\label{lst:formal:QRCU Initialization Process}
\end{listing}

\begin{fcvref}[ln:formal:promela:qrcu:init]
All that remains is the initialization block shown in
\cref{lst:formal:QRCU Initialization Process}.
This block simply initializes the counter pair on
\clnrefrange{i_ctr:b}{i_ctr:e},
spawns the reader processes on
\clnrefrange{spn_r:b}{spn_r:e}, and spawns the updater
processes on \clnrefrange{spn_u:b}{spn_u:e}.
This is all done within an atomic block to reduce state space.
\end{fcvref}

\subsubsection{Running the QRCU Example}
\label{sec:formal:Running the QRCU Example}

To run the QRCU example, combine the code fragments in the previous
section into a single file named \path{qrcu.spin}, and place the definitions
for \co{spin_lock()} and \co{spin_unlock()} into a file named
\path{lock.h}.
Then use the following commands to build and run the QRCU model:

\begin{VerbatimU}
spin -a qrcu.spin
cc -DSAFETY [-DCOLLAPSE] -o pan pan.c
./pan [-mN]
\end{VerbatimU}

\begin{table}
\centering
\begin{threeparttable}
\rowcolors{1}{}{lightgray}
\renewcommand*{\arraystretch}{1.2}
\footnotesize
\begin{tabular}{S[table-format = 1.0]S[table-format = 1.0]S[table-format = 9.0]
		S[table-format = 6.0]S[table-format = 5.1]}
	\toprule
	\multicolumn{1}{r}{updaters} &
	    \multicolumn{1}{r}{readers} &
		\multicolumn{1}{r}{\# states} &
		    \multicolumn{1}{r}{depth} &
			\multicolumn{1}{r}{memory (MB)\tnote{a}} \\
	\midrule
	1 & 1 &         376 &      95 &    128.7 \\
	1 & 2 &       6 177 &     218 &    128.9 \\
	1 & 3 &      99 728 &     385 &    132.6 \\
	2 & 1 &      29 399 &     859 &    129.8 \\
	2 & 2 &   1 071 181 &   2 352 &    169.6 \\
	2 & 3 &  33 866 736 &  12 857 &  1 540.8 \\
	3 & 1 &   2 749 453 &  53 809 &    236.6 \\
	3 & 2 & 186 202 860 & 328 014 & 10 483.7 \\
	\bottomrule
\end{tabular}
\begin{tablenotes}
	\item [a] Obtained with the compiler flag \co{-DCOLLAPSE}
		specified.
\end{tablenotes}
\end{threeparttable}
\caption{Memory Usage of QRCU Model}
\label{tab:advsync:Memory Usage of QRCU Model}
\end{table}

The output shows that this model passes all of the cases shown in
\cref{tab:advsync:Memory Usage of QRCU Model}.
It would be nice to run three readers and three
updaters, however, simple extrapolation indicates that this will
require about half a terabyte of memory.
What to do?

It turns out that \co{./pan} gives advice when it runs out of memory,
for example, when attempting to run three readers and three updaters:

\begin{VerbatimU}
hint: to reduce memory, recompile with
  -DCOLLAPSE # good, fast compression, or
  -DMA=96   # better/slower compression, or
  -DHC # hash-compaction, approximation
  -DBITSTATE # supertrace, approximation
\end{VerbatimU}

Let's try the suggested compiler flag \co{-DMA=N},
which generates code for aggressive compression of the
state space at the cost of greatly increased search overhead.
The required commands are as follows:

\begin{VerbatimU}
spin -a qrcu.spin
cc -DSAFETY -DMA=96 -O2 -o pan pan.c
./pan -m20000000
\end{VerbatimU}

Here, the depth limit of 20,000,000 is an order of magnitude
larger than the expected depth deduced from simple extrapolation.
Although this increases up-front memory usage, it avoids wasting
a long run due to incomplete search resulting from a too-tight
depth limit.
This run took a little more than 3~days on a \Power{9} server.
The result is shown in
\cref{lst:formal:spinhint:3 Readers 3 Updaters QRCU Spin Output with -DMA=96}.
This Spin run completed successfully with a total memory
usage of only 6.5\,GB, which is almost two orders of magnitude
lower than the \co{-DCOLLAPSE} usage of about half a terabyte.

\begin{listing}
\begin{Verbatim}
(Spin Version 6.4.6 -- 2 December 2016)
        + Partial Order Reduction
        + Graph Encoding (-DMA=96)

Full statespace search for:
        never claim             - (none specified)
        assertion violations    +
        cycle checks            - (disabled by -DSAFETY)
        invalid end states      +

State-vector 96 byte, depth reached 2055621, errors: 0
MA stats: -DMA=84 is sufficient
Minimized Automaton:    56420520 nodes and 1.75128e+08 edges
9.6647071e+09 states, stored
9.7503813e+09 states, matched
1.9415088e+10 transitions (= stored+matched)
7.2047951e+09 atomic steps

Stats on memory usage (in Megabytes):
1142905.887     equivalent memory usage for states
                  (stored*(State-vector + overhead))
 5448.879       actual memory usage for states
                  (compression: 0.48%)
 1068.115       memory used for DFS stack (-m20000000)
    1.619       memory lost to fragmentation
 6515.375       total actual memory usage

unreached in proctype qrcu_reader
        (0 of 18 states)
unreached in proctype qrcu_updater
        qrcu.spin:102, state 82, "-end-"
        (1 of 82 states)
unreached in init
        (0 of 23 states)

pan: elapsed time 2.72e+05 seconds
pan: rate 35500.523 states/second
\end{Verbatim}
\vspace*{-9pt}
\caption{3 Readers 3 Updaters QRCU Spin Output with \co{-DMA=96}}
\label{lst:formal:spinhint:3 Readers 3 Updaters QRCU Spin Output with -DMA=96}
\end{listing}

\QuickQuiz{
	A compression rate of 0.48\,\% corresponds to a 200-to-1 decrease
	in memory occupied by the states!
	Is the state-space search \emph{really} exhaustive???
}\QuickQuizAnswer{
	According to Spin's documentation, yes, it is.

\begin{listing}
\begin{Verbatim}
@@ -1,6 +1,6 @@
 (Spin Version 6.4.6 -- 2 December 2016)
         + Partial Order Reduction
-        + Compression
+        + Graph Encoding (-DMA=88)

 Full statespace search for:
         never claim             - (none specified)
@@ -9,27 +9,22 @@
         invalid end states      +

 State-vector 88 byte, depth reached 328014, errors: 0
+MA stats: -DMA=77 is sufficient
+Minimized Automaton:    2084798 nodes and 6.38445e+06 edges
 1.8620286e+08 states, stored
 1.7759831e+08 states, matched
 3.6380117e+08 transitions (= stored+matched)
 1.3724093e+08 atomic steps
-hash conflicts: 1.1445626e+08 (resolved)

 Stats on memory usage (in Megabytes):
 20598.919       equivalent memory usage for states
                   (stored*(State-vector + overhead))
- 8418.559       actual memory usage for states
-                  (compression: 40.87%)
-                state-vector as stored =
-                  19 byte + 28 byte overhead
- 2048.000       memory used for hash table (-w28)
+  204.907       actual memory usage for states
+                  (compression: 0.99%)
    17.624       memory used for DFS stack (-m330000)
-    1.509       memory lost to fragmentation
-10482.675       total actual memory usage
+  222.388       total actual memory usage

-nr of templates: [ 0:globals 1:chans 2:procs ]
-collapse counts: [ 0:1021 2:32 3:1869 4:2 ]
 unreached in proctype qrcu_reader
         (0 of 18 states)
 unreached in proctype qrcu_updater
@@ -38,5 +33,5 @@
 unreached in init
         (0 of 23 states)

-pan: elapsed time 369 seconds
-pan: rate 505107.58 states/second
+pan: elapsed time 2.68e+03 seconds
+pan: rate 69453.282 states/second
\end{Verbatim}
\vspace*{-9pt}
\caption{Spin Output Diff of \co{-DCOLLAPSE} and \co{-DMA=88}}
\label{lst:formal:promela:Spin Output Diff of -DCOLLAPSE and -DMA=88}
\end{listing}

	As an indirect evidence, let's compare the results of
	runs with \co{-DCOLLAPSE} and with \co{-DMA=88}
	(two readers and three updaters).
	The diff of outputs from those runs is shown in
	\cref{lst:formal:promela:Spin Output Diff of -DCOLLAPSE and -DMA=88}.
	As you can see, they agree on the numbers of states
	(stored and matched).
}\QuickQuizEnd

\begin{table*}
\rowcolors{6}{}{lightgray}
\renewcommand*{\arraystretch}{1.2}
\IfEbookSize{\scriptsize}{\footnotesize}
\centering
\OneColumnHSpace{-0.7in}%
\ebresizewidth{
\begin{tabular}{S[table-format = 1.0]S[table-format = 1.0]S[table-format = 9.0]
		S[table-format = 9.0]S[table-format = 2.0]S[table-format = 5.2]
		S[table-format = 4.2]S[table-format = 2.0]S[table-format = 4.2]
		S[table-format = 6.2]}
	\toprule
	\multicolumn{4}{r}{} & \multicolumn{3}{c}{\tco{-DCOLLAPSE}} &
					\multicolumn{3}{c}{\tco{-DMA=N}} \\
	\cmidrule(l){5-7} \cmidrule(l){8-10}
	\multicolumn{1}{r}{updaters} &
	    \multicolumn{1}{r}{readers} &
		\multicolumn{1}{r}{\# states} &
		    \multicolumn{1}{r}{depth reached} &
			\multicolumn{1}{r}{\tco{-wN}} &
			    \multicolumn{1}{r}{memory (MB)} &
				\multicolumn{1}{r}{runtime (s)} &
				    \multicolumn{1}{r}{\tco{N}} &
					\multicolumn{1}{r}{memory (MB)} &
					    \multicolumn{1}{r}{runtime (s)} \\
	\cmidrule{1-4} \cmidrule(l){5-7} \cmidrule(l){8-10}
	1 & 1 &           376 &         95 & 12 &     0.10 & 0.00 &
		40 &    0.29 &      0.00 \\
	1 & 2 &         6 177 &        218 & 12 &     0.39 & 0.01 &
		47 &    0.59 &      0.02 \\
	1 & 3 &        99 728 &        385 & 16 &     4.60 & 0.14 &
		54 &    3.04 &      0.45 \\
        2 & 1 &        29 399 &        859 & 16 &     2.30 & 0.03 &
		55 &    0.70 &      0.13 \\
        2 & 2 &     1 071 181 &      2 352 & 20 &    49.24 & 1.45 &
		62 &    7.77 &      5.76 \\
        2 & 3 &    33 866 736 &     12 857 & 24 & 1 540.70 & 62.5 &
		69 &  111.66 &    326    \\
        3 & 1 &     2 749 453 &     53 809 & 21 &   125.25 & 4.01 &
		70 &   11.41 &     19.5  \\
        3 & 2 &   186 202 860 &    328 014 & 28 & 10 482.51 & 390 &
		77 &  222.26 &   2560    \\
	3 & 3 & 9 664 707 100 &  2 055 621 &    &          &      &
		84 & 5557.02 & 266000    \\
	\bottomrule
\end{tabular}
}
\caption{QRCU Spin Result Summary}
\label{tab:formal:promela:QRCU Spin Result Summary}
\end{table*}

For reference, \cref{tab:formal:promela:QRCU Spin Result Summary}
summarizes the Spin results with \co{-DCOLLAPSE} and \co{-DMA=N}
compiler flags.
The memory usage is obtained with minimal sufficient
search depths and \co{-DMA=N} parameters shown in the table.
Hashtable sizes for \co{-DCOLLAPSE} runs are tweaked by
the \co{-wN} option of \co{./pan} to avoid using too much
memory hashing small state spaces.
Hence the memory usage is smaller than what is shown in
\cref{tab:advsync:Memory Usage of QRCU Model}, where the
hashtable size starts from the default of \co{-w24}.
The runtime is from a \Power{9} server, which shows that \co{-DMA=N}
suffers up to about an order of magnitude higher CPU overhead
than does \co{-DCOLLAPSE}, but on the other hand reduces memory overhead
by well over an order of magnitude.

So far so good.
But adding a few more updaters or readers would exhaust memory, even
with \co{-DMA=N}.\footnote{
	Alternatively, the CPU consumption would become excessive.}
So what to do?
Here are some possible approaches:

\begin{enumerate}
\item	See whether a smaller number of readers and updaters suffice
	to prove the general case.
\item	Manually construct a proof of correctness.
\item	Use a more capable tool.
\item	Divide and conquer.
\end{enumerate}

The following sections discuss each of these approaches.

\subsubsection{How Many Readers and Updaters Are Really Needed?}
\label{sec:formal:How Many Readers and Updaters Are Really Needed?}

One approach is to look carefully at the Promela code for
\co{qrcu_updater()} and notice that the only global state
change is happening under the lock.
Therefore, only one updater at a time can possibly be modifying
state visible to either readers or other updaters.
This means that any sequences of state changes can be carried
out serially by a single updater due to the fact that Promela does a full
state-space search.
Therefore, at most two updaters are required:
One to change state and a second to become confused.

The situation with the readers is less clear-cut, as each reader
does only a single read-side critical section then terminates.
It is possible to argue that the useful number of readers is limited,
due to the fact that the fastpath must see at most a zero and a one
in the counters.
This is a fruitful avenue of investigation, in fact, it leads to
the full proof of correctness described in the next section.

\subsubsection{Alternative Approach:
				     Proof of Correctness}
\label{sec:formal:Alternative Approach: Proof of Correctness}

An informal proof~\cite{PaulMcKenney2007QRCUpatch}
follows:

\begin{enumerate}
\item	For \co{synchronize_qrcu()} to exit too early, then
	by definition there must have been at least one reader
	present during \co{synchronize_qrcu()}'s full
	execution.
\item	The counter corresponding to this reader will have been
	at least 1 during this time interval.
\item	The \co{synchronize_qrcu()} code forces at least one
	of the counters to be at least 1 at all times.
\item	The above two items imply that if the counter corresponding
	to this reader is exactly one, then the other counter must be
	greater than or equal to one.
	Similarly, if the other counter is equal to zero, then the counter
	corresponding to the reader must be greater than or equal to two.
\item	Therefore, at any given point in time, either one of the
	counters will be at least 2, or both of the counters will
	be at least one.
\item	However, the \co{synchronize_qrcu()} fastpath code
	can read only one of the counters at a given time.
	It is therefore possible for the fastpath code to fetch
	the first counter while zero, but to race with a counter
	flip so that the second counter is seen as one.
\item	There can be at most one reader persisting through such
	a race condition, as otherwise the sum would be two or
	greater, which would cause the updater to take the slowpath.
\item	But if the race occurs on the fastpath's first read of the
	counters, and then again on its second read, there have
	to have been two counter flips.
\item	Because a given updater flips the counter only once, and
	because the update-side lock prevents a pair of updaters
	from concurrently flipping the counters, the only way that
	the fastpath code can race with a flip twice is if the
	first updater completes.
\item	But the first updater will not complete until after all
	pre-existing readers have completed.
\item	Therefore, if the fastpath races with a counter flip
	twice in succession, all pre-existing readers must have
	completed, so that it is safe to take the fastpath.
\end{enumerate}

Of course, not all parallel algorithms have such simple proofs.
In such cases, it may be necessary to enlist more capable tools.

\subsubsection{Alternative Approach:
				     More Capable Tools}
\label{sec:formal:Alternative Approach: More Capable Tools}

Although Promela and Spin are quite useful,
much more capable tools are available, particularly for verifying
hardware.
This means that if it is possible to translate your algorithm
to the hardware-design VHDL language, as it often will be for
low-level parallel algorithms, then it is possible to apply these
tools to your code (for example, this was done for the first
realtime RCU algorithm).
However, such tools can be quite expensive.

Although the advent of commodity multiprocessing
might eventually result in powerful free-software model-checkers
featuring fancy state-space-reduction capabilities,
this does not help much in the here and now.

As an aside, there are Spin features that support approximate searches
that require fixed amounts of memory, however, I have never been able
to bring myself to trust approximations when verifying parallel
algorithms.

Another approach might be to divide and conquer.

\subsubsection{Alternative Approach:
				     Divide and Conquer}
\label{sec:formal:Alternative Approach: Divide and Conquer}

It is often possible to break down a larger parallel algorithm into
smaller pieces, which can then be proven separately.
For example, a 10-billion-state model might be broken into a pair
of 100,000-state models.
Taking this approach not only makes it easier for tools such as
Promela to verify your algorithms, it can also make your algorithms
easier to understand.

\subsubsection{Is QRCU Really Correct?}
\label{sec:formal:Is QRCU Really Correct?}

Is QRCU really correct?
We have a Promela-based mechanical proof and a by-hand proof that both
say that it is.
However, a paper by \pplsur{Jade}{Alglave} et al.~\cite{JadeAlglave2013-cav}
says otherwise (see Section~5.1 of the paper at the bottom of page~12).
Which is it?

It turns out that both are correct!
When QRCU was added to a suite of formal-verification benchmarks,
its memory barriers were omitted, thus resulting in a buggy version
of QRCU\@.
So the real news here is that a number of formal-verification tools
incorrectly proved this buggy QRCU correct.
And this is why formal-verification tools themselves should be tested
using bug-injected versions of the code being verified.
If a given tool cannot find the injected bugs, then that tool is
clearly untrustworthy.

\QuickQuiz{
	But different formal-verification tools are often designed to
	locate particular classes of bugs.
	For example, very few formal-verification tools will find
	an error in the specification.
	So isn't this ``clearly untrustworthy'' judgment a bit harsh?
}\QuickQuizAnswer{
	It is certainly true that many formal-verification tools are
	specialized in some way.
	For example, Promela does not handle realistic memory models
	(though they can be programmed into
	\IX{Promela}~\cite{Desnoyers:2013:MSM:2506164.2506174}),
	\IXacr{cbmc}~\cite{EdmundClarke2004CBMC} does not detect probabilistic
	hangs and deadlocks, and
	\IX{Nidhugg}~\cite{CarlLeonardsson2014Nidhugg} does not detect
	bugs involving data nondeterminism.
	But this means that these tools cannot be trusted to find
	bugs that they are not designed to locate.

	And therefore people creating formal-verification tools should
	``tell the truth on the label'', clearly calling out what
	classes of bugs their tools can and cannot detect.
	Otherwise, the first time a practitioner finds a tool
	failing to detect a bug, that practitioner is likely to
	make extremely harsh and extremely public denunciations
	of that tool.
	Yes, yes, there is something to be said for putting your
	best foot forward, but putting it too far forward without
	appropriate disclaimers can easily trigger a land mine of
	negative reaction that your tool might or might not be able
	to recover from.

	You have been warned!
}\QuickQuizEnd

Therefore, if you do intend to use QRCU, please take care.
Its proofs of correctness might or might not themselves be correct.
Which is one reason why formal verification is unlikely to
completely replace testing, as Donald Knuth pointed out so long ago.

\QuickQuiz{
	Given that we have two independent proofs of correctness for
	the QRCU algorithm described herein, and given that the
	proof of incorrectness covers what is known to be a different
	algorithm, why is there any room for doubt?
}\QuickQuizAnswer{
	There is always room for doubt.
	In this case, it is important to keep in mind that the two proofs
	of correctness preceded the formalization of real-world memory
	models, raising the possibility that these two proofs are based
	on incorrect memory-ordering assumptions.
	Furthermore, since both proofs were constructed by the same person,
	it is quite possible that they contain a common error.
	Again, there is always room for doubt.
}\QuickQuizEnd

% formal/dyntickrcu.tex
% mainfile: ../perfbook.tex
% SPDX-License-Identifier: CC-BY-SA-3.0

% Disable frame around VerbatimN in two-column layout
\IfTwoColumn{
\RecustomVerbatimEnvironment{VerbatimN}{Verbatim}%
{numbers=left,numbersep=5pt,xleftmargin=10pt,xrightmargin=0pt,frame=none}
\setlength{\lnlblraise}{0pt}
}{}

\subsection{Promela Parable:
			     {dynticks} and Preemptible RCU}
\label{sec:formal:Promela Parable: dynticks and Preemptible RCU}

In early 2008, a preemptible variant of RCU was accepted into
mainline Linux in support of real-time workloads,
a variant similar to the RCU implementations in
the \rt\ patchset~\cite{IngoMolnar05a}
since August 2005.
Preemptible RCU is needed for real-time workloads because older
RCU implementations disable preemption across RCU read-side
critical sections, resulting in excessive real-time latencies.

However, one disadvantage of the older \rt\ implementation
was that each \IX{grace period}
requires work to be done on each CPU, even if that CPU is in a low-power
``dynticks-idle'' state,
and thus incapable of executing RCU read-side critical sections.
The idea behind the dynticks-idle state is that idle CPUs
should be physically powered down in order to conserve energy.
In short, preemptible RCU can disable a valuable energy-conservation
feature of recent Linux kernels.
Although Josh Triplett and Paul McKenney
had discussed some approaches for allowing
CPUs to remain in low-power state throughout an RCU grace period
(thus preserving the Linux kernel's ability to conserve energy), matters
did not come to a head until Steve Rostedt integrated a new dyntick
implementation with preemptible RCU in the \rt\ patchset.

This combination caused one of Steve's systems to hang on boot, so in
October, Paul coded up a dynticks-friendly modification to preemptible RCU's
grace-period processing.
Steve coded up \co{rcu_irq_enter()} and \co{rcu_irq_exit()}
interfaces called from the
\co{irq_enter()} and \co{irq_exit()} interrupt
entry/exit functions.
These \co{rcu_irq_enter()} and \co{rcu_irq_exit()}
functions are needed to allow RCU to reliably handle situations where
a dynticks-idle CPU is momentarily powered up for an interrupt
handler containing RCU read-side critical sections.
With these changes in place, Steve's system booted reliably,
but Paul continued inspecting the code periodically on the assumption
that we could not possibly have gotten the code right on the first try.

Paul reviewed the code repeatedly from October 2007 to February 2008,
and almost always found at least one bug.
In one case, Paul even coded and tested a fix before realizing that the
bug was illusory, and in fact in all cases, the ``bug'' turned out to be
illusory.

Near the end of February, Paul grew tired of this game.
He therefore decided to enlist the aid of
Promela and Spin.
The following presents a series of seven increasingly realistic
Promela models, the last of which passes, consuming about
40\,GB of main memory for the state space.

More important, Promela and Spin did find a very subtle bug for me!

\QuickQuiz{
	Yeah, that's just great!
	Now, just what am I supposed to do if I don't happen to have a
	machine with 40\,GB of main memory???
}\QuickQuizAnswer{
	Relax, there are a number of lawful answers to
	this question:
	\begin{enumerate}
	\item	Try compiler flags \co{-DCOLLAPSE} and \co{-DMA=N}
		to reduce memory consumption.
		See \cref{sec:formal:Running the QRCU Example}.
	\item	Further optimize the model, reducing its memory consumption.
	\item	Work out a pencil-and-paper proof, perhaps starting with the
		comments in the code in the Linux kernel.
	\item	Devise careful torture tests, which, though they cannot prove
		the code correct, can find hidden bugs.
	\item	There is some movement towards tools that do model
		checking on clusters of smaller machines.
		However, please note that we have not actually used such
		tools myself, courtesy of some large machines that Paul has
		occasional access to.
	\item	Wait for memory sizes of affordable systems to expand
		to fit your problem.
	\item	Use one of a number of cloud-computing services to rent
		a large system for a short time period.
	\end{enumerate}
}\QuickQuizEnd

Still better would be to come up with a simpler and faster algorithm
that has a smaller state space.
Even better would be an algorithm so simple that its correctness was
obvious to the casual observer!

\Crefrange{sec:formal:Introduction to Preemptible RCU and dynticks}
{sec:formal:Grace-Period Interface}
give an overview of preemptible RCU's dynticks interface,
followed by
\cref{sec:formal:Validating Preemptible RCU and dynticks}'s
discussion of the validation of the interface.

\subsubsection{Introduction to Preemptible RCU and dynticks}
\label{sec:formal:Introduction to Preemptible RCU and dynticks}

The per-CPU \co{dynticks_progress_counter} variable is
central to the interface between dynticks and preemptible RCU\@.
This variable has an even value whenever the corresponding CPU
is in dynticks-idle mode, and an odd value otherwise.
A CPU exits dynticks-idle mode for the following three reasons:

\begin{enumerate}
\item	To start running a task,
\item	When entering the outermost of a possibly nested set of interrupt
	handlers, and
\item	When entering an \IXacr{nmi} handler.
\end{enumerate}

Preemptible RCU's grace-period machinery samples the value of
the \co{dynticks_progress_counter} variable in order to
determine when a dynticks-idle CPU may safely be ignored.

The following three sections give an overview of the task
interface, the interrupt/NMI interface, and the use of
the \co{dynticks_progress_counter} variable by the
grace-period machinery as of Linux kernel v2.6.25-rc4.

\subsubsection{Task Interface}
\label{sec:formal:Task Interface}

When a given CPU enters dynticks-idle mode because it has no more
tasks to run, it invokes \co{rcu_enter_nohz()}:

\begin{VerbatimN}
static inline void rcu_enter_nohz(void)
{
	mb();
	__get_cpu_var(dynticks_progress_counter)++;
	WARN_ON(__get_cpu_var(dynticks_progress_counter) &
	        0x1);
}
\end{VerbatimN}

This function simply increments \co{dynticks_progress_counter} and
checks that the result is even, but first executing a memory barrier
to ensure that any other CPU that sees the new value of
\co{dynticks_progress_counter} will also see the completion
of any prior RCU read-side critical sections.

Similarly, when a CPU that is in dynticks-idle mode prepares to
start executing a newly runnable task, it invokes
\co{rcu_exit_nohz()}:

\begin{VerbatimN}
static inline void rcu_exit_nohz(void)
{
	__get_cpu_var(dynticks_progress_counter)++;
	mb();
	WARN_ON(!(__get_cpu_var(dynticks_progress_counter) &
	          0x1));
}
\end{VerbatimN}

This function again increments \co{dynticks_progress_counter},
but follows it with a memory barrier to ensure that if any other CPU
sees the result of any subsequent RCU read-side critical section,
then that other CPU will also see the incremented value of
\co{dynticks_progress_counter}.
Finally, \co{rcu_exit_nohz()} checks that the result of the
increment is an odd value.

The \co{rcu_enter_nohz()} and \co{rcu_exit_nohz()}
functions handle the case where a CPU enters and exits dynticks-idle
mode due to task execution, but does not handle interrupts, which are
covered in the following section.

\subsubsection{Interrupt Interface}
\label{sec:formal:Interrupt Interface}

The \co{rcu_irq_enter()} and \co{rcu_irq_exit()}
functions handle interrupt/NMI entry and exit, respectively.
Of course, nested interrupts must also be properly accounted for.
The possibility of nested interrupts is handled by a second per-CPU
variable, \co{rcu_update_flag}, which is incremented upon
entry to an interrupt or NMI handler (in \co{rcu_irq_enter()})
and is decremented upon exit (in \co{rcu_irq_exit()}).
In addition, the pre-existing \co{in_interrupt()} primitive is
used to distinguish between an outermost or a nested interrupt/NMI\@.

Interrupt entry is handled by the \co{rcu_irq_enter()}
shown below:

\begin{fcvlabel}[ln:formal:dyntickrcu:rcu_irq_enter]
\begin{VerbatimN}[commandchars=\\\[\]]
void rcu_irq_enter(void)
{
	int cpu = smp_processor_id();	\lnlbl[fetch]

	if (per_cpu(rcu_update_flag, cpu))	\lnlbl[inc:b]
		per_cpu(rcu_update_flag, cpu)++; \lnlbl[inc:e]
	if (!in_interrupt() &&			\lnlbl[chk_lv:b]
	    (per_cpu(dynticks_progress_counter,
	             cpu) & 0x1) == 0) {	\lnlbl[chk_lv:e]
		per_cpu(dynticks_progress_counter, cpu)++; \lnlbl[inc_cnt]
		smp_mb();			\lnlbl[mb]
		per_cpu(rcu_update_flag, cpu)++;\lnlbl[inc_flg]
	}
}
\end{VerbatimN}
\end{fcvlabel}

\begin{fcvref}[ln:formal:dyntickrcu:rcu_irq_enter]
\Clnref{fetch} fetches the current CPU's number, while \clnref{inc:b,inc:e}
increment the \co{rcu_update_flag} nesting counter if it
is already non-zero.
\Clnrefrange{chk_lv:b}{chk_lv:e} check to see whether we are
the outermost level of
interrupt, and, if so, whether \co{dynticks_progress_counter}
needs to be incremented.
If so, \clnref{inc_cnt} increments \co{dynticks_progress_counter},
\clnref{mb} executes a memory barrier, and \clnref{inc_flg} increments
\co{rcu_update_flag}.
As with \co{rcu_exit_nohz()}, the memory barrier ensures that
any other CPU that sees the effects of an RCU read-side critical section
in the interrupt handler (following the \co{rcu_irq_enter()}
invocation) will also see the increment of
\co{dynticks_progress_counter}.
\end{fcvref}

\QuickQuizSeries{%
\QuickQuizB{
	Why not simply increment \co{rcu_update_flag}, and then only
	increment \co{dynticks_progress_counter} if the old value
	of \co{rcu_update_flag} was zero???
}\QuickQuizAnswerB{
	This fails in presence of NMIs.
	To see this, suppose an NMI was received just after
	\co{rcu_irq_enter()} incremented \co{rcu_update_flag},
	but before it incremented \co{dynticks_progress_counter}.
	The instance of \co{rcu_irq_enter()} invoked by the NMI
	would see that the original value of \co{rcu_update_flag}
	was non-zero, and would therefore refrain from incrementing
	\co{dynticks_progress_counter}.
	This would leave the RCU grace-period machinery no clue that the
	NMI handler was executing on this CPU, so that any RCU read-side
	critical sections in the NMI handler would lose their RCU protection.

	The possibility of NMI handlers, which, by definition cannot
	be masked, does complicate this code.
}\QuickQuizEndB
\QuickQuizE{
	\begin{fcvref}[ln:formal:dyntickrcu:rcu_irq_enter]
	But if \clnref{chk_lv:b} finds that we are the outermost interrupt,
	wouldn't we \emph{always} need to increment
	\co{dynticks_progress_counter}?
	\end{fcvref}
}\QuickQuizAnswerE{
	Not if we interrupted a running task!
	In that case, \co{dynticks_progress_counter} would
	have already been incremented by \co{rcu_exit_nohz()},
	and there would be no need to increment it again.
}\QuickQuizEndE
}

Interrupt exit is handled similarly by
\co{rcu_irq_exit()}:

\begin{fcvlabel}[ln:formal:dyntickrcu:rcu_irq_exit]
\begin{VerbatimN}[commandchars=\\\[\]]
void rcu_irq_exit(void)
{
	int cpu = smp_processor_id();	\lnlbl[fetch]

	if (per_cpu(rcu_update_flag, cpu)) { \lnlbl[chk_flg]
		if (--per_cpu(rcu_update_flag, cpu)) \lnlbl[dec_flg]
			return;
		WARN_ON(in_interrupt());	\lnlbl[verify]
		smp_mb();			\lnlbl[mb]
		per_cpu(dynticks_progress_counter, cpu)++; \lnlbl[inc_cnt]
		WARN_ON(per_cpu(dynticks_progress_counter, \lnlbl[vrf_even:b]
		                cpu) & 0x1); \lnlbl[vrf_even:e]
	}
}
\end{VerbatimN}
\end{fcvlabel}

\begin{fcvref}[ln:formal:dyntickrcu:rcu_irq_exit]
\Clnref{fetch} fetches the current CPU's number, as before.
\Clnref{chk_flg} checks to see if the \co{rcu_update_flag} is
non-zero, returning immediately (via falling off the end of the
function) if not.
Otherwise, \clnrefthro{dec_flg}{vrf_even:e} come into play.
\Clnref{dec_flg} decrements \co{rcu_update_flag}, returning
if the result is not zero.
\Clnref{verify} verifies that we are indeed leaving the outermost
level of nested interrupts, \clnref{mb} executes a memory barrier,
\clnref{inc_cnt} increments \co{dynticks_progress_counter},
and \clnref{vrf_even:b,vrf_even:e} verify that this
variable is now even.
As with \co{rcu_enter_nohz()}, the memory barrier ensures that
any other CPU that sees the increment of
\co{dynticks_progress_counter}
will also see the effects of an RCU read-side critical section
in the interrupt handler (preceding the \co{rcu_irq_exit()}
invocation).
\end{fcvref}

These two sections have described how the
\co{dynticks_progress_counter} variable is maintained during
entry to and exit from dynticks-idle mode, both by tasks and by
interrupts and NMIs.
The following section describes how this variable is used by
preemptible RCU's grace-period machinery.

\subsubsection{Grace-Period Interface}
\label{sec:formal:Grace-Period Interface}

\begin{figure}
\centering
\resizebox{2.5in}{!}{\includegraphics{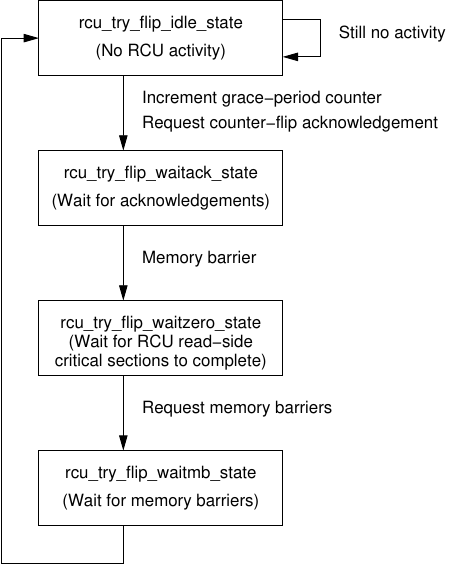}}
\caption{Preemptible RCU State Machine}
\label{fig:formal:Preemptible RCU State Machine}
\end{figure}

Of the four preemptible RCU grace-period states shown in
\cref{fig:formal:Preemptible RCU State Machine},
only the \co{rcu_try_flip_waitack_state}
and \co{rcu_try_flip_waitmb_state} states need to wait
for other CPUs to respond.

Of course, if a given CPU is in dynticks-idle state, we shouldn't
wait for it.
Therefore, just before entering one of these two states,
the preceding state takes a snapshot of each CPU's
\co{dynticks_progress_counter} variable, placing the
snapshot in another per-CPU variable,
\co{rcu_dyntick_snapshot}.
This is accomplished by invoking
\co{dyntick_save_progress_counter()}, shown below:

\begin{VerbatimN}
static void dyntick_save_progress_counter(int cpu)
{
	per_cpu(rcu_dyntick_snapshot, cpu) =
		per_cpu(dynticks_progress_counter, cpu);
}
\end{VerbatimN}

The \co{rcu_try_flip_waitack_state} state invokes
\co{rcu_try_flip_waitack_needed()}, shown below:

\begin{fcvlabel}[ln:formal:dyntickrcu:rcu_try_flip_waitack_needed]
\begin{VerbatimN}[samepage=true,commandchars=\\\[\]]
static inline int
rcu_try_flip_waitack_needed(int cpu)
{
	long curr;
	long snap;

	curr = per_cpu(dynticks_progress_counter, cpu); \lnlbl[curr]
	snap = per_cpu(rcu_dyntick_snapshot, cpu); \lnlbl[snap]
	smp_mb();				\lnlbl[mb]
	if ((curr == snap) && ((curr & 0x1) == 0)) \lnlbl[chk_remain]
		return 0;			\lnlbl[ret_0_r]
	if ((curr - snap) > 2 || (snap & 0x1) == 0) \lnlbl[chk_idle]
		return 0;			\lnlbl[ret_0_i]
	return 1;				\lnlbl[ret_1]
}
\end{VerbatimN}
\end{fcvlabel}

\begin{fcvref}[ln:formal:dyntickrcu:rcu_try_flip_waitack_needed]
\Clnref{curr,snap} pick up current and snapshot versions of
\co{dynticks_progress_counter}, respectively.
The memory barrier on \clnref{mb} ensures that the counter checks
in the later \co{rcu_try_flip_waitzero_state} follow
the fetches of these counters.
\Clnref{chk_remain,ret_0_r} return zero
(meaning no communication with the
specified CPU is required) if that CPU has remained in dynticks-idle
state since the time that the snapshot was taken.
Similarly, \clnref{chk_idle,ret_0_i} return zero
if that CPU was initially
in dynticks-idle state or if it has completely passed through a
dynticks-idle state.
In both these cases, there is no way that the CPU could have retained
the old value of the grace-period counter.
If neither of these conditions hold, \clnref{ret_1} returns one, meaning
that the CPU needs to explicitly respond.
\end{fcvref}

For its part, the \co{rcu_try_flip_waitmb_state} state
invokes \co{rcu_try_flip_waitmb_needed()}, shown below:

\begin{fcvlabel}[ln:formal:dyntickrcu:rcu_try_flip_waitmb_needed]
\begin{VerbatimN}[commandchars=\\\[\]]
static inline int
rcu_try_flip_waitmb_needed(int cpu)
{
	long curr;
	long snap;

	curr = per_cpu(dynticks_progress_counter, cpu);
	snap = per_cpu(rcu_dyntick_snapshot, cpu);
	smp_mb();
	if ((curr == snap) && ((curr & 0x1) == 0))
		return 0;
	if (curr != snap)		\lnlbl[chk_to_from]
		return 0;		\lnlbl[ret_0]
	return 1;
}
\end{VerbatimN}
\end{fcvlabel}

\begin{fcvref}[ln:formal:dyntickrcu:rcu_try_flip_waitmb_needed]
This is quite similar to \co{rcu_try_flip_waitack_needed()},
the difference being in \clnref{chk_to_from,ret_0}, because any transition
either to or from dynticks-idle state executes the memory barrier
needed by the \co{rcu_try_flip_waitmb_state} state.
\end{fcvref}

We now have seen all the code involved in the interface between
RCU and the dynticks-idle state.
The next section builds up the Promela model used to verify this
code.

\QuickQuiz{
	Can you spot any bugs in any of the code in this section?
}\QuickQuizAnswer{
	Read the next section to see if you were correct.
}\QuickQuizEnd

\subsection{Validating Preemptible RCU and dynticks}
\label{sec:formal:Validating Preemptible RCU and dynticks}

This section develops a Promela model for the interface between
dynticks and RCU step by step, with each of
\crefrange{sec:formal:Basic Model}{sec:formal:Validating NMI Handlers}
illustrating one step, starting with the process-level code,
adding assertions, interrupts, and finally NMIs.

\Cref{sec:formal:Lessons (Re)Learned} lists
lessons (re)learned during this effort, and
\crefrange{sec:formal:Simplicity Avoids Formal Verification}{sec:formal:Discussion}
present a simpler solution to RCU's dynticks problem.

\subsubsection{Basic Model}
\label{sec:formal:Basic Model}

This section translates the process-level dynticks entry/exit
code and the grace-period processing into
Promela~\cite{Holzmann03a}.
We start with \co{rcu_exit_nohz()} and
\co{rcu_enter_nohz()}
from the 2.6.25-rc4 kernel, placing these in a single Promela
process that models exiting and entering dynticks-idle mode in
a loop as follows:

\input{CodeSamples/formal/promela/dyntick/dyntickRCU-base=dyntick_nohz.fcv}

\begin{fcvref}[ln:formal:promela:dyntick:dyntickRCU-base:dyntick_nohz]
\Clnref{do,od} define a loop.
\Clnref{break} exits the loop once the loop counter \co{i}
has exceeded the limit \co{MAX_DYNTICK_LOOP_NOHZ}.
\Clnref{kick_loop} tells the loop construct to execute
\clnrefrange{ex_inc:b}{inc_i}
for each pass through the loop.
Because the conditionals on \clnref{break,kick_loop}
are exclusive of
each other, the normal Promela random selection of true conditions
is disabled.
\Clnref{ex_inc:b,ex_inc:e} model
\co{rcu_exit_nohz()}'s non-atomic
increment of \co{dynticks_progress_counter}, while
\clnref{ex_assert} models the \co{WARN_ON()}.
The \co{atomic} construct simply reduces the Promela state space,
given that the \co{WARN_ON()} is not strictly speaking part
of the algorithm.
\Clnrefrange{ent_inc:b}{ent_inc:e} similarly model the increment and
\co{WARN_ON()} for \co{rcu_enter_nohz()}.
Finally, \clnref{inc_i} increments the loop counter.
\end{fcvref}

Each pass through the loop therefore models a CPU exiting
dynticks-idle mode (for example, starting to execute a task), then
re-entering dynticks-idle mode (for example, that same task blocking).

\QuickQuizSeries{%
\QuickQuizB{
	Why isn't the memory barrier in \co{rcu_exit_nohz()}
	and \co{rcu_enter_nohz()} modeled in Promela?
}\QuickQuizAnswerB{
	Promela assumes sequential consistency, so
	it is not necessary to model memory barriers.
	In fact, one must instead explicitly model lack of memory barriers,
	for example, as shown in
	\cref{lst:formal:QRCU Unordered Summation} on
	\cpageref{lst:formal:QRCU Unordered Summation}.
}\QuickQuizEndB
\QuickQuizE{
	Isn't it a bit strange to model \co{rcu_exit_nohz()}
	followed by \co{rcu_enter_nohz()}?
	Wouldn't it be more natural to instead model entry before exit?
}\QuickQuizAnswerE{
	It probably would be more natural, but we will need
	this particular order for the liveness checks that we will add later.
}\QuickQuizEndE
}

The next step is to model the interface to RCU's grace-period
processing.
For this, we need to model
\co{dyntick_save_progress_counter()},
\co{rcu_try_flip_waitack_needed()},
\co{rcu_try_flip_waitmb_needed()},
as well as portions of
\co{rcu_try_flip_waitack()} and
\co{rcu_try_flip_waitmb()}, all from the 2.6.25-rc4 kernel.
The following \co{grace_period()} Promela process models
these functions as they would be invoked during a single pass
through preemptible RCU's grace-period processing.

\input{CodeSamples/formal/promela/dyntick/dyntickRCU-base=grace_period.fcv}

\begin{fcvref}[ln:formal:promela:dyntick:dyntickRCU-base:grace_period]
\Clnrefrange{print:b}{print:e} print out the loop limit
(but only into the ``.trail'' file
in case of error) and models a line of code
from \co{rcu_try_flip_idle()} and its call to
\co{dyntick_save_progress_counter()}, which takes a
snapshot of the current CPU's \co{dynticks_progress_counter}
variable.
These two lines are executed atomically to reduce state space.

\Clnrefrange{do1}{od1} model the relevant code in
\co{rcu_try_flip_waitack()} and its call to
\co{rcu_try_flip_waitack_needed()}.
This loop is modeling the grace-period state machine waiting for
a counter-flip acknowledgement from each CPU, but only that part
that interacts with dynticks-idle CPUs.

\Clnref{snap} models a line from \co{rcu_try_flip_waitzero()}
and its call to \co{dyntick_save_progress_counter()}, again
taking a snapshot of the CPU's \co{dynticks_progress_counter}
variable.

Finally, \clnrefrange{do2}{od2} model the relevant code in
\co{rcu_try_flip_waitack()} and its call to
\co{rcu_try_flip_waitack_needed()}.
This loop is modeling the grace-period state-machine waiting for
each CPU to execute a memory barrier, but again only that part
that interacts with dynticks-idle CPUs.
\end{fcvref}

\QuickQuiz{
	Wait a minute!
	In the Linux kernel, both \co{dynticks_progress_counter} and
	\co{rcu_dyntick_snapshot} are per-CPU variables.
	So why are they instead being modeled as single global variables?
}\QuickQuizAnswer{
	Because the grace-period code processes each
	CPU's \co{dynticks_progress_counter} and
	\co{rcu_dyntick_snapshot} variables separately,
	we can collapse the state onto a single CPU\@.
	If the grace-period code were instead to do something special
	given specific values on specific CPUs, then we would indeed need
	to model multiple CPUs.
	But fortunately, we can safely confine ourselves to two CPUs, the
	one running the grace-period processing and the one entering and
	leaving dynticks-idle mode.
}\QuickQuizEnd

The resulting model (\path{dyntickRCU-base.spin}),
when run with the
\path{runspin.sh} script,
generates 691 states and
passes without errors, which is not at all surprising given that
it completely lacks the assertions that could find failures.
The next section therefore adds safety assertions.

\subsubsection{Validating Safety}
\label{sec:formal:Validating Safety}

A safe RCU implementation must never permit a grace period to
complete before the completion of any RCU readers that started
before the start of the grace period.
This is modeled by a \co{grace_period_state} variable that
can take on three states as follows:

\input{CodeSamples/formal/promela/dyntick/dyntickRCU-base-s=grace_period_state.fcv}

The \co{grace_period()} process sets this variable as it
progresses through the grace-period phases, as shown below:

\input{CodeSamples/formal/promela/dyntick/dyntickRCU-base-s=grace_period.fcv}

\begin{fcvref}[ln:formal:promela:dyntick:dyntickRCU-base-s:grace_period]
\Clnref{upd_gps1,upd_gps2,upd_gps3,upd_gps4,upd_gps5,upd_gps6}
update this variable (combining
atomically with algorithmic operations where feasible) to
allow the \co{dyntick_nohz()} process to verify the basic
RCU safety property.
The form of this verification is to assert that the value of the
\co{grace_period_state} variable cannot jump from
\co{GP_IDLE} to \co{GP_DONE} during a time period
over which RCU readers could plausibly persist.
\end{fcvref}

\QuickQuiz{
	\begin{fcvref}[ln:formal:promela:dyntick:dyntickRCU-base-s:grace_period]
	Given there are a pair of back-to-back changes to
	\co{grace_period_state} on \clnref{upd_gps3,upd_gps4},
	how can we be sure that \clnref{upd_gps3}'s changes won't be lost?
	\end{fcvref}
}\QuickQuizAnswer{
	Recall that Promela and Spin trace out
	every possible sequence of state changes.
	Therefore, timing is irrelevant:
	Promela/Spin will be quite happy to jam the entire rest of
	the model between those two statements unless some state
	variable specifically prohibits doing so.
}\QuickQuizEnd

The \co{dyntick_nohz()} Promela process implements
this verification as shown below:

\input{CodeSamples/formal/promela/dyntick/dyntickRCU-base-s=dyntick_nohz.fcv}

\begin{fcvref}[ln:formal:promela:dyntick:dyntickRCU-base-s:dyntick_nohz]
\Clnref{new_flg} sets a new \co{old_gp_idle} flag if the
value of the \co{grace_period_state} variable is
\co{GP_IDLE} at the beginning of task execution,
and the assertion at \clnref{assert:b,assert:e}
fire if the \co{grace_period_state}
variable has advanced to \co{GP_DONE} during task execution,
which would be illegal given that a single RCU read-side critical
section could span the entire intervening time period.
\end{fcvref}

The resulting
model (\path{dyntickRCU-base-s.spin}),
when run with the \path{runspin.sh} script,
generates 964 states and passes without errors, which is reassuring.
That said, although safety is critically important, it is also quite
important to avoid indefinitely stalling grace periods.
The next section therefore covers verifying liveness.

\subsubsection{Validating Liveness}
\label{sec:formal:Validating Liveness}

\begin{fcvref}[ln:formal:promela:dyntick:dyntickRCU-base-sl-busted:dyntick_nohz]
Although liveness can be difficult to prove, there is a simple
trick that applies here.
The first step is to make \co{dyntick_nohz()} indicate that
it is done via a \co{dyntick_nohz_done} variable, as shown on
\clnref{done} of the following:
\end{fcvref}

\input{CodeSamples/formal/promela/dyntick/dyntickRCU-base-sl-busted=dyntick_nohz.fcv}

With this variable in place, we can add assertions to
\co{grace_period()} to check for unnecessary blockage
as follows:

\input{CodeSamples/formal/promela/dyntick/dyntickRCU-base-sl-busted=grace_period.fcv}

\begin{fcvref}[ln:formal:promela:dyntick:dyntickRCU-base-sl-busted:grace_period]
We have added the \co{shouldexit} variable on \clnref{shex},
which we initialize to zero on \clnref{init_shex}.
\Clnref{assert_shex} asserts that \co{shouldexit} is not set, while
\clnref{set_shex} sets \co{shouldexit} to the \co{dyntick_nohz_done}
variable maintained by \co{dyntick_nohz()}.
This assertion will therefore trigger if we attempt to take more than
one pass through the wait-for-counter-flip-acknowledgement
loop after \co{dyntick_nohz()} has completed
execution.
After all, if \co{dyntick_nohz()} is done, then there cannot be
any more state changes to force us out of the loop, so going through twice
in this state means an infinite loop, which in turn means no end to the
grace period.

\Clnref{init_shex2,assert_shex2,set_shex2} operate in a similar manner
for the second (memory-barrier) loop.

However, running this
model (\path{dyntickRCU-base-sl-busted.spin})
results in failure, as \clnref{chk_2} is checking that
the wrong variable
is even.
Upon failure, \co{spin} writes out a
``trail'' file
(\path{dyntickRCU-base-sl-busted.spin.trail}),
which records the sequence of states that lead to the failure.
Use the ``\co{spin -t -p -g -l\ }%
\path{dyntickRCU-base-sl-busted.spin}''
command to cause \co{spin} to retrace this sequence of states,
printing the statements executed and the values of variables
(\path{dyntickRCU-base-sl-busted.spin.trail.txt}).
Note that the line numbers do not match the listing above due to
the fact that \co{spin} takes both functions in a single file.
However, the line numbers \emph{do} match the full
model (\path{dyntickRCU-base-sl-busted.spin}).

We see that the \co{dyntick_nohz()} process completed
at step 34 (search for ``34:''), but that the
\co{grace_period()} process nonetheless failed to exit the loop.
The value of \co{curr} is \co{6} (see step 35)
and that the value of \co{snap} is \co{5} (see step 17).
Therefore the first condition on \clnref{chk_1} above
does not hold because
\qco{curr != snap}, and the second condition on \clnref{chk_2}
does not hold either because \co{snap} is odd and because
\co{curr} is only one greater than \co{snap}.
\end{fcvref}

So one of these two conditions has to be incorrect.
Referring to the comment block in \co{rcu_try_flip_waitack_needed()}
for the first condition:

\begin{quote}
	If the CPU remained in dynticks mode for the entire time
	and didn't take any interrupts, NMIs, SMIs, or whatever,
	then it cannot be in the middle of an \co{rcu_read_lock()}, so
	the next \co{rcu_read_lock()} it executes must use the new value
	of the counter.
	So we can safely pretend that this CPU already acknowledged
	the counter.
\end{quote}

The first condition does match this, because if \qco{curr == snap}
and if \co{curr} is even, then the corresponding CPU has been
in dynticks-idle mode the entire time, as required.
So let's look at the comment block for the second condition:

\begin{quote}
	If the CPU passed through or entered a dynticks idle phase with
	no active irq handlers, then, as above, we can safely pretend
	that this CPU already acknowledged the counter.
\end{quote}

The first part of the condition is correct, because if \co{curr}
and \co{snap} differ by two, there will be at least one even
number in between, corresponding to having passed completely through
a dynticks-idle phase.
However, the second part of the condition corresponds to having
\emph{started} in dynticks-idle mode, not having \emph{finished}
in this mode.
We therefore need to be testing \co{curr} rather than
\co{snap} for being an even number.

The corrected C code is as follows:

\begin{fcvlabel}[ln:formal:dyntickrcu:rcu_try_flip_waitack_needed_fixed]
\begin{VerbatimN}[commandchars=\\\[\]]
static inline int
rcu_try_flip_waitack_needed(int cpu)
{
	long curr;
	long snap;

	curr = per_cpu(dynticks_progress_counter, cpu);
	snap = per_cpu(rcu_dyntick_snapshot, cpu);
	smp_mb();
	if ((curr == snap) && ((curr & 0x1) == 0)) \lnlbl[if:b]
		return 0;
	if ((curr - snap) > 2 || (curr & 0x1) == 0)
		return 0;			\lnlbl[if:e]
	return 1;
}
\end{VerbatimN}
\end{fcvlabel}

\begin{fcvref}[ln:formal:dyntickrcu:rcu_try_flip_waitack_needed_fixed]
\Clnrefrange{if:b}{if:e} can now be combined and simplified,
resulting in the following.
A similar simplification can be applied to
\co{rcu_try_flip_waitmb_needed()}.
\end{fcvref}

\begin{VerbatimN}[commandchars=\\\[\]]
static inline int
rcu_try_flip_waitack_needed(int cpu)
{
	long curr;
	long snap;

	curr = per_cpu(dynticks_progress_counter, cpu);
	snap = per_cpu(rcu_dyntick_snapshot, cpu);
	smp_mb();
	if ((curr - snap) >= 2 || (curr & 0x1) == 0)
		return 0;
	return 1;
}
\end{VerbatimN}

Making the corresponding correction in the
model (\path{dyntickRCU-base-sl.spin})
results in a correct verification with 661 states that passes without
errors.
However, it is worth noting that the first version of the liveness
verification failed to catch this bug, due to a bug in the liveness
verification itself.
This liveness-verification bug was located by inserting an infinite
loop in the \co{grace_period()} process, and noting that
the liveness-verification code failed to detect this problem!

We have now successfully verified both safety and liveness
conditions, but only for processes running and blocking.
We also need to handle interrupts, a task taken up in the next section.

\subsubsection{Interrupts}
\label{sec:formal:Interrupts}

There are a couple of ways to model interrupts in Promela:
\begin{enumerate}
\item	Using C-preprocessor tricks to insert the interrupt handler
	between each and every statement of the \co{dynticks_nohz()}
	process, or
\item	Modeling the interrupt handler with a separate process.
\end{enumerate}

A bit of thought indicated that the second approach would have a
smaller state space, though it requires that the interrupt handler
somehow run atomically with respect to the \co{dynticks_nohz()}
process, but not with respect to the \co{grace_period()}
process.

Fortunately, it turns out that Promela permits you to branch
out of atomic statements.
This trick allows us to have the interrupt handler set a flag, and
recode \co{dynticks_nohz()} to atomically check this flag
and execute only when the flag is not set.
This can be accomplished with a C-preprocessor macro that takes
a label and a Promela statement as follows:

\input{CodeSamples/formal/promela/dyntick/dyntickRCU-irqnn-ssl=EXECUTE_MAINLINE.fcv}

One might use this macro as follows:

\begin{VerbatimU}
EXECUTE_MAINLINE(stmt1,
                 tmp = dynticks_progress_counter)
\end{VerbatimU}

\begin{fcvref}[ln:formal:promela:dyntick:dyntickRCU-irqnn-ssl:EXECUTE_MAINLINE]
\Clnref{label} of the macro creates the specified statement label.
\Clnrefrange{atm:b}{atm:e} are an atomic block that tests
the \co{in_dyntick_irq}
variable, and if this variable is set (indicating that the interrupt
handler is active), branches out of the atomic block back to the
label.
Otherwise, \clnref{else} executes the specified statement.
The overall effect is that mainline execution stalls any time an interrupt
is active, as required.
\end{fcvref}

\subsubsection{Validating Interrupt Handlers}
\label{sec:formal:Validating Interrupt Handlers}

The first step is to convert \co{dyntick_nohz()} to
\co{EXECUTE_MAINLINE()} form, as follows:

\input{CodeSamples/formal/promela/dyntick/dyntickRCU-irqnn-ssl=dyntick_nohz.fcv}

\begin{fcvref}[ln:formal:promela:dyntick:dyntickRCU-irqnn-ssl:dyntick_nohz]
It is important to note that when a group of statements is passed
to \co{EXECUTE_MAINLINE()}, as in \clnrefrange{stmt2:b}{stmt2:e}, all
statements in that group execute atomically.
\end{fcvref}

\QuickQuizSeries{%
\QuickQuizB{
	But what would you do if you needed the statements in a single
	\co{EXECUTE_MAINLINE()} group to execute non-atomically?
}\QuickQuizAnswerB{
	The easiest thing to do would be to put
	each such statement in its own \co{EXECUTE_MAINLINE()}
	statement.
}\QuickQuizEndB
\QuickQuizE{
	But what if the \co{dynticks_nohz()} process had
	\qco{if} or \qco{do} statements with conditions,
	where the statement bodies of these constructs
	needed to execute non-atomically?
}\QuickQuizAnswerE{
	One approach, as we will see in a later section,
	is to use explicit labels and \qco{goto} statements.
	For example, the construct:

\begin{VerbatimU}
	if
	:: i == 0 -> a = -1;
	:: else -> a = -2;
	fi;
\end{VerbatimU}
	could be modeled as something like:

\begin{VerbatimU}
	EXECUTE_MAINLINE(stmt1,
	                 if
	                 :: i == 0 -> goto stmt1_then;
	                 :: else -> goto stmt1_else;
	                 fi)
	stmt1_then: skip;
	EXECUTE_MAINLINE(stmt1_then1, a = -1; goto stmt1_end)
	stmt1_else: skip;
	EXECUTE_MAINLINE(stmt1_then1, a = -2)
	stmt1_end: skip;
\end{VerbatimU}

	However, it is not clear that the macro is helping much in the case
	of the \qco{if} statement, so these sorts of situations will
	be open-coded in the following sections.
}\QuickQuizEndE
}

The next step is to write a \co{dyntick_irq()} process
to model an interrupt handler:

\input{CodeSamples/formal/promela/dyntick/dyntickRCU-irqnn-ssl=dyntick_irq.fcv}

\begin{fcvref}[ln:formal:promela:dyntick:dyntickRCU-irqnn-ssl:dyntick_irq]
The loop from \clnrefrange{do}{od} models up to \co{MAX_DYNTICK_LOOP_IRQ}
interrupts, with \clnref{cond1,cond2} forming the loop condition and
\clnref{inc_i} incrementing the control variable.
\Clnref{in_irq} tells \co{dyntick_nohz()} that an interrupt handler
is running, and \clnref{clr_in_irq} tells \co{dyntick_nohz()} that this
handler has completed.
\Clnref{irq_done} is used for liveness verification, just like the corresponding
line of \co{dyntick_nohz()}.
\end{fcvref}

\QuickQuiz{
	\begin{fcvref}[ln:formal:promela:dyntick:dyntickRCU-irqnn-ssl:dyntick_irq]
	Why are \clnref{clr_in_irq,inc_i} (the \qtco{in_dyntick_irq = 0;}
	and the \qco{i++;}) executed atomically?
	\end{fcvref}
}\QuickQuizAnswer{
	These lines of code pertain to controlling the
	model, not to the code being modeled, so there is no reason to
	model them non-atomically.
	The motivation for modeling them atomically is to reduce the size
	of the state space.
}\QuickQuizEnd

\begin{fcvref}[ln:formal:promela:dyntick:dyntickRCU-irqnn-ssl:dyntick_irq]
\Clnrefrange{enter:b}{enter:e} model \co{rcu_irq_enter()}, and
\clnref{add_prmt_cnt:b,add_prmt_cnt:e} model the relevant snippet
of \co{__irq_enter()}.
\Clnrefrange{vrf_safe:b}{vrf_safe:e} verify safety in much the same manner
as do the corresponding lines of \co{dynticks_nohz()}.
\Clnref{irq_exit:b,irq_exit:e} model the relevant snippet of \co{__irq_exit()},
and finally \clnrefrange{exit:b}{exit:e} model \co{rcu_irq_exit()}.
\end{fcvref}

\QuickQuiz{
	What property of interrupts is this \co{dynticks_irq()}
	process unable to model?
}\QuickQuizAnswer{
	One such property is nested interrupts,
	which are handled in the following section.
}\QuickQuizEnd

The \co{grace_period()} process then becomes as follows:

\input{CodeSamples/formal/promela/dyntick/dyntickRCU-irqnn-ssl=grace_period.fcv}

\begin{fcvref}[ln:formal:promela:dyntick:dyntickRCU-irqnn-ssl:grace_period]
The implementation of \co{grace_period()} is very similar
to the earlier one.
The only changes are the addition of \clnref{MDLI} to add the new
interrupt-count parameter, changes to
\clnref{edit1,edit3} to add the new \co{dyntick_irq_done} variable
to the liveness checks, and of course the optimizations on \clnref{edit2,edit4}.
\end{fcvref}

This model (\path{dyntickRCU-irqnn-ssl.spin})
results in a correct verification with roughly half a million
states, passing without errors.
However, this version of the model does not handle nested
interrupts.
This topic is taken up in the next section.

\subsubsection{Validating Nested Interrupt Handlers}
\label{sec:formal:Validating Nested Interrupt Handlers}

Nested interrupt handlers may be modeled by splitting the body of
the loop in \co{dyntick_irq()} as follows:

\input{CodeSamples/formal/promela/dyntick/dyntickRCU-irq-ssl=dyntick_irq.fcv}

\begin{fcvref}[ln:formal:promela:dyntick:dyntickRCU-irq-ssl:dyntick_irq]
This is similar to the earlier \co{dynticks_irq()} process.
It adds a second counter variable \co{j} on \clnref{j}, so that
\co{i} counts entries to interrupt handlers and \co{j}
counts exits.
The \co{outermost} variable on \clnref{om} helps determine
when the \co{grace_period_state} variable needs to be sampled
for the safety checks.
The loop-exit check on \clnref{chk_ex:b,chk_ex:e} is updated to require that the
specified number of interrupt handlers are exited as well as entered,
and the increment of \co{i} is moved to \clnref{inc_i}, which is
the end of the interrupt-entry model.
\Clnrefrange{atm1:b}{atm1:e} set the \co{outermost} variable to indicate
whether this is the outermost of a set of nested interrupts and to
set the \co{in_dyntick_irq} variable that is used by the
\co{dyntick_nohz()} process.
\Clnrefrange{atm2:b}{atm2:e} capture the state of the \co{grace_period_state}
variable, but only when in the outermost interrupt handler.

\Clnref{cnd_ex} has the do-loop conditional for interrupt-exit modeling:
As long as we have exited fewer interrupts than we have entered, it is
legal to exit another interrupt.
\Clnrefrange{atm3:b}{atm3:e}
check the safety criterion, but only if we are exiting
from the outermost interrupt level.
Finally, \clnrefrange{atm4:b}{atm4:e} increment the interrupt-exit count \co{j}
and, if this is the outermost interrupt level, clears
\co{in_dyntick_irq}.
\end{fcvref}

This model (\path{dyntickRCU-irq-ssl.spin})
results in a correct verification with a bit more than half a million
states, passing without errors.
However, this version of the model does not handle NMIs,
which are taken up in the next section.

\subsubsection{Validating NMI Handlers}
\label{sec:formal:Validating NMI Handlers}

We take the same general approach for NMIs as we do for interrupts,
keeping in mind that NMIs do not nest.
This results in a \co{dyntick_nmi()} process as follows:

\input{CodeSamples/formal/promela/dyntick/dyntickRCU-irq-nmi-ssl=dyntick_nmi.fcv}

Of course, the fact that we have NMIs requires adjustments in
the other components.
For example, the \co{EXECUTE_MAINLINE()} macro now needs to
pay attention to the NMI handler (\co{in_dyntick_nmi}) as well
as the interrupt handler (\co{in_dyntick_irq}) by checking
the \co{dyntick_nmi_done} variable as follows:

\input{CodeSamples/formal/promela/dyntick/dyntickRCU-irq-nmi-ssl=EXECUTE_MAINLINE.fcv}

We will also need to introduce an \co{EXECUTE_IRQ()}
macro that checks \co{in_dyntick_nmi} in order to allow
\co{dyntick_irq()} to exclude \co{dyntick_nmi()}:

\input{CodeSamples/formal/promela/dyntick/dyntickRCU-irq-nmi-ssl=EXECUTE_IRQ.fcv}

It is further necessary to convert \co{dyntick_irq()}
to \co{EXECUTE_IRQ()} as follows:

\input{CodeSamples/formal/promela/dyntick/dyntickRCU-irq-nmi-ssl=dyntick_irq.fcv}

\begin{fcvref}[ln:formal:promela:dyntick:dyntickRCU-irq-nmi-ssl:dyntick_irq]
Note that we have open-coded the \qco{if} statements
(for example, \clnrefrange{stmt1:b}{stmt1:e}).
In addition, statements that process strictly local state
(such as \clnref{inc_i}) need not exclude \co{dyntick_nmi()}.
\end{fcvref}

Finally, \co{grace_period()} requires only a few changes:

\input{CodeSamples/formal/promela/dyntick/dyntickRCU-irq-nmi-ssl=grace_period.fcv}

\begin{fcvref}[ln:formal:promela:dyntick:dyntickRCU-irq-nmi-ssl:grace_period]
We have added the \co{printf()} for the new
\co{MAX_DYNTICK_LOOP_NMI} parameter on \clnref{MDL_NMI} and
added \co{dyntick_nmi_done} to the \co{shouldexit}
assignments on \clnref{nmi_done1,nmi_done2}.
\end{fcvref}

The model (\path{dyntickRCU-irq-nmi-ssl.spin})
results in a correct verification with several hundred million
states, passing without errors.

\QuickQuiz{
	Does Paul \emph{always} write his code in this painfully incremental
	manner?
}\QuickQuizAnswer{
	Not always, but more and more frequently.
	In this case, Paul started with the smallest slice of code that
	included an interrupt handler, because he was not sure how best
	to model interrupts in Promela.
	Once he got that working, he added other features.
	(But if he was doing it again, he would start with a ``toy'' handler.
	For example, he might have the handler increment a variable twice and
	have the mainline code verify that the value was always even.)

	Why the incremental approach?
	Consider the following, attributed to Brian W. Kernighan:

	\begin{quote}
		Debugging is twice as hard as writing the code in the first
		place.
		Therefore, if you write the code as cleverly as possible,
		you are, by definition, not smart enough to debug it.
	\end{quote}

	This means that any attempt to optimize the production of code should
	place at least 66\,\% of its emphasis on optimizing the debugging process,
	even at the expense of increasing the time and effort spent coding.
	Incremental coding and testing is one way to optimize the debugging
	process, at the expense of some increase in coding effort.
	Paul uses this approach because he rarely has the luxury of
	devoting full days (let alone weeks) to coding and debugging.
}\QuickQuizEnd

\subsubsection{Lessons (Re)Learned}
\label{sec:formal:Lessons (Re)Learned}

This effort provided some lessons (re)learned:

\begin{enumerate}
\item	{\bf Promela and Spin can verify interrupt\slash NMI\-/handler
	interactions.}
\item	{\bf Documenting code can help locate bugs.}
	In this case, the documentation effort located
	a misplaced memory barrier in
	\co{rcu_enter_nohz()} and \co{rcu_exit_nohz()},
	as shown by the following patch~\cite{PaulEMcKenney2008commit:ae66be9b71b1}.

\begin{VerbatimU}
 static inline void rcu_enter_nohz(void)
 {
+       mb();
        __get_cpu_var(dynticks_progress_counter)++;
-       mb();
 }

 static inline void rcu_exit_nohz(void)
 {
-       mb();
        __get_cpu_var(dynticks_progress_counter)++;
+       mb();
 }
\end{VerbatimU}

\item	{\bf Validate your code early, often, and up to the point
	of destruction.}
	This effort located one subtle bug in
	\co{rcu_try_flip_waitack_needed()}
	that would have been quite difficult to test or debug, as
	shown by the following patch~\cite{PaulEMcKenney2008commit:d7c0651390b6}.

\begin{VerbatimU}
-       if ((curr - snap) > 2 || (snap & 0x1) == 0)
+       if ((curr - snap) > 2 || (curr & 0x1) == 0)
\end{VerbatimU}

\item	{\bf Always verify your verification code.}
	The usual way to do this is to insert a deliberate bug
	and verify that the verification code catches it.
	Of course, if the verification code fails to catch this bug,
	you may also need to verify the bug itself, and so on,
	recursing infinitely.
	However, if you find yourself in this position,
	getting a good night's sleep
	can be an extremely effective debugging technique.
	You will then see that the obvious verify-the-verification
	technique is to deliberately insert bugs in the code being
	verified.
	If the verification fails to find them, the verification clearly
	is buggy.
\item	{\bf Use of atomic instructions can simplify verification.}
	Unfortunately, use of the \co{cmpxchg} atomic instruction
	would also slow down the critical \IXacr{irq} fastpath, so they
	are not appropriate in this case.
\item	{\bf The need for complex formal verification often indicates
	a need to re-think your design.}
\end{enumerate}

To this last point, it turns out that there is a much simpler solution to
the dynticks problem, which is presented in the next section.

\subsubsection{Simplicity Avoids Formal Verification}
\label{sec:formal:Simplicity Avoids Formal Verification}

The complexity of the dynticks interface for preemptible RCU is primarily
due to the fact that both \IRQ s and NMIs use the same code path and the
same state variables.
This leads to the notion of providing separate code paths and variables
for \IRQ s and NMIs, as has been done for
hierarchical RCU~\cite{PaulEMcKenney2008HierarchicalRCU}
as indirectly suggested by
Manfred Spraul~\cite{ManfredSpraul2008StateMachineRCU}.
This work was pulled into mainline kernel during the v2.6.29
development cycle~\cite{PaulEMcKenney2008commit:64db4cfff99c}.

\subsubsection{State Variables for Simplified Dynticks Interface}
\label{sec:formal:State Variables for Simplified Dynticks Interface}

\Cref{lst:formal:Variables for Simple Dynticks Interface}
shows the new per-CPU state variables.
These variables are grouped into structs to allow multiple independent
RCU implementations (e.g., \co{rcu} and \co{rcu_bh}) to conveniently
and efficiently share dynticks state.
In what follows, they can be thought of as independent per-CPU variables.

\begin{listing}
\begin{VerbatimL}
struct rcu_dynticks {
	int dynticks_nesting;
	int dynticks;
	int dynticks_nmi;
};

struct rcu_data {
	...
	int dynticks_snap;
	int dynticks_nmi_snap;
	...
};
\end{VerbatimL}
\caption{Variables for Simple Dynticks Interface}
\label{lst:formal:Variables for Simple Dynticks Interface}
\end{listing}

The \co{dynticks_nesting}, \co{dynticks}, and \co{dynticks_snap} variables
are for the \IRQ\ code paths, and the \co{dynticks_nmi} and
\co{dynticks_nmi_snap} variables are for the NMI code paths, although
the NMI code path will also reference (but not modify) the
\co{dynticks_nesting} variable.
These variables are used as follows:

\begin{description}[style=nextline]
\item	[\tco{dynticks_nesting}]
	This counts the number of reasons that the corresponding
	CPU should be monitored for RCU read-side critical sections.
	If the CPU is in dynticks-idle mode, then this counts the
	\IRQ\ nesting level, otherwise it is one greater than the
	\IRQ\ nesting level.
\item	[\tco{dynticks}]
	This counter's value is even if the corresponding CPU is
	in dynticks-idle mode and there are no \IRQ\ handlers currently
	running on that CPU, otherwise the counter's value is odd.
	In other words, if this counter's value is odd, then the
	corresponding CPU might be in an RCU read-side critical section.
\item	[\tco{dynticks_nmi}]
	This counter's value is odd if the corresponding CPU is
	in an NMI handler, but only if the NMI arrived while this
	CPU was in dyntick-idle mode with no \IRQ\ handlers running.
	Otherwise, the counter's value will be even.
\item	[\tco{dynticks_snap}]
	This will be a snapshot of the \co{dynticks} counter, but
	only if the current RCU grace period has extended for too
	long a duration.
\item	[\tco{dynticks_nmi_snap}]
	This will be a snapshot of the \co{dynticks_nmi} counter, but
	again only if the current RCU grace period has extended for too
	long a duration.
\end{description}

If both \co{dynticks} and \co{dynticks_nmi} have taken on an even
value during a given time interval, then the corresponding CPU has
passed through a \IX{quiescent state} during that interval.

\QuickQuiz{
	But what happens if an NMI handler starts running before
	an \IRQ\ handler completes, and if that NMI handler continues
	running until a second \IRQ\ handler starts?
}\QuickQuizAnswer{
	This cannot happen within the confines of a single CPU\@.
	The first \IRQ\ handler cannot complete until the NMI handler
	returns.
	Therefore, if each of the \co{dynticks} and \co{dynticks_nmi}
	variables have taken on an even value during a given time
	interval, the corresponding CPU really was in a quiescent
	state at some time during that interval.
}\QuickQuizEnd

\subsubsection{Entering and Leaving Dynticks-Idle Mode}
\label{sec:formal:Entering and Leaving Dynticks-Idle Mode}

\Cref{lst:formal:Entering and Exiting Dynticks-Idle Mode}
shows the \co{rcu_enter_nohz()} and \co{rcu_exit_nohz()},
which enter and exit dynticks-idle mode, also known as ``nohz'' mode.
These two functions are invoked from process context.

\begin{listing}
\begin{fcvlabel}[ln:formal:Entering and Exiting Dynticks-Idle Mode]
\begin{VerbatimL}[commandchars=\\\[\]]
void rcu_enter_nohz(void)
{
	unsigned long flags;
	struct rcu_dynticks *rdtp;

	smp_mb();			\lnlbl[mb]
	local_irq_save(flags);		\lnlbl[irq_sv]
	rdtp = &__get_cpu_var(rcu_dynticks); \lnlbl[get_ptr]
	rdtp->dynticks++;		\lnlbl[inc_cnt]
	rdtp->dynticks_nesting--;	\lnlbl[dec_nst]
	WARN_ON(rdtp->dynticks & 0x1);
	local_irq_restore(flags);	\lnlbl[irq_rs]
}

void rcu_exit_nohz(void)
{
	unsigned long flags;
	struct rcu_dynticks *rdtp;

	local_irq_save(flags);
	rdtp = &__get_cpu_var(rcu_dynticks);
	rdtp->dynticks++;
	rdtp->dynticks_nesting++;
	WARN_ON(!(rdtp->dynticks & 0x1));
	local_irq_restore(flags);
	smp_mb();
}
\end{VerbatimL}
\end{fcvlabel}
\caption{Entering and Exiting Dynticks-Idle Mode}
\label{lst:formal:Entering and Exiting Dynticks-Idle Mode}
\end{listing}

\begin{fcvref}[ln:formal:Entering and Exiting Dynticks-Idle Mode]
\Clnref{mb} ensures that any prior memory accesses (which might
include accesses from RCU read-side critical sections) are seen
by other CPUs before those marking entry to dynticks-idle mode.
\Clnref{irq_sv,irq_rs} disable and reenable \IRQ s.
\Clnref{get_ptr} acquires a pointer to the current CPU's \co{rcu_dynticks}
structure, and
\clnref{inc_cnt} increments the current CPU's \co{dynticks} counter, which
should now be even, given that we are entering dynticks-idle mode
in process context.
Finally, \clnref{dec_nst} decrements \co{dynticks_nesting},
which should now be zero.
\end{fcvref}

The \co{rcu_exit_nohz()} function is quite similar, but increments
\co{dynticks_nesting} rather than decrementing it and checks for
the opposite \co{dynticks} polarity.

\subsubsection{NMIs From Dynticks-Idle Mode}
\label{sec:formal:NMIs From Dynticks-Idle Mode}

\begin{fcvref}[ln:formal:NMIs From Dynticks-Idle Mode]
\Cref{lst:formal:NMIs From Dynticks-Idle Mode}
shows the \co{rcu_nmi_enter()} and \co{rcu_nmi_exit()} functions,
which inform RCU of NMI entry and exit, respectively, from dynticks-idle
mode.
However, if the NMI arrives during an \IRQ\ handler, then RCU will already
be on the lookout for RCU read-side critical sections from this CPU,
so \clnref{chk_ext1,ret1} of \co{rcu_nmi_enter()} and \clnref{chk_ext2,ret2}
of \co{rcu_nmi_exit()} silently return if \co{dynticks} is odd.
Otherwise, the two functions increment \co{dynticks_nmi}, with
\co{rcu_nmi_enter()} leaving it with an odd value and \co{rcu_nmi_exit()}
leaving it with an even value.
Both functions execute memory barriers between this increment
and possible RCU read-side critical sections on \clnref{mb1,mb2},
respectively.
\end{fcvref}

\begin{listing}
\begin{fcvlabel}[ln:formal:NMIs From Dynticks-Idle Mode]
\begin{VerbatimL}[commandchars=\\\[\]]
void rcu_nmi_enter(void)
{
	struct rcu_dynticks *rdtp;

	rdtp = &__get_cpu_var(rcu_dynticks);
	if (rdtp->dynticks & 0x1)	\lnlbl[chk_ext1]
		return;			\lnlbl[ret1]
	rdtp->dynticks_nmi++;
	WARN_ON(!(rdtp->dynticks_nmi & 0x1));
	smp_mb();			\lnlbl[mb1]
}

void rcu_nmi_exit(void)
{
	struct rcu_dynticks *rdtp;

	rdtp = &__get_cpu_var(rcu_dynticks);
	if (rdtp->dynticks & 0x1)	\lnlbl[chk_ext2]
		return;			\lnlbl[ret2]
	smp_mb();			\lnlbl[mb2]
	rdtp->dynticks_nmi++;
	WARN_ON(rdtp->dynticks_nmi & 0x1);
}
\end{VerbatimL}
\end{fcvlabel}
\caption{NMIs From Dynticks-Idle Mode}
\label{lst:formal:NMIs From Dynticks-Idle Mode}
\end{listing}

\subsubsection{Interrupts From Dynticks-Idle Mode}
\label{sec:formal:Interrupts From Dynticks-Idle Mode}

\begin{fcvref}[ln:formal:Interrupts From Dynticks-Idle Mode]
\Cref{lst:formal:Interrupts From Dynticks-Idle Mode}
shows \co{rcu_irq_enter()} and \co{rcu_irq_exit()}, which
inform RCU of entry to and exit from, respectively, \IRQ\ context.
\Clnref{inc_nst} of \co{rcu_irq_enter()} increments \co{dynticks_nesting},
and if this variable was already non-zero, \clnref{ret1} silently returns.
Otherwise, \clnref{inc_dynt1} increments \co{dynticks},
which will then have
an odd value, consistent with the fact that this CPU can now
execute RCU read-side critical sections.
\Clnref{mb1} therefore executes a memory barrier to ensure that
the increment of \co{dynticks} is seen before any
RCU read-side critical sections that the subsequent \IRQ\ handler
might execute.

\begin{listing}
\begin{fcvlabel}[ln:formal:Interrupts From Dynticks-Idle Mode]
\begin{VerbatimL}[commandchars=\\\[\]]
void rcu_irq_enter(void)
{
	struct rcu_dynticks *rdtp;

	rdtp = &__get_cpu_var(rcu_dynticks);
	if (rdtp->dynticks_nesting++)		\lnlbl[inc_nst]
		return;				\lnlbl[ret1]
	rdtp->dynticks++;			\lnlbl[inc_dynt1]
	WARN_ON(!(rdtp->dynticks & 0x1));
	smp_mb();				\lnlbl[mb1]
}

void rcu_irq_exit(void)
{
	struct rcu_dynticks *rdtp;

	rdtp = &__get_cpu_var(rcu_dynticks);
	if (--rdtp->dynticks_nesting)		\lnlbl[dec_nst]
		return;				\lnlbl[ret2]
	smp_mb();				\lnlbl[mb2]
	rdtp->dynticks++;			\lnlbl[inc_dynt2]
	WARN_ON(rdtp->dynticks & 0x1);		\lnlbl[chk_even]
	if (__get_cpu_var(rcu_data).nxtlist ||	\lnlbl[chk_cb:b]
	    __get_cpu_var(rcu_bh_data).nxtlist)
		set_need_resched();		\lnlbl[chk_cb:e]
}
\end{VerbatimL}
\end{fcvlabel}
\caption{Interrupts From Dynticks-Idle Mode}
\label{lst:formal:Interrupts From Dynticks-Idle Mode}
\end{listing}

\Clnref{dec_nst} of \co{rcu_irq_exit()} decrements
\co{dynticks_nesting}, and
if the result is non-zero, \clnref{ret2} silently returns.
Otherwise, \clnref{mb2} executes a memory barrier to ensure that the
increment of \co{dynticks} on \clnref{inc_dynt2} is seen after any RCU
read-side critical sections that the prior \IRQ\ handler might have executed.
\Clnref{chk_even} verifies that \co{dynticks} is now even, consistent with
the fact that no RCU read-side critical sections may appear in
dynticks-idle mode.
\Clnrefrange{chk_cb:b}{chk_cb:e} check to see
if the prior \IRQ\ handlers enqueued any
RCU callbacks, forcing this CPU out of dynticks-idle mode via
a reschedule API if so.
\end{fcvref}

\subsubsection{Checking For Dynticks Quiescent States}
\label{sec:formal:Checking For Dynticks Quiescent States}

\begin{fcvref}[ln:formal:Saving Dyntick Progress Counters]
\Cref{lst:formal:Saving Dyntick Progress Counters}
shows \co{dyntick_save_progress_counter()}, which takes a snapshot
of the specified CPU's \co{dynticks} and \co{dynticks_nmi}
counters.
\Clnref{snap,snapn} snapshot these two variables to locals, \clnref{mb}
executes a memory barrier to pair with the memory barriers in the functions in
\cref{lst:formal:Entering and Exiting Dynticks-Idle Mode,%
lst:formal:NMIs From Dynticks-Idle Mode,%
lst:formal:Interrupts From Dynticks-Idle Mode}.
\Clnref{rec_snap,rec_snapn} record the snapshots for later calls to
\co{rcu_implicit_dynticks_qs()},
and \clnref{chk_prgs} checks to see if the CPU is in dynticks-idle mode with
neither \IRQ s nor NMIs in progress (in other words, both snapshots
have even values), hence in an extended quiescent state.
If so, \clnref{cnt:b,cnt:e} count this event, and \clnref{ret} returns
true if the CPU was in a quiescent state.
\end{fcvref}

\begin{listing}
\begin{fcvlabel}[ln:formal:Saving Dyntick Progress Counters]
\begin{VerbatimL}[commandchars=\\\[\]]
static int
dyntick_save_progress_counter(struct rcu_data *rdp)
{
	int ret;
	int snap;
	int snap_nmi;

	snap = rdp->dynticks->dynticks;		\lnlbl[snap]
	snap_nmi = rdp->dynticks->dynticks_nmi;	\lnlbl[snapn]
	smp_mb();				\lnlbl[mb]
	rdp->dynticks_snap = snap;		\lnlbl[rec_snap]
	rdp->dynticks_nmi_snap = snap_nmi;	\lnlbl[rec_snapn]
	ret = ((snap & 0x1) == 0) && ((snap_nmi & 0x1) == 0); \lnlbl[chk_prgs]
	if (ret)				\lnlbl[cnt:b]
		rdp->dynticks_fqs++;		\lnlbl[cnt:e]
	return ret;				\lnlbl[ret]
}
\end{VerbatimL}
\end{fcvlabel}
\caption{Saving Dyntick Progress Counters}
\label{lst:formal:Saving Dyntick Progress Counters}
\end{listing}

\begin{fcvref}[ln:formal:Checking Dyntick Progress Counters]
\Cref{lst:formal:Checking Dyntick Progress Counters}
shows \co{rcu_implicit_dynticks_qs()}, which is called to check
whether a CPU has entered dyntick-idle mode subsequent to a call
to \co{dynticks_save_progress_counter()}.
\Clnref{curr,currn} take new snapshots of the corresponding CPU's
\co{dynticks} and \co{dynticks_nmi} variables, while
\clnref{snap,snapn} retrieve the snapshots saved earlier by
\co{dynticks_save_progress_counter()}.
\Clnref{mb} then
executes a memory barrier to pair with the memory barriers in
the functions in
\cref{lst:formal:Entering and Exiting Dynticks-Idle Mode,%
lst:formal:NMIs From Dynticks-Idle Mode,%
lst:formal:Interrupts From Dynticks-Idle Mode}.
\Clnrefrange{chk_q:b}{chk_q:e}
then check to see if the CPU is either currently in
a quiescent state (\co{curr} and \co{curr_nmi} having even values) or
has passed through a quiescent state since the last call to
\co{dynticks_save_progress_counter()} (the values of
\co{dynticks} and \co{dynticks_nmi} having changed).
If these checks confirm that the CPU has passed through a dyntick-idle
quiescent state, then \clnref{cnt} counts that fact and
\clnref{ret_1} returns an indication of this fact.
Either way, \clnref{chk_race}
checks for race conditions that can result in RCU
waiting for a CPU that is offline.
\end{fcvref}

\begin{listing}
\begin{fcvlabel}[ln:formal:Checking Dyntick Progress Counters]
\begin{VerbatimL}[commandchars=\\\[\]]
static int
rcu_implicit_dynticks_qs(struct rcu_data *rdp)
{
	long curr;
	long curr_nmi;
	long snap;
	long snap_nmi;

	curr = rdp->dynticks->dynticks;		\lnlbl[curr]
	snap = rdp->dynticks_snap;		\lnlbl[snap]
	curr_nmi = rdp->dynticks->dynticks_nmi;	\lnlbl[currn]
	snap_nmi = rdp->dynticks_nmi_snap;	\lnlbl[snapn]
	smp_mb();				\lnlbl[mb]
	if ((curr != snap || (curr & 0x1) == 0) && \lnlbl[chk_q:b]
	    (curr_nmi != snap_nmi || (curr_nmi & 0x1) == 0)) { \lnlbl[chk_q:e]
		rdp->dynticks_fqs++;		\lnlbl[cnt]
		return 1;			\lnlbl[ret_1]
	}
	return rcu_implicit_offline_qs(rdp);	\lnlbl[chk_race]
}
\end{VerbatimL}
\end{fcvlabel}
\caption{Checking Dyntick Progress Counters}
\label{lst:formal:Checking Dyntick Progress Counters}
\end{listing}

\QuickQuiz{
	This is still pretty complicated.
	Why not just have a \co{cpumask_t} with per-CPU bits, clearing
	the bit when entering an \IRQ\ or NMI handler, and setting it
	upon exit?
}\QuickQuizAnswer{
	Although this approach would be functionally correct, it
	would result in excessive \IRQ\ entry/exit overhead on
	large machines.
	In contrast, the approach laid out in this section allows
	each CPU to touch only per-CPU data on \IRQ\ and NMI entry/exit,
	resulting in much lower \IRQ\ entry/exit overhead, especially
	on large machines.
}\QuickQuizEnd

Linux-kernel RCU's dyntick-idle code has since been rewritten yet again
based on a suggestion from
Andy Lutomirski~\cite{PaulMcKenney2015dyntickAndyNMI},
but it is time to sum up and move on to other topics.

\subsubsection{Discussion}
\label{sec:formal:Discussion}

A slight shift in viewpoint resulted in a substantial simplification
of the dynticks interface for RCU\@.
The key change leading to this simplification was minimizing of
sharing between \IRQ\ and NMI contexts.
The only sharing in this simplified interface is references from NMI
context to \IRQ\ variables (the \co{dynticks} variable).
This type of sharing is benign, because the NMI functions never update
this variable, so that its value remains constant through the lifetime
of the NMI handler.
This limitation of sharing allows the individual functions to be
understood one at a time, in happy contrast to the situation
described in
\cref{sec:formal:Promela Parable: dynticks and Preemptible RCU},
where an NMI might change shared state at any point during execution of
the \IRQ\ functions.

Verification can be a good thing, but simplicity is even better.

% Restore frame around VerbatimN in two-column layout
\IfTwoColumn{
\RecustomVerbatimEnvironment{VerbatimN}{Verbatim}%
{numbers=left,numbersep=3pt,xleftmargin=5pt,xrightmargin=5pt,frame=single}
\setlength{\lnlblraise}{6pt}
}{}

% formal/ppcmem.tex
% mainfile: ../perfbook.tex
% SPDX-License-Identifier: CC-BY-SA-3.0

\section{Special-Purpose State-Space Search}
\label{sec:formal:Special-Purpose State-Space Search}
\epigraph{Jack of all trades, master of none.}{Unknown}

Although Promela and Spin allow you to verify pretty much any (smallish)
algorithm, their very generality can sometimes be a curse.
For example, Promela does not understand memory models or any sort
of reordering semantics.
This section therefore describes some state-space search tools that
understand memory models used by production systems, greatly simplifying the
verification of weakly ordered code.

For example,
\cref{sec:formal:Promela Example: QRCU}
showed how to convince Promela to account for weak memory ordering.
Although this approach can work well, it requires that the developer
fully understand the system's memory model.
Unfortunately, few (if any) developers fully understand the complex
memory models of modern CPUs.

Therefore, another approach is to use a tool that already understands
this memory ordering, such as the PPCMEM tool produced by
\ppl{Peter}{Sewell} and \ppl{Susmit}{Sarkar} at the University of Cambridge,
\ppl{Luc}{Maranget}, \ppl{Francesco Zappa}{Nardelli}, and
\ppl{Pankaj}{Pawan} at INRIA, and \ppl{Jade}{Alglave} at Oxford University,
in cooperation with \ppl{Derek}{Williams} of
IBM~\cite{JadeAlglave2011ppcmem}.
This group formalized the memory models of Power, \ARM, x86, as well
as that of the C/C++11 standard~\cite{RichardSmith2019N4800}, and
produced the PPCMEM tool based on the Power and \ARM\ formalizations.

\QuickQuiz{
	But x86 has strong memory ordering, so why formalize its memory
	model?
}\QuickQuizAnswer{
	Actually, academics consider the x86 memory model to be weak
	because it can allow prior stores to be reordered with
	subsequent loads.
	From an academic viewpoint, a strong memory model is one
	that allows absolutely no reordering, so that all threads
	agree on the order of all operations visible to them.

	Plus it really is the case that developers are sometimes confused
	about x86 memory ordering.
}\QuickQuizEnd

The PPCMEM tool takes \emph{litmus tests} as input.
A sample litmus test is presented in
\cref{sec:formal:Anatomy of a Litmus Test}.
\Cref{sec:formal:What Does This Litmus Test Mean?}
relates this litmus test to the equivalent C-language program,
\cref{sec:formal:Running a Litmus Test} describes how to
apply PPCMEM to this litmus test, and
\cref{sec:formal:PPCMEM Discussion}
discusses the implications.

\subsection{Anatomy of a Litmus Test}
\label{sec:formal:Anatomy of a Litmus Test}

An example PowerPC litmus test for PPCMEM is shown in
\cref{lst:formal:PPCMEM Litmus Test}.
The ARM interface works the same way, but with \ARM\ instructions
substituted for the Power instructions and with the initial \qco{PPC}
replaced by \qco{ARM}.

\begin{listing}
\begin{fcvlabel}[ln:formal:PPCMEM Litmus Test]
\begin{VerbatimL}[commandchars=\@\[\]]
PPC SB+lwsync-RMW-lwsync+isync-simple		@lnlbl[type]
""						@lnlbl[altname]
{						@lnlbl[init:b]
0:r2=x; 0:r3=2; 0:r4=y; 0:r10=0; 0:r11=0; 0:r12=z; @lnlbl[init:0]
1:r2=y; 1:r4=x;					@lnlbl[init:1]
}						@lnlbl[init:e]
 P0                 | P1           ;		@lnlbl[procid]
 li r1,1            | li r1,1      ;		@lnlbl[reginit]
 stw r1,0(r2)       | stw r1,0(r2) ;		@lnlbl[stw]
 lwsync             | sync         ; @lnlbl[P0lwsync] @lnlbl[P1sync]
                    | lwz r3,0(r4) ; @lnlbl[P0empty]  @lnlbl[P1lwz]
 lwarx  r11,r10,r12 | ;		@lnlbl[P0lwarx] @lnlbl[P1empty:b]
 stwcx. r11,r10,r12 | ;		@lnlbl[P0stwcx]
 bne Fail1          | ;		@lnlbl[P0bne]
 isync              | ;		@lnlbl[P0isync]
 lwz r3,0(r4)       | ;		@lnlbl[P0lwz]
 Fail1:             | ;		@lnlbl[P0fail1] @lnlbl[P1empty:e]

exists						@lnlbl[assert:b]
(0:r3=0 /\ 1:r3=0)				@lnlbl[assert:e]
\end{VerbatimL}
\end{fcvlabel}
\caption{PPCMEM Litmus Test}
\label{lst:formal:PPCMEM Litmus Test}
\end{listing}

\begin{fcvref}[ln:formal:PPCMEM Litmus Test]
In the example, \clnref{type} identifies the type of system (\qco{ARM} or
\qco{PPC}) and contains the title for the model.
\Clnref{altname} provides a place for an
alternative name for the test, which you will usually want to leave
blank as shown in the above example.
Comments can be inserted between
\clnref{altname,init:b} using the OCaml (or Pascal) syntax of \nbco{(* *)}.

\Clnrefrange{init:b}{init:e} give initial values for all registers;
each is of the form
\co{P:R=V}, where \co{P} is the process identifier, \co{R} is the register
identifier, and \co{V} is the value.
For example, process~0's register \co{r3} initially contains the value~2.
If the value is a variable (\co{x}, \co{y}, or \co{z} in the example)
then the register is initialized to the address of the variable.
It is also possible to initialize the contents of variables, for example,
\co{x=1} initializes the value of \co{x} to~1.
Uninitialized variables default to the value zero, so that in the
example, \co{x}, \co{y}, and~\co{z} are all initially zero.

\Clnref{procid} provides identifiers for the two processes, so that
the \co{0:r3=2} on \clnref{init:0} could instead have been written
\co{P0:r3=2}.
\Clnref{procid} is required, and the identifiers must be of the form
\co{Pn}, where \co{n} is the column number, starting from zero for
the left-most column.
This may seem unnecessarily strict, but it does prevent considerable
confusion in actual use.
\end{fcvref}

\QuickQuiz{
	\begin{fcvref}[ln:formal:PPCMEM Litmus Test]
	Why does \clnref{reginit} of \cref{lst:formal:PPCMEM Litmus Test}
	initialize the registers?
	Why not instead initialize them on \clnref{init:0,init:1}?
	\end{fcvref}
}\QuickQuizAnswer{
	Either way works.
	However, in general, it is better to use initialization than
	explicit instructions.
	The explicit instructions are used in this example to demonstrate
	their use.
	In addition, many of the litmus tests available on the tool's
	web site (\url{https://www.cl.cam.ac.uk/~pes20/ppcmem/}) were
	automatically generated, which generates explicit
	initialization instructions.
}\QuickQuizEnd

\begin{fcvref}[ln:formal:PPCMEM Litmus Test]
\Clnrefrange{reginit}{P0fail1} are the lines of code for each process.
A given process can have empty lines, as is the case for P0's
\clnref{P0empty} and P1's \clnrefrange{P1empty:b}{P1empty:e}.
Labels and branches are permitted, as demonstrated by the branch
on \clnref{P0bne} to the label on \clnref{P0fail1}.
That said, too-free use of branches will expand the state space.
Use of loops is a particularly good way to explode your state space.

\Clnrefrange{assert:b}{assert:e} show the assertion, which in this case
indicates that we are interested in whether P0's and P1's \co{r3} registers
can both contain zero after both threads complete execution.
This assertion is important because there are a number of use cases
that would fail miserably if both P0 and P1 saw zero in their
respective \co{r3} registers.

This should give you enough information to construct simple litmus tests.
Some additional documentation is available, though much of this
additional documentation is intended for a different research tool that
runs tests on actual hardware.
Perhaps more importantly, a large number of pre-existing litmus tests
are available with the online tool (available via the ``Select ARM Test''
and ``Select POWER Test'' buttons at
\url{https://www.cl.cam.ac.uk/~pes20/ppcmem/}).
It is quite likely that one of these pre-existing litmus tests will
answer your Power or \ARM\ memory-ordering question.

\subsection{What Does This Litmus Test Mean?}
\label{sec:formal:What Does This Litmus Test Mean?}

P0's \clnref{reginit,stw} are equivalent to the C statement \co{x=1}
because \clnref{init:0} defines P0's register \co{r2} to be the address
of~\co{x}.
P0's \clnref{P0lwarx,P0stwcx} are the mnemonics for load-linked
(``load register exclusive'' in \ARM\ parlance and ``load reserve''
in Power parlance) and store-conditional (``store register exclusive''
in \ARM\ parlance), respectively.
When these are used together, they form an atomic instruction sequence,
roughly similar to the \IXacrml{cas} sequences exemplified by the
x86 \co{lock;cmpxchg} instruction.
Moving to a higher level of abstraction, the sequence from
\clnrefrange{P0lwsync}{P0isync}
is equivalent to the Linux kernel's \co{atomic_add_return(&z, 0)}.
Finally, \clnref{P0lwz} is roughly equivalent to the C statement \co{r3=y}.

P1's \clnref{reginit,stw} are equivalent to the C statement \co{y=1},
\clnref{P1sync}
is a memory barrier, equivalent to the Linux kernel statement \co{smp_mb()},
and \clnref{P1lwz} is equivalent to the C statement \co{r3=x}.
\end{fcvref}

\QuickQuiz{
	\begin{fcvref}[ln:formal:PPCMEM Litmus Test]
	But whatever happened to \clnref{P0fail1} of
	\cref{lst:formal:PPCMEM Litmus Test},
	the one that is the \co{Fail1:} label?
	\end{fcvref}
}\QuickQuizAnswer{
	The implementation of PowerPC version of \co{atomic_add_return()}
	loops when the \co{stwcx} instruction fails, which it communicates
	by setting non-zero status in the condition-code register,
	which in turn is tested by the \co{bne} instruction.
	Because actually modeling the loop would result in state-space
	explosion, we instead branch to the \co{Fail1:} label,
	terminating the model with the initial value of~2 in P0's \co{r3}
	register, which will not trigger the exists assertion.

	There is some debate about whether this trick is universally
	applicable, but I have not seen an example where it fails.
}\QuickQuizEnd

\begin{listing}
\begin{VerbatimL}
void P0(void)
{
	int r3;

	x = 1; /* Lines 8 and 9 */
	atomic_add_return(&z, 0); /* Lines 10-15 */
	r3 = y; /* Line 16 */
}

void P1(void)
{
	int r3;

	y = 1; /* Lines 8-9 */
	smp_mb(); /* Line 10 */
	r3 = x; /* Line 11 */
}
\end{VerbatimL}
\caption{Meaning of PPCMEM Litmus Test}
\label{lst:formal:Meaning of PPCMEM Litmus Test}
\end{listing}

Putting all this together, the C-language equivalent to the entire litmus
test is as shown in
\cref{lst:formal:Meaning of PPCMEM Litmus Test}.
The key point is that if \co{atomic_add_return()} acts as a full
memory barrier (as the Linux kernel requires it to), 
then it should be impossible for \co{P0()}'s and \co{P1()}'s \co{r3}
variables to both be zero after execution completes.

The next section describes how to run this litmus test.

\subsection{Running a Litmus Test}
\label{sec:formal:Running a Litmus Test}

As noted earlier, litmus tests may be run interactively via
\url{https://www.cl.cam.ac.uk/~pes20/ppcmem/}, which can help build an
understanding of the memory model.
However, this approach requires that the user manually carry out the
full state-space search.
Because it is very difficult to be sure that you have checked every
possible sequence of events, a separate tool is provided for this
purpose~\cite{PaulEMcKenney2011ppcmem}.

\begin{listing}
\begin{VerbatimL}[numbers=none,xleftmargin=0pt]
./ppcmem -model lwsync_read_block \
         -model coherence_points filename.litmus
...
States 6
0:r3=0; 1:r3=0;
0:r3=0; 1:r3=1;
0:r3=1; 1:r3=0;
0:r3=1; 1:r3=1;
0:r3=2; 1:r3=0;
0:r3=2; 1:r3=1;
Ok
Condition exists (0:r3=0 /\ 1:r3=0)
Hash=e2240ce2072a2610c034ccd4fc964e77
Observation SB+lwsync-RMW-lwsync+isync Sometimes 1
\end{VerbatimL}
\caption{PPCMEM Detects an Error}
\label{lst:formal:PPCMEM Detects an Error}
\end{listing}

Because the litmus test shown in
\cref{lst:formal:PPCMEM Litmus Test}
contains read-modify-write instructions, we must add \co{-model}
arguments to the command line.
If the litmus test is stored in \co{filename.litmus},
this will result in the output shown in
\cref{lst:formal:PPCMEM Detects an Error},
where the \co{...} stands for voluminous making-progress output.
The list of states includes \co{0:r3=0; 1:r3=0;}, indicating once again
that the old PowerPC implementation of \co{atomic_add_return()} does
not act as a full barrier.
The ``Sometimes'' on the last line confirms this:
The assertion triggers for some executions, but not all of the time.

\begin{listing}
\begin{VerbatimL}[numbers=none,xleftmargin=0pt]
./ppcmem -model lwsync_read_block \
         -model coherence_points filename.litmus
...
States 5
0:r3=0; 1:r3=1;
0:r3=1; 1:r3=0;
0:r3=1; 1:r3=1;
0:r3=2; 1:r3=0;
0:r3=2; 1:r3=1;
No (allowed not found)
Condition exists (0:r3=0 /\ 1:r3=0)
Hash=77dd723cda9981248ea4459fcdf6097d
Observation SB+lwsync-RMW-lwsync+sync Never 0 5
\end{VerbatimL}
\caption{PPCMEM on Repaired Litmus Test}
\label{lst:formal:PPCMEM on Repaired Litmus Test}
\end{listing}

The fix to this Linux-kernel bug is to replace P0's \co{isync} with
\co{sync}, which results in the output shown in
\cref{lst:formal:PPCMEM on Repaired Litmus Test}.
As you can see, \co{0:r3=0; 1:r3=0;} does not appear in the list of states,
and the last line calls out ``Never''.
Therefore, the model predicts that the offending execution sequence
cannot happen.

\QuickQuizSeries{%
\QuickQuizB{
	Does the \ARM\ Linux kernel have a similar bug?
}\QuickQuizAnswer{
	\ARM\ does not have this particular bug because it places
	\co{smp_mb()} before and after the \co{atomic_add_return()}
	function's assembly-language implementation.
	PowerPC no longer has this bug; it has long since been
	fixed~\cite{BenjaminHerrenschmidt2011:powerpc:atomic_return}.
}\QuickQuizEndB
\QuickQuizE{
	\begin{fcvref}[ln:formal:PPCMEM Litmus Test]
	Does the \co{lwsync} on \clnref{P0lwsync} in
	\cref{lst:formal:PPCMEM Litmus Test} provide sufficient ordering?
	\end{fcvref}
}\QuickQuizAnswerE{
	It depends on the semantics required.
	The rest of this answer assumes that the assembly language
	for \co{P0} in
	\cref{lst:formal:PPCMEM Litmus Test}
	is supposed to implement a value-returning atomic operation.

	As is discussed in
	\cref{chp:Advanced Synchronization: Memory Ordering},
	Linux kernel's memory consistency model requires
	value-returning atomic RMW operations to be fully ordered
	on both sides.
	The ordering provided by \co{lwsync} is insufficient for this
	purpose, and so \co{sync} should be used instead.
	This change has since been
	made~\cite{BoqunFeng2015:powerpc:value-returning-atomics}
	in response to an email thread discussing a couple of other litmus
	tests~\cite{Paulmck2015:powerpc:value-returning-atomics}.
	Finding any other bugs that the Linux kernel might have is left
	as an exercise for the reader.

	In other enviroments providing weaker semantics, \co{lwsync}
	might be sufficient.
	But not for the Linux kernel's value-returning atomic operations!
}\QuickQuizEndE
}

\subsection{PPCMEM Discussion}
\label{sec:formal:PPCMEM Discussion}

These tools promise to be of great help to people working on low-level
parallel primitives that run on \ARM\ and on Power.
These tools do have some intrinsic limitations:

\begin{enumerate}
\item	These tools are research prototypes, and as such are unsupported.
\item	These tools do not constitute official statements by IBM or \ARM\
	on their respective CPU architectures.
	For example, both corporations reserve the right to report a bug
	at any time against any version of any of these tools.
	These tools are therefore not a substitute for careful stress
	testing on real hardware.
	Moreover, both the tools and the model that they are based on
	are under active development and might change at any time.
	On the other hand, this model was developed in consultation
	with the relevant hardware experts, so there is good reason to be
	confident that it is a robust representation of the architectures.
\item	These tools currently handle a subset of the instruction set.
	This subset has been sufficient for my purposes, but your mileage
	may vary.
	In particular, the tool handles only word-sized accesses (32 bits),
	and the words accessed must be properly aligned.\footnote{
		But recent work focuses on mixed-size
		accesses~\cite{Flur:2017:MCA:3093333.3009839}.}
	In addition, the tool does not handle some of the weaker variants
	of the \ARM\ memory-barrier instructions, nor does it handle
	arithmetic.
\item	The tools are restricted to small loop-free code fragments
	running on small numbers of threads.
	Larger examples result
	in state-space explosion, just as with similar tools such as
	Promela and Spin.
\item	The full state-space search does not give any indication of how
	each offending state was reached.
	That said, once you realize that the state is in fact reachable,
	it is usually not too hard to find that state using the
	interactive tool.
\item	These tools are not much good for complex data structures, although
	it is possible to create and traverse extremely simple linked
	lists using initialization statements of the form
	\qco{x=y; y=z; z=42;}.
\item	These tools do not handle memory mapped I/O or device registers.
	Of course, handling such things would require that they be
	formalized, which does not appear to be in the offing.
\item	The tools will detect only those problems for which you code an
	assertion.
	This weakness is common to all formal methods, and is yet another
	reason why testing remains important.
	In the immortal words of Donald Knuth quoted at the beginning of
	this chapter, ``Beware of bugs in the above code;
	I have only proved it correct, not tried it.''
\end{enumerate}

That said, one strength of these tools is that they are designed to
model the full range of behaviors allowed by the architectures, including
behaviors that are legal, but which current hardware implementations do
not yet inflict on unwary software developers.
Therefore, an algorithm that is vetted by these tools likely has some
additional safety margin when running on real hardware.
Furthermore, testing on real hardware can only find bugs; such testing
is inherently incapable of proving a given usage correct.
To appreciate this, consider that the researchers routinely ran in excess
of 100 billion test runs on real hardware to validate their model.
In one case, behavior that is allowed by the architecture did not occur,
despite 176 billion runs~\cite{JadeAlglave2011ppcmem}.
In contrast, the
full-state-space search allows the tool to prove code fragments correct.

It is worth repeating that formal methods and tools are no substitute for
testing.
The fact is that producing large reliable concurrent software artifacts,
the Linux kernel for example, is quite difficult.
Developers must therefore be prepared to apply every tool at their
disposal towards this goal.
The tools presented in this chapter are able to locate bugs that are
quite difficult to produce (let alone track down) via testing.
On the other hand, testing can be applied to far larger bodies of software
than the tools presented in this chapter are ever likely to handle.
As always, use the right tools for the job!

Of course, it is always best to avoid the need to work at this level
by designing your parallel code to be easily partitioned and then
using higher-level primitives (such as locks, sequence counters, atomic
operations, and RCU) to get your job done more straightforwardly.
And even if you absolutely must use low-level memory barriers and
read-modify-write instructions to get your job done, the more
conservative your use of these sharp instruments, the easier your life
is likely to be.

% formal/axiomatic.tex
% mainfile: ../perfbook.tex
% SPDX-License-Identifier: CC-BY-SA-3.0

\section{Axiomatic Approaches}
\label{sec:formal:Axiomatic Approaches}
\OriginallyPublished{Section}{sec:formal:Axiomatic Approaches}{Axiomatic Approaches}{Linux Weekly News}{PaulEMcKenney2014weakaxiom}
\epigraph{Theory helps us to bear our ignorance of facts.}
	{George Santayana}

Although the PPCMEM tool can solve the famous ``independent reads of
independent writes'' (IRIW) litmus test shown in
\cref{lst:formal:IRIW Litmus Test}, doing so requires no less than
fourteen CPU hours and generates no less than ten gigabytes of state space.
That said, this situation is a great improvement over that before the advent
of PPCMEM, where solving this problem required perusing volumes of
reference manuals, attempting proofs, discussing with experts, and
being unsure of the final answer.
Although fourteen hours can seem like a long time, it is much shorter
than weeks or even months.

\begin{listing}
\begin{fcvlabel}[ln:formal:IRIW Litmus Test]
\begin{VerbatimL}[commandchars=\%\@\$]
PPC IRIW.litmus
""
(* Traditional IRIW. *)
{
0:r1=1; 0:r2=x;
1:r1=1;         1:r4=y;
        2:r2=x; 2:r4=y;
        3:r2=x; 3:r4=y;
}
P0           | P1           | P2           | P3           ;
stw r1,0(r2) | stw r1,0(r4) | lwz r3,0(r2) | lwz r3,0(r4) ;
             |              | sync         | sync         ;
             |              | lwz r5,0(r4) | lwz r5,0(r2) ;

exists
(2:r3=1 /\ 2:r5=0 /\ 3:r3=1 /\ 3:r5=0)
\end{VerbatimL}
\end{fcvlabel}
\caption{IRIW Litmus Test}
\label{lst:formal:IRIW Litmus Test}
\end{listing}

However, the time required is a bit surprising given the simplicity
of the litmus test, which has two threads storing to two separate variables
and two other threads loading from these two variables in opposite
orders.
The assertion triggers if the two loading threads disagree on the order
of the two stores.
Even by the standards of memory-order litmus tests, this is quite simple.

One reason for the amount of time and space consumed is that PPCMEM does
a trace-based full-state-space search, which means that it must generate
and evaluate all possible orders and combinations of events at the
architectural level.
At this level, both loads and stores correspond to ornate sequences
of events and actions, resulting in a very large state space that must
be completely searched, in turn resulting in large memory and CPU
consumption.

Of course, many of the traces are quite similar to one another, which
suggests that an approach that treated similar traces as one might
improve performace.
One such approach is the axiomatic approach of
\pplsur{Jade}{Alglave} et al.~\cite{Alglave:2014:HCM:2594291.2594347},
which creates a set of axioms to represent the memory model and then
converts litmus tests to theorems that might be proven or disproven
over this set of axioms.
The resulting tool, called \qco{herd},  conveniently takes as input the
same litmus tests as PPCMEM, including the IRIW litmus test shown in
\cref{lst:formal:IRIW Litmus Test}.

\begin{listing}
\begin{fcvlabel}[ln:formal:Expanded IRIW Litmus Test]
\begin{VerbatimL}[commandchars=\%\@\$]
PPC IRIW5.litmus
""
(* Traditional IRIW, but with five stores instead of *)
(* just one.                                         *)
{
0:r1=1; 0:r2=x;
1:r1=1;         1:r4=y;
        2:r2=x; 2:r4=y;
        3:r2=x; 3:r4=y;
}
P0           | P1           | P2           | P3           ;
stw r1,0(r2) | stw r1,0(r4) | lwz r3,0(r2) | lwz r3,0(r4) ;
addi r1,r1,1 | addi r1,r1,1 | sync         | sync         ;
stw r1,0(r2) | stw r1,0(r4) | lwz r5,0(r4) | lwz r5,0(r2) ;
addi r1,r1,1 | addi r1,r1,1 |              |              ;
stw r1,0(r2) | stw r1,0(r4) |              |              ;
addi r1,r1,1 | addi r1,r1,1 |              |              ;
stw r1,0(r2) | stw r1,0(r4) |              |              ;
addi r1,r1,1 | addi r1,r1,1 |              |              ;
stw r1,0(r2) | stw r1,0(r4) |              |              ;

exists
(2:r3=1 /\ 2:r5=0 /\ 3:r3=1 /\ 3:r5=0)
\end{VerbatimL}
\end{fcvlabel}
\caption{Expanded IRIW Litmus Test}
\label{lst:formal:Expanded IRIW Litmus Test}
\end{listing}

However, where PPCMEM requires 14 CPU hours to solve IRIW, \co{herd} does so
in 17 milliseconds, which represents a speedup of more than six orders
of magnitude.
That said, the problem is exponential in nature, so we should expect
\co{herd} to exhibit exponential slowdowns for larger problems.
And this is exactly what happens, for example, if we add four more writes
per writing CPU as shown in
\cref{lst:formal:Expanded IRIW Litmus Test},
\co{herd} slows down by a factor of more than 50,000, requiring more than
15 \emph{minutes} of CPU time.
Adding threads also results in exponential
slowdowns~\cite{PaulEMcKenney2014weakaxiom}.

Despite their exponential nature, both PPCMEM and \co{herd} have proven quite
useful for checking key parallel algorithms, including the queued-lock
handoff on x86 systems.
The weaknesses of the \co{herd} tool are similar to those of PPCMEM, which
were described in
\cref{sec:formal:PPCMEM Discussion}.
There are some obscure (but very real) cases for which the PPCMEM and
\co{herd} tools disagree, and as of 2021 many but not all of these disagreements
was resolved.

It would be helpful if the litmus tests could be written in C
(as in \cref{lst:formal:Meaning of PPCMEM Litmus Test})
rather than assembly
(as in \cref{lst:formal:PPCMEM Litmus Test}).
This is now possible, as will be described in the following sections.

\subsection{Axiomatic Approaches and Locking}
\label{sec:formal:Axiomatic Approaches and Locking}

Axiomatic approaches may also be applied to higher-level
languages and also to higher-level synchronization primitives, as
exemplified by the lock-based litmus test shown in
\cref{lst:formal:Locking Example} (\path{C-Lock1.litmus}).
This litmus test can be modeled by
the \IXacrf{lkmm}~\cite{Alglave:2018:FSC:3173162.3177156,LucMaranget2018lock.cat}.
As expected, the \co{herd} tool's output features the string \co{Never},
correctly indicating that \co{P1()} cannot see \co{x} having a value
of one.\footnote{
	The output of the \co{herd} tool is compatible with that
	of PPCMEM, so feel free to look at
	\cref{lst:formal:PPCMEM Detects an Error,%
	lst:formal:PPCMEM on Repaired Litmus Test}
	for examples showing the output format.}

\begin{listing}
\input{CodeSamples/formal/herd/C-Lock1=whole.fcv}
\caption{Locking Example}
\label{lst:formal:Locking Example}
\end{listing}

\QuickQuiz{
	What do you have to do to run \co{herd} on litmus tests like
	that shown in \cref{lst:formal:Locking Example}?
}\QuickQuizAnswer{
	Get version v4.17 (or later) of the Linux-kernel source code,
	then follow the instructions in \path{tools/memory-model/README}
	to install the needed tools.
	Then follow the further instructions to run these tools on the
	litmus test of your choice.
}\QuickQuizEnd

\begin{listing}
\input{CodeSamples/formal/herd/C-Lock2=whole.fcv}
\caption{Broken Locking Example}
\label{lst:formal:Broken Locking Example}
\end{listing}

Of course, if \co{P0()} and \co{P1()} use different locks, as shown in
\cref{lst:formal:Broken Locking Example} (\path{C-Lock2.litmus}),
then all bets are off.
And in this case, the \co{herd} tool's output features the string
\co{Sometimes}, correctly indicating that use of different locks allows
\co{P1()} to see \co{x} having a value of one.

\QuickQuiz{
	Why bother modeling locking directly?
	Why not simply emulate locking with atomic operations?
}\QuickQuizAnswer{
	In a word, performance, as can be seen in
	\cref{tab:formal:Locking: Modeling vs. Emulation Time (s)}.
	The first column shows the number of \co{herd} processes modeled.
	The second column shows the \co{herd} runtime when modeling
	\co{spin_lock()} and \co{spin_unlock()} directly in \co{herd}'s
	cat language.
	The third column shows the \co{herd} runtime when emulating
	\co{spin_lock()} with \co{cmpxchg_acquire()} and \co{spin_unlock()}
	with \co{smp_store_release()}, using the \co{herd} \co{filter}
	clause to reject executions that fail to acquire the lock.
	The fourth column is like the third, but using \co{xchg_acquire()}
	instead of \co{cmpxchg_acquire()}.
	The fifth and sixth columns are like the third and fourth,
	but instead using the \co{herd} \co{exists} clause to reject
	executions that fail to acquire the lock.

\begin{table}
\rowcolors{10}{}{lightgray}
\small
\centering
\newcommand{\lockfml}[1]{\multicolumn{1}{c}{\begin{picture}(6,8)(0,0)\rotatebox{90}{#1}\end{picture}}}
\begin{tabular}{rrrrrr}
	\toprule
	& Model & \multicolumn{4}{c}{Emulate} \\
	\cmidrule(l){2-2} \cmidrule(l){3-6}
	& & \multicolumn{2}{c}{\tco{filter}} & \multicolumn{2}{c}{\tco{exists}} \\
	\cmidrule(l){3-4} \cmidrule(l){5-6}
	\lockfml{\# Proc.}
	&
	& \tco{cmpxchg}
	& \tco{xchg}
	& \tco{cmpxchg}
	& \tco{xchg}
	\\
	\cmidrule{1-1} \cmidrule(l){2-2} \cmidrule(l){3-4} \cmidrule(l){5-6}
	2 & 0.004 &  0.022 &  0.027 &   0.039 &   0.058 \\
	3 & 0.041 &  0.743 &  0.968 &   1.653 &   3.203 \\
	4 & 0.374 & 59.565 & 74.818 & 151.962 & 500.960 \\
	5 & 4.905 \\
	\bottomrule
\end{tabular}
\caption{Locking:
		  Modeling vs.\@ Emulation Time (s)}
\label{tab:formal:Locking: Modeling vs. Emulation Time (s)}
\end{table}

	Note also that use of the \co{filter} clause is about twice
	as fast as is use of the \co{exists} clause.
	This is no surprise because the \co{filter} clause allows
	early abandoning of excluded executions, where the executions
	that are excluded are the ones in which the lock is concurrently
	held by more than one process.

	More important, modeling \co{spin_lock()} and \co{spin_unlock()}
	directly ranges from five times faster to more than two orders
	of magnitude faster than modeling emulated locking.
	This should also be no surprise, as direct modeling raises
	the level of abstraction, thus reducing the number of events
	that \co{herd} must model.
	Because almost everything that \co{herd} does is of exponential
	computational complexity, modest reductions in the number of
	events produces exponentially large reductions in runtime.

	Thus, in formal verification even more than in parallel
	programming itself, divide and conquer!!!
}\QuickQuizEnd

But locking is not the only synchronization primitive that can be
modeled directly:
The next section looks at RCU\@.

\subsection{Axiomatic Approaches and RCU}
\label{sec:formal:Axiomatic Approaches and RCU}

\begin{fcvref}[ln:formal:C-RCU-remove:whole]
Axiomatic approaches can also analyze litmus tests involving
RCU~\cite{Alglave:2018:FSC:3173162.3177156}.
To that end,
\cref{lst:formal:Canonical RCU Removal Litmus Test}
(\path{C-RCU-remove.litmus})
shows a litmus test corresponding to the canonical RCU-mediated
removal from a linked list.
As with the locking litmus test, this RCU litmus test can be
modeled by LKMM, with similar performance advantages compared
to modeling emulations of RCU\@.
\Clnref{head} shows \co{x} as the list head, initially
referencing \co{y}, which in turn is initialized to the value
\co{2} on \clnref{tail:1}.

\begin{listing}
\input{CodeSamples/formal/herd/C-RCU-remove=whole.fcv}
\caption{Canonical RCU Removal Litmus Test}
\label{lst:formal:Canonical RCU Removal Litmus Test}
\end{listing}

\co{P0()} on \clnrefrange{P0start}{P0end}
removes element \co{y} from the list by replacing it with element \co{z}
(\clnref{assignnewtail}),
waits for a \IX{grace period} (\clnref{sync}),
and finally zeroes \co{y} to emulate \co{free()} (\clnref{free}).
\co{P1()} on \clnrefrange{P1start}{P1end}
executes within an RCU read-side critical section
(\clnrefrange{rl}{rul}),
picking up the list head (\clnref{rderef}) and then
loading the next element (\clnref{read}).
The next element should be non-zero, that is, not yet freed
(\clnref{exists_}).
Several other variables are output for debugging purposes
(\clnref{locations_}).

The output of the \co{herd} tool when running this litmus test features
\co{Never}, indicating that \co{P0()} never accesses a freed element,
as expected.
Also as expected, removing \clnref{sync} results in \co{P0()}
accessing a freed element, as indicated by the \co{Sometimes} in
the \co{herd} output.
\end{fcvref}

\begin{listing}
\ebresizeverb{.98}{\input{CodeSamples/formal/herd/C-RomanPenyaev-list-rcu-rr=whole.fcv}}
\caption{Complex RCU Litmus Test}
\label{lst:formal:Complex RCU Litmus Test}
\end{listing}

\begin{fcvref}[ln:formal:C-RomanPenyaev-list-rcu-rr:whole]
A litmus test for a more complex example proposed by
\ppl{Roman}{Penyaev}~\cite{RomanPenyaev2018rrRCU} is shown in
\cref{lst:formal:Complex RCU Litmus Test}
(\path{C-RomanPenyaev-list-rcu-rr.litmus}).
In this example, readers (modeled by \co{P0()} on
\clnrefrange{P0start}{P0end}) access a linked list
in a round-robin fashion by ``leaking'' a pointer to the last
list element accessed into variable \co{c}.
Updaters (modeled by \co{P1()} on \clnrefrange{P1start}{P1end})
remove an element, taking care to avoid disrupting current or future
readers.

\QuickQuiz{
	Wait!!!
	Isn't leaking pointers out of an RCU read-side critical
	section a critical bug???
}\QuickQuizAnswer{
	Yes, it usually is a critical bug.
	However, in this case, the updater has been cleverly constructed
	to properly handle such pointer leaks.
	But please don't make a habit of doing this sort of thing, and
	especially don't do this without having put a lot of thought
	into making some more conventional approach work.
}\QuickQuizEnd

\Clnrefrange{listtail}{listhead} define the initial linked
list, tail first.
In the Linux kernel, this would be a doubly linked circular list,
but \co{herd} is currently incapable of modeling such a beast.
The strategy is instead to use a singly linked linear list that
is long enough that the end is never reached.
\Clnref{rrcache} defines variable \co{c}, which is used to
cache the list pointer between successive RCU read-side critical
sections.

Again, \co{P0()} on \clnrefrange{P0start}{P0end} models readers.
This process models a pair of successive readers traversing round-robin
through the list, with the first reader on \clnrefrange{rl1}{rul1}
and the second reader on \clnrefrange{rl2}{rul2}.
\Clnref{rdcache} fetches the pointer cached in \co{c}, and if
\clnref{rdckcache} sees that the pointer was \co{NULL},
\clnref{rdinitcache} restarts at the beginning of the list.
In either case, \clnref{rdnext} advances to the next list element,
and \clnref{rdupdcache} stores a pointer to this element back into
variable \co{c}.
\Clnrefrange{rl2}{rul2} repeat this process, but using
registers \co{r3} and \co{r4} instead of \co{r1} and \co{r2}.
As with
\cref{lst:formal:Canonical RCU Removal Litmus Test},
this litmus test stores zero to emulate \co{free()}, so
\clnref{exists_} checks for any of these four registers being
\co{NULL}, also known as zero.

Because \co{P0()} leaks an RCU-protected pointer from its first
RCU read-side critical section to its second, \co{P1()} must carry
out its update (removing \co{x}) very carefully.
\Clnref{updremove} removes \co{x} by linking \co{w} to \co{y}.
\Clnref{updsync1} waits for readers, after which no subsequent reader
has a path to \co{x} via the linked list.
\Clnref{updrdcache} fetches \co{c}, and if \clnref{updckcache}
determines that \co{c} references the newly removed \co{x},
\clnref{updinitcache} sets \co{c} to \co{NULL}
and \clnref{updsync2} again waits for readers, after which no
subsequent reader can fetch \co{x} from \co{c}.
In either case, \clnref{updfree} emulates \co{free()} by storing
zero to \co{x}.

\QuickQuiz{
	\begin{fcvref}[ln:formal:C-RomanPenyaev-list-rcu-rr:whole]
	In \cref{lst:formal:Complex RCU Litmus Test},
	why couldn't a reader fetch \co{c} just before \co{P1()}
	zeroed it on \clnref{updinitcache}, and then later
	store this same value back into \co{c} just after it was
	zeroed, thus defeating the zeroing operation?
	\end{fcvref}
}\QuickQuizAnswer{
	\begin{fcvref}[ln:formal:C-RomanPenyaev-list-rcu-rr:whole]
	Because the reader advances to the next element on
	\clnref{rdnext}, thus avoiding storing a pointer to the
	same element as was fetched.
	\end{fcvref}
}\QuickQuizEnd

The output of the \co{herd} tool when running this litmus test features
\co{Never}, indicating that \co{P0()} never accesses a freed element,
as expected.
Also as expected, removing either \co{synchronize_rcu()} results
in \co{P1()} accessing a freed element, as indicated by \co{Sometimes}
in the \co{herd} output.
\end{fcvref}

\QuickQuizSeries{%
\QuickQuizB{
	\begin{fcvref}[ln:formal:C-RomanPenyaev-list-rcu-rr:whole]
	In \cref{lst:formal:Complex RCU Litmus Test},
	why not have just one call to \co{synchronize_rcu()}
	immediately before \clnref{updfree}?
	\end{fcvref}
}\QuickQuizAnswerB{
	\begin{fcvref}[ln:formal:C-RomanPenyaev-list-rcu-rr:whole]
	Because this results in \co{P0()} accessing a freed element.
	But don't take my word for this, try it out in \co{herd}!
	\end{fcvref}
}\QuickQuizEndB
\QuickQuizE{
	\begin{fcvref}[ln:formal:C-RomanPenyaev-list-rcu-rr:whole]
	Also in \cref{lst:formal:Complex RCU Litmus Test},
	can't \clnref{updfree} be \co{WRITE_ONCE()} instead
	of \co{smp_store_release()}?
	\end{fcvref}
}\QuickQuizAnswerE{
	\begin{fcvref}[ln:formal:C-RomanPenyaev-list-rcu-rr:whole]
	That is an excellent question.
	As of late 2021, the answer is ``no one knows''.
	Much depends on the semantics of \ARMv8's conditional-move
	instruction.
	While awaiting clarity on these semantics, \co{smp_store_release()}
	is the safe choice.
	\end{fcvref}
}\QuickQuizEndE
}

These sections have shown how axiomatic approaches can successfully
model synchronization primitives such as locking and RCU in C-language
litmus tests.
Longer term, the hope is that the axiomatic approaches will model
even higher-level software artifacts, producing exponential
verification speedups.
This could potentially allow axiomatic verification of much larger
software systems, perhaps incorporating spatial-synchronization techniques
from separation
logic~\cite{AlexeyGotsman2013ESOPRCU,PeterWOHearn2001SeparationLogic}.
Another alternative is to press the axioms of boolean logic into service,
as described in the next section.

% formal/sat.tex
% mainfile: ../perfbook.tex
% SPDX-License-Identifier: CC-BY-SA-3.0

\section{SAT Solvers}
\label{sec:formal:SAT Solvers}
\epigraph{Live by the heuristic, die by the heuristic.}{Unknown}

Any finite program with bounded loops and recursion can be converted
into a logic expression, which might express that program's assertions
in terms of its inputs.
Given such a logic expression, it would be quite interesting to know
whether any possible combinations of inputs could result in one of
the assertions triggering.
If the inputs are expressed as combinations of boolean variables,
this is simply SAT, also known as the satisfiability problem.
SAT solvers are heavily used in verification of hardware, which has
motivated great advances.
A world-class early 1990s SAT solver might be able to handle a logic
expression with 100 distinct boolean variables, but by the early 2010s
million-variable SAT solvers were readily
available~\cite{Kroening:2008:DPA:1391237}.

\begin{figure}
\centering
\resizebox{2in}{!}{\includegraphics{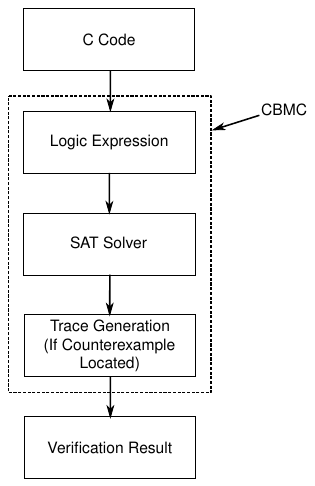}}
\caption{\IXacr{cbmc} Processing Flow}
\label{fig:formal:CBMC Processing Flow}
\end{figure}

In addition, front-end programs for SAT solvers can automatically translate
C code into logic expressions, taking assertions into account and generating
assertions for error conditions such as array-bounds errors.
One example is the \IXacrl{cbmc}, or \co{cbmc}, which is
available as part of many Linux distributions.
This tool is quite easy to use, with \co{cbmc test.c} sufficing to
validate \path{test.c}, resulting in the processing flow shown in
\cref{fig:formal:CBMC Processing Flow}.
This ease of use is exceedingly important because it opens the door
to formal verification being incorporated into regression-testing
frameworks.
In contrast, the traditional tools that require non-trivial translation
to a special-purpose language are confined to design-time verification.

More recently, SAT solvers have appeared that handle parallel code.
These solvers operate by converting the input code into single static
assignment (SSA) form, then generating all permitted access orders.
This approach seems promising, but it remains to be seen how well
it works in practice.
One encouraging sign is work in 2016 applying \co{cbmc} to Linux-kernel
RCU~\cite{LihaoLiang2016VerifyTreeRCU,Liang:2018:VTB,LanceRoy2017CBMC-SRCU}.
This work used minimal configurations of RCU, and verified scenarios
using small numbers of threads, but nevertheless successfully ingested
Linux-kernel C code and produced a useful result.
The logic expressions generated from the C code had up to 90~million
variables, 450~million clauses, occupied tens of gigabytes of memory,
and required up to 80~hours of CPU time for the SAT solver to produce
the correct result.

Nevertheless, a Linux-kernel hacker might be justified in feeling skeptical
of a claim that his or her code had been automatically verified, and
such hackers would find many fellow skeptics going back
decades~\cite{DeMillo:1979:SPP:359104.359106}.
One way to productively express such skepticism is to provide bug-injected
versions of the allegedly verified code.
If the formal-verification tool finds all the injected bugs, our hacker
might gain more confidence in the tool's capabilities.
Of course, tools that find valid bugs of which the hacker was not yet aware
will likely engender even more confidence.
And this is exactly why there is a \co{git} archive with a 20-branch set
of mutations, with each branch potentially containing a bug injected
into Linux-kernel RCU~\cite{PaulEMcKenney2017VerificationChallenge6}.
Anyone with a formal-verification tool is cordially invited to try that
tool out on this set of verification challenges.

Currently, \co{cbmc} is able to find a number of injected bugs,
however, it has not yet been able to locate a bug that RCU's
maintainer was not already aware of.
Nevertheless, there is some reason to hope that SAT solvers will someday
be useful for finding concurrency bugs in parallel code.

% formal/stateless.tex
% mainfile: ../perfbook.tex
% SPDX-License-Identifier: CC-BY-SA-3.0

\section{Stateless Model Checkers}
\label{sec:formal:Stateless Model Checkers}
\epigraph{He's making a list, he's permuting it twice\dots}
	{with apologies to Haven Gillespie and J. Fred Coots}

The SAT-solver approaches described in the previous section are quite
convenient and powerful, but the full tracking of all possible
executions, including state, can incur substantial overhead.
In fact, the memory and CPU-time overheads can sharply limit the size
of programs that can be feasibly verified, which raises the question
of whether less-exact approaches might find bugs in larger programs.

\begin{figure}
\centering
\resizebox{2.1in}{!}{\includegraphics{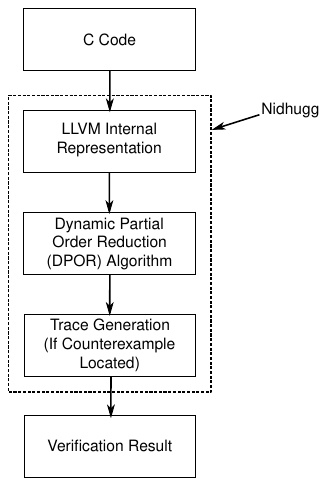}}
\caption{Nidhugg Processing Flow}
\label{fig:formal:Nidhugg Processing Flow}
\end{figure}

Although the jury is still out on this question, stateless model
checkers such as \IX{Nidhugg}~\cite{CarlLeonardsson2014Nidhugg} have in
some cases handled larger programs~\cite{SMC-TreeRCU}, and with
similar ease of use, as illustrated by
\cref{fig:formal:Nidhugg Processing Flow}.
In addition, Nidhugg was more than an order of magnitude faster than
was \co{cbmc} for some Linux-kernel RCU verification scenarios.
Of course, Nidhugg's speed and scalability advantages are tied to
the fact that it does not handle data non-determinism, but this
was not a factor in these particular verification scenarios.

Nevertheless, as with \co{cbmc}, Nidhugg has not yet been able to
locate a bug that Linux-kernel RCU's maintainer was not already
aware of.
However, it was able to demonstrate that one historical bug in
Linux-kernel RCU was fixed by a different commit than the maintainer
thought, which gives some additional hope that stateless model checkers
like Nidhugg might someday be useful for finding concurrency bugs in
parallel code.

\section{Summary}
\label{sec:formal:Summary}
\epigraph{Western thought has focused on True-False;
	  it is high time to shift to Robust-Fragile.}
	 {Nassim Nicholas Taleb, summarized}
% Full quote:
% Since Plato, Western thought and the theory of knowledge has focused on
% the notions of True-False; as commendable as that was, it is high time
% to shift the concern to Robust-Fragile, and social epistemology to the
% more serious problem of Sucker-Nonsucker.

The formal-verification techniques described in this chapter
are very powerful tools for validating small
parallel algorithms, but they should not be the only tools in your toolbox.
Despite decades of focus on formal verification, testing remains the
validation workhorse for large parallel software
systems~\cite{JonathanCorbet2006lockdep,DaveJones2011Trinity,PaulEMcKenney2016Formal}.

It is nevertheless quite possible that this will not always be the case.
To see this, consider that there is estimated to be more than twenty
billion instances of the Linux kernel as of 2017.
Suppose that the Linux kernel has a bug that manifests on average every million
years of runtime.
As noted at the end of the preceding chapter, this bug will be appearing
more than 50 times \emph{per day} across the installed base.
But the fact remains that most formal validation techniques can be used
only on very small codebases.
So what is a concurrency coder to do?

Think in terms of finding the first bug, the first relevant bug, the
last relevant bug, and the last bug.

The first bug is normally found via inspection or compiler diagnostics.
Although the increasingly sophisticated compiler diagnostics comprise
a lightweight sort of formal verification, it is not common to think of
them in those terms.
This is in part due to an odd practitioner prejudice which says ``If I am
using it, it cannot be formal verification'' on the one hand, and a large
gap between compiler diagnostics and verification research on the other.

Although the first relevant bug might be located via inspection or
compiler diagnostics, it is not unusual for these two steps to find
only typos and false positives.
Either way, the bulk of the relevant bugs, that is, those bugs that
might actually be encountered in production, will often be found via testing.

When testing is driven by anticipated or real use cases, it is not
uncommon for the last relevant bug to be located by testing.
This situation might motivate a complete rejection of formal verification,
however, irrelevant bugs have an annoying habit of suddenly becoming relevant
at the least convenient moment possible, courtesy of black-hat attacks.
For security-critical software, which appears to be a continually
increasing fraction of the total, there can thus be strong motivation
to find and fix the last bug.
Testing is demonstrably unable to find the last bug, so there is a
possible role for formal verification, assuming, that is, that
formal verification proves capable of growing into that role.
As this chapter has shown, current formal verification systems are
extremely limited.

\QuickQuiz{
	But shouldn't sufficiently low-level software be for all intents
	and purposes immune to being exploited by black hats?
}\QuickQuizAnswer{
	Unfortunately, no.

	At one time, Paul E. McKenny felt that Linux-kernel RCU
	was immune to such exploits, but the advent of Row Hammer
	showed him otherwise.
	After all, if the black hats can hit the system's DRAM,
	they can hit any and all low-level software, even including RCU\@.

	And in 2018, this possibility passed from the realm of
	theoretical speculation into the hard and fast realm of
	objective reality~\cite{McKenney:2019:CRS:3319647.3325836}.
}\QuickQuizEnd

Please note that formal verification is often much harder to use than
is testing.
This is in part a cultural statement, and there is reason to hope
that formal verification will be perceived to be easier with increased
familiarity.
That said, very simple test harnesses can find significant bugs in arbitrarily
large software systems.
In contrast, the effort required to apply formal verification seems to
increase dramatically as the system size increases.

I have nevertheless made occasional use of formal verification
for almost 30 years by playing to formal verification's strengths,
namely design-time verification of small complex portions of the overarching
software construct.
The larger overarching software construct is of course validated by testing.

\QuickQuiz{
	In light of the full verification of the L4 microkernel,
	isn't this limited view of formal verification just a little
	bit obsolete?
}\QuickQuizAnswer{
	Unfortunately, no.

	The first full verification of the L4 microkernel was a tour de force,
	with a large number of Ph.D.~students hand-verifying code at a
	very slow per-student rate.
	This level of effort could not be applied to most software projects
	because the rate of change is just too great.
	Furthermore, although the L4 microkernel is a large software
	artifact from the viewpoint of formal verification, it is tiny
	compared to a great number of projects, including LLVM,
	\GCC, the Linux kernel, Hadoop, MongoDB, and a great many others.
	In addition, this verification did have limits, as the researchers
	freely admit, to their credit:
	\url{https://docs.sel4.systems/projects/sel4/frequently-asked-questions.html\#does-sel4-have-zero-bugs}.

	Although formal verification is finally starting to show some
	promise, including more-recent L4 verifications involving greater
	levels of automation, it currently has no chance of completely
	displacing testing in the foreseeable future.
	And although I would dearly love to be proven wrong on this point,
	please note that such proof will be in the form of a real tool
	that verifies real software, not in the form of a large body of
	rousing rhetoric.

	Perhaps someday formal verification will be used heavily for
	validation, including for what is now known as regression testing.
	\Cref{sec:future:Formal Regression Testing?} looks at
	what would be required to make this possibility a reality.
}\QuickQuizEnd

One final approach is to consider the following two definitions from
\cref{sec:debugging:Required Mindset}
and the consequence that they imply:

\begin{description}[itemsep=0pt,labelindent=1em]
\item[Definition:]	Bug-free programs are trivial programs.
\item[Definition:]	Reliable programs have no known bugs.
\item[Consequence:]	Any non-trivial reliable program contains at least
			one as-yet-unknown bug.
\end{description}

From this viewpoint, any advances in validation and verification can
have but two effects:
\begin{enumerate*}[(1)]
\item An increase in the number of trivial programs or
\item A decrease in the number of reliable programs.
\end{enumerate*}
Of course, the human race's increasing reliance on multicore systems and
software provides extreme motivation for a very sharp increase in the
number of trivial programs.

However, if your code is so complex that you find yourself
relying too heavily on formal-verification
tools, you should carefully rethink your design, especially if your
formal-verification tools require your code to be hand-translated
to a special-purpose language.
For example, a complex implementation of the dynticks interface for
preemptible RCU that was presented in
\cref{sec:formal:Promela Parable: dynticks and Preemptible RCU}
turned out to
have a much simpler alternative implementation, as discussed in
\cref{sec:formal:Simplicity Avoids Formal Verification}.
All else being equal, a simpler implementation is much better than
a proof of correctness for a complex implementation.

And the open challenge to those working on formal verification techniques
and systems is to prove this summary wrong!
To assist in this task, Verification Challenge~6 is now
available~\cite{PaulEMcKenney2017VerificationChallenge6}.
Have at it!!!

\section{Choosing a Validation Plan}
\label{sec:formal:Choosing a Validation Plan}
\epigraph{Science is a first-rate piece of furniture for one's upper
	  chamber, but only given common sense on the ground floor.}
	 {Oliver Wendell Holmes, updated}

What sort of validation should you use for your project?

As is often the case in software in particular and in engineering
in general, the answer is ``it depends''.

Note that neither running a test nor undertaking formal verification
will change your project.
At best, such effort have an indirect effect by locating a bug that
is later fixed.
Nevertheless, fixing a bug might prevent inconvenience, monetary loss,
property damage, or even loss of life.
Clearly, this sort of indirect effect can be extremely valuable.

Unfortunately, as we have seen, it is difficult to predict whether or
not a given validation effort will find important bugs.
It is therefore all too easy to invest too little---or even to fail
to invest at all, especially if development estimates proved overly
optimistic or budgets unexpectedly tight, conditions which almost
always come into play in real-world software projects.

The decision to nevertheless invest in validation is often forced by
experienced people with forceful personalities.
But this is no guarantee, given that other stakeholders might also
have forceful personalities.
Worse yet, these other stakeholders might bring stories of expensive
validation efforts that nevertheless allowed embarrassing bugs to
escape to the end users.
So although a scarred, grey-haired, and grouchy veteran might carry
the day, a more organized approach would perhaps be more useful.

Fortunately, there is a strictly financial analog to investments in
validation, and that is the insurance policy.

\IfTwoColumn{
\begin{figure*}
\centering
\resizebox{6in}{!}{\includegraphics{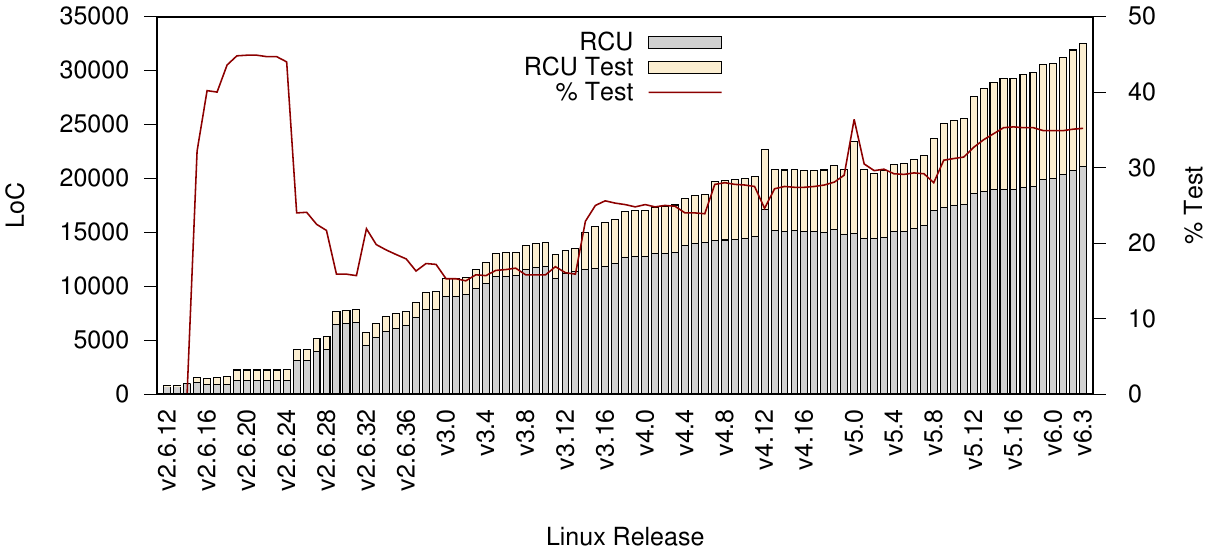}}
\caption{Linux-Kernel RCU Test Code}
\label{fig:formal:Linux-Kernel RCU Test Code}
\end{figure*}
}{}

Both insurance policies and validation efforts require consistent
up-front investments, and both defend against disasters that might
or might not ever happen.
Furthermore, both have exclusions of various types.
For example, insurance policies for coastal areas might exclude
damages due to tidal waves, while on the other hand we have seen
that there is not yet any validation methodology that can find
each and every bug.

In addition, it is possible to over-invest in both insurance and
in validation.
For but one example, a validation plan that consumed the entire
development budget would be just as pointless as would an insurance
policy that covered the Sun going nova.

One approach is to devote a given fraction of the software budget to
validation, with that fraction depending on the criticality of the
software, so that safety-critical avionics software might grant a
larger fraction of its budget to validation than would a homework
assignment.
Where available, experience from prior similar projects should be
brought to bear.
However, it is necessary to structure the project so that the validation
investment starts when the project does, otherwise the inevitable overruns
in spending on coding will crowd out the validation effort.

Staffing start-up projects with experienced people can result in
overinvestment in validation efforts.
Just as it is possible to go broke buying too much insurance, it is
possible to kill a project by investing too much in testing.
This is especially the case for first-of-a-kind projects where it is
not yet clear which use cases will be important, in which case testing
for all possible use cases will be a possibly fatal waste of time,
energy, and funding.

However, as the tasks supported by a start-up project become more routine,
users often become less forgiving of failures, thus increasing the need
for validation.
Managing this shift in investment can be extremely challenging,
especially in the all-too-common case where the users are unwilling
or unable to disclose the exact nature of their use case.
It then becomes critically important to reverse-engineer the
use cases from bug reports and from discussions with the users.
As these use cases are better understood, use of continuous integration
can help reduce the cost of finding and fixing any bugs located.

\IfTwoColumn{}{
\begin{figure}
\centering
\IfEbookSize{
\resizebox{\onecolumntextwidth}{!}{\includegraphics{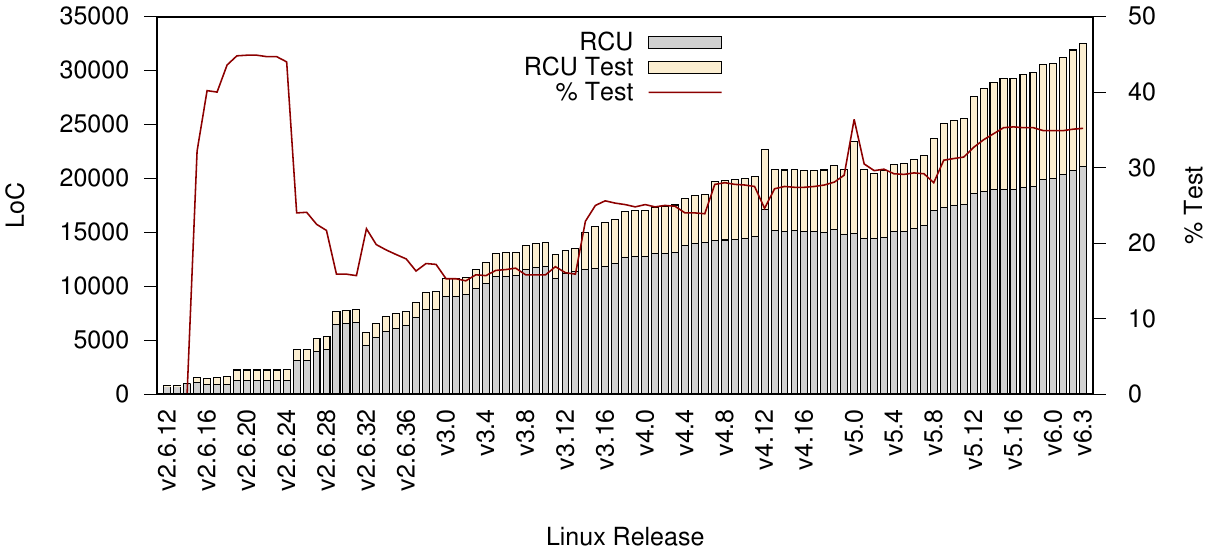}}
}{
\resizebox{4.5in}{!}{\includegraphics{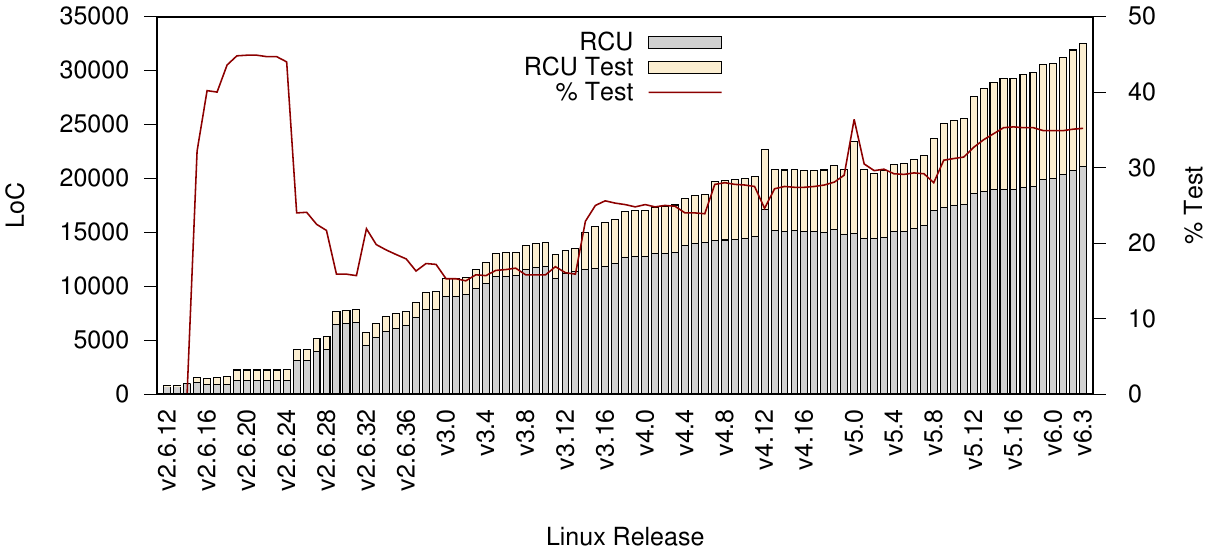}}
}
\caption{Linux-Kernel RCU Test Code}
\label{fig:formal:Linux-Kernel RCU Test Code}
\end{figure}
}

One example evolution of a software project's use of validation is
shown in
\cref{fig:formal:Linux-Kernel RCU Test Code}.
As can be seen in the \lcnamecref{fig:formal:Linux-Kernel RCU Test Code},
Linux-kernel RCU didn't have any validation code whatsoever until Linux
kernel v2.6.15, which was released more than two years after RCU was
accepted into the kernel.
The test suite achieved its peak fraction of the total lines of code
in Linux kernel v2.6.19--v2.6.21.
This fraction decreased sharply with the acceptance of preemptible RCU
for real-time applications in v2.6.25.
This decrease was due to the fact that the RCU API was identical
in the preemptible and non-preemptible variants of RCU\@.
This in turn meant that the existing test suite applied to both variants,
so that even though the Linux-kernel RCU code expanded significantly,
there was no need to expand the tests.

Subsequent bars in \cref{fig:formal:Linux-Kernel RCU Test Code} show
that the RCU code base expanded significantly, but that the
corresponding validation code expanded even more dramatically.
Linux kernel v3.5 added tests for the \co{rcu_barrier()} API, closing
a long-standing hole in test coverage.
Linux kernel v3.14 added automated testing and analysis of test results,
moving RCU towards continuous integration.
Linux kernel v4.7 added a performance validation suite for RCU's update-side
primitives.
Linux kernel v4.12 added Tree SRCU, featuring improved update-side
scalability, and v4.13 removed the old less-scalable SRCU implementation.
Linux kernel v5.0 briefly hosted the \path{nolibc} library within
the rcutorture scripting directory before it moved to its long-term
home in \path{tools/include/nolibc}.
Linux kernel v5.8 added the Tasks Trace and Rude flavors of RCU\@.
Linux kernel v5.9 added the \path{refscale.c} suite of read-side performance
tests.
Linux kernels v5.12 and v5.13 started adding the ability to change a
given CPU's callback-offloading status at runtime and also added the
\path{torture.sh} acceptance-test script.
Linux kernel v5.14 added distributed rcutorture.
Linux kernel v5.15 added demonic vCPU placement in rcutorture testing,
which was successful in locating a number of race conditions.\footnote{
	The trick is to place one pair of vCPUs within the same core on
	one socket, while placing another pair within the same core on
	some other socket.
	As you might expect from \cref{chp:Hardware and its Habits},
	this produces different memory latencies between different
	pairs of vCPUs
	(\url{https://paulmck.livejournal.com/62071.html}).}
Linux kernel v5.17 removed the \co{RCU_FAST_NO_HZ} Kconfig option.
Numerous other changes may be found in the Linux kernel's \co{git} archives.
% rcutorture
% v2.6.15: First torture test
% v2.6.19: SRCU: Plugin architecture avoids test-code explosion.
% v2.6.19-21: Peak test fraction.
% v2.6.25: preemptible RCU, consistent API avoids added test code.
% v3.4: Add tests for RCU CPU stall warnings.
% v3.5: Add tests for rcu_barrier(). *
% v3.14: Add rcutorture scripting automating tests and results analysis. *
% v3.15: Add support for multiple torture-tests suites for locktorture.
% v3.16: Add support for conditional grace-period primitives.
% v4.7: Add update-side performance validation suite. *
% v4.12: Added Tree SRCU.
% v4.13: Removed non-Tree SRCU.
% v5.0: nolibc was briefly in the rcutorture scripting directory.
% v5.8: Added Tasks Trace RCU and Rude RCU.
% v5.9: Added refscale.c.
% v5.12: Added torture.sh.  Added runtime adjustment of rcu_nocbs CPUs.
% v5.13: More rcu_nocbs CPUs, polls grace-period primitives, remote rcutorture.
% v5.14: Added demonic affinity to rcutorture scripting.
% v5.17: Added RCU callback (de)offloading optimizations and per-CPU
%	 tracking of RCU Tasks callbacks.
% v5.18: Improved RCU priority boosting and \co{rcu_barrier()} no longer
%	 blocks CPU-hotplug operations.
% v5.19: Enabled milliseconds-scale real-time response from
%	 \co{synchronize_rcu_expedited()} and also dynamically allocated
%	 and sized Tree SRCU's \co{srcu_node} array.  This last saves
%	 significant memory on kernels sized for thousands of CPUs
%	 that typically run on much smaller systems.

We have established that the validation budget varies from one project
to the next, and also over the lifetime of any given project.
But how should the validation investment be split between testing and
formal verification?

This question is being answered naturally as compilers adopt increasingly
aggressive formal-verification techniques into their diagnostics and
as formal-verification tools continue to mature.
In addition, the Linux-kernel lockdep and KCSAN tools illustrate the
advantages of combining formal verification techniques with run-time
analysis, as discussed in \cref{sec:debugging:Assertions}.
Other combined techniques analyze traces gathered from
executions~\cite{DanielBristot2019RTtrace}.
For the time being, the best practice is to focus first on testing and to
reserve explicit work on formal verification for those portions of the
project that are not well-served by testing, and that have exceptional
needs for robustness.
For example, Linux-kernel RCU relies primarily on testing, but has
made occasional use of formal verification as discussed in this chapter.

In short, choosing a validation plan for concurrent software remains
more an art than a science, let alone a field of engineering.
However, there is every reason to expect that increasingly rigorous
approaches will continue to become more prevalent.

	\setlength{\parskip}{0.0pt plus 1ex}
	\input{qqzformal}
	\setlength{\parskip}{0.0pt plus 1.0pt}% return to default

% together/together.tex
% mainfile: ../perfbook.tex
% SPDX-License-Identifier: CC-BY-SA-3.0

\QuickQuizChapter{chp:Putting It All Together}{Putting It All Together}{qqztogether}
\Epigraph{You don't learn how to shoot and then learn how to launch
	  and then learn to do a controlled spin---you learn to
	  launch-shoot-spin.}{\emph{Ender's Shadow}, Orson Scott Card}

% And the paragraph preceding this is also instructive:
% ``I may be pissed off, but that doesn't mean I can't learn.''

This \lcnamecref{chp:Putting It All Together}
gives some hints on concurrent-programming puzzles.
\Cref{sec:together:Counter Conundrums}
considers counter conundrums,
\cref{sec:together:Refurbish Reference Counting}
refurbishes reference counting,
\cref{sec:together:Hazard-Pointer Helpers}
helps with hazard pointers,
\cref{sec:together:Sequence-Locking Specials}
surmises on sequence-locking specials,
and finally
\cref{sec:together:RCU Rescues}
reflects on RCU rescues.

% together/count.tex
% mainfile: ../perfbook.tex
% SPDX-License-Identifier: CC-BY-SA-3.0

\section{Counter Conundrums}
\label{sec:together:Counter Conundrums}
\epigraph{Ford carried on counting quietly.
	  This is about the most aggressive thing you can do to a
	  computer, the equivalent of going up to a human being and saying
	  ``Blood \dots blood \dots blood \dots blood \dots''}
	 {Douglas Adams}

This \lcnamecref{sec:together:Counter Conundrums}
outlines solutions to counter conundrums.

\subsection{Counting Updates}
\label{sec:together:Counting Updates}

Suppose that Schr\"odinger (see
\cref{sec:datastruct:Motivating Application})
wants to count the number of updates for each animal,
and that these updates are synchronized using a per-data-element lock.
How can this counting best be done?

Of course, any number of counting algorithms from \cref{chp:Counting}
might qualify, but the optimal approach is quite simple.
Just place a counter in each data element, and increment it under the
protection of that element's lock!

If readers access the count locklessly, then updaters should use
\co{WRITE_ONCE()} to update the counter and lockless readers should
use \co{READ_ONCE()} to load it.

\subsection{Counting Lookups}
\label{sec:together:Counting Lookups}

Suppose that Schr\"odinger also wants to count the number of lookups for
each animal, where lookups are protected by RCU\@.
How can this counting best be done?

One approach would be to protect a lookup counter with the per-element
lock, as discussed in \cref{sec:together:Counting Updates}.
Unfortunately, this would require all lookups to acquire this lock,
which would be a severe bottleneck on large systems.

Another approach is to ``just say no'' to counting, following the example
of the \co{noatime} mount option.
If this approach is feasible, it is clearly the best:
After all, nothing is faster than doing nothing.
If the lookup count cannot be dispensed with, read on!

Any of the counters from \cref{chp:Counting}
could be pressed into service, with the statistical counters described in
\cref{sec:count:Statistical Counters} being perhaps the most common choice.
However, this results in a large memory footprint:
The number of counters required is the number of data elements multiplied
by the number of threads.

If this memory overhead is excessive, then one approach is to keep
per-core or even per-socket counters rather than per-CPU counters,
with an eye to the hash-table performance results depicted in
\cref{fig:datastruct:Read-Only Hash-Table Performance For Schroedinger's Zoo; 448 CPUs}.
This will require that the counter increments be atomic operations,
especially for user-mode execution where a given thread could migrate
to another CPU at any time.

If some elements are looked up very frequently, there are a number
of approaches that batch updates by maintaining a per-thread log,
where multiple log entries for a given element can be merged.
After a given log entry has a sufficiently large increment or after
sufficient time has passed, the log entries may be applied to the
corresponding data elements.
Silas Boyd-Wickizer has done some work formalizing this
notion~\cite{SilasBoydWickizerPhD}.

% together/refcnt.tex
% mainfile: ../perfbook.tex
% SPDX-License-Identifier: CC-BY-SA-3.0

\section{Refurbish Reference Counting}
\label{sec:together:Refurbish Reference Counting}
\epigraph{Counting is the religion of this generation.
	  It is its hope and its salvation.}
	 {Gertrude Stein}

Although reference counting is a conceptually simple technique,
many devils hide in the details when it is applied to concurrent
software.
After all, if the object was not subject to premature disposal,
there would be no need for the
\IXalt{reference counter}{reference count} in the first place.
But if the object can be disposed of, what prevents disposal during
the reference-acquisition process itself?

There are a number of ways to refurbish reference counters for
use in concurrent software, including:

\begin{enumerate}
\item	A lock residing outside of the object must be held while
	manipulating the reference count.
\item	The object is created with a non-zero reference count, and new
	references may be acquired only when the current value of
	the reference counter is non-zero.
	If a thread does not have a reference to a given object, it
	might seek help from another thread that already has a reference.
\item	In some cases, hazard pointers may be used as a drop-in
	replacement for reference counters.
\item	An \IX{existence guarantee} is provided for the object, thus preventing
	it from being freed while some other entity might be attempting
	to acquire a reference.
	Existence guarantees are often provided by automatic
	garbage collectors, and, as is seen in
	\cref{sec:defer:Hazard Pointers,sec:defer:Read-Copy Update (RCU)},
	by hazard pointers and RCU, respectively.
\item	A \IXalt{type-safety guarantee}{type-safe memory}
	is provided for the object.
	An additional identity check must be performed once
	the reference is acquired.
	Type-safety guarantees can be provided by special-purpose
	memory allocators, for example, by the
	\co{SLAB_TYPESAFE_BY_RCU} feature within the Linux kernel,
	as is seen in \cref{sec:defer:Read-Copy Update (RCU)}.
\end{enumerate}

Of course, any mechanism that provides existence guarantees
by definition also provides type-safety guarantees.
This results in four general categories of reference-acquisition
protection:
Reference counting, hazard pointers, sequence locking, and RCU\@.

\QuickQuiz{
	Why not implement reference-acquisition using
	a simple \IXacrml{cas} operation that only
	acquires a reference if the reference counter is
	non-zero?
}\QuickQuizAnswer{
	Although this can resolve the race between the release of
	the last reference and acquisition of a new reference,
	it does absolutely nothing to prevent the data structure
	from being freed and reallocated, possibly as some completely
	different type of structure.
	It is quite likely that the ``simple \acrml{cas}
	operation'' would give undefined results if applied to the
	differently typed structure.

	In short, use of atomic operations such as \acrml{cas}
	absolutely requires either type-safety or existence guarantees.

	But what if it is absolutely necessary to let the type change?

	One approach is for each such type to have the reference counter
	at the same location, so that as long as the reallocation results
	in an object from this group of types, all is well.
	If you do this in C, make sure you comment the reference counter
	in each structure in which it appears.
	In C++, use inheritance and templates.
}\QuickQuizEnd

\begin{table}
\renewcommand*{\arraystretch}{1.25}
\rowcolors{3}{}{lightgray}
\small
\centering
\begin{tabular}{lcccc}
	\toprule
	& \multicolumn{4}{c}{Release} \\
	\cmidrule(l){2-5}
	Acquisition	& Locks
				& \parbox[c]{.5in}{Reference\\Counts}
					& \parbox[c]{.5in}{Hazard\\Pointers}
						& RCU \\
	\cmidrule{1-1} \cmidrule(l){2-5}
	Locks		& $-$	& CAM	& M	& CA  \\
	\parbox[c][6ex]{.6in}{Reference\\Counts}
			& A	& AM    & M	& A   \\
	\parbox[c][6ex]{.6in}{Hazard\\Pointers}
			& M	& M	& M	& M   \\
	RCU		& CA	& MCA	& M	& CA  \\
	\bottomrule
\end{tabular}
\caption{Synchronizing Reference Counting}
\label{tab:together:Synchronizing Reference Counting}
\end{table}

Given that the key reference-counting issue
is synchronization between acquisition
of a reference and freeing of the object, we have nine possible
combinations of mechanisms, as shown in
\cref{tab:together:Synchronizing Reference Counting}.
This table
divides reference-counting mechanisms into the following broad categories:
\begin{enumerate}
\item	Simple counting with neither atomic operations,
	\IXpl{memory barrier}, nor alignment constraints (``$-$'').
\item	Atomic counting without memory barriers (``A'').
\item	Atomic counting, with memory barriers required only on release
	(``AM'').
\item	Atomic counting with a check combined with the atomic acquisition
	operation, and with memory barriers required only on release
	(``CAM'').
\item	Atomic counting with a check combined with the atomic acquisition
	operation (``CA'').
\item	Simple counting with a check combined with full memory barriers
	(``M'').
\item	Atomic counting with a check combined with the atomic acquisition
	operation, and with memory barriers also required on acquisition
	(``MCA'').
\end{enumerate}
However, because all Linux-kernel atomic operations that return a
value are defined to contain memory barriers,\footnote{
	With \co{atomic_read()} and \co{ATOMIC_INIT()} being the
	exceptions that prove the rule.}
all release operations
contain memory barriers, and all checked acquisition operations also
contain memory barriers.
Therefore, cases ``CA'' and ``MCA'' are equivalent to ``CAM'', so that
there are sections below for only the first four cases and the sixth case:
``$-$'', ``A'', ``AM'', ``CAM'', and ``M\@''.
Later sections describe optimizations that can improve performance
if reference acquisition and release is very frequent, and the
reference count need be checked for zero only very rarely.

\subsection{Implementation of Reference-Counting Categories}
\label{sec:together:Implementation of Reference-Counting Categories}

Simple counting protected by locking (``$-$'') is described in
\cref{sec:together:Simple Counting},
atomic counting with no memory barriers (``A'') is described in
\cref{sec:together:Atomic Counting},
atomic counting with acquisition memory barrier (``AM'') is described in
\cref{sec:together:Atomic Counting With Release Memory Barrier},
and
atomic counting with check and release memory barrier (``CAM'') is described in
\cref{sec:together:Atomic Counting With Check and Release Memory Barrier}.
Use of hazard pointers is described in
\cref{sec:defer:Hazard Pointers}
on \cpageref{sec:defer:Hazard Pointers}
and in
\cref{sec:together:Hazard-Pointer Helpers}.

\subsubsection{Simple Counting}
\label{sec:together:Simple Counting}

Simple counting, with neither atomic operations nor memory barriers,
can be used when the reference-counter acquisition and release are
both protected by the same lock.
In this case, it should be clear that the reference count itself
may be manipulated non-atomically, because the lock provides any
necessary exclusion, memory barriers, atomic instructions, and disabling
of compiler optimizations.
This is the method of choice when the lock is required to protect
other operations in addition to the reference count, but where
a reference to the object must be held after the lock is released.
\Cref{lst:together:Simple Reference-Count API} shows a simple
API that might be used to implement simple non-atomic reference
counting---although simple reference counting is almost always
open-coded instead.

\begin{listing}
\begin{fcvlabel}[ln:together:Simple Reference-Count API]
\begin{VerbatimL}[commandchars=\\\[\]]
struct sref {
	int refcount;
};

void sref_init(struct sref *sref)
{
	sref->refcount = 1;
}

void sref_get(struct sref *sref)
{
	sref->refcount++;
}

int sref_put(struct sref *sref,
             void (*release)(struct sref *sref))
{
	WARN_ON(release == NULL);
	WARN_ON(release == (void (*)(struct sref *))kfree);

	if (--sref->refcount == 0) {
		release(sref);
		return 1;
	}
	return 0;
}
\end{VerbatimL}
\end{fcvlabel}
\caption{Simple Reference-Count API}
\label{lst:together:Simple Reference-Count API}
\end{listing}

\subsubsection{Atomic Counting}
\label{sec:together:Atomic Counting}

Simple atomic counting may be used in cases where any CPU acquiring
a reference must already hold a reference.
This style is used when a single CPU creates an object for its own private
use, but must allow for accesses from other CPUs, tasks, timer handlers,
and so on.
Any CPU that hands the object off must first acquire a new reference on
behalf of the recipient on the one hand, or refrain from further accesses
after the handoff on the other.
In the Linux kernel, the \co{kref} primitives are used to implement
this style of reference counting, as shown in
\cref{lst:together:Linux Kernel kref API}.\footnote{
	As of Linux v4.10.
	Linux v4.11 introduced a \co{refcount_t} API that improves
	efficiency weakly ordered platforms, but which is functionally
	equivalent to the \co{atomic_t} that it replaced.}

Atomic counting is required in this case because locking does not protect
all reference-count operations, which means that two different CPUs
might concurrently manipulate the reference count.
If normal increment and decrement were used, a pair of CPUs might both
fetch the reference count concurrently, perhaps both obtaining
the value ``3''.
If both of them increment their value, they will both obtain ``4'',
and both will store this value back into the counter.
Since the new value of the counter should instead be ``5'', one
of the increments has been lost.
Therefore, atomic operations must be used both for counter increments
and for counter decrements.

If releases are guarded by locking, hazard pointers, or RCU,
memory barriers are \emph{not} required, but for different reasons.
In the case of locking, the locks provide any needed memory barriers
(and disabling of compiler optimizations), and the locks also
prevent a pair of releases from running concurrently.
In the case of hazard pointers and RCU, cleanup will be deferred,
and any needed memory barriers or disabling of compiler optimizations
will be provided by the hazard-pointers or RCU infrastructure.
Therefore, if two CPUs release the final two references concurrently, the
actual cleanup will be deferred until both CPUs have released their hazard
pointers or exited their RCU read-side critical sections, respectively.

\QuickQuiz{
	Why isn't it necessary to guard against cases where one CPU
	acquires a reference just after another CPU releases the last
	reference?
}\QuickQuizAnswer{
	Because a CPU must already hold a reference in order to legally
	acquire another reference.
	Therefore, if one CPU releases the last reference, there had
	better not be any CPU acquiring a new reference!
}\QuickQuizEnd

\begin{listing}
\begin{fcvlabel}[ln:together:Linux Kernel kref API]
\begin{VerbatimL}[commandchars=\\\[\]]
struct kref {						\lnlbl[kref:b]
	atomic_t refcount;
};							\lnlbl[kref:e]

void kref_init(struct kref *kref)			\lnlbl[init:b]
{
	atomic_set(&kref->refcount, 1);
}							\lnlbl[init:e]

void kref_get(struct kref *kref)			\lnlbl[get:b]
{
	WARN_ON(!atomic_read(&kref->refcount));
	atomic_inc(&kref->refcount);
}							\lnlbl[get:e]

static inline int					\lnlbl[sub:b]
kref_sub(struct kref *kref, unsigned int count,
         void (*release)(struct kref *kref))
{
	WARN_ON(release == NULL);

	if (atomic_sub_and_test((int) count,		\lnlbl[check]
	                        &kref->refcount)) {
		release(kref);				\lnlbl[rel]
		return 1;				\lnlbl[ret:1]
	}
	return 0;
}							\lnlbl[sub:e]
\end{VerbatimL}
\end{fcvlabel}
\caption{Linux Kernel \tco{kref} API}
\label{lst:together:Linux Kernel kref API}
\end{listing}

\begin{fcvref}[ln:together:Linux Kernel kref API]
The \co{kref} structure itself, consisting of a single atomic
data item, is shown in \clnrefrange{kref:b}{kref:e} of
\cref{lst:together:Linux Kernel kref API}.
The \co{kref_init()} function on \clnrefrange{init:b}{init:e}
initializes the counter to the value ``1''.
Note that the \co{atomic_set()} primitive is a simple
assignment, the name stems from the data type of \co{atomic_t}
rather than from the operation.
The \co{kref_init()} function must be invoked during object creation,
before the object has been made available to any other CPU\@.

The \co{kref_get()} function on \clnrefrange{get:b}{get:e}
unconditionally atomically increments the counter.
The \co{atomic_inc()} primitive does not necessarily explicitly
disable compiler
optimizations on all platforms, but the fact that the \co{kref}
primitives are in a separate module and that the Linux kernel build
process does no cross-module optimizations has the same effect.

The \co{kref_sub()} function on \clnrefrange{sub:b}{sub:e}
atomically decrements the
counter, and if the result is zero, \clnref{rel} invokes the specified
\co{release()} function and \clnref{ret:1} returns, informing the caller
that \co{release()} was invoked.
Otherwise, \co{kref_sub()} returns zero, informing the caller that
\co{release()} was not called.
\end{fcvref}

\QuickQuizSeries{%
\QuickQuizB{
	\begin{fcvref}[ln:together:Linux Kernel kref API]
	Suppose that just after the \co{atomic_sub_and_test()}
	on \clnref{check} of
	\cref{lst:together:Linux Kernel kref API} is invoked,
	that some other CPU invokes \co{kref_get()}.
	Doesn't this result in that other CPU now having an illegal
	reference to a released object?
	\end{fcvref}
}\QuickQuizAnswerB{
	This cannot happen if these functions are used correctly.
	It is illegal to invoke \co{kref_get()} unless you already
	hold a reference, in which case the \co{kref_sub()} could
	not possibly have decremented the counter to zero.
}\QuickQuizEndB
\QuickQuizM{
	Suppose that \co{kref_sub()} returns zero, indicating that
	the \co{release()} function was not invoked.
	Under what conditions can the caller rely on the continued
	existence of the enclosing object?
}\QuickQuizAnswerM{
	The caller cannot rely on the continued existence of the
	object unless it knows that at least one reference will
	continue to exist.
	Normally, the caller will have no way of knowing this, and
	must therefore carefully avoid referencing the object after
	the call to \co{kref_sub()}.

	Interested readers are encouraged to work around this limitation
	using RCU, in particular, \co{call_rcu()}.
}\QuickQuizEndM
\QuickQuizE{
	Why not just pass \co{kfree()} as the release function?
}\QuickQuizAnswerE{
	Because the \co{kref} structure normally is embedded in
	a larger structure, and it is necessary to free the entire
	structure, not just the \co{kref} field.
	This is normally accomplished by defining a wrapper function
	that does a \co{container_of()} and then a \co{kfree()}.
}\QuickQuizEndE
}

\subsubsection{Atomic Counting With Release Memory Barrier}
\label{sec:together:Atomic Counting With Release Memory Barrier}

Atomic reference counting with release memory barriers is used by the
Linux kernel's networking layer to track the destination caches that
are used in packet routing.
The actual implementation is quite a bit more involved; this section
focuses on the aspects of \co{struct dst_entry} reference-count
handling that matches this use case,
shown in \cref{lst:together:Linux Kernel dst-clone API}.\footnote{
	As of Linux v4.13.
	Linux v4.14 added a level of indirection to permit more
	comprehensive debugging checks, but the overall effect in the
	absence of bugs is identical.}

\begin{listing}
\begin{fcvlabel}[ln:together:Linux Kernel dst-clone API]
\begin{VerbatimL}[commandchars=\\\[\]]
static inline
struct dst_entry * dst_clone(struct dst_entry * dst)
{
	if (dst)
		atomic_inc(&dst->__refcnt);
	return dst;
}

static inline
void dst_release(struct dst_entry * dst)
{
	if (dst) {
		WARN_ON(atomic_read(&dst->__refcnt) < 1);
		smp_mb__before_atomic_dec();		\lnlbl[mb]
		atomic_dec(&dst->__refcnt);
	}
}
\end{VerbatimL}
\end{fcvlabel}
\caption{Linux Kernel \tco{dst_clone} API}
\label{lst:together:Linux Kernel dst-clone API}
\end{listing}

The \co{dst_clone()} primitive may be used if the caller
already has a reference to the specified \co{dst_entry},
in which case it obtains another reference that may be handed off
to some other entity within the kernel.
Because a reference is already held by the caller, \co{dst_clone()}
need not execute any memory barriers.
The act of handing the \co{dst_entry} to some other entity might
or might not require a memory barrier, but if such a memory barrier
is required, it will be embedded in the mechanism used to hand the
\co{dst_entry} off.

\begin{fcvref}[ln:together:Linux Kernel dst-clone API]
The \co{dst_release()} primitive may be invoked from any environment,
and the caller might well reference elements of the \co{dst_entry}
structure immediately prior to the call to \co{dst_release()}.
The \co{dst_release()} primitive therefore contains a memory
barrier on \clnref{mb} preventing both the compiler and the CPU
from misordering accesses.
\end{fcvref}

Please note that the programmer making use of \co{dst_clone()} and
\co{dst_release()} need not be aware of the memory barriers, only
of the rules for using these two primitives.

\subsubsection{Atomic Counting With Check and Release Memory Barrier}
\label{sec:together:Atomic Counting With Check and Release Memory Barrier}

Consider a situation where the caller must be able to acquire a new
reference to an object to which it does not already hold a reference,
but where that object's existence is guaranteed.
The fact that initial reference-count acquisition can now run concurrently
with reference-count release adds further complications.
Suppose that a reference-count release finds that the new
value of the reference count is zero, signaling that it is
now safe to clean up the reference-counted object.
We clearly cannot allow a reference-count acquisition to
start after such clean-up has commenced, so the acquisition
must include a check for a zero reference count.
This check must be part of the atomic increment operation,
as shown below.

\QuickQuiz{
	Why can't the check for a zero reference count be
	made in a simple \qco{if} statement with an atomic
	increment in its \qco{then} clause?
}\QuickQuizAnswer{
	Suppose that the \qco{if} condition completed, finding
	the reference counter value equal to one.
	Suppose that a release operation executes, decrementing
	the reference counter to zero and therefore starting
	cleanup operations.
	But now the \qco{then} clause can increment the counter
	back to a value of one, allowing the object to be
	used after it has been cleaned up.

	This use-after-cleanup bug is every bit as bad as a
	full-fledged use-after-free bug.
}\QuickQuizEnd

The Linux kernel's \co{fget()} and \co{fput()} primitives
use this style of reference counting.
Simplified versions of these functions are shown in
\cref{lst:together:Linux Kernel fget/fput API}.\footnote{
	As of Linux v2.6.38.
	Additional \co{O_PATH} functionality was added in v2.6.39,
	refactoring was applied in v3.14, and \co{mmap_sem} contention
	was reduced in v4.1.}

\begin{listing}
\begin{fcvlabel}[ln:together:Linux Kernel fget/fput API]
\begin{VerbatimL}[commandchars=\\\@\$]
struct file *fget(unsigned int fd)
{
	struct file *file;
	struct files_struct *files = current->files;	\lnlbl@fetch$

	rcu_read_lock();				\lnlbl@rrl$
	file = fcheck_files(files, fd);			\lnlbl@lookup$
	if (file) {
		if (!atomic_inc_not_zero(&file->f_count)) { \lnlbl@acq$
			rcu_read_unlock();		\lnlbl@rru1$
			return NULL;			\lnlbl@ret:null$
		}
	}
	rcu_read_unlock();				\lnlbl@rru2$
	return file;					\lnlbl@ret:file$
}

struct file *
fcheck_files(struct files_struct *files, unsigned int fd)
{
	struct file * file = NULL;
	struct fdtable *fdt = rcu_dereference((files)->fdt);  \lnlbl@deref:fdt$

	if (fd < fdt->max_fds)				\lnlbl@range$
		file = rcu_dereference(fdt->fd[fd]);	\lnlbl@deref:fd$
	return file;					\lnlbl@ret:file2$
}

void fput(struct file *file)
{
	if (atomic_dec_and_test(&file->f_count))	\lnlbl@dec$
		call_rcu(&file->f_u.fu_rcuhead, file_free_rcu);  \lnlbl@call$
}

static void file_free_rcu(struct rcu_head *head)
{
	struct file *f;

	f = container_of(head, struct file, f_u.fu_rcuhead); \lnlbl@obtain$
	kmem_cache_free(filp_cachep, f);		\lnlbl@free$
}
\end{VerbatimL}
\end{fcvlabel}
\caption{Linux Kernel \tco{fget}/\tco{fput} API}
\label{lst:together:Linux Kernel fget/fput API}
\end{listing}

\begin{fcvref}[ln:together:Linux Kernel fget/fput API]
\Clnref{fetch} of \co{fget()} fetches the pointer to the current
process's file-descriptor table, which might well be shared
with other processes.
\Clnref{rrl} invokes \co{rcu_read_lock()}, which
enters an RCU read-side critical section.
The callback function from any subsequent \co{call_rcu()} primitive
will be deferred until a matching \co{rcu_read_unlock()} is reached
(\clnref{rru1} or~\lnref{rru2} in this example).
\Clnref{lookup} looks up the file structure corresponding to the file
descriptor specified by the \co{fd} argument, as will be
described later.
If there is an open file corresponding to the specified file descriptor,
then \clnref{acq} attempts to atomically acquire a reference count.
If it fails to do so, \clnrefrange{rru1}{ret:null} exit the RCU read-side critical
section and report failure.
Otherwise, if the attempt is successful, \clnrefrange{rru2}{ret:file}
exit the read-side
critical section and return a pointer to the file structure.

The \co{fcheck_files()} primitive is a helper function for
\co{fget()}.
\Clnref{deref:fdt} uses \co{rcu_dereference()} to safely fetch an
RCU-protected pointer to this task's current file-descriptor table,
and \clnref{range} checks to see if the specified file descriptor is in range.
If so, \clnref{deref:fd} fetches the pointer to the file structure, again using
the \co{rcu_dereference()} primitive.
\Clnref{ret:file2} then returns a pointer to the file structure or \co{NULL}
in case of failure.

The \co{fput()} primitive releases a reference to a file structure.
\Clnref{dec} atomically decrements the reference count, and, if the
result was zero, \clnref{call} invokes the \co{call_rcu()} primitives
in order to free up the file structure (via the \co{file_free_rcu()}
function specified in \co{call_rcu()}'s second argument), but only after
all currently-executing RCU read-side critical sections complete, that
is, after an RCU \IX{grace period} has elapsed.

Once the grace period completes, the \co{file_free_rcu()} function
obtains a pointer to the file structure on \clnref{obtain}, and frees it
on \clnref{free}.
\end{fcvref}

This code fragment thus demonstrates how RCU can be used to guarantee
existence while an in-object reference count is being incremented.

\subsection{Counter Optimizations}
\label{sec:together:Counter Optimizations}

In some cases where increments and decrements are common, but checks
for zero are rare, it makes sense to maintain per-CPU or per-task
counters, as was discussed in \cref{chp:Counting}.
For example, see the paper on sleepable read-copy update (SRCU), which
applies this technique to RCU~\cite{PaulEMcKenney2006c}.
This approach eliminates the need for atomic instructions or memory
barriers on the increment and decrement primitives, but still requires
that code-motion compiler optimizations be disabled.
In addition, the primitives such as \co{synchronize_srcu()}
that check for the aggregate reference
count reaching zero can be quite slow.
This underscores the fact that these techniques are designed
for situations where the references are frequently acquired and
released, but where it is rarely necessary to check for a zero
reference count.

However, it is usually the case that use of reference counts requires
writing (often atomically) to a data structure that is otherwise
read only.
In this case, reference counts are imposing expensive cache misses
on readers.

It is therefore worthwhile to look into synchronization mechanisms
that do not require readers to write to the data structure being
traversed.
One possibility is the hazard pointers covered in
\cref{sec:defer:Hazard Pointers}
and another is RCU, which is covered in
\cref{sec:defer:Read-Copy Update (RCU)}.

% together/hazptr.tex
% mainfile: ../perfbook.tex
% SPDX-License-Identifier: CC-BY-SA-3.0

\section{Hazard-Pointer Helpers}
\label{sec:together:Hazard-Pointer Helpers}
\epigraph{It's the little things that count, hundreds of them.}
	 {Cliff Shaw}

This section looks at some issues that can be addressed with the
help of hazard pointers.
In addition, hazard pointers can sometimes be used to address the
issues called out in \cref{sec:together:RCU Rescues}, and vice versa.

\subsection{Scalable Reference Count}
\label{sec:together:Scalable Reference Count}

Suppose a reference count is becoming a performance or scalability
bottleneck.
What can you do?

One approach is to instead use \IXpl{hazard pointer}.

There are some differences, perhaps most notably that with
hazard pointers it is extremely expensive to determine when
the corresponding reference count has reached zero.

One way to work around this problem is to split the load between
reference counters and hazard pointers.
Each data element has a reference counter that tracks the number
of other data elements referencing this element on the one hand,
and readers use hazard pointers on the other.

Making this arrangement work both efficiently and correctly can be
quite challenging, and so interested readers are invited to examine
the UnboundedQueue and ConcurrentHashMap data structures implemented in
Folly open-source library.\footnote{
	\url{https://github.com/facebook/folly}}

\subsection{Long-Duration Accesses}
\label{sec:together:Long-Duration Accesses}

Suppose a reader-writer-locking reader is holding the lock for so
long that updates are excessively delayed.
If that reader can reasonably be converted to use reference counting
instead of reader-writer locking, but if performance and scalability
considerations prevent use of actual reference counters, then hazard
pointers provides a scalable variant of reference counting.

The key point is that where reader-writer locking readers block all
updates for that lock, hazard pointers instead simply hang onto the
data that is actually needed, while still allowing updates to proceed.

If the reader cannot be reasonably be converted to use reference
counting, the tricks in \cref{sec:together:Long-Duration Accesses Two}
might be helpful.

% @@@ papers to maybe cite: OrcGC, ThreadScan, Fast and Robust Memory...

% @@@ Generalized hazard-pointer link counts, if and when.

% @@@ Representative hazard pointer for list, so that nothing
% @@@ in list gets freed until list's hazard pointer is released.
% @@@ Midpoint between hazard pointers and RCU, in fact, you
% @@@ could argue that Tasks Trace RCU with read-side memory
% @@@ barriers is sort of a per-CPU hazard pointers implementing RCU.
% @@@ Except no re-checking because CPUs cannot be freed.

% together/seqlock.tex
% mainfile: ../perfbook.tex
% SPDX-License-Identifier: CC-BY-SA-3.0

\section{Sequence-Locking Specials}
\label{sec:together:Sequence-Locking Specials}
\epigraph{The girl who can't dance says the band can't play.}
	 {Yiddish proverb}

This section looks at some special uses of sequence locks.

\subsection{Dueling Sequence Locks}
\label{sec:together:Dueling Sequence Locks}

The classic sequence-locking use case enables a reader to see a consistent
snapshot of a small collection of variables, for example, calibration
constants for timekeeping.
This works quite well in practice because calibration constants are
rarely updated and, when updated, are updated quickly.
Readers therefore almost never need to retry.

However, if the updater is delayed during the update, readers will
also be delayed.
Such delays might be due to interrupts, NMIs, or even
virtual-CPU preemption.

One way to prevent updater delays from causing reader delays is to
maintain two sets of calibration constants.
Each set is updated in turn, but frequently enough that readers can
make good use of either set.
Each set has its own sequence lock (\co{seqlock_t} structure).

The updater alternates between the two sets, so that an delayed updater
delays readers of at most one of the sets.

Each reader attempts to access the first set, but upon retry attempts
to access the second set.
If the second set also forces a retry, the reader repeats starting
again from the first set.
If the updater is stuck, only one of the two sets will force
readers to retry, and therefore readers will succeed as soon as
they attempt to access the other set.

\QuickQuiz{
	Why don't all sequence-locking use cases replicate the
	data in this fashion?
}\QuickQuizAnswer{
	Such replication is impractical if the data is too
	large, as it might be in the Schr\"odinger's-zoo example
	described in
	\cref{sec:together:Correlated Data Elements}.

	Such replication is unnecessary if delays are prevented,
	for example, when updaters disable interrupts when running
	on bare-metal hardware (that is, without the use of
	a vCPU-preemption-prone hypervisor).

	Alternatively, if readers can tolerate the occasional delay,
	then replication is again unnecessary.
	Consider the example of reader-writer locking, where
	writers always delay readers and vice versa.

	However, if the data to be replicated is reasonably
	small, if delays are possible, and if readers cannot
	tolerate these delays, replicating the data is an
	excellent approach.
}\QuickQuizEnd

\subsection{Correlated Data Elements}
\label{sec:together:Correlated Data Elements}

Suppose we have a hash table where we need correlated views of two or
more of the elements.
These elements are updated together, and we do not want to see an old
version of the first element along with new versions of the other
elements.
For example, Schr\"odinger decided to add his extended family to his
in-memory database along with all his animals.
Although Schr\"odinger understands that marriages and divorces do not
happen instantaneously, he is also a traditionalist.
As such, he absolutely does not want his database ever to show that the
bride is now married, but the groom is not, and vice versa.
Plus, if you think Schr\"odinger is a traditionalist, you just
try conversing with some of his family members!
In other words, Schr\"odinger wants to be able to carry out a
wedlock-consistent traversal of his database.

One approach is to use sequence locks
(see \cref{sec:defer:Sequence Locks}),
so that wedlock-related updates are carried out under the
protection of \co{write_seqlock()}, while reads requiring
wedlock consistency are carried out within
a \co{read_seqbegin()} / \co{read_seqretry()} loop.
Note that sequence locks are not a replacement for RCU protection:
Sequence locks protect against concurrent modifications, but RCU
is still needed to protect against concurrent deletions.

This approach works quite well when the number of correlated elements is
small, the time to read these elements is short, and the update rate is
low.
Otherwise, updates might happen so quickly that readers might never complete.
Although Schr\"odinger does not expect that even his least-sane relatives
will marry and divorce quickly enough for this to be a problem,
he does realize that this problem could well arise in other situations.
One way to avoid this reader-starvation problem is to have the readers
use the update-side primitives if there have been too many retries,
but this can degrade both performance and scalability.
Another way to avoid \IX{starvation} is to have multiple sequence locks,
in Schr\"odinger's case, perhaps one per species.

In addition, if the update-side primitives are used too frequently,
poor performance and scalability will result due to lock contention.
One way to avoid this is to maintain a per-element sequence lock,
and to hold both spouses' locks when updating their marital status.
Readers can do their retry looping on either of the spouses' locks
to gain a stable view of any change in marital status involving both
members of the pair.
This avoids contention due to high marriage and divorce rates, but
complicates gaining a stable view of all marital statuses during a
single scan of the database.

If the element groupings are well-defined and persistent, which marital
status is hoped to be,
then one approach is to add pointers to the data elements to link
together the members of a given group.
Readers can then traverse these pointers to access all the data elements
in the same group as the first one located.

This technique is used heavily in the Linux kernel, perhaps most
notably in the dcache subsystem~\cite{NeilBrown2015RCUwalk}.
Note that it is likely that similar schemes also work with hazard
pointers.

This approach provides \IXh{sequential}{consistency} to successful readers,
each of which will either see the effects of a given update or not,
with any partial updates resulting in a read-side retry.
Sequential consistency is an extremely strong guarantee, incurring equally
strong restrictions and equally high overheads.
In this case, we saw that readers might be starved on the one hand, or
might need to acquire the update-side lock on the other.
Although this works very well in cases where updates are infrequent,
it unnecessarily forces read-side retries even when the update does not
affect any of the data that a retried reader accesses.
\Cref{sec:together:Correlated Fields} therefore covers a much weaker form
of consistency that not only avoids reader starvation, but also avoids
any form of read-side retry.
The next section instead presents a weaker form of consistency that
can be provided with much lower probabilities of reader starvation.

\subsection{Atomic Move}
\label{sec:together:Atomic Move}

Suppose that individual data elements are moved from one data structure
to another, and that readers look up only single data structures.
However, when a data element moves, readers must must never see it as
being in both structures at the same time and must also never see it as
missing from both structures at the same time.
At the same time, any reader seeing the element in its new location
must never subsequently see it in its old location.
In addition, the move may be implemented by inserting a new copy of the
old data element into the destination location.

For example, consider a hash table that supports an atomic-to-readers
rename operation.
Expanding on Schr\"odinger's zoo, suppose that an animal's name changes,
for example, each of the brides in Schr\"odinger's traditionalist family
might change their last name to match that of their groom.

But changing their name might change the hash value, and might also
require that the bride's element move from one hash chain to another.
The consistency set forth above requires that if a reader successfully
looks up the new name, then any subsequent lookup of the old name by
that reader must result in failure.
Similarly, if a reader's lookup of the old name results in lookup failure,
then any subsequent lookup of the new name by that reader must succeed.
In short, a given reader should not see a bride momentarily blinking out
of existence, nor should that reader lookup a bride under her new name
and then later lookup that bride under her old name.

This consistency guarantee could be enforced with a single global sequence
lock as described in \cref{sec:together:Correlated Data Elements}, but
this can result in reader starvation even for readers that are not looking
up a bride who is currently undergoing a name change.
This guarantee could also be enforced by requiring that readers acquire
a per-hash-chain lock, but reviewing
\cref{fig:datastruct:Read-Only Hash-Table Performance For Schroedinger's Zoo}
shows that this results in poor performance and scalabilty, even for
single-socket systems.

Another more reader-friendly way to implement this is to use RCU and to
place a sequence lock on each element.
Readers looking up a given element act as sequence-lock readers across
their full set of accesses to that element.
Note that these sequence-lock operations will order each reader's lookups.

Renaming an element can then proceed roughly as follows:

\begin{enumerate}
\item	Acquire a global lock protecting rename operations.
\item	Allocate and initialize a copy of the element with the new name.
\item	Write-acquire the sequence lock on the element with the old name,
	which has the side effect of ordering this acquisition with the
	following insertion.
	Concurrent lookups of the old name will now repeatedly retry.
\item	Insert the copy of the element with the new name.
	Lookups of the new name will now succeed.
\item	Execute \co{smp_wmb()} to order the prior insertion with the
	subsequent removal.
\item	Remove the element with the old name.
	Concurrent lookups of the old name will now fail.
\item	Write-release the sequence lock if necessary, for example, if
	required by lock dependency checkers.
\item	Release the global lock.
\end{enumerate}

Thus, readers looking up the old name will retry until the new name
is available, at which point their final retry will fail.
Any subsequent lookups of the new name will succeed.
Any reader succeeding in looking up the new name is guaranteed that
any subsequent lookup of the old name will fail, perhaps after a series
of retries.

\QuickQuizSeries{%
\QuickQuizB{
	Is it possible to write-acquire the sequence lock on
	the new element before it is inserted instead of acquiring
	that of the old element before it is removed?
}\QuickQuizAnswerB{
	Yes, and the details are left as an exercise to the reader.

	The term \emph{tombstone} is sometimes used to refer to the
	element with the old name after its sequence lock is acquired.
	Similarly, the term \emph{birthstone} is sometimes used to refer
	to the element with the new name while its sequence lock is
	still held.
}\QuickQuizEndB

\QuickQuizE{
	Is it possible to avoid the global lock?
}\QuickQuizAnswerE{
	Yes, and one way to do this would be to use per-hash-chain locks.
	The updater could acquire lock(s) corresponding to both the old
	and the new element, acquiring them in address order.
	In this case, the insertion and removal operations would of
	course need to refrain from acquiring and releasing these
	same per-hash-chain locks.
	This complexity can be worthwhile if rename operations are
	frequent, and of course can allow rename operations to execute
	concurrently.
}\QuickQuizEndE
}% End of \QuickQuizSeries

It is of course possible to instead implement this procedure somewhat
more efficiently using simple flags.
However, this can be thought of as a simplified variant of sequence
locking that relies on the fact that a given element's sequence lock is
never write-acquired more than once.

% @@@ Reference TBD flag-for-deletion section.

\subsection{Upgrade to Writer}
\label{sec:together:Upgrade to Writer}

As discussed in
\cref{sec:defer:Quasi Reader-Writer Lock},
RCU permits readers to upgrade to writers.
This capability can be quite useful when a reader scanning an
RCU-protected data structure notices that the current element
needs to be updated.
What happens when you try this trick with sequence locking?

It turns out that this sequence-locking trick is actually used in
the Linux kernel, for example, by the \co{sdma_flush()} function in
\path{drivers/infiniband/hw/hfi1/sdma.c}.
The effect is to doom the enclosing reader to retry.
This trick is therefore used when the reader detects some condition
that requires a retry.

% together/applyrcu.tex
% mainfile: ../perfbook.tex
% SPDX-License-Identifier: CC-BY-SA-3.0

\section{RCU Rescues}
\label{sec:together:RCU Rescues}
\epigraph{With great doubts comes great understanding, with little doubts
	  comes little understanding.}
	 {Chinese proverb}

This section shows how to apply RCU to some examples discussed earlier
in this book.
In some cases, RCU provides simpler code, in other cases better
performance and scalability, and in still other cases, both.

\subsection{RCU and Per-Thread-Variable-Based Statistical Counters}
\label{sec:together:RCU and Per-Thread-Variable-Based Statistical Counters}

\Cref{sec:count:Per-Thread-Variable-Based Implementation}
described an implementation of statistical counters that provided
excellent
performance, roughly that of simple increment (as in the C \co{++}
operator), and linear scalability---but only for incrementing
via \co{inc_count()}.
Unfortunately, threads needing to read out the value via \co{read_count()}
were required to acquire a global
lock, and thus incurred high overhead and suffered poor scalability.
The code for the lock-based implementation is shown in
\cref{lst:count:Per-Thread Statistical Counters} on
\cpageref{lst:count:Per-Thread Statistical Counters}.

\QuickQuiz{
	Why on earth did we need that global lock in the first place?
}\QuickQuizAnswer{
	A given thread's \co{__thread} variables vanish when that
	thread exits.
	It is therefore necessary to synchronize any operation that
	accesses other threads' \co{__thread} variables with
	thread exit.
	Without such synchronization, accesses to \co{__thread} variable
	of a just-exited thread will result in segmentation faults.
}\QuickQuizEnd

\subsubsection{Design}

The hope is to use RCU rather than \co{final_mutex} to protect the
thread traversal in \co{read_count()} in order to obtain excellent
performance and scalability from \co{read_count()}, rather than just
from \co{inc_count()}.
However, we do not want to give up any accuracy in the computed sum.
In particular, when a given thread exits, we absolutely cannot
lose the exiting thread's count, nor can we double-count it.
Such an error could result in inaccuracies equal to the full
precision of the result, in other words, such an error would
make the result completely useless.
And in fact, one of the purposes of \co{final_mutex} is to
ensure that threads do not come and go in the middle of \co{read_count()}
execution.

Therefore, if we are to dispense with \co{final_mutex}, we will need
to come up with some other method for ensuring consistency.
One approach is to place the total count for all previously exited
threads and the array of pointers to the per-thread counters into a single
structure.
Such a structure, once made available to \co{read_count()}, is
held constant, ensuring that \co{read_count()} sees consistent data.

\subsubsection{Implementation}

\begin{fcvref}[ln:count:count_end_rcu:whole]
\Clnrefrange{struct:b}{struct:e} of
\cref{lst:together:RCU and Per-Thread Statistical Counters}
show the \co{countarray} structure, which contains a
\co{->total} field for the count from previously exited threads,
and a \co{counterp[]} array of pointers to the per-thread
\co{counter} for each currently running thread.
This structure allows a given execution of \co{read_count()}
to see a total that is consistent with the indicated set of running
threads.

\begin{listing}
\ebresizeverb{.71}{\input{CodeSamples/count/count_end_rcu=whole.fcv}}
\caption{RCU and Per-Thread Statistical Counters}
\label{lst:together:RCU and Per-Thread Statistical Counters}
\end{listing}

\Clnrefrange{perthread:b}{perthread:e}
contain the definition of the per-thread \co{counter}
variable, the global pointer \co{countarrayp} referencing
the current \co{countarray} structure, and
the \co{final_mutex} spinlock.

\Clnrefrange{inc:b}{inc:e} show \co{inc_count()}, which is unchanged from
\cref{lst:count:Per-Thread Statistical Counters}.
\end{fcvref}

\begin{fcvref}[ln:count:count_end_rcu:whole:read]
\Clnrefrange{b}{e} show \co{read_count()}, which has changed significantly.
\Clnref{rrl,rru} substitute \co{rcu_read_lock()} and
\co{rcu_read_unlock()} for acquisition and release of \co{final_mutex}.
\Clnref{deref} uses \co{rcu_dereference()} to snapshot the
current \co{countarray} structure into local variable \co{cap}.
Proper use of RCU will guarantee that this \co{countarray} structure
will remain with us through at least the end of the current RCU
read-side critical section at \clnref{rru}.
\Clnref{init} initializes \co{sum} to \co{cap->total}, which is the
sum of the counts of threads that have previously exited.
\Clnrefrange{add:b}{add:e} add up the per-thread counters corresponding
to currently
running threads, and, finally, \clnref{ret} returns the sum.
\end{fcvref}

\begin{fcvref}[ln:count:count_end_rcu:whole:init]
The initial value for \co{countarrayp} is
provided by \co{count_init()} on \clnrefrange{b}{e}.
This function runs before the first thread is created, and its job
is to allocate
and zero the initial structure, and then assign it to \co{countarrayp}.
\end{fcvref}

\begin{fcvref}[ln:count:count_end_rcu:whole:reg]
\Clnrefrange{b}{e} show the \co{count_register_thread()} function, which
is invoked by each newly created thread.
\Clnref{idx} picks up the current thread's index, \clnref{acq} acquires
\co{final_mutex}, \clnref{set} installs a pointer to this thread's
\co{counter}, and \clnref{rel} releases \co{final_mutex}.
\end{fcvref}

\QuickQuiz{
	\begin{fcvref}[ln:count:count_end_rcu:whole:reg]
	Hey!!!
	\Clnref{set} of
	\cref{lst:together:RCU and Per-Thread Statistical Counters}
	modifies a value in a pre-existing \co{countarray} structure!
	Didn't you say that this structure, once made available to
	\co{read_count()}, remained constant???
	\end{fcvref}
}\QuickQuizAnswer{
	Indeed I did say that.
	And it would be possible to make \co{count_register_thread()}
	allocate a new structure, much as \co{count_unregister_thread()}
	currently does.

	But this is unnecessary.
	Recall the derivation of the error bounds of \co{read_count()}
	that was based on the snapshots of memory.
	Because new threads start with initial \co{counter} values of
	zero, the derivation holds even if we add a new thread partway
	through \co{read_count()}'s execution.
	So, interestingly enough, when adding a new thread, this
	implementation gets the effect of allocating a new structure,
	but without actually having to do the allocation.
}\QuickQuizEnd

\begin{fcvref}[ln:count:count_end_rcu:whole:unreg]
\Clnrefrange{b}{e} show \co{count_unregister_thread()}, which is invoked
by each thread just before it exits.
\Clnrefrange{alloc:b}{alloc:e} allocate a new \co{countarray} structure,
\clnref{acq} acquires \co{final_mutex} and \clnref{rel} releases it.
\Clnref{copy} copies the contents of the current \co{countarray} into
the newly allocated version, \clnref{add} adds the exiting thread's \co{counter}
to new structure's \co{->total}, and \clnref{null} \co{NULL}s the exiting thread's
\co{counterp[]} array element.
\Clnref{retain} then retains a pointer to the current (soon to be old)
\co{countarray} structure, and \clnref{assign} uses \co{rcu_assign_pointer()}
to install the new version of the \co{countarray} structure.
\Clnref{sync} waits for a grace period to elapse, so that any threads that
might be concurrently executing in \co{read_count()}, and thus might
have references to the old \co{countarray} structure, will be allowed
to exit their RCU read-side critical sections, thus dropping any such
references.
\Clnref{free} can then safely free the old \co{countarray} structure.
\end{fcvref}

\QuickQuiz{
	Given the fixed-size \co{counterp} array, exactly how does this
	code avoid a fixed upper bound on the number of threads???
}\QuickQuizAnswer{
	You are quite right, that array does in fact reimpose the fixed
	upper limit.
	This limit may be avoided by tracking threads with a linked list,
	as is done in userspace RCU~\cite{MathieuDesnoyers2012URCU}.
	Doing something similar for this code is left as an exercise for
	the reader.
}\QuickQuizEnd

\subsubsection{Discussion}

\EQuickQuiz{
	Wow!
	\Cref{lst:together:RCU and Per-Thread Statistical Counters}
	contains 70 lines of code, compared to only 42 in
	\cref{lst:count:Per-Thread Statistical Counters}.
	Is this extra complexity really worth it?
}\EQuickQuizAnswer{
	This of course needs to be decided on a case-by-case basis.
	If you need an implementation of \co{read_count()} that
	scales linearly, then the lock-based implementation shown in
	\cref{lst:count:Per-Thread Statistical Counters}
	simply will not work for you.
	On the other hand, if calls to \co{read_count()} are sufficiently
	rare, then the lock-based version is simpler and might thus be
	better, although much of the size difference is due
	to the structure definition, memory allocation, and \co{NULL}
	return checking.

	Of course, a better question is ``Why doesn't the language
	implement cross-thread access to \co{__thread} variables?''
	After all, such an implementation would make both the locking
	and the use of RCU unnecessary.
	This would in turn enable an implementation that
	was even simpler than the one shown in
	\cref{lst:count:Per-Thread Statistical Counters}, but
	with all the scalability and performance benefits of the
	implementation shown in
	\cref{lst:together:RCU and Per-Thread Statistical Counters}!
}\EQuickQuizEnd

Use of RCU enables exiting threads to wait until other threads are
guaranteed to be done using the exiting threads' \co{__thread} variables.
This allows the \co{read_count()} function to dispense with locking,
thereby providing
excellent performance and scalability for both the \co{inc_count()}
and \co{read_count()} functions.
However, this performance and scalability come at the cost of some increase
in code complexity.
It is hoped that compiler and library writers employ user-level
RCU~\cite{MathieuDesnoyers2009URCU} to provide safe cross-thread
access to \co{__thread} variables, greatly reducing the
complexity seen by users of \co{__thread} variables.

\subsection{RCU and Counters for Removable I/O Devices}
\label{sec:together:RCU and Counters for Removable I/O Devices}

\Cref{sec:count:Applying Exact Limit Counters}
showed a fanciful pair of code fragments for dealing with counting
I/O accesses to removable devices.
These code fragments suffered from high overhead on the fastpath
(starting an I/O) due to the need to acquire a reader-writer
lock.

This section shows how RCU may be used to avoid this overhead.

The code for performing an I/O is quite similar to the original, with
an RCU read-side critical section being substituted for the reader-writer
lock read-side critical section in the original:

\begin{VerbatimN}[tabsize=8]
rcu_read_lock();
if (removing) {
	rcu_read_unlock();
	cancel_io();
} else {
	add_count(1);
	rcu_read_unlock();
	do_io();
	sub_count(1);
}
\end{VerbatimN}
\vspace{5pt}

The RCU read-side primitives have minimal overhead, thus speeding up
the fastpath, as desired.

The updated code fragment removing a device is as follows:

\begin{fcvlabel}[ln:together:applyrcu:Removing Device]
\begin{VerbatimN}[tabsize=8,commandchars=\\\[\]]
spin_lock(&mylock);
removing = 1;
sub_count(mybias);
spin_unlock(&mylock);
synchronize_rcu();
while (read_count() != 0) {	\lnlbl[nextofsync]
	poll(NULL, 0, 1);
}
remove_device();
\end{VerbatimN}
\end{fcvlabel}

\begin{fcvref}[ln:together:applyrcu:Removing Device]
Here we replace the reader-writer lock with an exclusive spinlock and
add a \co{synchronize_rcu()} to wait for all of the RCU read-side
critical sections to complete.
Because of the \co{synchronize_rcu()},
once we reach \clnref{nextofsync},
we know that all remaining I/Os have been accounted for.

Of course, the overhead of \co{synchronize_rcu()} can be large,
but given that device removal is quite rare, this is usually a good
tradeoff.
\end{fcvref}

\FloatBarrier
\subsection{Array and Length}
\label{sec:together:Array and Length}

Suppose we have an RCU-protected variable-length array, as shown in
\cref{lst:together:RCU-Protected Variable-Length Array}.
The length of the array \co{->a[]} can change dynamically, and at any
given time, its length is given by the field \co{->length}.
Of course, this introduces the following \IX{race condition}:

\begin{listing}
\begin{VerbatimL}[tabsize=8]
struct foo {
	int length;
	char *a;
};
\end{VerbatimL}
\caption{RCU-Protected Variable-Length Array}
\label{lst:together:RCU-Protected Variable-Length Array}
\end{listing}

\begin{enumerate}
\item	The array is initially 16 characters long, and thus \co{->length}
	is equal to 16.
\item	CPU~0 loads the value of \co{->length}, obtaining the value 16.
\item	CPU~1 shrinks the array to be of length 8, and assigns a pointer
	to a new 8-character block of memory into \co{->a[]}.
\item	CPU~0 picks up the new pointer from \co{->a[]}, and stores a
	new value into element 12.
	Because the array has only 8 characters, this results in
	a SEGV or (worse yet) memory corruption.
\end{enumerate}

How can we prevent this?

One approach is to make careful use of memory barriers, which are
covered in \cref{chp:Advanced Synchronization: Memory Ordering}.
This works, but incurs read-side overhead and, perhaps worse, requires
use of explicit memory barriers.

\begin{listing}
\begin{VerbatimL}[tabsize=8]
struct foo_a {
	int length;
	char a[0];
};

struct foo {
	struct foo_a *fa;
};
\end{VerbatimL}
\caption{Improved RCU-Protected Variable-Length Array}
\label{lst:together:Improved RCU-Protected Variable-Length Array}
\end{listing}

A better approach is to put the value and the array into the same structure,
as shown in
\cref{lst:together:Improved RCU-Protected Variable-Length Array}~\cite{Arcangeli03}.
Allocating a new array (\co{foo_a} structure) then automatically provides
a new place for the array length.
This means that if any CPU picks up a reference to \co{->fa}, it is
guaranteed that the \co{->length} will match the \co{->a[]}.

\begin{enumerate}
\item	The array is initially 16 characters long, and thus \co{->length}
	is equal to 16.
\item	CPU~0 loads the value of \co{->fa}, obtaining a pointer to
	the structure containing the value 16 and the 16-byte array.
\item	CPU~0 loads the value of \co{->fa->length}, obtaining the value 16.
\item	CPU~1 shrinks the array to be of length 8, and assigns a pointer
	to a new \co{foo_a} structure containing an 8-character block
	of memory into \co{->fa}.
\item	CPU~0 picks up the new pointer from \co{->a[]}, and stores a
	new value into element 12.
	But because CPU~0 is still referencing the old \co{foo_a}
	structure that contains the 16-byte array, all is well.
\end{enumerate}

Of course, in both cases, CPU~1 must wait for a grace period before
freeing the old array.

A more general version of this approach is presented in the next section.

\subsection{Correlated Fields}
\label{sec:together:Correlated Fields}
\OriginallyPublished{Section}{sec:together:Correlated Fields}{Correlated Fields}{Oregon Graduate Institute}{PaulEdwardMcKenneyPhD}

Suppose that each of Sch\"odinger's animals is represented by the
data element shown in
\cref{lst:together:Uncorrelated Measurement Fields}.
The \co{meas_1}, \co{meas_2}, and \co{meas_3} fields are a set
of correlated measurements that are updated periodically.
It is critically important that readers see these three values from
a single measurement update:
If a reader sees an old value of \co{meas_1} but new values of
\co{meas_2} and \co{meas_3}, that reader will become fatally confused.
How can we guarantee that readers will see coordinated sets of these
three values?\footnote{
	This situation is similar to that described in
	\cref{sec:together:Correlated Data Elements},
	except that here readers need only see a consistent view of a
	given single data element, not the consistent view of a
	group of data elements that was required in that earlier
	\lcnamecref{sec:together:Correlated Data Elements}.}

\begin{listing}
\begin{VerbatimL}[tabsize=8]
struct animal {
	char name[40];
	double age;
	double meas_1;
	double meas_2;
	double meas_3;
	char photo[0]; /* large bitmap. */
};
\end{VerbatimL}
\caption{Uncorrelated Measurement Fields}
\label{lst:together:Uncorrelated Measurement Fields}
\end{listing}

One approach would be to allocate a new \co{animal} structure,
copy the old structure into the new structure, update the new
structure's \co{meas_1}, \co{meas_2}, and \co{meas_3} fields,
and then replace the old structure with a new one by updating
the pointer.
This does guarantee that all readers see coordinated sets of
measurement values, but it requires copying a large structure due
to the \co{->photo[]} field.
This copying might incur unacceptably large overhead.

\begin{listing}
\begin{VerbatimL}[tabsize=8]
struct measurement {
	double meas_1;
	double meas_2;
	double meas_3;
};

struct animal {
	char name[40];
	double age;
	struct measurement *mp;
	char photo[0]; /* large bitmap. */
};
\end{VerbatimL}
\caption{Correlated Measurement Fields}
\label{lst:together:Correlated Measurement Fields}
\end{listing}

Another approach is to impose a level of indirection, as shown in
\cref{lst:together:Correlated Measurement Fields}~\cite[Section 5.3.4]{PaulEdwardMcKenneyPhD}.
When a new measurement is taken, a new \co{measurement} structure
is allocated, filled in with the measurements, and the \co{animal}
structure's \co{->mp} field is updated to point to this new
\co{measurement} structure using \co{rcu_assign_pointer()}.
After a grace period elapses, the old \co{measurement} structure
can be freed.

\QuickQuiz{
	But cant't the approach shown in
	\cref{lst:together:Correlated Measurement Fields}
	result in extra cache misses, in turn resulting in additional
	read-side overhead?
}\QuickQuizAnswer{
	Indeed it can.

\begin{listing}
\begin{VerbatimL}[tabsize=8]
struct measurement {
	double meas_1;
	double meas_2;
	double meas_3;
};

struct animal {
	char name[40];
	double age;
	struct measurement *mp;
        struct measurement meas;
	char photo[0]; /* large bitmap. */
};
\end{VerbatimL}
\caption{Localized Correlated Measurement Fields}
\label{lst:together:Localized Correlated Measurement Fields}
\end{listing}

	One way to avoid this cache-miss overhead is shown in
	\cref{lst:together:Localized Correlated Measurement Fields}:
	Simply embed an instance of a \co{measurement} structure
	named \co{meas}
	into the \co{animal} structure, and point the \co{->mp}
	field at this \co{->meas} field.

	Measurement updates can then be carried out as follows:

	\begin{enumerate}
	\item	Allocate a new \co{measurement} structure and place
		the new measurements into it.
	\item	Use \co{rcu_assign_pointer()} to point \co{->mp} to
		this new structure.
	\item	Wait for a grace period to elapse, for example using
		either \co{synchronize_rcu()} or \co{call_rcu()}.
	\item	Copy the measurements from the new \co{measurement}
		structure into the embedded \co{->meas} field.
	\item	Use \co{rcu_assign_pointer()} to point \co{->mp}
		back to the old embedded \co{->meas} field.
	\item	After another grace period elapses, free up the
		new \co{measurement} structure.
	\end{enumerate}

	This approach uses a heavier weight update procedure to eliminate
	the extra cache miss in the common case.
	The extra cache miss will be incurred only while an update is
	actually in progress.
}\QuickQuizEnd

This approach enables readers to see correlated values for selected
fields, but while incurring minimal read-side overhead.
This per-data-element consistency suffices in the common case where
a reader looks only at a single data element.

% Flag for deletion (if not already covered in the defer chapter).
% @@@ Issaquah Challenge.
% @@@ RLU & MV-RLU (Eventually the corresponding patents.)

\subsection{Update-Friendly Traversal}
\label{sec:together:Update-Friendly Traversal}

Suppose that a statistical scan of all elements in a hash table is
required.
For example, Schr\"odinger might wish to compute the average
length-to-weight ratio over all of his animals.\footnote{
	Why would such a quantity be useful?
	Beats me!
	But group statistics are often useful.}
Suppose further that Schr\"odinger is willing to ignore slight
errors due to animals being added to and removed from the hash
table while this statistical scan is being carried out.
What should Schr\"odinger do to control concurrency?

One approach is to enclose the statistical scan in an RCU read-side
critical section.
This permits updates to proceed concurrently without unduly impeding
the scan.
In particular, the scan does not block the updates and vice versa,
which allows scan of hash tables containing very large numbers of
elements to be supported gracefully, even in the face of very high
update rates.

\QuickQuiz{
	But how does this scan work while a resizable hash table
	is being resized?
	In that case, neither the old nor the new hash table is
	guaranteed to contain all the elements in the hash table!
}\QuickQuizAnswer{
	True, resizable hash tables as described in
	\cref{sec:datastruct:Non-Partitionable Data Structures}
	cannot be fully scanned while being resized.
	One simple way around this is to acquire the
	\co{hashtab} structure's \co{->ht_lock} while scanning,
	but this prevents more than one scan from proceeding
	concurrently.

	Another approach is for updates to mutate the old hash
	table as well as the new one while resizing is in
	progress.
	This would allow scans to find all elements in the old
	hash table.
	Implementing this is left as an exercise for the reader.
}\QuickQuizEnd

\subsection{Scalable Reference Count Two}
\label{sec:together:Scalable Reference Count Two}

Suppose a \IX{reference count} is becoming a performance or scalability
bottleneck.
What can you do?

One approach is to use per-CPU counters for each reference count,
somewhat similar to the algorithms in \cref{chp:Counting}, in particular,
the exact limit counters described in
\cref{sec:count:Exact Limit Counters}.
The need to switch between per-CPU and global modes for these counters
results either in expensive increments and decrements on the one hand
(\cref{sec:count:Atomic Limit Counter Implementation})
or in the use of POSIX signals on the other
(\cref{sec:count:Signal-Theft Limit Counter Design}).

Another approach is to use RCU to mediate the switch between per-CPU
and global counting modes.
Each update is carried out within an RCU read-side critical section,
and each update checks a flag to determine whether to update the
per-CPU counters on the one hand or the global on the other.
To switch modes, update the flag, wait for a grace period, and then
move any remaining counts from the per-CPU counters to the global
counter or vice versa.

The Linux kernel uses this RCU-mediated approach in its \co{percpu_ref}
style of reference counter.
Code using this reference counter must initialize the \co{percpu_ref}
structure using \co{percpu_ref_init()}, which takes as arguments
a pointer to the structure, a pointer to a function to invoke when
the reference count reaches zero, a set of mode flags, and a set
of \co{kmalloc()} \co{GFP_} flags.
After normal initialization, the structure has one reference and
is in per-CPU mode.

The mode flags are usually zero, but can include the
\co{PERCPU_REF_INIT_ATOMIC} bit if the counter is to start in slow
non-per-CPU (that is, atomic) mode.
There is also a \co{PERCPU_REF_ALLOW_REINIT} bit that allows
a given \co{percpu_ref} counter to be reused via a call to
\co{percpu_ref_reinit()} without needing to be freed and reallocated.
Regardless of how the \co{percpu_ref} structure is initialized,
\co{percpu_ref_get()} may be used to acquire a reference and
\co{percpu_ref_put()} may be used to release a reference.

When in per-CPU mode, the \co{percpu_ref} structure cannot determine whether
or not its value has reached zero.
When such a determination is necessary, \co{percpu_ref_kill()} may
be invoked.
This function switches the structure into atomic mode and removes the
initial reference installed by the call to \co{percpu_ref_init()}.
Of course, when in atomic mode, calls to \co{percpu_ref_get()} and
\co{percpu_ref_put()} are quite expensive, but \co{percpu_ref_put()}
can tell when the value reaches zero.

Readers desiring more \co{percpu_ref} information are referred to
the Linux-kernel documentation and source code.

\subsection{Retriggered Grace Periods}
\label{sec:together:Retriggered Grace Periods}

There is no \co{call_rcu_cancel()}, so once an \co{rcu_head} structure
is passed to \co{call_rcu()}, there is no calling it back.
It must be left alone until the callback is invoked.
In the common case, this is as it should be because the \co{rcu_head}
structure is on a one-way journey to deallocation.

However, there are use cases that combine RCU and explicit \co{open()}
and \co{close()} calls.
After a \co{close()} call, readers are not supposed to begin new accesses
to the data structure, but there might well be readers completing their
traversal.
This situation can be handled in the usual manner:
Wait for a grace period following the \co{close()} call before freeing
the data structures.

But what if \co{open()} is called before the grace period ends?

Again, there is no \co{call_rcu_cancel()}, so another approach is to
set a flag that is checked by the callback function, which can opt out
of actually freeing anything.
Problem solved!

But what if \co{open()} and then another \co{close()} are both called
before the grace period ends?

One approach is to have a second value for the flag that causes the
callback to requeue itself.

But what if there is not only a \co{open()} and then another \co{close()},
but also another \co{open()} before the grace period ends?

In this case, the callback needs to set state to reflect that last
\co{open()} still being in effect.

\begin{figure}
\centering
\resizebox{2.5in}{!}{\includegraphics{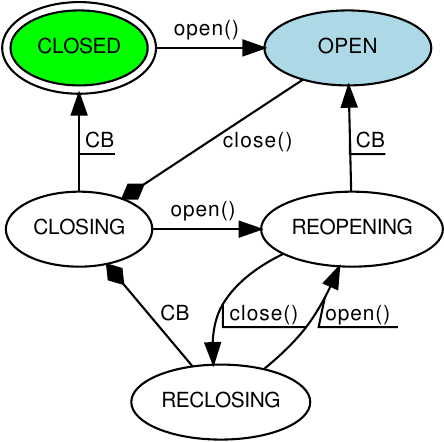}}
\caption{Retrigger-Grace-Period State Machine}
\label{fig:count:Retrigger-Grace-Period State Machine}
\end{figure}

Continuing this line of thought leads us to the state machine
shown in \cref{fig:count:Retrigger-Grace-Period State Machine}.
The initial state is CLOSED and the operational state is OPEN\@.
The diamond-shaped arrowheads denote \co{call_rcu()} invocation, while
the arrows labeled ``CB'' denote callback invocation.

The normal path through this state machine traverses the states CLOSED,
OPEN, CLOSING (with an invocation of \co{call_rcu()}), and back to CLOSED
once the callback has been invoked.
If \co{open()} is invoked before the grace period completes, the
state machine traverses the cycle OPEN, CLOSING (with
an invocation of \co{call_rcu()}), REOPENING, and back to OPEN once the
callback has been invoked.
If \co{open()} and then \co{close()} are invoked before the grace period
completes, the state machine traverses the cycle OPEN, CLOSING (with
an invocation of \co{call_rcu()}), REOPENING, RECLOSING, and back to CLOSING once the
callback has been invoked.

Given an indefinite alternating sequence of \co{close()} and \co{open()}
invocations, the state machine would traverse OPEN, and CLOSING (with
an invocation of \co{call_rcu()}), followed by alternating sojourns in
the REOPENING and RECLOSING states.
Once the grace period ends, the state machine would transition to
either of the CLOSING or the OPEN state, depending on which of the
RECLOSING or REOPENING states the callback was invoked in.

\begin{listing}
\ebresizeverb{.88}{\input{CodeSamples/together/retrigger-gp=whole.fcv}}
\caption{Retriggering a Grace Period (Pseudocode)}
\label{lst:together:Retriggering a Grace Period}
\end{listing}

Rough pseudocode of this state machine is shown in
\cref{lst:together:Retriggering a Grace Period}.
\begin{fcvref}[ln:together:retrigger-gp:whole]
The five states are shown on \clnrefrange{states:b}{states:e},
the current state is held in \co{rtrg_status} on \clnref{status},
which is protected by the lock on \clnref{lock}.

The three CB transitions (emanating from states CLOSING, REOPENING,
and RECLOSING) are implemented by the \co{close_cb()} function shown
on \clnrefrange{close_cb:b}{close_cb:e}.
\Clnref{cleanup} invokes a user-supplied \co{close_cleanup()} to
take any final cleanup actions such as freeing memory when
transitioning to the CLOSED state.
\Clnref{call_rcu1} contains the \co{call_rcu()} invocation that
causes a later transition to the CLOSED state.

The \co{open()} function on \clnrefrange{open:b}{open:e} implements
the transitions to the OPEN, CLOSING, and REOPENING states, with
\clnref{do_open} invoking a \co{do_open()} function to implement
any allocation and initialization of any needed data structures.

The \co{close()} function on \clnrefrange{close:b}{close:e}
implements the transitions to the CLOSING and RECLOSING states,
with \clnref{do_close} invoking a \co{do_close()} function to take
any actions that might be required to finalize this transition,
for example, causing later read-only traversals to return errors.
\Clnref{call_rcu2} contains the \co{call_rcu()} invocation that
causes a later transition to the CLOSED state.
\end{fcvref}

This state machine and pseudocode shows how to get the effect of a
\co{call_rcu_cancel()} in those rare situations needing such semantics.

\subsection{Long-Duration Accesses Two}
\label{sec:together:Long-Duration Accesses Two}

Suppose a reader-writer-locking reader is holding the lock for so
long that updates are excessively delayed.
Suppose further that this reader cannot reasonably be converted to
use reference counting
(otherwise, see \cref{sec:together:Long-Duration Accesses}).

If that reader can be reasonably converted to use RCU, that might
solve the problem.
The reason is that RCU readers do not completely block updates, but
rather block only the cleanup portions of those updates (including
memory reclamation).
Therefore, if the system has ample memory, converting the reader-writer
lock to RCU may suffice.

However, converting to RCU does not always suffice.
For example, the code might traverse an extremely large linked data
structure within a single RCU read-side critical section, which might
so greatly extend the RCU grace period that the system runs out of memory.
These situations can be handled in a couple of different ways:
\begin{enumerate*}[(1)]
\item	Use SRCU instead of RCU and
\item	Acquire a reference to exit the RCU reader.
\end{enumerate*}

\subsubsection{Use SRCU}
\label{sec:together:Use SRCU}

In the Linux kernel, RCU is global.
In other words, any long-running RCU reader anywhere in the kernel will
delay the current RCU grace period.
If the long-running RCU reader is traversing a small data structure,
that small amount of data is delaying freeing of all other data structures,
which can result in memory exhaustion.

One way to avoid this problem is to use SRCU for that long-running RCU
reader's data structure, with its own \co{srcu_struct} structure.
The resulting long-running SRCU readers will then delay only that
\co{srcu_struct} structure's grace periods, and not those of RCU,
thus avoiding memory exhaustion.
For more details, see the SRCU API in \cref{sec:defer:RCU Linux-Kernel API}.

Unfortunately, this approach does have some drawbacks.
For one thing, SRCU readers are not subject to priority boosting, which can
result in additional delays to low-priority SRCU readers on busy
systems.
Worse yet, defining a separate \co{srcu_struct} structure reduces the
number of RCU updaters, which in turn increases the grace-period
overhead per updater.
This means that giving each current Linux-kernel RCU use case its own
\co{srcu_struct} structure could multiply system-wide grace-period
overhead by the number of such structures.

Therefore, it is often better to acquire some sort of non-RCU reference
on the needed data to permit a momentary exit from the RCU read-side
critical section, as described in the next section.

\subsubsection{Acquire a Reference}
\label{sec:together:Acquire a Reference}

If the RCU read-side critical section is too long, shorten it!

In some cases, this can be done trivially.
For example, code that scans all of the hash chains of a statically
allocated array of hash buckets can just as easily scan each hash chain
within its own critical section.

This works because hash chains are normally quite short, and by design.
When traversing long linked structures, it is necessary to have some
way of stopping in the middle and resuming later.

For example, in Linux kernel v5.16, the \co{khugepaged_scan_file()}
function checks to see if some other task needs the current CPU
using \co{need_resched()}, and if so invokes \co{xas_pause()} to
adjust the traversal's iterator appropriately, and then invokes
\co{cond_resched_rcu()} to yield the CPU\@.
In turn, the \co{cond_resched_rcu()} function invokes \co{rcu_read_unlock()},
\co{cond_resched()}, and finally \co{rcu_read_lock()} to drop out of
the RCU read-side critical section in order to yield the CPU.

Of course, where feasible, another approach would be to switch to a
data structure such as a hash table that is more friendly to momentarily
dropping out of an RCU read-side critical section.

\QuickQuiz{
	But how would this work with a resizable hash table, such
	as the one described in
	\cref{sec:datastruct:Non-Partitionable Data Structures}?
}\QuickQuizAnswer{
	In this case, more care is required because the hash table
	might well be resized during the time that we momentarily
	exited the RCU read-side critical section.
	Worse yet, the resize operation can be expected to free the
	old hash buckets, leaving us pointing to the freelist.

	But it is not sufficient to prevent the old hash buckets
	from being freed.
	It is also necessary to ensure that those buckets continue
	to be updated.

	One way to handle this is to have a reference count on each
	set of buckets, which is initially set to the value one.
	A full-table scan would acquire a reference at the beginning of
	the scan (but only if the reference is non-zero) and release it
	at the end of the scan.
	The resizing would populate the new buckets, release the
	reference, wait for a grace period, and then wait for the
	reference to go to zero.
	Once the reference was zero, the resizing could let updaters
	forget about the old hash buckets and then free it.

	Actual implementation is left to the interested reader, who will
	gain much insight from this task.
}\QuickQuizEnd

% @@@ RCU link counts

% batching to trade off latency for perf/scale.

	\setlength{\parskip}{0.0pt plus 1ex}
	\input{qqztogether}
	\setlength{\parskip}{0.0pt plus 1.0pt}% return to default

% advsync/advsync.tex
% mainfile: ../perfbook.tex
% SPDX-License-Identifier: CC-BY-SA-3.0

\QuickQuizChapter{sec:advsync:Advanced Synchronization}{Advanced Synchronization}{qqzadvsync}
\Epigraph{If a little knowledge is a dangerous thing, just think what
	  you could do with a lot of knowledge!}{Unknown}

This chapter covers synchronization techniques used for lockless
algorithms and parallel real-time systems.

Although lockless algorithms can be quite helpful when faced with
extreme requirements, they are no panacea.
For example, as noted at the end of \cref{chp:Counting},
you should thoroughly apply partitioning, batching, and
well-tested packaged weak APIs
(see \cref{chp:Data Ownership,chp:Deferred Processing})
before even thinking about lockless algorithms.

But after doing all that, you still might find yourself needing the
advanced techniques described in this chapter.
To that end,
\cref{sec:advsync:Avoiding Locks}
summarizes techniques used thus far for avoiding locks and
\cref{sec:advsync:Non-Blocking Synchronization}
gives a brief overview of non-blocking synchronization.
Memory ordering is also quite important, but it warrants its own
\lcnamecref{chp:Advanced Synchronization: Memory Ordering}, namely
\cref{chp:Advanced Synchronization: Memory Ordering}.

The second form of advanced synchronization provides the stronger
forward-progress guarantees needed for parallel real-time computing,
which is the topic of
\cref{sec:advsync:Parallel Real-Time Computing}.

\section{Avoiding Locks}
\label{sec:advsync:Avoiding Locks}
\epigraph{We are confronted with insurmountable opportunities.}
	 {Walt Kelly}

Although locking is the workhorse of parallelism in production, in
many situations performance, scalability, and real-time response can
all be greatly improved through use of lockless techniques.
A particularly impressive example of such a lockless technique is
the statistical counters described in
\cref{sec:count:Statistical Counters},
which avoids not only locks, but also read-modify-write atomic operations,
memory barriers, and even cache misses for counter increments.
Other examples we have covered include:

\begin{enumerate}
\item	The fastpaths through a number of other counting algorithms
	in \cref{chp:Counting}.
\item	The fastpath through resource allocator caches in
	\cref{sec:SMPdesign:Resource Allocator Caches}.
\item	The maze solver in \cref{sec:SMPdesign:Beyond Partitioning}.
\item	The data-ownership techniques in \cref{chp:Data Ownership}.
\item	The reference-counting, hazard-pointer, and RCU techniques
	in \cref{chp:Deferred Processing}.
\item	The lookup code paths in \cref{chp:Data Structures}.
\item	Many of the techniques in \cref{chp:Putting It All Together}.
\end{enumerate}

In short, lockless techniques are quite useful and are heavily used.
However, it is best if lockless techniques are hidden behind a
well-defined API, such as the \co{inc_count()}, \co{memblock_alloc()},
\co{rcu_read_lock()}, and so on.
The reason for this is that undisciplined use of lockless techniques
is a good way to create difficult bugs.
If you believe that finding and fixing such bugs is easier than avoiding
them, please re-read
\cref{chp:Validation,chp:Formal Verification}.

\section{Non-Blocking Synchronization}
\label{sec:advsync:Non-Blocking Synchronization}
\epigraph{Never worry about theory as long as the machinery does what
	  it's supposed to do.}
	 {Robert A. Heinlein}

The term \IXBacrfst{nbs}~\cite{MauriceHerlihy90a}
describes eight classes of \IX{linearizable} algorithms with differing
\emph{\IXBpl{forward-progress guarantee}}~\cite{DanAlitarh2013PracticalProgress},
which are as follows:

\begin{enumerate}
\item	\emph{\IXalth{Bounded population-oblivious wait-free}{bounded population-oblivious}{wait free} synchronization}:
	Every thread will make progress within a specific finite period
	of time, where this period of time is independent of the number
	of threads~\cite{HerlihyShavit2008Textbook}.
	This level is widely considered to be even less achievable than
	bounded wait-free synchronization.
\item	\emph{\IXalth{Bounded wait-free}{bounded}{wait free} synchronization}:
	Every thread will make progress within
	a specific finite period of time~\cite{Herlihy91}.
	This level is widely considered to be unachievable, which might be why
	Alitarh et al.\ omitted it~\cite{DanAlitarh2013PracticalProgress}.
\item	\emph{\IXalt{Wait-free}{wait free} synchronization}:
	Every thread will make progress
	in finite time~\cite{Herlihy93}.
\item	\emph{\IXalt{Lock-free}{lock free} synchronization}:
	At least one thread will
	make progress in finite time~\cite{Herlihy93}.
\item	\emph{\IXalt{Obstruction-free}{obstruction free} synchronization}:
	Every thread will make progress in finite time in the absence of
	contention~\cite{HerlihyLM03}.
\item	\emph{\IXalt{Clash-free}{clash free} synchronization}:
	At least one thread will make progress in finite time in the absence of
	contention~\cite{DanAlitarh2013PracticalProgress}.
\item	\emph{\IXalt{Starvation-free}{starvation free} synchronization}:
	Every thread will make progress in finite time in the absence of
	failures~\cite{DanAlitarh2013PracticalProgress}.
\item	\emph{\IXalt{Deadlock-free}{deadlock free} synchronization}:
	At least one thread will make progress in finite time in the absence of
	failures~\cite{DanAlitarh2013PracticalProgress}.
\end{enumerate}

NBS class~1 was formulated some time before 2015,
classes~2, 3, and~4 were first formulated in the early 1990s,
class~5 was first formulated in the early 2000s,
and class~6 was first formulated in 2013.
The final two classes have seen informal use for a great many decades,
but were reformulated in 2013.

\QuickQuiz{
	Given that there will always be a sharply limited number of
	CPUs available, is population obliviousness really useful?
}\QuickQuizAnswer{
	Given the surprisingly limited scalability of any number of
	NBS algorithms, population obliviousness can be surprisingly
	useful.
	Nevertheless, the overall point of the question is valid.
	It is not normally helpful for an algorithm to scale beyond the
	size of the largest system it is ever going to run on.
}\QuickQuizEnd

In theory, any parallel algorithm can be cast into wait-free form,
but there are a relatively small subset of NBS algorithms that are
in common use.
A few of these are listed in the following section.

\subsection{Simple NBS}
\label{sec:advsync:Simple NBS}

Perhaps the simplest NBS algorithm is atomic update of an integer
counter using fetch-and-add (\co{atomic_add_return()}) primitives.
This section lists a few additional commonly used NBS algorithms in
roughly increasing order of complexity.

\subsubsection{NBS Sets}
\label{sec:advsync:NBS Sets}

One simple NBS algorithm implements a set of integers in an array.
Here the array index indicates a value that might be a member of the set
and the array element indicates whether or not that value actually is
a set member.
The linearizability criterion for NBS algorithms requires that reads from
and updates to the array either use atomic instructions or be accompanied
by memory barriers, but in the not-uncommon case where linearizability
is not important, simple volatile loads and stores suffice, for example,
using \co{READ_ONCE()} and \co{WRITE_ONCE()}.

An NBS set may also be implemented using a bitmap, where each value that
might be a member of the set corresponds to one bit.
Reads and updates must normally be carried out via atomic bit-manipulation
instructions, although compare-and-swap (\co{cmpxchg()} or CAS)
instructions can also be used.

\subsubsection{NBS Counters}
\label{sec:advsync:NBS Counters}

The statistical counters algorithm discussed in
\cref{sec:count:Statistical Counters}
can be considered to be bounded-wait-free, but only by using a cute
definitional trick in which the sum is considered to be approximate
rather than exact.\footnote{
	Citation needed.
	I heard of this trick verbally from Mark Moir.}
Given sufficiently wide error bounds that are a function of the length
of time that the \co{read_count()} function takes to sum the counters,
it is not possible to prove that any non-linearizable behavior occurred.
This definitely (if a bit artificially) classifies the statistical-counters
algorithm as bounded wait-free.
This algorithm is probably the most heavily used NBS algorithm in the
Linux kernel.

\subsubsection{Half-NBS Queue}
\label{sec:advsync:Half-NBS Queue}

\begin{fcvref}[ln:advsync:NBS Enqueue Algorithm]
Another common NBS algorithm is the atomic queue where elements are
enqueued using an atomic exchange instruction~\cite{MagedMichael1993JPDC},
followed by a store into the \co{->next} pointer of the new element's
predecessor, as shown in \cref{lst:advsync:NBS Enqueue Algorithm},
which shows the userspace-RCU library
implementation~\cite{MathieuDesnoyers2009URCU}.
\Clnref{tail} updates the tail pointer to reference the new element while
returning a reference to its predecessor, which is stored in
local variable \co{old_tail}.
\Clnref{pred} then updates the predecessor's \co{->next} pointer to
reference the newly added element, and finally \clnref{ret}
returns an indication as to whether or not the queue was initially
empty.

\begin{listing}
\begin{fcvlabel}[ln:advsync:NBS Enqueue Algorithm]
\begin{VerbatimL}[commandchars=\\\[\]]
static inline bool
___cds_wfcq_append(struct cds_wfcq_head *head,
                   struct cds_wfcq_tail *tail,
                   struct cds_wfcq_node *new_head,
                   struct cds_wfcq_node *new_tail)
{
	struct cds_wfcq_node *old_tail;

	old_tail = uatomic_xchg(&tail->p, new_tail);	\lnlbl[tail]
	CMM_STORE_SHARED(old_tail->next, new_head);     \lnlbl[pred]
	return old_tail != &head->node;			\lnlbl[ret]
}

static inline bool
_cds_wfcq_enqueue(struct cds_wfcq_head *head,
                  struct cds_wfcq_tail *tail,
                  struct cds_wfcq_node *new_tail)
{
	return ___cds_wfcq_append(head, tail,
	                          new_tail, new_tail);
}
\end{VerbatimL}
\end{fcvlabel}
\caption{NBS Enqueue Algorithm}
\label{lst:advsync:NBS Enqueue Algorithm}
\end{listing}

Although mutual exclusion is required to dequeue a single element
(so that dequeue is blocking), it is possible to carry out a non-blocking
removal of the entire contents of the queue.
What is not possible is to dequeue any given element in a non-blocking
manner:
The enqueuer might have failed between \clnref{tail,pred} of the
listing, so that the element in question is only partially enqueued.
This results in a half-NBS algorithm where enqueues are NBS but
dequeues are blocking.
This algorithm is nevertheless heavily used in practice, in part because
most production software is not required to tolerate arbitrary fail-stop
errors.
\end{fcvref}

\QuickQuiz{
	Wait!
	In order to dequeue all elements, both the \co{->head} and
	\co{->tail} pointers must be changed, which cannot be done
	atomically on typical computer systems.
	So how is this supposed to work???
}\QuickQuizAnswer{
	One pointer at a time!

	First, atomically exchange the \co{->head} pointer with \co{NULL}.
	If the return value from the atomic exchange operation is \co{NULL},
	the queue was empty and you are done.
	And if someone else attempts a dequeue-all at this point,
	they will get back a \co{NULL} pointer.

	Otherwise, atomically exchange the \co{->tail} pointer with a
	pointer to the now-\co{NULL} \co{->head} pointer.
	The return value from the atomic exchange operation is a pointer
	to the \co{->next} field of the eventual last element on the list.

	Producing and testing actual code is left as an exercise for the
	interested and enthusiastic reader, as are strategies for handling
	half-enqueued elements.
}\QuickQuizEnd

\subsubsection{NBS Stack}
\label{sec:advsync:NBS Stack}

\begin{fcvref}[ln:advsync:lifo_push:whole]
\Cref{lst:advsync:NBS Stack Algorithm}
shows the LIFO push algorithm, which boasts lock-free push and
bounded wait-free pop (\path{lifo-push.c}), forming an NBS stack.
The origins of this algorithm are unknown, but it was referred to in
a patent granted in 1975~\cite{PaulJBrown1975LIFOpush}.
This patent was filed in 1973, a few months before your editor
saw his first computer, which had but one CPU\@.

\begin{listing}
\input{CodeSamples/advsync/lifo_push=whole.fcv}
\caption{NBS Stack Algorithm}
\label{lst:advsync:NBS Stack Algorithm}
\end{listing}

\Clnrefrange{struct:b}{struct:e} show the \co{node_t} structure,
which contains an arbitrary value and a pointer to the next structure
on the stack and
\clnref{top} shows the top-of-stack pointer.

The \co{list_push()} function spans \clnrefrange{push:b}{push:e}.
\Clnref{push:alloc} allocates a new node and
\clnref{push:initialize} initializes it.
\Clnref{push:next} initializes the newly allocated node's \co{->next}
pointer, and \clnref{push:cmpxchg} attempts to push it on the stack.
If \clnref{push:check} detects \co{cmpxchg()} failure, another pass
through the loop retries.
Otherwise, the new node has been successfully pushed, and this function
returns to its caller.
Note that \clnref{push:check} resolves races in which two concurrent
instances of \co{list_push()} attempt to push onto the stack.
The \co{cmpxchg()} will succeed for one and fail for the other,
causing the other to retry, thereby selecting an arbitrary order for
the two node on the stack.

The \co{list_pop_all()} function spans \clnrefrange{popall:b}{popall:e}.
The \co{xchg()} statement on \clnref{popall:xchg} atomically removes
all nodes on the stack, placing the head of the resulting list in local
variable \co{p} and setting \co{top} to \co{NULL}.
This atomic operation serializes concurrent calls to \co{list_pop_all()}:
One of them will get the list, and the other a \co{NULL} pointer, at
least assuming that there were no concurrent calls to \co{list_push()}.

An instance of \co{list_pop_all()} that obtains a non-empty list in
\co{p} processes this list in the loop spanning
\clnrefrange{popall:loop:b}{popall:loop:e}.
\Clnref{popall:next} prefetches the \co{->next} pointer,
\clnref{popall:foo} invokes the function referenced by \co{foo()} on the
current node,
\clnref{popall:free} frees the current node, and
\clnref{popall:pnext} sets up \co{p} for the next pass through the loop.

But suppose that a pair of \co{list_push()} instances run concurrently
with a \co{list_pop_all()} with a list initially containing a single
\Node{A}.
Here is one way that this scenario might play out:

\begin{enumerate}
\item	The first \co{list_push()} instance pushes a new \Node{B},
	executing through \clnref{push:next}, having just stored
	a pointer to \Node{A} into \Node{B}'s \co{->next} pointer.
\item	The \co{list_pop_all()} instance runs to completion,
	setting \co{top} to \co{NULL} and freeing \Node{A}.
\item	The second \co{list_push()} instance runs to completion,
	pushing a new \Node{C}, but happens to allocate the memory
	that used to belong to \Node{A}.
\item	The first \co{list_push()} instance executes the \co{cmpxchg()}
	on \clnref{push:cmpxchg}.
	Because new \Node{C} has the same address as the newly freed \Node{A},
	this \co{cmpxchg()} succeeds and this \co{list_push()} instance
	runs to completion.
\end{enumerate}

Note that both pushes and the popall all ran successfully despite the
reuse of \Node{A}'s memory.
This is an unusual property:
Most data structures require protection against what is often called
the ABA problem.

But this property holds only for algorithm written in assembly
language.
The sad fact is that most languages (including C and C++) do not support
pointers to lifetime-ended objects, such as the pointer to the old \Node{A}
contained in \Node{B}'s \co{->next} pointer.
In fact, compilers are within their rights to assume that if two pointers
(call them \co{p} and \co{q}) were returned from two different calls to
\co{malloc()}, then those pointers must not be equal.
Real compilers really will generate the constant \co{false} in
response to a \co{p==q} comparison.
A pointer to an object that has been freed, but whose memory has been
reallocated for a compatibly typed object is termed a \emph{zombie pointer}.

Many concurrent applications avoid this problem by carefully hiding the
memory allocator from the compiler, thus preventing the compiler from
making inappropriate assumptions.
This obfuscatory approach currently works in practice, but might well
one day fall victim to increasingly aggressive optimizers.
There is work underway in both the C and C++ standards committees
to address this
problem~\cite{PaulEMcKenney2019PointerLifetimeEndZap,PaulEMcKenney2020PointerLifetimeEndZapCpp}.
In the meantime, please exercise great care when coding ABA-tolerant
algorithms.
\end{fcvref}

\QuickQuiz{
	So why not ditch antique languages like C and C++ for something
	more modern?
}\QuickQuizAnswer{
	That won't help unless the more-modern languages proponents
	are energetic enough to write their own compiler backends.
	The usual practice of re-using existing backends also reuses
	charming properties such as refusal to support pointers to
	lifetime-ended objects.
}\QuickQuizEnd

\subsection{Applicability of NBS Benefits}
\label{sec:advsync:Applicability of NBS Benefits}

The most heavily cited NBS benefits stem from its forward-progress
guarantees, its tolerance of fail-stop bugs, and from its linearizability.
Each of these is discussed in one of the following sections.

\subsubsection{NBS Forward Progress Guarantees}
\label{sec:advsync:NBS Forward Progress Guarantees}

NBS's forward-progress guarantees have caused many to suggest its use in
real-time systems, and NBS algorithms are in fact used in a great many
such systems.
However, it is important to note that forward-progress guarantees are
largely orthogonal to those that form the basis of real-time programming:

\begin{enumerate}
\item	Real-time forward-progress guarantees usually have some
	definite time associated with them, for example,
	``\IXh{scheduling}{latency} must be less than 100 microseconds.''
	In contrast, the most popular forms of NBS only guarantees
	that progress will be made in finite time, with no definite
	bound.
\item	Real-time forward-progress guarantees are often
	probabilistic, as in the soft-real-time guarantee that
	``at least 99.9\,\% of the time, scheduling latency must
	be less than 100 microseconds.''
	In contrast, many of NBS's forward-progress guarantees are
	unconditional.
\item	Real-time forward-progress guarantees are often conditioned on
	environmental constraints, for example, only being honored:
	\begin{enumerate*}[(1)]
	\item For the highest-priority tasks,
	\item When each CPU spends at least a certain fraction of its time idle,
	and
	\item When I/O rates are below some specified maximum.
	\end{enumerate*}
	In contrast, NBS's forward-progress
	guarantees are often unconditional, although recent NBS work
	accommodates conditional
	guarantees~\cite{DanAlitarh2013PracticalProgress}.
\item	An important component of a real-time program's environment
	is the scheduler.
	NBS algorithms assume a worst-case \emph{demonic scheduler},
	though for whatever reason, not a scheduler so demonic that
	it simply refuses to ever run the application housing the NBS
	algorithm.
	In contrast, real-time systems assume that the scheduler is
	doing its level best to satisfy any scheduling constraints
	it knows about, and, in the absence of such constraints,
	its level best to honor process priorities and to provide
	fair scheduling to processes of the same priority.
	Non-demonic schedulers allow real-time programs to use simpler
	algorithms than those required for
	NBS~\cite{DanAlitarh2013PracticalProgress,BjoernBrandenburgPhD}.
\item	NBS forward-progress guarantee classes assume that a number
	of underlying operations are lock-free or even wait-free,
	when in fact these operations are blocking on common-case
	computer systems.
\item	NBS forward-progress guarantees are often achieved by subdividing
	operations.
	For example, in order to avoid a blocking dequeue operation,
	an NBS algorithm might substitute a non-blocking polling
	operation.
	This is fine in theory, but not helpful in practice to real-world
	programs that require an element to propagate through the queue
	in a timely fashion.
\item	Real-time forward-progress guarantees usually apply only
	in the absence of software bugs.
	In contrast, many classes of NBS guarantees apply even in the
	face of fail-stop bugs.
\item	NBS forward-progress guarantee classes imply linearizability.
	In contrast, real-time forward progress guarantees are often
	independent of ordering constraints such as linearizability.
\end{enumerate}

\QuickQuiz{
	Why does anyone care about demonic schedulers?
}\QuickQuizAnswer{
	A demonic scheduler is one way to model an insanely overloaded
	system.
	After all, if you have an algorithm that you can prove runs
	reasonably given a demonic scheduler, mere overload should
	be no problem, right?

	On the other hand, it is only reasonable to ask if a demonic
	scheduler is really the best way to model overload conditions.
	And perhaps it is time for more accurate models.
	For one thing, a system might be overloaded in any of a number
	of ways.
	After all, an NBS algorithm that works fine on a demonic scheduler
	might or might not do well in out-of-memory conditions, when
	mass storage fills, or when the network is congested.

	Except that systems' core counts have been increasing, which
	means that an overloaded system is quite likely to be running
	more than one concurrent program.\footnote{
		As a point of reference, back in the mid-1990s, Paul
		witnessed a 16-CPU system running about 20 instances
		of a certain high-end proprietary database.}
	In that case, even if a demonic scheduler is not so demonic
	as to inject idle cycles while there are runnable tasks,
	it is easy to imagine such a scheduler consistently favoring
	the other program over yours.
	If both programs could consume all available CPU, then this
	scheduler might not run your program at all.

	One way to avoid these issues is to simply avoid overload
	conditions.
	This is often the preferred approach in production, where load
	balancers direct traffic away from overloaded systems.
	And if all systems are overloaded, it is not unheard of to simply
	\emph{shed load}, that is, to drop the low-priority incoming requests.
	Nor is this approach limited to computing, as those who have
	suffered through a rolling blackout can attest.
	But load-shedding is often considered a bad thing by those
	whose load is being shed.

	As always, choose wisely!
}\QuickQuizEnd

To reiterate, despite these differences, a number of NBS algorithms are
extremely useful in real-time programs.

\subsubsection{NBS Underlying Operations}
\label{sec:advsync:NBS Underlying Operations}

An NBS algorithm can be truly non-blocking only if the underlying
operations that it uses are also non-blocking.
In a surprising number of cases, this is not the case in practice.

For example, non-blocking algorithms often allocate memory.
In theory, this is fine, given the existence of lock-free
memory allocators~\cite{MagedMichael04NBSmalloc}.
But in practice, most environments must eventually obtain memory from
operating-system kernels, which commonly use locking.
Therefore, unless all the memory that will ever be needed is somehow
preallocated, a ``non-blocking'' algorithm that allocates memory will not
be non-blocking when running on common-case real-world computer systems.

This same point clearly also applies to algorithms performing I/O
operations or otherwise interacting with their environment.

Perhaps surprisingly, this point also applies to ostensibly non-blocking
algorithms that do only plain loads and stores, such as the counters
discussed in \cref{sec:advsync:NBS Counters}.
And at first glance, those loads and stores that can be compiled into
single load and store instructions, respectively, would seem to be
not just non-blocking, but bounded population-oblivious wait free.

Except that load and store instructions are not necessarily either
fast or deterministic.
For example, as noted in \cref{chp:Hardware and its Habits}, cache misses
can consume thousands of CPU cycles.
Worse yet, the measured cache-miss latencies can be a function of the
number of CPUs, as illustrated in
\cref{fig:count:Atomic Increment Scalability on x86}.
It is only reasonable to assume that these latencies also depend on
the details of the system's interconnect.
In addition, given that hardware vendors generally do not publish
upper bounds for cache-miss latencies, it seems brave to assume that
memory-reference instructions are in fact wait-free in modern computer
systems.
And the antique systems for which such bounds are available suffer from
profound overall slowness.

Furthermore, hardware is not the only source of slowness for
memory-reference instructions.
For example, when running on typical computer systems, both loads and
stores can result in page faults.
Which cause in-kernel page-fault handlers to be invoked.
Which might acquire locks, or even do I/O, potentially even using
something like network file system (NFS)\@.
All of which are most emphatically blocking operations.

Nor are page faults the only kernel-induced hazard.
A given CPU might be interrupted at any time, and the interrupt
handler might run for some time.
During this time, the user-mode ostensibly non-blocking algorithm
will not be running at all.
This situation raises interesting questions about the forward-progress
guarantees provided by system calls relying on interrupts, for example,
the \co{membarrier()} system call.

Things do look bleak, but the non-blocking nature of such algorithms
can be at least partially redeemed using a number of approaches:

\begin{enumerate}
\item	Run on bare metal, with paging disabled.
	If you are both brave and confident that you can write code that
	is free of wild-pointer bugs, this approach might be for you.
\item	Run on a non-blocking operating-system kernel~\cite{Cheriton96a}.
	Such kernels are quite rare, in part because they have
	traditionally completely failed to provide the hoped-for
	performance and scalability advantages over lock-based kernels.
	But perhaps you should write one.
\item	Use facilities such as \co{mlockall()} to avoid page faults,
	while also ensuring that your program preallocates all the
	memory it will ever need at boot time.
	This can work well, but at the expense of severe common-case
	underutilization of memory.
	In environments that are cost-constrained or power-limited,
	this approach is not likely to be feasible.
\item	Use facilities such as the Linux kernel's
	\co{NO_HZ_FULL} tickless mode~\cite{JonCorbet2013NO-HZ-FULL}.
	In recent versions of the Linux kernel, this mode directs
	interrupts away from a designated set of CPUs.
	However, this can sharply limit throughput for applications that
	are I/O bound during even part of their operation.
\end{enumerate}

Given these considerations, it is no surprise that non-blocking
synchronization is far more important in theory than it is in practice.

\subsubsection{NBS Subdivided Operations}
\label{sec:advsync:NBS Subdivided Operations}

One common trick that provides a given algorithm a loftier place on
the NBS ranking is to replace blocking operations with a polling API\@.
For example, instead of having a reliable dequeue operation that might be
merely lock-free or even blocking, instead provide a dequeue operation
that will spuriously fail in a wait-free manner rather than exhibiting
dreaded lock-free or blocking behaviors.

This can work well in theory, but a common effect in practice is to
merely move the lock-free or blocking behavior out of that specific
algorithm and into the hapless code making use of that algorithm.
In such cases, not only has nothing has been gained by this trick, but
this trick has increased the complexity of all users of this algorithm.

With concurrent algorithms as elsewhere, maximizing a specific metric
is no substitute for thinking carefully about the needs of one's users.

\subsubsection{NBS Fail-Stop Tolerance}
\label{sec:advsync:NBS Fail-Stop Tolerance}

Of the classes of NBS algorithms, wait-free synchronization (bounded or
otherwise), lock-free synchronization, obstruction-free synchronization,
and clash-free synchronization guarantee forward progress even in the
presence of fail-stop bugs.
An example fail-stop bug might cause some thread to be preempted indefinitely.
As we will see, this fail-stop-tolerant property can be useful, but the
fact is that composing a set of fail-stop-tolerant mechanisms does not
necessarily result in a fail-stop-tolerant system.
To see this, consider a system made up of a series of wait-free queues,
where an element is removed from one queue in the series, processed,
and then added to the next queue.

If a thread is preempted in the midst of a queuing operation, in theory
all is well because the wait-free nature of the queue will guarantee
forward progress.
But in practice, the element being processed is lost because the
fail-stop-tolerant nature of the wait-free queues does not extend to
the code using those queues.

Nevertheless, there are a few applications where NBS's rather limited
fail-stop-tolerance is useful.
For example, in some network-based or web applications, a fail-stop
event will eventually result in a retransmission, which will restart
any work that was lost due to the fail-stop event.
Systems running such applications can therefore be heavily loaded, even
to the point where the scheduler can no longer provide any reasonable
fairness guarantee.
In constrast, if a thread fail-stops while holding a lock, the application
might need to be restarted.
Nevertheless, NBS is not a panacea even within this restricted area,
due to the possibility of spurious retransmissions due to pure scheduling
delays.
In some cases, it may be more efficient to reduce the load to avoid
queueing delays, which will also improve the scheduler's ability to
provide fair access, reducing or even eliminating the fail-stop events,
thus reducing the number of retry operations, in turn further reducing
the load.

\subsubsection{NBS Linearizability}
\label{sec:advsync:NBS Linearizability}

It is important to note that linearizability can be quite useful,
especially when analyzing concurrent code made up of strict locking
and fully ordered atomic operations.\footnote{
	For example, the Linux kernel's value-returning atomic operations.}
Furthermore, this handling of fully ordered atomic operations
automatically covers simple NBS algorithms.

However, the linearization points of a complex NBS algorithms are often
buried deep within that algorithm, and thus not visible to users of
a library function implementing a part of such an algorithm.
Therefore, any claims that users benefit from the linearizability properties
of complex NBS algorithms should be regarded with deep
suspicion~\cite{AndreasHaas2012FIFOisnt}.

It is sometimes asserted that linearizability is necessary for developers
to produce proofs of correctness for their concurrent code.
However, such proofs are the exception rather than the rule, and modern
developers who do produce proofs often use modern proof techniques that
do not depend on linearizability.
Furthermore, developers frequently use modern proof techniques that do
not require a full specification, given that developers often learn
their specification after the fact, one bug at a time.
A few such proof techniques were discussed in
\cref{chp:Formal Verification}.\footnote{
	A memorable verbal discussion with an advocate of linearizability
	resulted in question:
	``So the reason linearizability is important is to rescue 1980s
	proof techniques?''
	The advocate immediately replied in the affirmative, then spent
	some time disparaging a particular modern proof technique.
	Oddly enough, that technique was one of those successfully
	applied to Linux-kernel RCU\@.}

It is often asserted that linearizability maps well to sequential
specifications, which are said to be more natural than are concurrent
specifications~\cite{SergioRajsbaum2020HistoryLinearizability}.
But this assertion fails to account for our highly concurrent objective
universe.
This universe can only be expected to select for ability to cope with
concurrency, especially for those participating in team sports or
overseeing small children.
In addition, given that the teaching of sequential
computing is still believed to be somewhat of a black
art~\cite{ElizabethPatitsas2020GradesNotBimodal}, it is reasonable
to expect that teaching of concurrent computing is in a similar state
of disarray.
Therefore, focusing on only one proof technique is unlikely to be a
good way forward.

Again, please understand that linearizability is quite useful in many
situations.
Then again, so is that venerable tool, the hammer.
But there comes a point in the field of computing where one should put
down the hammer and pick up a keyboard.
Similarly, it appears that there are times when linearizability is not
the best tool for the job.

To their credit, there are some linearizability advocates who are aware
of some of its shortcomings~\cite{SergioRajsbaum2020HistoryLinearizability}.
There are also proposals to extend linearizability, for example,
interval-linearizability, which is intended to handle the common case
of operations that require non-zero time to
complete~\cite{10.1145/3266457}.
It remains to be seen whether these proposals will result in theories
able to handle modern concurrent software artifacts, especially given
that several of the proof techniques discussed in \cref{chp:Formal
Verification} already handle many modern concurrent software artifacts.

\subsection{NBS Discussion}
\label{sec:advsync:NBS Discussion}

It is possible to create fully non-blocking queues~\cite{MichaelScott96},
however, such queues are much more complex than the half-NBS algorithm
outlined above.
The lesson here is to carefully consider your actual requirements.
Relaxing irrelevant requirements can often result in great
improvements in simplicity, performance, and scalability.

Recent research points to another important way to relax requirements.
It turns out that systems providing fair scheduling can enjoy most
of the benefits of wait-free synchronization even when running
algorithms that provide only non-blocking
synchronization, both in theory~\cite{DanAlitarh2013PracticalProgress}
and in practice~\cite{SamyAlBahra2013NBS}.
Because most schedulers used in production do in fact provide fairness,
the more-complex algorithms providing wait-free synchronization usually
provide no practical advantages over simpler and faster non-wait-free
algorithms.

Interestingly enough, fair scheduling is but one beneficial
constraint that is often respected in practice.
Other sets of constraints can permit blocking algorithms to
achieve deterministic real-time response.
For example, given:
\begin{enumerate*}[(1)]
\item Fair locks granted in FIFO order within a given priority level,
\item Priority inversion avoidance (for example, priority
inheritance~\cite{Takada:1995:RSN:527074.828566,Cai-DongWang1996PrioInherLock}
or priority ceiling),
\item A bounded number of threads,
\item Bounded critical section durations,
\item Bounded load,
and
\item Absence of fail-stop bugs,
\end{enumerate*}
lock-based applications can provide deterministic
response times~\cite{BjoernBrandenburgPhD,DipankarSarma2004OLSscalability}.
This approach of course blurs the distinction between blocking and wait-free
synchronization, which is all to the good.
Hopefully theoretical frameworks will continue to improve their ability
to describe software actually used in practice.

Those who feel that theory should lead the way are referred to the
inimitable Peter Denning, who said of operating systems:
``Theory follows practice''~\cite{Denning:2015:POF:2830903.2830904},
or to the eminent Tony Hoare, who said of the whole of engineering:
``In all branches of engineering science, the engineering starts before
the science; indeed, without the early products of engineering, there
would be nothing for the scientist to
study!''~\cite{RichardMorris2007TonyHoareInterview}.
Of course, once an appropriate body of theory becomes available, it is
wise to make use of it.
However, note well that the first \emph{appropriate} body of theory
is often one thing and the first \emph{proposed} body of theory quite
another.

\QuickQuiz{
	It seems like the various members of the NBS hierarchy are
	rather useless.
	So why bother with them at all???
}\QuickQuizAnswer{
	One advantage of the members of the NBS hierarchy is that they
	are reasonably simple to define and use from a theoretical viewpoint.
	We can hope that work done in the NBS arena will help lay the
	groundwork for analysis of real-world forward-progress guarantees
	for concurrent real-time programs.
	However, as of 2022 it appears that trace-based methodologies
	are in the lead~\cite{DanielBristot2019RTtrace}.

	So why bother learning about NBS at all?

	Because a great many people know of it, and are vaguely aware that
	it is somehow related to real-time computing.
	Their response to your carefully designed real-time constraints
	might well be of the form ``Bah, just use wait-free algorithms!''.
	In the all-too-common case where they are very convincing to your
	management, you will need to understand NBS in order to bring
	the discussion back to reality.
	I hope that this section has provided you with the required
	depth of understanding.

	Another thing to note is that learning about the NBS hierarchy
	is probably no more harmful than learning about transfinite
	numbers of the computational-complexity hierarchy.
	In all three cases, it is important to avoid over-applying
	the theory.
	Which is in and of itself good practice!
}\QuickQuizEnd

Proponents of NBS algorithms sometimes call out real-time computing as
an important NBS beneficiary.
The next section looks more deeply at the forward-progress needs of
real-time systems.

% advsync/rt.tex
% mainfile: ../perfbook.tex
% SPDX-License-Identifier: CC-BY-SA-3.0

\section{Parallel Real-Time Computing}
\label{sec:advsync:Parallel Real-Time Computing}
\epigraph{One always has time enough if one applies it well.}
	 {Johann Wolfgang von G\"othe}
% \epigraph{The difference between you and me is that I was right in time.}
% 	 {Konrad Adenauer}
% No support for this quote found.

An important emerging area in computing is that of parallel real-time
computing.
\Cref{sec:advsync:What is Real-Time Computing?}
looks at a number of definitions of ``real-time computing'', moving
beyond the usual sound bites to more meaningful criteria.
\Cref{sec:advsync:Who Needs Real-Time?}
surveys the sorts of applications that need real-time response.
\Cref{sec:advsync:Who Needs Parallel Real-Time?}
notes that parallel real-time computing is upon us, and discusses
when and why parallel real-time computing can be useful.
\Cref{sec:advsync:Implementing Parallel Real-Time Systems}
gives a brief overview of how parallel real-time systems may be implemented,
with
\cref{sec:advsync:Implementing Parallel Real-Time Operating Systems,%
sec:advsync:Implementing Parallel Real-Time Applications}
focusing on operating systems and applications, respectively.
Finally,
\cref{sec:advsync:Real Time vs. Real Fast: How to Choose?}
outlines how to decide whether or not your application needs real-time
facilities.

\subsection{What is Real-Time Computing?}
\label{sec:advsync:What is Real-Time Computing?}

One traditional way of classifying real-time computing is into the
categories of \emph{hard real time} and \emph{soft real time}, where
the macho hard real-time applications never miss their deadlines, but
the wimpy soft real-time applications miss their deadlines quite often.

\subsubsection{Soft Real Time}
\label{sec:Soft Real Time}

It should be easy to see problems with this definition of soft real time.
For one thing, by this definition, \emph{any} piece of software could be
said to be a soft real-time application:
``My application computes million-point Fourier transforms in half a
picosecond.''
``No way!!!
The clock cycle on this system is more than \emph{three hundred} picoseconds!''
``Ah, but it is a \emph{soft} real-time application!''
If the term ``soft real time'' is to be of any use whatsoever, some limits
are clearly required.

We might therefore say that a given soft real-time application must meet
its response-time requirements at least some fraction of the time, for
example, we might say that it must execute in less than 20 microseconds
99.9\,\% of the time.

This of course raises the question of what is to be done when the application
fails to meet its response-time requirements.
The answer varies with the application, but one possibility
is that the system being controlled has sufficient stability and inertia
to render harmless the occasional late control action.
Another possibility is that the application has two ways of computing
the result, a fast and deterministic but inaccurate method on the
one hand and
a very accurate method with unpredictable compute time on the other.
One reasonable approach would be to start
both methods in parallel, and if the accurate method fails to finish
in time, kill it and use the answer from the fast but inaccurate method.
One candidate for the fast but inaccurate method is to take
no control action during the current time period, and another candidate is
to take the same control action as was taken during the preceding time
period.

In short, it does not make sense to talk about soft real time without
some measure of exactly how soft it is.

\subsubsection{Hard Real Time}
\label{sec:Hard Real Time}

In contrast, the definition of hard real time is quite definite.
After all, a given system either always meets its deadlines or it
doesn't.

\begin{figure}
\centering
\resizebox{3in}{!}{\includegraphics{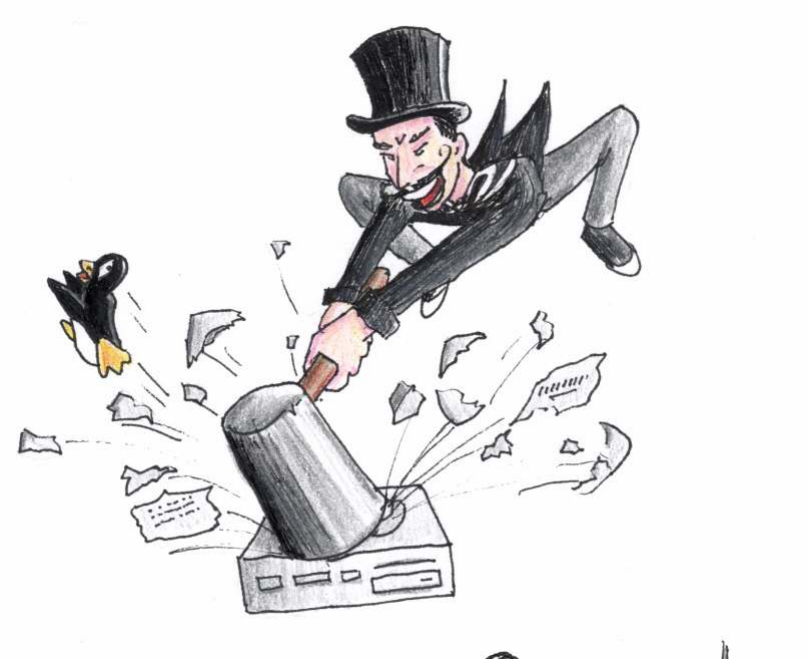}}
\caption{Real-Time Response, Meet Hammer}
\ContributedBy{Figure}{fig:advsync:Hard Real-Time Response; Meet Hammer}{Melissa Broussard}
\end{figure}

Unfortunately, a strict application of this definition would mean that
there can never be any hard real-time systems.
The reason for this is fancifully depicted in
\cref{fig:advsync:Hard Real-Time Response; Meet Hammer}.
And although you can always construct a more robust system, perhaps with
redundancy, your adversary can always get a bigger hammer.
But don't take \emph{my} word for it:
Ask the dinosaurs.

\begin{figure}
\centering
\resizebox{3in}{!}{\includegraphics{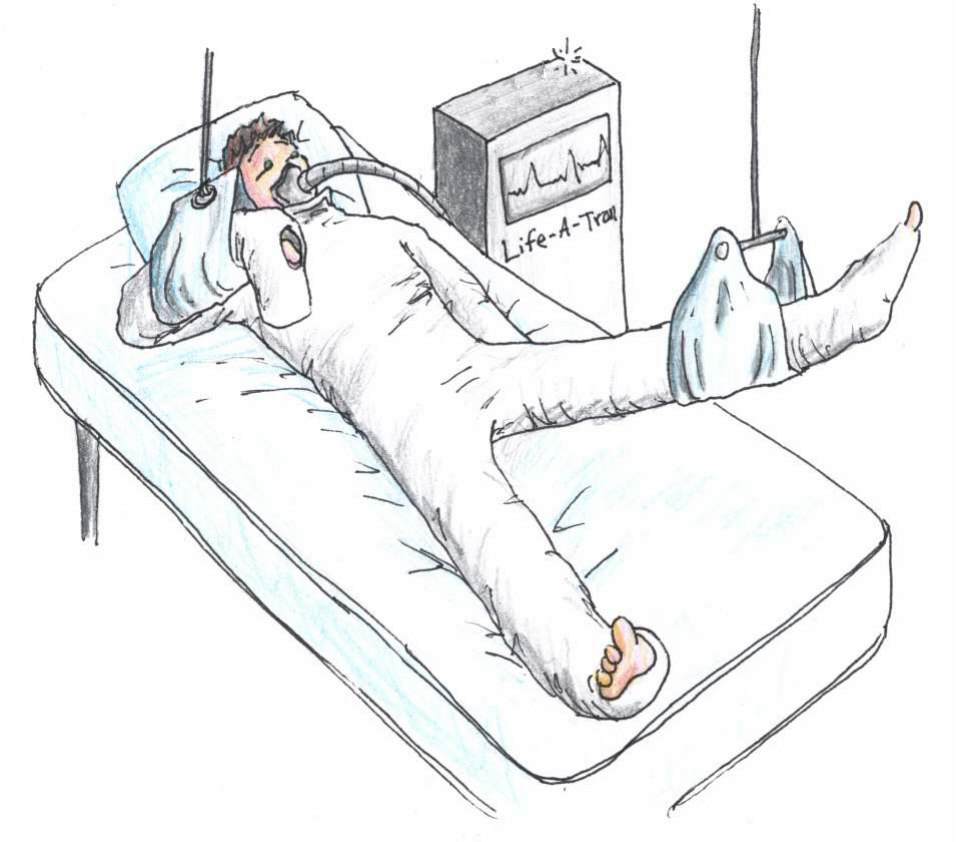}}
\caption{Real-Time Response:
			     Hardware Matters}
\ContributedBy{Figure}{fig:advsync:Real-Time Response: Hardware Matters}{Melissa Broussard}
\end{figure}

Then again, perhaps it is unfair to blame the software for what is clearly
not just a hardware problem, but a bona fide big-iron hardware problem
at that.\footnote{
	Or, given modern hammers, a big-steel problem.}
This suggests that we define hard real-time software as software that
will always meet its deadlines, but only in the absence of a hardware
failure.
Unfortunately, failure is not always an option, as fancifully depicted in
\cref{fig:advsync:Real-Time Response: Hardware Matters}.
We simply cannot expect the poor gentleman depicted in that figure to be
reassured our saying ``Rest assured that if a missed deadline results
in your tragic death, it most certainly will not have been due to a
software problem!''
Hard real-time response is a property of the entire system, not
just of the software.

But if we cannot demand perfection, perhaps we can make do with
notification, similar to the soft real-time approach noted earlier.
Then if the Life-a-Tron in
\cref{fig:advsync:Real-Time Response: Hardware Matters}
is about to miss its deadline,
it can alert the hospital staff.

\begin{figure*}
\centering
\resizebox{\onecolumntextwidth}{!}{\rotatebox{90}{\includegraphics{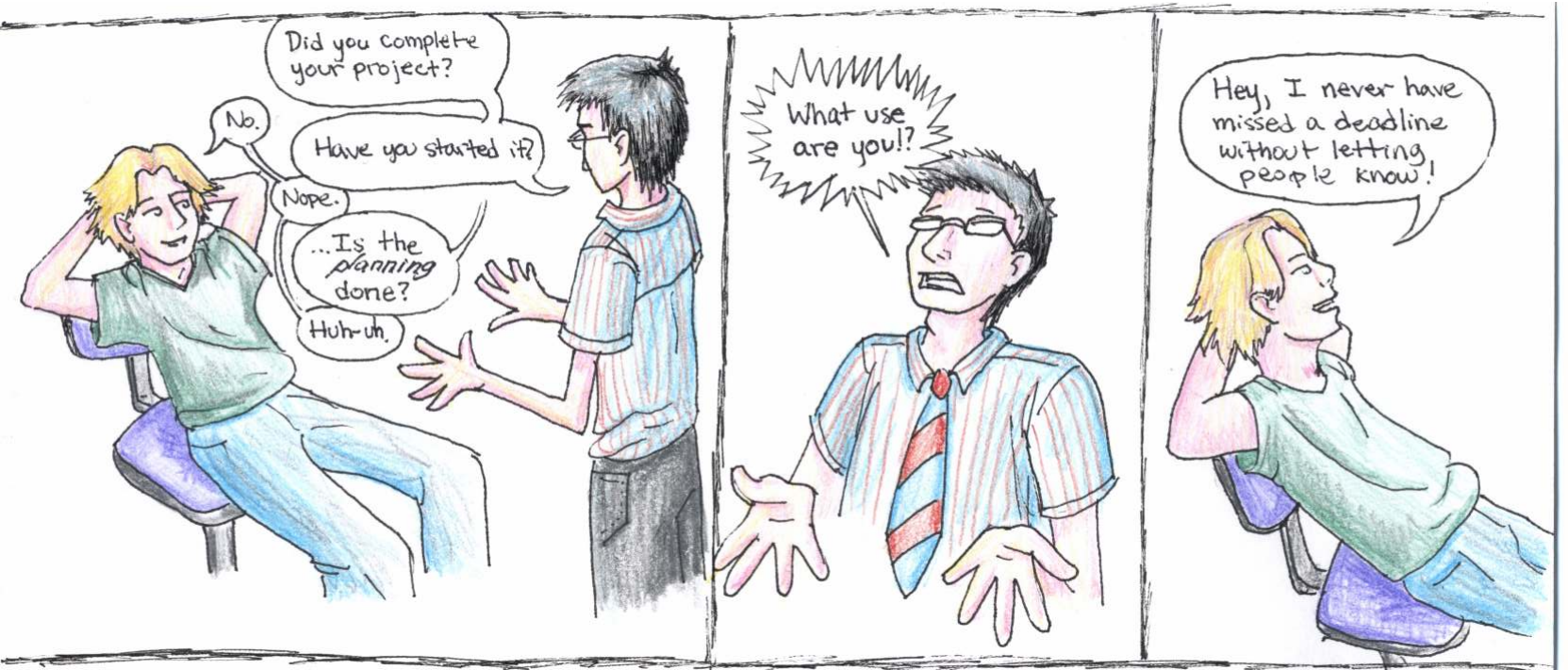}}}
\caption{Real-Time Response:
			     Notification Insufficient}
\ContributedBy{Figure}{fig:advsync:Real-Time Response: Notification Insufficient}{Melissa Broussard}
\end{figure*}

Unfortunately, this approach has the trivial solution fancifully depicted in
\cref{fig:advsync:Real-Time Response: Notification Insufficient}.
A system that always immediately issues a notification that it won't
be able to meet its deadline complies with the letter of the law,
but is completely useless.
There clearly must also be a requirement that the system meets its deadline
some fraction of the time, or perhaps that it be prohibited from missing
its deadlines on more than a certain number of consecutive operations.

We clearly cannot take a sound-bite approach to either hard or soft
real time.
The next section therefore takes a more real-world approach.

\subsubsection{Real-World Real Time}
\label{sec:advsync:Real-World Real Time}

Although sentences like ``Hard real-time systems \emph{always} meet
their deadlines!\@'' are catchy and easy to memorize, something else is
needed for real-world real-time systems.
Although the resulting specifications are
harder to memorize, they can simplify construction of a real-time
system by imposing constraints on the environment, the workload, and
the real-time application itself.

\paragraph{Environmental Constraints}
\label{sec:advsync:Environmental Constraints}

Constraints on the environment address the objection to open-ended
promises of response times implied by ``hard real time''.
These constraints might specify permissible operating temperatures,
air quality, levels and types of electromagnetic radiation, and, to
\cref{fig:advsync:Hard Real-Time Response; Meet Hammer}'s
point, levels of shock and vibration.

Of course, some constraints are easier to meet than others.
Any number of people have learned the hard way that
commodity computer components often refuse to operate at sub-freezing
temperatures, which suggests a set of climate-control requirements.

An old college friend once had the challenge of operating
a real-time system in an atmosphere featuring some rather aggressive
chlorine compounds, a challenge that he wisely handed off to his
colleagues designing the hardware.
In effect, my colleague imposed an atmospheric-composition constraint
on the environment immediately surrounding the computer, a constraint
that the hardware designers met through use of physical seals.

Another old college friend worked on a computer-controlled system
that sputtered ingots of titanium using an industrial-strength arc
in a vacuum.
From time to time, the arc would decide that it was bored with its path
through the ingot of titanium and choose a far shorter and more
entertaining path to ground.
As we all learned in our physics classes, a sudden shift in the flow of
electrons creates an electromagnetic wave, with larger shifts in larger
flows creating higher-power electromagnetic waves.
And in this case, the resulting electromagnetic pulses were sufficient
to induce a quarter of a volt potential difference in the leads of
a small ``rubber ducky'' antenna located more than 400 meters away.
This meant that nearby conductors experienced higher voltages, courtesy
of the inverse-square law.
This included those conductors making up the computer controlling the
sputtering process.
In particular, the voltage induced on that computer's reset line was
sufficient to actually reset the computer, mystifying everyone involved.
This situation was addressed using hardware, including some elaborate
shielding and a fiber-optic network with the lowest bitrate I have ever
heard of, namely 9600 baud.
Less spectacular electromagnetic environments can often be handled by
software through use of error detection and correction codes.
That said, it is important to remember that although error detection and
correction codes can reduce failure rates, they normally cannot reduce
them all the way down to zero, which can present yet another obstacle
to achieving hard real-time response.

There are also situations where a minimum level of energy
is required, for example, through the power leads of the system and
through the devices through which the system is to communicate with
that portion of the outside world that is to be monitored or controlled.

\QuickQuiz{
	But what about battery-powered systems?
	They don't require energy flowing into the system as a whole.
}\QuickQuizAnswer{
	Sooner or later, the battery must be recharged, which requires
	energy to flow into the system.
}\QuickQuizEnd

A number of systems are intended to operate in environments with impressive
levels of shock and vibration, for example, engine control systems.
More strenuous requirements may be found when we move away from
continuous vibrations to intermittent shocks.
For example, during my undergraduate studies, I encountered an old Athena
ballistics computer, which was designed to continue operating normally even if
a hand grenade went off nearby.\footnote{
	Decades later, the acceptance tests for some types of computer
	systems involve large detonations, and some types of
	communications networks must deal with what is delicately
	termed ``ballistic jamming.''}
And finally, the ``black boxes'' used in airliners must continue operating
before, during, and after a crash.

Of course, it is possible to make hardware more robust against
environmental shocks and insults.
Any number of ingenious mechanical shock-absorbing devices can reduce the
effects of shock and vibration, multiple layers of shielding can reduce
the effects of low-energy electromagnetic radiation, error-correction
coding can reduce the effects of high-energy radiation, various potting
and sealing techniques can reduce the effect of air quality, and any
number of heating and cooling systems can counter the effects of temperature.
In extreme cases, triple modular redundancy can reduce the probability that
a fault in one part of the system will result in incorrect behavior from
the overall system.
However, all of these methods have one thing in common:
Although they can reduce the probability of failure, they cannot reduce
it to zero.

These environmental challenges are often met via robust hardware, however,
the workload and application constraints in the next two sections are
often handled in software.

\paragraph{Workload Constraints}
\label{sec:advsync:Workload Constraints}

Just as with people, it is often possible to prevent a real-time system
from meeting its deadlines by overloading it.
For example, if the system is being interrupted too frequently, it might
not have sufficient CPU bandwidth to handle its real-time application.
A hardware solution to this problem might limit the rate at which
interrupts were delivered to the system.
Possible software solutions include disabling interrupts for some time if
they are being received too frequently,
resetting the device generating too-frequent interrupts,
or even avoiding interrupts altogether in favor of polling.

Overloading can also degrade response times due to queueing effects,
so it is not unusual for real-time systems to overprovision CPU bandwidth,
so that a running system has (say) 80\,\% idle time.
This approach also applies to storage and networking devices.
In some cases, separate storage and networking hardware might be reserved
for the sole use of high-priority portions of the real-time application.
In short, it is not unusual for this hardware to be mostly idle, given
that response time is more important than throughput in
real-time systems.

\QuickQuiz{
	But given the results from queueing theory, won't low utilization
	merely improve the average response time rather than improving
	the worst-case response time?
	And isn't worst-case response time all that most
	real-time systems really care about?
}\QuickQuizAnswer{
	Yes, but \ldots

	Those queueing-theory results assume infinite ``calling populations'',
	which in the Linux kernel might correspond to an infinite number
	of tasks.
	As of early 2021, no real system supports an infinite number of
	tasks, so results assuming infinite calling populations should
	be expected to have less-than-infinite applicability.

	Other queueing-theory results have \emph{finite}
	calling populations, which feature sharply bounded response
	times~\cite{Hillier86}.
	These results better model real systems, and these models do
	predict reductions in both average and worst-case response
	times as utilizations decrease.
	These results can be extended to model concurrent systems that use
	synchronization mechanisms such as
	locking~\cite{BjoernBrandenburgPhD,DipankarSarma2004OLSscalability}.

	In short, queueing-theory results that accurately describe
	real-world real-time systems show that worst-case response
	time decreases with decreasing utilization.
}\QuickQuizEnd

Of course, maintaining sufficiently low utilization requires great
discipline throughout the design and implementation.
There is nothing quite like a little feature creep to destroy deadlines.

\paragraph{Application Constraints}
\label{sec:advsync:Application Constraints}

It is easier to provide bounded response time for some operations than
for others.
For example, it is quite common to see response-time specifications for
interrupts and for wake-up operations, but quite rare for (say)
filesystem unmount operations.
One reason for this is that it is quite difficult to bound the amount
of work that a filesystem-unmount operation might need to do, given that
the unmount is required to flush all of that filesystem's in-memory
data to mass storage.

This means that real-time applications must be confined to operations
for which bounded latencies can reasonably be provided.
Other operations must either be pushed out into the non-real-time portions
of the application or forgone entirely.

There might also be constraints on the non-real-time portions of the
application.
For example, is the non-real-time application permitted to use the CPUs
intended for the real-time portion?
Are there time periods during which the real-time portion of the application
is expected to be unusually busy, and if so, is the non-real-time portion
of the application permitted to run at all during those times?
Finally, by what amount is the real-time portion of the application permitted
to degrade the throughput of the non-real-time portion?

\paragraph{Real-World Real-Time Specifications}
\label{sec:advsync:Real-World Real-Time Specifications}

As can be seen from the preceding sections, a real-world real-time
specification needs to include constraints on the environment,
on the workload, and on the application itself.
In addition, for the operations that the real-time portion of the
application is permitted to make use of, there must be constraints
on the hardware and software implementing those operations.

For each such operation, these constraints might include a maximum
response time (and possibly also a minimum response time) and a
probability of meeting that response time.
A probability of 100\,\% indicates that the corresponding operation
must provide hard real-time service.

In some cases, both the response times and the required probabilities of
meeting them might vary depending on the parameters to the operation in
question.
For example, a network operation over a local LAN would be much more likely
to complete in (say) 100~microseconds than would that same network operation
over a transcontinental WAN\@.
Furthermore, a network operation over a copper or fiber
LAN might have an extremely
high probability of completing without time-consuming retransmissions,
while that same networking operation over a lossy WiFi network might
have a much higher probability of missing tight deadlines.
Similarly, a read from a tightly coupled solid-state disk (SSD) could be
expected to complete much more quickly than that same read to an old-style
USB-connected rotating-rust disk drive.\footnote{
	Important safety tip:
	Worst-case response times from USB devices can be extremely long.
	Real-time systems should therefore take care to place any USB
	devices well away from critical paths.}

Some real-time applications pass through different phases of operation.
For example, a real-time system controlling a plywood lathe that peels
a thin sheet of wood (called ``veneer'') from a spinning log must:
\begin{enumerate*}[(1)]
\item Load the log into the lathe,
\item Position the log on the lathe's chucks so as to expose the largest
cylinder contained within that log to the blade,
\item Start spinning the log,
\item Continuously vary the knife's position so as to peel the log into veneer,
\item Remove the remaining core of the log that is too small to peel, and
\item Wait for the next log.
\end{enumerate*}
Each of these six phases of operation might well have its own set of
deadlines and environmental constraints,
for example, one would expect phase~4's deadlines to be much more severe
than those of phase~6, as in milliseconds rather than seconds.
One might therefore expect that low-priority work would be performed in
phase~6 rather than in phase~4.
In any case, careful choices of hardware, drivers, and software
configuration would be required to support phase~4's more severe
requirements.

A key advantage of this phase-by-phase approach is that the \IX{latency}
budgets can be broken down, so that the application's various components
can be developed independently, each with its own latency budget.
Of course, as with any other kind of budget, there will likely be the
occasional conflict as to which component gets which fraction of the
overall budget, and as with any other kind of budget, strong leadership
and a sense of shared goals can help to resolve these conflicts in
a timely fashion.
And, again as with other kinds of technical budget, a strong validation
effort is required in order to ensure proper focus on latencies and to
give early warning of latency problems.
A successful validation effort will almost always include a good test
suite, which might be unsatisfying to the theorists, but has the virtue
of helping to get the job done.
As a point of fact, as of early 2021, most real-world real-time system
use an acceptance test rather than formal proofs.

However, the widespread use of test suites to validate real-time systems
does have a very real disadvantage, namely that real-time software is
validated only on specific configurations of hardware and software.
Adding additional configurations requires additional costly and
time-consuming testing.
Perhaps the field of formal verification will advance sufficiently to
change this situation, but as of early 2021, rather
large advances are required.

\QuickQuiz{
	Formal verification is already quite capable, benefiting from
	decades of intensive study.
	Are additional advances \emph{really} required, or is this just
	a practitioner's excuse to continue to lazily ignore the awesome
	power of formal verification?
}\QuickQuizAnswer{
	Perhaps this situation is just a theoretician's excuse to avoid
	diving into the messy world of real software?
	Perhaps more constructively, the following advances are required:

	\begin{enumerate}
	\item	Formal verification needs to handle larger software
		artifacts.
		The largest verification efforts have been for systems
		of only about 10,000 lines of code, and those have been
		verifying much simpler properties than real-time latencies.
	\item	Hardware vendors will need to publish formal timing
		guarantees.
		This used to be common practice back when hardware was
		much simpler, but today's complex hardware results in
		excessively complex expressions for worst-case performance.
		Unfortunately, energy-efficiency concerns are pushing
		vendors in the direction of even more complexity.
	\item	Timing analysis needs to be integrated into development
		methodologies and IDEs.
	\end{enumerate}

	All that said, there is hope, given recent work formalizing
	the memory models of real computer
	systems~\cite{JadeAlglave2011ppcmem,Alglave:2013:SVW:2450268.2450306}.
	On the other hand, formal verification has just as much trouble
	as does testing with the astronomical number of variants of the
	Linux kernel that can be constructed from different combinations
	of its tens of thousands of Kconfig options.
	Sometimes life is hard!
}\QuickQuizEnd

In addition to latency requirements for the real-time portions of the
application, there will likely be performance and scalability requirements
for the non-real-time portions of the application.
These additional requirements reflect the fact that ultimate real-time
latencies are often attained by degrading scalability and average performance.

Software-engineering requirements can also be important, especially for
large applications that must be developed and maintained by large teams.
These requirements often favor increased modularity and fault isolation.

This is a mere outline of the work that would be required to specify
deadlines and environmental constraints for a production real-time system.
It is hoped that this outline clearly demonstrates the inadequacy of
the sound-bite-based approach to real-time computing.

\subsection{Who Needs Real-Time?}
\label{sec:advsync:Who Needs Real-Time?}

It is possible to argue that all computing is in fact real-time computing.
For one example, when you purchase a birthday gift online, you expect
the gift to arrive before the recipient's birthday.
And in fact even turn-of-the-millennium web services observed sub-second
response constraints~\cite{KristofferBohmann2001a}, and requirements have
not eased with the passage of time~\cite{DeCandia:2007:DAH:1323293.1294281}.
It is nevertheless useful to focus on those real-time applications
whose response-time requirements cannot be achieved straightforwardly
by non-real-time systems and applications.
Of course, as hardware costs decrease and bandwidths and memory sizes
increase, the line between real-time and non-real-time will continue
to shift, but such progress is by no means a bad thing.

\QuickQuiz{
	Differentiating real-time from non-real-time based on what can
	``be achieved straightforwardly by non-real-time systems and
	applications'' is a travesty!
	There is absolutely no theoretical basis for such a distinction!!!
	Can't we do better than that???
}\QuickQuizAnswer{
	This distinction is admittedly unsatisfying from a strictly
	theoretical perspective.
	But on the other hand, it is exactly what the developer needs
	in order to decide whether the application can be cheaply and
	easily developed using standard non-real-time approaches, or
	whether the more difficult and expensive real-time approaches
	are required.
	In other words, although theory is quite important, for those of
	us called upon to complete practical projects, theory supports
	practice, never the other way around.
}\QuickQuizEnd

Real-time computing is used in industrial-control applications, ranging from
manufacturing to avionics;
scientific applications, perhaps most spectacularly in the adaptive
optics used by
large Earth-bound telescopes to de-twinkle starlight;
military applications, including the afore-mentioned avionics;
and financial-services applications, where the first computer to recognize
an opportunity is likely to reap most of the profit.
These four areas could be characterized as ``in search of production'',
``in search of life'', ``in search of death'', and ``in search of money''.

Financial-services applications differ subtly from applications in
the other three categories in that money is non-material, meaning that
non-computational latencies are quite small.
In contrast, mechanical delays inherent in the other three categories
provide a very real point of diminishing returns beyond which further
reductions in the application's real-time response provide little or
no benefit.
This means that financial-services applications, along with other
real-time information-processing applications, face an arms race,
where the application with the lowest latencies normally wins.
Although the resulting latency requirements can still be specified
as described in
\pararef{sec:advsync:Real-World Real-Time Specifications},
the unusual nature of these requirements has led some to refer to
financial and information-processing applications as ``low latency''
rather than ``real time''.

Regardless of exactly what we choose to call it, there is substantial
need for real-time
computing~\cite{JeremyWPeters2006NYTDec11,BillInmon2007a}.

\subsection{Who Needs Parallel Real-Time?}
\label{sec:advsync:Who Needs Parallel Real-Time?}

It is less clear who really needs parallel real-time computing, but
the advent of low-cost multicore systems has brought it to the fore
regardless.
Unfortunately, the traditional mathematical basis for real-time
computing assumes single-CPU systems, with a few exceptions that
prove the rule~\cite{BjoernBrandenburgPhD}.
Fortunately, there are a couple of ways of squaring modern computing
hardware to fit the real-time mathematical circle, and a few Linux-kernel
hackers have been encouraging academics to make this
transition~\cite{DanielBristot2019RTtrace,ThomasGleixner2010AcademiaVsReality}.

\begin{figure}
\centering
\resizebox{3in}{!}{\includegraphics{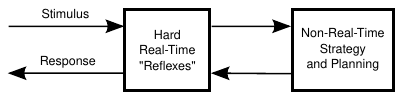}}
\caption{Real-Time Reflexes}
\label{fig:advsync:Real-Time Reflexes}
\end{figure}

One approach is to recognize the fact that many real-time systems
resemble biological nervous systems, with responses ranging from
real-time reflexes to non-real-time strategizing and planning,
as depicted in
\cref{fig:advsync:Real-Time Reflexes}.
The hard real-time reflexes, which read from sensors and control
actuators, run real-time on a single CPU or on special-purpose hardware
such as an FPGA\@.
The non-real-time strategy and planning portion of the application runs
on the remaining CPUs.
Strategy and planning activities might include statistical analysis,
periodic calibration, user interface, supply-chain activities, and
preparation.
For an example of high-compute-load preparation activities, think back
to the veneer-peeling application discussed in
\pararef{sec:advsync:Real-World Real-Time Specifications}.
While one CPU is attending to the high-speed real-time computations
required to peel one log, the other CPUs might be analyzing the size
and shape of the next log in order to determine how to position the
next log so as to obtain the largest cylinder of high-quality wood.
It turns out that many applications have non-real-time and real-time
components~\cite{RobertBerry2008IBMSysJ}, so this approach can
often be used to allow traditional real-time analysis to be combined
with modern multicore hardware.

Another trivial approach is to shut off all but one hardware thread so as
to return to the settled mathematics of uniprocessor real-time
computing.
However, this approach gives up potential cost and energy-efficiency
advantages.
That said, obtaining these advantages requires overcoming the parallel
performance obstacles covered in
\cref{chp:Hardware and its Habits},
and not merely on average, but instead in the worst case.

Implementing parallel real-time systems can therefore be quite a
challenge.
Ways of meeting this challenge are outlined in the following section.

\subsection{Implementing Parallel Real-Time Systems}
\label{sec:advsync:Implementing Parallel Real-Time Systems}

We will look at two major styles of real-time systems, event-driven and
polling.
An event-driven real-time system remains idle much of the time, responding
in real time to events passed up through the operating system to the
application.
Alternatively, the system could instead be running a background
non-real-time workload.
A polling real-time system features a real-time thread that is CPU
bound, running in a tight loop that polls inputs and updates outputs on
each pass.
This tight polling loop often executes entirely in user mode, reading from
and writing to hardware registers that have been mapped into the user-mode
application's address space.
Alternatively, some applications place the polling loop into the kernel,
for example, using loadable kernel modules.

\begin{figure}
\centering
\resizebox{3in}{!}{\includegraphics{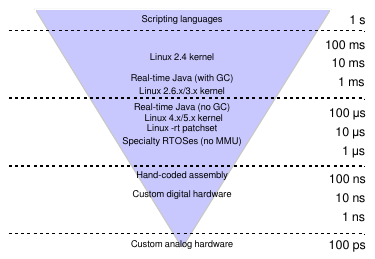}}
\caption{Real-Time Response Regimes}
\label{fig:advsync:Real-Time Response Regimes}
\end{figure}

Regardless of the style chosen, the approach used to implement a real-time
system will depend on the deadlines, for example, as shown in
\cref{fig:advsync:Real-Time Response Regimes}.
Starting from the top of this figure, if you can live with response times in
excess of one second, you might well be able to use scripting languages
to implement your real-time application---and scripting languages are
in fact used surprisingly often, not that I necessarily recommend this
practice.
If the required latencies exceed several tens of milliseconds,
old 2.4 versions of the Linux kernel can be used, not that I necessarily
recommend this practice, either.
Special real-time Java implementations can provide real-time response
latencies of a few milliseconds, even when the garbage collector is
used.
The Linux 2.6.x and 3.x kernels can provide real-time latencies of
a few hundred microseconds if painstakingly configured, tuned, and run
on real-time-friendly hardware.
Special real-time Java implementations can provide real-time latencies
below 100 microseconds if use of the garbage collector is carefully avoided.
(But note that avoiding the garbage collector means also avoiding
Java's large standard libraries, thus also avoiding Java's productivity
advantages.)
The Linux 4.x and 5.x kernels can provide sub-hundred-microsecond
latencies, but with all the same caveats as for the 2.6.x and 3.x kernels.
A Linux kernel incorporating the \rt\ patchset can provide latencies
well below 20 microseconds, and specialty real-time operating systems (RTOSes)
running without MMUs can provide sub-ten-microsecond
latencies.
Achieving sub-microsecond latencies typically requires hand-coded assembly
or even special-purpose hardware.

Of course, careful configuration and tuning are required all the way down
the stack.
In particular, if the hardware or firmware fails to provide real-time
latencies, there is nothing that the software can do to make up for the
lost time.
Worse yet, high-performance hardware sometimes sacrifices worst-case behavior
to obtain greater throughput.
In fact, timings from tight loops run with interrupts disabled can
provide the basis for a high-quality random-number
generator~\cite{PeterOkech2009InherentRandomness}.
Furthermore, some firmware does cycle-stealing to carry out various
housekeeping tasks, in some cases attempting to cover its tracks by
reprogramming the victim CPU's hardware clocks.
Of course, cycle stealing is expected behavior in virtualized
environment, but people are nevertheless working towards real-time
response in virtualized
environments~\cite{ThomasGleixner2012KVMrealtime,JanKiszka2014virtRT}.
It is therefore critically important to evaluate your hardware's and
firmware's real-time capabilities.

But given competent real-time hardware and firmware, the next
layer up the stack is the operating system, which is covered in
the next section.

\subsection{Implementing Parallel Real-Time Operating Systems}
\label{sec:advsync:Implementing Parallel Real-Time Operating Systems}

\begin{figure}
\centering
\resizebox{2.2in}{!}{\includegraphics{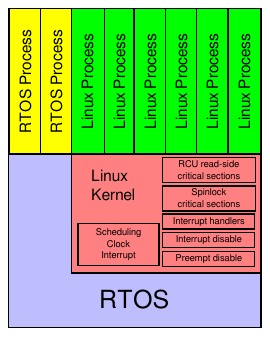}}
\caption{Linux Ported to RTOS}
\label{fig:advsync:Linux Ported to RTOS}
\end{figure}

There are a number of strategies that may be used to implement a
real-time system.
One approach is to port a general-purpose non-real-time OS on top
of a special purpose real-time operating system (RTOS), as shown in
\cref{fig:advsync:Linux Ported to RTOS}.
The green ``Linux Process'' boxes represent non-real-time processes
running on the Linux kernel, while the yellow ``RTOS Process''
boxes represent real-time processes running on the RTOS\@.

This was a very popular approach before the Linux kernel gained
real-time capabilities, and is still in
use~\cite{Xenomai2014,VictorYodaiken2004a}.
However, this approach requires that the application be split into
one portion that runs on the RTOS and another that runs on Linux.
Although it is possible to make the two environments look similar,
for example, by forwarding POSIX system calls from the RTOS to a
utility thread running on Linux, there are invariably rough edges.

In addition, the RTOS must interface to both the hardware and to
the Linux kernel, thus requiring significant maintenance with
changes in both hardware and kernel.
Furthermore, each such RTOS often has its own system-call interface
and set of system libraries, which can balkanize both ecosystems and
developers.
In fact, these problems seem to be what drove the combination of
RTOSes with Linux, as this approach allowed access to the full real-time
capabilities of the RTOS, while allowing the application's non-real-time
code full access to Linux's open-source ecosystem.

\begin{figure*}
\centering
\IfEbookSize{
\resizebox{.87\onecolumntextwidth}{!}{\includegraphics{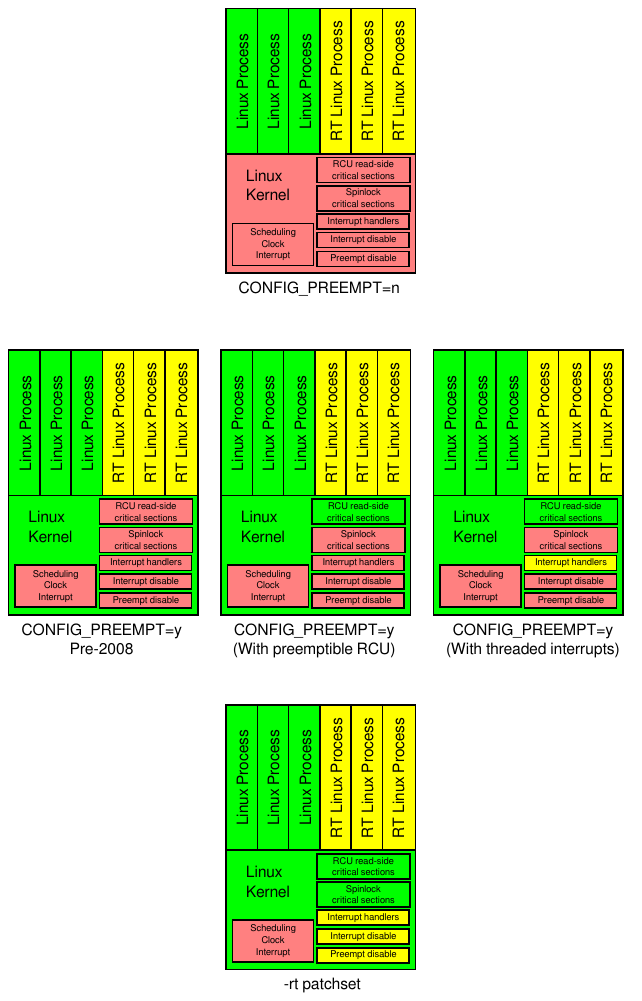}}
}{
\resizebox{4.4in}{!}{\includegraphics{advsync/preemption}}
}
\caption{Linux-Kernel Real-Time Implementations}
\label{fig:advsync:Linux-Kernel Real-Time Implementations}
\end{figure*}

Although pairing RTOSes with the Linux kernel was a clever and useful
short-term response during the time that the Linux kernel had minimal
real-time capabilities, it also motivated adding real-time capabilities
to the Linux kernel.
Progress towards this goal is shown in
\cref{fig:advsync:Linux-Kernel Real-Time Implementations}.
The upper row shows a diagram of the Linux kernel with preemption disabled,
thus having essentially no real-time capabilities.
The middle row shows a set of diagrams showing the increasing real-time
capabilities of the mainline Linux kernel with preemption enabled.
Finally, the bottom row shows a diagram of the Linux kernel with the
\rt\ patchset applied, maximizing real-time capabilities.
Functionality from the \rt\ patchset is added to mainline,
hence the increasing capabilities of the mainline Linux kernel over time.
Nevertheless, the most demanding real-time applications continue to use
the \rt\ patchset.

The non-preemptible kernel shown at the top of
\cref{fig:advsync:Linux-Kernel Real-Time Implementations}
is built with \co{CONFIG_PREEMPT=n}, so that execution within the Linux
kernel cannot be preempted.
This means that the kernel's real-time response latency is bounded below
by the longest code path in the Linux kernel, which is indeed long.
However, user-mode execution is preemptible, so that one of the
real-time Linux processes shown in the upper right may preempt any of the
non-real-time Linux processes shown in the upper left anytime the
non-real-time process is executing in user mode.

The middle row of
\cref{fig:advsync:Linux-Kernel Real-Time Implementations}
shows three stages (from left to right) in the development of Linux's
preemptible kernels.
In all three stages, most process-level code within the Linux kernel
can be preempted.
This of course greatly improves real-time response latency, but
preemption is still disabled
within RCU read-side critical sections,
spinlock critical sections,
interrupt handlers,
interrupt-disabled code regions, and
preempt-disabled code regions, as indicated by the red boxes in the
left-most diagram in the middle row of the figure.
The advent of preemptible RCU allowed RCU read-side critical sections
to be preempted, as shown in the central diagram,
and the advent of threaded interrupt handlers allowed device-interrupt
handlers to be preempted, as shown in the right-most diagram.
Of course, a great deal of other real-time functionality was added
during this time, however, it cannot be as easily represented on this
diagram.
It will instead be discussed in
\cref{sec:advsync:Event-Driven Real-Time Support}.

The bottom row of
\cref{fig:advsync:Linux-Kernel Real-Time Implementations}
shows the \rt\ patchset, which features threaded (and thus preemptible)
interrupt handlers for many devices, which also allows the corresponding
``interrupt-disabled'' regions of these drivers to be preempted.
These drivers instead use locking to coordinate the process-level
portions of each driver with its threaded interrupt handlers.
Finally, in some cases, disabling of preemption is replaced by
disabling of migration.
These measures result in excellent response times in many systems running
the \rt\ patchset~\cite{Reghenzani:2019:RLK:3309872.3297714,DanielBristot2019RTtrace}.

\begin{figure}
\centering
\resizebox{2.5in}{!}{\includegraphics{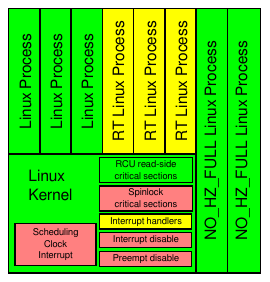}}
\caption{CPU Isolation}
\label{fig:advsync:CPU Isolation}
\end{figure}

A final approach is simply to get everything out of the way of the
real-time process, clearing all other processing off of any CPUs that
this process needs, as shown in \cref{fig:advsync:CPU Isolation}.
This was implemented in the 3.10 Linux kernel via the \co{CONFIG_NO_HZ_FULL}
Kconfig parameter~\cite{JonCorbet2013NO-HZ-FULL,FredericWeisbecker2013nohz}.
It is important to note that this approach requires at least one
\emph{housekeeping CPU} to do background processing, for example running
kernel daemons.
However, when there is only one runnable task on a given non-housekeeping CPU,
scheduling-clock interrupts are shut off on that CPU, removing an important
source of interference and \emph{OS jitter}.
With a few exceptions, the kernel does not force other processing off of the
non-housekeeping CPUs, but instead simply provides better performance
when only one runnable task is present on a given CPU\@.
Any number of userspace tools may be used to force a given CPU to have
no more that one runnable task.
If configured properly, a non-trivial undertaking, \co{CONFIG_NO_HZ_FULL}
offers real-time threads levels of performance that come close to those of
bare-metal systems~\cite{AbdullahAljuhni2018nohzfull}.
\ppl{Fr\'{e}d\'{e}ric}{Weisbecker} produced a practical guide to
\co{CONFIG_NO_HZ_FULL} configuration~\cite{
	FredericWeisbecker2022nohzIntro,
	FredericWeisbecker2022nohzFullDynticksInternals,
	FredericWeisbecker2022nohzfull,
	FredericWeisbecker2022HousekeepingTradeoffs,
	FredericWeisbecker2022practicalExample,
	FredericWeisbecker2022nohzfullTSC}.

There has of course been much debate over which of these approaches
is best for real-time systems, and this debate has been going on for
quite some
time~\cite{JonCorbet2004RealTimeLinuxPart1,JonCorbet2004RealTimeLinuxPart2}.
As usual, the answer seems to be ``It depends,'' as discussed in the
following sections.
\Cref{sec:advsync:Event-Driven Real-Time Support}
considers event-driven real-time systems, and
\cref{sec:advsync:Polling-Loop Real-Time Support}
considers real-time systems that use a CPU-bound polling loop.

\subsubsection{Event-Driven Real-Time Support}
\label{sec:advsync:Event-Driven Real-Time Support}

The operating-system support required for event-driven real-time
applications is quite extensive, however, this section will focus
on only a few items, namely
timers,
threaded interrupts,
priority inheritance,
preemptible RCU,
and
preemptible spinlocks.

\paragraph{Timers} are clearly critically important for real-time
operations.
After all, if you cannot specify that something be done at a specific
time, how are you going to respond by that time?
Even in non-real-time systems, large numbers of timers are generated,
so they must be handled extremely efficiently.
Example uses include retransmit timers for TCP connections (which are
almost always canceled before they have a chance to fire),\footnote{
	At least assuming reasonably low packet-loss rates!}
timed delays (as in \co{sleep(1)}, which are rarely canceled),
and timeouts for the \co{poll()} system call (which are often
canceled before they have a chance to fire).
A good data structure for such timers would therefore be a priority queue
whose addition and deletion primitives were fast and $\O{1}$ in the number
of timers posted.

The classic data structure for this purpose is the \emph{calendar queue},
which in the Linux kernel is called the \co{timer wheel}.
This age-old data structure is also heavily used in discrete-event
simulation.
The idea is that time is quantized, for example, in the Linux kernel,
the duration of the time quantum is the period of the scheduling-clock
interrupt.
A given time can be represented by an integer, and any attempt to post
a timer at some non-integral time will be rounded to a convenient nearby
integral time quantum.

One straightforward implementation would be to allocate a single array,
indexed by the low-order bits of the time.
This works in theory, but in practice systems create large numbers of
long-duration timeouts (for example, the two-hour keepalive timeouts for TCP
sessions) that are almost always canceled.
These long-duration timeouts cause problems for small arrays because
much time is wasted skipping timeouts that have not yet expired.
On the other hand, an array that is large enough to gracefully accommodate
a large number of long-duration timeouts would consume too much memory,
especially given that performance and scalability concerns require one
such array for each and every CPU\@.

\begin{figure}
\centering
\resizebox{2.0in}{!}{\includegraphics{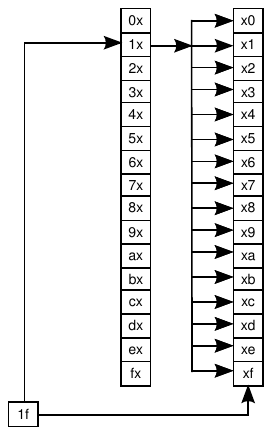}}
\caption{Timer Wheel}
\label{fig:advsync:Timer Wheel}
\end{figure}

A common approach for resolving this conflict is to provide multiple
arrays in a hierarchy.
At the lowest level of this hierarchy, each array element represents
one unit of time.
At the second level, each array element represents $N$ units of time,
where $N$ is the number of elements in each array.
At the third level, each array element represents $N^2$ units of time,
and so on up the hierarchy.
This approach allows the individual arrays to be indexed by different
bits, as illustrated by
\cref{fig:advsync:Timer Wheel}
for an unrealistically small eight-bit clock.
Here, each array has 16 elements, so the low-order four bits of the time
(currently \co{0xf}) index the low-order (rightmost) array, and the
next four bits (currently \co{0x1}) index the next level up.
Thus, we have two arrays each with 16 elements, for a total of 32 elements,
which, taken together, is much smaller than the 256-element array that
would be required for a single array.

This approach works extremely well for throughput-based systems.
Each timer operation is $\O{1}$ with small constant, and each timer
element is touched at most $m+1$ times, where $m$ is the number of
levels.

\begin{figure}
\centering
\resizebox{3.0in}{!}{\includegraphics{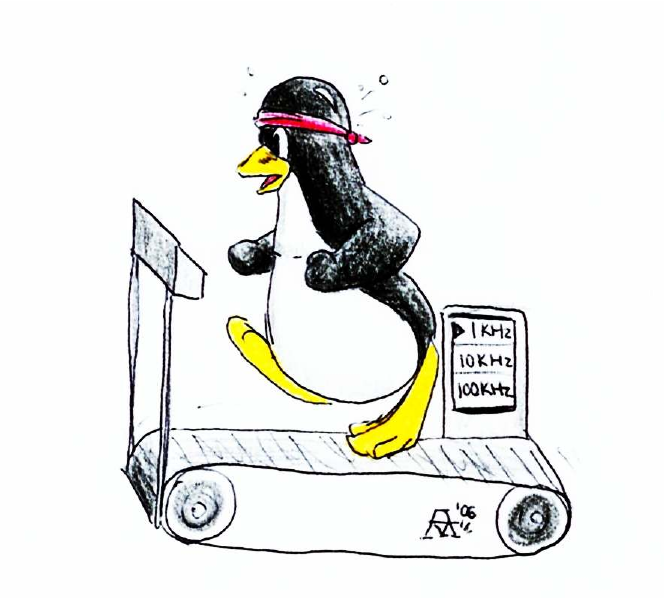}}
\caption{Timer Wheel at 1\,kHz}
\ContributedBy{Figure}{fig:advsync:Timer Wheel at 1kHz}{Melissa Broussard}
\end{figure}

\begin{figure}
\centering
\resizebox{3.0in}{!}{\includegraphics{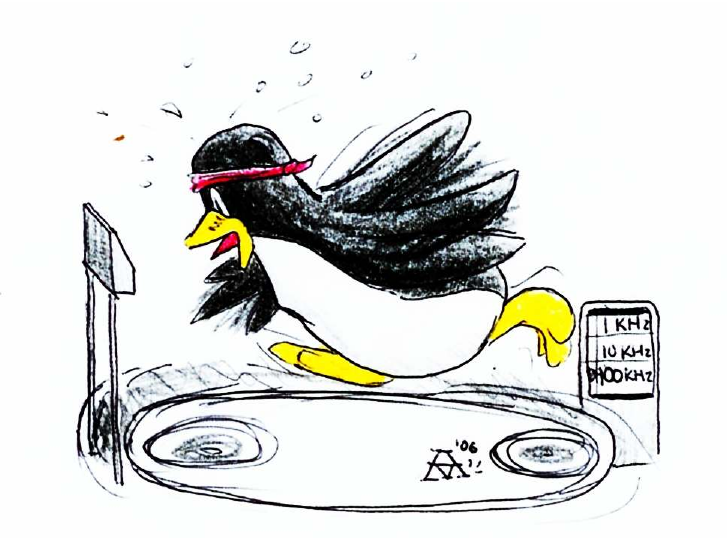}}
\caption{Timer Wheel at 100\,kHz}
\ContributedBy{Figure}{fig:advsync:Timer Wheel at 100kHz}{Melissa Broussard}
\end{figure}

Unfortunately, timer wheels do not work well for real-time systems, and for
two reasons.
The first reason is that there is a harsh tradeoff between timer
accuracy and timer overhead, which is fancifully illustrated by
\cref{fig:advsync:Timer Wheel at 1kHz,fig:advsync:Timer Wheel at 100kHz}.
In \cref{fig:advsync:Timer Wheel at 1kHz},
timer processing happens only once per millisecond, which keeps overhead
acceptably low for many (but not all!\@) workloads, but which also means
that timeouts cannot be set for finer than one-millisecond granularities.
On the other hand, \cref{fig:advsync:Timer Wheel at 100kHz}
shows timer processing taking place every ten microseconds, which
provides acceptably fine timer granularity for most (but not all!\@)
workloads, but which processes timers so frequently that the system
might well not have time to do anything else.

The second reason is the need to cascade timers from higher levels to
lower levels.
Referring back to \cref{fig:advsync:Timer Wheel},
we can see that any timers enqueued on element \co{1x} in the upper
(leftmost) array must be cascaded down to the lower (rightmost)
array so that may be invoked when their time arrives.
Unfortunately, there could be a large number of timeouts
waiting to be cascaded, especially for timer wheels with larger numbers
of levels.
The power of statistics causes this cascading to be a non-problem for
throughput-oriented systems, but cascading can result in problematic
degradations of latency in real-time systems.

\IfTwoColumn{
\begin{figure*}
\centering
\resizebox{1.4\twocolumnwidth}{!}{\includegraphics{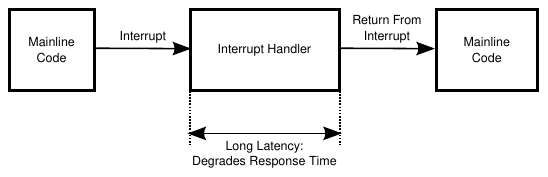}}
\caption{Non-Threaded Interrupt Handler}
\label{fig:advsync:Non-Threaded Interrupt Handler}
\end{figure*}
}{}

\IfTwoColumn{
\begin{figure*}
\centering
\resizebox{1.4\twocolumnwidth}{!}{\includegraphics{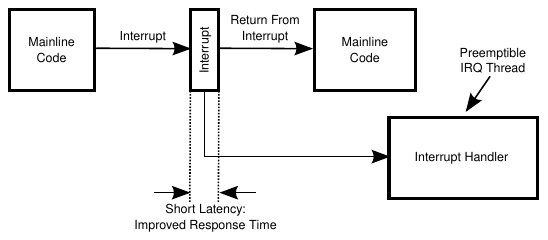}}
\caption{Threaded Interrupt Handler}
\label{fig:advsync:Threaded Interrupt Handler}
\end{figure*}
}{}

Of course, real-time systems could simply choose a different data
structure, for example, some form of heap or tree, giving up
$\O{1}$ bounds on insertion and deletion operations to gain $\O{\log n}$
limits on data-structure-maintenance operations.
This can be a good choice for special-purpose RTOSes, but is inefficient
for general-purpose systems such as Linux, which routinely support
extremely large numbers of timers.

The solution chosen for the Linux kernel's \rt\ patchset is to differentiate
between timers that schedule later activity and timeouts that schedule
error handling for low-probability errors such as TCP packet losses.
One key observation is that error handling is normally not particularly
time-critical, so that a timer wheel's millisecond-level granularity
is good and sufficient.
Another key observation is that error-handling timeouts are normally
canceled very early, often before they can be cascaded.
In addition, systems commonly have many more error-handling timeouts
than they do timer events, so that an $\O{\log n}$ data structure should
provide acceptable performance for timer events.

However, it is possible to do better, namely by simply refusing to
cascade timers.
Instead of cascading, the timers that would otherwise have been cascaded
all the way down the calendar queue are handled in place.
This does result in up to a few percent error for the time duration,
but the few situations where this is a problem can instead use tree-based
high-resolution timers (hrtimers).

In short, the Linux kernel's \rt\ patchset uses timer wheels for
error-handling timeouts and a tree for timer events, providing each
category the required quality of service.

\IfTwoColumn{}{
\begin{figure}
\centering
\IfEbookSize{
\resizebox{\onecolumntextwidth}{!}{\includegraphics{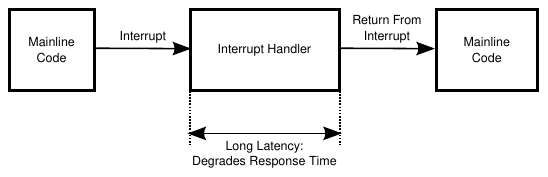}}
}{
\resizebox{1.4\twocolumnwidth}{!}{\includegraphics{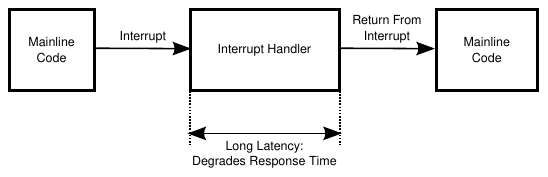}}
}
\caption{Non-Threaded Interrupt Handler}
\label{fig:advsync:Non-Threaded Interrupt Handler}
\end{figure}
}

\paragraph{Threaded interrupts}
are used to address a significant source of degraded real-time latencies,
namely long-running interrupt handlers,
as shown in \cref{fig:advsync:Non-Threaded Interrupt Handler}.
These latencies can be especially problematic for devices that can
deliver a large number of events with a single interrupt, which means
that the interrupt handler will run for an extended period of time
processing all of these events.
Worse yet are devices that can deliver new events to a still-running
interrupt handler, as such an interrupt handler might well run
indefinitely, thus indefinitely degrading real-time latencies.

\IfTwoColumn{}{
\begin{figure}
\centering
\IfEbookSize{
\resizebox{\onecolumntextwidth}{!}{\includegraphics{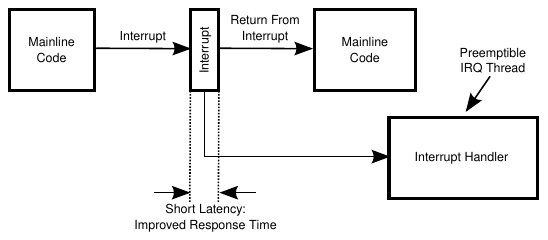}}
}{
\resizebox{1.4\twocolumnwidth}{!}{\includegraphics{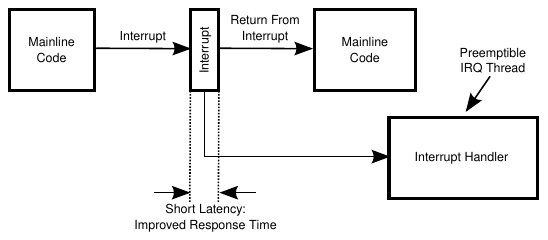}}
}
\caption{Threaded Interrupt Handler}
\label{fig:advsync:Threaded Interrupt Handler}
\end{figure}
}

One way of addressing this problem is the use of threaded interrupts shown in
\cref{fig:advsync:Threaded Interrupt Handler}.
Interrupt handlers run in the context of a preemptible \IXacr{irq} thread,
which runs at a configurable priority.
The device interrupt handler then runs for only a short time, just
long enough to make the \IRQ\ thread aware of the new event.
As shown in the figure, threaded interrupts can greatly improve
real-time latencies, in part because interrupt handlers running in
the context of the \IRQ\ thread may be preempted by high-priority real-time
threads.

However, there is no such thing as a free lunch, and there are downsides
to threaded interrupts.
One downside is increased interrupt latency.
Instead of immediately running the interrupt handler, the handler's execution
is deferred until the \IRQ\ thread gets around to running it.
Of course, this is not a problem unless the device generating the interrupt
is on the real-time application's critical path.

Another downside is that poorly written high-priority real-time code
might starve the interrupt handler, for example, preventing networking
code from running, in turn making it very difficult to debug the problem.
Developers must therefore take great care when writing high-priority
real-time code.
This has been dubbed the \emph{Spiderman principle}:
With great power comes great responsibility.

\paragraph{Priority inheritance} is used to handle priority inversion,
which can be caused by, among other things, locks acquired by
preemptible interrupt handlers~\cite{LuiSha1990PriorityInheritance}.
Suppose that a low-priority thread holds a lock, but is preempted by
a group of medium-priority threads, at least one such thread per CPU\@.
If an interrupt occurs, a high-priority \IRQ\ thread will preempt one
of the medium-priority threads, but only until it decides to acquire
the lock held by the low-priority thread.
Unfortunately, the low-priority thread cannot release the lock until
it starts running, which the medium-priority threads prevent it from
doing.
So the high-priority \IRQ\ thread cannot acquire the lock until after one
of the medium-priority threads releases its CPU\@.
In short, the medium-priority threads are indirectly blocking the
high-priority \IRQ\ threads, a classic case of priority inversion.

Note that this priority inversion could not happen with non-threaded
interrupts because the low-priority thread would have to disable interrupts
while holding the lock, which would prevent the medium-priority
threads from preempting it.

In the priority-inheritance solution, the high-priority thread attempting
to acquire the lock donates its priority to the low-priority thread holding
the lock until such time as the lock is released, thus preventing long-term
priority inversion.

\begin{figure}
\centering
\resizebox{3.4in}{!}{\includegraphics{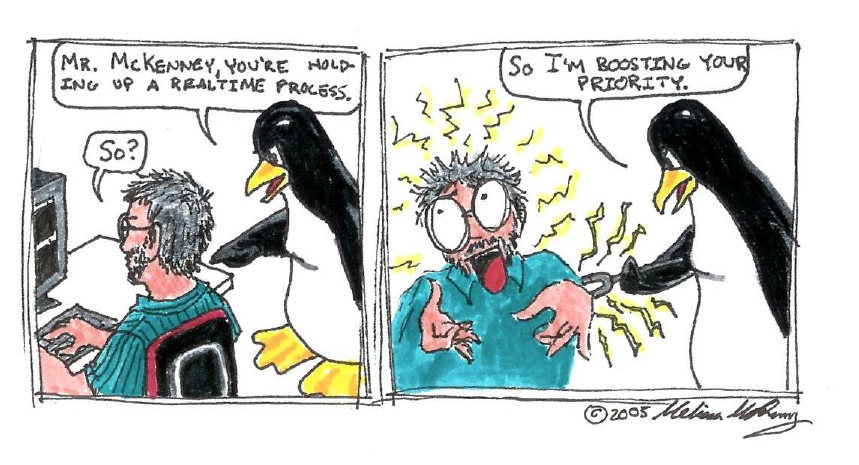}}
\caption{Priority Inversion and User Input}
\ContributedBy{Figure}{fig:advsync:Priority Inversion and User Input}{Melissa Broussard}
\end{figure}

Of course, priority inheritance does have its limitations.
For example, if you can design your application to avoid priority
inversion entirely, you will likely obtain somewhat better
latencies~\cite{VictorYodaiken2004a}.
This should be no surprise, given that priority inheritance adds
a pair of context switches to the worst-case latency.
That said, priority inheritance can convert indefinite postponement
into a limited increase in latency, and the software-engineering
benefits of priority inheritance may outweigh its latency costs in
many applications.

Another limitation is that it addresses only lock-based priority
inversions within the context of a given operating system.
One priority-inversion scenario that it cannot address is a high-priority
thread waiting on a network socket for a message that is to be written
by a low-priority process that is preempted by a set of CPU-bound
medium-priority processes.
In addition, a potential disadvantage of applying priority inheritance
to user input is fancifully depicted in
\cref{fig:advsync:Priority Inversion and User Input}.

A final limitation involves reader-writer locking.
Suppose that we have a very large number of low-priority threads, perhaps
even thousands of them, each
of which read-holds a particular reader-writer lock.
Suppose that all of these threads are preempted by a set of medium-priority
threads, with at least one medium-priority thread per CPU\@.
Finally, suppose that a high-priority thread awakens and attempts to
write-acquire this same reader-writer lock.
No matter how vigorously we boost the priority of the threads read-holding
this lock, it could well be a good long time before the high-priority
thread can complete its write-acquisition.

There are a number of possible solutions to this reader-writer lock
priority-inversion conundrum:

\begin{enumerate}
\item	Only allow one read-acquisition of a given reader-writer lock
	at a time.
	(This is the approach traditionally taken by the Linux
	kernel's \rt\ patchset.)
\item	Only allow $N$ read-acquisitions of a given reader-writer lock
	at a time, where $N$ is the number of CPUs.
\item	Only allow $N$ read-acquisitions of a given reader-writer lock
	at a time, where $N$ is a number specified somehow by the
	developer.
\item	Prohibit high-priority threads from write-acquiring reader-writer
	locks that are ever read-acquired by threads running at lower
	priorities.
	(This is a variant of the \emph{priority ceiling}
	protocol~\cite{LuiSha1990PriorityInheritance}.)
\end{enumerate}

\QuickQuiz{
	But if you only allow one reader at a time to read-acquire
	a reader-writer lock, isn't that the same as an exclusive
	lock???
}\QuickQuizAnswer{
	Indeed it is, other than the API\@.
	And the API is important because it allows the Linux kernel
	to offer real-time capabilities without having the \rt\ patchset
	grow to ridiculous sizes.

	However, this approach clearly and severely limits read-side
	scalability.
	The Linux kernel's \rt\ patchset was long able to live with this
	limitation for several reasons:
	\begin{enumerate*}[(1)]
	\item Real-time systems have traditionally been relatively small,
	\item Real-time systems have generally focused on process control,
	thus being unaffected by scalability limitations in the
	I/O subsystems, and
	\item Many of the Linux kernel's reader-writer locks have been
	converted to RCU\@.
	\end{enumerate*}

	However, the day came when it was absolutely necessary to
	permit concurrent readers, as described in the text following
	this quiz.
}\QuickQuizEnd

The no-concurrent-readers restriction eventually became intolerable, so
the \rt\ developers looked more carefully at how the Linux kernel uses
reader-writer spinlocks.
They learned that time-critical code rarely uses those parts of the
kernel that write-acquire reader-writer locks, so that the prospect
of writer \IX{starvation} was not a show-stopper.
They therefore constructed a real-time reader-writer lock in which
write-side acquisitions use priority inheritance among each other,
but where read-side acquisitions take absolute priority over
write-side acquisitions.
This approach appears to be working well in practice, and is another
lesson in the importance of clearly understanding what your users
really need.

One interesting detail of this implementation is that both the
\co{rt_read_lock()} and the \co{rt_write_lock()} functions enter an RCU
read-side critical section and both the \co{rt_read_unlock()} and the
\co{rt_write_unlock()} functions exit that critical section.
This is necessary because non-realtime kernels' reader-writer locking
functions disable preemption across their critical sections, and
there really are reader-writer locking use cases that rely on the fact
that \co{synchronize_rcu()} will therefore wait for all pre-existing
reader-writer-lock critical sections to complete.
Let this be a lesson to you:
Understanding what your users really need is critically important to
correct operation, not just to performance.
Not only that, but what your users really need changes over time.

This has the side-effect that all of a \rt\ kernel's reader-writer locking
critical sections are subject to RCU priority boosting.
This provides at least a partial solution to the problem of reader-writer
lock readers being preempted for extended periods of time.

It is also possible to avoid reader-writer lock priority inversion by
converting the reader-writer lock to RCU, as briefly discussed in the
next section.

\paragraph{Preemptible RCU}
can sometimes be used as a replacement for reader-writer
locking~\cite{PaulEMcKenney2007WhatIsRCUFundamentally,PaulMcKenney2012RCUUsage,PaulEMcKenney2014RCUAPI},
as was discussed in \cref{sec:defer:Read-Copy Update (RCU)}.
Where it can be used, it permits readers and updaters to run concurrently,
which prevents low-priority readers from inflicting any sort of
priority-inversion scenario on high-priority updaters.
However, for this to be useful, it is necessary to be able to preempt
long-running RCU read-side critical
sections~\cite{DinakarGuniguntala2008IBMSysJ}.
Otherwise, long RCU read-side critical sections would result in
excessive real-time latencies.

\begin{listing}
\begin{fcvlabel}[ln:advsync:Preemptible Linux-Kernel RCU]
\begin{VerbatimL}[commandchars=\\\[\]]
void __rcu_read_lock(void)			\lnlbl[lock:b]
{
	current->rcu_read_lock_nesting++;	\lnlbl[lock:inc]
	barrier();				\lnlbl[lock:bar]
}						\lnlbl[lock:e]

void __rcu_read_unlock(void)		        \lnlbl[unl:b]
{
	barrier();				\lnlbl[unl:bar1]
	if (!--current->rcu_read_lock_nesting)	\lnlbl[unl:decchk]
		barrier();			\lnlbl[unl:bar2]
		if (READ_ONCE(current->rcu_read_unlock_special.s)) { \lnlbl[unl:chks]
			rcu_read_unlock_special(t); \lnlbl[unl:unls]
		}				\lnlbl[unl:if:e]
}						\lnlbl[unl:e]
\end{VerbatimL}
\end{fcvlabel}
\caption{Preemptible Linux-Kernel RCU}
\label{lst:advsync:Preemptible Linux-Kernel RCU}
\end{listing}

A preemptible RCU implementation was therefore added to the Linux kernel.
This implementation avoids the need to individually track the state of
each and every task in the kernel by keeping lists of tasks that have
been preempted within their current RCU read-side critical sections.
A \IX{grace period} is permitted to end:
\begin{enumerate*}[(1)]
\item Once all CPUs have completed any RCU read-side critical sections
that were in effect before the start of the current grace period and
\item Once all tasks that were preempted
while in one of those pre-existing critical sections have removed
themselves from their lists.
\end{enumerate*}
A simplified version of this implementation is shown in
\cref{lst:advsync:Preemptible Linux-Kernel RCU}.
\begin{fcvref}[ln:advsync:Preemptible Linux-Kernel RCU]
The \co{__rcu_read_lock()} function spans \clnrefrange{lock:b}{lock:e} and
the \co{__rcu_read_unlock()} function spans \clnrefrange{unl:b}{unl:e}.
\end{fcvref}

\begin{fcvref}[ln:advsync:Preemptible Linux-Kernel RCU:lock]
\Clnref{inc} of \co{__rcu_read_lock()} increments a per-task count of the
number of nested \co{rcu_read_lock()} calls, and
\clnref{bar} prevents the compiler from reordering the subsequent code in the
RCU read-side critical section to precede the \co{rcu_read_lock()}.
\end{fcvref}

\begin{fcvref}[ln:advsync:Preemptible Linux-Kernel RCU:unl]
\Clnref{bar1} of \co{__rcu_read_unlock()} prevents the compiler from
reordering the code in the critical section with the remainder of
this function.
\Clnref{decchk} decrements the nesting count and checks to see if it
has become zero, in other words, if this corresponds to the outermost
\co{rcu_read_unlock()} of a nested set.
If so, \clnref{bar2} prevents the compiler from reordering this nesting
update with \clnref{chks}'s check for special handling.
If special handling is required, then the call to
\co{rcu_read_unlock_special()} on \clnref{unls} carries it out.

There are several types of special handling that can be required, but
we will focus on that required when the RCU read-side critical section
has been preempted.
In this case, the task must remove itself from the list that it was
added to when it was first preempted within its
RCU read-side critical section.
However, it is important to note that these lists are protected by locks,
which means that \co{rcu_read_unlock()} is no longer lockless.
However, the highest-priority threads will not be preempted, and therefore,
for those highest-priority threads, \co{rcu_read_unlock()} will never
attempt to acquire any locks.
In addition, if implemented carefully, locking can be used to synchronize
real-time software~\cite{BjoernBrandenburgPhD,DipankarSarma2004OLSscalability}.
\end{fcvref}

\QuickQuiz{
	\begin{fcvref}[ln:advsync:Preemptible Linux-Kernel RCU:unl]
	Suppose that preemption occurs just after the load from
	\co{t->rcu_read_unlock_special.s} on \clnref{chks} of
	\cref{lst:advsync:Preemptible Linux-Kernel RCU}.
	Mightn't that result in the task failing to invoke
	\co{rcu_read_unlock_special()}, thus failing to remove itself
	from the list of tasks blocking the current grace period,
	in turn causing that grace period to extend indefinitely?
	\end{fcvref}
}\QuickQuizAnswer{
	That is a real problem, and it is solved in RCU's scheduler hook.
	If that scheduler hook sees that the value of
	\co{t->rcu_read_lock_nesting} is negative, it invokes
	\co{rcu_read_unlock_special()} if needed before allowing
	the context switch to complete.
}\QuickQuizEnd

Another important real-time feature of RCU, whether preemptible or
not, is the ability to offload RCU callback execution to a kernel
thread.
To use this, your kernel must be built with \co{CONFIG_RCU_NOCB_CPU=y}
and booted with the \co{rcu_nocbs=} kernel boot parameter specifying
which CPUs are to be offloaded.
Alternatively, any CPU specified by the \co{nohz_full=} kernel boot parameter
described in
\cref{sec:advsync:Polling-Loop Real-Time Support}
will also have its RCU callbacks offloaded.

In short, this preemptible RCU implementation enables real-time response for
read-mostly data structures without the delays inherent to priority
boosting of large numbers of readers, and also without delays due to
callback invocation.

\paragraph{Preemptible spinlocks}
are an important part of the \rt\ patchset due to the long-duration
spinlock-based critical sections in the Linux kernel.
This functionality has not yet reached mainline:
Although they are a conceptually simple substitution of sleeplocks
for spinlocks, they have proven relatively controversial.
In addition the real-time functionality that is already in the mainline
Linux kernel suffices for a great many use cases, which slowed the \rt\
patchset's development rate in the early
2010s~\cite{JakeEdge2013Future-rtLinux,JakeEdge2014Future-rtLinux}.
However, preemptible spinlocks are absolutely necessary to the task of
achieving real-time latencies down in the tens of microseconds.
Fortunately, Linux Foundation organized an effort to fund moving the
remaining code from the \rt\ patchset to mainline.

\paragraph{Per-CPU variables}\ are used heavily in the Linux kernel
for performance reasons.
Unfortunately for real-time applications, many use cases for per-CPU
variables require coordinated update of multiple such variables,
which is normally provided by disabling preemption, which in turn
degrades real-time latencies.
Real-time applications clearly need some other way of coordinating
per-CPU variable updates.

One alternative is to supply per-CPU spinlocks, which as noted above
are actually sleeplocks, so that their critical sections can be
preempted and so that priority inheritance is provided.
In this approach, code updating groups of per-CPU variables must
acquire the current CPU's spinlock, carry out the update, then
release whichever lock is acquired, keeping in mind that a preemption
might have resulted in a migration to some other CPU\@.
However, this approach introduces both overhead and \IXpl{deadlock}.

Another alternative, which is used in the \rt\ patchset as of early 2021,
is to convert preemption disabling to migration disabling.
This ensures that a given kernel thread remains on its CPU through
the duration of the per-CPU-variable update, but could also allow some
other kernel thread to intersperse its own update of those same variables,
courtesy of preemption.
There are cases such as statistics gathering where this is not a problem.
In the surprisingly rare case where such mid-update preemption is a problem,
the use case at hand must properly synchronize the updates, perhaps through
a set of per-CPU locks specific to that use case.
Although introducing locks again introduces the possibility of deadlock,
the per-use-case nature of these locks makes any such deadlocks easier
to manage and avoid.

\paragraph{Closing event-driven remarks.}
There are of course any number of other Linux-kernel components that are
critically important to achieving world-class real-time latencies,
for example, deadline
scheduling~\cite{DanielBristot2018deadlinesched-1,DanielBristot2018deadlinesched-2},
however, those listed in this section give a good feeling for the workings
of the Linux kernel augmented by the \rt\ patchset.

\subsubsection{Polling-Loop Real-Time Support}
\label{sec:advsync:Polling-Loop Real-Time Support}

At first glance, use of a polling loop might seem to avoid all possible
operating-system interference problems.
After all, if a given CPU never enters the kernel, the kernel is
completely out of the picture.
And the traditional approach to keeping the kernel out of the way is
simply not to have a kernel, and many real-time applications do
indeed run on bare metal, particularly those running on eight-bit
microcontrollers.

One might hope to get bare-metal performance on a modern operating-system
kernel simply by running a single CPU-bound user-mode thread on a
given CPU, avoiding all causes of interference.
Although the reality is of course more complex, it is becoming
possible to do just that,
courtesy of the \co{NO_HZ_FULL} implementation led by
Frederic Weisbecker~\cite{JonCorbet2013NO-HZ-FULL,FredericWeisbecker2013nohz}
that was accepted into version 3.10 of the Linux kernel.
Nevertheless, considerable care is required to properly set up such
an environment, as it is necessary to control a number of possible
sources of OS jitter.
The discussion below covers the control of several sources of OS
jitter, including device interrupts, kernel threads and daemons,
scheduler real-time throttling (this is a feature, not a bug!),
timers, non-real-time device drivers, in-kernel global synchronization,
scheduling-clock interrupts, page faults, and finally, non-real-time
hardware and firmware.

Interrupts are an excellent source of large amounts of OS jitter.
Unfortunately, in most cases interrupts are absolutely required in order
for the system to communicate with the outside world.
One way of resolving this conflict between OS jitter and maintaining
contact with the outside world is to reserve a small number of
housekeeping CPUs, and to force all interrupts to these CPUs.
The \path{Documentation/IRQ-affinity.txt} file in the Linux source tree
describes how to direct device interrupts to specified CPUs,
which as of early 2021 involves something like the following:

\begin{VerbatimU}
$ echo 0f > /proc/irq/44/smp_affinity
\end{VerbatimU}

This command would confine interrupt \#44 to CPUs~0--3.
Note that scheduling-clock interrupts require special handling, and are
discussed later in this section.

A second source of OS jitter is due to kernel threads and daemons.
Individual kernel threads, such as RCU's grace-period kthreads
(\co{rcu_bh}, \co{rcu_preempt}, and \co{rcu_sched}), may be forced
onto any desired CPUs using the \co{taskset} command, the
\co{sched_setaffinity()} system call, or \co{cgroups}.

Per-CPU kthreads are often more challenging, sometimes constraining
hardware configuration and workload layout.
Preventing OS jitter from these kthreads requires either that certain
types of hardware
not be attached to real-time systems, that all interrupts and I/O
initiation take place on housekeeping CPUs, that special kernel
Kconfig or boot parameters be selected in order to direct work away from
the worker CPUs, or that worker CPUs never enter the kernel.
Specific per-kthread advice may be found in the Linux kernel source
\path{Documentation} directory at \path{kernel-per-CPU-kthreads.txt}.

A third source of OS jitter in the Linux kernel for CPU-bound threads
running at real-time priority is the scheduler itself.
This is an intentional debugging feature, designed to ensure that
important non-realtime work is allotted at least 50 milliseconds
out of each second, even if there is an infinite-loop bug in
your real-time application.
However, when you are running a polling-loop-style real-time application,
you will need to disable this debugging feature.
This can be done as follows:

\begin{VerbatimU}
$ echo -1 > /proc/sys/kernel/sched_rt_runtime_us
\end{VerbatimU}

You will of course need to be running as root to execute this command,
and you will also need to carefully consider the aforementioned Spiderman
principle.
One way to minimize the risks is to offload interrupts and
kernel threads/daemons from all CPUs running CPU-bound real-time
threads, as described in the paragraphs above.
In addition, you should carefully read the material in the
\path{Documentation/scheduler} directory.
The material in the \path{sched-rt-group.rst} file is particularly
important, especially if you are using the \co{cgroups} real-time features
enabled by the \co{CONFIG_RT_GROUP_SCHED} Kconfig parameter.

A fourth source of OS jitter comes from timers.
In most cases, keeping a given CPU out of the kernel will prevent
timers from being scheduled on that CPU\@.
One important exception are recurring timers, where a given timer
handler posts a later occurrence of that same timer.
If such a timer gets started on a given CPU for any reason, that
timer will continue to run periodically on that CPU, inflicting
OS jitter indefinitely.
One crude but effective way to offload recurring timers is to
use CPU hotplug to offline all worker CPUs that are to run CPU-bound
real-time application threads, online these same CPUs, then start
your real-time application.

A fifth source of OS jitter is provided by device drivers that were
not intended for real-time use.
For an old canonical example, in 2005, the VGA driver would blank
the screen by zeroing the frame buffer with interrupts disabled,
which resulted in tens of milliseconds of OS jitter.
One way of avoiding device-driver-induced OS jitter is to carefully
select devices that have been used heavily in real-time systems,
and which have therefore had their real-time bugs fixed.
Another way is to confine the device's interrupts and all code using
that device to designated housekeeping CPUs.
A third way is to test the device's ability to support real-time
workloads and fix any real-time bugs.\footnote{
	If you take this approach, please submit your fixes upstream
	so that others can benefit.
	After all, when you need to port your application to
	a later version of the Linux kernel, \emph{you} will be one of those
	``others''.}

A sixth source of OS jitter is provided by some in-kernel
full-system synchronization algorithms, perhaps most notably
the global TLB-flush algorithm.
This can be avoided by avoiding memory-unmapping operations, and especially
avoiding unmapping operations within the kernel.
As of early 2021, the way to avoid in-kernel
unmapping operations is to avoid unloading kernel modules.

A seventh source of OS jitter is provided by
scheduling-clock interrrupts and RCU callback invocation.
These may be avoided by building your kernel with the
\co{NO_HZ_FULL} Kconfig parameter enabled, and then booting
with the \co{nohz_full=} parameter specifying the list of
worker CPUs that are to run real-time threads.
For example, \co{nohz_full=2-7} would designate CPUs~2, 3, 4, 5, 6, and~7
as worker CPUs, thus leaving CPUs~0 and~1 as housekeeping CPUs.
The worker CPUs would not incur scheduling-clock interrupts as long
as there is no more than one runnable task on each worker CPU,
and each worker CPU's RCU callbacks would be invoked on one of the
housekeeping CPUs.
A CPU that has suppressed scheduling-clock interrupts due to there
only being one runnable task on that CPU is said to be in
\emph{adaptive ticks mode} or in \co{nohz_full} mode.
It is important to ensure that you have designated enough
housekeeping CPUs to handle the housekeeping load imposed by the
rest of the system, which requires careful benchmarking and tuning.

An eighth source of OS jitter is page faults.
Because most Linux implementations use an MMU for memory protection,
real-time applications running on these systems can be subject
to page faults.
Use the \co{mlock()} and \co{mlockall()} system calls to pin your
application's pages into memory, thus avoiding major page faults.
Of course, the Spiderman principle applies, because locking down
too much memory may prevent the system from getting other work done.

A ninth source of OS jitter is unfortunately the hardware and firmware.
It is therefore important to use systems that have been designed for
real-time use.

\begin{listing}
\begin{fcvlabel}[ln:advsync:Locating Sources of OS Jitter]
\begin{VerbatimL}[commandchars=\\\[\]]
cd /sys/kernel/debug/tracing
echo 1 > max_graph_depth		\lnlbl[echo1]
echo function_graph > current_tracer
# run workload
cat per_cpu/cpuN/trace			\lnlbl[cat]
\end{VerbatimL}
\end{fcvlabel}
\caption{Locating Sources of OS Jitter}
\label{lst:advsync:Locating Sources of OS Jitter}
\end{listing}

\begin{fcvref}[ln:advsync:Locating Sources of OS Jitter]
Unfortunately, this list of OS-jitter sources can never be complete,
as it will change with each new version of the kernel.
This makes it necessary to be able to track down additional sources
of OS jitter.
Given a CPU $N$ running a CPU-bound usermode thread, the
commands shown in
\cref{lst:advsync:Locating Sources of OS Jitter}
will produce a list of all the times that this CPU entered the kernel.
Of course, the \co{N} on \clnref{cat} must be replaced with the
number of the CPU in question, and the \co{1} on \clnref{echo1} may be
increased
to show additional levels of function call within the kernel.
The resulting trace can help track down the source of the OS jitter.
\end{fcvref}

As always, there is no free lunch, and \co{NO_HZ_FULL} is no exception.
As noted earlier,
\co{NO_HZ_FULL} makes kernel/user transitions more expensive due to the
need for delta process accounting and the need to inform kernel subsystems
(such as RCU) of the transitions.
As a rough rule of thumb, \co{NO_HZ_FULL} helps with many types of
real-time and heavy-compute workloads, but hurts other workloads
that feature high rates of system calls and
I/O~\cite{AbdullahAljuhni2018nohzfull}.
Additional limitations, tradeoffs, and configuration advice may be
found in \path{Documentation/timers/no_hz.rst}.

As you can see, obtaining bare-metal performance when running
CPU-bound real-time threads on a general-purpose OS such as Linux
requires painstaking attention to detail.
Automation would of course help, and some automation has been applied,
but given the relatively small number of users, automation can be
expected to appear relatively slowly.
Nevertheless, the ability to gain near-bare-metal performance while
running a general-purpose operating system promises to ease construction
of some types of real-time systems.

\subsection{Implementing Parallel Real-Time Applications}
\label{sec:advsync:Implementing Parallel Real-Time Applications}

Developing real-time applications is a wide-ranging topic, and this
section can only touch on a few aspects.
To this end,
\cref{sec:advsync:Real-Time Components}
looks at a few software components commonly used in real-time applications,
\cref{sec:advsync:Polling-Loop Applications}
provides a brief overview of how polling-loop-based applications may
be implemented,
\cref{sec:advsync:Streaming Applications}
gives a similar overview of streaming applications, and
\cref{sec:advsync:Event-Driven Applications}
briefly covers event-based applications.

\subsubsection{Real-Time Components}
\label{sec:advsync:Real-Time Components}

As in all areas of engineering, a robust set of components is essential
to \IX{productivity} and \IX{reliability}.
This section is not a full catalog of real-time software components---such
a catalog would fill multiple books---but rather a brief overview of the
types of components available.

A natural place to look for real-time software components would be
algorithms offering wait-free
synchronization~\cite{Herlihy91}, and in fact lockless
algorithms are very important to real-time computing.
However, wait-free synchronization only guarantees forward progress in
finite time.
Although a century is finite, this is unhelpful when your deadlines are
measured in microseconds, let alone milliseconds.

Nevertheless, there are some important wait-free algorithms that do
provide bounded response time, including atomic test and set,
atomic exchange,
atomic fetch-and-add,
single-producer/single-consumer FIFO queues based on circular arrays,
and numerous per-thread partitioned algorithms.
In addition, recent research has confirmed the observation that
algorithms with lock-free guarantees\footnote{
	Wait-free algorithms guarantee that all threads make progress in
	finite time, while lock-free algorithms only guarantee that at
	least one thread will make progress in finite time.
	See \cref{sec:advsync:Non-Blocking Synchronization} for more details.}
also provide the same latencies in practice (in the wait-free sense),
assuming a stochastically fair scheduler and absence of fail-stop
bugs~\cite{DanAlitarh2013PracticalProgress}.
This means that many non-wait-free stacks and queues are nevertheless
appropriate for real-time use.

\QuickQuiz{
	But isn't correct operation despite fail-stop bugs
	a valuable fault-tolerance property?
}\QuickQuizAnswer{
	Yes and no.

	Yes in that non-blocking algorithms can provide fault tolerance
	in the face of fail-stop bugs, but no in that this is grossly
	insufficient for practical fault tolerance.
	For example, suppose you had a wait-free queue, and further
	suppose that a thread has just dequeued an element.
	If that thread now succumbs to a fail-stop bug, the element
	it has just dequeued is effectively lost.
	True fault tolerance requires way more than mere non-blocking
	properties, and is beyond the scope of this book.
}\QuickQuizEnd

In practice, locking is often used in real-time programs, theoretical
concerns notwithstanding.
However, under more severe constraints, lock-based
algorithms can also provide bounded latencies~\cite{BjoernBrandenburgPhD}.
These constraints include:

\begin{enumerate}
\item	Fair scheduler.
	In the common case of a fixed-priority scheduler, the bounded
	latencies are provided only to the highest-priority threads.
\item	Sufficient bandwidth to support the workload.
	An implementation rule supporting this constraint might be
	``There will be at least 50\,\% idle time on all CPUs
	during normal operation,''
	or, more formally, ``The offered load will be sufficiently low
	to allow the workload to be schedulable at all times.''
\item	No fail-stop bugs.
\item	FIFO locking primitives with bounded acquisition, handoff,
	and release latencies.
	Again, in the common case of a locking primitive that is FIFO
	within priorities, the bounded latencies are provided only
	to the highest-priority threads.
\item	Some way of preventing unbounded priority inversion.
	The priority-ceiling and priority-inheritance disciplines
	mentioned earlier in this chapter suffice.
\item	Bounded nesting of lock acquisitions.
	We can have an unbounded number of locks, but only as long as a
	given thread never acquires more than a few of them (ideally only
	one of them) at a time.
\item	Bounded number of threads.
	In combination with the earlier constraints, this constraint means
	that there will be a bounded number of threads waiting on any
	given lock.
\item	Bounded time spent in any given critical section.
	Given a bounded number of threads waiting on any given lock and
	a bounded critical-section duration, the wait time will be bounded.
\end{enumerate}

\QuickQuiz{
	I couldn't help but spot the word ``include'' before this list.
	Are there other constraints?
}\QuickQuizAnswer{
	Indeed there are, and lots of them.
	However, they tend to be specific to a given situation,
	and many of them can be thought of as refinements of some of
	the constraints listed above.
	For example, the many constraints on choices of data structure
	will help meeting the ``Bounded time spent in any given critical
	section'' constraint.
}\QuickQuizEnd

This result opens a vast cornucopia of algorithms and data structures
for use in real-time software---and validates long-standing real-time practice.

Of course, a careful and simple application design is also extremely
important.
The best real-time components in the world cannot make up for a
poorly thought-out design.
For parallel real-time applications, synchronization overheads clearly
must be a key component of the design.

\subsubsection{Polling-Loop Applications}
\label{sec:advsync:Polling-Loop Applications}

Many real-time applications consist of a single CPU-bound loop that
reads sensor data, computes a control law, and writes control output.
If the hardware registers providing sensor data and taking control
output are mapped into the application's address space, this loop
might be completely free of system calls.
But beware of the Spiderman principle:
With great power comes great responsibility, in this case the
responsibility to avoid bricking the hardware by making inappropriate
references to the hardware registers.

This arrangement is often run on bare metal, without the benefits of
(or the interference from) an operating system.
However, increasing hardware capability and increasing levels of
automation motivates increasing software functionality, for example,
user interfaces, logging, and reporting, all of which can benefit from
an operating system.

One way of gaining much of the benefit of running on bare metal while
still having access to the full features and functions of a
general-purpose operating system is to use the Linux kernel's
\co{NO_HZ_FULL} capability, described in
\cref{sec:advsync:Polling-Loop Real-Time Support}.

\subsubsection{Streaming Applications}
\label{sec:advsync:Streaming Applications}

One type of big-data real-time application takes input from numerous
sources, processes it internally, and outputs alerts and summaries.
These \emph{streaming applications} are often highly parallel, processing
different information sources concurrently.

One approach for implementing streaming applications is to use
dense-array circular FIFOs to connect different processing
steps~\cite{AdrianSutton2013LCA:Disruptor}.
Each such FIFO has only a single thread producing into it and a
(presumably different) single thread consuming from it.
Fan-in and fan-out points use threads rather than data structures,
so if the output of several FIFOs needed to be merged, a separate
thread would input from them and output to another FIFO for which
this separate thread was the sole producer.
Similarly, if the output of a given FIFO needed to be split, a separate
thread would input from this FIFO and output to several FIFOs as needed.

This discipline might seem restrictive, but it allows communication
among threads with minimal synchronization overhead, and minimal
synchronization overhead is important when attempting to meet
tight latency constraints.
This is especially true when the amount of processing for each step
is small, so that the synchronization overhead is significant compared
to the processing overhead.

The individual threads might be CPU-bound, in which case the advice in
\cref{sec:advsync:Polling-Loop Applications} applies.
On the other hand, if the individual threads block waiting for
data from their input FIFOs, the advice of the next section applies.

\subsubsection{Event-Driven Applications}
\label{sec:advsync:Event-Driven Applications}

We will use fuel injection into a mid-sized industrial engine as a
fanciful example for event-driven applications.
Under normal operating conditions, this engine requires that the fuel
be injected within a one-degree interval surrounding top dead center.
If we assume a 1,500-RPM rotation rate, we have 25 rotations per second,
or about 9,000 degrees of rotation per second, which translates to
111 microseconds per degree.
We therefore need to schedule the fuel injection to within a time
interval of about 100 microseconds.

\begin{listing}
\begin{fcvlabel}[ln:advsync:Timed-Wait Test Program]
\begin{VerbatimL}
if (clock_gettime(CLOCK_REALTIME, &timestart) != 0) {
	perror("clock_gettime 1");
	exit(-1);
}
if (nanosleep(&timewait, NULL) != 0) {
	perror("nanosleep");
	exit(-1);
}
if (clock_gettime(CLOCK_REALTIME, &timeend) != 0) {
	perror("clock_gettime 2");
	exit(-1);
}
\end{VerbatimL}
\end{fcvlabel}
\caption{Timed-Wait Test Program}
\label{lst:advsync:Timed-Wait Test Program}
\end{listing}

Suppose that a timed wait was to be used to initiate fuel injection,
although if you are building an engine, I hope you supply a rotation
sensor.
We need to test the timed-wait functionality, perhaps using the test program
shown in
\cref{lst:advsync:Timed-Wait Test Program}.
Unfortunately, if we run this program, we can get unacceptable timer
jitter, even in a \rt\ kernel.

One problem is that POSIX \co{CLOCK_REALTIME} is, oddly enough, not intended
for real-time use.
Instead, it means ``realtime'' as opposed to the amount of CPU time
consumed by a process or thread.
For real-time use, you should instead use \co{CLOCK_MONOTONIC}.
However, even with this change, results are still unacceptable.

Another problem is that the thread must be raised to a real-time
priority by using the \co{sched_setscheduler()} system call.
But even this change is insufficient, because we can still see
page faults.
We also need to use the \co{mlockall()} system call to pin the
application's memory, preventing page faults.
With all of these changes, results might finally be acceptable.

In other situations, further adjustments might be needed.
It might be necessary to affinity time-critical threads onto their
own CPUs, and it might also be necessary to affinity interrupts
away from those CPUs.
It might be necessary to carefully select hardware and drivers,
and it will very likely be necessary to carefully select kernel
configuration.

As can be seen from this example, real-time computing can be quite
unforgiving.

\subsubsection{The Role of RCU}
\label{sec:advsync:The Role of RCU}

Suppose that you are writing a parallel real-time application that needs
to access
data that is subject to gradual change, perhaps due to changes in
temperature, humidity, and barometric pressure.
The real-time response constraints on this program are so severe that
it is not permissible to spin or block, thus ruling out locking,
nor is it permissible to use a retry loop, thus ruling out sequence locks
and \IXpl{hazard pointer}.
Fortunately, the temperature and pressure are normally controlled,
so that a default hard-coded set of data is usually sufficient.

However, the temperature, humidity, and pressure occasionally deviate too far
from the defaults, and in such situations it is necessary to provide
data that replaces the defaults.
Because the temperature, humidity, and pressure change gradually,
providing the updated values is not a matter of urgency, though
it must happen within a few minutes.
The program is to use a global pointer imaginatively named \co{cur_cal}
that normally references \co{default_cal}, which is a statically allocated
and initialized structure that contains the default calibration values
in fields imaginatively named \co{a}, \co{b}, and \co{c}.
Otherwise, \co{cur_cal} points to a dynamically allocated
structure providing the current calibration values.

\begin{listing}
\begin{fcvlabel}[ln:advsync:Real-Time Calibration Using RCU]
\begin{VerbatimL}[commandchars=\\\[\]]
struct calibration {
	short a;
	short b;
	short c;
};
struct calibration default_cal = { 62, 33, 88 };
struct calibration cur_cal = &default_cal;

short calc_control(short t, short h, short press)	\lnlbl[calc:b]
{
	struct calibration *p;

	p = rcu_dereference(cur_cal);
	return do_control(t, h, press, p->a, p->b, p->c);
}							\lnlbl[calc:e]

bool update_cal(short a, short b, short c)		\lnlbl[upd:b]
{
	struct calibration *p;
	struct calibration *old_p;

	old_p = rcu_dereference(cur_cal);
	p = malloc(sizeof(*p);
	if (!p)
		return false;
	p->a = a;
	p->b = b;
	p->c = c;
	rcu_assign_pointer(cur_cal, p);
	if (old_p == &default_cal)
		return true;
	synchronize_rcu();
	free(old_p);
	return true;
}							\lnlbl[upd:e]
\end{VerbatimL}
\end{fcvlabel}
\caption{Real-Time Calibration Using RCU}
\label{lst:advsync:Real-Time Calibration Using RCU}
\end{listing}

\begin{fcvref}[ln:advsync:Real-Time Calibration Using RCU]
\Cref{lst:advsync:Real-Time Calibration Using RCU}
shows how RCU can be used to solve this problem.
Lookups are deterministic, as shown in \co{calc_control()}
on \clnrefrange{calc:b}{calc:e}, consistent with real-time requirements.
Updates are more complex, as shown by \co{update_cal()}
on \clnrefrange{upd:b}{upd:e}.
\end{fcvref}

\QuickQuizSeries{%
\QuickQuizB{
	Given that real-time systems are often used for safety-critical
	applications, and given that runtime memory allocation is
	forbidden in many safety-critical situations, what is with
	the call to \co{malloc()}???
}\QuickQuizAnswerB{
	In early 2016, projects forbidding runtime memory allocation
	were also not at all interested in multithreaded computing.
	So the runtime memory allocation is not an additional
	obstacle to safety criticality.

	However, by 2020 runtime memory allocation in multi-core
	real-time systems was gaining some traction.
}\QuickQuizEndB
\QuickQuizE{
	Don't you need some kind of synchronization to protect
	\co{update_cal()}?
}\QuickQuizAnswerE{
	Indeed you do, and you could use any of a number of techniques
	discussed earlier in this book.
	One of those techniques is use of a single updater thread,
	which would result in exactly the code shown in \co{update_cal()}
	in \cref{lst:advsync:Real-Time Calibration Using RCU}.
}\QuickQuizEndE
}

This example shows how RCU can provide deterministic read-side
data-structure access to real-time programs.

\subsection{Real Time vs.~Real Fast:
				     How to Choose?}
\label{sec:advsync:Real Time vs. Real Fast: How to Choose?}

The choice between real-time and real-fast computing can be a difficult one.
Because real-time systems often inflict a throughput penalty on
non-real-time computing, using real-time when it is not required is
unwise, as fancifully depicted by
\cref{fig:advsync:The Dark Side of Real-Time Computing}.

\begin{figure}
\centering
\resizebox{3.2in}{!}{\includegraphics{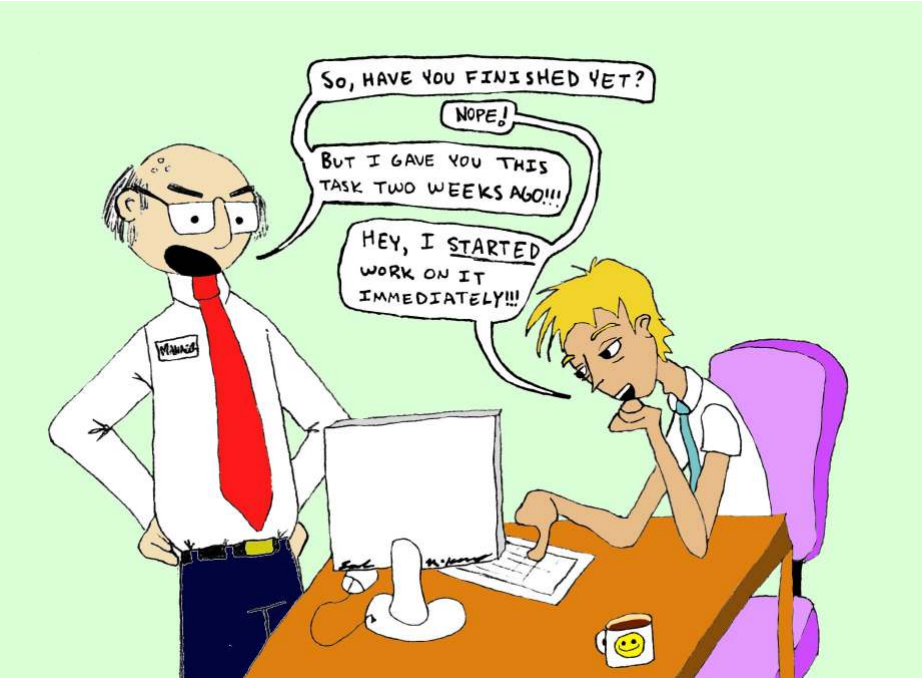}}
\caption{The Dark Side of Real-Time Computing}
\ContributedBy{Figure}{fig:advsync:The Dark Side of Real-Time Computing}{Sarah McKenney}
\end{figure}

On the other hand, failing to use real-time when it \emph{is} required
can also cause problems, as fancifully depicted by
\cref{fig:advsync:The Dark Side of Real-Fast Computing}.
It is almost enough to make you feel sorry for the boss!

\begin{figure}
\centering
\resizebox{3.2in}{!}{\includegraphics{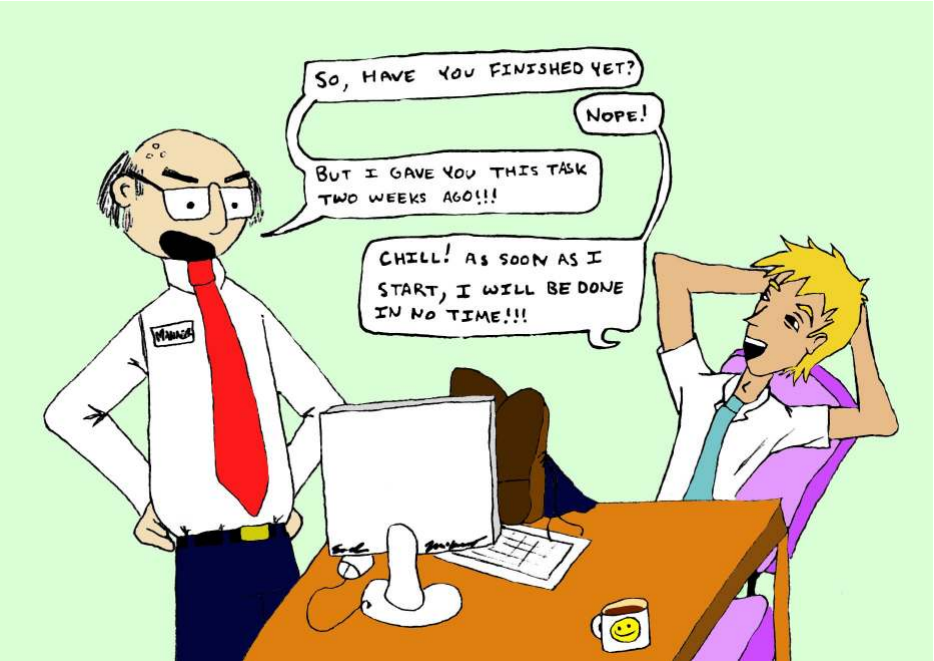}}
\caption{The Dark Side of Real-Fast Computing}
\ContributedBy{Figure}{fig:advsync:The Dark Side of Real-Fast Computing}{Sarah McKenney}
\end{figure}

One rule of thumb uses the following four questions to help you choose:

\begin{enumerate}
\item	Is average long-term throughput the only goal?
\item	Is it permissible for heavy loads to degrade response times?
\item	Is there high memory pressure, ruling out use of
	the \co{mlockall()} system call?
\item	Does the basic work item of your application take more than
	100 milliseconds to complete?
\end{enumerate}

If the answer to any of these questions is ``yes'', you should choose
real-fast over real-time, otherwise, real-time might be for you.

Choose wisely, and if you do choose real-time, make sure that your
hardware, firmware, and operating system are up to the job!

	\setlength{\parskip}{0.0pt plus 1ex}
	\input{qqzadvsync}
	\setlength{\parskip}{0.0pt plus 1.0pt}% return to default

% memorder/memorder.tex
% mainfile: ../perfbook.tex
% SPDX-License-Identifier: CC-BY-SA-3.0

\QuickQuizChapter{chp:Advanced Synchronization: Memory Ordering}{Advanced Synchronization: Memory Ordering}{qqzmemorder}
\OriginallyPublished{Chapter}{chp:Advanced Synchronization: Memory Ordering}{Advanced Synchronization: Memory Ordering}{the Linux kernel}{Howells2009membartxt}
\OriginallyPublished{Chapter}{chp:Advanced Synchronization: Memory Ordering}{Advanced Synchronization: Memory Ordering}{Linux Weekly News}{JadeAlglave2017LWN-LKMM-1,JadeAlglave2017LWN-LKMM-2}
\OriginallyPublished{Chapter}{chp:Advanced Synchronization: Memory Ordering}{Advanced Synchronization: Memory Ordering}{ASPLOS '18}{Alglave:2018:FSC:3173162.3177156}
\Epigraph{The art of progress is to preserve order amid change and to preserve change amid order.}{Alfred North Whitehead}

Causality and sequencing are deeply intuitive, and hackers often
have a strong grasp of these concepts.
These intuitions can be quite helpful when writing, analyzing, and
debugging not only sequential code, but also parallel code that makes
use of standard mutual-exclusion mechanisms such as locking.
Unfortunately, these intuitions break down completely in code that
instead uses weakly ordered atomic operations and memory barriers.
One example of such code implements the standard mutual-exclusion
mechanisms themselves, while another example implements fast
paths that use weaker synchronization.
Insults to intuition notwithstanding, some argue that weakness is a
virtue~\cite{JadeAlglave2013-WeaknessIsVirtue}.
Virtue or vice, this chapter will help you gain an understanding of
memory ordering, that, with practice, will be sufficient to implement
synchronization primitives and performance-critical fast paths.

\Cref{sec:memorder:Ordering: Why and How?}
will demonstrate that real computer systems can reorder memory references,
give some reasons why they do so, and provide some information on how
to prevent undesired reordering.
\Cref{sec:memorder:Tricks and Traps,%
sec:memorder:Compile-Time Consternation}
will cover the types of pain that hardware and compilers, respectively,
can inflict on unwary parallel programmers.
\Cref{sec:memorder:Higher-Level Primitives}
gives an overview of the benefits of modeling memory ordering at
higher levels of abstraction.
\Cref{sec:memorder:Hardware Specifics}
follows up with more detail on a few representative hardware platforms.
Finally, \cref{sec:memorder:Memory-Model Intuitions}
provides some reliable intuitions and useful rules of thumb.

\QuickQuiz{
	This chapter has been rewritten since the first edition,
	and heavily edited since the second edition.
	Did memory ordering change all \emph{that} since 2014,
	let alone 2021?
}\QuickQuizAnswer{
	The earlier memory-ordering section had its roots in a pair of
	Linux Journal articles~\cite{PaulMcKenney2005i,PaulMcKenney2005j}
	dating back to 2005.
	Since then, the C and C++ memory models~\cite{PeteBecker2011N3242}
	have been formalized
	(and critiqued~\cite{MarkBatty2013OOTA-WorkingNote,Boehm:2014:OGA:2618128.2618134,Vafeiadis:2015:CCO:2775051.2676995,conf/esop/BattyMNPS15,Lahav:2017:RSC:3140587.3062352,OlivierGiroux2017-P0668R1}),
	executable formal memory models for computer systems have become the
	norm~\cite{Maranget2012TutorialARMPower,PaulEMcKenney2011ppcmem,test6-pdf,JadeAlglave2011ppcmem,Alglave:2013:SVW:2450268.2450306,JadeAlglave2013-cav,Alglave:2014:HCM:2594291.2594347,PaulEMcKenney2014weakaxiom,Flur:2017:MCA:3093333.3009839,ARMv8A:2017},
	and there is even a memory model for the Linux
	kernel~\cite{JadeAlglave2017LWN-LKMM-1,JadeAlglave2017LWN-LKMM-2,Alglave:2018:FSC:3173162.3177156},
	along with a paper describing differences between the C11 and
	Linux memory models~\cite{PaulEMcKenney2016P0124R6-LKMM}.

	The \IXacrf{kcsan}~\cite{MarcoElver2020FearDataRaceDetector1,MarcoElver2020FearDataRaceDetector2},
	based in part on
	RacerD~\cite{SamBlackshear2018RacerD}
	and implementing \IXacr{lkmm}, has also been added to the Linux kernel
	and is now heavily used.

	Finally, there are now better ways of describing LKMM.

	Given all this progress, substantial change was required.
}\QuickQuizEnd

\section{Ordering:
		   Why and How?}
\label{sec:memorder:Ordering: Why and How?}
\epigraph{Nothing is orderly till people take hold of it.
	  Everything in creation lies around loose.}
	 {Henry Ward Beecher, updated}

One motivation for memory ordering can be seen in the trivial-seeming
litmus test in
\cref{lst:memorder:Memory Misordering: Store-Buffering Litmus Test}
(\path{C-SB+o-o+o-o.litmus}),
which at first glance might
appear to guarantee that the \co{exists} clause never triggers.\footnote{
	Purists would instead insist that the \co{exists} clause is
	never \emph{satisfied}, but we use ``trigger'' here by
	analogy with assertions.}
After all, if \nbco{0:r2=0} as shown in the \co{exists} clause,\footnote{
	That is, Thread~\co{P0()}'s instance of local variable \co{r2}
	equals zero.
	See \cref{sec:formal:Anatomy of a Litmus Test}
	for documentation of litmus-test nomenclature.}
we might hope that Thread~\co{P0()}'s load from~\co{x1} into \co{r2}
must have happened before Thread~\co{P1()}'s store to~\co{x1}, which
might raise further hopes that Thread~\co{P1()}'s load from~\co{x0}
into \co{r2} must happen after Thread~\co{P0()}'s store to~\co{x0},
so that \nbco{1:r2=2}, thus never triggering the \co{exists} clause.
The example is symmetric, so similar reasoning might lead
us to hope that \nbco{1:r2=0} guarantees that \nbco{0:r2=2}.
Unfortunately, the lack of memory barriers dashes these hopes.
The CPU is within its rights to reorder
the statements within both Thread~\co{P0()} and Thread~\co{P1()},
even on relatively strongly ordered systems such as x86.

\begin{listing}
\input{CodeSamples/formal/litmus/C-SB+o-o+o-o=whole.fcv}
\caption{Memory Misordering:
			     Store-Buffering Litmus Test}
\label{lst:memorder:Memory Misordering: Store-Buffering Litmus Test}
\end{listing}

\QuickQuiz{
	The compiler can also reorder Thread~\co{P0()}'s and
	Thread~\co{P1()}'s memory accesses in
	\cref{lst:memorder:Memory Misordering: Store-Buffering Litmus Test},
	right?
}\QuickQuizAnswer{
	In general, compiler optimizations carry out more extensive
	and profound reorderings than CPUs can.
	However, in this case, the volatile accesses in
	\co{READ_ONCE()} and \co{WRITE_ONCE()}
	prevent the compiler from reordering.
	And also from doing much else as well, so the examples in this
	section will be making heavy use of
	\co{READ_ONCE()} and \co{WRITE_ONCE()}.
	See \cref{sec:memorder:Compile-Time Consternation}
	for more detail on the need for \co{READ_ONCE()} and \co{WRITE_ONCE()}.
}\QuickQuizEnd

This willingness to reorder can be confirmed using tools such as
\co{litmus7}~\cite{Alglave:2014:HCM:2594291.2594347},
which found that the counter-intuitive ordering happened
314 times out of 100,000,000 trials on an x86 laptop.
Oddly enough, the perfectly legal outcome where both loads return the
value 2 occurred less frequently, in this case, only 167 times.\footnote{
	Please note that results are sensitive to the exact hardware
	configuration,
	how heavily the system is loaded, and much else besides.
	So why not try it out on your own system?}
The lesson here is clear:
Increased counter-intuitivity does not necessarily imply decreased probability!
% Run on June 23, 2017:
% litmus7 -r 1000 -carch X86 C-SB+o-o+o-o.litmus
% Test C-SB+o-o+o-o Allowed
% Histogram (4 states)
% 314   *>0:r2=0; 1:r2=0;
% 49999625:>0:r2=2; 1:r2=0;
% 49999894:>0:r2=0; 1:r2=2;
% 167   :>0:r2=2; 1:r2=2;

The following sections show exactly how this intuition breaks down,
and then put forward some mental models of memory ordering that can help
you avoid these pitfalls.

\Cref{sec:memorder:Why Hardware Misordering?}
gives a brief overview of why hardware misorders memory accesses, and then
\cref{sec:memorder:How to Force Ordering?}
gives an equally brief overview of how you can thwart such misordering.
Finally, \cref{sec:memorder:Basic Rules of Thumb}
lists some basic rules of thumb, which will be further refined in
later sections.
These sections focus on hardware reordering, but rest assured that compilers
reorder much more aggressively than hardware ever dreamed of doing.
Compiler-induced reordering will be taken up in
\cref{sec:memorder:Compile-Time Consternation}.

\subsection{Why Hardware Misordering?}
\label{sec:memorder:Why Hardware Misordering?}

But why does memory misordering happen in the first place?
Can't CPUs keep track of ordering on their own?
Isn't that why we have computers in the first place, to keep track of things?

\begin{figure*}
\centering
\resizebox{\textwidth}{!}{\includegraphics{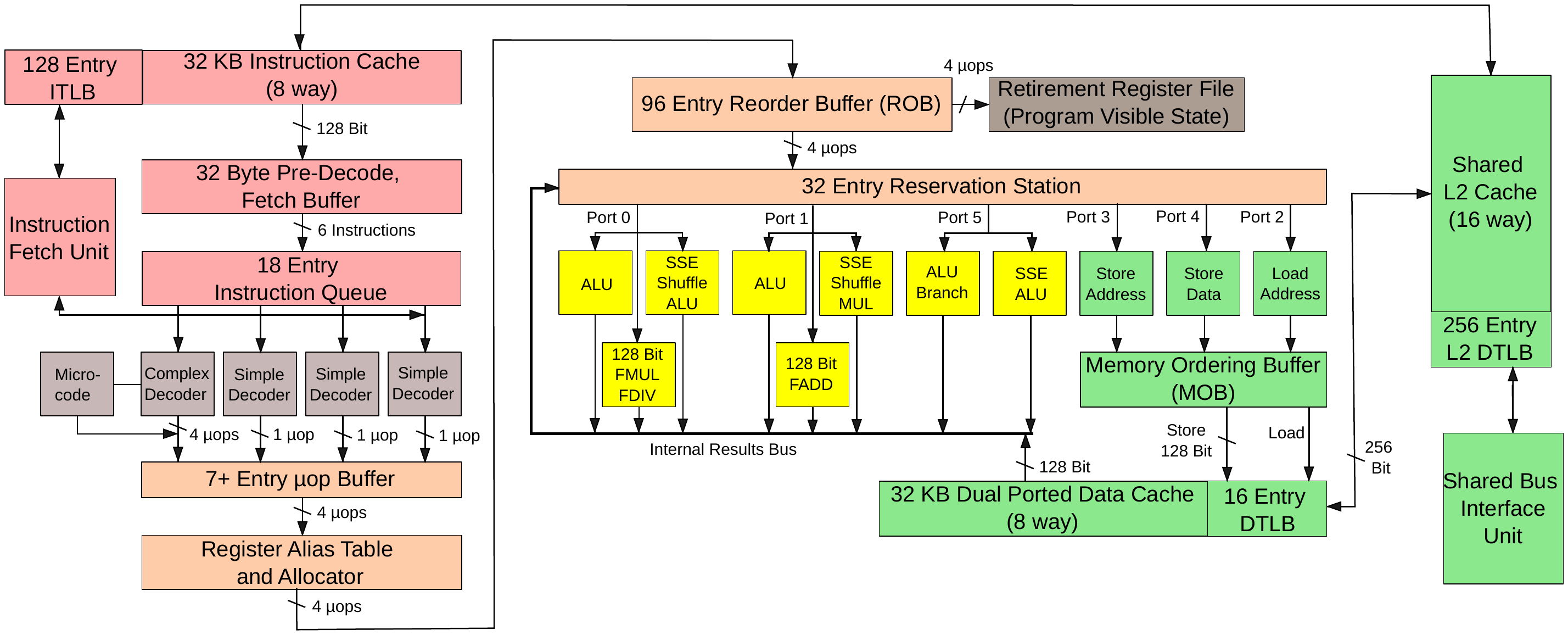}}
\caption{Intel Core 2 Architecture}
\ContributedBy{Figure}{fig:memorder:Intel Core 2 Architecture}{Wikipedia user “I, Appaloosa” CC BY-SA 3.0, reformatted}
\end{figure*}

Many people do indeed expect their computers to keep track of things,
but many also insist that they keep track of things quickly.
In fact, so intense is the focus on performance that modern CPUs
are extremely complex, as can be seen in the simplified block diagram in
\cref{fig:memorder:Intel Core 2 Architecture}.
Those needing to squeeze the last few percent of performance from their
systems will in turn need to pay close attention to the fine details of this
figure when tuning their software.
Except that this close attention to detail means that when a given CPU
degrades with age, the software will no longer run quickly on it.
For example, if the leftmost ALU fails, software tuned to take full
advantage of all of the ALUs might well run more slowly than untuned
software.
One solution to this problem is to take systems out of service as soon
as any of their CPUs start degrading.

\begin{figure*}
\centering
\resizebox{\textwidth}{!}{\includegraphics{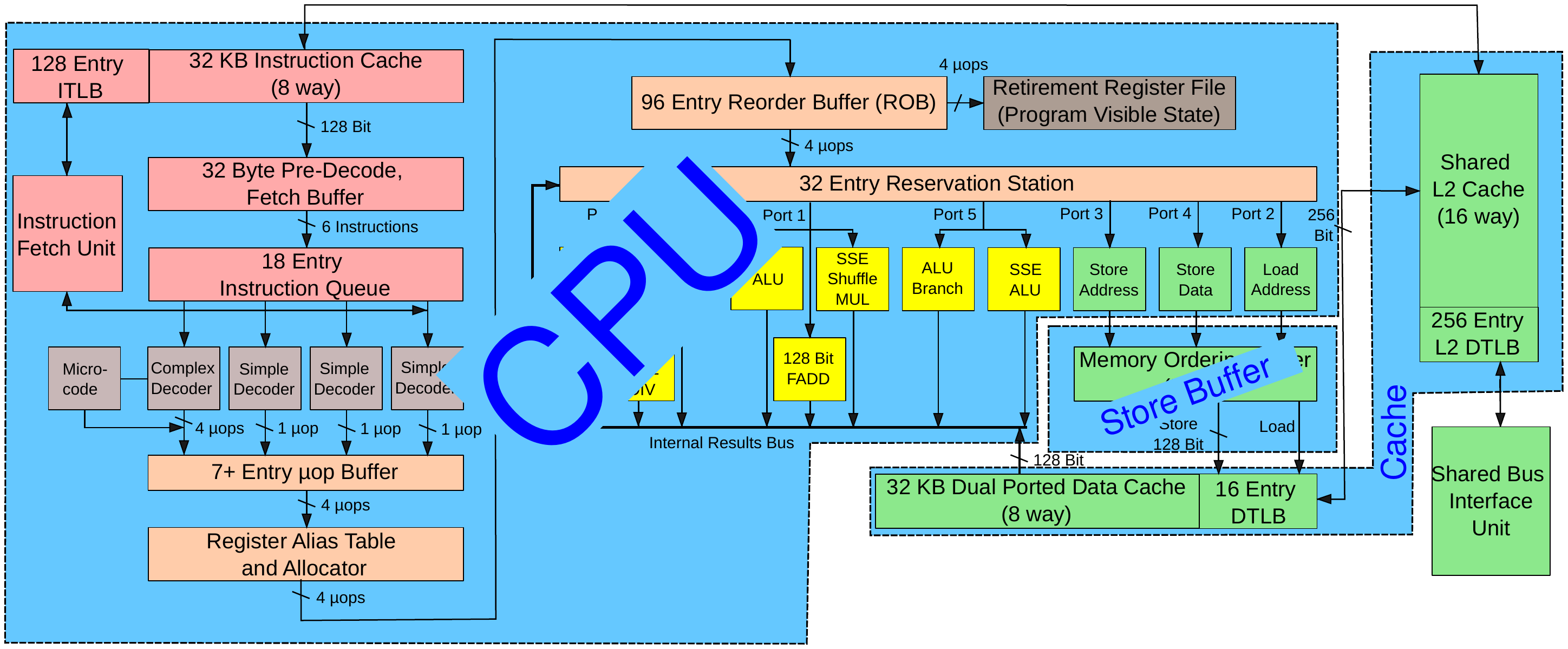}}
\caption{Intel Core 2 Architecture Simplified}
\ContributedBy{Figure}{fig:memorder:Intel Core 2 Architecture Simplified}{Wikipedia user “I, Appaloosa” CC BY-SA 3.0, reformatted}
\end{figure*}

Another option is to recall the lessons of
\cref{chp:Hardware and its Habits},
especially the lesson that for many important workloads, main memory
cannot keep up with modern CPUs, which can execute hundreds of
instructions in the time required to fetch a single variable from memory.
For such workloads, the detailed internal structure of the CPU is
irrelevant, and the CPU can instead be approximated by the blue shapes in
\cref{fig:memorder:Intel Core 2 Architecture Simplified}
labeled CPU, store buffer, and cache.

Because of these data-intensive workloads, CPUs sport increasingly large
caches, as was seen back in
\cref{fig:cpu:System Hardware Architecture},
which means that although the first load by a given CPU from a given
variable will result in an expensive \emph{cache miss} as was discussed in
\cref{sec:cpu:Cache Misses}, subsequent
repeated loads from that variable by that CPU might execute
very quickly because the initial cache miss will have loaded that
variable into that CPU's cache.

However, it is also necessary to accommodate frequent concurrent stores
from multiple CPUs to a set of shared variables.
In cache-coherent systems, if the caches hold multiple copies of a given
variable, all the copies of that variable must have the same value.
This works extremely well for concurrent loads, but not so well for
concurrent stores:
Each store must do something about all copies of the old value
(another cache miss!), which, given the finite speed of light and
the atomic nature of matter, will be slower than impatient software
hackers would like.
And these strings of stores are the reason for the blue block
labelled store buffer in
\cref{fig:memorder:Intel Core 2 Architecture Simplified}.

\begin{figure}
\centering
\resizebox{2.5in}{!}{\includegraphics{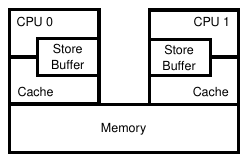}}
\caption{System Architecture With Store Buffers}
\label{fig:memorder:System Architecture With Store Buffers}
\end{figure}

Removing the internal CPU complexity from
\cref{fig:memorder:Intel Core 2 Architecture Simplified},
adding a second CPU, and showing main memory results in
\cref{fig:memorder:System Architecture With Store Buffers}.
When a given CPU stores to a variable
not present in that CPU's cache, then the new value
is instead placed in that CPU's store buffer.
The CPU can then proceed immediately, without having to wait for the
store to do something about all the old values of that variable
residing in other CPUs' caches.

\begin{figure}
\centering
\resizebox{2.4in}{!}{\includegraphics{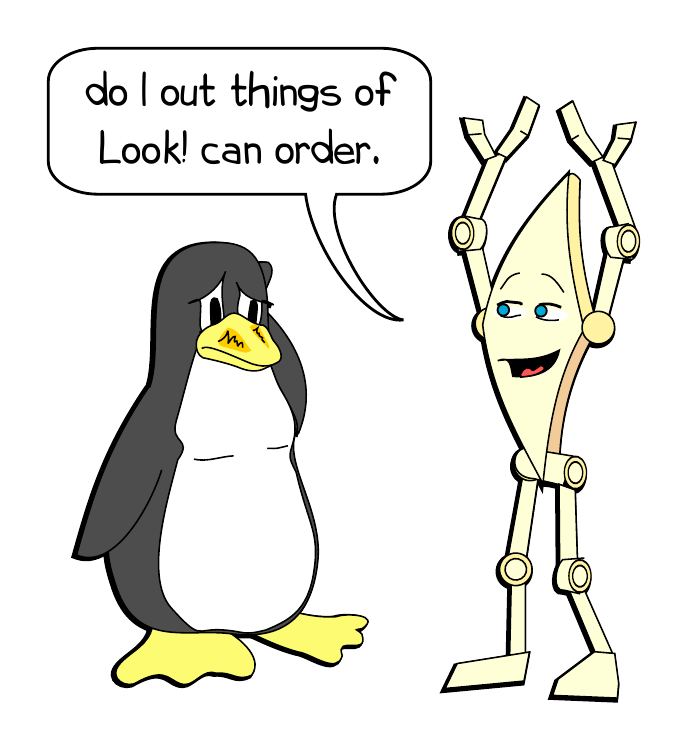}}
\caption{CPUs Can Do Things Out of Order}
\ContributedBy{Figure}{fig:memorder:CPUs Can Do Things Out of Order}{Melissa Broussard}
\end{figure}

Although store buffers can greatly increase performance, they can cause
instructions and memory references to execute out of order, which can
in turn cause serious confusion, as fancifully illustrated in
\cref{fig:memorder:CPUs Can Do Things Out of Order}.

\begin{table*}
\rowcolors{6}{}{lightgray}
\renewcommand*{\arraystretch}{1.1}
\small
\centering\OneColumnHSpace{-0.1in}
\ebresizewidth{
\begin{tabular}{rllllll}
	\toprule
	& \multicolumn{3}{c}{CPU 0} & \multicolumn{3}{c}{CPU 1} \\
	\cmidrule(l){2-4} \cmidrule(l){5-7}
	& Instruction & Store Buffer & Cache &
		Instruction & Store Buffer & Cache \\
	\cmidrule{1-1} \cmidrule(l){2-4} \cmidrule(l){5-7}
	1 & (Initial state) & & \tco{x1==0} &
		(Initial state) & & \tco{x0==0} \\
	2 & \tco{x0 = 2;} & \tco{x0==2} & \tco{x1==0} &
		\tco{x1 = 2;} & \tco{x1==2} & \tco{x0==0} \\
	3 & \tco{r2 = x1;} (0) & \tco{x0==2} & \tco{x1==0} &
		\tco{r2 = x0;} (0) & \tco{x1==2} & \tco{x0==0} \\
	4 & (Read-invalidate) & \tco{x0==2} & \tco{x0==0} &
		(Read-invalidate) & \tco{x1==2} & \tco{x1==0} \\
	5 & (Finish store) & & \tco{x0==2} &
		(Finish store) & & \tco{x1==2} \\
	\bottomrule
\end{tabular}
}
\caption{Memory Misordering:
			     Store-Buffering Sequence of Events}
\label{tab:memorder:Memory Misordering: Store-Buffering Sequence of Events}
\end{table*}

In particular, store buffers cause the memory misordering
illustrated by
\cref{lst:memorder:Memory Misordering: Store-Buffering Litmus Test}.
\Cref{tab:memorder:Memory Misordering: Store-Buffering Sequence of Events}
shows the steps leading to this misordering.
Row~1 shows the initial state, where CPU~0 has \co{x1} in its cache
and CPU~1 has \co{x0} in its cache, both variables having a value of zero.
\begin{fcvref}[ln:formal:C-SB+o-o+o-o:whole]
Row~2 shows the state change due to each CPU's store (\clnref{st0,st1} of
\cref{lst:memorder:Memory Misordering: Store-Buffering Litmus Test}).
Because neither CPU has the stored-to variable in its cache, both CPUs
record their stores in their respective store buffers.
\end{fcvref}

\QuickQuiz{
	But wait!!!
	On row~2 of
	\cref{tab:memorder:Memory Misordering: Store-Buffering Sequence of Events}
	both \co{x0} and \co{x1} each have two values at the same time,
	namely zero and two.
	How can that possibly work???
}\QuickQuizAnswer{
	There is an underlying cache-coherence protocol that straightens
	things out, which are discussed in
	\cref{sec:app:whymb:Cache-Coherence Protocols}.
	But if you think that a given variable having two values at
	the same time is surprising, just wait until you get to
	\cref{sec:memorder:Variables With Multiple Values}!
}\QuickQuizEnd

\begin{fcvref}[ln:formal:C-SB+o-o+o-o:whole]
Row~3 shows the two loads (\clnref{ld0,ld1} of
\cref{lst:memorder:Memory Misordering: Store-Buffering Litmus Test}).
Because the variable being loaded by each CPU is in that CPU's cache,
each load immediately returns the cached value, which in both cases
is zero.
\end{fcvref}

But the CPUs are not done yet:
Sooner or later, they must empty their store buffers.
Because caches move data around in relatively large blocks called
\emph{cachelines}, and because each cacheline can hold several
variables, each CPU must get the cacheline into its own cache so
that it can update the portion of that cacheline corresponding
to the variable in its store buffer, but without disturbing any
other part of the cacheline.
Each CPU must also ensure that the cacheline is not present in any other
CPU's cache, for which a read-invalidate operation is used.
As shown on row~4, after both read-invalidate operations complete,
the two CPUs have traded cachelines, so that CPU~0's cache now contains
\co{x0} and CPU~1's cache now contains \co{x1}.
Once these two variables are in their new homes, each CPU can flush
its store buffer into the corresponding \IX{cache line}, leaving each
variable with its final value as shown on row~5.

\QuickQuiz{
	But don't the values also need to be flushed from the cache
	to main memory?
}\QuickQuizAnswer{
	Perhaps surprisingly, not necessarily!
	On some systems,
	if the two variables are being used heavily, they might
	be bounced back and forth between the CPUs' caches and never
	land in main memory.
}\QuickQuizEnd

In summary, store buffers are needed to allow CPUs to handle
store instructions efficiently, but they can result in
counter-intuitive memory misordering.

But what do you do if your algorithm really needs its memory
references to be ordered?
For example, suppose that you are communicating with a driver using
a pair of flags, one that says whether or not the driver is running
and the other that says whether there is a request pending for that
driver.
The requester needs to set the request-pending flag, then check
the driver-running flag, and if false, wake the driver.
Once the driver has serviced all the pending requests that it knows about,
it needs to clear its driver-running flag, then check the request-pending
flag to see if it needs to restart.
This very reasonable approach cannot work unless there is some way
to make sure that the hardware processes the stores and loads in order.
This is the subject of the next section.

\subsection{How to Force Ordering?}
\label{sec:memorder:How to Force Ordering?}

It turns out that there are compiler directives and synchronization
primitives (such as locking and RCU) that are responsible for maintaining
the illusion of ordering through use of \emph{\IXBpl{memory barrier}} (for
example, \co{smp_mb()} in the Linux kernel).
These memory barriers can be explicit instructions, as they are on
\ARM, \Power{}, Itanium, and Alpha, or they can be implied by other instructions,
as they often are on x86.
Since these standard synchronization primitives preserve the illusion of
ordering, your path of least resistance is to simply use these primitives,
thus allowing you to stop reading this section.

\begin{listing}
\input{CodeSamples/formal/litmus/C-SB+o-mb-o+o-mb-o=whole.fcv}
\caption{Memory Ordering:
			  Store-Buffering Litmus Test}
\label{lst:memorder:Memory Ordering: Store-Buffering Litmus Test}
\end{listing}

However, if you need to implement the synchronization primitives
themselves, or if you are simply interested in understanding how memory
ordering works, read on!
The first stop on the journey is
\cref{lst:memorder:Memory Ordering: Store-Buffering Litmus Test}
(\path{C-SB+o-mb-o+o-mb-o.litmus}),
which places an \co{smp_mb()} Linux-kernel \IXh{full}{memory barrier} between
the store and load in both \co{P0()} and \co{P1()}, but is otherwise
identical to
\cref{lst:memorder:Memory Misordering: Store-Buffering Litmus Test}.
% Test C-SB+o-mb-o+o-mb-o Allowed
% Histogram (3 states)
% 49553298:>0:r2=2; 1:r2=0;
% 49636449:>0:r2=0; 1:r2=2;
% 810253:>0:r2=2; 1:r2=2;
% No
These barriers prevent the counter-intuitive outcome from happening
on 100,000,000 trials on my x86 laptop.
Interestingly enough, the added overhead due to these barriers causes the
legal outcome where both loads return the value two to happen more
than 800,000 times, as opposed to only 167 times for the
barrier-free code in
\cref{lst:memorder:Memory Misordering: Store-Buffering Litmus Test}.

\begin{table*}
\rowcolors{6}{}{lightgray}
\renewcommand*{\arraystretch}{1.1}
\small
\centering\OneColumnHSpace{-0.1in}
\ebresizewidth{
\begin{tabular}{rllllll}
	\toprule
	& \multicolumn{3}{c}{CPU 0} & \multicolumn{3}{c}{CPU 1} \\
	\cmidrule(l){2-4} \cmidrule(l){5-7}
	& Instruction & Store Buffer & Cache &
		Instruction & Store Buffer & Cache \\
	\cmidrule{1-1} \cmidrule(l){2-4} \cmidrule(l){5-7}
	1 & (Initial state) & & \tco{x1==0} &
		(Initial state) & & \tco{x0==0} \\
	2 & \tco{x0 = 2;} & \tco{x0==2} & \tco{x1==0} &
		\tco{x1 = 2;} & \tco{x1==2} & \tco{x0==0} \\
	3 & \tco{smp_mb();} & \tco{x0==2} & \tco{x1==0} &
		\tco{smp_mb();} & \tco{x1==2} & \tco{x0==0} \\
	4 & (Read-invalidate) & \tco{x0==2} & \tco{x0==0} &
		(Read-invalidate) & \tco{x1==2} & \tco{x1==0} \\
	5 & (Finish store) & & \tco{x0==2} &
		(Finish store) & & \tco{x1==2} \\
	6 & \tco{r2 = x1;} (2) & & \tco{x1==2} &
		\tco{r2 = x0;} (2) & & \tco{x0==2} \\
	\bottomrule
\end{tabular}
}
\caption{Memory Ordering:
			  Store-Buffering Sequence of Events}
\label{tab:memorder:Memory Ordering: Store-Buffering Sequence of Events}
\end{table*}

These barriers have a profound effect on ordering, as can be seen in
\cref{tab:memorder:Memory Ordering: Store-Buffering Sequence of Events}.
Although the first two rows are the same as in
\cref{tab:memorder:Memory Misordering: Store-Buffering Sequence of Events}
and although the \co{smp_mb()} instructions on row~3
do not change state
in and of themselves, they do cause the stores to complete
(rows~4 and~5) before the
loads (row~6), which rules out the counter-intuitive outcome shown in
\cref{tab:memorder:Memory Misordering: Store-Buffering Sequence of Events}.
Note that variables \co{x0} and \co{x1} each still have more than one
value on row~2, however, as promised earlier, the \co{smp_mb()}
invocations straighten things out in the end.

Although full barriers such as \co{smp_mb()} have extremely strong
ordering guarantees, their strength comes at a high price in terms
of foregone hardware and compiler optimizations.
A great many situations can be handled with much weaker ordering guarantees
that use much cheaper memory-ordering instructions, or, in some case, no
memory-ordering instructions at all.

\begin{table*}
\small
\centering\OneColumnHSpace{-0.7in}
\renewcommand*{\arraystretch}{1.1}
\rowcolors{7}{lightgray}{}
\ebresizewidth{
\begin{tabular}{lcccccccccccc}\toprule
	& & \multicolumn{4}{c}{Prior Ordered Operation} &
		\multicolumn{7}{c}{Subsequent Ordered Operation} \\
	\cmidrule(l){3-6} \cmidrule(l){7-13}
	Operation Providing Ordering & C &
		Self & R & W & RMW & Self & R & W & DR & DW & RMW & SV \\
	\cmidrule(r){1-1} \cmidrule{2-2} \cmidrule(l){3-6} \cmidrule(l){7-13}
	Store, for example, \tco{WRITE_ONCE()} &  &
		   Y &   &   &     &      &   &   &    &    &     &  Y \\
	Load, for example, \tco{READ_ONCE()} &  &
		   Y &   &   &     &      &   &   &  Y &  Y &     &  Y \\
	\tco{_relaxed()} RMW operation &  &
		   Y &   &   &     &      &   &   &  Y &  Y &     &  Y \\
	\tco{*_dereference()} &  &
		   Y &   &   &     &      &   &   &  Y &  Y &     &  Y \\
	Successful \tco{*_acquire()} &   &
		   R &   &   &     &      & Y & Y &  Y &  Y &   Y &  Y \\
	Successful \tco{*_release()} & C &
		     & Y & Y &   Y &    W &   &   &    &    &     &  Y \\
	\tco{smp_rmb()} &   &
		     & Y &   &   R &      & Y &   &  Y &    &   R &    \\
	\tco{smp_wmb()} &   &
		     &   & Y &   W &      &   & Y &    &  Y &   W &    \\
	\tco{smp_mb()} and \tco{synchronize_rcu()} & CP &
		     & Y & Y &   Y &      & Y & Y &  Y &  Y &   Y &    \\
	Successful full-strength non-\tco{void} RMW & CP &
		   Y & Y & Y &   Y &    Y & Y & Y &  Y &  Y &   Y &  Y \\
	\tco{smp_mb__before_atomic()} & CP &
		     & Y & Y &   Y &      & a & a & a  & a  &   Y &    \\
	\tco{smp_mb__after_atomic()} & CP &
		     & a & a &   Y &      & Y & Y &  Y &  Y &   Y &    \\
	\bottomrule
\end{tabular}
}

\vspace{5pt}\hfill
\ebresizeverb{.8}{
\framebox[\width]{\footnotesize\setlength{\tabcolsep}{3pt}
\rowcolors{1}{}{}
\begin{tabular}{lrl}
	Key:	& C: & Ordering is cumulative \\
		& P: & Ordering propagates \\
		& R: & Read, for example, \tco{READ_ONCE()}, or read portion of RMW \\
		& W: & Write, for example, \tco{WRITE_ONCE()}, or write portion of RMW \\
		& Y: & Provides the specified ordering \\
		& a: & Provides specified ordering given intervening RMW atomic operation \\
		& DR: & Dependent read (address dependency, \cref{sec:memorder:Address Dependencies}) \\
		& DW: & Dependent write (address, data, or control dependency, \crefrange{sec:memorder:Address Dependencies}{sec:memorder:Control Dependencies}) \\
		& RMW: & \IX{Atomic read-modify-write operation} \\
		& Self: & Orders self, as opposed to accesses both before
			  and after \\
		& SV: & Orders later accesses to the same variable \\
	\multicolumn{3}{l}{\emph{Applies to Linux kernel v4.15 and later.}} \\
\end{tabular}
}\OneColumnHSpace{-0.9in}
}
\caption{Linux-Kernel Memory-Ordering Cheat Sheet}
\label{tab:memorder:Linux-Kernel Memory-Ordering Cheat Sheet}
\end{table*}

\Cref{tab:memorder:Linux-Kernel Memory-Ordering Cheat Sheet}
provides a cheatsheet of the Linux kernel's ordering primitives and their
guarantees.
Each row corresponds to a primitive or category of primitives that might
or might not provide ordering, with the columns labeled
``Prior Ordered Operation'' and ``Subsequent Ordered Operation''
being the operations that might (or might not) be ordered against.
Cells containing ``Y'' indicate that ordering is supplied unconditionally,
while other characters indicate that ordering is supplied only partially or
conditionally.
Blank cells indicate that no ordering is supplied.

The ``Store'' row also covers the store portion of an
\IXalt{atomic RMW operation}{atomic read-modify-write operation}.
In addition, the ``Load'' row covers the load
component of a successful value-returning \co{_relaxed()} RMW atomic
operation, although the combined ``\co{_relaxed()} RMW operation''
line provides a convenient combined reference in the value-returning case.
A CPU executing unsuccessful value-returning atomic RMW operations must
invalidate the corresponding variable from all other CPUs' caches.
Therefore, unsuccessful value-returning atomic RMW operations have many
of the properties of a store, which means that the ``\co{_relaxed()}
RMW operation'' line also applies to unsuccessful value-returning atomic
RMW operations.

The \co{*_acquire} row covers \co{smp_load_acquire()},
\co{cmpxchg_acquire()}, \co{xchg_acquire()}, and so on; the \co{*_release}
row covers \co{smp_store_release()}, \co{rcu_assign_pointer()},
\co{cmpxchg_release()}, \co{xchg_release()}, and so on; and
the ``Successful full-strength non-\co{void} RMW'' row covers
\co{atomic_add_return()}, \co{atomic_add_unless()}, \co{atomic_dec_and_test()},
\co{cmpxchg()}, \co{xchg()}, and so on.
The ``Successful'' qualifiers apply to primitives such as
\co{atomic_add_unless()}, \co{cmpxchg_acquire()}, and \co{cmpxchg_release()},
which have no effect on either memory or on ordering when they indicate
failure, as indicated by the earlier ``\co{_relaxed()} RMW operation'' row.

Column ``C'' indicates cumulativity and propagation, as explained in
\cref{sec:memorder:Cumulativity,sec:memorder:Propagation}.
In the meantime, this column can usually be ignored when there
are at most two threads involved.

\QuickQuizSeries{%
\QuickQuizB{
	The rows in
	\cref{tab:memorder:Linux-Kernel Memory-Ordering Cheat Sheet}
	seem quite random and confused.
	Whatever is the conceptual basis of this table???
}\QuickQuizAnswerB{
	The rows correspond roughly to hardware mechanisms of increasing
	power and overhead.

	The \co{WRITE_ONCE()} row captures the fact that accesses to
	a single variable are always fully ordered, as indicated by
	the ``SV''column.
	Note that all other operations providing ordering against accesses
	to multiple variables also provide this same-variable ordering.

	The \co{READ_ONCE()} row captures the fact that (as of 2021) compilers
	and CPUs do not indulge in user-visible speculative stores, so that
	any store whose address, data, or execution depends on a prior load
	is guaranteed to happen after that load completes.
	However, this guarantee assumes that these dependencies have
	been constructed carefully, as described in
	\cref{sec:memorder:Address- and Data-Dependency Difficulties,%
	sec:memorder:Control-Dependency Calamities}.

	The ``\co{_relaxed()} RMW operation'' row captures the fact
	that a value-returning \co{_relaxed()} RMW has done a load and a
	store, which are every bit as good as a \co{READ_ONCE()} and a
	\co{WRITE_ONCE()}, respectively.

	The \co{*_dereference()} row captures the address and data
	dependency ordering provided by \co{rcu_dereference()} and friends.
	Again, these dependencies must been constructed carefully,
	as described in
	\cref{sec:memorder:Address- and Data-Dependency Difficulties}.

	The ``Successful \co{*_acquire()}'' row captures the fact that many
	CPUs have special ``acquire'' forms of loads and of atomic RMW
	instructions,
	and that many other CPUs have lightweight memory-barrier
	instructions that order prior loads against subsequent loads
	and stores.

	The ``Successful \co{*_release()}'' row captures the fact that many
	CPUs have special ``release'' forms of stores and of atomic RMW
	instructions, and that many other CPUs have lightweight memory-barrier
	instructions that order prior loads and stores against
	subsequent stores.

	The \co{smp_rmb()} row captures the fact that many CPUs have
	lightweight memory-barrier instructions that order prior loads against
	subsequent loads.
	Similarly,
	the \co{smp_wmb()} row captures the fact that many CPUs have
	lightweight memory-barrier instructions that order prior stores against
	subsequent stores.

	None of the ordering operations thus far require prior stores to be
	ordered against subsequent loads, which means that these operations
	need not interfere with store buffers, whose main purpose in life
	is in fact to reorder prior stores against subsequent loads.
	The lightweight nature of these operations is precisely due to
	their policy of store-buffer non-interference.
	However, as noted earlier, it is sometimes necessary to interfere
	with the store buffer in order to prevent prior stores from being
	reordered against later stores, which brings us to the remaining
	rows in this table.

	The \co{smp_mb()} row corresponds to the \IXh{full}{memory barrier}
	available on most platforms, with Itanium being the exception
	that proves the rule.
	However, even on Itanium, \co{smp_mb()} provides full ordering
	with respect to \co{READ_ONCE()} and \co{WRITE_ONCE()},
	as discussed in \cref{sec:memorder:Itanium}.

	The ``Successful full-strength non-\co{void} RMW'' row captures
	the fact that on some platforms (such as x86) atomic RMW instructions
	provide full ordering both before and after.
	The Linux kernel therefore requires that full-strength non-\co{void}
	atomic RMW operations provide full ordering in cases where these
	operations succeed.
	(Full-strength atomic RMW operation's names do not end in
	\co{_relaxed}, \co{_acquire}, or \co{_release}.)
	As noted earlier, the case where these operations do not succeed
	is covered by the ``\co{_relaxed()} RMW operation'' row.

	However, the Linux kernel does not require that either \co{void}
	or \co{_relaxed()} atomic RMW operations provide any ordering
	whatsoever, with the canonical example being \co{atomic_inc()}.
	Therefore, these operations, along with failing non-\co{void}
	atomic RMW operations may be preceded by \co{smp_mb__before_atomic()}
	and followed by \co{smp_mb__after_atomic()} to provide full
	ordering for any accesses preceding or following both.
	No ordering need be provided for accesses between the
	\co{smp_mb__before_atomic()} (or, similarly, the
	\co{smp_mb__after_atomic()}) and the atomic RMW operation, as
	indicated by the ``a'' entries on the \co{smp_mb__before_atomic()}
	and \co{smp_mb__after_atomic()} rows of the table.

	In short, the structure of this table is dictated by the
	properties of the underlying hardware, which are constrained by
	nothing other than the laws of physics, which were covered back in
	\cref{chp:Hardware and its Habits}.
	That is, the table is not random, although it is quite possible
	that you are confused.
}\QuickQuizEndB
\QuickQuizE{
	Why is
	\cref{tab:memorder:Linux-Kernel Memory-Ordering Cheat Sheet}
	missing \co{smp_mb__after_unlock_lock()} and
	\co{smp_mb__after_spinlock()}?
}\QuickQuizAnswerE{
	These two primitives are rather specialized, and at present
	seem difficult to fit into
	\cref{tab:memorder:Linux-Kernel Memory-Ordering Cheat Sheet}.
	The \co{smp_mb__after_unlock_lock()} primitive is intended to be placed
	immediately after a lock acquisition, and ensures that all CPUs
	see all accesses in prior critical sections as happening before
	all accesses following the \co{smp_mb__after_unlock_lock()}
	and also before all accesses in later critical sections.
	Here ``all CPUs'' includes those CPUs not holding that lock,
	and ``prior critical sections'' includes all prior critical sections
	for the lock in question as well as all prior critical sections
	for all other locks that were released by the same CPU that executed
	the  \co{smp_mb__after_unlock_lock()}.

	The \co{smp_mb__after_spinlock()} provides the same guarantees
	as does \co{smp_mb__after_unlock_lock()}, but also provides
	additional visibility guarantees for other accesses performed
	by the CPU that executed the \co{smp_mb__after_spinlock()}.
	Given any store S performed prior to any earlier lock acquisition
	and any load L performed after the \co{smp_mb__after_spinlock()},
	all CPUs will see S as happening before~L\@.
	In other words, if a CPU performs a store S, acquires a lock,
	executes an \co{smp_mb__after_spinlock()}, then performs a
	load L, all CPUs will see S as happening before~L\@.
}\QuickQuizEndE
}

It is important to note that this table is just a cheat sheet,
and is therefore in no way a replacement for a good understanding
of memory ordering.
To begin building such an understanding, the next section will
present some basic rules of thumb.

\subsection{Basic Rules of Thumb}
\label{sec:memorder:Basic Rules of Thumb}

This section presents some basic rules of thumb that are ``good and
sufficient'' for a great many situations.
In fact, you could write a great deal of concurrent code having
excellent performance and scalability without needing anything more
than these rules of thumb.
More sophisticated rules of thumb will be presented in
\cref{sec:memorder:Memory-Model Intuitions}.

\QuickQuiz{
	But how can I know that a given project can be designed
	and coded within the confines of these rules of thumb?
}\QuickQuizAnswer{
	Much of the purpose of the remainder of this chapter is
	to answer exactly that question!
}\QuickQuizEnd

\paragraph{A given thread sees its own accesses in order.}
This rule assumes that loads and stores from/to shared variables use
\co{READ_ONCE()} and \co{WRITE_ONCE()}, respectively.
Otherwise, the compiler can profoundly scramble\footnote{
	Many compiler writers prefer the word ``optimize''.}
your code, and sometimes the CPU can do a bit of scrambling as well,
as discussed in \cref{sec:memorder:Itanium}.

\paragraph{Interrupts and signal handlers are part of a thread.}
Both interrupt and signal handlers happen between a pair of adjacent
instructions in a thread.
This means that a given handler appears to execute atomically
from the viewpoint of the interrupted thread, at least at the
assembly-language level.
However, the C and C++ languages do not define the results of handlers
and interrupted threads sharing plain variables.
Instead, such shared variables must be \co{sig_atomic_t}, lock-free
atomics, or \co{volatile}.

On the other hand, because the handler executes within the interrupted
thread's context, the memory ordering used to synchronize communication
between the handler and the thread can be extremely lightweight.
For example, the counterpart of an acquire load is a \co{READ_ONCE()}
followed by a \co{barrier()} compiler directive and the counterpart
of a release store is a \co{barrier()} followed by a \co{WRITE_ONCE()}.
The counterpart of a full memory barrier is \co{barrier()}.
Finally, disabling interrupts or signals (as the case may be) within
the thread excludes handlers.

% @@@ Using flags to avoid expensive interrupt-disabling instructions
% @@@ and signal-blocking system calls.  Harder than it looks!

\begin{figure}
\centering
\resizebox{3in}{!}{\includegraphics{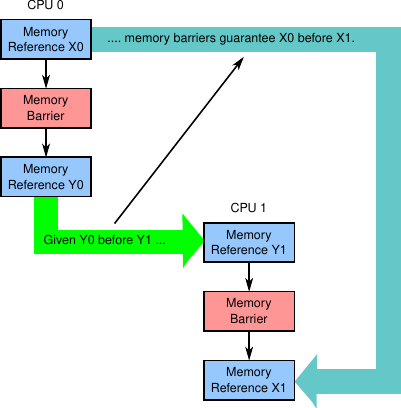}}
\caption{Memory Barriers Provide Conditional If-Then Ordering}
\label{fig:memorder:Memory Barriers Provide Conditional If-Then Ordering}
\end{figure}

\paragraph{Ordering has conditional if-then semantics.}
\Cref{fig:memorder:Memory Barriers Provide Conditional If-Then Ordering}
illustrates this for memory barriers.
Assuming that both memory barriers are strong enough, if CPU~1's access
Y1 happens after CPU~0's access Y0, then CPU~1's access X1 is guaranteed
to happen after CPU~0's access X0.
When in doubt as to which memory barriers are strong enough, \co{smp_mb()}
will always do the job, albeit at a price.

\QuickQuiz{
	How can you tell which memory barriers are strong enough for
	a given use case?
}\QuickQuizAnswer{
	Ah, that is a deep question whose answer requires most of the
	rest of this chapter.
	But the short answer is that \co{smp_mb()} is almost always
	strong enough, albeit at some cost.
}\QuickQuizEnd

\begin{fcvref}[ln:formal:C-SB+o-mb-o+o-mb-o:whole]
\Cref{lst:memorder:Memory Ordering: Store-Buffering Litmus Test}
is a case in point.
The \co{smp_mb()} on \clnref{P0:mb,P1:mb} serve as the barriers,
the store to \co{x0} on \clnref{P0:st} as X0, the load from \co{x1}
on \clnref{P0:ld} as Y0, the store to \co{x1} on \clnref{P1:st} as Y1,
and the load from \co{x0} on \clnref{P1:ld} as X1.
Applying the if-then rule step by step, we know that the store to
\co{x1} on \clnref{P1:st} happens after the load from \co{x1} on \clnref{P0:ld} if
\co{P0()}'s local variable \co{r2} is set to the value zero.
The if-then rule would then state that the load from \co{x0} on
\clnref{P1:ld} happens after the store to \co{x0} on \clnref{P0:st}.
In other words,
\co{P1()}'s local variable \co{r2} is guaranteed
to end up with the value two \emph{only if}
\co{P0()}'s local variable \co{r2} ends up with the value zero.
This underscores the point that memory ordering guarantees are
conditional, not absolute.
\end{fcvref}

Although
\cref{fig:memorder:Memory Barriers Provide Conditional If-Then Ordering}
specifically mentions memory barriers, this same if-then rule applies
to the rest of the Linux kernel's ordering operations.

\paragraph{Ordering operations must be paired.}
If you carefully order the operations in one thread, but then fail to do
so in another thread, then there is no ordering.
Both threads must provide ordering for the if-then rule to apply.\footnote{
	In \cref{sec:memorder:Propagation}, pairing will be
	generalized to cycles.}

\paragraph{Ordering operations almost never speed things up.}
If you find yourself tempted to add a memory barrier in an attempt
to force a prior store to be flushed to memory faster, resist!
Adding ordering usually slows things down.
Of course, there are situations where adding instructions speeds things
up, as was shown by
\cref{fig:defer:Pre-BSD Routing Table Protected by RCU QSBR} on
\cpageref{fig:defer:Pre-BSD Routing Table Protected by RCU QSBR},
but careful benchmarking is required in such cases.
And even then, it is quite possible that although you sped things up
a little bit on \emph{your} system, you might well have slowed things
down significantly on your users' systems.
Or on your future system.

\paragraph{Ordering operations are not magic.}
When your program is failing due to some \IX{race condition}, it is often
tempting to toss in a few memory-ordering operations in an attempt
to barrier your bugs out of existence.
A far better reaction is to use higher-level primitives in a carefully
designed manner.
With concurrent programming, it is almost always better to design your
bugs out of existence than to hack them down to lower probabilities.

\paragraph{These are only rough rules of thumb.}
Although these rules of thumb cover the vast majority of situations
seen in actual practice, as with any set of rules of thumb, they
do have their limits.
The next section will demonstrate some of these limits by introducing
trick-and-trap litmus tests that are intended to insult your
intuition while increasing your understanding.
These litmus tests will also illuminate many of the concepts
represented by the Linux-kernel memory-ordering cheat sheet shown in
\cref{tab:memorder:Linux-Kernel Memory-Ordering Cheat Sheet},
and can be automatically analyzed given proper
tooling~\cite{Alglave:2018:FSC:3173162.3177156}.
\Cref{sec:memorder:Memory-Model Intuitions} will
circle back to this cheat sheet, presenting a more sophisticated set of
rules of thumb in light of learnings from all the intervening tricks
and traps.

\QuickQuiz{
	Wait!!!
	Where do I find this tooling that automatically analyzes
	litmus tests???
}\QuickQuizAnswer{
	Get version v4.17 (or later) of the Linux-kernel source code,
	then follow the instructions in \path{tools/memory-model/README}
	to install the needed tools.
	Then follow the further instructions to run these tools on the
	litmus test of your choice.
}\QuickQuizEnd

\section{Tricks and Traps}
\label{sec:memorder:Tricks and Traps}
\epigraph{Knowing where the trap is---that's the first step in evading it.}
	 {Duke Leto Atreides, \emph{Dune}, Frank Herbert}

Now that you know that hardware can reorder memory accesses and that you
can prevent it from doing so, the next step is to get you to admit
that your intuition has a problem.
This painful task is taken up by
\cref{sec:memorder:Variables With Multiple Values},
which presents some code demonstrating that scalar variables can
take on multiple values simultaneously,
and by
\crefthro{sec:memorder:Memory-Reference Reordering}
{sec:memorder:Multicopy Atomicity},
which show a series of intuitively correct code fragments that fail miserably
on real hardware.
Once your intuition has made it through the grieving process, later
sections will summarize the basic rules that memory ordering follows.

But first, let's take a quick look at just how many values a single
variable might have at a single point in time.

\subsection{Variables With Multiple Values}
\label{sec:memorder:Variables With Multiple Values}

It is natural to think of a variable as taking on a well-defined
sequence of values in a well-defined, global order.
Unfortunately, the next stop on the journey says ``goodbye'' to this comforting fiction.
Hopefully, you already started to say ``goodbye'' in response to row~2 of
\cref{tab:memorder:Memory Misordering: Store-Buffering Sequence of Events,%
tab:memorder:Memory Ordering: Store-Buffering Sequence of Events},
and if so, the purpose of this section is to drive this point home.

\begin{fcvref}[ln:memorder:Software Logic Analyzer]
To this end, consider the program fragment shown in
\cref{lst:memorder:Software Logic Analyzer}.
This code fragment is executed in parallel by several CPUs.
\Clnref{setid} sets a shared variable to the current CPU's ID, \clnref{init}
initializes several variables from a \co{gettb()} function that
delivers the value of a fine-grained hardware ``timebase'' counter that is
synchronized among all CPUs (not available from all CPU architectures,
unfortunately!), and the loop from \clnrefrange{loop:b}{loop:e}
records the length of
time that the variable retains the value that this CPU assigned to it.
Of course, one of the CPUs will ``win'', and would thus never exit
the loop if not for the check on \clnrefrange{iftmout}{break}.
\end{fcvref}

\QuickQuiz{
	What assumption is the code fragment
	in \cref{lst:memorder:Software Logic Analyzer}
	making that might not be valid on real hardware?
}\QuickQuizAnswer{
	The code assumes that as soon as a given CPU stops
	seeing its own value, it will immediately see the
	final agreed-upon value.
	On real hardware, some of the CPUs might well see several
	intermediate results before converging on the final value.
	The actual code used to produce the data in the figures
	discussed later in this section was therefore somewhat more
	complex.
}\QuickQuizEnd

\begin{listing}
\begin{fcvlabel}[ln:memorder:Software Logic Analyzer]
\begin{VerbatimL}[commandchars=\\\[\]]
state.variable = mycpu;			\lnlbl[setid]
lasttb = oldtb = firsttb = gettb();	\lnlbl[init]
while (state.variable == mycpu) {	\lnlbl[loop:b]
	lasttb = oldtb;
	oldtb = gettb();
	if (lasttb - firsttb > 1000)	\lnlbl[iftmout]
		break;			\lnlbl[break]
}					\lnlbl[loop:e]
\end{VerbatimL}
\end{fcvlabel}
\caption{Software Logic Analyzer}
\label{lst:memorder:Software Logic Analyzer}
\end{listing}

\begin{figure}
\centering
\resizebox{3in}{!}{\includegraphics{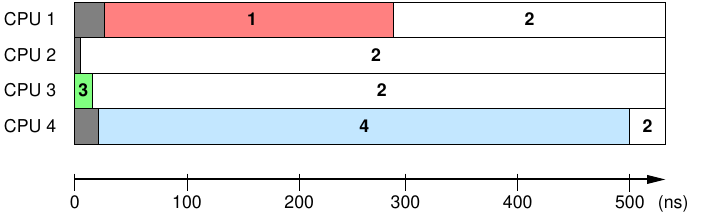}}
\caption{A Variable With Multiple Simultaneous Values}
\label{fig:memorder:A Variable With Multiple Simultaneous Values}
\end{figure}

Upon exit from the loop, \co{firsttb} will hold a timestamp
taken shortly after the assignment and \co{lasttb} will hold
a timestamp taken before the last sampling of the shared variable
that still retained the assigned value, or a value equal to \co{firsttb}
if the shared variable had changed before entry into the loop.
This allows us to plot each CPU's view of the value of \co{state.variable}
over a 532-nanosecond time period, as shown in
\cref{fig:memorder:A Variable With Multiple Simultaneous Values}.
This data was collected in 2006 on 1.5\,GHz \Power{5} system with 8 cores,
each containing a pair of hardware threads.
CPUs~1, 2, 3, and~4 recorded the values, while CPU~0 controlled the test.
The timebase counter period was about 5.32\,ns, sufficiently fine-grained
to allow observations of intermediate cache states.

Each horizontal bar represents the observations of a given CPU over time,
with the gray regions to the left indicating the time before the
corresponding CPU's first measurement.
During the first 5\,ns, only CPU~3 has an opinion about the value of the
variable.
During the next 10\,ns, CPUs~2 and~3 disagree on the value of the variable,
but thereafter agree that the value is~``2'', which is in fact
the final agreed-upon value.
However, CPU~1 believes that the value is~``1'' for almost 300\,ns, and
CPU~4 believes that the value is~``4'' for almost 500\,ns.

\QuickQuizSeries{%
\QuickQuizB{
	How could CPUs possibly have different views of the
	value of a single variable \emph{at the same time?}
}\QuickQuizAnswerB{
	As discussed in
	\cref{sec:memorder:Why Hardware Misordering?},
	many CPUs have store buffers that record the values of
	recent stores, which do not become globally visible until
	the corresponding cache line makes its way to the CPU\@.
	Therefore, it is quite possible for each CPU to see its own value
	for a given variable (in its own store buffer) at a single point
	in time---and for main memory to hold yet another value.
	One of the reasons that memory barriers were invented was
	to allow software to deal gracefully with situations like
	this one.

	Fortunately, software rarely cares about the fact that multiple
	CPUs might see multiple values for the same variable.
}\QuickQuizEndB
\QuickQuizE{
	Why do CPUs~2 and~3 come to agreement so quickly, when it
	takes so long for CPUs~1 and~4 to come to the party?
}\QuickQuizAnswerE{
	CPUs~2 and~3 are a pair of hardware threads on the same
	core, sharing the same cache hierarchy, and therefore have
	very low communications latencies.
	This is a \IXacr{numa}, or, more accurately, a \IXacr{nuca} effect.

	This leads to the question of why CPUs~2 and~3 ever disagree
	at all.
	One possible reason is that they each might have a small amount
	of private cache in addition to a larger shared cache.
	Another possible reason is instruction reordering, given the
	short 10-nanosecond duration of the disagreement and the
	total lack of memory-ordering operations in the code fragment.
}\QuickQuizEndE
}

And if you think that the situation with four CPUs was intriguing, consider
\cref{fig:memorder:A Variable With More Simultaneous Values},
which shows the same situation, but with 15~CPUs each assigning their
number to a single shared variable at time~$t=0$. Both diagrams in the
figure are drawn in the same way as
\cref{fig:memorder:A Variable With Multiple Simultaneous Values}.
The only difference is that the unit of horizontal axis is timebase ticks,
with each tick lasting about 5.3~nanoseconds.
The entire sequence therefore lasts a bit longer than the events recorded in
\cref{fig:memorder:A Variable With Multiple Simultaneous Values},
consistent with the increase in number of CPUs.
The upper diagram shows the overall picture, while the lower one zooms
in on the first 50~timebase ticks.
Again, CPU~0 coordinates the test, so does not record any values.

\begin{figure*}
\centering
\IfEbookSize{
\resizebox{\onecolumntextwidth}{!}{\includegraphics{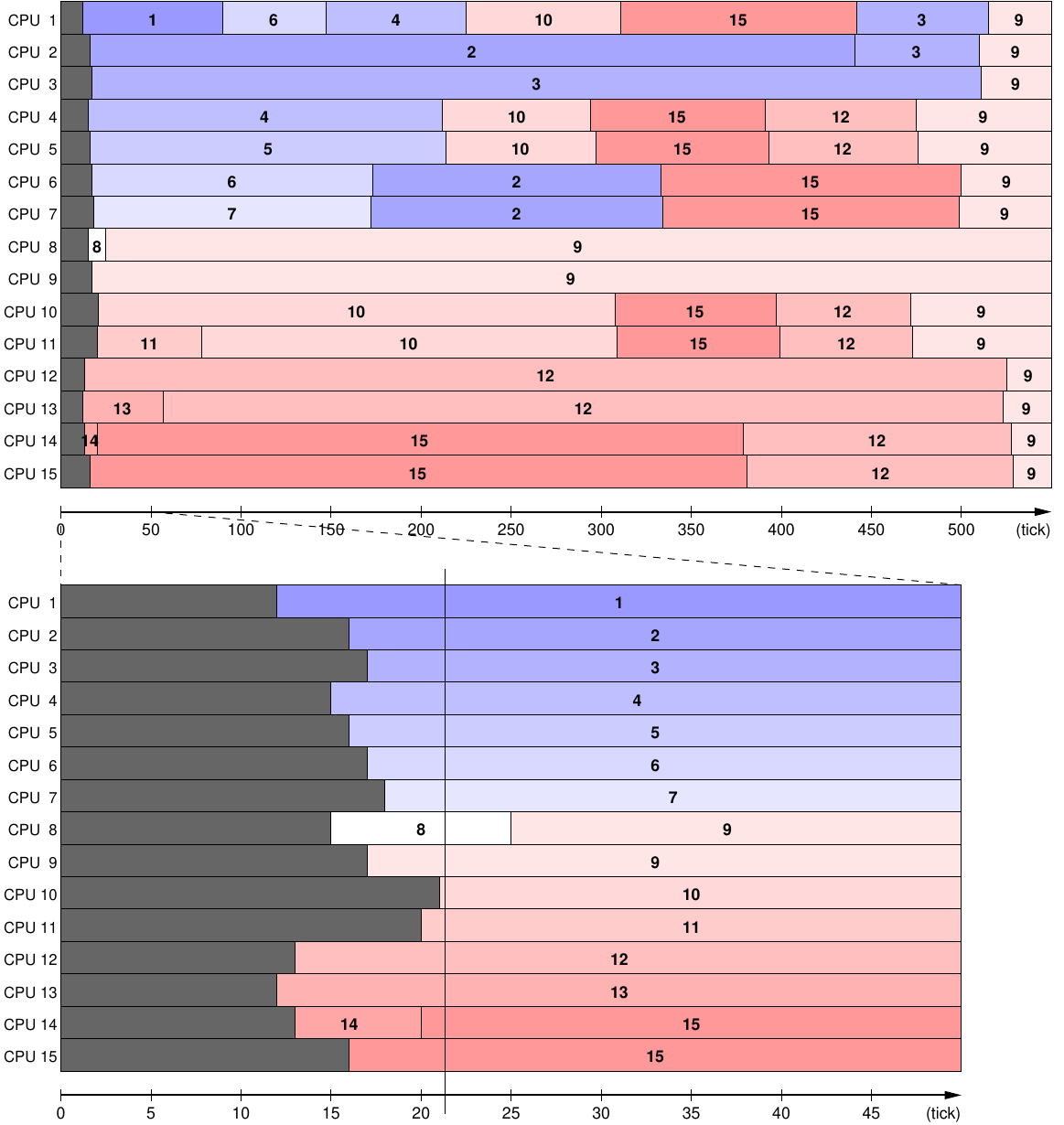}}
}{
\resizebox{5in}{!}{\includegraphics{memorder/MoreThanOneValue-15CPU}}
}
\caption{A Variable With More Simultaneous Values}
\ContributedBy{Figure}{fig:memorder:A Variable With More Simultaneous Values}{Akira Yokosawa}
\end{figure*}

\begin{figure}
\centering
\resizebox{2.0in}{!}{\includegraphics{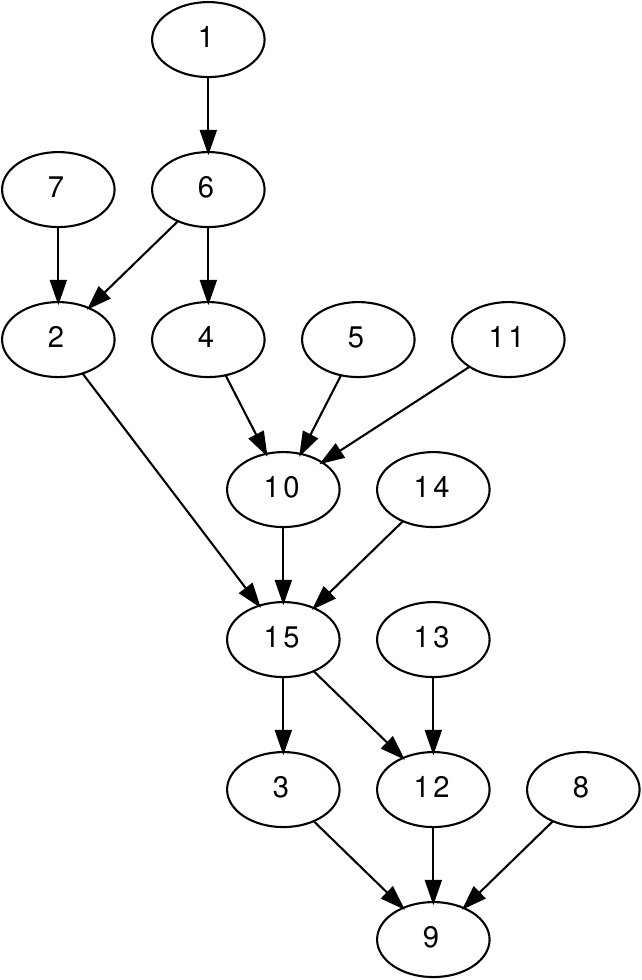}}
\caption{Possible Global Orders With More Simultaneous Values}
\label{fig:memorder:Possible Global Orders With More Simultaneous Values}
\end{figure}

All CPUs eventually agree on the final value of~9, but not before
the values~15 and~12 take early leads.
Note that there are fourteen different opinions on the variable's value
at time~21 indicated by the vertical line in the lower diagram.
Note also that all CPUs see sequences whose orderings are consistent with
the directed graph shown in
\cref{fig:memorder:Possible Global Orders With More Simultaneous Values}.
Nevertheless, these figures underscore the importance of
proper use of memory-ordering operations.

How many values can a single variable take on at a single point in
time?
As many as one per store buffer in the system!
We have therefore entered a regime where we must bid a fond farewell to
comfortable intuitions about values of variables and the passage of time.
This is the regime where memory-ordering operations are needed.

But remember well the lessons from
\cref{chp:Hardware and its Habits,%
chp:Partitioning and Synchronization Design}.
Having all CPUs store concurrently to the same variable
is no way to design a parallel program, at least
not if performance and scalability are at all important to you.

Unfortunately, memory ordering has many other ways of insulting your
intuition, and not all of these ways conflict with performance and
scalability.
The next section overviews reordering of unrelated memory reference.

\subsection{Memory-Reference Reordering}
\label{sec:memorder:Memory-Reference Reordering}

\Cref{sec:memorder:Why Hardware Misordering?}
showed that even relatively strongly ordered systems like x86
can reorder prior stores with later loads, at least when the
store and load are to different variables.
This section builds on that result, looking at the other combinations of
loads and stores.

\begin{listing}
\input{CodeSamples/formal/litmus/C-MP+o-wmb-o+o-o=whole.fcv}
\caption{Message-Passing Litmus Test (No Ordering)}
\label{lst:memorder:Message-Passing Litmus Test (No Ordering)}
\end{listing}

\subsubsection{Load Followed By Load}
\label{sec:memorder:Load Followed By Load}

\begin{fcvref}[ln:formal:C-MP+o-wmb-o+o-o:whole]
\Cref{lst:memorder:Message-Passing Litmus Test (No Ordering)}
(\path{C-MP+o-wmb-o+o-o.litmus})
shows the classic \emph{message-passing} litmus test, where \co{x0} is
the message and \co{x1} is a flag indicating whether or not a message
is available.
In this test, the \co{smp_wmb()} forces \co{P0()} stores to be ordered,
but no ordering is specified for the loads.
Relatively strongly ordered architectures, such as x86, do enforce ordering.
However, weakly ordered architectures often do
not~\cite{JadeAlglave2011ppcmem}.
Therefore, the \co{exists} clause on \clnref{exists} of the listing \emph{can}
trigger.
\end{fcvref}

One rationale for reordering loads from different locations is that doing
so allows execution to proceed when an earlier load misses the cache,
but the values for later loads are already present.

\QuickQuiz{
	But why make load-load reordering visible to the user?
	Why not just use speculative execution to allow execution to
	proceed in the common case where there are no intervening
	stores, in which case the reordering cannot be visible anyway?
}\QuickQuizAnswer{
	They can and many do, otherwise systems containing
	strongly ordered CPUs would be slow indeed.
	However, speculative execution does have its downsides, especially
	if speculation must be rolled back frequently, particularly
	on battery-powered systems.
	Speculative execution can also introduce side channels, which
	might in turn be exploited to exfiltrate information.
	But perhaps future systems will be able to overcome these
	disadvantages.
	Until then, we can expect vendors to continue producing
	weakly ordered CPUs.
}\QuickQuizEnd

\begin{listing}
\input{CodeSamples/formal/litmus/C-MP+o-wmb-o+o-rmb-o=whole.fcv}
\caption{Enforcing Order of Message-Passing Litmus Test}
\label{lst:memorder:Enforcing Order of Message-Passing Litmus Test}
\end{listing}

\begin{fcvref}[ln:formal:C-MP+o-wmb-o+o-rmb-o:whole]
Thus, portable code relying on ordered loads must
add explicit ordering, for example, the \co{smp_rmb()} shown on
\clnref{rmb} of
\cref{lst:memorder:Enforcing Order of Message-Passing Litmus Test}
(\path{C-MP+o-wmb-o+o-rmb-o.litmus}), which prevents
the \co{exists} clause from triggering.
\end{fcvref}

\begin{listing}
\input{CodeSamples/formal/litmus/C-LB+o-o+o-o=whole.fcv}
\caption{Load-Buffering Litmus Test (No Ordering)}
\label{lst:memorder:Load-Buffering Litmus Test (No Ordering)}
\end{listing}

\subsubsection{Load Followed By Store}
\label{sec:memorder:Load Followed By Store}

\begin{fcvref}[ln:formal:C-LB+o-o+o-o:whole]
\Cref{lst:memorder:Load-Buffering Litmus Test (No Ordering)}
(\path{C-LB+o-o+o-o.litmus})
shows the classic \emph{load-buffering} litmus test.
Although relatively strongly ordered systems such as x86
or the IBM Mainframe do not reorder prior loads with subsequent stores,
many weakly ordered architectures really do allow such
reordering~\cite{JadeAlglave2011ppcmem}.
Therefore, the \co{exists} clause on \clnref{exists} really can trigger.
\end{fcvref}

\begin{listing}
\input{CodeSamples/formal/litmus/C-LB+o-r+a-o=whole.fcv}
\caption{Enforcing Ordering of Load-Buffering Litmus Test}
\label{lst:memorder:Enforcing Ordering of Load-Buffering Litmus Test}
\end{listing}

\begin{fcvref}[ln:formal:C-LB+o-r+a-o:whole]
Although it is rare for actual hardware to
exhibit this reordering~\cite{LucMaranget2017aarch64},
one situation where it might be desirable to do so is when a load
misses the cache, the store buffer is nearly full, and the cacheline for
a subsequent store is ready at hand.
Therefore, portable code must enforce any required ordering, for example,
as shown in
\cref{lst:memorder:Enforcing Ordering of Load-Buffering Litmus Test}
(\path{C-LB+o-r+a-o.litmus}).
The \co{smp_store_release()} and \co{smp_load_acquire()} guarantee that
the \co{exists} clause on \clnref{exists} never triggers.
\end{fcvref}

\begin{listing}
\input{CodeSamples/formal/litmus/C-MP+o-o+o-rmb-o=whole.fcv}
\caption{Message-Passing Litmus Test, No Writer Ordering (No Ordering)}
\label{lst:memorder:Message-Passing Litmus Test; No Writer Ordering (No Ordering)}
\end{listing}

\subsubsection{Store Followed By Store}
\label{sec:memorder:Store Followed By Store}

\Cref{lst:memorder:Message-Passing Litmus Test; No Writer Ordering (No Ordering)}
(\path{C-MP+o-o+o-rmb-o.litmus})
once again shows the classic message-passing litmus test, with the
\co{smp_rmb()} providing ordering for \co{P1()}'s loads, but without
any ordering for \co{P0()}'s stores.
Again, the relatively strongly ordered architectures do enforce ordering,
but weakly ordered architectures do not necessarily do
so~\cite{JadeAlglave2011ppcmem}, which means that the
\co{exists} clause can trigger.
One situation in which such reordering could be beneficial is when
the store buffer is full, another store is ready to execute, but the
cacheline needed by the oldest store is not yet available.
In this situation, allowing stores to complete out of order would
allow execution to proceed.
Therefore, portable code must explicitly order the stores, for
example, as shown in
\cref{lst:memorder:Enforcing Order of Message-Passing Litmus Test},
thus preventing the \co{exists} clause from triggering.

\QuickQuiz{
	Why should strongly ordered systems pay the performance price
	of unnecessary \co{smp_rmb()} and \co{smp_wmb()} invocations?
	Shouldn't weakly ordered systems shoulder the full cost of
	their misordering choices???
}\QuickQuizAnswer{
	That is in fact exactly what happens.
	On strongly ordered systems, \co{smp_rmb()} and \co{smp_wmb()}
	emit no instructions, but instead just constrain the compiler.
	Thus, in this case, weakly ordered systems do in fact shoulder
	the full cost of their memory-ordering choices.
}\QuickQuizEnd

\subsection{Address Dependencies}
\label{sec:memorder:Address Dependencies}

An \emph{address dependency} occurs when the value returned by a load
instruction is used to compute the address used by a later memory-reference
instruction.
This means that the exact same sequence of instructions used to traverse
a linked data structure in single-threaded code provides weak but extremely
useful ordering in concurrent code.

\begin{listing}
\input{CodeSamples/formal/litmus/C-MP+o-wmb-o+o-addr-o=whole.fcv}
\caption{Message-Passing Address-Dependency Litmus Test (No Ordering Before v4.15)}
\label{lst:memorder:Message-Passing Address-Dependency Litmus Test (No Ordering Before v4.15)}
\end{listing}

\begin{fcvref}[ln:formal:C-MP+o-wmb-o+o-addr-o:whole]
\Cref{lst:memorder:Message-Passing Address-Dependency Litmus Test (No Ordering Before v4.15)}
(\path{C-MP+o-wmb-o+o-addr-o.litmus})
shows a linked variant of the message-passing pattern.
The head pointer is \co{x1}, which initially
references the \co{int} variable \co{y} (\clnref{init:x1}), which is in turn
initialized to the value $1$ (\clnref{init:y}).
\co{P0()} updates head pointer \co{x1} to reference \co{x0} (\clnref{P0:x1}),
but only after initializing it to $2$ (\clnref{P0:x0}) and forcing ordering
(\clnref{P0:wmb}).
\co{P1()} picks up the head pointer \co{x1} (\clnref{P1:x1}), and then loads
the referenced value (\clnref{P1:ref}).
There is thus an address dependency from the load on \clnref{P1:x1} to the
load on \clnref{P1:ref}.
In this case, the value returned by \clnref{P1:x1} is exactly the address
used by \clnref{P1:ref}, but many variations are possible,
\end{fcvref}
including field access using the C-language \co{->} operator,
addition, subtraction, and array indexing.\footnote{
	But note that in the Linux kernel, the address dependency must
	be carried through the pointer to the array, not through the
	array index.}

\begin{fcvref}[ln:formal:C-MP+o-wmb-o+o-addr-o:whole]
One might hope that \clnref{P1:x1}'s load from the head pointer would be ordered
before \clnref{P1:ref}'s dereference, which is in fact the case on Linux v4.15
and later.
However, prior to v4.15, this was not the case on DEC Alpha, which could
in effect use a speculated value for the dependent load, as described
in more detail in \cref{sec:memorder:Alpha}.
Therefore, on older versions of Linux,
\cref{lst:memorder:Message-Passing Address-Dependency Litmus Test (No Ordering Before v4.15)}'s
\co{exists} clause \emph{can} trigger.
\end{fcvref}

\begin{listing}
\begin{fcvlabel}[ln:memorder:Enforced Ordering of Message-Passing Address-Dependency Litmus Test (Before v4.15)]
\begin{VerbatimL}[commandchars=\@\[\]]
C C-MP+o-wmb-o+ld-addr-o

{
y=1;
x1=y;
}

P0(int* x0, int** x1) {
	WRITE_ONCE(*x0, 2);
	smp_wmb();
	WRITE_ONCE(*x1, x0);
}

P1(int** x1) {
	int *r2;
	int r3;

	r2 = lockless_dereference(*x1); // Obsolete @lnlbl[deref]
	r3 = READ_ONCE(*r2);			    @lnlbl[read]
}

exists (1:r2=x0 /\ 1:r3=1)
\end{VerbatimL}
\end{fcvlabel}
\caption{Enforced Ordering of Message-Passing Address-Dependency Litmus Test (Before v4.15)}
\label{lst:memorder:Enforced Ordering of Message-Passing Address-Dependency Litmus Test (Before v4.15)}
\end{listing}

\begin{fcvref}[ln:formal:C-MP+o-wmb-o+o-addr-o:whole]
\Cref{lst:memorder:Enforced Ordering of Message-Passing Address-Dependency Litmus Test (Before v4.15)}
% \path{C-MP+o-wmb-o+ld-addr-o.litmus} available at commit bc4b1c3f3b35
% ("styleguide: Loosen restriction on comment in litmus test")
shows how to make this work reliably on pre-v4.15 Linux kernels running on
DEC Alpha, by replacing \co{READ_ONCE()} on \clnref{P1:x1} of
\cref{lst:memorder:Message-Passing Address-Dependency Litmus Test (No Ordering Before v4.15)}
with \apikh{lockless_dereference()},\footnote{
	Note that \co{lockless_dereference()} is not needed on v4.15 and
	later, and therefore is not available in these later Linux kernels.
	Nor is it needed in versions of this book containing this footnote.}
which acts like \co{READ_ONCE()} on all platforms other than DEC Alpha,
where it acts like a \co{READ_ONCE()} followed by an \co{smp_mb()},
thereby forcing the required ordering on all platforms, in turn
preventing the \co{exists} clause from triggering.
\end{fcvref}

\begin{listing}
\input{CodeSamples/formal/litmus/C-S+o-wmb-o+o-addr-o=whole.fcv}
\caption{S Address-Dependency Litmus Test}
\label{lst:memorder:S Address-Dependency Litmus Test}
\end{listing}

\begin{fcvref}[ln:formal:C-S+o-wmb-o+o-addr-o:whole]
But what happens if the dependent operation is a store rather than
a load, for example, in the \emph{S}
litmus test~\cite{JadeAlglave2011ppcmem} shown in
\cref{lst:memorder:S Address-Dependency Litmus Test}
(\path{C-S+o-wmb-o+o-addr-o.litmus})?
Because no production-quality platform speculates stores,
it is not possible for the \co{WRITE_ONCE()} on \clnref{P0:x0} to overwrite
the \co{WRITE_ONCE()} on \clnref{P1:r2}, meaning that the \co{exists}
clause on \clnref{exists} cannot trigger, even on DEC Alpha, even
in pre-v4.15 Linux kernels.
\end{fcvref}

\QuickQuizSeries{%
\QuickQuizB{
	But how do we know that \emph{all} platforms really avoid
	triggering the \co{exists} clauses in
	\cref{lst:memorder:Enforced Ordering of Message-Passing Address-Dependency Litmus Test (Before v4.15),%
	lst:memorder:S Address-Dependency Litmus Test}?
}\QuickQuizAnswerB{
	Answering this requires identifying three major groups of platforms:
	\begin{enumerate*}[(1)]
	\item Total-store-order (TSO) platforms,
	\item Weakly ordered platforms, and
	\item DEC Alpha.
	\end{enumerate*}

	\begin{fcvref}[ln:memorder:Enforced Ordering of Message-Passing Address-Dependency Litmus Test (Before v4.15)]
	The TSO platforms order all pairs of memory references except for
	prior stores against later loads.
	Because the address dependency on \clnref{deref,read} of
	\cref{lst:memorder:Enforced Ordering of Message-Passing Address-Dependency Litmus Test (Before v4.15)}
	is instead a load followed by another load, TSO platforms preserve
	this address dependency.
	\end{fcvref}
	\begin{fcvref}[ln:formal:C-S+o-wmb-o+o-addr-o:whole]
	They also preserve the address dependency on \clnref{P1:x1,P1:r2} of
	\cref{lst:memorder:S Address-Dependency Litmus Test}
	because this is a load followed by a store.
	Because address dependencies must start with a load, TSO platforms
	implicitly but completely respect them, give or take compiler
	optimizations, hence the need for \co{READ_ONCE()}.
	\end{fcvref}

	Weakly ordered platforms don't necessarily maintain ordering of
	unrelated accesses.
	However, the address dependencies in
	\cref{lst:memorder:Enforced Ordering of Message-Passing Address-Dependency Litmus Test (Before v4.15),%
	lst:memorder:S Address-Dependency Litmus Test} are not unrelated:
	There is an address dependency.
	The hardware tracks dependencies and maintains the needed
	ordering.

	\begin{fcvref}[ln:memorder:Enforced Ordering of Message-Passing Address-Dependency Litmus Test (Before v4.15)]
	There is one (famous) exception to this rule for weakly ordered
	platforms, and that exception is DEC Alpha for load-to-load
	address dependencies.
	And this is why, in Linux kernels predating v4.15, DEC Alpha
	requires the explicit memory barrier supplied for it by the
	now-obsolete \apikh{lockless_dereference()} on \clnref{deref} of
	\cref{lst:memorder:Enforced Ordering of Message-Passing Address-Dependency Litmus Test (Before v4.15)}.
	\end{fcvref}
	\begin{fcvref}[ln:formal:C-S+o-wmb-o+o-addr-o:whole]
	However, DEC Alpha does track load-to-store address dependencies,
	which is why \clnref{P1:x1} of
	\cref{lst:memorder:S Address-Dependency Litmus Test}
	does not need a \co{lockless_dereference()}, even in Linux
	kernels predating v4.15.
	\end{fcvref}

	To sum up, current platforms either respect address dependencies
	implicitly, as is the case for TSO platforms (x86, mainframe,
	SPARC,~\dots), have hardware tracking for address dependencies
	(\ARM, PowerPC, MIPS,~\dots), have the required memory barriers
	supplied by \co{READ_ONCE()} (DEC Alpha in Linux kernel v4.15 and
	later), or supplied by
	\co{rcu_dereference()} (DEC Alpha in Linux kernel v4.14 and earlier).
}\QuickQuizEndB
\QuickQuizM{
	Why the use of \co{smp_wmb()} in
	\cref{lst:memorder:Enforced Ordering of Message-Passing Address-Dependency Litmus Test (Before v4.15),lst:memorder:S Address-Dependency Litmus Test}?
	Wouldn't \co{smp_store_release()} be a better choice?
}\QuickQuizAnswerM{
	In most cases, \co{smp_store_release()} is indeed a better choice.
	However, \co{smp_wmb()} was there first in the Linux kernel,
	so it is still good to understand how to use it.
}\QuickQuizEndM
\QuickQuizE{
	SP, MP, LB, and now~S\@.
	Where do all these litmus-test abbreviations come from and
	how can anyone keep track of them?
}\QuickQuizAnswerE{
	The best scorecard is the infamous
	\co{test6.pdf}~\cite{test6-pdf}.
	Unfortunately, not all of the abbreviations have catchy
	expansions like SB (store buffering), MP (message passing),
	and LB (load buffering), but at least the list of abbreviations
	is readily available.
}\QuickQuizEndE
}

However, it is important to note that address dependencies can
be fragile and easily broken by compiler optimizations, as discussed in
\cref{sec:memorder:Address- and Data-Dependency Difficulties}.

\subsection{Data Dependencies}
\label{sec:memorder:Data Dependencies}

A \emph{data dependency} occurs when the value returned by a load
instruction is used to compute the data stored by a later store
instruction.
Note well the ``data'' above:
If the value returned by a load was instead used to compute the address
used by a later store instruction, that would instead be an address
dependency, which was covered in
\cref{sec:memorder:Address Dependencies}.
However, the existence of data dependencies means that the exact same
sequence of instructions used to update a linked data structure in
single-threaded code provides weak but extremely useful ordering in
concurrent code.

\begin{listing}
\input{CodeSamples/formal/litmus/C-LB+o-r+o-data-o=whole.fcv}
\caption{Load-Buffering Data-Dependency Litmus Test}
\label{lst:memorder:Load-Buffering Data-Dependency Litmus Test}
\end{listing}

\begin{fcvref}[ln:formal:C-LB+o-r+o-data-o:whole]
\Cref{lst:memorder:Load-Buffering Data-Dependency Litmus Test}
(\path{C-LB+o-r+o-data-o.litmus})
is similar to
\cref{lst:memorder:Enforcing Ordering of Load-Buffering Litmus Test},
except that \co{P1()}'s ordering between \clnref{ld,st} is
enforced not by an \IX{acquire load}, but instead by a data dependency:
The value loaded by \clnref{ld} is what \clnref{st} stores.
The ordering provided by this data dependency is sufficient to prevent
the \co{exists} clause from triggering.
\end{fcvref}

Just as with address dependencies, data dependencies are
fragile and can be easily broken by compiler optimizations, as discussed in
\cref{sec:memorder:Address- and Data-Dependency Difficulties}.
In fact, data dependencies can be even more fragile than are address
dependencies.
The reason for this is that address dependencies normally involve
pointer values.
In contrast, as shown in
\cref{lst:memorder:Load-Buffering Data-Dependency Litmus Test},
it is tempting to carry data dependencies through integral values,
which the compiler has much more freedom to optimize into nonexistence.
For but one example, if the integer loaded was multiplied by the constant
zero, the compiler would know that the result was zero, and could therefore
substitute the constant zero for the value loaded, thus breaking
the dependency.

\QuickQuiz{
	\begin{fcvref}[ln:formal:C-LB+o-r+o-data-o:whole]
	But wait!!!
	\Clnref{ld} of
	\cref{lst:memorder:Load-Buffering Data-Dependency Litmus Test}
	uses \co{READ_ONCE()}, which marks the load as volatile,
	which means that the compiler absolutely must emit the load
	instruction even if the value is later multiplied by zero.
	So how can the compiler possibly break this data dependency?
	\end{fcvref}
}\QuickQuizAnswer{
	Yes, the compiler absolutely must emit a load instruction for
	a volatile load.
	But if you multiply the value loaded by zero, the compiler is
	well within its rights to substitute a constant zero for the
	result of that multiplication, which will break the data
	dependency on many platforms.

	Worse yet, if the dependent store does not use \co{WRITE_ONCE()},
	the compiler could hoist it above the load, which would cause
	even TSO platforms to fail to provide ordering.
}\QuickQuizEnd

In short, you can rely on data dependencies only if you prevent the
compiler from breaking them.

\subsection{Control Dependencies}
\label{sec:memorder:Control Dependencies}

A \emph{control dependency} occurs when the value returned by a load
instruction is tested to determine whether or not a later store instruction
is executed.
In other words, a simple conditional branch or conditional-move
instruction can act as a weak but low-overhead memory-barrier instruction.
However, note well the ``later store instruction'':
Although all platforms respect load-to-store dependencies, many platforms
do \emph{not} respect load-to-load control dependencies.

\begin{listing}
\input{CodeSamples/formal/litmus/C-LB+o-r+o-ctrl-o=whole.fcv}
\caption{Load-Buffering Control-Dependency Litmus Test}
\label{lst:memorder:Load-Buffering Control-Dependency Litmus Test}
\end{listing}

\begin{fcvref}[ln:formal:C-LB+o-r+o-ctrl-o:whole]
\Cref{lst:memorder:Load-Buffering Control-Dependency Litmus Test}
(\path{C-LB+o-r+o-ctrl-o.litmus})
shows another load-buffering example, this time using a control
dependency (\clnref{if}) to order the load on \clnref{ld} and the store on
\clnref{st}.
The ordering is sufficient to prevent the \co{exists} from triggering.
\end{fcvref}

However, control dependencies are even more susceptible to being optimized
out of existence than are data dependencies, and
\cref{sec:memorder:Control-Dependency Calamities}
describes some of the rules that must be followed in order to prevent
your compiler from breaking your control dependencies.

\begin{listing}
\input{CodeSamples/formal/litmus/C-MP+o-r+o-ctrl-o=whole.fcv}
\caption{Message-Passing Control-Dependency Litmus Test (No Ordering)}
\label{lst:memorder:Message-Passing Control-Dependency Litmus Test (No Ordering)}
\end{listing}

\begin{fcvref}[ln:formal:C-MP+o-r+o-ctrl-o:whole]
It is worth reiterating that control dependencies provide ordering only
from loads to stores.
Therefore, the load-to-load control dependency shown on \clnrefrange{ld1}{ld2} of
\cref{lst:memorder:Message-Passing Control-Dependency Litmus Test (No Ordering)}
(\path{C-MP+o-r+o-ctrl-o.litmus})
does \emph{not} provide ordering, and therefore does \emph{not}
prevent the \co{exists} clause from triggering.
\end{fcvref}

In summary, control dependencies can be useful, but they are
high-maintenance items.
You should therefore use them only when performance considerations
permit no other solution.

\QuickQuiz{
	Wouldn't control dependencies be more robust if they were
	mandated by language standards???
}\QuickQuizAnswer{
	But of course!
	And perhaps in the fullness of time they will be so mandated.
}\QuickQuizEnd

\subsection{Cache Coherence}
\label{sec:memorder:Cache Coherence}

On cache-coherent platforms, all CPUs agree on the order of loads and
stores to a given variable.
Fortunately, when \co{READ_ONCE()} and \co{WRITE_ONCE()} are used,
almost all platforms are cache-coherent, as indicated by the ``SV''
column of the cheat sheet shown in
\cref{tab:memorder:Linux-Kernel Memory-Ordering Cheat Sheet}.
Unfortunately, this property is so popular that it has been named
multiple times, with ``single-variable SC'',\footnote{
	Recall that SC stands for sequentially consistent.}
``single-copy atomic''~\cite{Stone:1995:SP:623262.623912},
and just plain ``coherence''~\cite{JadeAlglave2011ppcmem}
having seen use.
Rather than further compound the confusion by inventing yet another term
for this concept, this book uses ``\IX{cache coherence}'' and ``coherence''
interchangeably.

\begin{listing}
\input{CodeSamples/formal/litmus/C-CCIRIW+o+o+o-o+o-o=whole.fcv}
\caption{Cache-Coherent IRIW Litmus Test}
\label{lst:memorder:Cache-Coherent IRIW Litmus Test}
\end{listing}

\begin{fcvref}[ln:formal:C-CCIRIW+o+o+o-o+o-o:whole]
\Cref{lst:memorder:Cache-Coherent IRIW Litmus Test}
(\path{C-CCIRIW+o+o+o-o+o-o.litmus})
shows a litmus test that tests for cache coherence,
where ``IRIW'' stands
for ``independent reads of independent writes''.
Because this litmus test uses only one variable,
\co{P2()} and \co{P3()} must agree
on the order of \co{P0()}'s and \co{P1()}'s stores.
In other words, if \co{P2()} believes that \co{P0()}'s store
came first, then \co{P3()} had better not believe that
\co{P1()}'s store came first.
And in fact the \co{exists} clause on \clnref{exists} will trigger if this
situation arises.
\end{fcvref}

\QuickQuiz{
	But in
	\cref{lst:memorder:Cache-Coherent IRIW Litmus Test},
	wouldn't be just as bad if \co{P2()}'s \co{r1} and \co{r2}
	obtained the values 2 and 1, respectively, while \co{P3()}'s
	\co{r3} and \co{r4} obtained the values 1 and 2, respectively?
}\QuickQuizAnswer{
	Yes, it would.
	Feel free to modify the \co{exists} clause to
	check for that outcome and see what happens.
}\QuickQuizEnd

It is tempting to speculate that different-sized overlapping loads
and stores to a single region of memory (as might be set up using
the C-language \co{union} keyword) would provide similar ordering
guarantees.
However, Flur et al.~\cite{Flur:2017:MCA:3093333.3009839} discovered some
surprisingly simple litmus tests that demonstrate that such guarantees
can be violated on real hardware.
It is therefore necessary to restrict code to non-overlapping
same-sized aligned accesses to a given variable, at least if portability
is a consideration.\footnote{
	There is reason to believe that using atomic RMW operations
	(for example, \co{xchg()}) for all the stores will
	provide sequentially consistent ordering, but this has not
	yet been proven either way.}

Adding more variables and threads increases the scope for reordering
and other counter-intuitive behavior, as discussed in the next section.

\subsection{Multicopy Atomicity}
\label{sec:memorder:Multicopy Atomicity}

\begin{figure}
\centering
\resizebox{3.0in}{!}{\includegraphics{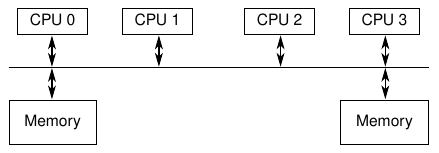}}
\caption{Global System Bus And Multi-Copy Atomicity}
\label{fig:memorder:Global System Bus And Multi-Copy Atomicity}
\end{figure}

Threads running on a fully
\emph{multicopy atomic}~\cite{Stone:1995:SP:623262.623912}
platform are guaranteed
to agree on the order of stores, even to different variables.
A useful mental model of such a system is the single-bus architecture
shown in
\cref{fig:memorder:Global System Bus And Multi-Copy Atomicity}.
If each store resulted in a message on the bus, and if the bus could
accommodate only one store at a time, then any pair of CPUs would
agree on the order of all stores that they observed.
Unfortunately, building a computer system as shown in the figure, without
store buffers or even caches, would result in glacially slow computation.
Most CPU vendors interested in providing multicopy atomicity therefore
instead provide the slightly weaker
\emph{other-multicopy atomicity}~\cite[Section B2.3]{ARMv8A:2017},
which excludes the CPU doing a given store from the requirement that all
CPUs agree on the order of all stores.\footnote{
	As of early 2021, \ARMv8 and x86 provide other-multicopy atomicity,
	IBM mainframe provides full multicopy atomicity, and PPC
	provides no multicopy atomicity at all.
	More detail is shown in
	\cref{tab:memorder:Summary of Memory Ordering}
	on
	\cpageref{tab:memorder:Summary of Memory Ordering}.}
This means that if only a subset of CPUs are doing stores, the
other CPUs will agree on the order of stores, hence the ``other''
in ``other-multicopy atomicity''.
Unlike multicopy-atomic platforms, within other-multicopy-atomic platforms,
the CPU doing the store is permitted to observe its
store early, which allows its later loads to obtain the newly stored
value directly from the store buffer, which improves performance.

\QuickQuiz{
	Can you give a specific example showing different behavior for
	multicopy atomic on the one hand and other-multicopy atomic
	on the other?
}\QuickQuizAnswer{
	\Cref{lst:memorder:Litmus Test Distinguishing Multicopy Atomic From Other Multicopy Atomic}
	(\path{C-MP-OMCA+o-o-o+o-rmb-o.litmus})
	shows such a test.

\begin{listing}
\input{CodeSamples/formal/litmus/C-MP-OMCA+o-o-o+o-rmb-o=whole.fcv}
\caption{Litmus Test Distinguishing Multicopy Atomic From Other Multicopy Atomic}
\label{lst:memorder:Litmus Test Distinguishing Multicopy Atomic From Other Multicopy Atomic}
\end{listing}

	\begin{fcvref}[ln:formal:C-MP-OMCA+o-o-o+o-rmb-o:whole]
	On a multicopy-atomic platform, \co{P0()}'s store to \co{x} on
	\clnref{P0:st} must become visible to both \co{P0()} and \co{P1()}
	simultaneously.
	Because this store becomes visible to \co{P0()} on \clnref{P0:ld}, before
	\co{P0()}'s store to \co{y} on \clnref{P0:y}, \co{P0()}'s store to
	\co{x} must become visible before its store to \co{y} everywhere,
	including \co{P1()}.
	Therefore, if \co{P1()}'s load from \co{y} on \clnref{P1:y} returns the
	value 1, so must its load from \co{x} on \clnref{P1:x}, given that
	the \co{smp_rmb()} on \clnref{P1:rmb} forces these two loads to execute
	in order.
	Therefore, the \co{exists} clause on \clnref{exists} cannot trigger on a
	multicopy-atomic platform.
	\end{fcvref}

	In contrast, on an other-multicopy-atomic platform, \co{P0()}
	could see its own store early, so that there would be no constraint
	on the order of visibility of the two stores from \co{P1()},
	which in turn allows the \co{exists} clause to trigger.
}\QuickQuizEnd

Perhaps there will come a day when all platforms provide some flavor
of multi-copy atomicity, but
in the meantime, non-multicopy-atomic platforms do exist, and so software
must deal with them.

\begin{listing}
\input{CodeSamples/formal/litmus/C-WRC+o+o-data-o+o-rmb-o=whole.fcv}
\caption{WRC Litmus Test With Dependencies (No Ordering)}
\label{lst:memorder:WRC Litmus Test With Dependencies (No Ordering)}
\end{listing}

\begin{fcvref}[ln:formal:C-WRC+o+o-data-o+o-rmb-o:whole]
\Cref{lst:memorder:WRC Litmus Test With Dependencies (No Ordering)}
(\path{C-WRC+o+o-data-o+o-rmb-o.litmus})
demonstrates multicopy atomicity, that is, on a multicopy-atomic platform,
the \co{exists} clause on \clnref{exists} cannot trigger.
In contrast, on a non-multicopy-atomic
platform this \co{exists} clause can trigger, despite
\co{P1()}'s accesses being ordered by a data dependency and \co{P2()}'s
accesses being ordered by an \co{smp_rmb()}.
Recall that the definition of multicopy atomicity requires that all
threads agree on the order of stores, which can be thought of as
all stores reaching all threads at the same time.
Therefore, a non-multicopy-atomic platform can have a store reach
different threads at different times.
In particular, \co{P0()}'s store might reach \co{P1()} long before it
reaches \co{P2()}, which raises the possibility that \co{P1()}'s store
might reach \co{P2()} before \co{P0()}'s store does.
\end{fcvref}

\begin{figure}
\centering
\resizebox{3.0in}{!}{\includegraphics{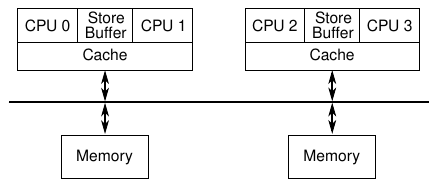}}
\caption{Shared Store Buffers And Multi-Copy Atomicity}
\label{fig:memorder:Shared Store Buffers And Multi-Copy Atomicity}
\end{figure}

This leads to the question of why a real system constrained by the
usual laws of physics would ever trigger the \co{exists} clause of
\cref{lst:memorder:WRC Litmus Test With Dependencies (No Ordering)}.
The cartoonish diagram of a such a real system is shown in
\cref{fig:memorder:Shared Store Buffers And Multi-Copy Atomicity}.
CPU~0 and CPU~1 share a store buffer, as do CPUs~2 and~3.
This means that CPU~1 can load a value out of the store buffer, thus
potentially immediately seeing a value stored by CPU~0.
In contrast, CPUs~2 and~3 will have to wait for the corresponding \IX{cache
line} to carry this new value to them.

\QuickQuiz{
	Then who would even \emph{think} of designing a system with shared
	store buffers???
}\QuickQuizAnswer{
	This is in fact a very natural design for any system having
	multiple hardware threads per core.
	Natural from a hardware point of view, that is!
}\QuickQuizEnd

\begin{table*}
\small
\centering\OneColumnHSpace{-0.8in}
\renewcommand*{\arraystretch}{1.1}
\rowcolors{10}{}{lightgray}
\ebresizewidth{
\begin{tabular}{rlllllll}\toprule
	& \multicolumn{1}{c}{\tco{P0()}} & \multicolumn{2}{c}{\tco{P0()} \& \tco{P1()}} &
		\multicolumn{1}{c}{\tco{P1()}} & \multicolumn{3}{c}{\tco{P2()}} \\
	\cmidrule(l){2-2} \cmidrule(l){3-4} \cmidrule(lr){5-5} \cmidrule(l){6-8}
	& Instruction & Store Buffer & Cache & Instruction &
			Instruction & Store Buffer & Cache \\
	\cmidrule{1-1} \cmidrule(l){2-2} \cmidrule(l){3-4}
		\cmidrule(lr){5-5} \cmidrule(l){6-8}
	1 & (Initial state) & & \tco{y==0} &
		(Initial state) &
			(Initial state) & & \tco{x==0} \\
	2 & \tco{x = 1;} & \tco{x==1} & \tco{y==0} &
		 & & & \tco{x==0} \\
	3 & (Read-Invalidate \tco{x}) & \tco{x==1} & \tco{y==0} & \tco{r1 = x} (1)
		 & & & \tco{x==0} \\
	4 &  & \tco{x==1} \tco{y==1} & \tco{y==0} & \tco{y = r1}
		 & \tco{r2 = y} & & \tco{x==0} \\
	5 &  & \tco{x==1} & \tco{y==1} & (Finish store)
		 & (Read \tco{y}) & & \tco{x==0} \\
	6 & (Respond \tco{y}) & \tco{x==1} & \tco{y==1} &
		 & (\tco{r2==1}) & & \tco{x==0} \tco{y==1} \\
	7 & & \tco{x==1} & \tco{y==1} &
		 & \tco{smp_rmb()} & & \tco{x==0} \tco{y==1} \\
	8 & & \tco{x==1} & \tco{y==1} &
		 & \tco{r3 = x (0)} & & \tco{x==0} \tco{y==1} \\
	9 & & \tco{x==1} & \tco{x==0} \tco{y==1} &
		 & (Respond \tco{x}) & & \tco{y==1} \\
	10 & (Finish store) & & \tco{x==1} \tco{y==1} &
		 &  & & \tco{y==1} \\
	\bottomrule
\end{tabular}
}
\caption{Memory Ordering:
			  WRC Sequence of Events}
\label{tab:memorder:Memory Ordering: WRC Sequence of Events}
\end{table*}

\Cref{tab:memorder:Memory Ordering: WRC Sequence of Events}
shows one sequence of events that can result in the \co{exists} clause in
\cref{lst:memorder:WRC Litmus Test With Dependencies (No Ordering)}
triggering.
This sequence of events will depend critically on \co{P0()} and
\co{P1()} sharing both cache and a store buffer in the manner shown in
\cref{fig:memorder:Shared Store Buffers And Multi-Copy Atomicity}.

\QuickQuiz{
	But just how is it fair that \co{P0()} and \co{P1()} must share a store
	buffer and a cache, but \co{P2()} gets one each of its very own???
}\QuickQuizAnswer{
	Presumably there is a \co{P3()}, as is in fact shown in
	\cref{fig:memorder:Shared Store Buffers And Multi-Copy Atomicity},
	that shares \co{P2()}'s store buffer and cache.
	But not necessarily.
	Some platforms allow different cores to disable different numbers
	of threads, allowing the hardware to adjust to the needs of the
	workload at hand.
	For example, a single-threaded critical-path portion of the workload
	might be assigned to a core with only one thread enabled, thus
	allowing the single thread running that portion of the workload
	to use the entire capabilities of that core.
	Other more highly parallel but cache-miss-prone portions of the
	workload might be assigned to cores with all hardware threads
	enabled to provide improved throughput.
	This improved throughput could be due to the fact that while one
	hardware thread is stalled on a cache miss, the other hardware
	threads can make forward progress.

	In such cases, performance requirements override quaint human
	notions of fairness.
}\QuickQuizEnd

\begin{fcvref}[ln:formal:C-WRC+o+o-data-o+o-rmb-o:whole]
Row~1 shows the initial state, with the initial value of \co{y} in
\co{P0()}'s and \co{P1()}'s shared cache, and the initial value of \co{x} in
\co{P2()}'s cache.

Row~2 shows the immediate effect of \co{P0()} executing its store on \clnref{P0:x}.
Because the cacheline containing \co{x} is not in \co{P0()}'s and \co{P1()}'s
shared cache, the new value (\co{1}) is stored in the shared store buffer.

Row~3 shows two transitions.
First, \co{P0()} issues a read-invalidate operation to fetch the cacheline
containing \co{x} so that it can flush the new value for \co{x} out of
the shared store buffer.
Second, \co{P1()} loads from \co{x} (\clnref{P1:x}), an operation that completes
immediately because the new value of \co{x} is immediately available
from the shared store buffer.

Row~4 also shows two transitions.
First, it shows the immediate effect of \co{P1()} executing its store to
\co{y} (\clnref{P1:y}), placing the new value into the shared store buffer.
Second, it shows the start of \co{P2()}'s load from \co{y} (\clnref{P2:y}).

Row~5 continues the tradition of showing two transitions.
First, it shows \co{P1()} complete its store to \co{y}, flushing
from the shared store buffer to the cache.
Second, it shows \co{P2()} request the cacheline containing \co{y}.

Row~6 shows \co{P2()} receive the cacheline containing \co{y}, allowing
it to finish its load into \co{r2}, which takes on the value \co{1}.

Row~7 shows \co{P2()} execute its \co{smp_rmb()} (\clnref{P2:rmb}), thus keeping
its two loads ordered.

Row~8 shows \co{P2()} execute its load from \co{x}, which immediately
returns with the value zero from \co{P2()}'s cache.

Row~9 shows \co{P2()} \emph{finally} responding to \co{P0()}'s request for
the cacheline containing \co{x}, which was made way back up on row~3.

Finally, row~10 shows \co{P0()} finish its store, flushing its value of
\co{x} from the shared store buffer to the shared cache.

Note well that the \co{exists} clause on \clnref{exists} has triggered.
The values of \co{r1} and \co{r2} are both the value one, and
the final value of \co{r3} the value zero.
This strange result occurred because \co{P0()}'s new value of \co{x} was
communicated to \co{P1()} long before it was communicated to \co{P2()}.
\end{fcvref}

\QuickQuiz{
	\begin{fcvref}[ln:formal:C-WRC+o+o-data-o+o-rmb-o:whole]
	Referring to
	\cref{tab:memorder:Memory Ordering: WRC Sequence of Events},
	why on earth would \co{P0()}'s store take so long to complete when
	\co{P1()}'s store complete so quickly?
	In other words, does the \co{exists} clause on \clnref{exists} of
	\cref{lst:memorder:WRC Litmus Test With Dependencies (No Ordering)}
	really trigger on real systems?
	\end{fcvref}
}\QuickQuizAnswer{
	You need to face the fact that it really can trigger.
	Akira Yokosawa used the \co{litmus7} tool to run this litmus test
	on a \Power{8} system.
	Out of 1,000,000,000 runs, 4 triggered the \co{exists} clause.
	Thus, triggering the \co{exists} clause is not merely a one-in-a-million
	occurrence, but rather a one-in-a-hundred-million occurrence.
	But it nevertheless really does trigger on real systems.
}\QuickQuizEnd

This counter-intuitive result happens because although dependencies
do provide ordering, they provide it only within the confines of their
own thread.
This three-thread example requires stronger ordering, which
is the subject of
\crefthro{sec:memorder:Cumulativity}
{sec:memorder:Release-Acquire Chains}.

\subsubsection{Cumulativity}
\label{sec:memorder:Cumulativity}

The three-thread example shown in
\cref{lst:memorder:WRC Litmus Test With Dependencies (No Ordering)}
requires \emph{cumulative} ordering, or \emph{cumulativity}.
A cumulative memory-ordering operation orders not just any given
access preceding it, but also earlier accesses by any thread to that
same variable.

\begin{listing}
\input{CodeSamples/formal/litmus/C-WRC+o+o-r+a-o=whole.fcv}
\caption{WRC Litmus Test With Release}
\label{lst:memorder:WRC Litmus Test With Release}
\end{listing}

Dependencies do not provide cumulativity,
which is why the ``C'' column is blank for the \co{READ_ONCE()} row
of \cref{tab:memorder:Linux-Kernel Memory-Ordering Cheat Sheet}
on
\cpageref{tab:memorder:Linux-Kernel Memory-Ordering Cheat Sheet}.
However, as indicated by the ``C'' in their ``C'' column,
release operations do provide cumulativity.
Therefore,
\cref{lst:memorder:WRC Litmus Test With Release}
(\path{C-WRC+o+o-r+a-o.litmus})
substitutes a release operation for
\cref{lst:memorder:WRC Litmus Test With Dependencies (No Ordering)}'s
data dependency.
\begin{fcvref}[ln:formal:C-WRC+o+o-r+a-o:whole]
Because the release operation is cumulative, its ordering applies not only to
\cref{lst:memorder:WRC Litmus Test With Release}'s
load from \co{x} by \co{P1()} on \clnref{P1:x}, but also to the store to \co{x}
by \co{P0()} on \clnref{P0:x}---but only if that load returns the value stored,
which matches the \co{1:r1=1} in the \co{exists} clause on \clnref{exists}.
This means that \co{P2()}'s load-acquire suffices to force the
load from \co{x} on \clnref{P2:x} to happen after the store on \clnref{P0:x}, so
the value returned is one, which does not match \co{2:r3=0}, which
in turn prevents the \co{exists} clause from triggering.
\end{fcvref}

\begin{figure*}
\centering
\includegraphics{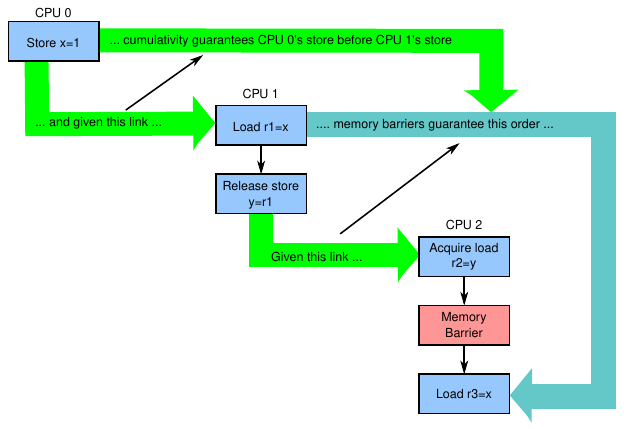}
\caption{Cumulativity}
\label{fig:memorder:Cumulativity}
\end{figure*}

\begin{fcvref}[ln:formal:C-WRC+o+o-r+a-o:whole]
These ordering constraints are depicted graphically in
\cref{fig:memorder:Cumulativity}.
Note also that cumulativity is not limited to a single step back in time.
If there was another load from \co{x} or store to \co{x} from any thread
that came before the store on \clnref{P0:x}, that prior load or store would also
be ordered before the load on \clnref{P2:x}, though only if both \co{r1} and
\co{r2} both end up containing the value \co{1}.
\end{fcvref}

In short, use of cumulative ordering operations can suppress
non-multicopy-atomic behaviors in some situations.
Cumulativity nevertheless has limits, which are examined in the next section.

\subsubsection{Propagation}
\label{sec:memorder:Propagation}

\begin{fcvref}[ln:formal:C-W+RWC+o-r+a-o+o-mb-o:whole]
\Cref{lst:memorder:W+RWC Litmus Test With Release (No Ordering)}
(\path{C-W+RWC+o-r+a-o+o-mb-o.litmus})
shows the limitations of cumulativity and store-release,
even with a full memory barrier.
The problem is that although the \co{smp_store_release()} on
\clnref{P0:sr} has cumulativity, and although that cumulativity does
order \co{P2()}'s load on \clnref{P2:ld}, the \co{smp_store_release()}'s
ordering cannot propagate through the combination of \co{P1()}'s
load (\clnref{P1:ld}) and \co{P2()}'s store (\clnref{P2:st}).
This means that the \co{exists} clause on \clnref{exists} really can trigger.
\end{fcvref}

\begin{listing}
\input{CodeSamples/formal/litmus/C-W+RWC+o-r+a-o+o-mb-o=whole.fcv}
\caption{W+RWC Litmus Test With Release (No Ordering)}
\label{lst:memorder:W+RWC Litmus Test With Release (No Ordering)}
\end{listing}

\QuickQuiz{
	But it is not necessary to worry about propagation unless
	there are at least three threads in the litmus test, right?
}\QuickQuizAnswer{
	Wrong.

\begin{listing}
\input{CodeSamples/formal/litmus/C-R+o-wmb-o+o-mb-o=whole.fcv}
\caption{R Litmus Test With Write Memory Barrier (No Ordering)}
\label{lst:memorder:R Litmus Test With Write Memory Barrier (No Ordering)}
\end{listing}

	\begin{fcvref}[ln:formal:C-R+o-wmb-o+o-mb-o:whole]
	\Cref{lst:memorder:R Litmus Test With Write Memory Barrier (No Ordering)}
	(\path{C-R+o-wmb-o+o-mb-o.litmus})
	shows a two-thread litmus test that requires propagation due to
	the fact that it only has store-to-store and load-to-store
	links between its pair of threads.
	Even though \co{P0()} is fully ordered by the \co{smp_wmb()} and
	\co{P1()} is fully ordered by the \co{smp_mb()}, the
	counter-temporal nature of the links means that
	the \co{exists} clause on \clnref{exists} really can trigger.
	To prevent this triggering, the \co{smp_wmb()} on \clnref{wmb}
	must become an \co{smp_mb()}, bringing propagation into play
	twice, once for each non-temporal link.
	\end{fcvref}
}\QuickQuizEnd

\QuickQuizLabel{\MemorderQQLitmusTestR}

\begin{figure}
\centering
\resizebox{\twocolumnwidth}{!}{\includegraphics{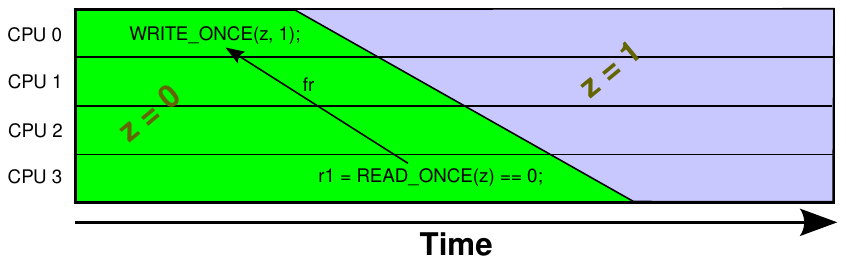}}
\caption{Load-to-Store is Counter-Temporal}
\label{fig:memorder:Load-to-Store is Counter-Temporal}
\end{figure}

This situation might seem completely counter-intuitive, but keep
in mind that the speed of light is finite and computers are of
non-zero size.
It therefore takes time for the effect of the \co{P2()}'s store to
\co{z} to propagate to \co{P1()}, which in turn means that it is possible
that \co{P1()}'s read from \co{z} happens much later in time, but
nevertheless still sees the old value of zero.
This situation is depicted in
\cref{fig:memorder:Load-to-Store is Counter-Temporal}:
Just because a load sees the old value does \emph{not} mean that
this load executed at an earlier time than did the store of the
new value.

Note that
\cref{lst:memorder:W+RWC Litmus Test With Release (No Ordering)}
also shows the limitations of memory-barrier pairing, given that
there are not two but three processes.
These more complex litmus tests can instead be said to have \emph{cycles},
where memory-barrier pairing is the special case of a two-thread cycle.
\begin{fcvref}[ln:formal:C-W+RWC+o-r+a-o+o-mb-o:whole]
The cycle in
\cref{lst:memorder:W+RWC Litmus Test With Release (No Ordering)}
goes through \co{P0()} (\clnref{P0:st,P0:sr}), \co{P1()} (\clnref{P1:la,P1:ld}),
\co{P2()} (\clnref{P2:st,P2:mb,P2:ld}), and back to \co{P0()} (\clnref{P0:st}).
The \co{exists} clause delineates this cycle:
The \co{1:r1=1} indicates that the \co{smp_load_acquire()} on \clnref{P1:la}
returned the value stored by the \co{smp_store_release()} on \clnref{P0:sr},
the \co{1:r2=0} indicates that the \co{WRITE_ONCE()} on \clnref{P2:st} came
too late to affect the value returned by the \co{READ_ONCE()} on \clnref{P1:ld},
and finally the \co{2:r3=0} indicates that the
\co{WRITE_ONCE()} on \clnref{P0:st} came too late to affect the value returned
by the \co{READ_ONCE()} on \clnref{P2:ld}.
In this case, the fact that the \co{exists} clause can trigger means that
the cycle is said to be \emph{allowed}.
In contrast, in cases where the \co{exists} clause cannot trigger,
the cycle is said to be \emph{prohibited}.
\end{fcvref}

\begin{listing}
\input{CodeSamples/formal/litmus/C-W+RWC+o-mb-o+a-o+o-mb-o=whole.fcv}
\caption{W+WRC Litmus Test With More Barriers}
\label{lst:memorder:W+WRC Litmus Test With More Barriers}
\end{listing}

\begin{fcvref}[ln:formal:C-W+RWC+o-r+a-o+o-mb-o:whole]
But what if we need to prohibit the cycle corresponding to the \co{exists}
clause on \clnref{exists} of
\cref{lst:memorder:W+RWC Litmus Test With Release (No Ordering)}?
One solution is to replace \co{P0()}'s \co{smp_store_release()}
with an \co{smp_mb()}, which
\cref{tab:memorder:Linux-Kernel Memory-Ordering Cheat Sheet}
shows to have not only cumulativity, but also propagation.
\end{fcvref}
The result is shown in
\cref{lst:memorder:W+WRC Litmus Test With More Barriers}
(\path{C-W+RWC+o-mb-o+a-o+o-mb-o.litmus}).

\QuickQuiz{
	\begin{fcvref}[ln:formal:C-W+RWC+o-r+a-o+o-mb-o:whole]
	But given that \co{smp_mb()} has the propagation property,
	why doesn't the \co{smp_mb()} on \clnref{P2:mb} of
	\cref{lst:memorder:W+RWC Litmus Test With Release (No Ordering)}
	prevent the \co{exists} clause from triggering?
	\end{fcvref}
}\QuickQuizAnswer{
	\begin{fcvref}[ln:formal:C-W+RWC+o-r+a-o+o-mb-o:whole]
	As a rough rule of thumb, the \co{smp_mb()} barrier's
	propagation property is sufficient to maintain ordering
	through only one load-to-store link between
	processes.
	Unfortunately,
	\cref{lst:memorder:W+RWC Litmus Test With Release (No Ordering)}
	has not one but two load-to-store links, with the
	first being from the \co{READ_ONCE()} on \clnref{P1:ld} to the
	\co{WRITE_ONCE()} on \clnref{P2:st} and the second being from
	the \co{READ_ONCE()} on \clnref{P2:ld} to the \co{WRITE_ONCE()}
	on \clnref{P0:st}.
	Therefore, preventing the \co{exists} clause from triggering
	should be expected to require not one but two
	instances of \co{smp_mb()}.
	\end{fcvref}

	As a special exception to this rule of thumb, a release-acquire
	chain can have one load-to-store link between processes
	and still prohibit the cycle.
}\QuickQuizEnd

\begin{figure}
\centering
\resizebox{\twocolumnwidth}{!}{\includegraphics{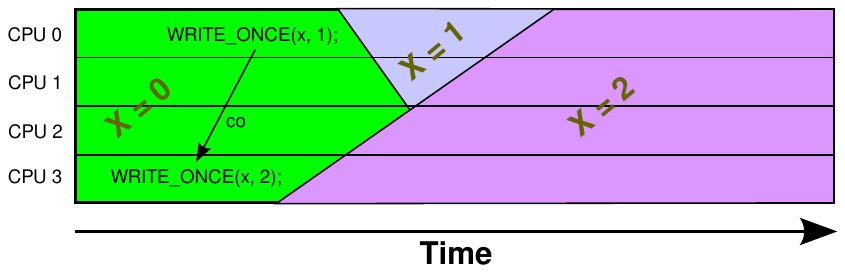}}
\caption{Store-to-Store is Counter-Temporal}
\label{fig:memorder:Store-to-Store is Counter-Temporal}
\end{figure}

For completeness,
\cref{fig:memorder:Store-to-Store is Counter-Temporal}
shows that the ``winning'' store among a group of stores to the
same variable is not necessarily the store that started last.
This should not come as a surprise to anyone who carefully examined
\cref{fig:memorder:A Variable With More Simultaneous Values}
on
\cpageref{fig:memorder:A Variable With More Simultaneous Values}.
One way to rationalize the counter-temporal properties of both
load-to-store and store-to-store ordering is to clearly distinguish
between the temporal order in which the store instructions executed on
the one hand, and the order in which the corresponding cacheline visited
the CPUs that executed those instructions on the other.
It is the cacheline-visitation order that defines the externally
visible ordering of the actual stores.
This cacheline-visitation order is not directly visible to the code
executing the store instructions, which results in the counter-intuitive
counter-temporal nature of load-to-store and store-to-store ordering.\footnote{
	In some hardware-multithreaded systems, the store would become
	visible to other CPUs in that same core as soon as the store
	reached the shared store buffer.
	As a result, such systems are non-multicopy atomic.}

\begin{listing}
\input{CodeSamples/formal/litmus/C-2+2W+o-wmb-o+o-wmb-o=whole.fcv}
\caption{2+2W Litmus Test With Write Barriers}
\label{lst:memorder:2+2W Litmus Test With Write Barriers}
\end{listing}

\QuickQuiz{
	But for litmus tests having only ordered stores, as shown in
	\cref{lst:memorder:2+2W Litmus Test With Write Barriers}
	(\path{C-2+2W+o-wmb-o+o-wmb-o.litmus}),
	research shows that the cycle is prohibited, even in weakly
	ordered systems such as \ARM\ and Power~\cite{test6-pdf}.
	Given that, are store-to-store really \emph{always}
	counter-temporal???
}\QuickQuizAnswer{
	This litmus test is indeed a very interesting curiosity.
	Its ordering apparently occurs naturally given typical
	weakly ordered hardware design, which would normally be
	considered a great gift from the relevant laws of physics
	and cache-coherency-protocol mathematics.

	\begin{fcvref}[ln:formal:C-2+2W+o-wmb-o+o-wmb-o:whole]
	Unfortunately, no one has been able to come up with a software use
	case for this gift that does not have a much better alternative
	implementation.
	Therefore, neither the C11 nor the Linux kernel memory models
	provide any guarantee corresponding to
	\cref{lst:memorder:2+2W Litmus Test With Write Barriers}.
	This means that the \co{exists} clause on \clnref{exists} can
	trigger.
	\end{fcvref}

\begin{listing}
\input{CodeSamples/formal/litmus/C-2+2W+o-o+o-o=whole.fcv}
\caption{2+2W Litmus Test (No Ordering)}
\label{lst:memorder:2+2W Litmus Test (No Ordering)}
\end{listing}

	Of course, without the barrier, there are no ordering
	guarantees, even on real weakly ordered hardware, as shown in
	\cref{lst:memorder:2+2W Litmus Test (No Ordering)}
	(\path{C-2+2W+o-o+o-o.litmus}).
}\QuickQuizEnd

But sometimes time really is on our side.
Read on!

\subsubsection{Happens-Before}
\label{sec:memorder:Happens-Before}

\begin{figure}
\centering
\resizebox{\twocolumnwidth}{!}{\includegraphics{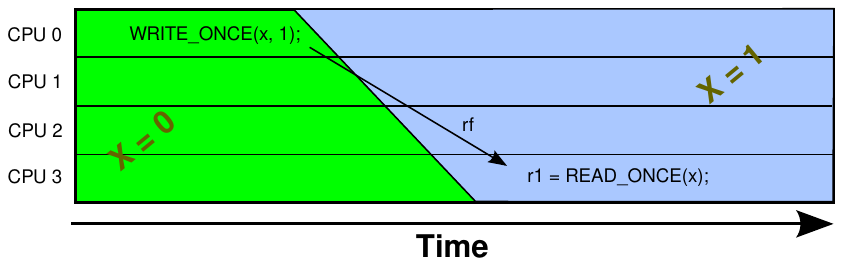}}
\caption{Store-to-Load is Temporal}
\label{fig:memorder:Store-to-Load is Temporal}
\end{figure}

As shown in
\cref{fig:memorder:Store-to-Load is Temporal},
on platforms without user-visible speculation, if a load returns the value
from a particular store, then, courtesy of the finite speed of light and
the non-zero size of modern computing systems, the store absolutely has
to have executed at an earlier time than did the load.
This means that carefully constructed programs can rely on the
passage of time itself as a memory-ordering operation.

\begin{listing}
\input{CodeSamples/formal/litmus/C-LB+a-o+o-data-o+o-data-o=whole.fcv}
\caption{LB Litmus Test With One Acquire}
\label{lst:memorder:LB Litmus Test With One Acquire}
\end{listing}

Of course, just the passage of time by itself is not enough, as
was seen in
\cref{lst:memorder:Load-Buffering Litmus Test (No Ordering)}
on
\cpageref{lst:memorder:Load-Buffering Litmus Test (No Ordering)},
which has nothing but store-to-load links and, because it provides
absolutely no ordering, still can trigger its \co{exists} clause.
However, as long as each thread provides even the weakest possible
ordering, \co{exists} clause would not be able to trigger.
For example,
\cref{lst:memorder:LB Litmus Test With One Acquire}
(\path{C-LB+a-o+o-data-o+o-data-o.litmus})
shows \co{P0()} ordered with an \co{smp_load_acquire()} and
both \co{P1()} and \co{P2()} ordered with data dependencies.
These orderings, which are close to the top of
\cref{tab:memorder:Linux-Kernel Memory-Ordering Cheat Sheet},
suffice to prevent the \co{exists} clause from triggering.

\QuickQuiz{
	Can you construct a litmus test like that in
	\cref{lst:memorder:LB Litmus Test With One Acquire}
	that uses \emph{only} dependencies?
}\QuickQuizAnswer{
	\Cref{lst:memorder:LB Litmus Test With No Acquires}
	shows a somewhat nonsensical but very real example.
	Creating a more useful (but still real) litmus test is left
	as an exercise for the reader.

\begin{listing}
\input{CodeSamples/formal/litmus/C-LB+o-data-o+o-data-o+o-data-o=whole.fcv}
\caption{LB Litmus Test With No Acquires}
\label{lst:memorder:LB Litmus Test With No Acquires}
\end{listing}
}\QuickQuizEnd

An important use of time for ordering memory accesses is covered in the
next section.

\subsubsection{Release-Acquire Chains}
\label{sec:memorder:Release-Acquire Chains}

A minimal release-acquire chain was shown in
\cref{lst:memorder:Enforcing Ordering of Load-Buffering Litmus Test}
on
\cpageref{lst:memorder:Enforcing Ordering of Load-Buffering Litmus Test},
but these chains can be much longer, as shown in
\cref{lst:memorder:Long LB Release-Acquire Chain}
(\path{C-LB+a-r+a-r+a-r+a-r.litmus}).
The longer the release-acquire chain, the more ordering is gained
from the passage of time, so that no matter how many threads are
involved, the corresponding \co{exists} clause cannot trigger.

\begin{listing}
\input{CodeSamples/formal/litmus/C-LB+a-r+a-r+a-r+a-r=whole.fcv}
\caption{Long LB Release-Acquire Chain}
\label{lst:memorder:Long LB Release-Acquire Chain}
\end{listing}

Although release-acquire chains are inherently store-to-load creatures,
it turns out that they can tolerate one load-to-store step, despite
such steps being counter-temporal, as shown in
\cref{fig:memorder:Load-to-Store is Counter-Temporal}
on
\cpageref{fig:memorder:Load-to-Store is Counter-Temporal}.
For example,
\cref{lst:memorder:Long ISA2 Release-Acquire Chain}
(\path{C-ISA2+o-r+a-r+a-r+a-o.litmus})
shows a three-step release-acquire chain, but where \co{P3()}'s
final access is a \co{READ_ONCE()} from \co{x0}, which is
accessed via \co{WRITE_ONCE()} by \co{P0()}, forming a non-temporal
load-to-store link between these two processes.
\begin{fcvref}[ln:formal:litmus:C-ISA2+o-r+a-r+a-r+a-o:whole]
However, because \co{P0()}'s \co{smp_store_release()} (\clnref{P0:rel})
is cumulative, if \co{P3()}'s \co{READ_ONCE()} returns zero,
this cumulativity will force the \co{READ_ONCE()} to be ordered
before \co{P0()}'s \co{smp_store_release()}.
In addition, the release-acquire chain
(\clnref{P0:rel,P1:acq,P1:rel,P2:acq,P2:rel,P3:acq})
forces \co{P3()}'s \co{READ_ONCE()} to be ordered after \co{P0()}'s
\co{smp_store_release()}.
Because \co{P3()}'s \co{READ_ONCE()} cannot be both before and after
\co{P0()}'s \co{smp_store_release()}, either or both of two things must
be true:
\end{fcvref}

\begin{listing}
\input{CodeSamples/formal/litmus/C-ISA2+o-r+a-r+a-r+a-o=whole.fcv}
\caption{Long ISA2 Release-Acquire Chain}
\label{lst:memorder:Long ISA2 Release-Acquire Chain}
\end{listing}

\begin{enumerate}
\item	\co{P3()}'s \co{READ_ONCE()} came after \co{P0()}'s
	\co{WRITE_ONCE()}, so that the \co{READ_ONCE()} returned
	the value two, so that the \co{exists} clause's \co{3:r2=0}
	is false.
\item	The release-acquire chain did not form, that is, one or more
	of the \co{exists} clause's \co{1:r2=2}, \co{2:r2=2}, or \co{3:r1=2}
	is false.
\end{enumerate}

Either way, the \co{exists} clause cannot trigger, despite this litmus
test containing a notorious load-to-store link between
\co{P3()} and \co{P0()}.
But never forget that release-acquire chains can tolerate only one
load-to-store link, as was seen in
\cref{lst:memorder:W+RWC Litmus Test With Release (No Ordering)}.

\begin{listing}
\input{CodeSamples/formal/litmus/C-Z6.2+o-r+a-r+a-r+a-o=whole.fcv}
\caption{Long Z6.2 Release-Acquire Chain}
\label{lst:memorder:Long Z6.2 Release-Acquire Chain}
\end{listing}

Release-acquire chains can also tolerate a single store-to-store step,
as shown in
\cref{lst:memorder:Long Z6.2 Release-Acquire Chain}
(\path{C-Z6.2+o-r+a-r+a-r+a-o.litmus}).
\begin{fcvref}[ln:formal:C-Z6.2+o-r+a-r+a-r+a-o:whole]
As with the previous example, \co{smp_store_release()}'s cumulativity
combined with the temporal nature of the release-acquire chain
prevents the \co{exists} clause on \clnref{exists} from triggering.
\end{fcvref}

\begin{listing}
\input{CodeSamples/formal/litmus/C-Z6.2+o-r+a-o+o-mb-o=whole.fcv}
\caption{Z6.2 Release-Acquire Chain (Ordering?)}
\label{lst:memorder:Z6.2 Release-Acquire Chain (Ordering?)}
\end{listing}

\QuickQuiz{
	Suppose we have a short release-acquire chain along with one
	load-to-store link and one store-to-store link, like that shown in
	\cref{lst:memorder:Z6.2 Release-Acquire Chain (Ordering?)}.
	Given that there is only one of each type of non-store-to-load
	link, the \co{exists} cannot trigger, right?
}\QuickQuizAnswer{
	Wrong.
	It is the number of non-store-to-load links that matters.
	If there is only one non-store-to-load link, a release-acquire
	chain can prevent the \co{exists} clause from triggering.
	However, if there is more than one non-store-to-load link,
	be they store-to-store, load-to-store, or any combination
	thereof, it is necessary to have at least one full barrier
	(\co{smp_mb()} or better) between each non-store-to-load link.
	In
	\cref{lst:memorder:Z6.2 Release-Acquire Chain (Ordering?)},
	preventing the \co{exists} clause from triggering therefore requires
	an additional full barrier between either \co{P0()}'s or
	\co{P1()}'s accesses.
}\QuickQuizEnd

\begin{listing}
\input{CodeSamples/formal/litmus/C-MP+o-r+a-o=whole.fcv}
\caption{A Release-Acquire Chain Ordering Multiple Accesses}
\label{lst:memorder:A Release-Acquire Chain Ordering Multiple Accesses}
\end{listing}

\begin{listing}
\input{CodeSamples/formal/litmus/C-MPO+o-r+a-o+o=whole.fcv}
\caption{A Release-Acquire Chain With Added Store (Ordering?)}
\label{lst:memorder:A Release-Acquire Chain With Added Store (Ordering?)}
\end{listing}

But beware:
Adding a second store-to-store link allows the correspondingly updated
\co{exists} clause to trigger.
To see this, review \cref{lst:memorder:A Release-Acquire Chain Ordering Multiple Accesses,lst:memorder:A Release-Acquire Chain With Added Store (Ordering?)},
which have identical \co{P0()} and \co{P1()} processes.
The only code difference is that
\cref{lst:memorder:A Release-Acquire Chain With Added Store (Ordering?)}
has an additional \co{P2()} that does an \co{smp_store_release()} to
the \co{x2} variable that \co{P0()} releases and \co{P1()} acquires.
The \co{exists} clause is also adjusted to exclude executions in which
\co{P2()}'s \co{smp_store_release()} precedes that of \co{P0()}.

Running the litmus test in
\cref{lst:memorder:A Release-Acquire Chain With Added Store (Ordering?)}
shows that the addition of \co{P2()} can totally destroy the
ordering from the release-acquire chain.
Therefore, when constructing release-acquire chains, please take care
to construct them properly.

\QuickQuiz{
	There are store-to-load links, load-to-store links, and
	store-to-store links.
	But what about load-to-load links?
}\QuickQuizAnswer{
	The problem with the concept of load-to-load links is that
	if the two loads from the same variable return the same
	value, there is no way to determine their ordering.
	The only way to determine their ordering is if they return
	different values, in which case there had to have been an
	intervening store.
	And that intervening store means that there is no load-to-load
	link, but rather a load-to-store link followed by a
	store-to-load link.
}\QuickQuizEnd

In short, properly constructed release-acquire chains form a peaceful
island of intuitive bliss surrounded by a strongly counter-intuitive
sea of more complex memory-ordering constraints.

\subsection{A Counter-Intuitive Case Study}
\label{sec:memorder:A Counter-Intuitive Case Study}

This section will revisit
\cref{lst:memorder:R Litmus Test With Write Memory Barrier (No Ordering)}
on \cpageref{lst:memorder:R Litmus Test With Write Memory Barrier (No Ordering)},
which was presented in the answer to
\QuickQuizARef{\MemorderQQLitmusTestR}.
This litmus test has only two threads, with the stores in \co{P0()}
being ordered by \co{smp_wmb()} and the accesses in \co{P1()} being
ordered by \co{smp_mb()}.
Despite this litmus test's small size and heavy ordering, the
counter-intuitive outcome shown in the \co{exists} clause is in fact
allowed.

One way to look at this was presented in the answer to
\QuickQuizARef{\MemorderQQLitmusTestR}, namely that the link from
\co{P0()} to \co{P1()} is a store-to-store link, and that back
from \co{P1()} to \co{P0()} is a store-to-store link.
Both links are counter-temporal, thus requiring full memory barriers
in both processes.
Revisiting
\cref{fig:memorder:Store-to-Store is Counter-Temporal,fig:memorder:Store-to-Load is Temporal}
shows that these counter-temporal links give the hardware considerable
latitude.

But that raises the question of exactly how hardware would go about using
this latitude to satisfy the \co{exists} clause in
\cref{lst:memorder:R Litmus Test With Write Memory Barrier (No Ordering)}.
There is no known ``toy'' hardware implementation that can do this, so
let us instead study the sequence of steps that the PowerPC architecture
goes through to make this happen.

The first step in this study is to translate
\cref{lst:memorder:R Litmus Test With Write Memory Barrier (No Ordering)}
to a PowerPC assembly language litmus test
(\cref{sec:formal:Anatomy of a Litmus Test} on
\cpageref{sec:formal:Anatomy of a Litmus Test}):

\begin{fcvlabel}[ln:memorder:inline:ppcasmr]
\begin{VerbatimN}[samepage=true,commandchars=\@\[\]]
PPC R+lwsync+sync
{
0:r1=1; 0:r2=x; 0:r4=y;		@lnlbl[init0]
1:r1=2; 1:r2=y; 1:r4=x;		@lnlbl[init1]
}
 P0           | P1           ;	@lnlbl[procs]
 stw r1,0(r2) | stw r1,0(r2) ;	@lnlbl[stores]
 lwsync       | sync         ;	@lnlbl[barriers]
 stw r1,0(r4) | lwz r3,0(r4) ;	@lnlbl[storeload]
exists (y=2 /\ 1:r3=0)		@lnlbl[exists]
\end{VerbatimN}
\end{fcvlabel}

\begin{fcvref}[ln:memorder:inline:ppcasmr]
The first line identifies the type of test (\co{PPC}) and gives
the test's name.
\Clnref{init0,init1} initialize \co{P0()}'s and \co{P1()}'s registers,
respectively.
\Clnrefrange{procs}{storeload} show the PowerPC assembly
statements corresponding to the C code from
\cref{lst:memorder:R Litmus Test With Write Memory Barrier (No Ordering)},
with the first column being the code for \co{P0()} and the second column
being the code for \co{P1()}.
\Clnref{stores} shows the initial \co{WRITE_ONCE()} calls in both columns;
the columns of \clnref{barriers} show the \co{smp_wmb()} and \co{smp_mb()}
for \co{P0()} and \co{P1()}, respectively;
the columns of \clnref{storeload} shows \co{P0()}'s \co{WRITE_ONCE()} and
\co{P1()}'s \co{READ_ONCE()}, respectively;
and finally \clnref{exists} shows the \co{exists} clause.
\end{fcvref}

In order for this \co{exists} clause to be satisfied, \co{P0()}'s
\co{stw} to \co{y} must precede that of \co{P1()}, but \co{P1()}'s
later \co{lwz} from \co{x} must precede \co{P0()}'s \co{stw} to \co{x}.
Seeing how this can happen requires a rough understanding of the
following PowerPC terminology.

\begin{description}[style=nextline]

\item[Instruction commit:]
This can be thought of as the execution of that instruction as opposed
to the memory-system consequences of having executed that instruction.

\item[Write reaching coherence point:]
This can be thought of as the value written being deposited into the
corresponding cache line.

\item[Partial coherence commit:]
This can be thought of as the system having worked out the order in which
a pair of values written will be deposited into the corresponding cache
line, but potentially well before that cache line arrives.
Some might argue that the data in
\cref{fig:memorder:A Variable With More Simultaneous Values}
suggests that real PowerPC hardware does in fact use partial coherence
commits to handle concurrent stores by multiple hardware threads within
a single core.

\item[Write propagate to thread:]
This occurs when a second hardware thread becomes aware of the first
hardware thread's write.
The time at which a write propagates to a given thread might not have
any relation to cache-line movement.
For example, if a pair of threads share a store buffer, they might see
each others' writes long before the cache line gets involved.
On the other hand, if a pair of hardware threads are widely separated,
the first thread's write's value might have been deposited into the
corresponding cache line long before the second thread learns of that
write.

\item[Barrier propagate to thread:]
Hardware threads make each other aware of memory-barrier instructions
as needed by propagating them to each other.

\item[Acknowledge \tco{sync}:]
The PowerPC \co{sync} instruction implements the Linux kernel's
\co{smp_mb()} full barrier.
And one reason that the \co{sync} instruction provides such strong
ordering is that each \co{sync} is not only propagated to other hardware
threads, but these other threads must also acknowledge each \co{sync}.
This two-way communication allows the hardware threads to cooperate
to produce the required strong global ordering.

\end{description}

\begin{figure*}[tbp]
\centering
\resizebox{\textwidth}{!}{\includegraphics{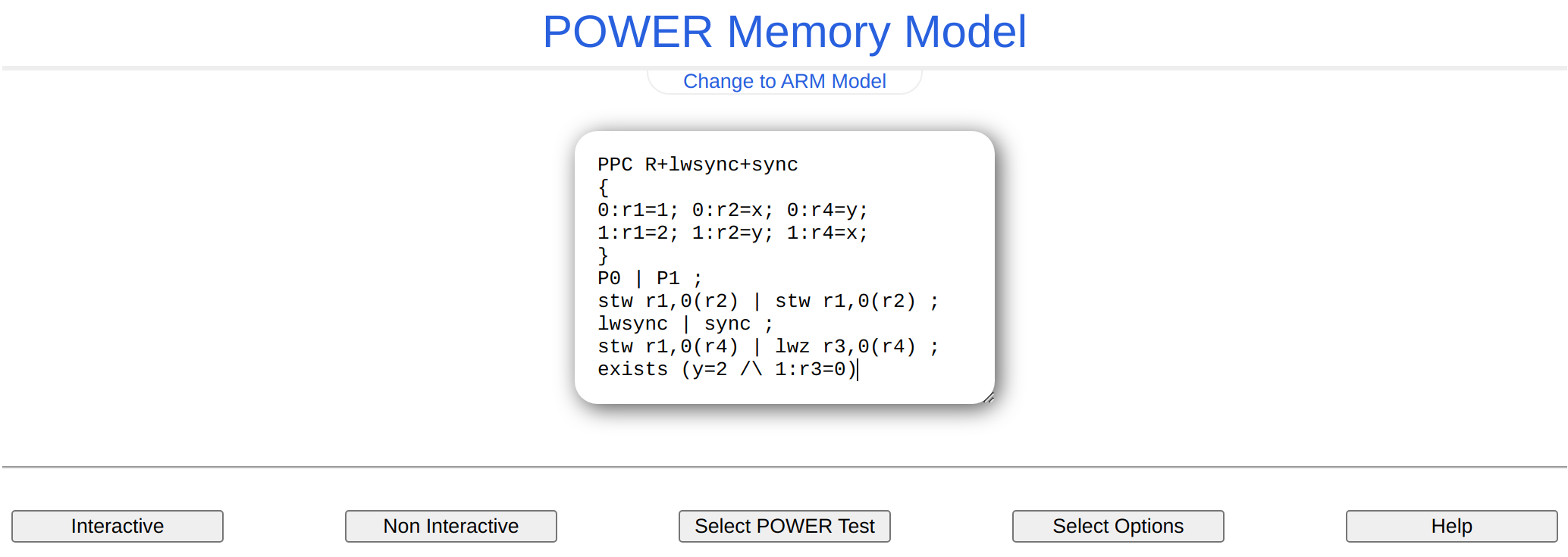}}
\caption{PPCMEM Initial R State}
\label{fig:memorder:PPCMEM Initial R State}
\end{figure*}

\begin{figure*}[tbp]
\centering
\resizebox{\textwidth}{!}{\includegraphics{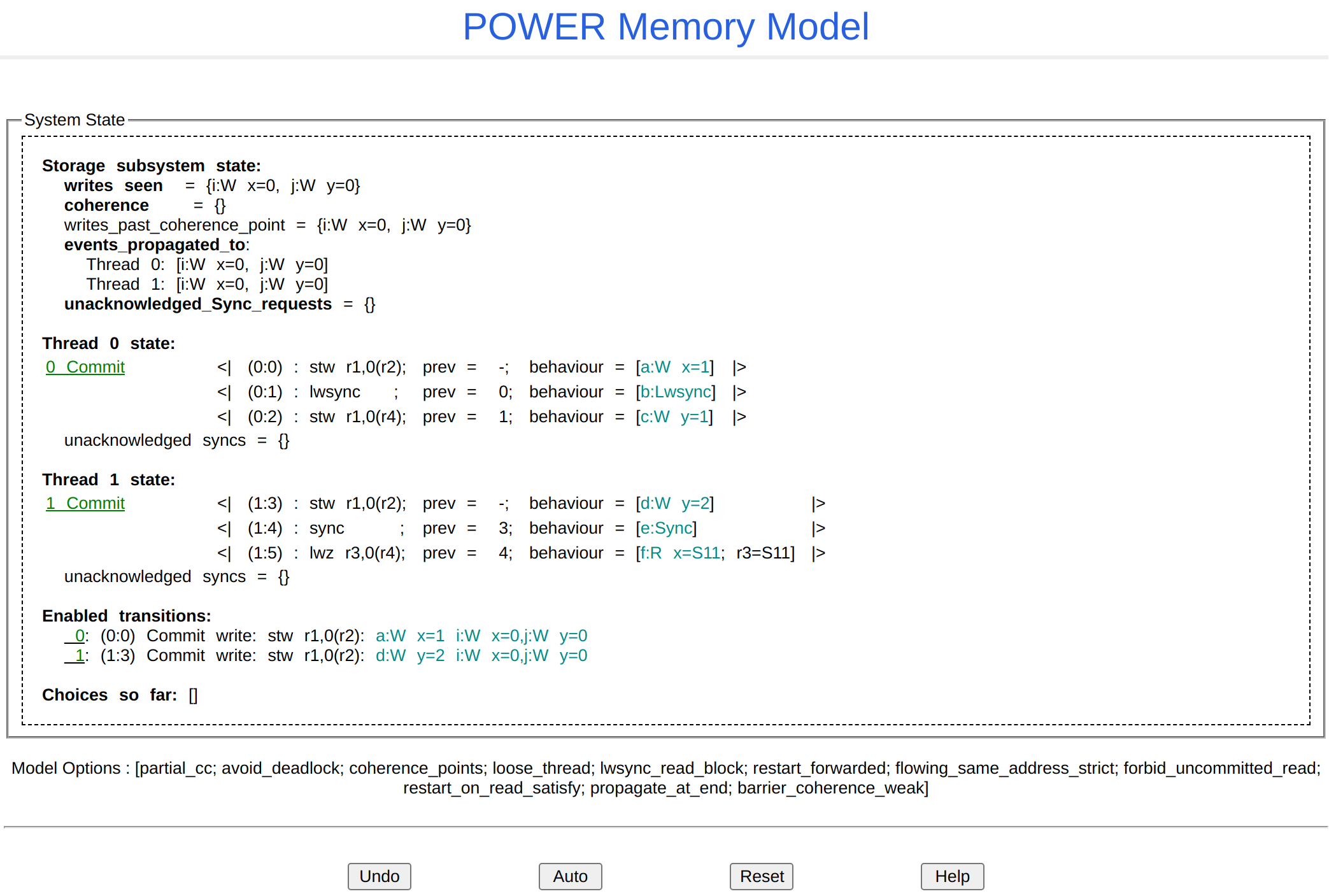}}
\caption{PPCMEM First R Step}
\label{fig:memorder:PPCMEM First R Step}
\end{figure*}

We are now ready to step through the PowerPC sequence of events that
satisfies the above \co{exists} clause.

To best understand this, please follow along at
\url{https://www.cl.cam.ac.uk/~pes20/ppcmem/index.html},
carefully copying the above assembly-language litmus test into the pane.
The result should look as shown in
\cref{fig:memorder:PPCMEM Initial R State}, give or take space characters.
Click on the ``Interactive'' button in the lower left, which, after a
short delay, should produce a display as shown in
\cref{fig:memorder:PPCMEM First R Step}.
If the ``Interactive'' button refuses to do anything, this usually means
that there is a syntax error, for example, a spurious newline character
might have been introduced during the copy-paste operation.

This display has one clickable link in each section displaying thread
state, and as the ``Commit'' in each link suggests, these links commit
each thread's first \co{stw} instruction.
If you prefer, you can instead click on the corresponding links listed
under ``Enabled transitions'' near the bottom of the screen.
Note well that some of the later memory-system transitions will appear
in the upper ``Storage subsystem state'' section of this display.

The following sequence of clicks demonstrates how the \co{exists} clause
can be satisfied:

\begin{enumerate}
\item	Commit \co{P0()}'s first \co{stw} instruction (to \co{x}).
\item	Commit \co{P1()}'s \co{stw} instruction.
\item	Commit \co{P0()}'s \co{lwsync} instruction.
\item	Commit \co{P0()}'s second \co{stw} instruction (to \co{y}).
\item	Commit \co{P1()}'s \co{sync} instruction.
\item	At this point, there should be no clickable links in either of
	the two sections displaying thread state, but there should be
	quite a few of them up in the ``Storage subsystem state''.
	The following steps tell you which of them to click on.
\item	\co{Partial coherence commit: c:W y=1 -> d:W y=2}.
	This commits the system to processing \co{P0()}'s store to
	\co{y} before \co{P1()}'s store even though neither store
	has reached either the coherence point or any other thread.
	One might imagine partial coherence commits happening within a
	store buffer that is shared by multiple hardware threads
	that are writing to the same variable.
\item	\co{Write propagate to thread: d:W y=2 to Thread 0}.
	This is necessary to allow \co{P1()}'s \co{sync} instruction
	to propagate to \co{P0()}.
\item	\co{Barrier propagate to thread: e:Sync  to Thread 0}.
\item	\co{Write reaching coherence point: a:W x=1}.
\item	\co{Write reaching coherence point: c:W y=1}.
\item	\co{Write reaching coherence point: d:W y=2}.
	These three operations were required in order to allow \co{P0()}
	to acknowledge \co{P1()}'s \co{sync} instruction.
\item	\co{Acknowledge sync: Sync e:Sync}.
\item	Back down in thread \co{P1()}'s state, click on \co{Read i:W
	x=0}, which loads the value zero, thus satisfying the \co{exists}
	clause.
	All that remains is cleanup, which can be carried out in any order.
\item	Commit \co{P1()}'s \co{lwz} instruction.
\item	\co{Write propagate to thread: a:W x=1 to Thread 1}.
\item	\co{Barrier propagate to thread: b:Lwsync  to Thread 1}.
\end{enumerate}

\begin{figure*}[tbp]
\centering
\resizebox{\textwidth}{!}{\includegraphics{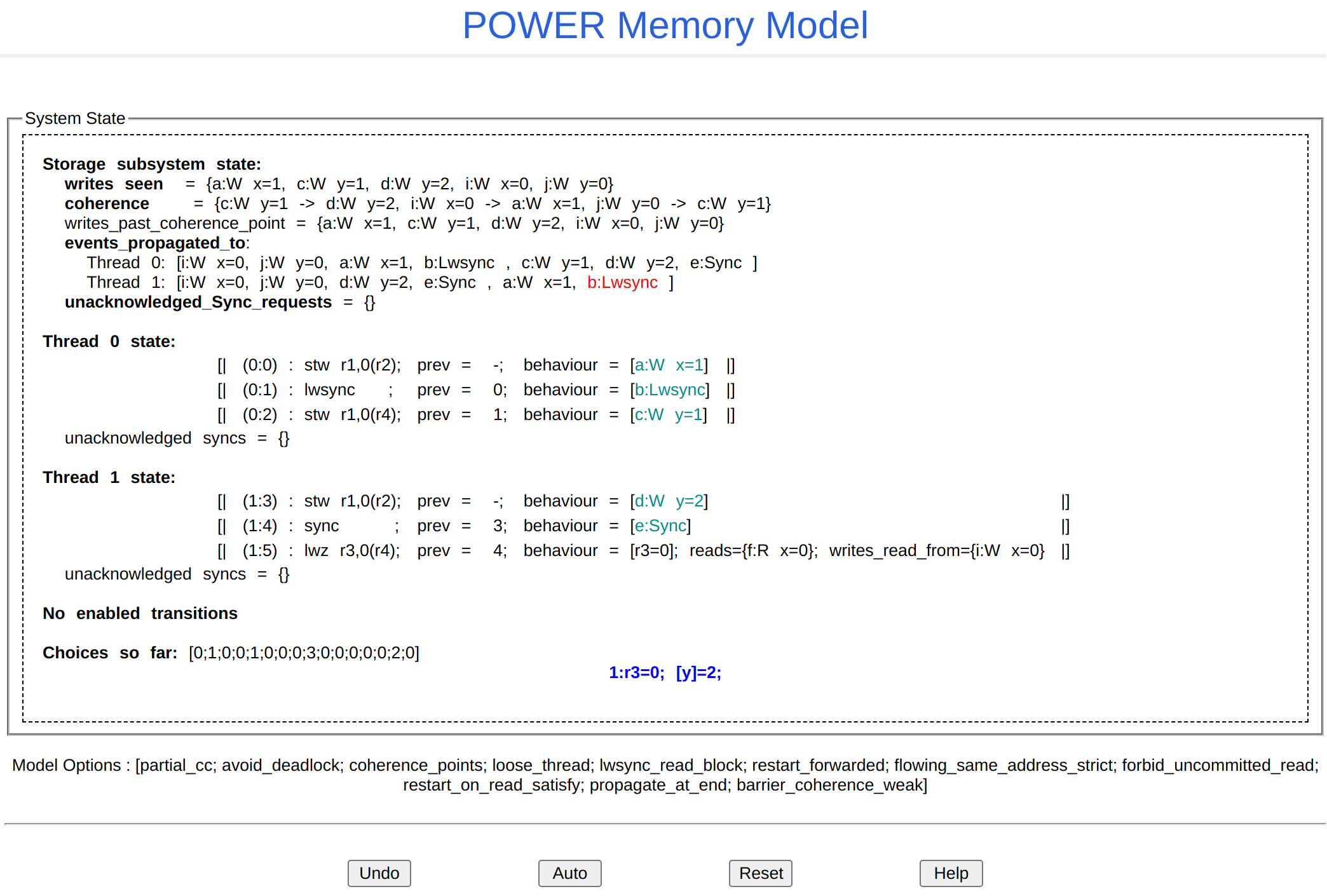}}
\caption{PPCMEM Final R State}
\label{fig:memorder:PPCMEM Final R State}
\end{figure*}

At this point, you should see something like
\cref{fig:memorder:PPCMEM Final R State}.
Note that the satisified \co{exists} clause is shown in blue near the
bottom, confirming that this counter-intuitive really can happen.
If you wish, you can click on ``Undo'' to explore other options or
click on ``Reset'' to start over.
It can be very helpful to carry out these steps in different orders
to better understand how a non-multicopy-atomic architecture operates.

\QuickQuiz{
	What happens if that \co{lwsync} instruction is instead a
	\co{sync} instruction?
}\QuickQuizAnswer{
	The counter-intuitive outcome cannot happen.
	(Try it!)
}\QuickQuizEnd

Although a full understanding of how this counter-intuitive outcome
happens would require hardware details that are beyond the scope of
this book, this exercise should provide some helpful intuitions.
Or perhaps more accurately, destroy some counter-productive intuitions.

% @@@ Exercises?
% @@@ Hardware details from Appendix?

\section{Compile-Time Consternation}
\label{sec:memorder:Compile-Time Consternation}
\epigraph{Science increases our power in proportion as it lowers our pride.}
	 {Claude Bernard}

Most languages, including C, were developed on uniprocessor systems
by people with little or no parallel-programming experience.
As a result, unless explicitly told otherwise, these languages assume
that the current CPU is the only thing that is reading or writing memory.
This in turn means that these languages' compilers' optimizers
are ready, willing, and oh so able to make dramatic changes to the
order, number, and sizes of memory references that your program
executes.
In fact, the reordering carried out by hardware can seem quite tame
by comparison.

This section will help you tame your compiler, thus avoiding a great
deal of compile-time consternation.
\Cref{sec:memorder:Memory-Reference Restrictions}
describes how to keep the compiler from destructively optimizing
your code's memory references,
\cref{sec:memorder:Address- and Data-Dependency Difficulties}
describes how to protect address and data dependencies,
and finally,
\cref{sec:memorder:Control-Dependency Calamities}
describes how to protect those delicate control dependencies.

\subsection{Memory-Reference Restrictions}
\label{sec:memorder:Memory-Reference Restrictions}

As noted in \cref{sec:toolsoftrade:Accessing Shared Variables},
unless told otherwise, compilers assume that nothing else
is affecting the variables that the code is accessing.
Furthermore, this assumption is not simply some design error, but is
instead enshrined in various standards.\footnote{
	Or perhaps it is a standardized design error.}
It is worth summarizing this material in preparation for the following
sections.

\IXplx{Plain access}{es}, as in plain-access C-language assignment statements such
as \qco{r1 = a} or \qco{b = 1} are subject to the
shared-variable shenanigans described in
\cref{sec:toolsoftrade:Shared-Variable Shenanigans}.
Ways of avoiding these shenanigans are described in
\crefrange{sec:toolsoftrade:A Volatile Solution}{sec:toolsoftrade:Avoiding Data Races}
starting on
\cpageref{sec:toolsoftrade:A Volatile Solution}:

\begin{enumerate}
\item	Plain accesses can tear, for example, the compiler could choose
	to access an eight-byte pointer one byte at a time.
	Tearing of aligned machine-sized accesses can be prevented by
	using \co{READ_ONCE()} and \co{WRITE_ONCE()}.
\item	Plain loads can fuse, for example, if the results of an earlier
	load from that same object are still in a machine register,
	the compiler might opt to reuse the value in that register
	instead of reloading from memory.
	Load fusing can be prevented by using \co{READ_ONCE()} or by
	enforcing ordering between the two loads using \co{barrier()},
	\co{smp_rmb()}, and other means shown in
	\cref{tab:memorder:Linux-Kernel Memory-Ordering Cheat Sheet}.
\item	Plain stores can fuse, so that a store can be omitted entirely
	if there is a later store to that same variable.
	Store fusing can be prevented by using \co{WRITE_ONCE()} or by
	enforcing ordering between the two stores using \co{barrier()},
	\co{smp_wmb()}, and other means shown in
	\cref{tab:memorder:Linux-Kernel Memory-Ordering Cheat Sheet}.
\item	Plain accesses can be reordered in surprising ways by modern
	optimizing compilers.
	This reordering can be prevented by enforcing ordering as
	called out above.
\item	Plain loads can be invented, for example, register pressure might
	cause the compiler to discard a previously loaded value from
	its register, and then reload it later on.
	Invented loads can be prevented by using \co{READ_ONCE()} or by
	enforcing ordering as called out above between the load and a
	later use of its value using \co{barrier()}.
\item	Stores can be invented before a plain store, for example, by
	using the stored-to location as temporary storage.
	This can be prevented by use of \co{WRITE_ONCE()}.
\item	Stores can be transformed into a load-check-store sequence,
	which can defeat control dependencies.
	This can be prevented by use of \co{smp_load_acquire()}.
\end{enumerate}

\QuickQuiz{
	Why not place a \co{barrier()} call immediately before
	a plain store to prevent the compiler from inventing stores?
}\QuickQuizAnswer{
	Because it would not work.
	Although the compiler would be prevented from inventing a
	store prior to the \co{barrier()}, nothing would prevent
	it from inventing a store between that \co{barrier()} and
	the plain store.
}\QuickQuizEnd

Please note that all of these shared-memory shenanigans can instead be
avoided by avoiding \IXpl{data race} on plain accesses, as described in
\cref{sec:toolsoftrade:Avoiding Data Races}.
After all, if there are no data races, then each and every one of the
compiler optimizations mentioned above is perfectly safe.
But for code containing data races, this list is subject to change
without notice as compiler optimizations continue becoming increasingly
aggressive.

In short, use of \co{READ_ONCE()}, \co{WRITE_ONCE()}, \co{barrier()},
\co{volatile}, and other primitives called out in
\cref{tab:memorder:Linux-Kernel Memory-Ordering Cheat Sheet}
on
\cpageref{tab:memorder:Linux-Kernel Memory-Ordering Cheat Sheet}
are valuable tools in preventing the compiler from
optimizing your parallel algorithm out of existence.
Compilers are starting to provide other mechanisms for avoiding
load and store tearing, for example, \co{memory_order_relaxed}
atomic loads and stores, however, work is still
needed~\cite{JonathanCorbet2016C11atomics}.
In addition, compiler issues aside, \co{volatile} is still needed
to avoid fusing and invention of accesses, including C11 atomic accesses.

Please note that, it is possible to overdo use of \co{READ_ONCE()} and
\co{WRITE_ONCE()}.
For example, if you have prevented a given variable from changing
(perhaps by holding the lock guarding all updates to that
variable), there is no point in using \co{READ_ONCE()}.
Similarly, if you have prevented any other CPUs or threads from
reading a given variable (perhaps because you are initializing
that variable before any other CPU or thread has access to it),
there is no point in using \co{WRITE_ONCE()}.
However, in my experience, developers need to use things like
\co{READ_ONCE()} and \co{WRITE_ONCE()} more often than they think that
they do, and the overhead of unnecessary uses is quite low.
In contrast, the penalty for failing to use them when needed can be quite high.

\subsection{Address- and Data-Dependency Difficulties}
\label{sec:memorder:Address- and Data-Dependency Difficulties}
\OriginallyPublished{Section}{sec:memorder:Address- and Data-Dependency Difficulties}{Address- and Data-Dependency Difficulties}{the Linux kernel}{PaulEMcKenney2014rcu-dereference}

The low overheads of the address and data dependencies discussed in
\cref{sec:memorder:Address Dependencies,sec:memorder:Data Dependencies},
respectively, makes their use extremely attractive.
Unfortunately, compilers do not understand either address or data
dependencies, although there are efforts underway to teach them, or at
the very least, standardize the process of teaching
them~\cite{PaulEMcKennneyConsumeP0190R4,PaulEMcKenney2017markconsumeP0462R1}.
In the meantime, it is necessary to be very careful in order to prevent
your compiler from breaking your dependencies.

\subsubsection{Give your dependency chain a good start}
The load that heads your dependency chain must use proper
ordering, for example \co{rcu_dereference()} or \co{READ_ONCE()}.
Failure to follow this rule can have serious side effects:

\begin{enumerate}
\item	On DEC Alpha, a dependent load might not be ordered with
	the load heading the dependency chain, as described in
	\cref{sec:memorder:Alpha}.
\item	If the load heading the dependency chain is a
	C11 non-volatile \co{memory_order_relaxed} load,
	the compiler could omit the load, for example, by using a value
	that it loaded in the past.
\item	If the load heading the dependency chain is a plain load,
	the compiler can omit the load, again by using a value
	that it loaded in the past.
	Worse yet, it could load twice instead of once, so that
	different parts of your code use different values---and
	compilers really do this, especially when under register
	pressure.
\item	The value loaded by the head of the dependency chain must
	be a pointer.
	In theory, yes, you could load an integer, perhaps to use
	it as an array index.
	In practice, the compiler knows too much about integers,
	and thus has way too many opportunities to break your
	dependency chain~\cite{PaulEMcKennneyConsumeP0190R4}.
\end{enumerate}

\subsubsection{Avoid arithmetic dependency breakage}
Although it is just fine to do some arithmetic operations on a pointer in
your dependency chain, you need to be careful to avoid giving the
compiler too much information.
After all, if the compiler learns enough to determine the exact value
of the pointer, it can use that exact value instead of the pointer itself.
As soon as the compiler does that, the dependency is broken and all
ordering is lost.

\begin{listing}
\begin{fcvlabel}[ln:memorder:Breakable Dependencies With Comparisons]
\begin{VerbatimL}[commandchars=\\\[\]]
int reserve_int;
int *gp;
int *p;

p = rcu_dereference(gp);
if (p == &reserve_int)		\lnlbl[cmp]
	handle_reserve(p);	\lnlbl[handle]
do_something_with(*p); /* buggy! */
\end{VerbatimL}
\end{fcvlabel}
\caption{Breakable Dependencies With Comparisons}
\label{lst:memorder:Breakable Dependencies With Comparisons}
\end{listing}

\begin{listing}
\begin{fcvlabel}[ln:memorder:Broken Dependencies With Comparisons]
\begin{VerbatimL}[commandchars=\\\[\]]
int reserve_int;
int *gp;
int *p;

p = rcu_dereference(gp);	\lnlbl[deref1]
if (p == &reserve_int) {
	handle_reserve(&reserve_int);
	do_something_with(reserve_int); /* buggy! */ \lnlbl[deref2]
} else {
	do_something_with(*p); /* OK! */
}
\end{VerbatimL}
\end{fcvlabel}
\caption{Broken Dependencies With Comparisons}
\label{lst:memorder:Broken Dependencies With Comparisons}
\end{listing}

\begin{enumerate}
\item	Although it is permissible to compute offsets from a
	pointer, these offsets must not result in total cancellation.
	For example, given a \co{char} pointer \co{cp},
	\co{cp-(uintptr_t)cp} will cancel and can allow the compiler
	to break your dependency chain.
	On the other hand, canceling offset values with each other
	is perfectly safe and legal.
	For example, if \co{a} and \co{b} are equal, \co{cp+a-b}
	is an identity function, including preserving the dependency.
\item	Comparisons can break dependencies.
	\Cref{lst:memorder:Breakable Dependencies With Comparisons}
	shows how this can happen.
	Here global pointer \co{gp} points to a dynamically allocated
	integer, but if memory is low, it might instead point to
	the \co{reserve_int} variable.
	\begin{fcvref}[ln:memorder:Breakable Dependencies With Comparisons]
	This \co{reserve_int} case might need special handling, as
	shown on \clnref{cmp,handle} of the listing.
	\end{fcvref}
	\begin{fcvref}[ln:memorder:Broken Dependencies With Comparisons]
	But the compiler could reasonably transform this code into
	the form shown in
	\cref{lst:memorder:Broken Dependencies With Comparisons},
	especially on systems where instructions with absolute
	addresses run faster than instructions using addresses
	supplied in registers.
	However, there is clearly no ordering between the pointer
	load on \clnref{deref1} and the dereference on \clnref{deref2}.
	Please note that this is simply an example:
	There are a great many other ways to break dependency chains
	with comparisons.
	\end{fcvref}
\end{enumerate}

\QuickQuizSeries{%
\QuickQuizB{
	\begin{fcvref}[ln:memorder:Breakable Dependencies With Comparisons]
	Why can't you simply dereference the pointer before comparing it
	to \co{&reserve_int} on \clnref{cmp} of
	\cref{lst:memorder:Breakable Dependencies With Comparisons}?
	\end{fcvref}
}\QuickQuizAnswerB{
	For first, it might be necessary to invoke
	\co{handle_reserve()} before \co{do_something_with()}.

	But more relevant to memory ordering, the compiler is often within
	its rights to hoist the comparison ahead of the dereferences,
	which would allow the compiler to use \co{&reserve_int} instead
	of the variable \co{p} that the hardware has tagged with
	a dependency.
}\QuickQuizEndB
\QuickQuizE{
	But it should be safe to compare two pointer variables, right?
	After all, the compiler doesn't know the value
	of either, so how can it possibly learn anything from the
	comparison?
}\QuickQuizAnswerE{
\begin{listing}
\begin{fcvlabel}[ln:memorder:Breakable Dependencies With Non-Constant Comparisons]
\begin{VerbatimL}
int *gp1;
int *p;
int *q;

p = rcu_dereference(gp1);
q = get_a_pointer();
if (p == q)
	handle_equality(p);
do_something_with(*p);
\end{VerbatimL}
\end{fcvlabel}
\caption{Breakable Dependencies With Non-Constant Comparisons}
\label{lst:memorder:Breakable Dependencies With Non-Constant Comparisons}
\end{listing}%
\begin{listing}
\begin{fcvlabel}[ln:memorder:Broken Dependencies With Non-Constant Comparisons]
\begin{VerbatimL}[commandchars=\\\[\]]
int *gp1;
int *p;
int *q;

p = rcu_dereference(gp1);		\lnlbl[p]
q = get_a_pointer();
if (p == q) {
	handle_equality(q);
	do_something_with(*q);		\lnlbl[q]
} else {
	do_something_with(*p);
}
\end{VerbatimL}
\end{fcvlabel}
\caption{Broken Dependencies With Non-Constant Comparisons}
\label{lst:memorder:Broken Dependencies With Non-Constant Comparisons}
\end{listing}%
	Unfortunately, the compiler really can learn enough to
	break your dependency chain, for example, as shown in
	\cref{lst:memorder:Breakable Dependencies With Non-Constant Comparisons}.
	The compiler is within its rights to transform this code
	into that shown in
	\cref{lst:memorder:Broken Dependencies With Non-Constant Comparisons},
	and might well make this transformation due to register pressure
	if \co{handle_equality()} was inlined and needed a lot of registers.
	\begin{fcvref}[ln:memorder:Broken Dependencies With Non-Constant Comparisons]
	\Clnref{q} of this transformed code uses \co{q}, which although
	equal to \co{p}, is not necessarily tagged by the hardware as
	carrying a dependency.
	Therefore, this transformed code does not necessarily guarantee
	that \clnref{q} is ordered after \clnref{p}.\footnote{
		Kudos to \ppl{Linus}{Torvalds} for providing this example.}
	\end{fcvref}
}\QuickQuizEndE
}

Note that a series of inequality comparisons might, when taken together,
give the compiler enough information to determine the exact value of
the pointer, at which point the dependency is broken.
Furthermore, the compiler might be able to combine information from
even a single inequality comparison with other information to learn
the exact value, again breaking the dependency.
Pointers to elements in arrays are especially susceptible to this latter
form of dependency breakage.

\subsubsection{Safe comparison of dependent pointers}
It turns out that there are several safe ways to compare dependent
pointers:

\begin{enumerate}
\item	Comparisons against the \co{NULL} pointer.
	In this case, all the compiler can learn is that the pointer
	is \co{NULL}, in which case you are not allowed to
	dereference it anyway.
\item	The dependent pointer is never dereferenced, whether before or
	after the comparison.
\item	The dependent pointer is compared to a pointer that references
	objects that were last modified a very long time ago, where
	the only unconditionally safe value of ``a very long time ago'' is
	``at compile time''.
	The key point is that something other than the address or data
	dependency guarantees ordering.
\item	Comparisons between two pointers, each of which carries
	an appropriate dependency.
	For example, you have a pair of pointers, each carrying a
	dependency, to data structures each containing a lock, and you
	want to avoid \IX{deadlock} by acquiring the locks in address order.
\item	The comparison is not-equal, and the compiler does not have
	enough other information to deduce the value of the
	pointer carrying the dependency.
\end{enumerate}

\begin{listing}
\begin{fcvlabel}[ln:memorder:Broken Dependencies With Pointer Comparisons]
\begin{VerbatimL}[commandchars=\\\[\]]
struct foo {		\lnlbl[foo:b]
	int a;
	int b;
	int c;
};                      \lnlbl[foo:e]
struct foo *gp1;	\lnlbl[gp1]
struct foo *gp2;	\lnlbl[gp2]

void updater(void)		\lnlbl[upd:b]
{
	struct foo *p;

	p = malloc(sizeof(*p));		\lnlbl[upd:alloc]
	BUG_ON(!p);			\lnlbl[upd:bug]
	p->a = 42;			\lnlbl[upd:init:a]
	p->b = 43;
	p->c = 44;			\lnlbl[upd:init:c]
	rcu_assign_pointer(gp1, p);	\lnlbl[upd:assign1]
	WRITE_ONCE(p->b, 143);		\lnlbl[upd:upd:b]
	WRITE_ONCE(p->c, 144);		\lnlbl[upd:upd:c]
	rcu_assign_pointer(gp2, p);	\lnlbl[upd:assign2]
}				\lnlbl[upd:e]

void reader(void)		\lnlbl[read:b]
{
	struct foo *p;
	struct foo *q;
	int r1, r2 = 0;

	p = rcu_dereference(gp2);	\lnlbl[read:gp2]
	if (p == NULL)			\lnlbl[read:nulchk]
		return;			\lnlbl[read:nulret]
	r1 = READ_ONCE(p->b);		\lnlbl[read:pb]
	q = rcu_dereference(gp1);	\lnlbl[read:gp1]
	if (p == q) {			\lnlbl[read:equ]
		r2 = READ_ONCE(p->c);	\lnlbl[read:pc]
	}
	do_something_with(r1, r2);
}				\lnlbl[read:e]
\end{VerbatimL}
\end{fcvlabel}
\caption{Broken Dependencies With Pointer Comparisons}
\label{lst:memorder:Broken Dependencies With Pointer Comparisons}
\end{listing}

Pointer comparisons can be quite tricky, and so it is well worth working
through the example shown in
\cref{lst:memorder:Broken Dependencies With Pointer Comparisons}.
\begin{fcvref}[ln:memorder:Broken Dependencies With Pointer Comparisons]
This example uses a simple \co{struct foo} shown on \clnrefrange{foo:b}{foo:e}
and two global pointers, \co{gp1} and \co{gp2}, shown on \clnref{gp1,gp2},
respectively.
This example uses two threads, namely \co{updater()} on
\clnrefrange{upd:b}{upd:e} and \co{reader()} on \clnrefrange{read:b}{read:e}.
\end{fcvref}

\begin{fcvref}[ln:memorder:Broken Dependencies With Pointer Comparisons:upd]
The \co{updater()} thread allocates memory on \clnref{alloc}, and complains
bitterly on \clnref{bug} if none is available.
\Clnrefrange{init:a}{init:c} initialize the newly allocated structure,
and then \clnref{assign1} assigns the pointer to \co{gp1}.
\Clnref{upd:b,upd:c} then update two of the structure's fields, and does
so \emph{after} \clnref{assign1} has made those fields visible to readers.
Please note that unsynchronized update of reader-visible fields
often constitutes a bug.
Although there are legitimate use cases doing just this, such use cases
require more care than is exercised in this example.

Finally, \clnref{assign2} assigns the pointer to \co{gp2}.
\end{fcvref}

\begin{fcvref}[ln:memorder:Broken Dependencies With Pointer Comparisons:read]
The \co{reader()} thread first fetches \co{gp2} on \clnref{gp2}, with
\clnref{nulchk,nulret} checking for \co{NULL} and returning if so.
\Clnref{pb} fetches field \co{->b} and
\clnref{gp1} fetches \co{gp1}.
If \clnref{equ} sees that the pointers fetched on \clnref{gp2,gp1}
are equal, \clnref{pc} fetches \co{p->c}.
Note that \clnref{pc} uses pointer \co{p} fetched on \clnref{gp2}, not
pointer \co{q} fetched on \clnref{gp1}.

But this difference might not matter.
An equals comparison on \clnref{equ} might lead the compiler to (incorrectly)
conclude that both pointers are equivalent, when in fact they carry
different dependencies.
This means that the compiler might well transform \clnref{pc} to instead
be \co{r2 = READ_ONCE(q->c)}, which might well cause the value 44 to be loaded
instead of the expected value 144.
\end{fcvref}

\QuickQuiz{
	\begin{fcvref}[ln:memorder:Broken Dependencies With Pointer Comparisons:read]
	But doesn't the condition in \clnref{equ} supply a control dependency
	that would keep \clnref{pc} ordered after \clnref{gp1}?
	\end{fcvref}
}\QuickQuizAnswer{
	\begin{fcvref}[ln:memorder:Broken Dependencies With Pointer Comparisons:read]
	Yes, but no.
	Yes, there is a control dependency, but control dependencies do
	not order later loads, only later stores.
	If you really need ordering, you could place an \co{smp_rmb()}
	between \clnref{equ,pc}.
	Or better yet, have \co{updater()}
	allocate two structures instead of reusing the structure.
	For more information, see
	\cref{sec:memorder:Control-Dependency Calamities}.
	\end{fcvref}
}\QuickQuizEnd

In short, great care is required to ensure that dependency
chains in your source code are still dependency chains in the
compiler-generated assembly code.

\subsection{Control-Dependency Calamities}
\label{sec:memorder:Control-Dependency Calamities}

The control dependencies described in
\cref{sec:memorder:Control Dependencies}
are attractive due to their low overhead, but are also especially
tricky because current compilers do not understand them and can easily
break them.
The rules and examples in this section are intended to help you
prevent your compiler's ignorance from breaking your code.

A load-load control dependency requires a full \IXh{read}{memory barrier},
not simply a data dependency barrier.
Consider the following bit of code:

\begin{VerbatimN}
q = READ_ONCE(x);
if (q) {
	<data dependency barrier>
	q = READ_ONCE(y);
}
\end{VerbatimN}

This will not have the desired effect because there is no actual data
dependency, but rather a control dependency that the CPU may short-circuit
by attempting to predict the outcome in advance, so that other CPUs see
the load from~\co{y} as having happened before the load from~\co{x}.
In such a case what's actually required is:

\begin{VerbatimN}
q = READ_ONCE(x);
if (q) {
	<read barrier>
	q = READ_ONCE(y);
}
\end{VerbatimN}

However, stores are not speculated.
This means that ordering \emph{is} provided for load-store control
dependencies, as in the following example:

\begin{VerbatimN}
q = READ_ONCE(x);
if (q)
	WRITE_ONCE(y, 1);
\end{VerbatimN}

Control dependencies pair normally with other types of ordering operations.
That said, please note that neither \co{READ_ONCE()} nor \co{WRITE_ONCE()}
are optional!
Without the \co{READ_ONCE()}, the compiler might fuse the load
from~\co{x} with other loads from~\co{x}.
Without the \co{WRITE_ONCE()}, the compiler might fuse the store
to~\co{y} with other stores to~\co{y}, or, worse yet, read the
value, compare it, and only conditionally do the store.
Any of these can result in highly counter-intuitive effects on ordering.

Worse yet, if the compiler is able to prove (say) that the value of
variable~\co{x} is always non-zero, it would be well within its rights
to optimize the original example by eliminating the \qco{if} statement
as follows:

\begin{VerbatimN}
q = READ_ONCE(x);
WRITE_ONCE(y, 1); /* BUG: CPU can reorder!!! */
\end{VerbatimN}

\QuickQuiz{
	But there is a \co{READ_ONCE()}, so how can the compiler
	prove anything about the value of \co{q}?
}\QuickQuizAnswer{
	Given the simple \co{if} statement comparing against zero,
	it is hard to imagine the compiler proving anything.
	But suppose that later code executed a division by \co{q}.
	Because division by zero is undefined behavior, as of 2023,
	many compilers will assume that the value of \co{q} must
	be non-zero, and will thus remove that \co{if} statement,
	thus unconditionally executing the \co{WRITE_ONCE()}, in
	turn destroying the control dependency.

	There are some who argue (correctly, in Paul's view) that
	back-propagating undefined behavior across volatile accesses
	constitutes a compiler bug, but many compiler writers insist
	that this is not a bug, but rather a valuable optimization.
}\QuickQuizEnd

It is tempting to try to enforce ordering on identical stores on both
branches of the \qco{if} statement as follows:

\begin{VerbatimN}
q = READ_ONCE(x);
if (q) {
	barrier();
	WRITE_ONCE(y, 1);
	do_something();
} else {
	barrier();
	WRITE_ONCE(y, 1);
	do_something_else();
}
\end{VerbatimN}

Unfortunately, current compilers will transform this as follows at high
optimization levels:

\begin{VerbatimN}
q = READ_ONCE(x);
barrier();
WRITE_ONCE(y, 1);  /* BUG: No ordering!!! */
if (q) {
	do_something();
} else {
	do_something_else();
}
\end{VerbatimN}

Now there is no conditional between the load from~\co{x} and the store
to~\co{y}, which means that the CPU is within its rights to reorder them:
The conditional is absolutely required, and must be present in the
assembly code even after all compiler optimizations have been applied.
Therefore, if you need ordering in this example, you need explicit
memory-ordering operations, for example, a \IX{release store}:

\begin{VerbatimN}
q = READ_ONCE(x);
if (q) {
	smp_store_release(&y, 1);
	do_something();
} else {
	smp_store_release(&y, 1);
	do_something_else();
}
\end{VerbatimN}

The initial \co{READ_ONCE()} is still required to prevent the compiler from
guessing the value of~\co{x}.
In addition, you need to be careful what you do with the local variable~%
\co{q},
otherwise the compiler might be able to guess its value and again remove
the needed conditional.
For example:

\begin{VerbatimN}
q = READ_ONCE(x);
if (q % MAX) {
	WRITE_ONCE(y, 1);
	do_something();
} else {
	WRITE_ONCE(y, 2);
	do_something_else();
}
\end{VerbatimN}

If \co{MAX} is defined to be~1, then the compiler knows that \co{(q\%MAX)} is
equal to zero, in which case the compiler is within its rights to
transform the above code into the following:

\begin{VerbatimN}
q = READ_ONCE(x);
WRITE_ONCE(y, 2);
do_something_else();
\end{VerbatimN}

Given this transformation, the CPU is not required to respect the ordering
between the load from variable~\co{x} and the store to variable~\co{y}.
It is tempting to add a \co{barrier()} to constrain the compiler,
but this does not help.
The conditional is gone, and the \co{barrier()} won't bring it back.
Therefore, if you are relying on this ordering, you should make sure
that \co{MAX} is greater than one, perhaps as follows:

\begin{VerbatimN}
q = READ_ONCE(x);
BUILD_BUG_ON(MAX <= 1);
if (q % MAX) {
	WRITE_ONCE(y, 1);
	do_something();
} else {
	WRITE_ONCE(y, 2);
	do_something_else();
}
\end{VerbatimN}

Please note once again that the stores to~\co{y} differ.
If they were identical, as noted earlier, the compiler could pull this
store outside of the \qco{if} statement.

You must also avoid excessive reliance on boolean short-circuit evaluation.
Consider this example:

\begin{VerbatimN}
q = READ_ONCE(x);
if (q || 1 > 0)
	WRITE_ONCE(y, 1);
\end{VerbatimN}

Because the first condition cannot fault and the second condition is
always true, the compiler can transform this example as following,
defeating the control dependency:

\begin{VerbatimN}
q = READ_ONCE(x);
WRITE_ONCE(y, 1);
\end{VerbatimN}

This example underscores the need to ensure that the compiler cannot
out-guess your code.
Never forget that, although \co{READ_ONCE()} does force
the compiler to actually emit code for a given load, it does not force
the compiler to use the value loaded.

In addition, control dependencies apply only to the then-clause and
else-clause of the if-statement in question.
In particular, it does
not necessarily apply to code following the if-statement:

\begin{VerbatimN}
q = READ_ONCE(x);
if (q) {
	WRITE_ONCE(y, 1);
} else {
	WRITE_ONCE(y, 2);
}
WRITE_ONCE(z, 1);  /* BUG: No ordering. */
\end{VerbatimN}

It is tempting to argue that there in fact is ordering because the
compiler cannot reorder volatile accesses and also cannot reorder
the writes to~\co{y} with the condition.
Unfortunately for this line
of reasoning, the compiler might compile the two writes to~\co{y} as
conditional-move instructions, as in this fanciful pseudo-assembly
language:

\begin{VerbatimN}
ld r1,x
cmp r1,$0
cmov,ne r4,$1
cmov,eq r4,$2
st r4,y
st $1,z
\end{VerbatimN}

A weakly ordered CPU would have no dependency of any sort between the load
from~\co{x} and the store to~\co{z}.
The control dependencies would extend
only to the pair of \co{cmov} instructions and the store depending on them.
In short, control dependencies apply only to the stores in the \qco{then}
and \qco{else} of the \qco{if} in question (including functions invoked by
those two clauses), and not necessarily to code following that \qco{if}.

Finally, control dependencies do \emph{not} provide cumulativity.\footnote{
	Refer to \cref{sec:memorder:Cumulativity} for
	the meaning of cumulativity.}
This is demonstrated by two related litmus tests, namely
\cref{lst:memorder:LB Litmus Test With Control Dependency,%
lst:memorder:WWC Litmus Test With Control Dependency (Cumulativity?)}
with the initial values
of~\co{x} and~\co{y} both being zero.

\begin{listing}
\input{CodeSamples/formal/litmus/C-LB+o-cgt-o+o-cgt-o=whole.fcv}
\caption{LB Litmus Test With Control Dependency}
\label{lst:memorder:LB Litmus Test With Control Dependency}
\end{listing}

The \co{exists} clause in the two-thread example of
\cref{lst:memorder:LB Litmus Test With Control Dependency}
(\path{C-LB+o-cgt-o+o-cgt-o.litmus})
will never trigger.
If control dependencies guaranteed cumulativity (which they do
not), then adding a thread to the example as in
\cref{lst:memorder:WWC Litmus Test With Control Dependency (Cumulativity?)}
(\path{C-WWC+o-cgt-o+o-cgt-o+o.litmus})
would guarantee the related \co{exists} clause never to trigger.

\begin{listing}
\input{CodeSamples/formal/litmus/C-WWC+o-cgt-o+o-cgt-o+o=whole.fcv}
\caption{WWC Litmus Test With Control Dependency (Cumulativity?)}
\label{lst:memorder:WWC Litmus Test With Control Dependency (Cumulativity?)}
\end{listing}

But because control dependencies do \emph{not} provide cumulativity, the
\co{exists} clause in the three-thread litmus test can trigger.
If you need the three-thread example to provide ordering, you will need
\co{smp_mb()} between the load and store in \co{P0()},
that is, just before or just after the \qco{if} statements.
Furthermore, the original two-thread example is very fragile and should be avoided.

\QuickQuiz{
	Can't you instead add an \co{smp_mb()} to \co{P1()} in
	\cref{lst:memorder:WWC Litmus Test With Control Dependency (Cumulativity?)}?
}\QuickQuizAnswer{
	Not given the Linux kernel memory model.
	(Try it!)
	However, you can instead replace \co{P0()}'s
	\co{WRITE_ONCE()} with \co{smp_store_release()},
	which usually has less overhead than does adding an \co{smp_mb()}.
}\QuickQuizEnd

The following list of rules summarizes the lessons of this section:

\begin{enumerate}
\item	Compilers do not understand control dependencies, so it is
	your job to make sure that the compiler cannot break your code.

\item	Control dependencies can order prior loads against later stores.
	However, they do \emph{not} guarantee any other sort of ordering:
	Not prior loads against later loads, nor prior stores against
	later anything.
	If you need these other forms of ordering, use \co{smp_rmb()},
	\co{smp_wmb()}, or, in the case of prior stores and later loads,
	\co{smp_mb()}.

\item	If both legs of the \qco{if} statement begin with identical stores
	to the same variable, then the control dependency will not order
	those stores,
	If ordering is needed, precede both of them with \co{smp_mb()} or
	use \co{smp_store_release()}.
	Please note that it is \emph{not} sufficient to use \co{barrier()}
	at beginning of each leg of the \qco{if} statement because, as shown
	by the example above, optimizing compilers can destroy the control
	dependency while respecting the letter of the \co{barrier()} law.

\item	Control dependencies require at least one run-time conditional
	between the prior load and the subsequent store, and this
	conditional must involve the prior load.
	If the compiler is able to optimize the conditional away, it
	will have also optimized away the ordering.
	Careful use of \co{READ_ONCE()} and \co{WRITE_ONCE()} can help
	to preserve the needed conditional.

\item	Control dependencies require that the compiler avoid reordering
	the dependency into nonexistence.
	Careful use of \co{READ_ONCE()}, \co{atomic_read()}, or
	\co{atomic64_read()} can help to preserve your control
	dependency.

\item	Control dependencies apply only to the \qco{then} and
	\qco{else} of the \qco{if} containing the control
	dependency, including any functions that these two clauses call.
	Control dependencies do \emph{not} apply to code following the
	end of the \qco{if} statement containing the control dependency.

\item	Control dependencies pair normally with other types of
	memory-ordering operations.

\item	Control dependencies do \emph{not} provide cumulativity.
	If you need cumulativity, use something that provides it,
	such as \co{smp_store_release()} or \co{smp_mb()}.
\end{enumerate}

Again, many popular languages were designed with single-threaded use
in mind.
Successful multithreaded use of these languages requires you to pay
special attention to your memory references and dependencies.

\section{Higher-Level Primitives}
\label{sec:memorder:Higher-Level Primitives}
\epigraph{Method will teach you to win time.}
	 {Johann Wolfgang von Goethe}

The answer to one of the quick quizzes in
\cref{sec:formal:Axiomatic Approaches and Locking}
demonstrated exponential speedups due to verifying programs
modeled at higher levels of abstraction.
This section will look into how higher levels of abstraction can
also provide a deeper understanding of the synchronization primitives
themselves.
\Cref{sec:memorder:Memory Allocation}
takes a look at memory allocation,
\cref{sec:memorder:Locking}
examines the surprisingly varied semantics of locking, and
\cref{sec:memorder:RCU}
digs more deeply into RCU\@.

\subsection{Memory Allocation}
\label{sec:memorder:Memory Allocation}

\Cref{sec:SMPdesign:Parallel Fastpath for Resource Allocation}
touched upon memory allocation, and this section expands upon the relevant
memory-ordering issues.

The key requirement is that any access executed on a given block of
memory before freeing that block must be ordered before any access
executed after that same block is reallocated.
It would after all be a cruel and unusual memory-allocator bug if a store
preceding the free were to be reordered after another store following
the reallocation!
However, it would also be cruel and unusual to require developers to use
\co{READ_ONCE()} and \co{WRITE_ONCE()} to access dynamically allocated
memory.
Full ordering must therefore be provided for plain accesses, in spite of
all the shared-variable shenanigans called out in
\cref{sec:toolsoftrade:Shared-Variable Shenanigans}.

Of course, each CPU sees its own accesses in order and the compiler
always has fully accounted for intra-CPU shenanigans, give or take
the occasional compiler bug.
These facts are what enables the lockless fastpaths in
\co{memblock_alloc()} and \co{memblock_free()}, which are shown in
\cref{lst:SMPdesign:Allocator-Cache Allocator Function,%
lst:SMPdesign:Allocator-Cache Free Function},
respectively.
However, this is also why the developer is responsible for providing
appropriate ordering (for example, by using \co{smp_store_release()})
when publishing a pointer to a newly allocated block of memory.
After all, in the CPU-local case, the allocator has not necessarily
provided any cross-CPU ordering.

This means that the allocator must provide ordering when rebalancing
its per-thread pools.
This ordering is provided by the calls to \co{spin_lock()} and
\co{spin_unlock()} from \co{memblock_alloc()} and \co{memblock_free()}.
For any block that has migrated from one thread to another, the old
thread will have executed \co{spin_unlock(&globalmem.mutex)} after
placing the block in the \co{globalmem} pool, and the new thread will
have executed \co{spin_lock(&globalmem.mutex)} before moving that
block to its per-thread pool.
This \co{spin_unlock()} and \co{spin_lock()} ensures that both the
old and new threads see the old thread's accesses as having happened
before those of the new thread.

\QuickQuiz{
	But doesn't PowerPC have weak unlock-lock ordering properties
	within the Linux kernel, allowing a write before the unlock to
	be reordered with a read after the lock?
}\QuickQuizAnswer{
	Yes, but only from the perspective of a third thread not holding
	that lock.
	In contrast, memory allocators need only concern themselves with
	the two threads migrating the memory.
	It is after all the developer's responsibility to properly
	synchronize with any other threads that need access to the newly
	migrated block of memory.
}\QuickQuizEnd

Therefore, the ordering required by conventional uses of memory allocation
can be provided solely by non-fastpath locking, allowing the fastpath to
remain synchronization-free.

\subsection{Locking}
\label{sec:memorder:Locking}

Locking is a well-known synchronization primitive with which the
parallel-programming community has had decades of experience.
As such, locking's semantics are quite simple.

That is, they are quite simple until you start trying to mathematically
model them.

The simple part is that any CPU or thread holding a given lock is
guaranteed to see any accesses executed by CPUs or threads while they
were previously holding that same lock.
Similarly, any CPU or thread holding a given lock is guaranteed not
to see accesses that will be executed by other CPUs or threads while
subsequently holding that same lock.
And what else is there?

As it turns out, quite a bit:

\begin{enumerate}
\item	Are CPUs, threads, or compilers allowed to pull memory accesses
	into a given lock-based critical section?
\item	Will a CPU or thread holding a given lock also be guaranteed
	to see accesses executed by CPUs and threads before they last
	acquired that same lock, and vice versa?
\item	Suppose that a given CPU or thread executes one access
	(call it ``A''), releases a lock, reacquires that same lock,
	then executes another access (call it ``B'')\@.
	Is some other CPU or thread not holding that lock guaranteed to
	see A and B in order?
\item	As above, but with the lock reacquisition carried out by some
	other CPU or thread?
\item	As above, but with the lock reacquisition being some other lock?
\item	What ordering guarantees are provided by \co{spin_is_locked()}?
\end{enumerate}

The reaction to some or even all of these questions might well be ``Why
would anyone do \emph{that}?''
However, any complete mathematical definition of locking must have
answers to all of these questions.
Therefore, the following sections address these questions in the context
of the Linux kernel.

\subsubsection{Accesses Into Critical Sections?}
\label{sec:memorder:Accesses Into Critical Sections?}

Can memory accesses be reordered into lock-based critical sections?

\begin{listing}
\input{CodeSamples/formal/herd/C-Lock-before-into=whole.fcv}
\caption{Prior Accesses Into Critical Section (Ordering?)}
\label{lst:memorder:Prior Accesses Into Critical Section (Ordering?)}
\end{listing}

\begin{listing}
\input{CodeSamples/formal/herd/C-Lock-after-into=whole.fcv}
\caption{Subsequent Accesses Into Critical Section (Ordering?)}
\label{lst:memorder:Subsequent Accesses Into Critical Section (Ordering?)}
\end{listing}

Within the context of the Linux-kernel memory model, the simple answer
is ``yes''.
This may be verified by running the litmus tests shown in
\cref{lst:memorder:Prior Accesses Into Critical Section (Ordering?),lst:memorder:Subsequent Accesses Into Critical Section (Ordering?)}
(\path{C-Lock-before-into.litmus} and \path{C-Lock-after-into.litmus},
respectively), both of which will yield the \co{Sometimes} result.
This result indicates that the \co{exists} clause can be satisfied, that
is, that the final value of both \co{P0()}'s and \co{P1()}'s \co{r1} variable
can be zero.
This means that neither \co{spin_lock()} nor \co{spin_unlock()}
are required to act as a \IXh{full}{memory barrier}.

However, other environments might make other choices.
For example, locking implementations that run only on the x86 CPU
family will have lock-acquisition primitives that fully order the lock
acquisition with any prior and any subsequent accesses.
Therefore, on such systems the ordering shown in
\cref{lst:memorder:Prior Accesses Into Critical Section (Ordering?)}
comes for free.
There are x86 lock-release implementations that are weakly ordered,
thus failing to provide the ordering shown in
\cref{lst:memorder:Subsequent Accesses Into Critical Section (Ordering?)},
but an implementation could nevertheless choose to guarantee this ordering.

For their part, weakly ordered systems might well choose to execute
the memory-barrier instructions required to guarantee both orderings,
possibly simplifying code making advanced use of combinations of locked
and lockless accesses.
However, as noted earlier, \IXacr{lkmm} chooses not to provide these additional
orderings, in part to avoid imposing performance penalties on the simpler
and more prevalent locking use cases.
Instead, the \co{smp_mb__after_spinlock()} and \co{smp_mb__after_unlock_lock()}
primitives are provided for those more complex use cases, as discussed
in \cref{sec:memorder:Hardware Specifics}.

Thus far, this section has discussed only hardware reordering.
Can the compiler also reorder memory references into lock-based
critical sections?

The answer to this question in the context of the Linux kernel is a
resounding ``No!''
One reason for this otherwise inexplicable favoring of hardware reordering
over compiler optimizations is that the hardware will avoid reordering
a page-faulting access into a lock-based critical section.
In contrast, compilers have no clue about page faults, and would
therefore happily reorder a page fault into a critical section, which
could crash the kernel.
The compiler is also unable to reliably determine which accesses
will result in cache misses, so that compiler reordering into critical
sections could also result in excessive lock contention.
Therefore, the Linux kernel prohibits the compiler (but not the CPU)
from moving accesses into lock-based critical sections.

\subsubsection{Accesses Outside of Critical Section?}
\label{sec:memorder:Accesses Outside of Critical Section?}

If a given CPU or thread holds a given lock, it is guaranteed to see
accesses executed during all prior critical sections for that same
lock.
Similarly, such a CPU or thread is guaranteed not to see accesses
that will be executed during all subsequent critical sections for
that same lock.

\begin{listing}
\input{CodeSamples/formal/herd/C-Lock-outside-across=whole.fcv}
\caption{Accesses Outside of Critical Sections}
\label{lst:memorder:Accesses Outside of Critical Sections}
\end{listing}

But what about accesses preceding prior critical sections and
following subsequent critical sections?

This question can be answered for the Linux kernel by referring to
\cref{lst:memorder:Accesses Outside of Critical Sections}
(\path{C-Lock-outside-across.litmus}).
Running this litmus test yields the \co{Never} result,
which means that accesses in code leading up to a prior critical section
is also visible to the current CPU or thread holding that same lock.
Similarly, code that is placed after a subsequent critical section
is never visible to the current CPU or thread holding that same lock.

As a result, the Linux kernel cannot allow accesses to be moved
across the entirety of a given critical section.
Other environments might well wish to allow such code motion, but please
be advised that doing so is likely to yield profoundly counter-intuitive
results.

In short, the ordering provided by \co{spin_lock()} extends not only
throughout the critical section, but also indefinitely beyond the end
of that critical section.
Similarly, the ordering provided by \co{spin_unlock()} extends not
only throughout the critical section, but also indefinitely beyond the
beginning of that critical section.

\subsubsection{Ordering for Non-Lock Holders?}
\label{sec:memorder:Ordering for Non-Lock Holders?}

Does a CPU or thread that is not holding a given lock see that lock's
critical sections as being ordered?

\begin{listing}
\input{CodeSamples/formal/herd/C-Lock-across-unlock-lock-1=whole.fcv}
\caption{Accesses Between Same-CPU Critical Sections (Ordering?)}
\label{lst:memorder:Accesses Between Same-CPU Critical Sections (Ordering?)}
\end{listing}

This question can be answered for the Linux kernel by referring to
\cref{lst:memorder:Accesses Between Same-CPU Critical Sections (Ordering?)}
(\path{C-Lock-across-unlock-lock-1.litmus}), which
shows an example where \co{P(0)} places its write and read in two
different critical sections for the same lock.
Running this litmus test shows that the \co{exists} can be satisfied,
which means that the answer is ``no'', and that CPUs can reorder accesses
across consecutive critical sections.
In other words, not only are \co{spin_lock()} and \co{spin_unlock()}
weaker than a full barrier when considered separately, they are also
weaker than a full barrier when taken together.

If the ordering of a given lock's critical sections are to be observed,
then either the observer must hold that lock on the one hand or either
\co{smp_mb__after_spinlock()} or \co{smp_mb__after_unlock_lock()}
must be executed just after the second lock acquisition on the other.

But what if the two critical sections run on different CPUs or threads?

\begin{listing}
\input{CodeSamples/formal/herd/C-Lock-across-unlock-lock-2=whole.fcv}
\caption{Accesses Between Different-CPU Critical Sections (Ordering?)}
\label{lst:memorder:Accesses Between Different-CPU Critical Sections (Ordering?)}
\end{listing}

This question is answered for the Linux kernel by referring to
\cref{lst:memorder:Accesses Between Different-CPU Critical Sections (Ordering?)}
(\path{C-Lock-across-unlock-lock-2.litmus}),
in which the first lock acquisition is executed by \co{P0()} and the
second lock acquisition is executed by \co{P1()}.
Note that \co{P1()} must read \co{x} to reject executions in which
\co{P1()} executes before \co{P0()} does.
Running this litmus test shows that the \co{exists} can be satisfied,
which means that the answer is ``no'', and that CPUs can reorder accesses
across consecutive critical sections, even if each of those critical
sections runs on a different CPU or thread.

\QuickQuiz{
	But if there are three critical sections, isn't it true that
	CPUs not holding the lock will observe the accesses from the
	first and the third critical section as being ordered?
}\QuickQuizAnswer{
	No.

\begin{listing}
\input{CodeSamples/formal/herd/C-Lock-across-unlock-lock-3=whole.fcv}
\caption{Accesses Between Multiple Different-CPU Critical Sections}
\label{lst:memorder:Accesses Between Multiple Different-CPU Critical Sections}
\end{listing}

	\Cref{lst:memorder:Accesses Between Multiple Different-CPU Critical Sections}
	shows an example three-critical-section chain
	(\path{Lock-across-unlock-lock-3.litmus}).
	Running this litmus test shows that the \co{exists} clause can
	still be satisfied, so this additional critical section is still
	not sufficient to force ordering.

	However, as the reader can verify, placing an
	\co{smp_mb__after_spinlock()} after either \co{P1()}'s or
	\co{P2()}'s lock acquisition does suffice to force ordering.
}\QuickQuizEnd

As before, if the ordering of a given lock's critical sections are to
be observed, then either the observer must hold that lock or either
\co{smp_mb__after_spinlock()} or \co{smp_mb__after_unlock_lock()} must
be executed just after \co{P1()}'s lock acquisition.

Given that ordering is not guaranteed when both critical sections are
protected by the same lock, there is no hope of any ordering guarantee
when different locks are used.
However, readers are encouraged to construct the corresponding litmus
test and see this for themselves.

This situation can seem counter-intuitive, but it is rare for code to
care.
This approach also allows certain weakly ordered systems to implement
locks more efficiently.

\subsubsection{Ordering for \tco{spin_is_locked()}?}
\label{sec:memorder:Ordering for spin-is-locked()?}

The Linux kernel's \co{spin_is_locked()} primitive returns
\co{true} if the specified lock is held and \co{false} otherwise.
Note that \co{spin_is_locked()} returns \co{true} when some other
CPU or thread holds the lock, not just when the current CPU or thread
holds that lock.
This raises the question of what ordering guarantees \co{spin_is_locked()}
might provide.

In the Linux kernel, the answer has varied over time.
Initially, \co{spin_is_locked()} was unordered, but a few interesting
use cases motivated strong ordering.
Later discussions surrounding the Linux-kernel memory model concluded
that \co{spin_is_locked()} should be used only for debugging.
Part of the reason for this is that even a fully ordered
\co{spin_is_locked()} might return \co{true} because some other CPU or
thread was just about to release the lock in question.
In this case, there is little that can be learned from that return value
of \co{true}, which means that reliable use of \co{spin_is_locked()}
is surprisingly complex.
Other approaches almost always work better, for example, use of explicit
shared variables or the \co{spin_trylock()} primitive.

This situation resulted in the current state, namely that
\co{spin_is_locked()} provides no ordering guarantees, except that if
it returns \co{false}, the current CPU or thread cannot be holding the
corresponding lock.

\QuickQuiz{
	But if \co{spin_is_locked()} returns \co{false}, don't we also
	know that no other CPU or thread is holding the corresponding
	lock?
}\QuickQuizAnswer{
	No.
	By the time that the code inspects the return value from
	\co{spin_is_locked()}, some other CPU or thread might well have
	acquired the corresponding lock.
}\QuickQuizEnd

\subsubsection{Why Mathematically Model Locking?}
\label{sec:memorder:Why Mathematically Model Locking?}

Given all these possible choices, why model locking in general?
Why not simply model a simple implementation?

One reason is modeling performance, as shown in
\cref{tab:formal:Locking: Modeling vs. Emulation Time (s)}
on
\cpageref{tab:formal:Locking: Modeling vs. Emulation Time (s)}.
Directly modeling locking in general is orders of magnitude faster
than emulating even a trivial implementation.
This should be no surprise, given the combinatorial explosion experienced
by present-day formal-verification tools with increases in the number of
memory accesses executed by the code being modeled.
Splitting the modeling at API boundaries can therefore result in
combinatorial implosion.

Another reason is that a trivial implementation might needlessly constrain
either real implementations or real use cases.
In contrast, modeling a platonic lock allows the widest variety of
implementations while providing specific guidance to locks' users.

\subsection{RCU}
\label{sec:memorder:RCU}

\begin{figure}
\centering
\resizebox{3in}{!}{\includegraphics{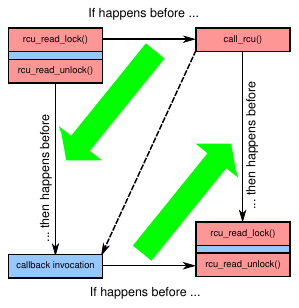}}
\caption{RCU Grace-Period Ordering Guarantees}
\label{fig:memorder:RCU Grace-Period Ordering Guarantees}
\end{figure}

As described in
\cref{sec:defer:RCU Fundamentals},
the fundamental property of RCU grace periods is this straightforward
two-part guarantee:
\begin{enumerate*}[(1)]
\item If any part of a given RCU read-side critical section precedes
the beginning of a given \IX{grace period}, then the entirety of that
critical section precedes the end of that grace period.
\item If any part of a given RCU read-side critical section follows
the end of a given grace period, then the entirety of that
critical section follows the beginning of that grace period.
\end{enumerate*}
These guarantees are summarized in
\cref{fig:memorder:RCU Grace-Period Ordering Guarantees},
where the grace period is denoted by the dashed arrow between the
\co{call_rcu()} invocation in the upper right and the corresponding
RCU callback invocation in the lower left.\footnote{
	For more detail, please see
	\crefrange{fig:defer:RCU Reader and Later Grace Period}{fig:defer:RCU Reader Within Grace Period}
	starting on
	\cpageref{fig:defer:RCU Reader and Later Grace Period}.}

\begin{listing}
\input{CodeSamples/formal/herd/C-SB+o-rcusync-o+rl-o-o-rul=whole.fcv}
\caption{RCU Fundamental Property}
\label{lst:memorder:RCU Fundamental Property}
\end{listing}

\begin{listing}
\input{CodeSamples/formal/herd/C-SB+o-rcusync-o+i-rl-o-o-rul=whole.fcv}
\caption{RCU Fundamental Property and Reordering}
\label{lst:memorder:RCU Fundamental Property and Reordering}
\end{listing}

In short, an RCU read-side critical section is guaranteed never to
completely overlap an RCU grace period, as demonstrated by
\cref{lst:memorder:RCU Fundamental Property}
(\path{C-SB+o-rcusync-o+rl-o-o-rul.litmus}).
Either or neither of the \co{r2} registers can have the final value of zero,
but at least one of them must be non-zero (that is, the cycle identified
by the \co{exists} clause is prohibited), courtesy of RCU's fundamental
grace-period guarantee, as can be seen by running \co{herd} on this litmus test.
Note that this guarantee is insensitive to the ordering of the accesses
within \co{P1()}'s critical section, so the litmus test shown in
\cref{lst:memorder:RCU Fundamental Property and Reordering}\footnote{
	Dependencies can of course limit the ability to reorder accesses
	within RCU read-side critical sections.}
also forbids this same cycle.

However, this definition is incomplete, as can be seen from the following
list of questions:\footnote{
	Several of which were introduced to Paul by \ppl{Jade}{Alglave} during
	early work on LKMM, and a few more of which came from other
	LKMM participants~\cite{Alglave:2018:FSC:3173162.3177156}.}

\begin{enumerate}
\item	What ordering is provided by \co{rcu_read_lock()}
	and \co{rcu_read_unlock()}, independent of RCU grace periods?
\item	What ordering is provided by \co{synchronize_rcu()}
	and \co{synchronize_rcu_expedited()}, independent of RCU read-side
	critical sections?
\item	If the entirety of a given RCU read-side critical section
	precedes the end of a given RCU grace period, what about
	accesses preceding that critical section?
\item	If the entirety of a given RCU read-side critical section
	follows the beginning of a given RCU grace period, what about
	accesses following that critical section?
\item	What happens in situations involving more than one RCU read-side
	critical section and/or more than one RCU grace period?
\item	What happens when RCU is combined with other memory-ordering
	mechanisms?
\end{enumerate}

These questions are addressed in the following sections.

\subsubsection{RCU Read-Side Ordering}
\label{sec:memorder:RCU Read-Side Ordering}

On their own, RCU's read-side primitives \co{rcu_read_lock()} and
\co{rcu_read_unlock()} provide no ordering whatsoever.
In particular, despite their names, they do not act like locks, as can
be seen in
\cref{lst:memorder:RCU Readers Provide No Lock-Like Ordering}
(\path{C-LB+rl-o-o-rul+rl-o-o-rul.litmus}).
This litmus test's cycle is allowed:
Both instances of the \co{r1} register can have final values of 1.

\begin{listing}
\input{CodeSamples/formal/herd/C-LB+rl-o-o-rul+rl-o-o-rul=whole.fcv}
\caption{RCU Readers Provide No Lock-Like Ordering}
\label{lst:memorder:RCU Readers Provide No Lock-Like Ordering}
\end{listing}

Nor do these primitives have barrier-like ordering properties,
at least not unless there is a grace period in the mix, as can be seen in
\cref{lst:memorder:RCU Readers Provide No Barrier-Like Ordering}
(\path{C-LB+o-rl-rul-o+o-rl-rul-o.litmus}).
This litmus test's cycle is also allowed.
(Try it!)

\begin{listing}
\input{CodeSamples/formal/herd/C-LB+o-rl-rul-o+o-rl-rul-o=whole.fcv}
\caption{RCU Readers Provide No Barrier-Like Ordering}
\label{lst:memorder:RCU Readers Provide No Barrier-Like Ordering}
\end{listing}

Of course, lack of ordering in both these litmus tests should be absolutely
no surprise, given that both \co{rcu_read_lock()} and \co{rcu_read_unlock()}
are no-ops in the \IXacr{qsbr} implementation of RCU\@.

\subsubsection{RCU Update-Side Ordering}
\label{sec:memorder:RCU Update-Side Ordering}

In contrast with RCU readers, the RCU update-side functions
\co{synchronize_rcu()} and \co{synchronize_rcu_expedited()}
provide memory ordering at least as strong as \co{smp_mb()},\footnote{
	And also way more expensive!}
as can be seen by running \co{herd} on the litmus test shown in
\cref{lst:memorder:RCU Updaters Provide Full Ordering}.
This test's cycle is prohibited, just as it would with \co{smp_mb()}.
This should be no surprise given the information presented in
\cref{tab:memorder:Linux-Kernel Memory-Ordering Cheat Sheet}.

\begin{listing}
\input{CodeSamples/formal/herd/C-SB+o-rcusync-o+o-rcusync-o=whole.fcv}
\caption{RCU Updaters Provide Full Ordering}
\label{lst:memorder:RCU Updaters Provide Full Ordering}
\end{listing}

\subsubsection{RCU Readers:
			    Before and After}
\label{sec:memorder:RCU Readers: Before and After}

Before reading this section, it would be well to reflect on the distinction
between guarantees that are available and guarantees that maintainable
software should rely on.
Keeping that firmly in mind, this section presents a few of the
more exotic RCU guarantees.

\begin{listing}
\input{CodeSamples/formal/herd/C-SB+o-rcusync-o+o-rl-o-rul=whole.fcv}
\caption{What Happens Before RCU Readers?}
\label{lst:memorder:What Happens Before RCU Readers?}
\end{listing}

\Cref{lst:memorder:What Happens Before RCU Readers?}
(\path{C-SB+o-rcusync-o+o-rl-o-rul.litmus})
shows a litmus test similar to that in
\cref{lst:memorder:RCU Fundamental Property},
but with the RCU reader's first access preceding the RCU read-side critical
section, rather than the more conventional (and maintainable!\@) approach of
being contained within it.
Perhaps surprisingly, running \co{herd} on this litmus test gives the
same result as for that in
\cref{lst:memorder:RCU Fundamental Property}:
The cycle is forbidden.

Why would this be the case?

Because both of \co{P1()}'s accesses are volatile,
as discussed in
\cref{sec:toolsoftrade:A Volatile Solution},
the compiler is not permitted to reorder them.
This means that the code emitted for \co{P1()}'s \co{WRITE_ONCE()} will
precede that of \co{P1()}'s \co{READ_ONCE()}.
Therefore, RCU implementations that place memory-barrier instructions in
\co{rcu_read_lock()} and \co{rcu_read_unlock()} will preserve the ordering
of \co{P1()}'s two accesses all the way down to the hardware level.
On the other hand, RCU implementations that rely on interrupt-based
state machines will also fully preserve this ordering
\emph{relative to the grace period} due to the fact that interrupts take
place at a precise location in the execution of the interrupted code.

This in turn means that if the \co{WRITE_ONCE()} follows the end of a
given RCU grace period, then the accesses within \emph{and following}
that RCU read-side critical section must follow the beginning of that
same grace period.
Similarly, if the \co{READ_ONCE()} precedes the beginning of the grace
period, everything within \emph{and preceding} that critical section
must precede the end of that same grace period.

\begin{listing}
\input{CodeSamples/formal/herd/C-SB+o-rcusync-o+rl-o-rul-o=whole.fcv}
\caption{What Happens After RCU Readers?}
\label{lst:memorder:What Happens After RCU Readers?}
\end{listing}

\Cref{lst:memorder:What Happens After RCU Readers?}
(\path{C-SB+o-rcusync-o+rl-o-rul-o.litmus})
is similar, but instead looks at accesses after the RCU read-side
critical section.
This test's cycle is also forbidden, as can be checked with the \co{herd}
tool.
The reasoning is similar to that for
\cref{lst:memorder:What Happens Before RCU Readers?},
and is left as an exercise for the reader.

\begin{listing}
\input{CodeSamples/formal/herd/C-SB+o-rcusync-o+o-rl-rul-o=whole.fcv}
\caption{What Happens With Empty RCU Readers?}
\label{lst:memorder:What Happens With Empty RCU Readers?}
\end{listing}

\Cref{lst:memorder:What Happens With Empty RCU Readers?}
(\path{C-SB+o-rcusync-o+o-rl-rul-o.litmus})
takes things one step farther, moving \co{P1()}'s \co{WRITE_ONCE()}
to precede the RCU read-side critical section and moving
\co{P1()}'s \co{READ_ONCE()} to follow it, resulting in an
empty RCU read-side critical section.

Perhaps surprisingly, despite the empty critical section, RCU nevertheless
still manages to forbid the cycle.
This can again be checked using the \co{herd} tool.
Furthermore, the reasoning is once again similar to that for
\cref{lst:memorder:What Happens Before RCU Readers?},
Recapping, if \co{P1()}'s \co{WRITE_ONCE()} follows the end of a given
grace period, then \co{P1()}'s RCU read-side critical section---and
everything following it---must follow the beginning of that same grace
period.
Similarly, if \co{P1()}'s \co{READ_ONCE()} precedes the beginning of a
given grace period, then \co{P1()}'s RCU read-side critical section---and
everything preceding it---must precede the end of that same grace period.
In both cases, the critical section's emptiness is irrelevant.

\QuickQuiz{
	Wait a minute!
	In QSBR implementations of RCU, no code is emitted for
	\co{rcu_read_lock()} and \co{rcu_read_unlock()}.
	This means that the RCU read-side critical section in
	\cref{lst:memorder:What Happens With Empty RCU Readers?}
	isn't just empty, it is completely nonexistent!!!
	So how can something that doesn't exist at all possibly have
	any effect whatsoever on ordering???
}\QuickQuizAnswer{
	Because in QSBR, RCU read-side critical sections don't
	actually disappear.
	Instead, they are extended in both directions until a quiescent
	state is encountered.
	For example, in the Linux kernel, the critical section might
	be extended back to the most recent \co{schedule()} call and
	ahead to the next \co{schedule()} call.
	Of course, in non-QSBR implementations, \co{rcu_read_lock()}
	and \co{rcu_read_unlock()} really do emit code, which can clearly
	provide ordering.
	And within the Linux kernel, even the QSBR implementation
	has a compiler \co{barrier()} in \co{rcu_read_lock()} and
	\co{rcu_read_unlock()}, which is necessary to prevent
	the compiler from moving memory accesses that might result
	in page faults into the RCU read-side critical section.

	Therefore, strange though it might seem, empty RCU read-side
	critical sections really can and do provide some degree of
	ordering.
}\QuickQuizEnd

\begin{listing}
\input{CodeSamples/formal/herd/C-SB+o-rcusync-o+o-o=whole.fcv}
\caption{What Happens With No RCU Readers?}
\label{lst:memorder:What Happens With No RCU Readers?}
\end{listing}

This situation leads to the question of what happens if
\co{rcu_read_lock()} and \co{rcu_read_unlock()} are omitted
entirely, as shown in
\cref{lst:memorder:What Happens With No RCU Readers?}
(\path{C-SB+o-rcusync-o+o-o.litmus}).
As can be checked with \co{herd}, this litmus test's cycle is allowed,
that is, both instances of \co{r2} can have final values of zero.

This might seem strange in light of the fact that empty RCU
read-side critical sections can provide ordering.
And it is true that QSBR implementations of RCU would in fact forbid
this outcome, due to the fact that there is no quiescent state anywhere
in \co{P1()}'s function body, so that \co{P1()} would run
within an implicit RCU read-side critical section.
However, RCU also has non-QSBR implementations, which have no implied
RCU read-side critical section, and in turn no way for RCU to enforce
ordering.
Therefore, this litmus test's cycle is allowed.

\QuickQuiz{
	Can \co{P1()}'s accesses be reordered in the litmus tests shown in
	\cref{lst:memorder:What Happens Before RCU Readers?,%
	lst:memorder:What Happens After RCU Readers?,%
	lst:memorder:What Happens With Empty RCU Readers?}
	in the same way that they were reordered going from
	\cref{lst:memorder:RCU Fundamental Property}
	to
	\cref{lst:memorder:RCU Fundamental Property and Reordering}?
}\QuickQuizAnswer{
	No, because none of these later litmus tests have more than one
	access within their RCU read-side critical sections.
	But what about swapping the accesses, for example, in
	\cref{lst:memorder:What Happens Before RCU Readers?},
	placing \co{P1()}'s \co{WRITE_ONCE()} within its critical
	section and the \co{READ_ONCE()} before its critical section?

	Swapping the accesses allows both instances of \co{r2} to
	have a final value of zero, in other words, although RCU read-side
	critical sections' ordering properties can extend outside of
	those critical sections, the same is not true of their
	reordering properties.
	Checking this with \co{herd} and explaining why is left as an
	exercise for the reader.
}\QuickQuizEnd

\subsubsection{Multiple RCU Readers and Updaters}
\label{sec:memorder:Multiple RCU Readers and Updaters}

Because \co{synchronize_rcu()} has ordering semantics that are at least
as strong as \co{smp_mb()}, no matter how many processes there are in
an SB litmus test
(such as \cref{lst:memorder:RCU Updaters Provide Full Ordering}),
placing \co{synchronize_rcu()} between each process's
accesses prohibits the cycle.
In addition, the cycle is prohibited in an SB test where one process
uses \co{synchronize_rcu()} and the other uses \co{rcu_read_lock()} and
\co{rcu_read_unlock()}, as shown by
\cref{lst:memorder:RCU Fundamental Property}.
However, if both processes use \co{rcu_read_lock()} and
\co{rcu_read_unlock()}, the cycle will be allowed, as shown by
\cref{lst:memorder:RCU Readers Provide No Lock-Like Ordering}.

Is it possible to say anything general about which RCU-protected
litmus tests will be prohibited and which will be allowed?
This section takes up that question.

\begin{listing}
\input{CodeSamples/formal/herd/C-SB+o-rcusync-o+rl-o-o-rul+rl-o-o-rul=whole.fcv}
\caption{One RCU Grace Period and Two Readers}
\label{lst:memorder:One RCU Grace Period and Two Readers}
\end{listing}

\begin{listing}
\input{CodeSamples/formal/herd/C-SB+o-rcusync-o+o-rcusync-o+rl-o-o-rul+rl-o-o-rul=whole.fcv}
\caption{Two RCU Grace Periods and Two Readers}
\label{lst:memorder:Two RCU Grace Periods and Two Readers}
\end{listing}

More specifically, what if the litmus test has one RCU grace
period and two RCU readers, as shown in
\cref{lst:memorder:One RCU Grace Period and Two Readers}?
The \co{herd} tool says that this cycle is allowed, but it would be
good to know \emph{why}.\footnote{
	Especially given that Paul changed his mind several times about
	this particular litmus test when working with \ppl{Jade}{Alglave} to
	generalize RCU ordering semantics.}

\begin{figure*}
\centering
\resizebox{0.75\onecolumntextwidth}{!}{\includegraphics{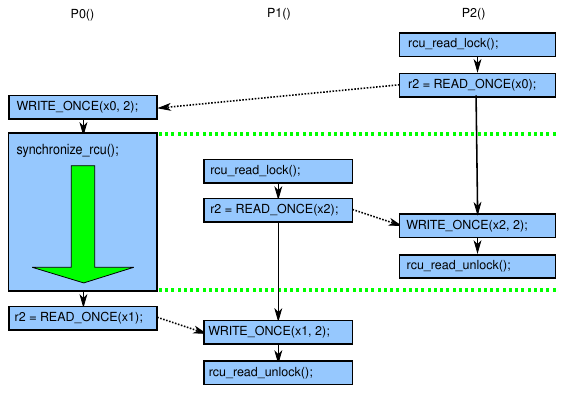}}
\caption{Cycle for One RCU Grace Period and Two RCU Readers}
\label{fig:memorder:Cycle for One RCU Grace Period and Two RCU Readers}
\end{figure*}

\begin{figure*}
\centering
\resizebox{\onecolumntextwidth}{!}{\includegraphics{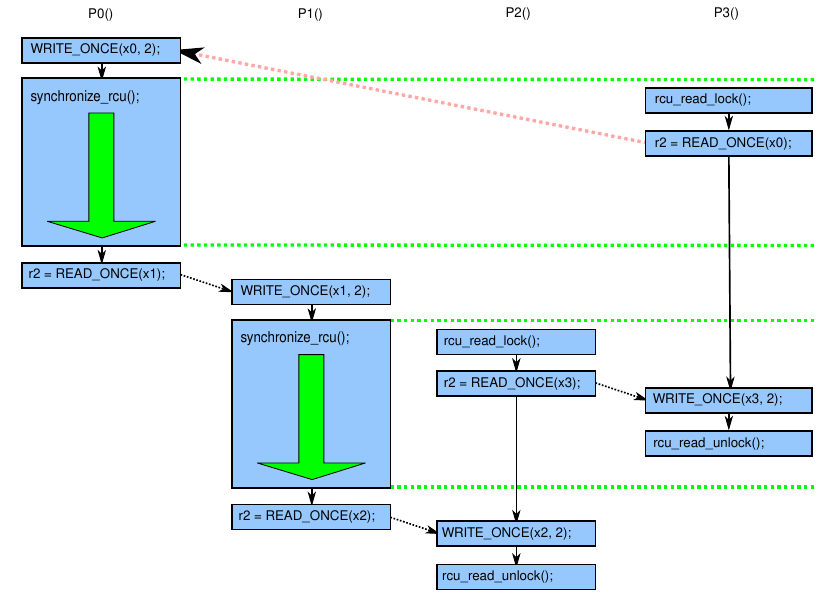}}
\caption{No Cycle for Two RCU Grace Periods and Two RCU Readers}
\label{fig:memorder:No Cycle for Two RCU Grace Periods and Two RCU Readers}
\end{figure*}

The key point is that even strongly ordered CPUs such as x86 can
and will reorder \co{P1()}'s and \co{P2()}'s \co{WRITE_ONCE()} and
\co{READ_ONCE()}.
With that reordering,
\cref{fig:memorder:Cycle for One RCU Grace Period and Two RCU Readers}
shows how the cycle forms:

\begin{enumerate}
\item	\co{P0()}'s read from \co{x1} precedes \co{P1()}'s write, as
	depicted by the dashed arrow near the bottom of the diagram.
\item	Because \co{P1()}'s write follows the end of \co{P0()}'s grace period,
	\co{P1()}'s read from \co{x2} cannot precede the beginning of
	\co{P0()}'s grace period.
\item	\co{P1()}'s read from \co{x2} precedes \co{P2()}'s write.
\item	Because \co{P2()}'s write to \co{x2} precedes the end of
	\co{P0()}'s grace period, it is completely legal for \co{P2()}'s
	read from \co{x0} to precede the beginning of \co{P0()}'s grace period.
\item	Therefore, \co{P2()}'s read from \co{x0} can precede \co{P0()}'s
	write, thus allowing the cycle to form.
\end{enumerate}

But what happens when another grace period is added?
This situation is shown in
\cref{lst:memorder:Two RCU Grace Periods and Two Readers},
an SB litmus test in which \co{P0()} and \co{P1()} have RCU grace periods
and \co{P2()} and \co{P3()} have RCU readers.
Again, the CPUs can reorder the accesses within RCU read-side critical
sections, as shown in
\cref{fig:memorder:No Cycle for Two RCU Grace Periods and Two RCU Readers}.
For this cycle to form, \co{P2()}'s critical section must
end after \co{P1()}'s grace period and \co{P3()}'s must end after the
beginning of that same grace period, which happens to also be after the
end of \co{P0()}'s grace period.
Therefore, \co{P3()}'s critical section must start after the beginning
of \co{P0()}'s grace period, which in turn means that \co{P3()}'s
read from \co{x0} cannot possibly precede \co{P0()}'s write.
Therefore, the cycle is forbidden because RCU read-side critical sections
cannot span full RCU grace periods.

However, a closer look at
\cref{fig:memorder:No Cycle for Two RCU Grace Periods and Two RCU Readers}
makes it clear that adding a third reader would allow the cycle.
This is because this third reader could end before the end of \co{P0()}'s
grace period, and thus start before the beginning of that same grace
period.
This in turn suggests the general rule, which is:
In these sorts of RCU-only litmus tests, if there are at least as many
RCU grace periods as there are RCU read-side critical sections,
the cycle is forbidden.\footnote{
	Interestingly enough, Alan Stern proved that within the context
	of LKMM, the two-part fundamental property of RCU expressed
	in \cref{sec:defer:RCU Fundamentals} actually implies
	this seemingly more general result, which is called the RCU
	axiom~\cite{Alglave:2018:FSC:3173162.3177156}.}

\subsubsection{RCU and Other Ordering Mechanisms}
\label{sec:memorder:RCU and Other Ordering Mechanisms}

But what about litmus tests that combine RCU with other ordering
mechanisms?

The general rule is that it takes only one mechanism to forbid a cycle.

For example, refer back to
\cref{lst:memorder:RCU Readers Provide No Lock-Like Ordering}.
Applying the general rule from the previous section, because this litmus
test has two RCU read-side critical sections and no RCU grace periods,
the cycle is allowed.
But what if \co{P0()}'s \co{WRITE_ONCE()} is replaced by an
\co{smp_store_release()} and \co{P1()}'s \co{READ_ONCE()} is replaced
by an \co{smp_load_acquire()}?

RCU would still allow the cycle, but the release-acquire pair would
forbid it.
Because it only takes one mechanism to forbid a cycle, the release-acquire
pair would prevail, thus forbidding the cycle.

\begin{figure*}
\centering
\resizebox{0.75\onecolumntextwidth}{!}{\includegraphics{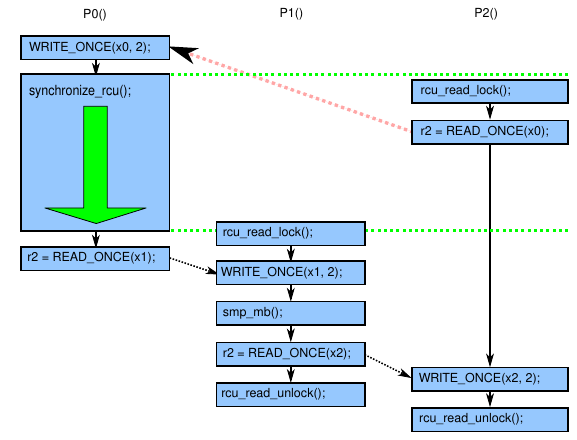}}
\caption{Cycle for One RCU Grace Period, Two RCU Readers, and Memory Barrier}
\label{fig:memorder:Cycle for One RCU Grace Period; Two RCU Readers; and Memory Barrier}
\end{figure*}

For another example, refer back to
\cref{lst:memorder:One RCU Grace Period and Two Readers}.
Because this litmus test has two RCU readers but only one grace period,
its cycle is allowed.
But suppose that an \co{smp_mb()} was placed between \co{P1()}'s
pair of accesses.
In this new litmus test, because of the addition of the \co{smp_mb()},
\co{P2()}'s as well as \co{P1()}'s critical sections would extend beyond the
end of \co{P0()}'s grace period, which in turn would prevent \co{P2()}'s
read from \co{x0} from preceding \co{P0()}'s write, as depicted by the
red dashed arrow in
\cref{fig:memorder:Cycle for One RCU Grace Period; Two RCU Readers; and Memory Barrier}.
In this case, RCU and the \IXh{full}{memory barrier} work together to forbid
the cycle, with RCU preserving ordering between \co{P0()} and both
\co{P1()} and \co{P2()}, and with the \co{smp_mb()} preserving
ordering between \co{P1()} and \co{P2()}.

\QuickQuiz{
	What would happen if the \co{smp_mb()} was instead added between
	\co{P2()}'s accesses in
	\cref{lst:memorder:One RCU Grace Period and Two Readers}?
}\QuickQuizAnswer{
	The cycle would again be forbidden.
	Further analysis is left as an exercise for the reader.
}\QuickQuizEnd

In short, where RCU's semantics were once purely pragmatic, they are
now fully
formalized~\cite{PaulMcKenney2005RCUSemantics,MathieuDesnoyers2012URCU,AlexeyGotsman2013ESOPRCU,Alglave:2018:FSC:3173162.3177156}.

% \subsection{SRCU}
% \label{sec:memorder:SRCU}
% @@@ After LWN article

% Nesting vs. value passed from \co{srcu_read_lock()} to
% \co{srcu_read_unlock()}.

% When augmented by \co{smp_mb__after_srcu_read_unlock()}.

\subsection{Higher-Level Primitives:
	    Discussion}
\label{sec:memorder:Higher-Level Primitives: Discussion}

It is quite beneficial to verify code in terms of a higher-level primitive
instead of in terms of the low-level memory accesses used in a particular
implementation of that primitive.
First, this allows code using`those primitives to be
verified against an abstract representation of those primitives,
thus making that code less vulnerable to implementation changes.
Second, partitioning the verification at API boundaries results in
combinatorial implosion, greatly reducing the overhead of formal
verification.

It is hoped that verifying against detailed semantics for higher-level
primitives will greatly increase the effectiveness of static analysis
and model checking.

\section{Hardware Specifics}
\label{sec:memorder:Hardware Specifics}
\OriginallyPublished{Section}{sec:memorder:Hardware Specifics}{Memory-Barrier Instructions For Specific CPUs}{Linux Journal}{PaulMcKenney2005i,PaulMcKenney2005j}
\epigraph{Rock beats paper!}{Derek Williams}

Each CPU family has its own peculiar approach to memory ordering, which
can make portability a challenge, as you can see in
\cref{tab:memorder:Summary of Memory Ordering}.

\begin{table*}[tb] % @@@ Omitting 'p' prevents unordered floats in 2c builds
\rowcolors{4}{}{lightgray}
\small
\centering
\newcommand{\cpufml}[1]{\begin{picture}(6,50)(0,0)\rotatebox{90}{#1}\end{picture}}
\renewcommand*{\arraystretch}{1.2}\OneColumnHSpace{-.35in}
\ebresizewidth{
\begin{tabular}{llccccccccc}
	\toprule
	\multicolumn{2}{l}{~} & \multicolumn{9}{c}{CPU Family} \\
	\cmidrule{3-11}
	\multicolumn{2}{c}{\raisebox{.5ex}{Property}}
	& \cpufml{Alpha}
	& \cpufml{\ARMv7-A/R}
	& \cpufml{\ARMv8}
	& \cpufml{Itanium}
	& \cpufml{MIPS}
	& \cpufml{\Power{}}
	& \cpufml{SPARC TSO}
	& \cpufml{x86}
	& \cpufml{z~Systems}
	\\
	\cmidrule(r){1-2} \cmidrule{3-11}
%		 Alpha ARMv8 ARMv7 Itanium MIPS PPC SPARC x86 z Systems
\cellcolor{white}
	Memory Ordering
	& Loads Reordered After Loads or Stores?
		 & Y   & Y   & Y   & Y     & Y  & Y & ~   & ~ & ~ \\
	& Stores Reordered After Stores?
		 & Y   & Y   & Y   & Y     & Y  & Y & ~   & ~ & ~ \\
\cellcolor{white}
	& Stores Reordered After Loads?
		 & Y   & Y   & Y   & Y     & Y  & Y & Y   & Y & Y \\
	& \parbox[c][6ex]{2in}{\raggedright Atomic Instructions Reordered With\par Loads or Stores?}
		 & Y   & Y   & Y   & ~     & Y  & Y & ~   & ~ & ~ \\
\cellcolor{white}
	& Dependent Loads Reordered?
		 & Y   & ~   & ~   & ~     & ~  & ~ & ~   & ~ & ~ \\
	& Dependent Stores Reordered?
		 & ~   & ~   & ~   & ~     & ~  & ~ & ~   & ~ & ~ \\
\cellcolor{white}
	& Non-Sequentially Consistent?
		 & Y   & Y   & Y   & Y     & Y  & Y & Y   & Y & Y \\
	& Non-Multicopy Atomic?
		 & Y   & Y   & Y   & Y     & Y  & Y & Y   & Y & ~ \\
\cellcolor{white}
	& Non-Other-Multicopy Atomic?
		 & Y   & Y   & ~   & Y     & Y  & Y & ~   & ~ & ~ \\
	& Non-Cache Coherent?
		 & ~   & ~   & ~   & Y     & ~  & ~ & ~   & ~ & ~ \\
	\cmidrule(r){1-2} \cmidrule{3-11}
\cellcolor{white}
	Instructions
	& Load-Acquire/Store-Release?
		 & F   & F   & i   & I     & F  & b & ~   & ~ & ~ \\
	& Atomic RMW Instruction Type?
		 & L   & L   & L   & C     & L  & L & C   & C & C \\
\cellcolor{white}
	& Incoherent Instruction Cache/Pipeline?
		 & Y   & Y   & Y   & Y     & Y  & Y & Y   & Y & Y \\
	\bottomrule
\end{tabular}
}

\vspace{5pt}\hfill
\ebresizeverb{.6}{
\framebox[\width]{\footnotesize\setlength{\tabcolsep}{3pt}
\rowcolors{1}{}{}
\renewcommand*{\arraystretch}{1}
\begin{tabular}{llcl}
	{ \bf Key: }
	  & \multicolumn{3}{l}{Load-Acquire/Store-Release?} \\
	~ & ~ & b: & Lightweight memory barrier \\
	~ & ~ & F: & Full memory barrier \\
	~ & ~ & i: & Instruction with lightweight ordering \\
	~ & ~ & I: & Instruction with heavyweight ordering \\
	~ &\multicolumn{3}{l}{Atomic RMW Instruction Type?} \\
	~ & ~ & C: & Compare-and-exchange instruction \\
	~ & ~ & L: & Load-linked/store-conditional instruction \\
\end{tabular}
}
}\OneColumnHSpace{-0.4in}
\caption{Summary of Memory Ordering}
\label{tab:memorder:Summary of Memory Ordering}
\end{table*}

In fact, some software environments simply prohibit
direct use of memory-ordering operations, restricting the programmer
to mutual-exclusion primitives that incorporate them to the extent that
they are required.
Please note that this section is not intended to be a reference manual
covering all (or even most) aspects of each CPU family, but rather
a high-level overview providing a rough comparison.
For full details, see the reference manual for the CPU of interest.

Getting back to
\cref{tab:memorder:Summary of Memory Ordering},
the first group of rows look at memory-ordering
properties and the second group looks at instruction properties.
Please note that these properties hold at the machine-instruction
level.
Compilers can and do reorder far more aggressively than does hardware.
Use marked accesses such as \co{READ_ONCE()} and \co{WRITE_ONCE()}
to constrain the compiler's optimizations and prevent undesireable
reordering.

The first three rows indicate whether a given CPU allows the four
possible combinations of loads and stores to be reordered, as discussed
in
\cref{sec:memorder:Ordering: Why and How?} and
\crefrange{sec:memorder:Load Followed By Load}{sec:memorder:Store Followed By Store}.
The next row (``Atomic Instructions Reordered With Loads or Stores?\@'')
indicates whether a given CPU allows loads and stores
to be reordered with atomic instructions.

The fifth and sixth rows cover reordering and dependencies,
which was covered in
\crefrange{sec:memorder:Address Dependencies}{sec:memorder:Control Dependencies}
and which is explained in more detail in
\cref{sec:memorder:Alpha}.
The short version is that Alpha requires memory barriers for readers
as well as updaters of linked data structures, however, these memory
barriers are provided by the Alpha architecture-specific code in
v4.15 and later Linux kernels.

The next row, ``Non-Sequentially Consistent'', indicates whether
the CPU's normal load and store instructions are constrained by
sequential consistency.
Performance considerations have dictated that no modern mainstream
system is sequentially consistent.

The next three rows cover multicopy atomicity, which was defined in
\cref{sec:memorder:Multicopy Atomicity}.
The first is full-up (and rare) multicopy atomicity, the second is the
weaker other-multicopy atomicity, and the third is the weakest
non-multicopy atomicity.

The next row, ``Non-Cache Coherent'', covers accesses from multiple
threads to a single variable, which was discussed in
\cref{sec:memorder:Cache Coherence}.

The final three rows cover instruction-level choices and issues.
The first row indicates how each CPU implements load-acquire
and store-release, the second row classifies CPUs by atomic-instruction
type, and the third and final row
indicates whether a given CPU has an incoherent
instruction cache and pipeline.
Such CPUs require special instructions be executed for self-modifying
code.

%Parenthesized CPU names indicate modes that are architecturally allowed,
%but rarely used in practice.

The common ``just say no'' approach to memory-ordering operations
can be eminently reasonable where it applies,
but there are environments, such as the Linux kernel, where direct
use of memory-ordering operations is required.
Therefore,
Linux provides a carefully chosen least-common-denominator
set of memory-ordering primitives, which are as follows:
\begin{description}
\item	[\tco{smp_mb()}] (\IXh{full}{memory barrier}) that orders both loads and
	stores.
	This means that loads and stores preceding the memory barrier
	will be committed to memory before any loads and stores
	following the memory barrier.
\item	[\tco{smp_rmb()}] (\IXh{read}{memory barrier}) that orders only loads.
\item	[\tco{smp_wmb()}] (\IXh{write}{memory barrier}) that orders only stores.
\item	[\tco{smp_mb__before_atomic()}] that forces ordering of accesses
	preceding the \co{smp_mb__before_atomic()} against accesses following
	a later RMW atomic operation.
	This is a noop on systems that fully order atomic RMW operations.
\item	[\tco{smp_mb__after_atomic()}] that forces ordering of accesses
	preceding an earlier RMW atomic operation against accesses
	following the \co{smp_mb__after_atomic()}.
	This is also a noop on systems that fully order atomic RMW operations.
\item	[\tco{smp_mb__after_spinlock()}] that forces ordering of accesses
	preceding a lock acquisition against accesses
	following the \co{smp_mb__after_spinlock()}.
	This is also a noop on systems that fully order lock acquisitions.
\item	[\tco{mmiowb()}] that forces ordering on MMIO writes that are guarded
	by global spinlocks, and is more thoroughly described
	in a 2016 LWN article on MMIO~\cite{PaulEMcKenney2016LinuxKernelMMIO}.
\end{description}
The \co{smp_mb()}, \co{smp_rmb()}, and \co{smp_wmb()}
primitives also force
the compiler to eschew any optimizations that would have the effect
of reordering memory optimizations across the barriers.

\QuickQuiz{
	What happens to code between an atomic operation and an
	\co{smp_mb__after_atomic()}?
}\QuickQuizAnswer{
	First, please don't do this!

	But if you do, this intervening code will either be ordered
	after the atomic operation or before the
	\co{smp_mb__after_atomic()}, depending on the architecture,
	but not both.
	This also applies to \co{smp_mb__before_atomic()} and
	\co{smp_mb__after_spinlock()}, that is, both the uncertain
	ordering of the intervening code and the plea to avoid such code.
}\QuickQuizEnd

These primitives generate code only in SMP kernels, however, several
have UP versions ({\tt mb()}, {\tt rmb()}, and {\tt wmb()},
respectively) that generate a memory barrier even in UP kernels.
The \co{smp_} versions should be used in most cases.
However, these latter primitives are useful when writing drivers,
because MMIO accesses must remain ordered even in UP kernels.
In absence of memory-ordering operations, both CPUs and compilers would
happily rearrange these accesses, which at best would make the device
act strangely, and could crash your kernel or even damage your hardware.

So most kernel programmers need not worry about the memory-ordering
peculiarities of each and every CPU, as long as they stick to these
interfaces and to the fully ordered atomic operations.\footnote{
	For a full list, expand the patterns in
	\path{Documentation/atomic_t.txt}.}
If you are working deep in a given CPU's architecture-specific code,
of course, all bets are off.

Furthermore,
all of Linux's locking primitives (spinlocks, reader-writer locks,
semaphores, RCU, \ldots) include any needed ordering primitives.
So if you are working with code that uses these primitives properly,
you need not worry about Linux's memory-ordering primitives.

That said, deep knowledge of each CPU's memory-consistency model
can be very helpful when debugging, to say nothing of when writing
architecture-specific code or synchronization primitives.

Besides, they say that a little knowledge is a very dangerous thing.
Just imagine the damage you could do with a lot of knowledge!
For those who wish to understand more about individual CPUs'
\IXh{memory}{consistency} models, the next sections describe those of a few
popular and prominent CPUs.
Although there is no substitute for actually reading a given CPU's
documentation, these sections do give a good overview.

\subsection{Alpha}
\label{sec:memorder:Alpha}

It may seem strange to say much of anything about a CPU whose end of life
has long since passed, but Alpha is interesting because it is the only
mainstream CPU that reorders dependent loads, and has thus had outsized
influence on concurrency APIs, including within the Linux kernel.
The need for core Linux-kernel code to accommodate Alpha ended
with version v4.15 of the Linux kernel, and all traces of this
accommodation were removed in v5.9 with the removal of the
\co{smp_read_barrier_depends()} and \co{read_barrier_depends()} APIs.
This section is nevertheless retained in the Third Edition
because here in early 2023 there are still a few Linux kernel hackers
still working on pre-v4.15 versions of the Linux kernel.
In addition, the modifications to \co{READ_ONCE()} that permitted
these APIs to be removed have not necessarily propagated to all
userspace projects that might still support Alpha.

\begin{fcvref}[ln:memorder:Insert and Lock-Free Search (No Ordering)]
The dependent-load difference between Alpha and the other CPUs is
illustrated by the code shown in
\cref{lst:memorder:Insert and Lock-Free Search (No Ordering)}.
This \co{smp_store_release()}
guarantees that the element initialization
in \clnrefrange{init:b}{init:e} is executed before the element is added to the
list on \clnref{add}, so that the lock-free search will work correctly.
That is, it makes this guarantee on all CPUs {\em except} Alpha.
\end{fcvref}

\begin{listing}
\begin{fcvlabel}[ln:memorder:Insert and Lock-Free Search (No Ordering)]
\begin{VerbatimL}[commandchars=\\\[\]]
struct el *insert(long key, long data)
{
	struct el *p;
	p = kmalloc(sizeof(*p), GFP_ATOMIC);
	spin_lock(&mutex);
	p->next = head.next;		\lnlbl[init:b]
	p->key = key;
	p->data = data;			\lnlbl[init:e]
	smp_store_release(&head.next, p); \lnlbl[add]
	spin_unlock(&mutex);
}

struct el *search(long searchkey)
{
	struct el *p;
	p = READ_ONCE_OLD(head.next);	\lnlbl[h:next]
	while (p != &head) {
		/* Prior to v4.15, BUG ON ALPHA!!! */ \lnlbl[BUG]
		if (p->key == searchkey) {	\lnlbl[key]
			return (p);
		}
		p = READ_ONCE_OLD(p->next);	\lnlbl[next]
	};
	return (NULL);
}
\end{VerbatimL}
\end{fcvlabel}
\caption{Insert and Lock-Free Search (No Ordering)}
\label{lst:memorder:Insert and Lock-Free Search (No Ordering)}
\end{listing}

\begin{figure}
\centering
\resizebox{\twocolumnwidth}{!}{\includegraphics{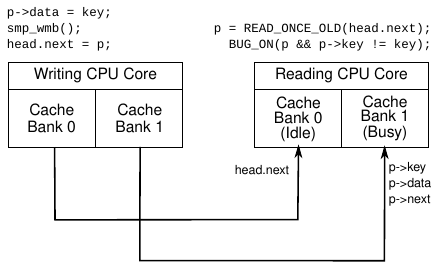}}
\caption{Why \tco{smp_read_barrier_depends()} is Required in Pre-v4.15 Linux Kernels}
\label{fig:memorder:fig:memorder:Why smp-read-barrier-depends() is Required in Pre-v4.15 Linux Kernels}
\end{figure}

\begin{fcvref}[ln:memorder:Insert and Lock-Free Search (No Ordering)]
Given the pre-v4.15 implementation of \co{READ_ONCE()}, indicated by
\co{READ_ONCE_OLD()} in the listing, Alpha actually allows the code on
\clnref{key} of
\cref{lst:memorder:Insert and Lock-Free Search (No Ordering)}
to see the old garbage values that were present before the initialization
on \clnrefrange{init:b}{init:e}.

\Cref{fig:memorder:fig:memorder:Why smp-read-barrier-depends() is Required in Pre-v4.15 Linux Kernels}
shows how this can happen on
an aggressively parallel machine with partitioned caches, so that
alternating \IXpl{cache line} are processed by the different partitions
of the caches.
For example, the load of \co{head.next} on \clnref{h:next} of
\cref{lst:memorder:Insert and Lock-Free Search (No Ordering)}
might access cache bank~0,
and the load of \co{p->key} on \clnref{key} and of \co{p->next} on \clnref{next}
might access cache bank~1.
On Alpha, the \co{smp_store_release()} will guarantee that the cache
invalidations performed by \clnrefrange{init:b}{init:e} of
\cref{lst:memorder:Insert and Lock-Free Search (No Ordering)}
(for \co{p->next}, \co{p->key}, and \co{p->data}) will reach
the interconnect before that of \clnref{add} (for \co{head.next}), but
makes absolutely no guarantee about the order of
propagation through the reading CPU's cache banks.
For example, it is possible that the reading CPU's cache bank~1 is very
busy, but cache bank~0 is idle.
This could result in the cache invalidations for the new element
(\co{p->next}, \co{p->key}, and \co{p->data}) being
delayed, so that the reading CPU loads the new value for \co{head.next},
but loads the old cached values for \co{p->key} and \co{p->next}.
Yes, this does mean that Alpha can in effect fetch
the data pointed to {\em before} it fetches the pointer itself, strange
but true.
\end{fcvref}
See the documentation~\cite{Compaq01,WilliamPugh2000Gharachorloo}
called out earlier for more information,
or if you think that I am just making all this up.\footnote{
	Of course, the astute reader will have already recognized that
	Alpha is nowhere near as mean and nasty as it could be,
	the (thankfully) mythical architecture in
	\cref{sec:app:whymb:Ordering-Hostile Architecture}
	being a case in point.}
The benefit of this unusual approach to ordering is that Alpha can use
simpler cache hardware, which in turn permitted higher clock frequencies
in Alpha's heyday.

One could place an \co{smp_rmb()} primitive
between the pointer fetch and dereference in order to force Alpha
to order the pointer fetch with the later dependent load.
However, this imposes unneeded overhead on systems (such as \ARM,
Itanium, PPC, and SPARC) that respect data dependencies on the read side.
A \co{smp_read_barrier_depends()} primitive was therefore added to the
Linux kernel to eliminate overhead on these systems, but was removed
in v5.9 of the Linux kernel in favor of augmenting Alpha's definition
of \co{READ_ONCE()}.
Thus, as of v5.9, core kernel code no longer needs to concern itself
with this aspect of DEC Alpha.
\begin{fcvref}[ln:memorder:Insert and Lock-Free Search (No Ordering)]
\end{fcvref}
\begin{fcvref}[ln:memorder:Safe Insert and Lock-Free Search]
However, it is better to use \co{rcu_dereference()}
as shown on \clnref{deref1,deref2} of
\cref{lst:memorder:Safe Insert and Lock-Free Search},
which works safely and efficiently for all recent kernel versions.
\end{fcvref}

It is also possible to implement a software mechanism
that could be used in place of \co{smp_store_release()} to force
all reading CPUs to see the writing CPU's writes in order.
This software barrier could be implemented by sending \IXacrfpl{ipi}
to all other CPUs.
Upon receipt of such an IPI, a CPU would execute a memory-barrier
instruction, implementing a system-wide memory barrier similar to that
provided by the Linux kernel's \co{sys_membarrier()} system call.
Additional logic is required to avoid deadlocks.
Of course, CPUs that respect data dependencies would define such a barrier
to simply be \co{smp_store_release()}.
However, this approach was deemed by the Linux community
to impose excessive overhead~\cite{McKenney01f}, and to their point would
be completely inappropriate for systems having
aggressive real-time response requirements.

\begin{listing}
\begin{fcvlabel}[ln:memorder:Safe Insert and Lock-Free Search]
\begin{VerbatimL}[commandchars=\\\[\]]
struct el *insert(long key, long data)
{
	struct el *p;
	p = kmalloc(sizeof(*p), GFP_ATOMIC);
	spin_lock(&mutex);
	p->next = head.next;
	p->key = key;
	p->data = data;
	smp_store_release(&head.next, p);
	spin_unlock(&mutex);
}

struct el *search(long searchkey)
{
	struct el *p;
	p = rcu_dereference(head.next);		\lnlbl[deref1]
	while (p != &head) {
		if (p->key == searchkey) {
			return (p);
		}
		p = rcu_dereference(p->next);	\lnlbl[deref2]
	};
	return (NULL);
}
\end{VerbatimL}
\end{fcvlabel}
\caption{Safe Insert and Lock-Free Search}
\label{lst:memorder:Safe Insert and Lock-Free Search}
\end{listing}

The Linux memory-barrier primitives took their names from the Alpha
instructions, so \co{smp_mb()} is {\tt mb}, \co{smp_rmb()} is {\tt rmb},
and \co{smp_wmb()} is {\tt wmb}.
Alpha is the only CPU whose \co{READ_ONCE()} includes an \co{smp_mb()}.

\QuickQuizSeries{%
\QuickQuizB{
	Why does Alpha's \co{READ_ONCE()} include an
	\co{mb()} rather than \co{rmb()}?
}\QuickQuizAnswerB{
	Alpha has only \co{mb} and \co{wmb} instructions,
	so \co{smp_rmb()} would be implemented by the Alpha \co{mb}
	instruction in either case.
	In addition, at the time that the Linux kernel started relying on
	dependency ordering, it was not clear that Alpha ordered dependent
	stores, and thus \co{smp_mb()} was therefore the safe choice.

	However, given the aforementioned v5.9 changes to \co{READ_ONCE()}
	and a few of Alpha's atomic read-modify-write operations,
	no Linux-kernel core code need concern itself with DEC Alpha,
	thus greatly reducing Paul E.~McKenney's incentive to remove
	Alpha support from the kernel.
}\QuickQuizEndB
\QuickQuizE{
	Isn't DEC Alpha significant as having the weakest possible
	memory ordering?
}\QuickQuizAnswerE{
	Although DEC Alpha does take considerable flak, it does avoid
	reordering reads from the same CPU to the same variable.
	It also avoids the out-of-thin-air problem that plagues
	the Java and C11 memory
	models~\cite{Boehm:2014:OGA:2618128.2618134,conf/esop/BattyMNPS15,MarkBatty2013OOTA-WorkingNote,HansBoehm2020ConcurrentUB,DavidGoldblatt2019NoElegantOOTAfix,AlanJeffrey2014JavaDRF,PaulEMcKenney2020RelaxedGuideRelaxed,PaulEMcKenney2016OOTA,Sevcik:2011:SOS:1993316.1993534,Vafeiadis:2015:CCO:2775051.2676995}.
}\QuickQuizEndE
}

For more on Alpha, see its reference manual~\cite{ALPHA2002}.

\subsection{\ARMv7-A/R}
\label{sec:memorder:ARMv7-A/R}

The \ARM\ family of CPUs is popular in deep embedded applications,
particularly for power-constrained microcontrollers.
Its memory model is similar to that of \Power{}
(see \cref{sec:memorder:POWER / PowerPC}), but \ARM\ uses a
different set of memory-barrier instructions~\cite{ARMv7A:2010}:

\begin{description}
\item	[\tco{DMB}] (data memory barrier) causes the specified type of
	operations to \emph{appear} to have completed before any
	subsequent operations of the same type.
	The ``type'' of operations can be all operations or can be
	restricted to only writes (similar to the Alpha \co{wmb}
	and the \Power{} \co{eieio} instructions).
	In addition, \ARM\ allows \IX{cache coherence} to have one of three
	scopes:
	Single processor, a subset of the processors
	(``inner'') and global (``outer'').
\item	[\tco{DSB}] (data synchronization barrier) causes the specified
	type of operations to actually complete before any subsequent
	operations (of any type) are executed.
	The ``type'' of operations is the same as that of \co{DMB}.
	The \co{DSB} instruction was called \co{DWB} (drain write buffer
	or data write barrier, your choice) in early versions of the
	\ARM\ architecture.
\item	[\tco{ISB}] (instruction synchronization barrier) flushes the CPU
	pipeline, so that all instructions following the \co{ISB}
	are fetched only after the \co{ISB} completes.
	For example, if you are writing a self-modifying program
	(such as a JIT), you should execute an \co{ISB} between
	generating the code and executing it.
\end{description}

None of these instructions exactly match the semantics of Linux's
\co{rmb()} primitive, which must therefore be implemented as a full
\co{DMB}.
The \co{DMB} and \co{DSB} instructions have a recursive definition
of accesses ordered before and after the barrier, which has an effect
similar to that of \Power{}'s cumulativity, both of which are
stronger than \IXacr{lkmm}'s cumulativity described in
\cref{sec:memorder:Cumulativity}.

\ARM\ also implements control dependencies, so that if a conditional
branch depends on a load, then any store executed after that conditional
branch will be ordered after the load.
However, loads following the conditional branch will \emph{not}
be guaranteed to be ordered unless there is an \co{ISB}
instruction between the branch and the load.
Consider the following example:

\begin{fcvlabel}[ln:memorder:ARM:load-store control dependency]
\begin{VerbatimN}[commandchars=\\\[\]]
r1 = x;			\lnlbl[x]
if (r1 == 0)		\lnlbl[if]
	nop();		\lnlbl[nop]
y = 1;			\lnlbl[y]
r2 = z;			\lnlbl[z1]
ISB();			\lnlbl[isb]
r3 = z;			\lnlbl[z2]
\end{VerbatimN}
\end{fcvlabel}

\begin{fcvref}[ln:memorder:ARM:load-store control dependency]
In this example, load-store control dependency ordering causes
the load from \co{x} on \clnref{x} to be ordered before the store to
\co{y} on \clnref{y}.
However, \ARM\ does not respect load-load control dependencies, so that
the load on \clnref{x} might well happen \emph{after} the
load on \clnref{z1}.
On the other hand, the combination of the conditional branch on \clnref{if}
and the \co{ISB} instruction on \clnref{isb} ensures that
the load on \clnref{z2} happens after the load on \clnref{x}.
Note that inserting an additional \co{ISB} instruction somewhere between
\clnref{if,z1} would enforce ordering between \clnref{x,z1}.
\end{fcvref}

\subsection{\ARMv8}
\label{sec:memorder:ARMv8}

\begin{figure}
\centering
\resizebox{2in}{!}{\includegraphics{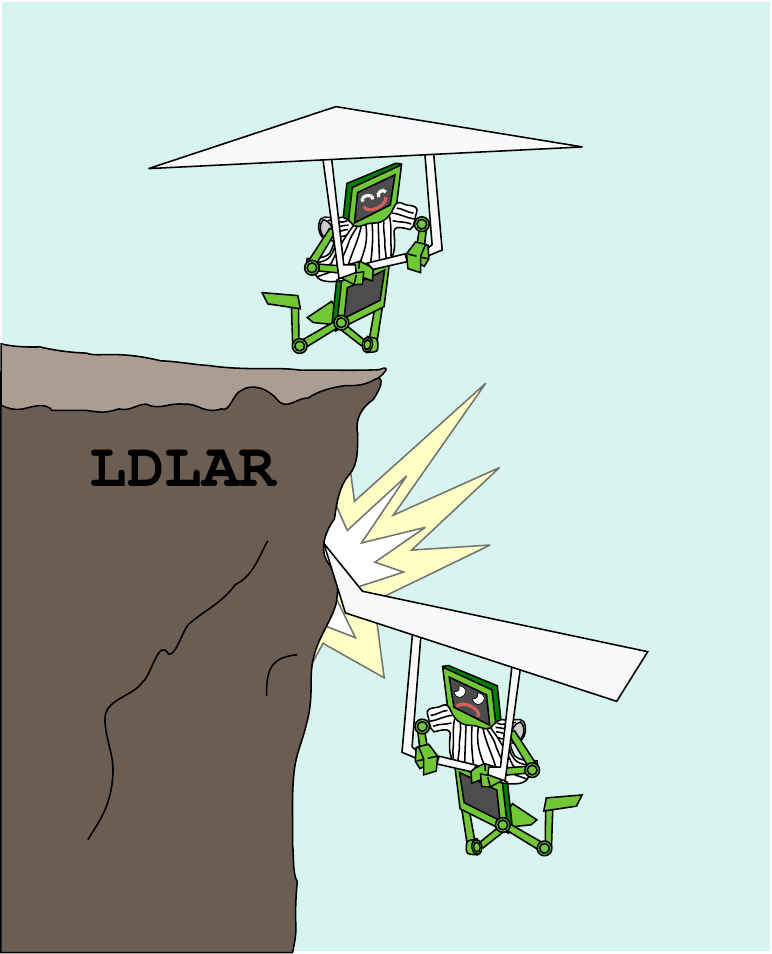}}
\caption{Half Memory Barrier}
\ContributedBy{Figure}{fig:memorder:Half Memory Barrier}{Melissa Brossard}
\end{figure}

\ARM's \ARMv8 CPU family~\cite{ARMv8A:2017}
includes 64-bit capabilities,
in contrast to their 32-bit-only CPU described in
\cref{sec:memorder:ARMv7-A/R}.
\ARMv8's memory model closely resembles its \ARMv7 counterpart,
but adds load-acquire (\co{LDLARB}, \co{LDLARH}, and \co{LDLAR})
and store-release (\co{STLLRB}, \co{STLLRH}, and \co{STLLR})
instructions.
These instructions act as ``half memory barriers'', so that
\ARMv8 CPUs can reorder previous accesses with a later \co{LDLAR}
instruction, but are prohibited from reordering an earlier \co{LDLAR}
instruction with later accesses, as fancifully depicted in
\cref{fig:memorder:Half Memory Barrier}.
Similarly, \ARMv8 CPUs can reorder an earlier \co{STLLR} instruction with
a subsequent access, but are prohibited from reordering
previous accesses with a later \co{STLLR} instruction.
As one might expect, this means that these instructions directly support
the C11 notion of load-acquire and store-release.

However, \ARMv8 goes well beyond the C11 memory model by mandating that
the combination of a store-release and load-acquire act as a full
barrier under certain circumstances.
For example, in \ARMv8, given a store followed by a store-release followed
a load-acquire followed by a load, all to different variables and all from
a single CPU, all CPUs
would agree that the initial store preceded the final load.
Interestingly enough, most TSO architectures (including x86 and the
mainframe) do not make this guarantee, as the two loads could be
reordered before the two stores.

\ARMv8 is one of only two architectures that needs the
\co{smp_mb__after_spinlock()} primitive to be a full barrier,
due to its relatively weak lock-acquisition implementation in
the Linux kernel.

\ARMv8 also has the distinction of being the first CPU whose vendor publicly
defined its memory ordering with an executable formal model~\cite{ARMv8A:2017}.

\subsection{Itanium}
\label{sec:memorder:Itanium}

Itanium offers a \IXh{weak}{consistency}
model, so that in absence of explicit
memory-barrier instructions or dependencies, Itanium is within its rights
to arbitrarily reorder memory references~\cite{IntelItanium02v2}.
Itanium has a memory-fence instruction named {\tt mf}, but also has
``half-memory fence'' modifiers to loads, stores, and to some of its atomic
instructions~\cite{IntelItanium02v3}.
The {\tt acq} modifier prevents subsequent memory-reference instructions
from being reordered before the {\tt acq}, but permits
prior memory-reference instructions to be reordered after the {\tt acq},
similar to the \ARMv8 load-acquire instructions.
Similarly, the {\tt rel} modifier prevents prior memory-reference
instructions from being reordered after the {\tt rel}, but allows
subsequent memory-reference instructions to be reordered before
the {\tt rel}.

These half-memory fences are useful for critical sections, since
it is safe to push operations into a critical section, but can be
fatal to allow them to bleed out.
However, as one of the few CPUs with this property, Itanium at one
time defined Linux's semantics of memory ordering associated with lock
acquisition and release.\footnote{
	PowerPC is now the architecture with this dubious privilege.}
Oddly enough, actual Itanium hardware is rumored to implement
both load-acquire and store-release instructions as full barriers.
Nevertheless, Itanium was the first mainstream CPU to introduce the concept
(if not the reality) of load-acquire and store-release into its
instruction set.

\QuickQuiz{
	Given that hardware can have a half memory barrier, why don't
	locking primitives allow the compiler to move memory-reference
	instructions into lock-based critical sections?
}\QuickQuizAnswer{
	In fact, as we saw in \cref{sec:memorder:ARMv8} and will
	see in \cref{sec:memorder:POWER / PowerPC}, hardware really does
	implement partial memory-ordering instructions and it also turns
	out that these really are used to construct locking primitives.
	However, these locking primitives use full compiler barriers,
	thus preventing the compiler from reordering memory-reference
	instructions both out of and into the corresponding critical
	section.

\begin{listing}
\begin{fcvlabel}[ln:memorder:synchronize-rcu]
\begin{VerbatimL}[commandchars=\@\[\]]
static inline int rcu_gp_ongoing(unsigned long *ctr)
{
	unsigned long v;

	v = LOAD_SHARED(*ctr);@lnlbl[load]
	return v && (v != rcu_gp_ctr);
}

static void update_counter_and_wait(void)
{
	struct rcu_reader *index;

	STORE_SHARED(rcu_gp_ctr, rcu_gp_ctr + RCU_GP_CTR);
	barrier();
	list_for_each_entry(index, &registry, node) {@lnlbl[loop]
		while (rcu_gp_ongoing(&index->ctr))@lnlbl[call2]
			msleep(10);
	}
}

void synchronize_rcu(void)
{
	unsigned long was_online;

	was_online = rcu_reader.ctr;
	smp_mb();
	if (was_online)@lnlbl[if]
		STORE_SHARED(rcu_reader.ctr, 0);@lnlbl[store]
	mutex_lock(&rcu_gp_lock);@lnlbl[acqmutex]
	update_counter_and_wait();@lnlbl[call1]
	mutex_unlock(&rcu_gp_lock);
	if (was_online)
		STORE_SHARED(rcu_reader.ctr, LOAD_SHARED(rcu_gp_ctr));
	smp_mb();
}
\end{VerbatimL}
\end{fcvlabel}
\caption{Userspace RCU Code Reordering}
\label{lst:memorder:Userspace RCU Code Reordering}
\end{listing}

	To see why the compiler is forbidden from doing reordering that
	is permitted by hardware, consider the following sample code
	in \cref{lst:memorder:Userspace RCU Code Reordering}.
	This code is based on the userspace RCU update-side
	code~\cite[Supplementary Materials Figure 5]{MathieuDesnoyers2012URCU}.

\begin{fcvref}[ln:memorder:synchronize-rcu]
	Suppose that the compiler reordered \clnref{if,store} into
	the critical section starting at \clnref{acqmutex}.
	Now suppose that two updaters start executing \co{synchronize_rcu()}
	at about the same time.
	Then consider the following sequence of events:
	\begin{enumerate}
	\item	CPU~0 acquires the lock at \clnref{acqmutex}.
	\item	\Clnref{if} determines that CPU~0 was online, so it clears
		its own counter at \clnref{store}.
		(Recall that \clnref{if,store} have been reordered by the
		compiler to follow \clnref{acqmutex}).
	\item	CPU~0 invokes \co{update_counter_and_wait()} from
		\clnref{call1}.
	\item	CPU~0 invokes \co{rcu_gp_ongoing()} on itself at
		\clnref{call2}, and \clnref{load} sees that CPU~0 is
		in a quiescent state.
		Control therefore returns to \co{update_counter_and_wait()},
		and \clnref{loop} advances to CPU~1.
	\item	CPU~1 invokes \co{synchronize_rcu()}, but because CPU~0
		already holds the lock, CPU~1 blocks waiting for this
		lock to become available.
		Because the compiler reordered \clnref{if,store} to follow
		\clnref{acqmutex}, CPU~1 does not clear its own counter,
		despite having been online.
	\item	CPU~0 invokes \co{rcu_gp_ongoing()} on CPU~1 at
		\clnref{call2}, and \clnref{load} sees that CPU~1 is
		not in a quiescent state.
		The \co{while} loop at \clnref{call2} therefore never
		exits.
	\end{enumerate}

	So the compiler's reordering results in a deadlock.
	In contrast, hardware reordering is temporary, so that CPU~1
	might undertake its first attempt to acquire the mutex on
	\clnref{acqmutex} before executing \clnref{if,store}, but it
	will eventually execute \clnref{if,store}.
	Because hardware reordering only results in a short delay, it
	can be tolerated.
	On the other hand, because compiler reordering results in a
	deadlock, it must be prohibited.

	Some research efforts have used hardware transactional memory
	to allow compilers to safely reorder more aggressively, but
	the overhead of hardware transactions has thus far made
	such optimizations unattractive.
	% @@@ Citation for compilers use of HTM in this manner?
\end{fcvref}
}\QuickQuizEnd

The Itanium {\tt mf} instruction is used for the \co{smp_rmb()},
\co{smp_mb()}, and \co{smp_wmb()} primitives in the Linux kernel.
Despite persistent rumors to the contrary, the \qco{mf} mnemonic stands
for ``memory fence''.

Itanium also offers a global total order for release operations,
including the \co{mf} instruction.
This provides the notion of transitivity, where if a given code fragment
sees a given access as having happened, any later code fragment will
also see that earlier access as having happened.
Assuming, that is, that all the code fragments involved correctly use
memory barriers.

Finally, Itanium is the only architecture supporting the Linux kernel
that can reorder normal loads to the same variable.
The Linux kernel avoids this issue because \co{READ_ONCE()} emits
a \co{volatile} load, which is compiled as a \co{ld,acq} instruction,
which forces ordering of all \co{READ_ONCE()} invocations by a given
CPU, including those to the same variable.

\subsection{MIPS}

The MIPS memory model~\cite[page~479]{MIPSvII-A-2016}
appears to resemble that of \ARM, Itanium, and \Power{},
being weakly ordered by default, but respecting dependencies.
MIPS has a wide variety of memory-barrier instructions, but ties them
not to hardware considerations, but rather to the use cases provided
by the Linux kernel and the C++11 standard~\cite{RichardSmith2019N4800}
in a manner similar to the \ARMv8 additions:

\begin{description}[style=nextline]
\item[\tco{SYNC}]
	Full barrier for a number of hardware operations in addition
	to memory references, which is used to implement the v4.13
	Linux kernel's \co{smp_mb()} for OCTEON systems.
\item[\tco{SYNC_WMB}]
	Write memory barrier, which can be used on OCTEON systems
	to implement the
	\co{smp_wmb()} primitive in the v4.13 Linux kernel via the
	\co{syncw} mnemonic.
	Other systems use plain \co{sync}.
\item[\tco{SYNC_MB}]
	Full memory barrier, but only for memory operations.
	This may be used to implement the
	C++ \co{atomic_thread_fence(memory_order_seq_cst)}.
\item[\tco{SYNC_ACQUIRE}]
	Acquire memory barrier, which could be used to implement
	C++'s \co{atomic_thread_fence(memory_order_acquire)}.
	In theory, it could also be used to implement the v4.13 Linux-kernel
	\co{smp_load_acquire()} primitive, but in practice
	\co{sync} is used instead.
\item[\tco{SYNC_RELEASE}]
	Release memory barrier, which may be used to implement
	C++'s \co{atomic_thread_fence(memory_order_release)}.
	In theory, it could also be used to implement the v4.13 Linux-kernel
	\co{smp_store_release()} primitive, but in practice
	\co{sync} is used instead.
\item[\tco{SYNC_RMB}]
	Read memory barrier, which could in theory be used to implement the
	\co{smp_rmb()} primitive in the Linux kernel, except that current
	MIPS implementations supported by the v4.13 Linux kernel do not
	need an explicit instruction to force ordering.
	Therefore, \co{smp_rmb()} instead simply constrains the compiler.
\item[\tco{SYNCI}]
	Instruction-cache synchronization, which is used in conjunction with
	other instructions to allow self-modifying code, such as that produced
	by just-in-time (JIT) compilers.
\end{description}

Informal discussions with MIPS architects indicates that MIPS has a
definition of transitivity or cumulativity similar to that of
\ARM\ and \Power{}\@.
However, it appears that different MIPS implementations can have
different memory-ordering properties, so it is important to consult
the documentation for the specific MIPS implementation you are using.

\subsection{\Power{} / PowerPC}
\label{sec:memorder:POWER / PowerPC}

The \Power{} and PowerPC CPU families have a wide variety of memory-barrier
instructions~\cite{PowerPC94,MichaelLyons05a}:
\begin{description}
\item	[\tco{sync}] causes all preceding operations to {\em appear to have}
	completed before any subsequent operations are started.
	This instruction is therefore quite expensive.
\item	[\tco{lwsync}] (lightweight sync) orders loads with respect to
	subsequent loads and stores, and also orders stores.
	However, it does {\em not} order stores with respect to subsequent
	loads.
	The \co{lwsync} instruction may be used to implement
	load-acquire and store-release operations.
	Interestingly enough, the {\tt lwsync} instruction enforces
	the same within-CPU ordering as does x86, z~Systems, and coincidentally,
	SPARC TSO\@.
	However, placing the \co{lwsync} instruction between each
	pair of memory-reference instructions will \emph{not}
	result in x86, z~Systems, or SPARC TSO memory ordering.
	On these other systems, if a pair of CPUs independently execute
	stores to different variables, all other CPUs will agree on the
	order of these stores.
	Not so on PowerPC, even with an \co{lwsync} instruction between each
	pair of memory-reference instructions, because PowerPC is
	non-multicopy atomic.
\item	[\tco{eieio}] (enforce in-order execution of I/O, in case you
	were wondering) causes all preceding cacheable stores to appear
	to have completed before all subsequent stores.
	However, stores to cacheable memory are ordered separately from
	stores to non-cacheable memory, which means that {\tt eieio}
	will not force an MMIO store to precede a spinlock release.
	This instruction may well be unique in having a five-vowel mnemonic.
\item	[\tco{isync}] forces all preceding instructions to appear to have
	completed before any subsequent instructions start execution.
	This means that the preceding instructions must have progressed
	far enough that any traps they might generate have either happened
	or are guaranteed not to happen, and that any side-effects of
	these instructions (for example, page-table changes) are seen by the
	subsequent instructions.
	However, it does \emph{not} force all memory references to be
	ordered, only the actual execution of the instruction itself.
	Thus, the loads might return old still-cached values and the
	\co{isync} instruction does not force values previously stored
	to be flushed from the store buffers.
\end{description}

Unfortunately, none of these instructions line up exactly with Linux's
\co{wmb()} primitive, which requires \emph{all} stores to be ordered,
but does not require the other high-overhead actions of the \co{sync}
instruction.
The \co{rmb()} primitive doesn't have a matching light-weight instruction
either.
But there is no choice:
{ppc64} versions of \co{wmb()}, \co{rmb()}, and \co{mb()} are defined
to be the heavyweight \co{sync} instruction.
However, Linux's \co{smp_wmb()} primitive is never used for MMIO
(since a driver must carefully order MMIOs in UP as well as
SMP kernels, after all), so it is defined to be the lighter weight
\co{eieio} or \co{lwsync} instruction~\cite{PaulEMcKenney2016LinuxKernelMMIO}.
The \co{smp_mb()} primitive is also defined to be the \co{sync}
instruction, while \co{smp_rmb()} is defined to be the lighter-weight
\co{lwsync} instruction.

\Power{} features ``cumulativity'', which can be used to obtain
transitivity.
When used properly, any code seeing the results of an earlier
code fragment will also see the accesses that this earlier code
fragment itself saw.
Much more detail is available from
McKenney and Silvera~\cite{PaulEMcKenneyN2745r2009}.

\Power{} respects control dependencies in much the same way that \ARM\
does, with the exception that the \Power{} \co{isync} instruction
is substituted for the \ARM\ \co{ISB} instruction.

Like \ARMv8, \Power{} requires \co{smp_mb__after_spinlock()} to be
a full memory barrier.
In addition, \Power{} is the only architecture requiring
\co{smp_mb__after_unlock_lock()} to be a full memory barrier.
In both cases, this is because of the weak ordering properties
of \Power{}'s locking primitives, due to the use of the \co{lwsync}
instruction to provide ordering for both acquisition and release.

Many members of the \Power{} architecture have incoherent instruction
caches, so that a store to memory will not necessarily be reflected
in the instruction cache.
Thankfully, few people write self-modifying code these days, but JITs
and compilers do it all the time.
Furthermore, recompiling a recently run program looks just like
self-modifying code from the CPU's viewpoint.
The {\tt icbi} instruction (instruction cache block invalidate)
invalidates a specified cache line from
the instruction cache, and may be used in these situations.

\subsection{SPARC TSO}

Although SPARC's TSO (total-store order) is used by both Linux and
Solaris, the architecture also defines PSO (partial store order) and RMO
(relaxed-memory order).
Any program that runs in RMO will also run in either PSO or TSO, and similarly,
a program that runs in PSO will also run in TSO\@.
Moving a shared-memory parallel program in the other direction may
require careful insertion of memory barriers.

Although SPARC's PSO and RMO modes are not used much these days, they
did give rise to a very flexible memory-barrier instruction~\cite{SPARC94}
that permits fine-grained control of ordering:
\begin{description}
\item	[\tco{StoreStore}] orders preceding stores before subsequent stores.
	(This option is used by the Linux \co{smp_wmb()} primitive.)
\item	[\tco{LoadStore}] orders preceding loads before subsequent stores.
\item	[\tco{StoreLoad}] orders preceding stores before subsequent loads.
\item	[\tco{LoadLoad}] orders preceding loads before subsequent loads.
	(This option is used by the Linux \co{smp_rmb()} primitive.)
\item	[\tco{Sync}] fully completes all preceding operations before starting
	any subsequent operations.
\item	[\tco{MemIssue}] completes preceding memory operations before subsequent
	memory operations, important for some instances of memory-mapped
	I/O.
\item	[\tco{Lookaside}] does the same as MemIssue,
	but only applies to preceding stores
	and subsequent loads, and even then only for stores and loads that
	access the same memory location.
\end{description}

So, why is \qco{membar #MemIssue} needed?
Because a \qco{membar #StoreLoad} could permit a subsequent
load to get its value from a store buffer, which would be
disastrous if the write was to an MMIO register that induced side effects
on the value to be read.
In contrast, \qco{membar #MemIssue} would wait until the store buffers
were flushed before permitting the loads to execute,
thereby ensuring that the load actually gets its value from the MMIO register.
Drivers could instead use \qco{membar #Sync}, but the lighter-weight
\qco{membar #MemIssue} is preferred in cases where the additional function
of the more-expensive \qco{membar #Sync} are not required.

The \qco{membar #Lookaside} is a lighter-weight version of
\qco{membar #MemIssue}, which is useful when writing to a given MMIO register
affects the value that will next be read from that register.
However, the heavier-weight \qco{membar #MemIssue} must be used when
a write to a given MMIO register affects the value that will next be
read from {\em some other} MMIO register.

SPARC requires a {\tt flush} instruction be used between the time that
the instruction stream is modified and the time that any of these
instructions are executed~\cite{SPARC94}.
This is needed to flush any prior value for that location from
the SPARC's instruction cache.
Note that {\tt flush} takes an address, and will flush only that address
from the instruction cache.
On SMP systems, all CPUs' caches are flushed, but there is no
convenient way to determine when the off-CPU flushes complete,
though there is a reference to an implementation note.

But again, the Linux kernel runs SPARC in TSO mode, so
all of the above \co{membar} variants are strictly of historical
interest.
In particular, the \co{smp_mb()} primitive only needs to use \co{#StoreLoad}
because the other three reorderings are prohibited by TSO\@.

\subsection{x86}

Historically, the x86 CPUs provided ``process ordering'' so that all CPUs
agreed on the order of a given CPU's writes to memory.
This allowed the \co{smp_wmb()}
primitive to be a no-op for the CPU~\cite{IntelXeonV3-96a}.
Of course, a compiler directive was also required to prevent optimizations
that would reorder across the \co{smp_wmb()} primitive.
In ancient times, certain x86 CPUs gave no ordering guarantees for loads, so
the \co{smp_mb()} and \co{smp_rmb()} primitives expanded to {\tt lock;addl}.
This atomic instruction acts as a barrier to both loads and stores.

But those were ancient times.
More recently, Intel has published a memory model for
x86~\cite{Intelx86MemoryOrdering2007}.
It turns out that Intel's modern CPUs enforce tighter ordering than was
claimed in the previous specifications, so this model simply mandates
this modern behavior.
Even more recently, Intel published an updated memory model for
x86~\cite[Section 8.2]{Intel64IA32v3A2011}, which mandates a total global order
for stores, although individual CPUs are still permitted to see their
own stores as having happened earlier than this total global order
would indicate.
This exception to the total ordering is needed to allow important
hardware optimizations involving store buffers.
In addition, x86 provides other-multicopy atomicity, for example,
so that if CPU~0 sees a store by CPU~1, then CPU~0 is guaranteed to see
all stores that CPU~1 saw prior to its store.
Software may use atomic operations to override these hardware optimizations,
which is one reason that atomic operations tend to be more expensive
than their non-atomic counterparts.

It is also important to note that atomic instructions operating
on a given memory location should all be of the same
size~\cite[Section 8.1.2.2]{Intel64IA32v3A2016}.
For example, if you write a program where one CPU atomically increments
a byte while another CPU executes a 4-byte atomic increment on
that same location, you are on your own.

Some SSE instructions are weakly ordered ({\tt clflush}
and non-temporal move instructions~\cite{IntelXeonV2b-96a}).
Code that uses these non-temporal move instructions
can also use {\tt mfence} for \co{smp_mb()},
{\tt lfence} for \co{smp_rmb()}, and {\tt sfence} for \co{smp_wmb()}.
A few older variants of the x86 CPU have a mode bit that enables out-of-order
stores, and for these CPUs, \co{smp_wmb()} must also be defined to
be {\tt lock;addl}.

Although newer x86 implementations accommodate self-modifying code
without any special instructions, to be fully compatible with
past and potential future x86 implementations, a given CPU must
execute a jump instruction or a serializing instruction (e.g., \co{cpuid})
between modifying the code and executing
it~\cite[Section 8.1.3]{Intel64IA32v3A2011}.

\subsection{z Systems}

The z~Systems machines make up the IBM mainframe family, previously
known as the 360, 370, 390 and zSeries~\cite{IBMzSeries04a}.
Parallelism came late to z~Systems, but given that these mainframes first
shipped in the mid 1960s, this is not saying much.
The \qco{bcr 15,0} instruction is used for the Linux \co{smp_mb()} primitives,
but compiler constraints suffices for both the
\co{smp_rmb()} and \co{smp_wmb()} primitives.
It also has strong memory-ordering semantics, as shown in
\cref{tab:memorder:Summary of Memory Ordering}.
In particular, all CPUs will agree on the order of unrelated stores from
different CPUs, that is, the z~Systems CPU family is fully multicopy
atomic, and is the only commercially available system with this property.

As with most CPUs, the z~Systems architecture does not guarantee a
cache-coherent instruction stream, hence,
self-modifying code must execute a serializing instruction between updating
the instructions and executing them.
That said, many actual z~Systems machines do in fact accommodate self-modifying
code without serializing instructions.
The z~Systems instruction set provides a large set of serializing instructions,
including compare-and-swap, some types of branches (for example, the
aforementioned \qco{bcr 15,0} instruction), and test-and-set.

\subsection{Hardware Specifics:
	    Discussion}
\label{sec:memorder:Hardware Specifics: Discussion}

There is considerable variation among these CPU families, and this section
only scratched the surface of a few families that are either heavily used
or historically significant.
Those wishing more detail are invited to consult the reference manuals.

But a big benefit of the Linux-kernel memory model is that you can
ignore these details when writing architecture-independent Linux-kernel
code.

\section{Memory-Model Intuitions}
\label{sec:memorder:Memory-Model Intuitions}
\epigraph{Almost all people are intelligent.
	  It is method that they lack.}
	 {F. W. Nichol}

This section revisits
\cref{tab:memorder:Linux-Kernel Memory-Ordering Cheat Sheet}
and \cref{sec:memorder:Basic Rules of Thumb},
summarizing the intervening discussion with some appeals to transitive
intuitions and with more sophisticated rules of thumb.

But first, it is necessary to review the temporal and non-temporal
nature of communication from one thread to another when using memory
as the communications medium, as was discussed in detail in
\cref{sec:memorder:Multicopy Atomicity}.
The key point is that although loads and stores are conceptually simple,
on real multicore hardware significant periods of time are required for
their effects to become visible to all other threads.

The simple and intuitive case occurs when one thread loads a value that
some other thread stored.
This straightforward cause-and-effect case exhibits temporal behavior, so
that the software can safely assume that the store instruction completed
before the load instruction started.
In real life, the load instruction might well have started quite some
time before the store instruction did, but all modern hardware must
carefully hide such cases from the software.
Software will thus see the expected temporal cause-and-effect
behavior when one thread loads a value that some other thread stores,
as discussed in \cref{sec:memorder:Happens-Before}.

This temporal behavior provides the basis for the next section's
transitive intuitions.

\subsection{Transitive Intuitions}
\label{sec:memorder:Transitive Intuitions}

This section summarizes intuitions regarding single threads or variables,
locking, release-acquire chains, RCU, and fully ordered code.

\subsubsection{Singular Intuitive Bliss}
\label{sec:memorder:Singular Intuitive Bliss}

A program that has only one variable or only one thread will see
all accesses in order.
There is quite a bit of code that can attain adequate performance
when running single-threaded on modern computer systems, but
this book is primarily about software that needs multiple CPUs.
On, then, to the next section.

\subsubsection{Locking Intuitions}
\label{sec:memorder:Locking Intuitions}

Another transitive intuition involves that much-maligned workhorse,
locking, described in more detail in
\cref{sec:memorder:Locking},
to say nothing of
\cref{chp:Locking}.
This section contains a graphical description followed by a verbal
description.

\begin{figure*}
\centering
\resizebox{\textwidth}{!}{\includegraphics{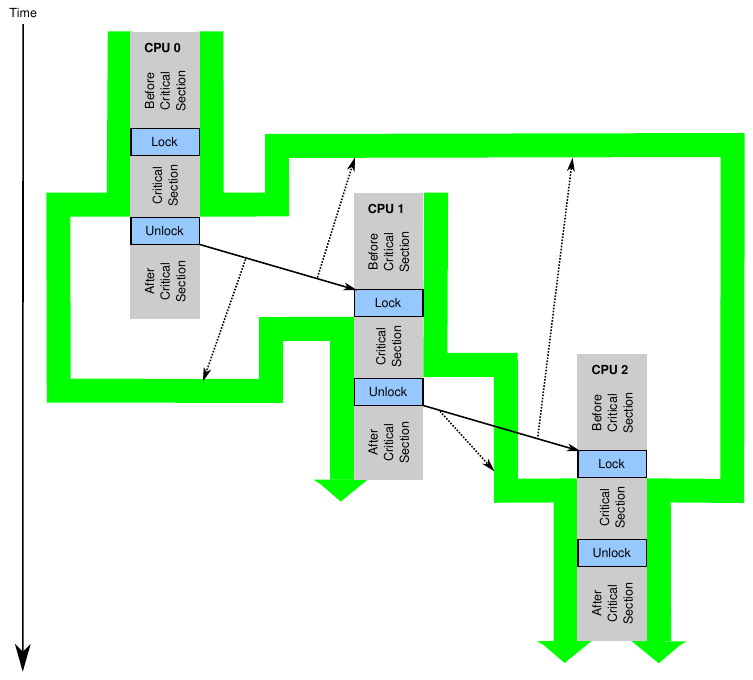}}
\caption{Locking Intuitions}
\label{fig:memorder:Locking Intuitions}
\end{figure*}

The graphical description is shown in
\cref{fig:memorder:Locking Intuitions},
which shows a lock being acquired and released by CPUs~0, 1, and~2
in that order.
The solid black arrows depict the unlock-lock ordering.
The dotted lines emanating from them to the wide green arrows
show the effects on ordering.
In particular:

\begin{enumerate}
\item	The fact that CPU~0's unlock precedes CPU~1's lock ensures that
	any access executed by CPU~0 within or before its critical
	section will be seen by accesses executed by CPU~1 within
	and after its critical section.
\item	The fact that CPU~0's unlock precedes CPU~2's lock ensures that
	any access executed by CPU~0 within or before its critical
	section will be seen by accesses executed by CPU~2 within
	and after its critical section.
\item	The fact that CPU~1's unlock precedes CPU~2's lock ensures that
	any access executed by CPU~1 within or before its critical
	section will be seen by accesses executed by CPU~2 within and
	after its critical section.
\end{enumerate}

In short, lock-based ordering is transitive through CPUs~0, 1, and~2.
A key point is that this ordering extends beyond the critical sections,
so that everything before an earlier lock release is seen by everything
after a later lock acquisition.

For those who prefer words to diagrams, code holding a given lock will
see the accesses in all prior critical sections for that same lock,
transitively.
And if such code sees the accesses in a given critical section, it will
also see the accesses in all of that CPU's code preceding
that critical section.
In other words, when a CPU releases a given lock, all
of that lock's subsequent critical sections will see the accesses in
all of that CPU's code preceding that lock release.

Inversely, code holding a given lock will be protected from seeing the
accesses in any subsequent critical sections for that same lock, again,
transitively.
And if such code is protected against seeing the accesses in a given
critical section, it will also be protected against seeing the accesses
in all of that CPU's code following that critical section.
In other words, when a CPU acquires a given lock, all of
that lock's previous critical sections will be protected from seeing
the accesses in all of that CPU's code following that lock
acquisition.

But what does it mean to ``see accesses'' and exactly what accesses
are seen?

To start, an access is either a load or a store, possibly occurring as part
of a read-modify-write operation.

If a CPU's code prior to its release of a given lock contains
an access A to a given variable, then for an access B to that same variable
contained in any CPU's code following a later acquisition
of that same lock:

\begin{enumerate}
\item	If A and B are both loads, then B will return either the same
	value that A did or some later value.
\item	If A is a load and B is a store, then B will overwrite either the
	value loaded by A or some later value.
\item	If A is a store and B is a load, then B will return either the
	value stored by A or some later value.
\item	If A and B are both stores, then B will overwrite either the value
	stored by A or some later value.
\end{enumerate}

Here, ``some later value'' is shorthand for ``the value stored by some
intervening access''.

Locking is strongly intuitive, which is one reason why it has survived
so many attempts to eliminate it.
This is also one reason why you should use it where it applies.

\subsubsection{Release-Acquire Intuitions}
\label{sec:memorder:Release-Acquire Intuitions}

Release-acquire chains also behave in a transitively intuitive manner
not unlike that of locking.
This section also contains a graphical description followed by a verbal
description.

\begin{figure*}
\centering
\resizebox{\textwidth}{!}{\includegraphics{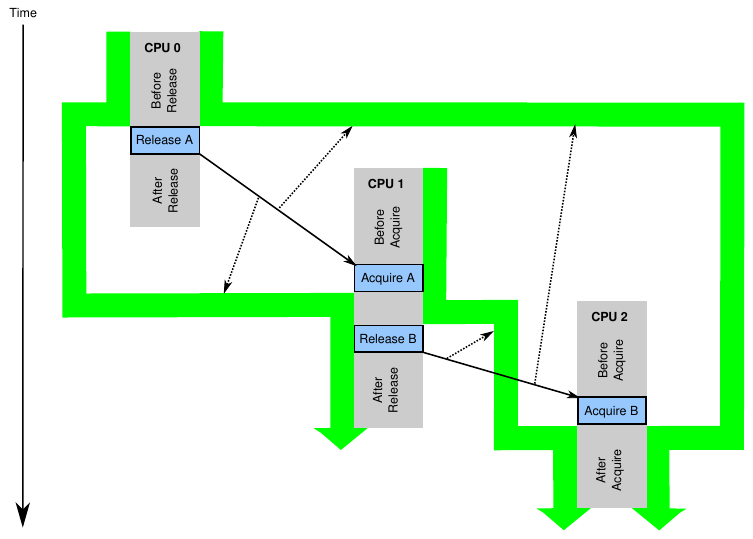}}
\caption{Release-Acquire Intuitions}
\label{fig:memorder:Release-Acquire Intuitions}
\end{figure*}

The graphical description is shown in
\cref{fig:memorder:Release-Acquire Intuitions},
which shows a release-acquire chain extending through CPUs~0, 1, and~2.
The solid black arrows depict the release-acquire ordering.
The dotted lines emanating from them to the wide green arrows show the
effects on ordering.

\begin{enumerate}
\item	The fact that CPU~0's release of A is read by CPU~1's acquire of A
	ensures that any accesses executed by CPU~0 prior to its release
	will be seen by any accesses executed by CPU~1 after its acquire.
\item	The fact that CPU~1's release of B is read by CPU~2's acquire of B
	ensures that any accesses executed by CPU~1 prior to its release
	will be seen by any accesses executed by CPU~2 after its acquire.
\item	Note also that CPU~0's release of A is read by CPU~1's acquire of
	A, which precedes CPU 1's release of B, which is read by CPU~2's
	acquire of B\@.
	Taken together, all this ensures that any accesses executed by
	CPU~0 prior to its release will be seen by any accesses executed
	by CPU~2 after its acquire.
\end{enumerate}

This illustrates that properly constructed release-acquire ordering is
transitive through CPUs~0, 1, and~2, and in fact may be extended through
as many CPUs as needed.\footnote{
	But please note that stray stores to either A or B can break
	the release-acquire chain, as illustrated by
	\cref{lst:memorder:A Release-Acquire Chain With Added Store (Ordering?)}.}

For those who prefer words to diagrams, when an acquire loads the value
stored by a release, discussed in
\cref{sec:memorder:Release-Acquire Chains},
then the code following that release will see all accesses preceding
the acquire.
More precisely, if CPU~0 does an acquire that loads the value stored by
CPU~1's release, than all the subsequent accesses executed by CPU~0 will
see the all of CPU~1's accesses prior to its release.

Similarly, the accesses preceding that release access will be protected
from seeing the accesses following the acquire access.
(More precision is left as an exercise to the reader.)

Releases and acquires can be chained, for example CPU~0's release
stores the value loaded by CPU~1's acquire, a later release by CPU~1
stores the value loaded by CPU~2's acquire, and so on.
The accesses following a given acquire will see the accesses preceding
each prior release in the chain, and, inversely, the accesses preceding a
given release will be protected from seeing the accesses following each
later acquire in the chain.
Some long-chain examples are illustrated by
\cref{lst:memorder:Long LB Release-Acquire Chain,lst:memorder:Long ISA2 Release-Acquire Chain,lst:memorder:Long Z6.2 Release-Acquire Chain}.

The seeing and not seeing of accesses works the same way as described in
\cref{sec:memorder:Locking Intuitions}.

However, as illustrated by
\cref{lst:memorder:A Release-Acquire Chain With Added Store (Ordering?)},
the acquire access must load exactly what was stored by the release access.
Any intervening store that is not itself part of that same release-acquire
chain will break the chain.

Nevertheless, properly constructed release-acquire chains are transitive,
intuitive, and useful.

\subsubsection{RCU Intuitions}
\label{sec:memorder:RCU Intuitions}

As noted in \cref{sec:defer:RCU Fundamentals} on
\cpageref{sec:defer:RCU Fundamentals}, RCU provides a number
of ordering guarantees.

The first is the publish-subscribe mechanism described in
\cref{sec:defer:Publish-Subscribe Mechanism}
on
\cpageref{sec:defer:Publish-Subscribe Mechanism}.
This resembles the acquire-release chains discussed in the previous
section, but substitutes a member of the \co{rcu_dereference()} family
of primitives for the \co{smp_load_acquire()}.
Unlike \co{smp_load_acquire()}, the ordering implied by
\co{rcu_dereference()} applies only to subsequent accesses that
dereference the pointer returned by that \co{rcu_dereference()}
as shown in
\cref{fig:defer:Publication/Subscription Constraints}
on
\cpageref{fig:defer:Publication/Subscription Constraints}.

The second guarantee says that if any part of an RCU read-side
critical section precedes the beginning of a grace period, then
the entirety of that critical section precedes the end of that
grace period, as shown in
\cref{fig:defer:RCU Reader and Later Grace Period}
on
\cpageref{fig:defer:RCU Reader and Later Grace Period}.

The third guarantee says that if any part of an RCU read-side critical
section follows the end of a grace period, then the entirety of that
critical section follows the beginning of that grace period, as shown in
\cref{fig:defer:RCU Reader and Earlier Grace Period}
on
\cpageref{fig:defer:RCU Reader and Earlier Grace Period}.

Both of these two guarantees are discussed in
\cref{sec:defer:Wait For Pre-Existing RCU Readers}
on
\cpageref{sec:defer:Wait For Pre-Existing RCU Readers},
with more examples shown in
\cref{fig:defer:RCU Reader Within Grace Period,fig:defer:Summary of RCU Grace-Period Ordering Guarantees}
on
\cpageref{fig:defer:RCU Reader Within Grace Period,fig:defer:Summary of RCU Grace-Period Ordering Guarantees}.
These two guarantees have further version-maintenance consequences that
are discussed in
\cref{sec:defer:Maintain Multiple Versions of Recently Updated Objects}
on
\cpageref{sec:defer:Maintain Multiple Versions of Recently Updated Objects}.

These guarantees are discussed somewhat more formally in
\cref{sec:memorder:RCU}.

Much of the sophistication of RCU lies not in its guarantees, but in its
use cases, which are the subject of
\cref{sec:defer:RCU Usage}
starting on
\cpageref{sec:defer:RCU Usage}.

\subsubsection{Fully Ordered Intuitions}
\label{sec:memorder:Fully Ordered Intuitions}

A more extreme example of transitivity places at least one \co{smp_mb()}
between each pair of accesses.
All accesses seen by any given access will also be seen by all later
accesses.

The resulting program will be fully ordered, if somewhat slow.
Such programs will be sequentially consistent and much loved by
formal-verification experts who specialize in tried-and-true 1980s
proof techniques.
But slow or not, \co{smp_mb()} is always there when you need it!

Nevertheless, there are situations that cannot be addressed by these
intuitive approaches.
The next section therefore presents a more complete, if less transitive,
set of rules of thumb.

\subsection{Rules of Thumb}
\label{sec:memorder:Rules of Thumb}

The transitive intuitions presented in the previous section are
very appealing, at least as memory models go.
Unfortunately, hardware is under no obligation to provide temporal
cause-and-effect illusions when one thread's store overwrites a value either
loaded or stored by some other thread.
It is quite possible that, from the software's viewpoint, an earlier store
will overwrite a later store's value, but only if those two stores were
executed by different threads, as illustrated by
\cref{fig:memorder:Store-to-Store is Counter-Temporal}.
Similarly, a later load might well read a value overwritten by an
earlier store, but again only if that load and store were executed by
different threads, as illustrated by
\cref{fig:memorder:Load-to-Store is Counter-Temporal}.
This counter-intuitive behavior occurs due to the need to buffer
stores in order to achieve adequate performance, as discussed in
\cref{sec:memorder:Propagation}.

As a result, situations where one thread reads a value written by some
other thread can make do with far weaker ordering than can situations
where one thread overwrites a value loaded or stored by some other thread.
These differences are captured by the following rules of thumb.

The first rule of thumb is that memory-ordering operations are only
required where there is a possibility of interaction between at least
two variables shared among at least two threads, which underlies the
singular intuitive bliss presented in
\cref{sec:memorder:Singular Intuitive Bliss}.
In light of the intervening material, this single sentence encapsulates much of
\cref{sec:memorder:Basic Rules of Thumb}'s basic rules of thumb,
for example, keeping in mind that ``memory-barrier pairing'' is a
two-thread special case of ``cycle''.
And, as always, if a single-threaded program will provide sufficient
performance, why bother with parallelism?\footnote{
	Hobbyists and researchers should of course feel free to ignore
	this and many other cautions.}
After all, avoiding parallelism also avoids the added cost and complexity
of memory-ordering operations.

The second rule of thumb involves load-buffering situations:
If all thread-to-thread communication in a given cycle use store-to-load
links (that is, the next thread's load returns the value stored by
the previous thread), minimal ordering suffices.
Minimal ordering includes dependencies and acquires as well as all stronger
ordering operations.
Because a lock acquisition must load the lock-word value stored by any
prior release of that lock, this rule of thumb underlies the locking
intuitions presented in
\cref{sec:memorder:Locking Intuitions}.

The third rule of thumb involves release-acquire chains:
If all but one of the links in a given cycle is a store-to-load
link, it is sufficient to use release-acquire pairs for each of
those store-to-load links, as illustrated by
\cref{lst:memorder:Long ISA2 Release-Acquire Chain,%
lst:memorder:Long Z6.2 Release-Acquire Chain}.
This rule underlies the release-acquire intuitions presented in
\cref{sec:memorder:Release-Acquire Intuitions}.

You can replace a given acquire with a dependency in environments permitting
this, keeping in mind that the C11 standard's memory model does \emph{not}
fully respect dependencies.
Therefore, a dependency leading to a load must be headed by
a \co{READ_ONCE()} or an \co{rcu_dereference()}:
A plain C-language load is not sufficient.
In addition, carefully review
\cref{sec:memorder:Address- and Data-Dependency Difficulties,%
sec:memorder:Control-Dependency Calamities}, because
a dependency broken by your compiler will not order anything.
The two threads sharing the sole non-store-to-load link can
sometimes substitute \co{WRITE_ONCE()} plus \co{smp_wmb()} for
\co{smp_store_release()} on the one hand,
and \co{READ_ONCE()} plus \co{smp_rmb()} for \co{smp_load_acquire()}
on the other.
However, the wise developer will check such substitutions carefully,
for example, using the herd tool as described in
\cref{sec:formal:Axiomatic Approaches}.

\QuickQuiz{
	Why is it necessary to use heavier-weight ordering for
	load-to-store and store-to-store links, but not for
	store-to-load links?
	What on earth makes store-to-load links so special???
}\QuickQuizAnswer{
	Recall that load-to-store and store-to-store links can be
	counter-temporal, as illustrated by
	\cref{fig:memorder:Load-to-Store is Counter-Temporal,%
	fig:memorder:Store-to-Store is Counter-Temporal} in
	\cref{sec:memorder:Propagation}.
	This counter-temporal nature of load-to-store and store-to-store
	links necessitates strong ordering.

	In constrast, store-to-load links are temporal, as illustrated by
	\cref{lst:memorder:Load-Buffering Data-Dependency Litmus Test,%
	lst:memorder:Load-Buffering Control-Dependency Litmus Test}.
	This temporal nature of store-to-load links permits use of
	minimal ordering.
}\QuickQuizEnd

The fourth and final rule of thumb identifies where full memory barriers
(or stronger) are required:
If a given cycle contains two or more non-store-to-load links (that is, a
total of two or more links that are either load-to-store or store-to-store
links), you will need at least one full barrier between each pair of
non-store-to-load links in that cycle, as illustrated by
\cref{lst:memorder:W+WRC Litmus Test With More Barriers}
as well as in the answer to
\QuickQuizARef{\MemorderQQLitmusTestR}.
Full barriers include \co{smp_mb()}, successful full-strength non-\co{void}
atomic RMW operations, and other atomic RMW operations in conjunction with
either \co{smp_mb__before_atomic()} or \co{smp_mb__after_atomic()}.
Any of RCU's grace-period-wait primitives (\co{synchronize_rcu()} and
friends) also act as full barriers, but at far greater expense than
\co{smp_mb()}.
With strength comes expense, though full barriers
usually hurt performance more than they hurt scalability.
The extreme logical endpoint of this rule of thumb underlies the
fully ordered intuitions presented in
\cref{sec:memorder:Fully Ordered Intuitions}.

Recapping the rules:

\begin{enumerate}
\item	Memory-ordering operations are required only if at least
	two variables are shared by at least two threads.
\item	If all links in a cycle are store-to-load links, then
	minimal ordering suffices.
\item	If all but one of the links in a cycle are store-to-load links,
	then each store-to-load link may use a release-acquire pair.
\item	Otherwise, at least one full barrier is required between
	each pair of non-store-to-load links.
\end{enumerate}

Note that an architecture is permitted to provide stronger guarantees, as
discussed in \cref{sec:memorder:Hardware Specifics}, but these guarantees
may only be relied upon in code that runs only for that architecture.
In addition, more accurate memory models~\cite{Alglave:2018:FSC:3173162.3177156}
may give stronger guarantees with lower-overhead operations than do
these rules of thumb, albeit at the expense of greater complexity.
In these more formal memory-ordering papers, a store-to-load link is an
example of a reads-from (rf) link, a load-to-store link is an example
of a from-reads (fr) link, and a store-to-store link is an example of
a coherence (co) link.

One final word of advice:
Use of raw memory-ordering primitives is a last resort.
It is almost always better to use existing primitives, such as locking
or RCU, thus letting those primitives do the memory ordering for you.

	\setlength{\parskip}{0.0pt plus 1ex}
	\input{qqzmemorder}
	\setlength{\parskip}{0.0pt plus 1.0pt}% return to default

% easy/easy.tex
% mainfile: ../perfbook.tex
% SPDX-License-Identifier: CC-BY-SA-3.0

\QuickQuizChapter{chp:Ease of Use}{Ease of Use}{qqzeasy}
\Epigraph{Creating a perfect API is like committing the perfect crime.
	  There are at least fifty things that can go wrong, and if you are
	  a genius, you might be able to anticipate twenty-five of them.}
	 {With apologies to any Kathleen Turner fans who might
	  still be alive.}

\section{What is Easy?}
\label{sec:easy:What is Easy?}
\epigraph{When someone says ``I want a programming language in which I
	  need only say what I wish done,'' give them a lollipop.}
	 {Alan J.~ Perlis, updated}
% http://www.cs.yale.edu/homes/perlis-alan/quotes.html

If you are tempted to look down on ease-of-use requirements, please
consider that an ease-of-use bug in Linux-kernel RCU resulted in an
exploitable Linux-kernel security bug in a use of
RCU~\cite{McKenney:2019:CRS:3319647.3325836}.
It is therefore clearly important that even in-kernel APIs be easy to use.

Unfortunately, ``easy'' is a relative term.
For example, many people would consider a 15-hour airplane flight to be
a bit of an ordeal---unless they stopped to consider alternative modes
of transportation, especially swimming.
This means that creating an easy-to-use API requires that you understand
your intended users well enough to know what is easy for them.
Which might or might not have anything to do with what is easy for you.

The following question illustrates this point:
``Given a randomly chosen person among everyone alive today, what one
change would improve that person's life?''

There is no single change that would be guaranteed to help everyone's life.
After all, there is an extremely wide range of people, with a correspondingly
wide range of needs, wants, desires, and aspirations.
A starving person might need food, but additional food might well hasten
the death of a morbidly obese person.
The high level of excitement so fervently desired by many young people
might well be fatal to someone recovering from a heart attack.
Information critical to the success of one person might contribute to
the failure of someone suffering from information overload.
In short, if you are working on a software project that is intended to
help people you know nothing about, you should not be surprised when
those people find fault with your project.

If you really want to help a given group of people, there is simply no
substitute for working closely with them over an extended period of time,
as in years.
Nevertheless, there are some simple things that you can do to increase
the odds of your users being happy with your software, and some of these
things are covered in the next section.

\section{Rusty Scale for API Design}
\label{sec:easy:Rusty Scale for API Design}
\epigraph{Finding the appropriate measurement is thus not a mathematical
	  exercise.
	  It is a risk-taking judgment.}
	 {Peter Drucker}
% http://billhennessy.com/simple-strategies/2015/09/09/i-wish-drucker-never-said-it
% Rusty is OK with this: July 19, 2006.

This section is adapted from portions of Rusty Russell's 2003 Ottawa Linux
Symposium keynote address~\cite[Slides~39--57]{RustyRussell2003OLSkeynote}.
Rusty's key point is that the goal should not be merely to make an API
easy to use, but rather to make the API hard to misuse.
To that end, Rusty proposed his ``Rusty Scale'' in decreasing order
of this important hard-to-misuse property.

The following list attempts to generalize the Rusty Scale beyond the
Linux kernel:

\begin{enumerate}
\item	It is impossible to get wrong.
	Although this is the standard to which all API designers should
	strive, only the mythical \co{dwim()}\footnote{
		The \co{dwim()} function is an acronym that expands to
		``do what I mean''.}
	command manages to come close.
\item	The compiler or linker won't let you get it wrong.
\item	The compiler or linker will warn you if you get it wrong.
	\co{BUILD_BUG_ON()} is your users' friend.
\item	The simplest use is the correct one.
\item	The name tells you how to use it.
	But names can be two-edged swords.
	Although \co{rcu_read_lock()} is plain enough for someone
	converting code from reader-writer locking, it might cause
	some consternation for someone converting code from
	reference counting.
\item	Do it right or it will always break at runtime.
	\co{WARN_ON_ONCE()} is your users' friend.
\item	Follow common convention and you will get it right.
	The \co{malloc()} library function is a good example.
	Although it is easy to get memory allocation wrong, a
	great many projects do manage to get it right, at least most
	of the time.
	Using \co{malloc()} in conjunction with
	Valgrind~\cite{ValgrindHomePage} moves \co{malloc()}
	almost up to the ``do it right or it will always break at runtime''
	point on the scale.
\item	Read the documentation and you will get it right.
\item	Read the implementation and you will get it right.
\item	Read the right mailing-list archive and you will get it right.
\item	Read the right mailing-list archive and you will get it wrong.
\item	Read the implementation and you will get it wrong.
	The original non-\co{CONFIG_PREEMPT} implementation of
	\co{rcu_read_lock()}~\cite{PaulEMcKenney2007PreemptibleRCU}
	is an infamous example of this point on the scale.
\item	Read the documentation and you will get it wrong.
	For example, the DEC Alpha \co{wmb} instruction's
	documentation~\cite{ALPHA2002} fooled a
	number of developers into thinking that this instruction
	had much stronger memory-order semantics than it actually does.
	Later documentation clarified this
	point~\cite{Compaq01,WilliamPugh2000Gharachorloo},
	moving the \co{wmb} instruction up to the
	``read the documentation and you will get it right'' point on
	the scale.
\item	Follow common convention and you will get it wrong.
	The \co{printf()} statement is an example of this point on the
	scale because
	developers almost always fail to check \co{printf()}'s error return.
\item	Do it right and it will break at runtime.
\item	The name tells you how not to use it.
\item	The obvious use is wrong.
	The Linux kernel \co{smp_mb()} function is an example of
	this point on the scale.
	Many developers assume that this function has much
	stronger ordering semantics than it actually possesses.
	\Cref{chp:Advanced Synchronization: Memory Ordering} contains the
	information needed to avoid this mistake, as does the
	Linux-kernel source tree's \path{Documentation} and
	\path{tools/memory-model} directories.
\item	The compiler or linker will warn you if you get it right.
\item	The compiler or linker won't let you get it right.
\item	It is impossible to get right.
	The \co{gets()} function is a famous example of this point on
	the scale.
	In fact, \co{gets()} can perhaps best be described as
	an unconditional buffer-overflow security hole.
\end{enumerate}

\section{Shaving the Mandelbrot Set}
\label{sec:easy:Shaving the Mandelbrot Set}
\epigraph{Simplicity does not precede complexity, \\ but follows it.}
	 {Alan J.~Perlis}

\begin{figure}
\centering
\resizebox{2.5in}{!}{\includegraphics{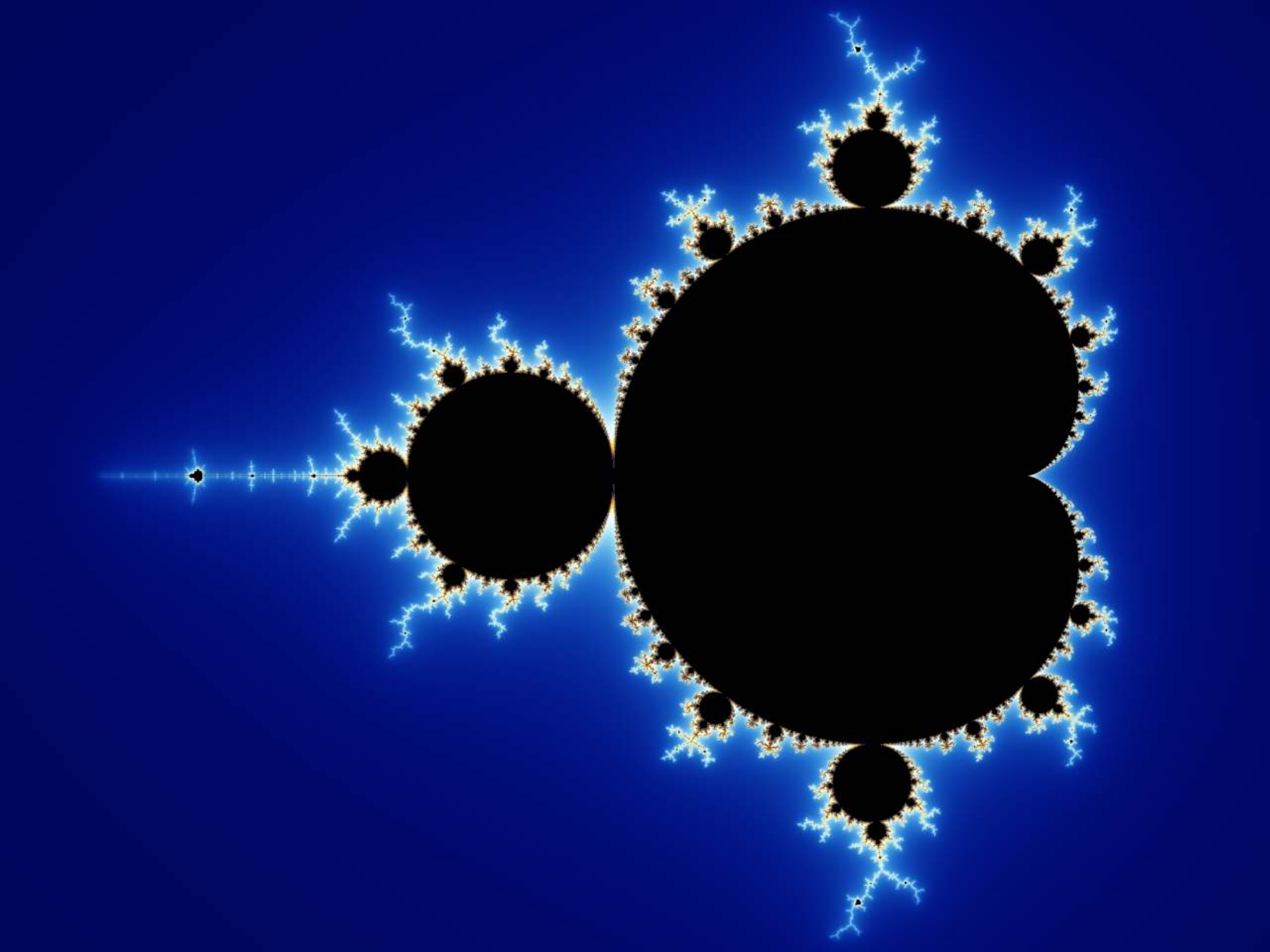}}
\caption{Mandelbrot Set (Courtesy of Wikipedia)}
\label{fig:easy:Mandelbrot Set}
\end{figure}

The set of useful programs resembles the Mandelbrot set
(shown in \cref{fig:easy:Mandelbrot Set})
in that it does
not have a clear-cut smooth boundary---if it did, the halting problem
would be solvable.
But we need APIs that real people can use, not ones that require a
Ph.D. dissertation be completed for each and every potential use.
So, we ``shave the Mandelbrot set'',\footnote{
	Due to Josh Triplett.}
restricting the use of the
API to an easily described subset of the full set of potential uses.

Such shaving may seem counterproductive.
After all, if an algorithm works, why shouldn't it be used?

To see why at least some shaving is absolutely necessary, consider
a locking design that avoids \IX{deadlock}, but in perhaps the worst possible way.
This design uses a circular doubly linked list, which contains one
element for each thread in the system along with a header element.
When a new thread is spawned, the parent thread must insert a new
element into this list, which requires some sort of synchronization.

One way to protect the list is to use a global lock.
However, this might be a bottleneck if threads were being created and
deleted frequently.\footnote{
	Those of you with strong operating-system backgrounds, please
	suspend disbelief.
	Those unable to suspend disbelief are encouraged to provide
	better examples.}
Another approach would be to use a hash table and to lock the individual
hash buckets, but this can perform poorly when scanning the list in order.

A third approach is to lock the individual list elements, and to require
the locks for both the predecessor and successor to be held during the
insertion.
Since both locks must be acquired, we need to decide which order to
acquire them in.
Two conventional approaches would be to acquire the locks in address
order, or to acquire them in the order that they appear in the list,
so that the header is always acquired first when it is one of the two
elements being locked.
However, both of these methods require special checks and branches.

The to-be-shaven solution is to unconditionally acquire the locks in
list order.
But what about deadlock?

Deadlock cannot occur.

To see this, number the elements in the list starting with zero for the
header up to $N$ for the last element in the list (the one preceding the
header, given that the list is circular).
Similarly, number the threads from zero to $N-1$.
If each thread attempts to lock some consecutive pair of elements,
at least one of the threads is guaranteed to be able to acquire both
locks.

Why?

Because there are not enough threads to reach all the way around the list.
Suppose thread~0 acquires element~0's lock.
To be blocked, some other thread must have already acquired element~1's
lock, so let us assume that thread~1 has done so.
Similarly, for thread~1 to be blocked, some other thread must have acquired
element~2's lock, and so on, up through thread~$N-1$, who acquires
element~$N-1$'s lock.
For thread~$N-1$ to be blocked, some other thread must have acquired
element~$N$'s lock.
But there are no more threads, and so thread~$N-1$ cannot be blocked.
Therefore, deadlock cannot occur.

So why should we prohibit use of this delightful little algorithm?

The fact is that if you \emph{really} want to use it, we cannot stop you.
We \emph{can}, however, recommend against such code being included
in any project that we care about.

But, before you use this algorithm, please think through the following
Quick Quiz.

\QuickQuiz{
	Can a similar algorithm be used when deleting elements?
}\QuickQuizAnswer{
	Yes.
	However, since each thread must hold the locks of three
	consecutive elements to delete the middle one, if there
	are $N$ threads, there must be $2N+1$ elements (rather than
	just $N+1$) in order to avoid deadlock.
}\QuickQuizEnd

The fact is that this algorithm is extremely specialized (it only works
on certain sized lists), and also quite fragile.
Any bug that accidentally failed to add a node to the list could result
in deadlock.
In fact, simply adding the node a bit too late could result in deadlock,
as could increasing the number of threads.

In addition, the other algorithms described above are ``good and sufficient''.
For example, simply acquiring the locks in address order is fairly simple
and quick, while allowing the use of lists of any size.
Just be careful of the special cases presented by empty lists and lists
containing only one element!

\QuickQuiz{
	Yetch!
	What ever possessed someone to come up with an algorithm
	that deserves to be shaved as much as this one does???
}\QuickQuizAnswer{
	That would be Paul.

	He was considering the \emph{Dining Philosopher's Problem}, which
	involves a rather unsanitary spaghetti dinner attended by
	five philosophers.
	Given that there are five plates and but five forks on the table, and
	given that each philosopher requires two forks at a time to eat,
	one is supposed to come up with a fork-allocation algorithm that
	avoids deadlock.
	Paul's response was ``Sheesh!
			      Just get five more forks!''

	This in itself was OK, but Paul then applied this same solution to
	circular linked lists.

	This would not have been so bad either, but he had to go and tell
	someone about it!
}\QuickQuizEnd

\begin{figure}
\centering
\resizebox{2.5in}{!}{\includegraphics{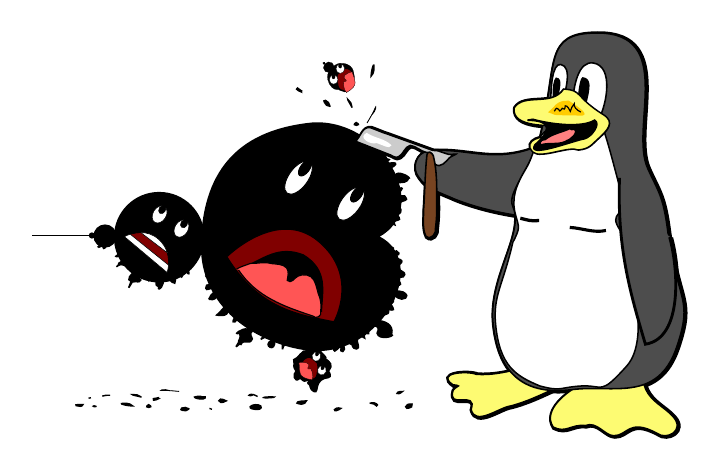}}
\caption{Shaving the Mandelbrot Set}
\ContributedBy{Figure}{fig:easy:Shaving the Mandelbrot Set}{Melissa Broussard}
\end{figure}

In summary, we do not use algorithms simply because they happen to work.
We instead restrict ourselves to algorithms that are useful enough to
make it worthwhile learning about them.
The more difficult and complex the algorithm, the more generally useful
it must be in order for the pain of learning it and fixing its bugs to
be worthwhile.

\QuickQuiz{
	Give an exception to this rule.
}\QuickQuizAnswer{
	One exception would be a difficult and complex algorithm that
	was the only one known to work in a given situation.
	Another exception would be a difficult and complex algorithm
	that was nonetheless the simplest of the set known to work in
	a given situation.
	However, even in these cases, it may be very worthwhile to spend
	a little time trying to come up with a simpler algorithm!
	After all, if you managed to invent the first algorithm
	to do some task, it shouldn't be that hard to go on to
	invent a simpler one.
}\QuickQuizEnd

Exceptions aside, we must continue to shave the software ``Mandelbrot
set'' so that our programs remain maintainable, as shown in
\cref{fig:easy:Shaving the Mandelbrot Set}.

	\setlength{\parskip}{0.0pt plus 1ex}
	\input{qqzeasy}
	\setlength{\parskip}{0.0pt plus 1.0pt}% return to default

% future/future.tex
% mainfile: ../perfbook.tex
% SPDX-License-Identifier: CC-BY-SA-3.0

\QuickQuizChapter{chp:Conflicting Visions of the Future}{Conflicting Visions of the Future}{qqzfuture}
\Epigraph{Prediction is very difficult, especially about the future.}
	 {Niels Bohr}

This chapter presents some conflicting visions of the future of parallel
programming.
It is not clear which of these will come to pass, in fact, it is not
clear that any of them will.
They are nevertheless important because each vision has its devoted
adherents, and if enough people believe in something fervently enough,
you will need to deal with that thing's existence in the form of its
influence on the thoughts, words, and deeds of its adherents.
Besides which, one or more of these visions will actually come to pass.
But most are bogus.
Tell which is which and you'll be rich~\cite{KeithRSpitz1977}!

Therefore, the following sections give an overview of transactional
memory, hardware transactional memory,
formal verification in regression testing, and
parallel functional programming.
But first, a cautionary tale on prognostication taken from the early 2000s.

% future/cpu.tex
% mainfile: ../perfbook.tex
% SPDX-License-Identifier: CC-BY-SA-3.0

\section{The Future of CPU Technology Ain't What it Used to Be}
\label{sec:future:The Future of CPU Technology Ain't What it Used to Be}
\epigraph{A great future behind him.}{David Maraniss}

Years past always seem so simple and innocent when viewed through the
lens of many years of experience.
And the early 2000s were for the most part innocent of the impending
failure of \IXr{Moore's Law} to continue delivering the then-traditional
increases in CPU clock frequency.
Oh, there were the occasional warnings about the limits of technology,
but such warnings had been sounded for decades.
With that in mind, consider the following scenarios:

\begin{figure}
\centering
\resizebox{3in}{!}{\includegraphics{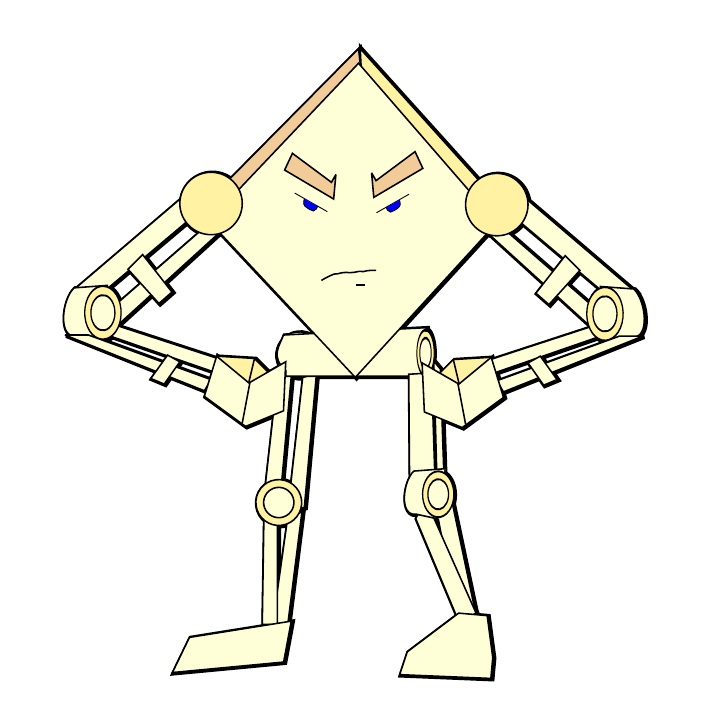}}
\caption{Uniprocessor \"Uber Alles}
\ContributedBy{Figure}{fig:future:Uniprocessor \"Uber Alles}{Melissa Broussard}
\end{figure}

\begin{figure}
\centering
\resizebox{2.6in}{!}{\includegraphics{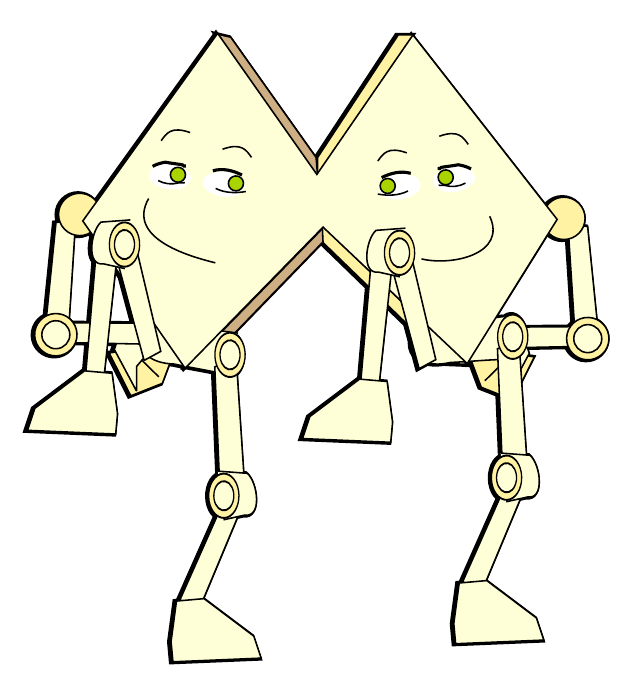}}
\caption{Multithreaded Mania}
\ContributedBy{Figure}{fig:future:Multithreaded Mania}{Melissa Broussard}
\end{figure}

\begin{figure}
\centering
\resizebox{2.5in}{!}{\includegraphics{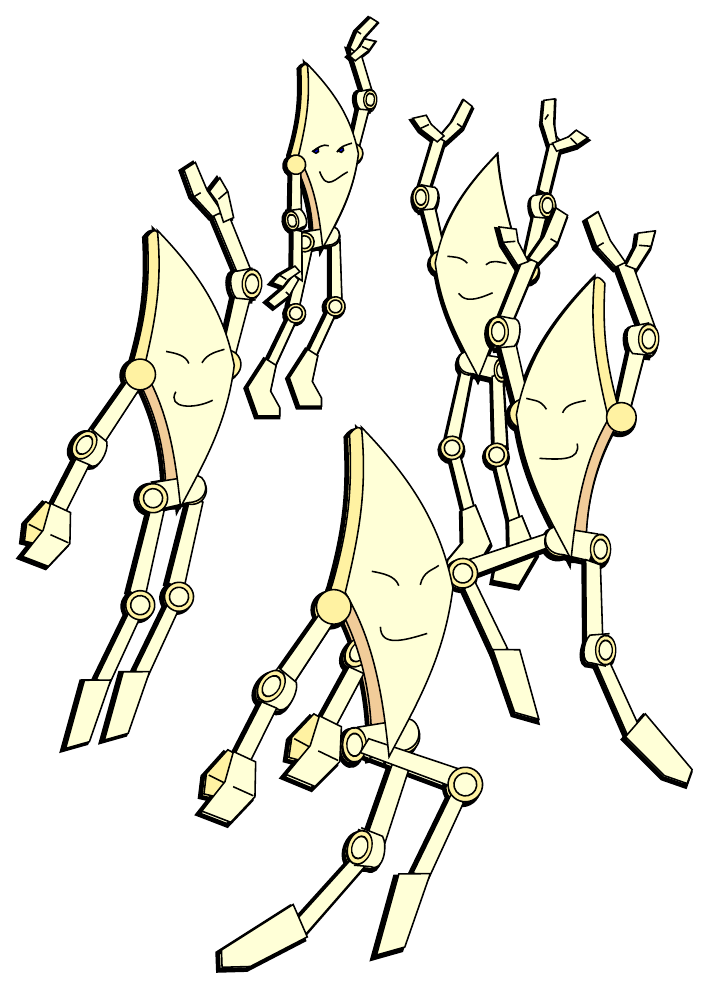}}
\caption{More of the Same}
\ContributedBy{Figure}{fig:future:More of the Same}{Melissa Broussard}
\end{figure}

\begin{figure}
\centering
\resizebox{3in}{!}{\includegraphics{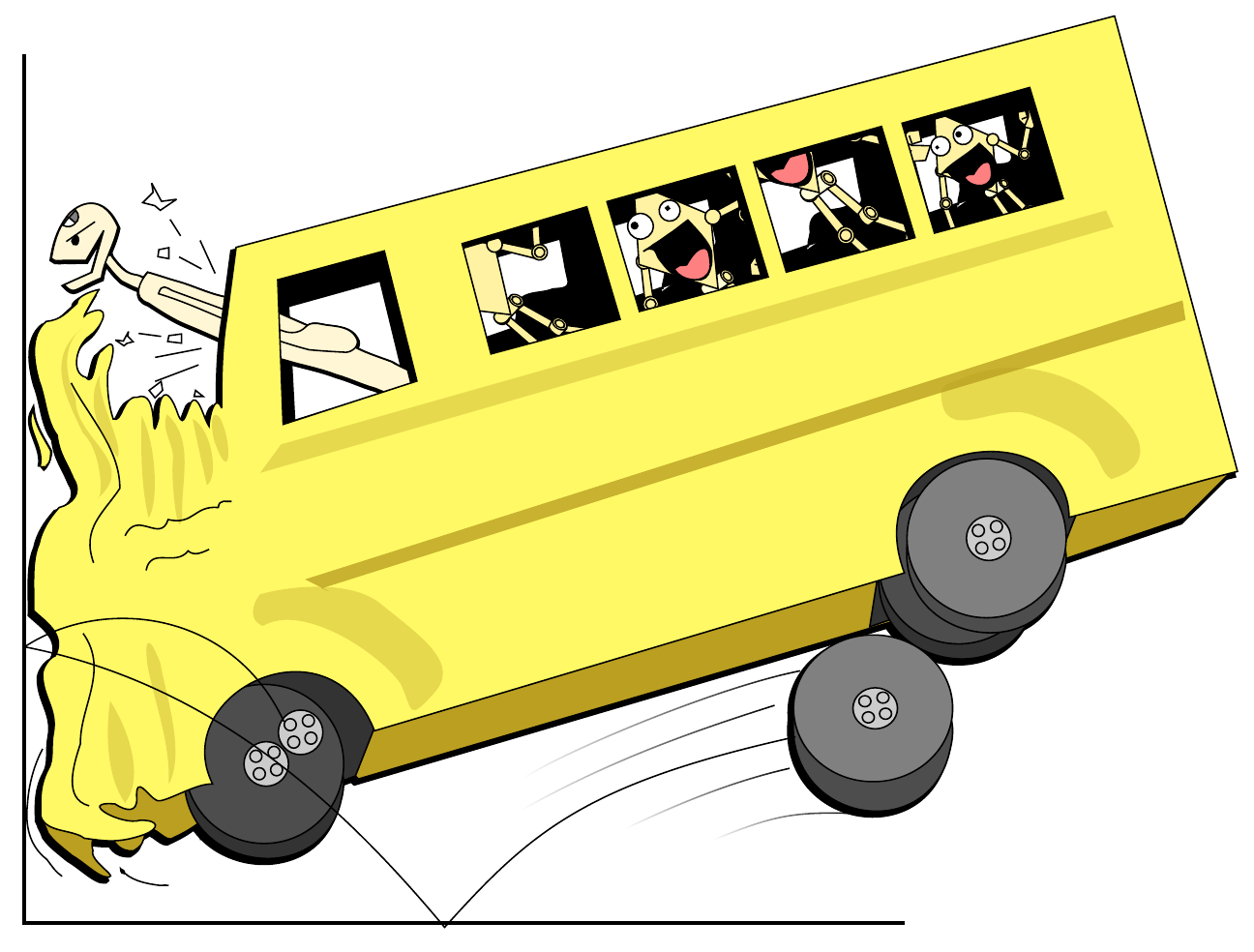}}
\caption{Crash Dummies Slamming into the Memory Wall}
\ContributedBy{Figure}{fig:future:Crash Dummies Slamming into the Memory Wall}{Melissa Broussard}
\end{figure}

\begin{figure}
\centering
\resizebox{3in}{!}{\includegraphics{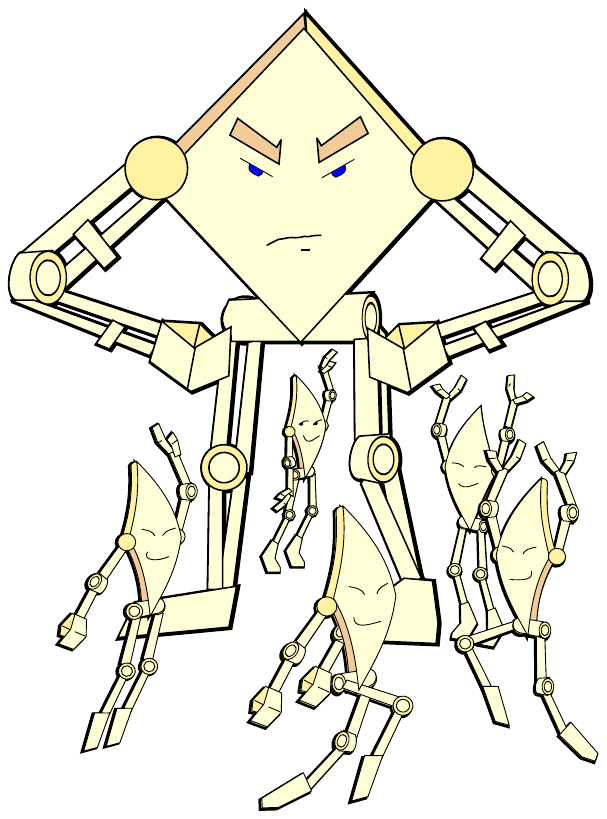}}
\caption{Astounding Accelerators}
\ContributedBy{Figure}{fig:future:Astounding Accelerators}{Melissa Broussard, remixed}
\end{figure}

\begin{enumerate}
\item	Uniprocessor \"Uber Alles
	(\cref{fig:future:Uniprocessor \"Uber Alles}),
\item	Multithreaded Mania
	(\cref{fig:future:Multithreaded Mania}),
\item	More of the Same
	(\cref{fig:future:More of the Same}), and
\item	Crash Dummies Slamming into the Memory Wall
	(\cref{fig:future:Crash Dummies Slamming into the Memory Wall}).
\item	Astounding Accelerators
	(\cref{fig:future:Astounding Accelerators}).
\end{enumerate}

Each of these scenarios is covered in the following sections.

\subsection{Uniprocessor \"Uber Alles}
\label{sec:future:Uniprocessor \"Uber Alles}

As was said in 2004~\cite{PaulEdwardMcKenneyPhD}:

\begin{quote}
	In this scenario, the combination of \IXaltr{Moore's-Law}{Moore's Law}
	increases in CPU
	clock rate and continued progress in horizontally scaled computing
	render SMP systems irrelevant.
	This scenario is therefore dubbed ``Uniprocessor \"Uber
	Alles'', literally, uniprocessors above all else.

	These uniprocessor systems would be subject only to instruction
	overhead, since \IXpl{memory barrier}, cache thrashing, and contention
	do not affect single-CPU systems.
	In this scenario, RCU is useful only for niche applications, such
	as interacting with \IXacrpl{nmi}.
	It is not clear that an operating system lacking RCU would see
	the need to adopt it, although operating
	systems that already implement RCU might continue to do so.

	However, recent progress with multithreaded CPUs seems to indicate
	that this scenario is quite unlikely.
\end{quote}

Unlikely indeed!
But the larger software community was reluctant to accept the fact that
they would need to embrace parallelism, and so it was some time before
this community concluded that the ``free lunch'' of
\IXaltr{Moore's-Law}{Moore's Law}-induced
CPU core-clock frequency increases was well and truly finished.
Never forget:
Belief is an emotion, not necessarily the result of a rational technical
thought process!

\subsection{Multithreaded Mania}
\label{sec:future:Multithreaded Mania}

Also from 2004~\cite{PaulEdwardMcKenneyPhD}:

\begin{quote}
	A less-extreme variant of Uniprocessor \"Uber Alles features
	uniprocessors with hardware multithreading, and in fact
	multithreaded CPUs are now standard for many desktop and laptop
	computer systems.
	The most aggressively multithreaded CPUs share all levels of
	cache hierarchy, thereby eliminating CPU-to-CPU \IXh{memory}{latency},
	in turn greatly reducing the performance penalty for traditional
	synchronization mechanisms.
	However, a multithreaded CPU would still incur overhead due to
	contention and to pipeline stalls caused by memory barriers.
	Furthermore, because all hardware threads share all levels
	of cache, the cache available to a given hardware thread is a
	fraction of what it would be on an equivalent single-threaded
	CPU, which can degrade performance for applications with large
	cache footprints.
	There is also some possibility that the restricted amount of cache
	available will cause RCU-based algorithms to incur performance
	penalties due to their grace-period-induced additional memory
	consumption.
	Investigating this possibility is future work.

	However, in order to avoid such performance degradation, a number
	of multithreaded CPUs and multi-CPU chips partition at least
	some of the levels of cache on a per-hardware-thread basis.
	This increases the amount of cache available to each hardware
	thread, but re-introduces memory latency for cachelines that
	are passed from one hardware thread to another.
\end{quote}

And we all know how this story has played out, with multiple multi-threaded
cores on a single die plugged into a single socket, with varying degrees
of optimization for lower numbers of active threads per core.
The question then becomes whether or not future shared-memory systems will
always fit into a single socket.

\subsection{More of the Same}
\label{sec:meas:More of the Same}

Again from 2004~\cite{PaulEdwardMcKenneyPhD}:

\begin{quote}
	The More-of-the-Same scenario assumes that the memory-latency
	ratios will remain roughly where they are today.

	This scenario actually represents a change, since to have more
	of the same, interconnect performance must begin keeping up
	with the \IXaltr{Moore's-Law}{Moore's Law} increases in core CPU performance.
	In this scenario, overhead due to pipeline stalls, memory latency,
	and contention remains significant, and RCU retains the high
	level of applicability that it enjoys today.
\end{quote}

And the change has been the ever-increasing levels of integration
that \IXr{Moore's Law} is still providing.
But longer term, which will it be?
More CPUs per die?
Or more I/O, cache, and memory?

Servers seem to be choosing the former, while embedded systems on a chip
(SoCs) continue choosing the latter.

\subsection{Crash Dummies Slamming into the Memory Wall}
\label{sec:future:Crash Dummies Slamming into the Memory Wall}

\begin{figure}
\centering
\epsfxsize=3in
\epsfbox{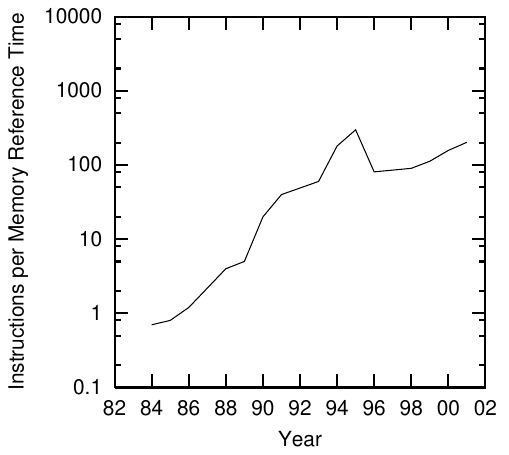}
% from Ph.D. thesis: related/latencytrend.eps
\caption{Instructions per Local Memory Reference for Sequent Computers}
\label{fig:future:Instructions per Local Memory Reference for Sequent Computers}
\end{figure}

\begin{figure}
\centering
\epsfxsize=3in
\epsfbox{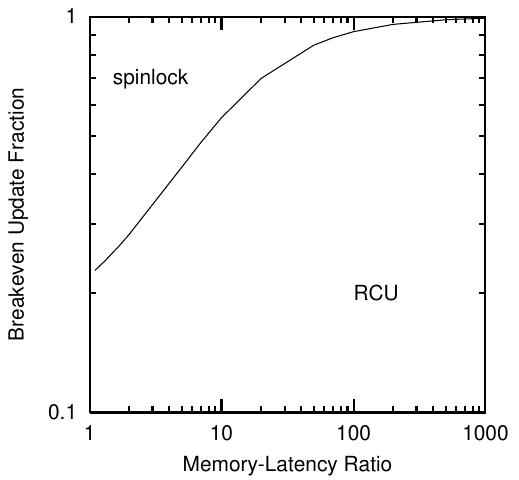}
% from Ph.D. thesis: an/plots/be-lb-n4-rf-all.eps
\caption{Breakevens vs.\@ $r$, $\lambda$ Large, Four CPUs}
\label{fig:future:Breakevens vs. r; lambda Large; Four CPUs}
\end{figure}

\begin{figure}
\centering
\epsfxsize=3in
\epsfbox{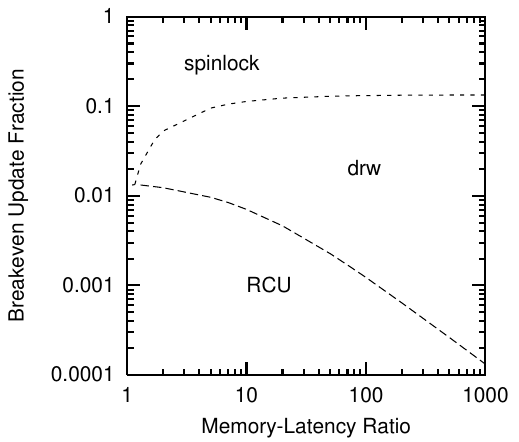}
% from Ph.D. thesis: an/plots/be-lw-n4-rf-all.eps
\caption{Breakevens vs.\@ $r$, $\lambda$ Small, Four CPUs}
\label{fig:future:Breakevens vs. r; Worst-Case lambda; Four CPUs}
\end{figure}

And one more quote from 2004~\cite{PaulEdwardMcKenneyPhD}:

\begin{quote}
	If the memory-latency trends shown in
	\cref{fig:future:Instructions per Local Memory Reference for Sequent Computers}
	continue, then memory latency will continue to grow relative
	to instruction-execution overhead.
	Systems such as Linux that have significant use of RCU will find
	additional use of RCU to be profitable, as shown in
	\cref{fig:future:Breakevens vs. r; lambda Large; Four CPUs}.
	As can be seen in this figure, if RCU is heavily used, increasing
	memory-latency ratios give RCU an increasing advantage over other
	synchronization mechanisms.
	In contrast, systems with minor
	use of RCU will require increasingly high degrees of read intensity
	for use of RCU to pay off, as shown in
	\cref{fig:future:Breakevens vs. r; Worst-Case lambda; Four CPUs}.
	As can be seen in this figure, if RCU is lightly used,
	increasing memory-latency ratios
	put RCU at an increasing disadvantage compared to other synchronization
	mechanisms.
	Since Linux has been observed with over 1,600 callbacks per \IX{grace
	period} under heavy load~\cite{Sarma04c},
	it seems safe to say that Linux falls into the former category.
\end{quote}

On the one hand, this passage failed to anticipate the cache-warmth
issues that RCU can suffer from in workloads with significant update
intensity, in part because it seemed unlikely that RCU would really
be used for such workloads.
In the event, the \co{SLAB_TYPESAFE_BY_RCU} has been pressed into
service in a number of instances where these cache-warmth issues would
otherwise be problematic, as has sequence locking.
On the other hand, this passage also failed to anticipate that
RCU would be used to reduce scheduling latency or for security.

Much of the data generated for this book was collected on an eight-socket
system with 28 cores per socket and two hardware threads per core, for
a total of 448 hardware threads.
The idle-system memory latencies are less than one microsecond, which
are no worse than those of similar-sized systems of the year 2004.
Some claim that these latencies approach a microsecond only because of
the x86 CPU family's relatively strong memory ordering, but it may be
some time before that particular argument is settled.

\subsection{Astounding Accelerators}
\label{sec:future:Astounding Accelerators}

The potential of hardware accelerators was not quite as clear in 2004
as it is in 2021, so this section has no quote.
However, the November 2020 Top 500 list~\cite{Top500} features a great
many accelerators, so one could argue that this section is a view of
the present rather than of the future.
The same could be said of most of the preceding sections.

Hardware accelerators are being put to many other uses, including
encryption, compression, machine learning.

In short, beware of prognostications, including those in the remainder
of this chapter.

\IfTwoColumn{}{\FloatBarrier}
% future/tm.tex
% mainfile: ../perfbook.tex
% SPDX-License-Identifier: CC-BY-SA-3.0

\section{Transactional Memory}
\label{sec:future:Transactional Memory}
\epigraph{Everything should be as simple as it can be, but not simpler.}
	 {Albert Einstein, by way of Louis Zukofsky}

The idea of using transactions outside of databases goes back many
decades~\cite{DBLomet1977SIGSOFT,Knight:1986:AMF:319838.319854,Herlihy93a},
with the key difference between
database and non-database transactions being that non-database transactions
drop the ``D'' in the ``ACID''\footnote{
	Atomicity, consistency, isolation, and durability.}
properties defining database transactions.
The idea of supporting memory-based transactions, or ``transactional memory''
(TM), in hardware
is more recent~\cite{Herlihy93a}, but unfortunately, support for such
transactions in commodity hardware was not immediately forthcoming,
despite other somewhat similar proposals being put forward~\cite{JMStone93}.
Not long after, Shavit and Touitou proposed a software-only implementation
of transactional memory (STM) that was capable of running on commodity
hardware, give or take memory-ordering issues~\cite{Shavit95}.
This proposal languished for many years, perhaps due to the fact that
the research community's attention was absorbed by \IXacrl{nbs}
(see \cref{sec:advsync:Non-Blocking Synchronization}).

But by the turn of the century, TM started receiving
more attention~\cite{Martinez01a,Rajwar01a}, and by the middle of the
decade, the level of interest can only be termed
``incandescent''~\cite{MauriceHerlihy2005-TM-manifesto.pldi,
DanGrossman2007TMGCAnalogy}, with only a few voices of
caution~\cite{Blundell2005DebunkTM,McKenney2007PLOSTM}.

The basic idea behind TM is to execute a section of
code atomically, so that other threads see no intermediate state.
As such, the semantics of TM could be implemented
by simply replacing each transaction with a recursively acquirable
global lock acquisition and release, albeit with abysmal performance
and scalability.
Much of the complexity inherent in TM implementations, whether hardware
or software, is efficiently detecting when concurrent transactions can safely
run in parallel.
Because this detection is done dynamically, conflicting transactions
can be aborted or ``rolled back'', and in some implementations, this
failure mode is visible to the programmer.

Because transaction roll-back is increasingly unlikely as transaction
size decreases, TM might become quite attractive for small memory-based
operations,
such as linked-list manipulations used for stacks, queues, hash tables,
and search trees.
However, it is currently much more difficult to make the case for large
transactions, particularly those containing non-memory operations such
as I/O and process creation.
The following sections look at current challenges to the grand vision of
``Transactional Memory Everywhere''~\cite{PaulEMcKenney2009TMeverywhere}.
\Cref{sec:future:Outside World} examines the challenges faced
interacting with the outside world,
\cref{sec:future:Process Modification} looks at interactions
with process modification primitives,
\cref{sec:future:Synchronization} explores interactions with
other synchronization primitives, and finally
\cref{sec:future:Discussion} closes with some discussion.

\subsection{Outside World}
\label{sec:future:Outside World}

In the wise words of \ppl{Donald}{Knuth}:

\begin{quote}
	Many computer users feel that input and output are not actually part
	of ``real programming,'' they are merely things that (unfortunately)
	must be done in order to get information in and out of the machine.
\end{quote}

Whether or not we believe that input and output are ``real programming'',
the fact is that software absolutely must deal with the outside world.
This section therefore critiques transactional memory's outside-world
capabilities, focusing on I/O operations, time delays, and persistent
storage.

\subsubsection{I/O Operations}
\label{sec:future:I/O Operations}

One can execute I/O operations within a lock-based critical section,
while holding a \IX{hazard pointer}, within a sequence-locking read-side
critical section, and from within a userspace-RCU read-side critical
section, and even all at the same time, if need be.
What happens when you attempt to execute an I/O operation from within
a transaction?

The underlying problem is that transactions may be rolled back, for
example, due to conflicts.
Roughly speaking, this requires that all operations within any given
transaction be revocable, so that executing the operation twice has
the same effect as executing it once.
Unfortunately, I/O is in general the prototypical irrevocable
operation, making it difficult to include general I/O operations in
transactions.
In fact, general I/O is irrevocable:
Once you have pushed the proverbial button launching the nuclear warheads,
there is no turning back.

Here are some options for handling of I/O within transactions:

\begin{enumerate}
\item	Restrict I/O within transactions to buffered I/O with in-memory
	buffers.
	These buffers may then be included in the transaction in the
	same way that any other memory location might be included.
	This seems to be the mechanism of choice, and it does work
	well in many common cases of situations such as stream I/O and
	mass-storage I/O\@.
	However, special handling is required in cases where multiple
	record-oriented output streams are merged onto a single file
	from multiple processes, as might be done using the ``a+''
	option to \co{fopen()} or the \co{O_APPEND}  flag to \co{open()}.
	In addition, as will be seen in the next section, common
	networking operations cannot be handled via buffering.
\item	Prohibit I/O within transactions, so that any attempt to execute
	an I/O operation aborts the enclosing transaction (and perhaps
	multiple nested transactions).
	This approach seems to be the conventional TM approach for
	unbuffered I/O, but requires that TM interoperate with other
	synchronization primitives tolerating I/O.
\item	Prohibit I/O within transactions, but enlist the compiler's aid
	in enforcing this prohibition.
\item	Permit only one special
	\emph{irrevocable} transaction~\cite{SpearMichaelScott2008InevitableSTM}
	to proceed
	at any given time, thus allowing irrevocable transactions to
	contain I/O operations.\footnote{
		In earlier literature, irrevocable transactions are
		termed \emph{inevitable} transactions.}
	This works in general, but severely limits the scalability and
	performance of I/O operations.
	Given that scalability and performance is a first-class goal of
	parallelism, this approach's generality seems a bit self-limiting.
	Worse yet, use of irrevocability to tolerate I/O operations
	seems to greatly restrict use of manual transaction-abort
	operations.\footnote{
		This difficulty was pointed out by Michael Factor.
		To see the problem, think through what TM should do
		in response to an attempt to abort a transaction after
		it has executed an irrevocable operation.}
	Finally, if there is an irrevocable transaction manipulating
	a given data item, any other transaction manipulating that
	same data item cannot have non-blocking semantics.
\item	Create new hardware and protocols such that I/O operations can
	be pulled into the transactional substrate.
	In the case of input operations, the hardware would need to
	correctly predict the result of the operation, and to abort the
	transaction if the prediction failed.
\end{enumerate}

I/O operations are a well-known weakness of TM, and it is not clear
that the problem of supporting I/O in transactions has a reasonable
general solution, at least if ``reasonable'' is to include usable
performance and scalability.
Nevertheless, continued time and attention to this problem will likely
produce additional progress.

\subsubsection{RPC Operations}
\label{sec:future:RPC Operations}

One can execute RPCs within a lock-based critical section, while holding
a hazard pointer, within a sequence-locking read-side critical section,
and from within a userspace-RCU read-side critical section, and even
all at the same time, if need be.
What happens when you attempt to execute an RPC from within a transaction?

If both the RPC request and its response are to be contained within the
transaction, and if some part of the transaction depends on the result
returned by the response, then it is not possible to use the memory-buffer
tricks that can be used in the case of buffered I/O\@.
Any attempt to
take this buffering approach would deadlock the transaction, as the
request could not be transmitted until the transaction was guaranteed
to succeed, but the transaction's success might not be knowable until
after the response is received, as is the case in the following example:

\begin{VerbatimN}[samepage=true]
begin_trans();
rpc_request();
i = rpc_response();
a[i]++;
end_trans();
\end{VerbatimN}

The transaction's memory footprint cannot be determined until after the
RPC response is received, and until the transaction's memory footprint
can be determined, it is impossible to determine whether the transaction
can be allowed to commit.
The only action consistent with transactional semantics is therefore to
unconditionally abort the transaction, which is, to say the least,
unhelpful.

Here are some options available to TM:

\begin{enumerate}
\item	Prohibit RPC within transactions, so that any attempt to execute
	an RPC operation aborts the enclosing transaction (and perhaps
	multiple nested transactions).
	Alternatively, enlist the compiler to enforce RPC-free
	transactions.
	This approach does work, but will require TM to
	interact with other synchronization primitives.
\item	Permit only one special
	irrevocable transaction~\cite{SpearMichaelScott2008InevitableSTM}
	to proceed at any given time, thus allowing irrevocable
	transactions to contain RPC operations.
	This works in general, but severely limits the scalability and
	performance of RPC operations.
	Given that scalability and performance is a first-class goal of
	parallelism, this approach's generality seems a bit self-limiting.
	Furthermore, use of irrevocable transactions to permit RPC
	operations restricts manual transaction-abort operations
	once the RPC operation has started.
	Finally, if there is an irrevocable transaction manipulating
	a given data item, any other transaction manipulating that
	same data item must have blocking semantics.
\item	Identify special cases where the success of the transaction may
	be determined before the RPC response is received, and
	automatically convert these to irrevocable transactions immediately
	before sending the RPC request.
	Of course, if several concurrent transactions attempt RPC calls
	in this manner, it might be necessary to roll all but one of them
	back, with consequent degradation of performance and scalability.
	This approach nevertheless might be valuable given long-running
	transactions ending with an RPC\@.
	This approach must still restrict  manual transaction-abort
	operations.
\item	Identify special cases where the RPC response may be moved out
	of the transaction, and then proceed using techniques similar
	to those used for buffered I/O.
\item	Extend the transactional substrate to include the RPC server as
	well as its client.
	This is in theory possible, as has been demonstrated by
	distributed databases.
	However, it is unclear whether the requisite performance and
	scalability requirements can be met by distributed-database
	techniques, given that memory-based TM has no slow disk drives
	behind which to hide such latencies.
	Of course, given the advent of solid-state disks, it is also quite
	possible that databases will need to redesign their approach to
	latency hiding.
\end{enumerate}

As noted in the prior section, I/O is a known weakness of TM, and RPC
is simply an especially problematic case of I/O.

\subsubsection{Time Delays}
\label{sec:future:Time Delays}

An important special case of interaction with extra-transactional accesses
involves explicit time delays within a transaction.
Of course, the idea of a time delay within a transaction flies in the
face of TM's atomicity property, but this sort of thing is arguably what
weak atomicity is all about.
Furthermore, correct interaction with memory-mapped I/O sometimes requires
carefully controlled timing, and applications often use time delays
for varied purposes.
Finally, one can execute time delays within a lock-based critical section,
while holding a hazard pointer, within a sequence-locking read-side
critical section, and from within a userspace-RCU read-side critical
section, and even all at the same time, if need be.
Doing so might not be wise from a contention or scalability viewpoint,
but then again, doing so does not raise any fundamental conceptual issues.

So, what can TM do about time delays within transactions?

\begin{enumerate}
\item	Ignore time delays within transactions.
	This has an appearance of elegance, but like too many other
	``elegant'' solutions, fails to survive first contact with
	legacy code.
	Such code, which might well have important time delays in critical
	sections, would fail upon being transactionalized.
\item	Abort transactions upon encountering a time-delay operation.
	This is attractive, but it is unfortunately not always possible
	to automatically detect a time-delay operation.
	Is that tight loop carrying out a critical computation, or is it
	simply waiting for time to elapse?
\item	Enlist the compiler to prohibit time delays within transactions.
\item	Let the time delays execute normally.
	Unfortunately, some TM implementations publish modifications only
	at commit time, which could defeat the purpose of the time delay.
\end{enumerate}

It is not clear that there is a single correct answer.
TM implementations featuring weak atomicity that publish changes
immediately within the transaction (rolling these changes back upon abort)
might be reasonably well served by the last alternative.
Even in this case, the code (or possibly even hardware) at the other
end of the transaction may require a substantial redesign to tolerate
aborted transactions.
This need for redesign would make it more difficult to apply transactional
memory to legacy code.

\subsubsection{Persistence}
\label{sec:future:Persistence}

There are many different types of locking primitives.
One interesting distinction is persistence, in other words, whether the
lock can exist independently of the address space of the process using
the lock.

Non-persistent locks include \co{pthread_mutex_lock()},
\co{pthread_rwlock_rdlock()}, and most kernel-level locking primitives.
If the memory locations instantiating a non-persistent lock's data
structures disappear, so does the lock.
For typical use of \co{pthread_mutex_lock()}, this means that when the
process exits, all of its locks vanish.
This property can be exploited in order to trivialize lock cleanup
at program shutdown time, but makes it more difficult for unrelated
applications to share locks, as such sharing requires the applications
to share memory.

\QuickQuiz{
	But suppose that an application exits while holding a
	\co{pthread_mutex_lock()} that happens to be located in a
	file-mapped region of memory?
}\QuickQuizAnswer{
	Indeed, in this case the lock would persist, much to the
	consternation of other processes attempting to acquire this
	lock that is held by a process that no longer exists.
	Which is why great care is required when using \co{pthread_mutex}
	objects located in file-mapped memory regions.
}\QuickQuizEnd

Persistent locks help avoid the need to share memory among unrelated
applications.
Persistent locking APIs include the flock family, \co{lockf()}, System
V semaphores, or the \co{O_CREAT} flag to \co{open()}.
These persistent APIs can be used to protect large-scale operations
spanning runs of multiple applications, and, in the case of \co{O_CREAT}
even surviving operating-system reboot.
If need be, locks can even span multiple computer systems via distributed
lock managers and distributed filesystems---and persist across reboots
of any or all of those computer systems.

Persistent locks can be used by any application, including applications
written using multiple languages and software environments.
In fact, a persistent lock might well be acquired by an application written
in C and released by an application written in Python.

How could a similar persistent functionality be provided for TM?

\begin{enumerate}
\item	Restrict persistent transactions to special-purpose environments
	designed to support them, for example, SQL\@.
	This clearly works, given the decades-long history of database
	systems, but does not provide the same degree of flexibility
	provided by persistent locks.
\item	Use snapshot facilities provided by some storage devices and/or
	filesystems.
	Unfortunately, this does not handle network communication,
	nor does it handle I/O to devices that do not provide snapshot
	capabilities, for example, memory sticks.
\item	Build a time machine.
\item	Avoid the problem entirely by using existing persistent facilities,
	presumably avoiding such use within transactions.
\end{enumerate}

Of course, the fact that it is called transactional \emph{memory}
should give us pause, as the name itself conflicts with the concept of
a persistent transaction.
It is nevertheless worthwhile to consider this possibility as an important
test case probing the inherent limitations of transactional memory.

\subsection{Process Modification}
\label{sec:future:Process Modification}

Processes are not eternal:
They are created and destroyed, their memory mappings are modified,
they are linked to dynamic libraries, and they are debugged.
These sections look at how transactional memory can handle an
ever-changing execution environment.

\subsubsection{Multithreaded Transactions}
\label{sec:future:Multithreaded Transactions}

It is perfectly legal to create processes and threads while holding
a lock or, for that matter, while holding a hazard pointer, within
a sequence-locking read-side critical section, and from within a
userspace-RCU read-side critical section, and even all at the same time,
if need be.
Not only is it legal, but it is quite simple, as can be seen from the
following code fragment:

\begin{VerbatimN}
pthread_mutex_lock(...);
for (i = 0; i < ncpus; i++)
	pthread_create(&tid[i], ...);
for (i = 0; i < ncpus; i++)
	pthread_join(tid[i], ...);
pthread_mutex_unlock(...);
\end{VerbatimN}

This pseudo-code fragment uses \co{pthread_create()} to spawn one thread
per CPU, then uses \co{pthread_join()} to wait for each to complete,
all under the protection of \co{pthread_mutex_lock()}.
The effect is to execute a lock-based critical section in parallel,
and one could obtain a similar effect using \co{fork()} and \co{wait()}.
Of course, the critical section would need to be quite large to justify
the thread-spawning overhead, but there are many examples of large
critical sections in production software.

What might TM do about thread spawning within a transaction?

\begin{enumerate}
\item	Declare \co{pthread_create()} to be illegal within transactions,
	preferably by aborting the transaction.
	Alternatively, enlist the compiler to enforce
	\co{pthread_create()}-free transactions.
\item	Permit \co{pthread_create()} to be executed within a
	transaction, but only the parent thread will be considered to
	be part of the transaction.
	This approach seems to be reasonably compatible with existing and
	posited TM implementations, but seems to be a trap for the unwary.
	This approach raises further questions, such as how to handle
	conflicting child-thread accesses.
\item	Convert the \co{pthread_create()}s to function calls.
	This approach is also an attractive nuisance, as it does not
	handle the not-uncommon cases where the child threads communicate
	with one another.
	In addition, it does not permit concurrent execution of the body
	of the transaction.
\item	Extend the transaction to cover the parent and all child threads.
	This approach raises interesting questions about the nature of
	conflicting accesses, given that the parent and children are
	presumably permitted to conflict with each other, but not with
	other threads.
	It also raises interesting questions as to what should happen
	if the parent thread does not wait for its children before
	committing the transaction.
	Even more interesting, what happens if the parent conditionally
	executes \co{pthread_join()} based on the values of variables
	participating in the transaction?
	The answers to these questions are reasonably straightforward
	in the case of locking.
	The answers for TM are left as an exercise for the reader.
\end{enumerate}

Given that parallel execution of transactions is commonplace in the
database world, it is perhaps surprising that current TM proposals do
not provide for it.
On the other hand, the example above is a fairly sophisticated use
of locking that is not normally found in simple textbook examples,
so perhaps its omission is to be expected.
That said, some researchers are using transactions to autoparallelize
code~\cite{ArunRaman2010MultithreadedTransactions},
and there are rumors that other TM researchers are investigating
fork/join parallelism within transactions, so perhaps this topic will
soon be addressed more thoroughly.

\subsubsection{The \tco{exec()} System Call}
\label{sec:future:The exec System Call}

One can execute an \co{exec()} system call within a lock-based critical
section, while holding a hazard pointer, within a sequence-locking
read-side critical section, and from within a userspace-RCU read-side
critical section, and even all at the same time, if need be.
The exact semantics depends on the type of primitive.

In the case of non-persistent primitives (including
\co{pthread_mutex_lock()}, \co{pthread_rwlock_rdlock()}, and userspace RCU),
if the \co{exec()} succeeds, the whole address space vanishes, along
with any locks being held.
Of course, if the \co{exec()} fails, the address space still lives,
so any associated locks would also still live.
A bit strange perhaps, but well defined.

On the other hand, persistent primitives (including the flock family,
\co{lockf()}, System V semaphores, and the \co{O_CREAT} flag to
\co{open()}) would survive regardless of whether the \co{exec()}
succeeded or failed, so that the \co{exec()}ed program might well
release them.

\QuickQuiz{
	What about non-persistent primitives represented by data
	structures in \co{mmap()} regions of memory?
	What happens when there is an \co{exec()} within a critical
	section of such a primitive?
}\QuickQuizAnswer{
	If the \co{exec()}ed program maps those same regions of
	memory, then this program could in principle simply release
	the lock.
	The question as to whether this approach is sound from a
	software-engineering viewpoint is left as an exercise for
	the reader.
}\QuickQuizEnd

What happens when you attempt to execute an \co{exec()} system call
from within a transaction?

\begin{enumerate}
\item	Disallow \co{exec()} within transactions, so that the enclosing
	transactions abort upon encountering the \co{exec()}.
	This is well defined, but clearly requires non-TM synchronization
	primitives for use in conjunction with \co{exec()}.
\item	Disallow \co{exec()} within transactions, with the compiler
	enforcing this prohibition.
	There is a draft specification for TM in C++ that takes
	this approach, allowing functions to be decorated with
	the \co{transaction_safe} and \co{transaction_unsafe}
	attributes.\footnote{
		Thanks to Mark Moir for pointing me at this spec, and
		to Michael Wong for having pointed me at an earlier
		revision some time back.}
	This approach has some advantages over aborting the transaction
	at runtime, but again requires non-TM synchronization primitives
	for use in conjunction with \co{exec()}.
	One disadvantage is the need to decorate a great many library
	functions with \co{transaction_safe} and \co{transaction_unsafe}
	attributes.
\item	Treat the transaction in a manner similar to non-persistent
	locking primitives, so that the transaction survives if \co{exec()}
	fails, and silently commits if the \co{exec()} succeeds.
	The case where only some of the variables affected by the
	transaction reside in \co{mmap()}ed memory (and thus could
	survive a successful \co{exec()} system call) is left as an
	exercise for the reader.
\item	Abort the transaction (and the \co{exec()} system call) if the
	\co{exec()} system call would have succeeded, but allow the
	transaction to continue if the \co{exec()} system call would
	fail.
	This is in some sense the ``correct'' approach, but it would
	require considerable work for a rather unsatisfying result.
\end{enumerate}

The \co{exec()} system call is perhaps the strangest example of an
obstacle to universal TM applicability, as it is not completely clear
what approach makes sense, and some might argue that this is merely a
reflection of the perils of real-life interaction with \co{exec()}.
That said, the two options prohibiting \co{exec()} within transactions
are perhaps the most logical of the group.

Similar issues surround the \co{exit()} and \co{kill()} system calls,
as well as a \co{longjmp()} or an exception that would exit the transaction.
(Where did the \co{longjmp()} or exception come from?)

\subsubsection{Dynamic Linking and Loading}
\label{sec:future:Dynamic Linking and Loading}

Lock-based critical section, code holding a hazard pointer,
sequence-locking read-side critical sections, and userspace-RCU read-side
critical sections can (separately or in combination) legitimately contain
code that invokes dynamically linked and loaded functions, including C/C++
shared libraries and Java class libraries.
Of course, the code contained in these libraries is by definition
unknowable at compile time.
So, what happens if a dynamically loaded function is invoked within
a transaction?

This question has two parts:
\begin{enumerate*}[(a)]
\item How do you dynamically link and load a function within a transaction
and
\item What do you do about the unknowable nature of the code within
this function?
\end{enumerate*}
To be fair, item (b) poses some challenges for locking and userspace-RCU
as well, at least in theory.
For example, the dynamically linked function might introduce a \IX{deadlock}
for locking or might (erroneously) introduce a \IX{quiescent state} into a
userspace-RCU read-side critical section.
The difference is that while the class of operations permitted in locking
and userspace-RCU critical sections is well-understood, there appears
to still be considerable uncertainty in the case of TM\@.
In fact, different implementations of TM seem to have different restrictions.

So what can TM do about dynamically linked and loaded library functions?
Options for part (a), the actual loading of the code, include the following:

\begin{enumerate}
\item	Treat the dynamic linking and loading in a manner similar to a
	page fault, so that the function is loaded and linked, possibly
	aborting the transaction in the process.
	If the transaction is aborted, the retry will find the function
	already present, and the transaction can thus be expected to
	proceed normally.
\item	Disallow dynamic linking and loading of functions from within
	transactions.
\end{enumerate}

Options for part (b), the inability to detect TM-unfriendly operations
in a not-yet-loaded function, possibilities include the following:

\begin{enumerate}
\item	Just execute the code:
	If there are any TM-unfriendly operations in the function,
	simply abort the transaction.
	Unfortunately, this approach makes it impossible for the compiler
	to determine whether a given group of transactions may be safely
	composed.
	One way to permit composability regardless is irrevocable
	transactions, however, current implementations permit only a
	single irrevocable transaction to proceed at any given time,
	which can severely limit performance and scalability.
	Irrevocable transactions also to restrict use of manual
	transaction-abort operations.
	Finally, if there is an irrevocable transaction manipulating
	a given data item, any other transaction manipulating that
	same data item cannot have non-blocking semantics.
\item	Decorate the function declarations indicating which functions
	are TM-friendly.
	These decorations can then be enforced by the compiler's type system.
	Of course, for many languages, this requires language extensions
	to be proposed, standardized, and implemented, with the
	corresponding time delays, and also with the corresponding
	decoration of a great many otherwise uninvolved library functions.
	That said, the standardization effort is already in
	progress~\cite{Ali-Reza-Adl-Tabatabai2009CppTM}.
\item	As above, disallow dynamic linking and loading of functions from
	within transactions.
\end{enumerate}

I/O operations are of course a known weakness of TM, and dynamic linking
and loading can be thought of as yet another special case of I/O\@.
Nevertheless, the proponents of TM must either solve this problem, or
resign themselves to a world where TM is but one tool of several in the
parallel programmer's toolbox.
(To be fair, a number of TM proponents have long since resigned themselves
to a world containing more than just TM.)

\subsubsection{Memory-Mapping Operations}
\label{sec:future:Memory-Mapping Operations}

It is perfectly legal to execute memory-mapping operations (including
\co{mmap()}, \co{shmat()}, and \co{munmap()}~\cite{TheOpenGroup1997SUS})
within a lock-based critical section, while holding a hazard pointer,
within a sequence-locking read-side critical section, and from within a
userspace-RCU read-side critical section, and even all at the same time,
if need be.
What happens when you attempt to execute such an operation from within
a transaction?
More to the point, what happens if the memory region being remapped
contains some variables participating in the current thread's transaction?
And what if this memory region contains variables participating in some
other thread's transaction?

It should not be necessary to consider cases where the TM system's
metadata is remapped, given that most locking primitives do not define
the outcome of remapping their lock variables.

Here are some TM memory-mapping options:

\begin{enumerate}
\item	Memory remapping is illegal within a transaction, and will result
	in all enclosing transactions being aborted.
	This does simplify things somewhat, but also requires that TM
	interoperate with synchronization primitives that do tolerate
	remapping from within their critical sections.
\item	Memory remapping is illegal within a transaction, and the
	compiler is enlisted to enforce this prohibition.
\item	Memory mapping is legal within a transaction, but aborts all
	other transactions having variables in the region mapped over.
\item	Memory mapping is legal within a transaction, but the mapping
	operation will fail if the region being mapped overlaps with
	the current transaction's footprint.
\item	All memory-mapping operations, whether within or outside a
	transaction, check the region being mapped against the memory
	footprint of all transactions in the system.
	If there is overlap, then the memory-mapping operation fails.
\item	The effect of memory-mapping operations that overlap the memory
	footprint of any transaction in the system is determined by the
	TM conflict manager, which might dynamically determine whether
	to fail the memory-mapping operation or abort any conflicting
	transactions.
\end{enumerate}

It is interesting to note that \co{munmap()} leaves the relevant region
of memory unmapped, which could have additional interesting
implications.\footnote{
	This difference between mapping and unmapping was noted by
	Josh Triplett.}

\subsubsection{Debugging}
\label{sec:future:Debugging}

The usual debugging operations such as breakpoints work normally within
lock-based critical sections and from usespace-RCU read-side critical sections.
However, in initial transactional-memory hardware
implementations~\cite{DaveDice2009ASPLOSRockHTM} an exception within
a transaction will abort that transaction, which in turn means that
breakpoints abort all enclosing transactions.

So how can transactions be debugged?

\begin{enumerate}
\item	Use software emulation techniques within transactions containing
	breakpoints.
	Of course, it might be necessary to emulate all transactions
	any time a breakpoint is set within the scope of any transaction.
	If the runtime system is unable to determine whether or not a
	given breakpoint is within the scope of a transaction, then it
	might be necessary to emulate all transactions just to be on
	the safe side.
	However, this approach might impose significant overhead, which
	might in turn obscure the bug being pursued.
\item	Use only hardware TM implementations that are capable of
	handling breakpoint exceptions.
	Unfortunately, as of this writing (March 2021), all such
	implementations are research prototypes.
\item	Use only software TM implementations, which are
	(very roughly speaking) more tolerant of exceptions than are
	the simpler of the hardware TM implementations.
	Of course, software TM tends to have higher overhead than hardware
	TM, so this approach may not be acceptable in all situations.
\item	Program more carefully, so as to avoid having bugs in the
	transactions in the first place.
	As soon as you figure out how to do this, please do let everyone
	know the secret!
\end{enumerate}

There is some reason to believe that transactional memory will deliver
\IX{productivity} improvements compared to other synchronization mechanisms,
but it does seem quite possible that these improvements could easily
be lost if traditional debugging techniques cannot be applied to
transactions.
This seems especially true if transactional memory is to be used by
novices on large transactions.
In contrast, macho ``top-gun'' programmers might be able to dispense with
such debugging aids, especially for small transactions.

Therefore, if transactional memory is to deliver on its productivity
promises to novice programmers, the debugging problem does need to
be solved.

\subsection{Synchronization}
\label{sec:future:Synchronization}

If transactional memory someday proves that it can be everything to everyone,
it will not need to interact with any other synchronization mechanism.
Until then, it will need to work with synchronization mechanisms that
can do what it cannot, or that work more naturally in a given situation.
The following sections outline the current challenges in this area.

\subsubsection{Locking}
\label{sec:future:Locking}

It is commonplace to acquire locks while holding other locks, which works
quite well, at least as long as the usual well-known software-engineering
techniques are employed to avoid deadlock.
It is not unusual to acquire locks from within RCU read-side critical
sections, which eases deadlock concerns because RCU read-side primitives
cannot participate in lock-based deadlock cycles.
It is also possible to acquire locks while holding hazard pointers and
within sequence-lock read-side critical sections.
But what happens when you attempt to acquire a lock from within a transaction?

In theory, the answer is trivial:
Simply manipulate the data structure representing the lock as part of
the transaction, and everything works out perfectly.
In practice, a number of non-obvious complications~\cite{Volos2008TRANSACT}
can arise, depending on implementation details of the TM system.
These complications can be resolved, but at the cost of a 45\,\% increase in
overhead for locks acquired outside of transactions and a 300\,\% increase
in overhead for locks acquired within transactions.
Although these overheads might be acceptable for transactional
programs containing small amounts of locking, they are often completely
unacceptable for production-quality lock-based programs wishing to use
the occasional transaction.

\begin{enumerate}
\item	Use only locking-friendly TM implementations.
	Unfortunately, the locking-unfriendly implementations have some
	attractive properties, including low overhead for successful
	transactions and the ability to accommodate extremely large
	transactions.
\item	Use TM only ``in the small'' when introducing TM to lock-based
	programs, thereby accommodating the limitations of
	locking-friendly TM implementations.
\item	Set aside locking-based legacy systems entirely, re-implementing
	everything in terms of transactions.
	This approach has no shortage of advocates, but this requires
	that all the issues described in this series be resolved.
	During the time it takes to resolve these issues, competing
	synchronization mechanisms will of course also have the
	opportunity to improve.
\item	Use TM strictly as an optimization in lock-based systems, as was
	done by the TxLinux~\cite{ChistopherJRossbach2007a} group and
	by a great many transactional lock elision
	projects~\cite{MartinPohlack2011HTM2TLE,Kleen:2014:SEL:2566590.2576793,PascalFelber2016rwlockElision,SeongJaePark2020HTMRCUlock}.
	This approach seems sound, but leaves the locking design
	constraints (such as the need to avoid deadlock) firmly in place.
\item	Strive to reduce the overhead imposed on locking primitives.
\end{enumerate}

The fact that there could possibly be a problem interfacing TM and locking
came as a surprise to many, which underscores the need to try out new
mechanisms and primitives in real-world production software.
Fortunately, the advent of open source means that a huge quantity of
such software is now freely available to everyone, including researchers.

\subsubsection{Reader-Writer Locking}
\label{sec:future:Reader-Writer Locking}

It is commonplace to read-acquire reader-writer locks while holding
other locks, which just works, at least as long as the usual well-known
software-engineering techniques are employed to avoid deadlock.
Read-acquiring reader-writer locks from within RCU read-side critical
sections also works, and doing so eases deadlock concerns because RCU
read-side primitives cannot participate in lock-based deadlock cycles.
It is also possible to acquire locks while holding hazard pointers and
within sequence-lock read-side critical sections.
But what happens when you attempt to read-acquire a reader-writer lock
from within a transaction?

Unfortunately, the straightforward approach to read-acquiring the
traditional counter-based reader-writer lock within a transaction defeats
the purpose of the reader-writer lock.
To see this, consider a pair of transactions concurrently attempting to
read-acquire the same reader-writer lock.
Because read-acquisition involves modifying the reader-writer lock's
data structures, a conflict will result, which will roll back one of
the two transactions.
This behavior is completely inconsistent with the reader-writer lock's
goal of allowing concurrent readers.

Here are some options available to TM:

\begin{enumerate}
\item	Use per-CPU or per-thread reader-writer
	locking~\cite{WilsonCHsieh92a}, which allows a
	given CPU (or thread, respectively) to manipulate only local
	data when read-acquiring the lock.
	This would avoid the conflict between the two transactions
	concurrently read-acquiring the lock, permitting both to proceed,
	as intended.
	Unfortunately, (1)~the write-acquisition overhead of
	per-CPU/thread locking can be extremely high, (2)~the memory
	overhead of per-CPU/thread locking can be prohibitive, and
	(3)~this transformation is available only when you have access to
	the source code in question.
	Other more-recent scalable
	reader-writer locks~\cite{YossiLev2009SNZIrwlock}
	might avoid some or all of these problems.
\item	Use TM only ``in the small'' when introducing TM to lock-based
	programs, thereby avoiding read-acquiring reader-writer locks
	from within transactions.
\item	Set aside locking-based legacy systems entirely, re-implementing
	everything in terms of transactions.
	This approach has no shortage of advocates, but this requires
	that \emph{all} the issues described in this series be resolved.
	During the time it takes to resolve these issues, competing
	synchronization mechanisms will of course also have the
	opportunity to improve.
\item	Use TM strictly as an optimization in lock-based systems, as was
	done by the TxLinux~\cite{ChistopherJRossbach2007a} group,
	and as has been done by more recent work using TM to elide
	reader writer locks~\cite{PascalFelber2016rwlockElision}.
	This approach seems sound, at least on \Power{8}
	CPUs~\cite{Le:2015:TMS:3266491.3266500}, but leaves the locking
	design constraints (such as the need to avoid deadlock) firmly
	in place.
\end{enumerate}

Of course, there might well be other non-obvious issues surrounding
combining TM with reader-writer locking, as there in fact were with
exclusive locking.

\subsubsection{Deferred Reclamation}
\label{sec:future:Deferred Reclamation}

This section focuses mainly on RCU\@.
Similar issues and possible resolutions arise when combining TM with
other deferred-reclamation mechanisms such as
\IXalt{reference counters}{reference count} and
hazard pointers.
In the text below, known differences are specifically called out.

Reference counting, hazard pointers, and RCU are all heavily used, as noted in
\cref{sec:defer:RCU Related Work,sec:defer:Which to Choose? (Production Use)}.
This means that any TM implementation that chooses not to surmount each
and every challenge called out in this section needs to interoperate
cleanly and efficiently with all of these synchronization mechanisms.

The TxLinux group from the University of Texas at Austin appears to be
the group to take on the challenge of RCU/TM
interoperability~\cite{ChistopherJRossbach2007a}.
Because they applied TM to the Linux 2.6 kernel, which uses RCU, they
had no choice but to integrate TM and RCU, with TM taking the place of
locking for RCU updates.
Unfortunately, although the paper does state that the RCU implementation's
locks (e.g., \co{rcu_ctrlblk.lock}) were converted to transactions,
it is silent about what was done with those locks used by RCU-based updates
(for example, \co{dcache_lock}).

More recently, \ppl{Dimitrios}{Siakavaras} et al.~have applied
HTM and RCU to search trees~\cite{Siakavaras2017CombiningHA,DimitriosSiakavaras2020RCU-HTM-B+Trees},
\ppl{Christina}{Giannoula} et al.~have used HTM and RCU to color
graphs~\cite{ChristinaGiannoula2018HTM-RCU-graphcoloring},
and
\ppl{SeongJae}{Park} et al.~have used HTM and RCU to optimize high-contention
locking on \IXacr{numa} systems~\cite{SeongJaePark2020HTMRCUlock}.

It is important to note that RCU permits readers and updaters to run
concurrently, further permitting RCU readers to access data that is in
the act of being updated.
Of course, this property of RCU, whatever its performance, scalability,
and real-time-response benefits might be, flies in the face of the
underlying atomicity properties of TM, although the \Power{8} CPU family's
suspended-transaction facility~\cite{Le:2015:TMS:3266491.3266500} makes
it an exception to this rule.

So how should TM-based updates interact with concurrent RCU readers?
Some possibilities are as follows:

\begin{enumerate}
\item	RCU readers abort concurrent conflicting TM updates.
	This is in fact the approach taken by the TxLinux project.
	This approach does preserve RCU semantics, and also preserves
	RCU's read-side performance, scalability, and real-time-response
	properties, but it does have the unfortunate side-effect of
	unnecessarily aborting conflicting updates.
	In the worst case, a long sequence of RCU readers could
	potentially starve all updaters, which could in theory result
	in system hangs.
	In addition, not all TM implementations offer the strong atomicity
	required to implement this approach, and for good reasons.
\item	RCU readers that run concurrently with conflicting TM updates
	get old (pre-transaction) values from any conflicting RCU loads.
	This preserves RCU semantics and performance, and also prevents
	RCU-update \IX{starvation}.
	However, not all TM implementations can provide timely access
	to old values of variables that have been tentatively updated
	by an in-flight transaction.
	In particular, log-based TM implementations that maintain
	old values in the log (thus providing excellent TM commit
	performance) are not likely to be happy with this approach.
	Perhaps the \co{rcu_dereference()} primitive can be leveraged
	to permit RCU to access the old values within a greater range
	of TM implementations, though performance might still be an issue.
	Nevertheless, there are popular TM implementations that have
	been integrated with RCU in this
	manner~\cite{DonaldEPorter2007TRANSACT,PhilHoward2011RCUTMRBTree,
	PhilipWHoward2013RCUrbtree}.
\item	If an RCU reader executes an access that conflicts with an
	in-flight transaction, then that RCU access is delayed until
	the conflicting transaction either commits or aborts.
	This approach preserves RCU semantics, but not RCU's performance
	or real-time response, particularly in presence of long-running
	transactions.
	In addition, not all TM implementations are capable of delaying
	conflicting accesses.
	Nevertheless, this approach seems eminently reasonable for hardware
	TM implementations that support only small transactions.
\item	RCU readers are converted to transactions.
	This approach pretty much guarantees that RCU is compatible with
	any TM implementation, but it also imposes TM's rollbacks on RCU
	read-side critical sections, destroying RCU's real-time response
	guarantees, and also degrading RCU's read-side performance.
	Furthermore, this approach is infeasible in cases where any of
	the RCU read-side critical sections contains operations that
	the TM implementation in question is incapable of handling.
	This approach is more difficult to apply to hazard pointers and
	reference counters, which do not have a sharply defined notion
	of a reader as a section of code.
\item	Many update-side uses of RCU modify a single pointer to publish
	a new data structure.
	In some of these cases, RCU can safely be permitted to see a
	transactional pointer update that is subsequently rolled back,
	as long as the transaction respects memory ordering and as long
	as the roll-back process uses \co{call_rcu()} to free up the
	corresponding structure.
	Unfortunately, not all TM implementations respect memory barriers
	within a transaction.
	Apparently, the thought is that because transactions are supposed
	to be atomic, the ordering of the accesses within the transaction
	is not supposed to matter.
\item	Prohibit use of TM in RCU updates.
	This is guaranteed to work, but restricts use of TM.
\end{enumerate}

It seems likely that additional approaches will be uncovered, especially
given the advent of user-level RCU and hazard-pointer
implementations.\footnote{
	Kudos to the TxLinux group, Maged Michael, and Josh Triplett
	for coming up with a number of the above alternatives.}
It is interesting to note that many of the better performing and
scaling STM implementations make use of RCU-like techniques
internally~\cite{UCAM-CL-TR-579,KeirFraser2007withoutLocks,Gu:2019:PSE:3358807.3358885,Kim:2019:MSR:3297858.3304040}.

\QuickQuiz{
	MV-RLU looks pretty good!
	Doesn't it beat RCU hands down?
}\QuickQuizAnswer{
	One might get that impression from a quick read of the abstract,
	but more careful readers will notice the ``for a wide range of
	workloads'' phrase in the last sentence.
	It turns out that this phrase is quite important:

	\begin{enumerate}
	\item	Their RCU evaluation uses synchronous grace periods, which
		needlessly throttle updates, as noted in their
		Section~6.2.1.
		See \cref{fig:datastruct:Read-Side RCU-Protected Hash-Table Performance For Schroedinger's Zoo in the Presence of Updates}
		\cpageref{fig:datastruct:Read-Side RCU-Protected Hash-Table Performance For Schroedinger's Zoo in the Presence of Updates}
		of this book to see that the venerable asynchronous
		\co{call_rcu()} primitive enables RCU to perform and
		scale quite well with large numbers of updaters.
		Furthermore, in Section~3.7 of their paper, the authors
		admit that asynchronous grace periods are important to
		MV-RLU scalability.
		A fair comparison would also allow RCU the benefits of
		asynchrony.
	\item	They use a poorly tuned 1,000-bucket hash table containing
		10,000~elements.
		In addition, their 448~hardware threads need considerably
		more than 1,000~buckets to avoid the lock contention
		that they correctly state limits RCU performance in
		their benchmarks.
		A useful comparison would feature a properly tuned
		hash table.
	\item	Their RCU hash table used per-bucket locks, which they
		call out as a bottleneck, which is not a surprise given
		the long hash chains and small ratio of buckets to threads.
		A number of their competing mechanisms instead use
		lockfree techniques, thus avoiding the per-bucket-lock
		bottleneck, which cynics might claim sheds some light
		on the authors' otherwise inexplicable choice of poorly
		tuned hash tables.
		The first graph in the middle row of the authors'
		Figure~4 show what RCU can achieve if not hobbled by
		artificial bottlenecks, as does the first portion of
		the second graph in that same row.
	\item	Their linked-list operation permits RLU to do concurrent
		modifications of different elements in the list, while
		RCU is forced to serialize updates.
		Again, RCU has always worked just fine in conjunction
		with lockless updaters, a fact that has been set forth
		in academic literature that the authors
		cited~\cite{MathieuDesnoyers2012URCU}.
		A fair comparison would use the same style of update
		for RCU as it does for MV-RLU.
	\item	The authors fail to consider combining RCU and sequence
		locking, which is used in the Linux kernel to give
		readers coherent views of multi-pointer updates.
	\item	The authors fail to consider RCU-based solutions to the
		Issaquah Challenge~\cite{PaulEMcKenney2016IssaquahCPPCON},
		which also gives readers a coherent view of multi-pointer
		updates, albeit with a weaker view of ``coherent''.
	\end{enumerate}

	It is surprising that the anonymous reviewers of this paper did
	not demand an apples-to-apples comparison of MV-RLU and RCU\@.
	Nevertheless, the authors should be congratulated on producing
	an academic paper that presents an all-too-rare example of good
	scalability combined with strong read-side coherence.
	They are also to be congratulated on overcoming the traditional
	academic prejudice against asynchronous grace periods,
	which greatly aided their scalability.

	Interestingly enough, RLU and RCU take different approaches to avoid
	the inherent limitations of STM noted by \ppl{Hagit}{Attiya} et
	al.~\cite{Attiya:2009:STMReadOnlyLimits}.
	RCU avoids providing strict serializability and RLU avoids providing
	invisible read-only transactions, both thus avoiding the
	limitations.
}\QuickQuizEnd

\subsubsection{Extra-Transactional Accesses}
\label{sec:future:Extra-Transactional Accesses}

Within a lock-based critical section, it is perfectly legal to manipulate
variables that are concurrently accessed or even modified outside that
lock's critical section, with one common example being statistical
counters.
The same thing is possible within RCU read-side critical
sections, and is in fact the common case.

Given mechanisms such as the so-called ``dirty reads'' that are
prevalent in production database systems, it is not surprising
that extra-transactional accesses have received serious attention
from the proponents of TM, with the concept of weak
atomicity~\cite{Blundell2006TMdeadlock} being but one case in point.

Here are some extra-transactional options:

\begin{enumerate}
\item	Conflicts due to extra-transactional accesses always abort
	transactions.
	This is strong atomicity.
\item	Conflicts due to extra-transactional accesses are ignored,
	so only conflicts among transactions can abort transactions.
	This is weak atomicity.
\item	Transactions are permitted to carry out non-transactional
	operations in special cases, such as when allocating memory or
	interacting with lock-based critical sections.
\item	Produce hardware extensions that permit some operations
	(for example, addition) to be carried out concurrently on a
	single variable by multiple transactions.
\item	Introduce weak semantics to transactional memory.
	One approach is the combination with RCU described in
	\cref{sec:future:Deferred Reclamation},
	while Gramoli and Guerraoui
	survey a number of other weak-transaction
	approaches~\cite{Gramoli:2014:DTP:2541883.2541900}, for example,
	restricted partitioning of large
	``elastic'' transactions into smaller transactions, thus
	reducing conflict probabilities (albeit with tepid performance
	and scalability).
	Perhaps further experience will show that some uses of
	extra-transactional accesses can be replaced by weak
	transactions.
\end{enumerate}

It appears that transactions were conceived in a vacuum, with no
interaction required with any other synchronization mechanism.
If so, it is no surprise that much confusion and complexity arises when
combining transactions with non-transactional accesses.
But unless transactions are to be confined to small updates to isolated
data structures, or alternatively to be confined to new programs
that do not interact with the huge body of existing parallel code,
then transactions absolutely must be so combined if they are to have
large-scale practical impact in the near term.

% @@@ Huge transactions.  Or perhaps conflict handling.
% Contention managers: WilliamNSchererIII2005
% Unbounded transactional memory: CScottAnanian2006 (UTM)
%	KevinEMoore2006 (LogTM).
%	SanjeevKumar2006 (Hybrid HTM/STM).
%	BratinSaha2006MICRO ("mark bits", similar to LL/SC markings).
%	DonaldEPorter2007TRANSACT (Need to loosen TM consistency).
%	YujieLiu2011ToxicTransactions (Toxic Transactions).

\subsection{Discussion}
\label{sec:future:Discussion}

The obstacles to universal TM adoption lead to the following
conclusions:

\begin{enumerate}
\item	One interesting property of TM is the fact that transactions are
	subject to rollback and retry.
	This property underlies TM's difficulties with irreversible
	operations, including unbuffered I/O, RPCs, memory-mapping
	operations, time delays, and the \co{exec()} system call.
	This property also has the unfortunate consequence of introducing
	all the complexities inherent in the possibility of failure,
	often in a developer-visible manner.
\item	Another interesting property of TM, noted by
	Shpeisman et al.~\cite{TatianaShpeisman2009CppTM}, is that TM
	intertwines the synchronization with the data it protects.
	This property underlies TM's issues with I/O, memory-mapping
	operations, extra-transactional accesses, and debugging
	breakpoints.
	In contrast, conventional synchronization primitives, including
	locking and RCU, maintain a clear separation between the
	synchronization primitives and the data that they protect.
\item	One of the stated goals of many workers in the TM area is to
	ease parallelization of large sequential programs.
	As such, individual transactions are commonly expected to
	execute serially, which might do much to explain TM's issues
	with multithreaded transactions.
\end{enumerate}

\QuickQuiz{
	Given things like \co{spin_trylock()}, how does it make any
	sense at all to claim that TM introduces the concept of failure???
}\QuickQuizAnswer{
	When using locking, \co{spin_trylock()} is a choice, with a
	corresponding failure-free choice being \co{spin_lock()},
	which is used in the common case, as in there are more than
	100 times as many calls to \co{spin_lock()} than to
	\co{spin_trylock()} in the v5.11 Linux kernel.
	When using TM, the only failure-free choice is the irrevocable
	transaction, which is not used in the common case.
	In fact, the irrevocable transaction is not even available
	in all TM implementations.
}\QuickQuizEnd

What should TM researchers and developers do about all of this?

One approach is to focus on TM in the small, focusing on small
transactions where hardware assist potentially provides substantial
advantages over other synchronization primitives and on small programs
where there is some evidence for increased productivity for a combined
TM-locking approach~\cite{VPankratius2011TMvsLockingProductivity}.
Sun took the small-transaction approach with its Rock research
CPU~\cite{DaveDice2009ASPLOSRockHTM}.
Some TM researchers seem to agree with these two small-is-beautiful
approaches~\cite{JMStone93}, others have much higher hopes for TM, and yet others
hint that high TM aspirations might be TM's worst
enemy~\cite[Section 6]{Attiya:2010:ICT:1835698.1835699}.
It is nonetheless quite possible that TM will be able to take on larger
problems, and this section has listed a few of the issues that must be
resolved if TM is to achieve this lofty goal.

Of course, everyone involved should treat this as a learning experience.
It would seem that TM researchers have great deal to learn from
practitioners who have successfully built large software systems using
traditional synchronization primitives.

And vice versa.

\QuickQuiz{
	What is to learn?
	Why not just use TM for memory-based data structures and locking
	for those rare cases featuring the many silly corner cases listed
	in this silly section???
}\QuickQuizAnswer{
	The year 2005 just called, and it says that it wants its
	incandescent TM marketing hype back.

	In the year 2021, TM still has significant proving to do,
	even with the advent of HTM, which is covered in the
	upcoming
	\cref{sec:future:Hardware Transactional Memory}.
}\QuickQuizEnd

\begin{figure}
\centering
\resizebox{3in}{!}{\includegraphics{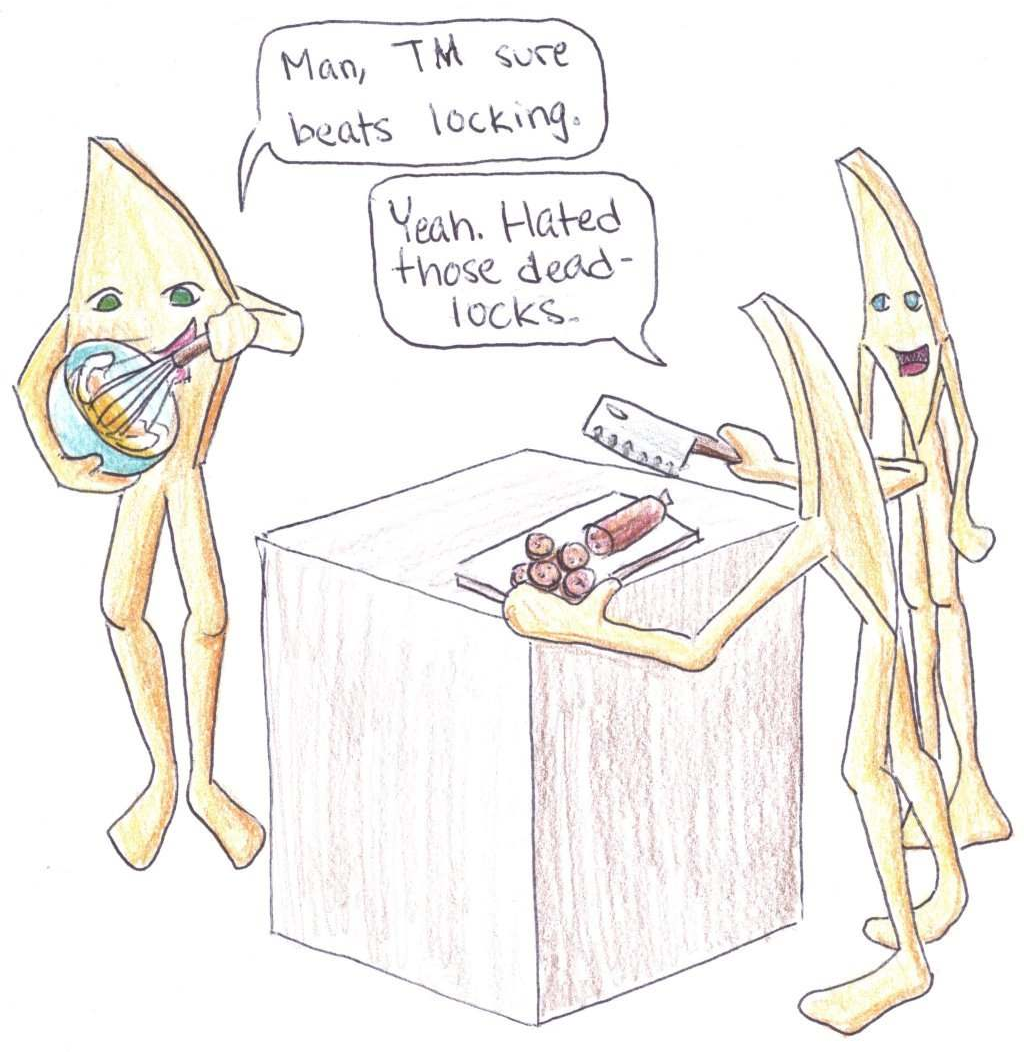}}
\caption{The STM Vision}
\ContributedBy{Figure}{fig:future:The STM Vision}{Melissa Broussard}
\end{figure}

\begin{figure}
\centering
\resizebox{2.7in}{!}{\includegraphics{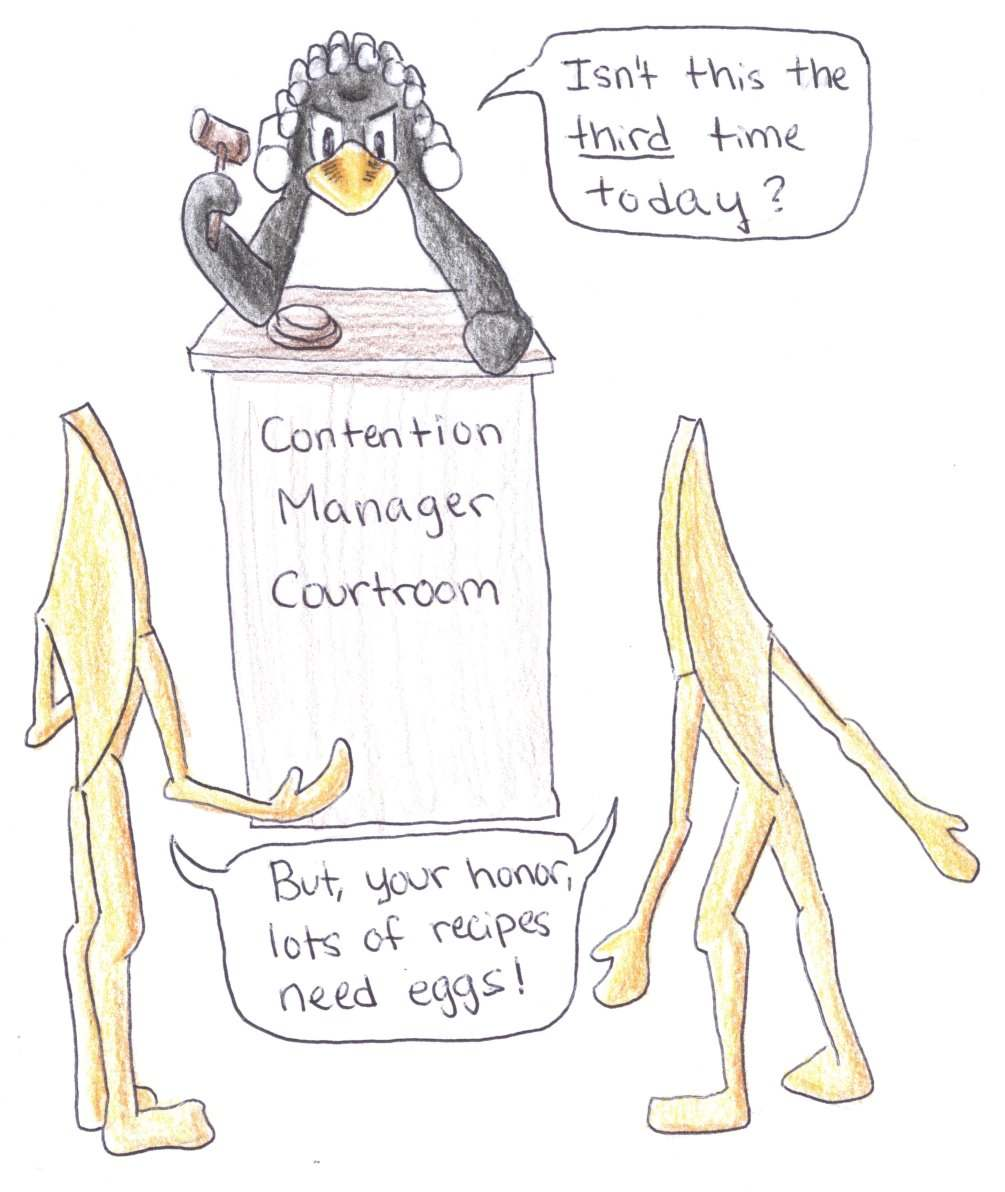}}
\caption{The STM Reality:
			  Conflicts}
\ContributedBy{Figure}{fig:future:The STM Reality: Conflicts}{Melissa Broussard}
\end{figure}

\begin{figure}
\centering
\resizebox{3in}{!}{\includegraphics{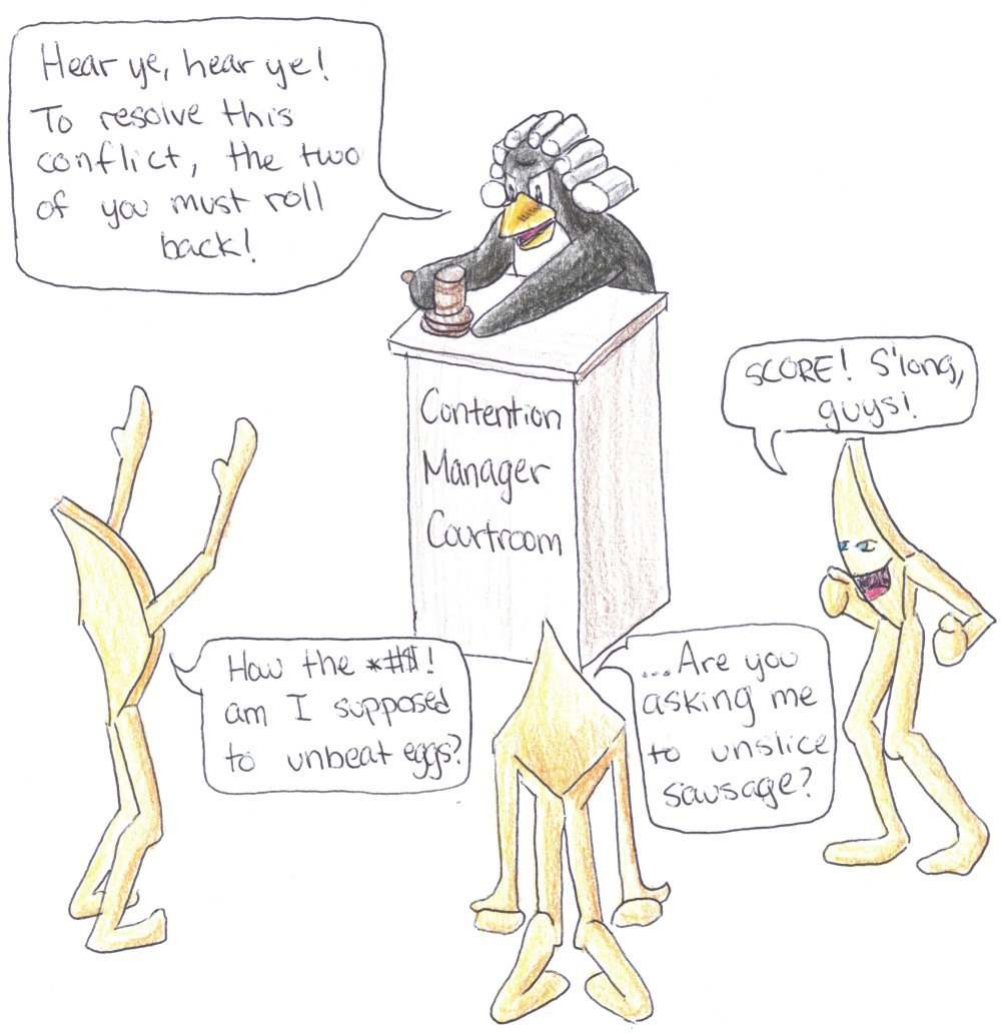}}
\caption{The STM Reality:
			  Irrevocable Operations}
\ContributedBy{Figure}{fig:future:The STM Reality: Irrevocable Operations}{Melissa Broussard}
\end{figure}

\begin{figure}
\centering
\resizebox{2.7in}{!}{\includegraphics{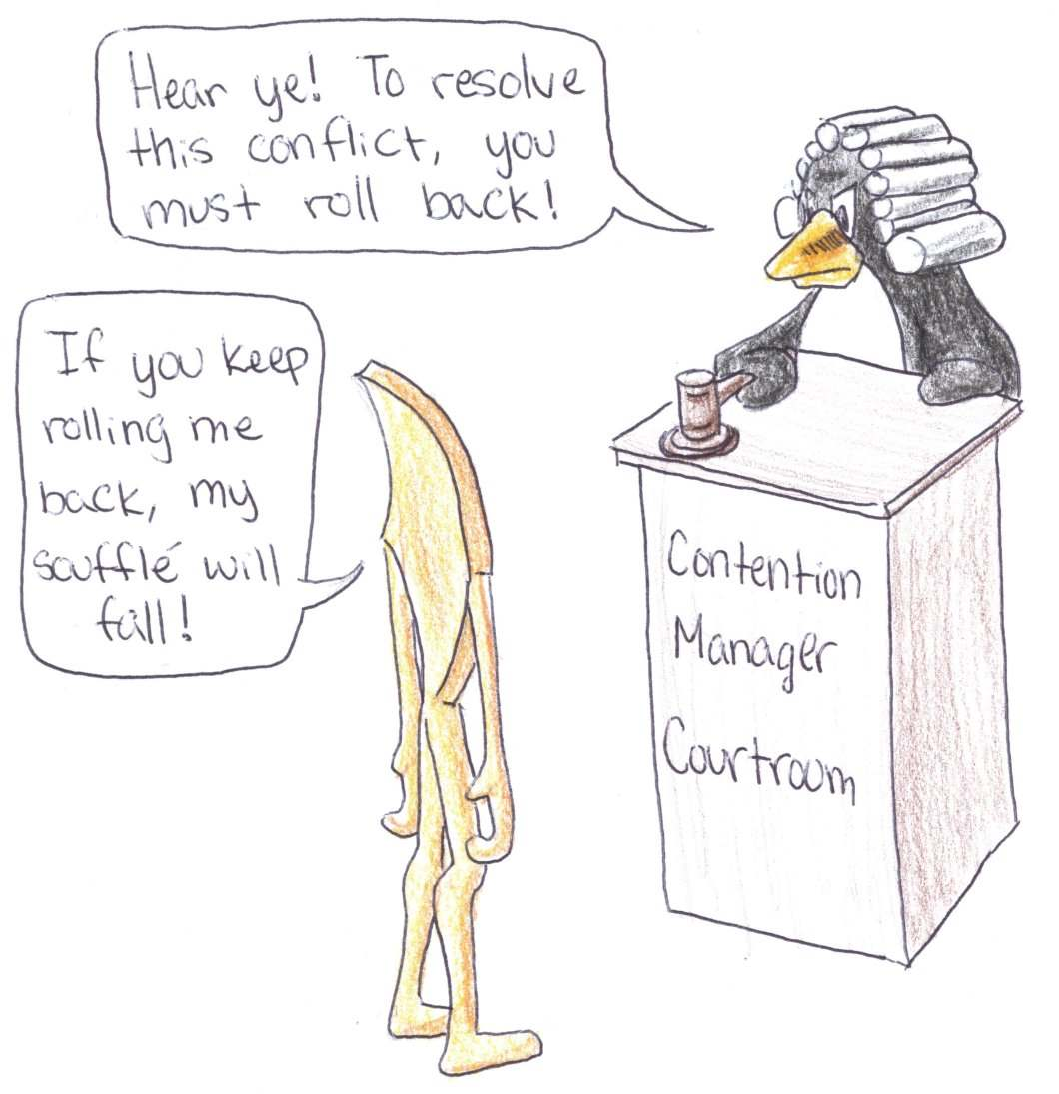}}
\caption{The STM Reality:
			  Realtime Response}
\ContributedBy{Figure}{fig:future:The STM Reality: Realtime Response}{Melissa Broussard}
\end{figure}

But for the moment, the current state of STM
can best be summarized with a series of cartoons.
First,
\cref{fig:future:The STM Vision}
shows the STM vision.
As always, the reality is a bit more nuanced, as fancifully depicted by
\cref{fig:future:The STM Reality: Conflicts,%
fig:future:The STM Reality: Irrevocable Operations,%
fig:future:The STM Reality: Realtime Response}.\footnote{
	Recent academic work-in-progress has investigated lock-based STM
	systems for real-time use~\cite{JimAnderson2019STMRT,CatherineNemitz2018LockSTMrealtime},
	albeit without any performance results, and with some indications
	that real-time hybrid STM/HTM systems must choose between fast
	common-case performance and worst-case forward-progress
	guarantees~\cite{DBLP:journals/corr/AlistarhKKRS14,MartinSchoeberl2010realtimeTM}.}
Less fanciful STM retrospectives are also
available~\cite{JoeDuffy2010RetroTM,JoeDuffy2010RetroTM2}.

Some commercially available hardware supports restricted variants of
HTM, which are addressed in the following section.

\IfTwoColumn{}{\FloatBarrier}
% future/htm.tex
% mainfile: ../perfbook.tex
% SPDX-License-Identifier: CC-BY-SA-3.0

\section{Hardware Transactional Memory}
\label{sec:future:Hardware Transactional Memory}
\epigraph{Make sure your report system is reasonably clean and efficient
	  before you automate.
	  Otherwise, your new computer will just speed up the mess.}
	 {Robert Townsend}
% If at first you do succeed---try to hide your astonishment.
% Harry F.~Banks

As of 2021, \IXacrf{htm} has been available for many
years on several types of commercially available commodity computer
systems~\cite{Yoo:2013:PEI:2503210.2503232,RickMerrit2011PowerTM,ChristianJacobi2012MainframeTM,TimothyHayes2020ARM-HTM}.
This section makes an attempt to identify HTM's place in the parallel
programmer's toolbox.

From a conceptual viewpoint, HTM uses processor caches and speculative
execution to make a designated group of statements (a ``transaction'')
take effect atomically
from the viewpoint of any other transactions running on other processors.
This transaction is initiated by a
begin-transaction machine instruction and completed by a commit-transaction
machine instruction.
There is typically also an abort-transaction machine instruction, which
squashes the speculation (as if the begin-transaction instruction and
all following instructions had not executed) and commences execution
at a failure handler.
The location of the failure handler is typically specified by the
begin-transaction instruction, either as an explicit failure-handler
address or via a condition code set by the instruction itself.
Each transaction executes atomically with respect to all other transactions.

HTM has a number of important benefits, including automatic
dynamic partitioning of data structures, reducing synchronization-primitive
cache misses, and supporting a fair number of practical applications.

However, it always pays to read the fine print, and HTM is no exception.
A major point of this section is determining under what conditions HTM's
benefits outweigh the complications hidden in its fine print.
To this end, \cref{sec:future:HTM Benefits WRT Locking}
describes HTM's benefits and
\cref{sec:future:HTM Weaknesses WRT Locking} describes its weaknesses.
This is the same approach used in earlier
papers~\cite{McKenney2007PLOSTM,PaulEMcKenney2010OSRGrassGreener}
and also in the previous section.\footnote{
	I gratefully acknowledge many stimulating
	discussions with the other authors, Maged Michael, Josh Triplett,
	and Jonathan Walpole, as well as with Andi Kleen.}

\Cref{sec:future:HTM Weaknesses WRT Locking When Augmented} then describes
HTM's weaknesses with respect to the combination of synchronization
primitives used in the Linux kernel (and in many user-space applications).
\Cref{sec:future:Where Does HTM Best Fit In?} looks at where HTM
might best fit into the parallel programmer's toolbox, and
\cref{sec:future:Potential Game Changers} lists some events that might
greatly increase HTM's scope and appeal.
Finally, \cref{sec:future:Conclusions}
presents concluding remarks.

\subsection{HTM Benefits WRT Locking}
\label{sec:future:HTM Benefits WRT Locking}

The primary benefits of HTM are
(1)~its avoidance of the cache misses that are often incurred by
other synchronization primitives,
(2)~its ability to dynamically partition
data structures,
and (3)~the fact that it has
a fair number of practical applications.
I break from TM tradition by not listing ease of use separately
for two reasons.
First, ease of use should stem from HTM's primary benefits,
which this section focuses on.
Second, there has been considerable controversy surrounding attempts to
test for raw programming
talent~\cite{RichardBornat2006SheepGoats,SaeedDehnadi2009SheepGoats,ElizabethPatitsas2020GradesNotBimodal}
and even around the use of small programming exercises in job
interviews~\cite{RegBraithwaite2007FizzBuzz}.
This indicates that we really do not have a firm grasp on what makes
programming easy or hard.
Therefore, the remainder of this section focuses on the three benefits
listed above.

\subsubsection{Avoiding Synchronization Cache Misses}
\label{sec:future:Avoiding Synchronization Cache Misses}

Most synchronization mechanisms are based on data structures that are
operated on by atomic instructions.
Because these atomic instructions normally operate by first causing
the relevant \IX{cache line} to be owned by the CPU that they are running on,
a subsequent execution
of the same instance of that synchronization primitive on some other
CPU will result in a cache miss.
These communications cache misses severely degrade both the performance and
scalability of conventional synchronization
mechanisms~\cite[Section 4.2.3]{Anderson97}.

In contrast, HTM synchronizes by using the CPU's cache, avoiding the need
for a separate synchronization data structure and resultant cache misses.
HTM's advantage is greatest in cases where a lock data structure is
placed in a separate cache line, in which case, converting a given
critical section to an HTM transaction can reduce that critical section's
overhead by a full cache miss.
These savings can be quite significant for the common case of short
critical sections, at least for those situations where the elided lock
does not share a cache line with an oft-written variable protected by
that lock.

\QuickQuiz{
	Why would it matter that oft-written variables shared the cache
	line with the lock variable?
}\QuickQuizAnswer{
	If the lock is in the same cacheline as some of the variables
	that it is protecting, then writes to those variables by one CPU
	will invalidate that cache line for all the other CPUs.
	These \IXpl{invalidation} will
	generate large numbers of conflicts and retries, perhaps even
	degrading performance and scalability compared to locking.
}\QuickQuizEnd

\subsubsection{Dynamic Partitioning of Data Structures}
\label{sec:future:Dynamic Partitioning of Data Structures}

A major obstacle to the use of some conventional synchronization mechanisms
is the need to statically partition data structures.
There are a number of data structures that are trivially
partitionable, with the most prominent example being hash tables,
where each hash chain constitutes a partition.
Allocating a lock for each hash chain then trivially parallelizes
the hash table for operations confined to a given chain.\footnote{
	And it is also easy to extend this scheme to operations accessing
	multiple hash chains by having such operations acquire the
	locks for all relevant chains in hash order.}
Partitioning is similarly trivial for arrays, radix trees, skiplists, and
several other data structures.

However, partitioning for many types of trees and graphs is quite
difficult, and the results are often quite complex~\cite{Ellis80}.
Although it is possible to use two-phased locking and hashed arrays
of locks to partition general data structures, other techniques
have proven preferable~\cite{DavidSMiller2006HashedLocking},
as will be discussed in
\cref{sec:future:HTM Weaknesses WRT Locking When Augmented}.
Given its avoidance of synchronization cache misses,
HTM is therefore a very real possibility for large non-partitionable
data structures, at least assuming relatively small updates.

\QuickQuiz{
	Why are relatively small updates important to \IXacr{htm} performance
	and scalability?
}\QuickQuizAnswer{
	The larger the updates, the greater the probability of conflict,
	and thus the greater probability of retries, which degrade
	performance.
}\QuickQuizEnd

\subsubsection{Practical Value}
\label{sec:future:Practical Value}

Some evidence of HTM's practical value has been demonstrated in a number
of hardware platforms, including
Sun Rock~\cite{DaveDice2009ASPLOSRockHTM},
Azul Vega~\cite{CliffClick2009AzulHTM},
IBM Blue Gene/Q~\cite{RickMerrit2011PowerTM},
Intel Haswell TSX~\cite{RaviRajwar2012TSX}, and
IBM System z~\cite{ChristianJacobi2012MainframeTM}.

Expected practical benefits include:

\begin{enumerate}
\item	Lock elision for in-memory data access and
	update~\cite{Martinez01a,Rajwar02a}.
\item	Concurrent access and small random updates to large non-partitionable
	data structures.
\end{enumerate}

However, HTM also has some very real shortcomings, which will be discussed
in the next section.

\subsection{HTM Weaknesses WRT Locking}
\label{sec:future:HTM Weaknesses WRT Locking}

The concept of HTM is quite simple:
A group of accesses and updates to memory occurs atomically.
However, as is the case with many simple ideas, complications arise
when you apply it to real systems in the real world.
These complications are as follows:

\begin{enumerate}
\item	Transaction-size limitations.
\item	Conflict handling.
\item	Aborts and rollbacks.
\item	Lack of forward-progress guarantees.
\item	Irrevocable operations.
\item	Semantic differences.
\end{enumerate}

Each of these complications is covered in the following sections,
followed by a summary.

\subsubsection{Transaction-Size Limitations}
\label{sec:future:Transaction-Size Limitations}

The transaction-size limitations of current HTM implementations
stem from the use of the processor caches to hold the data
affected by the transaction.
Although this allows a given CPU to make the transaction appear atomic to
other CPUs by executing the transaction within the confines of its cache,
it also means that any transaction that does not fit cannot commit.
Furthermore, events that change execution context, such as interrupts,
system calls, exceptions, traps, and context switches either must
abort any ongoing transaction on the CPU in question or must further
restrict transaction size due to the cache footprint of the other
execution context.

Of course, modern CPUs tend to have large caches, and the data required
for many transactions would fit easily in a one-megabyte cache.
Unfortunately, with caches, sheer size is not all that matters.
The problem is that most caches
can be thought of hash tables implemented in hardware.
However, hardware caches do not chain their buckets (which are normally
called \emph{sets}), but rather
provide a fixed number of cachelines per set.
The number of elements provided for each set in a given cache
is termed that cache's \emph{\IXalt{associativity}{cache associativity}}.

Although cache associativity varies, the eight-way associativity of
the level-0 cache on the laptop I am typing this on is not unusual.
What this means is that if a given transaction needed to touch
nine cache lines, and if all nine cache lines mapped to the same
set, then that transaction cannot possibly complete, never mind how
many megabytes of additional space might be available in that cache.
Yes, given randomly selected data elements in a given data structure,
the probability of that transaction being able to commit is quite
high, but there can be no guarantee~\cite{PaulEMcKenney2012HTMCacheGeometry}.

There has been some research work to alleviate this limitation.
Fully associative \emph{victim caches} would alleviate the associativity
constraints, but there are currently stringent performance and
energy-efficiency constraints on the sizes of victim caches.
That said, HTM victim caches for unmodified cache lines can be quite
small, as they need to retain only the address:
The data itself can be written to memory or shadowed by other caches,
while the address itself is sufficient to detect a conflicting
write~\cite{RaviRajwar2012TSX}.

\IXAcrmfst{utm}
schemes~\cite{CScottAnanian2006,KevinEMoore2006}
use DRAM as an extremely large victim cache, but integrating such schemes
into a production-quality
\IXalt{cache-coherence}{cache coherence} mechanism is still an unsolved
problem.
In addition, use of DRAM as a victim cache may have unfortunate
performance and energy-efficiency consequences, particularly
if the victim cache is to be
\IXalth{fully associative}{fully associative}{cache}.
Finally, the ``unbounded'' aspect of UTM assumes that all of DRAM
could be used as a victim cache, while in reality
the large but still fixed amount of DRAM assigned to a given CPU
would limit the size of that CPU's transactions.
Other schemes use a combination of hardware and software transactional
memory~\cite{SanjeevKumar2006} and one could imagine using \IXacr{stm} as a
fallback mechanism for HTM\@.

However, to the best of my knowledge, with the exception of abbreviating
representation of TM read sets, currently available systems do not
implement any of these research ideas, and perhaps for good reason.

\subsubsection{Conflict Handling}
\label{sec:future:Conflict Handling}

The first complication is the possibility of \emph{conflicts}.
For example, suppose that transactions~A and~B are defined as follows:

\begin{VerbatimU}
Transaction A       Transaction B

x = 1;              y = 2;
y = 3;              x = 4;
\end{VerbatimU}

Suppose that each transaction executes concurrently on its own processor.
If transaction~A stores to \co{x} at the same time that transaction~B
stores to \co{y}, neither transaction can progress.
To see this, suppose that transaction~A executes its store to \co{y}.
Then transaction~A will be interleaved within transaction~B, in violation
of the requirement that transactions execute atomically with respect to
each other.
Allowing transaction~B to execute its store to \co{x} similarly violates
the atomic-execution requirement.
This situation is termed a \emph{conflict}, which happens whenever two
concurrent transactions access the same variable where at least one of
the accesses is a store.
The system is therefore obligated to abort one or both of the transactions
in order to allow execution to progress.
The choice of exactly which transaction to abort is an interesting topic
that will very likely retain the ability to generate Ph.D. dissertations for
some time to come, see for
example~\cite{EgeAkpinar2011HTM2TLE}.\footnote{
	Liu's and Spear's paper entitled ``Toxic
	Transactions''~\cite{YujieLiu2011ToxicTransactions} is
	particularly instructive.}
For the purposes of this section, we can assume that the system makes
a random choice.

Another complication is conflict detection, which is comparatively
straightforward, at least in the simplest case.
When a processor is executing a transaction, it marks every cache line
touched by that transaction.
If the processor's cache receives a request involving a cache line that
has been marked as touched by the current transaction, a potential
conflict has occurred.
More sophisticated systems might try to order the current processors'
transaction to precede that of the processor sending the request, and
optimizing this process will likely also retain the ability to generate
Ph.D. dissertations for quite some time.
However this section assumes a very simple conflict-detection strategy.

However, for HTM to work effectively, the probability of conflict must
be quite low, which in turn requires that the data structures
be organized so as to maintain a sufficiently low probability of conflict.
For example, a red-black tree with simple insertion, deletion, and search
operations fits this description, but a red-black
tree that maintains an accurate count of the number of elements in
the tree does not.\footnote{
	The need to update the count would result in additions to and
	deletions from the tree conflicting with each other, resulting
	in strong non-commutativity~\cite{HagitAttiya2011LawsOfOrder,Attiya:2011:LOE:1925844.1926442,PaulEMcKenney2011SNC}.}
For another example, a red-black tree that enumerates all elements in
the tree in a single transaction will have high conflict probabilities,
degrading performance and scalability.
As a result, many serial programs will require some restructuring before
HTM can work effectively.
In some cases, practitioners will prefer to take the extra steps
(in the red-black-tree case, perhaps switching to a partitionable
data structure such as a radix tree or a hash table), and just
use locking, particularly until such time as HTM is readily available
on all relevant
architectures~\cite{CliffClick2009AzulHTM}.

\QuickQuiz{
	How could a red-black tree possibly efficiently enumerate all
	elements of the tree regardless of choice of synchronization
	mechanism???
}\QuickQuizAnswer{
	In many cases, the enumeration need not be exact.
	In these cases, hazard pointers or \IXacr{rcu} may be used to protect
	readers, which provides low probability of conflict with any
	given insertion or deletion.
}\QuickQuizEnd

Furthermore, the potential for conflicting accesses among concurrent
transactions can result in failure.
Handling such failure is discussed in the next section.

\subsubsection{Aborts and Rollbacks}
\label{sec:future:Aborts and Rollbacks}

Because any transaction might be aborted at any time, it is important
that transactions contain no statements that cannot be rolled back.
This means that transactions cannot do I/O, system calls, or debugging
breakpoints (no single stepping in the debugger for HTM transactions!!!).
Instead, transactions must confine themselves to accessing normal
cached memory.
Furthermore, on some systems, interrupts, exceptions, traps,
TLB misses, and other events will also abort transactions.
Given the number of bugs that have resulted from improper handling
of error conditions, it is fair to ask what impact aborts and rollbacks
have on ease of use.

\QuickQuiz{
	But why can't a debugger emulate single stepping by setting
	breakpoints at successive lines of the transaction, relying
	on the retry to retrace the steps of the earlier instances
	of the transaction?
}\QuickQuizAnswer{
	This scheme might work with reasonably high probability, but it
	can fail in ways that would be quite surprising to most users.
	To see this, consider the following transaction:

\begin{fcvlabel}[ln:future:htm:debug rollbacks]
\begin{VerbatimN}[commandchars=\\\[\]]
begin_trans();
if (a) {
	do_one_thing();
	do_another_thing();	\lnlbl[another]
} else {
	do_a_third_thing();
	do_a_fourth_thing();
}
end_trans();
\end{VerbatimN}
\end{fcvlabel}

	\begin{fcvref}[ln:future:htm:debug rollbacks]
	Suppose that the user sets a breakpoint at \clnref{another},
	which triggers,
	aborting the transaction and entering the debugger.
	\end{fcvref}
	Suppose that between the time that the breakpoint triggers
	and the debugger gets around to stopping all the threads, some
	other thread sets the value of \co{a} to zero.
	When the poor user attempts to single-step the program, surprise!
	The program is now in the else-clause instead of the then-clause.

	This is \emph{not} what I call an easy-to-use debugger.
}\QuickQuizEnd

Of course, aborts and rollbacks raise the question of whether HTM can
be useful for hard real-time systems.
Do the performance benefits of HTM outweigh the costs of the aborts
and rollbacks, and if so under what conditions?
Can transactions use priority boosting?
Or should transactions for high-priority threads instead preferentially
abort those of low-priority threads?
If so, how is the hardware efficiently informed of priorities?
The literature on real-time use of HTM is quite sparse, perhaps
because there are more than enough problems in making HTM work well in
non-real-time environments.

Because current HTM implementations might deterministically abort a
given transaction, software must provide fallback code.
This fallback code must use some other form of synchronization, for
example, locking.
If a lock-based fallback is ever used, then all the limitations of locking,
including the possibility of \IX{deadlock}, reappear.
One can of course hope that the fallback isn't used often, which might
allow simpler and less deadlock-prone locking designs to be used.
But this raises the question of how the system transitions from using
the lock-based fallbacks back to transactions.\footnote{
	The possibility of an application getting stuck in fallback
	mode has been termed the ``lemming effect'', a term that
	Dave Dice has been credited with coining.}
One approach is to use a test-and-test-and-set discipline~\cite{Martinez02a},
so that everyone holds off until the lock is released, allowing the
system to start from a clean slate in transactional mode at that point.
However, this could result in quite a bit of spinning, which might not
be wise if the lock holder has blocked or been preempted.
Another approach is to allow transactions to proceed in parallel with
a thread holding a lock~\cite{Martinez02a}, but this raises difficulties
in maintaining atomicity, especially if the reason that the thread is
holding the lock is because the corresponding transaction would not fit
into cache.

Finally, dealing with the possibility of aborts and rollbacks seems to
put an additional burden on the developer, who must correctly handle
all combinations of possible error conditions.

It is clear that users of HTM must put considerable validation effort
into testing both the fallback code paths and transition from fallback
code back to transactional code.
Nor is there any reason to believe that the validation requirements of
HTM hardware are any less daunting.

\subsubsection{Lack of Forward-Progress Guarantees}
\label{sec:future:Lack of Forward-Progress Guarantees}

Even though transaction size, conflicts, and aborts/rollbacks can all
cause transactions to abort, one might hope that sufficiently small and
short-duration transactions could be guaranteed to eventually succeed.
This would permit a transaction to be unconditionally retried, in the
same way that \IXacrmf{cas} and load-linked/store-conditional
(LL/SC) operations are unconditionally retried in code that uses these
instructions to implement atomic operations.

Unfortunately, other than low-clock-rate academic research
prototypes~\cite{MartinSchoeberl2010realtimeTM},
currently available HTM implementations refuse to make any
sort of forward-progress guarantee.
As noted earlier, HTM therefore cannot be used to avoid deadlock on
those systems.
Hopefully future implementations of HTM will provide some sort of
forward-progress guarantees.
Until that time, HTM must be used with extreme caution in real-time
applications.

The one exception to this gloomy picture as of 2021 is
the IBM mainframe, which provides
\emph{constrained transactions}~\cite{ChristianJacobi2012MainframeTM}.
The constraints are quite severe, and are presented in
\cref{sec:future:Forward-Progress Guarantees}.
It will be interesting to see if HTM forward-progress guarantees migrate
from the mainframe to commodity CPU families.

\subsubsection{Irrevocable Operations}
\label{sec:future:Irrevocable Operations}

Another consequence of aborts and rollbacks is that HTM transactions
cannot accommodate irrevocable operations.
Current HTM implementations typically enforce this limitation by
requiring that all of the accesses in the transaction be to cacheable
memory (thus prohibiting MMIO accesses) and aborting transactions on
interrupts, traps, and exceptions (thus prohibiting system calls).

Note that buffered I/O can be accommodated by HTM transactions as
long as the buffer fill/flush operations occur extra-transactionally.
The reason that this works is that adding data to and removing data
from the buffer is revocable:
Only the actual buffer fill/flush operations are irrevocable.
Of course, this buffered-I/O approach has the effect of including the I/O
in the transaction's footprint, increasing the size of the transaction
and thus increasing the probability of failure.

\subsubsection{Semantic Differences}
\label{sec:future:Semantic Differences}

Although HTM can in many cases be used as a drop-in replacement for locking
(hence the name \IXacrfst{tle}~\cite{DaveDice2008TransactLockElision}),
there are subtle differences in semantics.
A particularly nasty example involving coordinated lock-based critical
sections that results in deadlock or \IX{livelock} when executed transactionally
was given by Blundell~\cite{Blundell2006TMdeadlock}, but a much simpler
example is the empty critical section.

In a lock-based program, an empty critical section will guarantee
that all processes that had previously been holding that lock have
now released it.
This idiom was used by the 2.4 Linux kernel's networking stack to
coordinate changes in configuration.
But if this empty critical section is translated to a transaction,
the result is a no-op.
The guarantee that all prior critical sections have terminated is
lost.
In other words, transactional lock elision preserves the data-protection
semantics of locking, but loses locking's time-based messaging semantics.

\QuickQuizSeries{%
\QuickQuizB{
	But why would \emph{anyone} need an empty lock-based critical
	section???
}\QuickQuizAnswerB{
	See the answer to \QuickQuizARef{\QlockingQemptycriticalsection} in
	\cref{sec:locking:Exclusive Locks}.

	However, it is claimed that given a strongly atomic \IXacr{htm}
	implementation without forward-progress guarantees, any
	memory-based locking design based on empty critical sections
	will operate correctly in the presence of transactional
	lock elision.
	Although I have not seen a proof of this statement, there
	is a straightforward rationale for this claim.
	The main idea is that in a strongly atomic HTM implementation,
	the results of a given transaction are not visible until
	after the transaction completes successfully.
	Therefore, if you can see that a transaction has started,
	it is guaranteed to have already completed, which means
	that a subsequent empty lock-based critical section will
	successfully ``wait'' on it---after all, there is no waiting
	required.

	This line of reasoning does not apply to weakly atomic
	systems (including many \IXacr{stm} implementation), and it also
	does not apply to lock-based programs that use means other
	than memory to communicate.
	One such means is the passage of time (for example, in
	hard real-time systems) or flow of priority (for example,
	in soft real-time systems).

	Locking designs that rely on priority boosting are of particular
	interest.
}\QuickQuizEndB
\QuickQuizM{
	Can't transactional lock elision trivially handle locking's
	time-based messaging semantics
	by simply choosing not to elide empty lock-based critical sections?
}\QuickQuizAnswerM{
	It could do so, but this would be both unnecessary and
	insufficient.

	It would be unnecessary in cases where the empty critical section
	was due to conditional compilation.
	Here, it might well be that the only purpose of the lock was to
	protect data, so eliding it completely would be the right thing
	to do.
	In fact, leaving the empty lock-based critical section would
	degrade performance and scalability.

	On the other hand, it is possible for a non-empty lock-based
	critical section to be relying on both the data-protection
	and time-based and messaging semantics of locking.
	Using transactional lock elision in such a case would be
	incorrect, and would result in bugs.
}\QuickQuizEndM
\QuickQuizE{
	Given modern hardware~\cite{PeterOkech2009InherentRandomness},
	how can anyone possibly expect parallel software relying
	on timing to work?
}\QuickQuizAnswerE{
	The short answer is that on commonplace commodity hardware,
	synchronization designs based on any sort of fine-grained
	timing are foolhardy and cannot be expected to operate correctly
	under all conditions.

	That said, there are systems designed for hard real-time use
	that are much more deterministic.
	In the (very unlikely) event that you are using such a system,
	here is a toy example showing how time-based synchronization can
	work.
	Again, do \emph{not} try this on commodity microprocessors,
	as they have highly nondeterministic performance characteristics.

	This example uses multiple worker threads along with a control
	thread.
	Each worker thread corresponds to an outbound data feed, and
	records the current time (for example, from the
	\co{clock_gettime()} system call) in a per-thread
	\co{my_timestamp} variable after executing each unit
	of work.
	The real-time nature of this example results in the following
	set of constraints:

	\begin{enumerate}
	\item	It is a fatal error for a given worker thread to fail
		to update its timestamp for a time period of more than
		\co{MAX_LOOP_TIME}.
	\item	Locks are used sparingly to access and update global
		state.
	\item	Locks are granted in strict FIFO order within
		a given thread priority.
	\end{enumerate}

	When worker threads complete their feed, they must disentangle
	themselves from the rest of the application and place a status
	value in a per-thread \co{my_status} variable that is initialized
	to $-1$.
	Threads do not exit; they instead are placed on a thread pool
	to accommodate later processing requirements.
	The control thread assigns (and re-assigns) worker threads as
	needed, and also maintains a histogram of thread statuses.
	The control thread runs at a real-time priority no higher than
	that of the worker threads.

	Worker threads' code is as follows:

\begin{VerbatimN}
	int my_status = -1;  /* Thread local. */

	while (continue_working()) {
		enqueue_any_new_work();
		wp = dequeue_work();
		do_work(wp);
		my_timestamp = clock_gettime(...);
	}

	acquire_lock(&departing_thread_lock);

	/*
	 * Disentangle from application, might
	 * acquire other locks, can take much longer
	 * than MAX_LOOP_TIME, especially if many
	 * threads exit concurrently.
	 */
	my_status = get_return_status();
	release_lock(&departing_thread_lock);

	/* thread awaits repurposing. */
\end{VerbatimN}

	The control thread's code is as follows:

\begin{fcvlabel}[ln:future:htm:control thread]
\begin{VerbatimN}[commandchars=\\\@\$]
	for (;;) {
		for_each_thread(t) {
			ct = clock_gettime(...);
			d = ct - per_thread(my_timestamp, t);
			if (d >= MAX_LOOP_TIME) {	\lnlbl@if$
				/* thread departing. */	\lnlbl@dep:b$
				acquire_lock(&departing_thread_lock); \lnlbl@acq$
				release_lock(&departing_thread_lock); \lnlbl@rel$
				i = per_thread(my_status, t);
				status_hist[i]++; /* Bug if TLE! */ \lnlbl@dep:e$
			}
		}
		/* Repurpose threads as needed. */
	}
\end{VerbatimN}
\end{fcvlabel}

	\begin{fcvref}[ln:future:htm:control thread]
	\Clnref{if} uses the passage of time to deduce that the thread
	has exited, executing \clnref{dep:b,dep:e} if so.
	The empty lock-based critical section on \clnref{acq,rel}
	guarantees that any thread in the process of exiting
	completes (remember that locks are granted in FIFO order!).
	\end{fcvref}

	Once again, do not try this sort of thing on commodity
	microprocessors.
	After all, it is difficult enough to get this right on systems
	specifically designed for hard real-time use!
}\QuickQuizEndE
}

One important semantic difference between locking and transactions
is the priority boosting that is used to avoid priority inversion
in lock-based real-time programs.
One way in which priority inversion can occur is when a
low-priority thread holding a lock
is preempted by a medium-priority CPU-bound thread.
If there is at least one such medium-priority thread per CPU, the
low-priority thread will never get a chance to run.
If a high-priority thread now attempts to acquire the lock,
it will block.
It cannot acquire the lock until the low-priority thread releases it,
the low-priority thread cannot release the lock until it gets a chance
to run, and it cannot get a chance to run until one of the medium-priority
threads gives up its CPU\@.
Therefore, the medium-priority threads are in effect blocking the
high-priority process, which is the rationale for the name ``priority
inversion.''

\begin{listing}
\begin{fcvlabel}[ln:future:Exploiting Priority Boosting]
\begin{VerbatimL}[commandchars=\\\@\$]
void boostee(void)		\lnlbl@low:b$
{
	int i = 0;

	acquire_lock(&boost_lock[i]);	\lnlbl@1stacq$
	for (;;) {
		acquire_lock(&boost_lock[!i]);
		release_lock(&boost_lock[i]);
		i = i ^ 1;
		do_something();
	}
}				\lnlbl@low:e$

void booster(void)		\lnlbl@high:b$
{
	int i = 0;

	for (;;) {
		usleep(500); /* sleep 0.5 ms. */
		acquire_lock(&boost_lock[i]);	\lnlbl@acq$
		release_lock(&boost_lock[i]);	\lnlbl@rel$
		i = i ^ 1;
	}
}                               \lnlbl@high:e$
\end{VerbatimL}
\end{fcvlabel}
\caption{Exploiting Priority Boosting}
\label{lst:future:Exploiting Priority Boosting}
\end{listing}

One way to avoid priority inversion is \emph{priority inheritance},
in which a high-priority thread blocked on a lock temporarily donates
its priority to the lock's holder, which is also called \emph{priority
boosting}.
However, priority boosting can be used for things other than avoiding
priority inversion, as shown in
\cref{lst:future:Exploiting Priority Boosting}.
\begin{fcvref}[ln:future:Exploiting Priority Boosting]
\Clnrefrange{low:b}{low:e} of this listing show a low-priority process that must
nevertheless run every millisecond or so, while \clnrefrange{high:b}{high:e} of
this same listing show a high-priority process that uses priority
boosting to ensure that \co{boostee()} runs periodically as needed.

The \co{boostee()} function arranges this by always holding one of
the two \co{boost_lock[]} locks, so that \clnrefrange{acq}{rel} of
\co{booster()} can boost priority as needed.
\end{fcvref}

\QuickQuiz{
	But the \co{boostee()} function in
	\cref{lst:future:Exploiting Priority Boosting}
	alternatively acquires its locks in reverse order!
	Won't this result in deadlock?
}\QuickQuizAnswer{
	No deadlock will result.
	To arrive at deadlock, two different threads must each
	acquire the two locks in opposite orders, which does not
	happen in this example.
	However, deadlock detectors such as
	lockdep~\cite{JonathanCorbet2006lockdep}
	will flag this as a false positive.
}\QuickQuizEnd

\begin{fcvref}[ln:future:Exploiting Priority Boosting]
This arrangement requires that \co{boostee()} acquire its first
lock on \clnref{1stacq} before the system becomes busy, but this is easily
arranged, even on modern hardware.

Unfortunately, this arrangement can break down in presence of transactional
lock elision.
The \co{boostee()} function's overlapping critical sections become
one infinite transaction, which will sooner or later abort,
for example, on the first time that the thread running
the \co{boostee()} function is preempted.
At this point, \co{boostee()} will fall back to locking, but given
its low priority and that the quiet initialization period is now
complete (which after all is why \co{boostee()} was preempted),
this thread might never again get a chance to run.

And if the \co{boostee()} thread is not holding the lock, then
the \co{booster()} thread's empty critical section on \clnref{acq,rel} of
\cref{lst:future:Exploiting Priority Boosting}
will become an empty transaction that has no effect, so that
\co{boostee()} never runs.
This example illustrates some of the subtle consequences of
transactional memory's rollback-and-retry semantics.
\end{fcvref}

Given that experience will likely uncover additional subtle semantic
differences, application of HTM-based lock elision to large programs
should be undertaken with caution.
That said, where it does apply, HTM-based lock elision can eliminate
the cache misses associated with the lock variable, which has resulted
in tens of percent performance increases in large real-world software
systems as of early 2015.
We can therefore expect to see substantial use of this technique on
hardware providing reliable support for it.

\QuickQuiz{
	So a bunch of people set out to supplant locking, and they
	mostly end up just optimizing locking???
}\QuickQuizAnswer{
	At least they accomplished something useful!
	And perhaps there will continue to be additional \IXacr{htm} progress
	over time~\cite{Siakavaras2017CombiningHA,DimitriosSiakavaras2020RCU-HTM-B+Trees,ChristinaGiannoula2018HTM-RCU-graphcoloring,SeongJaePark2020HTMRCUlock}.
}\QuickQuizEnd

\subsubsection{Summary}
\label{sec:future:HTM Weaknesses WRT Locking: Summary}

% future/HTMtable.tex
% SPDX-License-Identifier: CC-BY-SA-3.0

\begin{table*}
\centering
% \scriptsize
\small
% future/HTMtableColor.tex
% SPDX-License-Identifier: CC-BY-SA-3.0

\definecolor{plus}{cmyk}{0.1,0,0,0}
\definecolor{minus}{cmyk}{0,0.05,0.2,0.05}
\definecolor{down}{cmyk}{0,0.15,0.15,0.1}
\newcommand{\Pl}{\cellcolor{plus}}
\newcommand{\Mn}{\cellcolor{minus}}
\newcommand{\Dw}{\cellcolor{down}}

\setlength{\tabcolsep}{4pt}\OneColumnHSpace{-.9in}
\ebresizewidth{
\begin{tabularx}{6.5in}{p{0.95in}cXcX}
\toprule
  &
    & \multicolumn{1}{c}{Locking} &
      & \multicolumn{1}{c}{Hardware Transactional Memory} \\
\midrule
  Basic Idea &
    & Allow only one thread at a time to access a given set of objects. &
      & Cause a given operation over a set of objects to execute atomically. \\
\midrule
  Scope &
    & \Pl Handles all operations. &
      & \Pl Handles revocable operations. \\
\addlinespace[4pt]
  &
    & &
      & \Mn Irrevocable operations force fallback (typically to locking). \\
\midrule
  Composability &
    & \Dw Limited by deadlock. &
      & \Dw Limited by irrevocable operations, transaction size,
        and deadlock (assuming lock-based fallback code). \\
\midrule
  Scalability \& Performance &
    & \Mn Data must be partitionable to avoid lock contention. &
      & \Mn Data must be partitionable to avoid conflicts. \\
\cmidrule{3-5}
  &
    & \Dw Partioning must typically be fixed at design time. &
      & \Pl Dynamic adjustment of partitioning carried out automatically down
        to cacheline boundaries. \\
\addlinespace[4pt]
  &
    & &
      & \Mn Partitioning required for fallbacks (less important for rare
        fallbacks). \\
\cmidrule{3-5}
  &
    & \Dw Locking primitives typically result in expensive cache misses
      and memory-barrier instructions. &
      & \Mn Transactions begin/end instructions typically do not result in cache
        misses, but do have memory-ordering and overhead consequences. \\
\cmidrule{3-5}
  &
    & \Pl Contention effects are focused on acquisition and release, so
      that the critical section runs at full speed. &
      & \Mn Contention aborts conflicting transactions, even if they have been
        running for a long time. \\
\cmidrule{3-5}
  &
    & \Pl Privatization operations are simple, intuitive, performant,
      and scalable. &
      & \Mn Privatized data contributes to transaction size. \\
\midrule
  Hardware Support &
    & \Pl Commodity hardware suffices. &
      & \Mn New hardware required (and is starting to become available). \\
\cmidrule{3-5}
  &
    & \Pl Performance is insensitive to cache-geometry details. &
      & \Mn Performance depends critically on cache geometry. \\
\midrule
  Software Support &
    & \Pl APIs exist, large body of code and experience, debuggers operate
      naturally. &
      & \Mn APIs emerging, little experience outside of DBMS, breakpoints
        mid-transaction can be problematic. \\
\midrule
  Interaction With Other Mechanisms &
    & \Pl Long experience of successful interaction. &
      & \Dw Just beginning investigation of interaction. \\
\midrule
  Practical Apps &
    & \Pl Yes. &
      & \Pl Yes. \\
\midrule
  Wide Applicability &
    & \Pl Yes. &
      & \Mn Jury still out. \\
\bottomrule
\end{tabularx}
}
\IfTwoColumn{
\caption{Comparison of Locking and HTM (\colorbox{plus}{Advantage},
  \colorbox{minus}{Disadvantage}, \colorbox{down}{Strong Disadvantage})}
}{
\caption{Comparison of Locking and HTM (\colorbox{plus}{Advantage},
  \colorbox{minus}{Disadvantage},
  \colorbox{down}{Strong} \colorbox{down}{Disadvantage})}
}
\label{tab:future:Comparison of Locking and HTM}
\end{table*}

Although it seems likely that HTM will have compelling use cases,
current implementations have serious transaction-size limitations,
conflict-handling complications, abort-and-rollback issues, and
semantic differences that will require careful handling.
HTM's current situation relative to locking is summarized in
\cref{tab:future:Comparison of Locking and HTM}.
As can be seen, although the current state of HTM alleviates some
serious shortcomings of locking,\footnote{
	In fairness, it is important to emphasize that locking's shortcomings
	do have well-known and heavily used engineering solutions, including
	deadlock detectors~\cite{JonathanCorbet2006lockdep}, a wealth
	of data structures that have been adapted to locking, and
	a long history of augmentation, as discussed in
	\cref{sec:future:HTM Weaknesses WRT Locking When Augmented}.
	In addition, if locking really were as horrible as a quick skim
	of many academic papers might reasonably lead one to believe,
	where did all the large lock-based parallel programs (both
	FOSS and proprietary) come from, anyway?}
it does so by introducing a significant
number of shortcomings of its own.
These shortcomings are acknowledged by leaders in the TM
community~\cite{AlexanderMatveev2012PessimisticTM}.\footnote{
	In addition, in early 2011, I was invited to deliver a critique of
	some of the assumptions underlying transactional
	memory~\cite{PaulEMcKenney2011Verico}.
	The audience was surprisingly non-hostile, though perhaps they
	were taking it easy on me due to the fact that I was heavily
	jet-lagged while giving the presentation.}

In addition, this is not the whole story.
Locking is not normally used by itself, but is instead typically
augmented by other synchronization mechanisms,
including reference counting, atomic operations, non-blocking data structures,
\IXpl{hazard pointer}~\cite{MagedMichael04a,HerlihyLM02},
and RCU~\cite{McKenney98,McKenney01a,ThomasEHart2007a,PaulEMcKenney2012ELCbattery}.
The next section looks at how such augmentation changes the equation.

\subsection{HTM Weaknesses WRT Locking When Augmented}
\label{sec:future:HTM Weaknesses WRT Locking When Augmented}

% future/HTMtableRCU.tex
% SPDX-License-Identifier: CC-BY-SA-3.0

\begin{table*}
\centering
\small
% future/HTMtableColor.tex
% SPDX-License-Identifier: CC-BY-SA-3.0

\definecolor{plus}{cmyk}{0.1,0,0,0}
\definecolor{minus}{cmyk}{0,0.05,0.2,0.05}
\definecolor{down}{cmyk}{0,0.15,0.15,0.1}
\newcommand{\Pl}{\cellcolor{plus}}
\newcommand{\Mn}{\cellcolor{minus}}
\newcommand{\Dw}{\cellcolor{down}}

\setstretch{0.95}
\setlength{\tabcolsep}{4pt}\OneColumnHSpace{-.9in}
\ebresizewidth{
\resizebox{6.5in}{!}{
\begin{tabularx}{6.8in}{p{0.95in}cXcX}
\toprule
  &
    & \multicolumn{1}{c}{Locking with Userspace RCU or Hazard Pointers} &
      & \multicolumn{1}{c}{Hardware Transactional Memory} \\
\midrule
  Basic Idea &
    & Allow only one thread at a time to access a given set of objects. &
      & Cause a given operation over a set of objects to execute atomically. \\
\midrule
  Scope &
    & \Pl Handles all operations. &
      & \Pl Handles revocable operations. \\
\addlinespace[4pt]
  &
    & &
      & \Mn Irrevocable operations force fallback (typically to locking). \\
\midrule
  Composability &
    & \Pl Readers limited only by grace-period-wait operations. &
      & \Dw Limited by irrevocable operations, transaction size, and deadlock.
        (Assuming lock-based fallback code.)\\
\addlinespace[4pt]
  &
    & \Mn Updaters limited by deadlock.  Readers reduce deadlock. &
      & \\
\midrule
  Scalability \& Performance &
    & \Mn Data must be partitionable to avoid lock contention among updaters. &
      & \Mn Data must be partitionable to avoid conflicts. \\
\addlinespace[4pt]
  &
    & \Pl Partitioning not needed for readers. &
      & \\
\cmidrule{3-5}
  &
    & \Dw Partitioning for updaters must typically be fixed at design time. &
      & \Pl Dynamic adjustment of partitioning carried out automatically down
        to cacheline boundaries. \\
\cmidrule{3-5}
  &
    & \Pl Partitioning not needed for readers. &
      & \Mn Partitioning required for fallbacks (less important for rare
        fallbacks). \\
\cmidrule{3-5}
  &
    & \Dw Updater locking primitives typically result in expensive cache misses
      and memory-barrier instructions. &
      & \Mn Transactions begin/end instructions typically do not result in
        cache misses, but do have memory-ordering and overhead consequences. \\
\cmidrule{3-5}
  &
    & \Pl Update-side contention effects are focused on acquisition and release,
      so that the critical section runs at full speed. &
      & \Mn Contention aborts conflicting transactions, even if they have been
        running for a long time. \\
\addlinespace[4pt]
  &
    & \Pl Readers do not contend with updaters or with each other. &
      & \\
\cmidrule{3-5}
  &
    & \Pl Read-side primitives are typically bounded wait-free with low
      overhead.
      (Lock-free with low overhead for hazard pointers.) &
      & \Mn Read-only transactions subject to conflicts and rollbacks.
        No forward-progress guarantees other than those supplied by fallback
        code. \\
\cmidrule{3-5}
  &
    & \Pl Privatization operations are simple, intuitive, performant, and
      scalable when data is visible only to updaters. &
      & \Mn Privatized data contributes to transaction size. \\
\addlinespace[4pt]
  &
    & \Mn Privatization operations are expensive (though still intuitive
      and scalable) for reader-visible data. &
      & \\
\midrule
  Hardware Support &
    & \Pl Commodity hardware suffices. &
      & \Mn New hardware required (and is starting to become available). \\
\cmidrule{3-5}
  &
    & \Pl Performance is insensitive to cache-geometry details. &
      & \Mn Performance depends critically on cache geometry. \\
\midrule
  Software Support &
    & \Pl APIs exist, large body of code and experience, debuggers operate
      naturally. &
      & \Mn APIs emerging, little experience outside of DBMS, breakpoints
        mid-transaction can be problematic. \\
\midrule
  Interaction With Other Mechanisms &
    & \Pl Long experience of successful interaction. &
      & \Dw Just beginning investigation of interaction. \\
\midrule
  Practical Apps &
    & \Pl Yes. &
      & \Pl Yes. \\
\midrule
  Wide Applicability &
    & \Pl Yes. &
      & \Mn Jury still out. \\
\bottomrule
\end{tabularx}
}}
\caption{Comparison of Locking (Augmented by RCU or Hazard Pointers) and HTM
  (\colorbox{plus}{Advantage}, \colorbox{minus}{Disadvantage},
  \colorbox{down}{Strong Disadvantage})}
\label{tab:future:Comparison of Locking (Augmented by RCU or Hazard Pointers) and HTM}
\end{table*}

Practitioners have long used reference counting, atomic operations,
non-blocking data structures, hazard pointers, and RCU to avoid some
of the shortcomings of locking.
For example, deadlock can be avoided in many cases by using reference
counts, hazard pointers, or RCU to protect data structures,
particularly for read-only critical
sections~\cite{MagedMichael04a,HerlihyLM02,MathieuDesnoyers2012URCU,DinakarGuniguntala2008IBMSysJ,ThomasEHart2007a}.
These approaches also reduce the need to partition data
structures, as was seen in \cref{chp:Data Structures}.
RCU further provides contention-free bounded wait-free read-side
primitives~\cite{McKenney98,MathieuDesnoyers2012URCU}, while hazard pointers
provides lock-free read-side
primitives~\cite{Michael02a,HerlihyLM02,MagedMichael04a}.
Adding these considerations to
\cref{tab:future:Comparison of Locking and HTM}
results in the updated comparison between augmented locking and HTM
shown in
\cref{tab:future:Comparison of Locking (Augmented by RCU or Hazard Pointers) and HTM}.
A summary of the differences between the two tables is as follows:

\begin{enumerate}
\item	Use of non-blocking read-side mechanisms alleviates deadlock issues.
\item	Read-side mechanisms such as hazard pointers and RCU can operate
	efficiently on non-partitionable data.
\item	Hazard pointers and RCU do not contend with each other or with
	updaters, allowing excellent performance and scalability for
	read-mostly workloads.
\item	Hazard pointers and RCU provide forward-progress guarantees
	(lock freedom and bounded wait-freedom, respectively).
\item	Privatization operations for hazard pointers and RCU are
	straightforward.
\end{enumerate}

\IfEbookSize{}{
% future/HTMtableFull.tex
% SPDX-License-Identifier: CC-BY-SA-3.0

\begin{sidewaystable*}[htbp]
% future/HTMtableColor.tex
% SPDX-License-Identifier: CC-BY-SA-3.0

\definecolor{plus}{cmyk}{0.1,0,0,0}
\definecolor{minus}{cmyk}{0,0.05,0.2,0.05}
\definecolor{down}{cmyk}{0,0.15,0.15,0.1}
\newcommand{\Pl}{\cellcolor{plus}}
\newcommand{\Mn}{\cellcolor{minus}}
\newcommand{\Dw}{\cellcolor{down}}

\centering
\caption{Comparison of Locking (Plain and Augmented) and HTM
  (\colorbox{plus}{Advantage}, \colorbox{minus}{Disadvantage},
  \colorbox{down}{Strong Disadvantage})}
\label{tab:future:Comparison of Locking (Plain and Augmented) and HTM}
\footnotesize
\setstretch{0.95}
\setlength{\tabcolsep}{3pt}
\resizebox{8in}{!}{
\begin{tabularx}{8.5in}{p{.85in}cXcXcX}
\toprule
  &
    & \multicolumn{1}{c}{Locking} &
      & \multicolumn{1}{c}{Locking with Userspace RCU or Hazard Pointers} &
        & \multicolumn{1}{c}{Hardware Transactional Memory} \\
\midrule
  Basic Idea &
    & Allow only one thread at a time to access a given set of objects. &
      & Allow only one thread at a time to access a given set of objects. &
        & Cause a given operation over a set of objects to execute atomically. \\
\midrule
  Scope &
    & \Pl Handles all operations. &
      & \Pl Handles all operations. &
        & \Pl Handles revocable operations. \\
\addlinespace[4pt]
  &
    & &
      & &
        & \Mn Irrevocable operations force fallback (typically to locking). \\
\midrule
  Composability &
    & \Dw Limited by deadlock. &
      & \Pl Readers limited only by grace-period-wait operations. &
        & \Dw Limited by irrevocable operations, transaction size, and deadlock.
          (Assuming lock-based fallback code.) \\
\addlinespace[4pt]
  &
    & &
      & \Mn Updaters limited by deadlock. Readers reduce deadlock. &
        & \\
\midrule
  Scalability \& Performance &
    & \Mn Data must be partitionable to avoid lock contention. &
      & \Mn Data must be partitionable to avoid lock contention among updaters. &
        & \Mn Data must be partitionable to avoid conflicts. \\
\addlinespace[4pt]
  &
    & &
      & \Pl Partitioning not needed for readers. &
        & \\
\cmidrule{3-7}
  &
    & \Dw Partitioning must typically be fixed at design time. &
      & \Dw Partitioning for updaters must typically be fixed at design time. &
        & \Pl Dynamic adjustment of partitioning carried out automatically
          down to cacheline boundaries. \\
\addlinespace[4pt]
  &
    & &
      & \Pl Partitioning not needed for readers. &
        & \Mn Partitioning required for fallbacks (less important for rare
          fallbacks). \\
\cmidrule{3-7}
  &
    & \Dw Locking primitives typically result in expensive cache misses and
      memory-barrier instructions.&
      & \Dw Updater locking primitives typically result in expensive cache
        misses and memory-barrier instructions. &
        & \Mn Transactions begin/end instructions typically do not result in
          cache misses, but do have memory-ordering and overhead
	  consequences. \\
\cmidrule{3-7}
  &
    & \Pl Contention effects are focused on acquisition and release, so that
      the critical section runs at full speed. &
      & \Pl Update-side contention effects are focused on acquisition and
        release, so that the critical section runs at full speed. &
        & \Mn Contention aborts conflicting transactions, even if they have been
          running for a long time. \\
\addlinespace[4pt]
  &
    & &
      & \Pl Readers do not contend with updaters or with each other. &
        & \\
\cmidrule{3-7}
  &
    & &
      & \Pl Read-side primitives are typically bounded wait-free with
        low overhead.
        (Lock-free with low overhead for hazard pointers.) &
        & \Mn Read-only transactions subject to conflicts and rollbacks. No
          forward-progress guarantees other than those supplied by fallback
          code. \\
\cmidrule{3-7}
  &
    & \Pl Privatization operations are simple, intuitive, performant,
      and scalable. &
      & \Pl Privatization operations are simple, intuitive, performant,
        and scalable when data is visible only to updaters. &
        & \Mn Privatized data contributes to transaction size. \\
\addlinespace[4pt]
  &
    & &
      & \Mn Privatization operations are expensive (though still intuitive
        and scalable) for reader-visible data. &
        & \\
\midrule
  Hardware Support &
    & \Pl Commodity hardware suffices. &
      & \Pl Commodity hardware suffices. &
        & \Mn New hardware required (and is starting to become available). \\
\cmidrule{3-7}
  &
    & \Pl Performance is insensitive to cache-geometry details. &
      & \Pl Performance is insensitive to cache-geometry details. &
        & \Mn Performance depends critically on cache geometry. \\
\midrule
  Software Support &
    & \Pl APIs exist, large body of code and experience, debuggers operate
      naturally. &
      & \Pl APIs exist, large body of code and experience, debuggers operate
        naturally. &
        & \Mn APIs emerging, little experience outside of DBMS, breakpoints
              mid-transaction can be problematic. \\
\midrule
  Interaction With Other Mechanisms &
    & \Pl Long experience of successful interaction. &
      & \Pl Long experience of successful interaction. &
        & \Dw Just beginning investigation of interaction. \\
\midrule
  Practical Apps &
    & \Pl Yes. &
      & \Pl Yes. &
        & \Pl Yes. \\
\midrule
  Wide Applicability &
    & \Pl Yes. &
      & \Pl Yes. &
        & \Mn Jury still out. \\
\bottomrule
\end{tabularx}
}
\end{sidewaystable*}

For those with good eyesight,
\cref{tab:future:Comparison of Locking (Plain and Augmented) and HTM}
combines
\cref{tab:future:Comparison of Locking and HTM,%
tab:future:Comparison of Locking (Augmented by RCU or Hazard Pointers) and HTM}.
}

\QuickQuiz{
	\Cref{tab:future:Comparison of Locking and HTM,tab:future:Comparison of Locking (Augmented by RCU or Hazard Pointers) and HTM}
	state that hardware is only starting to become available.
	But hasn't HTM hardware support been widely available
	for almost a full decade?
}\QuickQuizAnswer{
	Yes and no.
	It appears that implementing even the HTM subset of TM in real
	hardware is a bit trickier than it
	appears~\cite{ChristianJacobi2012MainframeTM,ScottWasson2014HaswellDisableTSX,Intel2020HaswellTSXordering,Intel2021perfTSXordering,MichaelLarabel2021DisableTSX}.
	Therefore, the sad fact is that ``starting to become available'' is
	all too accurate as of 2021.
	In fact, vendors are beginning to deprecate their HTM
	implementations~\cite[Book III Appendix A]{PowerISA3.1-2020}.
}\QuickQuizEnd

Of course, it is also possible to augment HTM,
as discussed in the next section.

\subsection{Where Does HTM Best Fit In?}
\label{sec:future:Where Does HTM Best Fit In?}

Although it will likely be some time before HTM's area of applicability
can be as crisply delineated as that shown for RCU in
\cref{fig:defer:RCU Areas of Applicability} on
\cpageref{fig:defer:RCU Areas of Applicability}, that is no reason not to
start moving in that direction.

HTM seems best suited to update-heavy workloads involving relatively
small changes to disparate portions of relatively large in-memory
data structures running on large multiprocessors,
as this meets the size restrictions of current HTM implementations while
minimizing the probability of conflicts and attendant aborts and
rollbacks.
This scenario is also one that is relatively difficult to handle given
current synchronization primitives.

Use of locking in conjunction with HTM seems likely to overcome HTM's
difficulties with irrevocable operations, while use of RCU or
hazard pointers might alleviate HTM's transaction-size limitations
for read-only operations that traverse large fractions of the data
structure~\cite{SeongJaePark2020HTMRCUlock}.
Current HTM implementations unconditionally abort an update transaction
that conflicts with an RCU or hazard-pointer reader, but perhaps future
HTM implementations will interoperate more smoothly with these
synchronization mechanisms.
In the meantime, the probability of an update conflicting with a
large RCU or hazard-pointer read-side critical section should be
much smaller than the probability of conflicting with the equivalent
read-only transaction.\footnote{
	It is quite ironic that strictly transactional mechanisms are
	appearing in shared-memory systems at just about the time
	that NoSQL databases are relaxing the traditional
	database-application reliance on strict transactions.
	Nevertheless, HTM has in fact realized the ease-of-use promise
	of TM, albeit for black-hat attacks on the Linux kernel's
	address-space randomization defense
	mechanism~\cite{YeongjinJang2016TSXbreakKASLR,Jang:2016:BKA:2976749.2978321}.}
Nevertheless, it is quite possible that a steady stream of RCU or
hazard-pointer readers might starve updaters due to a corresponding
steady stream of conflicts.
This vulnerability could be eliminated (at significant
hardware cost and complexity) by giving extra-transactional
reads the pre-transaction copy of the memory location being loaded.

The fact that HTM transactions must have fallbacks might in some cases
force static partitionability of data structures back onto HTM\@.
This limitation might be alleviated if future HTM implementations
provide forward-progress guarantees, which might eliminate the need
for fallback code in some cases, which in turn might allow HTM to
be used efficiently in situations with higher conflict probabilities.

In short, although HTM is likely to have important uses and applications,
it is another tool in the parallel programmer's toolbox, not a replacement
for the toolbox in its entirety.

\subsection{Potential Game Changers}
\label{sec:future:Potential Game Changers}

Game changers that could greatly increase the need for HTM include
the following:

\begin{enumerate}
\item	Forward-progress guarantees.
\item	Transaction-size increases.
\item	Improved debugging support.
\item	Weak atomicity.
\end{enumerate}

These are expanded upon in the following sections.

\subsubsection{Forward-Progress Guarantees}
\label{sec:future:Forward-Progress Guarantees}

As was discussed in
\cref{sec:future:Lack of Forward-Progress Guarantees},
current HTM implementations lack forward-progress guarantees, which requires
that fallback software is available to handle HTM failures.
Of course, it is easy to demand guarantees, but not always easy
to provide them.
In the case of HTM, obstacles to guarantees can include cache size and
associativity, TLB size and associativity, transaction duration and
interrupt frequency, and scheduler implementation.

Cache size and associativity was discussed in
\cref{sec:future:Transaction-Size Limitations},
along with some research intended to work around current limitations.
However, HTM forward-progress guarantees would
come with size limits, large though these limits might one day be.
So why don't current HTM implementations provide forward-progress
guarantees for small transactions, for example, limited to the
associativity of the cache?
One potential reason might be the need to deal with hardware failure.
For example, a failing cache SRAM cell might be handled by deactivating
the failing cell, thus reducing the associativity of the cache and
therefore also the maximum size of transactions that can be guaranteed
forward progress.
Given that this would simply decrease the guaranteed transaction size,
it seems likely that other reasons are at work.
Perhaps providing forward progress guarantees on production-quality
hardware is more difficult than one might think, an entirely plausible
explanation given the difficulty of making forward-progress guarantees
in software.
Moving a problem from software to hardware does not necessarily make
it easier to solve~\cite{ChristianJacobi2012MainframeTM}.

Given a physically tagged and indexed cache, it is not enough for the
transaction to fit in the cache.
Its address translations must also fit in the TLB\@.
Any forward-progress guarantees must therefore also take TLB size
and associativity into account.

Given that interrupts, traps, and exceptions abort transactions in current
HTM implementations, it is necessary that the execution duration of
a given transaction be shorter than the expected interval between
interrupts.
No matter how little data a given transaction touches, if it runs too
long, it will be aborted.
Therefore, any forward-progress guarantees must be conditioned not only
on transaction size, but also on transaction duration.

Forward-progress guarantees depend critically on the ability to determine
which of several conflicting transactions should be aborted.
It is all too easy to imagine an endless series of transactions, each
aborting an earlier transaction only to itself be aborted by a later
transactions, so that none of the transactions actually commit.
The complexity of conflict handling is
evidenced by the large number of HTM conflict-resolution policies
that have been proposed~\cite{EgeAkpinar2011HTM2TLE,YujieLiu2011ToxicTransactions}.
Additional complications are introduced by extra-transactional accesses,
as noted by Blundell~\cite{Blundell2006TMdeadlock}.
It is easy to blame the extra-transactional accesses for all of these
problems, but the folly of this line of thinking is easily demonstrated
by placing each of the extra-transactional accesses into its own
single-access transaction.
It is the pattern of accesses that is the issue, not whether or not they
happen to be enclosed in a transaction.

Finally, any forward-progress guarantees for transactions also
depend on the scheduler, which must let the thread executing the
transaction run long enough to successfully commit.

So there are significant obstacles to HTM vendors offering forward-progress
guarantees.
However, the impact of any of them doing so would be enormous.
It would mean that HTM transactions would no longer need software
fallbacks, which would mean that HTM could finally deliver on the
TM promise of deadlock elimination.

However, in late 2012, the IBM Mainframe announced an HTM implementation
that includes \emph{constrained transactions} in addition to the usual
best-effort HTM
implementation~\cite{ChristianJacobi2012MainframeTM}.
A constrained transaction starts with the \co{tbeginc} instruction
instead of the \co{tbegin} instruction that is used for best-effort
transactions.
Constrained transactions are guaranteed to always complete (eventually),
so if a transaction aborts, rather than branching to a fallback path
(as is done for best-effort transactions), the hardware instead restarts
the transaction at the \co{tbeginc} instruction.

The Mainframe architects needed to take extreme measures to deliver on
this forward-progress guarantee.
If a given constrained transaction repeatedly fails, the CPU
might disable branch prediction, force in-order execution, and even
disable pipelining.
If the repeated failures are due to high contention, the CPU might
disable speculative fetches, introduce random delays, and even
serialize execution of the conflicting CPUs.
``Interesting'' forward-progress scenarios involve as few as two CPUs
or as many as one hundred CPUs.
Perhaps these extreme measures provide some insight as to why other CPUs
have thus far refrained from offering constrained transactions.

As the name implies, constrained transactions are in fact severely constrained:

\begin{enumerate}
\item	The maximum data footprint is four blocks of memory,
	where each block can be no larger than 32 bytes.
\item	The maximum code footprint is 256 bytes.
\item	If a given 4K page contains a constrained transaction's code,
	then that page may not contain that transaction's data.
\item	The maximum number of assembly instructions that may be executed
	is 32.
\item	Backwards branches are forbidden.
\end{enumerate}

Nevertheless, these constraints support a number of important data structures,
including linked lists, stacks, queues, and arrays.
Constrained HTM therefore seems likely to become an important tool in
the parallel programmer's toolbox.

Note that these forward-progress guarantees need not be absolute.
For example, suppose that a use of HTM uses a global lock as fallback.
Assuming that the fallback mechanism has been carefully designed to
avoid the ``lemming effect'' discussed in
\cref{sec:future:Aborts and Rollbacks},
then if HTM rollbacks are sufficiently infrequent, the global lock
will not be a bottleneck.
That said, the larger the system, the longer the critical sections,
and the longer the time required to recover from the ``lemming effect'',
the more rare ``sufficiently infrequent'' needs to be.

\subsubsection{Transaction-Size Increases}
\label{sec:future:Transaction-Size Increases}

Forward-progress guarantees are important, but as we saw, they will
be conditional guarantees based on transaction size and duration.
There has been some progress, for example, some commercially available
HTM implementations use approximation techniques to support extremely
large HTM read sets~\cite{RaviRajwar2012TSX}.
For another example, \Power{8} HTM supports suspended transations, which
avoid adding irrelevant accesses to the suspended transation's read and
write sets~\cite{Le:2015:TMS:3266491.3266500}.
This capability has been used to produce a high performance
reader-writer lock~\cite{PascalFelber2016rwlockElision}.

It is important to note that even small-sized guarantees will be
quite useful.
For example,
a guarantee of two cache lines is sufficient for a stack, queue, or dequeue.
However, larger data structures require larger guarantees, for example,
traversing a tree in order requires a guarantee equal to the number
of nodes in the tree.
Therefore, even modest increases in the size of the guarantee also
increases the usefulness of HTM, thereby increasing the need for CPUs
to either provide it or provide good-and-sufficient workarounds.

\subsubsection{Improved Debugging Support}
\label{sec:future:Improved Debugging Support}

Another inhibitor to transaction size is the need to debug the transactions.
The problem with current mechanisms is that a single-step exception
aborts the enclosing transaction.
There are a number of workarounds for this issue, including emulating
the processor (slow!), substituting STM for HTM (slow and slightly
different semantics!),
playback techniques using repeated retries to emulate forward
progress (strange failure modes!), and
full support of debugging HTM transactions (complex!).

Should one of the HTM vendors produce an HTM system that allows
straightforward use of classical debugging techniques within
transactions, including breakpoints, single stepping, and
print statements, this will make HTM much more compelling.
Some transactional-memory researchers started to recognize this
problem in 2013, with at least one proposal involving hardware-assisted
debugging facilities~\cite{JustinGottschlich2013TMdebug}.
Of course, this proposal depends on readily available hardware gaining such
facilities~\cite{TimothyHayes2020ARM-HTM,Intel2020TSXdevguide}.
Worse yet, some cutting-edge debugging facilities are incompatible
with HTM~\cite{RobertOCallahan2020DebuggingHTM}.

\subsubsection{Weak Atomicity}
\label{sec:future:Weak Atomicity}

Given that HTM is likely to face some sort of size limitations for the
foreseeable future, it will be necessary for HTM to interoperate
smoothly with other mechanisms.
HTM's interoperability with read-mostly mechanisms such as hazard pointers
and RCU would be improved if extra-transactional reads did not
unconditionally abort transactions with conflicting writes---instead,
the read could simply be provided with the pre-transaction value.
In this way, hazard pointers and RCU could be used to allow HTM to handle
larger data structures and to reduce conflict probabilities.

This is not necessarily simple, however.
The most straightforward way of implementing this requires an additional
state in each cache line and on the bus, which is a non-trivial added
expense.
The benefit that goes along with this expense is permitting
large-footprint readers without the risk of starving updaters due
to continual conflicts.
An alternative approach, applied to great effect to binary search trees
by Siakavaras et al.~\cite{Siakavaras2017CombiningHA},
is to use RCU for read-only traversals and HTM
only for the actual updates themselves.
This combination outperformed other transactional-memory techniques by
up to 220\,\%, a speedup similar to that observed by
Howard and Walpole~\cite{PhilHoward2011RCUTMRBTree}
when they combined RCU with STM\@.
In both cases, the weak atomicity is implemented in software rather than
in hardware.
It would nevertheless be interesting to see what additional speedups
could be obtained by implementing weak atomicity in both hardware and
software.

\subsection{Conclusions}
\label{sec:future:Conclusions}

Although current HTM implementations have delivered real performance
benefits in some situations, they also have significant shortcomings.
The most significant shortcomings appear to be
limited transaction sizes,
the need for conflict handling, the need for aborts and rollbacks,
the lack of forward-progress guarantees,
the inability to handle irrevocable operations,
and subtle semantic differences
from locking.
There are also reasons for lingering concerns surrounding HTM-implementation
reliability~\cite{ChristianJacobi2012MainframeTM,ScottWasson2014HaswellDisableTSX,Intel2020HaswellTSXordering,Intel2021perfTSXordering,MichaelLarabel2021DisableTSX,PowerISA3.1-2020}.

Some of these shortcomings might be alleviated in future implementations,
but it appears that there will continue to be a strong need to make
HTM work well with the many other types of synchronization mechanisms,
as noted earlier~\cite{McKenney2007PLOSTM,PaulEMcKenney2010OSRGrassGreener}.
Although there has been some work using HTM with
RCU~\cite{Siakavaras2017CombiningHA,DimitriosSiakavaras2020RCU-HTM-B+Trees,ChristinaGiannoula2018HTM-RCU-graphcoloring,SeongJaePark2020HTMRCUlock},
there has been little evidence of progress towards HTM work better with
RCU and with other deferred-reclamation mechanisms.

In short, current HTM implementations appear to be welcome and useful
additions to the parallel programmer's toolbox, and much interesting
and challenging work is required to make use of them.
However, they cannot be
considered to be a magic wand with which to wave away all parallel-programming
problems.

% future/formalregress.tex
% mainfile: ../perfbook.tex
% SPDX-License-Identifier: CC-BY-SA-3.0

\section{Formal Regression Testing?}
\label{sec:future:Formal Regression Testing?}
\epigraph{Theory without experiments:
	  Have we gone too far?}
	 {Michael Mitzenmacher}

Formal verification has long proven useful in a number of production
environments~\cite{JamesRLarus2004RightingSoftware,AlBessey2010BillionLoCLater,ByronCook2018FormalAmazon,CaitlinSadowski2018staticAnalysisGoogle,DinoDistefano2019FBstaticAnalysis}.
However, it is an open question as to whether hard-core formal verification
will ever be included in the automated regression-test suites used for
continuous integration within complex concurrent codebases, such as the
Linux kernel.
Although there is already a proof of concept for Linux-kernel
SRCU~\cite{LanceRoy2017CBMC-SRCU}, this test is for a small portion
of one of the simplest RCU implementations, and has proven difficult
to keep it current with the ever-changing Linux kernel.
It is therefore worth asking what would be required to incorporate
formal verification as first-class members of the Linux kernel's
regression tests.

The following list is a good
start~\cite[slide 34]{PaulEMcKenney2015DagstuhlVerification}:

\begin{enumerate}
\item	Any required translation must be automated.
\item	The environment (including memory ordering) must be correctly
	handled.
\item	The memory and CPU overhead must be acceptably modest.
\item	Specific information leading to the location of the bug
	must be provided.
\item	Information beyond the source code and inputs must be
	modest in scope.
\item	The bugs located must be relevant to the code's users.
\end{enumerate}

This list builds on, but is somewhat more modest than, Richard Bornat's
dictum:
``Formal-verification researchers should verify the code that
developers write, in the language they write it in, running in the
environment that it runs in, as they write it.''
The following sections discuss each of the above requirements, followed
by a section presenting a scorecard of how well a few tools stack up
against these requirements.

\QuickQuiz{
	This list is ridiculously utopian!
	Why not stick to the current state of the formal-verification art?
}\QuickQuizAnswer{
	You are welcome to your opinion on what is and is not utopian,
	but I will be paying more attention to people actually making
	progress on the items in that list than to anyone who might be
	objecting to them.
	This might have something to do with my long experience with
	people attempting to talk me out of specific things that their
	favorite tools cannot handle.

	In the meantime, please feel free to read the papers written by
	the people who are actually making progress, for example, this
	one~\cite{DinoDistefano2019FBstaticAnalysis}.
}\QuickQuizEnd

\subsection{Automatic Translation}
\label{sec:future:Automatic Translation}

Although Promela and \co{spin}
are invaluable design aids, if you need to formally regression-test
your C-language program, you must hand-translate to Promela each time
you would like to re-verify your code.
If your code happens to be in the Linux kernel, which releases every
60--90 days, you will need to hand-translate from four to six times
each year.
Over time, human error will creep in, which means that the verification
won't match the source code, rendering the verification useless.
Repeated verification clearly requires either that the formal-verification
tooling input your code directly, or that there be bug-free automatic
translation of your code to the form required for verification.

PPCMEM and \co{herd} can in theory directly input assembly language
and C++ code, but these tools work only on very small litmus tests,
which normally means that you must extract the core of your
mechanism---by hand.
As with Promela and \co{spin}, both PPCMEM and \co{herd} are
extremely useful, but they are not well-suited for regression suites.

In contrast, \IXaltacr{\co{cbmc}}{cbmc} and \IX{Nidhugg} can input
C programs of reasonable
(though still quite limited) size, and if their capabilities continue
to grow, could well become excellent additions to regression suites.
The Coverity static-analysis tool also inputs C programs, and of very
large size, including the Linux kernel.
Of course, Coverity's static analysis is quite simple compared to that
of \co{cbmc} and Nidhugg.
On the other hand, Coverity had an all-encompassing definition of
``C program'' that posed special challenges~\cite{AlBessey2010BillionLoCLater}.
Amazon Web Services uses a variety of formal-verification tools,
including \co{cbmc}, and applies some of these tools to regression
testing~\cite{ByronCook2018FormalAmazon}.
Google uses a number of relatively simple static analysis tools directly
on large Java code bases, which are arguably less diverse than C code
bases~\cite{CaitlinSadowski2018staticAnalysisGoogle}.
Facebook uses more aggressive forms of formal verification against its
code bases, including analysis of concurrency~\cite{DinoDistefano2019FBstaticAnalysis,PeterWOHearn2019incorrectnessLogic},
though not yet on the Linux kernel.
Finally, Microsoft has long used static analysis on its code
bases~\cite{JamesRLarus2004RightingSoftware}.

Given this list, it is clearly possible to create sophisticated
formal-verification tools that directly consume production-quality
source code.

However, one shortcoming of taking C code as input is that it assumes
that the compiler is correct.
An alternative approach is to take the binary produced by the C compiler
as input, thereby accounting for any relevant compiler bugs.
This approach has been used in a number of verification efforts,
perhaps most notably by the SEL4
project~\cite{ThomasSewell2013L4binaryVerification}.

\QuickQuiz{
	Given the groundbreaking nature of the various verifiers used
	in the SEL4 project, why doesn't this chapter cover them in
	more depth?
}\QuickQuizAnswer{
	There can be no doubt that the verifiers used by the SEL4
	project are quite capable.
	However, SEL4 started as a single-CPU project.
	And although SEL4 has gained multi-processor
	capabilities, it is currently using very coarse-grained
	locking that is similar to the Linux kernel's old
	Big Kernel Lock (BKL)\@.
	There will hopefully come a day when it makes sense to add
	SEL4's verifiers to a book on parallel programming, but
	this is not yet that day.
}\QuickQuizEnd

However, verifying directly from either the source or binary both have the
advantage of eliminating human translation errors, which is critically
important for reliable regression testing.

This is not to say that tools with special-purpose languages are useless.
On the contrary, they can be quite helpful for design-time verification,
as was discussed in
\cref{chp:Formal Verification}.
However, such tools are not particularly helpful for automated regression
testing, which is in fact the topic of this section.

\subsection{Environment}
\label{sec:future:Environment}

It is critically important that formal-verification tools correctly
model their environment.
One all-too-common omission is the memory model, where a great
many formal-verification tools, including Promela/spin, are
restricted to \IXh{sequential}{consistency}.
The QRCU experience related in
\cref{sec:formal:Is QRCU Really Correct?}
is an important cautionary tale.

Promela and \co{spin} assume sequential consistency, which is not a
good match for modern computer systems, as was seen in
\cref{chp:Advanced Synchronization: Memory Ordering}.
In contrast, one of the great strengths of PPCMEM and \co{herd}
is their detailed modeling of various CPU families memory models,
including x86, \ARM, Power, and, in the case of \co{herd},
a Linux-kernel memory model~\cite{Alglave:2018:FSC:3173162.3177156},
which was accepted into Linux-kernel version v4.17.

The \co{cbmc} and Nidhugg tools provide some ability to select
memory models, but do not provide the variety that PPCMEM and
\co{herd} do.
However, it is likely that the larger-scale tools will adopt
a greater variety of memory models as time goes on.

In the longer term, it would be helpful for formal-verification
tools to include I/O~\cite{PaulEMcKenney2016LinuxKernelMMIO},
but it may be some time before this comes to pass.

Nevertheless, tools that fail to match the environment can still
be useful.
For example, a great many concurrency bugs would still be bugs on
a mythical sequentially consistent system, and these bugs could
be located by a tool that over-approximates the system's memory model
with sequential consistency.
Nevertheless, these tools will fail to find bugs involving missing
memory-ordering directives, as noted in the aforementioned
cautionary tale of
\cref{sec:formal:Is QRCU Really Correct?}.

\subsection{Overhead}
\label{sec:future:Overhead}

Almost all hard-core formal-verification tools are exponential
in nature, which might seem discouraging until you consider that
many of the most interesting software questions are in fact undecidable.
However, there are differences in degree, even among exponentials.

PPCMEM by design is unoptimized, in order to provide greater assurance
that the memory models of interest are accurately represented.
The \co{herd} tool optimizes more aggressively, as described in
\cref{sec:formal:Axiomatic Approaches}, and is thus orders of magnitude
faster than PPCMEM\@.
Nevertheless, both PPCMEM and \co{herd} target very small litmus tests
rather than larger bodies of code.

In contrast, Promela/\co{spin}, \co{cbmc}, and Nidhugg are designed for
(somewhat) larger bodies of code.
Promela/\co{spin} was used to verify the Curiosity rover's
filesystem~\cite{DBLP:journals/amai/GroceHHJX14} and, as noted earlier,
both \co{cbmc} and Nidhugg were appled to Linux-kernel RCU\@.

If advances in heuristics continue at the rate of the past three
decades, we can look forward to large reductions in overhead for
formal verification.
That said, combinatorial explosion is still combinatorial explosion,
which would be expected to sharply limit the size of programs that
could be verified, with or without continued improvements in
heuristics.

However, the flip side of combinatorial explosion is Philip II of
Macedon's timeless advice:
``Divide and rule.''
If a large program can be divided and the pieces verified, the result
can be combinatorial \emph{implosion}~\cite{PaulEMcKenney2011Verico}.
One natural place to divide is on API boundaries, for example, those
of locking primitives.
One verification pass can then verify that the locking implementation
is correct, and additional verification passes can verify correct
use of the locking APIs.

\begin{listing}
\input{CodeSamples/formal/herd/C-SB+l-o-o-u+l-o-o-u-C=whole.fcv}
\caption{Emulating Locking with \tco{cmpxchg_acquire()}}
\label{lst:future:Emulating Locking with cmpxchg}
\end{listing}

\begin{table}
\rowcolors{1}{}{lightgray}
\renewcommand*{\arraystretch}{1.1}
\small
\centering
\begin{tabular}{S[table-format=1.0]S[table-format=1.3]S[table-format=2.3]}
	\toprule
	\multicolumn{1}{c}{\# Threads} & \multicolumn{1}{c}{Locking} &
			\multicolumn{1}{c}{\tco{cmpxchg_acquire}} \\
	\midrule
	2 & 0.004 &  0.022 \\
	3 & 0.041 &  0.743 \\
	4 & 0.374 & 59.565 \\
	5 & 4.905 &        \\
	\bottomrule
\end{tabular}
\caption{Emulating Locking:
			    Performance (s)}
\label{tab:future:Emulating Locking: Performance (s)}
\end{table}

The performance benefits of this approach can be demonstrated using
the Linux-kernel memory
model~\cite{Alglave:2018:FSC:3173162.3177156}.
This model provides \co{spin_lock()} and \co{spin_unlock()}
primitives, but these primitives can also be emulated using
\co{cmpxchg_acquire()} and \co{smp_store_release()}, as shown in
\cref{lst:future:Emulating Locking with cmpxchg}
(\path{C-SB+l-o-o-u+l-o-o-*u.litmus} and \path{C-SB+l-o-o-u+l-o-o-u*-C.litmus}).
\Cref{tab:future:Emulating Locking: Performance (s)}
compares the performance and scalability of using the model's
\co{spin_lock()} and \co{spin_unlock()} against emulating these
primitives as shown in the listing.
The difference is not insignificant:
At four processes, the model is more than two orders of magnitude
faster than emulation!

\QuickQuiz{
\begin{fcvref}[ln:future:formalregress:C-SB+l-o-o-u+l-o-o-u-C:whole]
	Why bother with a separate \co{filter} command on \clnref{filter_} of
	\cref{lst:future:Emulating Locking with cmpxchg}
	instead of just adding the condition to the \co{exists} clause?
	And wouldn't it be simpler to use \co{xchg_acquire()} instead
	of \co{cmpxchg_acquire()}?
\end{fcvref}
}\QuickQuizAnswer{
	The \co{filter} clause causes the \co{herd} tool to discard
	executions at an earlier stage of processing than does
	the \co{exists} clause, which provides significant speedups.

\begin{table}
\rowcolors{7}{lightgray}{}
\renewcommand*{\arraystretch}{1.1}
\small
\centering
\begin{tabular}{S[table-format=1.0]S[table-format=1.3]S[table-format=2.3]
		S[table-format=3.3]S[table-format=2.3]S[table-format=3.3]}
	\toprule
	& & \multicolumn{2}{c}{\tco{cmpxchg_acquire()}}
		& \multicolumn{2}{c}{\tco{xchg_acquire()}} \\
	\cmidrule(l){3-4} \cmidrule(l){5-6}
	\multicolumn{1}{c}{\#} & \multicolumn{1}{c}{Lock}
		& \multicolumn{1}{c}{\tco{filter}}
			& \multicolumn{1}{c}{\tco{exists}}
				& \multicolumn{1}{c}{\tco{filter}}
					& \multicolumn{1}{c}{\tco{exists}} \\
	\cmidrule{1-1} \cmidrule(l){2-2} \cmidrule(l){3-4} \cmidrule(l){5-6}
	2 & 0.004 &  0.022 &   0.039 &  0.027 &  0.058 \\
	3 & 0.041 &  0.743 &   1.653 &  0.968 &  3.203 \\
	4 & 0.374 & 59.565 & 151.962 & 74.818 & 500.96 \\
	5 & 4.905 &        &         &        &        \\
	\bottomrule
\end{tabular}
\caption{Emulating Locking:
			    Performance Comparison (s)}
\label{tab:future:Emulating Locking: Performance Comparison (s)}
\end{table}

	As for \co{xchg_acquire()}, this atomic operation will do a
	write whether or not lock acquisition succeeds, which means
	that a model using \co{xchg_acquire()} will have more operations
	than one using \co{cmpxchg_acquire()}, which won't do a write
	in the failed-acquisition case.
	More writes means more combinatorial to explode, as shown in
	\cref{tab:future:Emulating Locking: Performance Comparison (s)}
	(\path{C-SB+l-o-o-u+l-o-o-*u.litmus},
	\path{C-SB+l-o-o-u+l-o-o-u*-C.litmus},
	\path{C-SB+l-o-o-u+l-o-o-u*-CE.litmus},
	\path{C-SB+l-o-o-u+l-o-o-u*-X.litmus}, and
	\path{C-SB+l-o-o-u+l-o-o-u*-XE.litmus}).
	This table clearly shows that \co{cmpxchg_acquire()}
	outperforms \co{xchg_acquire()} and that use of the
	\co{filter} clause outperforms use of the \co{exists} clause.
}\QuickQuizEnd

It would of course be quite useful for tools to automatically divide
up large programs, verify the pieces, and then verify the combinations
of pieces.
In the meantime, verification of large programs will require significant
manual intervention.
This intervention will preferably mediated by scripting, the better to
reliably carry out repeated verifications on each release, and
preferably eventually in a manner well-suited for continuous integration.
And Facebook's Infer tool has taken important steps towards doing just
that, via compositionality and
abstraction~\cite{SamBlackshear2018RacerD,DinoDistefano2019FBstaticAnalysis}.

In any case, we can expect formal-verification capabilities to continue
to increase over time, and any such increases will in turn increase
the applicability of formal verification to regression testing.

\subsection{Locate Bugs}
\label{sec:future:Locate Bugs}

Any software artifact of any size contains bugs.
Therefore, a formal-verification tool that reports only the
presence or absence of bugs is not particularly useful.
What is needed is a tool that gives at least \emph{some} information
as to where the bug is located and the nature of that bug.

The \co{cbmc} output includes a traceback mapping back to the source
code, similar to Promela/spin's, as does Nidhugg.
Of course, these tracebacks can be quite long, and analyzing them
can be quite tedious.
However, doing so is usually quite a bit faster
and more pleasant than locating bugs the old-fashioned way.

In addition, one of the simplest tests of formal-verification tools is
bug injection.
After all, not only could any of us write
\co{printf("VERIFIED\\n")}, but the plain fact is that
developers of formal-verification tools are just as bug-prone as
are the rest of us.
Therefore, formal-verification tools that just proclaim that a
bug exists are fundamentally less trustworthy because it is
more difficult to verify them on real-world code.

All that aside, people writing formal-verification tools are
permitted to leverage existing tools.
For example, a tool designed to determine only the presence
or absence of a serious but rare bug might leverage bisection.
If an old version of the program under test did not contain the bug,
but a new version did, then bisection could be used to quickly
locate the commit that inserted the bug, which might be
sufficient information to find and fix the bug.
Of course, this sort of strategy would not work well for common
bugs because in this case bisection would fail due to all commits
having at least one instance of the common bug.

Therefore, the execution traces provided
by many formal-verification tools will continue to be valuable,
particularly for complex and difficult-to-understand bugs.
In addition, recent work applies \emph{incorrectness-logic}
formalism reminiscent of the traditional Hoare logic used for
full-up correctness proofs, but with the sole purpose of finding
bugs~\cite{PeterWOHearn2019incorrectnessLogic}.

\subsection{Minimal Scaffolding}
\label{sec:future:Minimal Scaffolding}

In the old days, formal-verification researchers demanded a full
specification against which the software would be verified.
Unfortunately, a mathematically rigorous specification might well
be larger than the actual code, and each line of specification
is just as likely to contain bugs as is each line of code.
A formal verification effort proving that the code faithfully implemented
the specification would be a proof of bug-for-bug compatibility between
the two, which might not be all that helpful.

Worse yet, the requirements for a number of software artifacts,
including Linux-kernel RCU, are empirical in
nature~\cite{PaulEMcKenney2015RCUreqts1,PaulEMcKenney2015RCUreqts2,PaulEMcKenney2015RCUreqts3}.\footnote{
	Or, in formal-verification parlance, Linux-kernel RCU has an
	\emph{incomplete specification}.}
For this common type of software, a complete specification is a
polite fiction.
Nor are complete specifications any less fictional for hardware,
as was made clear by the late-2017 Meltdown and Spectre side-channel
attacks~\cite{JannHorn2018MeltdownSpectre}.

This situation might cause one to give up all hope of formal verification
of real-world software and hardware artifacts, but it turns out that there is
quite a bit that can be done.
For example, design and coding rules can act as a partial specification,
as can assertions contained in the code.
And in fact formal-verification tools such as \co{cbmc} and Nidhugg
both check for assertions that can be triggered, implicitly treating
these assertions as part of the specification.
However, the assertions are also part of the code, which makes it less
likely that they will become obsolete, especially if the code is
also subjected to stress tests.\footnote{
	And you \emph{do} stress-test your code, don't you?}
The \co{cbmc} tool also checks for array-out-of-bound references,
thus implicitly adding them to the specification.
The aforementioned incorrectness logic can also be thought of as using
an implicit bugs-not-present
specification~\cite{PeterWOHearn2019incorrectnessLogic}.

This implicit-specification approach makes quite a bit of sense, particularly
if you look at formal verification not as a full proof of correctness,
but rather an alternative form of validation with a different set of
strengths and weaknesses than the common case, that is, testing.
From this viewpoint, software will always have bugs, and therefore any
tool of any kind that helps to find those bugs is a very good thing
indeed.

\subsection{Relevant Bugs}
\label{sec:future:Relevant Bugs}

Finding bugs---and fixing them---is of course the whole point of any
type of validation effort.
Clearly, false positives are to be avoided.
But even in the absence of false positives, there are bugs and there are bugs.

For example, suppose that a software artifact had exactly 100 remaining
bugs, each of which manifested on average once every million years
of runtime.
Suppose further that an omniscient formal-verification tool located
all 100 bugs, which the developers duly fixed.
What happens to the reliability of this software artifact?

The answer is that the reliability \emph{decreases}.

To see this, keep in mind that historical experience indicates that
about 7\,\% of fixes introduce a new bug~\cite{RexBlack2012SQA}.
Therefore, fixing the 100 bugs, which had a combined mean time to failure
(MTBF) of about 10,000 years, will introduce seven more bugs.
Historical statistics indicate that each new bug will have an MTBF
much less than 70,000 years.
This in turn suggests that the combined MTBF of these seven new bugs
will most likely be much less than 10,000 years, which in turn means
that the well-intentioned fixing of the original 100 bugs actually
decreased the reliability of the overall software.

\QuickQuizSeries{%
\QuickQuizB{
	How do we know that the MTBFs of known bugs is a good estimate
	of the MTBFs of bugs that have not yet been located?
}\QuickQuizAnswerB{
	We don't, but it does not matter.

	To see this, note that the 7\,\% figure only applies to injected
	bugs that were subsequently located:
	It necessarily ignores any injected bugs that were never found.
	Therefore, the MTBF statistics of known bugs is likely to be
	a good approximation of that of the injected bugs that are
	subsequently located.

	A key point in this whole section is that we should be more
	concerned about bugs that inconvenience users than about
	other bugs that never actually manifest.
	This of course is \emph{not} to say that we should completely
	ignore bugs that have not yet inconvenienced users, just that
	we should properly prioritize our efforts so as to fix the
	most important and urgent bugs first.
}\QuickQuizEndB
\QuickQuizE{
	But the formal-verification tools should immediately find all the
	bugs introduced by the fixes, so why is this a problem?
}\QuickQuizAnswerE{
	It is a problem because real-world formal-verification tools
	(as opposed to those that exist only in the imaginations of
	the more vociferous proponents of formal verification) are
	not omniscient, and thus are only able to locate certain types
	of bugs.
	For but one example, formal-verification tools are unlikely to
	spot a bug corresponding to an omitted assertion or, equivalently,
	a bug corresponding to an undiscovered portion of the specification.
}\QuickQuizEndE
}

Worse yet, imagine another software artifact with one bug that fails
once every day on average and 99 more that fail every million years
each.
Suppose that a formal-verification tool located the 99 million-year
bugs, but failed to find the one-day bug.
Fixing the 99 bugs located will take time and effort, decrease
reliability, and do nothing at all about the pressing each-day failure
that is likely causing embarrassment and perhaps much worse besides.

Therefore, it would be best to have a validation tool that
preferentially located the most troublesome bugs.
However, as noted in
\cref{sec:future:Locate Bugs},
it is permissible to leverage additional tools.
One powerful tool is none other than plain old testing.
Given knowledge of the bug, it should be possible to construct
specific tests for it, possibly also using some of the techniques
described in
\cref{sec:debugging:Hunting Heisenbugs}
to increase the probability of the bug manifesting.
These techniques should allow calculation of a rough estimate of the
bug's raw failure rate, which could in turn be used to prioritize
bug-fix efforts.

\QuickQuiz{
	But many formal-verification tools can only find one bug at
	a time, so that each bug must be fixed before the tool can
	locate the next.
	How can bug-fix efforts be prioritized given such a tool?
}\QuickQuizAnswer{
	One approach is to provide a simple fix that might not be
	suitable for a production environment, but which allows
	the tool to locate the next bug.
	Another approach is to restrict configuration or inputs
	so that the bugs located thus far cannot occur.
	There are a number of similar approaches, but the common theme
	is that fixing the bug from the tool's viewpoint is usually much
	easier than constructing and validating a production-quality fix,
	and the key point is to prioritize the larger efforts required
	to construct and validate the production-quality fixes.
}\QuickQuizEnd

There has been some recent formal-verification work that prioritizes
executions having fewer preemptions, under that reasonable assumption
that smaller numbers of preemptions are more likely.

Identifying relevant bugs might sound like too much to ask, but it is what
is really required if we are to actually increase software reliability.

\subsection{Formal Regression Scorecard}
\label{sec:future:Formal Regression Scorecard}

\begin{table*}
% \rowcolors{6}{}{lightgray}
%\renewcommand*{\arraystretch}{1.1}
\small
\centering
\setlength{\tabcolsep}{2pt}
\begin{tabular}{lcccccccccc}
	\toprule
	& & Promela & & PPCMEM & & \tco{herd} & & \tco{cbmc} & & Nidhugg \\
	\midrule
	(1) Automated &
		& \cellcolor{red!50} &
			& \cellcolor{orange!50} &
				& \cellcolor{orange!50} &
					& \cellcolor{blue!50} &
						& \cellcolor{blue!50} \\
	\addlinespace[3pt]
	(2) Environment &
		& \cellcolor{red!50} (MM) &
			& \cellcolor{green!50} &
				& \cellcolor{blue!50} &
					& \cellcolor{yellow!50} (MM) &
						& \cellcolor{orange!50} (MM) \\
	\addlinespace[3pt]
	(3) Overhead &
		& \cellcolor{yellow!50} &
			& \cellcolor{red!50} &
				& \cellcolor{yellow!50} &
					& \cellcolor{yellow!50} (SAT) &
						& \cellcolor{green!50} \\
	\addlinespace[3pt]
	(4) Locate Bugs &
		& \cellcolor{yellow!50} &
			& \cellcolor{yellow!50} &
				& \cellcolor{yellow!50} &
					& \cellcolor{green!50} &
						& \cellcolor{green!50} \\
	\addlinespace[3pt]
	(5) Minimal Scaffolding &
		& \cellcolor{green!50} &
			& \cellcolor{yellow!50} &
				& \cellcolor{yellow!50} &
					& \cellcolor{blue!50} &
						& \cellcolor{blue!50} \\
	\addlinespace[3pt]
	(6) Relevant Bugs &
		& \cellcolor{yellow!50} ??? &
			& \cellcolor{yellow!50} ??? &
				& \cellcolor{yellow!50} ??? &
					& \cellcolor{yellow!50} ??? &
						& \cellcolor{yellow!50} ??? \\
	\bottomrule
\end{tabular}
\caption{Formal Regression Scorecard}
\label{tab:future:Formal Regression Scorecard}
\end{table*}

\Cref{tab:future:Formal Regression Scorecard}
shows a rough-and-ready scorecard for the formal-verification tools
covered in this chapter.
Shorter wavelengths are better than longer wavelengths.

Promela requires hand translation and supports only sequential
consistency, so its first two cells are red.
It has reasonable overhead (for formal verification, anyway)
and provides a traceback, so its next two cells are yellow.
Despite requiring hand translation, Promela handles assertions
in a natural way, so its fifth cell is green.

PPCMEM usually requires hand translation due to the small size of litmus
tests that it supports, so its first cell is orange.
It handles several memory models, so its second cell is green.
Its overhead is quite high, so its third cell is red.
It provides a graphical display of relations among operations, which
is not as helpful as a traceback, but is still quite useful, so its
fourth cell is yellow.
It requires constructing an \co{exists} clause and cannot take
intra-process assertions, so its fifth cell is also yellow.

The \co{herd} tool has size restrictions similar to those of PPCMEM,
so \co{herd}'s first cell is also orange.
It supports a wide variety of memory models, so its second cell is blue.
It has reasonable overhead, so its third cell is yellow.
Its bug-location and assertion capabilities are quite similar to those
of PPCMEM, so \co{herd} also gets yellow for the next two cells.

The \co{cbmc} tool inputs C code directly, so its first cell is blue.
It supports a few memory models, so its second cell is yellow.
It has reasonable overhead, so its third cell is also yellow, however,
perhaps SAT-solver performance will continue improving.
It provides a traceback, so its fourth cell is green.
It takes assertions directly from the C code, so its fifth cell is blue.

Nidhugg also inputs C code directly, so its first cell is also blue.
It supports only a couple of memory models, so its second cell is orange.
Its overhead is quite low (for formal-verification), so its
third cell is green.
It provides a traceback, so its fourth cell is green.
It takes assertions directly from the C code, so its fifth cell is blue.

So what about the sixth and final row?
It is too early to tell how any of the tools do at finding the right bugs,
so they are all yellow with question marks.

\QuickQuizSeries{%
\QuickQuizB{
	How would testing stack up in the scorecard shown in
	\cref{tab:future:Formal Regression Scorecard}?
}\QuickQuizAnswerB{
	It would be blue all the way down, with the possible
	exception of the third row (overhead) which might well
	be marked down for testing's difficulty finding
	improbable bugs.

	On the other hand, improbable bugs are often also
	irrelevant bugs, so your mileage may vary.

	Much depends on the size of your installed base.
	If your code is only ever going to run on (say) 10,000
	systems, Murphy can actually be a really nice guy.
	Everything that can go wrong, will.
	Eventually.
	Perhaps in geologic time.

	But if your code is running on 20~billion systems,
	like the Linux kernel was said to be by late 2017,
	Murphy can be a real jerk!
	Everything that can go wrong, will, and it can go wrong
	really quickly!!!
}\QuickQuizEndB
\QuickQuizE{
	But aren't there a great many more formal-verification systems
	than are shown in
	\cref{tab:future:Formal Regression Scorecard}?
}\QuickQuizAnswerE{
	Indeed there are!
	This table focuses on those that Paul has used, but others are
	proving to be useful.
	Formal verification has been heavily used in the seL4
	project~\cite{ThomasSewell2013L4binaryVerification},
	and its tools can now handle modest levels of concurrency.
	More recently, Catalin Marinas used Lamport's
	TLA tool~\cite{Lamport:2002:SST:579617} to locate some
	forward-progress bugs in the Linux kernel's queued spinlock
	implementation.
	Will Deacon fixed these bugs~\cite{WillDeacon2018qspinlock},
	and Catalin verified Will's
	fixes~\cite{CatalinMarinas2018qspinlockTLA}.

	Lighter-weight formal verification tools have been used heavily
	in production~\cite{JamesRLarus2004RightingSoftware,AlBessey2010BillionLoCLater,ByronCook2018FormalAmazon,CaitlinSadowski2018staticAnalysisGoogle,DinoDistefano2019FBstaticAnalysis}.
}\QuickQuizEndE
}

Once again, please note that this table rates these tools for use in
regression testing.
Just because many of them are a poor fit for regression testing does
not at all mean that they are useless, in fact,
many of them have proven their worth many times over.\footnote{
	For but one example, Promela was used to verify the file system
	of none other than the Curiosity Rover.
	Was \emph{your} formal verification tool used on software that
	currently runs on Mars???}
Just not for regression testing.

However, this might well change.
After all, formal verification tools made impressive strides in the 2010s.
If that progress continues, formal verification might well become an
indispensable tool in the parallel programmer's validation toolbox.

% future/fp.tex
% mainfile: ../perfbook.tex
% SPDX-License-Identifier: CC-BY-SA-3.0

\section{Functional Programming for Parallelism}
\label{sec:future:Functional Programming for Parallelism}
\epigraph{The curious failure of functional programming for parallel
	  applications.}
	 {Malte Skarupke}

When I took my first-ever functional-programming class in the early 1980s,
the professor asserted that the side-effect-free functional-programming
style was well-suited to trivial parallelization and analysis.
Thirty years later, this assertion remains, but mainstream production
use of parallel functional languages is minimal, a state of affairs that
might not be entirely unrelated to professor's additional
assertion that programs should neither maintain state nor do I/O\@.
There is niche use of functional languages such as Erlang, and
multithreaded support has been added to several other functional languages,
but mainstream production usage remains the province of procedural
languages such as C, C++, Java, and Fortran (usually augmented with
OpenMP, MPI, or coarrays).

This situation naturally leads to the question ``If analysis is the goal,
why not transform the procedural language into a functional language before
doing the analysis?''
There are of course a number of objections to this approach, of which
I list but three:

\begin{enumerate}
\item	Procedural languages often make heavy use of global variables,
	which can be updated independently by different
	functions, or, worse yet, by multiple threads.
	Note that Haskell's \emph{monads} were invented to deal with
	single-threaded global state, and that multi-threaded access to
	global state inflicts additional violence on the functional model.
\item	Multithreaded procedural languages often use synchronization
	primitives such as locks, atomic operations, and transactions,
	which inflict added violence upon the functional model.
\item	Procedural languages can \emph{alias} function arguments,
	for example, by passing a pointer to the same structure via two
	different arguments to the same invocation of a given function.
	This can result in the function unknowingly updating that
	structure via two different (and possibly overlapping) code
	sequences, which greatly complicates analysis.
\end{enumerate}

Of course, given the importance of global state, synchronization
primitives, and aliasing, clever functional-programming experts have
proposed any number of attempts to reconcile the function programming
model to them, monads being but one case in point.

Another approach is to compile the parallel procedural program into
a functional program, then to use functional-programming tools to analyze
the result.
But it is possible to do much better than this, given that any real
computation is a large finite-state machine with finite input that
runs for a finite time interval.
This means that any real program can be transformed into an expression,
possibly albeit an impractically large one~\cite{VijayDSilva2012-sas}.

However, a number of the low-level kernels of parallel algorithms transform
into expressions that are small enough to fit easily into the memories
of modern computers.
If such an expression is coupled with an assertion, checking to see if
the assertion would ever fire becomes a satisfiability problem.
Even though satisfiability problems are NP-complete, they can often
be solved in much less time than would be required to generate the
full state space.
In addition, the solution time appears to be only weakly dependent on
the underlying memory model, so that algorithms running on weakly ordered
systems can also be checked~\cite{JadeAlglave2013-cav}.

The general approach is to transform the program into single-static-assignment
(SSA) form, so that each assignment to a variable creates a separate
version of that variable.
This applies to assignments from all the active threads, so that the
resulting expression embodies all possible executions of the code
in question.
The addition of an assertion entails asking whether any combination of
inputs and initial values can result in the assertion firing, which,
as noted above, is exactly the satisfiability problem.

One possible objection is that it does not gracefully handle arbitrary
looping constructs.
However, in many cases, this can be handled by unrolling the loop a
finite number of times.
In addition, perhaps some loops will also prove amenable to collapse
via inductive methods.

Another possible objection is that spinlocks involve arbitrarily long
loops, and any finite unrolling would fail to capture the full behavior
of the spinlock.
It turns out that this objection is easily overcome.
Instead of modeling a full spinlock, model a trylock that attempts to
obtain the lock, and aborts if it fails to immediately do so.
The assertion must then be crafted so as to avoid firing in cases
where a spinlock aborted due to the lock not being immediately available.
Because the logic expression is independent of time, all possible
concurrency behaviors will be captured via this approach.

A final objection is that this technique is unlikely to be able to handle
a full-sized software artifact such as the millions of lines of code making
up the Linux kernel.
This is likely the case, but the fact remains that exhaustive validation
of each of the much smaller parallel primitives within the Linux kernel
would be quite valuable.
And in fact the researchers spearheading this approach have applied it
to non-trivial real-world code, including the Tree RCU implementation in
the Linux
kernel~\cite{LihaoLiang2016VerifyTreeRCU,MichalisKokologiannakis2017NidhuggRCU}.

It remains to be seen how widely applicable this technique is, but
it is one of the more interesting innovations in the field of
formal verification.
Although it might well be that the functional-programming advocates
are at long last correct in their assertion of the inevitable
dominance of functional programming, it is clearly the case
that this long-touted methodology is starting to see credible
competition on its formal-verification home turf.
There is therefore continued reason to doubt the inevitability of
functional-programming dominance.

% @@@ Maybe add quantum computing back in, but heavily summarized.

\section{Summary}
\label{sec:future:Summary}

This chapter has taken a quick tour of a number of possible futures,
including multicore, transactional memory, formal verification as
a regression test, and concurrent functional programming.
Any of these futures might come true, but it is more likely that, as in
the past, the future will be far stranger than we can possibly imagine.

	\setlength{\parskip}{0.0pt plus 1ex}
	\input{qqzfuture}
	\setlength{\parskip}{0.0pt plus 1.0pt}% return to default

% summary.tex
% mainfile: perfbook.tex
% SPDX-License-Identifier: CC-BY-SA-3.0

\chapter{Looking Forward and Back}
\label{chp:Looking Forward and Back}
\Epigraph{History is the sum total of things that could have been avoided.}
	  {Konrad Adenauer}

You have arrived at the end of this book, well done!
I~hope that your journey was a pleasant but challenging and worthwhile
one.

For your editor and contributors, this is the end of the journey to the
Second Edition, but for those willing to join in, it is also the start
of the journey to the Third Edition.
Either way, it is good to recap this past journey.

\Cref{chp:How To Use This Book} covered what this book is about, along
with some alternatives for those interested in something other than
low-level parallel programming.

\Cref{chp:Introduction} covered parallel-programming challenges and
high-level approaches for addressing them.
It also touched on ways of avoiding these challenges while nevertheless
still gaining most of the benefits of parallelism.

\Cref{chp:Hardware and its Habits} gave a high-level overview of multicore
hardware, especially those aspects that pose challenges for concurrent
software.
This \lcnamecref{chp:Hardware and its Habits} puts the blame for these
challenges where it belongs, very much on the laws of physics and rather
less on intransigent hardware architects and designers.
However, there might be some things that hardware architects and engineers
can do, and this \lcnamecref{chp:Hardware and its Habits} discusses a few of
them.
In the meantime, software architects and engineers must do their part
to meet these challenges, as discussed in the rest of the book.

\Cref{chp:Tools of the Trade}
gave a quick overview of the tools of the low-level concurrency trade.
\Cref{chp:Counting} then demonstrated use of those tools---and, more
importantly, use of parallel-programming design techniques---on the
simple but surprisingly challenging task of concurrent counting.
So challenging, in fact, that a number of concurrent counting algorithms
are in common use, each specialized for a different use case.

\Cref{chp:Partitioning and Synchronization Design} dug more deeply
into the most important parallel-programming design technique, namely
partitioning the problem at the highest possible level.
This \lcnamecref{chp:Partitioning and Synchronization Design} also
overviewed a number of points in this design space.

\Cref{chp:Locking} expounded on that parallel-programming workhorse
(and villain), locking.
This \lcnamecref{chp:Locking} covered a number of types of locking
and presented some engineering solutions to many well-known and
aggressively advertised shortcomings of locking.

\Cref{chp:Data Ownership} discussed the uses of data ownership, where
synchronization is supplied by the association of a given data item
with a specific thread.
Where it applies, this approach combines excellent performance and
scalability with profound simplicity.

\Cref{chp:Deferred Processing} showed how a little procrastination can
greatly improve performance and scalability, while in a surprisingly large
number of cases also simplifying the code.
A number of the mechanisms presented in this
\lcnamecref{chp:Deferred Processing}
take advantage of the ability of CPU caches to replicate read-only data,
thus sidestepping the laws of physics that cruelly limit the speed of
light and the smallness of atoms.
\Cref{chp:Data Structures} looked at concurrent data structures, with
emphasis on hash tables, which have a long and honorable history in
parallel programs.

\Cref{chp:Validation} dug into code-review and testing methods, and
\cref{chp:Formal Verification} overviewed formal verification.
Whichever side of the formal-verification/\-testing divide you might
be on, if code has not been thoroughly validated, it does not work.
And that goes at least double for concurrent code.

\Cref{chp:Putting It All Together} presented a number of situations
where combining concurrency mechanisms with each other or with other
design tricks can greatly ease parallel programmers' lives.
\Cref{sec:advsync:Advanced Synchronization} looked at advanced
synchronization methods, including lockless programming, \IXacrl{nbs},
and parallel real-time computing.
\Cref{chp:Advanced Synchronization: Memory Ordering} dug into the
critically important topic of memory ordering, presenting techniques
and tools to help you not only solve memory-ordering problems, but
also to avoid them completely.
\Cref{chp:Ease of Use} presented a brief overview of the surprisingly
important topic of ease of use.

Last, but definitely not least, \cref{chp:Conflicting Visions of the Future}
expounded on a number of conflicting visions of the future, including
CPU-technology trends, transactional memory, hardware transactional
memory, use of formal verification in regression testing, and the
long-standing prediction that the future of parallel programming belongs
to functional-programming languages.

But now that we have recapped the contents of this Second Edition, how did
this book get started?

Paul's parallel-programming journey started in earnest in 1990, when
he joined Sequent Computer Systems, Inc.
Sequent used an apprenticeship-like program in which newly hired engineers
were placed in cubicles surrounded by experienced engineers, who mentored
them, reviewed their code, and gave copious quantities of advice on
a variety of topics.
A few of the newly hired engineers were greatly helped by the fact that
there were no on-chip caches in those days, which meant that
logic analyzers could easily display a given CPU's instruction stream
and memory accesses, complete with accurate timing information.
Of course, the downside of this transparency was that CPU core clock
frequencies were 100 times slower than those of the twenty-first
century.
Between apprenticeship and hardware performance transparency, these newly
hired engineers became productive parallel programmers within two or three
months, and some were doing ground-breaking work within a couple of years.

Sequent understood that its ability to quickly train new engineers in the
mysteries of parallelism was unusual, so it produced a slim volume that
crystalized the company's parallel-programming wisdom~\cite{SQNTParallel},
which joined a pair of groundbreaking papers that had been written a
few years earlier~\cite{Beck85,Inman85}.
People already steeped in these mysteries saluted this book and these
papers, but novices were usually unable to benefit much from them,
invariably making highly creative and quite destructive errors that
were not explicitly prohibited by either the book or the papers.\footnote{
	``But why on earth would you do \emph{that}???''
	``Well, why not?''}
This situation of course caused Paul to start thinking in terms of
writing an improved book, but his efforts during this time were limited
to internal training materials and to published papers.

By the time Sequent was acquired by IBM in 1999, many of the world's
largest database instances ran on Sequent hardware.
But times change, and by 2001 many of Sequent's parallel programmers
had shifted their focus to the Linux kernel.
After some initial reluctance, the Linux kernel community embraced
concurrency both enthusiastically and
effectively~\cite{SilasBoydWickizer2010LinuxScales48,McKenney:2012:BEP:2414729.2414734},
with many excellent innovations and improvements from throughout the
community.
The thought of writing a book occurred to Paul from time to time, but
life was flowing fast, so he made no progress on this project.

In 2006, Paul was invited to a conference on Linux scalability, and was
granted the privilege of asking the last question of panel of esteemed
parallel-programming experts.
Paul began his question by noting that in the 15~years from 1991 to 2006,
the price of a parallel system had dropped from that of a house to that
of a mid-range bicycle, and it was clear that there was much more room
for additional dramatic price decreases over the next 15~years
extending to the year 2021.
He also noted that decreasing price should result in greater familiarity
and faster progress in solving parallel-programming problems.
This led to his question:
``In the year 2021, why wouldn't parallel programming have become routine?''

The first panelist seemed quite disdainful of anyone who would ask such
an absurd question, and quickly responded with a soundbite answer.
To which Paul gave a soundbite response.
They went back and forth for some time, for example, the panelist's
sound-bite answer ``Deadlock'' provoked Paul's sound-bite response ``Lock
dependency checker''.

The panelist eventually ran out of soundbites, improvising a final
``People like you should be hit over the head with a hammer!''

Paul's response was of course ``You will have to get in line for that!''

Paul turned his attention to the next panelist, who seemed torn between
agreeing with the first panelist and not wishing to have to deal with
Paul's series of responses.
He therefore have a short non-committal speech.
And so it went through the rest of the panel.

Until it was the turn of the last panelist, who was someone you might have
heard of who goes by the name of Linus Torvalds.
Linus noted that three years earlier (that is, 2003), the initial version
of any concurrency-related patch was usually quite poor, having design
flaws and many bugs.
And even when it was cleaned up enough to be accepted, bugs still
remained.
Linus contrasted this with the then-current situation in 2006, in which
he said that it was not unusual for the first version of a concurrency-related
patch to be well-designed with few or even no bugs.
He then suggested that \emph{if} tools continued to improve, then \emph{maybe}
parallel programming would become routine by the year 2021.\footnote{
	Tools have in fact continued to improve, including fuzzers,
	lock dependency checkers, static analyzers, formal verification,
	memory models, and code-modification tools such as coccinelle.
	Therefore, those who wish to assert that year-2021
	parallel programming is not routine should refer to
	\cref{chp:Introduction}'s epigraph.}

The conference then concluded.
Paul was not surprised to be given wide berth by many audience members,
especially those who saw the world in the same way as did the first panelist.
Paul was also not surprised that a few audience members thanked him for
the question.
However, he was quite surprised when one man came up to say ``thank
you'' with tears streaming down his face, sobbing so hard that he could
barely speak.

You see, this man had worked several years at Sequent, and thus very
well understood parallel programming.
Furthermore, he was currently assigned to a group whose job it was to
write parallel code.
Which was not going well.
You see, it wasn't that they had trouble understanding his explanations
of parallel programming.

It was that they refused to listen to him \emph{at all}.

In short, his group was treating this man in the same way that the first
panelist attempted to treat Paul.
And so in that moment, Paul went from ``I should write a book some day''
to ``I will do whatever it takes to write this book''.
Paul is embarrassed to admit that he does not remember the man's name,
if in fact he ever knew it.

This book is nevertheless for that man.

\begin{figure}
\centering
\resizebox{3in}{!}{\includegraphics{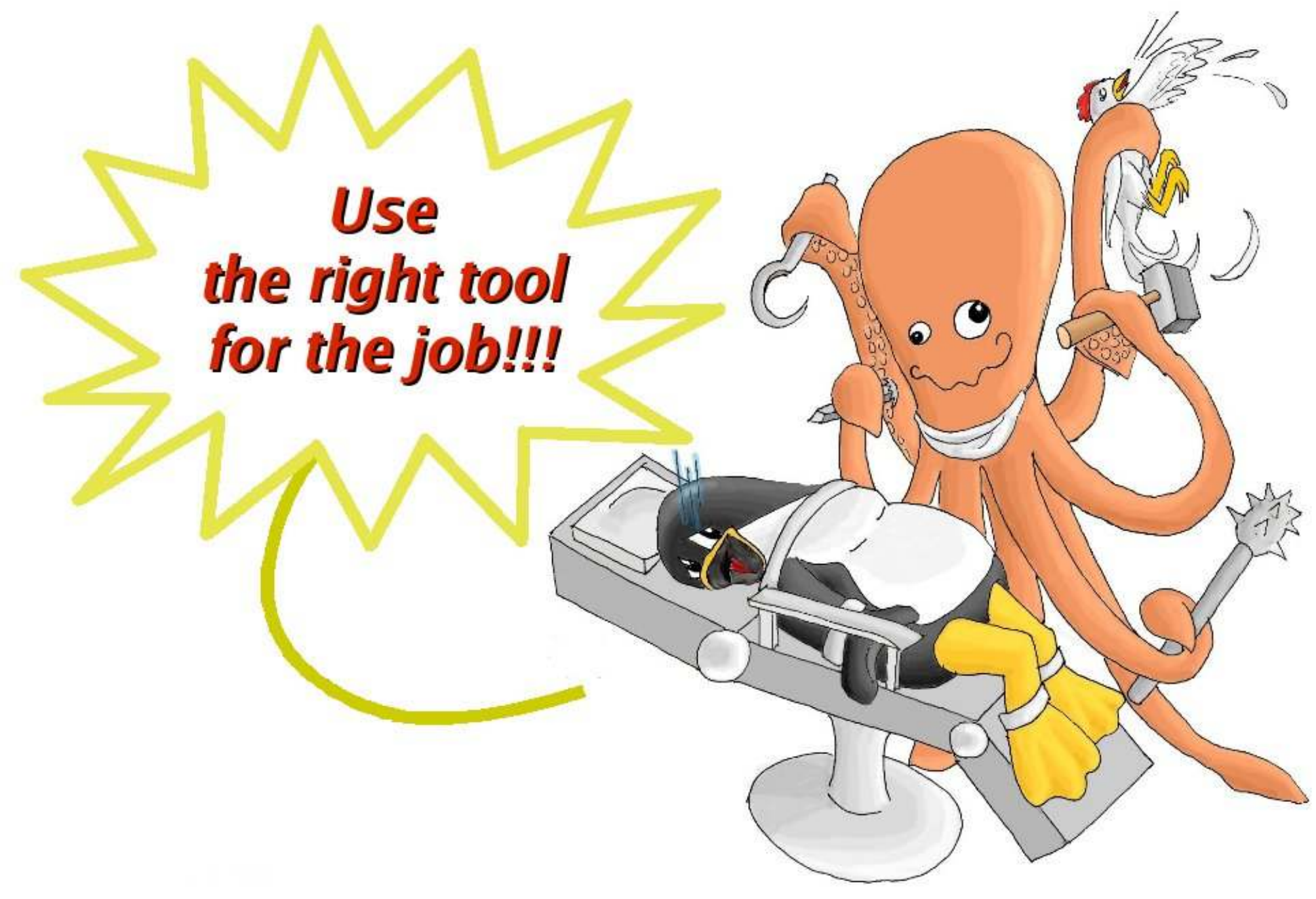}}
\caption{The Most Important Lesson}
\ContributedBy{Figure}{fig:summary:The Most Important Lesson}{Melissa Broussard}
\end{figure}

And this book is also for everyone else who would like to add low-level
concurrency to their skillset.
If you remember nothing else from this book, let it be the lesson of
\cref{fig:summary:The Most Important Lesson}.

And this book is also a salute to that unnamed panelist's unnamed employer.
Some years later, this employer choose to appoint someone with more
useful experience and fewer sound bites.
That someone was also on a panel, and during that session he looked
directly at me when he stated that parallel programming was perhaps 5\%
more difficult than sequential programming.

For the rest of us, when someone tries to show us a solution to pressing
problem, perhaps we should at the very least do them the courtesy of
listening!

\appendix

% appendix/appendix.tex
% mainfile: ../perfbook.tex
% SPDX-License-Identifier: CC-BY-SA-3.0

% appendix/questions/questions.tex
% mainfile: ../../perfbook.tex
% SPDX-License-Identifier: CC-BY-SA-3.0

\QuickQuizChapter{chp:app:Important Questions}{Important Questions}{qqzquestions}
\Epigraph{Ask me no questions, and I'll tell you no fibs.}
	 {\emph{She Stoops to Conquer}, Oliver Goldsmith}

The following sections discuss some important questions relating to
SMP programming.
Each section also shows how to avoid worrying about
the corresponding question, which can be extremely important if
your goal is to simply get your SMP code working as quickly and
painlessly as possible---which is an excellent goal, by the way!

Although the answers to these questions are often less
intuitive than they would be in a single-threaded setting,
with a bit of work, they are not that difficult to understand.
If you managed to master recursion, there is nothing here that should
pose an overwhelming challenge.

With that, here are the questions:

\begin{enumerate}
\item	Why aren't parallel programs always faster?
	(\cref{sec:app:questions:Why Aren't Parallel Programs Always Faster?})
\item	Why not remove locking?
	(\cref{sec:app:questions:Why Not Remove Locking?})
\item	What time is it?
	(\cref{sec:app:questions:What Time Is It?})
\item	What does ``after'' mean?
	(\cref{sec:app:questions:What Does ``After'' Mean?})
\item	How much ordering is needed?
	(\cref{sec:app:questions:How Much Ordering Is Needed?})
\item	What is the difference between ``concurrent'' and ``parallel''?
	(\cref{sec:app:questions:What is the Difference Between ``Concurrent'' and ``Parallel''?})
\item	Why is software buggy?
	(\cref{sec:app:questions:Why Is Software Buggy?})
\end{enumerate}

Read on to learn some answers.
Improve upon these answers if you can!

% appendix/questions/parallelfaster.tex
% mainfile: ../../perfbook.tex
% SPDX-License-Identifier: CC-BY-SA-3.0

\section{Why Aren't Parallel Programs Always Faster?}
\label{sec:app:questions:Why Aren't Parallel Programs Always Faster?}

The short answer is ``because parallel execution often requires
communication, and communication is not free''.

For more information on this question, see
\cref{chp:Hardware and its Habits},
\cref{sec:count:Why Isn't Concurrent Counting Trivial?},
and especially
\cref{chp:Partitioning and Synchronization Design},
each of which present ways of slowing down your code by ineptly
parallelizing it.
Of course, much of this book deals with ways of ensuring that your
parallel programs really are faster than their sequential counterparts.

However, never forget that parallel programs can be quite fast while at
the same time being quite simple, with the example in
\cref{sec:toolsoftrade:Scripting Languages}
being a case in point.
Also never forget that parallel execution is but one optimization of many,
and there are programs for which other optimizations produce better results.

% appendix/questions/removelocking.tex
% mainfile: ../../perfbook.tex
% SPDX-License-Identifier: CC-BY-SA-3.0

\section{Why Not Remove Locking?}
\label{sec:app:questions:Why Not Remove Locking?}

There can be no doubt that many have cast locking as the evil villain
of parallel programming, and not entirely without reason.
And there are important examples where lockless code does much better
than its locked counterpart, a few of which are discussed in
\cref{sec:advsync:Non-Blocking Synchronization}.

However, lockless algorithms are not guaranteed to perform and scale
well, as shown by
\cref{fig:count:Atomic Increment Scalability on x86} on
\cpageref{fig:count:Atomic Increment Scalability on x86}.
Furthermore, as a general rule, the more complex the algorithm,
the greater the advantage of combining locking with selected
lockless techniques, even with significant hardware support,
as shown in
\IfEbookSize{
\cref{tab:future:Comparison of Locking and HTM,%
tab:future:Comparison of Locking (Augmented by RCU or Hazard Pointers) and HTM}
on
\cpageref{tab:future:Comparison of Locking and HTM,%
tab:future:Comparison of Locking (Augmented by RCU or Hazard Pointers) and HTM}.
}{
\cref{tab:future:Comparison of Locking (Plain and Augmented) and HTM}
on
\cpageref{tab:future:Comparison of Locking (Plain and Augmented) and HTM}.
}
\Cref{sec:advsync:Non-Blocking Synchronization}
looks more deeply at \IXacrl{nbs}, which is a popular
lockless methodology.

As a more general rule, a sound-bite approach to parallel programming
is not likely to end well.
Some would argue that this is also true of many other fields of endeavor.

% appendix/questions/time.tex
% mainfile: ../../perfbook.tex
% SPDX-License-Identifier: CC-BY-SA-3.0

\section{What Time Is It?}
\label{sec:app:questions:What Time Is It?}

\begin{figure}
\centering
\resizebox{2.6in}{!}{\includegraphics{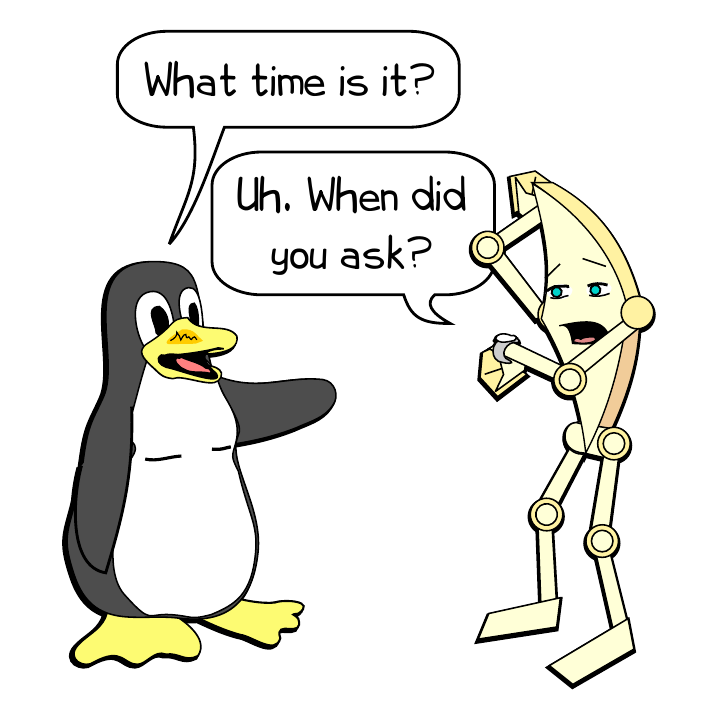}}
\caption{What Time Is It?}
\ContributedBy{Figure}{fig:app:questions:What Time Is It?}{Melissa Broussard}
\end{figure}

A key issue with timekeeping on multicore computer systems is illustrated
by \cref{fig:app:questions:What Time Is It?}.
One problem is that it takes time to read out the time.
An instruction might read from a hardware clock, and might
have to go off-core (or worse yet, off-socket) to complete
this read operation.
It might also be necessary to do some computation on the value read out,
for example, to convert it to the desired format, to apply network time
protocol (NTP) adjustments, and so on.
So does the time eventually returned correspond to the beginning of
the resulting time interval, the end, or somewhere in between?

Worse yet, the thread reading the time might be interrupted or preempted.
Furthermore, there will likely be some computation between reading out
the time and the actual use of the time that has been read out.
Both of these possibilities further extend the interval of uncertainty.

One approach is to read the time twice, and take the arithmetic mean
of the two readings, perhaps one on each side of the operation being
timestamped.
The difference between the two readings is then a measure of uncertainty
of the time at which the intervening operation occurred.

Of course, in many cases, the exact time is not necessary.
For example, when printing the time for the benefit of a human user,
we can rely on slow human reflexes to render internal hardware and
software delays irrelevant.
Similarly, if a server needs to timestamp the response to a client, any
time between the reception of the request and the transmission of the
response will do equally well.

There is an old saying that those who have but one clock always
know the time, but those who have several clocks can never be sure.
And there was a time when the typical low-end computer's sole
software-visible clock was its program counter, but those days are
long gone.
This is not a bad thing, considering that on modern computer systems,
the program counter is a truly horrible
clock~\cite{PeterOkech2009InherentRandomness}.

In addition, different clocks provide different tradeoffs of performance,
accuracy, precision, and ordering.
For example, in the Linux kernel, the \co{jiffies} counter\footnote{
	The \co{jiffies} variable is a location in normal memory that
	is incremented by software in response to events such as the
	scheduling-clock interrupt.}
provides high-speed access to a course-grained counter (at best
one-millisecond accuracy and precision) that imposes very little ordering
on either the compiler or the hardware.
In contrast, the x86 HPET hardware provides an accurate and
precise clock, but at the price of slow access.
The x86 time-stamp counter (TSC) has a checkered past, but is more
recently held out as providing a good combination of precision, accuracy,
and performance.
Unfortunately, for all of these counters, ordering against all effects
of prior and subsequent code requires expensive memory-barrier instructions.
And this expense appears to be an unavoidable consequence of the
complex superscalar nature of modern computer systems.

\begin{figure}
\centering
\resizebox{3in}{!}{\includegraphics{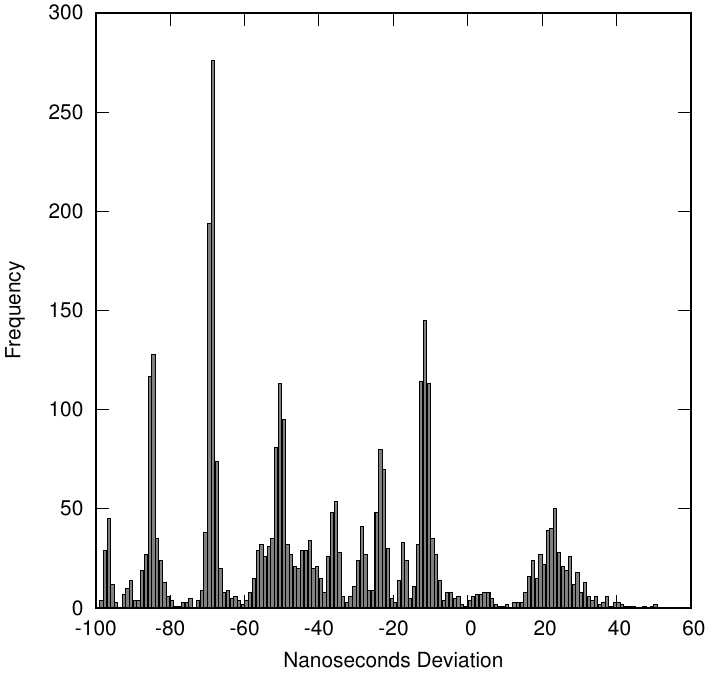}}
\caption{\tco{clock_gettime(CLOCK_REALTIME)} Deviation From Immediately Preceding \tco{clock_gettime(CLOCK_MONOTONIC)}}
\label{fig:app:questions:clock-gettime(CLOCK-REALTIME) Deviation From Immediately Preceding clock-gettime(CLOCK-MONOTONIC)}
\end{figure}

In addition, each clock source provides its own timebase.
\Cref{fig:app:questions:clock-gettime(CLOCK-REALTIME) Deviation From Immediately Preceding clock-gettime(CLOCK-MONOTONIC)}
shows a histogram of the value returned by a call to
\co{clock_gettime(CLOCK_MONOTONIC)} subtracted from that returned by an
immediately following \co{clock_gettime(CLOCK_REALTIME)}
(\path{timeskew.c}).
Because some time passes between these two function calls, it is no
surprise that there are positive deviations, but the negative deviations
should give us some pause.
Nevertheless, such deviations are possible, if for no other reason than
the machinations of network time protocol
(NTP)~\cite{FredericWeisbecker2022nohzfullTSC}.

Worse yet, identical clocksources on different systems
are not necessarily compatible with that of another.
For example, the \co{jiffies} counters on a pair of systems very likely
started counting at different times, and worse yet might well be counting
at different rates.
This brings up the topic of synchronizing a given system's counters
with some real-world notion of time such as the aforementioned NTP,
but that topic is beyond the scope of this book.

In short, time is a slippery topic that causes untold confusion to
parallel programmers and to their code.

% appendix/questions/after.tex
% mainfile: ../../perfbook.tex
% SPDX-License-Identifier: CC-BY-SA-3.0

\section{What Does ``After'' Mean?}
\label{sec:app:questions:What Does ``After'' Mean?}

``After'' is an intuitive, but surprisingly difficult concept.
An important non-intuitive issue is that code can be delayed at
any point for any amount of time.
Consider a producing and a consuming thread that communicate using
a global struct with a timestamp ``t'' and integer fields ``a'', ``b'',
and ``c''.
The producer loops recording the current time
(in seconds since 1970 in decimal),
then updating the values of ``a'', ``b'', and ``c'',
as shown in \cref{lst:app:questions:After Producer Function}.
The consumer code loops, also recording the current time, but also
copying the producer's timestamp along with the fields ``a'',
``b'', and ``c'', as shown in
\cref{lst:app:questions:After Consumer Function}.
At the end of the run, the consumer outputs a list of anomalous recordings,
e.g., where time has appeared to go backwards.

\begin{listing}
\input{CodeSamples/api-pthreads/QAfter/time=producer.fcv}
\caption{``After'' Producer Function}
\label{lst:app:questions:After Producer Function}
\end{listing}

\begin{listing}
\ebresizeverb{.96}{\input{CodeSamples/api-pthreads/QAfter/time=consumer.fcv}}
\caption{``After'' Consumer Function}
\label{lst:app:questions:After Consumer Function}
\end{listing}

\QuickQuiz{
	What SMP coding errors can you see in these examples?
	See \path{time.c} for full code.
}\QuickQuizAnswer{
	Here are errors you might have found:

	\begin{enumerate}
	\item	Missing barrier() or volatile on tight loops.
	\item	Missing memory barriers on update side.
	\item	Lack of synchronization between producer and consumer.
	\end{enumerate}
}\QuickQuizEnd

One might intuitively expect that the difference between the producer
and consumer timestamps would be quite small, as it should not take
much time for the producer to record the timestamps or the values.
An excerpt of some sample output on a dual-core 1\,GHz x86 is shown in
\cref{tab:app:questions:After Program Sample Output}.
Here, the ``seq'' column is the number of times through the loop,
the ``time'' column is the time of the anomaly in seconds, the ``delta''
column is the number of seconds the consumer's timestamp follows that
of the producer (where a negative value indicates that the consumer
has collected its timestamp before the producer did), and the
columns labelled ``a'', ``b'', and ``c'' show the amount that these
variables increased since the prior snapshot collected by the consumer.

\begin{table}
\rowcolors{1}{}{lightgray}
\renewcommand*{\arraystretch}{1.2}
\sisetup{group-digits=false}
\centering
\scriptsize
\begin{tabular}{rS[table-format=7.6]rS[table-format=3.0]S[table-format=3.0]S[table-format=3.0]}
\toprule
seq    & \multicolumn{1}{c}{time (seconds)} & delta~  &  a &  b &  c \\
\midrule
17563: & 1152396.251585 & ($-16.928$) & 27 & 27 & 27 \\
18004: & 1152396.252581 & ($-12.875$) & 24 & 24 & 24 \\
18163: & 1152396.252955 & ($-19.073$) & 18 & 18 & 18 \\
18765: & 1152396.254449 & ($-148.773$) & 216 & 216 & 216 \\
19863: & 1152396.256960 & ($-6.914$) & 18 & 18 & 18 \\
21644: & 1152396.260959 & ($-5.960$) & 18 & 18 & 18 \\
23408: & 1152396.264957 & ($-20.027$) & 15 & 15 & 15 \\
\bottomrule
\end{tabular}
\caption{``After'' Program Sample Output}
\label{tab:app:questions:After Program Sample Output}
\end{table}

Why is time going backwards?
The number in parentheses is the difference in microseconds, with
a large number exceeding 10 microseconds, and one exceeding even
100 microseconds!
Please note that this CPU can potentially execute more than 100,000
instructions in that time.

\begin{figure}
\centering
\includegraphics{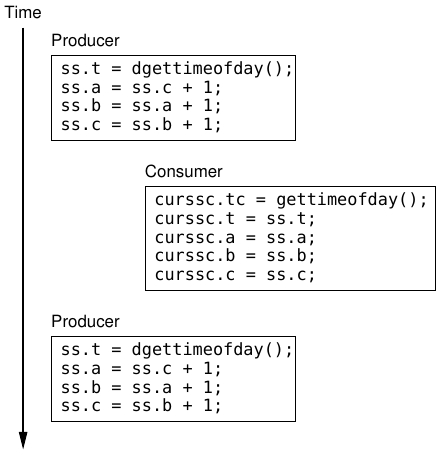}
\caption{Effect of Locking on Snapshot Collection}
\label{fig:app:questions:Effect of Locking on Snapshot Collection}
\end{figure}

\begin{fcvref}[ln:api-pthreads:QAfter:time]
One possible reason is given by the following sequence of events:
\begin{enumerate}
\item	Consumer obtains timestamp
	(\cref{lst:app:questions:After Consumer Function},
	\clnref{consumer:tod}).
\item	Consumer is preempted.
\item	An arbitrary amount of time passes.
\item	Producer obtains timestamp
	(\cref{lst:app:questions:After Producer Function},
	\clnref{producer:tod}).
\item	Consumer starts running again, and picks up the producer's
	timestamp
	(\cref{lst:app:questions:After Consumer Function},
	\clnref{consumer:prodtod}).
\end{enumerate}

In this scenario, the producer's timestamp might be an arbitrary
amount of time after the consumer's timestamp.

How do you avoid agonizing over the meaning of ``after'' in your
SMP code?

Simply use SMP primitives as designed.

In this example, the easiest fix is to use locking, for example,
acquire a lock in the producer before \clnref{producer:tod} in
\cref{lst:app:questions:After Producer Function} and in
the consumer before \clnref{consumer:tod} in
\cref{lst:app:questions:After Consumer Function}.
This lock must also be released after \clnref{producer:upd:c} in
\cref{lst:app:questions:After Producer Function} and
after \clnref{consumer:upd:c} in
\cref{lst:app:questions:After Consumer Function}.
These locks cause the code segments in
\clnrefrange{producer:tod}{producer:upd:c} of
\cref{lst:app:questions:After Producer Function} and in
\clnrefrange{consumer:tod}{consumer:upd:c} of
\cref{lst:app:questions:After Consumer Function} to {\em exclude}
each other, in other words, to run atomically with respect to each other.
This is represented in
\cref{fig:app:questions:Effect of Locking on Snapshot Collection}:
The locking prevents any of the boxes of code from overlapping in time, so
that the consumer's timestamp must be collected after the prior
producer's timestamp.
The segments of code in each box in this figure are termed
``critical sections''; only one such critical section may be executing
at a given time.
\end{fcvref}

This addition of locking results in output as shown in
\cref{fig:app:questions:Locked After Program Sample Output}.
Here there are no instances of time going backwards, instead,
there are only cases with more than 1,000 counts difference between
consecutive reads by the consumer.

\begin{table}
\renewcommand*{\arraystretch}{1.2}
\sisetup{group-digits=false}
\centering
\scriptsize
\begin{tabular}{rS[table-format=7.6]rS[table-format=4.0]S[table-format=4.0]S[table-format=4.0]}
\toprule
seq    & \multicolumn{1}{c}{time (seconds)} & delta~  &  a &  b &  c \\
\midrule
58597:  & 1156521.556296 & ($3.815$) & 1485 & 1485 & 1485 \\
403927: & 1156523.446636 & ($2.146$) & 2583 & 2583 & 2583 \\
\bottomrule
\end{tabular}
\caption{Locked ``After'' Program Sample Output}
\label{fig:app:questions:Locked After Program Sample Output}
\end{table}

\QuickQuiz{
	How could there be such a large gap between successive
	consumer reads?
	See \path{timelocked.c} for full code.
}\QuickQuizAnswer{
	Here are a few reasons for such gaps:

	\begin{enumerate}
	\item	The consumer might be preempted for long time periods.
	\item	A long-running interrupt might delay the consumer.
	\item	Cache misses might delay the consumer.
	\item	The producer might also be running on a faster CPU than is the
		consumer (for example, one of the CPUs might have had to
		decrease its
		clock frequency due to heat-dissipation or power-consumption
		constraints).
	\end{enumerate}
}\QuickQuizEnd

In summary, if you acquire an \IXh{exclusive}{lock}, you {\em know} that
anything you do while holding that lock will appear to happen after
anything done by any prior holder of that lock, at least give or
take \IXacrl{tle}
(see \cref{sec:future:Semantic Differences}).
No need to worry about which CPU did or did not execute a \IX{memory
barrier}, no need to worry about the CPU or compiler reordering
operations---life is simple.
Of course, the fact that this locking prevents these two pieces of
code from running concurrently might limit the program's ability
to gain increased performance on multiprocessors, possibly resulting
in a ``safe but slow'' situation.
\Cref{chp:Partitioning and Synchronization Design} describes ways of
gaining performance and scalability in many situations.

In short, in many parallel programs, the really important definition
of ``after'' is ordering of operations, which is covered in dazzling
detail in
\cref{chp:Advanced Synchronization: Memory Ordering}.

However, in most cases, if you find yourself worrying about what happens
before or after a given piece of code, you should take this as a hint to
make better use of the standard primitives.
Let these primitives do the worrying for you.

% appendix/questions/ordering.tex
% mainfile: ../../perfbook.tex
% SPDX-License-Identifier: CC-BY-SA-3.0

\section{How Much Ordering Is Needed?}
\label{sec:app:questions:How Much Ordering Is Needed?}

Perhaps you have carefully constructed a strongly ordered concurrent
system, only to find that it neither performs nor scales well.
Or perhaps you threw caution to the wind, only to find that your
brilliantly fast and scalable software is also unreliable.
Is there a happy medium with both robust reliability on the one
hand and powerful performance augmented by scintellating scalability on
the other?

The answer, as is so often the case, is ``it depends''.

One approach is to construct a strongly ordered system, then examine
its performance and scalability.
If these suffice, the system is good and sufficient, and no more need
be done.
Otherwise, undertake careful analysis
(see \cref{sec:debugging:Performance Estimation})
and attack each bottleneck until the system's performance is good and
sufficient.

This approach can work very well, especially in contrast to the
all-too-common approach of optimizing random components of the system
in the hope of achieving significant system-wide benefits.
However, starting with strong ordering can also be quite wasteful,
given that weakening ordering of the system's bottleneck can require
that large portions of the rest of the system be redesigned and
rewritten to accommodate the weakening.
Worse yet, eliminating one bottleneck often exposes another, which
in turn needs to be weakened and which in turn can result in wholesale
redesigns and rewrites of other parts of the system.
Perhaps even worse is the approach, also common, of starting with a
fast but unreliable system and then playing whack-a-mole with an endless
succession of concurrency bugs, though in the latter case,
\cref{chp:Validation,chp:Formal Verification}
are always there for you.

It would be better to have design-time tools to determine which portions
of the system could use weak ordering, and at the same time, which
portions actually benefit from weak ordering.
These tasks are taken up by the following sections.

\subsection{Where is the Defining Data?}
\label{sec:app:questions:Where is the Defining Data?}

One way to do this is to keep firmly in mind that the region of
consistency engendered by strong ordering cannot extend out past the
boundaries of the system.\footnote{
	Which might well be a distributed system.}
Portions of the system whose role is to track the state of the outside
world can usually feature weak ordering, given that speed-of-light delays
will force the within-system state to lag that of the outside world.
There is often no point in incurring large overheads to force a consistent
view of data that is inherently out of date.
In these cases, the methods of \cref{chp:Deferred Processing} can be
quite helpful, as can some of the data structures described in
\cref{chp:Data Structures}.

Nevertheless, it is wise to adopt some meaningful semantics that are
visible to those accessing the data, for example, a given function's
return value might be:

\begin{enumerate}
\item	Some value between the conceptual value at the time of the call
	to the function and the conceptual value at the time of the
	return from that function.
	For example, see the statistical counters discussed in
	\cref{sec:count:Statistical Counters}, keeping in mind that such
	counters are normally monotonic, at least between consecutive
	overflows.
\item	The actual value at some time between the call to and the return
	from that function.
	For example, see the single-variable atomic counter shown in
	\cref{lst:count:Just Count Atomically!}.
\item	If the values used by that function remain unchanged during the
	time between that function's call and return, the expected
	value, otherwise some approximation to the expected value.
	Precise specification of the bounds on the approximation can
	be quite challenging.
	For example, consider a function combining values from
	different elements of an RCU-protected linked data structure,
	as described in \cref{sec:datastruct:Read-Mostly Data Structures}.
\end{enumerate}

Weaker ordering usually implies weaker semantics, and you should be
able to give some sort of promise to your users as to how this weakening
affects them.
At the same time, unless the caller holds a lock across both the
function call and the use of any values computed by that function,
even fully ordered implementations normally cannot do any better
than the semantics given by the options above.

\QuickQuiz{
	But if fully ordered implementations cannot offer stronger
	guarantees than the better performing and more scalable weakly
	ordered implementations, why bother with full ordering?
}\QuickQuizAnswer{
	Because strongly ordered implementations are sometimes
	able to provide greater consistency among sets of calls to
	functions accessing a given data structure.
	For example, compare the atomic counter of
	\cref{lst:count:Just Count Atomically!}
	to the statistical counter of
	\cref{sec:count:Statistical Counters}.
	Suppose that one thread is adding the value~3 and another is
	adding the value~5, while two other threads are concurrently
	reading the counter's value.
	With atomic counters, it is not possible for one of the readers
	to obtain the value~3 while the other obtains the value~5.
	With statistical counters, this outcome really can happen.
	In fact, in some computing environments, this outcome can happen
	even on relatively strongly ordered hardware such as x86.

	Therefore, if your user happen to need this admittedly
	unusual level of consistency, you should avoid weakly ordered
	statistical counters.
}\QuickQuizEnd

Some might argue that useful computing deals only with the outside world,
and therefore that all computing can use weak ordering.
Such arguments are incorrect.
For example, the value of your bank account is defined within your
bank's computers, and people often prefer exact computations involving
their account balances, especially those who might suspect that any such
approximations would be in the bank's favor.

In short, although data tracking external state can be an attractive
candidate for weakly ordered access, please think carefully about
exactly what is being tracked and what is doing the tracking.

\subsection{Consistent Data Used Consistently?}
\label{sec:app:questions:Consistent Data Used Consistently?}

Another hint that weakening is safe can appear in the guise of data
that is computed while holding a lock, but then used after the lock
is released.
The computed result clearly becomes at best an approximation as soon as
the lock is released, which suggests computing an approximate result
in the first place, possibly permitting use of weaker ordering.
To this end, \cref{chp:Counting} covers numerous approximate methods
for counting.

Great care is required, however.
Is the use of data following lock release a hint that weak-ordering
optimizations might be helpful?
Or is instead a bug in which the lock was released too soon?

\subsection{Is the Problem Partitionable?}
\label{sec:app:questions:Is the Problem Partitionable?}

Suppose that the system holds the defining instance of the data,
or that using a computed value past lock release proved to be a bug.
What then?

One approach is to partition the system, as discussed in
\cref{chp:Partitioning and Synchronization Design}.
Partititioning can provide excellent scalability and in its more
extreme form, per-CPU performance rivaling that of a sequential program,
as discussed in \cref{chp:Data Ownership}.
Partial partitioning is often mediated by locking, which is the subject of
\cref{chp:Locking}.

\subsection{None of the Above?}
\label{sec:app:questions:None of the Above?}

The previous sections described the easier ways to gain performance
and scalability, sometimes using weaker ordering and sometimes not.
But the plain fact is that multicore systems are under no compunction
to make life easy.
But perhaps the advanced topics covered in
\cref{sec:advsync:Advanced Synchronization,%
chp:Advanced Synchronization: Memory Ordering}
will prove helpful.

But please proceed with care, as it is all too easy to destabilize
your codebase optimizing non-bottlenecks.
Once again, \cref{sec:debugging:Performance Estimation} can help.
It might also be worth your time to review other portions of this
book, as it contains much information on handling a number of tricky
situations.

% appendix/questions/concurrentparallel.tex
% mainfile: ../../perfbook.tex
% SPDX-License-Identifier: CC-BY-SA-3.0

\section{What is the Difference Between ``Concurrent'' and ``Parallel''?}
\label{sec:app:questions:What is the Difference Between ``Concurrent'' and ``Parallel''?}

From a classic computing perspective, ``\IX{concurrent}'' and ``\IX{parallel}''
are clearly synonyms.
However, this has not stopped many people from drawing distinctions
between the two, and it turns out that these distinctions can be
understood from a couple of different perspectives.

The first perspective treats ``parallel'' as an abbreviation for
``data parallel'', and treats ``concurrent'' as pretty much everything
else.
From this perspective, in parallel computing, each partition of the
overall problem can proceed completely independently, with no
communication with other partitions.
In this case, little or no coordination among partitions is required.
In contrast, concurrent computing might well have tight interdependencies,
in the form of contended locks, transactions, or other synchronization
mechanisms.

\QuickQuiz{
	Suppose a portion of a program uses RCU read-side primitives
	as its only synchronization mechanism.
	Is this parallelism or concurrency?
}\QuickQuizAnswer{
	Yes.
}\QuickQuizEnd

This of course begs the question of why such a distinction matters,
which brings us to the second perspective, that of the underlying scheduler.
Schedulers come in a wide range of complexities and capabilities, and
as a rough rule of thumb, the more tightly and irregularly a set of
parallel processes communicate, the higher the level of sophistication
required from the scheduler.
As such, parallel computing's avoidance of interdependencies means that
parallel-computing programs run well on the least-capable schedulers.
In fact, a pure parallel-computing program can run successfully after
being arbitrarily subdivided and interleaved onto a uniprocessor.\footnote{
	Yes, this does mean that data-parallel-computing programs are
	best-suited for sequential execution.
	Why did you ask?}
In contrast, concurrent-computing programs might well require extreme
subtlety on the part of the scheduler.

One could argue that we should simply demand a reasonable level of
competence from the scheduler, so that we could simply ignore any
distinctions between parallelism and concurrency.
Although this is often a good strategy,
there are important situations where \IX{efficiency},
\IX{performance}, and \IX{scalability} concerns sharply limit the level
of competence that the scheduler can reasonably offer.
One important example is when the scheduler is implemented in
hardware, as it often is in SIMD units or GPGPUs.
Another example is a workload where the units of work are quite
short, so that even a software-based scheduler must make hard choices
between subtlety on the one hand and efficiency on the other.

Now, this second perspective can be thought of as making the workload
match the available scheduler, with parallel workloads able to
use simple schedulers and concurrent workloads requiring
sophisticated schedulers.

Unfortunately, this perspective does not always align with the
dependency-based distinction put forth by the first perspective.
For example, a highly interdependent lock-based workload
with one thread per CPU can make do with a trivial scheduler
because no scheduler decisions are required.
In fact, some workloads of this type can even be run one after another
on a sequential machine.
Therefore, such a workload would be labeled ``concurrent'' by the first
perspective and ``parallel'' by many taking the second perspective.

\QuickQuiz{
	In what part of the second (scheduler-based) perspective would
	the lock-based single-thread-per-CPU workload be considered
	``concurrent''?
}\QuickQuizAnswer{
	The people who would like to arbitrarily subdivide and interleave
	the workload.
	Of course, an arbitrary subdivision might end up separating
	a lock acquisition from the corresponding lock release, which
	would prevent any other thread from acquiring that lock.
	If the locks were pure spinlocks, this could even result in
	deadlock.
}\QuickQuizEnd

Which is just fine.
No rule that humankind writes carries any weight against the objective
universe, not even rules dividing multiprocessor programs into categories
such as ``concurrent'' and ``parallel''.

This categorization failure does not mean such rules are useless,
but rather that you should take on a suitably skeptical frame of mind when
attempting to apply them to new situations.
As always, use such rules where they apply and ignore them otherwise.

In fact, it is likely that new categories will arise in addition
to parallel, concurrent, map-reduce, task-based, and so on.
Some will stand the test of time, but good luck guessing which!

% appendix/questions/buggy.tex
% mainfile: ../../perfbook.tex
% SPDX-License-Identifier: CC-BY-SA-3.0

\section{Why Is Software Buggy?}
\label{sec:app:questions:Why Is Software Buggy?}

The short answer is ``because it was written by humans, and to err
is human''.
This does not necessarily mean that automated code generation is
the answer, because the program that does the code generation will
have been written by humans.
In addition, one of the biggest problems in producing software is
working out what that software is supposed to do, and this task
has thus far proven rather resistant to automation.

Nevertheless, automation is an important part of the process of reducing
the number of bugs in software.
For but one example, despite their many flaws, it is almost always better
to use a compiler than to write in assembly language.

Furthermore, careful validation can be very helpful in finding bugs,
as discussed in
\crefrange{chp:Validation}{chp:Formal Verification}.

% @@@ How much documentation is enough?

	\setlength{\parskip}{0.0pt plus 1ex}
	\input{qqzquestions}
	\setlength{\parskip}{0.0pt plus 1.0pt}% return to default

% appendix/toyrcu/toyrcu.tex
% mainfile: ../../perfbook.tex
% SPDX-License-Identifier: CC-BY-SA-3.0

\QuickQuizChapter{chp:app:``Toy'' RCU Implementations}{``Toy'' RCU Implementations}{qqztoyrcu}
\Epigraph{The only difference between men and boys is the price of their toys.}
	 {M. H\'ebert}
% https://www.ncbi.nlm.nih.gov/pubmed/11548147

The toy RCU implementations in this appendix are designed not for
high performance, practicality, or any kind of production use,\footnote{
	However, production-quality user-level RCU implementations
	are available~\cite{MathieuDesnoyers2009URCU,MathieuDesnoyers2012URCU}.}
but rather for clarity.
Nevertheless, you will need a thorough understanding of
\cref{chp:Introduction,%
chp:Hardware and its Habits,%
chp:Tools of the Trade,%
chp:Partitioning and Synchronization Design,%
chp:Deferred Processing}
for even these toy RCU implementations to be easily understandable.

This appendix provides a series of RCU implementations in order of
increasing sophistication, from the viewpoint of solving the
existence-guarantee problem.
\Cref{sec:app:toyrcu:Lock-Based RCU} presents a rudimentary
RCU implementation based on simple locking, while
\crefthro{sec:app:toyrcu:Per-Thread Lock-Based RCU}
{sec:app:toyrcu:RCU Based on Quiescent States}
present a series of
simple RCU implementations based on locking,
\IXalt{reference counters}{reference count},
and free-running counters.
Finally, \cref{sec:app:toyrcu:Summary of Toy RCU Implementations}
provides a summary and a list of desirable RCU properties.

\section{Lock-Based RCU}
\label{sec:app:toyrcu:Lock-Based RCU}

Perhaps the simplest RCU implementation leverages locking, as
shown in
\cref{lst:app:toyrcu:Lock-Based RCU Implementation}
(\path{rcu_lock.h} and \path{rcu_lock.c}).

\begin{listing}
\input{CodeSamples/defer/rcu_lock=lock_unlock.fcv}\vspace*{-11pt}\fvset{firstnumber=last}
\input{CodeSamples/defer/rcu_lock=synchronize.fcv}\fvset{firstnumber=auto}
\caption{Lock-Based RCU Implementation}
\label{lst:app:toyrcu:Lock-Based RCU Implementation}
\end{listing}

In this implementation, \co{rcu_read_lock()} acquires a global
spinlock, \co{rcu_read_unlock()} releases it, and
\co{synchronize_rcu()} acquires it then immediately releases it.

Because \co{synchronize_rcu()} does not return until it has acquired
(and released) the lock, it cannot return until all prior RCU read-side
critical sections have completed, thus faithfully implementing
RCU semantics.
Of course, only one RCU reader may be in its read-side critical section
at a time, which almost entirely defeats the purpose of RCU\@.
In addition, the lock operations in \co{rcu_read_lock()} and
\co{rcu_read_unlock()} are extremely heavyweight,
with read-side overhead ranging from about 100~nanoseconds on a single \Power{5}
CPU up to more than 17~\emph{microseconds} on a 64-CPU system.
Worse yet,
these same lock operations permit \co{rcu_read_lock()}
to participate in \IXpl{deadlock cycle}.
Furthermore, in absence of recursive locks,
RCU read-side critical sections cannot be nested, and, finally,
although concurrent RCU updates could in principle be satisfied by
a common \IX{grace period}, this implementation serializes grace periods,
preventing grace-period sharing.

\QuickQuizSeries{%
\QuickQuizB{
	Why wouldn't any deadlock in the RCU implementation in
	\cref{lst:app:toyrcu:Lock-Based RCU Implementation}
	also be a deadlock in any other RCU implementation?
}\QuickQuizAnswerB{
	\begin{fcvref}[ln:app:toyrcu:Deadlock in Lock-Based RCU Implementation]
	Suppose the functions \co{foo()} and \co{bar()} in
	\cref{lst:app:toyrcu:Deadlock in Lock-Based RCU Implementation}
	are invoked concurrently from different CPUs.
	Then \co{foo()} will acquire \co{my_lock()} on \clnref{foo:acq},
	while \co{bar()} will acquire \co{rcu_gp_lock} on
	\clnref{bar:rrl}.
	\end{fcvref}

\begin{listing}
\begin{fcvlabel}[ln:app:toyrcu:Deadlock in Lock-Based RCU Implementation]
\begin{VerbatimL}[commandchars=\\\[\]]
void foo(void)
{
	spin_lock(&my_lock);		\lnlbl[foo:acq]
	rcu_read_lock();		\lnlbl[foo:rrl]
	do_something();
	rcu_read_unlock();
	do_something_else();
	spin_unlock(&my_lock);
}

void bar(void)
{
	rcu_read_lock();		\lnlbl[bar:rrl]
	spin_lock(&my_lock);		\lnlbl[bar:acq]
	do_some_other_thing();
	spin_unlock(&my_lock);
	do_whatever();
	rcu_read_unlock();
}
\end{VerbatimL}
\end{fcvlabel}
\caption{Deadlock in Lock-Based RCU Implementation}
\label{lst:app:toyrcu:Deadlock in Lock-Based RCU Implementation}
\end{listing}

	\begin{fcvref}[ln:app:toyrcu:Deadlock in Lock-Based RCU Implementation]
	When \co{foo()} advances to \clnref{foo:rrl}, it will attempt to
	acquire \co{rcu_gp_lock}, which is held by \co{bar()}.
	Then when \co{bar()} advances to \clnref{bar:acq}, it will attempt
	to acquire \co{my_lock}, which is held by \co{foo()}.
	\end{fcvref}

	Each function is then waiting for a lock that the other
	holds, a classic deadlock.

	Other RCU implementations neither spin nor block in
	\co{rcu_read_lock()}, hence avoiding deadlocks.
}\QuickQuizEndB
\QuickQuizE{
	Why not simply use reader-writer locks in the RCU implementation
	in
	\cref{lst:app:toyrcu:Lock-Based RCU Implementation}
	in order to allow RCU readers to proceed in parallel?
}\QuickQuizAnswerE{
	One could in fact use reader-writer locks in this manner.
	However, textbook reader-writer locks suffer from memory
	contention, so that the RCU read-side critical sections would
	need to be quite long to actually permit parallel
	execution~\cite{McKenney03a}.

	On the other hand, use of a reader-writer lock that is
	read-acquired in \co{rcu_read_lock()} would avoid the
	deadlock condition noted above.
}\QuickQuizEndE
}

It is hard to imagine this implementation being useful
in a production setting, though it does have the virtue
of being implementable in almost any user-level application.
Furthermore, similar implementations having one lock per CPU
or using reader-writer locks have been used in production
in the 2.4 Linux kernel.

A modified version of this one-lock-per-CPU approach, but instead using
one lock per thread, is described
in the next section.

\section{Per-Thread Lock-Based RCU}
\label{sec:app:toyrcu:Per-Thread Lock-Based RCU}

\Cref{lst:app:toyrcu:Per-Thread Lock-Based RCU Implementation}
(\path{rcu_lock_percpu.h} and \path{rcu_lock_percpu.c})
shows an implementation based on one lock per thread.
The \co{rcu_read_lock()} and \co{rcu_read_unlock()} functions
acquire and release, respectively, the current thread's lock.
The \co{synchronize_rcu()} function acquires and releases each thread's
lock in turn.
Therefore, all RCU read-side critical sections running
when \co{synchronize_rcu()} starts must have completed before
\co{synchronize_rcu()} can return.

\begin{listing}
\input{CodeSamples/defer/rcu_lock_percpu=lock_unlock.fcv}\vspace*{-11pt}\fvset{firstnumber=last}
\input{CodeSamples/defer/rcu_lock_percpu=sync.fcv}\fvset{firstnumber=auto}
\caption{Per-Thread Lock-Based RCU Implementation}
\label{lst:app:toyrcu:Per-Thread Lock-Based RCU Implementation}
\end{listing}

This implementation does have the virtue of permitting concurrent
RCU readers, and does avoid the deadlock condition that can arise
with a single global lock.
Furthermore, the read-side overhead, though high at roughly 140 nanoseconds,
remains at about 140 nanoseconds regardless of the number of CPUs.
However, the update-side overhead ranges from about 600 nanoseconds
on a single \Power{5} CPU
up to more than 100 \emph{microseconds} on 64 CPUs.

\QuickQuizSeries{%
\QuickQuizB{
	\begin{fcvref}[ln:defer:rcu_lock_percpu:sync:loop]
	Wouldn't it be cleaner to acquire all the locks, and then
	release them all in the loop from \clnrefrange{b}{e} of
	\cref{lst:app:toyrcu:Per-Thread Lock-Based RCU Implementation}?
	After all, with this change, there would be a point in time
	when there were no readers, simplifying things greatly.
	\end{fcvref}
}\QuickQuizAnswerB{
	Making this change would re-introduce the deadlock, so
	no, it would not be cleaner.
}\QuickQuizEndB
\QuickQuizM{
	Is the implementation shown in
	\cref{lst:app:toyrcu:Per-Thread Lock-Based RCU Implementation}
	free from deadlocks?
	Why or why not?
}\QuickQuizAnswerM{
	One deadlock is where a lock is
	held across \co{synchronize_rcu()}, and that same lock is
	acquired within an RCU read-side critical section.
	However, this situation could deadlock any correctly designed
	RCU implementation.
	After all, the \co{synchronize_rcu()} primitive must wait for all
	pre-existing RCU read-side critical sections to complete,
	but if one of those critical sections is spinning on a lock
	held by the thread executing the \co{synchronize_rcu()},
	we have a deadlock inherent in the definition of RCU\@.

	Another deadlock happens when attempting to nest RCU read-side
	critical sections.
	This deadlock is peculiar to this implementation, and might
	be avoided by using recursive locks, or by using reader-writer
	locks that are read-acquired by \co{rcu_read_lock()} and
	write-acquired by \co{synchronize_rcu()}.

	However, if we exclude the above two cases,
	this implementation of RCU does not introduce any deadlock
	situations.
	This is because only time some other thread's lock is acquired is when
	executing \co{synchronize_rcu()}, and in that case, the lock
	is immediately released, prohibiting a deadlock cycle that
	does not involve a lock held across the \co{synchronize_rcu()}
	which is the first case above.
}\QuickQuizEndM
\QuickQuizE{
	Isn't one advantage of the RCU algorithm shown in
	\cref{lst:app:toyrcu:Per-Thread Lock-Based RCU Implementation}
	that it uses only primitives that are widely available,
	for example, in POSIX pthreads?
}\QuickQuizAnswerE{
	This is indeed an advantage, but do not forget that
	\co{rcu_dereference()} and \co{rcu_assign_pointer()}
	are still required, which means \co{volatile} manipulation
	for \co{rcu_dereference()} and memory barriers for
	\co{rcu_assign_pointer()}.
	Of course, many Alpha CPUs require memory barriers for both
	primitives.
}\QuickQuizEndE
}

This approach could be useful in some situations, given that a similar
approach was used in the
Linux 2.4 kernel~\cite{Molnar00a}.

The counter-based RCU implementation described next overcomes some of
the shortcomings of the lock-based implementation.

\section{Simple Counter-Based RCU}
\label{sec:app:toyrcu:Simple Counter-Based RCU}

A slightly more sophisticated RCU implementation is shown in
\cref{lst:app:toyrcu:RCU Implementation Using Single Global Reference Counter}
(\path{rcu_rcg.h} and \path{rcu_rcg.c}).
\begin{fcvref}[ln:defer:rcu_rcg]
This implementation makes use of a global reference counter
\co{rcu_refcnt} defined on \clnref{lock_unlock:grc}.
The \co{rcu_read_lock()} primitive atomically increments this
counter, then executes a memory barrier to ensure that the
RCU read-side critical section is ordered after the atomic
increment.
Similarly, \co{rcu_read_unlock()} executes a memory barrier to
confine the RCU read-side critical section, then atomically
decrements the counter.
The \co{synchronize_rcu()} primitive spins waiting for the reference
counter to reach zero, surrounded by \IXpl{memory barrier}.
The \co{poll()} on \clnref{sync:poll} merely provides pure delay, and from
a pure RCU-semantics point of view could be omitted.
Again, once \co{synchronize_rcu()} returns, all prior
RCU read-side critical sections are guaranteed to have completed.
\end{fcvref}

\begin{listing}
\input{CodeSamples/defer/rcu_rcg=lock_unlock.fcv}\vspace*{-11pt}\fvset{firstnumber=last}
\input{CodeSamples/defer/rcu_rcg=sync.fcv}\fvset{firstnumber=auto}
\caption{RCU Implementation Using Single Global Reference Counter}
\label{lst:app:toyrcu:RCU Implementation Using Single Global Reference Counter}
\end{listing}

In happy contrast to the lock-based implementation shown in
\cref{sec:app:toyrcu:Lock-Based RCU}, this implementation
allows parallel execution of RCU read-side critical sections.
In happy contrast to the per-thread lock-based implementation shown in
\cref{sec:app:toyrcu:Per-Thread Lock-Based RCU},
it also allows them to be nested.
In addition, the \co{rcu_read_lock()} primitive cannot possibly
participate in \IXpl{deadlock cycle}, as it never spins nor blocks.

\QuickQuiz{
	But what if you hold a lock across a call to
	\co{synchronize_rcu()}, and then acquire that same lock within
	an RCU read-side critical section?
}\QuickQuizAnswer{
	Indeed, this would deadlock any legal RCU implementation.
	But is \co{rcu_read_lock()} \emph{really} participating in
	the deadlock cycle?
	If you believe that it is, then please
	ask yourself this same question when looking at the
	RCU implementation in
	\cref{sec:app:toyrcu:RCU Based on Quiescent States}.
}\QuickQuizEnd

However, this implementation still has some serious shortcomings.
First, the atomic operations in \co{rcu_read_lock()} and
\co{rcu_read_unlock()} are still quite  heavyweight,
with read-side overhead ranging from about 100~nanoseconds on
a single \Power{5} CPU up to almost 40~\emph{microseconds}
on a 64-CPU system.
This means that the RCU read-side critical sections
have to be extremely long in order to get any real
read-side parallelism.
On the other hand, in the absence of readers, grace periods elapse
in about 40~\emph{nanoseconds}, many orders of magnitude faster
than production-quality implementations in the Linux kernel.

\QuickQuiz{
	How can the grace period possibly elapse in 40 nanoseconds when
	\co{synchronize_rcu()} contains a 10-millisecond delay?
}\QuickQuizAnswer{
	The update-side test was run in absence of readers, so the
	\co{poll()} system call was never invoked.
	In addition, the actual code has this \co{poll()}
	system call commented out, the better to evaluate the
	true overhead of the update-side code.
	Any production uses of this code would be better served by
	using the \co{poll()} system call, but then again,
	production uses would be even better served by other implementations
	shown later in this section.
}\QuickQuizEnd

Second, if there are many concurrent \co{rcu_read_lock()}
and \co{rcu_read_unlock()} operations, there will
be extreme memory contention on \co{rcu_refcnt},
resulting in expensive cache misses.
Both of these first two shortcomings largely defeat a major purpose of
RCU, namely to provide low-overhead read-side synchronization primitives.

Finally, a large number of RCU readers with long read-side
critical sections could prevent \co{synchronize_rcu()}
from ever completing, as the global counter might
never reach zero.
This could result in \IX{starvation} of RCU updates, which
is of course unacceptable in production settings.

\QuickQuiz{
	Why not simply make \co{rcu_read_lock()} wait when a concurrent
	\co{synchronize_rcu()} has been waiting too long in
	the RCU implementation in
	\cref{lst:app:toyrcu:RCU Implementation Using Single Global Reference Counter}?
	Wouldn't that prevent \co{synchronize_rcu()} from starving?
}\QuickQuizAnswer{
	Although this would in fact eliminate the starvation, it would
	also mean that \co{rcu_read_lock()} would spin or block waiting
	for the writer, which is in turn waiting on readers.
	If one of these readers is attempting to acquire a lock that
	the spinning/blocking \co{rcu_read_lock()} holds, we again
	have deadlock.

	In short, the cure is worse than the disease.
	See \cref{sec:app:toyrcu:Starvation-Free Counter-Based RCU}
	for a proper cure.
}\QuickQuizEnd

Therefore, it is still hard to imagine this implementation being useful
in a production setting, though it has a bit more potential
than the lock-based mechanism, for example, as an RCU implementation
suitable for a high-stress debugging environment.
The next section describes a variation on the reference-counting
scheme that is more favorable to writers.

\section{Starvation-Free Counter-Based RCU}
\label{sec:app:toyrcu:Starvation-Free Counter-Based RCU}

\Cref{lst:app:toyrcu:RCU Read-Side Using Global Reference-Count Pair}
(\path{rcu_rcpg.h})
shows the read-side primitives of an RCU implementation that uses a pair
of reference counters (\co{rcu_refcnt[]}),
along with a global index that
selects one counter out of the pair (\co{rcu_idx}),
a per-thread nesting counter \co{rcu_nesting},
a per-thread snapshot of the global index (\co{rcu_read_idx}),
and a global lock (\co{rcu_gp_lock}),
which are themselves shown in
\cref{lst:app:toyrcu:RCU Global Reference-Count Pair Data}.

\begin{listing}
\input{CodeSamples/defer/rcu_rcpg=define.fcv}
\caption{RCU Global Reference-Count Pair Data}
\label{lst:app:toyrcu:RCU Global Reference-Count Pair Data}
\end{listing}

\begin{listing}
\input{CodeSamples/defer/rcu_rcpg=r.fcv}
\caption{RCU Read-Side Using Global Reference-Count Pair}
\label{lst:app:toyrcu:RCU Read-Side Using Global Reference-Count Pair}
\end{listing}

\paragraph{Design}

It is the two-element \co{rcu_refcnt[]} array that provides the freedom
from starvation.
The key point is that \co{synchronize_rcu()} is only required to wait
for pre-existing readers.
If a new reader starts after a given instance of \co{synchronize_rcu()}
has already begun execution, then that instance of \co{synchronize_rcu()}
need not wait on that new reader.
At any given time, when a given reader enters its RCU read-side critical
section via \co{rcu_read_lock()},
it increments the element of the \co{rcu_refcnt[]} array indicated by
the \co{rcu_idx} variable.
When that same reader exits its RCU read-side critical section via
\co{rcu_read_unlock()}, it decrements whichever element it incremented,
ignoring any possible subsequent changes to the \co{rcu_idx} value.

This arrangement means that \co{synchronize_rcu()} can avoid starvation
by complementing the value of \co{rcu_idx}, as in \co{rcu_idx = !rcu_idx}.
Suppose that the old value of \co{rcu_idx} was zero, so that the new
value is one.
New readers that arrive after the complement operation will increment
\co{rcu_refcnt[1]}, while the old readers that previously incremented
\co{rcu_refcnt[0]} will decrement \co{rcu_refcnt[0]} when they exit their
RCU read-side critical sections.
This means that the value of \co{rcu_refcnt[0]} will no longer be incremented,
and thus will be monotonically decreasing.\footnote{
	There is a \IX{race condition} that this ``monotonically decreasing''
	statement ignores.
	This race condition will be dealt with by the code for
	\co{synchronize_rcu()}.
	In the meantime, I suggest suspending disbelief.}
This means that all that \co{synchronize_rcu()} need do is wait for the
value of \co{rcu_refcnt[0]} to reach zero.

With the background, we are ready to look at the implementation of the
actual primitives.

\paragraph{Implementation}

The \co{rcu_read_lock()} primitive atomically increments the member of the
\co{rcu_refcnt[]} pair indexed by \co{rcu_idx}, and keeps a
snapshot of this index in the per-thread variable \co{rcu_read_idx}.
The \co{rcu_read_unlock()} primitive then atomically decrements
whichever counter of the pair that the corresponding \co{rcu_read_lock()}
incremented.
However, because only one value of \co{rcu_idx} is remembered per thread,
additional measures must be taken to permit nesting.
These additional measures use the per-thread \co{rcu_nesting} variable
to track nesting.

\begin{fcvref}[ln:defer:rcu_rcpg:r:lock]
To make all this work, \clnref{pick} of \co{rcu_read_lock()} in
\cref{lst:app:toyrcu:RCU Read-Side Using Global Reference-Count Pair}
picks up the
current thread's instance of \co{rcu_nesting}, and if \clnref{if} finds
that this is the outermost \co{rcu_read_lock()},
then \clnrefrange{cur:b}{cur:e} pick up the current value of
\co{rcu_idx}, save it in this thread's instance of \co{rcu_read_idx},
and atomically increment the selected element of \co{rcu_refcnt}.
Regardless of the value of \co{rcu_nesting}, \clnref{inc} increments it.
\Clnref{mb} executes a memory barrier to ensure that the RCU read-side
critical section does not bleed out before the \co{rcu_read_lock()} code.
\end{fcvref}

\begin{fcvref}[ln:defer:rcu_rcpg:r:unlock]
Similarly, the \co{rcu_read_unlock()} function executes a memory barrier
at \clnref{mb}
to ensure that the RCU read-side critical section does not bleed out
after the \co{rcu_read_unlock()} code.
\Clnref{nest} picks up this thread's instance of \co{rcu_nesting}, and if
\clnref{if} finds that this is the outermost \co{rcu_read_unlock()},
then \clnref{idx,atmdec} pick up this thread's instance of \co{rcu_read_idx}
(saved by the outermost \co{rcu_read_lock()}) and atomically decrements
the selected element of \co{rcu_refcnt}.
Regardless of the nesting level, \clnref{decnest} decrements this thread's
instance of \co{rcu_nesting}.
\end{fcvref}

\begin{listing}
\input{CodeSamples/defer/rcu_rcpg=sync.fcv}
\caption{RCU Update Using Global Reference-Count Pair}
\label{lst:app:toyrcu:RCU Update Using Global Reference-Count Pair}
\end{listing}

\begin{fcvref}[ln:defer:rcu_rcpg:sync]
\Cref{lst:app:toyrcu:RCU Update Using Global Reference-Count Pair}
(\path{rcu_rcpg.c})
shows the corresponding \co{synchronize_rcu()} implementation.
\Clnref{acq,rel} acquire and release \co{rcu_gp_lock} in order to
prevent more than one concurrent instance of \co{synchronize_rcu()}.
\Clnref{pick,compl} pick up the value of \co{rcu_idx} and complement it,
respectively, so that subsequent instances of \co{rcu_read_lock()}
will use a different element of \co{rcu_refcnt} than did preceding
instances.
\Clnrefrange{while:b}{while:e}
then wait for the prior element of \co{rcu_refcnt} to
reach zero, with the memory barrier on \clnref{mb2} ensuring that the check
of \co{rcu_refcnt} is not reordered to precede the complementing of
\co{rcu_idx}.
\Clnrefrange{mb3}{while2:e} repeat this process, and
\clnref{mb5} ensures that any
subsequent reclamation operations are not reordered to precede the
checking of \co{rcu_refcnt}.
\end{fcvref}

\QuickQuizSeries{%
\QuickQuizB{
	\begin{fcvref}[ln:defer:rcu_rcpg:sync]
	Why the memory barrier on \clnref{mb1} of \co{synchronize_rcu()} in
	\cref{lst:app:toyrcu:RCU Update Using Global Reference-Count Pair}
	given that there is a spin-lock acquisition immediately after?
	\end{fcvref}
}\QuickQuizAnswerB{
	The spin-lock acquisition only guarantees that the spin-lock's
	critical section will not ``bleed out'' to precede the
	acquisition.
	It in no way guarantees that code preceding the spin-lock
	acquisition won't be reordered into the critical section.
	Such reordering could cause a removal from an RCU-protected
	list to be reordered to follow the complementing of
	\co{rcu_idx}, which could allow a newly starting RCU
	read-side critical section to see the recently removed
	data element.

	Exercise for the reader:
	Use a tool such as Promela/spin to determine which (if any) of
	the memory barriers in
	\cref{lst:app:toyrcu:RCU Update Using Global Reference-Count Pair}
	are really needed.
	See \cref{chp:Formal Verification}
	for information on using these tools.
	The first correct and complete response will be credited.
}\QuickQuizEndB
\QuickQuizE{
	Why is the counter flipped twice in
	\cref{lst:app:toyrcu:RCU Update Using Global Reference-Count Pair}?
	Shouldn't a single flip-and-wait cycle be sufficient?
}\QuickQuizAnswerE{
	\begin{fcvref}[ln:defer:rcu_rcpg]
	Both flips are absolutely required.
	To see this, consider the following sequence of events:
	\begin{sequence}
	\item	\Clnref{r:lock:cur:b} of \co{rcu_read_lock()} in
		\cref{lst:app:toyrcu:RCU Read-Side Using Global Reference-Count Pair}
		picks up \co{rcu_idx}, finding its value to be zero.
	\item	\Clnref{sync:compl} of \co{synchronize_rcu()} in
		\cref{lst:app:toyrcu:RCU Update Using Global Reference-Count Pair}
		complements the value of \co{rcu_idx}, setting its
		value to one.
	\item	\Clnrefrange{sync:while:b}{sync:while:e}
		of \co{synchronize_rcu()} find that the
		value of \co{rcu_refcnt[0]} is zero, and thus
		returns.
		(Recall that the question is asking what happens if
		\clnrefrange{sync:mb3}{sync:mb5} are omitted.)
	\item	\Clnref{r:lock:set,r:lock:cur:e}
		of \co{rcu_read_lock()} store the
		value zero to this thread's instance of \co{rcu_read_idx}
		and increments \co{rcu_refcnt[0]}, respectively.
		Execution then proceeds into the RCU read-side critical
		section.
		\label{sec:app:toyrcu:rcu_rcgp:RCU Read Side Start}
	\item	Another instance of \co{synchronize_rcu()} again complements
		\co{rcu_idx}, this time setting its value to zero.
		Because \co{rcu_refcnt[1]} is zero, \co{synchronize_rcu()}
		returns immediately.
		(Recall that \co{rcu_read_lock()} incremented
		\co{rcu_refcnt[0]}, not \co{rcu_refcnt[1]}!)
		\label{sec:app:toyrcu:rcu_rcgp:RCU Grace Period Start}
	\item	The grace period that started in
		\cref{sec:app:toyrcu:rcu_rcgp:RCU Grace Period Start}
		has been allowed to end, despite
		the fact that the RCU read-side critical section
		that started beforehand in
		\cref{sec:app:toyrcu:rcu_rcgp:RCU Read Side Start}
		has not completed.
		This violates RCU semantics, and could allow the update
		to free a data element that the RCU read-side critical
		section was still referencing.
	\end{sequence}

	Exercise for the reader:
	What happens if \co{rcu_read_lock()} is preempted for a very long
	time (hours!\@) just after \clnref{r:lock:cur:b}?
	Does this implementation operate correctly in that case?
	Why or why not?
	The first correct and complete response will be credited.
	\end{fcvref}
}\QuickQuizEndE
}

This implementation avoids the update-starvation issues that could
occur in the single-counter implementation shown in
\cref{lst:app:toyrcu:RCU Implementation Using Single Global Reference Counter}.

\paragraph{Discussion}

There are still some serious shortcomings.
First, the atomic operations in \co{rcu_read_lock()}
and \co{rcu_read_unlock()}
are still quite heavyweight.
In fact, they are more complex than those
of the single-counter variant shown in
\cref{lst:app:toyrcu:RCU Implementation Using Single Global Reference Counter},
with the read-side primitives consuming about 150~nanoseconds on a single
\Power{5} CPU and almost 40~\emph{microseconds} on a 64-CPU system.
The update-side \co{synchronize_rcu()} primitive is more costly as
well, ranging from about 200~nanoseconds on a single \Power{5} CPU to
more than 40~\emph{microseconds} on a 64-CPU system.
This means that the RCU read-side critical sections
have to be extremely long in order to get any real
read-side parallelism.

Second, if there are many concurrent \co{rcu_read_lock()}
and \co{rcu_read_unlock()} operations, there will
be extreme memory contention on the \co{rcu_refcnt}
elements, resulting in expensive cache misses.
This further extends the RCU read-side critical-section
duration required to provide parallel read-side access.
These first two shortcomings defeat the purpose of RCU in most
situations.

Third, the need to flip \co{rcu_idx} twice imposes substantial
overhead on updates, especially if there are large
numbers of threads.

Finally, despite the fact that concurrent RCU updates could in principle be
satisfied by a common grace period, this implementation
serializes grace periods, preventing grace-period
sharing.

\QuickQuiz{
	\begin{fcvref}[ln:defer:rcu_rcpg:r]
	Given that atomic increment and decrement are so expensive,
	why not just use non-atomic increment on \clnref{lock:cur:e} and a
	non-atomic decrement on \clnref{unlock:atmdec} of
	\cref{lst:app:toyrcu:RCU Read-Side Using Global Reference-Count Pair}?
	\end{fcvref}
}\QuickQuizAnswer{
	Using non-atomic operations would cause increments and decrements
	to be lost, in turn causing the implementation to fail.
	See \cref{sec:app:toyrcu:Scalable Counter-Based RCU}
	for a safe way to use non-atomic operations in
	\co{rcu_read_lock()} and \co{rcu_read_unlock()}.
}\QuickQuizEnd

Despite these shortcomings, one could imagine this variant
of RCU being used on small tightly coupled multiprocessors,
perhaps as a memory-conserving implementation that maintains
API compatibility with more complex implementations.
However, it would not likely scale well beyond a few CPUs.

The next section describes yet another variation on the reference-counting
scheme that provides greatly improved read-side performance and scalability.

\section{Scalable Counter-Based RCU}
\label{sec:app:toyrcu:Scalable Counter-Based RCU}

\Cref{lst:app:toyrcu:RCU Read-Side Using Per-Thread Reference-Count Pair}
(\path{rcu_rcpl.h})
shows the read-side primitives of an RCU implementation that uses per-thread
pairs of reference counters.
This implementation is quite similar to that shown in
\cref{lst:app:toyrcu:RCU Read-Side Using Global Reference-Count Pair},
the only difference being that \co{rcu_refcnt} is now a per-thread
array (as shown in
\cref{lst:app:toyrcu:RCU Per-Thread Reference-Count Pair Data}).
As with the algorithm in the previous section, use of this two-element
array prevents readers from starving updaters.
One benefit of per-thread \co{rcu_refcnt[]} array is that the
\co{rcu_read_lock()} and \co{rcu_read_unlock()} primitives no longer
perform atomic operations.

\begin{listing}
\input{CodeSamples/defer/rcu_rcpl=define.fcv}
\caption{RCU Per-Thread Reference-Count Pair Data}
\label{lst:app:toyrcu:RCU Per-Thread Reference-Count Pair Data}
\end{listing}

\begin{listing}
\input{CodeSamples/defer/rcu_rcpl=r.fcv}
\caption{RCU Read-Side Using Per-Thread Reference-Count Pair}
\label{lst:app:toyrcu:RCU Read-Side Using Per-Thread Reference-Count Pair}
\end{listing}

\QuickQuiz{
	Come off it!
	We can see the \co{atomic_read()} primitive in
	\co{rcu_read_lock()}!!!
	So why are you trying to pretend that \co{rcu_read_lock()}
	contains no atomic operations???
}\QuickQuizAnswer{
	The \co{atomic_read()} primitives does not actually execute
	atomic machine instructions, but rather does a normal load
	from an \co{atomic_t}.
	Its sole purpose is to keep the compiler's type-checking happy.
	If the Linux kernel ran on 8-bit CPUs, it would also need to
	prevent ``store tearing'', which could happen due to the need
	to store a 16-bit pointer with two eight-bit accesses on some
	8-bit systems.
	But thankfully, it seems that no one runs Linux on 8-bit systems.
}\QuickQuizEnd

\begin{listing}
\input{CodeSamples/defer/rcu_rcpl=u.fcv}
\caption{RCU Update Using Per-Thread Reference-Count Pair}
\label{lst:app:toyrcu:RCU Update Using Per-Thread Reference-Count Pair}
\end{listing}

\Cref{lst:app:toyrcu:RCU Update Using Per-Thread Reference-Count Pair}
(\path{rcu_rcpl.c})
shows the implementation of \co{synchronize_rcu()}, along with a helper
function named \co{flip_counter_and_wait()}.
\begin{fcvref}[ln:defer:rcu_rcpl:u:sync]
The \co{synchronize_rcu()} function resembles that shown in
\cref{lst:app:toyrcu:RCU Update Using Global Reference-Count Pair},
except that the repeated counter flip is replaced by a pair of calls
on \clnref{flip1,flip2} to the new helper function.
\end{fcvref}

\begin{fcvref}[ln:defer:rcu_rcpl:u:flip]
The new \co{flip_counter_and_wait()} function updates the
\co{rcu_idx} variable on \clnref{atmset},
executes a memory barrier on \clnref{mb1},
then \clnrefrange{loop:b}{loop:e}
spin on each thread's prior \co{rcu_refcnt} element,
waiting for it to go to zero.
Once all such elements have gone to zero,
it executes another memory barrier on \clnref{mb2} and returns.
\end{fcvref}

This RCU implementation imposes important new requirements on its
software environment, namely, (1) that it be possible to declare
per-thread variables, (2) that these per-thread variables be accessible
from other threads, and (3) that it is possible to enumerate all threads.
These requirements can be met in almost all software environments,
but often result in fixed upper bounds on the number of threads.
More-complex implementations might avoid such bounds, for example, by using
expandable hash tables.
Such implementations might dynamically track threads, for example, by
adding them on their first call to \co{rcu_read_lock()}.

\QuickQuiz{
	Great, if we have $N$ threads, we can have $2N$ ten-millisecond
	waits (one set per \co{flip_counter_and_wait()} invocation,
	and even that assumes that we wait only once for each thread).
	Don't we need the grace period to complete \emph{much} more quickly?
}\QuickQuizAnswer{
	Keep in mind that we only wait for a given thread if that thread
	is still in a pre-existing RCU read-side critical section,
	and that waiting for one hold-out thread gives all the other
	threads a chance to complete any pre-existing RCU read-side
	critical sections that they might still be executing.
	So the only way that we would wait for $2N$ intervals
	would be if the last thread still remained in a pre-existing
	RCU read-side critical section despite all the waiting for
	all the prior threads.
	In short, this implementation will not wait unnecessarily.

	However, if you are stress-testing code that uses RCU, you
	might want to comment out the \co{poll()} statement in
	order to better catch bugs that incorrectly retain a reference
	to an RCU-protected data element outside of an RCU
	read-side critical section.
}\QuickQuizEnd

This implementation still has several shortcomings.
First, the need to flip \co{rcu_idx} twice imposes substantial overhead
on updates, especially if there are large numbers of threads.

Second, \co{synchronize_rcu()} must now examine a number of variables
that increases linearly with the number of threads, imposing substantial
overhead on applications with large numbers of threads.

Third, as before, although concurrent RCU updates could in principle
be satisfied by a common grace period, this implementation serializes
grace periods, preventing grace-period sharing.

Finally, as noted in the text, the need for per-thread variables
and for enumerating threads may be problematic in some software
environments.

That said, the read-side primitives scale very nicely, requiring about
115~nanoseconds regardless of whether running on a single-CPU or a 64-CPU
\Power{5} system.
As noted above, the \co{synchronize_rcu()} primitive does not scale,
ranging in overhead from almost a microsecond on a single \Power{5} CPU
up to almost 200~microseconds on a 64-CPU system.
This implementation could conceivably form the basis for a
production-quality user-level RCU implementation.

The next section describes an algorithm permitting more efficient
concurrent RCU updates.

\section{Scalable Counter-Based RCU With Shared Grace Periods}
\label{sec:app:toyrcu:Scalable Counter-Based RCU With Shared Grace Periods}

\Cref{lst:app:toyrcu:RCU Read-Side Using Per-Thread Reference-Count Pair and Shared Update}
(\path{rcu_rcpls.h})
shows the read-side primitives for an RCU implementation using per-thread
reference count pairs, as before, but permitting updates to share
grace periods.
\begin{fcvref}[ln:defer:rcu_rcpls:r]
The main difference from the earlier implementation shown in
\cref{lst:app:toyrcu:RCU Read-Side Using Per-Thread Reference-Count Pair}
is that \co{rcu_idx} is now a \co{long} that counts freely,
so that \clnref{lock:idx} of
\cref{lst:app:toyrcu:RCU Read-Side Using Per-Thread Reference-Count Pair and Shared Update}
must mask off the low-order bit.
We also switched from using \co{atomic_read()} and \co{atomic_set()}
to using \co{READ_ONCE()}.
The data is also quite similar, as shown in
\cref{lst:app:toyrcu:RCU Read-Side Using Per-Thread Reference-Count Pair and Shared Update Data},
with \co{rcu_idx} now being a \co{long} instead of an
\co{atomic_t}.
\end{fcvref}

\begin{listing}
\input{CodeSamples/defer/rcu_rcpls=define.fcv}
\caption{RCU Read-Side Using Per-Thread Reference-Count Pair and Shared Update Data}
\label{lst:app:toyrcu:RCU Read-Side Using Per-Thread Reference-Count Pair and Shared Update Data}
\end{listing}

\begin{listing}
\input{CodeSamples/defer/rcu_rcpls=r.fcv}
\caption{RCU Read-Side Using Per-Thread Reference-Count Pair and Shared Update}
\label{lst:app:toyrcu:RCU Read-Side Using Per-Thread Reference-Count Pair and Shared Update}
\end{listing}

\Cref{lst:app:toyrcu:RCU Shared Update Using Per-Thread Reference-Count Pair}
(\path{rcu_rcpls.c})
shows the implementation of \co{synchronize_rcu()} and its helper
function \co{flip_counter_and_wait()}.
These are similar to those in
\cref{lst:app:toyrcu:RCU Update Using Per-Thread Reference-Count Pair}.
The differences in \co{flip_counter_and_wait()} include:
\begin{fcvref}[ln:defer:rcu_rcpls:u:flip]
\begin{enumerate}
\item	\Clnref{inc} uses \co{WRITE_ONCE()} instead of \co{atomic_set()},
	and increments rather than complementing.
\item	A new \clnref{mask} masks the counter down to its bottom bit.
\end{enumerate}
\end{fcvref}

\begin{listing}
\input{CodeSamples/defer/rcu_rcpls=u.fcv}
\caption{RCU Shared Update Using Per-Thread Reference-Count Pair}
\label{lst:app:toyrcu:RCU Shared Update Using Per-Thread Reference-Count Pair}
\end{listing}

\begin{fcvref}[ln:defer:rcu_rcpls:u:sync]
The changes to \co{synchronize_rcu()} are more pervasive:
\begin{enumerate}
\item	There is a new \co{oldctr} local variable that captures
	the pre-lock-acquisition value of \co{rcu_idx} on
	\clnref{oldctr}.
\item	\Clnref{idx} uses \co{READ_ONCE()} instead of \co{atomic_read()}.
\item	\Clnrefrange{if:b}{ret} check to see if at least three counter flips were
	performed by other threads while the lock was being acquired,
	and, if so, releases the lock, does a memory barrier, and returns.
	In this case, there were two full waits for the counters to
	go to zero, so those other threads already did all the required work.
\item	At \clnrefrange{ifpair}{flip2}, \co{flip_counter_and_wait()} is only
	invoked a second time if there were fewer than two counter flips
	while the lock was being acquired.
	On the other hand, if there were two counter flips, some other
	thread did one full wait for all the counters to go to zero,
	so only one more is required.
\end{enumerate}
\end{fcvref}

With this approach, if an arbitrarily large number of threads invoke
\co{synchronize_rcu()} concurrently, with one CPU for each thread, there
will be a total of only three waits for counters to go to zero.

Despite the improvements, this implementation of RCU still
has a few shortcomings.
First, as before, the need to flip \co{rcu_idx} twice imposes substantial
overhead on updates, especially if there are large
numbers of threads.

Second, each updater still acquires \co{rcu_gp_lock}, even if there
is no work to be done.
This can result in a severe scalability limitation
if there are large numbers of concurrent updates.
There are ways of avoiding this, as was done in a
production-quality real-time implementation of RCU for the Linux
kernel~\cite{PaulEMcKenney2007PreemptibleRCU}.

Third, this implementation requires per-thread variables
and the ability to enumerate threads, which again can be
problematic in some software environments.

Finally, on 32-bit machines, a given update thread might be
preempted long enough for the \co{rcu_idx}
counter to overflow.
This could cause such a thread to force an unnecessary
pair of counter flips.
However, even if each grace period took only one
microsecond, the offending thread would need to be
preempted for more than an hour, in which case an
extra pair of counter flips is likely the least of
your worries.

As with the implementation described in
\cref{sec:app:toyrcu:Simple Counter-Based RCU},
the read-side primitives scale extremely well, incurring roughly
115~nanoseconds of overhead regardless of the number of CPUs.
The \co{synchronize_rcu()} primitive is still expensive,
ranging from about one microsecond up to about 16~microseconds.
This is nevertheless much cheaper than the roughly 200~microseconds
incurred by the implementation in
\cref{sec:app:toyrcu:Scalable Counter-Based RCU}.
So, despite its shortcomings, one could imagine this
RCU implementation being used in production in real-life applications.

\QuickQuiz{
	All of these toy RCU implementations have either atomic operations
	in \co{rcu_read_lock()} and \co{rcu_read_unlock()},
	or \co{synchronize_rcu()}
	overhead that increases linearly with the number of threads.
	Under what circumstances could an RCU implementation enjoy
	lightweight implementations for all three of these primitives,
	all having deterministic ($\O{1}$) overheads and latencies?
}\QuickQuizAnswer{
	Special-purpose uniprocessor implementations of RCU can attain
	this ideal~\cite{PaulEMcKenney2009BloatwatchRCU}.
}\QuickQuizEnd

Referring back to
\cref{lst:app:toyrcu:RCU Read-Side Using Per-Thread Reference-Count Pair and Shared Update},
we see that there is one global-variable access and no fewer than four
accesses to thread-local variables.
Given the relatively high cost of thread-local accesses on systems
implementing POSIX threads, it is tempting to collapse the three
thread-local variables into a single structure, permitting
\co{rcu_read_lock()} and \co{rcu_read_unlock()} to access their
thread-local data with a single thread-local-storage access.
However, an even better approach would be to reduce the number of
thread-local accesses to one, as is done in the next section.

\section{RCU Based on Free-Running Counter}
\label{sec:app:toyrcu:RCU Based on Free-Running Counter}

\Cref{lst:app:toyrcu:Free-Running Counter Using RCU}
(\path{rcu.h} and \path{rcu.c})
shows an RCU implementation based on a single global free-running counter
that takes on only even-numbered values, with data shown in
\cref{lst:app:toyrcu:Data for Free-Running Counter Using RCU}.

\begin{listing}
\input{CodeSamples/defer/rcu=define.fcv}
\caption{Data for Free-Running Counter Using RCU}
\label{lst:app:toyrcu:Data for Free-Running Counter Using RCU}
\end{listing}

\begin{listing}
\input{CodeSamples/defer/rcu=read_lock_unlock.fcv}\vspace*{-11pt}\fvset{firstnumber=last}
\input{CodeSamples/defer/rcu=synchronize.fcv}\fvset{firstnumber=auto}
\caption{Free-Running Counter Using RCU}
\label{lst:app:toyrcu:Free-Running Counter Using RCU}
\end{listing}

The resulting \co{rcu_read_lock()} implementation is extremely
straightforward.
\begin{fcvref}[ln:defer:rcu:read_lock_unlock:lock]
\Clnref{gp1,gp2} simply
add the value one to the global free-running \co{rcu_gp_ctr}
variable and stores the resulting odd-numbered value into the
\co{rcu_reader_gp} per-thread variable.
\Clnref{mb} executes a \IX{memory barrier} to prevent the content of the
subsequent RCU read-side critical section from ``leaking out''.
\end{fcvref}

\begin{fcvref}[ln:defer:rcu:read_lock_unlock:unlock]
The \co{rcu_read_unlock()} implementation is similar.
\Clnref{mb} executes a memory barrier, again to prevent the prior RCU
read-side critical section from ``leaking out''.
\Clnref{gp1,gp2} then copy the \co{rcu_gp_ctr} global variable to the
\co{rcu_reader_gp} per-thread variable, leaving this per-thread
variable with an even-numbered value so that a concurrent instance
of \co{synchronize_rcu()} will know to ignore it.
\end{fcvref}

\QuickQuiz{
	\begin{fcvref}[ln:defer:rcu:read_lock_unlock:unlock]
	If any even value is sufficient to tell \co{synchronize_rcu()}
	to ignore a given task, why don't \clnref{gp1,gp2} of
	\cref{lst:app:toyrcu:Free-Running Counter Using RCU}
	simply assign zero to \co{rcu_reader_gp}?
	\end{fcvref}
}\QuickQuizAnswer{
	Assigning zero (or any other even-numbered constant)
	would in fact work, but assigning the value of
	\co{rcu_gp_ctr} can provide a valuable debugging aid,
	as it gives the developer an idea of when the corresponding
	thread last exited an RCU read-side critical section.
}\QuickQuizEnd

\begin{fcvref}[ln:defer:rcu:synchronize:syn]
Thus, \co{synchronize_rcu()} could wait for all of the per-thread
\co{rcu_reader_gp} variables to take on even-numbered values.
However, it is possible to do much better than that because
\co{synchronize_rcu()} need only wait on \emph{pre-existing}
RCU read-side critical sections.
\Clnref{mb1} executes a memory barrier to prevent prior manipulations
of RCU-protected data structures from being reordered (by either
the CPU or the compiler) to follow the increment on
\clnref{increasegp}.
\Clnref{spinlock} acquires the \co{rcu_gp_lock}
(and \clnref{spinunlock} releases it)
in order to prevent multiple
\co{synchronize_rcu()} instances from running concurrently.
\Clnref{increasegp} then increments the global \co{rcu_gp_ctr} variable by
two, so that all pre-existing RCU read-side critical sections will
have corresponding per-thread \co{rcu_reader_gp} variables with
values less than that of \co{rcu_gp_ctr}, modulo the machine's
word size.
Recall also that threads with even-numbered values of \co{rcu_reader_gp}
are not in an RCU read-side critical section,
so that \clnrefrange{scan:b}{scan:e}
scan the \co{rcu_reader_gp} values until they
all are either even (\clnref{even}) or are greater than the global
\co{rcu_gp_ctr} (\clnrefrange{gt1}{gt2}).
\Clnref{poll} blocks for a short period of time to wait for a
pre-existing RCU read-side critical section, but this can be replaced with
a spin-loop if \IXh{grace-period}{latency} is of the essence.
Finally, the memory barrier at \clnref{mb3} ensures that any subsequent
destruction will not be reordered into the preceding loop.
\end{fcvref}

\QuickQuiz{
	\begin{fcvref}[ln:defer:rcu:synchronize:syn]
	Why are the memory barriers on \clnref{mb1,mb3} of
	\cref{lst:app:toyrcu:Free-Running Counter Using RCU}
	needed?
	Aren't the memory barriers inherent in the locking
	primitives on \clnref{spinlock,spinunlock} sufficient?
	\end{fcvref}
}\QuickQuizAnswer{
	These memory barriers are required because the locking
	primitives are only guaranteed to confine the critical
	section.
	The locking primitives are under absolutely no obligation
	to keep other code from bleeding in to the critical section.
	The pair of memory barriers are therefore requires to prevent
	this sort of code motion, whether performed by the compiler
	or by the CPU\@.
}\QuickQuizEnd

This approach achieves much better read-side performance, incurring
roughly 63~nanoseconds of overhead regardless of the number of
\Power{5} CPUs.
Updates incur more overhead, ranging from about 500~nanoseconds on
a single \Power{5} CPU to more than 100~\emph{microseconds} on 64
such CPUs.

\QuickQuiz{
	Couldn't the update-side batching optimization described in
	\cref{sec:app:toyrcu:Scalable Counter-Based RCU With Shared Grace Periods}
	be applied to the implementation shown in
	\cref{lst:app:toyrcu:Free-Running Counter Using RCU}?
}\QuickQuizAnswer{
	Indeed it could, with a few modifications.
	This work is left as an exercise for the reader.
}\QuickQuizEnd

\begin{fcvref}[ln:defer:rcu:read_lock_unlock:lock]
This implementation suffers from some serious shortcomings in
addition to the high update-side overhead noted earlier.
First, it is no longer permissible to nest RCU read-side critical
sections, a topic that is taken up in the next section.
Second, if a reader is preempted at \clnref{gp1} of
\cref{lst:app:toyrcu:Free-Running Counter Using RCU} after fetching from
\co{rcu_gp_ctr} but before storing to \co{rcu_reader_gp},
and if the \co{rcu_gp_ctr} counter then runs through more than half
but less than all of its possible values, then \co{synchronize_rcu()}
will ignore the subsequent RCU read-side critical section.
Third and finally, this implementation requires that the enclosing software
environment be able to enumerate threads and maintain per-thread
variables.
\end{fcvref}

\QuickQuiz{
	\begin{fcvref}[ln:defer:rcu:read_lock_unlock:lock]
	Is the possibility of readers being preempted in
	\clnrefrange{gp1}{gp2} of
	\cref{lst:app:toyrcu:Free-Running Counter Using RCU}
	a real problem, in other words, is there a real sequence
	of events that could lead to failure?
	If not, why not?
	If so, what is the sequence of events, and how can the
	failure be addressed?
	\end{fcvref}
}\QuickQuizAnswer{
	It is a real problem, there is a sequence of events leading to
	failure, and there are a number of possible ways of
	addressing it.
	For more details, see the Quick Quizzes near the end of
	\cref{sec:app:toyrcu:Nestable RCU Based on Free-Running Counter}.
	The reason for locating the discussion there is to (1) give you
	more time to think about it, and (2) because the nesting support
	added in that section greatly reduces the time required to
	overflow the counter.
}\QuickQuizEnd

\section{Nestable RCU Based on Free-Running Counter}
\label{sec:app:toyrcu:Nestable RCU Based on Free-Running Counter}

\Cref{lst:app:toyrcu:Nestable RCU Using a Free-Running Counter}
(\path{rcu_nest.h} and \path{rcu_nest.c})
shows an RCU implementation based on a single global free-running counter,
but that permits nesting of RCU read-side critical sections.
This nestability is accomplished by reserving the low-order bits of the
global \co{rcu_gp_ctr} to count nesting, using the definitions shown in
\cref{lst:app:toyrcu:Data for Nestable RCU Using a Free-Running Counter}.
This is a generalization of the scheme in
\cref{sec:app:toyrcu:RCU Based on Free-Running Counter},
which can be thought of as having a single low-order bit reserved
for counting nesting depth.
Two C-preprocessor macros are used to arrange this,
\co{RCU_GP_CTR_NEST_MASK} and
\co{RCU_GP_CTR_BOTTOM_BIT}.
These are related: \co{RCU_GP_CTR_NEST_MASK=RCU_GP_CTR_BOTTOM_BIT-1}.
The \co{RCU_GP_CTR_BOTTOM_BIT} macro contains a single bit that is
positioned just above the bits reserved for counting nesting,
and the \co{RCU_GP_CTR_NEST_MASK} has all one bits covering the
region of \co{rcu_gp_ctr} used to count nesting.
Obviously, these two C-preprocessor macros must reserve enough
of the low-order bits of the counter to permit the maximum required
nesting of RCU read-side critical sections, and this implementation
reserves seven bits, for a maximum RCU read-side critical-section
nesting depth of 127, which should be well in excess of that needed
by most applications.

\begin{listing}
\input{CodeSamples/defer/rcu_nest=define.fcv}
\caption{Data for Nestable RCU Using a Free-Running Counter}
\label{lst:app:toyrcu:Data for Nestable RCU Using a Free-Running Counter}
\end{listing}

\begin{listing}
\input{CodeSamples/defer/rcu_nest=read_lock_unlock.fcv}\vspace*{-11pt}\fvset{firstnumber=last}
\input{CodeSamples/defer/rcu_nest=synchronize.fcv}\fvset{firstnumber=auto}
\caption{Nestable RCU Using a Free-Running Counter}
\label{lst:app:toyrcu:Nestable RCU Using a Free-Running Counter}
\end{listing}

\begin{fcvref}[ln:defer:rcu_nest:read_lock_unlock:lock]
The resulting \co{rcu_read_lock()} implementation is still reasonably
straightforward.
\Clnref{readgp} places a pointer to
this thread's instance of \co{rcu_reader_gp}
into the local variable \co{rrgp}, minimizing the number of expensive
calls to the pthreads thread-local-state API\@.
\Clnref{wtmp1} records the current value of \co{rcu_reader_gp}
into another local variable \co{tmp}, and \clnref{checktmp} checks
to see if the low-order bits are zero, which would indicate that
this is the outermost \co{rcu_read_lock()}.
If so, \clnref{wtmp2} places the global \co{rcu_gp_ctr}
into \co{tmp} because the current value previously fetched by
\clnref{wtmp1} is likely to be obsolete.
In either case, \clnref{inctmp} increments the nesting depth,
which you will recall is stored in the seven low-order bits of the counter.
\Clnref{writegp} stores the updated counter back into this thread's
instance of \co{rcu_reader_gp}, and,
finally, \clnref{mb1} executes a memory barrier
to prevent the RCU read-side critical section from bleeding out
into the code preceding the call to \co{rcu_read_lock()}.
\end{fcvref}

In other words, this implementation of \co{rcu_read_lock()} picks up a copy
of the global \co{rcu_gp_ctr} unless the current invocation of
\co{rcu_read_lock()} is nested within an RCU read-side critical section,
in which case it instead fetches the contents of the current thread's
instance of \co{rcu_reader_gp}.
Either way, it increments whatever value it fetched in order to record
an additional nesting level, and stores the result in the current
thread's instance of \co{rcu_reader_gp}.

\begin{fcvref}[ln:defer:rcu_nest:read_lock_unlock:unlock]
Interestingly enough, despite their \co{rcu_read_lock()} differences,
the implementation of \co{rcu_read_unlock()}
is broadly similar to that shown in
\cref{sec:app:toyrcu:RCU Based on Free-Running Counter}.
\Clnref{mb1} executes a memory barrier
in order to prevent the RCU read-side
critical section from bleeding out into code following the call
to \co{rcu_read_unlock()}, and
\clnref{decgp} decrements this thread's instance of \co{rcu_reader_gp},
which has the effect of decrementing the nesting count contained in
\co{rcu_reader_gp}'s low-order bits.
Debugging versions of this primitive would check (before decrementing!\@)
that these low-order bits were non-zero.
\end{fcvref}

\begin{fcvref}[ln:defer:rcu_nest:synchronize:syn]
The implementation of \co{synchronize_rcu()} is quite similar to
that shown in
\cref{sec:app:toyrcu:RCU Based on Free-Running Counter}.
There are two differences.
The first is that \clnref{incgp1,incgp2}
adds \co{RCU_GP_CTR_BOTTOM_BIT} to the global \co{rcu_gp_ctr}
instead of adding the constant ``2'',
and the second is that the comparison on \clnref{ongoing}
has been abstracted out to a separate function,
where it checks the bits indicated by \co{RCU_GP_CTR_NEST_MASK}
instead of unconditionally checking the low-order bit.
\end{fcvref}

This approach achieves read-side performance almost equal to that
shown in
\cref{sec:app:toyrcu:RCU Based on Free-Running Counter}, incurring
roughly 65~nanoseconds of overhead regardless of the number of
\Power{5} CPUs.
Updates again incur more overhead, ranging from about 600~nanoseconds on
a single \Power{5} CPU to more than 100~\emph{microseconds} on 64
such CPUs.

\QuickQuiz{
	Why not simply maintain a separate per-thread nesting-level
	variable, as was done in previous section, rather than having
	all this complicated bit manipulation?
}\QuickQuizAnswer{
	The apparent simplicity of the separate per-thread variable
	is a red herring.
	This approach incurs much greater complexity in the guise
	of careful ordering of operations, especially if signal
	handlers are to be permitted to contain RCU read-side
	critical sections.
	But don't take my word for it, code it up and see what you
	end up with!
}\QuickQuizEnd

This implementation suffers from the same shortcomings as does that of
\cref{sec:app:toyrcu:RCU Based on Free-Running Counter}, except that
nesting of RCU read-side critical sections is now permitted.
In addition, on 32-bit systems, this approach shortens the time
required to overflow the global \co{rcu_gp_ctr} variable.
The following section shows one way to greatly increase the time
required for overflow to occur, while greatly reducing read-side
overhead.

\QuickQuizSeries{%
\QuickQuizB{
	Given the algorithm shown in
	\cref{lst:app:toyrcu:Nestable RCU Using a Free-Running Counter},
	how could you double the time required to overflow the global
	\co{rcu_gp_ctr}?
}\QuickQuizAnswerB{
	\begin{fcvref}[ln:defer:rcu_nest:synchronize:syn]
	One way would be to replace the magnitude comparison on
	\clnref{lt1,lt2} with an inequality check of
	the per-thread \co{rcu_reader_gp} variable against
	\co{rcu_gp_ctr+RCU_GP_CTR_BOTTOM_BIT}.
	\end{fcvref}
}\QuickQuizEndB
\QuickQuizE{
	Again, given the algorithm shown in
	\cref{lst:app:toyrcu:Nestable RCU Using a Free-Running Counter},
	is counter overflow fatal?
	Why or why not?
	If it is fatal, what can be done to fix it?
}\QuickQuizAnswerE{
	It can indeed be fatal.
	To see this, consider the following sequence of events:
	\begin{enumerate}
	\item	Thread~0 enters \co{rcu_read_lock()}, determines
		that it is not nested, and therefore fetches the
		value of the global \co{rcu_gp_ctr}.
		Thread~0 is then preempted for an extremely long time
		(before storing to its per-thread \co{rcu_reader_gp}
		variable).
	\item	Other threads repeatedly invoke \co{synchronize_rcu()},
		so that the new value of the global \co{rcu_gp_ctr}
		is now \co{RCU_GP_CTR_BOTTOM_BIT}
		less than it was when thread~0 fetched it.
	\item	Thread~0 now starts running again, and stores into
		its per-thread \co{rcu_reader_gp} variable.
		The value it stores is
		\co{RCU_GP_CTR_BOTTOM_BIT+1}
		greater than that of the global \co{rcu_gp_ctr}.
	\item	Thread~0 acquires a reference to RCU-protected data
		element~A.
	\item	Thread~1 now removes the data element~A that thread~0
		just acquired a reference to.
	\item	Thread~1 invokes \co{synchronize_rcu()}, which
		increments the global \co{rcu_gp_ctr} by
		\co{RCU_GP_CTR_BOTTOM_BIT}.
		It then checks all of the per-thread \co{rcu_reader_gp}
		variables, but thread~0's value (incorrectly) indicates
		that it started after thread~1's call to
		\co{synchronize_rcu()}, so thread~1 does not wait
		for thread~0 to complete its RCU read-side critical
		section.
	\item	Thread~1 then frees up data element~A, which thread~0
		is still referencing.
	\end{enumerate}

	Note that scenario can also occur in the implementation presented in
	\cref{sec:app:toyrcu:RCU Based on Free-Running Counter}.

	One strategy for fixing this problem is to use 64-bit
	counters so that the time required to overflow them would exceed
	the useful lifetime of the computer system.
	Note that non-antique members of the 32-bit x86 CPU family
	allow atomic manipulation of 64-bit counters via the
	\co{cmpxchg64b} instruction.

	Another strategy is to limit the rate at which grace periods are
	permitted to occur in order to achieve a similar effect.
	For example, \co{synchronize_rcu()} could record the last time
	that it was invoked, and any subsequent invocation would then
	check this time and block as needed to force the desired
	spacing.
	For example, if the low-order four bits of the counter were
	reserved for nesting, and if grace periods were permitted to
	occur at most ten times per second, then it would take more
	than 300 days for the counter to overflow.
	However, this approach is not helpful if there is any possibility
	that the system will be fully loaded with CPU-bound high-priority
	real-time threads for the full 300 days.
	(A remote possibility, perhaps, but best to consider it ahead
	of time.)

	A third approach is to administratively abolish real-time threads
	from the system in question.
	In this case, the preempted process will age up in priority,
	thus getting to run long before the counter had a chance to
	overflow.
	Of course, this approach is less than helpful for real-time
	applications.

	A fourth approach would be for \co{rcu_read_lock()} to recheck
	the value of the global \co{rcu_gp_ctr} after storing to its
	per-thread \co{rcu_reader_gp} counter, retrying if the new
	value of the global \co{rcu_gp_ctr} is inappropriate.
	This works, but introduces non-deterministic execution time
	into \co{rcu_read_lock()}.
	On the other hand, if your application is being preempted long
	enough for the counter to overflow, you have no hope of
	deterministic execution time in any case!

	A fifth approach is for the grace period process to wait for
	all readers to become aware of the new grace period.
	This works nicely in theory, but hangs if a reader blocks
	indefinitely outside of an RCU read-side critical section.

	A final approach is, oddly enough, to use a single-bit
	grace-period counter and for each call to \co{synchronize_rcu()}
	to take two passes through its algorithm.
	This is the approached use by userspace
	RCU~\cite{MathieuDesnoyers2009URCU}, and is described in
	detail in the journal article and supplementary
	materials~\cite[Appendix D]{MathieuDesnoyers2012URCU}.
}\QuickQuizEndE
}

\section{RCU Based on Quiescent States}
\label{sec:app:toyrcu:RCU Based on Quiescent States}

\begin{fcvref}[ln:defer:rcu_qs:read_lock_unlock]
\Cref{lst:app:toyrcu:Quiescent-State-Based RCU Read Side}
(\path{rcu_qs.h})
shows the read-side primitives used to construct a user-level
implementation of RCU based on \IXpl{quiescent state}, with the data shown in
\cref{lst:app:toyrcu:Data for Quiescent-State-Based RCU}.
As can be seen from \clnrefrange{lock:b}{unlock:e} in the listing,
the \co{rcu_read_lock()}
and \co{rcu_read_unlock()} primitives do nothing, and can in fact
be expected to be inlined and optimized away, as they are in
server builds of the Linux kernel.
This is due to the fact that quiescent-state-based RCU implementations
\emph{approximate} the extents of RCU read-side critical sections
using the aforementioned quiescent states.
Each of these quiescent states contains a call to
\co{rcu_quiescent_state()}, which is shown from
\clnrefrange{qs:b}{qs:e} in the listing.
Threads entering extended quiescent states (for example, when blocking)
may instead call \co{rcu_thread_offline()}
(\clnrefrange{offline:b}{offline:e}) when entering
an extended quiescent state and then call
\co{rcu_thread_online()}
(\clnrefrange{online:b}{online:e}) when leaving it.
As such, \co{rcu_thread_online()} is analogous to \co{rcu_read_lock()}
and \co{rcu_thread_offline()} is analogous to \co{rcu_read_unlock()}.
In addition, \co{rcu_quiescent_state()} can be thought of as a
\co{rcu_thread_online()} immediately followed by a
\co{rcu_thread_offline()}.\footnote{
	Although the code in the listing is consistent with
	\co{rcu_quiescent_state()}
	being the same as \co{rcu_thread_online()} immediately followed by
	\co{rcu_thread_offline()}, this relationship is obscured by
	performance optimizations.}
It is illegal to invoke \co{rcu_quiescent_state()}, \co{rcu_thread_offline()},
or \co{rcu_thread_online()} from an RCU read-side critical section.
\end{fcvref}

\begin{listing}
\input{CodeSamples/defer/rcu_qs=define.fcv}
\caption{Data for Quiescent-State-Based RCU}
\label{lst:app:toyrcu:Data for Quiescent-State-Based RCU}
\end{listing}

\begin{listing}
\input{CodeSamples/defer/rcu_qs=read_lock_unlock.fcv}
\caption{Quiescent-State-Based RCU Read Side}
\label{lst:app:toyrcu:Quiescent-State-Based RCU Read Side}
\end{listing}

\begin{fcvref}[ln:defer:rcu_qs:read_lock_unlock:qs]
In \co{rcu_quiescent_state()}, \clnref{mb1} executes a memory barrier
to prevent any code prior to the quiescent state (including possible
RCU read-side critical sections) from being reordered
into the quiescent state.
\Clnrefrange{gp1}{gp2} pick up
a copy of the global \co{rcu_gp_ctr}, using
\co{READ_ONCE()} to ensure that the compiler does not employ any
optimizations that would result in \co{rcu_gp_ctr} being fetched
more than once,
and then adds one to the value fetched and stores it into
the per-thread \co{rcu_reader_qs_gp} variable, so that any concurrent
instance of \co{synchronize_rcu()} will see an odd-numbered value,
thus becoming aware that a new RCU read-side critical section has started.
Instances of \co{synchronize_rcu()} that are waiting on older
RCU read-side critical sections will thus know to ignore this new one.
Finally, \clnref{mb2} executes a memory barrier, which prevents subsequent
code (including a possible RCU read-side critical section) from being
re-ordered with the \clnrefrange{gp1}{gp2}.
\end{fcvref}

\QuickQuiz{
	\begin{fcvref}[ln:defer:rcu_qs:read_lock_unlock:qs]
	Doesn't the additional memory barrier shown on \clnref{mb2} of
	\cref{lst:app:toyrcu:Quiescent-State-Based RCU Read Side}
	greatly increase the overhead of \co{rcu_quiescent_state}?
	\end{fcvref}
}\QuickQuizAnswer{
	\begin{fcvref}[ln:defer:rcu_qs:read_lock_unlock:qs]
	Indeed it does!
	An application using this implementation of RCU should therefore
	invoke \co{rcu_quiescent_state} sparingly, instead using
	\co{rcu_read_lock()} and \co{rcu_read_unlock()} most of the
	time.

	However, this memory barrier is absolutely required so that
	other threads will see the store on
	\clnrefrange{gp1}{gp2} before any
	subsequent RCU read-side critical sections executed by the
	caller.
	\end{fcvref}
}\QuickQuizEnd

Some applications might use RCU only occasionally, but use it very heavily
when they do use it.
Such applications might choose to use \co{rcu_thread_online()} when
starting to use RCU and \co{rcu_thread_offline()} when no longer
using RCU\@.
The time between a call to \co{rcu_thread_offline()} and a subsequent
call to \co{rcu_thread_online()} is an extended quiescent state,
so that RCU will not expect explicit quiescent states to be registered
during this time.

The \co{rcu_thread_offline()} function simply sets the
per-thread \co{rcu_reader_qs_gp} variable to the current value of
\co{rcu_gp_ctr}, which has an even-numbered value.
Any concurrent instances of \co{synchronize_rcu()} will thus know to
ignore this thread.

\QuickQuiz{
	\begin{fcvref}[ln:defer:rcu_qs:read_lock_unlock:qs]
	Why are the two memory barriers on \clnref{mb1,mb2} of
	\cref{lst:app:toyrcu:Quiescent-State-Based RCU Read Side}
	needed?
	\end{fcvref}
}\QuickQuizAnswer{
	\begin{fcvref}[ln:defer:rcu_qs:read_lock_unlock:qs]
	The memory barrier on \clnref{mb1} prevents any RCU read-side
	critical sections that might precede the
	call to \co{rcu_thread_offline()} won't be reordered by either
	the compiler or the CPU to follow the assignment on
	\clnrefrange{gp1}{gp2}.
	The memory barrier on \clnref{mb2} is, strictly speaking, unnecessary,
	as it is illegal to have any RCU read-side critical sections
	following the call to \co{rcu_thread_offline()}.
	\end{fcvref}
}\QuickQuizEnd

The \co{rcu_thread_online()} function simply invokes
\co{rcu_quiescent_state()}, thus marking the end of the extended
quiescent state.

\begin{listing}
\input{CodeSamples/defer/rcu_qs=synchronize.fcv}
\caption{RCU Update Side Using Quiescent States}
\label{lst:app:toyrcu:RCU Update Side Using Quiescent States}
\end{listing}

\Cref{lst:app:toyrcu:RCU Update Side Using Quiescent States}
(\path{rcu_qs.c})
shows the implementation of \co{synchronize_rcu()}, which is
quite similar to that of the preceding sections.

This implementation has blazingly fast read-side primitives, with
an \co{rcu_read_lock()}--\co{rcu_read_unlock()} round trip incurring
an overhead of roughly 50~\emph{picoseconds}.
The \co{synchronize_rcu()} overhead ranges from about 600~nanoseconds
on a single-CPU \Power{5} system up to more than 100~microseconds on
a 64-CPU system.

\QuickQuiz{
	To be sure, the clock frequencies of \Power{}
	systems in 2008 were quite high, but even a 5\,GHz clock
	frequency is insufficient to allow
	loops to be executed in 50~picoseconds!
	What is going on here?
}\QuickQuizAnswer{
	Since the measurement loop contains a pair of empty functions,
	the compiler optimizes it away.
	The measurement loop takes 1,000 passes between each call to
	\co{rcu_quiescent_state()}, so this measurement is roughly
	one thousandth of the overhead of a single call to
	\co{rcu_quiescent_state()}.
}\QuickQuizEnd

However, this implementation requires that each thread either
invoke \co{rcu_quiescent_state()} periodically or to invoke
\co{rcu_thread_offline()} for extended quiescent states.
The need to invoke these functions periodically can make this
implementation difficult to use in some situations, such as for
certain types of library functions.

\QuickQuizSeries{%
\QuickQuizB{
	Why would the fact that the code is in a library make
	any difference for how easy it is to use the RCU
	implementation shown in
	\cref{lst:app:toyrcu:Quiescent-State-Based RCU Read Side,%
	lst:app:toyrcu:RCU Update Side Using Quiescent States}?
}\QuickQuizAnswerB{
	A library function has absolutely no control over the caller,
	and thus cannot force the caller to invoke \co{rcu_quiescent_state()}
	periodically.
	On the other hand, a library function that made many references
	to a given RCU-protected data structure might be able to invoke
	\co{rcu_thread_online()} upon entry,
	\co{rcu_quiescent_state()} periodically, and
	\co{rcu_thread_offline()} upon exit.
}\QuickQuizEndB
\QuickQuizE{
	But what if you hold a lock across a call to
	\co{synchronize_rcu()}, and then acquire that same lock within
	an RCU read-side critical section?
	This should be a deadlock, but how can a primitive that
	generates absolutely no code possibly participate in a
	deadlock cycle?
}\QuickQuizAnswerE{
	Please note that the RCU read-side critical section is in
	effect extended beyond the enclosing
	\co{rcu_read_lock()} and \co{rcu_read_unlock()}, out to
	the previous and next call to \co{rcu_quiescent_state()}.
	This \co{rcu_quiescent_state} can be thought of as an
	\co{rcu_read_unlock()} immediately followed by an
	\co{rcu_read_lock()}.

	Even so, the actual deadlock itself will involve the lock
	acquisition in the RCU read-side critical section and
	the \co{synchronize_rcu()}, never the \co{rcu_quiescent_state()}.
}\QuickQuizEndE
}

In addition, this implementation does not permit concurrent calls
to \co{synchronize_rcu()} to share grace periods.
That said, one could easily imagine a production-quality RCU
implementation based on this version of RCU\@.

\section{Summary of Toy RCU Implementations}
\label{sec:app:toyrcu:Summary of Toy RCU Implementations}

If you made it this far, congratulations!
You should now have a much clearer understanding
not only of RCU itself, but also of the requirements of enclosing
software environments and applications.
Those wishing an even deeper understanding are invited to read
descriptions of production-quality RCU
implementations~\cite{MathieuDesnoyers2012URCU,PaulEMcKenney2007PreemptibleRCU,PaulEMcKenney2008HierarchicalRCU,PaulEMcKenney2009BloatwatchRCU}.

The preceding sections listed some desirable properties of the
various RCU primitives.
The following list is provided for easy reference for those wishing to
create a new RCU implementation.

\begin{enumerate}
\item	There must be read-side primitives (such as \co{rcu_read_lock()}
	and \co{rcu_read_unlock()}) and grace-period primitives
	(such as \co{synchronize_rcu()} and \co{call_rcu()}), such
	that any RCU read-side critical section in existence at the
	start of a grace period has completed by the end of the
	grace period.
\item	RCU read-side primitives should have minimal overhead.
	In particular, expensive operations such as cache misses,
	atomic instructions, memory barriers, and branches should
	be avoided.
\item	RCU read-side primitives should have $\O{1}$ computational
	complexity to enable real-time use.
	(This implies that readers run concurrently with updaters.)
\item	RCU read-side primitives should be usable in all contexts
	(in the Linux kernel, they are permitted everywhere except in
	the idle loop).
	An important special case is that RCU read-side primitives be
	usable within an RCU read-side critical section, in other words,
	that it be possible to nest RCU read-side critical sections.
\item	RCU read-side primitives should be unconditional, with no
	failure returns.
	This property is extremely important, as failure checking
	increases complexity and complicates testing and validation.
\item	Any operation other than a quiescent state (and thus a grace
	period) should be permitted in an RCU read-side critical section.
	In particular, irrevocable operations such as I/O should be
	permitted.
\item	It should be possible to update an RCU-protected data structure
	while executing within an RCU read-side critical section.
\item	Both RCU read-side and update-side primitives should be independent
	of memory allocator design and implementation, in other words,
	the same RCU implementation should be able to protect a given
	data structure regardless of how the data elements are allocated
	and freed.
\item	RCU grace periods should not be blocked by threads that
	halt outside of RCU read-side critical sections.
	(But note that most quiescent-state-based implementations
	violate this desideratum.)
\end{enumerate}

\QuickQuiz{
	Given that grace periods are prohibited within RCU read-side
	critical sections, how can an RCU data structure possibly be
	updated while in an RCU read-side critical section?
}\QuickQuizAnswer{
	This situation is one reason for the existence of asynchronous
	grace-period primitives such as \co{call_rcu()}.
	This primitive may be invoked within an RCU read-side critical
	section, and the specified RCU callback will in turn be invoked
	at a later time, after a grace period has elapsed.

	The ability to perform an RCU update while within an RCU read-side
	critical section can be extremely convenient, and is analogous
	to a (mythical) unconditional read-to-write upgrade for
	reader-writer locking.
}\QuickQuizEnd

	\setlength{\parskip}{0.0pt plus 1ex}
	\input{qqztoyrcu}
	\setlength{\parskip}{0.0pt plus 1.0pt}% return to default

% appendix/whymb/whymemorybarriers.tex
% mainfile: ../../perfbook.tex
% SPDX-License-Identifier: CC-BY-SA-3.0

\QuickQuizChapter{chp:app:whymb:Why Memory Barriers?}{Why Memory Barriers?}{qqzwhymb}
\Epigraph{Order!
	  Order in the court!}
	 {Unknown}

So what possessed CPU designers to cause them to inflict \IXBpl{memory barrier}
on poor unsuspecting SMP software designers?

In short, because reordering memory references allows much better performance,
courtesy of the finite speed of light and the non-zero size of atoms
noted in \cref{sec:cpu:Overheads}, and particularly in the
hardware-performance question posed by \QuickQuizRef{\QspeedOfLightAtoms}.
Therefore, memory barriers are needed to force ordering in things like
synchronization primitives whose correct operation depends on ordered
memory references.

Getting a more detailed answer to this question requires a good understanding
of how CPU caches work, and especially what is required to make
caches really work well.
The following sections:
\begin{enumerate}
\item	Present the structure of a cache,
\item	Describe how cache-coherency protocols ensure that CPUs agree
	on the value of each location in memory, and, finally,
\item	Outline how store buffers and invalidate queues help
	caches and cache-coherency protocols achieve high performance.
\end{enumerate}
We will see that memory barriers are a necessary evil that is required
to enable good performance and scalability, an evil that stems from
the fact that CPUs are orders of magnitude faster than are both the
interconnects between them and the memory they are attempting to access.

\section{Cache Structure}
\label{sec:app:whymb:Cache Structure}

\begin{figure}
\centering
\resizebox{3in}{!}{\includegraphics{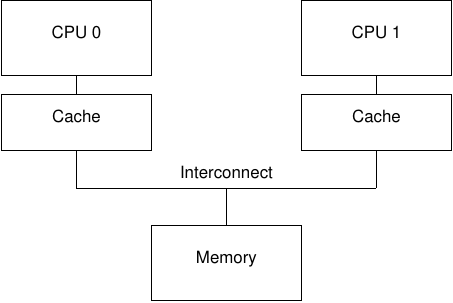}}
\caption{Modern Computer System Cache Structure}
\label{fig:app:whymb:Modern Computer System Cache Structure}
\end{figure}

Modern CPUs are much faster than are modern memory systems.
A 2006 CPU might be capable of executing ten instructions per nanosecond,
but will require many tens of nanoseconds to fetch a data item from
main memory.
This disparity in speed---more than two orders of magnitude---has
resulted in the multi-megabyte caches found on modern CPUs.
These caches are associated with the CPUs as shown in
\cref{fig:app:whymb:Modern Computer System Cache Structure},
and can typically be accessed in a few cycles.\footnote{
	It is standard practice to use multiple levels of cache,
	with a small level-one cache close to the CPU with
	single-cycle access time, and a larger level-two cache
	with a longer access time, perhaps roughly ten clock cycles.
	Higher-performance CPUs often have three or even four levels
	of cache.}

Data flows among the CPUs' caches and memory in fixed-length blocks
called ``\IXpl{cache line}'', which are normally a power of two in size,
ranging from 16 to 256 bytes.
When a given data item is first accessed by a given CPU, it will
be absent from that CPU's cache, meaning that a ``cache miss''
(or, more specifically, a ``startup'' or ``warmup'' cache miss)
has occurred.
The cache miss means that the CPU will
have to wait (or be ``stalled'') for hundreds of cycles while the
item is fetched from memory.
However, the item will be loaded into that CPU's cache, so that
subsequent accesses will find it in the cache and therefore run
at full speed.

After some time, the CPU's cache will fill, and subsequent
misses will likely need to eject an item from the cache in order
to make room for the newly fetched item.
Such a cache miss is termed a ``\IXalth{capacity miss}
{capacity}{cache miss}'', because it is caused
by the cache's limited capacity.
However, most caches can be forced to eject an old item to make room
for a new item even when they are not yet full.
This is due to the fact that large caches are implemented as hardware
hash tables with fixed-size hash buckets (or ``sets'', as CPU designers
call them) and no chaining, as shown in
\cref{fig:app:whymb:CPU Cache Structure}.

This cache has sixteen ``sets'' and two ``ways'' for a total of 32
``lines'', each entry containing a single 256-byte ``cache line'',
which is a 256-byte-aligned block of memory.
This cache line size is a little on the large size, but makes the hexadecimal
arithmetic much simpler.
In hardware parlance, this is a two-way set-associative cache, and
is analogous to a software hash table with
sixteen buckets, where each bucket's hash chain is limited to
at most two elements.
The size (32 cache lines in this case) and the
\IXalt{associativity}{cache associativity} (two in
this case) are collectively called the cache's
``\IXalt{geometry}{cache geometry}''.
Since this cache is implemented in hardware, the hash function is
extremely simple:
Extract four bits from the memory address.

\begin{figure}
\centering
\small
\begin{picture}(170,170)(0,0)

	% Addresses

	\put(0,0){\makebox(20,10){\tt 0xF}}
	\put(0,10){\makebox(20,10){\tt 0xE}}
	\put(0,20){\makebox(20,10){\tt 0xD}}
	\put(0,30){\makebox(20,10){\tt 0xC}}
	\put(0,40){\makebox(20,10){\tt 0xB}}
	\put(0,50){\makebox(20,10){\tt 0xA}}
	\put(0,60){\makebox(20,10){\tt 0x9}}
	\put(0,70){\makebox(20,10){\tt 0x8}}
	\put(0,80){\makebox(20,10){\tt 0x7}}
	\put(0,90){\makebox(20,10){\tt 0x6}}
	\put(0,100){\makebox(20,10){\tt 0x5}}
	\put(0,110){\makebox(20,10){\tt 0x4}}
	\put(0,120){\makebox(20,10){\tt 0x3}}
	\put(0,130){\makebox(20,10){\tt 0x2}}
	\put(0,140){\makebox(20,10){\tt 0x1}}
	\put(0,150){\makebox(20,10){\tt 0x0}}

	% Way 0

	\put(20,163){\makebox(80,10){Way 0}}
	\put(20,0){\framebox(80,10){\tt }}
	\put(20,10){\framebox(80,10){\tt 0x12345E00}}
	\put(20,20){\framebox(80,10){\tt 0x12345D00}}
	\put(20,30){\framebox(80,10){\tt 0x12345C00}}
	\put(20,40){\framebox(80,10){\tt 0x12345B00}}
	\put(20,50){\framebox(80,10){\tt 0x12345A00}}
	\put(20,60){\framebox(80,10){\tt 0x12345900}}
	\put(20,70){\framebox(80,10){\tt 0x12345800}}
	\put(20,80){\framebox(80,10){\tt 0x12345700}}
	\put(20,90){\framebox(80,10){\tt 0x12345600}}
	\put(20,100){\framebox(80,10){\tt 0x12345500}}
	\put(20,110){\framebox(80,10){\tt 0x12345400}}
	\put(20,120){\framebox(80,10){\tt 0x12345300}}
	\put(20,130){\framebox(80,10){\tt 0x12345200}}
	\put(20,140){\framebox(80,10){\tt 0x12345100}}
	\put(20,150){\framebox(80,10){\tt 0x12345000}}

	% Way 1

	\put(100,163){\makebox(80,10){Way 1}}
	\put(100,0){\framebox(80,10){\tt }}
	\put(100,10){\framebox(80,10){\tt 0x43210E00}}
	\put(100,20){\framebox(80,10){\tt }}
	\put(100,30){\framebox(80,10){\tt }}
	\put(100,40){\framebox(80,10){\tt }}
	\put(100,50){\framebox(80,10){\tt }}
	\put(100,60){\framebox(80,10){\tt }}
	\put(100,70){\framebox(80,10){\tt }}
	\put(100,80){\framebox(80,10){\tt }}
	\put(100,90){\framebox(80,10){\tt }}
	\put(100,100){\framebox(80,10){\tt }}
	\put(100,110){\framebox(80,10){\tt }}
	\put(100,120){\framebox(80,10){\tt }}
	\put(100,130){\framebox(80,10){\tt }}
	\put(100,140){\framebox(80,10){\tt }}
	\put(100,150){\framebox(80,10){\tt }}

\end{picture}
\caption{CPU Cache Structure}
\label{fig:app:whymb:CPU Cache Structure}
\end{figure}

In \cref{fig:app:whymb:CPU Cache Structure},
each box corresponds to a cache entry, which
can contain a 256-byte cache line.
However, a cache entry can be empty, as indicated by the empty boxes
in the figure.
The rest of the boxes are flagged with the memory address of the cache line
that they contain.
Since the cache lines must be 256-byte aligned, the low eight bits of
each address are
zero, and the choice of hardware hash function means that the next-higher
four bits match the hash line number.

The situation depicted in the figure might arise if the program's code
were located at address 0x43210E00 through 0x43210EFF, and this program
accessed data sequentially from 0x12345000 through 0x12345EFF\@.
Suppose that the program were now to access location 0x12345F00.
This location hashes to line 0xF, and both ways of this line are
empty, so the corresponding 256-byte line can be accommodated.
If the program were to access location 0x1233000, which hashes to line
0x0, the corresponding 256-byte cache line can be accommodated in
way 1.
However, if the program were to access location 0x1233E00, which hashes
to line 0xE, one of the existing lines must be ejected from the cache
to make room for the new cache line.
If this ejected line were accessed later, a cache miss would result.
Such a cache miss is termed an ``\IXalth{associativity miss}
{associativity}{cache miss}''.

Thus far, we have been considering only cases where a CPU reads
a data item.
What happens when it does a write?
Because it is important that all CPUs agree on the value of a given
data item, before a given CPU writes to that data item, it must first
cause it to be removed, or ``invalidated'', from other CPUs' caches.
Once this \IX{invalidation} has completed, the CPU may safely modify the
data item.
If the data item was present in this CPU's cache, but was read-only,
this process is termed a ``\IXalth{write miss}{write}{cache miss}''.
Once a given CPU has completed invalidating a given data item from other
CPUs' caches, that CPU may repeatedly write (and read) that data item.

Later, if one of the other CPUs attempts to access the data item, it
will incur a cache miss, this time because the first CPU invalidated
the item in order to write to it.
This type of cache miss is termed a ``\IXalth{communication miss}
{communication}{cache miss}'', since it
is usually due to several CPUs using the data items to communicate
(for example, a lock is a data item that is used to communicate among
CPUs using a mutual-exclusion algorithm).

Clearly, much care must be taken to ensure that all CPUs maintain
a coherent view of the data.
With all this fetching, invalidating, and writing, it is easy to
imagine data being lost or (perhaps worse) different CPUs having
conflicting values for the same data item in their respective
caches.
These problems are prevented by ``cache-coherency protocols'',
described in the next section.

\section{Cache-Coherence Protocols}
\label{sec:app:whymb:Cache-Coherence Protocols}

\IXpl{Cache-coherence protocol} manage cache-line states so as to prevent
inconsistent or lost data.
These protocols can be quite complex, with many tens
of states,\footnote{
	See Culler et al.~\cite{DavidECuller1999} pages 670 and 671
	for the nine-state and 26-state diagrams for SGI Origin2000
	and Sequent (now IBM) NUMA-Q, respectively.
	Both diagrams are significantly simpler than real life.}
but for our purposes we need only concern ourselves with the
four-state \IXaltr{MESI cache-coherence protocol}{MESI protocol}.

\subsection{MESI States}
\label{sec:app:whymb:MESI States}

MESI stands for ``modified'', ``exclusive'', ``shared'', and ``invalid'',
the four states a given cache line can take on using this
protocol.
Caches using this protocol therefore maintain a two-bit state ``tag'' on each
cache line in addition to that line's physical address and data.
% cite Schimmel's book on virtual caches.

A line in the ``modified'' state has been subject to a recent memory store
from the corresponding CPU, and the corresponding memory is guaranteed
not to appear in any other CPU's cache.
Cache lines in the ``modified'' state can thus be said to be ``owned''
by the CPU\@.
Because this cache holds the only up-to-date copy of the data, this
cache is ultimately responsible for either writing it back to memory
or handing it off to some other cache, and must do so before reusing
this line to hold other data.

The ``exclusive'' state is very similar to the ``modified'' state,
the single exception being that the cache line has not yet been
modified by the corresponding CPU, which in turn means that the
copy of the cache line's data that resides in memory is up-to-date.
However, since the CPU can store to this line at any time, without
consulting other CPUs, a line in the ``exclusive'' state can still
be said to be owned by the corresponding CPU\@.
That said, because the corresponding value in memory is up to date,
this cache can discard this data without writing it back to memory
or handing it off to some other CPU\@.

A line in the ``shared'' state might be replicated in at least
one other CPU's cache, so that this CPU is not permitted to store
to the line without first consulting with other CPUs.
As with the ``exclusive'' state, because the corresponding value
in memory is up to date,
this cache can discard this data without writing it back to memory
or handing it off to some other CPU\@.

A line in the ``invalid'' state is empty, in other words, it holds
no data.
When new data enters the cache, it is placed into a
cache line that was in the ``invalid'' state if possible.
This approach is preferred because replacing a line in any other
state could result in an expensive cache miss should the replaced
line be referenced in the future.

Since all CPUs must maintain a coherent view of the data carried in
the cache lines, the cache-coherence protocol provides messages
that coordinate the movement of cache lines through the system.

\subsection{MESI Protocol Messages}
\label{sec:app:whymb:MESI Protocol Messages}

Many of the transitions described in the previous section require
communication among the CPUs.
If the CPUs are on a single shared bus, the following messages suffice:
\begin{description}[style=nextline]
\item	[Read:]
	The ``read'' message contains the physical address of the cache line
	to be read.
\item	[Read Response:]
	The ``read response'' message contains the data requested by an
	earlier ``read'' message.
	This ``read response'' message might be supplied either by
	memory or by one of the other caches.
	For example, if one of the caches has the desired data in
	``modified'' state, that cache must supply the ``read response''
	message.
\item	[Invalidate:]
	The ``invalidate'' message contains the physical address of the
	cache line to be invalidated.
	All other caches must remove the corresponding data from their
	caches and respond.
\item	[Invalidate Acknowledge:]
	A CPU receiving an ``invalidate'' message must respond with an
	``invalidate acknowledge'' message after removing the specified
	data from its cache.
\item	[Read Invalidate:]
	The ``read invalidate'' message contains the physical address
	of the cache line to be read, while at the same time directing
	other caches to remove the data.
	Hence, it is a combination of a ``read'' and an ``invalidate'',
	as indicated by its name.
	A ``read invalidate'' message requires both a ``read response''
	and a set of ``invalidate acknowledge'' messages in reply.
\item	[Writeback:]
	The ``writeback'' message contains both the address and the
	data to be written back to memory (and perhaps ``snooped''
	into other CPUs' caches along the way).
	This message permits caches to eject lines in the ``modified''
	state as needed to make room for other data.
\end{description}

\QuickQuiz{
	Where does a writeback message originate from and where does
	it go to?
}\QuickQuizAnswer{
	The writeback message originates from a given CPU, or in some
	designs from a given level of a given CPU's cache---or even
	from a cache that might be shared among several CPUs.
	The key point is that a given cache does not have room for
	a given data item, so some other piece of data must be ejected
	from the cache to make room.
	If there is some other piece of data that is duplicated in some
	other cache or in memory, then that piece of data may be simply
	discarded, with no writeback message required.

	On the other hand, if every piece of data that might be ejected
	has been modified so that the only up-to-date copy is in this
	cache, then one of those data items must be copied somewhere
	else.
	This copy operation is undertaken using a ``writeback message''.

	The destination of the writeback message has to be something
	that is able to store the new value.
	This might be main memory, but it also might be some other cache.
	If it is a cache, it is normally a higher-level cache for the
	same CPU, for example, a level-1 cache might write back to a
	level-2 cache.
	However, some hardware designs permit cross-CPU writebacks,
	so that CPU~0's cache might send a writeback message to CPU~1.
	This would normally be done if CPU~1 had somehow indicated
	an interest in the data, for example, by having recently
	issued a read request.

	In short, a writeback message is sent from some part of the
	system that is short of space, and is received by some other
	part of the system that can accommodate the data.
}\QuickQuizEnd

Interestingly enough, a shared-memory multiprocessor system really
is a message-passing computer under the covers.
This means that clusters of SMP machines that use distributed shared memory
are using message passing to implement shared memory at two different
levels of the system architecture.

\QuickQuizSeries{%
\QuickQuizB{
	What happens if two CPUs attempt to invalidate the
	same cache line concurrently?
}\QuickQuizAnswerB{
	One of the CPUs gains access to the shared bus first,
	and that CPU ``wins''.
	The other CPU must invalidate its copy of the cache line and
	transmit an ``invalidate acknowledge'' message to the other CPU\@.

	Of course, the losing CPU can be expected to immediately issue a
	``read invalidate'' transaction, so the winning CPU's victory will
	be quite ephemeral.
}\QuickQuizEndB
\QuickQuizM{
	When an ``invalidate'' message appears in a large multiprocessor,
	every CPU must give an ``invalidate acknowledge'' response.
	Wouldn't the resulting ``storm'' of ``invalidate acknowledge''
	responses totally saturate the system bus?
}\QuickQuizAnswerM{
	It might, if large-scale multiprocessors were in fact implemented
	that way.
	Larger multiprocessors, particularly NUMA machines,
	tend to use so-called ``directory-based'' cache-coherence
	protocols to avoid this and other problems.
}\QuickQuizEndM
\QuickQuizE{
	If SMP machines are really using message passing
	anyway, why bother with SMP at all?
}\QuickQuizAnswerE{
	There has been quite a bit of controversy on this topic over
	the past few decades.
	One answer is that the cache-coherence
	protocols are quite simple, and therefore can be implemented
	directly in hardware, gaining bandwidths and latencies
	unattainable by software message passing.
	Another answer is that
	the real truth is to be found in economics due to the relative
	prices of large SMP machines and that of clusters of smaller
	SMP machines.
	A third answer is that the SMP programming model is easier to
	use than that of distributed systems, but a rebuttal might note
	the appearance of HPC clusters and MPI\@.
	And so the argument continues.
}\QuickQuizEndE
}

\subsection{MESI State Diagram}
\label{sec:app:whymb:MESI State Diagram}

\begin{figure}
\centering
% \resizebox{3in}{!}{\includegraphics{appendix/whymb/MESI}}
\includegraphics{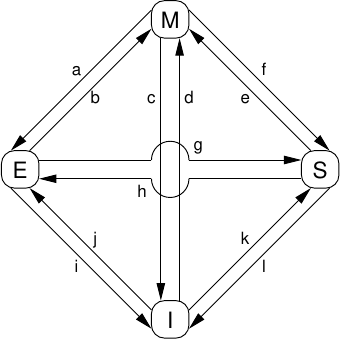}
\caption{MESI Cache-Coherency State Diagram}
\label{fig:app:whymb:MESI Cache-Coherency State Diagram}
\end{figure}

A given cache line's state changes
as protocol messages are sent and received, as
shown in \cref{fig:app:whymb:MESI Cache-Coherency State Diagram}.

The transition arcs in this figure are as follows:
\begin{description}[style=nextline]
\item	[Transition (a):]
	A cache line is written back to memory, but the CPU retains
	it in its cache and further retains the right to modify it.
	This transition requires a ``writeback'' message.
\item	[Transition (b):]
	The CPU writes to the cache line that it already had exclusive
	access to.
	This transition does not require any messages to be sent or
	received.
\item	[Transition (c):]
	The CPU receives a ``read invalidate'' message for a cache line
	that it has modified.
	The CPU must invalidate its local copy, then respond with both a
	``read response'' and an ``invalidate acknowledge'' message,
	both sending the data to the requesting CPU and indicating
	that it no longer has a local copy.
\item	[Transition (d):]
	The CPU does an \IX{atomic read-modify-write operation} on a data item
	that was not present in its cache.
	It transmits a ``read invalidate'', receiving the data via
	a ``read response''.
	The CPU can complete the transition once it has also received a
	full set of ``invalidate acknowledge'' responses.
\item	[Transition (e):]
	The CPU does an atomic read-modify-write operation on a data item
	that was previously read-only in its cache.
	It must transmit ``invalidate'' messages, and must wait for a
	full set of ``invalidate acknowledge'' responses before completing
	the transition.
\item	[Transition (f):]
	Some other CPU reads the cache line, and it is supplied from
	this CPU's cache, which retains a read-only copy, possibly also
	writing it back to memory.
	This transition is initiated by the reception of a ``read''
	message, and this CPU responds with a ``read response'' message
	containing the requested data.
\item	[Transition (g):]
	Some other CPU reads a data item in this cache line,
	and it is supplied either from this CPU's cache or from memory.
	In either case, this CPU retains a read-only copy.
	This transition is initiated by the reception of a ``read''
	message, and this CPU responds with a ``read response'' message
	containing the requested data.
\item	[Transition (h):]
	This CPU realizes that it will soon need to write to some data
	item in this cache line, and thus transmits an ``invalidate'' message.
	The CPU cannot complete the transition until it receives a full
	set of ``invalidate acknowledge'' responses, indicating that
	no other CPU has this cacheline in its cache.
	In other words, this CPU is the only CPU caching it.
\item	[Transition (i):]
	Some other CPU does an atomic read-modify-write operation on
	a data item in a cache line held only in this CPU's cache,
	so this CPU invalidates it from its cache.
	This transition is initiated by the reception of a ``read invalidate''
	message, and this CPU responds with both a ``read response''
	and an ``invalidate acknowledge'' message.
\item	[Transition (j):]
	This CPU does a store to a data item in a cache line that was not
	in its cache, and thus transmits a ``read invalidate'' message.
	The CPU cannot complete the transition until it receives the
	``read response'' and a full set of ``invalidate acknowledge''
	messages.
	The cache line will presumably transition to ``modified'' state via
	transition (b) as soon as the actual store completes.
\item	[Transition (k):]
	This CPU loads a data item in a cache line that was not
	in its cache.
	The CPU transmits a ``read'' message, and completes the
	transition upon receiving the corresponding ``read response''.
\item	[Transition (l):]
	Some other CPU does a store to
	a data item in this cache line, but holds this cache line in read-only
	state due to its being held in other CPUs' caches (such as the
	current CPU's cache).
	This transition is initiated by the reception of an ``invalidate''
	message, and this CPU responds with
	an ``invalidate acknowledge'' message.
\end{description}

\QuickQuiz{
	How does the hardware handle the delayed transitions
	described above?
}\QuickQuizAnswer{
	Usually by adding additional states, though these additional
	states need not be actually stored with the cache line, due to
	the fact that only a few lines at a time will be transitioning.
	The need to delay transitions is but one issue that results in
	real-world cache coherence protocols being much more complex than
	the over-simplified MESI protocol described in this appendix.
	Hennessy and Patterson's classic introduction to computer
	architecture~\cite{Hennessy95a} covers many of these issues.
}\QuickQuizEnd

\subsection{MESI Protocol Example}
\label{sec:app:whymb:MESI Protocol Example}

Let's now look at this from the perspective of a cache line's worth
of data, initially residing in memory at address~0,
as it travels through the various single-line \IXhpl{direct-mapped}{cache}
in a four-CPU system.
\Cref{tab:app:whymb:Cache Coherence Example}
shows this flow of data, with the first column showing the sequence
of operations, the second the CPU performing the operation,
the third the operation being performed, the next four the state
of each CPU's cache line (memory address followed by MESI state),
and the final two columns whether the corresponding memory contents
are up to date (``V'') or not (``I'').

Initially, the CPU cache lines in which the data would reside are
in the ``invalid'' state, and the data is valid in memory.
When CPU~0 loads the data at address~0, it enters the ``shared'' state in
CPU~0's cache, and is still valid in memory.
CPU~3 also loads the data at address~0, so that it is in the
``shared'' state in both CPUs' caches, and is still valid in memory.
Next CPU~0 loads some other cache line (at address~8),
which forces the data at address~0 out of its cache via an invalidation,
replacing it with the data at address~8.
CPU~2 now does a load from address~0, but this CPU realizes that it will
soon need to store to it, and so it uses a ``read invalidate'' message
in order to gain an exclusive copy, invalidating
it from CPU~3's cache (though the copy in memory remains up to date).
Next CPU~2 does its anticipated store, changing the state to ``modified''.
The copy of the data in memory is now out of date.
CPU~1 does an atomic increment, using a ``read invalidate'' to snoop
the data from CPU~2's cache
and invalidate it, so that the copy in CPU~1's cache is in the ``modified''
state (and the copy in memory remains out of date).
Finally, CPU~1 reads the cache line at address~8, which uses a
``writeback'' message to push address~0's data back out to memory.

\begin{table*}
\small
\centering
\renewcommand*{\arraystretch}{1.2}
\rowcolors{6}{}{lightgray}
\begin{tabular}{rclcccccc}
	\toprule
	& & & \multicolumn{4}{c}{CPU Cache} & \multicolumn{2}{c}{Memory} \\
	\cmidrule(lr){4-7} \cmidrule(l){8-9}
	Sequence \# & CPU \# & Operation & 0 & 1 & 2 & 3 & 0 & 8 \\
	\cmidrule(r){1-3} \cmidrule(lr){4-7} \cmidrule(l){8-9}
%	Seq CPU Operation	------------- CPU -------------   - Memory -
%				   0	   1	   2	   3	    0   8
	0 &   & Initial State	& $-$/I & $-$/I & $-$/I & $-$/I   & V & V \\
	1 & 0 & Load		& 0/S &   $-$/I & $-$/I & $-$/I   & V & V \\
	2 & 3 & Load		& 0/S &   $-$/I & $-$/I & 0/S     & V & V \\
	3 & 0 & Invalidation	& 8/S &   $-$/I & $-$/I & 0/S     & V & V \\
	4 & 2 & RMW		& 8/S &   $-$/I & 0/E &   $-$/I   & V & V \\
	5 & 2 & Store		& 8/S &   $-$/I & 0/M &   $-$/I   & I & V \\
	6 & 1 & Atomic Inc	& 8/S &   0/M &   $-$/I & $-$/I   & I & V \\
	7 & 1 & Writeback	& 8/S &   8/S &   $-$/I & $-$/I   & V & V \\
	\bottomrule
\end{tabular}
\caption{Cache Coherence Example}
\label{tab:app:whymb:Cache Coherence Example}
\end{table*}

Note that we end with data in some of the CPU's caches.

\QuickQuiz{
	What sequence of operations would put the CPUs' caches
	all back into the ``invalid'' state?
}\QuickQuizAnswer{
	There is no such sequence, at least in absence of special
	``flush my cache'' instructions in the CPU's instruction set.
	Most CPUs do have such instructions.
}\QuickQuizEnd

\section{Stores Result in Unnecessary Stalls}
\label{sec:app:whymb:Stores Result in Unnecessary Stalls}

\begin{figure}
\centering
% \resizebox{3in}{!}{\includegraphics{appendix/whymb/cacheSCwrite}}
\includegraphics{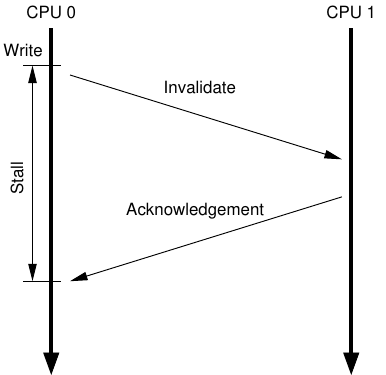}
\caption{Writes See Unnecessary Stalls}
\label{fig:app:whymb:Writes See Unnecessary Stalls}
\end{figure}

Although the cache structure shown in
\cref{fig:app:whymb:Modern Computer System Cache Structure}
provides good performance for repeated reads and writes from a given CPU
to a given item of data, its performance for the first write to
a given cache line is quite poor.
To see this, consider
\cref{fig:app:whymb:Writes See Unnecessary Stalls},
which shows a timeline of a write by CPU~0 to a cacheline held in
CPU~1's cache.
Since CPU~0 must wait for the cache line to arrive before it can
write to it, CPU~0 must stall for an extended period of time.\footnote{
	The time required to transfer a cache line from one CPU's cache
	to another's is typically a few orders of magnitude more than
	that required to execute a simple register-to-register instruction.}

But there is no real reason to force CPU~0 to stall for so long---after
all, regardless of what data happens to be in the cache line that CPU~1
sends it, CPU~0 is going to unconditionally overwrite it.

\subsection{Store Buffers}
\label{sec:app:whymb:Store Buffers}

\begin{figure}
\centering
\resizebox{3in}{!}{\includegraphics{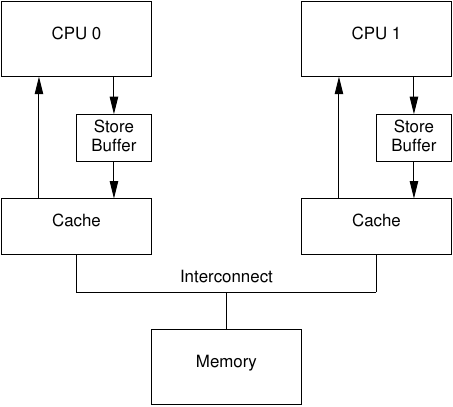}}
\caption{Caches With Store Buffers}
\label{fig:app:whymb:Caches With Store Buffers}
\end{figure}

One way to prevent this unnecessary stalling of writes is to add
``store buffers'' between each CPU and its cache, as shown in
\cref{fig:app:whymb:Caches With Store Buffers}.
With the addition of these store buffers, CPU~0 can simply record
its write in its store buffer and continue executing.
When the cache line does finally make its way from CPU~1 to CPU~0,
the data will be moved from the store buffer to the cache line.

\QuickQuiz{
	But then why do uniprocessors also have store buffers?
}\QuickQuizAnswer{
	Because the purpose of store buffers is not just to hide
	acknowledgement latencies in multiprocessor cache-coherence protocols,
	but to hide memory latencies in general.
	Because memory is much slower than is cache on uniprocessors,
	store buffers on uniprocessors can help to hide write-miss
	memory latencies.
}\QuickQuizEnd

Please note that the store buffer does not necessarily operate on
full cache lines.
The reason for this is that a given store-buffer entry need only contain
the value stored, not the other data contained in the corresponding
cache line.
Which is a good thing, because the CPU doing the store has no idea
what that other data might be!
But once the corresponding cache line arrives, any values from the
store buffer that update that cache line can be merged into it,
and the corresponding entries can then be removed from the store buffer.
Any other data in that cache line is of course left intact.

\QuickQuiz{
	So store-buffer entries are variable length?
	Isn't that difficult to implement in hardware?
}\QuickQuizAnswer{
	Here are two ways for hardware to easily handle variable-length
	stores.

	First, each store-buffer entry could be a single byte wide.
	Then an 64-bit store would consume eight store-buffer entries.
	This approach is simple and flexible, but one disadvantage is
	that each entry would need to replicate much of the address that
	was stored to.

	Second, each store-buffer entry could be double the size of a
	cache line, with half of the bits containing the values stored,
	and the other half indicating which bits had been stored to.
	So, assuming a 32-bit cache line, a single-byte store of 0x5a
	to the low-order byte of a given cache line would result in
	\co{0xXXXXXX5a} for the first half and \co{0x000000ff} for the
	second half, where the values labeled \co{X} are arbitrary
	because they would be ignored.
	This approach allows multiple consecutive stores corresponding to
	a given cache line to be merged into a single store-buffer entry,
	but is space-inefficient for random stores of single bytes.

	Much more complex and efficient schemes are of course used
	by actual hardware designers.
}\QuickQuizEnd

These store buffers are local to a given CPU or, on systems with
hardware multithreading, local to a given core.
Either way, a given CPU is permitted to access only the store buffer
assigned to it.
For example, in
\cref{fig:app:whymb:Caches With Store Buffers}, CPU~0 cannot
access CPU~1's store buffer and vice versa.
This restriction simplifies the hardware by separating concerns:
The store buffer improves performance for consecutive writes, while
the responsibility for communicating among CPUs (or cores, as the
case may be) is fully shouldered by the cache-coherence protocol.
However, even given this restriction, there are complications that must
be addressed, which are covered in the next two sections.

\subsection{Store Forwarding}
\label{sec:app:whymb:Store Forwarding}

To see the first complication, a violation of self-consistency,
consider the following code with variables ``a'' and ``b'' both initially
zero, and with the cache line containing variable ``a'' initially
owned by CPU~1 and that containing ``b'' initially owned by CPU~0:

\begin{VerbatimN}[fontsize=\footnotesize,samepage=true]
a = 1;
b = a + 1;
assert(b == 2);
\end{VerbatimN}

One would not expect the assertion to fail.
However, if one were foolish enough to use the very simple architecture
shown in
\cref{fig:app:whymb:Caches With Store Buffers},
one would be surprised.
Such a system could potentially see the following sequence of events:
\begin{sequence}
\item	CPU~0 starts executing the \co{a = 1}.
\item	CPU~0 looks ``a'' up in the cache, and finds that it is missing.
\item	CPU~0 therefore sends a ``read invalidate'' message in order to
	get exclusive ownership of the cache line containing ``a''.
\item	CPU~0 records the store to ``a'' in its store buffer.
\item	CPU~1 receives the ``read invalidate'' message, and responds
	by transmitting the cache line and removing that cacheline from
	its cache.
\item	CPU~0 starts executing the \co{b = a + 1}.
\item	CPU~0 receives the cache line from CPU~1, which still has
	a value of zero for ``a''.
\item	CPU~0 loads ``a'' from its cache, finding the value zero.
	\label{item:app:whymb:Need Store Buffer}
\item	CPU~0 applies the entry from its store buffer to the newly
	arrived cache line, setting the value of ``a'' in its cache
	to one.
\item	CPU~0 adds one to the value zero loaded for ``a'' above,
	and stores it into the cache line containing ``b''
	(which we will assume is already owned by CPU~0).
\item	CPU~0 executes \co{assert(b == 2)}, which fails.
\end{sequence}

The problem is that we have two copies of ``a'', one in the cache and
the other in the store buffer.

\begin{figure}
\centering
\resizebox{3in}{!}{\includegraphics{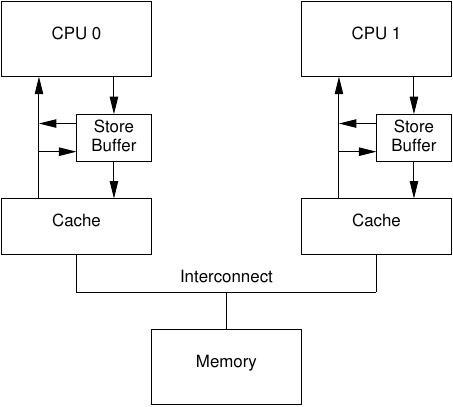}}
\caption{Caches With Store Forwarding}
\label{fig:app:whymb:Caches With Store Forwarding}
\end{figure}

This example breaks a very important guarantee, namely that each CPU
will always see its own operations as if they happened in program order.
Breaking this guarantee is violently counter-intuitive to software types,
so much so
that the hardware guys took pity and implemented ``store forwarding'',
where each CPU refers to (or ``snoops'') its store buffer as well
as its cache when performing loads, as shown in
\cref{fig:app:whymb:Caches With Store Forwarding}.
In other words, a given CPU's stores are directly forwarded to its
subsequent loads, without having to pass through the cache.

With store forwarding in place, item~\ref{item:app:whymb:Need Store Buffer}
in the above sequence would have found the correct value of 1 for ``a'' in
the store buffer, so that the final value of ``b'' would have been 2,
as one would hope.

\subsection{Store Buffers and Memory Barriers}
\label{sec:app:whymb:Store Buffers and Memory Barriers}

To see the second complication, a violation of global memory ordering,
consider the following code sequences
with variables ``a'' and ``b'' initially zero:

\begin{VerbatimN}[fontsize=\footnotesize,samepage=true]
void foo(void)
{
	a = 1;
	b = 1;
}

void bar(void)
{
	while (b == 0) continue;
	assert(a == 1);
}
\end{VerbatimN}

Suppose CPU~0 executes foo() and CPU~1 executes bar().
Suppose further that the cache line containing ``a'' resides only in CPU~1's
cache, and that the cache line containing ``b'' is owned by CPU~0.
Then the sequence of operations might be as follows:
\begin{sequence}
\item	CPU~0 executes \co{a = 1}.
	The cache line is not in CPU~0's cache, so CPU~0 places the new
	value of ``a'' in its store buffer and transmits a ``read
	invalidate'' message.
	\label{seq:app:whymb:Store Buffers and Memory Barriers}
\item	CPU~1 executes \co{while (b == 0) continue}, but the cache line
	containing ``b'' is not in its cache.
	It therefore transmits a ``read'' message.
\item	CPU~0 executes \co{b = 1}.
	It already owns this cache line (in other words, the cache line
	is already in either the ``modified'' or the ``exclusive'' state),
	so it stores the new value of ``b'' in its cache line.
\item	CPU~0 receives the ``read'' message, and transmits the
	cache line containing the now-updated value of ``b''
	to CPU~1, also marking the line as ``shared'' in its own cache
	(but only after writing back the line containing ``b'' to main
	memory).
	\label{seq:app:whymb:Store Buffers and Memory Barriers store}
\item	CPU~1 receives the cache line containing ``b'' and installs
	it in its cache.
\item	CPU~1 can now finish executing \co{while (b == 0) continue},
	and since it finds that the value of ``b'' is 1, it proceeds
	to the next statement.
\item	CPU~1 executes the \co{assert(a == 1)}, and, since CPU~1 is
	working with the old value of ``a'', this assertion fails.
\item	CPU~1 receives the ``read invalidate'' message, and
	transmits the cache line containing ``a'' to CPU~0 and
	invalidates this cache line from its own cache.
	But it is too late.
\item	CPU~0 receives the cache line containing ``a'' and applies
	the buffered store just in time to fall victim to CPU~1's
	failed assertion.
	\label{seq:app:whymb:Store Buffers and Memory Barriers victim}
\end{sequence}

\EQuickQuiz{
	In \cref{seq:app:whymb:Store Buffers and Memory Barriers} above,
	why does CPU~0 need to issue a ``read invalidate''
	rather than a simple ``invalidate''?
	After all, \co{foo()} will overwrite the variable \co{a} in any
	case, so why should it care about the old value of \co{a}?
}\EQuickQuizAnswer{
	Because the cache line in question contains more data than just the
	variable \co{a}.
	Issuing ``invalidate'' instead of the needed ``read invalidate''
	would cause that other data to be lost, which would constitute
	a serious bug in the hardware.
}\EQuickQuizEnd

\EQuickQuiz{
	In \cref{seq:app:whymb:Store Buffers and Memory Barriers store}
	above, don't systems avoid that store to memory?
}\EQuickQuizAnswer{
	Yes, they do.
	But to do so, they add states beyond the MESI quadruple that
	this example is working within.
}\EQuickQuizEnd

\EQuickQuiz{
	In \cref{seq:app:whymb:Store Buffers and Memory Barriers victim}
	above, did \co{bar()} read a stale value from \co{a}, or did
	its reads of \co{b} and \co{a} get reordered?
}\EQuickQuizAnswer{
	It could be either, depending on the hardware implementation.
	And it really does not matter which.
	After all, the \co{bar()} function's \co{assert()} cannot tell
	the difference!
}\EQuickQuizEnd

The hardware designers cannot help directly here, since the CPUs have
no idea which variables are related, let alone how they might be related.
Therefore, the hardware designers provide memory-barrier instructions
to allow the software to tell the CPU about such relations.
The program fragment must be updated to contain the memory barrier:

\begin{VerbatimN}[fontsize=\footnotesize,samepage=true]
void foo(void)
{
	a = 1;
	smp_mb();
	b = 1;
}

void bar(void)
{
	while (b == 0) continue;
	assert(a == 1);
}
\end{VerbatimN}

The memory barrier \co{smp_mb()} will cause the CPU to flush its store
buffer before applying each subsequent store to its variable's cache line.
The CPU could either simply stall until the store buffer was empty
before proceeding, or it could use the store buffer to hold subsequent
stores until all of the prior entries in the store buffer had been
applied.

With this latter approach the sequence of operations might be as follows:
\begin{sequence}
\item	CPU~0 executes \co{a = 1}.
	The cache line is not in CPU~0's cache, so CPU~0 places the new
	value of ``a'' in its store buffer and transmits a ``read
	invalidate'' message.
\item	CPU~1 executes \co{while (b == 0) continue}, but the cache line
	containing ``b'' is not in its cache.
	It therefore transmits a ``read'' message.
\item	CPU~0 executes \co{smp_mb()}, and marks all current store-buffer
	entries (namely, the \co{a = 1}).
\item	CPU~0 executes \co{b = 1}.
	It already owns this cache line (in other words, the cache line
	is already in either the ``modified'' or the ``exclusive'' state),
	but there is a marked entry in the store buffer.
	Therefore, rather than store the new value of ``b'' in the
	cache line, it instead places it in the store buffer (but
	in an \emph{unmarked} entry).
\item	CPU~0 receives the ``read'' message, and transmits the
	cache line containing the original value of ``b''
	to CPU~1.
	It also marks its own copy of this cache line as ``shared''.
\item	CPU~1 receives the cache line containing ``b'' and installs
	it in its cache.
\item	CPU~1 can now load the value of ``b'',
	but since it finds that the value of ``b'' is still 0, it repeats
	the \co{while} statement.
	The new value of ``b'' is safely hidden in CPU~0's store buffer.
\item	CPU~1 receives the ``read invalidate'' message, and
	transmits the cache line containing ``a'' to CPU~0 and
	invalidates this cache line from its own cache.
\item	CPU~0 receives the cache line containing ``a'' and applies
	the buffered store, placing this line into the ``modified''
	state.
\item	Since the store to ``a'' was the only
	entry in the store buffer that was marked by the \co{smp_mb()},
	CPU~0 can also store the new value of ``b''---except for the
	fact that the cache line containing ``b'' is now in ``shared''
	state.
\item	CPU~0 therefore sends an ``invalidate'' message to CPU~1.
\item	CPU~1 receives the ``invalidate'' message, invalidates the
	cache line containing ``b'' from its cache, and sends an
	``acknowledgement'' message to CPU~0.
\item	CPU~1 executes \co{while (b == 0) continue}, but the cache line
	containing ``b'' is not in its cache.
	It therefore transmits a ``read'' message to CPU~0.
\item	CPU~0 receives the ``acknowledgement'' message, and puts
	the cache line containing ``b'' into the ``exclusive'' state.
	CPU~0 now stores the new value of ``b'' into the cache line.
\item	CPU~0 receives the ``read'' message, and transmits the
	cache line containing the new value of ``b''
	to CPU~1.
	It also marks its own copy of this cache line as ``shared''.%
	\label{seq:app:whymb:Store buffers: All copies shared}
\item	CPU~1 receives the cache line containing ``b'' and installs
	it in its cache.
\item	CPU~1 can now load the value of ``b'',
	and since it finds that the value of ``b'' is 1, it
	exits the \co{while} loop and proceeds
	to the next statement.
\item	CPU~1 executes the \co{assert(a == 1)}, but the cache line containing
	``a'' is no longer in its cache.
	Once it gets this cache from CPU~0, it will be
	working with the up-to-date value of ``a'', and the assertion
	therefore passes.
\end{sequence}

\QuickQuiz{
	After \cref{seq:app:whymb:Store buffers: All copies shared}
	in \cref{sec:app:whymb:Store Buffers and Memory Barriers} on
	\cpageref{seq:app:whymb:Store buffers: All copies shared},
	both CPUs might drop the cache line containing the new value of
	``b''.
	Wouldn't that cause this new value to be lost?
}\QuickQuizAnswer{
	It might, and that is why real hardware takes steps to avoid
	this problem.
	A traditional approach, pointed out by Vasilevsky Alexander,
	is to write this cache line back to main memory before marking
	the cache line as ``shared''.
	A more efficient (though more complex) approach is to use
	additional state to indicate whether or not the cache line
	is ``dirty'', allowing the writeback to happen.
	Year-2000 systems went further, using much more state in order to
	avoid redundant writebacks~\cite[Figure 8.42]{DavidECuller1999}.
	It would be reasonable to assume that complexity has not decreased
	in the meantime.
}\QuickQuizEnd

As you can see, this process involves no small amount of bookkeeping.
Even something intuitively simple, like ``load the value of a'' can
involve lots of complex steps in silicon.

\section{Store Sequences Result in Unnecessary Stalls}
\label{sec:app:whymb:Store Sequences Result in Unnecessary Stalls}

Unfortunately, each store buffer must be relatively small, which means
that a CPU executing a modest sequence of stores can fill its store
buffer (for example, if all of them result in cache misses).
At that point, the CPU must once again wait for \IXpl{invalidation} to complete
in order to drain its store buffer before it can continue executing.
This same situation can arise immediately after a memory barrier, when
\emph{all} subsequent store instructions must wait for invalidations to
complete, regardless of whether or not these stores result in cache misses.

This situation can be improved by making invalidate acknowledge
messages arrive more quickly.
One way of accomplishing this is to use per-CPU queues of
invalidate messages, or ``invalidate queues''.

\subsection{Invalidate Queues}
\label{sec:app:whymb:Invalidate Queues}

One reason that invalidate acknowledge messages can take so long
is that they must ensure that the corresponding cache line is
actually invalidated, and this invalidation can be delayed if
the cache is busy, for example, if the CPU is intensively loading
and storing data, all of which resides in the cache.
In addition, if a large number of invalidate messages arrive
in a short time period, a given CPU might fall behind in processing
them, thus possibly stalling all the other CPUs.

However, the CPU need not actually invalidate the cache line
before sending the acknowledgement.
It could instead queue the invalidate message with the understanding
that the message will be processed before the CPU sends any further
messages regarding that cache line.

\subsection{Invalidate Queues and Invalidate Acknowledge}
\label{sec:app:whymb:Invalidate Queues and Invalidate Acknowledge}

\begin{figure}
\centering
\resizebox{3in}{!}{\includegraphics{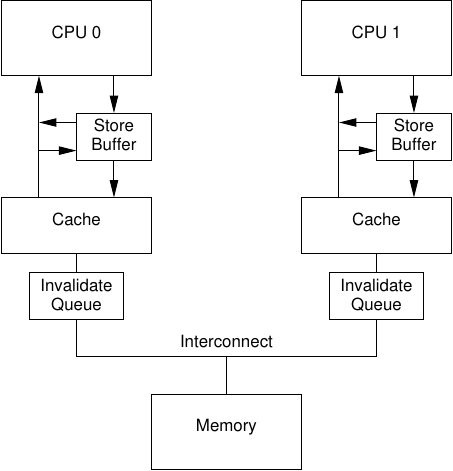}}
\caption{Caches With Invalidate Queues}
\label{fig:app:whymb:Caches With Invalidate Queues}
\end{figure}

\Cref{fig:app:whymb:Caches With Invalidate Queues}
shows a system with invalidate queues.
A CPU with an invalidate queue may acknowledge an invalidate message
as soon as it is placed in the queue, instead of having to wait until
the corresponding line is actually invalidated.
Of course, the CPU must refer to its invalidate queue when preparing
to transmit invalidation messages---if an entry for the corresponding
cache line is in the invalidate queue, the CPU cannot immediately
transmit the invalidate message; it must instead wait until the
invalidate-queue entry has been processed.

Placing an entry into the invalidate queue is essentially a promise
by the CPU to process that entry before transmitting any MESI protocol
messages regarding that cache line.
As long as the corresponding data structures are not highly contended,
the CPU will rarely be inconvenienced by such a promise.

However, the fact that invalidate messages can be buffered in the
invalidate queue provides additional opportunity for memory-misordering,
as discussed in the next section.

\subsection{Invalidate Queues and Memory Barriers}
\label{sec:app:whymb:Invalidate Queues and Memory Barriers}

Let us suppose that CPUs queue invalidation requests, but respond to
them immediately.
This approach minimizes the \IXh{cache-invalidation}{latency} seen by CPUs
doing stores, but can defeat memory barriers, as seen in the following
example.

Suppose the values of ``a'' and ``b'' are initially zero,
that ``a'' is replicated read-only (MESI ``shared'' state),
and that ``b''
is owned by CPU~0 (MESI ``exclusive'' or ``modified'' state).
Then suppose that CPU~0 executes \co{foo()} while CPU~1 executes
function \co{bar()} in the following code fragment:

\begin{fcvlabel}[ln:app:whymb:Breaking mb]
\begin{VerbatimN}[fontsize=\footnotesize,samepage=true,commandchars=\\\[\]]
void foo(void)
{
	a = 1;
	smp_mb();	\lnlbl[mb]
	b = 1;
}

void bar(void)
{
	while (b == 0) continue;
	assert(a == 1);
}
\end{VerbatimN}
\end{fcvlabel}

Then the sequence of operations might be as follows:
\begin{fcvref}[ln:app:whymb:Breaking mb]
\begin{sequence}
\item	CPU~0 executes \co{a = 1}.
	The corresponding cache line is read-only in CPU~0's cache, so
	CPU~0 places the new value of ``a'' in its store buffer and
	transmits an ``invalidate'' message in order to flush the
	corresponding cache line from CPU~1's cache.
	\label{seq:app:whymb:Invalidate Queues and Memory Barriers}
\item	CPU~1 executes \co{while (b == 0) continue}, but the cache line
	containing ``b'' is not in its cache.
	It therefore transmits a ``read'' message.
\item	CPU~1 receives CPU~0's ``invalidate'' message, queues it, and
	immediately responds to it.
\item	CPU~0 receives the response from CPU~1, and is therefore free
	to proceed past the \co{smp_mb()} on \clnref{mb} above, moving
	the value of ``a'' from its store buffer to its cache line.
\item	CPU~0 executes \co{b = 1}.
	It already owns this cache line (in other words, the cache line
	is already in either the ``modified'' or the ``exclusive'' state),
	so it stores the new value of ``b'' in its cache line.
\item	CPU~0 receives the ``read'' message, and transmits the
	cache line containing the now-updated value of ``b''
	to CPU~1, also marking the line as ``shared'' in its own cache.
\item	CPU~1 receives the cache line containing ``b'' and installs
	it in its cache.
\item	CPU~1 can now finish executing \co{while (b == 0) continue},
	and since it finds that the value of ``b'' is 1, it proceeds
	to the next statement.
\item	CPU~1 executes the \co{assert(a == 1)}, and, since the
	old value of ``a'' is still in CPU~1's cache,
	this assertion fails.
\item	Despite the assertion failure, CPU~1 processes the queued
	``invalidate'' message, and (tardily)
	invalidates the cache line containing ``a'' from its own cache.
\end{sequence}
\end{fcvref}

\QuickQuiz{
	In \cref{seq:app:whymb:Invalidate Queues and Memory Barriers}
	of the first scenario in
	\cref{sec:app:whymb:Invalidate Queues and Memory Barriers},
	why is an ``invalidate'' sent instead of a ''read invalidate''
	message?
	Doesn't CPU~0 need the values of the other variables that share
	this cache line with ``a''?
}\QuickQuizAnswer{
	CPU~0 already has the values of these variables, given that it
	has a read-only copy of the cache line containing ``a''.
	Therefore, all CPU~0 need do is to cause the other CPUs to discard
	their copies of this cache line.
	An ``invalidate'' message therefore suffices.
}\QuickQuizEnd

There is clearly not much point in accelerating invalidation responses
if doing so causes memory barriers to effectively be ignored.
However, the memory-barrier instructions can interact with
the invalidate queue, so that when a given CPU executes a memory
barrier, it marks all the entries currently in its invalidate queue,
and forces any subsequent load to wait until all marked entries
have been applied to the CPU's cache.
Therefore, we can add a memory barrier to function \co{bar} as follows:

\begin{fcvlabel}[ln:app:whymb:Add mb]
\begin{VerbatimN}[fontsize=\footnotesize,samepage=true,commandchars=\\\[\]]
void foo(void)
{
	a = 1;
	smp_mb();		\lnlbl[mb1]
	b = 1;
}

void bar(void)
{
	while (b == 0) continue;
	smp_mb();
	assert(a == 1);
}
\end{VerbatimN}
\end{fcvlabel}

\QuickQuiz{
	Say what???
	Why do we need a memory barrier here, given that the CPU cannot
	possibly execute the \co{assert()} until after the
	\co{while} loop completes?
}\QuickQuizAnswer{
	Suppose that memory barrier was omitted.

	Keep in mind that CPUs are free to speculatively execute later
	loads, which can have the effect of executing the assertion
	before the \co{while} loop completes.
	Furthermore, compilers assume that only the currently executing
	thread is updating the variables, and this assumption allows
	the compiler to hoist the load of \co{a} to precede the
	loop.

	In fact, some compilers would transform the loop to a branch
	around an infinite loop as follows:

\begin{VerbatimN}[fontsize=\footnotesize,samepage=true]
void foo(void)
{
	a = 1;
	smp_mb();
	b = 1;
}

void bar(void)
{
	if (b == 0)
		for (;;)
			continue;
	assert(a == 1);
}
\end{VerbatimN}

	Given this optimization, the code would behave in a completely
	different way than the original code.
	If \co{bar()} observed \qco{b == 0}, the assertion could of
	course not be reached at all due to the infinite loop.
	However, if \co{bar()} loaded the value \qco{1} just as
	\qco{foo()} stored it, the CPU might still have the old
	zero value of \qco{a} in its cache, which would cause
	the assertion to fire.
	You should of course use volatile casts (for example, those
	volatile casts implied by the C11 relaxed atomic load operation)
	to prevent the compiler from optimizing your parallel code
	into oblivion.
	But volatile casts would not prevent a weakly ordered CPU
	from loading the old value for \qco{a} from its cache, which
	means that this code also requires the explicit memory barrier
	in \qco{bar()}.

	In short, both compilers and CPUs aggressively apply
	code-reordering optimizations, so you must clearly communicate
	your constraints using the compiler directives and memory barriers
	provided for this purpose.
}\QuickQuizEnd

\begin{fcvref}[ln:app:whymb:Add mb]
With this change, the sequence of operations might be as follows:
\begin{sequence}
\item	CPU~0 executes \co{a = 1}.
	The corresponding cache line is read-only in CPU~0's cache,
	so CPU~0 places the new value of ``a'' in its store buffer and
	transmits an ``invalidate'' message in order to flush the
	corresponding cache line from CPU~1's cache.
\item	CPU~1 executes \co{while (b == 0) continue}, but the cache line
	containing ``b'' is not in its cache.
	It therefore transmits a ``read'' message.
\item	CPU~1 receives CPU~0's ``invalidate'' message, queues it, and
	immediately responds to it.
\item	CPU~0 receives the response from CPU~1, and is therefore free
	to proceed past the \co{smp_mb()} on \clnref{mb1} above, moving
	the value of ``a'' from its store buffer to its cache line.
\item	CPU~0 executes \co{b = 1}.
	It already owns this cache line (in other words, the cache line
	is already in either the ``modified'' or the ``exclusive'' state),
	so it stores the new value of ``b'' in its cache line.
\item	CPU~0 receives the ``read'' message, and transmits the
	cache line containing the now-updated value of ``b''
	to CPU~1, also marking the line as ``shared'' in its own cache.
\item	CPU~1 receives the cache line containing ``b'' and installs
	it in its cache.
\item	CPU~1 can now finish executing \co{while (b == 0) continue},
	and since it finds that the value of ``b'' is 1, it proceeds
	to the next statement, which is now a memory barrier.
\item	CPU~1 must now stall until it processes all pre-existing
	messages in its invalidation queue.
\item	CPU~1 now processes the queued
	``invalidate'' message, and
	invalidates the cache line containing ``a'' from its own cache.
\item	CPU~1 executes the \co{assert(a == 1)}, and, since the
	cache line containing ``a'' is no longer in CPU~1's cache,
	it transmits a ``read'' message.
\item	CPU~0 responds to this ``read'' message with the cache line
	containing the new value of ``a''.
\item	CPU~1 receives this cache line, which contains a value of 1 for
	``a'', so that the assertion does not trigger.
\end{sequence}
\end{fcvref}

With much passing of MESI messages, the CPUs arrive at the correct answer.
This section illustrates why CPU designers must be extremely careful
with their cache-coherence optimizations.
The key requirement is that the memory barriers provide the appearance
of ordering to the software.
As long as these appearances are maintained, the hardware can carry
out whatever queueing, buffering, marking, stallings, and flushing
optimizations it likes.

\QuickQuiz{
	Instead of all of this marking of invalidation-queue entries
	and stalling of loads, why not simply force an immediate flush
	of the invalidation queue?
}\QuickQuizAnswer{
	An immediate flush of the invalidation queue would do the trick.
	Except that the common-case super-scalar CPU is executing many
	instructions at once, and not necessarily even in the expected
	order.
	So what would ``immediate'' even mean?
	The answer is clearly ``not much''.

	Nevertheless, for simpler CPUs that execute instructions serially,
	flushing the invalidation queue might be a reasonable implementation
	strategy.
}\QuickQuizEnd

\section{Read and Write Memory Barriers}
\label{sec:app:whymb:Read and Write Memory Barriers}

In the previous section, memory barriers were used to mark entries in both
the store buffer and the invalidate queue.
But in our code fragment, \co{foo()} had no reason to do anything with the
invalidate queue, and \co{bar()} similarly had no reason to do anything
with the store buffer.

Many CPU architectures therefore provide weaker memory-barrier
instructions that do only one or the other of these two.
Roughly speaking, a ``\IXBh{read}{memory barrier}'' marks only the invalidate
queue (and snoops entries in the store buffer) and a ``\IXBh{write}{memory
barrier}'' marks only the store buffer, while a full-fledged memory
barrier does all of the above.

The software-visible effect of these hardware mechanisms is that a read
memory barrier orders only loads on the CPU that executes it, so that
all loads preceding the read memory barrier will appear to have completed
before any load following the read memory barrier.
Similarly, a write memory barrier orders only stores, again on the
CPU that executes it, and again so that all stores preceding the write
memory barrier will appear to have completed before any store following
the write memory barrier.
A full-fledged memory barrier orders both loads and stores, but again
only on the CPU executing the memory barrier.

\QuickQuiz{
	But can't full memory barriers impose global ordering?
	After all, isn't that needed to provide the ordering
	shown in \cref{lst:formal:IRIW Litmus Test}?
}\QuickQuizAnswer{
	Sort of.

	Note well that this litmus test has not one but two full
	memory-barrier instructions, namely the two \co{sync} instructions
	executed by \co{P2} and \co{P3}.

	It is the interaction of those two instructions that provides
	the global ordering, not just their individual execution.
	For example, each of those two \co{sync} instructions might stall
	waiting for all CPUs to process their invalidation queues before
	allowing subsequent instructions to execute.\footnote{
		Real-life hardware of course applies many optimizations
		to minimize the resulting stalls.}
}\QuickQuizEnd

If we update \co{foo} and \co{bar} to use read and write memory
barriers, they appear as follows:

\begin{VerbatimN}[fontsize=\footnotesize,samepage=true]
void foo(void)
{
	a = 1;
	smp_wmb();
	b = 1;
}

void bar(void)
{
	while (b == 0) continue;
	smp_rmb();
	assert(a == 1);
}
\end{VerbatimN}

Some computers have even more flavors of memory barriers, but
understanding these three variants will provide a good introduction
to memory barriers in general.

\section{Example Memory-Barrier Sequences}
\label{sec:app:whymb:Example Memory-Barrier Sequences}

This section presents some seductive but subtly broken uses of
memory barriers.
Although many of them will work most of the time, and some will
work all the time on some specific CPUs, these uses must be avoided
if the goal is to produce code that works reliably on all CPUs.
To help us better see the subtle breakage, we first need to focus
on an ordering-hostile architecture.

\subsection{Ordering-Hostile Architecture}
\label{sec:app:whymb:Ordering-Hostile Architecture}

A number of ordering-hostile computer systems have been produced over
the decades,
but the nature of the hostility has always been extremely subtle,
and understanding it has required detailed knowledge of the specific
hardware.
Rather than picking on a specific hardware vendor, and as a presumably
attractive alternative to dragging the reader through detailed
technical specifications, let us instead design a mythical but maximally
memory-ordering-hostile computer architecture.\footnote{
	Readers preferring a detailed look at real hardware
	architectures are encouraged to consult CPU vendors'
	manuals~\cite{ALPHA95,AMDOpteron02,IntelItanium02v2,PowerPC94,MichaelLyons05a,SPARC94,IntelXeonV3-96a,IntelXeonV2b-96a,IBMzSeries04a},
	Gharachorloo's dissertation~\cite{Gharachorloo95},
	Peter Sewell's work~\cite{PeterSewell2021weakmemory}, or
	the excellent hardware-oriented primer by
	Sorin, Hill, and Wood~\cite{DanielJSorin2011MemModel}.}

This hardware must obey the following ordering
constraints~\cite{PaulMcKenney2005i,PaulMcKenney2005j}:
\begin{enumerate}
\item	Each CPU will always perceive its own memory accesses
	as occurring in program order.
\item	CPUs will reorder a given operation with a store only
	if the two operations are referencing different locations.
\item	All of a given CPU's loads preceding a read memory barrier
	(\co{smp_rmb()}) will be perceived by all CPUs to precede
	any loads following that read memory barrier.
\item	All of a given CPU's stores preceding a write memory barrier
	(\co{smp_wmb()}) will be perceived by all CPUs to precede
	any stores following that write memory barrier.
\item	All of a given CPU's accesses (loads and stores) preceding a
	full memory barrier
	(\co{smp_mb()}) will be perceived by all CPUs to precede
	any accesses following that memory barrier.
\end{enumerate}

\QuickQuiz{
	Does the guarantee that each CPU sees its own memory accesses
	in order also guarantee that each user-level thread will see
	its own memory accesses in order?
	Why or why not?
}\QuickQuizAnswer{
	No.
	Consider the case where a thread migrates from one CPU to
	another, and where the destination CPU perceives the source
	CPU's recent memory operations out of order.
	To preserve user-mode sanity, kernel hackers must use memory
	barriers in the context-switch path.
	However, the locking already required to safely do a context
	switch should automatically provide the memory barriers needed
	to cause the user-level task to see its own accesses in order.
	That said, if you are designing a super-optimized scheduler,
	either in the kernel or at user level,
	please keep this scenario in mind!
}\QuickQuizEnd

\begin{figure}
\centering
\resizebox{3in}{!}{\includegraphics{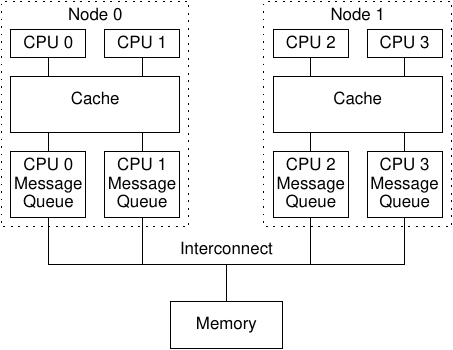}}
\caption{Example Ordering-Hostile Architecture}
\label{fig:app:whymb:Example Ordering-Hostile Architecture}
\end{figure}

Imagine a large \IXacrf{nuca} system that,
in order to provide fair allocation
of interconnect bandwidth to CPUs in a given node, provided per-CPU
queues in each node's interconnect interface, as shown in
\cref{fig:app:whymb:Example Ordering-Hostile Architecture}.
Although a given CPU's accesses are ordered as specified by memory
barriers executed by that CPU, however, the relative order of a
given pair of CPUs' accesses could be severely reordered,
as we will see.\footnote{
	Any real hardware architect or designer will no doubt be
	objecting strenuously,
	as they just might be a bit upset about the prospect of working
	out which queue should handle a message involving a cache line
	that both CPUs accessed, to say nothing of the many races that
	this example poses.
	All I can say is ``Give me a better example''.}

\subsection{Example 1}
\label{sec:app:whymb:Example 1}

\Cref{lst:app:whymb:Memory Barrier Example 1}
shows three code fragments, executed concurrently by CPUs~0, 1, and 2.
Each of ``a'', ``b'', and ``c'' are initially zero.

\floatstyle{plaintop}
\restylefloat{listing}

\begin{listing}
\scriptsize
\centering{\tt
\begin{tabular}{l|l|l}
	\multicolumn{1}{c|}{\nf{CPU~0}} &
		\multicolumn{1}{c|}{\nf{CPU~1}} &
			\multicolumn{1}{c}{\nf{CPU~2}} \\
	\hline
	\hline
	a = 1;		 &		& \\
	\tco{smp_wmb();} & while (b == 0); & \\
	b = 1;		 & c = 1;	& z = c; \\
			 &		& \tco{smp_rmb();} \\
			 &		& x = a; \\
			 &		& assert(z == 0 || x == 1); \\
\end{tabular}}
\caption{Memory Barrier Example 1}
\label{lst:app:whymb:Memory Barrier Example 1}
\end{listing}

Suppose CPU~0 recently experienced many cache misses, so that its
message queue is full, but that CPU~1 has been running exclusively within
the cache, so that its message queue is empty.
Then CPU~0's assignment to ``a'' and ``b'' will appear in Node~0's cache
immediately (and thus be visible to CPU~1), but will be blocked behind
CPU~0's prior traffic.
In contrast, CPU~1's assignment to ``c'' will sail through CPU~1's
previously empty queue.
Therefore, CPU~2 might well see CPU~1's assignment to ``c'' before
it sees CPU~0's assignment to ``a'', causing the assertion to fire,
despite the memory barriers.

Therefore, portable code cannot rely on this assertion not firing,
as both the compiler and the CPU can reorder the code so as to trip
the assertion.

\QuickQuiz{
	Could this code be fixed by inserting a memory barrier
	between CPU~1's ``while'' and assignment to ``c''?
	Why or why not?
}\QuickQuizAnswer{
	No.
	Such a memory barrier would only force ordering local to CPU~1.
	It would have no effect on the relative ordering of CPU~0's and
	CPU~1's accesses, so the assertion could still fail.
	However, all mainstream computer systems provide one mechanism
	or another to provide ``transitivity'', which provides
	intuitive causal ordering:
	If B saw the effects of A's accesses, and C saw the effects of
	B's accesses, then C must also see the effects of A's accesses.
	In short, hardware designers have taken at least a little pity
	on software developers.
}\QuickQuizEnd

\subsection{Example 2}
\label{sec:app:whymb:Example 2}

\Cref{lst:app:whymb:Memory Barrier Example 2}
shows three code fragments, executed concurrently by CPUs~0, 1, and 2.
Both ``a'' and ``b'' are initially zero.

\begin{listing}
\scriptsize
\centering{\tt
\begin{tabular}{l|l|l}
	\multicolumn{1}{c|}{\nf{CPU~0}} &
		\multicolumn{1}{c|}{\nf{CPU~1}} &
			\multicolumn{1}{c}{\nf{CPU~2}} \\
	\hline
	\hline
	a = 1;	     & while (a == 0); & \\
		     & \tco{smp_mb();}	& y = b; \\
		     & b = 1;		& \tco{smp_rmb();} \\
		     &			& x = a; \\
		     &			& assert(y == 0 || x == 1); \\
\end{tabular}}
\caption{Memory Barrier Example 2}
\label{lst:app:whymb:Memory Barrier Example 2}
\end{listing}

Again, suppose CPU~0 recently experienced many cache misses, so that its
message queue is full, but that CPU~1 has been running exclusively within
the cache, so that its message queue is empty.
Then CPU~0's assignment to ``a'' will appear in Node~0's cache
immediately (and thus be visible to CPU~1), but will be blocked behind
CPU~0's prior traffic.
In contrast, CPU~1's assignment to ``b'' will sail through CPU~1's
previously empty queue.
Therefore, CPU~2 might well see CPU~1's assignment to ``b'' before
it sees CPU~0's assignment to ``a'', causing the assertion to fire,
despite the memory barriers.

In theory, portable code should not rely on this example code fragment,
however, as before, in practice it actually does work on most
mainstream computer systems.

\subsection{Example 3}
\label{sec:app:whymb:Example 3}

\Cref{lst:app:whymb:Memory Barrier Example 3}
shows three code fragments, executed concurrently by CPUs~0, 1, and 2.
All variables are initially zero.

\begin{listing*}
\scriptsize
\centering{\tt
\begin{tabular}{r|l|l|l}
	& \multicolumn{1}{c|}{\nf{CPU~0}} &
		\multicolumn{1}{c|}{\nf{CPU~1}} &
			\multicolumn{1}{c}{\nf{CPU~2}} \\
	\hline
	\hline
 1 &	a = 1; &			& \\
 2 &	\tco{smp_wmb();}&		& \\
 3 &	b = 1;		& while (b == 0); & while (b == 0); \\
 4 &			& \tco{smp_mb();}& \tco{smp_mb();} \\
 5 &			& c = 1;	& d = 1; \\
 6 &	while (c == 0); &		& \\
 7 &	while (d == 0); &		& \\
 8 &	\tco{smp_mb();}	&		& \\
 9 &	e = 1; &			& assert(e == 0 || a == 1); \\
\end{tabular}}
\caption{Memory Barrier Example 3}
\label{lst:app:whymb:Memory Barrier Example 3}
\end{listing*}

\floatstyle{ruled}
\restylefloat{listing}

Note that neither CPU~1 nor CPU~2 can proceed to line~5 until they see
CPU~0's assignment to ``b'' on line~3.
Once CPU~1 and~2 have executed their memory barriers on line~4, they
are both guaranteed to see all assignments by CPU~0 preceding its memory
barrier on line~2.
Similarly, CPU~0's memory barrier on line~8 pairs with those of CPUs~1 and~2
on line~4, so that CPU~0 will not execute the assignment to ``e'' on
line~9 until after its assignment to ``b'' is visible to both of the
other CPUs.
Therefore, CPU~2's assertion on line~9 is guaranteed \emph{not} to fire.

\QuickQuizSeries{%
\QuickQuizB{
	Suppose that lines~3--5 for CPUs~1 and~2 in
	\cref{lst:app:whymb:Memory Barrier Example 3}
	are in an interrupt
	handler, and that the CPU~2's line~9 runs at process level.
	In other words, the code in all three columns of the table
	runs on the same CPU, but the first two columns run in an
	interrupt handler, and the third column runs at process
	level, so that the code in third column can be interrupted
	by the code in the first two columns.
	What changes, if any, are required to enable the code to work
	correctly, in other words, to prevent the assertion from firing?
}\QuickQuizAnswerB{
	The assertion must ensure that the load of
	``e'' precedes that of ``a''.
	In the Linux kernel, the \co{barrier()} primitive may be used to
	accomplish this in much the same way that the memory barrier was
	used in the assertions in the previous examples.
	For example, the assertion can be modified as follows:

\begin{VerbatimU}[fontsize=\footnotesize]
r1 = e;
barrier();
assert(r1 == 0 || a == 1);
\end{VerbatimU}

	No changes are needed to the code in the first two columns,
	because interrupt handlers run atomically from the perspective
	of the interrupted code.
}\QuickQuizEndB
\QuickQuizE{
	If CPU~2 executed an \co{assert(e==0||c==1)} in the example in
	\cref{lst:app:whymb:Memory Barrier Example 3},
	would this assert ever trigger?
}\QuickQuizAnswerE{
	The result depends on whether the CPU supports ``transitivity''.
	In other words, CPU~0 stored to ``e'' after seeing CPU~1's
	store to ``c'', with a memory barrier between CPU~0's load
	from ``c'' and store to ``e''.
	If some other CPU sees CPU~0's store to ``e'', is it also
	guaranteed to see CPU~1's store?

	All CPUs I am aware of claim to provide transitivity.
}\QuickQuizEndE
}

The Linux kernel's \co{synchronize_rcu()} primitive uses an algorithm
similar to that shown in this example.

\section{Are Memory Barriers Forever?}
\label{sec:app:whymb:Are Memory Barriers Forever?}

There have been a number of recent systems that are significantly less
aggressive about out-of-order execution in general and re-ordering
memory references in particular.
Will this trend continue to the point where memory barriers are a thing
of the past?

The argument in favor would cite proposed massively multi-threaded hardware
architectures, so that each thread would wait until memory was ready,
with tens, hundreds, or even thousands of other threads making progress
in the meantime.
In such an architecture, there would be no need for memory barriers,
because a given thread would simply wait for all outstanding operations
to complete before proceeding to the next instruction.
Because there would be potentially thousands of other threads, the
CPU would be completely utilized, so no CPU time would be wasted.

The argument against would cite the extremely limited number of applications
capable of scaling up to a thousand threads, as well as increasingly
severe realtime requirements, which are in the tens of microseconds
for some applications.
The realtime-response requirements are difficult enough to meet as is,
and would be even more difficult to meet given the extremely low
single-threaded throughput implied by the massive multi-threaded
scenarios.

Another argument in favor would cite increasingly sophisticated
latency-hiding hardware implementation techniques that might well allow
the CPU to provide the illusion of fully sequentially consistent
execution while still providing almost all of the performance advantages
of out-of-order execution.
A counter-argument would cite the increasingly severe power-efficiency
requirements presented both by battery-operated devices and by
environmental responsibility.

Who is right?
We have no clue, so we are preparing to live with either scenario.

\section{Advice to Hardware Designers}
\label{sec:app:whymb:Advice to Hardware Designers}

There are any number of things that hardware designers can do
to make the lives of software people difficult.
Here is a list of a few such things that we have encountered in
the past, presented here in the hope that it might help prevent
future such problems:
\begin{enumerate}
\item	I/O devices that ignore cache coherence.

	This charming misfeature can result in DMAs from memory
	missing recent changes to the output buffer, or, just as
	bad, cause input buffers to be overwritten by the contents
	of CPU caches just after the DMA completes.
	To make your system work in face of such misbehavior,
	you must carefully flush the CPU caches of any location
	in any DMA buffer before presenting that buffer to the
	I/O device.
	Otherwise, a store from one of the CPUs might not be
	accounted for in the data DMAed out through the device.
	This is a form of data corruption, which is an extremely
	serious bug.

	Similarly, you need to invalidate\footnote{
		Why not flush?
		If there is a difference, then a CPU must have incorrectly
		stored to the DMA buffer in the midst of the DMA operation.}
	the CPU caches corresponding to any location in any DMA buffer
	after DMA to that buffer completes.
	Otherwise, a given CPU might see the old data still residing in
	its cache instead of the newly DMAed data that it was supposed
	to see.
	This is another form of data corruption.

	And even then, you need to be \emph{very} careful to avoid
	pointer bugs, as even a misplaced read to an input buffer
	can result in corrupting the data input!
	One way to avoid this is to invalidate all of the caches of
	all of the CPUs once the DMA completes, but it is much easier
	and more efficient if the device DMA participates in the
	cache-coherence protocol, making all of this flushing and
	invalidating unnecessary.

\item	External busses that fail to transmit cache-coherence data.

	This is an even more painful variant of the above problem,
	but causes groups of devices---and even memory itself---to
	fail to respect cache coherence.
	It is my painful duty to inform you that as embedded systems
	move to multicore architectures, we will no doubt see a fair
	number of such problems arise.
	By the year 2021, there were some efforts to address
	these problems with new interconnect standards, with some
	debate as to how effective these standards will really
	be~\cite{WilliamGWong2019CCIX-CXL}.

\item	Device interrupts that ignore cache coherence.

	This might sound innocent enough---after all, interrupts
	aren't memory references, are they?
	But imagine a CPU with a split cache, one bank of which is
	extremely busy, therefore holding onto the last cacheline
	of the input buffer.
	If the corresponding I/O-complete interrupt reaches this
	CPU, then that CPU's memory reference to the last cache
	line of the buffer could return old data, again resulting
	in data corruption, but in a form that will be invisible
	in a later crash dump.
	By the time the system gets around to dumping the offending
	input buffer, the DMA will most likely have completed.

\item	\IXAcrfpl{ipi} that ignore cache coherence.

	This can be problematic if the IPI reaches its destination
	before all of the cache lines in the corresponding message
	buffer have been committed to memory.

\item	Context switches that get ahead of cache coherence.

	If memory accesses can complete too wildly out of order,
	then context switches can be quite harrowing.
	If the task flits from one CPU to another before all the
	memory accesses visible to the source CPU make it to the
	destination CPU, then the task could easily see the corresponding
	variables revert to prior values, which can fatally confuse
	most algorithms.

\item	Overly kind simulators and emulators.

	It is difficult to write simulators or emulators that force
	memory re-ordering, so software that runs just fine in
	these environments can get a nasty surprise when it first
	runs on the real hardware.
	Unfortunately, it is still the rule that the hardware is more
	devious than are the simulators and emulators, but we hope that
	this situation changes.
\end{enumerate}

Again, we encourage hardware designers to avoid these practices!

	\setlength{\parskip}{0.0pt plus 1ex}
	\input{qqzwhymb}
	\setlength{\parskip}{0.0pt plus 1.0pt}% return to default

% appendix/styleguide/styleguide.tex
% mainfile: ../../perfbook.tex
% SPDX-License-Identifier: CC-BY-SA-3.0

\chapter{Style Guide}
\label{chp:app:styleguide:Style Guide}
\Epigraph{De gustibus non est disputandum.}{Latin maxim}

This appendix is a collection of style guides which is intended
as a reference to improve consistency in perfbook.
It also contains several suggestions and their experimental examples.

\Cref{sec:app:styleguide:Paul's Conventions} describes basic
punctuation and spelling rules.
\Cref{sec:app:styleguide:NIST Style Guide} explains rules
related to unit symbols.
\Cref{sec:app:styleguide:LaTeX Conventions} summarizes
\LaTeX-specific conventions.

\section{Paul's Conventions}
\label{sec:app:styleguide:Paul's Conventions}

Following is the list of Paul's conventions assembled from his
answers to Akira's questions regarding perfbook's punctuation policy.

\begin{itemize}
\item (On punctuations and quotations)
  Despite being American myself, for this sort of book, the UK approach
  is better because it removes ambiguities like the following:
  \begin{quote}
    Type ``\nbco{ls -a},'' look for the file ``\co{.},''
    and file a bug if you don't see it.
  \end{quote}

  The following is much more clear:
  \begin{quote}
    Type ``\nbco{ls -a}'', look for the file ``\co{.}'',
    and file a bug if you don't see it.
  \end{quote}
\item American English spelling: ``color'' rather than ``colour''.
\item Oxford comma: ``a, b, and~c'' rather than ``a, b and~c''.
  This is arbitrary.
  Cases where the Oxford comma results in ambiguity should be reworded,
  for example, by introducing numbering:  ``a, b, and c and~d'' should
  be ``(1)~a, (2)~b, and (3)~c and~d''.
\item Italic for emphasis.
  Use sparingly.
\item \verb|\co{}| for identifiers, \verb|\url{}| for URLs,
  \verb|\path{}| for filenames.
\item Dates should use an unambiguous format.
  Never ``mm/dd/yy'' or ``dd/mm/yy'', but rather ``July 26, 2016''
  or ``26 July 2016'' or ``26-Jul-2016'' or ``2016/07/26''.
  I tend to use \path{yyyy.mm.ddA} for filenames, for example.
\item North American rules on periods and abbreviations.
  For example neither of the following can reasonably be interpreted
  as two sentences:
  \begin{itemize}
  \item Say hello, to Mr.~Jones.
  \item If it looks like she sprained her ankle, call Dr.~Smith and
    then tell her to keep the ankle iced and elevated.
  \end{itemize}

  An ambiguous example:
  \begin{quote}
    If I take the cow, the pig, the horse, etc. George will be upset.
  \end{quote}
  can be written with more words:
  \begin{quote}
    If I take the cow, the pig, the horse, or much of anything else,
    George will be upset.
  \end{quote}
  or:
  \begin{quote}
    If I take the cow, the pig, the horse, etc., George will be upset.
  \end{quote}
\item I don't like ampersand (``\&'') in headings, but will sometimes
  use it if doing so prevents a line break in that heading.
\item When mentioning words, I use quotations.
  When introducing a new word, I use \verb|\emph{}|.
\end{itemize}

Following is a convention regarding punctuation in \LaTeX\ sources.

\begin{itemize}
\item Place a newline after a colon (\co{:}) and the end of a sentence.
  This avoids the whole one-space/two-space food fight and also has
  the advantage of more clearly showing changes to single sentences
  in the middle of long paragraphs.
\end{itemize}

\section{NIST Style Guide}
\label{sec:app:styleguide:NIST Style Guide}

\subsection{Unit Symbol}
\label{sec:app:styleguide:Unit Symbol}

\subsubsection{SI Unit Symbol}
\label{sec:app:styleguide:SI Unit Symbol}

NIST style guide~\cite[Chapter 5]{NIST:SP:330:2019}
states the following rules (rephrased for perfbook).

\begin{itemize}
\item When SI unit symbols such as ``ns'', ``MHz'', and ``K'' (kelvin)
are used behind numerical values, narrow spaces should be placed between
the values and the symbols.

A narrow space can be coded in \LaTeX{} by the sequence of \qco{\\,}.
For example,
\begin{quote}
  ``2.4\,GHz'', rather then ``2.4GHz''.
\end{quote}

\item Even when the value is used in adjectival sense, a narrow space
  should be placed.
  For example,
\begin{quote}
  ``a~10\,ms interval'', rather than ``a~10\=/ms interval'' nor
  ``a~10ms interval''.
\end{quote}
\end{itemize}

The symbol of micro (\micro :$10^{-6}$) can be typeset easily by
the help of ``gensymb'' \LaTeX\ package.
A macro \qco{\\micro} can be used in both text and math modes.
To typeset the symbol of ``microsecond'', you can do
so by \qco{\\micro s}.
For example,
\begin{quote}
  10\,\micro s
\end{quote}

Note that math mode \qco{\\mu} is italic by default and should not
be used as a prefix.
An improper example:
\begin{quote}
  10\,$\mu $s (math mode \qco{\\mu})
\end{quote}

\subsubsection{Non-SI Unit Symbol}
\label{sec:app:styleguide:Non-SI Unit Symbol}

Although NIST style guide does not cover non-SI unit symbols
such as ``KB'', ``MB'', and ``GB'', the same rule should be followed.

Example:

\begin{quote}
  ``A~240\,GB hard drive'', rather than ``a~240\=/GB hard drive''
  nor ``a~240GB hard drive''.
\end{quote}

Strictly speaking, NIST guide requires us to use the binary prefixes
``Ki'', ``Mi'', or ``Gi'' to represent powers of~$2^{10}$.
However, we accept the JEDEC conventions to use ``K'', ``M'',
and ``G'' as binary prefixes in describing memory
capacity~\cite{JEDEC:dict:prefixmega}.

An acceptable example:
\begin{quote}
  ``8\,GB of main memory'', meaning ``8\,GiB of main memory''.
\end{quote}

Also, it is acceptable to use just ``K'', ``M'', or ``G'' as abbreviations
appended to a numerical value, e.g., ``4K~entries''.
In such cases, no space before an abbreviation is required.
For example,

\begin{quote}
  ``8K entries'', rather than ``8\,K entries''.
\end{quote}

If you put a space in between, the symbol looks like a unit symbol and
is confusing.
Note that ``K'' and ``k'' represent $2^{10}$ and $10^3$, respectively.
``M'' can represent either $2^{20}$ or $10^6$, and ``G'' can represent
either $2^{30}$ or $10^9$.
These ambiguities should not be confusing in discussing approximate order.

\subsubsection{Degree Symbol}
\label{sec:app:styleguide:Degree Symbol}

The angular-degree symbol (\degree) does not require any space in front
of it.
NIST style guide clearly states so.

The symbol of degree can also be typeset easily by the help of gensymb
package.
A macro \qco{\\degree} can be used in both text and math modes.

Example:

\begin{quote}
  $45\degree$, rather than $45\,\degree$.
\end{quote}

\subsubsection{Percent Symbol}
\label{sec:app:styleguide:Percent Symbol}

NIST style guide treats the percent symbol (\%) as the same as SI unit
symbols.

\begin{quote}
  50\,\% possibility, rather than 50\% possibility.
\end{quote}

\subsubsection{Font Style}
\label{sec:app:styleguide:Font Style}

Quote from NIST check list~\cite[\#6]{NIST:UnitCheckList}:

\begin{quote}
  Variables and quantity symbols are in italic type.
  Unit symbols are in roman type.
  Numbers should generally be written in roman type.
  These rules apply irrespective of the typeface used in the surrounding
  text.
\end{quote}

For example,
\begin{quote}
  {\textit e} (elementary charge)
\end{quote}

On the other hand, mathematical constants such as the base
of natural logarithms should be roman~\cite{NIST:TypeFaces}.
For example,

\begin{quote}
  $\mathrm{e}^x$
\end{quote}

%\footnote{
%  See \url{https://tex.stackexchange.com/questions/119248/}
%  for the historical reason.}

\subsection{NIST Guide Yet To Be Followed}
\label{sec:app:styleguide:NIST Guides Yet To Be Followed}

There are a few cases where NIST style guide is not followed.
Other English conventions are followed in such cases.

\subsubsection{Digit Grouping}
\label{sec:app:styleguide:Digit Grouping}

Quote from NIST checklist~\cite[\#16]{NIST:UnitCheckList}:

\begin{quote}
  The digits of numerical values having more than four digits on either
  side of the decimal marker are separated into groups of three using
  a thin, fixed space counting from both the left and right of the decimal
  marker.
  Commas are not used to separate digits into groups of three.
\end{quote}

\begin{quote}
\begin{tabular}{ll}
  NIST Example:& 15\,739.012\,53\,ms\\
  Our convention:& 15,739.01253\,ms\\
\end{tabular}
\end{quote}

In \LaTeX\ coding, it is cumbersome to place thin spaces as are recommended
in NIST guide.
The \verb|\num{}| command provided by the ``siunitx'' package would be
of help for us to follow this rule.
It would also help us overcome different conventions.
We can select a specific digit-grouping style as
a default in preamble, or specify an option to each \verb|\num{}|
command as is shown in
\cref{tab:app:styleguide:Digit-Grouping Style}.

\newcommand{\NumDigitGrpA}{12 345}
\newcommand{\NumDigitGrpB}{12.345}
\newcommand{\NumDigitGrp}{1 234 567.89}
\begin{table}
\small\centering
\begin{tabular}{lrrr}\toprule
  Style & \multicolumn{3}{c}{Outputs of \co{\\num\{\}}} \\
  \midrule
  NIST/SI (English) & \num[group-separator={\,},group-digits=integer]{\NumDigitGrpA} &
    \num[group-separator={\,},group-digits=integer]{\NumDigitGrpB} &
      \num[group-separator={\,},group-digits=integer]{\NumDigitGrp} \\
  SI (French) & \num[locale=FR,group-separator={\,}]{\NumDigitGrpA} &
    \num[locale=FR,group-separator={\,}]{\NumDigitGrpB} &
      \num[locale=FR,group-separator={\,}]{\NumDigitGrp} \\
  English & \num[group-separator={,},group-digits=integer]{\NumDigitGrpA} &
    \num[group-separator={,},group-digits=integer]{\NumDigitGrpB} &
      \num[group-separator={,},group-digits=integer]{\NumDigitGrp} \\
  French & \num[locale=FR,group-separator={\,}]{\NumDigitGrpA} &
    \num[locale=FR,group-separator={\,}]{\NumDigitGrpB} &
      \num[locale=FR,group-separator={\,}]{\NumDigitGrp} \\
  Other Europe & \num[group-separator={.},output-decimal-marker={,},group-digits=integer]{\NumDigitGrpA} &
    \num[group-separator={.},output-decimal-marker={,},group-digits=integer]{\NumDigitGrpB} &
      \num[group-separator={.},output-decimal-marker={,},group-digits=integer]{\NumDigitGrp} \\
\bottomrule
\end{tabular}
\caption{Digit-Grouping Style}
\label{tab:app:styleguide:Digit-Grouping Style}
\end{table}

As are evident in
\cref{tab:app:styleguide:Digit-Grouping Style},
periods and commas used as other than decimal markers are confusing
and should be avoided, especially in documents expecting global
audiences.

By marking up constant decimal values by \verb|\num{}| commands,
the \LaTeX\ source would be exempted from any particular conventions.

Because of its open-source policy, this approach should give
more ``portability'' to perfbook.

\section{\LaTeX\ Conventions}
\label{sec:app:styleguide:LaTeX Conventions}

Good looking \LaTeX\ documents require further considerations
on proper use of font styles, line break exceptions, etc.
This section summarizes guidelines specific to \LaTeX.

\subsection{Monospace Font}
\label{sec:app:styleguide:Monospace Font}

Monospace font (or typewriter font) is heavily used in this textbook.
First policy regarding monospace font in perfbook is to avoid
directly using \qco{\\texttt} or \qco{\\tt} macro.
It is highly recommended to use a macro or an environment
indicating the reason why you want the font.

This section explains the use cases of such macros and environments.

\subsubsection{Code Snippet}
\label{sec:app:styleguide:Code Snippet}

Because the \qco{verbatim} environment is a primitive way to include
listings, we have transitioned to a scheme which uses
the \qco{fancyvrb} package for code snippets.

The goal of the scheme is to extract \LaTeX\ sources of
code snippets directly from code samples under \path{CodeSamples}
directory.
It also makes it possible to embed line labels in the code samples,
which can be referenced within the \LaTeX\ sources.
This reduces the burden of keeping line numbers
in the text consistent with those in code snippets.

Code-snippet extraction is handled by a couple of
perl scripts and recipes in Makefile.
We use the escaping feature of the \co{fancyvrb} package
to embed line labels as comments.

We used to use the \qco{verbbox} environment provided
by the \qco{verbatimbox} package.
\Cref{sec:app:styleguide:Code Snippet (Obsolete)} describes how
\co{verbbox} can automatically generate line numbers,
but those line numbers cannot be referenced within the \LaTeX\ sources.

Let's start by looking at how code snippets are coded in the current scheme.
There are three customized environments of \qco{Verbatim}.
\qco{VerbatimL} is for floating snippets within the \qco{listing}
environment.
\qco{VerbatimN} is for inline snippets with line count enabled.
\qco{VerbatimU} is for inline snippets without line count.
They are defined in the preamble as shown below:

\begin{VerbatimU}
\DefineVerbatimEnvironment{VerbatimL}{Verbatim}%
  {fontsize=\scriptsize,numbers=left,numbersep=5pt,%
    xleftmargin=9pt,obeytabs=true,tabsize=2}
\AfterEndEnvironment{VerbatimL}{\vspace*{-9pt}}
\DefineVerbatimEnvironment{VerbatimN}{Verbatim}%
  {fontsize=\scriptsize,numbers=left,numbersep=3pt,%
    xleftmargin=5pt,xrightmargin=5pt,obeytabs=true,%
    tabsize=2,frame=single}
\DefineVerbatimEnvironment{VerbatimU}{Verbatim}%
  {fontsize=\scriptsize,numbers=none,xleftmargin=5pt,%
    xrightmargin=5pt,obeytabs=true,tabsize=2,%
    samepage=true,frame=single}
\end{VerbatimU}

% Another option would be the ``lstlisting'' environment provided
%  by the ``listings'' package. We are already using its ``lstinline''
%  command in the definition of \co{\\co\{\}} macro.

\begin{listing}
\fvset{fontsize=\scriptsize,numbers=left,numbersep=5pt,xleftmargin=9pt,obeytabs=true,tabsize=8,commandchars=\%\~\^}
\begin{fcvlabel}[ln:app:styleguide:LaTeX Source of Sample Code Snippet (Current)]
\begin{Verbatim}
\begin{listing}
\begin{fcvlabel}[ln:base1]			%lnlbl~beg:fcvlabel^
\begin{VerbatimL}[commandchars=\$\[\]]
/*
 * Sample Code Snippet
 */
#include <stdio.h>
int main(void)
{
	printf("Hello world!\n");	$lnlbl[printf]
	return 0;			$lnlbl[return]
}
\end{VerbatimL}
\end{fcvlabel}					%lnlbl~end:fcvlabel^
\caption{Sample Code Snippet}
\label{lst:app:styleguide:Sample Code Snippet}
\end{listing}
\end{Verbatim}
\end{fcvlabel}
\vspace*{-9pt}
\caption{\LaTeX\ Source of Sample Code Snippet (Current)}
\label{lst:app:styleguide:LaTeX Source of Sample Code Snippet (Current)}
\end{listing}

\begin{listing}
\begin{fcvlabel}[ln:base1]			%lnlbl~beg:fcvlabel^
\begin{VerbatimL}[commandchars=\$\[\]]
/*
 * Sample Code Snippet
 */
#include <stdio.h>
int main(void)
{
	printf("Hello world!\n");	$lnlbl[printf]
	return 0;			$lnlbl[return]
}
\end{VerbatimL}
\end{fcvlabel}					%lnlbl~end:fcvlabel^
\caption{Sample Code Snippet}
\label{lst:app:styleguide:Sample Code Snippet}
\end{listing}

The \LaTeX\ source of a sample code snippet is shown in
\cref{lst:app:styleguide:LaTeX Source of Sample Code Snippet (Current)}
and is typeset as shown in
\cref{lst:app:styleguide:Sample Code Snippet}.

Labels to lines are specified in \qco{$lnlbl[]} command.
The characters specified by \qco{commandchars} option to \co{VarbatimL}
environment are used by the \co{fancyvrb} package to substitute
\qco{\\lnlbl\{\}} for \qco{$lnlbl[]}.
Those characters should be selected so that they don't appear
elsewhere in the code snippet.

Labels \qco{printf} and \qco{return} in
\cref{lst:app:styleguide:Sample Code Snippet}
can be referred to as shown below:

\begin{VerbatimU}
\begin{fcvref}[ln:base1]
\Clnref{printf, return} can be referred
to from text.
\end{fcvref}
\end{VerbatimU}

Above code results in the paragraph below:

\begin{quote}
\begin{fcvref}[ln:base1]
\Clnref{printf, return} can be referred
to from text.
\end{fcvref}
\end{quote}

Macros \qco{\\lnlbl\{\}} and \qco{\\lnref\{\}} are defined in
the preamble as follows:

\begin{VerbatimU}
\newcommand{\lnlblbase}{}
\newcommand{\lnlbl}[1]{%
  \phantomsection\label{\lnlblbase:#1}}
\newcommand{\lnrefbase}{}
\newcommand{\lnref}[1]{\ref{\lnrefbase:#1}}
\end{VerbatimU}

Environments \qco{fcvlabel} and \qco{fcvref} are defined as
shown below:

\begin{VerbatimU}
\newenvironment{fcvlabel}[1][]{%
  \renewcommand{\lnlblbase}{#1}%
  \ignorespaces}{\ignorespacesafterend}
\newenvironment{fcvref}[1][]{%
  \renewcommand{\lnrefbase}{#1}%
  \ignorespaces}{\ignorespacesafterend}
\end{VerbatimU}

\begin{fcvref}[ln:app:styleguide:LaTeX Source of Sample Code Snippet (Current)]
The main part of \LaTeX\ source shown on
\clnrefrange{beg:fcvlabel}{end:fcvlabel} in
\cref{lst:app:styleguide:LaTeX Source of Sample Code Snippet (Current)}
can be extracted from a code sample of
\cref{lst:app:styleguide:Source of Code Sample} by a perl script
\path{utilities/fcvextract.pl}.
All the relevant rules of extraction are described as recipes in the
top level \path{Makefile} and a script to generate dependencies
(\path{utilities/gen_snippet_d.pl}).
\end{fcvref}

\begin{listing*}
\fvset{fontsize=\scriptsize,numbers=left,numbersep=5pt,xleftmargin=9pt,obeytabs=true,tabsize=8}
\begin{Verbatim}
//\begin{snippet}[labelbase=ln:base1,keepcomment=yes,commandchars=\$\[\]]
/*
 * Sample Code Snippet
 */
#include <stdio.h>
int main(void)
{
	printf("Hello world!\n");	//\lnlbl{printf}
	return 0;			//\lnlbl{return}
}
//\end{snippet}
\end{Verbatim}
\vspace*{-9pt}
\caption{Source of Code Sample with ``snippet'' Meta Command}
\label{lst:app:styleguide:Source of Code Sample}
\end{listing*}

As you can see, \cref{lst:app:styleguide:Source of Code Sample}
has meta commands in comments of C (C++ style).
Those meta commands are interpreted by \path{utilities/fcvextract.pl},
which distinguishes the type of comment style by the suffix of code
sample's file name.

Meta commands which can be used in code samples are listed below:

\begin{itemize}[noitemsep]
\item \co{\\begin\{snippet\}[<options>]}
\item \co{\\end\{snippet\}}
\item \co{\\lnlbl\{<label string>\}}
\item \co{\\fcvexclude}
\item \co{\\fcvblank}
\end{itemize}

\qco{<options>} to the \co{\\begin\{snippet\}} meta command
is a comma-spareted list of options shown below:

\begin{itemize}[noitemsep]
\item \co{labelbase=<label base string>}
\item \co{keepcomment=yes}
\item \co{gobbleblank=yes}
\item \co{commandchars=\\X\\Y\\Z}
\end{itemize}

The \qco{labelbase} option is mandatory and the string given to it
will be passed to the
``\co{\\begin\{fcvlabel\}[<label base string>]}'' command as shown on
\clnrefr{ln:app:styleguide:LaTeX Source of Sample Code Snippet (Current):beg:fcvlabel} of
\cref{lst:app:styleguide:LaTeX Source of Sample Code Snippet (Current)}.
The \qco{keepcomment=yes} option tells \co{fcvextract.pl} to keep
comment blocks.
Otherwise, comment blocks in C source code will be omitted.
The \qco{gobbleblank=yes} option will remove empty or blank lines
in the resulting snippet.
The \qco{commandchars} option is given to the \co{VerbatimL} environment
as is.
At the moment, it is also mandatory and must come at the end of options
listed above.
Other types of options, if any, are also passed to the \co{VerbatimL}
environment.

The \qco{\\lnlbl} commands are converted along the way to reflect
the escape-character choice.\footnote{
	Characters forming comments around the \qco{\\lnlbl} commands
	are also gobbled up regardless of the \qco{keepcomment} setting.
}
Source lines with \qco{\\fcvexclude} are removed.
\qco{\\fcvblank} can be used to keep blank lines when the
\qco{gobbleblank=yes} option is specified.

There can be multiple pairs of \co{\\begin\{snippet\}} and \co{\\end\{snippet\}}
as long as they have unique \qco{labelbase} strings.

Our naming scheme of \qco{labelbase} for unique name space is as follows:

\begin{VerbatimU}
ln:<Chapter/Subdirectory>:<File Name>:<Function Name>
\end{VerbatimU}

Litmus tests, which are handled by \qco{herdtools7} commands such as
\qco{litmus7} and \qco{herd7}, were problematic in this scheme.
Those commands have particular rules of where comments can be
placed and restriction on permitted characters in comments.
They also forbid a couple of tokens to appear in comments.
(Tokens in comments might sound strange, but they do have such restriction.)

For example, the first token in a litmus test must be one of
\qco{C}, \qco{PPC}, \qco{X86}, \qco{LISA}, etc., which indicates
the flavor of the test.
This means no comment is allowed at the beginning of a litmus test.

Similarly, several tokens such as \qco{exists}, \qco{filter},
and~\qco{locations} indicate the end of litmus test's body.
Once one of them appears in a litmus test, comments should be of
OCaml style (\qco{(* ... *)}).
Those tokens keep the same meaning even when they appear in comments!

The pair of characters \qco{\{} and \qco{\}} also have special
meaning in the C flavour tests.
They are used to separate portions in a litmus test.

First pair of \qco{\{} and \qco{\}} encloses initialization part.
Comments in this part should also be in the ocaml form.

You can't use \qco{\{} and \qco{\}} in comments in litmus tests, either.

Examples of disallowed comments in a litmus test are shown below:

\begin{fcvlabel}[ln:app:styleguide:Bad comments in Litmus Test]
\begin{VerbatimN}[tabsize=8]
// Comment at first
C C-sample
// Comment with { and } characters
{
x=2;  // C style comment in initialization
}

P0(int *x}
{
	int r1;

	r1 = READ_ONCE(*x);  // Comment with "exists"
}

[...]

exists (0:r1=0)  // C++ style comment after test body
\end{VerbatimN}
\end{fcvlabel}

To avoid parse errors, meta commands in litmus tests (C flavor) are embedded
in the following way.

\begin{fcvlabel}[ln:app:styleguide:Sample Source of Litmus Test]
\begin{VerbatimN}[tabsize=8]
C C-SB+o-o+o-o
//\begin[snippet][labelbase=ln:base,commandchars=\%\@\$]

{
1:r2=0			(*\lnlbl[initr2]*)
}

P0(int *x0, int *x1)		//\lnlbl[P0:b]
{
	int r2;

	WRITE_ONCE(*x0, 2);
	r2 = READ_ONCE(*x1);
}				//\lnlbl[P0:e]

P1(int *x0, int *x1)
{
	int r2;

	WRITE_ONCE(*x1, 2);
	r2 = READ_ONCE(*x0);
}

//\end[snippet]
exists (1:r2=0 /\ 0:r2=0)  (* \lnlbl[exists_] *)
\end{VerbatimN}
\end{fcvlabel}

Example above is converted to the following intermediate code
by a script \path{utilities/reorder_ltms.pl}.\footnote{
	Currently, only C flavor litmus tests are supported.
}
The intermediate code can be handled
by the common script \path{utilities/fcvextract.pl}.

\begin{fcvlabel}[ln:app:styleguide:Intermediate Source of Litmus Test]
\begin{VerbatimN}[tabsize=8]
// Do not edit!
// Generated by utillities/reorder_ltms.pl
//\begin{snippet}[labelbase=ln:base,commandchars=\%\@\$]
C C-SB+o-o+o-o

{
1:r2=0			//\lnlbl{initr2}
}

P0(int *x0, int *x1)		//\lnlbl{P0:b}
{
	int r2;

	WRITE_ONCE(*x0, 2);
	r2 = READ_ONCE(*x1);
}				//\lnlbl{P0:e}

P1(int *x0, int *x1)
{
	int r2;

	WRITE_ONCE(*x1, 2);
	r2 = READ_ONCE(*x0);
}

exists (1:r2=0 /\ 0:r2=0)  \lnlbl{exists_}
//\end{snippet}
\end{VerbatimN}
\end{fcvlabel}

Note that each litmus test's source file can contain at most one
pair of \co{\\begin[snippet]} and \co{\\end[snippet]} because of
the restriction of comments.

\subsubsection{Code Snippet (Obsolete)}
\label{sec:app:styleguide:Code Snippet (Obsolete)}

Sample \LaTeX\ source of a code snippet coded using
the \qco{verbatimbox} package is shown in
\cref{lst:app:styleguide:LaTeX Source of Sample Code Snippet (Obsolete)}
and is typeset as shown in
\cref{lst:app:styleguide:Sample Code Snippet (Obsolete)}.

\begin{listing}
\begin{fcvlabel}[ln:app:styleguide:samplecodesnippetlstlbl]
\fvset{fontsize=\scriptsize,numbers=left,numbersep=5pt,xleftmargin=9pt,commandchars=\%\@\$}
\begin{Verbatim}
\begin{listing}
{ \scriptsize
\begin{verbbox}[\LstLineNo] %lnlbl@lineno$
/*				%lnlbl@b$
 * Sample Code Snippet
 */
#include <stdio.h>
int main(void)
{
  printf("Hello world!\n");
  return 0;
}				%lnlbl@e$
\end{verbbox}
}
\centering
\theverbbox			%lnlbl@theverbbox$
\caption{Sample Code Snippet (Obsolete)}
\label{lst:app:styleguide:Sample Code Snippet (Obsolete)}
\end{listing}
\end{Verbatim}
\end{fcvlabel}
\vspace*{-9pt}
\caption{\LaTeX\ Source of Sample Code Snippet (Obsolete)}
\label{lst:app:styleguide:LaTeX Source of Sample Code Snippet (Obsolete)}
\end{listing}

\begin{listing}
{ \scriptsize
\begin{verbbox}[\LstLineNo]
/*
 * Sample Code Snippet
 */
#include <stdio.h>
int main(void)
{
  printf("Hello world!\n");
  return 0;
}
\end{verbbox}
}
\centering
\theverbbox
\caption{Sample Code Snippet (Obsolete)}
\label{lst:app:styleguide:Sample Code Snippet (Obsolete)}
\end{listing}

The auto\-/numbering feature of \co{verbbox} is enabled by
the ``\verb|\LstLineNo|'' macro specified in the option to verbbox
(\clnrefr{ln:app:styleguide:samplecodesnippetlstlbl:lineno} in
\cref{lst:app:styleguide:LaTeX Source of Sample Code Snippet (Obsolete)}).
The macro is defined in the preamble of \path{perfbook.tex}
as follows:

\begin{VerbatimU}
\newcommand{\LstLineNo}
  {\makebox[5ex][r]{\arabic{VerbboxLineNo}\hspace{2ex}}}
\end{VerbatimU}

The \qco{verbatim} environment is used for listings with too many lines
to fit in a column.
It is also used to avoid overwhelming \LaTeX\ with a lot of floating objects.
They are being converted to the scheme using the \co{VerbatimN} environment.

\subsubsection{Identifier}
\label{sec:app:styleguide:Identifier}

We use ``\verb|\co{}|'' macro for inline identifiers.
(``co'' stands for ``code''.)

By putting them into \verb|\co{}|, underscore characters in
their names are free of escaping in \LaTeX\ source.
It is convenient to search them in source files.
Also, \verb|\co{}| macro has a capability to permit line breaks
at particular sequences of letters.
Current definition permits a line break at an underscore (\tco{_}),
two consecutive underscores (\tco{__}), a white space, or an
operator \tco{->}.

\subsubsection{Identifier inside Table and Heading}
\label{sec:app:styleguide:Identifier inside Table and Heading}

Although \verb|\co{}| command is convenient for inlining within text,
it is fragile because of its capability of line break.
When it is used inside a \qco{tabular} environment or its derivative
such as \qco{tabularx}, it confuses column width
estimation of those environments.
Furthermore, \verb|\co{}| can not be safely used in section headings nor
description headings.

As a workaround, we use ``\verb|\tco{}|'' command
inside tables and headings.
It has no capability of line break at particular sequences, but
still frees us from escaping underscores.

When used in text, \verb|\tco{}| permits line breaks at
white spaces.

\subsubsection{Other Use Cases of Monospace Font}
\label{sec:app:styleguide:Other Use Cases of Monospace Font}

For URLs, we use ``\verb|\url{}|'' command provided by the
\qco{hyperref} package.
It will generate hyper references to the URLs.

For path names, we use ``\verb|\path{}|'' command.
It won't generate hyper references.

Both \verb|\url{}| and \verb|\path{}| permit line breaks
at \qco{/}, \qco{-}, and \qco{.}.\footnote{
  Overfill can be a problem if the URL or the path name contains
  long runs of unbreakable characters.
}

For short monospace statements not to be line broken, we use
the ``\verb|\nbco{}|'' (non-breakable co) macro.

\subsubsection{Limitations}
\label{sec:app:styleguide:Limitations}

There are a few cases where macros introduced in this section
do not work as expected.
\Cref{tab:app:styleguide:Limitation of Monospace Macro}
lists such limitations.

\begin{table}
\renewcommand*{\arraystretch}{1.2}\centering\footnotesize
\begin{tabular}{@{}lll@{}}\toprule
  Macro &  Need Escape & Should Avoid \\
  \midrule
  \co{\\co}, \co{\\nbco} & \co{\\}, \%, \{, \} & \\
  \co{\\tco}  & \# & \%, \{, \}, \co{\\} \\
  \bottomrule
\end{tabular}
\caption{Limitation of Monospace Macro}
\label{tab:app:styleguide:Limitation of Monospace Macro}
\end{table}

While \verb|\co{}| requires some characters to be escaped,
it can contain any character.

On the other hand, \verb|\tco{}| can not handle
\qco{\%}, \qco{\{}, \qco{\}}, nor \qco{\\} properly.
If they are escaped by a~\qco{\\},
they appear in the end result with the escape character.
The \qco{\\verb} command can be used in running text if you
need to use monospace font for a string which contains
many characters to escape.\footnote{
  The \co{\\verb} command is not almighty though.
  For example, you can't use it within a footnote.
  If you do so, you will see a fatal \LaTeX\ error.
  A workaround would be a macro named \co{\\VerbatimFootnotes}
  provided by the \co{fancyvrb} package.
  Unfortunately, perfbook can't employ it due to the interference
  with the \co{footnotebackref} package.
  }

\subsection{Cross-reference}
\label{sec:app:styleguide:Cross-Reference}

Cross-references to \namecrefs{chp:Introduction},
\namecrefs{sec:intro:Parallel Programming Goals},
\namecrefs{lst:app:styleguide:Source of Code Sample}, etc.\@ have
been expressed by combinations of names and bare \verb|\ref{}|
commands in the following way:

\begin{VerbatimN}
Chapter~\ref{chp:Introduction},
Table~\ref{tab:app:styleguide:Digit-Grouping Style}
\end{VerbatimN}

This is a traditional way of cross\-/referencing.
However, it is tedious and sometimes error-prone to put a name
manually on every cross\-/reference.
The \co{cleveref} package provides a nicer way of cross\-/referencing.
A few examples follow:

\begin{VerbatimN}
\Cref{chp:Introduction},
\cref{sec:intro:Parallel Programming Goals},
\cref{chp:app:styleguide:Style Guide},
\cref{tab:app:styleguide:Digit-Grouping Style}, and
\cref{lst:app:styleguide:Source of Code Sample} are
examples of cross\-/references.
\end{VerbatimN}

Above code is typeset as follows:

\begin{quote}
\Cref{chp:Introduction},
\cref{sec:intro:Parallel Programming Goals},
\cref{chp:app:styleguide:Style Guide},
\cref{tab:app:styleguide:Digit-Grouping Style}, and
\cref{lst:app:styleguide:Source of Code Sample} are
examples of cross\-/references.
\end{quote}

As you can see, naming of cross\-/references is automated.
Current setting generates capitalized names for both of
\verb|\Cref{}| and \verb|\cref{}|, but the former
should be used at the beginning of a sentence.

We are in the middle of conversion to
\co{cleveref}-style cross\-/referencing.

Cross-references to line numbers of code snippets
can be done in a similar way by using \verb|\Clnref{}| and
\verb|\clnref{}| macros, which mimic \co{cleveref}.
The former puts ``Line'' as the name of the reference
and the latter ``line''.

Please refer to \co{cleveref}'s documentation for further
info on its cleverness.

\subsection{Non Breakable Spaces}
\label{sec:app:styleguide:Non Breakable Spaces}

In \LaTeX\ conventions, proper use of non-breakable white spaces
is highly recommended.
They can prevent widowing and orphaning of single digit numbers
or short variable names, which would cause the text to be confusing
at first glance.

The thin space mentioned earlier to be placed in front of a unit
symbol is non breakable.

Other cases to use a non-breakable space (``\verb|~|'' in \LaTeX\
source, often referred to as ``nbsp'')
are the following (inexhaustive).

\begin{itemize}
\item Reference to a Chapter or a Section:
  \begin{quote}
    Please refer to \cref{sec:app:styleguide:NIST Style Guide}.
  \end{quote}
\item Calling out CPU number or Thread name:
  \begin{quote}
    After they load the pointer, CPUs~1 and~2 will see the stored
    value.
  \end{quote}
\item Short variable name:
  \begin{quote}
    The results will be stored in variables~\co{a} and~\co{b}.
  \end{quote}
\end{itemize}

\subsection{Hyphenation and Dashes}
\label{sec:app:styleguide:Hyphenation and Dashes}

\subsubsection{Hyphenation in Compound Word}
\label{sec:app:styleguide:Hyphenation in Compound Word}

In plain \LaTeX, compound words such as ``high-frequency''
can be hyphenated only at the hyphen.
This sometimes results in poor typesetting.
For example:

\begin{center}\begin{minipage}{2.6in}\vspace{0.6\baselineskip}
  High-frequency radio wave, high-frequency radio wave,
  high-frequency radio wave, high-frequency radio wave,
  high-frequency radio wave, high-frequency radio wave.
\vspace{0.6\baselineskip}\end{minipage}\end{center}

By using a shortcut \qco{\\-/} provided by the
``extdash'' package, hyphenation in elements of compound
words is enabled in perfbook.\footnote{
  In exchange for enabling the shortcut, we can't use plain
  \LaTeX's shortcut \qco{\\-} to specify hyphenation points.
  Use \path{pfhyphex.tex} to add such exceptions.
}

Example with \qco{\\-/}:

\begin{center}\begin{minipage}{2.6in}\vspace{0.6\baselineskip}
  High\-/frequency radio wave, high\-/frequency radio wave,
  high\-/frequency radio wave, high\-/frequency radio wave,
  high\-/frequency radio wave, high\-/frequency radio wave.
\vspace{0.6\baselineskip}\end{minipage}\end{center}

\subsubsection{Non Breakable Hyphen}
\label{sec:app:styleguide:Non Breakable Hyphen}

We want hyphenated compound terms such as ``x\=/coordinate'',
``y\=/coordinate'', etc.\@ not to be broken at the hyphen
following a single letter.

To make a hyphen unbreakable, we can use a short cut
\qco{\\=/} also provided by the ``extdash'' package.

Example without a shortcut:

\begin{center}\begin{minipage}{2.55in}\vspace{0.6\baselineskip}
x-, y-, and z-coordinates; x-, y-, and z-coordinates;
x-, y-, and z-coordinates; x-, y-, and z-coordinates;
x-, y-, and z-coordinates; x-, y-, and z-coordinates;
\vspace{0.6\baselineskip}\end{minipage}\end{center}

Example with \qco{\\-/}:

\begin{center}\begin{minipage}{2.55in}\vspace{0.6\baselineskip}
x-, y-, and z\-/coordinates; x-, y-, and z\-/coordinates;
x-, y-, and z\-/coordinates; x-, y-, and z\-/coordinates;
x-, y-, and z\-/coordinates; x-, y-, and z\-/coordinates;
\vspace{0.6\baselineskip}\end{minipage}\end{center}

Example with \qco{\\=/}:

\begin{center}\begin{minipage}{2.55in}\vspace{0.6\baselineskip}
x-, y-, and z\=/coordinates; x-, y-, and z\=/coordinates;
x-, y-, and z\=/coordinates; x-, y-, and z\=/coordinates;
x-, y-, and z\=/coordinates; x-, y-, and z\=/coordinates;
\vspace{0.6\baselineskip}\end{minipage}\end{center}

Note that \qco{\\=/} enables hyphenation in elements
of compound words as the same as \qco{\\-/} does.

\subsubsection{Em Dash}
\label{sec:app:styleguide:Em Dash}

Em dashes are used to indicate parenthetic expression.
In perfbook, em dashes are placed without spaces around it.
In \LaTeX\ source, an em dash is represented by \qco{---}.

Example (quote from \cref{sec:app:whymb:Cache Structure}):
\begin{quote}
  This disparity in speed---more than two orders of magnitude---has
  resulted in the multi-megabyte caches found on modern CPUs.
\end{quote}

\subsubsection{En Dash}
\label{sec:app;styleguide:En Dash}

In \LaTeX\ convention, en~dashes (\==) are used for ranges of (mostly)
numbers.
Past revisions of perfbook didn't follow this rule and used
plain dashes (\=/) for such cases.

Now that \co{\\clnrefrange}, \co{\\crefrange},
and their variants, which generate en~dashes, are used for ranges of
cross\-/references, the remaining couple of tens of simple dashes
of other types of ranges have been converted to en~dashes for
consistency.

Example with a simple dash:

\begin{quote}
\begin{fcvref}[ln:app:styleguide:samplecodesnippetlstlbl]
  Lines~\lnref{b}\=/\lnref{e} in
  \cref{lst:app:styleguide:LaTeX Source of Sample Code Snippet (Obsolete)}
  are the contents of the verbbox environment.
  The box is output by the \co{\\theverbbox} macro on \clnref{theverbbox}.
\end{fcvref}
\end{quote}

Example with an en dash:

\begin{quote}
\begin{fcvref}[ln:app:styleguide:samplecodesnippetlstlbl]
  Lines~\lnref{b}\==\lnref{e} in
  \cref{lst:app:styleguide:LaTeX Source of Sample Code Snippet (Obsolete)}
  are the contents of the verbbox environment.
  The box is output by the \co{\\theverbbox} macro on \clnref{theverbbox}.
\end{fcvref}
\end{quote}

\subsubsection{Numerical Minus Sign}
\label{sec:app:styleguide:Numerical Minus Sign}

Numerical minus signs should be coded as math mode minus signs,
namely \qco{$-$}.\footnote{This rule assumes that math mode uses the
  same upright glyph as text mode.
  Our default font choice meets the assumption.
\IfSansSerif{
  Experimental targets with sans-serif fonts such as
  ``1csf'' and ``ebsf'' \emph{do} use a different font
  for math mode figures as of August 2021.}{}
}
For example,

\begin{quote}
  $-30$, rather than -30.
\end{quote}

\subsection{Punctuation}
\label{sec:app:styleguide:Punctuation}

\subsubsection{Ellipsis}
\label{sec:app:styleguide:Ellipsis}

In monospace fonts, ellipses can be expressed by series of periods.
For example:

\begin{quote}
  \verb|Great ... So how do I fix it?|
\end{quote}

However, in proportional fonts, the series of periods is printed
with tight spaces as follows:

\begin{quote}
  Great ... So how do I fix it?
\end{quote}

Standard \LaTeX\ defines the \verb|\dots| macro for this purpose.
However, it has a kludge in the evenness of spaces.
The ``ellipsis'' package redefines the \verb|\dots| macro to fix
the issue.\footnote{To be exact, it is the \co{\\textellipsis} macro
  that is redefined.
  The behavior of \co{\\dots} macro in math mode is not affected.
  The ``amsmath'' package has another definition of \co{\\dots}.
  It is not used in perfbook at the moment.}
By using \verb|\dots|, the above example is typeset as the following:

\begin{quote}
  Great \dots So how do I fix it?
\end{quote}

Note that the ``xspace'' option specified to the ``ellipsis'' package
adjusts the spaces after ellipses depending on what follows them.

For example:

\begin{itemize}[itemsep=.2ex]
\item He said, ``I~\dots really don't remember~\dots''
\item Sequence A: (one, two, three, \dots)
\item Sequence B: (4, 5, \dots, $n$)
\end{itemize}

As you can see, extra space is placed before the comma.

\verb|\dots| macro can also be used in math mode:

\begin{itemize}[itemsep=.2ex]
\item Sequence C: $(1, 2, 3, 5, 8, \dots)$
\item Sequence D: $(10, 12, \dots, 20)$
\end{itemize}

The \verb|\ldots| macro behaves the same as the \verb|\dots| macro.

\subsubsection{Full Stop}
\label{sec:app:styleguide:Full Stop}

\LaTeX\ treats a full stop in front of a white space as an end of
a sentence and puts a slightly wider skip by default (double spacing).
There is an exception to this rule, i.e.\@ where the full stop is next
to a capital letter, \LaTeX\ assumes it represents an abbreviation
and puts a normal skip.

To make \LaTeX\ use proper skips, one need to annotate such exceptions.
For example, given the following \LaTeX\ source:

\begin{VerbatimU}
\begin{quote}
	Lock~1 is owned by CPU~A.
	Lock~2 is owned by CPU~B.  (Bad.)

	Lock~1 is owned by CPU~A\@.
	Lock~2 is owned by CPU~B\@.  (Good.)
\end{quote}
\end{VerbatimU}
the output will be as the following:

\begin{quote}
	Lock~1 is owned by CPU~A.
	Lock~2 is owned by CPU~B.  (Bad.)

	Lock~1 is owned by CPU~A\@.
	Lock~2 is owned by CPU~B\@.  (Good.)
\end{quote}

On the other hand, where a full stop is following a lower case
letter, e.g.\@ as in ``Mr.~Smith'', a wider skip will follow
in the output unless it is properly hinted.
Such hintings can be done in one of several ways.

Given the following source,

\begin{VerbatimU}
\begin{itemize}[nosep]
	\item Mr. Smith (bad)
	\item Mr.~Smith (good)
	\item Mr.\ Smith (good)
	\item Mr.\@ Smith (good)
\end{itemize}
\end{VerbatimU}

\noindent%
the result will look as follows:

\begin{itemize}[nosep]
	\item Mr. Smith (bad)
	\item Mr.~Smith (good)
	\item Mr.\ Smith (good)
	\item Mr.\@ Smith (good)
\end{itemize}

\subsection{Floating Object Format}
\label{sec:app:styleguide:Floating Object Format}

\subsubsection{Ruled Line in Table}
\label{sec:app:styleguide:Ruled Line in Table}

They say that tables drawn by using ruled lines of plain \LaTeX\
look ugly.\footnote{
  \url{https://www.inf.ethz.ch/personal/markusp/teaching/guides/guide-tables.pdf}
}
Vertical lines should be avoided and horizontal lines should be
used sparingly, especially in tables of simple structure.

\IfTblCpTop{}{
\floatstyle{plaintop}
\restylefloat{table}
\setlength{\abovetopsep}{-2pt}
\addtolength{\abovecaptionskip}{-2.5pt}
}

\newcommand{\TLo}{T_\mathrm{L}}
\newcommand{\THi}{T_\mathrm{H}}
\newcommand{\CPf}{C_\mathrm{P}}

\Cref{tab:app:styleguide:Refrigeration Power Consumption}
(corresponding to a table from a now-deleted section)
is drawn by using the features of ``booktabs'' and ``xcolor'' packages.
Note that ruled lines of booktabs can not be mixed with
vertical lines in a table.\footnote{
  There is another package named ``arydshln'' which provides dashed lines
  to be used in tables.
  A couple of experimental examples are presented in
  \cref{sec:app:styleguide:Table Layout Experiment}.
}

\begin{table}
\rowcolors{1}{}{lightgray}
\renewcommand*{\arraystretch}{1.2}\centering\small
\begin{tabular}{lrrr}\toprule
Situation
	& $T$ (K)
		& $\CPf$ & \parbox[b]{.75in}{\raggedleft Power per watt\par waste heat (W)} \\
\midrule
Dry Ice
	& $195$
		& $1.990$
			& 0.5 \\
Liquid N$_2$
	& $77$
		& $0.356$
			& 2.8 \\
Liquid H$_2$
	& $20$
		& $0.073$
			& 13.7 \\
Liquid He
	& $4$
		& $0.0138$
			& 72.3 \\
IBM~Q	& $0.015$
		& $0.000051$
			& 19,500.0 \\
\bottomrule
\end{tabular}
\caption{Refrigeration Power Consumption}
\label{tab:app:styleguide:Refrigeration Power Consumption}
\end{table}

\subsubsection{Position of Caption}
\label{sec:app:styleguide:Position of Caption}

In \LaTeX\ conventions, captions of tables are usually placed
above them.
The reason is the flow of your eye movement when you look at them.
Most tables have a row of heading at the top.
You naturally look at the top of a table at first.
Captions at the bottom of tables disturb this flow.
The same can be said of code snippets, which are read from
top to bottom.

For code snippets, the ``ruled'' style chosen for listing
environment places the caption at the top.
See \cref{lst:app:styleguide:Sample Code Snippet}
for an example.

As for tables, the position of caption is tweaked by
\verb|\floatstyle{}| and \verb|\restylefloat{}| macros
in preamble.

Vertical skips below captions are reduced by setting a smaller
value to the \verb|\abovecaptionskip| variable,
which would also affect captions to figures.

In the tables which use horizontal rules of ``booktabs'' package,
the vertical skips between captions and tables are further reduced
by setting a negative value to the \co{\\abovetopsep} variable,
which controls the behavior of \co{\\toprule}.

\subsection{Improvement Candidates}
\label{sec:app:styleguide:Improvement Candidates}

\begin{figure*}\centering
\ebresizewidth{
\begin{minipage}[t][][t]{2.1in}
\resizebox{2.1in}{!}{\includegraphics{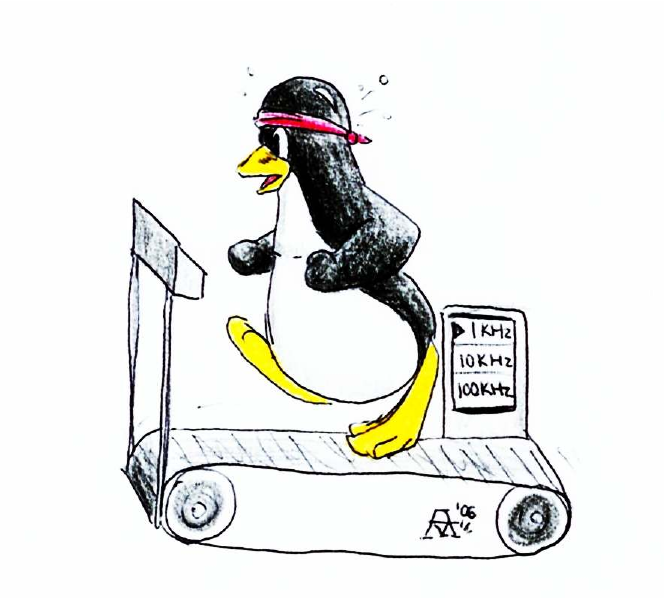}}
\caption{Timer Wheel at 1\,kHz}
\label{fig:app:styleguide:Timer Wheel at 1kHz}
\end{minipage}
\qquad
\begin{minipage}[t][][t]{2.3in}
\resizebox{2.3in}{!}{\includegraphics{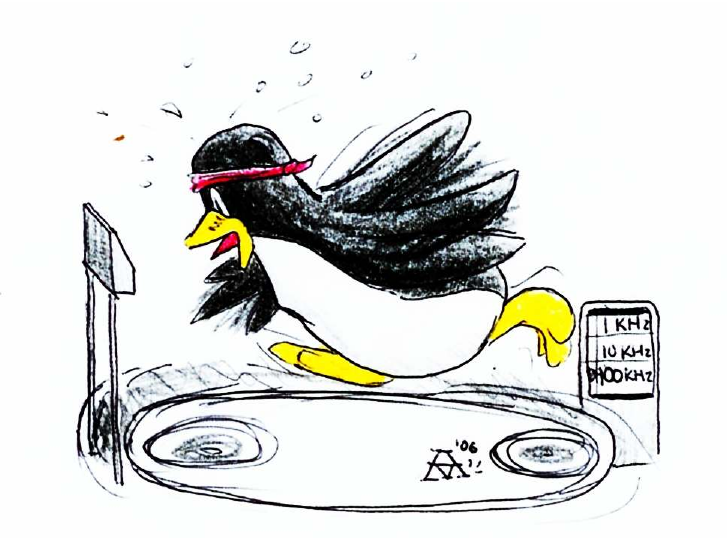}}
\caption{Timer Wheel at 100\,kHz}
\label{fig:app:styleguide:Timer Wheel at 100kHz}
\end{minipage}
}
\end{figure*}

\begin{listing*}%
\caption{Message-Passing Litmus Test (by subfig)}%
\label{lst:app:styleguide:Message-Passing Litmus Test (subfig)}%
{\scriptsize%
\begin{verbbox}[\LstLineNo]
C C-MP+o-wmb-o+o-o.litmus

{
}

P0(int* x0, int* x1) {

  WRITE_ONCE(*x0, 2);
  smp_wmb();
  WRITE_ONCE(*x1, 2);

}

P1(int* x0, int* x1) {

  int r2;
  int r3;

  r2 = READ_ONCE(*x1);
  r3 = READ_ONCE(*x0);

}

exists (1:r2=2 /\ 1:r3=0)
\end{verbbox}
}
\centering
\hspace*{\fill}
\subfloat[Not Enforcing Order]{
  \theverbbox
  \label{sublst:app:styleguide:Not Enforcing Order}
}
\hspace{\fill}
{\scriptsize%
\begin{verbbox}[\LstLineNo]
C C-MP+o-wmb-o+o-rmb-o.litmus

{
}

P0(int* x0, int* x1) {

  WRITE_ONCE(*x0, 2);
  smp_wmb();
  WRITE_ONCE(*x1, 2);

}

P1(int* x0, int* x1) {

  int r2;
  int r3;

  r2 = READ_ONCE(*x1);
  smp_rmb();
  r3 = READ_ONCE(*x0);

}

exists (1:r2=2 /\ 1:r3=0)
\end{verbbox}
}%
\subfloat[Enforcing Order]{%
  \theverbbox
  \label{sublst:app:styleguide:Enforcing Order}
}\hspace*{\fill}%
\end{listing*}

There are a few areas yet to be attempted in perfbook
which would further improve its appearance.
This section lists such candidates.

\subsubsection{Grouping Related Figures/Listings}
\label{sec:app:styleguide:Grouping Related Figures/Listings}

To prevent a pair of closely related figures or listings
from being placed in different pages, it is desirable to group
them into a single floating object.
The ``subfig'' package provides the features to do so.\footnote{
  One problem of grouping figures might be the complexity in
  \LaTeX\ source.}

Two floating objects can be placed side by side by using
\co{\\parbox} or \co{minipage}.
For example,
\cref{fig:advsync:Timer Wheel at 1kHz,fig:advsync:Timer Wheel at 100kHz}
can be grouped together by using a pair of \co{minipage}s
as shown in
\cref{fig:app:styleguide:Timer Wheel at 1kHz,%
fig:app:styleguide:Timer Wheel at 100kHz}.

By using subfig package,
\cref{lst:memorder:Message-Passing Litmus Test (No Ordering),%
lst:memorder:Enforcing Order of Message-Passing Litmus Test}
can be grouped together as shown in
\cref{lst:app:styleguide:Message-Passing Litmus Test (subfig)}
with sub\-/captions (with a minor change of blank line).

Note that they can not be grouped in the same way as
\cref{fig:app:styleguide:Timer Wheel at 1kHz,%
fig:app:styleguide:Timer Wheel at 100kHz}
because the ``ruled'' style prevents their captions
from being properly typeset.

The sub\-/caption can be cited by combining a \verb|\cref{}| macro
and a \verb|\subref{}| macro, for example,
``\cref{lst:app:styleguide:Message-Passing Litmus Test (subfig)}\,%
\subref{sublst:app:styleguide:Not Enforcing Order}''.

It can also be cited by a \verb|\cref{}| macro, for example,
``\cref{sublst:app:styleguide:Enforcing Order}''.
Note the difference in the resulting format.
For the citing by a \verb|\cref{}| to work, you need to place
the \verb|\label{}| macro of the combined floating object
ahead of the definition of subfloats.
Otherwise, the resulting caption number would be off by one
from the actual number.

\subsubsection{Table Layout Experiment}
\label{sec:app:styleguide:Table Layout Experiment}

This section presents some experimental tables
using booktabs, xcolors, and arydshln packages.
The corresponding tables in the text have been converted using one of
the format shown here.
The source of this section can be regarded as a reference to be
consulted when new tables are added in the text.

\begin{table}
\rowcolors{1}{}{lightgray}
\renewcommand*{\arraystretch}{1.1}
\sisetup{group-minimum-digits=4,group-separator={,}}
\centering\small
\begin{tabular}
  {
    l
    S[table-format = 9.1]
    S[table-format = 9.1]
  }
	\toprule
	Operation		& \multicolumn{1}{r}{Cost (ns)}
			& {\parbox[b]{.7in}{\raggedleft Ratio\\(cost/clock)}} \\
	\midrule
	Clock period		&           0.5	&           1.0 \\
	Best-case CAS		&           7.0	&          14.6 \\
	Best-case lock		&          15.4	&          32.3 \\
	Blind CAS		&           7.2	&          15.2 \\
	CAS			&          18.0	&          37.7 \\
	Blind CAS (off-core)	&          47.5	&          99.8 \\
	CAS (off-core)		&         101.9	&         214.0 \\
	Blind CAS (off-socket)	&         148.8	&         312.5 \\
	CAS (off-socket)	&         442.9	&         930.1 \\
	Comms Fabric		&       5 000	&      10 500	\\
	Global Comms		& 195 000 000	& 409 500 000   \\
	\bottomrule
\end{tabular}
\caption{CPU 0 View of Synchronization Mechanisms on 8-Socket System With Intel Xeon Platinum 8176 CPUs @ 2.10GHz}
\label{tab:app:styleguide:CPU 0 View of Synchronization Mechanisms on 8-Socket System With Intel Xeon Platinum 8176 CPUs @ 2.10GHz}
\end{table}

In
\cref{tab:app:styleguide:CPU 0 View of Synchronization Mechanisms on 8-Socket System With Intel Xeon Platinum 8176 CPUs @ 2.10GHz}
(corresponding to
\cref{tab:cpu:CPU 0 View of Synchronization Mechanisms on 8-Socket System With Intel Xeon Platinum 8176 CPUs at 2.10GHz}),
the ``S'' column specifiers provided
by the ``siunitx'' package are used to align numbers.

\Cref{tab:app:styleguide:Synchronization and Reference Counting}
(corresponding to
\cref{tab:together:Synchronizing Reference Counting})
is an example of table with a complex header.
In
\cref{tab:app:styleguide:Synchronization and Reference Counting},
the gap in the mid-rule corresponds to the distinction
which had been represented by double vertical rules before the conversion.
The legends in the frame box appended here explain the abbreviations used
in the matrix.
Two types of memory barrier are denoted by subscripts here.
The legends and subscripts are not present in
\cref{tab:together:Synchronizing Reference Counting}
since they are redundant there.

\begin{table}
\renewcommand*{\arraystretch}{1.25}
\rowcolors{3}{}{lightgray}
\small
\centering
\begin{tabular}{lcccc}
	\toprule
	& \multicolumn{4}{c}{Release} \\
	\cmidrule(l){2-5}
	Acquisition	& Locks
				& \parbox[c]{.5in}{Reference\\Counts}
					& \parbox[c]{.5in}{Hazard\\Pointers}
						& RCU \\
	\cmidrule{1-1} \cmidrule(l){2-5}
	Locks		& $-$	& CAM\textsubscript{R}	& M	& CA  \\
	\parbox[c][6ex]{.6in}{Reference\\Counts}
			& A	& AM\textsubscript{R}    & M	& A   \\
	\parbox[c][6ex]{.6in}{Hazard\\Pointers}
			& M	& M	& M	& M   \\
	RCU		& CA	& M\textsubscript{A}CA	& M	& CA  \\
	\bottomrule
\end{tabular}

\vspace{5pt}\hfill
\framebox[\width]{\footnotesize\setlength{\tabcolsep}{3pt}
\rowcolors{1}{}{}
  \begin{tabular}{lrp{2in}}
	Key:	& A: & Atomic counting \\
		& C: & Check combined with the atomic acquisition operation \\
		& M: & Full memory barriers required \\
		& M\textsubscript{R}: & Memory barriers required only on release \\
		& M\textsubscript{A}: & Memory barriers required on acquire \\
  \end{tabular}
}
\caption{Synchronization and Reference Counting}
\label{tab:app:styleguide:Synchronization and Reference Counting}
\end{table}

\Cref{tab:app:styleguide:Cache Coherence Example}
(corresponding to
\cref{tab:app:whymb:Cache Coherence Example})
is a sequence diagram drawn as a table.

\begin{table*}
\small
\centering
\renewcommand*{\arraystretch}{1.2}
\rowcolors{6}{}{lightgray}
% "6" is chosen due to disturbance of row count by cmidrule.
% The command definition is:
%     \rowcolors{<row>}{<odd-row color>}{<even-row color>}
% Here, <row> specifies the row count where the coloring start.
% In this table, the "Seq = 0" row is the 3rd row, so a "3" would
% be a right choice.
% However, because of the \cmidrule{} commands used in the heading,
% internal row count of the "Seq = 0" row becomes "6".
% This is why the 3rd row has the background color of <even-row color>.
%
% \cline of plain LaTeX also interfares the row count.
\ebresizewidth{
\begin{tabular}{rclcccccc}
	\toprule
	& & & \multicolumn{4}{c}{CPU Cache} & \multicolumn{2}{c}{Memory} \\
	\cmidrule(lr){4-7} \cmidrule(l){8-9}
	Sequence \# & CPU \# & Operation & 0 & 1 & 2 & 3 & 0 & 8 \\
	\cmidrule(r){1-3} \cmidrule(lr){4-7} \cmidrule(l){8-9}
%	Seq CPU Operation	------------- CPU -------------   - Memory -
%				   0	   1	   2	   3	    0   8
	0 &   & Initial State	& $-$/I & $-$/I & $-$/I & $-$/I   & V & V \\
	1 & 0 & Load		& 0/S &   $-$/I & $-$/I & $-$/I   & V & V \\
	2 & 3 & Load		& 0/S &   $-$/I & $-$/I & 0/S     & V & V \\
	3 & 0 & Invalidation	& 8/S &   $-$/I & $-$/I & 0/S     & V & V \\
	4 & 2 & RMW		& 8/S &   $-$/I & 0/E &   $-$/I   & V & V \\
	5 & 2 & Store		& 8/S &   $-$/I & 0/M &   $-$/I   & I & V \\
	6 & 1 & Atomic Inc	& 8/S &   0/M &   $-$/I & $-$/I   & I & V \\
	7 & 1 & Writeback	& 8/S &   8/S &   $-$/I & $-$/I   & V & V \\
	\bottomrule
\end{tabular}
}
\caption{Cache Coherence Example}
\label{tab:app:styleguide:Cache Coherence Example}
\end{table*}

\Cref{tab:app:styleguide:RCU Publish-Subscribe and Version Maintenance APIs}
is a tweaked version of
\cref{tab:defer:RCU Publish-Subscribe and Version Maintenance APIs}.
Here, the ``Category'' column in the original is removed
and the categories are indicated in rows of bold-face font
just below the mid-rules.
This change makes it easier for \verb|\rowcolors{}| command of
``xcolor'' package to work properly.

\Cref{tab:app:styleguide:RCU Publish-Subscribe and Version Maintenance APIs (colortbl)}
is another version which keeps original columns and colors rows only where
a category has multiple rows.
This is done by combining \verb|\rowcolors{}| of ``xcolor'' and
\verb|\cellcolor{}| commands of the ``colortbl'' package
(\verb|\cellcolor{}| overrides \verb|\rowcolors{}|).

In
\cref{tab:defer:RCU Publish-Subscribe and Version Maintenance APIs},
the latter layout without partial row coloring has been
chosen for simplicity.

\begin{table*}
\rowcolors{2}{}{blue!15}
\renewcommand*{\arraystretch}{1.1}
\footnotesize
\centering
\ebresizewidth{
\begin{tabular}{lll}
\toprule
	Primitives &
		Availability &
			Overhead \\
\midrule
	\multicolumn{3}{l}{\bfseries List traversal} \\
	\tco{list_for_each_entry_rcu()} &
		2.5.59 &
			Simple instructions (memory barrier on Alpha) \\
\midrule
	\multicolumn{3}{l}{\bfseries List update} \\
	\tco{list_add_rcu()} &
		2.5.44 &
			Memory barrier \\
	\rowcolor{lightgray}\tco{list_add_tail_rcu()} &
		2.5.44 &
			Memory barrier \\
	\tco{list_del_rcu()} &
		2.5.44 &
			Simple instructions \\
	\rowcolor{lightgray}\tco{list_replace_rcu()} &
		2.6.9 &
			Memory barrier \\
	\tco{list_splice_init_rcu()} &
		2.6.21 &
			Grace-period latency \\
\midrule
	\multicolumn{3}{l}{\bfseries Hlist traversal} \\
	\tco{hlist_for_each_entry_rcu()} &
		2.6.8 &
			Simple instructions (memory barrier on Alpha) \\
\midrule
	\multicolumn{3}{l}{\bfseries Hlist update} \\
	\tco{hlist_add_after_rcu()} &
		2.6.14 &
			Memory barrier \\
	\rowcolor{lightgray}\tco{hlist_add_before_rcu()} &
		2.6.14 &
			Memory barrier \\
	\tco{hlist_add_head_rcu()} &
		2.5.64 &
			Memory barrier \\
	\rowcolor{lightgray}\tco{hlist_del_rcu()} &
		2.5.64 &
			Simple instructions \\
	\tco{hlist_replace_rcu()} &
		2.6.15 &
			Memory barrier \\
\midrule
	\multicolumn{3}{l}{\bfseries Pointer traversal} \\
	\tco{rcu_dereference()} &
		2.6.9 &
			Simple instructions (memory barrier on Alpha) \\
\midrule
	\multicolumn{3}{l}{\bfseries Pointer update} \\
	\tco{rcu_assign_pointer()} &
		2.6.10 &
			Memory barrier \\
\bottomrule
\end{tabular}
}
\caption{RCU Publish-Subscribe and Version Maintenance APIs}
\label{tab:app:styleguide:RCU Publish-Subscribe and Version Maintenance APIs}
\end{table*}

\begin{table*}
\renewcommand*{\arraystretch}{1.2}
\rowcolors{3}{lightgray}{}
\footnotesize
\centering
\ebresizewidth{
\begin{tabular}{lllp{1.2in}}\toprule
Category &
	Primitives &
		Availability &
			Overhead \\
\midrule
List traversal &
	\tco{list_for_each_entry_rcu()} &
		2.5.59 &
			Simple instructions (memory barrier on Alpha) \\
\midrule
\cellcolor{white}List update &
	\tco{list_add_rcu()} &
		2.5.44 &
			Memory barrier \\
&
	\tco{list_add_tail_rcu()} &
		2.5.44 &
			Memory barrier \\
\cellcolor{white} &
	\tco{list_del_rcu()} &
		2.5.44 &
			Simple instructions \\
&
	\tco{list_replace_rcu()} &
		2.6.9 &
			Memory barrier \\
\cellcolor{white} &
	\tco{list_splice_init_rcu()} &
		2.6.21 &
			Grace-period latency \\
\midrule
Hlist traversal &
	\tco{hlist_for_each_entry_rcu()} &
		2.6.8 &
			Simple instructions (memory barrier on Alpha) \\
\midrule
\cellcolor{white}Hlist update &
	\tco{hlist_add_after_rcu()} &
		2.6.14 &
			Memory barrier \\
&
	\tco{hlist_add_before_rcu()} &
		2.6.14 &
			Memory barrier \\
\cellcolor{white} &
	\tco{hlist_add_head_rcu()} &
		2.5.64 &
			Memory barrier \\
&
	\tco{hlist_del_rcu()} &
		2.5.64 &
			Simple instructions \\
\cellcolor{white} &
	\tco{hlist_replace_rcu()} &
		2.6.15 &
			Memory barrier \\
\midrule\hiderowcolors
Pointer traversal &
	\tco{rcu_dereference()} &
		2.6.9 &
			Simple instructions (memory barrier on Alpha) \\
\midrule
Pointer update &
	\tco{rcu_assign_pointer()} &
		2.6.10 &
			Memory barrier \\
\bottomrule
\end{tabular}
}
\caption{RCU Publish-Subscribe and Version Maintenance APIs}
\label{tab:app:styleguide:RCU Publish-Subscribe and Version Maintenance APIs (colortbl)}
\end{table*}

\Cref{tab:app:styleguide:Memory Misordering: Store-Buffering Sequence of Events}
(corresponding to
\cref{tab:memorder:Memory Misordering: Store-Buffering Sequence of Events})
is also a sequence diagram drawn as a tabular object.

\begin{table*}
\rowcolors{6}{}{lightgray}
\renewcommand*{\arraystretch}{1.1}
\small
\centering\OneColumnHSpace{-0.1in}
\ebresizewidth{
\begin{tabular}{rllllll}
	\toprule
	& \multicolumn{3}{c}{CPU 0} & \multicolumn{3}{c}{CPU 1} \\
	\cmidrule(l){2-4} \cmidrule(l){5-7}
	& Instruction & Store Buffer & Cache &
		Instruction & Store Buffer & Cache \\
	\cmidrule{1-1} \cmidrule(l){2-4} \cmidrule(l){5-7}
	1 & (Initial state) & & \tco{x1==0} &
		(Initial state) & & \tco{x0==0} \\
	2 & \tco{x0 = 2;} & \tco{x0==2} & \tco{x1==0} &
		\tco{x1 = 2;} & \tco{x1==2} & \tco{x0==0} \\
	3 & \tco{r2 = x1;} (0) & \tco{x0==2} & \tco{x1==0} &
		\tco{r2 = x0;} (0) & \tco{x1==2} & \tco{x0==0} \\
	4 & (Read-invalidate) & \tco{x0==2} & \tco{x0==0} &
		(Read-invalidate) & \tco{x1==2} & \tco{x1==0} \\
	5 & (Finish store) & & \tco{x0==2} &
		(Finish store) & & \tco{x1==2} \\
	\bottomrule
\end{tabular}
}
\caption{Memory Misordering: Store-Buffering Sequence of Events}
\label{tab:app:styleguide:Memory Misordering: Store-Buffering Sequence of Events}
\end{table*}

\Cref{tab:app:styleguide:Refrigeration Power Consumption (arydshln)}
shows another version of
\cref{tab:app:styleguide:Refrigeration Power Consumption}
with dashed horizontal and vertical rules of the arydshln package.

\setlength\dashlinedash{.5pt}
\setlength\dashlinegap{1pt}

\begin{table}
\renewcommand*{\arraystretch}{1.2}\centering\small
\begin{tabular}{l:r:r:r}\toprule
Situation
	& $T$ (K)
		& $\CPf$ & \parbox[b]{.75in}{\raggedleft Power per watt\par waste heat (W)} \\
\hline
Dry Ice
	& $195$
		& $1.990$
			& 0.5 \\ \hdashline
Liquid N$_2$
	& $77$
		& $0.356$
			& 2.8 \\ \hdashline
Liquid H$_2$
	& $20$
		& $0.073$
			& 13.7 \\ \hdashline
Liquid He
	& $4$
		& $0.0138$
			& 72.3 \\ \hdashline
IBM~Q	& $0.015$
		& $0.000051$
			& 19,500.0 \\
\bottomrule
\end{tabular}
\caption{Refrigeration Power Consumption}
\label{tab:app:styleguide:Refrigeration Power Consumption (arydshln)}
\end{table}

In this case, the vertical dashed rules seems unnecessary.
The one without the vertical rules is shown in
\cref{tab:app:styleguide:Refrigeration Power Consumption (arydshln-2)}.

\begin{table}
\renewcommand*{\arraystretch}{1.2}\centering\small
\begin{tabular}{lrrr}\toprule
Situation
	& $T$ (K)
		& $\CPf$ & \parbox[b]{.75in}{\raggedleft Power per watt\par waste heat (W)} \\
\midrule
Dry Ice
	& $195$
		& $1.990$
			& 0.5 \\ \hdashline
Liquid N$_2$
	& $77$
		& $0.356$
			& 2.8 \\ \hdashline
Liquid H$_2$
	& $20$
		& $0.073$
			& 13.7 \\ \hdashline
Liquid He
	& $4$
		& $0.0138$
			& 72.3 \\ \hdashline
IBM~Q	& $0.015$
		& $0.000051$
			& 19,500.0 \\
\bottomrule
\end{tabular}
\caption{Refrigeration Power Consumption}
\label{tab:app:styleguide:Refrigeration Power Consumption (arydshln-2)}
\end{table}

\IfTblCpTop{}{
\floatstyle{plain}
\restylefloat{table}
\addtolength{\abovecaptionskip}{2.5pt}
\setlength{\abovetopsep}{0pt}
}

\FloatBarrier

\subsubsection{Miscellaneous Candidates}
\label{sec:app:styleguide:Miscellaneous Candidates}

Other improvement candidates are listed in the source of this
section as comments.

% Ugly line break by \co{}
%                                 __
%        atomic_store()
%
%                           seqlock_
%        t
%
%   Is there any way to prevent these breaks?
%   Maybe we need an on-the-fly script to convert such \co{}s
%   to couples of \co{}s.
%   Example:
%     \co{__atomic_store()} -> \co{__}\co{atomic_store()}
%     \co{seqlock_t} ->        \co{seqlock_}\co{t}

\renewcommand{\bottomtitlespace}{.08\textheight}
	\renewcommand*{\theHNum}{\arabic{section}.\arabic{quickquizctrC}}
	\chapter{Answers to Quick Quizzes}
	\label{chp:app:Answers to Quick Quizzes}
	\Epigraph{The Answer to the Ultimate Question of Life, The Universe,
		  and Everything.}
		 {\emph{The Hitchhikers Guide to the Galaxy}, Douglas~Adams}
	\setlength{\parskip}{0.0pt plus 1ex}
	% mainfile: perfbook.tex
\QuickQAC{chp:How To Use This Book}{How To Use This Book}{qqzhowto}
\QuickQ{}
	Where are the answers to the Quick Quizzes found?
\QuickA{}
	In \cref{chp:app:Answers to Quick Quizzes} starting on
	\cpageref{chp:app:Answers to Quick Quizzes}.
	Hey, I thought I owed you an easy one!
\QuickE{}
\QuickQ{}
	Some of the Quick Quiz questions seem to be from the viewpoint
	of the reader rather than the author.
	Is that really the intent?
\QuickA{}
	Indeed it is!
	Many are questions that Paul E.~McKenney would probably have
	asked if he was a novice student in a class covering this material.
	It is worth noting that Paul was taught most of this material by
	parallel hardware and software, not by professors.
	In Paul's experience, professors are much more likely to provide
	answers to verbal questions than are parallel systems, recent
	advances in voice-activated assistants notwithstanding.
	Of course, we could have a lengthy debate over which of professors
	or parallel systems provide the most useful answers to these sorts
	of questions,
	but for the time being let's just agree that usefulness of
	answers varies widely across the population both of professors
	and of parallel systems.

	Other quizzes are quite similar to actual questions that have been
	asked during conference presentations and lectures covering the
	material in this book.
	A few others are from the viewpoint of the author.
\QuickE{}
\QuickQ{}
	These Quick Quizzes are just not my cup of tea.
	What can I do about it?
\QuickA{}
Here are a few possible strategies:

\begin{enumerate}
\item	Just ignore the Quick Quizzes and read the rest of
	the book.
	You might miss out on the interesting material in
	some of the Quick Quizzes, but the rest of the book
	has lots of good material as well.
	This is an eminently reasonable approach if your main
	goal is to gain a general understanding of the material
	or if you are skimming through the book to find a
	solution to a specific problem.
\item	Look at the answer immediately rather than investing
	a large amount of time in coming up with your own
	answer.
	This approach is reasonable when a given Quick Quiz's
	answer holds the key to a specific problem you are
	trying to solve.
	This approach is also reasonable if you want a somewhat
	deeper understanding of the material, but when you do not
	expect to be called upon to generate parallel solutions given
	only a blank sheet of paper.
\item	If you find the Quick Quizzes distracting but impossible
	to ignore, you can always clone the \LaTeX{} source for
	this book from the git archive.
	You can then run the command \co{make nq}, which will
	produce a \co{perfbook-nq.pdf}.
	This PDF contains unobtrusive boxed tags where the Quick Quizzes
	would otherwise be, and gathers each chapter's Quick Quizzes
	at the end of that chapter in the classic textbook style.
\item	Learn to like (or at least tolerate) the Quick Quizzes.
	Experience indicates that quizzing yourself periodically
	while reading greatly increases comprehension and depth
	of understanding.
\end{enumerate}

Note that the quick quizzes are hyperlinked to the answers and vice versa.
Click either the ``Quick Quiz'' heading or the small black square
to move to the beginning of the answer.
From the answer, click on the heading or the small black square to
move to the beginning of the quiz, or, alternatively, click on the
small white square at the end of the answer to move to the end of the
corresponding quiz.
\QuickE{}
\QuickQ{}
	If passively reading this book doesn't get me full problem-solving
	and code-production capabilities, what on earth is the point???
\QuickA{}
	For those preferring analogies, coding concurrent software is
	similar to playing music in that there are good uses for many
	different levels of talent and skill.
	Not everyone needs to devote their entire live to becoming a
	concert pianist.
	In fact, for every such virtuoso, there are a great many lesser
	pianists whose of music is welcomed by their friends and families.
	But these lesser pianists are probably doing something else to
	support themselves, and so it is with concurrent coding.

	One potential benefit of passively reading this book is the ability
	to read and understand modern concurrent code.
	This ability might in turn permit you to:

	\begin{enumerate}
	\item	See what the kernel does so that you can check to see
		if a proposed use case is valid.
	\item	Chase down a kernel bug.
	\item	Use information in the kernel to more easily chase down
		a userspace bug.
	\item	Produce a fix for a kernel bug.
	\item	Create a straightforward kernel feature, whether from
		scratch or using the modern copy-pasta development
		methodology.
	\end{enumerate}

	If you are proficient with straightforward uses of locks and
	atomic operations, passively reading this book should enable
	you to successfully apply modern concurrency techniques.

	And finally, if your job is to coordinate the activities of
	developers making use of modern concurrency techniques, passively
	reading this book might help you understand what on earth they
	are talking about.
\QuickE{}
\QuickQAC{chp:Introduction}{Introduction}{qqzintro}
\QuickQ{}
	Come on now!!!
	Parallel programming has been known to be exceedingly
	hard for many decades.
	You seem to be hinting that it is not so hard.
	What sort of game are you playing?
\QuickA{}
	If you really believe that parallel programming is exceedingly
	hard, then you should have a ready answer to the question
	``Why is parallel programming hard?''
	One could list any number of reasons, ranging from deadlocks to
	race conditions to testing coverage, but the real answer is that
	{\em it is not really all that hard}.
	After all, if parallel programming was really so horribly difficult,
	how could a large number of open-source projects, ranging from Apache
	to MySQL to the Linux kernel, have managed to master it?

	A better question might be:
	``Why is parallel programming \emph{perceived} to be so difficult?''
	To see the answer, let's go back to the year 1991.
	Paul McKenney was walking across the parking lot to Sequent's
	benchmarking center carrying six dual-80486 Sequent Symmetry CPU
	boards, when he suddenly realized that he was carrying several
	times the price of the house he had just purchased.\footnote{
		Yes, this sudden realization {\em did} cause him to walk quite
		a bit more carefully.
		Why do you ask?}
	This high cost of parallel systems meant that
	parallel programming was restricted to a privileged few who
	worked for an employer who either manufactured or could afford to
	purchase machines costing upwards of \$100,000---in 1991 dollars US.

	In contrast, in 2020, Paul finds himself typing these words on a
	six-core x86 laptop.
	Unlike the dual-80486 CPU boards, this laptop also contains
	64\,GB of main memory, a 1\,TB solid-state disk, a display, Ethernet,
	USB ports, wireless, and Bluetooth.
	And the laptop is more than an order of magnitude cheaper than
	even one of those dual-80486 CPU boards, even before taking inflation
	into account.

	Parallel systems have truly arrived.
	They are no longer the sole domain of a privileged few, but something
	available to almost everyone.

	The earlier restricted availability of parallel hardware is
	the \emph{real} reason that parallel programming is considered
	so difficult.
	After all, it is quite difficult to learn to program even the simplest
	machine if you have no access to it.
	Since the age of rare and expensive parallel machines is for the most
	part behind us, the age during which
	parallel programming is perceived to be mind-crushingly difficult is
	coming to a close.\footnote{
		Parallel programming is in some ways more difficult than
		sequential programming, for example, parallel validation
		is more difficult.
		But no longer mind-crushingly difficult.}
\QuickE{}
\QuickQ{}
	How could parallel programming \emph{ever} be as easy
	as sequential programming?
\QuickA{}
	It depends on the programming environment.
	SQL~\cite{DIS9075SQL92} is an underappreciated success
	story, as it permits programmers who know nothing about parallelism
	to keep a large parallel system productively busy.
	We can expect more variations on this theme as parallel
	computers continue to become cheaper and more readily available.
	For example, one possible contender in the scientific and
	technical computing arena is MATLAB*P,
	which is an attempt to automatically parallelize common
	matrix operations.

	Finally, on Linux and UNIX systems, consider the following
	shell command:

	\begin{VerbatimU}
	get_input | grep "interesting" | sort
	\end{VerbatimU}

	This shell pipeline runs the \co{get_input}, \co{grep},
	and \co{sort} processes in parallel.
	There, that wasn't so hard, now was it?

	In short, parallel programming is just as easy as sequential
	programming---at least in those environments that hide the parallelism
	from the user!
\QuickE{}
\QuickQ{}
	Oh, really???
	What about correctness, maintainability, robustness, and so on?
\QuickA{}
	These are important goals, but they are just as important for
	sequential programs as they are for parallel programs.
	Therefore, important though they are, they do not belong on
	a list specific to parallel programming.
\QuickE{}
\QuickQ{}
	And if correctness, maintainability, and robustness don't
	make the list, why do productivity and generality?
\QuickA{}
	Given that parallel programming is perceived to be much harder
	than sequential programming, productivity is tantamount and
	therefore must not be omitted.
	Furthermore, high-productivity parallel-programming environments
	such as SQL serve a specific purpose, hence generality must
	also be added to the list.
\QuickE{}
\QuickQ{}
	Given that parallel programs are much harder to prove
	correct than are sequential programs, again, shouldn't
	correctness \emph{really} be on the list?
\QuickA{}
	From an engineering standpoint, the difficulty in proving
	correctness, either formally or informally, would be important
	insofar as it impacts the primary goal of productivity.
	So, in cases where correctness proofs are important, they
	are subsumed under the ``productivity'' rubric.
\QuickE{}
\QuickQ{}
	What about just having fun?
\QuickA{}
	Having fun is important as well, but, unless you are a hobbyist,
	would not normally be a \emph{primary} goal.
	On the other hand, if you \emph{are} a hobbyist, go wild!
\QuickE{}
\QuickQ{}
	Are there no cases where parallel programming is about something
	other than performance?
\QuickA{}
	There certainly are cases where the problem to be solved is
	inherently parallel, for example, Monte Carlo methods and
	some numerical computations.
	Even in these cases, however, there will be some amount of
	extra work managing the parallelism.

	Parallelism is also sometimes used for reliability.
	For but one example,
	triple-modulo redundancy has three systems run in parallel
	and vote on the result.
	In extreme cases, the three systems will be independently
	implemented using different algorithms and technologies.
\QuickE{}
\QuickQ{}
	Why not instead rewrite programs from inefficient scripting
	languages to C or C++?
\QuickA{}
	If the developers, budget, and time is available for such a
	rewrite, and if the result will attain the required levels
	of performance on a single CPU, this can be a reasonable
	approach.
\QuickE{}
\QuickQ{}
	Why all this prattling on about non-technical issues???
	And not just \emph{any} non-technical issue, but \emph{productivity}
	of all things?
	Who cares?
\QuickA{}
	If you are a pure hobbyist, perhaps you don't need to care.
	But even pure hobbyists will often care about how much they
	can get done, and how quickly.
	After all, the most popular hobbyist tools are usually those
	that are the best suited for the job, and an important part of
	the definition of ``best suited'' involves productivity.
	And if someone is paying you to write parallel code, they will
	very likely care deeply about your productivity.
	And if the person paying you cares about something, you would
	be most wise to pay at least some attention to it!

	Besides, if you \emph{really} didn't care about productivity,
	you would be doing it by hand rather than using a computer!
\QuickE{}
\QuickQ{}
	Given how cheap parallel systems have become, how can anyone
	afford to pay people to program them?
\QuickA{}
	There are a number of answers to this question:
	\begin{enumerate}
	\item	Given a large computational cluster of parallel machines,
		the aggregate cost of the cluster can easily justify
		substantial developer effort, because the development
		cost can be spread over the large number of machines.
	\item	Popular software that is run by tens of millions of users
		can easily justify substantial developer effort,
		as the cost of this development can be spread over the tens
		of millions of users.
		Note that this includes things like kernels and system
		libraries.
	\item	If the low-cost parallel machine is controlling the operation
		of a valuable piece of equipment, then the cost of this
		piece of equipment might easily justify substantial
		developer effort.
	\item	If the software for the low-cost parallel machine produces an
		extremely valuable result (e.g., energy savings),
		then this valuable result might again justify substantial
		developer cost.
	\item	Safety-critical systems protect lives, which can clearly
		justify very large developer effort.
	\item	Hobbyists and researchers might instead seek knowledge,
		experience, fun, or glory.
	\end{enumerate}
	So it is not the case that the decreasing cost of hardware renders
	software worthless, but rather that it is no longer possible to
	``hide'' the cost of software development within the cost of
	the hardware, at least not unless there are extremely large
	quantities of hardware.
\QuickE{}
\QuickQ{}
	This is a ridiculously unachievable ideal!
	Why not focus on something that is achievable in practice?
\QuickA{}
	This is eminently achievable.
	The cellphone is a computer that can be used to make phone
	calls and to send and receive text messages with little or
	no programming or configuration on the part of the end user.

	This might seem to be a trivial example at first glance,
	but if you consider it carefully you will see that it is
	both simple and profound.
	When we are willing to sacrifice generality, we can achieve
	truly astounding increases in productivity.
	Those who indulge in excessive generality will therefore fail to set
	the productivity bar high enough to succeed near the top of the
	software stack.
	This fact of life even has its own acronym:
	YAGNI, or ``You Ain't Gonna Need It.''
\QuickE{}
\QuickQ{}
	Wait a minute!
	Doesn't this approach simply shift the development effort from
	you to whoever wrote the existing parallel software you are using?
\QuickA{}
	Exactly!
	And that is the whole point of using existing software.
	One team's work can be used by many other teams, resulting in a
	large decrease in overall effort compared to all teams
	needlessly reinventing the wheel.
\QuickE{}
\QuickQ{}
	What other bottlenecks might prevent additional CPUs from
	providing additional performance?
\QuickA{}
	There are any number of potential bottlenecks:
	\begin{enumerate}
	\item	Main memory.
		If a single thread consumes all available
		memory, additional threads will simply page themselves
		silly.
	\item	Cache.
		If a single thread's cache footprint completely
		fills any shared CPU cache(s), then adding more threads
		will simply thrash those affected caches, as will be
		seen in \cref{chp:Data Structures}.
	\item	Memory bandwidth.
		If a single thread consumes all available
		memory bandwidth, additional threads will simply
		result in additional queuing on the system interconnect.
	\item	I/O bandwidth.
		If a single thread is I/O bound,
		adding more threads will simply result in them all
		waiting in line for the affected I/O resource.
	\end{enumerate}

	Specific hardware systems might have any number of additional
	bottlenecks.
	The fact is that every resource which is shared between
	multiple CPUs or threads is a potential bottleneck.
\QuickE{}
\QuickQ{}
	Other than CPU cache capacity, what might require limiting the
	number of concurrent threads?
\QuickA{}
	There are any number of potential limits on the number of
	threads:
	\begin{enumerate}
	\item	Main memory.
		Each thread consumes some memory
		(for its stack if nothing else), so that excessive
		numbers of threads can exhaust memory, resulting
		in excessive paging or memory-allocation failures.
	\item	I/O bandwidth.
		If each thread initiates a given
		amount of mass-storage I/O or networking traffic,
		excessive numbers of threads can result in excessive
		I/O queuing delays, again degrading performance.
		Some networking protocols may be subject to timeouts
		or other failures if there are so many threads that
		networking events cannot be responded to in a timely
		fashion.
	\item	Synchronization overhead.
		For many synchronization protocols, excessive numbers
		of threads can result in excessive spinning, blocking,
		or rollbacks, thus degrading performance.
	\end{enumerate}

	Specific applications and platforms may have any number of additional
	limiting factors.
\QuickE{}
\QuickQ{}
	Just what is ``explicit timing''???
\QuickA{}
	Where each thread is given access to some set of resources during
	an agreed-to slot of time.
	For example, a parallel program with eight threads might be
	organized into eight-millisecond time intervals, so that the
	first thread is given access during the first millisecond of
	each interval, the second thread during the second millisecond,
	and so on.
	This approach clearly requires carefully synchronized clocks
	and careful control of execution times, and therefore should
	be used with considerable caution.

	In fact, outside of hard realtime environments, you almost
	certainly want to use something else instead.
	Explicit timing is nevertheless worth a mention, as it is
	always there when you need it.
\QuickE{}
\QuickQ{}
	Are there any other obstacles to parallel programming?
\QuickA{}
	There are a great many other potential obstacles to parallel
	programming.
	Here are a few of them:
	\begin{enumerate}
	\item	The only known algorithms for a given project might
		be inherently sequential in nature.
		In this case, either avoid parallel programming
		(there being no law saying that your project \emph{has}
		to run in parallel) or invent a new parallel algorithm.
	\item	The project allows binary-only plugins that share the same
		address space, such that no one developer has access to
		all of the source code for the project.
		Because many parallel bugs, including deadlocks, are
		global in nature, such binary-only plugins pose a severe
		challenge to current software development methodologies.
		This might well change, but for the time being, all
		developers of parallel code sharing a given address space
		need to be able to see \emph{all} of the code running in
		that address space.
	\item	The project contains heavily used APIs that were designed
		without regard to
		parallelism~\cite{HagitAttiya2011LawsOfOrder,Clements:2013:SCR:2517349.2522712}.
		Some of the more ornate features of the System V
		message-queue API form a case in point.
		Of course, if your project has been around for a few
		decades, and its developers did not have access to
		parallel hardware, it undoubtedly has at least
		its share of such APIs.
	\item	The project was implemented without regard to parallelism.
		Given that there are a great many techniques that work
		extremely well in a sequential environment, but that
		fail miserably in parallel environments, if your project
		ran only on sequential hardware for most of its lifetime,
		then your project undoubtably has at least its share of
		parallel-unfriendly code.
	\item	The project was implemented without regard to good
		software-development practice.
		The cruel truth is that shared-memory parallel
		environments are often much less forgiving of sloppy
		development practices than are sequential environments.
		You may be well-served to clean up the existing design
		and code prior to attempting parallelization.
	\item	The people who originally did the development on your
		project have since moved on, and the people remaining,
		while well able to maintain it or add small features,
		are unable to make ``big animal'' changes.
		In this case, unless you can work out a very simple
		way to parallelize your project, you will probably
		be best off leaving it sequential.
		That said, there are a number of simple approaches that
		you might use
		to parallelize your project, including running multiple
		instances of it, using a parallel implementation of
		some heavily used library function, or making use of
		some other parallel project, such as a database.
	\end{enumerate}

	One can argue that many of these obstacles are non-technical
	in nature, but that does not make them any less real.
	In short, parallelization of a large body of code
	can be a large and complex effort.
	As with any large and complex effort, it makes sense to
	do your homework beforehand.
\QuickE{}
\QuickQAC{chp:Hardware and its Habits}{Hardware and its Habits}{qqzcpu}
\QuickQ{}
	Why should parallel programmers bother learning low-level
	properties of the hardware?
	Wouldn't it be easier, better, and more elegant to remain at
	a higher level of abstraction?
\QuickA{}
	It might well be easier to ignore the detailed properties of
	the hardware, but in most cases it would be quite foolish
	to do so.
	If you accept that the only purpose of parallelism is to
	increase performance, and if you further accept that
	performance depends on detailed properties of the hardware,
	then it logically follows that parallel programmers are going
	to need to know at least a few hardware properties.

	This is the case in most engineering disciplines.
	Would \emph{you} want to use a bridge designed by an
	engineer who did not understand the properties of
	the concrete and steel making up that bridge?
	If not, why would you expect a parallel programmer to be
	able to develop competent parallel software without at least
	\emph{some} understanding of the underlying hardware?
\QuickE{}
\QuickQ{}
	What types of machines would allow atomic operations on
	multiple data elements?
\QuickA{}
	One answer to this question is that it is often possible to
	pack multiple elements of data into a single machine word,
	which can then be manipulated atomically.

	A more trendy answer would be machines supporting transactional
	memory~\cite{DBLomet1977SIGSOFT,Knight:1986:AMF:319838.319854,Herlihy93a}.
	By early 2014, several mainstream systems provided limited
	hardware transactional memory implementations, which is covered
	in more detail in
	\cref{sec:future:Hardware Transactional Memory}.
	The jury is still out on the applicability of software transactional
	memory~\cite{McKenney2007PLOSTM,DonaldEPorter2007TRANSACT,
	ChistopherJRossbach2007a,CalinCascaval2008tmtoy,
	AleksandarDragovejic2011STMnotToy,AlexanderMatveev2012PessimisticTM},
	which is covered in \cref{sec:future:Transactional Memory}.
\QuickE{}
\QuickQ{}
	So have CPU designers also greatly reduced the overhead of
	cache misses?
\QuickA{}
	Unfortunately, not so much.
	There has been some reduction given constant numbers of CPUs,
	but the finite speed of light and the atomic nature of
	matter limits their ability to reduce cache-miss overhead
	for larger systems.
	\Cref{sec:cpu:Hardware Free Lunch?}
	discusses some possible avenues for possible future progress.
\QuickE{}
\QuickQ{}
	This is a \emph{simplified} sequence of events?
	How could it \emph{possibly} be any more complex?
\QuickA{}
	This sequence ignored a number of possible complications,
	including:

	\begin{enumerate}
	\item	Other CPUs might be concurrently attempting to perform
		memory-reference operations involving this same cacheline.
	\item	The cacheline might have been replicated read-only in
		several CPUs' caches, in which case, it would need to
		be flushed from their caches.
	\item	CPU~7 might have been operating on the cache line when
		the request for it arrived, in which case CPU~7 might
		need to hold off the request until its own operation
		completed.
	\item	CPU~7 might have ejected the cacheline from its cache
		(for example, in order to make room for other data),
		so that by the time that the request arrived, the
		cacheline was on its way to memory.
	\item	A correctable error might have occurred in the cacheline,
		which would then need to be corrected at some point before
		the data was used.
	\end{enumerate}

	Production-quality cache-coherence mechanisms are extremely
	complicated due to these sorts of
	considerations~\cite{Hennessy95a,DavidECuller1999,MiloMKMartin2012scale,DanielJSorin2011MemModel}.
\QuickE{}
\QuickQ{}
	Why is it necessary to flush the cacheline from CPU~7's cache?
\QuickA{}
	If the cacheline was not flushed from CPU~7's cache, then
	CPUs~0 and~7 might have different values for the same set
	of variables in the cacheline.
	This sort of incoherence greatly complicates parallel software,
	which is why wise hardware architects avoid it.
\QuickE{}
\QuickQ{}
	\Cref{tab:cpu:CPU 0 View of Synchronization Mechanisms on 8-Socket System With Intel Xeon Platinum 8176 CPUs at 2.10GHz}
	shows CPU~0 sharing a core with CPU~224.
	Shouldn't that instead be CPU~1???
\QuickA{}
	It is easy to be sympathetic to this view, but the file
	\path{/sys/devices/system/cpu/cpu0/cache/index0/shared_cpu_list}
	really does contain the string \co{0,224}.
	Therefore, CPU~0's hyperthread twin really is CPU~224.
	Some people speculate that this numbering allows naive applications
	and schedulers to perform better, citing the fact that on many
	workloads the second hyperthread does not provide a huge
	amount of additional performance.
	This speculation assumes that naive applications and schedulers
	would utilize CPUs in numerical order, leaving aside the weaker
	hyperthread twin CPUs until all cores are in use.
\QuickE{}
\QuickQ{}
	Surely the hardware designers could be persuaded to improve
	this situation!
	Why have they been content with such abysmal performance
	for these single-instruction operations?
\QuickA{}
	The hardware designers \emph{have} been working on this
	problem, and have consulted with no less a luminary than
	the late physicist Stephen Hawking.
	Hawking's observation was that the hardware designers have
	two basic problems~\cite{BryanGardiner2007}:

	\begin{enumerate}
	\item	The finite speed of light, and
	\item	The atomic nature of matter.
	\end{enumerate}

\begin{table}
\renewcommand*{\arraystretch}{1.1}
\centering\small
\begin{tabular}
  {
    ll
    S[table-format = 9.1]
    S[table-format = 9.1]
  }
	\toprule
	\multicolumn{2}{l}{Operation}
			& \multicolumn{1}{r}{Cost (ns)}
			& {\parbox[b]{.7in}{\raggedleft Ratio\\(cost/clock)}} \\
	\midrule
	\multicolumn{2}{l}{Clock period}
			&           0.4	&           1.0 \\
        \midrule
	\multicolumn{2}{l}{Same-CPU}
			&		&		\\
	& CAS		&          12.2	&          33.8 \\
	& lock		&          25.6	&          71.2 \\
        \midrule
        \multicolumn{2}{l}{On-Core}
			&		&		\\
	& Blind CAS	&          12.9	&          35.8 \\
	& CAS		&           7.0	&          19.4 \\
	\midrule
        \multicolumn{2}{l}{Off-Core}
			&		&		\\
	& Blind CAS	&          31.2	&          86.6 \\
	& CAS		&          31.2	&          86.5 \\
	\midrule
	\multicolumn{2}{l}{Off-Socket}
			&		&		\\
	& Blind CAS	&          92.4	&         256.7 \\
	& CAS		&          95.9	&         266.4 \\
	\midrule
	\multicolumn{2}{l}{Off-System}
			&		&		\\
	& Comms Fabric	&       2 600   &       7 220   \\
	& Global Comms	& 195 000 000	& 542 000 000   \\
	\bottomrule
\end{tabular}
\caption{Performance of Synchronization Mechanisms on 16-CPU 2.8\,GHz Intel X5550 (Nehalem) System}
\label{tab:cpu:Performance of Synchronization Mechanisms on 16-CPU 2.8GHz Intel X5550 (Nehalem) System}
\end{table}

	The first problem limits raw speed, and the second limits
	miniaturization, which in turn limits frequency.
	And even this sidesteps the power-consumption issue that
	is currently limiting production frequencies to well below
	10\,GHz.

	In addition,
	\cref{tab:cpu:CPU 0 View of Synchronization Mechanisms on 8-Socket System With Intel Xeon Platinum 8176 CPUs at 2.10GHz}
	on
	\cpageref{tab:cpu:CPU 0 View of Synchronization Mechanisms on 8-Socket System With Intel Xeon Platinum 8176 CPUs at 2.10GHz}
	represents a reasonably large system with no fewer than 448~hardware
	threads.
	Smaller systems often achieve better latency, as may be seen in
	\cref{tab:cpu:Performance of Synchronization Mechanisms on 16-CPU 2.8GHz Intel X5550 (Nehalem) System},
	which represents a much smaller system with only 16~hardware threads.
	A similar view is provided by the rows of
	\cref{tab:cpu:CPU 0 View of Synchronization Mechanisms on 8-Socket System With Intel Xeon Platinum 8176 CPUs at 2.10GHz}
	down to and including the two ``Off-Core'' rows.

\begin{table}
\renewcommand*{\arraystretch}{1.1}
\centering\small
\tcresizewidth{
\begin{tabular}
  {
    ll
    S[table-format = 9.1]
    S[table-format = 9.1]
    r
  }
	\toprule
	\multicolumn{2}{l}{Operation}
		& \multicolumn{1}{r}{Cost (ns)}
			& {\parbox[b]{.7in}{\raggedleft Ratio\\(cost/clock)}}
			& CPUs \\
	\midrule
	\multicolumn{2}{l}{Clock period}
				     &   0.5 &    1.0 &			  \\
        \midrule
	\multicolumn{2}{l}{Same-CPU} &       &        &	0		  \\
	& CAS			     &   6.2 &   13.6 &			  \\
	& lock			     &  13.5 &   29.6 &			  \\
        \midrule
	\multicolumn{2}{l}{On-Core}  &       &        &	6		  \\
	& Blind CAS		     &   6.5 &   14.3 &			  \\
	& CAS			     &  16.2 &   35.6 &			  \\
        \midrule
	\multicolumn{2}{l}{Off-Core} &       &        &	1--5		  \\
	& Blind CAS		     &  22.2 &   48.8 & 7--11		  \\
	& CAS			     &  53.6 &  117.9 &			  \\
	\midrule
	\multicolumn{2}{l}{Off-System}&       &        &		  \\
	& Comms Fabric		      & 5 000 & 11 000 &		  \\
	& Global Comms		      & 195 000 000 & 429 000 000 &	  \\
	\bottomrule
\end{tabular}
}
\caption{CPU 0 View of Synchronization Mechanisms on 12-CPU Intel Core i7-8750H CPU @ 2.20\,GHz}
\label{tab:cpu:CPU 0 View of Synchronization Mechanisms on 12-CPU Intel Core i7-8750H CPU @ 2.20GHz}
\end{table}

	Furthermore, newer small-scale single-socket systems such
	as the laptop on which I am typing this also have more
	reasonable latencies, as can be seen in
	\cref{tab:cpu:CPU 0 View of Synchronization Mechanisms on 12-CPU Intel Core i7-8750H CPU @ 2.20GHz}.

	Alternatively, a 64-CPU system in the mid 1990s had
	cross-interconnect latencies in excess of five microseconds,
	so even the eight-socket 448-hardware-thread monster shown in
	\cref{tab:cpu:CPU 0 View of Synchronization Mechanisms on 8-Socket System With Intel Xeon Platinum 8176 CPUs at 2.10GHz}
	represents more than a five-fold improvement over its
	25-years-prior counterparts.

	Integration of hardware threads in a single core and multiple
	cores on a die have improved latencies greatly, at least within the
	confines of a single core or single die.
	There has been some improvement in overall system latency,
	but only by about a factor of two.
	Unfortunately, neither the speed of light nor the atomic nature
	of matter has changed much in the past few
	years~\cite{NoBugsHare2016CPUoperations}.
	Therefore, spatial and temporal locality are first-class concerns
	for concurrent software, even when running on relatively
	small systems.

	\Cref{sec:cpu:Hardware Free Lunch?}
	looks at what else hardware designers might be
	able to do to ease the plight of parallel programmers.
\QuickE{}
\QuickQ{}
	\Cref{tab:cpu:Performance of Synchronization Mechanisms on 16-CPU 2.8GHz Intel X5550 (Nehalem) System}
	in the answer to \QuickQuizARef{\QspeedOfLightAtoms} on
	\cpageref{tab:cpu:Performance of Synchronization Mechanisms on 16-CPU 2.8GHz Intel X5550 (Nehalem) System}
	says that on-core CAS is faster than both of same-CPU CAS and
	on-core blind CAS\@.
	What is happening there?
\QuickA{}
	I \emph{was} surprised by the data I obtained and did a rigorous
	check of their validity.
	I got the same result persistently.
	One theory that might explain the observation would be:
	The two threads in the core are able to overlap their accesses,
	while the single CPU must do everything sequentially.
	Unfortunately, there seems to be no public documentation explaining
	why the Intel X5550 (Nehalem) system behaved like that.
\QuickE{}
\QuickQ{}
	These numbers are insanely large!
	How can I possibly get my head around them?
\QuickA{}
	Get a roll of toilet paper.
	In the USA, each roll will normally have somewhere around
	350--500 sheets.
	Tear off one sheet to represent a single clock cycle, setting it aside.
	Now unroll the rest of the roll.

	The resulting pile of toilet paper will likely represent a single
	\IXacr{cas} cache miss.

	For the more-expensive inter-system communications latencies,
	use several rolls (or multiple cases) of toilet paper to represent
	the communications latency.

	Important safety tip:
	Make sure to account for the needs of those you live with when
	appropriating toilet paper, especially
	in 2020 or during a similar time when store shelves are free of
	toilet paper and much else besides.

	Furthermore, for those working on kernel code, a CPU disabling
	interrupts across a cache miss is analogous to you holding your
	breath while unrolling a roll of toilet paper.
	How many rolls of toilet paper can \emph{you} unroll while holding
	your breath?
	You might wish to avoid disabling interrupts across that many
	cache misses.\footnote{
		Kudos to Matthew Wilcox for this holding-breath analogy.}
\QuickE{}
\QuickQ{}
	But individual electrons don't move anywhere near that fast,
	even in conductors!!!
	The electron drift velocity in a conductor under semiconductor
	voltage levels is on the order of only one \emph{millimeter}
	per second.
	What gives???
\QuickA{}
	Electron drift velocity tracks the long-term movement of individual
	electrons.
	It turns out that individual electrons bounce around quite
	randomly, so that their instantaneous speed is very high, but
	over the long term, they don't move very far.
	In this, electrons resemble long-distance commuters, who
	might spend most of their time traveling at full highway
	speed, but over the long term go nowhere.
	These commuters' speed might be 70 miles per hour
	(113 kilometers per hour), but their long-term drift velocity
	relative to the planet's surface is zero.

	Therefore, we should pay attention not to the electrons'
	drift velocity, but to their instantaneous velocities.
	However, even their instantaneous velocities are nowhere near
	a significant fraction of the speed of light.
	Nevertheless, the measured velocity of electric waves
	in conductors \emph{is} a substantial fraction of the
	speed of light, so we still have a mystery on our hands.

	The other trick is that electrons interact with each other at
	significant distances (from an atomic perspective, anyway),
	courtesy of their negative charge.
	This interaction is carried out by photons, which \emph{do}
	move at the speed of light.
	So even with electricity's electrons, it is photons
	doing most of the fast footwork.

	Extending the commuter analogy, a driver might use a smartphone
	to inform other drivers of an accident or congestion, thus
	allowing a change in traffic flow to propagate much faster
	than the instantaneous velocity of the individual cars.
	Summarizing the analogy between electricity and traffic flow:

	\begin{enumerate}
	\item	The (very low) drift velocity of an electron is similar
		to the long-term velocity of a commuter, both being
		very nearly zero.
	\item	The (still rather low) instantaneous velocity of
		an electron is similar to the instantaneous velocity
		of a car in traffic.
		Both are much higher than the drift velocity, but
		quite small compared to the rate at which changes
		propagate.
	\item	The (much higher) propagation velocity of an electric
		wave is primarily due to photons transmitting
		electromagnetic force among the electrons.
		Similarly, traffic patterns can change quite quickly
		due to communication among drivers.
		Not that this is necessarily of much help to the
		drivers already stuck in traffic, any more than it
		is to the electrons already pooled in a given capacitor.
	\end{enumerate}

	Of course, to fully understand this topic, you should read
	up on electrodynamics.
\QuickE{}
\QuickQ{}
	Given that distributed-systems communication is so horribly
	expensive, why does anyone bother with such systems?
\QuickA{}
	There are a number of reasons:

	\begin{enumerate}
	\item	Shared-memory multiprocessor systems have strict size limits.
		If you need more than a few thousand CPUs, you have no
		choice but to use a distributed system.
	\item	Large shared-memory systems tend to be more expensive
		per unit computation than their smaller counterparts.
	\item	Large shared-memory systems tend to have much longer
		cache-miss latencies than do smaller system.
		To see this, compare
		\cref{tab:cpu:CPU 0 View of Synchronization Mechanisms on 8-Socket System With Intel Xeon Platinum 8176 CPUs at 2.10GHz}
		on \cpageref{tab:cpu:CPU 0 View of Synchronization Mechanisms on 8-Socket System With Intel Xeon Platinum 8176 CPUs at 2.10GHz}
		with
		\cref{tab:cpu:CPU 0 View of Synchronization Mechanisms on 12-CPU Intel Core i7-8750H CPU @ 2.20GHz}.
	\item	The distributed-systems communications operations do
		not necessarily use much CPU, so that computation can
		proceed in parallel with message transfer.
	\item	Many important problems are ``embarrassingly parallel'',
		so that extremely large quantities of processing may
		be enabled by a very small number of messages.
		SETI@HOME~\cite{SETIatHOME2008}
		was but one example of such an application.
		These sorts of applications can make good use of networks
		of computers despite extremely long communications
		latencies.
	\end{enumerate}

	Thus, large shared-memory systems tend to be used for applications
	that benefit from faster latencies than can be provided by
	distributed computing, and particularly for those applications
	that benefit from a large shared memory.

	It is likely that continued work on parallel applications will
	increase the number of embarrassingly parallel applications that
	can run well on machines and/or clusters having long communications
	latencies, reductions in cost being the driving force that it is.
	That said, greatly reduced hardware latencies would be an
	extremely welcome development, both for single-system and
	for distributed computing.
\QuickE{}
\QuickQ{}
	OK, if we are going to have to apply distributed-programming
	techniques to shared-memory parallel programs, why not just
	always use these distributed techniques and dispense with
	shared memory?
\QuickA{}
	Because it is often the case that only a small fraction of
	the program is performance-critical.
	Shared-memory parallelism allows us to focus distributed-programming
	techniques on that small fraction, allowing simpler shared-memory
	techniques to be used on the non-performance-critical bulk of
	the program.
\QuickE{}
\QuickQAC{chp:Tools of the Trade}{Tools of the Trade}{qqztoolsoftrade}
\QuickQ{}
	You call these tools???
	They look more like low-level synchronization primitives to me!
\QuickA{}
	They look that way because they are in fact low-level synchronization
	primitives.
	And they are in fact the fundamental tools for building low-level
	concurrent software.
\QuickE{}
\QuickQ{}
	But this silly shell script isn't a \emph{real} parallel program!
	Why bother with such trivia???
\QuickA{}
	Because you should \emph{never} forget the simple stuff!

	Please keep in mind that the title of this book is
	``Is Parallel Programming Hard, And, If So, What Can You Do About It?''.
	One of the most effective things you can do about it is to
	avoid forgetting the simple stuff!
	After all, if you choose to do parallel programming the hard
	way, you have no one but yourself to blame.
\QuickE{}
\QuickQ{}
	Is there a simpler way to create a parallel shell script?
	If so, how?
	If not, why not?
\QuickA{}
	One straightforward approach is the shell pipeline:

\begin{VerbatimU}
grep $pattern1 | sed -e 's/a/b/' | sort
\end{VerbatimU}

	For a sufficiently large input file,
	\co{grep} will pattern-match in parallel with \co{sed}
	editing and with the input processing of \co{sort}.
	See the file \path{parallel.sh} for a demonstration of
	shell-script parallelism and pipelining.
\QuickE{}
\QuickQ{}
	But if script-based parallel programming is so easy, why
	bother with anything else?
\QuickA{}
	In fact, it is quite likely that a very large fraction of
	parallel programs in use today are script-based.
	However, script-based parallelism does have its limitations:
	\begin{enumerate}
	\item	Creation of new processes is usually quite heavyweight,
		involving the expensive \co{fork()} and \co{exec()}
		system calls.
	\item	Sharing of data, including pipelining, typically involves
		expensive file I/O.
	\item	The reliable synchronization primitives available to
		scripts also typically involve expensive file I/O.
	\item	Scripting languages are often too slow, but are often
		quite useful when coordinating execution of long-running
		programs written in lower-level programming languages.
	\end{enumerate}
	These limitations require that script-based parallelism use
	coarse-grained parallelism, with each unit of work having
	execution time of at least tens of milliseconds, and preferably
	much longer.

	Those requiring finer-grained parallelism are well advised to
	think hard about their problem to see if it can be expressed
	in a coarse-grained form.
	If not, they should consider using other parallel-programming
	environments, such as those discussed in
	\cref{sec:toolsoftrade:POSIX Multiprocessing}.
\QuickE{}
\QuickQ{}
	Why does this \co{wait()} primitive need to be so complicated?
	Why not just make it work like the shell-script \co{wait} does?
\QuickA{}
	Some parallel applications need to take special action when
	specific children exit, and therefore need to wait for each
	child individually.
	In addition, some parallel applications need to detect the
	reason that the child died.
	As we saw in \cref{lst:toolsoftrade:Using the wait() Primitive},
	it is not hard to build a \co{waitall()} function out of
	the \co{wait()} function, but it would be impossible to
	do the reverse.
	Once the information about a specific child is lost, it is lost.
\QuickE{}
\QuickQ{}
	Isn't there a lot more to \co{fork()} and \co{wait()}
	than discussed here?
\QuickA{}
	Indeed there is, and
	it is quite possible that this section will be expanded in
	future versions to include messaging features (such as UNIX
	pipes, TCP/IP, and shared file I/O) and memory mapping
	(such as \co{mmap()} and \co{shmget()}).
	In the meantime, there are any number of textbooks that cover
	these primitives in great detail,
	and the truly motivated can read manpages, existing parallel
	applications using these primitives, as well as the
	source code of the Linux-kernel implementations themselves.

	It is important to note that the parent process in
	\cref{lst:toolsoftrade:Processes Created Via fork() Do Not Share Memory}
	waits until after the child terminates to do its \co{printf()}.
	Using \co{printf()}'s buffered I/O concurrently to the same file
	from multiple processes is non-trivial, and is best avoided.
	If you really need to do concurrent buffered I/O,
	consult the documentation for your OS\@.
	For UNIX/Linux systems, Stewart Weiss's lecture notes provide
	a good introduction with informative
	examples~\cite{StewartWeiss2013UNIX}.
\QuickE{}
\QuickQ{}
	If the \co{mythread()} function in
	\cref{lst:toolsoftrade:Threads Created Via pthread-create() Share Memory}
	can simply return, why bother with \co{pthread_exit()}?
\QuickA{}
	In this simple example, there is no reason whatsoever.
	However, imagine a more complex example, where \co{mythread()}
	invokes other functions, possibly separately compiled.
	In such a case, \co{pthread_exit()} allows these other functions
	to end the thread's execution without having to pass some sort
	of error return all the way back up to \co{mythread()}.
\QuickE{}
\QuickQ{}
	If the C language makes no guarantees in presence of a data
	race, then why does the Linux kernel have so many data races?
	Are you trying to tell me that the Linux kernel is completely
	broken???
\QuickA{}
	Ah, but the Linux kernel is written in a carefully selected
	superset of the C language that includes special GNU
	extensions, such as asms, that permit safe execution even
	in presence of data races.
	In addition, the Linux kernel does not run on a number of
	platforms where data races would be especially problematic.
	For an example, consider embedded systems with 32-bit pointers
	and 16-bit busses.
	On such a system, a data race involving a store to and a load
	from a given pointer might well result in the load returning the
	low-order 16 bits of the old value of the pointer concatenated
	with the high-order 16 bits of the new value of the pointer.

	Nevertheless, even in the Linux kernel, data races can be
	quite dangerous and should be avoided where
	feasible~\cite{JonCorbet2012ACCESS:ONCE}.
\QuickE{}
\QuickQ{}
	What if I want several threads to hold the same lock at the
	same time?
\QuickA{}
	The first thing you should do is to ask yourself why you would
	want to do such a thing.
	If the answer is ``because I have a lot of data that is read
	by many threads, and only occasionally updated'', then
	POSIX reader-writer locks might be what you are looking for.
	These are introduced in
	\cref{sec:toolsoftrade:POSIX Reader-Writer Locking}.

	Another way to get the effect of multiple threads holding
	the same lock is for one thread to acquire the lock, and
	then use \co{pthread_create()} to create the other threads.
	The question of why this would ever be a good idea is left
	to the reader.
\QuickE{}
\QuickQ{}
	Why not simply make the argument to \co{lock_reader()}
	on \clnrefr{ln:toolsoftrade:lock:reader_writer:reader:b} of
	\cref{lst:toolsoftrade:Demonstration of Exclusive Locks}
	be a pointer to a \co{pthread_mutex_t}?
\QuickA{}
	Because we will need to pass \co{lock_reader()} to
	\co{pthread_create()}.
	Although we could cast the function when passing it to
	\co{pthread_create()}, function casts are quite a bit
	uglier and harder to get right than are simple pointer casts.
\QuickE{}
\QuickQ{}
	\begin{fcvref}[ln:toolsoftrade:lock:reader_writer]
	What is the \apik{READ_ONCE()} on
	\clnref{reader:read_x,writer:inc} and the
	\apik{WRITE_ONCE()} on \clnref{writer:inc} of
	\cref{lst:toolsoftrade:Demonstration of Exclusive Locks}?
	\end{fcvref}
\QuickA{}
	These macros constrain the compiler so as to prevent it from
	carrying out optimizations that would be problematic for concurrently
	accessed shared variables.
	They don't constrain the CPU at all, other than by preventing
	reordering of accesses to a given single variable.
	Note that this single-variable constraint does apply to the
	code shown in
	\cref{lst:toolsoftrade:Demonstration of Exclusive Locks}
	because only the variable \co{x} is accessed.

	For more information on \co{READ_ONCE()} and \co{WRITE_ONCE()},
	please see
	\cref{sec:toolsoftrade:Atomic Operations (gcc Classic)}.
	For more information on ordering accesses to multiple variables
	by multiple threads, please see
	\cref{chp:Advanced Synchronization: Memory Ordering}.
	In the meantime, \co{READ_ONCE(x)} has much in common with
	the \GCC\  intrinsic \co{__atomic_load_n(&x, __ATOMIC_RELAXED)}
	and \co{WRITE_ONCE(x, v)} has much in common with the \GCC\
	intrinsic \co{__atomic_store_n(&x, v, __ATOMIC_RELAXED)}.
\QuickE{}
\QuickQ{}
	Writing four lines of code for each acquisition and release
	of a \co{pthread_mutex_t} sure seems painful!
	Isn't there a better way?
\QuickA{}
	Indeed!
	And for that reason, the \co{pthread_mutex_lock()} and
	\co{pthread_mutex_unlock()} primitives are normally wrapped
	in functions that do this error checking.
	Later on, we will wrap them with the Linux kernel
	\co{spin_lock()} and \co{spin_unlock()} APIs.
\QuickE{}
\QuickQ{}
	Is ``x = 0'' the only possible output from the code fragment
	shown in
	\cref{lst:toolsoftrade:Demonstration of Same Exclusive Lock}?
	If so, why?
	If not, what other output could appear, and why?
\QuickA{}
	No.
	The reason that ``x = 0'' was output was that \co{lock_reader()}
	acquired the lock first.
	Had \co{lock_writer()} instead acquired the lock first, then
	the output would have been \mbox{``x = 3''}.
	However, because the code fragment started \co{lock_reader()} first
	and because this run was performed on a multiprocessor,
	one would normally expect \co{lock_reader()} to acquire the
	lock first.
	Nevertheless, there are no guarantees, especially on a busy system.
\QuickE{}
\QuickQ{}
	Using different locks could cause quite a bit of confusion,
	what with threads seeing each others' intermediate states.
	So should well-written parallel programs restrict themselves
	to using a single lock in order to avoid this kind of confusion?
\QuickA{}
	Although it is sometimes possible to write a program using a
	single global lock that both performs and scales well, such
	programs are exceptions to the rule.
	You will normally need to use multiple locks to attain good
	performance and scalability.

	One possible exception to this rule is ``transactional memory'',
	which is currently a research topic.
	Transactional\-/memory semantics can be loosely thought of as those
	of a single global lock with optimizations permitted and
	with the addition of rollback~\cite{HansJBoehm2009HOTPAR}.
\QuickE{}
\QuickQ{}
	In the code shown in
	\cref{lst:toolsoftrade:Demonstration of Different Exclusive Locks},
	is \co{lock_reader()} guaranteed to see all the values produced
	by \co{lock_writer()}?
	Why or why not?
\QuickA{}
	No.
	On a busy system, \co{lock_reader()} might be preempted
	for the entire duration of \co{lock_writer()}'s execution,
	in which case it would not see \emph{any} of \co{lock_writer()}'s
	intermediate states for \co{x}.
\QuickE{}
\QuickQ{}
	Wait a minute here!!!
	\Cref{lst:toolsoftrade:Demonstration of Same Exclusive Lock}
	didn't initialize shared variable~\co{x},
	so why does it need to be initialized in
	\cref{lst:toolsoftrade:Demonstration of Different Exclusive Locks}?
\QuickA{}
	See \clnrefr{ln:toolsoftrade:lock:reader_writer:x} of
	\cref{lst:toolsoftrade:Demonstration of Exclusive Locks}.
	Because the code in
	\cref{lst:toolsoftrade:Demonstration of Same Exclusive Lock}
	ran first, it could rely on the compile-time initialization of~\co{x}.
	The code in
	\cref{lst:toolsoftrade:Demonstration of Different Exclusive Locks}
	ran next, so it had to re-initialize~\co{x}.
\QuickE{}
\QuickQ{}
	Instead of using \apik{READ_ONCE()} everywhere, why not just
	declare \co{goflag} as \co{volatile} on
	\clnrefr{ln:toolsoftrade:rwlockscale:reader:goflag:e} of
	\cref{lst:toolsoftrade:Measuring Reader-Writer Lock Scalability}?
\QuickA{}
	A \co{volatile} declaration is in fact a reasonable alternative in
	this particular case.
	However, use of \apik{READ_ONCE()} has the benefit of clearly
	flagging to the reader that \co{goflag} is subject to concurrent
	reads and updates.
	Note that \apik{READ_ONCE()} is especially useful in cases where
	most of the accesses are protected by a lock (and thus \emph{not}
	subject to change), but where a few of the accesses are made outside
	of the lock.
	Using a \co{volatile} declaration in this case would make it harder
	for the reader to note the special accesses outside of the lock,
	and would also make it harder for the compiler to generate good
	code under the lock.
\QuickE{}
\QuickQ{}
	\apik{READ_ONCE()} only affects the compiler, not the CPU\@.
	Don't we also need memory barriers to make sure
	that the change in \co{goflag}'s value propagates to the
	CPU in a timely fashion in
	\cref{lst:toolsoftrade:Measuring Reader-Writer Lock Scalability}?
\QuickA{}
	No, memory barriers are not needed and won't help here.
	Memory barriers only enforce ordering among multiple memory
	references:
	They absolutely do not guarantee to expedite the propagation
	of data from one part of the system to another.\footnote{
		There have been persistent rumors of hardware in which
		memory barriers actually do expedite propagation of data,
		but no confirmed sightings.}
	This leads to a quick rule of thumb:
	You do not need memory barriers unless you are using more
	than one variable to communicate between multiple threads.

	But what about \co{nreadersrunning}?
	Isn't that a second variable used for communication?
	Indeed it is, and there really are the needed memory-barrier
	instructions buried in \apig{__sync_fetch_and_add()},
	which make sure that the thread proclaims its presence
	before checking to see if it should start.
\QuickE{}
\QuickQ{}
	Would it ever be necessary to use \apik{READ_ONCE()} when accessing
	a per-thread variable, for example, a variable declared using
	\GCC's \apig{__thread} storage class?
\QuickA{}
	It depends.
	If the per-thread variable was accessed only from its thread,
	and never from a signal handler, then no.
	Otherwise, it is quite possible that \apik{READ_ONCE()} is needed.
	We will see examples of both situations in
	\cref{sec:count:Signal-Theft Limit Counter Implementation}.

	This leads to the question of how one thread can gain access to
	another thread's \apig{__thread} variable, and the answer is that
	the second thread must store a pointer to its \apig{__thread}
	variable somewhere that the first thread has access to.
	One common approach is to maintain a linked list with one
	element per thread, and to store the address of each thread's
	\apig{__thread} variable in the corresponding element.
\QuickE{}
\QuickQ{}
	Isn't comparing against single-CPU throughput a bit harsh?
\QuickA{}
	Not at all.
	In fact, this comparison was, if anything, overly lenient.
	A more balanced comparison would be against single-CPU
	throughput with the locking primitives commented out.
\QuickE{}
\QuickQ{}
	But one microsecond is not a particularly small size for
	a critical section.
	What do I do if I need a much smaller critical section, for
	example, one containing only a few instructions?
\QuickA{}
	If the data being read \emph{never} changes, then you do not
	need to hold any locks while accessing it.
	If the data changes sufficiently infrequently, you might be
	able to checkpoint execution, terminate all threads, change
	the data, then restart at the checkpoint.

	Another approach is to keep a single exclusive lock per
	thread, so that a thread read-acquires the larger aggregate
	reader-writer lock by acquiring its own lock, and write-acquires
	by acquiring all the per-thread locks~\cite{WilsonCHsieh92a}.
	This can work quite well for readers, but causes writers
	to incur increasingly large overheads as the number of threads
	increases.

	Some other ways of efficiently handling very small critical
	sections are described in \cref{chp:Deferred Processing}.
\QuickE{}
\QuickQ{}
	The system used is a few years old, and new hardware should
	be faster.
	So why should anyone worry about reader-writer locks being slow?
\QuickA{}
	In general, newer hardware is improving.
	However, it will need to improve several orders of magnitude
	to permit reader-writer lock to achieve ideal performance on
	448 CPUs.
	Worse yet, the greater the number of CPUs, the larger the
	required performance improvement.
	The performance problems of reader-writer locking are therefore
	very likely to be with us for quite some time to come.
\QuickE{}
\QuickQ{}
	Is it really necessary to have both sets of primitives?
\QuickA{}
	Strictly speaking, no.
	One could implement any member of the second set using the
	corresponding member of the first set.
	For example, one could implement \apig{__sync_nand_and_fetch()}
	in terms of \apig{__sync_fetch_and_nand()} as follows:

\begin{VerbatimU}
tmp = v;
ret = __sync_fetch_and_nand(p, tmp);
ret = ~ret & tmp;
\end{VerbatimU}

	It is similarly possible to implement \apig{__sync_fetch_and_add()},
	\apig{__sync_fetch_and_sub()}, and \apig{__sync_fetch_and_xor()}
	in terms of their post-value counterparts.

	However, the alternative forms can be quite convenient, both
	for the programmer and for the compiler/library implementor.
\QuickE{}
\QuickQ{}
	Given that these atomic operations will often be able to
	generate single atomic instructions that are directly
	supported by the underlying instruction set, shouldn't
	they be the fastest possible way to get things done?
\QuickA{}
	Unfortunately, no.
	See \cref{chp:Counting} for some stark counterexamples.
\QuickE{}
\QuickQ{}
	What happened to \apikh{ACCESS_ONCE()}?
\QuickA{}
	In the 2018 v4.15 release, the Linux kernel's \apikh{ACCESS_ONCE()} was
	replaced by \apik{READ_ONCE()} and \apik{WRITE_ONCE()} for reads and
	writes, respectively~\cite{JonCorbet2012ACCESS:ONCE,
	JonathanCorbet2014ACCESS:ONCEcompilerBugs,
	MarkRutland2017ACCESS:ONCE:remove}.
	\apikh{ACCESS_ONCE()} was introduced as a helper in RCU code, but was
	promoted to core API soon afterward~\cite{
	PaulEMcKenney2007ACCESS:ONCE:rcu,
	LinusTorvalds2008ACCESS:ONCE:move}.
	Linux kernel's \apik{READ_ONCE()} and \apik{WRITE_ONCE()} have
	evolved into complex forms that look quite different than
	the original \apikh{ACCESS_ONCE()} implementation due to the
	need to support access-once semantics for large structures,
	but with the possibility of load/store tearing if the structure
	cannot be loaded and stored with a single machine instruction.
\QuickE{}
\QuickQ{}
	What happened to the Linux-kernel equivalents to \apipx{fork()}
	and \apipx{wait()}?
\QuickA{}
	They don't really exist.
	All tasks executing within the Linux kernel share memory,
	at least unless you want to do a huge amount of memory-mapping
	work by hand.
\QuickE{}
\QuickQ{}
	What problems could occur if the variable \co{counter} were
	incremented without the protection of \co{mutex}?
\QuickA{}
	On CPUs with load-store architectures, incrementing \co{counter}
	might compile into something like the following:

\begin{VerbatimU}
LOAD counter,r0
INC r0
STORE r0,counter
\end{VerbatimU}

	On such machines, two threads might simultaneously load the
	value of \co{counter}, each increment it, and each store the
	result.
	The new value of \co{counter} will then only be one greater
	than before, despite two threads each incrementing it.
\QuickE{}
\QuickQ{}
	What is wrong with loading
	\cref{lst:toolsoftrade:Living Dangerously Early 1990s Style}'s
	\co{global_ptr} up to three times?
\QuickA{}
	Suppose that \co{global_ptr} is initially non-\co{NULL},
	but that some other thread sets \co{global_ptr} to \co{NULL}.
	\begin{fcvref}[ln:toolsoftrade:C Compilers Can Invent Loads]
	Suppose further that \clnref{if:a} of the transformed code
	(\cref{lst:toolsoftrade:C Compilers Can Invent Loads})
	executes just before \co{global_ptr} is set to \co{NULL} and
	\clnref{if:b} just after.
	Then \clnref{if:a} will conclude that
	\co{global_ptr} is non-\co{NULL},
	\clnref{if:b} will conclude that it is less than
	\co{high_address},
	so that \clnref{do_low} passes \co{do_low()} a \co{NULL} pointer,
	which \co{do_low()} just might not be prepared to deal with.
	\end{fcvref}

	Your editor made exactly this mistake in the DYNIX/ptx
	kernel's memory allocator in the early 1990s.
	Tracking down the bug consumed a holiday weekend not just
	for your editor, but also for several of his colleagues.
	In short, this is not a new problem, nor is it likely to
	go away on its own.
\QuickE{}
\QuickQ{}
	Why does it matter whether \co{do_something()} and
	\co{do_something_else()} in
	\cref{lst:toolsoftrade:C Compilers Can Fuse Non-Adjacent Loads}
	are inline functions?
\QuickA{}
	\begin{fcvref}[ln:toolsoftrade:C Compilers Can Fuse Non-Adjacent Loads]
	Because \co{gp} is not a static variable, if either
	\co{do_something()} or \co{do_something_else()} were separately
	compiled, the compiler would have to assume that either or both
	of these two functions might change the value of \co{gp}.
	This possibility would force the compiler to reload \co{gp}
	on \clnref{p3}, thus avoiding the \co{NULL}-pointer dereference.
	\end{fcvref}
\QuickE{}
\QuickQ{}
	Ouch!
	So can't the compiler invent a store to a normal variable pretty
	much any time it likes?
\QuickA{}
	Thankfully, the answer is no.
	This is because the compiler is forbidden from introducing data races.
	The case of inventing a store just before a normal store is
	quite special:
	It is not possible for some other entity, be it CPU, thread,
	signal handler, or interrupt handler, to be able to see the
	invented store unless the code already has a data race, even
	without the invented store.
	And if the code already has a data race, it already invokes
	the dreaded spectre of undefined behavior, which allows the
	compiler to generate pretty much whatever code it wants,
	regardless of the wishes of the developer.

	But if the original store is volatile, as in \apik{WRITE_ONCE()},
	for all the compiler knows, there might be a side effect
	associated with the store that could signal some other thread,
	allowing data-race-free access to the variable.
	By inventing the store, the compiler might be introducing a
	data race, which it is not permitted to do.

	Furthermore, in
	\cref{lst:toolsoftrade:Compiler Invents an Invited Store},
	the address of that variable is passed to
	\co{do_a_bunch_of_stuff()}.
	If the compiler can see this function's definition, and
	notes that \co{a} is unconditionally stored to without
	any synchronization operations, then the compiler can be
	quite sure that it is not introducing a data race in this
	case.

	In the case of \co{volatile} and atomic variables, the compiler
	is specifically forbidden from inventing writes.
\QuickE{}
\QuickQ{}
	But aren't full memory barriers very heavyweight?
	Isn't there a cheaper way to enforce the ordering needed in
	\cref{lst:toolsoftrade:Preventing Reordering}?
\QuickA{}
	As is often the case, the answer is ``it depends''.
	However, if only two threads are accessing the \co{status}
	and \co{other_task_ready} variables, then the
	\apik{smp_store_release()} and \apik{smp_load_acquire()}
	functions discussed in
	\cref{sec:toolsoftrade:Atomic Operations}
	will suffice.
\QuickE{}
\QuickQ{}
	What needs to happen if an interrupt or signal handler
	might itself be interrupted?
\QuickA{}
	Then that interrupt handler must follow the same rules that
	are followed by other interrupted code.
	Only those handlers that cannot be themselves interrupted
	or that access no variables shared with an interrupting handler
	may safely use plain accesses, and even then only if those
	variables cannot be concurrently accessed by some other CPU or
	thread.
\QuickE{}
\QuickQ{}
	How could you work around the lack of a per-thread-variable
	API on systems that do not provide it?
\QuickA{}
	One approach would be to create an array indexed by
	\apipf{smp_thread_id()}, and another would be to use a hash
	table to map from \apipf{smp_thread_id()} to an array
	index---which is in fact what this
	set of APIs does in pthread environments.

	Another approach would be for the parent to allocate a structure
	containing fields for each desired per-thread variable, then
	pass this to the child during thread creation.
	However, this approach can impose large software-engineering
	costs in large systems.
	To see this, imagine if all global variables in a large system
	had to be declared in a single file, regardless of whether or
	not they were C static variables!
\QuickE{}
\QuickQ{}
	What do you do if you need a per-thread (not per-CPU!) variable
	in the Linux kernel?
\QuickA{}
	First, needing a per-thread variable is less likely than
	you might think.
	Per-CPU variables can often do a per-thread variable's job.
	For example, if you only need to do addition, bitwise AND,
	bitwise OR, exchange, or compare-and-exchange, then the
	\co{this_cpu_add()},
	\co{this_cpu_add_return()},
	\co{this_cpu_and()},
	\co{this_cpu_or()},
	\co{this_cpu_xchg()},
	\co{this_cpu_cmpxchg()}, and
	\co{this_cpu_cmpxchg_double()}
	operations, respectively, will do the job cheaply and atomically
	with respect to context switches, interrupt handlers, and softirq
	handlers, but \emph{not} non-maskable interrupts.

	Second, within a preemption-disabled region of code, for
	example, one surrounded by the \co{preempt_disable()} and
	\co{preempt_enable()} macros, the current task is guaranteed to
	remain executing on the current CPU\@.
	Therefore, while within one such region, any series of accesses
	to per-CPU variables is atomic with respect to context switches,
	though not with respect to interrupt handlers, softirq handlers,
	and non-maskable interrupts.
	But please be aware that a preemption-disabled region of code
	that runs for more than a few microseconds will not be looked upon
	with favor by people attempting to construct real-time systems.

	Third, a field added to the \co{task_struct} structure acts
	as set of per-task variables.
	However, there are those who keep a close eye on the size of
	this structure, and these people are likely to ask hard
	questions about the need for any added fields.
	Therefore, if your field is being added for some facility
	that is only built into some kernels, you should definitely
	place your new \co{task_struct} fields under an appropriate
	\co{#ifdef}.

	Fourth and finally, your per-task variable might instead
	be located in some other structure and protected by some
	synchronization mechanism that is already in use.
	For example, if your code must hold a given lock, can accesses
	to this storage instead be protected by that lock?
	The fact that this is at the end of the list notwithstanding,
	you should look into this possibility first, not last!
\QuickE{}
\QuickQ{}
	Wouldn't the shell normally use \apipx{vfork()} rather than
	\apipx{fork()}?
\QuickA{}
	It might well do that, however, checking is left as an exercise
	for the reader.
	But in the meantime, I hope that we can agree that \apipx{vfork()}
	is a variant of \apipx{fork()}, so that we can use \apipx{fork()}
	as a generic term covering both.
\QuickE{}
\QuickQAC{chp:Counting}{Counting}{qqzcount}
\QuickQ{}
	Why should efficient and scalable counting be hard???
	After all, computers have special hardware for the sole purpose
	of doing counting!!!
\QuickA{}
	Because the straightforward counting algorithms, for example,
	atomic operations on a shared counter, either are slow and scale
	badly, or are inaccurate, as will be seen in
	\cref{sec:count:Why Isn't Concurrent Counting Trivial?}.
\QuickE{}
\QuickQ{}
	{\bfseries Network-packet counting problem.}
	Suppose that you need to collect statistics on the number
	of networking packets transmitted and received.
	Packets might be transmitted or received by any CPU on the system.
	Suppose further that your system is capable of
	handling millions of packets per second per CPU, and that
	a systems-monitoring package reads the count every five seconds.
	How would you implement this counter?
\QuickA{}
	Hint:
	The act of updating the counter must be blazingly fast, but
	because the counter is read out only about once in five million
	updates, the act of reading out the counter can be quite slow.
	In addition, the value read out normally need not be all that
	accurate---after all, since the counter is updated a thousand
	times per millisecond, we should be able to work with a value
	that is within a few thousand counts of the ``true value'',
	whatever ``true value'' might mean in this context.
	However, the value read out should maintain roughly the same
	absolute error over time.
	For example, a 1\,\% error might be just fine when the count
	is on the order of a million or so, but might be absolutely
	unacceptable once the count reaches a trillion.
	See \cref{sec:count:Statistical Counters}.
\QuickE{}
\QuickQ{}
	{\bfseries Approximate structure-allocation limit problem.}
	Suppose that you need to maintain a count of the number of
	structures allocated in order to fail any allocations
	once the number of structures in use exceeds a limit
	(say, 10,000).
	Suppose further that the structures are short-lived, the
	limit is rarely exceeded, and a ``sloppy'' approximate limit
	is acceptable.
\QuickA{}
	Hint:
	The act of updating the counter must again be blazingly
	fast, but the counter is read out each time that the
	counter is increased.
	However, the value read out need not be accurate
	\emph{except} that it must distinguish approximately
	between values below the limit and values greater than or
	equal to the limit.
	See \cref{sec:count:Approximate Limit Counters}.
\QuickE{}
\QuickQ{}
	{\bfseries Exact structure-allocation limit problem.}
	Suppose that you need to maintain a count of the number of
	structures allocated in order to fail any allocations
	once the number of structures in use exceeds an exact limit
	(again, say 10,000).
	Suppose further that these structures are short-lived,
	and that the limit is rarely exceeded, that there is almost
	always at least one structure in use, and suppose further
	still that it is necessary to know exactly when this counter reaches
	zero, for example, in order to free up some memory
	that is not required unless there is at least one structure
	in use.
\QuickA{}
	Hint:
	The act of updating the counter must once again be blazingly
	fast, but the counter is read out each time that the
	counter is increased.
	However, the value read out need not be accurate
	\emph{except} that it absolutely must distinguish perfectly
	between values between the limit and zero on the one hand,
	and values that either are less than or equal to zero or
	are greater than or equal to the limit on the other hand.
	See \cref{sec:count:Exact Limit Counters}.
\QuickE{}
\QuickQ{}
	{\bfseries Removable I/O device access-count problem.}
	Suppose that you need to maintain a \IX{reference count} on a
	heavily used removable mass-storage device, so that you
	can tell the user when it is safe to remove the device.
	As usual, the user indicates a desire to remove the device, and
	the system tells the user when it is safe to do so.
\QuickA{}
	Hint:
	Yet again, the act of updating the counter must be blazingly
	fast and scalable in order to avoid slowing down I/O operations,
	but because the counter is read out only when the
	user wishes to remove the device, the counter read-out
	operation can be extremely slow.
	Furthermore, there is no need to be able to read out
	the counter at all unless the user has already indicated
	a desire to remove the device.
	In addition, the value read out need not be accurate
	\emph{except} that it absolutely must distinguish perfectly
	between non-zero and zero values, and even then only when
	the device is in the process of being removed.
	However, once it has read out a zero value, it must act
	to keep the value at zero until it has taken some action
	to prevent subsequent threads from gaining access to the
	device being removed.
	See \cref{sec:count:Applying Exact Limit Counters}.
\QuickE{}
\QuickQ{}
	One thing that could be simpler is \co{++} instead of that
	concatenation of \co{READ_ONCE()} and \co{WRITE_ONCE()}.
	Why all that extra typing???
\QuickA{}
	See \cref{sec:toolsoftrade:Shared-Variable Shenanigans}
	on \cpageref{sec:toolsoftrade:Shared-Variable Shenanigans}
	for more information on how the compiler can cause trouble,
	as well as how \co{READ_ONCE()} and \co{WRITE_ONCE()} can avoid
	this trouble.
\QuickE{}
\QuickQ{}
	But can't a smart compiler prove that
	\clnrefr{ln:count:count_nonatomic:inc-read:inc}
	of
	\cref{lst:count:Just Count!}
	is equivalent to the \co{++} operator and produce an x86
	add-to-memory instruction?
	And won't the CPU cache cause this to be atomic?
\QuickA{}
	Although the \co{++} operator \emph{could} be atomic, there
	is no requirement that it be so unless it is applied to a
	C11 \co{_Atomic} variable.
	And indeed, in the absence of \co{_Atomic}, \GCC\ often
	chooses to load the value to a register, increment
	the register, then store the value to memory, which is
	decidedly non-atomic.

	Furthermore, note the volatile casts in
	\co{READ_ONCE()} and \co{WRITE_ONCE()}, which tell
	the compiler that the location might well be an MMIO
	device register.
	Because MMIO registers are not cached, it would be unwise for
	the compiler to assume that the increment operation is atomic.
\QuickE{}
\QuickQ{}
	The 8-figure accuracy on the number of failures indicates
	that you really did test this.
	Why would it be necessary to test such a trivial program,
	especially when the bug is easily seen by inspection?
\QuickA{}
	Not only are there very few
	trivial parallel programs, and most days I am
	not so sure that there are many trivial sequential programs, either.

	No matter how small or simple the program, if you haven't tested
	it, it does not work.
	And even if you have tested it, Murphy's Law says that there will
	be at least a few bugs still lurking.

	Furthermore, while proofs of correctness certainly do have their
	place, they never will replace testing, including the
	\path{counttorture.h} test setup used here.
	After all, proofs are only as good as the assumptions that they
	are based on.
	Finally, proofs can be every bit as buggy as are programs!
\QuickE{}
\QuickQ{}
	Why doesn't the horizontal dashed line on the x~axis meet the
	diagonal line at $x=1$?
\QuickA{}
	Because of the overhead of the atomic operation.
	The dashed line on the x~axis represents the overhead of
	a single \emph{non-atomic} increment.
	After all, an \emph{ideal} algorithm would not only scale
	linearly, it would also incur no performance penalty compared
	to single-threaded code.

	This level of idealism may seem severe, but if it is good
	enough for \ppl{Linus}{Torvalds}, it is good enough for you.
\QuickE{}
\QuickQ{}
	But atomic increment is still pretty fast.
	And incrementing a single variable in a tight loop sounds
	pretty unrealistic to me, after all, most of the program's
	execution should be devoted to actually doing work, not accounting
	for the work it has done!
	Why should I care about making this go faster?
\QuickA{}
	In many cases, atomic increment will in fact be fast enough
	for you.
	In those cases, you should by all means use atomic increment.
	That said, there are many real-world situations where
	more elaborate counting algorithms are required.
	The canonical example of such a situation is counting packets
	and bytes in highly optimized networking stacks, where it is
	all too easy to find much of the execution time going into
	these sorts of accounting tasks, especially on large
	multiprocessors.

	In addition, as noted at the beginning of this chapter,
	counting provides an excellent view of the
	issues encountered in shared-memory parallel programs.
\QuickE{}
\QuickQ{}
	But why can't CPU designers simply ship the addition operation to the
	data, avoiding the need to circulate the cache line containing
	the global variable being incremented?
\QuickA{}
	It might well be possible to do this in some cases.
	However, there are a few complications:
	\begin{enumerate}
	\item	If the value of the variable is required, then the
		thread will be forced to wait for the operation
		to be shipped to the data, and then for the result
		to be shipped back.
	\item	If the atomic increment must be ordered with respect
		to prior and/or subsequent operations, then the thread
		will be forced to wait for the operation to be shipped
		to the data, and for an indication that the operation
		completed to be shipped back.
	\item	Shipping operations among CPUs will likely require
		more lines in the system interconnect, which will consume
		more die area and more electrical power.
	\end{enumerate}
	But what if neither of the first two conditions holds?
	Then you should think carefully about the algorithms discussed
	in \cref{sec:count:Statistical Counters}, which achieve
	near-ideal performance on commodity hardware.

\begin{figure}
\centering
\resizebox{3in}{!}{\includegraphics{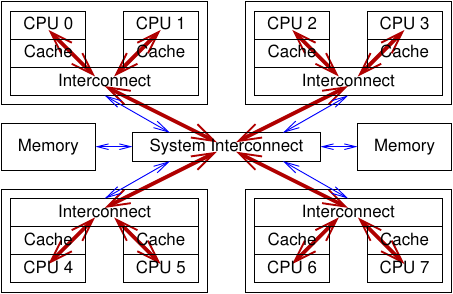}}
\caption{Data Flow For Global Combining-Tree Atomic Increment}
\label{fig:count:Data Flow For Global Combining-Tree Atomic Increment}
\end{figure}

	If either or both of the first two conditions hold, there is
	\emph{some} hope for improved hardware.
	One could imagine the hardware implementing a combining tree,
	so that the increment requests from multiple CPUs are combined
	by the hardware into a single addition when the combined request
	reaches the hardware.
	The hardware could also apply an order to the requests, thus
	returning to each CPU the return value corresponding to its
	particular atomic increment.
	This results in instruction latency that varies as $\O{\log N}$,
	where $N$ is the number of CPUs, as shown in
	\cref{fig:count:Data Flow For Global Combining-Tree Atomic Increment}.
	And CPUs with this sort of hardware optimization started to
	appear in 2011.

	This is a great improvement over the $\O{N}$ performance
	of current hardware shown in
	\cref{fig:count:Data Flow For Global Atomic Increment},
	and it is possible that hardware latencies might decrease
	further if innovations such as three-dimensional fabrication prove
	practical.
	Nevertheless, we will see that in some important special cases,
	software can do \emph{much} better.
\QuickE{}
\QuickQ{}
	But doesn't the fact that C's ``integers'' are limited in size
	complicate things?
\QuickA{}
	No, because modulo addition is still commutative and associative.
	At least as long as you use unsigned integers.
	Recall that in the C standard, overflow of signed integers results
	in undefined behavior, never mind the fact that machines that
	do anything other than wrap on overflow are quite rare these days.
	Unfortunately, compilers frequently carry out optimizations that
	assume that signed integers will not overflow, so if your code
	allows signed integers to overflow, you can run into trouble
	even on modern twos-complement hardware.

	That said, one potential source of additional complexity arises
	when attempting to gather (say) a 64-bit sum from 32-bit
	per-thread counters.
	Dealing with this added complexity is left as
	an exercise for the reader, for whom some of the techniques
	introduced later in this chapter could be quite helpful.
\QuickE{}
\QuickQ{}
	An array???
	But doesn't that limit the number of threads?
\QuickA{}
	It can, and in this toy implementation, it does.
	But it is not that hard to come up with an alternative
	implementation that permits an arbitrary number of threads,
	for example, using C11's \co{_Thread_local} facility,
	as shown in
	\cref{sec:count:Per-Thread-Variable-Based Implementation}.
\QuickE{}
\QuickQ{}
	What other nasty optimizations could \GCC\ apply?
\QuickA{}
	See \cref{sec:toolsoftrade:Shared-Variable Shenanigans,%
	sec:memorder:Compile-Time Consternation}
	for more information.
	One nasty optimization would be to apply common subexpression
	elimination to successive calls to the \co{read_count()} function,
	which might come as a surprise to code expecting changes in the
	values returned from successive calls to that function.
\QuickE{}
\QuickQ{}
	How does the per-thread \co{counter} variable in
	\cref{lst:count:Array-Based Per-Thread Statistical Counters}
	get initialized?
\QuickA{}
	The C standard specifies that the initial value of
	global variables is zero, unless they are explicitly initialized,
	thus implicitly initializing all the instances of \co{counter}
	to zero.
	Besides, in the common case where the user is interested only in
	differences between consecutive reads from statistical counters,
	the initial value is irrelevant.
\QuickE{}
\QuickQ{}
	How is the code in
	\cref{lst:count:Array-Based Per-Thread Statistical Counters}
	supposed to permit more than one counter?
\QuickA{}
	Indeed, this toy example does not support more than one counter.
	Modifying it so that it can provide multiple counters is left
	as an exercise to the reader.
\QuickE{}
\QuickQ{}
	The read operation takes time to sum up the per-thread values,
	and during that time, the counter could well be changing.
	This means that the value returned by
	\co{read_count()} in
	\cref{lst:count:Array-Based Per-Thread Statistical Counters}
	will not necessarily be exact.
	Assume that the counter is being incremented at rate
	$r$ counts per unit time, and that \co{read_count()}'s
	execution consumes $\Delta$ units of time.
	What is the expected error in the return value?
\QuickA{}
	Let's do worst-case analysis first, followed by a less
	conservative analysis.

	In the worst case, the read operation completes immediately,
	but is then delayed for $\Delta$ time units before returning,
	in which case the worst-case error is simply $r \Delta$.

	This worst-case behavior is rather unlikely, so let us instead
	consider the case where the reads from each of the $N$
	counters is spaced equally over the time period $\Delta$.
	There will be $N+1$ intervals of duration $\frac{\Delta}{N+1}$
	between the $N$ reads.
	The rate $r$ of increments is expected to be spread evenly over
	the $N$ counters, for $\frac{r}{N}$ increments per unit time
	for each individual counter.
	The error due to the delay after the read from the last thread's
	counter will be given by $\frac{r \Delta}{N \left( N + 1 \right)}$,
	the second-to-last thread's counter by
	$\frac{2 r \Delta}{N \left( N + 1 \right)}$,
	the third-to-last by
	$\frac{3 r \Delta}{N \left( N + 1 \right)}$,
	and so on.
	The total error is given by the sum of the errors due to the
	reads from each thread's counter, which is:

	\begin{equation}
		\frac{r \Delta}{N \left( N + 1 \right)}
			\sum_{i = 1}^N i
	\end{equation}

	Expressing the summation in closed form yields:

	\begin{equation}
		\frac{r \Delta}{N \left( N + 1 \right)}
			\frac{N \left( N + 1 \right)}{2}
	\end{equation}

	Canceling yields the intuitively expected result:

	\begin{equation}
		\frac{r \Delta}{2}
	\label{eq:count:CounterErrorAverage}
	\end{equation}

	It is important to remember that error continues accumulating
	as the caller executes code making use of the count returned
	by the read operation.
	For example, if the caller spends time $t$ executing some
	computation based on the result of the returned count, the
	worst-case error will have increased to $r \left(\Delta + t\right)$.

	The expected error will have similarly increased to:

	\begin{equation}
		r \left( \frac{\Delta}{2} + t \right)
	\end{equation}

	Of course, it is sometimes unacceptable for the counter to
	continue incrementing during the read operation.
	\Cref{sec:count:Applying Exact Limit Counters}
	discusses a way to handle this situation.

	Thus far, we have been considering a counter that is only
	increased, never decreased.
	If the counter value is being changed by $r$ counts per unit
	time, but in either direction, we should expect the error
	to reduce.
	However, the worst case is unchanged because although the
	counter \emph{could} move in either direction, the worst
	case is when the read operation completes immediately,
	but then is delayed for $\Delta$ time units, during which
	time all the changes in the counter's value move it in
	the same direction, again giving us an absolute error
	of $r \Delta$.

	There are a number of ways to compute the average error,
	based on a variety of assumptions about the patterns of
	increments and decrements.
	For simplicity, let's assume that the $f$ fraction of
	the operations are decrements, and that the error of interest
	is the deviation from the counter's long-term trend line.
	Under this assumption, if $f$ is less than or equal to 0.5,
	each decrement will be canceled by an increment, so that
	$2f$ of the operations will cancel each other, leaving
	$1-2f$ of the operations being uncanceled increments.
	On the other hand, if $f$ is greater than 0.5, $1-f$ of
	the decrements are canceled by increments, so that the
	counter moves in the negative direction by $-1+2\left(1-f\right)$,
	which simplifies to $1-2f$, so that the counter moves an average
	of $1-2f$ per operation in either case.
	Therefore, that the long-term
	movement of the counter is given by $\left( 1-2f \right) r$.
	Plugging this into
	\cref{eq:count:CounterErrorAverage} yields:

	\begin{equation}
		\frac{\left( 1 - 2 f \right) r \Delta}{2}
	\end{equation}

	All that aside, in most uses of statistical counters, the
	error in the value returned by \co{read_count()} is
	irrelevant.
	This irrelevance is due to the fact that the time required
	for \co{read_count()} to execute is normally extremely
	small compared to the time interval between successive
	calls to \co{read_count()}.
\QuickE{}
\QuickQ{}
	Doesn't that explicit \co{counterp} array in
	\cref{lst:count:Per-Thread Statistical Counters}
	reimpose an arbitrary limit on the number of threads?
	Why doesn't the C language provide a \co{per_thread()} interface, similar
	to the Linux kernel's \co{per_cpu()} primitive, to allow
	threads to more easily access each others' per-thread variables?
\QuickA{}
	Why indeed?

	To be fair, user-mode thread-local storage faces some challenges
	that the Linux kernel gets to ignore.
	When a user-level thread exits, its per-thread variables all
	disappear, which complicates the problem of per-thread-variable
	access, particularly before the advent of user-level RCU
	(see \cref{sec:defer:Read-Copy Update (RCU)}).
	In contrast, in the Linux kernel, when a CPU goes offline,
	that CPU's per-CPU variables remain mapped and accessible.

	Similarly, when a new user-level thread is created, its
	per-thread variables suddenly come into existence.
	In contrast, in the Linux kernel, all per-CPU variables are
	mapped and initialized at boot time, regardless of whether
	the corresponding CPU exists yet, or indeed, whether the
	corresponding CPU will ever exist.

	A key limitation that the Linux kernel imposes is a compile-time
	maximum bound on the number of CPUs, namely, \co{CONFIG_NR_CPUS},
	along with a typically tighter boot-time bound of \co{nr_cpu_ids}.
	In contrast, in user space, there is not necessarily a hard-coded
	upper limit on the number of threads.

	Of course, both environments must handle dynamically loaded
	code (dynamic libraries in user space, kernel modules in the
	Linux kernel), which increases the complexity of per-thread
	variables.

	These complications make it significantly harder for user-space
	environments to provide access to other threads' per-thread
	variables.
	Nevertheless, such access is highly useful, and it is hoped
	that it will someday appear.

	In the meantime, textbook examples such as this one can use
	arrays whose limits can be easily adjusted by the user.
	Alternatively, such arrays can be dynamically allocated and
	expanded as needed at runtime.
	Finally, variable-length data structures such as linked
	lists can be used, as is done in the userspace RCU
	library~\cite{MathieuDesnoyers2009URCU,MathieuDesnoyers2012URCU}.
	This last approach can also reduce \IX{false sharing} in some cases.
\QuickE{}
\QuickQ{}
	\begin{fcvref}[ln:count:count_end:whole:read]
	Doesn't the check for \co{NULL} on \clnref{check} of
	\cref{lst:count:Per-Thread Statistical Counters}
	add extra branch mispredictions?
	Why not have a variable set permanently to zero, and point
	unused counter-pointers to that variable rather than setting
	them to \co{NULL}?
	\end{fcvref}
\QuickA{}
	This is a reasonable strategy.
	Checking for the performance difference is left as an exercise
	for the reader.
	However, please keep in mind that the fastpath is not
	\co{read_count()}, but rather \co{inc_count()}.
\QuickE{}
\QuickQ{}
	Why on earth do we need something as heavyweight as a \emph{lock}
	guarding the summation in the function \co{read_count()} in
	\cref{lst:count:Per-Thread Statistical Counters}?
\QuickA{}
	Remember, when a thread exits, its per-thread variables disappear.
	Therefore, if we attempt to access a given thread's per-thread
	variables after that thread exits, we will get a segmentation
	fault.
	The lock coordinates summation and thread exit, preventing this
	scenario.

	Of course, we could instead read-acquire a reader-writer lock,
	but \cref{chp:Deferred Processing} will introduce even
	lighter-weight mechanisms for implementing the required coordination.

	Another approach would be to use an array instead of a per-thread
	variable, which, as Alexey Roytman notes, would eliminate
	the tests against \co{NULL}.
	However, array accesses are often slower than accesses to
	per-thread variables, and use of an array would imply a
	fixed upper bound on the number of threads.
	Also, note that neither tests nor locks are needed on the
	\co{inc_count()} fastpath.
\QuickE{}
\QuickQ{}
	Why on earth do we need to acquire the lock in
	\co{count_register_thread()} in
	\cref{lst:count:Per-Thread Statistical Counters}?
	It is a single properly aligned machine-word store to a location
	that no other thread is modifying, so it should be atomic anyway,
	right?
\QuickA{}
	This lock could in fact be omitted, but better safe than
	sorry, especially given that this function is executed only at
	thread startup, and is therefore not on any critical path.
	Now, if we were testing on machines with thousands of CPUs,
	we might need to omit the lock, but on machines with ``only''
	a hundred or so CPUs, there is no need to get fancy.
\QuickE{}
\QuickQ{}
	Fine, but the Linux kernel doesn't have to acquire a lock
	when reading out the aggregate value of per-CPU counters.
	So why should user-space code need to do this???
\QuickA{}
	Remember, the Linux kernel's per-CPU variables are always
	accessible, even if the corresponding CPU is offline---even
	if the corresponding CPU never existed and never will exist.

\begin{listing}
\begin{fcvlabel}[ln:count:count_tstat:whole]
\begin{VerbatimL}[commandchars=\\\@\$]
unsigned long __thread counter = 0;
unsigned long *counterp[NR_THREADS] = { NULL };
int finalthreadcount = 0;
DEFINE_SPINLOCK(final_mutex);

static __inline__ void inc_count(void)
{
	WRITE_ONCE(counter, counter + 1);
}

static __inline__ unsigned long read_count(void)
                  /* need to tweak counttorture! */
{
	int t;
	unsigned long sum = 0;

	for_each_thread(t) {
		if (READ_ONCE(counterp[t]) != NULL)
			sum += READ_ONCE(*counterp[t]);
	}
	return sum;
}

void count_register_thread(unsigned long *p)
{
	WRITE_ONCE(counterp[smp_thread_id()], &counter);
}

void count_unregister_thread(int nthreadsexpected)
{
	spin_lock(&final_mutex);
	finalthreadcount++;
	spin_unlock(&final_mutex);
	while (READ_ONCE(finalthreadcount) < nthreadsexpected)
		poll(NULL, 0, 1);
}
\end{VerbatimL}
\end{fcvlabel}
 \caption{Per-Thread Statistical Counters With Lockless Summation}
\label{lst:count:Per-Thread Statistical Counters With Lockless Summation}
\end{listing}

	One workaround is to ensure that each thread continues to exist
	until all threads are finished, as shown in
	\cref{lst:count:Per-Thread Statistical Counters With Lockless Summation}
	(\path{count_tstat.c}).
	Analysis of this code is left as an exercise to the reader,
	however, please note that it requires tweaks in the
	\path{counttorture.h} counter-evaluation scheme.
	(Hint:
		See \co{#ifndef KEEP_GCC_THREAD_LOCAL}.)
	\Cref{chp:Deferred Processing} will introduce
	synchronization mechanisms that handle this situation in a much
	more graceful manner.
\QuickE{}
\QuickQ{}
	Why doesn't \co{inc_count()} in
	\cref{lst:count:Array-Based Per-Thread Eventually Consistent Counters}
	need to use atomic instructions?
	After all, we now have multiple threads accessing the per-thread
	counters!
\QuickA{}
	Because one of the two threads only reads, and because the
	variable is aligned and machine-sized, non-atomic instructions
	suffice.
	That said, the \co{READ_ONCE()} macro is used to prevent
	compiler optimizations that might otherwise prevent the
	counter updates from becoming visible to
	\co{eventual()}.\footnote{
		A simple definition of \co{READ_ONCE()} is shown in
		\cref{lst:toolsoftrade:Compiler Barrier Primitive (for GCC)}.}

	An older version of this algorithm did in fact use atomic
	instructions, kudos to Ersoy Bayramoglu for pointing out that
	they are in fact unnecessary.
	However, note that on a 32-bit system,
	the per-thread \co{counter} variables
	might need to be limited to 32 bits in order to sum them accurately,
	but with a 64-bit \co{global_count} variable to avoid overflow.
	In this case, it is necessary to zero the per-thread
	\co{counter} variables periodically in order to avoid overflow,
	which does require atomic instructions.
	It is extremely important to note that this zeroing cannot
	be delayed too long or overflow of the smaller per-thread
	variables will result.
	This approach therefore imposes real-time requirements on the
	underlying system, and in turn must be used with extreme care.

	In contrast, if all variables are the same size, overflow
	of any variable is harmless because the eventual sum
	will be modulo the word size.
\QuickE{}
\QuickQ{}
	Won't the single global thread in the function \co{eventual()} of
	\cref{lst:count:Array-Based Per-Thread Eventually Consistent Counters}
	be just as severe a bottleneck as a global lock would be?
\QuickA{}
	In this case, no.
	What will happen instead is that as the number of threads increases,
	the estimate of the counter
	value returned by \co{read_count()} will become more inaccurate.
\QuickE{}
\QuickQ{}
	Won't the estimate returned by \co{read_count()} in
	\cref{lst:count:Array-Based Per-Thread Eventually Consistent Counters}
	become increasingly
	inaccurate as the number of threads rises?
\QuickA{}
	Yes.
	If this proves problematic, one fix is to provide multiple
	\co{eventual()} threads, each covering its own subset of
	the other threads.
	In more extreme cases, a tree-like hierarchy of
	\co{eventual()} threads might be required.
\QuickE{}
\QuickQ{}
	Given that in the eventually\-/consistent algorithm shown in
	\cref{lst:count:Array-Based Per-Thread Eventually Consistent Counters}
	both reads and updates have extremely low overhead
	and are extremely scalable, why would anyone bother with the
	implementation described in
	\cref{sec:count:Array-Based Implementation},
	given its costly read-side code?
\QuickA{}
	The thread executing \co{eventual()} consumes CPU time.
	As more of these eventually\-/consistent counters are added,
	the resulting \co{eventual()} threads will eventually
	consume all available CPUs.
	This implementation therefore suffers a different sort of
	scalability limitation, with the scalability limit being in
	terms of the number of eventually consistent counters rather
	than in terms of the number of threads or CPUs.

	Of course, it is possible to make other tradeoffs.
	For example, a single thread could be created to handle all
	eventually\-/consistent counters, which would limit the
	overhead to a single CPU, but would result in increasing
	update-to-read latencies as the number of counters increased.
	Alternatively, that single thread could track the update rates
	of the counters, visiting the frequently\-/updated counters
	more frequently.
	In addition, the number of threads handling the counters could
	be set to some fraction of the total number of CPUs, and
	perhaps also adjusted at runtime.
	Finally, each counter could specify its latency, and
	deadline\-/scheduling techniques could be used to provide
	the required latencies to each counter.

	There are no doubt many other tradeoffs that could be made.
\QuickE{}
\QuickQ{}
	What is the accuracy of the estimate returned by \co{read_count()} in
	\cref{lst:count:Array-Based Per-Thread Eventually Consistent Counters}?
\QuickA{}
	A straightforward way to evaluate this estimate is to use the
	analysis derived in \QuickQuizARef{\StatisticalCounterAccuracy},
	but set $\Delta$ to the interval between the beginnings of
	successive runs of the \co{eventual()} thread.
	Handling the case where a given counter has multiple \co{eventual()}
	threads is left as an exercise for the reader.
\QuickE{}
\QuickQ{}
	What fundamental difference is there between counting packets
	and counting the total number of bytes in the packets, given
	that the packets vary in size?
\QuickA{}
	When counting packets, the counter is only incremented by the
	value one.
	On the other hand, when counting bytes, the counter might
	be incremented by largish numbers.

	Why does this matter?
	Because in the increment-by-one case, the value returned will
	be exact in the sense that the counter must necessarily have
	taken on that value at some point in time, even if it is impossible
	to say precisely when that point occurred.
	In contrast, when counting bytes, two different threads might
	return values that are inconsistent with any global ordering
	of operations.

	To see this, suppose that thread~0 adds the value three to its
	counter, thread~1 adds the value five to its counter, and
	threads~2 and~3 sum the counters.
	If the system is ``weakly ordered'' or if the compiler
	uses aggressive optimizations, thread~2 might find the
	sum to be three and thread~3 might find the sum to be five.
	The only possible global orders of the sequence of values
	of the counter are 0,3,8 and 0,5,8, and neither order is
	consistent with the results obtained.

	If you missed this one, you are not alone.
	Michael Scott used this question to stump Paul E.~McKenney
	during Paul's Ph.D. defense.
\QuickE{}
\QuickQ{}
	Given that the reader must sum all the threads' counters, this
	counter-read operation could take a long time given large numbers
	of threads.
	Is there any way that the increment operation can remain
	fast and scalable while allowing readers to also enjoy
	not only reasonable performance and scalability, but also
	good accuracy?
\QuickA{}
	One approach would be to maintain a global approximation
	to the value, similar to the approach described in
	\cref{sec:count:Eventually Consistent Implementation}.
	Updaters would increment their per-thread variable, but when it
	reached some predefined limit, atomically add it to a global
	variable, then zero their per-thread variable.
	This would permit a tradeoff between average increment overhead
	and accuracy of the value read out.
	In particular, it would allow sharp bounds on the read-side
	inaccuracy.

	Another approach makes use of the fact that readers often care
	only about certain transitions in value, not in the exact value.
	This approach is examined in
	\cref{sec:count:Approximate Limit Counters}.

	The reader is encouraged to think up and try out other approaches,
	for example, using a combining tree.
\QuickE{}
\QuickQ{}
	Why does
	\cref{lst:count:Simple Limit Counter Add; Subtract; and Read}
	provide \co{add_count()} and \co{sub_count()} instead of the
	\co{inc_count()} and \co{dec_count()} interfaces show in
	\cref{sec:count:Statistical Counters}?
\QuickA{}
	Because structures come in different sizes.
	Of course, a limit counter corresponding to a specific size
	of structure might still be able to use
	\co{inc_count()} and \co{dec_count()}.
\QuickE{}
\QuickQ{}
	What is with the strange form of the condition on
	\clnrefr{ln:count:count_lim:add_sub_read:add:checklocal} of
	\cref{lst:count:Simple Limit Counter Add; Subtract; and Read}?
	Why not the more intuitive form of the fastpath shown in
	\cref{lst:count:Intuitive Fastpath}?
\QuickA{}
	Two words.
	``Integer overflow.''

	Try the formulation in \cref{lst:count:Intuitive Fastpath}
	with \co{counter} equal to~10 and
	\co{delta} equal to \co{ULONG_MAX}.
	Then try it again with the code shown in
	\cref{lst:count:Simple Limit Counter Add; Subtract; and Read}.

	A good understanding of integer overflow will be required for
	the rest of this example, so if you have never dealt with
	integer overflow before, please try several examples to get
	the hang of it.
	Integer overflow can sometimes be more difficult to get right
	than parallel algorithms!
\QuickE{}
\QuickQ{}
	Why does \co{globalize_count()} zero the per-thread variables,
	only to later call \co{balance_count()} to refill them in
	\cref{lst:count:Simple Limit Counter Add; Subtract; and Read}?
	Why not just leave the per-thread variables non-zero?
\QuickA{}
	That is in fact what an earlier version of this code did.
	But addition and subtraction are extremely cheap, and handling
	all of the special cases that arise is quite complex.
	Again, feel free to try it yourself, but beware of integer
	overflow!
\QuickE{}
\QuickQ{}
	Given that \co{globalreserve} counted against us in \co{add_count()},
	why doesn't it count for us in \co{sub_count()} in
	\cref{lst:count:Simple Limit Counter Add; Subtract; and Read}?
\QuickA{}
	The \co{globalreserve} variable tracks the sum of all threads'
	\co{countermax} variables.
	The sum of these threads' \co{counter} variables might be anywhere
	from zero to \co{globalreserve}.
	We must therefore take a conservative approach, assuming that
	all threads' \co{counter} variables are full in \co{add_count()}
	and that they are all empty in \co{sub_count()}.

	But remember this question, as we will come back to it later.
\QuickE{}
\QuickQ{}
	Suppose that one thread invokes \co{add_count()} shown in
	\cref{lst:count:Simple Limit Counter Add; Subtract; and Read},
	and then another thread invokes \co{sub_count()}.
	Won't \co{sub_count()} return failure even though the value of
	the counter is non-zero?
\QuickA{}
	Indeed it will!
	In many cases, this will be a problem, as discussed in
	\cref{sec:count:Simple Limit Counter Discussion}, and
	in those cases the algorithms from
	\cref{sec:count:Exact Limit Counters}
	will likely be preferable.
\QuickE{}
\QuickQ{}
	Why have both \co{add_count()} and \co{sub_count()} in
	\cref{lst:count:Simple Limit Counter Add; Subtract; and Read}?
	Why not simply pass a negative number to \co{add_count()}?
\QuickA{}
	Given that \co{add_count()} takes an \co{unsigned} \co{long}
	as its argument, it is going to be a bit tough to pass it a
	negative number.
	And unless you have some anti-matter memory, there is little
	point in allowing negative numbers when counting the number
	of structures in use!

	All kidding aside, it would of course be possible to combine
	\co{add_count()} and \co{sub_count()}, however, the \co{if}
	conditions on the combined function would be more complex
	than in the current pair of functions, which would in turn
	mean slower execution of these fast paths.
\QuickE{}
\QuickQ{}
	\begin{fcvref}[ln:count:count_lim:utility:balance]
	Why set \co{counter} to \co{countermax / 2} in \clnref{middle} of
	\cref{lst:count:Simple Limit Counter Utility Functions}?
	Wouldn't it be simpler to just take \co{countermax} counts?
	\end{fcvref}
\QuickA{}
	\begin{fcvref}[ln:count:count_lim:utility:balance]
	First, it really is reserving \co{countermax} counts
	(see \clnref{adjreserve}), however,
	it adjusts so that only half of these are actually in use
	by the thread at the moment.
	This allows the thread to carry out at least \co{countermax / 2}
	increments or decrements before having to refer back to
	\co{globalcount} again.

	Note that the accounting in \co{globalcount} remains accurate,
	thanks to the adjustment in \clnref{adjglobal}.
	\end{fcvref}
\QuickE{}
\QuickQ{}
	In \cref{fig:count:Schematic of Globalization and Balancing},
	even though a quarter of the remaining count up to the limit is
	assigned to thread~0, only an eighth of the remaining count is
	consumed, as indicated by the uppermost dotted line connecting
	the center and the rightmost configurations.
	Why is that?
\QuickA{}
	The reason this happened is that thread~0's \co{counter} was
	set to half of its \co{countermax}.
	Thus, of the quarter assigned to thread~0, half of that quarter
	(one eighth) came from \co{globalcount}, leaving the other half
	(again, one eighth) to come from the remaining count.

	There are two purposes for taking this approach:
	\begin{enumerate*}[(1)]
	\item To allow thread~0 to use the fastpath for decrements as
	well as increments and
	\item To reduce the inaccuracies if all threads are monotonically
	incrementing up towards the limit.
	\end{enumerate*}
	To see this last point, step through the algorithm and watch
	what it does.
\QuickE{}
\QuickQ{}
	Why is it necessary to atomically manipulate the thread's
	\co{counter} and \co{countermax} variables as a unit?
	Wouldn't it be good enough to atomically manipulate them
	individually?
\QuickA{}
	This might well be possible, but great care is required.
	Note that removing \co{counter} without first zeroing
	\co{countermax} could result in the corresponding thread
	increasing \co{counter} immediately after it was zeroed,
	completely negating the effect of zeroing the counter.

	The opposite ordering, namely zeroing \co{countermax} and then
	removing \co{counter}, can also result in a non-zero
	\co{counter}.
	To see this, consider the following sequence of events:

	\begin{enumerate}
	\item	Thread~A fetches its \co{countermax}, and finds that
		it is non-zero.
	\item	Thread~B zeroes Thread~A's \co{countermax}.
	\item	Thread~B removes Thread~A's \co{counter}.
	\item	Thread~A, having found that its \co{countermax}
		is non-zero, proceeds to add to its \co{counter},
		resulting in a non-zero value for \co{counter}.
	\end{enumerate}

	Again, it might well be possible to atomically manipulate
	\co{countermax} and \co{counter} as separate variables,
	but it is clear that great care is required.
	It is also quite likely that doing so will slow down the
	fastpath.

	Exploring these possibilities are left as exercises for
	the reader.
\QuickE{}
\QuickQ{}
	In what way does
	\clnrefr{ln:count:count_lim_atomic:var_access:var:CM_BITS} of
	\cref{lst:count:Atomic Limit Counter Variables and Access Functions}
	violate the C standard?
\QuickA{}
	It assumes eight bits per byte.
	This assumption does hold for all current commodity microprocessors
	that can be easily assembled into shared-memory multiprocessors,
	but certainly does not hold for all computer systems that have
	ever run C code.
	(What could you do instead in order to comply with the C standard?
	What drawbacks would it have?)
\QuickE{}
\QuickQ{}
	Given that there is only one \co{counterandmax} variable,
	why bother passing in a pointer to it on
	\clnrefr{ln:count:count_lim_atomic:var_access:split:func} of
	\cref{lst:count:Atomic Limit Counter Variables and Access Functions}?
\QuickA{}
	There is only one \co{counterandmax} variable \emph{per thread}.
	Later, we will see code that needs to pass other threads'
	\co{counterandmax} variables to \co{split_counterandmax()}.
\QuickE{}
\QuickQ{}
	Why does \co{merge_counterandmax()} in
	\cref{lst:count:Atomic Limit Counter Variables and Access Functions}
	return an \co{int} rather than storing directly into an
	\co{atomic_t}?
\QuickA{}
	Later, we will see that we need the \co{int} return to pass
	to the \co{atomic_cmpxchg()} primitive.
\QuickE{}
\QuickQ{}
	Yecch!
	Why the ugly \co{goto} on
	\clnrefr{ln:count:count_lim_atomic:add_sub:add:goto} of
	\cref{lst:count:Atomic Limit Counter Add and Subtract}?
	Haven't you heard of the \co{break} statement???
\QuickA{}
	Replacing the \co{goto} with a \co{break} would require keeping
	a flag to determine whether or not
	\clnrefr{ln:count:count_lim_atomic:add_sub:add:return:fs}
	should return, which
	is not the sort of thing you want on a fastpath.
	If you really hate the \co{goto} that much, your best bet would
	be to pull the fastpath into a separate function that returned
	success or failure, with ``failure'' indicating a need for the
	slowpath.
	This is left as an exercise for goto-hating readers.
\QuickE{}
\QuickQ{}
	\begin{fcvref}[ln:count:count_lim_atomic:add_sub:add]
	Why would the \co{atomic_cmpxchg()} primitive at
	\clnrefrange{atmcmpex}{loop:e} of
	\cref{lst:count:Atomic Limit Counter Add and Subtract}
	ever fail?
	After all, we picked up its old value on \clnref{split} and have not
	changed it!
	\end{fcvref}
\QuickA{}
	\begin{fcvref}[ln:count:count_lim_atomic:add_sub:add]
	Later, we will see how the \co{flush_local_count()} function in
	\cref{lst:count:Atomic Limit Counter Utility Functions 1}
	might update this thread's \co{counterandmax} variable concurrently
	with the execution of the fastpath on
	\clnrefrange{fast:b}{loop:e} of
	\cref{lst:count:Atomic Limit Counter Add and Subtract}.
	\end{fcvref}
\QuickE{}
\QuickQ{}
	What stops a thread from simply refilling its
	\co{counterandmax} variable immediately after
	\co{flush_local_count()} on
	\clnrefr{ln:count:count_lim_atomic:utility1:flush:b} of
	\cref{lst:count:Atomic Limit Counter Utility Functions 1}
	empties it?
\QuickA{}
	This other thread cannot refill its \co{counterandmax}
	until the caller of \co{flush_local_count()} releases the
	\co{gblcnt_mutex}.
	By that time, the caller of \co{flush_local_count()} will have
	finished making use of the counts, so there will be no problem
	with this other thread refilling---assuming that the value
	of \co{globalcount} is large enough to permit a refill.
\QuickE{}
\QuickQ{}
	What prevents concurrent execution of the fastpath of either
	\co{add_count()} or \co{sub_count()} from interfering with
	the \co{counterandmax} variable while
	\co{flush_local_count()} is accessing it on
	\clnrefr{ln:count:count_lim_atomic:utility1:flush:atmxchg} of
	\cref{lst:count:Atomic Limit Counter Utility Functions 1}?
\QuickA{}
	Nothing.
	Consider the following three cases:
	\begin{enumerate}
	\item	If \co{flush_local_count()}'s \co{atomic_xchg()} executes
		before the \co{split_counterandmax()} of either fastpath,
		then the fastpath will see a zero \co{counter} and
		\co{countermax}, and will thus transfer to the slowpath
		(unless of course \co{delta} is zero).
	\item	If \co{flush_local_count()}'s \co{atomic_xchg()} executes
		after the \co{split_counterandmax()} of either fastpath,
		but before that fastpath's \co{atomic_cmpxchg()},
		then the \co{atomic_cmpxchg()} will fail, causing the
		fastpath to restart, which reduces to case~1 above.
	\item	If \co{flush_local_count()}'s \co{atomic_xchg()} executes
		after the \co{atomic_cmpxchg()} of either fastpath,
		then the fastpath will (most likely) complete successfully
		before \co{flush_local_count()} zeroes the thread's
		\co{counterandmax} variable.
	\end{enumerate}
	Either way, the race is resolved correctly.
\QuickE{}
\QuickQ{}
	Given that the \co{atomic_set()} primitive does a simple
	store to the specified \co{atomic_t}, how can
	\clnrefr{ln:count:count_lim_atomic:utility2:balance:atmcset} of
	\co{balance_count()} in
	\cref{lst:count:Atomic Limit Counter Utility Functions 2}
	work correctly in face of concurrent \co{flush_local_count()}
	updates to this variable?
\QuickA{}
	The caller of both \co{balance_count()} and
	\co{flush_local_count()} hold \co{gblcnt_mutex}, so
	only one may be executing at a given time.
\QuickE{}
\QuickQ{}
	But signal handlers can be migrated to some other
	CPU while running.
	Doesn't this possibility require that atomic instructions
	and memory barriers are required to reliably communicate
	between a thread and a signal handler that interrupts that
	thread?
\QuickA{}
	No.
	If the signal handler is migrated to another CPU, then the
	interrupted thread is also migrated along with it.
\QuickE{}
\QuickQ{}
	In \cref{fig:count:Signal-Theft State Machine}, why is
	the REQ \co{theft} state colored red?
\QuickA{}
	To indicate that only the fastpath is permitted to change the
	\co{theft} state, and that if the thread remains in this
	state for too long, the thread running the slowpath will
	resend the POSIX signal.
\QuickE{}
\QuickQ{}
	In \cref{fig:count:Signal-Theft State Machine}, what is
	the point of having separate REQ and ACK \co{theft} states?
	Why not simplify the state machine by collapsing
	them into a single REQACK state?
	Then whichever of the signal handler or the fastpath gets there
	first could set the state to READY\@.
\QuickA{}
	Reasons why collapsing the REQ and ACK states would be a very
	bad idea include:
	\begin{enumerate}
	\item	The slowpath uses the REQ and ACK states to determine
		whether the signal should be retransmitted.
		If the states were collapsed, the slowpath would have
		no choice but to send redundant signals, which would
		have the unhelpful effect of needlessly slowing down
		the fastpath.
	\item	The following race would result:
		\begin{enumerate}
		\item	The slowpath sets a given thread's state to REQACK.
		\item	That thread has just finished its fastpath, and
			notes the REQACK state.
		\item	The thread receives the signal, which also notes
			the REQACK state, and, because there is no fastpath
			in effect, sets the state to READY\@.
		\item	The slowpath notes the READY state, steals the
			count, and sets the state to IDLE, and completes.
		\item	The fastpath sets the state to READY, disabling
			further fastpath execution for this thread.
		\end{enumerate}
		The basic problem here is that the combined REQACK state
		can be referenced by both the signal handler and the
		fastpath.
		The clear separation maintained by the four-state
		setup ensures orderly state transitions.
	\end{enumerate}
	That said, you might well be able to make a three-state setup
	work correctly.
	If you do succeed, compare carefully to the four-state setup.
	Is the three-state solution really preferable, and why or why not?
\QuickE{}
\QuickQ{}
	In \cref{lst:count:Signal-Theft Limit Counter Value-Migration Functions},
	doesn't \co{flush_local_count_sig()} need stronger memory barriers?
\QuickA{}
	No, that \co{smp_store_release()} suffices because this code
	communicates only with \co{flush_local_count()}, and there is
	no need for store-to-load ordering.
\QuickE{}
\QuickQ{}
	In \cref{lst:count:Signal-Theft Limit Counter Value-Migration Functions},
	why is it safe for
	\clnrefr{ln:count:count_lim_sig:migration:flush:checkmax}
	to directly access the other thread's
	\co{countermax} variable?
\QuickA{}
	Because the other thread is not permitted to change the value
	of its \co{countermax} variable unless it holds the
	\co{gblcnt_mutex} lock.
	But the caller has acquired this lock, so it is not possible
	for the other thread to hold it, and therefore the other thread
	is not permitted to change its \co{countermax} variable.
	We can therefore safely access it---but not change it.
\QuickE{}
\QuickQ{}
	In \cref{lst:count:Signal-Theft Limit Counter Value-Migration Functions},
	why doesn't
	\clnrefr{ln:count:count_lim_sig:migration:flush:signal}
	check for the current thread sending itself
	a signal?
\QuickA{}
	There is no need for an additional check.
	The caller of \co{flush_local_count()} has already invoked
	\co{globalize_count()}, so the check on
	\clnrefr{ln:count:count_lim_sig:migration:flush:checkmax}
	will have succeeded, skipping the later \co{pthread_kill()}.
\QuickE{}
\QuickQ{}
	The code shown in
	\cref{lst:count:Signal-Theft Limit Counter Data,%
	lst:count:Signal-Theft Limit Counter Value-Migration Functions}
	works with \GCC\ and POSIX\@.
	What would be required to make it also conform to the ISO C standard?
\QuickA{}
	The \co{theft} variable must be of type \co{sig_atomic_t}
	to guarantee that it can be safely shared between the signal
	handler and the code interrupted by the signal.
\QuickE{}
\QuickQ{}
	In \cref{lst:count:Signal-Theft Limit Counter Value-Migration Functions},
	why does \clnrefr{ln:count:count_lim_sig:migration:flush:signal2}
	resend the signal?
\QuickA{}
	Because many operating systems over several decades have
	had the property of losing the occasional signal.
	Whether this is a feature or a bug is debatable, but
	irrelevant.
	The obvious symptom from the user's viewpoint will not be
	a kernel bug, but rather a user application hanging.

	\emph{Your} user application hanging!
\QuickE{}
\QuickQ{}
	Not only are POSIX signals slow, sending one to each thread
	simply does not scale.
	What would you do if you had (say) 10,000 threads and needed
	the read side to be fast?
\QuickA{}
	One approach is to use the techniques shown in
	\cref{sec:count:Eventually Consistent Implementation},
	summarizing an approximation to the overall counter value in
	a single variable.
	Another approach would be to use multiple threads to carry
	out the reads, with each such thread interacting with a
	specific subset of the updating threads.
\QuickE{}
\QuickQ{}
	What if you want an exact limit counter to be exact only for
	its lower limit, but to allow the upper limit to be inexact?
\QuickA{}
	One simple solution is to overstate the upper limit by the
	desired amount.
	The limiting case of such overstatement results in the
	upper limit being set to the largest value that the counter is
	capable of representing.
\QuickE{}
\QuickQ{}
	What else had you better have done when using a biased counter?
\QuickA{}
	You had better have set the upper limit to be large enough
	accommodate the bias, the expected maximum number of accesses,
	and enough ``slop'' to allow the counter to work efficiently
	even when the number of accesses is at its maximum.
\QuickE{}
\QuickQ{}
	This is ridiculous!
	We are \emph{read}-acquiring a reader-writer lock to
	\emph{update} the counter?
	What are you playing at???
\QuickA{}
	Strange, perhaps, but true!
	Almost enough to make you think that the name
	``reader-writer lock'' was poorly chosen, isn't it?
\QuickE{}
\QuickQ{}
	What other issues would need to be accounted for in a real system?
\QuickA{}
	A huge number!

	Here are a few to start with:

	\begin{enumerate}
	\item	There could be any number of devices, so that the
		global variables are inappropriate, as are the
		lack of arguments to functions like \co{do_io()}.
	\item	Polling loops can be problematic in real systems,
		wasting CPU time and energy.
		In many cases, an event-driven design is far better,
		for example, where the last completing I/O wakes up the
		device-removal thread.
	\item	The I/O might fail, and so \co{do_io()} will likely
		need a return value.
	\item	If the device fails, the last I/O might never complete.
		In such cases, there might need to be some sort of
		timeout to allow error recovery.
	\item	Both \co{add_count()} and \co{sub_count()} can
		fail, but their return values are not checked.
	\item	Reader-writer locks do not scale well.
		One way of avoiding the high read-acquisition costs
		of reader-writer locks is presented in
		\cref{chp:Locking,chp:Deferred Processing}.
\QuickE{}
	\end{enumerate}
\QuickQ{}
	On the \path{count_stat.c} row of
	\cref{tab:count:Statistical/Limit Counter Performance on x86},
	we see that the read-side scales linearly with the number of
	threads.
	How is that possible given that the more threads there are,
	the more per-thread counters must be summed up?
\QuickA{}
	The read-side code must scan the entire fixed-size array, regardless
	of the number of threads, so there is no difference in performance.
	In contrast, in the last two algorithms, readers must do more
	work when there are more threads.
	In addition, the last two algorithms interpose an additional
	level of indirection because they map from integer thread ID
	to the corresponding \co{_Thread_local} variable.
\QuickE{}
\QuickQ{}
	Even on the fourth row of
	\cref{tab:count:Statistical/Limit Counter Performance on x86},
	the read-side performance of these statistical counter
	implementations is pretty horrible.
	So why bother with them?
\QuickA{}
	``Use the right tool for the job.''

	As can be seen from
	\cref{fig:count:Atomic Increment Scalability on x86},
	single-variable atomic increment need not apply for any job
	involving heavy use of parallel updates.
	In contrast, the algorithms shown in the top half of
	\cref{tab:count:Statistical/Limit Counter Performance on x86}
	do an excellent job of handling update-heavy situations.
	Of course, if you have a read-mostly situation, you should
	use something else, for example, an eventually consistent design
	featuring a single atomically incremented
	variable that can be read out using a single load,
	similar to the approach used in
	\cref{sec:count:Eventually Consistent Implementation}.
\QuickE{}
\QuickQ{}
	Given the performance data shown in the bottom half of
	\cref{tab:count:Statistical/Limit Counter Performance on x86},
	we should always prefer signals over atomic operations, right?
\QuickA{}
	That depends on the workload.
	Note that on a 64-core system, you need more than
	one hundred non-atomic operations (with roughly
	a 40-nanosecond performance gain) to make up for even one
	signal (with almost a 5-\emph{microsecond} performance loss).
	Although there are no shortage of workloads with far greater
	read intensity, you will need to consider your particular
	workload.

	In addition, although memory barriers have historically been
	expensive compared to ordinary instructions, you should
	check this on the specific hardware you will be running.
	The properties of computer hardware do change over time,
	and algorithms must change accordingly.
\QuickE{}
\QuickQ{}
	Can advanced techniques be applied to address the lock
	contention for readers seen in the bottom half of
	\cref{tab:count:Statistical/Limit Counter Performance on x86}?
\QuickA{}
	One approach is to give up some update-side performance, as is
	done with scalable non-zero indicators
	(SNZI)~\cite{FaithEllen:2007:SNZI}.
	There are a number of other ways one might go about this, and these
	are left as exercises for the reader.
	Any number of approaches that apply hierarchy, which replace
	frequent global-lock acquisitions with local lock acquisitions
	corresponding to lower levels of the hierarchy, should work quite well.
\QuickE{}
\QuickQ{}
	The \co{++} operator works just fine for 1,000-digit numbers!
	Haven't you heard of operator overloading???
\QuickA{}
	In the C++ language, you might well be able to use \co{++}
	on a 1,000-digit number, assuming that you had access to a
	class implementing such numbers.
	But as of 2021, the C language does not permit operator overloading.
\QuickE{}
\QuickQ{}
	But if we are going to have to partition everything, why bother
	with shared-memory multithreading?
	Why not just partition the problem completely and run as
	multiple processes, each in its own address space?
\QuickA{}
	Indeed, multiple processes with separate address spaces can be
	an excellent way to exploit parallelism, as the proponents of
	the fork-join methodology and the Erlang language would be very
	quick to tell you.
	However, there are also some advantages to shared-memory parallelism:
	\begin{enumerate}
	\item	Only the most performance-critical portions of the
		application must be partitioned, and such portions
		are usually a small fraction of the application.
	\item	Although cache misses are quite slow compared to
		individual register-to-register instructions,
		they are typically considerably faster than
		inter-process-communication primitives, which in
		turn are considerably faster than things like
		TCP/IP networking.
	\item	Shared-memory multiprocessors are readily available
		and quite inexpensive, so, in stark contrast to the
		1990s, there is little cost penalty for use of
		shared-memory parallelism.
	\end{enumerate}
	As always, use the right tool for the job!
\QuickE{}
\QuickQAC{chp:Partitioning and Synchronization Design}{Partitioning and Synchronization Design}{qqzSMPdesign}
\QuickQ{}
	Is there a better solution to the Dining
	Philosophers Problem?
\QuickA{}
	One such improved solution is shown in
	\cref{fig:SMPdesign:Dining Philosophers Problem; Fully Partitioned},
	where the philosophers are simply provided with an additional
	five forks.
	All five philosophers may now eat simultaneously, and there
	is never any need for philosophers to wait on one another.
	In addition, this approach offers greatly improved disease control.

\begin{figure}
\centering
\includegraphics[scale=.7]{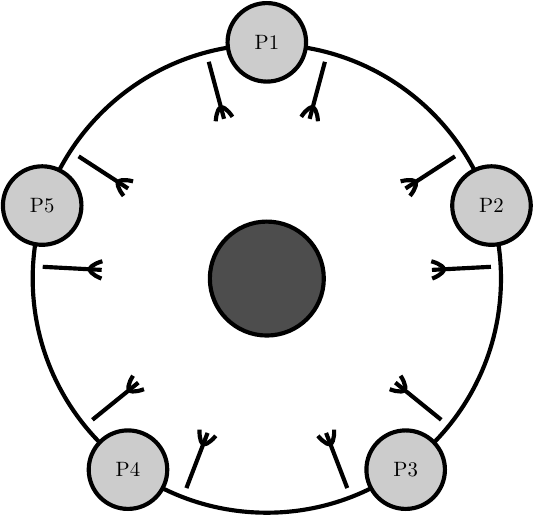}
\caption{Dining Philosophers Problem, Fully Partitioned}
\ContributedBy{Figure}{fig:SMPdesign:Dining Philosophers Problem; Fully Partitioned}{Kornilios Kourtis}
\end{figure}

	This solution might seem like cheating to some, but such
	``cheating'' is key to finding good solutions to many
	concurrency problems, as any hungry philosopher would agree.

	And this is one solution to the Dining Philosophers
	concurrent-consumption problem called out on
	\cpageref{sec:SMPdesign:Problems Dining Philosophers}.
\QuickE{}
\QuickQ{}
	How would you valididate an algorithm alleged to solve the Dining
	Philosophers Problem?
\QuickA{}
	Much depends on the details of the algorithm, but here are a
	couple of places to start.

	First, for algorithms in which picking up left-hand and right-hand
	forks are separate operations, start with all forks on the table.
	Then have all philosophers attempt to pick up their first fork.
	Once all philosophers either have their first fork or are waiting
	for someone to put down their first fork, have each non-waiting
	philosopher pick up their second fork.
	At this point in any starvation-free solution, at least one
	philosopher will be eating.
	If there were any waiting philosophers, repeat this test,
	preferably imposing random variations in timing.

	Second, create a stress test in which philosphers start and
	stop eating at random times.
	Generate starvation and fairness conditions and verify that
	these conditions are met.
	Here are a couple of example starvation and fairness conditions:

	\begin{enumerate}
	\item	If all other philosophers have stopped eating $N$ times
		since a given philosopher attempted to pick up a given
		fork, that philosopher should have succeeded in picking
		up that fork.
		For high-quality solutions using high-quality locking
		primitives (or high-quality atomic operations), $N=1$
		is doable.
	\item	Given an upper bound $T$ on the time any philosopher holds
		onto both forks before putting them down, the maximum
		waiting time for any philosopher should be bounded by
		$NT$ for some $N$ that is not hugely larger than the
		number of philosophers.
	\item	Generate some statistic representing the time from
		when philosophers attempt to pick up their first fork
		to the time when they start eating.
		The smaller this statistic, the better the solution.
		Mean, median, and maximum are all useful statistics,
		but examining the full distribution can also be
		enlightening.
	\end{enumerate}

	Readers are encouraged to actually try testing any of the
	solutions presented in this book, and especially testing solutions
	of their own devising.
\QuickE{}
\QuickQ{}
	And in just what sense can this ``horizontal parallelism'' be
	said to be ``horizontal''?
\QuickA{}
	Inman was working with protocol stacks, which are normally
	depicted vertically, with the application on top and the
	hardware interconnect on the bottom.
	Data flows up and down this stack.
	``Horizontal parallelism'' processes packets from different network
	connections in parallel, while ``vertical parallelism''
	handles different protocol-processing steps for a given
	packet in parallel.

	``Vertical parallelism'' is also called ``pipelining''.
\QuickE{}
\QuickQ{}
	In this compound double-ended queue implementation, what should
	be done if the queue has become non-empty while releasing
	and reacquiring the lock?
\QuickA{}
	In this case, simply dequeue an item from the non-empty
	queue, release both locks, and return.
\QuickE{}
\QuickQ{}
	Is the hashed double-ended queue a good solution?
	Why or why not?
\QuickA{}
	The best way to answer this is to run \path{lockhdeq.c} on
	a number of different multiprocessor systems, and you are
	encouraged to do so in the strongest possible terms.
	One reason for concern is that each operation on this
	implementation must acquire not one but two locks.

	The first well-designed performance study will be cited.\footnote{
		The studies by Dalessandro
		et al.~\cite{LukeDalessandro:2011:ASPLOS:HybridNOrecSTM:deque}
		and Dice et al.~\cite{DavidDice:2010:SCA:HTM:deque} are
		excellent starting points.}
	Do not forget to compare to a sequential implementation!
\QuickE{}
\QuickQ{}
	Move \emph{all} the elements to the queue that became empty?
	In what possible universe is this brain-dead solution in any
	way optimal???
\QuickA{}
	It is optimal in the case where data flow switches direction only
	rarely.
	It would of course be an extremely poor choice if the double-ended
	queue was being emptied from both ends concurrently.
	This of course raises another question, namely, in what possible
	universe emptying from both ends concurrently would be a reasonable
	thing to do.
	Work-stealing queues are one possible answer to this question.
\QuickE{}
\QuickQ{}
	Why can't the compound parallel double-ended queue
	implementation be symmetric?
\QuickA{}
	The need to avoid deadlock by imposing a lock hierarchy
	forces the asymmetry, just as it does in the fork-numbering
	solution to the Dining Philosophers Problem
	(see \cref{sec:SMPdesign:Dining Philosophers Problem}).
\QuickE{}
\QuickQ{}
	Why is it necessary to retry the right-dequeue operation
	on \clnrefr{ln:SMPdesign:locktdeq:pop_push:popr:deq:rr2} of
	\cref{lst:SMPdesign:Compound Parallel Double-Ended Queue Implementation}?
\QuickA{}
	\begin{fcvref}[ln:SMPdesign:locktdeq:pop_push:popr]
	This retry is necessary because some other thread might have
	enqueued an element between the time that this thread dropped
	\co{d->rlock} on \clnref{rel:r1} and the time that it reacquired this
	same lock on \clnref{acq:r2}.
	\end{fcvref}
\QuickE{}
\QuickQ{}
	Surely the left-hand lock must \emph{sometimes} be available!!!
	So why is it necessary that
	\clnrefr{ln:SMPdesign:locktdeq:pop_push:popr:rel:r1} of
	\cref{lst:SMPdesign:Compound Parallel Double-Ended Queue Implementation}
	unconditionally release the right-hand lock?
\QuickA{}
	It would be possible to use \co{spin_trylock()} to attempt
	to acquire the left-hand lock when it was available.
	However, the failure case would still need to drop the
	right-hand lock and then re-acquire the two locks in order.
	Making this transformation (and determining whether or not
	it is worthwhile) is left as an exercise for the reader.
\QuickE{}
\QuickQ{}
	But in the case where data is flowing in only one direction,
	the algorithm shown in
	\cref{lst:SMPdesign:Compound Parallel Double-Ended Queue Implementation}
	will have both ends attempting to acquire the same lock
	whenever the consuming end empties its underlying
	double-ended queue.
	Doesn't that mean that sometimes this algorithm fails to
	provide concurrent access to both ends of the queue even
	when the queue contains an arbitrarily large number of elements?
\QuickA{}
	Indeed it does!

	But the same is true of other algorithms claiming this property.
	For example, in solutions using software transactional memory
	mechanisms based on hashed arrays of locks,
	the leftmost and rightmost elements' addresses will sometimes
	happen to hash to the same lock.
	These hash collisions will also prevent concurrent access.
	For another example, solutions using hardware transactional
	memory mechanisms with software
	fallbacks~\cite{Yoo:2013:PEI:2503210.2503232,RickMerrit2011PowerTM,ChristianJacobi2012MainframeTM}
	often use locking within those software fallbacks, and thus
	suffer (albeit hopefully rarely) from whatever concurrency
	limitations that these locking solutions suffer from.

	Therefore, as of 2021, all practical solutions to the
	concurrent double-ended queue problem fail to provide full
	concurrency in at least some circumstances, including the
	compound double-ended queue.
\QuickE{}
\QuickQ{}
	Why are there not one but two solutions to the double-ended queue
	problem?
\QuickA{}
	There are actually at least three.
	The third, by Dominik Dingel, makes interesting use of
	reader-writer locking, and may be found in \path{lockrwdeq.c}.

	And so there is not one, but rather three solutions to the
	lock-based double-ended queue problem on
	\cpageref{sec:SMPdesign:Problems Double-Ended Queue}!
\QuickE{}
\QuickQ{}
	The tandem double-ended queue runs about twice as fast as
	the hashed double-ended queue, even when I increase the
	size of the hash table to an insanely large number.
	Why is that?
\QuickA{}
	The hashed double-ended queue's locking design only permits
	one thread at a time at each end, and further requires
	two lock acquisitions for each operation.
	The tandem double-ended queue also permits one thread at a time
	at each end, and in the common case requires only one lock
	acquisition per operation.
	Therefore, the tandem double-ended queue should be expected to
	outperform the hashed double-ended queue.

	Can you create a double-ended queue that allows multiple
	concurrent operations at each end?
	If so, how?
	If not, why not?
\QuickE{}
\QuickQ{}
	Is there a significantly better way of handling concurrency
	for double-ended queues?
\QuickA{}
	One approach is to transform the problem to be solved
	so that multiple double-ended queues can be used in parallel,
	allowing the simpler single-lock double-ended queue to be used,
	and perhaps also replace each double-ended queue with a pair of
	conventional single-ended queues.
	Without such ``horizontal scaling'', the speedup is limited
	to 2.0.
	In contrast, horizontal-scaling designs can achieve very large
	speedups, and are especially attractive if there are multiple threads
	working either end of the queue, because in this
	multiple-thread case the dequeue
	simply cannot provide strong ordering guarantees.
	After all, the fact that a given thread removed an item first
	in no way implies that it will process that item
	first~\cite{AndreasHaas2012FIFOisnt}.
	And if there are no guarantees, we may as well obtain the
	performance benefits that come with refusing to provide these
	guarantees.

	Regardless of whether or not the problem can be transformed
	to use multiple queues, it is worth asking whether work can
	be batched so that each enqueue and dequeue operation corresponds
	to larger units of work.
	This batching approach decreases contention on the queue data
	structures, which increases both performance and scalability,
	as will be seen in
	\cref{sec:SMPdesign:Synchronization Granularity}.
	After all, if you must incur high synchronization overheads,
	be sure you are getting your money's worth.

	Other researchers are working on other ways to take advantage
	of limited ordering guarantees in
	queues~\cite{ChristophMKirsch2012FIFOisntTR}.
\QuickE{}
\QuickQ{}
	Don't all these problems with critical sections mean that
	we should just always use
	non-blocking synchronization~\cite{MauriceHerlihy90a},
	which don't have critical sections?
\QuickA{}
	Although non-blocking synchronization can be very useful
	in some situations, it is no panacea, as discussed in
	\cref{sec:advsync:Non-Blocking Synchronization}.
	Also, non-blocking synchronization really does have
	critical sections, as noted by Josh Triplett.
	For example, in a non-blocking algorithm based on
	compare-and-swap operations, the code starting at the
	initial load and continuing to the compare-and-swap
	is analogous to a lock-based critical section.
\QuickE{}
\QuickQ{}
	What should you do to validate a hash table?
\QuickA{}
	Quite a bit, actually.

	See \cref{sec:datastruct:RCU-Protected Hash Table Validation}
	for a good starting point.
\QuickE{}
\QuickQ{}
	``Partitioning time''?
	Isn't that an odd turn of phrase?
\QuickA{}
	Perhaps so.

	But in the next section we will be partitioning space (that is,
	address space) as well as time.
	This nomenclature will permit us to partition spacetime, as
	opposed to (say) partitioning space but segmenting time.
\QuickE{}
\QuickQ{}
	What are some ways of preventing a structure from being freed while
	its lock is being acquired?
\QuickA{}
	Here are a few possible solutions to this
	\emph{\IX{existence guarantee}} problem:

	\begin{enumerate}
	\item	Provide a statically allocated lock that is held while
		the per-structure lock is being acquired, which is an
		example of hierarchical locking (see
		\cref{sec:SMPdesign:Hierarchical Locking}).
		Of course, using a single global lock for this purpose
		can result in unacceptably high levels of lock contention,
		dramatically reducing performance and scalability.
	\item	Provide an array of statically allocated locks, hashing
		the structure's address to select the lock to be acquired,
		as described in \cref{chp:Locking}.
		Given a hash function of sufficiently high quality, this
		avoids the scalability limitations of the single global
		lock, but in read-mostly situations, the lock-acquisition
		overhead can result in unacceptably degraded performance.
	\item	Use a garbage collector, in software environments providing
		them, so that a structure cannot be deallocated while being
		referenced.
		This works very well, removing the existence-guarantee
		burden (and much else besides) from the developer's
		shoulders, but imposes the overhead of garbage collection
		on the program.
		Although garbage-collection technology has advanced
		considerably in the past few decades, its overhead
		may be unacceptably high for some applications.
		In addition, some applications require that the developer
		exercise more control over the layout and placement of
		data structures than is permitted by most garbage collected
		environments.
	\item	As a special case of a garbage collector, use a global
		reference counter, or a global array of reference counters.
		These have strengths and limitations similar to those
		called out above for locks.
	\item	Use \emph{\IXpl{hazard pointer}}~\cite{MagedMichael04a}, which
		can be thought of as an inside-out reference count.
		Hazard-pointer-based algorithms maintain a per-thread list of
		pointers, so that the appearance of a given pointer on
		any of these lists acts as a reference to the corresponding
		structure.
		Hazard pointers are starting to see significant production use
		(see \cref{sec:defer:Production Uses of Hazard Pointers}).
	\item	Use transactional memory
		(TM)~\cite{Herlihy93a,DBLomet1977SIGSOFT,Shavit95},
		so that each reference and
		modification to the data structure in question is
		performed atomically.
		Although TM has engendered much excitement in recent years,
		and seems likely to be of some use in production software,
		developers should exercise some
		caution~\cite{Blundell2005DebunkTM,Blundell2006TMdeadlock,McKenney2007PLOSTM},
		particularly in performance-critical code.
		In particular, existence guarantees require that the
		transaction covers the full path from a global reference
		to the data elements being updated.
		For more on TM, including ways to overcome some of its
		weaknesses by combining it with other synchronization
		mechanisms, see
		\cref{sec:future:Transactional Memory,sec:future:Hardware Transactional Memory}.
	\item	Use RCU, which can be thought of as an extremely lightweight
		approximation to a garbage collector.
		Updaters are not permitted to free RCU-protected
		data structures that RCU readers might still be referencing.
		RCU is most heavily used for read-mostly data structures,
		and is discussed at length in
		\cref{sec:defer:Read-Copy Update (RCU)}.
	\end{enumerate}

	For more on providing existence guarantees, see
	\cref{chp:Locking,chp:Deferred Processing}.
\QuickE{}
\QuickQ{}
	But won't system boot and shutdown (or application startup
	and shutdown) be partitioning time, even for data ownership?
\QuickA{}
	You can indeed think in these terms.

	And if you are working on a persistent data store where state
	survives shutdown, thinking in these terms might even be useful.
\QuickE{}
\QuickQ{}
	How can a single-threaded 64-by-64 matrix multiple possibly
	have an efficiency of less than 1.0?
	Shouldn't all of the traces in
	\cref{fig:SMPdesign:Matrix Multiply Efficiency}
	have efficiency of exactly 1.0 when running on one thread?
\QuickA{}
	The \path{matmul.c} program creates the specified number of
	worker threads, so even the single-worker-thread case incurs
	thread-creation overhead.
	Making the changes required to optimize away thread-creation
	overhead in the single-worker-thread case is left as an
	exercise to the reader.
\QuickE{}
\QuickQ{}
	How are data-parallel techniques going to help with matrix
	multiply?
	It is \emph{already} data parallel!!!
\QuickA{}
	I am glad that you are paying attention!
	This example serves to show that although data parallelism can
	be a very good thing, it is not some magic wand that automatically
	wards off any and all sources of inefficiency.
	Linear scaling at full performance, even to ``only'' 64 threads,
	requires care at all phases of design and implementation.

	In particular, you need to pay careful attention to the
	size of the partitions.
	For example, if you split a 64-by-64 matrix multiply across
	64 threads, each thread gets only 64 floating-point multiplies.
	The cost of a floating-point multiply is minuscule compared to
	the overhead of thread creation, and cache-miss overhead
	also plays a role in spoiling the theoretically perfect scalability
	(and also in making the traces so jagged).
	The full 448~hardware threads would require a matrix with
	hundreds of thousands of rows and columns to attain good
	scalability, but by that point GPGPUs become quite attractive,
	especially from a price/performance viewpoint.

	Moral:
	If you have a parallel program with variable input,
	always include a check for the input size being too small to
	be worth parallelizing.
	And when it is not helpful to parallelize, it is not helpful
	to incur the overhead required to spawn a thread, now is it?
\QuickE{}
\QuickQ{}
	What did you do to validate this matrix multiply algorithm?
\QuickA{}
	For this simple approach, very little.

	However, the validation of production-quality matrix multiply
	requires great care and attention.
	Some cases require careful handling of floating-point rounding
	errors, others involve complex sparse-matrix data structures,
	and still others make use of special-purpose arithmetic hardware
	such as vector units or GPGPUs.
	Adequate tests for handling of floating-point rounding errors
	can be especially challenging.
\QuickE{}
\QuickQ{}
	In what situation would hierarchical locking work well?
\QuickA{}
	If the comparison on
	\clnrefr{ln:SMPdesign:Hierarchical-Locking Hash Table Search:retval} of
	\cref{lst:SMPdesign:Hierarchical-Locking Hash Table Search}
	were replaced by a much heavier-weight operation,
	then releasing \co{bp->bucket_lock} \emph{might} reduce lock
	contention enough to outweigh the overhead of the extra
	acquisition and release of \co{cur->node_lock}.
\QuickE{}
\QuickQ{}
	Doesn't this resource-allocator design resemble that of
	the approximate limit counters covered in
	\cref{sec:count:Approximate Limit Counters}?
\QuickA{}
	Indeed it does!
	We are used to thinking of allocating and freeing memory,
	but the algorithms in
	\cref{sec:count:Approximate Limit Counters}
	are taking very similar actions to allocate and free
	``count''.
\QuickE{}
\QuickQ{}
	In \cref{fig:SMPdesign:Allocator Cache Performance},
	there is a pattern of performance rising with increasing run
	length in groups of three samples, for example, for run lengths
	10, 11, and 12.
	Why?
\QuickA{}
	This is due to the per-CPU target value being three.
	A run length of 12 must acquire the global-pool lock twice,
	while a run length of 13 must acquire the global-pool lock
	three times.
\QuickE{}
\QuickQ{}
	Allocation failures were observed in the two-thread
	tests at run lengths of 19 and greater.
	Given the global-pool size of 40 and the per-thread target
	pool size $s$ of three, number of threads $n$ equal to two,
	and assuming that the per-thread pools are initially
	empty with none of the memory in use, what is the smallest allocation
	run length $m$ at which failures can occur?
	(Recall that each thread repeatedly allocates $m$ block of memory,
	and then frees the $m$ blocks of memory.)
	Alternatively, given $n$ threads each with pool size $s$, and
	where each thread repeatedly first allocates $m$ blocks of memory
	and then frees those $m$ blocks, how large must the global pool
	size be?
	\emph{Note:}
	Obtaining the correct answer will require you to
	examine the \path{smpalloc.c} source code, and very likely
	single-step it as well.
	You have been warned!
\QuickA{}
	This solution is adapted from one put forward by Alexey Roytman.
	It is based on the following definitions:

	\begin{description}
	\item[$g$]	Number of blocks globally available.
	\item[$i$]	Number of blocks left in the initializing thread's
			per-thread pool.
			(This is one reason you needed to look at the code!)
	\item[$m$]	Allocation/free run length.
	\item[$n$]	Number of threads, excluding the initialization thread.
	\item[$p$]	Per-thread maximum block consumption, including
			both the blocks actually allocated and the blocks
			remaining in the per-thread pool.
	\end{description}

	The values $g$, $m$, and $n$ are given.
	The value for $p$ is $m$ rounded up to the next multiple of $s$,
	as follows:

	\begin{equation}
		p = s \left \lceil \frac{m}{s} \right \rceil
	\label{sec:SMPdesign:p}
	\end{equation}

	The value for $i$ is as follows:

	\begin{equation}
		i = \left \{
			\begin{array}{l}
				g \pmod{2 s} = 0: 2 s \\
				g \pmod{2 s} \ne 0: g \pmod{2 s}
			\end{array}
		    \right .
	\label{sec:SMPdesign:i}
	\end{equation}

	\begin{figure}
	\centering
	\resizebox{3in}{!}{\includegraphics{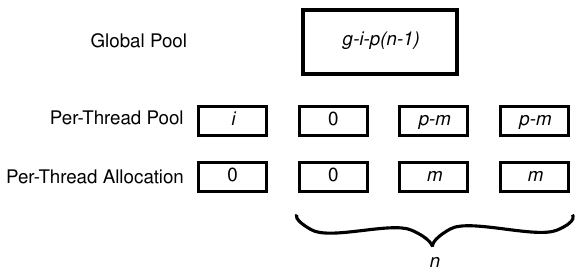}}
	\caption{Allocator Cache Run-Length Analysis}
	\label{fig:SMPdesign:Allocator Cache Run-Length Analysis}
	\end{figure}

	The relationships between these quantities are shown in
	\cref{fig:SMPdesign:Allocator Cache Run-Length Analysis}.
	The global pool is shown on the top of this figure, and
	the ``extra'' initializer thread's per-thread pool and
	per-thread allocations are the left-most pair of boxes.
	The initializer thread has no blocks allocated, but has
	$i$ blocks stranded in its per-thread pool.
	The rightmost two pairs of boxes are the per-thread pools and
	per-thread allocations of threads holding the maximum possible
	number of blocks, while the second-from-left pair of boxes
	represents the thread currently trying to allocate.

	The total number of blocks is $g$, and adding up the per-thread
	allocations and per-thread pools, we see that the global pool
	contains $g-i-p(n-1)$ blocks.
	If the allocating thread is to be successful, it needs at least
	$m$ blocks in the global pool, in other words:

	\begin{equation}
		g - i - p(n - 1) \ge m
	\label{sec:SMPdesign:g-vs-m}
	\end{equation}

	The question has $g=40$, $s=3$, and $n=2$.
	\Cref{sec:SMPdesign:i} gives $i=4$, and
	\cref{sec:SMPdesign:p} gives $p=18$ for $m=18$
	and $p=21$ for $m=19$.
	Plugging these into \cref{sec:SMPdesign:g-vs-m}
	shows that $m=18$ will not overflow, but that $m=19$ might
	well do so.

	The presence of $i$ could be considered to be a bug.
	After all, why allocate memory only to have it stranded in
	the initialization thread's cache?
	One way of fixing this would be to provide a \co{memblock_flush()}
	function that flushed the current thread's pool into the
	global pool.
	The initialization thread could then invoke this function
	after freeing all of the blocks.
\QuickE{}
\QuickQ{}
	Given that a 2D maze achieved 4x speedup on two CPUs, would
	a 3D maze achieve an 8x speedup on two CPUs?
\QuickA{}
	This is an excellent question that is left to a suitably
	interested and industrious reader.
\QuickE{}
\QuickQ{}
	Why place the third, fourth, and so on threads on the diagonal?
	Why not instead distribute them evenly around the maze?
\QuickA{}
	There are indeed a great many ways to distribute the extra
	threads.
	Evaluation of distribution strategies is left to a suitably
	interested and industrious reader.
\QuickE{}
\QuickQAC{chp:Locking}{Locking}{qqzlocking}
\QuickQ{}
	Just how can serving as a whipping boy be considered to be
	in any way honorable???
\QuickA{}
	The reason locking serves as a research-paper whipping boy is
	because it is heavily used in practice.
	In contrast, if no one used or cared about locking, most research
	papers would not bother even mentioning it.
\QuickE{}
\QuickQ{}
	But the definition of lock-based deadlock only said that each
	thread was holding at least one lock and waiting on another lock
	that was held by some thread.
	How do you know that there is a cycle?
\QuickA{}
	Suppose that there is no cycle in the graph.
	We would then have a directed acyclic graph (DAG), which would
	have at least one leaf node.

	If this leaf node was a lock, then we would have a thread
	that was waiting on a lock that wasn't held by any thread,
	counter to the definition.
	In this case the thread would immediately acquire the lock.

	On the other hand, if this leaf node was a thread, then
	we would have a thread that was not waiting on any lock,
	again counter to the definition.
	And in this case, the thread would either be running or
	be blocked on something that is not a lock.
	In the first case, in the absence of infinite-loop bugs,
	the thread will eventually release the lock.
	In the second case, in the absence of a failure-to-wake bug,
	the thread will eventually wake up and release the lock.\footnote{
		Of course, one type of failure-to-wake bug is a
		deadlock that involves not only locks, but also non-lock
		resources.
		But the question really did say ``lock-based deadlock''!}

	Therefore, given this definition of lock-based deadlock, there
	must be a cycle in the corresponding graph.
\QuickE{}
\QuickQ{}
	Are there any exceptions to this rule, so that there really could be
	a deadlock cycle containing locks from both the library and
	the caller, even given that the library code never invokes
	any of the caller's functions?
\QuickA{}
	Indeed there are!
	Here are a few of them:
	\begin{enumerate}
	\item	If one of the library function's arguments is a pointer
		to a lock that this library function acquires, and if
		the library function holds one of its locks while
		acquiring the caller's lock, then we could have a
		deadlock cycle involving both caller and library locks.
	\item	If one of the library functions returns a pointer to
		a lock that is acquired by the caller, and if the
		caller acquires one of its locks while holding the
		library's lock, we could again have a deadlock
		cycle involving both caller and library locks.
	\item	If one of the library functions acquires a lock and
		then returns while still holding it, and if the caller
		acquires one of its locks, we have yet another way
		to create a deadlock cycle involving both caller
		and library locks.
	\item	If the caller has a signal handler that acquires
		locks, then the deadlock cycle can involve both
		caller and library locks.
		In this case, however, the library's locks are
		innocent bystanders in the deadlock cycle.
		That said, please note that acquiring a lock from
		within a signal handler is a no-no in many
		environments---it is not just a bad idea, it
		is unsupported.
		But if you absolutely must acquire a lock in a signal
		handler, be sure to block that signal while holding that
		same lock in thread context, and also while holding any
		other locks acquired while that same lock is held.
\QuickE{}
	\end{enumerate}
\QuickQ{}
	But if \co{qsort()} releases all its locks before invoking
	the comparison function, how can it protect against races
	with other \co{qsort()} threads?
\QuickA{}
	By privatizing the data elements being compared
	(as discussed in \cref{chp:Data Ownership})
	or through use of deferral mechanisms such as
	reference counting (as discussed in
	\cref{chp:Deferred Processing}).
	Or through use of layered locking hierarchies, as described
	in \cref{sec:locking:Layered Locking Hierarchies}.

	On the other hand, changing a key in a list that is
	currently being sorted is at best rather brave.
\QuickE{}
\QuickQ{}
	So the iterating thread may or may not observe the added child.
	What is the big deal?
\QuickA{}
	There are at least two hazards in this situation.

	One is indeed that the number of children may or may not be
	observed to have changed.
	While that would be consistent with \co{tree_add()} being called
	either before or after the iterator started, it is better not
	left to the vagaries of the compiler.
	A more serious problem is that \co{realloc()} may not be able
	to extend the array in place, causing the heap to free the
	one used by the iterator and replace it with another block of
	memory.
	If the \co{children} pointer is not re-read then the iterating
	thread will access invalid memory (either free or reclaimed).
\QuickE{}
\QuickQ{}
	What do you mean ``cannot always safely invoke the scheduler''?
	Either \co{call_rcu()} can or cannot safely invoke the scheduler,
	right?
\QuickA{}
	It really does depend.

	The scheduler locks are always held with interrupts disabled.
	Therefore, if \co{call_rcu()} is invoked with interrupts
	enabled, no scheduler locks are held, and \co{call_rcu()}
	can safely call into the scheduler.
	Otherwise, if interrupts are disabled, one of the scheduler
	locks \emph{might} be held, so \co{call_rcu()} must play it
	safe and refrain from calling into the scheduler.
\QuickE{}
\QuickQ{}
	Name one common situation where a pointer to a lock is passed
	into a function.
\QuickA{}
	Locking primitives, of course!
\QuickE{}
\QuickQ{}
	Doesn't the fact that \co{pthread_cond_wait()} first releases the
	mutex and then re-acquires it eliminate the possibility of deadlock?
\QuickA{}
	Absolutely not!

	Consider a program that acquires \co{mutex_a}, and then
	\co{mutex_b}, in that order, and then passes \co{mutex_a}
	to \co{pthread_cond_wait()}.
	Now, \co{pthread_cond_wait()} will release \co{mutex_a}, but
	will re-acquire it before returning.
	If some other thread acquires \co{mutex_a} in the meantime
	and then blocks on \co{mutex_b}, the program will deadlock.
\QuickE{}
\QuickQ{}
	Can the transformation from
	\cref{lst:locking:Protocol Layering and Deadlock} to
	\cref{lst:locking:Avoiding Deadlock Via Conditional Locking}
	be applied universally?
\QuickA{}
	Absolutely not!

	This transformation assumes that the
	\co{layer_2_processing()} function is idempotent, given that
	it might be executed multiple times on the same packet when
	the \co{layer_1()} routing decision changes.
	Therefore, in real life, this transformation can become
	arbitrarily complex.
\QuickE{}
\QuickQ{}
	But the complexity in
	\cref{lst:locking:Avoiding Deadlock Via Conditional Locking}
	is well worthwhile given that it avoids deadlock, right?
\QuickA{}
	Maybe.

	If the routing decision in \co{layer_1()} changes often enough,
	the code will always retry, never making forward progress.
	This is termed ``\IX{livelock}'' if no thread makes any
	forward progress or ``\IX{starvation}''
	if some threads make forward progress but others do not
	(see \cref{sec:locking:Livelock and Starvation}).
\QuickE{}
\QuickQ{}
	When using the ``acquire needed locks first'' approach described in
	\cref{sec:locking:Acquire Needed Locks First},
	how can livelock be avoided?
\QuickA{}
	Provide an additional global lock.
	If a given thread has repeatedly tried and failed to acquire the needed
	locks, then have that thread unconditionally acquire the new
	global lock, and then unconditionally acquire any needed locks.
	(Suggested by Doug Lea.)
\QuickE{}
\QuickQ{}
	Suppose Lock~A is never acquired within a signal handler,
	but Lock~B is acquired both from thread context and by signal
	handlers.
	Suppose further that Lock~A is sometimes acquired with signals
	unblocked.
	Why is it illegal to acquire Lock~A holding Lock~B?
\QuickA{}
	Because this would lead to deadlock.
	Given that Lock~A is sometimes held outside of a signal
	handler without blocking signals, a signal might be handled while
	holding this lock.
	The corresponding signal handler might then acquire
	Lock~B, so that Lock~B is acquired while holding Lock~A\@.
	Therefore, if we also acquire Lock~A while holding Lock~B,
	we will have a deadlock cycle.
	Note that this problem exists even if signals are blocked while
	holding Lock~B.

	This is another reason to be very careful with locks that are
	acquired within interrupt or signal handlers.
	But the Linux kernel's lock dependency checker knows about this
	situation and many others as well, so please do make full use
	of it!
\QuickE{}
\QuickQ{}
	How can you legally block signals within a signal handler?
\QuickA{}
	One of the simplest and fastest ways to do so is to use
	the \co{sa_mask} field of the \co{struct sigaction} that
	you pass to \co{sigaction()} when setting up the signal.
\QuickE{}
\QuickQ{}
	If acquiring locks in signal handlers is such a bad idea, why
	even discuss ways of making it safe?
\QuickA{}
	Because these same rules apply to the interrupt handlers used in
	operating-system kernels and in some embedded applications.

	In many application environments, acquiring locks in signal
	handlers is frowned upon~\cite{OpenGroup1997pthreads}.
	However, that does not stop clever developers from (perhaps
	unwisely) fashioning home-brew locks out of atomic operations.
	And atomic operations are in many cases perfectly legal in
	signal handlers.
\QuickE{}
\QuickQ{}
	Given an object-oriented application that passes control freely
	among a group of objects such that there is no straightforward
	locking hierarchy,\footnote{
		Also known as ``object-oriented spaghetti code.''}
	layered or otherwise, how can this
	application be parallelized?
\QuickA{}
	There are a number of approaches:
	\begin{enumerate}
	\item	In the case of parametric search via simulation,
		where a large number of simulations will be run
		in order to converge on (for example) a good design
		for a mechanical or electrical device, leave the
		simulation single-threaded, but run many instances
		of the simulation in parallel.
		This retains the object-oriented design, and gains
		parallelism at a higher level, and likely also avoids
		both deadlocks and synchronization overhead.
	\item	Partition the objects into groups such that there
		is no need to operate on objects in
		more than one group at a given time.
		Then associate a lock with each group.
		This is an example of a single-lock-at-a-time
		design, which discussed in
		\cref{sec:locking:Single-Lock-at-a-Time Designs}.
	\item	Partition the objects into groups such that threads
		can all operate on objects in the groups in some
		groupwise ordering.
		Then associate a lock with each group, and impose a
		locking hierarchy over the groups.
	\item	Impose an arbitrarily selected hierarchy on the locks,
		and then use conditional locking if it is necessary
		to acquire a lock out of order, as was discussed in
		\cref{sec:locking:Conditional Locking}.
	\item	Before carrying out a given group of operations, predict
		which locks will be acquired, and attempt to acquire them
		before actually carrying out any updates.
		If the prediction turns out to be incorrect, drop
		all the locks and retry with an updated prediction
		that includes the benefit of experience.
		This approach was discussed in
		\cref{sec:locking:Acquire Needed Locks First}.
	\item	Use transactional memory.
		This approach has a number of advantages and disadvantages
		which will be discussed in
		\crefrange{sec:future:Transactional Memory}{sec:future:Hardware Transactional Memory}.
	\item	Refactor the application to be more concurrency-friendly.
		This would likely also have the side effect of making
		the application run faster even when single-threaded, but might
		also make it more difficult to modify the application.
	\item	Use techniques from later chapters in addition to locking.
\QuickE{}
	\end{enumerate}
\QuickQ{}
	How can the livelock shown in
	\cref{lst:locking:Abusing Conditional Locking}
	be avoided?
\QuickA{}
	\Cref{lst:locking:Avoiding Deadlock Via Conditional Locking}
	provides some good hints.
	In many cases, livelocks are a hint that you should revisit your
	locking design.
	Or visit it in the first place if your locking design
	``just grew''.

	That said, one good-and-sufficient approach due to Doug Lea
	is to use conditional locking as described in
	\cref{sec:locking:Conditional Locking}, but combine this
	with acquiring all needed locks first, before modifying shared
	data, as described in
	\cref{sec:locking:Acquire Needed Locks First}.
	If a given critical section retries too many times,
	unconditionally acquire
	a global lock, then unconditionally acquire all the needed locks.
	This avoids both deadlock and livelock, and scales reasonably
	assuming that the global lock need not be acquired too often.
\QuickE{}
\QuickQ{}
	What problems can you spot in the code in
	\cref{lst:locking:Conditional Locking and Exponential Backoff}?
\QuickA{}
	Here are a couple:
	\begin{enumerate}
	\item	A one-second wait is way too long for most uses.
		Wait intervals should begin with roughly the time
		required to execute the critical section, which will
		normally be in the microsecond or millisecond range.
	\item	The code does not check for overflow.
		On the other hand, this bug is nullified
		by the previous bug:
		32 bits worth of seconds is more than 50 years.
\QuickE{}
	\end{enumerate}
\QuickQ{}
	Wouldn't it be better just to use a good parallel design
	so that lock contention was low enough to avoid unfairness?
\QuickA{}
	It would be better in some sense, but there are situations
	where it can be appropriate to use
	designs that sometimes result in high lock contentions.

	For example, imagine a system that is subject to a rare error
	condition.
	It might well be best to have a simple error-handling design
	that has poor performance and scalability for the duration of
	the rare error condition, as opposed to a complex and
	difficult-to-debug design that is helpful only when one of
	those rare error conditions is in effect.

	That said, it is usually worth putting some effort into
	attempting to produce a design that both simple as well as
	efficient during error conditions, for example by partitioning
	the problem.
\QuickE{}
\QuickQ{}
	How might the lock holder be interfered with?
\QuickA{}
	If the data protected by the lock is in the same cache line
	as the lock itself, then attempts by other CPUs to acquire
	the lock will result in expensive cache misses on the part
	of the CPU holding the lock.
	This is a special case of \IX{false sharing}, which can also occur
	if a pair of variables protected by different locks happen
	to share a cache line.
	In contrast, if the lock is in a different cache line than
	the data that it protects, the CPU holding the lock will
	usually suffer a cache miss only on first access to a given
	variable.

	Of course, the downside of placing the lock and data into separate
	cache lines is that the code will incur two cache misses rather
	than only one in the uncontended case.
	As always, choose wisely!
\QuickE{}
\QuickQ{}
	Does it ever make sense to have an exclusive lock acquisition
	immediately followed by a release of that same lock, that is,
	an empty critical section?
\QuickA{}
	Empty lock-based critical sections are rarely used, but they
	do have their uses.
	The point is that the semantics of exclusive locks have two
	components:
	\begin{enumerate*}[(1)]
	\item The familiar data-protection semantic and
	\item A messaging semantic, where releasing a given lock notifies
	a waiting acquisition of that same lock.
	\end{enumerate*}
	An empty critical section uses the messaging component without
	the data-protection component.

	The rest of this answer provides some example uses of empty
	critical sections, however, these examples should be considered
	``gray magic.''\footnote{
		Thanks to Alexey Roytman for this description.}
	As such, empty critical sections are almost never used in practice.
	Nevertheless, pressing on into this gray area \ldots

	One historical use of empty critical sections appeared in the
	networking stack of the 2.4 Linux kernel through use of a
	read-side-scalable reader-writer lock called \co{brlock}
	for ``big reader lock''.
	This use case is a way of approximating the semantics of read-copy
	update (RCU), which is discussed in
	\cref{sec:defer:Read-Copy Update (RCU)}.
	And in fact this Linux-kernel use case has been replaced
	with RCU\@.

	The empty-lock-critical-section idiom can also be used to
	reduce lock contention in some situations.
	For example, consider a multithreaded user-space application where
	each thread processes units of work maintained in a per-thread
	list, where threads are prohibited from touching each others'
	lists~\cite{PaulEMcKenney2012EmptyLocks}.
	There could also be updates that require that all previously
	scheduled units of work have completed before the update can
	progress.
	One way to handle this is to schedule a unit of work on each
	thread, so that when all of these units of work complete, the
	update may proceed.

	In some applications, threads can come and go.
	For example, each thread might correspond to one user of the
	application, and thus be removed when that user logs out or
	otherwise disconnects.
	In many applications, threads cannot depart atomically:
	They must instead explicitly unravel themselves from various
	portions of the application using a specific sequence of actions.
	One specific action will be refusing to accept further requests
	from other threads, and another specific action will be disposing
	of any remaining units of work on its list, for example, by
	placing these units of work in a global work-item-disposal list
	to be taken by one of the remaining threads.
	(Why not just drain the thread's work-item list by executing
	each item?
	Because a given work item might generate more work items, so
	that the list could not be drained in a timely fashion.)

	If the application is to perform and scale well, a good locking
	design is required.
	One common solution is to have a global lock (call it \co{G})
	protecting the entire
	process of departing (and perhaps other things as well),
	with finer-grained locks protecting the
	individual unraveling operations.

	Now, a departing thread must clearly refuse to accept further
	requests before disposing of the work on its list, because
	otherwise additional work might arrive after the disposal action,
	which would render that disposal action ineffective.
	So simplified pseudocode for a departing thread might be as follows:

	\begin{enumerate}
	\item	Acquire lock \co{G}.
	\item	Acquire the lock guarding communications.
	\item	Refuse further communications from other threads.
	\item	Release the lock guarding communications.
	\item	Acquire the lock guarding the global work-item-disposal list.
	\item	Move all pending work items to the global
		work-item-disposal list.
	\item	Release the lock guarding the global work-item-disposal list.
	\item	Release lock \co{G}.
	\end{enumerate}

	Of course, a thread that needs to wait for all pre-existing work
	items will need to take departing threads into account.
	To see this, suppose that this thread starts waiting for all
	pre-existing work items just after a departing thread has refused
	further communications from other threads.
	How can this thread wait for the departing thread's work items
	to complete, keeping in mind that threads are not allowed to
	access each others' lists of work items?

	One straightforward approach is for this thread to acquire \co{G}
	and then the lock guarding the global work-item-disposal list, then
	move the work items to its own list.
	The thread then release both locks,
	places a work item on the end of its own list,
	and then wait for all of the work items that it placed on each thread's
	list (including its own) to complete.

	This approach does work well in many cases, but if special
	processing is required for each work item as it is pulled in
	from the global work-item-disposal list, the result could be
	excessive contention on \co{G}.
	One way to avoid that contention is to acquire \co{G} and then
	immediately release it.
	Then the process of waiting for all prior work items look
	something like the following:

	\begin{enumerate}
	\item	Set a global counter to one and initialize a condition
		variable to zero.
	\item	Send a message to all threads to cause them to atomically
		increment the global counter, and then to enqueue a
		work item.
		The work item will atomically decrement the global
		counter, and if the result is zero, it will set a
		condition variable to one.
	\item	Acquire \co{G}, which will wait on any currently departing
		thread to finish departing.
		Because only one thread may depart at a time, all the
		remaining threads will have already received the message
		sent in the preceding step.
	\item	Release \co{G}.
	\item	Acquire the lock guarding the global work-item-disposal list.
	\item	Move all work items from the global work-item-disposal list
		to this thread's list, processing them as needed along the way.
	\item	Release the lock guarding the global work-item-disposal list.
	\item	Enqueue an additional work item onto this thread's list.
		(As before, this work item will atomically decrement
		the global counter, and if the result is zero, it will
		set a condition variable to one.)
	\item	Wait for the condition variable to take on the value one.
	\end{enumerate}

	Once this procedure completes, all pre-existing work items are
	guaranteed to have completed.
	The empty critical sections are using locking for messaging as
	well as for protection of data.
\QuickE{}
\QuickQ{}
	Is there any other way for the VAX/VMS DLM to emulate
	a reader-writer lock?
\QuickA{}
	There are in fact several.
	One way would be to use the null, protected-read, and exclusive
	modes.
	Another way would be to use the null, protected-read, and
	concurrent-write modes.
	A third way would be to use the null, concurrent-read, and
	exclusive modes.
\QuickE{}
\QuickQ{}
	The code in
	\cref{lst:locking:Conditional Locking to Reduce Contention}
	is ridiculously complicated!
	Why not conditionally acquire a single global lock?
\QuickA{}
	Conditionally acquiring a single global lock does work very well,
	but only for relatively small numbers of CPUs.
	To see why it is problematic in systems with many hundreds of
	CPUs, look at
	\cref{fig:count:Atomic Increment Scalability on x86}.
\QuickE{}
\QuickQ{}
	\begin{fcvref}[ln:locking:Conditional Locking to Reduce Contention]
	Wait a minute!
	If we ``win'' the tournament on \clnref{flag_not_set} of
	\cref{lst:locking:Conditional Locking to Reduce Contention},
	we get to do all the work of \co{do_force_quiescent_state()}.
	Exactly how is that a win, really?
	\end{fcvref}
\QuickA{}
	How indeed?
	This just shows that in concurrency, just as in life, one
	should take care to learn exactly what winning entails before
	playing the game.
\QuickE{}
\QuickQ{}
	\begin{fcvref}[ln:locking:xchglock:lock_unlock]
	Why not rely on the C language's default initialization of
	zero instead of using the explicit initializer shown on
	\clnref{initval} of
	\cref{lst:locking:Sample Lock Based on Atomic Exchange}?
	\end{fcvref}
\QuickA{}
	Because this default initialization does not apply to locks
	allocated as auto variables within the scope of a function.
\QuickE{}
\QuickQ{}
	\begin{fcvref}[ln:locking:xchglock:lock_unlock:lock]
	Why bother with the inner loop on \clnrefrange{inner:b}{inner:e} of
	\cref{lst:locking:Sample Lock Based on Atomic Exchange}?
	Why not simply repeatedly do the atomic exchange operation
	on \clnref{atmxchg}?
	\end{fcvref}
\QuickA{}
	\begin{fcvref}[ln:locking:xchglock:lock_unlock:lock]
	Suppose that the lock is held and that several threads
	are attempting to acquire the lock.
	In this situation, if these threads all loop on the atomic
	exchange operation, they will ping-pong the cache line
	containing the lock among themselves, imposing load
	on the interconnect.
	In contrast, if these threads are spinning in the inner
	loop on \clnrefrange{inner:b}{inner:e},
	they will each spin within their own
	caches, placing negligible load on the interconnect.
	\end{fcvref}
\QuickE{}
\QuickQ{}
	\begin{fcvref}[ln:locking:xchglock:lock_unlock:unlock]
	Why not simply store zero into the lock word on \clnref{atmxchg} of
	\cref{lst:locking:Sample Lock Based on Atomic Exchange}?
	\end{fcvref}
\QuickA{}
	This can be a legitimate implementation, but only if
	this store is preceded by a memory barrier and makes use
	of \co{WRITE_ONCE()}.
	The memory barrier is not required when the \co{xchg()}
	operation is used because this operation implies a
	full memory barrier due to the fact that it returns
	a value.
\QuickE{}
\QuickQ{}
	How can you tell if one counter is greater than another,
	while accounting for counter wrap?
\QuickA{}
	In the C language, the following macro correctly handles this:

\begin{VerbatimU}
#define ULONG_CMP_LT(a, b) \
        (ULONG_MAX / 2 < (a) - (b))
\end{VerbatimU}

	Although it is tempting to simply subtract two signed integers,
	this should be avoided because signed overflow is undefined
	in the C language.
	For example, if the compiler knows that one of the values is
	positive and the other negative, it is within its rights to
	simply assume that the positive number is greater than the
	negative number, even though subtracting the negative number
	from the positive number might well result in overflow and
	thus a negative number.

	How could the compiler know the signs of the two numbers?
	It might be able to deduce it based on prior assignments
	and comparisons.
	In this case, if the per-CPU counters were signed, the compiler
	could deduce that they were always increasing in value, and
	then might assume that they would never go negative.
	This assumption could well lead the compiler to generate
	unfortunate code~\cite{PaulEMcKenney2012SignedOverflow,JohnRegehr2010UndefinedBehavior}.
\QuickE{}
\QuickQ{}
	Which is better, the counter approach or the flag approach?
\QuickA{}
	The flag approach will normally suffer fewer cache misses,
	but a better answer is to try both and see which works best
	for your particular workload.
\QuickE{}
\QuickQ{}
	How can relying on implicit existence guarantees result in
	a bug?
\QuickA{}
	Here are some bugs resulting from improper use of implicit
	existence guarantees:
	\begin{enumerate}
	\item	A program writes the address of a global variable to
		a file, then a later instance of that same program
		reads that address and attempts to dereference it.
		This can fail due to address-space randomization,
		to say nothing of recompilation of the program.
	\item	A module can record the address of one of its variables
		in a pointer located in some other module, then attempt
		to dereference that pointer after the module has
		been unloaded.
	\item	A function can record the address of one of its on-stack
		variables into a global pointer, which some other
		function might attempt to dereference after that function
		has returned.
	\end{enumerate}
	I am sure that you can come up with additional possibilities.
\QuickE{}
\QuickQ{}
	\begin{fcvref}[ln:locking:Per-Element Locking Without Existence Guarantees]
	What if the element we need to delete is not the first element
	of the list on \clnref{chk_first} of
	\cref{lst:locking:Per-Element Locking Without Existence Guarantees (Buggy!)}?
	\end{fcvref}
\QuickA{}
	This is a very simple hash table with no chaining, so the only
	element in a given bucket is the first element.
	The reader is invited to adapt this example to a hash table with
	full chaining.
\QuickE{}
\QuickQAC{chp:Data Ownership}{Data Ownership}{qqzowned}
\QuickQ{}
	What form of data ownership is extremely difficult
	to avoid when creating shared-memory parallel programs
	(for example, using pthreads) in C or C++?
\QuickA{}
	Use of auto variables in functions.
	By default, these are private to the thread executing the
	current function.
\QuickE{}
\QuickQ{}
	What synchronization remains in the example shown in
	\cref{sec:owned:Multiple Processes}?
\QuickA{}
	The creation of the threads via the \co{sh} \co{&} operator
	and the joining of thread via the \co{sh} \co{wait}
	command.

	Of course, if the processes explicitly share memory, for
	example, using the \co{shmget()} or \co{mmap()} system
	calls, explicit synchronization might well be needed when
	acccessing or updating the shared memory.
	The processes might also synchronize using any of the following
	interprocess communications mechanisms:
	\begin{enumerate}
	\item	System V semaphores.
	\item	System V message queues.
	\item	UNIX-domain sockets.
	\item	Networking protocols, including TCP/IP, UDP, and a whole
		host of others.
	\item	File locking.
	\item	Use of the \co{open()} system call with the
		\co{O_CREAT} and \co{O_EXCL} flags.
	\item	Use of the \co{rename()} system call.
	\end{enumerate}
	A complete list of possible synchronization mechanisms is left
	as an exercise to the reader, who is warned that it will be
	an extremely long list.
	A surprising number of unassuming system calls can be pressed
	into service as synchronization mechanisms.
\QuickE{}
\QuickQ{}
	Is there any shared data in the example shown in
	\cref{sec:owned:Multiple Processes}?
\QuickA{}
	That is a philosophical question.

	Those wishing the answer ``no'' might argue that processes by
	definition do not share memory.

	Those wishing to answer ``yes'' might list a large number of
	synchronization mechanisms that do not require shared memory,
	note that the kernel will have some shared state, and perhaps
	even argue that the assignment of process IDs (PIDs) constitute
	shared data.

	Such arguments are excellent intellectual exercise, and are
	also a wonderful way of feeling intelligent and scoring points
	against hapless classmates or colleagues, but are mostly a way
	of avoiding getting anything useful done.
\QuickE{}
\QuickQ{}
	Does it ever make sense to have partial data ownership where
	each thread reads only its own instance of a per-thread variable,
	but writes to other threads' instances?
\QuickA{}
	Amazingly enough, yes.
	One example is a simple message-passing system where threads post
	messages to other threads' mailboxes, and where each thread
	is responsible for removing any message it sent once that message
	has been acted on.
	Implementation of such an algorithm is left as an exercise for
	the reader, as is identifying other algorithms with similar
	ownership patterns.
\QuickE{}
\QuickQ{}
	What mechanisms other than POSIX signals may be used for function
	shipping?
\QuickA{}
	There is a very large number of such mechanisms, including:
	\begin{enumerate}
	\item	System V message queues.
	\item	Shared-memory dequeue (see
		\cref{sec:SMPdesign:Double-Ended Queue}).
	\item	Shared-memory mailboxes.
	\item	UNIX-domain sockets.
	\item	TCP/IP or UDP, possibly augmented by any number of
		higher-level protocols, including RPC, HTTP,
		XML, SOAP, and so on.
	\end{enumerate}
	Compilation of a complete list is left as an exercise to
	sufficiently single-minded readers, who are warned that the
	list will be extremely long.
\QuickE{}
\QuickQ{}
	\begin{fcvref}[ln:count:count_stat_eventual:whole:eventual]
	But none of the data in the \co{eventual()} function shown on
	\clnrefrange{b}{e} of
	\cref{lst:count:Array-Based Per-Thread Eventually Consistent Counters}
	is actually owned by the \co{eventual()} thread!
	In just what way is this data ownership???
	\end{fcvref}
\QuickA{}
	\begin{fcvref}[ln:count:count_stat_eventual:whole]
	The key phrase is ``owns the rights to the data''.
	In this case, the rights in question are the rights to access
	the per-thread \co{counter} variable defined on \clnref{per_thr_cnt}
	of the listing.
	This situation is similar to that described in
	\cref{sec:owned:Partial Data Ownership and pthreads}.

	However, there really is data that is owned by the \co{eventual()}
	thread, namely the \co{t} and \co{sum} variables defined on
	\clnref{t,sum} of the listing.

	For other examples of designated threads, look at the kernel
	threads in the Linux kernel, for example, those created by
	\co{kthread_create()} and \co{kthread_run()}.
	\end{fcvref}
\QuickE{}
\QuickQ{}
	Is it possible to obtain greater accuracy while still
	maintaining full privacy of the per-thread data?
\QuickA{}
	Yes.
	One approach is for \co{read_count()} to add the value
	of its own per-thread variable.
	This maintains full ownership and performance, but only
	a slight improvement in accuracy, particularly on systems
	with very large numbers of threads.

	Another approach is for \co{read_count()} to use function
	shipping, for example, in the form of per-thread signals.
	This greatly improves accuracy, but at a significant performance
	cost for \co{read_count()}.

	However, both of these methods have the advantage of eliminating
	cache thrashing for the common case of updating counters.
\QuickE{}
\QuickQAC{chp:Deferred Processing}{Deferred Processing}{qqzdefer}
\QuickQ{}
	Why bother with a use-after-free check?
\QuickA{}
	To greatly increase the probability of finding bugs.
	A small torture-test program
	(\path{routetorture.h}) that allocates and frees only
	one type of structure can tolerate a surprisingly
	large amount of use-after-free misbehavior.
	See \cref{fig:debugging:Number of Tests Required for 99 Percent Confidence Given Failure Rate}
	on \cpageref{fig:debugging:Number of Tests Required for 99 Percent Confidence Given Failure Rate}
	and the related discussion in
	\cref{sec:debugging:Hunting Heisenbugs}
	starting on
	\cpageref{sec:debugging:Hunting Heisenbugs}
	for more on the importance
	of increasing the probability of finding bugs.
\QuickE{}
\QuickQ{}
	Why doesn't \co{route_del()} in
	\cref{lst:defer:Reference-Counted Pre-BSD Routing Table Add/Delete}
	use reference counts to
	protect the traversal to the element to be freed?
\QuickA{}
	Because the traversal is already protected by the lock, so
	no additional protection is required.
\QuickE{}
\QuickQ{}
	Why the break in the ``ideal'' line  at 224 CPUs in
	\cref{fig:defer:Pre-BSD Routing Table Protected by Reference Counting}?
	Shouldn't it be a straight line?
\QuickA{}
	The break is due to hyperthreading.
	On this particular system, the first hardware thread in each
	core within a socket have consecutive CPU numbers,
	followed by the first hardware threads in each core for the
	other sockets,
	and finally followed by the second hardware thread in each core
	on all the sockets.
	On this particular system, CPU numbers 0--27 are the first
	hardware threads in each of the 28 cores in the first socket,
	numbers 28--55 are the first hardware threads in each of the
	28 cores in the second socket, and so on, so that numbers 196--223
	are the first hardware threads in each of the 28 cores in
	the eighth socket.
	Then CPU numbers 224--251 are the second hardware threads in each 
	of the 28 cores of the first socket, numbers 252--279 are the
	second hardware threads in each of the 28 cores of the second
	socket, and so on until numbers 420--447 are the second hardware
	threads in each of the 28 cores of the eighth socket.

	Why does this matter?

	Because the two hardware threads of a given core share resources,
	and this workload seems to allow a single hardware thread to
	consume more than half of the relevant resources within its core.
	Therefore, adding the second hardware thread of that core adds
	less than one might hope.
	Other workloads might gain greater benefit from each core's
	second hardware thread, but much depends on the details of both
	the hardware and the workload.
\QuickE{}
\QuickQ{}
	Shouldn't the refcnt trace in
	\cref{fig:defer:Pre-BSD Routing Table Protected by Reference Counting}
	be at least a little bit off of the x-axis???
\QuickA{}
	Define ``a little bit.''

\begin{figure}
\centering
\resizebox{2.5in}{!}{\includegraphics{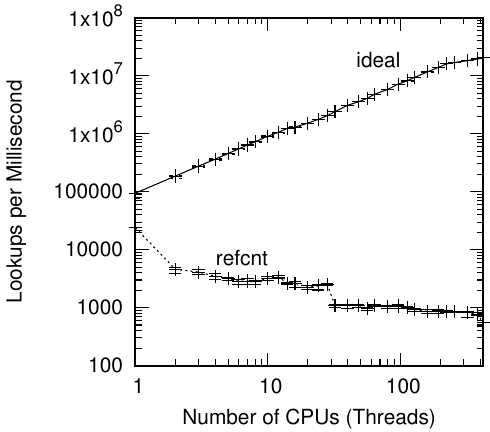}}
\caption{Pre-BSD Routing Table Protected by Reference Counting, Log Scale}
\label{fig:defer:Pre-BSD Routing Table Protected by Reference Counting; Log Scale}
\end{figure}

	\Cref{fig:defer:Pre-BSD Routing Table Protected by Reference Counting; Log Scale}
	shows the same data, but on a log-log plot.
	As you can see, the refcnt line drops below 5,000 at two CPUs.
	This means that the refcnt performance at two CPUs is more than
	one thousand times smaller than the first y-axis tick of
	$5 \times 10^6$ in
	\cref{fig:defer:Pre-BSD Routing Table Protected by Reference Counting}.
	Therefore, the depiction of the performance of reference counting
	shown in
	\cref{fig:defer:Pre-BSD Routing Table Protected by Reference Counting}
	is all too accurate.
\QuickE{}
\QuickQ{}
	If concurrency has ``most definitely reduced the usefulness
	of reference counting'', why are there so many reference
	counters in the Linux kernel?
\QuickA{}
	That sentence did say ``reduced the usefulness'', not
	``eliminated the usefulness'', now didn't it?

	Please see
	\cref{sec:together:Refurbish Reference Counting},
	which discusses some of the techniques that the Linux kernel
	uses to take advantage of reference counting in a highly
	concurrent environment.
\QuickE{}
\QuickQ{}
	Given that papers on hazard pointers use the bottom bits
	of each pointer to mark deleted elements, what is up with
	\co{HAZPTR_POISON}?
\QuickA{}
	The published implementations of hazard pointers used
	non-blocking synchronization techniques for insertion
	and deletion.
	These techniques require that readers traversing the
	data structure ``help'' updaters complete their updates,
	which in turn means that readers need to look at the successor
	of a deleted element.

	In contrast, we will be using locking to synchronize updates,
	which does away with the need for readers to help updaters
	complete their updates, which in turn allows us to leave
	pointers' bottom bits alone.
	This approach allows read-side code to be simpler and faster.
\QuickE{}
\QuickQ{}
	Why does \co{hp_try_record()} in
	\cref{lst:defer:Hazard-Pointer Recording and Clearing}
	take a double indirection to the data element?
	Why not \co{void *} instead of \co{void **}?
\QuickA{}
	Because \co{hp_try_record()} must check for concurrent modifications.
	To do that job, it needs a pointer to a pointer to the element,
	so that it can check for a modification to the pointer to the
	element.
\QuickE{}
\QuickQ{}
	Why bother with \co{hp_try_record()}?
	Wouldn't it be easier to just use the failure-immune
	\co{hp_record()} function?
\QuickA{}
	It might be easier in some sense, but as will be seen in the
	Pre-BSD routing example, there are situations for which
	\co{hp_record()} simply does not work.
\QuickE{}
\QuickQ{}
	Readers must ``typically'' restart?
	What are some exceptions?
\QuickA{}
	If the pointer emanates from a global variable or is otherwise
	not subject to being freed, then \co{hp_record()} may be
	used to repeatedly attempt to record the hazard pointer,
	even in the face of concurrent deletions.

	In certain cases, restart can be avoided by using link counting
	as exemplified by the UnboundedQueue and ConcurrentHashMap data
	structures implemented in Folly open-source library.\footnote{
		\url{https://github.com/facebook/folly}}
\QuickE{}
\QuickQ{}
	But don't these restrictions on hazard pointers also apply
	to other forms of reference counting?
\QuickA{}
	Yes and no.
	These restrictions apply only to reference-counting mechanisms whose
	reference acquisition can fail.
\QuickE{}
\QuickQ{}
	\Cref{fig:defer:Pre-BSD Routing Table Protected by Hazard Pointers}
	shows no sign of hyperthread-induced flattening at 224 threads.
	Why is that?
\QuickA{}
	Modern microprocessors are complicated beasts, so significant
	skepticism is appropriate for any simple answer.
	That aside, the most likely reason is the full memory barriers
	required by hazard-pointers readers.
	Any delays resulting from those memory barriers would make time
	available to the other hardware thread sharing the core, resulting
	in greater scalability at the expense of per-hardware-thread
	performance.
\QuickE{}
\QuickQ{}
	The paper ``Structured Deferral:
	Synchronization via
	Procrastination''~\cite{McKenney:2013:SDS:2483852.2483867}
	shows that hazard pointers have near-ideal performance.
	Whatever happened in
	\cref{fig:defer:Pre-BSD Routing Table Protected by Hazard Pointers}???
\QuickA{}
	First,
	\cref{fig:defer:Pre-BSD Routing Table Protected by Hazard Pointers}
	has a linear y-axis, while most of the graphs in the
	``Structured Deferral'' paper have logscale y-axes.
	Next, that paper uses lightly-loaded hash tables, while
	\cref{fig:defer:Pre-BSD Routing Table Protected by Hazard Pointers}'s
	uses a 10-element simple linked list, which means that hazard pointers
	face a larger memory-barrier penalty in this workload than in
	that of the ``Structured Deferral'' paper.
	Finally, that paper used an older modest-sized x86 system, while
	a much newer and larger system was used to generate the data
	shown in
	\cref{fig:defer:Pre-BSD Routing Table Protected by Hazard Pointers}.

	In addition, use of pairwise asymmetric
	barriers~\cite{Windows2008FlushProcessWriteBuffers,JonathanCorbet2010sys-membarrier,Linuxmanpage2018sys-membarrier}
	has been proposed to eliminate the read-side hazard-pointer
	memory barriers on systems supporting this notion~\cite{DavidGoldblatt2018asymmetricFences},
	which might improve the performance of hazard pointers beyond
	what is shown in the figure.

	As always, your mileage may vary.
	Given the difference in performance, it is clear that hazard
	pointers give you the best performance either for
	very large data structures (where the memory-barrier overhead
	will at least partially overlap cache-miss penalties) and
	for data structures such as hash tables where a lookup
	operation needs a minimal number of hazard pointers.
\QuickE{}
\QuickQ{}
	Why isn't this sequence-lock discussion in \cref{chp:Locking},
	you know, the one on \emph{locking}?
\QuickA{}
	The sequence-lock mechanism is really a combination of two
	separate synchronization mechanisms, sequence counts and
	locking.
	In fact, the sequence-count mechanism is available separately
	in the Linux kernel via the
	\co{write_seqcount_begin()} and \co{write_seqcount_end()}
	primitives.

	However, the combined \co{write_seqlock()} and
	\co{write_sequnlock()} primitives are used much more heavily
	in the Linux kernel.
	More importantly, many more people will understand what you
	mean if you say ``sequence lock'' than if you say
	``sequence count''.

	So this section is entitled ``Sequence Locks'' so that people
	will understand what it is about just from the title, and
	it appears in the ``Deferred Processing'' because (1) of the
	emphasis on the ``sequence count'' aspect of ``sequence locks''
	and (2) because a ``sequence lock'' is much more than merely
	a lock.
\QuickE{}
\QuickQ{}
	Why not have \co{read_seqbegin()} in
	\cref{lst:defer:Sequence-Locking Implementation}
	check for the low-order bit being set, and retry
	internally, rather than allowing a doomed read to start?
\QuickA{}
	That would be a legitimate implementation.
	However, if the workload is read-mostly, it would likely
	increase the overhead of the common-case successful read,
	which could be counter-productive.
	However, given a sufficiently large fraction of updates
	and sufficiently high-overhead readers, having the
	check internal to \co{read_seqbegin()} might be preferable.
\QuickE{}
\QuickQ{}
	Why is the \co{smp_mb()} on
	\clnrefr{ln:defer:seqlock:impl:read_seqretry:mb} of
	\cref{lst:defer:Sequence-Locking Implementation}
	needed?
\QuickA{}
	If it was omitted, both the compiler and the CPU would be
	within their rights to move the critical section preceding
	the call to \co{read_seqretry()} down below this function.
	This would prevent the sequence lock from protecting the
	critical section.
	The \co{smp_mb()} primitive prevents such reordering.
\QuickE{}
\QuickQ{}
	Can't weaker memory barriers be used in the code in
	\cref{lst:defer:Sequence-Locking Implementation}?
\QuickA{}
	In older versions of the Linux kernel, no.

	\begin{fcvref}[ln:defer:seqlock:impl]
	In very new versions of the Linux kernel,
	\clnref{read_seqbegin:fetch} could use
	\co{smp_load_acquire()} instead of \co{READ_ONCE()}, which
	in turn would allow the \co{smp_mb()} on
	\clnref{read_seqbegin:mb} to be dropped.
	Similarly, \clnref{write_sequnlock:inc} could use an
	\co{smp_store_release()}, for
	example, as follows:

\begin{VerbatimU}
smp_store_release(&slp->seq, READ_ONCE(slp->seq) + 1);
\end{VerbatimU}

	This would allow the \co{smp_mb()} on
	\clnref{write_sequnlock:mb} to be dropped.
	\end{fcvref}
\QuickE{}
\QuickQ{}
	What prevents sequence-locking updaters from starving readers?
\QuickA{}
	Nothing.
	This is one of the weaknesses of sequence locking, and as a
	result, you should use sequence locking only in read-mostly
	situations.
	Unless of course read-side starvation is acceptable in your
	situation, in which case, go wild with the sequence-locking updates!
\QuickE{}
\QuickQ{}
	What if something else serializes writers, so that the lock
	is not needed?
\QuickA{}
	In this case, the \co{->lock} field could be omitted, as it
	is in \co{seqcount_t} in the Linux kernel.
\QuickE{}
\QuickQ{}
	Why isn't \co{seq} on
	\clnrefr{ln:defer:seqlock:impl:typedef:seq} of
	\cref{lst:defer:Sequence-Locking Implementation}
	\co{unsigned} rather than \co{unsigned long}?
	After all, if \co{unsigned} is good enough for the Linux
	kernel, shouldn't it be good enough for everyone?
\QuickA{}
	Not at all.
	The Linux kernel has a number of special attributes that allow
	it to ignore the following sequence of events:
	\begin{enumerate}
	\item	Thread~0 executes \co{read_seqbegin()}, picking up
		\co{->seq} in
		\clnrefr{ln:defer:seqlock:impl:read_seqbegin:fetch},
		noting that the value is even,
		and thus returning to the caller.
	\item	Thread~0 starts executing its read-side critical section,
		but is then preempted for a long time.
	\item	Other threads repeatedly invoke \co{write_seqlock()} and
		\co{write_sequnlock()}, until the value of \co{->seq}
		overflows back to the value that Thread~0 fetched.
	\item	Thread~0 resumes execution, completing its read-side
		critical section with inconsistent data.
	\item	Thread~0 invokes \co{read_seqretry()}, which incorrectly
		concludes that Thread~0 has seen a consistent view of
		the data protected by the sequence lock.
	\end{enumerate}

	The Linux kernel uses sequence locking for things that are
	updated rarely, with time-of-day information being a case
	in point.
	This information is updated at most once per millisecond,
	so that seven weeks would be required to overflow the counter.
	If a kernel thread was preempted for seven weeks, the Linux
	kernel's soft-lockup code would be emitting warnings every two
	minutes for that entire time.

	In contrast, with a 64-bit counter, more than five centuries
	would be required to overflow, even given an update every
	\emph{nano}second.
	Therefore, this implementation uses a type for \co{->seq}
	that is 64 bits on 64-bit systems.
\QuickE{}
\QuickQ{}
	Can this bug be fixed?
	In other words, can you use sequence locks as the \emph{only}
	synchronization mechanism protecting a linked list supporting
	concurrent addition, deletion, and lookup?
\QuickA{}
	One trivial way of accomplishing this is to surround all
	accesses, including the read-only accesses, with
	\co{write_seqlock()} and \co{write_sequnlock()}.
	Of course, this solution also prohibits all read-side
	parallelism, resulting in massive lock contention,
	and furthermore could just as easily be implemented
	using simple locking.

	If you do come up with a solution that uses \co{read_seqbegin()}
	and \co{read_seqretry()} to protect read-side accesses, make
	sure that you correctly handle the following sequence of events:

	\begin{enumerate}
	\item	CPU~0 is traversing the linked list, and picks up a pointer
		to list element~A.
	\item	CPU~1 removes element~A from the list and frees it.
	\item	CPU~2 allocates an unrelated data structure, and gets
		the memory formerly occupied by element~A\@.
		In this unrelated data structure, the memory previously
		used for element~A's \co{->next} pointer is now occupied
		by a floating-point number.
	\item	CPU~0 picks up what used to be element~A's \co{->next}
		pointer, gets random bits, and therefore gets a
		segmentation fault.
	\end{enumerate}

	One way to protect against this sort of problem requires use
	of ``type-safe memory'', which will be discussed in
	\cref{sec:defer:Type-Safe Memory}.
	Roughly similar solutions are possible using the hazard pointers
	discussed in
	\cref{sec:defer:Hazard Pointers}.
	But in either case, you would be using some other synchronization
	mechanism in addition to sequence locks!
\QuickE{}
\QuickQ{}
	Why does
	\cref{fig:defer:Deletion With Concurrent Readers}
	use \co{smp_store_release()} given that it is storing
	a \co{NULL} pointer?
	Wouldn't \co{WRITE_ONCE()} work just as well in this case,
	given that there is no structure initialization to order
	against the store of the \co{NULL} pointer?
\QuickA{}
	Yes, it would.

	Because a \co{NULL} pointer is being assigned, there is nothing
	to order against, so there is no need for \co{smp_store_release()}.
	In contrast, when assigning a non-\co{NULL} pointer, it is
	necessary to use \co{smp_store_release()} in order to ensure
	that initialization of the pointed-to structure is carried
	out before assignment of the pointer.

	In short, \co{WRITE_ONCE()} would work, and would
	save a little bit of CPU time on some architectures.
	However, as we will see, software-engineering concerns
	will motivate use of a special \co{rcu_assign_pointer()}
	that is quite similar to \co{smp_store_release()}.
\QuickE{}
\QuickQ{}
	Readers running concurrently with each other and with the procedure
	outlined in
	\cref{fig:defer:Deletion With Concurrent Readers}
	can disagree on the value of \co{gptr}.
	Isn't that just a wee bit problematic???
\QuickA{}
	Not necessarily.

	As hinted at in
	\cref{sec:cpu:Hardware Optimizations,sec:cpu:Hardware Free Lunch?},
	speed-of-light delays mean that a computer's data is always
	stale compared to whatever external reality that data is intended
	to model.

	Real-world algorithms therefore absolutely must tolerate
	inconsistancies between external reality and the in-computer
	data reflecting that reality.
	Many of those algorithms are also able to tolerate some degree
	of inconsistency within the in-computer data.
	\Cref{sec:datastruct:RCU-Protected Hash Table Discussion}
	discusses this point in more detail.

	Please note that this need to tolerate inconsistent and stale
	data is not limited to RCU\@.
	It also applies to reference counting, hazard pointers, sequence
	locks, and even to some locking use cases.
	For example, if you compute some quantity while holding a lock,
	but use that quantity after releasing that lock,
	you might well be using stale data.
	After all, the data that quantity is based on might change
	arbitrarily as soon as the lock is released.

	So yes, RCU readers can see stale and inconsistent data, but no,
	this is not necessarily problematic.
	And, when needed, there are RCU usage patterns that avoid both
	staleness and inconsistency~\cite{Arcangeli03}.
\QuickE{}
\QuickQ{}
	What is an RCU-protected pointer?
\QuickA{}
	A pointer to \IXr{RCU-protected data}.
	RCU-protected data is in turn a block of dynamically allocated
	memory whose freeing will be deferred such that an RCU grace
	period will elapse between the time that there were no longer
	any RCU-reader-accessible pointers to that block and the time
	that that block is freed.
	This ensures that no RCU readers will have access to that block at
	the time that it is freed.

	RCU-protected pointers must be handled carefully.
	For example, any reader that intends to dereference an
	RCU-protected pointer must use \co{rcu_dereference()} (or
	stronger) to load that pointer.
	In addition, any updater must use \co{rcu_assign_pointer()}
	(or stronger) to store to that pointer.
\QuickE{}
\QuickQ{}
	What does \co{synchronize_rcu()} do if it starts at about
	the same time as an \co{rcu_read_lock()}?
\QuickA{}
	If a \co{synchronize_rcu()} cannot prove that it started
	before a given \co{rcu_read_lock()}, then it must wait
	for the corresponding \co{rcu_read_unlock()}.
\QuickE{}
\QuickQ{}
	In \cref{fig:defer:QSBR: Waiting for Pre-Existing Readers},
	the last of CPU~3's readers that could possibly have
	access to the old data item ended before the grace period
	even started!
	So why would anyone bother waiting until CPU~3's later context
	switch???
\QuickA{}
	Because that waiting is exactly what enables readers to use
	the same sequence of instructions that is appropriate for
	single-theaded situations.
	In other words, this additional ``redundant'' waiting enables
	excellent read-side performance, scalability, and real-time
	response.
\QuickE{}
\QuickQ{}
	What is the point of \co{rcu_read_lock()} and \co{rcu_read_unlock()} in
	\cref{lst:defer:Insertion and Deletion With Concurrent Readers}?
	Why not just let the quiescent states speak for themselves?
\QuickA{}
	Recall that readers are not permitted to pass through a quiescent
	state.
	For example, within the Linux kernel, RCU readers are not permitted
	to execute a context switch.
	Use of \co{rcu_read_lock()} and \co{rcu_read_unlock()} enables
	debug checks for improperly placed quiescent states, making it
	easy to find bugs that would otherwise be difficult to find,
	intermittent, and quite destructive.
\QuickE{}
\QuickQ{}
	What is the point of \co{rcu_dereference()}, \co{rcu_assign_pointer()}
	and \co{RCU_INIT_POINTER()} in
	\cref{lst:defer:Insertion and Deletion With Concurrent Readers}?
	Why not just use \co{READ_ONCE()}, \co{smp_store_release()}, and
	\co{WRITE_ONCE()}, respectively?
\QuickA{}
	The RCU-specific APIs do have similar semantics to the suggested
	replacements, but also enable static-analysis debugging checks
	that complain if an RCU-specific API is invoked on a non-RCU
	pointer and vice versa.
\QuickE{}
\QuickQ{}
	But what if the old structure needs to be freed, but the caller
	of \co{ins_route()} cannot block, perhaps due to performance
	considerations or perhaps because the caller is executing within
	an RCU read-side critical section?
\QuickA{}
	A \co{call_rcu()} function, which is described in
	\cref{sec:defer:Wait For Pre-Existing RCU Readers},
	permits asynchronous grace-period waits.
\QuickE{}
\QuickQ{}
	Doesn't \cref{sec:defer:Sequence Locks}'s seqlock
	also permit readers and updaters to make useful concurrent
	forward progress?
\QuickA{}
	Yes and no.
	Although seqlock readers can run concurrently with
	seqlock writers, whenever this happens, the \co{read_seqretry()}
	primitive will force the reader to retry.
	This means that any work done by a seqlock reader running concurrently
	with a seqlock updater will be discarded and then redone upon retry.
	So seqlock readers can \emph{run} concurrently with updaters,
	but they cannot actually get any work done in this case.

	In contrast, RCU readers can perform useful work even in presence
	of concurrent RCU updaters.

	However, both reference counters
	(\cref{sec:defer:Reference Counting})
	and hazard pointers
	(\cref{sec:defer:Hazard Pointers})
	really do permit useful concurrent forward progress for both
	updaters and readers, just at somewhat greater cost.
	Please see
	\cref{sec:defer:Which to Choose?}
	for a comparison of these different solutions to the
	deferred-reclamation problem.
\QuickE{}
\QuickQ{}
	Wouldn't use of data ownership for RCU updaters mean that
	the updates could use exactly the same sequence of instructions
	as would the corresponding single-threaded code?
\QuickA{}
	Sometimes, for example, on TSO systems such as x86 or the IBM
	mainframe where a store-release operation emits a single store
	instruction.
	However, weakly ordered systems must also emit a memory barrier
	or perhaps a store-release instruction.
	In addition, removing data requires quite a bit of additional
	work because it is necessary to wait for pre-existing readers
	before freeing the removed data.
\QuickE{}
\QuickQ{}
	But suppose that updaters are adding and removing multiple data
	items from a linked list while a reader is iterating over that
	same list.
	Specifically, suppose that a list initially contains elements
	A, B, and~C, and that an updater removes element A and then
	adds a new element D at the end of the list.
	The reader might well see \{A, B, C, D\}, when that sequence of
	elements never actually ever existed!
	In what alternate universe would that qualify as ``not disrupting
	concurrent readers''???
\QuickA{}
	In the universe where an iterating reader is only required to
	traverse elements that were present throughout the full duration
	of the iteration.
	In the example, that would be elements B and~C\@.
	Because elements A and~D were each present for only part of the
	iteration, the reader is permitted to iterate over them, but not
	obliged to.
	Note that this supports the common case where the reader is simply
	looking up a single item, and does not know or care about the
	presence or absence of other items.

	If stronger consistency is required, then higher-cost
	synchronization mechanisms are required, for example, sequence
	locking or reader-writer locking.
	But if stronger consistency is \emph{not} required (and it very often
	is not), then why pay the higher cost?
\QuickE{}
\QuickQ{}
	What other final values of \co{r1} and \co{r2} are possible in
	\cref{fig:defer:RCU Reader and Later Grace Period}?
\QuickA{}
	The \co{r1 == 0 && r2 == 0} possibility was called out in the text.
	Given that \co{r1 == 0} implies \co{r2 == 0}, we know that
	\co{r1 == 0 && r2 == 1} is forbidden.
	The following discussion will show that both
	\co{r1 == 1 && r2 == 1} and \co{r1 == 1 && r2 == 0} are possible.
\QuickE{}
\QuickQ{}
	What would happen if the order of \co{P0()}'s two accesses was
	reversed in
	\cref{fig:defer:RCU Reader and Earlier Grace Period}?
\QuickA{}
	Absolutely nothing would change.
	The fact that \co{P0()}'s loads from \co{x} and \co{y} are
	in the same RCU read-side critical section suffices;
	their order is irrelevant.
\QuickE{}
\QuickQ{}
	What would happen if \co{P0()}'s accesses in
	\crefrange{fig:defer:RCU Reader and Later Grace Period}{fig:defer:RCU Reader Within Grace Period}
	were stores?
\QuickA{}
	The exact same ordering rules would apply, that is,
	(1)~If any part of \co{P0()}'s RCU read-side critical section
	preceded the beginning of \co{P1()}'s grace period, all of
	\co{P0()}'s RCU read-side critical section would precede the
	end of \co{P1()}'s grace period, and
	(2)~If any part of \co{P0()}'s RCU read-side critical section
	followed the end of \co{P1()}'s grace period, all of \co{P0()}'s
	RCU read-side critical section would follow the beginning of
	\co{P1()}'s grace period.

	It might seem strange to have RCU read-side critical sections
	containing writes, but this capability is not only permitted,
	but also highly useful.
	For example, the Linux kernel frequently carries out an
	RCU-protected traversal of a linked data structure and then
	acquires a reference to the destination data element.
	Because this data element must not be freed in the meantime,
	that element's reference counter must necessarily be incremented
	within the traversal's RCU read-side critical section.
	However, that increment entails a write to memory.
	Therefore, it is a very good thing that memory writes are
	permitted within RCU read-side critical sections.

	If having writes in RCU read-side critical sections still seems
	strange, please review
	\cref{sec:count:Applying Exact Limit Counters},
	which presented a use case for writes in reader-writer locking
	read-side critical sections.
\QuickE{}
\QuickQ{}
	How would you modify the deletion example to permit more than two
	versions of the list to be active?
\QuickA{}
	One way of accomplishing this is as shown in
	\cref{lst:defer:Concurrent RCU Deletion}.

\begin{listing}
\begin{VerbatimL}
spin_lock(&mylock);
p = search(head, key);
if (p == NULL)
	spin_unlock(&mylock);
else {
	list_del_rcu(&p->list);
	spin_unlock(&mylock);
	synchronize_rcu();
	kfree(p);
}
\end{VerbatimL}
\caption{Concurrent RCU Deletion}
\label{lst:defer:Concurrent RCU Deletion}
\end{listing}

	Note that this means that multiple concurrent deletions might be
	waiting in \co{synchronize_rcu()}.
\QuickE{}
\QuickQ{}
	How many RCU versions of a given list can be
	active at any given time?
\QuickA{}
	That depends on the synchronization design.
	If a semaphore protecting the update is held across the grace period,
	then there can be at most two versions, the old and the new.

	However, suppose that only the search, the update, and the
	\co{list_replace_rcu()} were protected by a lock, so that
	the \co{synchronize_rcu()} was outside of that lock, similar
	to the code shown in
	\cref{lst:defer:Concurrent RCU Deletion}.
	Suppose further that a large number of threads undertook an
	RCU replacement at about the same time, and that readers
	are also constantly traversing the data structure.

	Then the following sequence of events could occur, starting from
	the end state of
	\cref{fig:defer:Multiple RCU Data-Structure Versions}:

	\begin{enumerate}
	\item	Thread~A traverses the list, obtaining a reference to
		Element~C.
	\item	Thread~B replaces Element~C with a new
		Element~F, then waits for its \co{synchronize_rcu()}
		call to return.
	\item	Thread~C traverses the list, obtaining a reference to
		Element~F.
	\item	Thread~D replaces Element~F with a new
		Element~G, then waits for its \co{synchronize_rcu()}
		call to return.
	\item	Thread~E traverses the list, obtaining a reference to
		Element~G.
	\item	Thread~F replaces Element~G with a new
		Element~H, then waits for its \co{synchronize_rcu()}
		call to return.
	\item	Thread~G traverses the list, obtaining a reference to
		Element~H.
	\item	And the previous two steps repeat quickly with additional
		new elements, so that all of them happen before any of
		the \co{synchronize_rcu()} calls return.
	\end{enumerate}

	Thus, there can be an arbitrary number of versions active,
	limited only by memory and by how many updates could be completed
	within a grace period.
	But please note that data structures that are updated so frequently
	are not likely to be good candidates for RCU\@.
	Nevertheless, RCU can handle high update rates when necessary.
\QuickE{}
\QuickQ{}
	How can the per-update overhead of RCU be reduced?
\QuickA{}
	The most effective way to reduce the per-update overhead
	of RCU is to increase the number of updates served by
	a given grace period.
	This works because the per-grace period overhead is nearly
	independent of the number of updates served by that
	grace period.

	One way to do this is to delay the start of a given grace
	period in the hope that more updates requiring that grace
	period appear in the meantime.
	Another way is to slow down execution of the grace period
	in the hope that more updates requiring an additional
	grace period will accumulate in the meantime.

	There are many other possible optimizations, and fanatically
	devoted readers are referred to the Linux-kernel RCU
	implementation.
\QuickE{}
\QuickQ{}
	How can RCU updaters possibly delay RCU readers, given that
	neither \co{rcu_read_lock()} nor \co{rcu_read_unlock()}
	spin or block?
\QuickA{}
	The modifications undertaken by a given RCU updater will cause the
	corresponding CPU to invalidate cache lines containing the data,
	forcing the CPUs running concurrent RCU readers to incur expensive
	cache misses.
	(Can you design an algorithm that changes a data structure
	\emph{without}
	inflicting expensive cache misses on concurrent readers?
	On subsequent readers?)
\QuickE{}
\QuickQ{}
	Why do some of the cells in
	\cref{tab:defer:RCU Wait-to-Finish APIs}
	have exclamation marks (``!'')?
\QuickA{}
	The API members with exclamation marks (\co{rcu_read_lock()},
	\co{rcu_read_unlock()}, and \co{call_rcu()}) were the
	only members of the Linux RCU API that Paul E. McKenney was aware
	of back in the mid-90s.
	During this timeframe, he was under the mistaken impression that
	he knew all that there is to know about RCU\@.
\QuickE{}
\QuickQ{}
	How do you prevent a huge number of RCU read-side critical
	sections from indefinitely blocking a \co{synchronize_rcu()}
	invocation?
\QuickA{}
	There is no need to do anything to prevent RCU read-side
	critical sections from indefinitely blocking a
	\co{synchronize_rcu()} invocation, because the
	\co{synchronize_rcu()} invocation need wait only for
	\emph{pre-existing} RCU read-side critical sections.
	So as long as each RCU read-side critical section is
	of finite duration, RCU grace periods will also remain
	finite.
\QuickE{}
\QuickQ{}
	The \co{synchronize_rcu()} API waits for all pre-existing
	interrupt handlers to complete, right?
\QuickA{}
	In v4.20 and later Linux kernels,
	yes~\cite{PaulEMcKenney2019RCUCVE,McKenney:2019:CRS:3319647.3325836}.

	But not in earlier kernels, and especially not when using
	preemptible RCU\@!
	You instead want \apik{synchronize_irq()}.
	Alternatively, you can place calls to \co{rcu_read_lock()}
	and \co{rcu_read_unlock()} in the specific interrupt handlers that
	you want \co{synchronize_rcu()} to wait for.
	But even then, be careful, as preemptible RCU will not be guaranteed
	to wait for that portion of the interrupt handler preceding the
	\co{rcu_read_lock()} or following the \co{rcu_read_unlock()}.
\QuickE{}
\QuickQ{}
	What is the difference between \co{synchronize_rcu()} and
	\co{rcu_barrier()}?
\QuickA{}
	They wait on different things.
	While \co{synchronize_rcu()} waits for pre-existing RCU read-side
	critical sections to complete, \co{rcu_barrier()} instead waits
	for callbacks from prior calls to \co{call_rcu()} to be invoked.

\begin{listing}
\begin{fcvlabel}[ln:defer:synchonize-rcu vs rcu-barrier]
\begin{VerbatimL}[commandchars=\\\@\$]
	do_something_1();			\lnlbl@ds1$
	rcu_read_lock();			\lnlbl@rrl$
	do_something_2();			\lnlbl@ds2$
	call_rcu(&p->rh, f);			\lnlbl@cr$
	do_something_3();			\lnlbl@ds3$
	rcu_read_unlock();			\lnlbl@rrul$
	do_something_4();			\lnlbl@ds4$
	// f(&p->rh) invoked			\lnlbl@cb$
	do_something_5();			\lnlbl@ds5$
\end{VerbatimL}
\end{fcvlabel}
\caption{\tco{synchronize_rcu()} vs. \tco{rcu_barrier()}}
\label{lst:defer:synchronize-rcu vs. rcu-barrier}
\end{listing}

	This distinction is illustrated by
	\cref{lst:defer:synchronize-rcu vs. rcu-barrier}, which
	shows code being executed by a given CPU\@.
	For simplicity, assume that no other CPU is executing
	\co{rcu_read_lock()}, \co{rcu_read_unlock()}, or
	\co{call_rcu()}.

\begin{table*}
\renewcommand*{\arraystretch}{1.2}
\centering
\small
\begin{fcvref}[ln:defer:synchonize-rcu vs rcu-barrier]
\begin{tabular}{lll}
\toprule
            & \multicolumn{2}{c}{Must Wait Until (in \cref{lst:defer:synchronize-rcu vs. rcu-barrier}):} \\
\cmidrule{2-3}
\multicolumn{1}{c}{Invoked at:} & \multicolumn{1}{c}{\tco{synchronize_rcu()}}
					& \multicolumn{1}{c}{\tco{rcu_barrier()}} \\
\cmidrule(r){1-1} \cmidrule{2-3}
\tco{do_something_1()} & 			 & \\
\tco{do_something_2()} & \tco{rcu_read_unlock()} (\clnref{rrul}) & \\
\tco{do_something_3()} & \tco{rcu_read_unlock()} (\clnref{rrul})
						 & \tco{f(&p->rh)} (\clnref{cb}) \\
\tco{do_something_4()} &			 & \tco{f(&p->rh)} (\clnref{cb}) \\
\tco{do_something_5()} & 			 & \\
\bottomrule
\end{tabular}
\end{fcvref}
\caption{\tco{synchronize_rcu()} vs. \tco{rcu_barrier()}}
\label{tab:defer:synchonize-rcu vs rcu-barrier}
\end{table*}

	\Cref{tab:defer:synchonize-rcu vs rcu-barrier}
	shows how long each primitive must wait if invoked
	concurrently with each of the \co{do_something_*()}
	functions, with empty cells indicating that no
	waiting is necessary.
	As you can see, \co{synchronize_rcu()} need not wait unless
	it is in an RCU read-side critical section, in which case
	it must wait for the \co{rcu_read_unlock()} that ends that
	critical section.
	In contrast, RCU read-side critical sections have no effect
	on \co{rcu_barrier()}.
	However, when \co{rcu_barrier()} executes after a
	\co{call_rcu()} invocation, it must wait until the
	corresponding RCU callback is invoked.

	All that said, there is a special case where each call to
	\co{rcu_barrier()} can be replaced by a direct call to
	\co{synchronize_rcu()}, and that is where \co{synchronize_rcu()}
	is implemented in terms of \co{call_rcu()} and where there is
	a single global list of callbacks.
	But please do not do this in portable code!!!
\QuickE{}
\QuickQ{}
	Under what conditions can \co{synchronize_srcu()} be safely
	used within an SRCU read-side critical section?
\QuickA{}
	In principle, you can use either \co{synchronize_srcu()} or
	\co{synchronize_srcu_expedited()} with a given \co{srcu_struct}
	within an SRCU read-side critical section that uses some other
	\co{srcu_struct}.
	In practice, however, doing this is almost certainly a bad idea.
	In particular, the code shown in
	\cref{lst:defer:Multistage SRCU Deadlocks}
	could still result in deadlock.
\QuickE{}

\begin{listing}
\begin{VerbatimL}
idx = srcu_read_lock(&ssa);
synchronize_srcu(&ssb);
srcu_read_unlock(&ssa, idx);

/* . . . */

idx = srcu_read_lock(&ssb);
synchronize_srcu(&ssa);
srcu_read_unlock(&ssb, idx);
\end{VerbatimL}
\caption{Multistage SRCU Deadlocks}
\label{lst:defer:Multistage SRCU Deadlocks}
\end{listing}
\QuickQ{}
	In a kernel built with \co{CONFIG_PREEMPT_NONE=y}, won't
	\co{synchronize_rcu()} wait for all trampolines, given
	that preemption is disabled and that trampolines never
	directly or indirectly invoke \co{schedule()}?
\QuickA{}
	You are quite right!

	In fact, in nonpreemptible kernels, \co{synchronize_rcu_tasks()}
	is a wrapper around \co{synchronize_rcu()}.
\QuickE{}
\QuickQ{}
	Normally, any pointer subject to \co{rcu_dereference()} \emph{must}
	always be updated using one of the pointer-publish functions in
	\cref{tab:defer:RCU Publish-Subscribe and Version Maintenance APIs},
	for example, \co{rcu_assign_pointer()}.

	What is an exception to this rule?
\QuickA{}
	One such exception is when a multi-element linked
	data structure is initialized as a unit while inaccessible to other
	CPUs, and then a single \co{rcu_assign_pointer()} is used
	to plant a global pointer to this data structure.
	The initialization-time pointer assignments need not use
	\co{rcu_assign_pointer()}, though any such assignments that
	happen after the structure is globally visible \emph{must} use
	\co{rcu_assign_pointer()}.

	However, unless this initialization code is on an impressively hot
	code-path, it is probably wise to use \co{rcu_assign_pointer()}
	anyway, even though it is in theory unnecessary.
	It is all too easy for a ``minor'' change to invalidate your cherished
	assumptions about the initialization happening privately.
\QuickE{}
\QuickQ{}
	Are there any downsides to the fact that these traversal and update
	primitives can be used with any of the RCU API family members?
\QuickA{}
	It can sometimes be difficult for automated
	code checkers such as ``sparse'' (or indeed for human beings) to
	work out which type of RCU read-side critical section a given
	RCU traversal primitive corresponds to.
	For example, consider the code shown in
	\cref{lst:defer:Diverse RCU Read-Side Nesting}.

\begin{listing}
\begin{VerbatimL}
rcu_read_lock();
preempt_disable();
p = rcu_dereference(global_pointer);

/* . . . */

preempt_enable();
rcu_read_unlock();
\end{VerbatimL}
\caption{Diverse RCU Read-Side Nesting}
\label{lst:defer:Diverse RCU Read-Side Nesting}
\end{listing}

	Is the \co{rcu_dereference()} primitive in a vanilla RCU critical
	section or an RCU Sched critical section?
	What would you have to do to figure this out?

	But perhaps after the consolidation of the RCU flavors in
	the v4.20 Linux kernel we no longer need to care!
\QuickE{}
\QuickQ{}
	But what if an \co{hlist_nulls} reader gets moved to some other
	bucket and then back again?
\QuickA{}
	One way to handle this is to always move nodes to the beginning
	of the destination bucket, ensuring that when the reader reaches
	the end of the list having a matching \co{NULL} pointer, it will
	have searched the entire list.

	Of course, if there are too many move operations in a hash table
	with many elements per bucket, the reader might never reach the
	end of a list.
	One way of avoiding this in the common case is to keep hash
	tables well-tuned, thus with short lists.
	One way of detecting the problem and handling it is for the
	reader to terminate the search after traversing some large
	number of nodes, acquire the update-side lock, and redo the
	search, but this might introduce deadlocks.
	Another way of avoiding the problem entirely is for readers to
	search within RCU read-side critical sections, and to wait for
	an RCU grace period between successive updates.
	An intermediate position might wait for an RCU grace period
	every $N$ updates, for some suitable value of $N$.
\QuickE{}
\QuickQ{}
	Why isn't there a \co{rcu_read_lock_tasks_held()} for Tasks RCU?
\QuickA{}
	Because Tasks RCU does not have read-side markers.
	Instead, Tasks RCU read-side critical sections are
	bounded by voluntary context switches.
\QuickE{}
\QuickQ{}
	Wait, what???
	How can RCU QSBR possibly be better than ideal?
	Just what rubbish definition of ideal would fail to be the best
	of all possible results???
\QuickA{}
	This is an excellent question, and the answer is that modern
	CPUs and compilers are extremely complex.
	But before getting into that, it is well worth noting that
	RCU QSBR's performance advantage appears only in the
	one-hardware-thread-per-core regime.
	Once the system is fully loaded, RCU QSBR's performance drops
	back to ideal.

	The RCU variant of the \co{route_lookup()} search loop actually
	has one more x86 instruction than does the sequential version,
	namely the \co{lea} in the sequence
	\co{cmp}, \co{je}, \co{mov}, \co{cmp}, \co{lea}, and \co{jne}.
	This extra instruction is due to the \co{rcu_head} structure
	at the beginning of the RCU variant's \co{route_entry} structure,
	so that, unlike the sequential variant, the RCU variant's
	\co{->re_next.next} pointer has a non-zero offset.
	Back in the 1980s, this additional \co{lea} instruction might
	have reliably resulted in the RCU variant being slower, but we
	are now in the 21\textsuperscript{st} century, and the 1980s
	are long gone.

	But those of you who read
	\cref{sec:cpu:Pipelined CPUs}
	carefully already knew all of this!

	These counter-intuitive results of course means that any
	performance result on modern microprocessors must be subject to
	some skepticism.
	In theory, it really does not make sense to obtain performance
	results that are better than ideal, but it really can happen
	on modern microprocessors.
	Such results can be thought of as similar to the celebrated
	super-linear speedups (see
	\cref{sec:SMPdesign:Beyond Partitioning}
	for one such example), that is, of interest but also of limited
	practical importance.
	Nevertheless, one of the strengths of RCU is that its read-side
	overhead is so low that tiny effects such as this one are visible
	in real performance measurements.

\begin{figure}
\centering
\resizebox{2.5in}{!}{\includegraphics{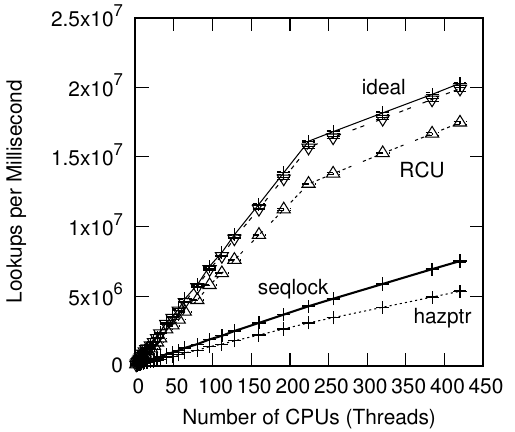}}
\caption{Pre-BSD Routing Table Protected by RCU QSBR With Non-Initial \tco{rcu_head}}
\label{fig:defer:Pre-BSD Routing Table Protected by RCU QSBR With Non-Initial rcu-head}
\end{figure}

	This raises the question as to what would happen if the
	\co{rcu_head} structure were to be moved so that RCU's
	\co{->re_next.next} pointer also had zero offset, just the
	same as the sequential variant.
	And the answer, as can be seen in
	\cref{fig:defer:Pre-BSD Routing Table Protected by RCU QSBR With Non-Initial rcu-head},
	is that this causes RCU QSBR's performance to decrease to where
	it is still very nearly ideal, but no longer super-ideal.
\QuickE{}
\QuickQ{}
	Given RCU QSBR's read-side performance, why bother with any
	other flavor of userspace RCU?
\QuickA{}
	Because RCU QSBR places constraints on the overall application
	that might not be tolerable,
	for example, requiring that each and every thread in the
	application regularly pass through a quiescent state.
	Among other things, this means that RCU QSBR is not helpful
	to library writers, who might be better served by other
	flavors of userspace RCU~\cite{PaulMcKenney2013LWNURCU}.
\QuickE{}
\QuickQ{}
	Suppose that the \co{nmi_profile()} function was preemptible.
	What would need to change to make this example work correctly?
\QuickA{}
	One approach would be to use
	\co{rcu_read_lock()} and \co{rcu_read_unlock()}
	in \co{nmi_profile()}, and to replace the
	\co{synchronize_sched()} with \co{synchronize_rcu()},
	perhaps as shown in
	\cref{lst:defer:Using RCU to Wait for Mythical Preemptible NMIs to Finish}.
\begin{listing}
\begin{VerbatimL}
struct profile_buffer {
	long size;
	atomic_t entry[0];
};
static struct profile_buffer *buf = NULL;

void nmi_profile(unsigned long pcvalue)
{
	struct profile_buffer *p;

	rcu_read_lock();
	p = rcu_dereference(buf);
	if (p == NULL) {
		rcu_read_unlock();
		return;
	}
	if (pcvalue >= p->size) {
		rcu_read_unlock();
		return;
	}
	atomic_inc(&p->entry[pcvalue]);
	rcu_read_unlock();
}

void nmi_stop(void)
{
	struct profile_buffer *p = buf;

	if (p == NULL)
		return;
	rcu_assign_pointer(buf, NULL);
	synchronize_rcu();
	kfree(p);
}
\end{VerbatimL}
\caption{Using RCU to Wait for Mythical Preemptible NMIs to Finish}
\label{lst:defer:Using RCU to Wait for Mythical Preemptible NMIs to Finish}
\end{listing}

	But why on earth would an NMI handler be preemptible???
\QuickE{}
\QuickQ{}
	What is the point of the second call to \co{synchronize_rcu()}
	in function
	\co{maint()} in \cref{lst:defer:Phased State Change for Maintenance Operations}?
	Isn't it OK for any \co{cco()} invocations in the clean-up
	phase to invoke either \co{cco_carefully()} or \co{cco_quickly()}?
\QuickA{}
	The problem is that there is no ordering between the \co{cco()}
	function's load from \co{be_careful} and any memory loads
	executed by the \co{cco_quickly()} function.
	Because there is no ordering, without that second call to
	\co{syncrhonize_rcu()}, memory ordering could cause loads
	in \co{cco_quickly()} to overlap with stores by \co{do_maint()}.

	Another alternative would be to compensate for the removal of
	that second call to \co{synchronize_rcu()} by changing the
	\co{READ_ONCE()} to \co{smp_load_acquire()} and the
	\co{WRITE_ONCE()} to \co{smp_store_release()}, thus restoring
	the needed ordering.
\QuickE{}
\QuickQ{}
	How can you be sure that the code shown in
	\co{maint()} in \cref{lst:defer:Phased State Change for Maintenance Operations}
	really works?\@
\QuickA{}
	By one popular school of thought, you cannot.

	But in this case, those willing to jump ahead to
	\cref{chp:Formal Verification}
	and
	\cref{chp:Advanced Synchronization: Memory Ordering}
	might find a couple of LKMM litmus tests to be interesting
	(\path{C-RCU-phased-state-change-1.litmus} and
	\path{C-RCU-phased-state-change-2.litmus}).
	These tests could be argued to demonstrate that this code
	and a variant of it really do work.
\QuickE{}
\QuickQ{}
	But what if there is an arbitrarily long series of RCU
	read-side critical sections in multiple threads, so that at
	any point in time there is at least one thread in the system
	executing in an RCU read-side critical section?
	Wouldn't that prevent any data from a \co{SLAB_TYPESAFE_BY_RCU}
	slab ever being returned to the system, possibly resulting
	in OOM events?
\QuickA{}
	There could certainly be an arbitrarily long period of time
	during which at least one thread is always in an RCU read-side
	critical section.
	However, the key words in the description in
	\cref{sec:defer:Type-Safe Memory}
	are ``in-use'' and ``pre-existing''.
	Keep in mind that a given RCU read-side critical section is
	conceptually only permitted to gain references to data elements
	that were visible to readers during that critical section.
	Furthermore, remember that a slab cannot be returned to the
	system until all of its data elements have been freed, in fact,
	the RCU grace period cannot start until after they have all been
	freed.

	Therefore, the slab cache need only wait for those RCU read-side
	critical sections that started before the freeing of the last element
	of the slab.
	This in turn means that any RCU grace period that begins after
	the freeing of the last element will do---the slab may be returned
	to the system after that grace period ends.
\QuickE{}
\QuickQ{}
	What if the element we need to delete is not the first element
	of the list on
	\clnrefr{ln:defer:Existence Guarantees Enable Per-Element Locking:chkkey} of
	\cref{lst:defer:Existence Guarantees Enable Per-Element Locking}?
\QuickA{}
	As with the (bug-ridden)
	\cref{lst:locking:Per-Element Locking Without Existence Guarantees (Buggy!)},
	this is a very simple hash table with no chaining, so the only
	element in a given bucket is the first element.
	The reader is again invited to adapt this example to a hash table with
	full chaining.
	Less energetic readers might wish to refer to
	\cref{chp:Data Structures}.
\QuickE{}
\QuickQ{}
	\begin{fcvref}[ln:defer:Existence Guarantees Enable Per-Element Locking]
	Why is it OK to exit the RCU read-side critical section on
	\clnref{rdunlock2} of
	\cref{lst:defer:Existence Guarantees Enable Per-Element Locking}
	before releasing the lock on \clnref{rel1}?
	\end{fcvref}
\QuickA{}
	\begin{fcvref}[ln:defer:Existence Guarantees Enable Per-Element Locking]
	First, please note that the second check on \clnref{chkkey2} is
	necessary because some other
	CPU might have removed this element while we were waiting
	to acquire the lock.
	However, the fact that we were in an RCU read-side critical section
	while acquiring the lock guarantees that this element could not
	possibly have been re-allocated and re-inserted into this
	hash table.
	Furthermore, once we acquire the lock, the lock itself guarantees
	the element's existence, so we no longer need to be in an
	RCU read-side critical section.

	The question as to whether it is necessary to re-check the
	element's key is left as an exercise to the reader.
	\end{fcvref}
\QuickE{}
\QuickQ{}
	\begin{fcvref}[ln:defer:Existence Guarantees Enable Per-Element Locking]
	Why not exit the RCU read-side critical section on
	\clnref{rdunlock3} of
	\cref{lst:defer:Existence Guarantees Enable Per-Element Locking}
	before releasing the lock on \clnref{rel2}?
	\end{fcvref}
\QuickA{}
	Suppose we reverse the order of these two lines.
	Then this code is vulnerable to the following sequence of
	events:
	\begin{enumerate}
	\begin{fcvref}[ln:defer:Existence Guarantees Enable Per-Element Locking]
	\item	CPU~0 invokes \co{delete()}, and finds the element
		to be deleted, executing through \clnref{rdunlock2}.
		It has not yet actually deleted the element, but
		is about to do so.
	\item	CPU~1 concurrently invokes \co{delete()}, attempting
		to delete this same element.
		However, CPU~0 still holds the lock, so CPU~1 waits
		for it at \clnref{acq}.
	\item	CPU~0 executes \clnref{remove,rel1},
		and blocks at \clnref{sync_rcu} waiting for CPU~1
		to exit its RCU read-side critical section.
	\item	CPU~1 now acquires the lock, but the test on \clnref{chkkey2}
		fails because CPU~0 has already removed the element.
		CPU~1 now executes \clnref{rel2}
		(which we switched with \clnref{rdunlock3}
		for the purposes of this Quick Quiz)
		and exits its RCU read-side critical section.
	\item	CPU~0 can now return from \co{synchronize_rcu()},
		and thus executes \clnref{kfree}, sending the element to
		the freelist.
	\item	CPU~1 now attempts to release a lock for an element
		that has been freed, and, worse yet, possibly
		reallocated as some other type of data structure.
		This is a fatal memory-corruption error.
	\end{fcvref}
\QuickE{}
	\end{enumerate}
\QuickQ{}
	The RCU-based algorithm shown in
	\cref{lst:defer:Existence Guarantees Enable Per-Element Locking}
	locks very similar to that in
	\cref{lst:locking:Per-Element Locking With Lock-Based Existence Guarantees},
	so why should the RCU-based approach be any better?
\QuickA{}
	\Cref{lst:defer:Existence Guarantees Enable Per-Element Locking}
	replaces the per-element \co{spin_lock()} and \co{spin_unlock()}
	shown in
	\cref{lst:locking:Per-Element Locking With Lock-Based Existence Guarantees}
	with a much cheaper \co{rcu_read_lock()} and \co{rcu_read_unlock()},
	thus greatly improving both performance and scalability.
	For more detail, please see
	\cref{sec:datastruct:RCU-Protected Hash Table Performance}.
\QuickE{}
\QuickQ{}
	WTF\@?
	How the heck do you expect me to believe that RCU can have less
	than a 300-picosecond overhead when the clock period at 2.10\,GHz
	is almost 500\,picoseconds?
\QuickA{}
	First, consider that the inner loop used to
	take this measurement is as follows:

\begin{VerbatimN}
	for (i = nloops; i >= 0; i--) {
		rcu_read_lock();
		rcu_read_unlock();
	}
\end{VerbatimN}

	Next, consider the effective definitions of \co{rcu_read_lock()}
	and \co{rcu_read_unlock()}:

\begin{VerbatimN}
#define rcu_read_lock()   barrier()
#define rcu_read_unlock() barrier()
\end{VerbatimN}

	These definitions constrain compiler code-movement optimizations
	involving memory references, but emit no instructions in and
	of themselves.
	However, if the loop variable is maintained in a register,
	the accesses to \co{i} will not count as memory references.
	Furthermore, the compiler can do loop unrolling,
	allowing the resulting code to ``execute'' multiple passes
	through the loop body simply by incrementing \co{i} by
	some value larger than the value 1.

	So the ``measurement'' of 267 picoseconds is simply the fixed
	overhead of the timing measurements divided by the number of
	passes through the inner loop containing the calls
	to \co{rcu_read_lock()} and \co{rcu_read_unlock()}, plus
	the code to manipulate \co{i} divided by the loop-unrolling
	factor.
	And therefore, this measurement really is in error, in fact,
	it exaggerates the overhead by an arbitrary number of orders
	of magnitude.
	After all, in terms of machine instructions emitted, the actual
	overheads of \co{rcu_read_lock()} and of \co{rcu_read_unlock()}
	are each precisely zero.

	It is not just every day that a timing measurement of 267
	picoseconds turns out to be an overestimate!
\QuickE{}
\QuickQ{}
	Didn't an earlier edition of this book show RCU read-side
	overhead way down in the sub-picosecond range?
	What happened???
\QuickA{}
	Excellent memory!!!
	The overhead in some early releases was in fact roughly
	100~femtoseconds.

	What happened was that RCU usage spread more broadly through the
	Linux kernel, including into code that takes page faults.
	Back at that time, \co{rcu_read_lock()} and \co{rcu_read_unlock()}
	were complete no-ops in \co{CONFIG_PREEMPT=n} kernels.
	Unfortunately, that situation allowed the compiler to reorder
	page-faulting memory accesses into RCU read-side critical
	sections.
	Of course, page faults can block, which destroys those critical
	sections.

	Nor was this a theoretical problem:
	A failure actually manifested in 2019.
	\ppl{Herbert}{Xu} tracked down this failure down and
	\ppl{Linus}{Torvalds}
	therefore queued a commit to upgrade \co{rcu_read_lock()} and
	\co{rcu_read_unlock()} to unconditionally include a call to
	\co{barrier()}~\cite{LinusTorvalds2019:RCUreader.barrier}.
	And although \co{barrier()} emits no code, it does constrain
	compiler optimizations.
	And so the price of widespread RCU usage is slightly higher
	\co{rcu_read_lock()} and \co{rcu_read_unlock()} overhead.
	As such, Linux-kernel RCU has proven to be a victim of its
	own success.

	Of course, it is also the case that the older results were obtained
	on a different system than were those shown in
	\cref{fig:defer:Performance Advantage of RCU Over Reader-Writer Locking}.
	So which change had the most effect, Linus's commit or the change in
	the system?
	This question is left as an exercise to the reader.
\QuickE{}
\QuickQ{}
	Why is there such large variation for the \co{RCU} trace in
	\cref{fig:defer:Performance Advantage of RCU Over Reader-Writer Locking}?
\QuickA{}
	Keep in mind that this is a log-log plot, so those large-seeming
	\co{RCU} variances in reality span only a few hundred picoseconds.
	And that is such a short time that anything could cause it.
	However, given that the variance decreases with both small and
	large numbers of CPUs, one hypothesis is that the variation is
	due to migrations from one CPU to another.

	Yes, these measurements were taken with interrupts disabled, but
	they were also taken within a guest OS, so that preemption was
	still possible at the hypervisor level.
	In addition, the system featured hyperthreading and a single
	hardware thread running this RCU workload is able to consume
	more than half of the core's resources.
	Therefore, the overall throughput varies depending on how many
	of a given guest OS's CPUs share cores.
	Attempting to reduce these variations by running the guest OSes
	at real-time priority (as suggested by Joel Fernandes) is left
	as an exercise for the reader.
\QuickE{}
\QuickQ{}
	Given that the system had no fewer than 448~hardware threads,
	why only 192~CPUs?
\QuickA{}
	Because the script (\path{rcuscale.sh}) that generates this data
	spawns a guest operating system for each set of points gathered,
	and on this particular system, both \co{qemu} and KVM limit the
	number of CPUs that may be configured into a given guest OS\@.
	Yes, it would have been possible to run a few more CPUs, but
	192 is a nice round number from a binary perspective, given
	that 256 is infeasible.
\QuickE{}
\QuickQ{}
	Why the larger error ranges for the submicrosecond durations in
	\cref{fig:defer:Comparison of RCU to Reader-Writer Locking as Function of Critical-Section Duration}?
\QuickA{}
	Because smaller disturbances result in greater relative errors
	for smaller measurements.
	Also, the Linux kernel's \co{ndelay()} nanosecond-scale primitive
	is (as of 2020) less accurate than is the \co{udelay()} primitive
	used for the data for durations of a microsecond or more.
	It is instructive to compare to the zero-length case shown in
	\cref{fig:defer:Performance Advantage of RCU Over Reader-Writer Locking}.
\QuickE{}
\QuickQ{}
	Is there an exception to this deadlock immunity, and if so,
	what sequence of events could lead to deadlock?
\QuickA{}
	One way to cause a deadlock cycle involving
	RCU read-side primitives is via the following (illegal) sequence
	of statements:

\begin{VerbatimU}
rcu_read_lock();
synchronize_rcu();
rcu_read_unlock();
\end{VerbatimU}

	The \co{synchronize_rcu()} cannot return until all
	pre-existing RCU read-side critical sections complete, but
	is enclosed in an RCU read-side critical section that cannot
	complete until the \co{synchronize_rcu()} returns.
	The result is a classic self-deadlock---you get the same
	effect when attempting to write-acquire a reader-writer lock
	while read-holding it.

	Note that this self-deadlock scenario does not apply to
	RCU QSBR, because the context switch performed by the
	\co{synchronize_rcu()} would act as a quiescent state
	for this CPU, allowing a grace period to complete.
	However, this is if anything even worse, because data used
	by the RCU read-side critical section might be freed as a
	result of the grace period completing.
	Plus Linux kernel's lockdep facility will yell at you.

	In short, do not invoke synchronous RCU update-side primitives, which
	are listed in
	\cref{tab:defer:RCU Wait-to-Finish APIs},
	from within an RCU read-side critical section.

	In addition, within the Linux kernel, RCU uses the scheduler
	and the scheduler uses RCU\@.
	In some cases, both RCU and the scheduler must take care to
	avoid deadlock.
\QuickE{}
\QuickQ{}
	Immunity to both deadlock and priority inversion???
	Sounds too good to be true.
	Why should I believe that this is even possible?
\QuickA{}
	It really does work.
	After all, if it didn't work, the Linux kernel would not run.
\QuickE{}
\QuickQ{}
	But how many other algorithms really tolerate stale and
	inconsistent data?
\QuickA{}
	Quite a few!

	Please keep in mind that the finite speed of light means that
	data reaching a given computer system is at least slightly stale
	at the time that it arrives, and extremely stale in the case
	of astronomical data.
	The finite speed of light also places a sharp limit on the
	consistency of data arriving from different sources of via
	different paths.

	You might as well face the fact that the laws of physics
	are incompatible with naive notions of perfect freshness and
	consistency.
\QuickE{}
\QuickQ{}
	If Tasks RCU Trace might someday be priority boosted, why
	not also Tasks RCU and Tasks RCU Rude?
\QuickA{}
	Maybe, but these are less likely.

	In the case of Tasks RCU, recall that the quiescent state is
	a voluntary context switch.
	Thus, all tasks not blocked after a voluntary context switch
	might need to be boosted, and the mechanics of deboosting would
	not likely be at all pretty.

	In the case of Tasks RCU Rude, as was the case with the old
	RCU Sched, any preemptible region of code is a quiescent state.
	Thus, the only tasks that might need boosting are those currently
	running with preemption disabled.
	But boosting the priority of a preemption-disabled task has no
	effect.
	It therefore seems doubly unlikely that priority boosting will
	ever be introduced to Tasks RCU Rude, at least in its current
	form.
\QuickE{}
\QuickQ{}
	But doesn't the RCU grace period start sometime after the
	call to \co{synchronize_rcu()} rather than in the middle
	of that \co{xchg()} statement?
\QuickA{}
	Which grace period, exactly?

	The updater is required to wait for at least one grace
	period that starts at or some time after the removal,
	in this case, the \co{xchg()}.
	So in
	\cref{fig:defer:RCU Spatial/Temporal Synchronization},
	the indicated grace period starts as early as theoretically
	possible and extends to the return from \co{synchronize_rcu()}.
	This is a perfectly legal grace period corresponding to the
	change carried out by that \co{xchg()} statement.
\QuickE{}
\QuickQ{}
	Is RCU the only synchronization mechanism that combines temporal
	and spatial synchronization in this way?
\QuickA{}
	Not at all.

	Hazard pointers can be considered to combine temporal and spatial
	synchronization in a similar manner.
	Referring to
	\cref{lst:defer:Hazard-Pointer Recording and Clearing},
	the \co{hp_record()} function's acquisition of a reference
	provides both spatial and temporal synchronization, subscribing
	to a version and marking the start of a reference, respectively.
	This function therefore combines the effects of RCU's
	\co{rcu_read_lock()} and \co{rcu_dereference()}.
	Referring now to
	\cref{lst:defer:Hazard-Pointer Scanning and Freeing},
	the \co{hp_clear()} function's release of a reference provides
	temporal synchronization marking the end of a reference, and is
	thus similar to RCU's \co{rcu_read_unlock()}.
	The \co{hazptr_free_later()} function's retiring of a
	hazard-pointer-protected object provides temporal synchronization,
	similar to RCU's \co{call_rcu()}.
	The primitives used to mutate a hazard-pointer-protected
	structure provide spatial synchronization, similar to RCU's
	\co{rcu_assign_pointer()}.

	Alternatively, one could instead come at hazard pointers by
	analogy with reference counting.
\QuickE{}
\QuickQ{}
	But wait!
	This is exactly the same code that might be used when thinking
	of RCU as a replacement for reader-writer locking!
	What gives?
\QuickA{}
	This is an effect of the Law of Toy Examples:
	Beyond a certain point, the code fragments look the same.
	The only difference is in how we think about the code.
	For example, what does an \co{atomic_inc()} operation do?
	It might be acquiring another explicit reference to an object
	to which we already have a reference, it might be incrementing
	an often-read/seldom-updated statistical counter, it might
	be checking into an HPC-style barrier, or any of a number of
	other things.

	However, these differences can be extremely important.
	For but one example of the importance, consider that if we think
	of RCU as a restricted reference counting scheme, we would never
	be fooled into thinking that the updates would exclude the RCU
	read-side critical sections.

	It nevertheless is often useful to think of RCU as a replacement
	for reader-writer locking, for example, when you are replacing
	reader-writer locking with RCU\@.
\QuickE{}
\QuickQ{}
	Which of these use cases best describes the Pre-BSD routing
	example in
	\cref{sec:defer:RCU for Pre-BSD Routing}?
\QuickA{}
	Pre-BSD routing could be argued to fit into either
	quasi reader-writer lock, quasi reference count, or
	quasi multi-version concurrency control.
	The code is the same either way.
	This is similar to things like \co{atomic_inc()}, another tool
	that can be put to a great many uses.
\QuickE{}
\QuickQ{}
	Garbage collectors?
	Passive serialization?
	System reference points?
	Quiescent states?
	Aging?
	Generations?
	Why on earth couldn't the knuckleheads working on these early
	papers bring themselves to agree on a common terminology???
\QuickA{}
	There were multiple independent inventions of mechanisms
	vaguely resembling RCU\@.
	Each group of inventors was unaware of the others, so each
	made up its own terminology as a matter of course.
	And the different terminology made it quite difficult for
	any one group to find any of the others.

	Sorry, but life is like that sometimes!
\QuickE{}
\QuickQ{}
	Why didn't Kung's and Lehman's paper result in immediate use
	of RCU?
\QuickA{}
	One reason is that Kung and Lehman were simply ahead of their
	time.
	Another reason was that their approach, ground-breaking though
	it was, did not take a number of software-engineering and
	performance issues into account.

	To see that they were ahead of their time, consider that three
	years after their paper was published, Paul was working on a
	PDP-11 system running BSD 2.8.
	This system lacked any sort of automatic configuration, which
	meant that any hardware modification, including adding a new
	disk drive, required hand-editing and rebuilding the kernel.
	Furthermore, this was a single-CPU system, which meant that
	full-system synchronization was a simple matter of disabling
	interrupts.

	Fast-forward a number of years, and multicore systems permitting
	runtime changes in hardware configuration were commonplace.
	This meant that the hardware configuration data that was implicitly
	represented in 1980s kernel source code was now a mutable
	data structure that was accessed on every I/O\@.
	Such data structures rarely change, but could change at any time.
	And this read-mostly property applies to many other new-age
	data structures, including those concerning networking (rare in
	the 1980s), security policies (physical locks in the 1980s),
	software configuration (immutable at runtime in the 1980s),
	and much else besides.
	There was thus much more opportunity for RCU to demonstrate its
	benefits in the 1990s and 2000s than there was in the 1980s.

	Kung's and Lehman's software-engineering sins included failing
	to mark readers (thus presenting debugging difficulties),
	failing to provide a clean RCU API (thus tying their mechanism
	to a specific data structure), and failing to allow for any
	post-grace-period operation other than freeing memory (thus
	disallowing a number of RCU use cases).

	Kung and Lehman presented two garbage-collection strategies.
	The first waited for all processes running at a given time
	to terminate, which represented another software-engineering
	sin that ruled out their mechanism's use in software that
	runs indefinitely.
	The second used per-object reference counting, which greatly
	complicates their read-side code (thus representing yet
	another software-engineering sin), and, on modern hardware,
	results in severe cache-miss overhead (thus representing a
	performance sin, see for example
	\cref{fig:defer:Performance of RCU vs. Reference Counting,fig:defer:Performance of Preemptible RCU vs. Reference Counting}).

	Despite this long list of software-engineering and performance
	sins, Kung's and Lehman's paper remains a truly impressive piece
	of work, especially considering that much of the later work
	(both independent and not) committed these same sins, plus others
	as well.
\QuickE{}
\QuickQ{}
	Why not just drop the lock before waiting for the grace
	period, or using something like \co{call_rcu()}
	instead of waiting for a grace period?
\QuickA{}
	The authors wished to support \IX{linearizable} tree
	operations, so that concurrent additions to, deletions
	from, and searches of the tree would appear to execute
	in some globally agreed-upon order.
	In their search trees, this requires holding locks
	across grace periods.
	(It is probably better to drop linearizability as a
	requirement in most cases, but linearizability is a
	surprisingly popular (and costly!\@) requirement.)
\QuickE{}
\QuickQ{}
	Why can't users dynamically allocate the hazard pointers as they
	are needed?
\QuickA{}
	They can, but at the expense of additional reader-traversal
	overhead and, in some environments, the need to handle
	memory-allocation failure.
\QuickE{}
\QuickQ{}
	But don't Linux-kernel \co{kref} reference counters allow
	guaranteed unconditional reference acquisition?
\QuickA{}
	Yes they do, but the guarantee only applies unconditionally
	in cases where a reference is already held.
	With this in mind, please review the paragraph at the beginning of
	\cref{sec:defer:Which to Choose?}, especially the part
	saying ``large enough that readers do not hold references from
	one traversal to another''.
\QuickE{}
\QuickQ{}
	But didn't the answer to one of the quick quizzes in
	\cref{sec:defer:Hazard Pointers}
	say that pairwise asymmetric barriers could eliminate the
	read-side \co{smp_mb()} from hazard pointers?
\QuickA{}
	Yes, it did.
	However, doing this could be argued to change hazard-pointers
	``Reclamation Forward Progress'' row (discussed later) from
	lock-free to blocking because a CPU spinning with interrupts
	disabled in the kernel would prevent the update-side portion of
	the asymmetric barrier from completing.
	In the Linux kernel, such blocking could in theory be prevented
	by building the kernel with \co{CONFIG_NO_HZ_FULL}, designating
	the relevant CPUs as \co{nohz_full} at boot time, ensuring that
	only one thread was ever runnable on a given CPU at a given
	time, and avoiding ever calling into the kernel.
	Alternatively, you could ensure that the kernel was free of any
	bugs that might cause CPUs to spin with interrupts disabled.

	Given that CPUs spinning in the Linux kernel with interrupts
	disabled seems to be rather rare, one might counter-argue that
	asymmetric-barrier hazard-pointer updates are non-blocking
	in practice, if not in theory.
\QuickE{}
\QuickQAC{chp:Data Structures}{Data Structures}{qqzdatastruct}
\QuickQ{}
	But chained hash tables are but one type of many.
	Why the focus on chained hash tables?
\QuickA{}
	Chained hash tables are completely partitionable, and thus
	well-suited to concurrent use.
	There are other completely-partitionable hash tables, for
	example, split-ordered list~\cite{OriShalev2006SplitOrderListHash},
	but they are considerably more complex.
	We therefore start with chained hash tables.
\QuickE{}
\QuickQ{}
	\begin{fcvref}[ln:datastruct:hash_bkt:lookup]
	But isn't the double comparison on
	\clnrefrange{hashmatch}{return} in
	\cref{lst:datastruct:Hash-Table Lookup} inefficient
	in the case where the key fits into an unsigned long?
	\end{fcvref}
\QuickA{}
	Indeed it is!
	However, hash tables quite frequently store information with
	keys such as character strings that do not necessarily fit
	into an unsigned long.
	Simplifying the hash-table implementation for the case where
	keys always fit into unsigned longs is left as an exercise
	for the reader.
\QuickE{}
\QuickQ{}
	Instead of simply increasing the number of hash buckets,
	wouldn't it be better to cache-align the existing hash buckets?
\QuickA{}
	The answer depends on a great many things.
	If the hash table has a large number of elements per bucket, it
	would clearly be better to increase the number of hash buckets.
	On the other hand, if the hash table is lightly loaded,
	the answer depends on the hardware, the effectiveness of the
	hash function, and the workload.
	Interested readers are encouraged to experiment.
\QuickE{}
\QuickQ{}
	Given the negative scalability of the Schr\"odinger's
	Zoo application across sockets, why not just run multiple
	copies of the application, with each copy having a subset
	of the animals and confined to run on a single socket?
\QuickA{}
	You can do just that!
	In fact, you can extend this idea to large clustered systems,
	running one copy of the application on each node of the cluster.
	This practice is called ``sharding'', and is heavily used in
	practice by large web-based
	retailers~\cite{DeCandia:2007:DAH:1323293.1294281}.

	However, if you are going to shard on a per-socket basis within
	a multisocket system, why not buy separate smaller and cheaper
	single-socket systems, and then run one shard of the database
	on each of those systems?
\QuickE{}
\QuickQ{}
	But if elements in a hash table can be removed concurrently
	with lookups, doesn't that mean that a lookup could return
	a reference to a data element that was removed immediately
	after it was looked up?
\QuickA{}
	Yes it can!
	This is why \co{hashtab_lookup()} must be invoked within an
	RCU read-side critical section, and it is why
	\co{hashtab_add()} and \co{hashtab_del()} must also use
	RCU-aware list-manipulation primitives.
	Finally, this is why the caller of \co{hashtab_del()} must
	wait for a grace period (e.g., by calling \co{synchronize_rcu()})
	before freeing the removed element.
	This will ensure that all RCU readers that might reference
	the newly removed element have completed before that element
	is freed.
\QuickE{}
\QuickQ{}
	The \path{hashtorture.h} file contains more than 1,000 lines!
	Is that a comprehensive test or what???
\QuickA{}
	What.

	The \path{hashtorture.h} tests are a good start and suffice
	for a textbook algorithm.
	If this code was to be used in production, much more testing
	would be required:

	\begin{enumerate}
	\item	Have some subset of elements that always reside in the
		table, and verify that lookups always find these elements
		regardless of the number and type of concurrent updates
		in flight.
	\item	Pair an updater with one or more readers, verifying that
		after an element is added, once a reader successfully
		looks up that element, all later lookups succeed.
		The definition of ``later'' will depend on the table's
		consistency requirements.
	\item	Pair an updater with one or more readers, verifying that
		after an element is deleted, once a reader's lookup
		of that element fails, all later lookups also fail.
	\end{enumerate}

	There are many more tests where those came from, the exact
	nature of which depend on the details of the requirements
	on your particular hash table.
\QuickE{}
\QuickQ{}
	How can we be so sure that the hash-table size is at fault here,
	especially given that
	\cref{fig:datastruct:Read-Only Hash-Table Performance For Schroedinger's Zoo; Varying Buckets}
	on \cpageref{fig:datastruct:Read-Only Hash-Table Performance For Schroedinger's Zoo; Varying Buckets}
	shows that varying hash-table size has almost
	no effect?
	Might the problem instead be something like \IX{false sharing}?
\QuickA{}
	Excellent question!

	False sharing requires writes, which are not featured in the
	unsynchronized and RCU runs of this lookup-only
	benchmark.
	The problem is therefore not false sharing.

\begin{figure}
\centering
\resizebox{3in}{!}{\includegraphics{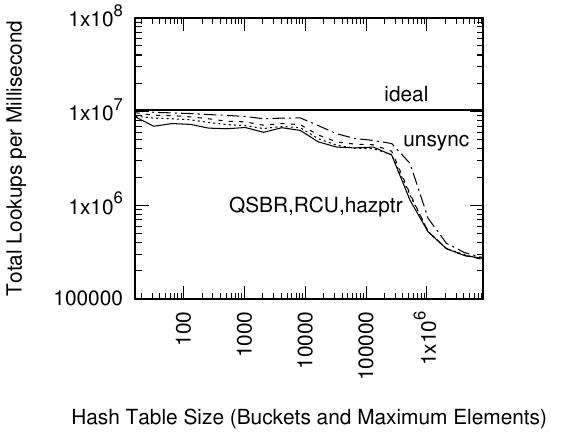}}
\caption{Read-Only RCU-Protected Hash-Table Performance For Schr\"odinger's Zoo at 448 CPUs, Varying Table Size}
\label{fig:datastruct:Read-Only RCU-Protected Hash-Table Performance For Schr\"odinger's Zoo at 448 CPUs; Varying Table Size}
\end{figure}

	Still unconvinced?
	Then look at the log-log plot in
	\cref{fig:datastruct:Read-Only RCU-Protected Hash-Table Performance For Schr\"odinger's Zoo at 448 CPUs; Varying Table Size},
	which shows performance for 448~CPUs as a function of the
	hash-table size, that is, number of buckets and maximum number
	of elements.
	A hash-table of size 1,024 has 1,024~buckets and contains
	at most 1,024~elements, with the average occupancy being
	512~elements.
	Because this is a read-only benchmark, the actual occupancy is
	always equal to the average occupancy.

	This figure shows near-ideal performance below about 8,000~elements,
	that is, when the hash table comprises less than 1\,MB of data.
	This near-ideal performance is consistent with that for the
	pre-BSD routing table shown in
	\cref{fig:defer:Pre-BSD Routing Table Protected by RCU}
	on \cpageref{fig:defer:Pre-BSD Routing Table Protected by RCU},
	even at 448~CPUs.
	However, the performance drops significantly (this is a log-log
	plot) at about 8,000~elements, which is where the 1,048,576-byte
	L2 cache overflows.
	Performance falls off a cliff (even on this log-log plot) at about
	300,000~elements, where the 40,370,176-byte L3 cache overflows.
	This demonstrates that the memory-system bottleneck is profound,
	degrading performance by well in excess of an order of magnitude
	for the large hash tables.
	This should not be a surprise, as the size-8,388,608 hash table
	occupies about 1\,GB of memory, overflowing the L3 caches by
	a factor of 25.

	The reason that
	\cref{fig:datastruct:Read-Only Hash-Table Performance For Schroedinger's Zoo; Varying Buckets}
	on \cpageref{fig:datastruct:Read-Only Hash-Table Performance For Schroedinger's Zoo; Varying Buckets}
	shows little effect is that its data was gathered from
	bucket-locked hash tables, where locking overhead and contention
	drowned out cache-capacity effects.
	In contrast, both RCU and hazard-pointers readers avoid stores
	to shared data, which means that the cache-capacity effects come
	to the fore.

	Still not satisfied?
	Find a multi-socket system and run this code, making use of
	whatever performance-counter hardware is available.
	This hardware should allow you to track down the precise cause
	of any slowdowns exhibited on your particular system.
	The experience gained by doing this exercise will be extremely
	valuable, giving you a significant advantage over those whose
	understanding of this issue is strictly theoretical.\footnote{
		Of course, a theoretical understanding beats no
		understanding.}
\QuickE{}
\QuickQ{}
	The memory system is a serious bottleneck on this big system.
	Why bother putting 448~CPUs on a system without giving them
	enough memory bandwidth to do something useful???
\QuickA{}
	It would indeed be a bad idea to use this large and expensive
	system for a workload consisting solely of simple hash-table
	lookups of small data elements.
	However, this system is extremely useful for a great many
	workloads that feature more processing and less memory accessing.
	For example, some in-memory databases run extremely well on
	this class of system, albeit when running much more complex
	sets of queries than performed by the benchmarks in this chapter.
	For example, such systems might be processing images or video
	streams stored in each element, providing further performance
	benefits due to the fact that the resulting sequential memory
	accesses will make better use of the available memory bandwidth
	than will a pure pointer-following workload.

	But let this be a lesson to you.
	Modern computer systems come in a great many shapes and sizes,
	and great care is frequently required to select one that suits
	your application.
	And perhaps even more frequently, significant care and work is
	required to adjust your application to the specific computer
	systems at hand.
\QuickE{}
\QuickQ{}
	The dangers of extrapolating from 28~CPUs to 448~CPUs was
	made quite clear in
	\cref{sec:datastruct:Hash-Table Performance}.
	Would extrapolating up from 448~CPUs be any safer?
\QuickA{}
	In theory, no, it isn't any safer, and a useful exercise would be
	to run these programs on larger systems.
	In practice, there are only a very few systems with more than
	448~CPUs, in contrast to the huge number having more than 28~CPUs.
	This means that although it is dangerous to extrapolate beyond
	448~CPUs, there is very little need to do so.

	In addition, other testing has shown that RCU read-side primitives
	offer consistent performance and scalability up to at least 1024~CPUs.
	However, it is useful to review
	\cref{fig:datastruct:Read-Only RCU-Protected Hash-Table Performance For Schr\"odinger's Zoo at 448 CPUs; Varying Table Size}
	and its associated commentary.
	You see, unlike the 448-CPU system that provided this data,
	the system enjoying linear scalability up to 1024~CPUs boasted
	excellent memory bandwidth.
\QuickE{}
\QuickQ{}
	How does the code in
	\cref{lst:datastruct:Resizable Hash-Table Bucket Selection}
	protect against the resizing process progressing past the
	selected bucket?
\QuickA{}
	It does not provide any such protection.
	That is instead the job of the update-side concurrency-control
	functions described next.
\QuickE{}
\QuickQ{}
	Suppose that one thread is inserting an element into the
	hash table during a resize operation.
	What prevents this insertion from being lost due to a subsequent
	resize operation completing before the insertion does?
\QuickA{}
	The second resize operation will not be able to move beyond
	the bucket into which the insertion is taking place due to
	the insertion holding the lock(s) on one or both of the hash
	buckets in the hash tables.
	Furthermore, the insertion operation takes place within an
	RCU read-side critical section.
	As we will see when we examine the \co{hashtab_resize()}
	function, this means that each resize operation uses
	\co{synchronize_rcu()} invocations to wait for the insertion's
	read-side critical section to complete.
\QuickE{}
\QuickQ{}
	The \co{hashtab_lookup()} function in
	\cref{lst:datastruct:Resizable Hash-Table Access Functions}
	ignores concurrent resize operations.
	Doesn't this mean that readers might miss an element that was
	previously added during a resize operation?
\QuickA{}
	No.
	As we will see soon,
	the \co{hashtab_add()} and \co{hashtab_del()} functions
	keep the old hash table up-to-date while a resize operation
	is in progress.
\QuickE{}
\QuickQ{}
	The \co{hashtab_add()} and \co{hashtab_del()} functions in
	\cref{lst:datastruct:Resizable Hash-Table Access Functions}
	can update two hash buckets while a resize operation is progressing.
	This might cause poor performance if the frequency of resize operation
	is not negligible.
	Isn't it possible to reduce the cost of updates in such cases?
\QuickA{}
	Yes, at least assuming that a slight increase in the cost of
	\co{hashtab_lookup()} is acceptable.
	One approach is shown in
	\cref{lst:datastruct:Resizable Hash-Table Access Functions (Fewer Updates),%
	lst:datastruct:Resizable Hash-Table Update-Side Locking Function (Fewer Updates)}
	(\path{hash_resize_s.c}).

\begin{listing}
\begin{fcvlabel}[ln:datastruct:hash_resize_s:access]
\begin{VerbatimL}[commandchars=\\\@\$]
struct ht_elem *\lnlbl@lkp:b$
hashtab_lookup(struct hashtab *htp_master, void *key)
{
	struct ht *htp;
	struct ht_elem *htep;

	htp = rcu_dereference(htp_master->ht_cur);\lnlbl@lkp:get_curtbl$
	htep = ht_search_bucket(htp, key);\lnlbl@lkp:get_curbkt$
	if (htep)\lnlbl@lkp:entchk$
		return htep;\lnlbl@lkp:ret_match$
	htp = rcu_dereference(htp->ht_new);\lnlbl@lkp:get_nxttbl$
	if (!htp)\lnlbl@lkp:htpchk$
		return NULL;\lnlbl@lkp:noresize$
	return ht_search_bucket(htp, key);\lnlbl@lkp:ret_nxtbkt$
}\lnlbl@lkp:e$

void hashtab_add(struct ht_elem *htep,\lnlbl@add:b$
                 struct ht_lock_state *lsp)
{
	struct ht_bucket *htbp = lsp->hbp[0];\lnlbl@add:htbp$
	int i = lsp->hls_idx[0];\lnlbl@add:i$

	htep->hte_next[!i].prev = NULL;\lnlbl@add:initp$
	cds_list_add_rcu(&htep->hte_next[i], &htbp->htb_head);\lnlbl@add:add$
}\lnlbl@add:e$

void hashtab_del(struct ht_elem *htep,\lnlbl@del:b$
                 struct ht_lock_state *lsp)
{
	int i = lsp->hls_idx[0];\lnlbl@del:i$

	if (htep->hte_next[i].prev) {\lnlbl@del:if$
		cds_list_del_rcu(&htep->hte_next[i]);\lnlbl@del:del$
		htep->hte_next[i].prev = NULL;\lnlbl@del:init$
	}
	if (lsp->hbp[1] && htep->hte_next[!i].prev) {\lnlbl@del:ifnew$
		cds_list_del_rcu(&htep->hte_next[!i]);\lnlbl@del:delnew$
		htep->hte_next[!i].prev = NULL;\lnlbl@del:initnew$
	}
}\lnlbl@del:e$
\end{VerbatimL}
\end{fcvlabel}
 \caption{Resizable Hash-Table Access Functions (Fewer Updates)}
\label{lst:datastruct:Resizable Hash-Table Access Functions (Fewer Updates)}
\end{listing}

\begin{listing}
\begin{fcvlabel}[ln:datastruct:hash_resize_s:lock_mod]
\begin{VerbatimL}[commandchars=\\\@\$]
static void\lnlbl@l:b$
hashtab_lock_mod(struct hashtab *htp_master, void *key,
                 struct ht_lock_state *lsp)
{
	long b;
	unsigned long h;
	struct ht *htp;
	struct ht_bucket *htbp;

	rcu_read_lock();\lnlbl@l:rcu_lock$
	htp = rcu_dereference(htp_master->ht_cur);\lnlbl@l:refhashtbl$
	htbp = ht_get_bucket(htp, key, &b, &h);\lnlbl@l:refbucket$
	spin_lock(&htbp->htb_lock);\lnlbl@l:acq_bucket$
	lsp->hbp[0] = htbp;\lnlbl@l:lsp0b$
	lsp->hls_idx[0] = htp->ht_idx;
	if (b > READ_ONCE(htp->ht_resize_cur)) {\lnlbl@l:ifresized$
		lsp->hbp[1] = NULL;\lnlbl@l:lsp1_1$
		return;\lnlbl@l:fastret1$
	}
	htp = rcu_dereference(htp->ht_new);\lnlbl@l:new_hashtbl$
	htbp = ht_get_bucket(htp, key, &b, &h);\lnlbl@l:get_newbkt$
	spin_lock(&htbp->htb_lock);\lnlbl@l:acq_newbkt$
	lsp->hbp[1] = lsp->hbp[0];\lnlbl@l:lsp1b$
	lsp->hls_idx[1] = lsp->hls_idx[0];
	lsp->hbp[0] = htbp;
	lsp->hls_idx[0] = htp->ht_idx;
}\lnlbl@l:e$
\end{VerbatimL}
\end{fcvlabel}
 \caption{Resizable Hash-Table Update-Side Locking Function (Fewer Updates)}
\label{lst:datastruct:Resizable Hash-Table Update-Side Locking Function (Fewer Updates)}
\end{listing}

	This version of \co{hashtab_add()} adds an element to
	either the old bucket if it is not resized yet, or to the new
	bucket if it has been resized, and \co{hashtab_del()} removes
	the specified element from any buckets into which it has been inserted.
	The \co{hashtab_lookup()} function searches the new bucket
	if the search of the old bucket fails, which has the disadvantage
	of adding overhead to the lookup fastpath.
	The alternative \co{hashtab_lock_mod()} returns the locking
	state of the new bucket in \co{->hbp[0]} and \co{->hls_idx[0]}
	if resize operation is in progress, instead of the perhaps
	more natural choice of \co{->hbp[1]} and \co{->hls_idx[1]}.
	However, this less-natural choice has the advantage of simplifying
	\co{hashtab_add()}.

	Further analysis of the code is left as an exercise for the reader.
\QuickE{}
\QuickQ{}
	\begin{fcvref}[ln:datastruct:hash_resize:resize]
	In the \co{hashtab_resize()} function in
	\cref{lst:datastruct:Resizable Hash-Table Resizing},
	what guarantees that the update to \co{->ht_new} on \clnref{set_newtbl}
	will be seen as happening before the update to \co{->ht_resize_cur}
	on \clnref{update_resize} from the perspective of
	\co{hashtab_add()} and \co{hashtab_del()}?
	In other words, what prevents \co{hashtab_add()}
	and \co{hashtab_del()} from dereferencing
	a \co{NULL} pointer loaded from \co{->ht_new}?
	\end{fcvref}
\QuickA{}
	\begin{fcvref}[ln:datastruct:hash_resize:resize]
	The \co{synchronize_rcu()} on \clnref{sync_rcu} of
	\cref{lst:datastruct:Resizable Hash-Table Resizing}
	ensures that all pre-existing RCU readers have completed between
	the time that we install the new hash-table reference on
	\clnref{set_newtbl} and the time that we update \co{->ht_resize_cur} on
	\clnref{update_resize}.
	This means that any reader that sees a non-negative value
	of \co{->ht_resize_cur} cannot have started before the
	assignment to \co{->ht_new}, and thus must be able to see
	the reference to the new hash table.

	And this is why the update-side \co{hashtab_add()} and
	\co{hashtab_del()} functions must be enclosed
	in RCU read-side critical sections, courtesy of
	\co{hashtab_lock_mod()} and \co{hashtab_unlock_mod()} in
	\cref{lst:datastruct:Resizable Hash-Table Update-Side Concurrency Control}.
	\end{fcvref}
\QuickE{}
\QuickQ{}
	\begin{fcvref}[ln:datastruct:hash_resize:resize]
	Why is there a \co{WRITE_ONCE()} on \clnref{update_resize}
	in \cref{lst:datastruct:Resizable Hash-Table Resizing}?
	\end{fcvref}
\QuickA{}
	\begin{fcvref}[ln:datastruct:hash_resize:lock_unlock_mod]
	Together with the \co{READ_ONCE()}
	on \clnref{l:ifresized} in \co{hashtab_lock_mod()}
	of \cref{lst:datastruct:Resizable Hash-Table Update-Side Concurrency Control},
	it tells the compiler that the non-initialization accesses
	to \co{->ht_resize_cur} must remain because reads
	from \co{->ht_resize_cur} really can race with writes,
	just not in a way to change the ``if'' conditions.
	\end{fcvref}
\QuickE{}
\QuickQ{}
	How much of the difference in performance between the large and
	small hash tables shown in
	\cref{fig:datastruct:Overhead of Resizing Hash Tables Between 262;144 and 524;288 Buckets vs. Total Number of Elements}
	was due to long hash chains and how much was due to
	memory-system bottlenecks?
\QuickA{}
	The easy way to answer this question is to do another run with
	2,097,152 elements, but this time also with 2,097,152 buckets,
	thus bringing the average number of elements per bucket back down
	to unity.

\begin{figure}
\centering
\resizebox{2.7in}{!}{\includegraphics{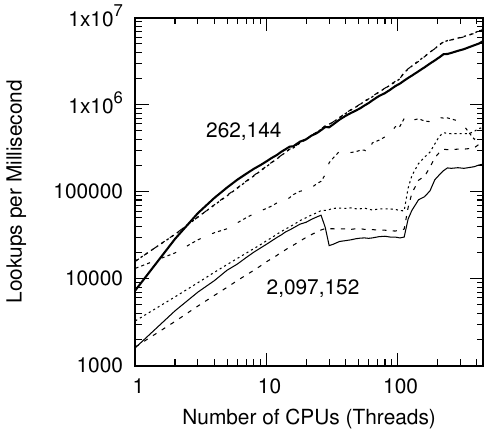}}
\caption{Effect of Memory-System Bottlenecks on Hash Tables}
\label{fig:datastruct:Effect of Memory-System Bottlenecks on Hash Tables}
\end{figure}

	The results are shown by the triple-dashed new trace in
	the middle of
	\cref{fig:datastruct:Effect of Memory-System Bottlenecks on Hash Tables}.
	The other six traces are identical to their counterparts in
	\cref{fig:datastruct:Overhead of Resizing Hash Tables Between 262;144 and 524;288 Buckets vs. Total Number of Elements}
	on \cpageref{fig:datastruct:Overhead of Resizing Hash Tables Between 262;144 and 524;288 Buckets vs. Total Number of Elements}.
	The gap between this new trace and the lower set of three
	traces is a rough measure of how much of the difference in
	performance was due to hash-chain length, and the gap between
	the new trace and the upper set of three traces is a rough measure
	of how much of that difference was due to memory-system bottlenecks.
	The new trace starts out slightly below its 262,144-element
	counterpart at a single CPU, showing that cache capacity is
	degrading performance slightly even on that single CPU\@.\footnote{
		Yes, as far as hardware architects are concerned,
		caches are part of the memory system.}
	This is to be expected, given that unlike its smaller counterpart,
	the 2,097,152-bucket hash table does not fit into the L3 cache.
	This new trace rises just past 28~CPUs, which is also to be
	expected.
	This rise is due to the fact that the 29\textsuperscript{th}
	CPU is on another socket, which brings with it an additional
	39\,MB of cache as well as additional memory bandwidth.

	But the large hash table's advantage over that of the hash table
	with 524,288~buckets (but still 2,097,152 elements) decreases
	with additional CPUs, which is consistent with the bottleneck
	residing in the memory system.
	Above about 400~CPUs, the 2,097,152-bucket hash table is
	actually outperformed slightly by the 524,288-bucket hash
	table.
	This should not be a surprise because the memory system is
	the bottleneck and the larger number of buckets increases this
	workload's memory footprint.

	The alert reader will have noted the word ``rough'' above
	and might be interested in a more detailed analysis.
	Such readers are invited to run similar benchmarks, using
	whatever performance counters or hardware-analysis tools
	they might have available.
	This can be a long and complex journey, but those brave enough
	to embark on it will be rewarded with detailed knowledge of
	hardware performance and its effect on software.
\QuickE{}
\QuickQ{}
	How much do these specializations really save?
	Are they really worth it?
\QuickA{}
	The answer to the first question is left as an exercise to
	the reader.
	Try specializing the resizable hash table and see how much
	performance improvement results.
	The second question cannot be answered in general, but must
	instead be answered with respect to a specific use case.
	Some use cases are extremely sensitive to performance and
	scalability, while others are less so.
\QuickE{}
\QuickQAC{chp:Validation}{Validation}{qqzdebugging}
\QuickQ{}
	When in computing is it necessary to follow a
	fragmentary plan?
\QuickA{}
	There are any number of situations, but perhaps the most important
	situation is when no one has ever created anything resembling
	the program to be developed.
	In this case, the only way to create a credible plan is to
	implement the program, create the plan, and implement it a
	second time.
	But whoever implements the program for the first time has no
	choice but to follow a fragmentary plan because any detailed
	plan created in ignorance cannot survive first contact with
	the real world.

	And perhaps this is one reason why evolution has favored insanely
	optimistic human beings who are happy to follow fragmentary plans!
\QuickE{}
\QuickQ{}
	Who cares about the organization?
	After all, it is the project that is important!
\QuickA{}
	Yes, projects are important, but if you like being paid for your
	work, you need organizations as well as projects.
\QuickE{}
\QuickQ{}
	Suppose that you are writing a script that processes the
	output of the \co{time} command, which looks as follows:

	\begin{center}
	\fbox{\BUseVerbatim[boxwidth=2in,baseline=c]{VerbDebuggingQQZ}}
	\end{center}

	The script is required to check its input for errors, and to
	give appropriate diagnostics if fed erroneous \co{time} output.
	What test inputs should you provide to this program to test it
	for use with \co{time} output generated by single-threaded programs?
\QuickA{}
	Can you say ``Yes'' to all the following questions?

	\begin{enumerate}
	\item	Do you have a test case in which all the time is
		consumed in user mode by a CPU-bound program?
	\item	Do you have a test case in which all the time is
		consumed in system mode by a CPU-bound program?
	\item	Do you have a test case in which all three times
		are zero?
	\item	Do you have a test case in which the \qco{user} and \qco{sys}
		times sum to more than the \qco{real} time?
		(This would of course be completely legitimate in
		a multithreaded program.)
	\item	Do you have a set of tests cases in which one of the
		times uses more than one second?
	\item	Do you have a set of tests cases in which one of the
		times uses more than ten seconds?
	\item	Do you have a set of test cases in which one of the
		times has non-zero minutes?
		(For example, \qco{15m36.342s}.)
	\item	Do you have a set of test cases in which one of the
		times has a seconds value of greater than 60?
	\item	Do you have a set of test cases in which one of the
		times overflows 32 bits of milliseconds?
		64 bits of milliseconds?
	\item	Do you have a set of test cases in which one of the
		times is negative?
	\item	Do you have a set of test cases in which one of the
		times has a positive minutes value but a negative
		seconds value?
	\item	Do you have a set of test cases in which one of the
		times omits the \qco{m} or the \qco{s}?
	\item	Do you have a set of test cases in which one of the
		times is non-numeric?
		(For example, \qco{Go Fish}.)
	\item	Do you have a set of test cases in which one of the
		lines is omitted?
		(For example, where there is a \qco{real} value and
		a \qco{sys} value, but no \qco{user} value.)
	\item	Do you have a set of test cases where one of the
		lines is duplicated?
		Or duplicated, but with a different time value for
		the duplicate?
	\item	Do you have a set of test cases where a given line
		has more than one time value?
		(For example, \qco{real 0m0.132s 0m0.008s}.)
	\item	Do you have a set of test cases containing random
		characters?
	\item	In all test cases involving invalid input, did you
		generate all permutations?
	\item	For each test case, do you have an expected outcome
		for that test?
	\end{enumerate}

	If you did not generate test data for a substantial number of
	the above cases, you will need to cultivate a more destructive
	attitude in order to have a chance of generating high-quality
	tests.

	Of course, one way to economize on destructiveness is to
	generate the tests with the to-be-tested source code at hand,
	which is called white-box testing (as opposed to black-box testing).
	However, this is no panacea:
	You will find that it is all too easy to find your thinking
	limited by what the program can handle, thus failing to generate
	truly destructive inputs.
\QuickE{}
\QuickQ{}
	You are asking me to do all this validation BS before
	I even start coding???
	That sounds like a great way to never get started!!!
\QuickA{}
	If it is your project, for example, a hobby, do what you like.
	Any time you waste will be your own, and you have no one else
	to answer to for it.
	And there is a good chance that the time will not be completely
	wasted.
	For example, if you are embarking on a first-of-a-kind project,
	the requirements are in some sense unknowable anyway.
	In this case, the best approach might be to quickly prototype
	a number of rough solutions, try them out, and see what works
	best.

	On the other hand, if you are being paid to produce a system that
	is broadly similar to existing systems, you owe it to your users,
	your employer, and your future self to validate early and often.
\QuickE{}
\QuickQ{}
	Are you actually suggesting that it is possible to test
	correctness into software???
	Everyone knows that is impossible!!!
\QuickA{}
	Please note that the text used the word ``validation'' rather
	than the word ``testing''.
	The word ``validation'' includes formal methods as well as
	testing, for more on which please see
	\cref{chp:Formal Verification}.

	But as long as we are bringing up things that everyone should
	know, let's remind ourselves that Darwinian evolution is
	not about correctness, but rather about survival.
	As is software.
	My goal as a developer is not that my software be attractive
	from a theoretical viewpoint, but rather that it survive
	whatever its users throw at it.

	Although the notion of correctness does have its uses, its
	fundamental limitation is that the specification against which
	correctness is judged will also have bugs.
	This means nothing more nor less than that traditional correctness
	proofs prove that the code in question contains the intended
	set of bugs!

	Alternative definitions of correctness instead focus on the
	lack of problematic properties, for example, proving that the
	software has no use-after-free bugs, no \co{NULL} pointer
	dereferences, no array-out-of-bounds references, and so on.
	Make no mistake, finding and eliminating such classes of bugs
	can be highly useful.
	But the fact remains that the lack of certain classes of bugs
	does nothing to demonstrate fitness for any specific purpose.

	Therefore, usage-driven validation remains critically important.

	Besides, it is also impossible to verify correctness into your
	software, especially given the problematic need to verify both
	the verifier and the specification.
\QuickE{}
\QuickQ{}
	How can you implement \co{WARN_ON_ONCE()}?
\QuickA{}
	If you don't mind \co{WARN_ON_ONCE()} sometimes warning more
	than once, simply maintain a static variable that is initialized
	to zero.
	If the condition triggers, check the variable, and
	if it is non-zero, return.
	Otherwise, set it to one, print the message, and return.

	If you really need the message to never appear more than once,
	you can use an atomic exchange operation in place of ``set it
	to one'' above.
	Print the message only if the atomic exchange operation returns
	zero.
\QuickE{}
\QuickQ{}
	Just what invalid assumptions are you accusing Linux kernel
	hackers of harboring???
\QuickA{}
	Those wishing a complete answer to this question are encouraged
	to search the Linux kernel \co{git} repository for commits
	containing the string \qco{Fixes:}.
	There were many thousands of them just in the year 2020, including
	fixes for the following invalid assumptions:

	\begin{enumerate}
	\item	Testing for a non-zero denominator will prevent
		divide-by-zero errors.
		(Hint:
		Suppose that the test uses 64-bit arithmetic
		but that the division uses 32-bit arithmetic.)
	\item	Userspace can be trusted to zero out versioned data
		structures used to communicate with the kernel.
		(Hint:
		Sometimes userspace has no idea how large the
		data structure is.)
	\item	Outdated TCP duplicate selective acknowledgement (D-SACK)
		packets can be completely ignored.
		(Hint:
		These packets might also contain other information.)
	\item	All CPUs are little-endian.
	\item	Once a data structure is no longer needed, all of its
		memory may be immediately freed.
	\item	All devices can be initialized while in standby mode.
	\item	Developers can be trusted to consistently do correct
		hexidecimal arithmetic.
	\end{enumerate}

	Those who look at these commits in greater detail will conclude
	that invalid assumptions are the rule, not the exception.
\QuickE{}
\QuickQ{}
	Why would anyone bother copying existing code in pen on paper???
	Doesn't that just increase the probability of transcription errors?
\QuickA{}
	If you are worried about transcription errors, please allow me
	to be the first to introduce you to a really cool tool named
	\co{diff}.
	In addition, carrying out the copying can be quite valuable:
	\begin{enumerate}
	\item	If you are copying a lot of code, you are probably failing
		to take advantage of an opportunity for abstraction.
		The act of copying code can provide great motivation
		for abstraction.
	\item	Copying the code gives you an opportunity to think about
		whether the code really works in its new setting.
		Is there some non-obvious constraint, such as the need
		to disable interrupts or to hold some lock?
	\item	Copying the code also gives you time to consider whether
		there is some better way to get the job done.
	\end{enumerate}
	So, yes, copy the code!
\QuickE{}
\QuickQ{}
	This procedure is ridiculously over-engineered!
	How can you expect to get a reasonable amount of software
	written doing it this way???
\QuickA{}
	Indeed, repeatedly copying code by hand is laborious and slow.
	However, when combined with heavy-duty stress testing and
	proofs of correctness, this approach is also extremely effective
	for complex parallel code where ultimate performance and
	reliability are required and where debugging is difficult.
	The Linux-kernel RCU implementation is a case in point.

	On the other hand, if you are writing a simple single-threaded
	shell script, then you would be best-served by a different
	methodology.
	For example, enter each command one at a time into an interactive
	shell with a test data set to make sure that it does what you
	want, then copy-and-paste the successful commands into your
	script.
	Finally, test the script as a whole.

	If you have a friend or colleague who is willing to help out,
	pair programming can work very well, as can any number of
	formal design- and code-review processes.

	And if you are writing code as a hobby, then do whatever you like.

	In short, different types of software need different development
	methodologies.
\QuickE{}
\QuickQ{}
	What do you do if, after all the pen-on-paper copying, you find
	a bug while typing in the resulting code?
\QuickA{}
	The answer, as is often the case, is ``it depends''.
	If the bug is a simple typo, fix that typo and continue typing.
	However, if the bug indicates a design flaw, go back to pen
	and paper.
\QuickE{}
\QuickQ{}
	Wait!
	Why on earth would an abstract piece of software fail only
	sometimes???
\QuickA{}
	Because complexity and concurrency can produce results that
	are indistinguishable from
	randomness~\cite{PeterOkech2009InherentRandomness}.
	For example, a bug in Linux-kernel RCU required the following
	to hold before that bug would manifest:
	\begin{enumerate}
	\item	The kernel was built for HPC or real-time use, so that
		a given CPU's RCU work could be offloaded to some other
		CPU\@.
	\item	An offloaded CPU went offline just after generating a
		large quantity of RCU work.
	\item	A special \co{rcu_barrier()} API was invoked just at
		this time.
	\item	The RCU work from the newly offlined CPU was still being
		processed after \co{rcu_barrier()} returned.
	\item	One of these remaining RCU work items was related to
		the code invoking the \co{rcu_barrier()}.
	\end{enumerate}
	Making this bug manifest therefore required considerable luck
	or great testing skill.
	But the testing skill could be effective only if the bug was
	known, which of course it was not.
	Therefore, the manifesting of this bug was very well modeled
	as a probabilistic process.
\QuickE{}
\QuickQ{}
	Suppose that you had a very large number of systems at your
	disposal.
	For example, at current cloud prices, you can purchase a
	huge amount of CPU time at low cost.
	Why not use this approach to get close enough to certainty
	for all practical purposes?
\QuickA{}
	This approach might well be a valuable addition to your
	validation arsenal.
	But it does have limitations that rule out ``for all practical
	purposes'':
	\begin{enumerate}
	\item	Some bugs have extremely low probabilities of occurrence,
		but nevertheless need to be fixed.
		For example, suppose that the Linux kernel's RCU
		implementation had a bug that is triggered only once
		per million years of machine time on average.
		A million years of CPU time is hugely expensive even on
		the cheapest cloud platforms, but we could expect
		this bug to result in more than 50 failures per day
		on the more than 20~billion Linux instances in the
		world as of 2017.
	\item	The bug might well have zero probability of occurrence
		on your particular cloud-computing test setup, which
		means that you won't see it no matter how much machine
		time you burn testing it.
		For but one example, there are RCU bugs that appear
		only in preemptible kernels, and also other RCU bugs
		that appear only in non-preemptible kernels.
	\end{enumerate}
	Of course, if your code is small enough, formal validation
	may be helpful, as discussed in
	\cref{chp:Formal Verification}.
	But beware:
	Formal validation of your code will not find errors in your
	assumptions, misunderstanding of the requirements,
	misunderstanding of the software or hardware primitives you use,
	or errors that you did not think to construct a proof for.
\QuickE{}
\QuickQ{}
	Say what???
	When I plug the earlier five-test 10\,\%-failure-rate example into
	the formula, I get 59,050\,\% and that just doesn't make sense!!!
\QuickA{}
	You are right, that makes no sense at all.

	Remember that a probability is a number between zero and one,
	so that you need to divide a percentage by 100 to get a
	probability.
	So 10\,\% is a probability of 0.1, which gets a probability
	of 0.4095, which rounds to 41\,\%, which quite sensibly
	matches the earlier result.
\QuickE{}
\QuickQ{}
	In \cref{eq:debugging:Binomial Number of Tests Required},
	are the logarithms base-10, base-2, or base-$\euler$?
\QuickA{}
	It does not matter.
	You will get the same answer no matter what base of logarithms
	you use because the result is a pure ratio of logarithms.
	The only constraint is that you use the same base for both
	the numerator and the denominator.
\QuickE{}
\QuickQ{}
	Suppose that a bug causes a test failure three times per hour
	on average.
	How long must the test run error-free to provide 99.9\,\%
	confidence that the fix significantly reduced the probability
	of failure?
\QuickA{}
	We set $n$ to $3$ and $P$ to $99.9$ in
	\cref{eq:debugging:Error-Free Test Duration}, resulting in:

	\begin{equation}
		T = - \frac{1}{3} \ln \frac{100 - 99.9}{100} = 2.3
	\end{equation}

	If the test runs without failure for 2.3 hours, we can be 99.9\,\%
	certain that the fix reduced the probability of failure.
\QuickE{}
\QuickQ{}
	Doing the summation of all the factorials and exponentials
	is a real pain.
	Isn't there an easier way?
\QuickA{}
	One approach is to use the open-source symbolic manipulation
	program named ``maxima''.
	Once you have installed this program, which is a part of many
	Linux distributions, you can run it and give the
	\co{load(distrib);} command followed by any number of
	\co{bfloat(cdf_poisson(m,l));} commands, where the \co{m} is
	replaced by the desired value of $m$ (the actual number of failures in
	actual test) and the \co{l} is replaced by the desired value of
	$\lambda$ (the expected number of failures in the actual test).

	In particular, the \co{bfloat(cdf_poisson(2,24));} command
	results in \co{1.181617112359357b-8}, which matches the value
	given by \cref{eq:debugging:Possion CDF}.

\begin{table}
\renewcommand*{\arraystretch}{1.25}
\rowcolors{3}{}{lightgray}
\small
\centering
\begin{tabular}{rrrr}
	\toprule
		& \multicolumn{3}{c}{Improvement} \\
		\cmidrule(l){2-4}
	Certainty (\%)
		& Any
			& 10x
				& 100x \\
	\cmidrule{1-1} \cmidrule(l){2-4}
	90.0	& 2.3	& 23.0	& 230.0  \\
	95.0	& 3.0	& 30.0	& 300.0  \\
	99.0	& 4.6	& 46.1	& 460.5  \\
	99.9	& 6.9	& 69.1	& 690.7  \\
	\bottomrule
\end{tabular}
\caption{Human-Friendly Poisson-Function Display}
\label{tab:debugging:Human-Friendly Poisson-Function Display}
\end{table}

	Another approach is to recognize that in this real world,
	it is not all that useful to compute (say) the duration of a test
	having two or fewer errors that would give a 76.8\,\% confidence
	of a 349.2x improvement in reliability.
	Instead, human beings tend to focus on specific values, for
	example, a 95\,\% confidence of a 10x improvement.
	People also greatly prefer error-free test runs, and so should
	you because doing so reduces your required test durations.
	Therefore, it is quite possible that the values in
	\cref{tab:debugging:Human-Friendly Poisson-Function Display}
	will suffice.
	Simply look up the desired confidence and degree of improvement,
	and the resulting number will give you the required
	error-free test duration in terms of the expected time for
	a single error to appear.
	So if your pre-fix testing suffered one failure per hour, and the
	powers that be require a 95\,\% confidence of a 10x improvement,
	you need a 30-hour error-free run.

	Alternatively, you can use the rough-and-ready method described in
	\cref{sec:debugging:Statistics Abuse for Discrete Testing}.
\QuickE{}
\QuickQ{}
	But wait!!!
	Given that there has to be \emph{some} number of failures
	(including the possibility of zero failures), shouldn't
	\cref{eq:debugging:Possion CDF}
	approach the value $1$ as $m$ goes to infinity?
\QuickA{}
	Indeed it should.
	And it does.

	To see this, note that $\euler^{-\lambda}$ does not depend on $i$,
	which means that it can be pulled out of the summation as follows:

	\begin{equation}
		\euler^{-\lambda} \sum_{i=0}^\infty \frac{\lambda^i}{i!}
	\end{equation}

	The remaining summation is exactly the Taylor series for
	$\euler^\lambda$, yielding:

	\begin{equation}
		\euler^{-\lambda} \euler^\lambda
	\end{equation}

	The two exponentials are reciprocals, and therefore cancel,
	resulting in exactly $1$, as required.
\QuickE{}
\QuickQ{}
	How is this approach supposed to help if the corruption affected some
	unrelated pointer, which then caused the corruption???
\QuickA{}
	Indeed, that can happen.
	Many CPUs have hardware-debugging facilities that can help you
	locate that unrelated pointer.
	Furthermore, if you have a core dump, you can search the core
	dump for pointers referencing the corrupted region of memory.
	You can also look at the data layout of the corruption, and
	check pointers whose type matches that layout.

	You can also step back and test the modules making up your
	program more intensively, which will likely confine the corruption
	to the module responsible for it.
	If this makes the corruption vanish, consider adding additional
	argument checking to the functions exported from each module.

	Nevertheless, this is a hard problem, which is why I used the
	words ``a bit of a dark art''.
\QuickE{}
\QuickQ{}
	But I did the bisection, and ended up with a huge commit.
	What do I do now?
\QuickA{}
	A huge commit?
	Shame on you!
	This is but one reason why you are supposed to keep the commits small.

	And that is your answer:
	Break up the commit into bite-sized pieces and bisect the pieces.
	In my experience, the act of breaking up the commit is often
	sufficient to make the bug painfully obvious.
\QuickE{}
\QuickQ{}
	Why don't conditional-locking primitives provide this
	spurious-failure functionality?
\QuickA{}
	There are locking algorithms that depend on conditional-locking
	primitives telling them the truth.
	For example, if conditional-lock failure signals that
	some other thread is already working on a given job,
	spurious failure might cause that job to never get done,
	possibly resulting in a hang.
\QuickE{}
\QuickQ{}
	That is ridiculous!!!
	After all, isn't getting the correct answer later than one would like
	better than getting an incorrect answer???
\QuickA{}
	This question fails to consider the option of choosing not to
	compute the answer at all, and in doing so, also fails to consider
	the costs of computing the answer.
	For example, consider short-term weather forecasting, for which
	accurate models exist, but which require large (and expensive)
	clustered supercomputers, at least if you want to actually run
	the model faster than the weather.

	And in this case, any performance bug that prevents the model from
	running faster than the actual weather prevents any forecasting.
	Given that the whole purpose of purchasing the large clustered
	supercomputers was to forecast weather, if you cannot run the
	model faster than the weather, you would be better off not running
	the model at all.

	More severe examples may be found in the area of safety-critical
	real-time computing.
\QuickE{}
\QuickQ{}
	But if you are going to put in all the hard work of parallelizing
	an application, why not do it right?
	Why settle for anything less than optimal performance and
	linear scalability?
\QuickA{}
	Although I do heartily salute your spirit and aspirations,
	you are forgetting that there may be high costs due to delays
	in the program's completion.
	For an extreme example, suppose that a 40\,\% performance shortfall
	from a single-threaded application is causing one person to die
	each day.
	Suppose further that in a day you could hack together a
	quick and dirty
	parallel program that ran 50\,\% faster on an eight-CPU system
	than the sequential version, but that an optimal parallel
	program would require four months of painstaking design, coding,
	debugging, and tuning.

	It is safe to say that more than 100 people would prefer the
	quick and dirty version.
\QuickE{}
\QuickQ{}
	But what about other sources of error, for example, due to
	interactions between caches and memory layout?
\QuickA{}
	Changes in memory layout can indeed result in unrealistic
	decreases in execution time.
	For example, suppose that a given microbenchmark almost
	always overflows the L0 cache's associativity, but with just the right
	memory layout, it all fits.
	If this is a real concern, consider running your microbenchmark
	using huge pages (or within the kernel or on bare metal) in
	order to completely control the memory layout.

	But note that there are many different possible memory-layout
	bottlenecks.
	Benchmarks sensitive to memory bandwidth (such as those involving
	matrix arithmetic) should spread the running threads across the
	available cores and sockets to maximize memory parallelism.
	They should also spread the data across \IXplr{NUMA node}, memory
	controllers, and DRAM chips to the extent possible.
	In contrast, benchmarks sensitive to memory latency (including
	most poorly scaling applications) should instead maximize
	locality, filling each core and socket in turn before adding
	another one.
\QuickE{}
\QuickQ{}
	Wouldn't the techniques suggested to isolate the code under
	test also affect that code's performance, particularly if
	it is running within a larger application?
\QuickA{}
	Indeed it might, although in most microbenchmarking efforts
	you would extract the code under test from the enclosing
	application.
	Nevertheless, if for some reason you must keep the code under
	test within the application, you will very likely need to use
	the techniques discussed in
	\cref{sec:debugging:Detecting Interference}.
\QuickE{}
\QuickQ{}
	This approach is just plain weird!
	Why not use means and standard deviations, like we were taught
	in our statistics classes?
\QuickA{}
	Because mean and standard deviation were not designed to do this job.
	To see this, try applying mean and standard deviation to the
	following data set, given a 1\,\% relative error in measurement:

	\begin{quote}
		49,548.4 49,549.4 49,550.2 49,550.9 49,550.9 49,551.0
		49,551.5 49,552.1 49,899.0 49,899.3 49,899.7 49,899.8
		49,900.1 49,900.4 52,244.9 53,333.3 53,333.3 53,706.3
		53,706.3 54,084.5
	\end{quote}

	The problem is that mean and standard deviation do not rest on
	any sort of measurement-error assumption, and they will therefore
	see the difference between the values near 49,500 and those near
	49,900 as being statistically significant, when in fact they are
	well within the bounds of estimated measurement error.

	Of course, it is possible to create a script similar to
	that in
	\cref{lst:debugging:Statistical Elimination of Interference}
	that uses standard deviation rather than absolute difference
	to get a similar effect,
	and this is left as an exercise for the interested reader.
	Be careful to avoid divide-by-zero errors arising from strings
	of identical data values!
\QuickE{}
\QuickQ{}
	But what if all the y-values in the trusted group of data
	are exactly zero?
	Won't that cause the script to reject any non-zero value?
\QuickA{}
	Indeed it will!
	But if your performance measurements often produce a value of
	exactly zero, perhaps you need to take a closer look at your
	performance-measurement code.

	Note that many approaches based on mean and standard deviation
	will have similar problems with this sort of dataset.
\QuickE{}
\QuickQAC{chp:Formal Verification}{Formal Verification}{qqzformal}
\QuickQ{}
	Why is there an unreached statement in locker?
	After all, isn't this a \emph{full} state-space
	search?
\QuickA{}
	The locker process is an infinite loop, so control
	never reaches the end of this process.
	However, since there are no monotonically increasing variables,
	Promela is able to model this infinite loop with a small
	number of states.
\QuickE{}
\QuickQ{}
	What are some Promela code-style issues with this example?
\QuickA{}
	There are several:
	\begin{enumerate}
	\item	The declaration of \co{sum} should be moved to within
		the init block, since it is not used anywhere else.
	\item	The assertion code should be moved outside of the
		initialization loop.
		The initialization loop can then be placed in an atomic
		block, greatly reducing the state space (by how much?).
	\item	The atomic block covering the assertion code should
		be extended to include the initialization of \co{sum}
		and \co{j}, and also to cover the assertion.
		This also reduces the state space (again, by how
		much?).
\QuickE{}
	\end{enumerate}
\QuickQ{}
	Is there a more straightforward way to code the \co{do-od} statement?
\QuickA{}
	Yes.
	Replace it with \co{if-fi} and remove the two \co{break} statements.
\QuickE{}
\QuickQ{}
	\begin{fcvref}[ln:formal:promela:qrcu:updater]
	Why are there atomic blocks at \clnrefrange{atm1:b}{atm1:e}
	and \clnrefrange{atm2:b}{atm2:e}, when the operations
	within those atomic
	blocks have no atomic implementation on any current
	production microprocessor?
	\end{fcvref}
\QuickA{}
	Because those operations are for the benefit of the
	assertion only.
	They are not part of the algorithm itself.
	There is therefore no harm in marking them atomic, and
	so marking them greatly reduces the state space that must
	be searched by the Promela model.
\QuickE{}
\QuickQ{}
	\begin{fcvref}[ln:formal:promela:qrcu:updater]
	Is the re-summing of the counters on
	\clnrefrange{reinvoke:b}{reinvoke:e}
	\emph{really} necessary?
	\end{fcvref}
\QuickA{}
	Yes.
	To see this, delete these lines and run the model.

	Alternatively, consider the following sequence of steps:

	\begin{enumerate}
	\item	One process is within its RCU read-side critical
		section, so that the value of \co{ctr[0]} is zero and
		the value of \co{ctr[1]} is two.
	\item	An updater starts executing, and sees that the sum of
		the counters is two so that the fastpath cannot be
		executed.
		It therefore acquires the lock.
	\item	A second updater starts executing, and fetches the value
		of \co{ctr[0]}, which is zero.
	\item	The first updater adds one to \co{ctr[0]}, flips
		the index (which now becomes zero), then subtracts
		one from \co{ctr[1]} (which now becomes one).
	\item	The second updater fetches the value of \co{ctr[1]},
		which is now one.
	\item	The second updater now incorrectly concludes that it
		is safe to proceed on the fastpath, despite the fact
		that the original reader has not yet completed.
\QuickE{}
	\end{enumerate}
\QuickQ{}
	A compression rate of 0.48\,\% corresponds to a 200-to-1 decrease
	in memory occupied by the states!
	Is the state-space search \emph{really} exhaustive???
\QuickA{}
	According to Spin's documentation, yes, it is.

\begin{listing}
\begin{Verbatim}
@@ -1,6 +1,6 @@
 (Spin Version 6.4.6 -- 2 December 2016)
         + Partial Order Reduction
-        + Compression
+        + Graph Encoding (-DMA=88)

 Full statespace search for:
         never claim             - (none specified)
@@ -9,27 +9,22 @@
         invalid end states      +

 State-vector 88 byte, depth reached 328014, errors: 0
+MA stats: -DMA=77 is sufficient
+Minimized Automaton:    2084798 nodes and 6.38445e+06 edges
 1.8620286e+08 states, stored
 1.7759831e+08 states, matched
 3.6380117e+08 transitions (= stored+matched)
 1.3724093e+08 atomic steps
-hash conflicts: 1.1445626e+08 (resolved)

 Stats on memory usage (in Megabytes):
 20598.919       equivalent memory usage for states
                   (stored*(State-vector + overhead))
- 8418.559       actual memory usage for states
-                  (compression: 40.87%)
-                state-vector as stored =
-                  19 byte + 28 byte overhead
- 2048.000       memory used for hash table (-w28)
+  204.907       actual memory usage for states
+                  (compression: 0.99%)
    17.624       memory used for DFS stack (-m330000)
-    1.509       memory lost to fragmentation
-10482.675       total actual memory usage
+  222.388       total actual memory usage

-nr of templates: [ 0:globals 1:chans 2:procs ]
-collapse counts: [ 0:1021 2:32 3:1869 4:2 ]
 unreached in proctype qrcu_reader
         (0 of 18 states)
 unreached in proctype qrcu_updater
@@ -38,5 +33,5 @@
 unreached in init
         (0 of 23 states)

-pan: elapsed time 369 seconds
-pan: rate 505107.58 states/second
+pan: elapsed time 2.68e+03 seconds
+pan: rate 69453.282 states/second
\end{Verbatim}
\vspace*{-9pt}
\caption{Spin Output Diff of \co{-DCOLLAPSE} and \co{-DMA=88}}
\label{lst:formal:promela:Spin Output Diff of -DCOLLAPSE and -DMA=88}
\end{listing}

	As an indirect evidence, let's compare the results of
	runs with \co{-DCOLLAPSE} and with \co{-DMA=88}
	(two readers and three updaters).
	The diff of outputs from those runs is shown in
	\cref{lst:formal:promela:Spin Output Diff of -DCOLLAPSE and -DMA=88}.
	As you can see, they agree on the numbers of states
	(stored and matched).
\QuickE{}
\QuickQ{}
	But different formal-verification tools are often designed to
	locate particular classes of bugs.
	For example, very few formal-verification tools will find
	an error in the specification.
	So isn't this ``clearly untrustworthy'' judgment a bit harsh?
\QuickA{}
	It is certainly true that many formal-verification tools are
	specialized in some way.
	For example, Promela does not handle realistic memory models
	(though they can be programmed into
	\IX{Promela}~\cite{Desnoyers:2013:MSM:2506164.2506174}),
	\IXacr{cbmc}~\cite{EdmundClarke2004CBMC} does not detect probabilistic
	hangs and deadlocks, and
	\IX{Nidhugg}~\cite{CarlLeonardsson2014Nidhugg} does not detect
	bugs involving data nondeterminism.
	But this means that these tools cannot be trusted to find
	bugs that they are not designed to locate.

	And therefore people creating formal-verification tools should
	``tell the truth on the label'', clearly calling out what
	classes of bugs their tools can and cannot detect.
	Otherwise, the first time a practitioner finds a tool
	failing to detect a bug, that practitioner is likely to
	make extremely harsh and extremely public denunciations
	of that tool.
	Yes, yes, there is something to be said for putting your
	best foot forward, but putting it too far forward without
	appropriate disclaimers can easily trigger a land mine of
	negative reaction that your tool might or might not be able
	to recover from.

	You have been warned!
\QuickE{}
\QuickQ{}
	Given that we have two independent proofs of correctness for
	the QRCU algorithm described herein, and given that the
	proof of incorrectness covers what is known to be a different
	algorithm, why is there any room for doubt?
\QuickA{}
	There is always room for doubt.
	In this case, it is important to keep in mind that the two proofs
	of correctness preceded the formalization of real-world memory
	models, raising the possibility that these two proofs are based
	on incorrect memory-ordering assumptions.
	Furthermore, since both proofs were constructed by the same person,
	it is quite possible that they contain a common error.
	Again, there is always room for doubt.
\QuickE{}
\QuickQ{}
	Yeah, that's just great!
	Now, just what am I supposed to do if I don't happen to have a
	machine with 40\,GB of main memory???
\QuickA{}
	Relax, there are a number of lawful answers to
	this question:
	\begin{enumerate}
	\item	Try compiler flags \co{-DCOLLAPSE} and \co{-DMA=N}
		to reduce memory consumption.
		See \cref{sec:formal:Running the QRCU Example}.
	\item	Further optimize the model, reducing its memory consumption.
	\item	Work out a pencil-and-paper proof, perhaps starting with the
		comments in the code in the Linux kernel.
	\item	Devise careful torture tests, which, though they cannot prove
		the code correct, can find hidden bugs.
	\item	There is some movement towards tools that do model
		checking on clusters of smaller machines.
		However, please note that we have not actually used such
		tools myself, courtesy of some large machines that Paul has
		occasional access to.
	\item	Wait for memory sizes of affordable systems to expand
		to fit your problem.
	\item	Use one of a number of cloud-computing services to rent
		a large system for a short time period.
\QuickE{}
	\end{enumerate}
\QuickQ{}
	Why not simply increment \co{rcu_update_flag}, and then only
	increment \co{dynticks_progress_counter} if the old value
	of \co{rcu_update_flag} was zero???
\QuickA{}
	This fails in presence of NMIs.
	To see this, suppose an NMI was received just after
	\co{rcu_irq_enter()} incremented \co{rcu_update_flag},
	but before it incremented \co{dynticks_progress_counter}.
	The instance of \co{rcu_irq_enter()} invoked by the NMI
	would see that the original value of \co{rcu_update_flag}
	was non-zero, and would therefore refrain from incrementing
	\co{dynticks_progress_counter}.
	This would leave the RCU grace-period machinery no clue that the
	NMI handler was executing on this CPU, so that any RCU read-side
	critical sections in the NMI handler would lose their RCU protection.

	The possibility of NMI handlers, which, by definition cannot
	be masked, does complicate this code.
\QuickE{}
\QuickQ{}
	\begin{fcvref}[ln:formal:dyntickrcu:rcu_irq_enter]
	But if \clnref{chk_lv:b} finds that we are the outermost interrupt,
	wouldn't we \emph{always} need to increment
	\co{dynticks_progress_counter}?
	\end{fcvref}
\QuickA{}
	Not if we interrupted a running task!
	In that case, \co{dynticks_progress_counter} would
	have already been incremented by \co{rcu_exit_nohz()},
	and there would be no need to increment it again.
\QuickE{}
\QuickQ{}
	Can you spot any bugs in any of the code in this section?
\QuickA{}
	Read the next section to see if you were correct.
\QuickE{}
\QuickQ{}
	Why isn't the memory barrier in \co{rcu_exit_nohz()}
	and \co{rcu_enter_nohz()} modeled in Promela?
\QuickA{}
	Promela assumes sequential consistency, so
	it is not necessary to model memory barriers.
	In fact, one must instead explicitly model lack of memory barriers,
	for example, as shown in
	\cref{lst:formal:QRCU Unordered Summation} on
	\cpageref{lst:formal:QRCU Unordered Summation}.
\QuickE{}
\QuickQ{}
	Isn't it a bit strange to model \co{rcu_exit_nohz()}
	followed by \co{rcu_enter_nohz()}?
	Wouldn't it be more natural to instead model entry before exit?
\QuickA{}
	It probably would be more natural, but we will need
	this particular order for the liveness checks that we will add later.
\QuickE{}
\QuickQ{}
	Wait a minute!
	In the Linux kernel, both \co{dynticks_progress_counter} and
	\co{rcu_dyntick_snapshot} are per-CPU variables.
	So why are they instead being modeled as single global variables?
\QuickA{}
	Because the grace-period code processes each
	CPU's \co{dynticks_progress_counter} and
	\co{rcu_dyntick_snapshot} variables separately,
	we can collapse the state onto a single CPU\@.
	If the grace-period code were instead to do something special
	given specific values on specific CPUs, then we would indeed need
	to model multiple CPUs.
	But fortunately, we can safely confine ourselves to two CPUs, the
	one running the grace-period processing and the one entering and
	leaving dynticks-idle mode.
\QuickE{}
\QuickQ{}
	\begin{fcvref}[ln:formal:promela:dyntick:dyntickRCU-base-s:grace_period]
	Given there are a pair of back-to-back changes to
	\co{grace_period_state} on \clnref{upd_gps3,upd_gps4},
	how can we be sure that \clnref{upd_gps3}'s changes won't be lost?
	\end{fcvref}
\QuickA{}
	Recall that Promela and Spin trace out
	every possible sequence of state changes.
	Therefore, timing is irrelevant:
	Promela/Spin will be quite happy to jam the entire rest of
	the model between those two statements unless some state
	variable specifically prohibits doing so.
\QuickE{}
\QuickQ{}
	But what would you do if you needed the statements in a single
	\co{EXECUTE_MAINLINE()} group to execute non-atomically?
\QuickA{}
	The easiest thing to do would be to put
	each such statement in its own \co{EXECUTE_MAINLINE()}
	statement.
\QuickE{}
\QuickQ{}
	But what if the \co{dynticks_nohz()} process had
	\qco{if} or \qco{do} statements with conditions,
	where the statement bodies of these constructs
	needed to execute non-atomically?
\QuickA{}
	One approach, as we will see in a later section,
	is to use explicit labels and \qco{goto} statements.
	For example, the construct:

\begin{VerbatimU}
	if
	:: i == 0 -> a = -1;
	:: else -> a = -2;
	fi;
\end{VerbatimU}
	could be modeled as something like:

\begin{VerbatimU}
	EXECUTE_MAINLINE(stmt1,
	                 if
	                 :: i == 0 -> goto stmt1_then;
	                 :: else -> goto stmt1_else;
	                 fi)
	stmt1_then: skip;
	EXECUTE_MAINLINE(stmt1_then1, a = -1; goto stmt1_end)
	stmt1_else: skip;
	EXECUTE_MAINLINE(stmt1_then1, a = -2)
	stmt1_end: skip;
\end{VerbatimU}

	However, it is not clear that the macro is helping much in the case
	of the \qco{if} statement, so these sorts of situations will
	be open-coded in the following sections.
\QuickE{}
\QuickQ{}
	\begin{fcvref}[ln:formal:promela:dyntick:dyntickRCU-irqnn-ssl:dyntick_irq]
	Why are \clnref{clr_in_irq,inc_i} (the \qtco{in_dyntick_irq = 0;}
	and the \qco{i++;}) executed atomically?
	\end{fcvref}
\QuickA{}
	These lines of code pertain to controlling the
	model, not to the code being modeled, so there is no reason to
	model them non-atomically.
	The motivation for modeling them atomically is to reduce the size
	of the state space.
\QuickE{}
\QuickQ{}
	What property of interrupts is this \co{dynticks_irq()}
	process unable to model?
\QuickA{}
	One such property is nested interrupts,
	which are handled in the following section.
\QuickE{}
\QuickQ{}
	Does Paul \emph{always} write his code in this painfully incremental
	manner?
\QuickA{}
	Not always, but more and more frequently.
	In this case, Paul started with the smallest slice of code that
	included an interrupt handler, because he was not sure how best
	to model interrupts in Promela.
	Once he got that working, he added other features.
	(But if he was doing it again, he would start with a ``toy'' handler.
	For example, he might have the handler increment a variable twice and
	have the mainline code verify that the value was always even.)

	Why the incremental approach?
	Consider the following, attributed to Brian W. Kernighan:

	\begin{quote}
		Debugging is twice as hard as writing the code in the first
		place.
		Therefore, if you write the code as cleverly as possible,
		you are, by definition, not smart enough to debug it.
	\end{quote}

	This means that any attempt to optimize the production of code should
	place at least 66\,\% of its emphasis on optimizing the debugging process,
	even at the expense of increasing the time and effort spent coding.
	Incremental coding and testing is one way to optimize the debugging
	process, at the expense of some increase in coding effort.
	Paul uses this approach because he rarely has the luxury of
	devoting full days (let alone weeks) to coding and debugging.
\QuickE{}
\QuickQ{}
	But what happens if an NMI handler starts running before
	an \IRQ\ handler completes, and if that NMI handler continues
	running until a second \IRQ\ handler starts?
\QuickA{}
	This cannot happen within the confines of a single CPU\@.
	The first \IRQ\ handler cannot complete until the NMI handler
	returns.
	Therefore, if each of the \co{dynticks} and \co{dynticks_nmi}
	variables have taken on an even value during a given time
	interval, the corresponding CPU really was in a quiescent
	state at some time during that interval.
\QuickE{}
\QuickQ{}
	This is still pretty complicated.
	Why not just have a \co{cpumask_t} with per-CPU bits, clearing
	the bit when entering an \IRQ\ or NMI handler, and setting it
	upon exit?
\QuickA{}
	Although this approach would be functionally correct, it
	would result in excessive \IRQ\ entry/exit overhead on
	large machines.
	In contrast, the approach laid out in this section allows
	each CPU to touch only per-CPU data on \IRQ\ and NMI entry/exit,
	resulting in much lower \IRQ\ entry/exit overhead, especially
	on large machines.
\QuickE{}
\QuickQ{}
	But x86 has strong memory ordering, so why formalize its memory
	model?
\QuickA{}
	Actually, academics consider the x86 memory model to be weak
	because it can allow prior stores to be reordered with
	subsequent loads.
	From an academic viewpoint, a strong memory model is one
	that allows absolutely no reordering, so that all threads
	agree on the order of all operations visible to them.

	Plus it really is the case that developers are sometimes confused
	about x86 memory ordering.
\QuickE{}
\QuickQ{}
	\begin{fcvref}[ln:formal:PPCMEM Litmus Test]
	Why does \clnref{reginit} of \cref{lst:formal:PPCMEM Litmus Test}
	initialize the registers?
	Why not instead initialize them on \clnref{init:0,init:1}?
	\end{fcvref}
\QuickA{}
	Either way works.
	However, in general, it is better to use initialization than
	explicit instructions.
	The explicit instructions are used in this example to demonstrate
	their use.
	In addition, many of the litmus tests available on the tool's
	web site (\url{https://www.cl.cam.ac.uk/~pes20/ppcmem/}) were
	automatically generated, which generates explicit
	initialization instructions.
\QuickE{}
\QuickQ{}
	\begin{fcvref}[ln:formal:PPCMEM Litmus Test]
	But whatever happened to \clnref{P0fail1} of
	\cref{lst:formal:PPCMEM Litmus Test},
	the one that is the \co{Fail1:} label?
	\end{fcvref}
\QuickA{}
	The implementation of PowerPC version of \co{atomic_add_return()}
	loops when the \co{stwcx} instruction fails, which it communicates
	by setting non-zero status in the condition-code register,
	which in turn is tested by the \co{bne} instruction.
	Because actually modeling the loop would result in state-space
	explosion, we instead branch to the \co{Fail1:} label,
	terminating the model with the initial value of~2 in P0's \co{r3}
	register, which will not trigger the exists assertion.

	There is some debate about whether this trick is universally
	applicable, but I have not seen an example where it fails.
\QuickE{}
\QuickQ{}
	Does the \ARM\ Linux kernel have a similar bug?
\QuickA{}
	\ARM\ does not have this particular bug because it places
	\co{smp_mb()} before and after the \co{atomic_add_return()}
	function's assembly-language implementation.
	PowerPC no longer has this bug; it has long since been
	fixed~\cite{BenjaminHerrenschmidt2011:powerpc:atomic_return}.
\QuickE{}
\QuickQ{}
	\begin{fcvref}[ln:formal:PPCMEM Litmus Test]
	Does the \co{lwsync} on \clnref{P0lwsync} in
	\cref{lst:formal:PPCMEM Litmus Test} provide sufficient ordering?
	\end{fcvref}
\QuickA{}
	It depends on the semantics required.
	The rest of this answer assumes that the assembly language
	for \co{P0} in
	\cref{lst:formal:PPCMEM Litmus Test}
	is supposed to implement a value-returning atomic operation.

	As is discussed in
	\cref{chp:Advanced Synchronization: Memory Ordering},
	Linux kernel's memory consistency model requires
	value-returning atomic RMW operations to be fully ordered
	on both sides.
	The ordering provided by \co{lwsync} is insufficient for this
	purpose, and so \co{sync} should be used instead.
	This change has since been
	made~\cite{BoqunFeng2015:powerpc:value-returning-atomics}
	in response to an email thread discussing a couple of other litmus
	tests~\cite{Paulmck2015:powerpc:value-returning-atomics}.
	Finding any other bugs that the Linux kernel might have is left
	as an exercise for the reader.

	In other enviroments providing weaker semantics, \co{lwsync}
	might be sufficient.
	But not for the Linux kernel's value-returning atomic operations!
\QuickE{}
\QuickQ{}
	What do you have to do to run \co{herd} on litmus tests like
	that shown in \cref{lst:formal:Locking Example}?
\QuickA{}
	Get version v4.17 (or later) of the Linux-kernel source code,
	then follow the instructions in \path{tools/memory-model/README}
	to install the needed tools.
	Then follow the further instructions to run these tools on the
	litmus test of your choice.
\QuickE{}
\QuickQ{}
	Why bother modeling locking directly?
	Why not simply emulate locking with atomic operations?
\QuickA{}
	In a word, performance, as can be seen in
	\cref{tab:formal:Locking: Modeling vs. Emulation Time (s)}.
	The first column shows the number of \co{herd} processes modeled.
	The second column shows the \co{herd} runtime when modeling
	\co{spin_lock()} and \co{spin_unlock()} directly in \co{herd}'s
	cat language.
	The third column shows the \co{herd} runtime when emulating
	\co{spin_lock()} with \co{cmpxchg_acquire()} and \co{spin_unlock()}
	with \co{smp_store_release()}, using the \co{herd} \co{filter}
	clause to reject executions that fail to acquire the lock.
	The fourth column is like the third, but using \co{xchg_acquire()}
	instead of \co{cmpxchg_acquire()}.
	The fifth and sixth columns are like the third and fourth,
	but instead using the \co{herd} \co{exists} clause to reject
	executions that fail to acquire the lock.

\begin{table}
\rowcolors{10}{}{lightgray}
\small
\centering
\newcommand{\lockfml}[1]{\multicolumn{1}{c}{\begin{picture}(6,8)(0,0)\rotatebox{90}{#1}\end{picture}}}
\begin{tabular}{rrrrrr}
	\toprule
	& Model & \multicolumn{4}{c}{Emulate} \\
	\cmidrule(l){2-2} \cmidrule(l){3-6}
	& & \multicolumn{2}{c}{\tco{filter}} & \multicolumn{2}{c}{\tco{exists}} \\
	\cmidrule(l){3-4} \cmidrule(l){5-6}
	\lockfml{\# Proc.}
	&
	& \tco{cmpxchg}
	& \tco{xchg}
	& \tco{cmpxchg}
	& \tco{xchg}
	\\
	\cmidrule{1-1} \cmidrule(l){2-2} \cmidrule(l){3-4} \cmidrule(l){5-6}
	2 & 0.004 &  0.022 &  0.027 &   0.039 &   0.058 \\
	3 & 0.041 &  0.743 &  0.968 &   1.653 &   3.203 \\
	4 & 0.374 & 59.565 & 74.818 & 151.962 & 500.960 \\
	5 & 4.905 \\
	\bottomrule
\end{tabular}
\caption{Locking:
		  Modeling vs.\@ Emulation Time (s)}
\label{tab:formal:Locking: Modeling vs. Emulation Time (s)}
\end{table}

	Note also that use of the \co{filter} clause is about twice
	as fast as is use of the \co{exists} clause.
	This is no surprise because the \co{filter} clause allows
	early abandoning of excluded executions, where the executions
	that are excluded are the ones in which the lock is concurrently
	held by more than one process.

	More important, modeling \co{spin_lock()} and \co{spin_unlock()}
	directly ranges from five times faster to more than two orders
	of magnitude faster than modeling emulated locking.
	This should also be no surprise, as direct modeling raises
	the level of abstraction, thus reducing the number of events
	that \co{herd} must model.
	Because almost everything that \co{herd} does is of exponential
	computational complexity, modest reductions in the number of
	events produces exponentially large reductions in runtime.

	Thus, in formal verification even more than in parallel
	programming itself, divide and conquer!!!
\QuickE{}
\QuickQ{}
	Wait!!!
	Isn't leaking pointers out of an RCU read-side critical
	section a critical bug???
\QuickA{}
	Yes, it usually is a critical bug.
	However, in this case, the updater has been cleverly constructed
	to properly handle such pointer leaks.
	But please don't make a habit of doing this sort of thing, and
	especially don't do this without having put a lot of thought
	into making some more conventional approach work.
\QuickE{}
\QuickQ{}
	\begin{fcvref}[ln:formal:C-RomanPenyaev-list-rcu-rr:whole]
	In \cref{lst:formal:Complex RCU Litmus Test},
	why couldn't a reader fetch \co{c} just before \co{P1()}
	zeroed it on \clnref{updinitcache}, and then later
	store this same value back into \co{c} just after it was
	zeroed, thus defeating the zeroing operation?
	\end{fcvref}
\QuickA{}
	\begin{fcvref}[ln:formal:C-RomanPenyaev-list-rcu-rr:whole]
	Because the reader advances to the next element on
	\clnref{rdnext}, thus avoiding storing a pointer to the
	same element as was fetched.
	\end{fcvref}
\QuickE{}
\QuickQ{}
	\begin{fcvref}[ln:formal:C-RomanPenyaev-list-rcu-rr:whole]
	In \cref{lst:formal:Complex RCU Litmus Test},
	why not have just one call to \co{synchronize_rcu()}
	immediately before \clnref{updfree}?
	\end{fcvref}
\QuickA{}
	\begin{fcvref}[ln:formal:C-RomanPenyaev-list-rcu-rr:whole]
	Because this results in \co{P0()} accessing a freed element.
	But don't take my word for this, try it out in \co{herd}!
	\end{fcvref}
\QuickE{}
\QuickQ{}
	\begin{fcvref}[ln:formal:C-RomanPenyaev-list-rcu-rr:whole]
	Also in \cref{lst:formal:Complex RCU Litmus Test},
	can't \clnref{updfree} be \co{WRITE_ONCE()} instead
	of \co{smp_store_release()}?
	\end{fcvref}
\QuickA{}
	\begin{fcvref}[ln:formal:C-RomanPenyaev-list-rcu-rr:whole]
	That is an excellent question.
	As of late 2021, the answer is ``no one knows''.
	Much depends on the semantics of \ARMv8's conditional-move
	instruction.
	While awaiting clarity on these semantics, \co{smp_store_release()}
	is the safe choice.
	\end{fcvref}
\QuickE{}
\QuickQ{}
	But shouldn't sufficiently low-level software be for all intents
	and purposes immune to being exploited by black hats?
\QuickA{}
	Unfortunately, no.

	At one time, Paul E. McKenny felt that Linux-kernel RCU
	was immune to such exploits, but the advent of Row Hammer
	showed him otherwise.
	After all, if the black hats can hit the system's DRAM,
	they can hit any and all low-level software, even including RCU\@.

	And in 2018, this possibility passed from the realm of
	theoretical speculation into the hard and fast realm of
	objective reality~\cite{McKenney:2019:CRS:3319647.3325836}.
\QuickE{}
\QuickQ{}
	In light of the full verification of the L4 microkernel,
	isn't this limited view of formal verification just a little
	bit obsolete?
\QuickA{}
	Unfortunately, no.

	The first full verification of the L4 microkernel was a tour de force,
	with a large number of Ph.D.~students hand-verifying code at a
	very slow per-student rate.
	This level of effort could not be applied to most software projects
	because the rate of change is just too great.
	Furthermore, although the L4 microkernel is a large software
	artifact from the viewpoint of formal verification, it is tiny
	compared to a great number of projects, including LLVM,
	\GCC, the Linux kernel, Hadoop, MongoDB, and a great many others.
	In addition, this verification did have limits, as the researchers
	freely admit, to their credit:
	\url{https://docs.sel4.systems/projects/sel4/frequently-asked-questions.html\#does-sel4-have-zero-bugs}.

	Although formal verification is finally starting to show some
	promise, including more-recent L4 verifications involving greater
	levels of automation, it currently has no chance of completely
	displacing testing in the foreseeable future.
	And although I would dearly love to be proven wrong on this point,
	please note that such proof will be in the form of a real tool
	that verifies real software, not in the form of a large body of
	rousing rhetoric.

	Perhaps someday formal verification will be used heavily for
	validation, including for what is now known as regression testing.
	\Cref{sec:future:Formal Regression Testing?} looks at
	what would be required to make this possibility a reality.
\QuickE{}
\QuickQAC{chp:Putting It All Together}{Putting It All Together}{qqztogether}
\QuickQ{}
	Why not implement reference-acquisition using
	a simple \IXacrml{cas} operation that only
	acquires a reference if the reference counter is
	non-zero?
\QuickA{}
	Although this can resolve the race between the release of
	the last reference and acquisition of a new reference,
	it does absolutely nothing to prevent the data structure
	from being freed and reallocated, possibly as some completely
	different type of structure.
	It is quite likely that the ``simple \acrml{cas}
	operation'' would give undefined results if applied to the
	differently typed structure.

	In short, use of atomic operations such as \acrml{cas}
	absolutely requires either type-safety or existence guarantees.

	But what if it is absolutely necessary to let the type change?

	One approach is for each such type to have the reference counter
	at the same location, so that as long as the reallocation results
	in an object from this group of types, all is well.
	If you do this in C, make sure you comment the reference counter
	in each structure in which it appears.
	In C++, use inheritance and templates.
\QuickE{}
\QuickQ{}
	Why isn't it necessary to guard against cases where one CPU
	acquires a reference just after another CPU releases the last
	reference?
\QuickA{}
	Because a CPU must already hold a reference in order to legally
	acquire another reference.
	Therefore, if one CPU releases the last reference, there had
	better not be any CPU acquiring a new reference!
\QuickE{}
\QuickQ{}
	\begin{fcvref}[ln:together:Linux Kernel kref API]
	Suppose that just after the \co{atomic_sub_and_test()}
	on \clnref{check} of
	\cref{lst:together:Linux Kernel kref API} is invoked,
	that some other CPU invokes \co{kref_get()}.
	Doesn't this result in that other CPU now having an illegal
	reference to a released object?
	\end{fcvref}
\QuickA{}
	This cannot happen if these functions are used correctly.
	It is illegal to invoke \co{kref_get()} unless you already
	hold a reference, in which case the \co{kref_sub()} could
	not possibly have decremented the counter to zero.
\QuickE{}
\QuickQ{}
	Suppose that \co{kref_sub()} returns zero, indicating that
	the \co{release()} function was not invoked.
	Under what conditions can the caller rely on the continued
	existence of the enclosing object?
\QuickA{}
	The caller cannot rely on the continued existence of the
	object unless it knows that at least one reference will
	continue to exist.
	Normally, the caller will have no way of knowing this, and
	must therefore carefully avoid referencing the object after
	the call to \co{kref_sub()}.

	Interested readers are encouraged to work around this limitation
	using RCU, in particular, \co{call_rcu()}.
\QuickE{}
\QuickQ{}
	Why not just pass \co{kfree()} as the release function?
\QuickA{}
	Because the \co{kref} structure normally is embedded in
	a larger structure, and it is necessary to free the entire
	structure, not just the \co{kref} field.
	This is normally accomplished by defining a wrapper function
	that does a \co{container_of()} and then a \co{kfree()}.
\QuickE{}
\QuickQ{}
	Why can't the check for a zero reference count be
	made in a simple \qco{if} statement with an atomic
	increment in its \qco{then} clause?
\QuickA{}
	Suppose that the \qco{if} condition completed, finding
	the reference counter value equal to one.
	Suppose that a release operation executes, decrementing
	the reference counter to zero and therefore starting
	cleanup operations.
	But now the \qco{then} clause can increment the counter
	back to a value of one, allowing the object to be
	used after it has been cleaned up.

	This use-after-cleanup bug is every bit as bad as a
	full-fledged use-after-free bug.
\QuickE{}
\QuickQ{}
	Why don't all sequence-locking use cases replicate the
	data in this fashion?
\QuickA{}
	Such replication is impractical if the data is too
	large, as it might be in the Schr\"odinger's-zoo example
	described in
	\cref{sec:together:Correlated Data Elements}.

	Such replication is unnecessary if delays are prevented,
	for example, when updaters disable interrupts when running
	on bare-metal hardware (that is, without the use of
	a vCPU-preemption-prone hypervisor).

	Alternatively, if readers can tolerate the occasional delay,
	then replication is again unnecessary.
	Consider the example of reader-writer locking, where
	writers always delay readers and vice versa.

	However, if the data to be replicated is reasonably
	small, if delays are possible, and if readers cannot
	tolerate these delays, replicating the data is an
	excellent approach.
\QuickE{}
\QuickQ{}
	Is it possible to write-acquire the sequence lock on
	the new element before it is inserted instead of acquiring
	that of the old element before it is removed?
\QuickA{}
	Yes, and the details are left as an exercise to the reader.

	The term \emph{tombstone} is sometimes used to refer to the
	element with the old name after its sequence lock is acquired.
	Similarly, the term \emph{birthstone} is sometimes used to refer
	to the element with the new name while its sequence lock is
	still held.
\QuickE{}
\QuickQ{}
	Is it possible to avoid the global lock?
\QuickA{}
	Yes, and one way to do this would be to use per-hash-chain locks.
	The updater could acquire lock(s) corresponding to both the old
	and the new element, acquiring them in address order.
	In this case, the insertion and removal operations would of
	course need to refrain from acquiring and releasing these
	same per-hash-chain locks.
	This complexity can be worthwhile if rename operations are
	frequent, and of course can allow rename operations to execute
	concurrently.
\QuickE{}
\QuickQ{}
	Why on earth did we need that global lock in the first place?
\QuickA{}
	A given thread's \co{__thread} variables vanish when that
	thread exits.
	It is therefore necessary to synchronize any operation that
	accesses other threads' \co{__thread} variables with
	thread exit.
	Without such synchronization, accesses to \co{__thread} variable
	of a just-exited thread will result in segmentation faults.
\QuickE{}
\QuickQ{}
	\begin{fcvref}[ln:count:count_end_rcu:whole:reg]
	Hey!!!
	\Clnref{set} of
	\cref{lst:together:RCU and Per-Thread Statistical Counters}
	modifies a value in a pre-existing \co{countarray} structure!
	Didn't you say that this structure, once made available to
	\co{read_count()}, remained constant???
	\end{fcvref}
\QuickA{}
	Indeed I did say that.
	And it would be possible to make \co{count_register_thread()}
	allocate a new structure, much as \co{count_unregister_thread()}
	currently does.

	But this is unnecessary.
	Recall the derivation of the error bounds of \co{read_count()}
	that was based on the snapshots of memory.
	Because new threads start with initial \co{counter} values of
	zero, the derivation holds even if we add a new thread partway
	through \co{read_count()}'s execution.
	So, interestingly enough, when adding a new thread, this
	implementation gets the effect of allocating a new structure,
	but without actually having to do the allocation.
\QuickE{}
\QuickQ{}
	Given the fixed-size \co{counterp} array, exactly how does this
	code avoid a fixed upper bound on the number of threads???
\QuickA{}
	You are quite right, that array does in fact reimpose the fixed
	upper limit.
	This limit may be avoided by tracking threads with a linked list,
	as is done in userspace RCU~\cite{MathieuDesnoyers2012URCU}.
	Doing something similar for this code is left as an exercise for
	the reader.
\QuickE{}
\QuickQ{}
	Wow!
	\Cref{lst:together:RCU and Per-Thread Statistical Counters}
	contains 70 lines of code, compared to only 42 in
	\cref{lst:count:Per-Thread Statistical Counters}.
	Is this extra complexity really worth it?
\QuickA{}
	This of course needs to be decided on a case-by-case basis.
	If you need an implementation of \co{read_count()} that
	scales linearly, then the lock-based implementation shown in
	\cref{lst:count:Per-Thread Statistical Counters}
	simply will not work for you.
	On the other hand, if calls to \co{read_count()} are sufficiently
	rare, then the lock-based version is simpler and might thus be
	better, although much of the size difference is due
	to the structure definition, memory allocation, and \co{NULL}
	return checking.

	Of course, a better question is ``Why doesn't the language
	implement cross-thread access to \co{__thread} variables?''
	After all, such an implementation would make both the locking
	and the use of RCU unnecessary.
	This would in turn enable an implementation that
	was even simpler than the one shown in
	\cref{lst:count:Per-Thread Statistical Counters}, but
	with all the scalability and performance benefits of the
	implementation shown in
	\cref{lst:together:RCU and Per-Thread Statistical Counters}!
\QuickE{}
\QuickQ{}
	But cant't the approach shown in
	\cref{lst:together:Correlated Measurement Fields}
	result in extra cache misses, in turn resulting in additional
	read-side overhead?
\QuickA{}
	Indeed it can.

\begin{listing}
\begin{VerbatimL}[tabsize=8]
struct measurement {
	double meas_1;
	double meas_2;
	double meas_3;
};

struct animal {
	char name[40];
	double age;
	struct measurement *mp;
        struct measurement meas;
	char photo[0]; /* large bitmap. */
};
\end{VerbatimL}
\caption{Localized Correlated Measurement Fields}
\label{lst:together:Localized Correlated Measurement Fields}
\end{listing}

	One way to avoid this cache-miss overhead is shown in
	\cref{lst:together:Localized Correlated Measurement Fields}:
	Simply embed an instance of a \co{measurement} structure
	named \co{meas}
	into the \co{animal} structure, and point the \co{->mp}
	field at this \co{->meas} field.

	Measurement updates can then be carried out as follows:

	\begin{enumerate}
	\item	Allocate a new \co{measurement} structure and place
		the new measurements into it.
	\item	Use \co{rcu_assign_pointer()} to point \co{->mp} to
		this new structure.
	\item	Wait for a grace period to elapse, for example using
		either \co{synchronize_rcu()} or \co{call_rcu()}.
	\item	Copy the measurements from the new \co{measurement}
		structure into the embedded \co{->meas} field.
	\item	Use \co{rcu_assign_pointer()} to point \co{->mp}
		back to the old embedded \co{->meas} field.
	\item	After another grace period elapses, free up the
		new \co{measurement} structure.
	\end{enumerate}

	This approach uses a heavier weight update procedure to eliminate
	the extra cache miss in the common case.
	The extra cache miss will be incurred only while an update is
	actually in progress.
\QuickE{}
\QuickQ{}
	But how does this scan work while a resizable hash table
	is being resized?
	In that case, neither the old nor the new hash table is
	guaranteed to contain all the elements in the hash table!
\QuickA{}
	True, resizable hash tables as described in
	\cref{sec:datastruct:Non-Partitionable Data Structures}
	cannot be fully scanned while being resized.
	One simple way around this is to acquire the
	\co{hashtab} structure's \co{->ht_lock} while scanning,
	but this prevents more than one scan from proceeding
	concurrently.

	Another approach is for updates to mutate the old hash
	table as well as the new one while resizing is in
	progress.
	This would allow scans to find all elements in the old
	hash table.
	Implementing this is left as an exercise for the reader.
\QuickE{}
\QuickQ{}
	But how would this work with a resizable hash table, such
	as the one described in
	\cref{sec:datastruct:Non-Partitionable Data Structures}?
\QuickA{}
	In this case, more care is required because the hash table
	might well be resized during the time that we momentarily
	exited the RCU read-side critical section.
	Worse yet, the resize operation can be expected to free the
	old hash buckets, leaving us pointing to the freelist.

	But it is not sufficient to prevent the old hash buckets
	from being freed.
	It is also necessary to ensure that those buckets continue
	to be updated.

	One way to handle this is to have a reference count on each
	set of buckets, which is initially set to the value one.
	A full-table scan would acquire a reference at the beginning of
	the scan (but only if the reference is non-zero) and release it
	at the end of the scan.
	The resizing would populate the new buckets, release the
	reference, wait for a grace period, and then wait for the
	reference to go to zero.
	Once the reference was zero, the resizing could let updaters
	forget about the old hash buckets and then free it.

	Actual implementation is left to the interested reader, who will
	gain much insight from this task.
\QuickE{}
\QuickQAC{sec:advsync:Advanced Synchronization}{Advanced Synchronization}{qqzadvsync}
\QuickQ{}
	Given that there will always be a sharply limited number of
	CPUs available, is population obliviousness really useful?
\QuickA{}
	Given the surprisingly limited scalability of any number of
	NBS algorithms, population obliviousness can be surprisingly
	useful.
	Nevertheless, the overall point of the question is valid.
	It is not normally helpful for an algorithm to scale beyond the
	size of the largest system it is ever going to run on.
\QuickE{}
\QuickQ{}
	Wait!
	In order to dequeue all elements, both the \co{->head} and
	\co{->tail} pointers must be changed, which cannot be done
	atomically on typical computer systems.
	So how is this supposed to work???
\QuickA{}
	One pointer at a time!

	First, atomically exchange the \co{->head} pointer with \co{NULL}.
	If the return value from the atomic exchange operation is \co{NULL},
	the queue was empty and you are done.
	And if someone else attempts a dequeue-all at this point,
	they will get back a \co{NULL} pointer.

	Otherwise, atomically exchange the \co{->tail} pointer with a
	pointer to the now-\co{NULL} \co{->head} pointer.
	The return value from the atomic exchange operation is a pointer
	to the \co{->next} field of the eventual last element on the list.

	Producing and testing actual code is left as an exercise for the
	interested and enthusiastic reader, as are strategies for handling
	half-enqueued elements.
\QuickE{}
\QuickQ{}
	So why not ditch antique languages like C and C++ for something
	more modern?
\QuickA{}
	That won't help unless the more-modern languages proponents
	are energetic enough to write their own compiler backends.
	The usual practice of re-using existing backends also reuses
	charming properties such as refusal to support pointers to
	lifetime-ended objects.
\QuickE{}
\QuickQ{}
	Why does anyone care about demonic schedulers?
\QuickA{}
	A demonic scheduler is one way to model an insanely overloaded
	system.
	After all, if you have an algorithm that you can prove runs
	reasonably given a demonic scheduler, mere overload should
	be no problem, right?

	On the other hand, it is only reasonable to ask if a demonic
	scheduler is really the best way to model overload conditions.
	And perhaps it is time for more accurate models.
	For one thing, a system might be overloaded in any of a number
	of ways.
	After all, an NBS algorithm that works fine on a demonic scheduler
	might or might not do well in out-of-memory conditions, when
	mass storage fills, or when the network is congested.

	Except that systems' core counts have been increasing, which
	means that an overloaded system is quite likely to be running
	more than one concurrent program.\footnote{
		As a point of reference, back in the mid-1990s, Paul
		witnessed a 16-CPU system running about 20 instances
		of a certain high-end proprietary database.}
	In that case, even if a demonic scheduler is not so demonic
	as to inject idle cycles while there are runnable tasks,
	it is easy to imagine such a scheduler consistently favoring
	the other program over yours.
	If both programs could consume all available CPU, then this
	scheduler might not run your program at all.

	One way to avoid these issues is to simply avoid overload
	conditions.
	This is often the preferred approach in production, where load
	balancers direct traffic away from overloaded systems.
	And if all systems are overloaded, it is not unheard of to simply
	\emph{shed load}, that is, to drop the low-priority incoming requests.
	Nor is this approach limited to computing, as those who have
	suffered through a rolling blackout can attest.
	But load-shedding is often considered a bad thing by those
	whose load is being shed.

	As always, choose wisely!
\QuickE{}
\QuickQ{}
	It seems like the various members of the NBS hierarchy are
	rather useless.
	So why bother with them at all???
\QuickA{}
	One advantage of the members of the NBS hierarchy is that they
	are reasonably simple to define and use from a theoretical viewpoint.
	We can hope that work done in the NBS arena will help lay the
	groundwork for analysis of real-world forward-progress guarantees
	for concurrent real-time programs.
	However, as of 2022 it appears that trace-based methodologies
	are in the lead~\cite{DanielBristot2019RTtrace}.

	So why bother learning about NBS at all?

	Because a great many people know of it, and are vaguely aware that
	it is somehow related to real-time computing.
	Their response to your carefully designed real-time constraints
	might well be of the form ``Bah, just use wait-free algorithms!''.
	In the all-too-common case where they are very convincing to your
	management, you will need to understand NBS in order to bring
	the discussion back to reality.
	I hope that this section has provided you with the required
	depth of understanding.

	Another thing to note is that learning about the NBS hierarchy
	is probably no more harmful than learning about transfinite
	numbers of the computational-complexity hierarchy.
	In all three cases, it is important to avoid over-applying
	the theory.
	Which is in and of itself good practice!
\QuickE{}
\QuickQ{}
	But what about battery-powered systems?
	They don't require energy flowing into the system as a whole.
\QuickA{}
	Sooner or later, the battery must be recharged, which requires
	energy to flow into the system.
\QuickE{}
\QuickQ{}
	But given the results from queueing theory, won't low utilization
	merely improve the average response time rather than improving
	the worst-case response time?
	And isn't worst-case response time all that most
	real-time systems really care about?
\QuickA{}
	Yes, but \ldots

	Those queueing-theory results assume infinite ``calling populations'',
	which in the Linux kernel might correspond to an infinite number
	of tasks.
	As of early 2021, no real system supports an infinite number of
	tasks, so results assuming infinite calling populations should
	be expected to have less-than-infinite applicability.

	Other queueing-theory results have \emph{finite}
	calling populations, which feature sharply bounded response
	times~\cite{Hillier86}.
	These results better model real systems, and these models do
	predict reductions in both average and worst-case response
	times as utilizations decrease.
	These results can be extended to model concurrent systems that use
	synchronization mechanisms such as
	locking~\cite{BjoernBrandenburgPhD,DipankarSarma2004OLSscalability}.

	In short, queueing-theory results that accurately describe
	real-world real-time systems show that worst-case response
	time decreases with decreasing utilization.
\QuickE{}
\QuickQ{}
	Formal verification is already quite capable, benefiting from
	decades of intensive study.
	Are additional advances \emph{really} required, or is this just
	a practitioner's excuse to continue to lazily ignore the awesome
	power of formal verification?
\QuickA{}
	Perhaps this situation is just a theoretician's excuse to avoid
	diving into the messy world of real software?
	Perhaps more constructively, the following advances are required:

	\begin{enumerate}
	\item	Formal verification needs to handle larger software
		artifacts.
		The largest verification efforts have been for systems
		of only about 10,000 lines of code, and those have been
		verifying much simpler properties than real-time latencies.
	\item	Hardware vendors will need to publish formal timing
		guarantees.
		This used to be common practice back when hardware was
		much simpler, but today's complex hardware results in
		excessively complex expressions for worst-case performance.
		Unfortunately, energy-efficiency concerns are pushing
		vendors in the direction of even more complexity.
	\item	Timing analysis needs to be integrated into development
		methodologies and IDEs.
	\end{enumerate}

	All that said, there is hope, given recent work formalizing
	the memory models of real computer
	systems~\cite{JadeAlglave2011ppcmem,Alglave:2013:SVW:2450268.2450306}.
	On the other hand, formal verification has just as much trouble
	as does testing with the astronomical number of variants of the
	Linux kernel that can be constructed from different combinations
	of its tens of thousands of Kconfig options.
	Sometimes life is hard!
\QuickE{}
\QuickQ{}
	Differentiating real-time from non-real-time based on what can
	``be achieved straightforwardly by non-real-time systems and
	applications'' is a travesty!
	There is absolutely no theoretical basis for such a distinction!!!
	Can't we do better than that???
\QuickA{}
	This distinction is admittedly unsatisfying from a strictly
	theoretical perspective.
	But on the other hand, it is exactly what the developer needs
	in order to decide whether the application can be cheaply and
	easily developed using standard non-real-time approaches, or
	whether the more difficult and expensive real-time approaches
	are required.
	In other words, although theory is quite important, for those of
	us called upon to complete practical projects, theory supports
	practice, never the other way around.
\QuickE{}
\QuickQ{}
	But if you only allow one reader at a time to read-acquire
	a reader-writer lock, isn't that the same as an exclusive
	lock???
\QuickA{}
	Indeed it is, other than the API\@.
	And the API is important because it allows the Linux kernel
	to offer real-time capabilities without having the \rt\ patchset
	grow to ridiculous sizes.

	However, this approach clearly and severely limits read-side
	scalability.
	The Linux kernel's \rt\ patchset was long able to live with this
	limitation for several reasons:
	\begin{enumerate*}[(1)]
	\item Real-time systems have traditionally been relatively small,
	\item Real-time systems have generally focused on process control,
	thus being unaffected by scalability limitations in the
	I/O subsystems, and
	\item Many of the Linux kernel's reader-writer locks have been
	converted to RCU\@.
	\end{enumerate*}

	However, the day came when it was absolutely necessary to
	permit concurrent readers, as described in the text following
	this quiz.
\QuickE{}
\QuickQ{}
	\begin{fcvref}[ln:advsync:Preemptible Linux-Kernel RCU:unl]
	Suppose that preemption occurs just after the load from
	\co{t->rcu_read_unlock_special.s} on \clnref{chks} of
	\cref{lst:advsync:Preemptible Linux-Kernel RCU}.
	Mightn't that result in the task failing to invoke
	\co{rcu_read_unlock_special()}, thus failing to remove itself
	from the list of tasks blocking the current grace period,
	in turn causing that grace period to extend indefinitely?
	\end{fcvref}
\QuickA{}
	That is a real problem, and it is solved in RCU's scheduler hook.
	If that scheduler hook sees that the value of
	\co{t->rcu_read_lock_nesting} is negative, it invokes
	\co{rcu_read_unlock_special()} if needed before allowing
	the context switch to complete.
\QuickE{}
\QuickQ{}
	But isn't correct operation despite fail-stop bugs
	a valuable fault-tolerance property?
\QuickA{}
	Yes and no.

	Yes in that non-blocking algorithms can provide fault tolerance
	in the face of fail-stop bugs, but no in that this is grossly
	insufficient for practical fault tolerance.
	For example, suppose you had a wait-free queue, and further
	suppose that a thread has just dequeued an element.
	If that thread now succumbs to a fail-stop bug, the element
	it has just dequeued is effectively lost.
	True fault tolerance requires way more than mere non-blocking
	properties, and is beyond the scope of this book.
\QuickE{}
\QuickQ{}
	I couldn't help but spot the word ``include'' before this list.
	Are there other constraints?
\QuickA{}
	Indeed there are, and lots of them.
	However, they tend to be specific to a given situation,
	and many of them can be thought of as refinements of some of
	the constraints listed above.
	For example, the many constraints on choices of data structure
	will help meeting the ``Bounded time spent in any given critical
	section'' constraint.
\QuickE{}
\QuickQ{}
	Given that real-time systems are often used for safety-critical
	applications, and given that runtime memory allocation is
	forbidden in many safety-critical situations, what is with
	the call to \co{malloc()}???
\QuickA{}
	In early 2016, projects forbidding runtime memory allocation
	were also not at all interested in multithreaded computing.
	So the runtime memory allocation is not an additional
	obstacle to safety criticality.

	However, by 2020 runtime memory allocation in multi-core
	real-time systems was gaining some traction.
\QuickE{}
\QuickQ{}
	Don't you need some kind of synchronization to protect
	\co{update_cal()}?
\QuickA{}
	Indeed you do, and you could use any of a number of techniques
	discussed earlier in this book.
	One of those techniques is use of a single updater thread,
	which would result in exactly the code shown in \co{update_cal()}
	in \cref{lst:advsync:Real-Time Calibration Using RCU}.
\QuickE{}
\QuickQAC{chp:Advanced Synchronization: Memory Ordering}{Advanced Synchronization: Memory Ordering}{qqzmemorder}
\QuickQ{}
	This chapter has been rewritten since the first edition,
	and heavily edited since the second edition.
	Did memory ordering change all \emph{that} since 2014,
	let alone 2021?
\QuickA{}
	The earlier memory-ordering section had its roots in a pair of
	Linux Journal articles~\cite{PaulMcKenney2005i,PaulMcKenney2005j}
	dating back to 2005.
	Since then, the C and C++ memory models~\cite{PeteBecker2011N3242}
	have been formalized
	(and critiqued~\cite{MarkBatty2013OOTA-WorkingNote,Boehm:2014:OGA:2618128.2618134,Vafeiadis:2015:CCO:2775051.2676995,conf/esop/BattyMNPS15,Lahav:2017:RSC:3140587.3062352,OlivierGiroux2017-P0668R1}),
	executable formal memory models for computer systems have become the
	norm~\cite{Maranget2012TutorialARMPower,PaulEMcKenney2011ppcmem,test6-pdf,JadeAlglave2011ppcmem,Alglave:2013:SVW:2450268.2450306,JadeAlglave2013-cav,Alglave:2014:HCM:2594291.2594347,PaulEMcKenney2014weakaxiom,Flur:2017:MCA:3093333.3009839,ARMv8A:2017},
	and there is even a memory model for the Linux
	kernel~\cite{JadeAlglave2017LWN-LKMM-1,JadeAlglave2017LWN-LKMM-2,Alglave:2018:FSC:3173162.3177156},
	along with a paper describing differences between the C11 and
	Linux memory models~\cite{PaulEMcKenney2016P0124R6-LKMM}.

	The \IXacrf{kcsan}~\cite{MarcoElver2020FearDataRaceDetector1,MarcoElver2020FearDataRaceDetector2},
	based in part on
	RacerD~\cite{SamBlackshear2018RacerD}
	and implementing \IXacr{lkmm}, has also been added to the Linux kernel
	and is now heavily used.

	Finally, there are now better ways of describing LKMM.

	Given all this progress, substantial change was required.
\QuickE{}
\QuickQ{}
	The compiler can also reorder Thread~\co{P0()}'s and
	Thread~\co{P1()}'s memory accesses in
	\cref{lst:memorder:Memory Misordering: Store-Buffering Litmus Test},
	right?
\QuickA{}
	In general, compiler optimizations carry out more extensive
	and profound reorderings than CPUs can.
	However, in this case, the volatile accesses in
	\co{READ_ONCE()} and \co{WRITE_ONCE()}
	prevent the compiler from reordering.
	And also from doing much else as well, so the examples in this
	section will be making heavy use of
	\co{READ_ONCE()} and \co{WRITE_ONCE()}.
	See \cref{sec:memorder:Compile-Time Consternation}
	for more detail on the need for \co{READ_ONCE()} and \co{WRITE_ONCE()}.
\QuickE{}
\QuickQ{}
	But wait!!!
	On row~2 of
	\cref{tab:memorder:Memory Misordering: Store-Buffering Sequence of Events}
	both \co{x0} and \co{x1} each have two values at the same time,
	namely zero and two.
	How can that possibly work???
\QuickA{}
	There is an underlying cache-coherence protocol that straightens
	things out, which are discussed in
	\cref{sec:app:whymb:Cache-Coherence Protocols}.
	But if you think that a given variable having two values at
	the same time is surprising, just wait until you get to
	\cref{sec:memorder:Variables With Multiple Values}!
\QuickE{}
\QuickQ{}
	But don't the values also need to be flushed from the cache
	to main memory?
\QuickA{}
	Perhaps surprisingly, not necessarily!
	On some systems,
	if the two variables are being used heavily, they might
	be bounced back and forth between the CPUs' caches and never
	land in main memory.
\QuickE{}
\QuickQ{}
	The rows in
	\cref{tab:memorder:Linux-Kernel Memory-Ordering Cheat Sheet}
	seem quite random and confused.
	Whatever is the conceptual basis of this table???
\QuickA{}
	The rows correspond roughly to hardware mechanisms of increasing
	power and overhead.

	The \co{WRITE_ONCE()} row captures the fact that accesses to
	a single variable are always fully ordered, as indicated by
	the ``SV''column.
	Note that all other operations providing ordering against accesses
	to multiple variables also provide this same-variable ordering.

	The \co{READ_ONCE()} row captures the fact that (as of 2021) compilers
	and CPUs do not indulge in user-visible speculative stores, so that
	any store whose address, data, or execution depends on a prior load
	is guaranteed to happen after that load completes.
	However, this guarantee assumes that these dependencies have
	been constructed carefully, as described in
	\cref{sec:memorder:Address- and Data-Dependency Difficulties,%
	sec:memorder:Control-Dependency Calamities}.

	The ``\co{_relaxed()} RMW operation'' row captures the fact
	that a value-returning \co{_relaxed()} RMW has done a load and a
	store, which are every bit as good as a \co{READ_ONCE()} and a
	\co{WRITE_ONCE()}, respectively.

	The \co{*_dereference()} row captures the address and data
	dependency ordering provided by \co{rcu_dereference()} and friends.
	Again, these dependencies must been constructed carefully,
	as described in
	\cref{sec:memorder:Address- and Data-Dependency Difficulties}.

	The ``Successful \co{*_acquire()}'' row captures the fact that many
	CPUs have special ``acquire'' forms of loads and of atomic RMW
	instructions,
	and that many other CPUs have lightweight memory-barrier
	instructions that order prior loads against subsequent loads
	and stores.

	The ``Successful \co{*_release()}'' row captures the fact that many
	CPUs have special ``release'' forms of stores and of atomic RMW
	instructions, and that many other CPUs have lightweight memory-barrier
	instructions that order prior loads and stores against
	subsequent stores.

	The \co{smp_rmb()} row captures the fact that many CPUs have
	lightweight memory-barrier instructions that order prior loads against
	subsequent loads.
	Similarly,
	the \co{smp_wmb()} row captures the fact that many CPUs have
	lightweight memory-barrier instructions that order prior stores against
	subsequent stores.

	None of the ordering operations thus far require prior stores to be
	ordered against subsequent loads, which means that these operations
	need not interfere with store buffers, whose main purpose in life
	is in fact to reorder prior stores against subsequent loads.
	The lightweight nature of these operations is precisely due to
	their policy of store-buffer non-interference.
	However, as noted earlier, it is sometimes necessary to interfere
	with the store buffer in order to prevent prior stores from being
	reordered against later stores, which brings us to the remaining
	rows in this table.

	The \co{smp_mb()} row corresponds to the \IXh{full}{memory barrier}
	available on most platforms, with Itanium being the exception
	that proves the rule.
	However, even on Itanium, \co{smp_mb()} provides full ordering
	with respect to \co{READ_ONCE()} and \co{WRITE_ONCE()},
	as discussed in \cref{sec:memorder:Itanium}.

	The ``Successful full-strength non-\co{void} RMW'' row captures
	the fact that on some platforms (such as x86) atomic RMW instructions
	provide full ordering both before and after.
	The Linux kernel therefore requires that full-strength non-\co{void}
	atomic RMW operations provide full ordering in cases where these
	operations succeed.
	(Full-strength atomic RMW operation's names do not end in
	\co{_relaxed}, \co{_acquire}, or \co{_release}.)
	As noted earlier, the case where these operations do not succeed
	is covered by the ``\co{_relaxed()} RMW operation'' row.

	However, the Linux kernel does not require that either \co{void}
	or \co{_relaxed()} atomic RMW operations provide any ordering
	whatsoever, with the canonical example being \co{atomic_inc()}.
	Therefore, these operations, along with failing non-\co{void}
	atomic RMW operations may be preceded by \co{smp_mb__before_atomic()}
	and followed by \co{smp_mb__after_atomic()} to provide full
	ordering for any accesses preceding or following both.
	No ordering need be provided for accesses between the
	\co{smp_mb__before_atomic()} (or, similarly, the
	\co{smp_mb__after_atomic()}) and the atomic RMW operation, as
	indicated by the ``a'' entries on the \co{smp_mb__before_atomic()}
	and \co{smp_mb__after_atomic()} rows of the table.

	In short, the structure of this table is dictated by the
	properties of the underlying hardware, which are constrained by
	nothing other than the laws of physics, which were covered back in
	\cref{chp:Hardware and its Habits}.
	That is, the table is not random, although it is quite possible
	that you are confused.
\QuickE{}
\QuickQ{}
	Why is
	\cref{tab:memorder:Linux-Kernel Memory-Ordering Cheat Sheet}
	missing \co{smp_mb__after_unlock_lock()} and
	\co{smp_mb__after_spinlock()}?
\QuickA{}
	These two primitives are rather specialized, and at present
	seem difficult to fit into
	\cref{tab:memorder:Linux-Kernel Memory-Ordering Cheat Sheet}.
	The \co{smp_mb__after_unlock_lock()} primitive is intended to be placed
	immediately after a lock acquisition, and ensures that all CPUs
	see all accesses in prior critical sections as happening before
	all accesses following the \co{smp_mb__after_unlock_lock()}
	and also before all accesses in later critical sections.
	Here ``all CPUs'' includes those CPUs not holding that lock,
	and ``prior critical sections'' includes all prior critical sections
	for the lock in question as well as all prior critical sections
	for all other locks that were released by the same CPU that executed
	the  \co{smp_mb__after_unlock_lock()}.

	The \co{smp_mb__after_spinlock()} provides the same guarantees
	as does \co{smp_mb__after_unlock_lock()}, but also provides
	additional visibility guarantees for other accesses performed
	by the CPU that executed the \co{smp_mb__after_spinlock()}.
	Given any store S performed prior to any earlier lock acquisition
	and any load L performed after the \co{smp_mb__after_spinlock()},
	all CPUs will see S as happening before~L\@.
	In other words, if a CPU performs a store S, acquires a lock,
	executes an \co{smp_mb__after_spinlock()}, then performs a
	load L, all CPUs will see S as happening before~L\@.
\QuickE{}
\QuickQ{}
	But how can I know that a given project can be designed
	and coded within the confines of these rules of thumb?
\QuickA{}
	Much of the purpose of the remainder of this chapter is
	to answer exactly that question!
\QuickE{}
\QuickQ{}
	How can you tell which memory barriers are strong enough for
	a given use case?
\QuickA{}
	Ah, that is a deep question whose answer requires most of the
	rest of this chapter.
	But the short answer is that \co{smp_mb()} is almost always
	strong enough, albeit at some cost.
\QuickE{}
\QuickQ{}
	Wait!!!
	Where do I find this tooling that automatically analyzes
	litmus tests???
\QuickA{}
	Get version v4.17 (or later) of the Linux-kernel source code,
	then follow the instructions in \path{tools/memory-model/README}
	to install the needed tools.
	Then follow the further instructions to run these tools on the
	litmus test of your choice.
\QuickE{}
\QuickQ{}
	What assumption is the code fragment
	in \cref{lst:memorder:Software Logic Analyzer}
	making that might not be valid on real hardware?
\QuickA{}
	The code assumes that as soon as a given CPU stops
	seeing its own value, it will immediately see the
	final agreed-upon value.
	On real hardware, some of the CPUs might well see several
	intermediate results before converging on the final value.
	The actual code used to produce the data in the figures
	discussed later in this section was therefore somewhat more
	complex.
\QuickE{}
\QuickQ{}
	How could CPUs possibly have different views of the
	value of a single variable \emph{at the same time?}
\QuickA{}
	As discussed in
	\cref{sec:memorder:Why Hardware Misordering?},
	many CPUs have store buffers that record the values of
	recent stores, which do not become globally visible until
	the corresponding cache line makes its way to the CPU\@.
	Therefore, it is quite possible for each CPU to see its own value
	for a given variable (in its own store buffer) at a single point
	in time---and for main memory to hold yet another value.
	One of the reasons that memory barriers were invented was
	to allow software to deal gracefully with situations like
	this one.

	Fortunately, software rarely cares about the fact that multiple
	CPUs might see multiple values for the same variable.
\QuickE{}
\QuickQ{}
	Why do CPUs~2 and~3 come to agreement so quickly, when it
	takes so long for CPUs~1 and~4 to come to the party?
\QuickA{}
	CPUs~2 and~3 are a pair of hardware threads on the same
	core, sharing the same cache hierarchy, and therefore have
	very low communications latencies.
	This is a \IXacr{numa}, or, more accurately, a \IXacr{nuca} effect.

	This leads to the question of why CPUs~2 and~3 ever disagree
	at all.
	One possible reason is that they each might have a small amount
	of private cache in addition to a larger shared cache.
	Another possible reason is instruction reordering, given the
	short 10-nanosecond duration of the disagreement and the
	total lack of memory-ordering operations in the code fragment.
\QuickE{}
\QuickQ{}
	But why make load-load reordering visible to the user?
	Why not just use speculative execution to allow execution to
	proceed in the common case where there are no intervening
	stores, in which case the reordering cannot be visible anyway?
\QuickA{}
	They can and many do, otherwise systems containing
	strongly ordered CPUs would be slow indeed.
	However, speculative execution does have its downsides, especially
	if speculation must be rolled back frequently, particularly
	on battery-powered systems.
	Speculative execution can also introduce side channels, which
	might in turn be exploited to exfiltrate information.
	But perhaps future systems will be able to overcome these
	disadvantages.
	Until then, we can expect vendors to continue producing
	weakly ordered CPUs.
\QuickE{}
\QuickQ{}
	Why should strongly ordered systems pay the performance price
	of unnecessary \co{smp_rmb()} and \co{smp_wmb()} invocations?
	Shouldn't weakly ordered systems shoulder the full cost of
	their misordering choices???
\QuickA{}
	That is in fact exactly what happens.
	On strongly ordered systems, \co{smp_rmb()} and \co{smp_wmb()}
	emit no instructions, but instead just constrain the compiler.
	Thus, in this case, weakly ordered systems do in fact shoulder
	the full cost of their memory-ordering choices.
\QuickE{}
\QuickQ{}
	But how do we know that \emph{all} platforms really avoid
	triggering the \co{exists} clauses in
	\cref{lst:memorder:Enforced Ordering of Message-Passing Address-Dependency Litmus Test (Before v4.15),%
	lst:memorder:S Address-Dependency Litmus Test}?
\QuickA{}
	Answering this requires identifying three major groups of platforms:
	\begin{enumerate*}[(1)]
	\item Total-store-order (TSO) platforms,
	\item Weakly ordered platforms, and
	\item DEC Alpha.
	\end{enumerate*}

	\begin{fcvref}[ln:memorder:Enforced Ordering of Message-Passing Address-Dependency Litmus Test (Before v4.15)]
	The TSO platforms order all pairs of memory references except for
	prior stores against later loads.
	Because the address dependency on \clnref{deref,read} of
	\cref{lst:memorder:Enforced Ordering of Message-Passing Address-Dependency Litmus Test (Before v4.15)}
	is instead a load followed by another load, TSO platforms preserve
	this address dependency.
	\end{fcvref}
	\begin{fcvref}[ln:formal:C-S+o-wmb-o+o-addr-o:whole]
	They also preserve the address dependency on \clnref{P1:x1,P1:r2} of
	\cref{lst:memorder:S Address-Dependency Litmus Test}
	because this is a load followed by a store.
	Because address dependencies must start with a load, TSO platforms
	implicitly but completely respect them, give or take compiler
	optimizations, hence the need for \co{READ_ONCE()}.
	\end{fcvref}

	Weakly ordered platforms don't necessarily maintain ordering of
	unrelated accesses.
	However, the address dependencies in
	\cref{lst:memorder:Enforced Ordering of Message-Passing Address-Dependency Litmus Test (Before v4.15),%
	lst:memorder:S Address-Dependency Litmus Test} are not unrelated:
	There is an address dependency.
	The hardware tracks dependencies and maintains the needed
	ordering.

	\begin{fcvref}[ln:memorder:Enforced Ordering of Message-Passing Address-Dependency Litmus Test (Before v4.15)]
	There is one (famous) exception to this rule for weakly ordered
	platforms, and that exception is DEC Alpha for load-to-load
	address dependencies.
	And this is why, in Linux kernels predating v4.15, DEC Alpha
	requires the explicit memory barrier supplied for it by the
	now-obsolete \apikh{lockless_dereference()} on \clnref{deref} of
	\cref{lst:memorder:Enforced Ordering of Message-Passing Address-Dependency Litmus Test (Before v4.15)}.
	\end{fcvref}
	\begin{fcvref}[ln:formal:C-S+o-wmb-o+o-addr-o:whole]
	However, DEC Alpha does track load-to-store address dependencies,
	which is why \clnref{P1:x1} of
	\cref{lst:memorder:S Address-Dependency Litmus Test}
	does not need a \co{lockless_dereference()}, even in Linux
	kernels predating v4.15.
	\end{fcvref}

	To sum up, current platforms either respect address dependencies
	implicitly, as is the case for TSO platforms (x86, mainframe,
	SPARC,~\dots), have hardware tracking for address dependencies
	(\ARM, PowerPC, MIPS,~\dots), have the required memory barriers
	supplied by \co{READ_ONCE()} (DEC Alpha in Linux kernel v4.15 and
	later), or supplied by
	\co{rcu_dereference()} (DEC Alpha in Linux kernel v4.14 and earlier).
\QuickE{}
\QuickQ{}
	Why the use of \co{smp_wmb()} in
	\cref{lst:memorder:Enforced Ordering of Message-Passing Address-Dependency Litmus Test (Before v4.15),lst:memorder:S Address-Dependency Litmus Test}?
	Wouldn't \co{smp_store_release()} be a better choice?
\QuickA{}
	In most cases, \co{smp_store_release()} is indeed a better choice.
	However, \co{smp_wmb()} was there first in the Linux kernel,
	so it is still good to understand how to use it.
\QuickE{}
\QuickQ{}
	SP, MP, LB, and now~S\@.
	Where do all these litmus-test abbreviations come from and
	how can anyone keep track of them?
\QuickA{}
	The best scorecard is the infamous
	\co{test6.pdf}~\cite{test6-pdf}.
	Unfortunately, not all of the abbreviations have catchy
	expansions like SB (store buffering), MP (message passing),
	and LB (load buffering), but at least the list of abbreviations
	is readily available.
\QuickE{}
\QuickQ{}
	\begin{fcvref}[ln:formal:C-LB+o-r+o-data-o:whole]
	But wait!!!
	\Clnref{ld} of
	\cref{lst:memorder:Load-Buffering Data-Dependency Litmus Test}
	uses \co{READ_ONCE()}, which marks the load as volatile,
	which means that the compiler absolutely must emit the load
	instruction even if the value is later multiplied by zero.
	So how can the compiler possibly break this data dependency?
	\end{fcvref}
\QuickA{}
	Yes, the compiler absolutely must emit a load instruction for
	a volatile load.
	But if you multiply the value loaded by zero, the compiler is
	well within its rights to substitute a constant zero for the
	result of that multiplication, which will break the data
	dependency on many platforms.

	Worse yet, if the dependent store does not use \co{WRITE_ONCE()},
	the compiler could hoist it above the load, which would cause
	even TSO platforms to fail to provide ordering.
\QuickE{}
\QuickQ{}
	Wouldn't control dependencies be more robust if they were
	mandated by language standards???
\QuickA{}
	But of course!
	And perhaps in the fullness of time they will be so mandated.
\QuickE{}
\QuickQ{}
	But in
	\cref{lst:memorder:Cache-Coherent IRIW Litmus Test},
	wouldn't be just as bad if \co{P2()}'s \co{r1} and \co{r2}
	obtained the values 2 and 1, respectively, while \co{P3()}'s
	\co{r3} and \co{r4} obtained the values 1 and 2, respectively?
\QuickA{}
	Yes, it would.
	Feel free to modify the \co{exists} clause to
	check for that outcome and see what happens.
\QuickE{}
\QuickQ{}
	Can you give a specific example showing different behavior for
	multicopy atomic on the one hand and other-multicopy atomic
	on the other?
\QuickA{}
	\Cref{lst:memorder:Litmus Test Distinguishing Multicopy Atomic From Other Multicopy Atomic}
	(\path{C-MP-OMCA+o-o-o+o-rmb-o.litmus})
	shows such a test.

\begin{listing}
\begin{fcvlabel}[ln:formal:C-MP-OMCA+o-o-o+o-rmb-o:whole]
\begin{VerbatimL}[commandchars=\@\[\]]
C C-MP-OMCA+o-o-o+o-rmb-o

{}

P0(int *x, int *y)
{
	int r0;

	WRITE_ONCE(*x, 1);@lnlbl[P0:st]
	r0 = READ_ONCE(*x);@lnlbl[P0:ld]
	WRITE_ONCE(*y, r0);@lnlbl[P0:y]
}

P1(int *x, int *y)
{
	int r1;
	int r2;

	r1 = READ_ONCE(*y);@lnlbl[P1:y]
	smp_rmb();@lnlbl[P1:rmb]
	r2 = READ_ONCE(*x);@lnlbl[P1:x]
}

exists (1:r1=1 /\ 1:r2=0)@lnlbl[exists]
\end{VerbatimL}
\end{fcvlabel}
 \caption{Litmus Test Distinguishing Multicopy Atomic From Other Multicopy Atomic}
\label{lst:memorder:Litmus Test Distinguishing Multicopy Atomic From Other Multicopy Atomic}
\end{listing}

	\begin{fcvref}[ln:formal:C-MP-OMCA+o-o-o+o-rmb-o:whole]
	On a multicopy-atomic platform, \co{P0()}'s store to \co{x} on
	\clnref{P0:st} must become visible to both \co{P0()} and \co{P1()}
	simultaneously.
	Because this store becomes visible to \co{P0()} on \clnref{P0:ld}, before
	\co{P0()}'s store to \co{y} on \clnref{P0:y}, \co{P0()}'s store to
	\co{x} must become visible before its store to \co{y} everywhere,
	including \co{P1()}.
	Therefore, if \co{P1()}'s load from \co{y} on \clnref{P1:y} returns the
	value 1, so must its load from \co{x} on \clnref{P1:x}, given that
	the \co{smp_rmb()} on \clnref{P1:rmb} forces these two loads to execute
	in order.
	Therefore, the \co{exists} clause on \clnref{exists} cannot trigger on a
	multicopy-atomic platform.
	\end{fcvref}

	In contrast, on an other-multicopy-atomic platform, \co{P0()}
	could see its own store early, so that there would be no constraint
	on the order of visibility of the two stores from \co{P1()},
	which in turn allows the \co{exists} clause to trigger.
\QuickE{}
\QuickQ{}
	Then who would even \emph{think} of designing a system with shared
	store buffers???
\QuickA{}
	This is in fact a very natural design for any system having
	multiple hardware threads per core.
	Natural from a hardware point of view, that is!
\QuickE{}
\QuickQ{}
	But just how is it fair that \co{P0()} and \co{P1()} must share a store
	buffer and a cache, but \co{P2()} gets one each of its very own???
\QuickA{}
	Presumably there is a \co{P3()}, as is in fact shown in
	\cref{fig:memorder:Shared Store Buffers And Multi-Copy Atomicity},
	that shares \co{P2()}'s store buffer and cache.
	But not necessarily.
	Some platforms allow different cores to disable different numbers
	of threads, allowing the hardware to adjust to the needs of the
	workload at hand.
	For example, a single-threaded critical-path portion of the workload
	might be assigned to a core with only one thread enabled, thus
	allowing the single thread running that portion of the workload
	to use the entire capabilities of that core.
	Other more highly parallel but cache-miss-prone portions of the
	workload might be assigned to cores with all hardware threads
	enabled to provide improved throughput.
	This improved throughput could be due to the fact that while one
	hardware thread is stalled on a cache miss, the other hardware
	threads can make forward progress.

	In such cases, performance requirements override quaint human
	notions of fairness.
\QuickE{}
\QuickQ{}
	\begin{fcvref}[ln:formal:C-WRC+o+o-data-o+o-rmb-o:whole]
	Referring to
	\cref{tab:memorder:Memory Ordering: WRC Sequence of Events},
	why on earth would \co{P0()}'s store take so long to complete when
	\co{P1()}'s store complete so quickly?
	In other words, does the \co{exists} clause on \clnref{exists} of
	\cref{lst:memorder:WRC Litmus Test With Dependencies (No Ordering)}
	really trigger on real systems?
	\end{fcvref}
\QuickA{}
	You need to face the fact that it really can trigger.
	Akira Yokosawa used the \co{litmus7} tool to run this litmus test
	on a \Power{8} system.
	Out of 1,000,000,000 runs, 4 triggered the \co{exists} clause.
	Thus, triggering the \co{exists} clause is not merely a one-in-a-million
	occurrence, but rather a one-in-a-hundred-million occurrence.
	But it nevertheless really does trigger on real systems.
\QuickE{}
\QuickQ{}
	But it is not necessary to worry about propagation unless
	there are at least three threads in the litmus test, right?
\QuickA{}
	Wrong.

\begin{listing}
\begin{fcvlabel}[ln:formal:C-R+o-wmb-o+o-mb-o:whole]
\begin{VerbatimL}[commandchars=\@\[\]]
C C-R+o-wmb-o+o-mb-o

{}

P0(int *x0, int *x1)
{
	WRITE_ONCE(*x0, 1);
	smp_wmb();@lnlbl[wmb]
	WRITE_ONCE(*x1, 1);
}

P1(int *x0, int *x1)
{
	int r2;

	WRITE_ONCE(*x1, 2);
	smp_mb();
	r2 = READ_ONCE(*x0);
}

exists (1:r2=0 /\ x1=2)@lnlbl[exists]
\end{VerbatimL}
\end{fcvlabel}
 \caption{R Litmus Test With Write Memory Barrier (No Ordering)}
\label{lst:memorder:R Litmus Test With Write Memory Barrier (No Ordering)}
\end{listing}

	\begin{fcvref}[ln:formal:C-R+o-wmb-o+o-mb-o:whole]
	\Cref{lst:memorder:R Litmus Test With Write Memory Barrier (No Ordering)}
	(\path{C-R+o-wmb-o+o-mb-o.litmus})
	shows a two-thread litmus test that requires propagation due to
	the fact that it only has store-to-store and load-to-store
	links between its pair of threads.
	Even though \co{P0()} is fully ordered by the \co{smp_wmb()} and
	\co{P1()} is fully ordered by the \co{smp_mb()}, the
	counter-temporal nature of the links means that
	the \co{exists} clause on \clnref{exists} really can trigger.
	To prevent this triggering, the \co{smp_wmb()} on \clnref{wmb}
	must become an \co{smp_mb()}, bringing propagation into play
	twice, once for each non-temporal link.
	\end{fcvref}
\QuickE{}
\QuickQ{}
	\begin{fcvref}[ln:formal:C-W+RWC+o-r+a-o+o-mb-o:whole]
	But given that \co{smp_mb()} has the propagation property,
	why doesn't the \co{smp_mb()} on \clnref{P2:mb} of
	\cref{lst:memorder:W+RWC Litmus Test With Release (No Ordering)}
	prevent the \co{exists} clause from triggering?
	\end{fcvref}
\QuickA{}
	\begin{fcvref}[ln:formal:C-W+RWC+o-r+a-o+o-mb-o:whole]
	As a rough rule of thumb, the \co{smp_mb()} barrier's
	propagation property is sufficient to maintain ordering
	through only one load-to-store link between
	processes.
	Unfortunately,
	\cref{lst:memorder:W+RWC Litmus Test With Release (No Ordering)}
	has not one but two load-to-store links, with the
	first being from the \co{READ_ONCE()} on \clnref{P1:ld} to the
	\co{WRITE_ONCE()} on \clnref{P2:st} and the second being from
	the \co{READ_ONCE()} on \clnref{P2:ld} to the \co{WRITE_ONCE()}
	on \clnref{P0:st}.
	Therefore, preventing the \co{exists} clause from triggering
	should be expected to require not one but two
	instances of \co{smp_mb()}.
	\end{fcvref}

	As a special exception to this rule of thumb, a release-acquire
	chain can have one load-to-store link between processes
	and still prohibit the cycle.
\QuickE{}
\QuickQ{}
	But for litmus tests having only ordered stores, as shown in
	\cref{lst:memorder:2+2W Litmus Test With Write Barriers}
	(\path{C-2+2W+o-wmb-o+o-wmb-o.litmus}),
	research shows that the cycle is prohibited, even in weakly
	ordered systems such as \ARM\ and Power~\cite{test6-pdf}.
	Given that, are store-to-store really \emph{always}
	counter-temporal???
\QuickA{}
	This litmus test is indeed a very interesting curiosity.
	Its ordering apparently occurs naturally given typical
	weakly ordered hardware design, which would normally be
	considered a great gift from the relevant laws of physics
	and cache-coherency-protocol mathematics.

	\begin{fcvref}[ln:formal:C-2+2W+o-wmb-o+o-wmb-o:whole]
	Unfortunately, no one has been able to come up with a software use
	case for this gift that does not have a much better alternative
	implementation.
	Therefore, neither the C11 nor the Linux kernel memory models
	provide any guarantee corresponding to
	\cref{lst:memorder:2+2W Litmus Test With Write Barriers}.
	This means that the \co{exists} clause on \clnref{exists} can
	trigger.
	\end{fcvref}

\begin{listing}
\begin{fcvlabel}[ln:formal:C-2+2W+o-o+o-o:whole]
\begin{VerbatimL}[commandchars=\@\[\]]
C C-2+2W+o-o+o-o

{}

P0(int *x0, int *x1)
{
	WRITE_ONCE(*x0, 1);
	WRITE_ONCE(*x1, 2);
}

P1(int *x0, int *x1)
{
	WRITE_ONCE(*x1, 1);
	WRITE_ONCE(*x0, 2);
}

exists (x0=1 /\ x1=1)
\end{VerbatimL}
\end{fcvlabel}
 \caption{2+2W Litmus Test (No Ordering)}
\label{lst:memorder:2+2W Litmus Test (No Ordering)}
\end{listing}

	Of course, without the barrier, there are no ordering
	guarantees, even on real weakly ordered hardware, as shown in
	\cref{lst:memorder:2+2W Litmus Test (No Ordering)}
	(\path{C-2+2W+o-o+o-o.litmus}).
\QuickE{}
\QuickQ{}
	Can you construct a litmus test like that in
	\cref{lst:memorder:LB Litmus Test With One Acquire}
	that uses \emph{only} dependencies?
\QuickA{}
	\Cref{lst:memorder:LB Litmus Test With No Acquires}
	shows a somewhat nonsensical but very real example.
	Creating a more useful (but still real) litmus test is left
	as an exercise for the reader.
\QuickE{}

\begin{listing}
\begin{fcvlabel}[ln:formal:C-LB+o-data-o+o-data-o+o-data-o:whole]
\begin{VerbatimL}[commandchars=\@\[\]]
C C-LB+o-data-o+o-data-o+o-data-o

{
x1=1;
x2=2;
}

P0(int *x0, int *x1)
{
	int r2;

	r2 = READ_ONCE(*x0);
	WRITE_ONCE(*x1, r2);
}

P1(int *x1, int *x2)
{
	int r2;

	r2 = READ_ONCE(*x1);
	WRITE_ONCE(*x2, r2);
}

P2(int *x2, int *x0)
{
	int r2;

	r2 = READ_ONCE(*x2);
	WRITE_ONCE(*x0, r2);
}

exists (0:r2=2 /\ 1:r2=0 /\ 2:r2=1)
\end{VerbatimL}
\end{fcvlabel}
 \caption{LB Litmus Test With No Acquires}
\label{lst:memorder:LB Litmus Test With No Acquires}
\end{listing}
\QuickQ{}
	Suppose we have a short release-acquire chain along with one
	load-to-store link and one store-to-store link, like that shown in
	\cref{lst:memorder:Z6.2 Release-Acquire Chain (Ordering?)}.
	Given that there is only one of each type of non-store-to-load
	link, the \co{exists} cannot trigger, right?
\QuickA{}
	Wrong.
	It is the number of non-store-to-load links that matters.
	If there is only one non-store-to-load link, a release-acquire
	chain can prevent the \co{exists} clause from triggering.
	However, if there is more than one non-store-to-load link,
	be they store-to-store, load-to-store, or any combination
	thereof, it is necessary to have at least one full barrier
	(\co{smp_mb()} or better) between each non-store-to-load link.
	In
	\cref{lst:memorder:Z6.2 Release-Acquire Chain (Ordering?)},
	preventing the \co{exists} clause from triggering therefore requires
	an additional full barrier between either \co{P0()}'s or
	\co{P1()}'s accesses.
\QuickE{}
\QuickQ{}
	There are store-to-load links, load-to-store links, and
	store-to-store links.
	But what about load-to-load links?
\QuickA{}
	The problem with the concept of load-to-load links is that
	if the two loads from the same variable return the same
	value, there is no way to determine their ordering.
	The only way to determine their ordering is if they return
	different values, in which case there had to have been an
	intervening store.
	And that intervening store means that there is no load-to-load
	link, but rather a load-to-store link followed by a
	store-to-load link.
\QuickE{}
\QuickQ{}
	What happens if that \co{lwsync} instruction is instead a
	\co{sync} instruction?
\QuickA{}
	The counter-intuitive outcome cannot happen.
	(Try it!)
\QuickE{}
\QuickQ{}
	Why not place a \co{barrier()} call immediately before
	a plain store to prevent the compiler from inventing stores?
\QuickA{}
	Because it would not work.
	Although the compiler would be prevented from inventing a
	store prior to the \co{barrier()}, nothing would prevent
	it from inventing a store between that \co{barrier()} and
	the plain store.
\QuickE{}
\QuickQ{}
	\begin{fcvref}[ln:memorder:Breakable Dependencies With Comparisons]
	Why can't you simply dereference the pointer before comparing it
	to \co{&reserve_int} on \clnref{cmp} of
	\cref{lst:memorder:Breakable Dependencies With Comparisons}?
	\end{fcvref}
\QuickA{}
	For first, it might be necessary to invoke
	\co{handle_reserve()} before \co{do_something_with()}.

	But more relevant to memory ordering, the compiler is often within
	its rights to hoist the comparison ahead of the dereferences,
	which would allow the compiler to use \co{&reserve_int} instead
	of the variable \co{p} that the hardware has tagged with
	a dependency.
\QuickE{}
\QuickQ{}
	But it should be safe to compare two pointer variables, right?
	After all, the compiler doesn't know the value
	of either, so how can it possibly learn anything from the
	comparison?
\QuickA{}
\begin{listing}
\begin{fcvlabel}[ln:memorder:Breakable Dependencies With Non-Constant Comparisons]
\begin{VerbatimL}
int *gp1;
int *p;
int *q;

p = rcu_dereference(gp1);
q = get_a_pointer();
if (p == q)
	handle_equality(p);
do_something_with(*p);
\end{VerbatimL}
\end{fcvlabel}
\caption{Breakable Dependencies With Non-Constant Comparisons}
\label{lst:memorder:Breakable Dependencies With Non-Constant Comparisons}
\end{listing}%
\begin{listing}
\begin{fcvlabel}[ln:memorder:Broken Dependencies With Non-Constant Comparisons]
\begin{VerbatimL}[commandchars=\\\[\]]
int *gp1;
int *p;
int *q;

p = rcu_dereference(gp1);		\lnlbl[p]
q = get_a_pointer();
if (p == q) {
	handle_equality(q);
	do_something_with(*q);		\lnlbl[q]
} else {
	do_something_with(*p);
}
\end{VerbatimL}
\end{fcvlabel}
\caption{Broken Dependencies With Non-Constant Comparisons}
\label{lst:memorder:Broken Dependencies With Non-Constant Comparisons}
\end{listing}%
	Unfortunately, the compiler really can learn enough to
	break your dependency chain, for example, as shown in
	\cref{lst:memorder:Breakable Dependencies With Non-Constant Comparisons}.
	The compiler is within its rights to transform this code
	into that shown in
	\cref{lst:memorder:Broken Dependencies With Non-Constant Comparisons},
	and might well make this transformation due to register pressure
	if \co{handle_equality()} was inlined and needed a lot of registers.
	\begin{fcvref}[ln:memorder:Broken Dependencies With Non-Constant Comparisons]
	\Clnref{q} of this transformed code uses \co{q}, which although
	equal to \co{p}, is not necessarily tagged by the hardware as
	carrying a dependency.
	Therefore, this transformed code does not necessarily guarantee
	that \clnref{q} is ordered after \clnref{p}.\footnote{
		Kudos to \ppl{Linus}{Torvalds} for providing this example.}
	\end{fcvref}
\QuickE{}
\QuickQ{}
	\begin{fcvref}[ln:memorder:Broken Dependencies With Pointer Comparisons:read]
	But doesn't the condition in \clnref{equ} supply a control dependency
	that would keep \clnref{pc} ordered after \clnref{gp1}?
	\end{fcvref}
\QuickA{}
	\begin{fcvref}[ln:memorder:Broken Dependencies With Pointer Comparisons:read]
	Yes, but no.
	Yes, there is a control dependency, but control dependencies do
	not order later loads, only later stores.
	If you really need ordering, you could place an \co{smp_rmb()}
	between \clnref{equ,pc}.
	Or better yet, have \co{updater()}
	allocate two structures instead of reusing the structure.
	For more information, see
	\cref{sec:memorder:Control-Dependency Calamities}.
	\end{fcvref}
\QuickE{}
\QuickQ{}
	But there is a \co{READ_ONCE()}, so how can the compiler
	prove anything about the value of \co{q}?
\QuickA{}
	Given the simple \co{if} statement comparing against zero,
	it is hard to imagine the compiler proving anything.
	But suppose that later code executed a division by \co{q}.
	Because division by zero is undefined behavior, as of 2023,
	many compilers will assume that the value of \co{q} must
	be non-zero, and will thus remove that \co{if} statement,
	thus unconditionally executing the \co{WRITE_ONCE()}, in
	turn destroying the control dependency.

	There are some who argue (correctly, in Paul's view) that
	back-propagating undefined behavior across volatile accesses
	constitutes a compiler bug, but many compiler writers insist
	that this is not a bug, but rather a valuable optimization.
\QuickE{}
\QuickQ{}
	Can't you instead add an \co{smp_mb()} to \co{P1()} in
	\cref{lst:memorder:WWC Litmus Test With Control Dependency (Cumulativity?)}?
\QuickA{}
	Not given the Linux kernel memory model.
	(Try it!)
	However, you can instead replace \co{P0()}'s
	\co{WRITE_ONCE()} with \co{smp_store_release()},
	which usually has less overhead than does adding an \co{smp_mb()}.
\QuickE{}
\QuickQ{}
	But doesn't PowerPC have weak unlock-lock ordering properties
	within the Linux kernel, allowing a write before the unlock to
	be reordered with a read after the lock?
\QuickA{}
	Yes, but only from the perspective of a third thread not holding
	that lock.
	In contrast, memory allocators need only concern themselves with
	the two threads migrating the memory.
	It is after all the developer's responsibility to properly
	synchronize with any other threads that need access to the newly
	migrated block of memory.
\QuickE{}
\QuickQ{}
	But if there are three critical sections, isn't it true that
	CPUs not holding the lock will observe the accesses from the
	first and the third critical section as being ordered?
\QuickA{}
	No.

\begin{listing}
\begin{fcvlabel}[ln:formal:C-Lock-across-unlock-lock-3:whole]
\begin{VerbatimL}[commandchars=\%\@\$]
C Lock-across-unlock-lock-3

{}

P0(int *x, spinlock_t *sp)
{
	spin_lock(sp);
	WRITE_ONCE(*x, 1);
	spin_unlock(sp);
}

P1(int *x, int *y, int *z, spinlock_t *sp)
{
	int r1;

	spin_lock(sp);
	r1 = READ_ONCE(*x);
	WRITE_ONCE(*z, 1);
	spin_unlock(sp);
}

P2(int *x, int *y, int *z, spinlock_t *sp)
{
	int r1;
	int r2;

	spin_lock(sp);
	r1 = READ_ONCE(*z);
	r2 = READ_ONCE(*y);
	spin_unlock(sp);
}

P3(int *x, int *y, spinlock_t *sp)
{
	int r1;

	WRITE_ONCE(*y, 1);
	smp_mb();
	r1 = READ_ONCE(*x);
}

exists (1:r1=1 /\ 2:r1=1 /\ 2:r2=0 /\ 3:r1=0)
\end{VerbatimL}
\end{fcvlabel}
 \caption{Accesses Between Multiple Different-CPU Critical Sections}
\label{lst:memorder:Accesses Between Multiple Different-CPU Critical Sections}
\end{listing}

	\Cref{lst:memorder:Accesses Between Multiple Different-CPU Critical Sections}
	shows an example three-critical-section chain
	(\path{Lock-across-unlock-lock-3.litmus}).
	Running this litmus test shows that the \co{exists} clause can
	still be satisfied, so this additional critical section is still
	not sufficient to force ordering.

	However, as the reader can verify, placing an
	\co{smp_mb__after_spinlock()} after either \co{P1()}'s or
	\co{P2()}'s lock acquisition does suffice to force ordering.
\QuickE{}
\QuickQ{}
	But if \co{spin_is_locked()} returns \co{false}, don't we also
	know that no other CPU or thread is holding the corresponding
	lock?
\QuickA{}
	No.
	By the time that the code inspects the return value from
	\co{spin_is_locked()}, some other CPU or thread might well have
	acquired the corresponding lock.
\QuickE{}
\QuickQ{}
	Wait a minute!
	In QSBR implementations of RCU, no code is emitted for
	\co{rcu_read_lock()} and \co{rcu_read_unlock()}.
	This means that the RCU read-side critical section in
	\cref{lst:memorder:What Happens With Empty RCU Readers?}
	isn't just empty, it is completely nonexistent!!!
	So how can something that doesn't exist at all possibly have
	any effect whatsoever on ordering???
\QuickA{}
	Because in QSBR, RCU read-side critical sections don't
	actually disappear.
	Instead, they are extended in both directions until a quiescent
	state is encountered.
	For example, in the Linux kernel, the critical section might
	be extended back to the most recent \co{schedule()} call and
	ahead to the next \co{schedule()} call.
	Of course, in non-QSBR implementations, \co{rcu_read_lock()}
	and \co{rcu_read_unlock()} really do emit code, which can clearly
	provide ordering.
	And within the Linux kernel, even the QSBR implementation
	has a compiler \co{barrier()} in \co{rcu_read_lock()} and
	\co{rcu_read_unlock()}, which is necessary to prevent
	the compiler from moving memory accesses that might result
	in page faults into the RCU read-side critical section.

	Therefore, strange though it might seem, empty RCU read-side
	critical sections really can and do provide some degree of
	ordering.
\QuickE{}
\QuickQ{}
	Can \co{P1()}'s accesses be reordered in the litmus tests shown in
	\cref{lst:memorder:What Happens Before RCU Readers?,%
	lst:memorder:What Happens After RCU Readers?,%
	lst:memorder:What Happens With Empty RCU Readers?}
	in the same way that they were reordered going from
	\cref{lst:memorder:RCU Fundamental Property}
	to
	\cref{lst:memorder:RCU Fundamental Property and Reordering}?
\QuickA{}
	No, because none of these later litmus tests have more than one
	access within their RCU read-side critical sections.
	But what about swapping the accesses, for example, in
	\cref{lst:memorder:What Happens Before RCU Readers?},
	placing \co{P1()}'s \co{WRITE_ONCE()} within its critical
	section and the \co{READ_ONCE()} before its critical section?

	Swapping the accesses allows both instances of \co{r2} to
	have a final value of zero, in other words, although RCU read-side
	critical sections' ordering properties can extend outside of
	those critical sections, the same is not true of their
	reordering properties.
	Checking this with \co{herd} and explaining why is left as an
	exercise for the reader.
\QuickE{}
\QuickQ{}
	What would happen if the \co{smp_mb()} was instead added between
	\co{P2()}'s accesses in
	\cref{lst:memorder:One RCU Grace Period and Two Readers}?
\QuickA{}
	The cycle would again be forbidden.
	Further analysis is left as an exercise for the reader.
\QuickE{}
\QuickQ{}
	What happens to code between an atomic operation and an
	\co{smp_mb__after_atomic()}?
\QuickA{}
	First, please don't do this!

	But if you do, this intervening code will either be ordered
	after the atomic operation or before the
	\co{smp_mb__after_atomic()}, depending on the architecture,
	but not both.
	This also applies to \co{smp_mb__before_atomic()} and
	\co{smp_mb__after_spinlock()}, that is, both the uncertain
	ordering of the intervening code and the plea to avoid such code.
\QuickE{}
\QuickQ{}
	Why does Alpha's \co{READ_ONCE()} include an
	\co{mb()} rather than \co{rmb()}?
\QuickA{}
	Alpha has only \co{mb} and \co{wmb} instructions,
	so \co{smp_rmb()} would be implemented by the Alpha \co{mb}
	instruction in either case.
	In addition, at the time that the Linux kernel started relying on
	dependency ordering, it was not clear that Alpha ordered dependent
	stores, and thus \co{smp_mb()} was therefore the safe choice.

	However, given the aforementioned v5.9 changes to \co{READ_ONCE()}
	and a few of Alpha's atomic read-modify-write operations,
	no Linux-kernel core code need concern itself with DEC Alpha,
	thus greatly reducing Paul E.~McKenney's incentive to remove
	Alpha support from the kernel.
\QuickE{}
\QuickQ{}
	Isn't DEC Alpha significant as having the weakest possible
	memory ordering?
\QuickA{}
	Although DEC Alpha does take considerable flak, it does avoid
	reordering reads from the same CPU to the same variable.
	It also avoids the out-of-thin-air problem that plagues
	the Java and C11 memory
	models~\cite{Boehm:2014:OGA:2618128.2618134,conf/esop/BattyMNPS15,MarkBatty2013OOTA-WorkingNote,HansBoehm2020ConcurrentUB,DavidGoldblatt2019NoElegantOOTAfix,AlanJeffrey2014JavaDRF,PaulEMcKenney2020RelaxedGuideRelaxed,PaulEMcKenney2016OOTA,Sevcik:2011:SOS:1993316.1993534,Vafeiadis:2015:CCO:2775051.2676995}.
\QuickE{}
\QuickQ{}
	Given that hardware can have a half memory barrier, why don't
	locking primitives allow the compiler to move memory-reference
	instructions into lock-based critical sections?
\QuickA{}
	In fact, as we saw in \cref{sec:memorder:ARMv8} and will
	see in \cref{sec:memorder:POWER / PowerPC}, hardware really does
	implement partial memory-ordering instructions and it also turns
	out that these really are used to construct locking primitives.
	However, these locking primitives use full compiler barriers,
	thus preventing the compiler from reordering memory-reference
	instructions both out of and into the corresponding critical
	section.

\begin{listing}
\begin{fcvlabel}[ln:memorder:synchronize-rcu]
\begin{VerbatimL}[commandchars=\@\[\]]
static inline int rcu_gp_ongoing(unsigned long *ctr)
{
	unsigned long v;

	v = LOAD_SHARED(*ctr);@lnlbl[load]
	return v && (v != rcu_gp_ctr);
}

static void update_counter_and_wait(void)
{
	struct rcu_reader *index;

	STORE_SHARED(rcu_gp_ctr, rcu_gp_ctr + RCU_GP_CTR);
	barrier();
	list_for_each_entry(index, &registry, node) {@lnlbl[loop]
		while (rcu_gp_ongoing(&index->ctr))@lnlbl[call2]
			msleep(10);
	}
}

void synchronize_rcu(void)
{
	unsigned long was_online;

	was_online = rcu_reader.ctr;
	smp_mb();
	if (was_online)@lnlbl[if]
		STORE_SHARED(rcu_reader.ctr, 0);@lnlbl[store]
	mutex_lock(&rcu_gp_lock);@lnlbl[acqmutex]
	update_counter_and_wait();@lnlbl[call1]
	mutex_unlock(&rcu_gp_lock);
	if (was_online)
		STORE_SHARED(rcu_reader.ctr, LOAD_SHARED(rcu_gp_ctr));
	smp_mb();
}
\end{VerbatimL}
\end{fcvlabel}
\caption{Userspace RCU Code Reordering}
\label{lst:memorder:Userspace RCU Code Reordering}
\end{listing}

	To see why the compiler is forbidden from doing reordering that
	is permitted by hardware, consider the following sample code
	in \cref{lst:memorder:Userspace RCU Code Reordering}.
	This code is based on the userspace RCU update-side
	code~\cite[Supplementary Materials Figure 5]{MathieuDesnoyers2012URCU}.

\begin{fcvref}[ln:memorder:synchronize-rcu]
	Suppose that the compiler reordered \clnref{if,store} into
	the critical section starting at \clnref{acqmutex}.
	Now suppose that two updaters start executing \co{synchronize_rcu()}
	at about the same time.
	Then consider the following sequence of events:
	\begin{enumerate}
	\item	CPU~0 acquires the lock at \clnref{acqmutex}.
	\item	\Clnref{if} determines that CPU~0 was online, so it clears
		its own counter at \clnref{store}.
		(Recall that \clnref{if,store} have been reordered by the
		compiler to follow \clnref{acqmutex}).
	\item	CPU~0 invokes \co{update_counter_and_wait()} from
		\clnref{call1}.
	\item	CPU~0 invokes \co{rcu_gp_ongoing()} on itself at
		\clnref{call2}, and \clnref{load} sees that CPU~0 is
		in a quiescent state.
		Control therefore returns to \co{update_counter_and_wait()},
		and \clnref{loop} advances to CPU~1.
	\item	CPU~1 invokes \co{synchronize_rcu()}, but because CPU~0
		already holds the lock, CPU~1 blocks waiting for this
		lock to become available.
		Because the compiler reordered \clnref{if,store} to follow
		\clnref{acqmutex}, CPU~1 does not clear its own counter,
		despite having been online.
	\item	CPU~0 invokes \co{rcu_gp_ongoing()} on CPU~1 at
		\clnref{call2}, and \clnref{load} sees that CPU~1 is
		not in a quiescent state.
		The \co{while} loop at \clnref{call2} therefore never
		exits.
	\end{enumerate}

	So the compiler's reordering results in a deadlock.
	In contrast, hardware reordering is temporary, so that CPU~1
	might undertake its first attempt to acquire the mutex on
	\clnref{acqmutex} before executing \clnref{if,store}, but it
	will eventually execute \clnref{if,store}.
	Because hardware reordering only results in a short delay, it
	can be tolerated.
	On the other hand, because compiler reordering results in a
	deadlock, it must be prohibited.

	Some research efforts have used hardware transactional memory
	to allow compilers to safely reorder more aggressively, but
	the overhead of hardware transactions has thus far made
	such optimizations unattractive.
\end{fcvref}
\QuickE{}
\QuickQ{}
	Why is it necessary to use heavier-weight ordering for
	load-to-store and store-to-store links, but not for
	store-to-load links?
	What on earth makes store-to-load links so special???
\QuickA{}
	Recall that load-to-store and store-to-store links can be
	counter-temporal, as illustrated by
	\cref{fig:memorder:Load-to-Store is Counter-Temporal,%
	fig:memorder:Store-to-Store is Counter-Temporal} in
	\cref{sec:memorder:Propagation}.
	This counter-temporal nature of load-to-store and store-to-store
	links necessitates strong ordering.

	In constrast, store-to-load links are temporal, as illustrated by
	\cref{lst:memorder:Load-Buffering Data-Dependency Litmus Test,%
	lst:memorder:Load-Buffering Control-Dependency Litmus Test}.
	This temporal nature of store-to-load links permits use of
	minimal ordering.
\QuickE{}
\QuickQAC{chp:Ease of Use}{Ease of Use}{qqzeasy}
\QuickQ{}
	Can a similar algorithm be used when deleting elements?
\QuickA{}
	Yes.
	However, since each thread must hold the locks of three
	consecutive elements to delete the middle one, if there
	are $N$ threads, there must be $2N+1$ elements (rather than
	just $N+1$) in order to avoid deadlock.
\QuickE{}
\QuickQ{}
	Yetch!
	What ever possessed someone to come up with an algorithm
	that deserves to be shaved as much as this one does???
\QuickA{}
	That would be Paul.

	He was considering the \emph{Dining Philosopher's Problem}, which
	involves a rather unsanitary spaghetti dinner attended by
	five philosophers.
	Given that there are five plates and but five forks on the table, and
	given that each philosopher requires two forks at a time to eat,
	one is supposed to come up with a fork-allocation algorithm that
	avoids deadlock.
	Paul's response was ``Sheesh!
			      Just get five more forks!''

	This in itself was OK, but Paul then applied this same solution to
	circular linked lists.

	This would not have been so bad either, but he had to go and tell
	someone about it!
\QuickE{}
\QuickQ{}
	Give an exception to this rule.
\QuickA{}
	One exception would be a difficult and complex algorithm that
	was the only one known to work in a given situation.
	Another exception would be a difficult and complex algorithm
	that was nonetheless the simplest of the set known to work in
	a given situation.
	However, even in these cases, it may be very worthwhile to spend
	a little time trying to come up with a simpler algorithm!
	After all, if you managed to invent the first algorithm
	to do some task, it shouldn't be that hard to go on to
	invent a simpler one.
\QuickE{}
\QuickQAC{chp:Conflicting Visions of the Future}{Conflicting Visions of the Future}{qqzfuture}
\QuickQ{}
	But suppose that an application exits while holding a
	\co{pthread_mutex_lock()} that happens to be located in a
	file-mapped region of memory?
\QuickA{}
	Indeed, in this case the lock would persist, much to the
	consternation of other processes attempting to acquire this
	lock that is held by a process that no longer exists.
	Which is why great care is required when using \co{pthread_mutex}
	objects located in file-mapped memory regions.
\QuickE{}
\QuickQ{}
	What about non-persistent primitives represented by data
	structures in \co{mmap()} regions of memory?
	What happens when there is an \co{exec()} within a critical
	section of such a primitive?
\QuickA{}
	If the \co{exec()}ed program maps those same regions of
	memory, then this program could in principle simply release
	the lock.
	The question as to whether this approach is sound from a
	software-engineering viewpoint is left as an exercise for
	the reader.
\QuickE{}
\QuickQ{}
	MV-RLU looks pretty good!
	Doesn't it beat RCU hands down?
\QuickA{}
	One might get that impression from a quick read of the abstract,
	but more careful readers will notice the ``for a wide range of
	workloads'' phrase in the last sentence.
	It turns out that this phrase is quite important:

	\begin{enumerate}
	\item	Their RCU evaluation uses synchronous grace periods, which
		needlessly throttle updates, as noted in their
		Section~6.2.1.
		See \cref{fig:datastruct:Read-Side RCU-Protected Hash-Table Performance For Schroedinger's Zoo in the Presence of Updates}
		\cpageref{fig:datastruct:Read-Side RCU-Protected Hash-Table Performance For Schroedinger's Zoo in the Presence of Updates}
		of this book to see that the venerable asynchronous
		\co{call_rcu()} primitive enables RCU to perform and
		scale quite well with large numbers of updaters.
		Furthermore, in Section~3.7 of their paper, the authors
		admit that asynchronous grace periods are important to
		MV-RLU scalability.
		A fair comparison would also allow RCU the benefits of
		asynchrony.
	\item	They use a poorly tuned 1,000-bucket hash table containing
		10,000~elements.
		In addition, their 448~hardware threads need considerably
		more than 1,000~buckets to avoid the lock contention
		that they correctly state limits RCU performance in
		their benchmarks.
		A useful comparison would feature a properly tuned
		hash table.
	\item	Their RCU hash table used per-bucket locks, which they
		call out as a bottleneck, which is not a surprise given
		the long hash chains and small ratio of buckets to threads.
		A number of their competing mechanisms instead use
		lockfree techniques, thus avoiding the per-bucket-lock
		bottleneck, which cynics might claim sheds some light
		on the authors' otherwise inexplicable choice of poorly
		tuned hash tables.
		The first graph in the middle row of the authors'
		Figure~4 show what RCU can achieve if not hobbled by
		artificial bottlenecks, as does the first portion of
		the second graph in that same row.
	\item	Their linked-list operation permits RLU to do concurrent
		modifications of different elements in the list, while
		RCU is forced to serialize updates.
		Again, RCU has always worked just fine in conjunction
		with lockless updaters, a fact that has been set forth
		in academic literature that the authors
		cited~\cite{MathieuDesnoyers2012URCU}.
		A fair comparison would use the same style of update
		for RCU as it does for MV-RLU.
	\item	The authors fail to consider combining RCU and sequence
		locking, which is used in the Linux kernel to give
		readers coherent views of multi-pointer updates.
	\item	The authors fail to consider RCU-based solutions to the
		Issaquah Challenge~\cite{PaulEMcKenney2016IssaquahCPPCON},
		which also gives readers a coherent view of multi-pointer
		updates, albeit with a weaker view of ``coherent''.
	\end{enumerate}

	It is surprising that the anonymous reviewers of this paper did
	not demand an apples-to-apples comparison of MV-RLU and RCU\@.
	Nevertheless, the authors should be congratulated on producing
	an academic paper that presents an all-too-rare example of good
	scalability combined with strong read-side coherence.
	They are also to be congratulated on overcoming the traditional
	academic prejudice against asynchronous grace periods,
	which greatly aided their scalability.

	Interestingly enough, RLU and RCU take different approaches to avoid
	the inherent limitations of STM noted by \ppl{Hagit}{Attiya} et
	al.~\cite{Attiya:2009:STMReadOnlyLimits}.
	RCU avoids providing strict serializability and RLU avoids providing
	invisible read-only transactions, both thus avoiding the
	limitations.
\QuickE{}
\QuickQ{}
	Given things like \co{spin_trylock()}, how does it make any
	sense at all to claim that TM introduces the concept of failure???
\QuickA{}
	When using locking, \co{spin_trylock()} is a choice, with a
	corresponding failure-free choice being \co{spin_lock()},
	which is used in the common case, as in there are more than
	100 times as many calls to \co{spin_lock()} than to
	\co{spin_trylock()} in the v5.11 Linux kernel.
	When using TM, the only failure-free choice is the irrevocable
	transaction, which is not used in the common case.
	In fact, the irrevocable transaction is not even available
	in all TM implementations.
\QuickE{}
\QuickQ{}
	What is to learn?
	Why not just use TM for memory-based data structures and locking
	for those rare cases featuring the many silly corner cases listed
	in this silly section???
\QuickA{}
	The year 2005 just called, and it says that it wants its
	incandescent TM marketing hype back.

	In the year 2021, TM still has significant proving to do,
	even with the advent of HTM, which is covered in the
	upcoming
	\cref{sec:future:Hardware Transactional Memory}.
\QuickE{}
\QuickQ{}
	Why would it matter that oft-written variables shared the cache
	line with the lock variable?
\QuickA{}
	If the lock is in the same cacheline as some of the variables
	that it is protecting, then writes to those variables by one CPU
	will invalidate that cache line for all the other CPUs.
	These \IXpl{invalidation} will
	generate large numbers of conflicts and retries, perhaps even
	degrading performance and scalability compared to locking.
\QuickE{}
\QuickQ{}
	Why are relatively small updates important to \IXacr{htm} performance
	and scalability?
\QuickA{}
	The larger the updates, the greater the probability of conflict,
	and thus the greater probability of retries, which degrade
	performance.
\QuickE{}
\QuickQ{}
	How could a red-black tree possibly efficiently enumerate all
	elements of the tree regardless of choice of synchronization
	mechanism???
\QuickA{}
	In many cases, the enumeration need not be exact.
	In these cases, hazard pointers or \IXacr{rcu} may be used to protect
	readers, which provides low probability of conflict with any
	given insertion or deletion.
\QuickE{}
\QuickQ{}
	But why can't a debugger emulate single stepping by setting
	breakpoints at successive lines of the transaction, relying
	on the retry to retrace the steps of the earlier instances
	of the transaction?
\QuickA{}
	This scheme might work with reasonably high probability, but it
	can fail in ways that would be quite surprising to most users.
	To see this, consider the following transaction:

\begin{fcvlabel}[ln:future:htm:debug rollbacks]
\begin{VerbatimN}[commandchars=\\\[\]]
begin_trans();
if (a) {
	do_one_thing();
	do_another_thing();	\lnlbl[another]
} else {
	do_a_third_thing();
	do_a_fourth_thing();
}
end_trans();
\end{VerbatimN}
\end{fcvlabel}

	\begin{fcvref}[ln:future:htm:debug rollbacks]
	Suppose that the user sets a breakpoint at \clnref{another},
	which triggers,
	aborting the transaction and entering the debugger.
	\end{fcvref}
	Suppose that between the time that the breakpoint triggers
	and the debugger gets around to stopping all the threads, some
	other thread sets the value of \co{a} to zero.
	When the poor user attempts to single-step the program, surprise!
	The program is now in the else-clause instead of the then-clause.

	This is \emph{not} what I call an easy-to-use debugger.
\QuickE{}
\QuickQ{}
	But why would \emph{anyone} need an empty lock-based critical
	section???
\QuickA{}
	See the answer to \QuickQuizARef{\QlockingQemptycriticalsection} in
	\cref{sec:locking:Exclusive Locks}.

	However, it is claimed that given a strongly atomic \IXacr{htm}
	implementation without forward-progress guarantees, any
	memory-based locking design based on empty critical sections
	will operate correctly in the presence of transactional
	lock elision.
	Although I have not seen a proof of this statement, there
	is a straightforward rationale for this claim.
	The main idea is that in a strongly atomic HTM implementation,
	the results of a given transaction are not visible until
	after the transaction completes successfully.
	Therefore, if you can see that a transaction has started,
	it is guaranteed to have already completed, which means
	that a subsequent empty lock-based critical section will
	successfully ``wait'' on it---after all, there is no waiting
	required.

	This line of reasoning does not apply to weakly atomic
	systems (including many \IXacr{stm} implementation), and it also
	does not apply to lock-based programs that use means other
	than memory to communicate.
	One such means is the passage of time (for example, in
	hard real-time systems) or flow of priority (for example,
	in soft real-time systems).

	Locking designs that rely on priority boosting are of particular
	interest.
\QuickE{}
\QuickQ{}
	Can't transactional lock elision trivially handle locking's
	time-based messaging semantics
	by simply choosing not to elide empty lock-based critical sections?
\QuickA{}
	It could do so, but this would be both unnecessary and
	insufficient.

	It would be unnecessary in cases where the empty critical section
	was due to conditional compilation.
	Here, it might well be that the only purpose of the lock was to
	protect data, so eliding it completely would be the right thing
	to do.
	In fact, leaving the empty lock-based critical section would
	degrade performance and scalability.

	On the other hand, it is possible for a non-empty lock-based
	critical section to be relying on both the data-protection
	and time-based and messaging semantics of locking.
	Using transactional lock elision in such a case would be
	incorrect, and would result in bugs.
\QuickE{}
\QuickQ{}
	Given modern hardware~\cite{PeterOkech2009InherentRandomness},
	how can anyone possibly expect parallel software relying
	on timing to work?
\QuickA{}
	The short answer is that on commonplace commodity hardware,
	synchronization designs based on any sort of fine-grained
	timing are foolhardy and cannot be expected to operate correctly
	under all conditions.

	That said, there are systems designed for hard real-time use
	that are much more deterministic.
	In the (very unlikely) event that you are using such a system,
	here is a toy example showing how time-based synchronization can
	work.
	Again, do \emph{not} try this on commodity microprocessors,
	as they have highly nondeterministic performance characteristics.

	This example uses multiple worker threads along with a control
	thread.
	Each worker thread corresponds to an outbound data feed, and
	records the current time (for example, from the
	\co{clock_gettime()} system call) in a per-thread
	\co{my_timestamp} variable after executing each unit
	of work.
	The real-time nature of this example results in the following
	set of constraints:

	\begin{enumerate}
	\item	It is a fatal error for a given worker thread to fail
		to update its timestamp for a time period of more than
		\co{MAX_LOOP_TIME}.
	\item	Locks are used sparingly to access and update global
		state.
	\item	Locks are granted in strict FIFO order within
		a given thread priority.
	\end{enumerate}

	When worker threads complete their feed, they must disentangle
	themselves from the rest of the application and place a status
	value in a per-thread \co{my_status} variable that is initialized
	to $-1$.
	Threads do not exit; they instead are placed on a thread pool
	to accommodate later processing requirements.
	The control thread assigns (and re-assigns) worker threads as
	needed, and also maintains a histogram of thread statuses.
	The control thread runs at a real-time priority no higher than
	that of the worker threads.

	Worker threads' code is as follows:

\begin{VerbatimN}
	int my_status = -1;  /* Thread local. */

	while (continue_working()) {
		enqueue_any_new_work();
		wp = dequeue_work();
		do_work(wp);
		my_timestamp = clock_gettime(...);
	}

	acquire_lock(&departing_thread_lock);

	/*
	 * Disentangle from application, might
	 * acquire other locks, can take much longer
	 * than MAX_LOOP_TIME, especially if many
	 * threads exit concurrently.
	 */
	my_status = get_return_status();
	release_lock(&departing_thread_lock);

	/* thread awaits repurposing. */
\end{VerbatimN}

	The control thread's code is as follows:

\begin{fcvlabel}[ln:future:htm:control thread]
\begin{VerbatimN}[commandchars=\\\@\$]
	for (;;) {
		for_each_thread(t) {
			ct = clock_gettime(...);
			d = ct - per_thread(my_timestamp, t);
			if (d >= MAX_LOOP_TIME) {	\lnlbl@if$
				/* thread departing. */	\lnlbl@dep:b$
				acquire_lock(&departing_thread_lock); \lnlbl@acq$
				release_lock(&departing_thread_lock); \lnlbl@rel$
				i = per_thread(my_status, t);
				status_hist[i]++; /* Bug if TLE! */ \lnlbl@dep:e$
			}
		}
		/* Repurpose threads as needed. */
	}
\end{VerbatimN}
\end{fcvlabel}

	\begin{fcvref}[ln:future:htm:control thread]
	\Clnref{if} uses the passage of time to deduce that the thread
	has exited, executing \clnref{dep:b,dep:e} if so.
	The empty lock-based critical section on \clnref{acq,rel}
	guarantees that any thread in the process of exiting
	completes (remember that locks are granted in FIFO order!).
	\end{fcvref}

	Once again, do not try this sort of thing on commodity
	microprocessors.
	After all, it is difficult enough to get this right on systems
	specifically designed for hard real-time use!
\QuickE{}
\QuickQ{}
	But the \co{boostee()} function in
	\cref{lst:future:Exploiting Priority Boosting}
	alternatively acquires its locks in reverse order!
	Won't this result in deadlock?
\QuickA{}
	No deadlock will result.
	To arrive at deadlock, two different threads must each
	acquire the two locks in opposite orders, which does not
	happen in this example.
	However, deadlock detectors such as
	lockdep~\cite{JonathanCorbet2006lockdep}
	will flag this as a false positive.
\QuickE{}
\QuickQ{}
	So a bunch of people set out to supplant locking, and they
	mostly end up just optimizing locking???
\QuickA{}
	At least they accomplished something useful!
	And perhaps there will continue to be additional \IXacr{htm} progress
	over time~\cite{Siakavaras2017CombiningHA,DimitriosSiakavaras2020RCU-HTM-B+Trees,ChristinaGiannoula2018HTM-RCU-graphcoloring,SeongJaePark2020HTMRCUlock}.
\QuickE{}
\QuickQ{}
	\Cref{tab:future:Comparison of Locking and HTM,tab:future:Comparison of Locking (Augmented by RCU or Hazard Pointers) and HTM}
	state that hardware is only starting to become available.
	But hasn't HTM hardware support been widely available
	for almost a full decade?
\QuickA{}
	Yes and no.
	It appears that implementing even the HTM subset of TM in real
	hardware is a bit trickier than it
	appears~\cite{ChristianJacobi2012MainframeTM,ScottWasson2014HaswellDisableTSX,Intel2020HaswellTSXordering,Intel2021perfTSXordering,MichaelLarabel2021DisableTSX}.
	Therefore, the sad fact is that ``starting to become available'' is
	all too accurate as of 2021.
	In fact, vendors are beginning to deprecate their HTM
	implementations~\cite[Book III Appendix A]{PowerISA3.1-2020}.
\QuickE{}
\QuickQ{}
	This list is ridiculously utopian!
	Why not stick to the current state of the formal-verification art?
\QuickA{}
	You are welcome to your opinion on what is and is not utopian,
	but I will be paying more attention to people actually making
	progress on the items in that list than to anyone who might be
	objecting to them.
	This might have something to do with my long experience with
	people attempting to talk me out of specific things that their
	favorite tools cannot handle.

	In the meantime, please feel free to read the papers written by
	the people who are actually making progress, for example, this
	one~\cite{DinoDistefano2019FBstaticAnalysis}.
\QuickE{}
\QuickQ{}
	Given the groundbreaking nature of the various verifiers used
	in the SEL4 project, why doesn't this chapter cover them in
	more depth?
\QuickA{}
	There can be no doubt that the verifiers used by the SEL4
	project are quite capable.
	However, SEL4 started as a single-CPU project.
	And although SEL4 has gained multi-processor
	capabilities, it is currently using very coarse-grained
	locking that is similar to the Linux kernel's old
	Big Kernel Lock (BKL)\@.
	There will hopefully come a day when it makes sense to add
	SEL4's verifiers to a book on parallel programming, but
	this is not yet that day.
\QuickE{}
\QuickQ{}
\begin{fcvref}[ln:future:formalregress:C-SB+l-o-o-u+l-o-o-u-C:whole]
	Why bother with a separate \co{filter} command on \clnref{filter_} of
	\cref{lst:future:Emulating Locking with cmpxchg}
	instead of just adding the condition to the \co{exists} clause?
	And wouldn't it be simpler to use \co{xchg_acquire()} instead
	of \co{cmpxchg_acquire()}?
\end{fcvref}
\QuickA{}
	The \co{filter} clause causes the \co{herd} tool to discard
	executions at an earlier stage of processing than does
	the \co{exists} clause, which provides significant speedups.

\begin{table}
\rowcolors{7}{lightgray}{}
\renewcommand*{\arraystretch}{1.1}
\small
\centering
\begin{tabular}{S[table-format=1.0]S[table-format=1.3]S[table-format=2.3]
		S[table-format=3.3]S[table-format=2.3]S[table-format=3.3]}
	\toprule
	& & \multicolumn{2}{c}{\tco{cmpxchg_acquire()}}
		& \multicolumn{2}{c}{\tco{xchg_acquire()}} \\
	\cmidrule(l){3-4} \cmidrule(l){5-6}
	\multicolumn{1}{c}{\#} & \multicolumn{1}{c}{Lock}
		& \multicolumn{1}{c}{\tco{filter}}
			& \multicolumn{1}{c}{\tco{exists}}
				& \multicolumn{1}{c}{\tco{filter}}
					& \multicolumn{1}{c}{\tco{exists}} \\
	\cmidrule{1-1} \cmidrule(l){2-2} \cmidrule(l){3-4} \cmidrule(l){5-6}
	2 & 0.004 &  0.022 &   0.039 &  0.027 &  0.058 \\
	3 & 0.041 &  0.743 &   1.653 &  0.968 &  3.203 \\
	4 & 0.374 & 59.565 & 151.962 & 74.818 & 500.96 \\
	5 & 4.905 &        &         &        &        \\
	\bottomrule
\end{tabular}
\caption{Emulating Locking:
			    Performance Comparison (s)}
\label{tab:future:Emulating Locking: Performance Comparison (s)}
\end{table}

	As for \co{xchg_acquire()}, this atomic operation will do a
	write whether or not lock acquisition succeeds, which means
	that a model using \co{xchg_acquire()} will have more operations
	than one using \co{cmpxchg_acquire()}, which won't do a write
	in the failed-acquisition case.
	More writes means more combinatorial to explode, as shown in
	\cref{tab:future:Emulating Locking: Performance Comparison (s)}
	(\path{C-SB+l-o-o-u+l-o-o-*u.litmus},
	\path{C-SB+l-o-o-u+l-o-o-u*-C.litmus},
	\path{C-SB+l-o-o-u+l-o-o-u*-CE.litmus},
	\path{C-SB+l-o-o-u+l-o-o-u*-X.litmus}, and
	\path{C-SB+l-o-o-u+l-o-o-u*-XE.litmus}).
	This table clearly shows that \co{cmpxchg_acquire()}
	outperforms \co{xchg_acquire()} and that use of the
	\co{filter} clause outperforms use of the \co{exists} clause.
\QuickE{}
\QuickQ{}
	How do we know that the MTBFs of known bugs is a good estimate
	of the MTBFs of bugs that have not yet been located?
\QuickA{}
	We don't, but it does not matter.

	To see this, note that the 7\,\% figure only applies to injected
	bugs that were subsequently located:
	It necessarily ignores any injected bugs that were never found.
	Therefore, the MTBF statistics of known bugs is likely to be
	a good approximation of that of the injected bugs that are
	subsequently located.

	A key point in this whole section is that we should be more
	concerned about bugs that inconvenience users than about
	other bugs that never actually manifest.
	This of course is \emph{not} to say that we should completely
	ignore bugs that have not yet inconvenienced users, just that
	we should properly prioritize our efforts so as to fix the
	most important and urgent bugs first.
\QuickE{}
\QuickQ{}
	But the formal-verification tools should immediately find all the
	bugs introduced by the fixes, so why is this a problem?
\QuickA{}
	It is a problem because real-world formal-verification tools
	(as opposed to those that exist only in the imaginations of
	the more vociferous proponents of formal verification) are
	not omniscient, and thus are only able to locate certain types
	of bugs.
	For but one example, formal-verification tools are unlikely to
	spot a bug corresponding to an omitted assertion or, equivalently,
	a bug corresponding to an undiscovered portion of the specification.
\QuickE{}
\QuickQ{}
	But many formal-verification tools can only find one bug at
	a time, so that each bug must be fixed before the tool can
	locate the next.
	How can bug-fix efforts be prioritized given such a tool?
\QuickA{}
	One approach is to provide a simple fix that might not be
	suitable for a production environment, but which allows
	the tool to locate the next bug.
	Another approach is to restrict configuration or inputs
	so that the bugs located thus far cannot occur.
	There are a number of similar approaches, but the common theme
	is that fixing the bug from the tool's viewpoint is usually much
	easier than constructing and validating a production-quality fix,
	and the key point is to prioritize the larger efforts required
	to construct and validate the production-quality fixes.
\QuickE{}
\QuickQ{}
	How would testing stack up in the scorecard shown in
	\cref{tab:future:Formal Regression Scorecard}?
\QuickA{}
	It would be blue all the way down, with the possible
	exception of the third row (overhead) which might well
	be marked down for testing's difficulty finding
	improbable bugs.

	On the other hand, improbable bugs are often also
	irrelevant bugs, so your mileage may vary.

	Much depends on the size of your installed base.
	If your code is only ever going to run on (say) 10,000
	systems, Murphy can actually be a really nice guy.
	Everything that can go wrong, will.
	Eventually.
	Perhaps in geologic time.

	But if your code is running on 20~billion systems,
	like the Linux kernel was said to be by late 2017,
	Murphy can be a real jerk!
	Everything that can go wrong, will, and it can go wrong
	really quickly!!!
\QuickE{}
\QuickQ{}
	But aren't there a great many more formal-verification systems
	than are shown in
	\cref{tab:future:Formal Regression Scorecard}?
\QuickA{}
	Indeed there are!
	This table focuses on those that Paul has used, but others are
	proving to be useful.
	Formal verification has been heavily used in the seL4
	project~\cite{ThomasSewell2013L4binaryVerification},
	and its tools can now handle modest levels of concurrency.
	More recently, Catalin Marinas used Lamport's
	TLA tool~\cite{Lamport:2002:SST:579617} to locate some
	forward-progress bugs in the Linux kernel's queued spinlock
	implementation.
	Will Deacon fixed these bugs~\cite{WillDeacon2018qspinlock},
	and Catalin verified Will's
	fixes~\cite{CatalinMarinas2018qspinlockTLA}.

	Lighter-weight formal verification tools have been used heavily
	in production~\cite{JamesRLarus2004RightingSoftware,AlBessey2010BillionLoCLater,ByronCook2018FormalAmazon,CaitlinSadowski2018staticAnalysisGoogle,DinoDistefano2019FBstaticAnalysis}.
\QuickE{}
\QuickQAC{chp:app:Important Questions}{Important Questions}{qqzquestions}
\QuickQ{}
	What SMP coding errors can you see in these examples?
	See \path{time.c} for full code.
\QuickA{}
	Here are errors you might have found:

	\begin{enumerate}
	\item	Missing barrier() or volatile on tight loops.
	\item	Missing memory barriers on update side.
	\item	Lack of synchronization between producer and consumer.
\QuickE{}
	\end{enumerate}
\QuickQ{}
	How could there be such a large gap between successive
	consumer reads?
	See \path{timelocked.c} for full code.
\QuickA{}
	Here are a few reasons for such gaps:

	\begin{enumerate}
	\item	The consumer might be preempted for long time periods.
	\item	A long-running interrupt might delay the consumer.
	\item	Cache misses might delay the consumer.
	\item	The producer might also be running on a faster CPU than is the
		consumer (for example, one of the CPUs might have had to
		decrease its
		clock frequency due to heat-dissipation or power-consumption
		constraints).
\QuickE{}
	\end{enumerate}
\QuickQ{}
	But if fully ordered implementations cannot offer stronger
	guarantees than the better performing and more scalable weakly
	ordered implementations, why bother with full ordering?
\QuickA{}
	Because strongly ordered implementations are sometimes
	able to provide greater consistency among sets of calls to
	functions accessing a given data structure.
	For example, compare the atomic counter of
	\cref{lst:count:Just Count Atomically!}
	to the statistical counter of
	\cref{sec:count:Statistical Counters}.
	Suppose that one thread is adding the value~3 and another is
	adding the value~5, while two other threads are concurrently
	reading the counter's value.
	With atomic counters, it is not possible for one of the readers
	to obtain the value~3 while the other obtains the value~5.
	With statistical counters, this outcome really can happen.
	In fact, in some computing environments, this outcome can happen
	even on relatively strongly ordered hardware such as x86.

	Therefore, if your user happen to need this admittedly
	unusual level of consistency, you should avoid weakly ordered
	statistical counters.
\QuickE{}
\QuickQ{}
	Suppose a portion of a program uses RCU read-side primitives
	as its only synchronization mechanism.
	Is this parallelism or concurrency?
\QuickA{}
	Yes.
\QuickE{}
\QuickQ{}
	In what part of the second (scheduler-based) perspective would
	the lock-based single-thread-per-CPU workload be considered
	``concurrent''?
\QuickA{}
	The people who would like to arbitrarily subdivide and interleave
	the workload.
	Of course, an arbitrary subdivision might end up separating
	a lock acquisition from the corresponding lock release, which
	would prevent any other thread from acquiring that lock.
	If the locks were pure spinlocks, this could even result in
	deadlock.
\QuickE{}
\QuickQAC{chp:app:``Toy'' RCU Implementations}{``Toy'' RCU Implementations}{qqztoyrcu}
\QuickQ{}
	Why wouldn't any deadlock in the RCU implementation in
	\cref{lst:app:toyrcu:Lock-Based RCU Implementation}
	also be a deadlock in any other RCU implementation?
\QuickA{}
	\begin{fcvref}[ln:app:toyrcu:Deadlock in Lock-Based RCU Implementation]
	Suppose the functions \co{foo()} and \co{bar()} in
	\cref{lst:app:toyrcu:Deadlock in Lock-Based RCU Implementation}
	are invoked concurrently from different CPUs.
	Then \co{foo()} will acquire \co{my_lock()} on \clnref{foo:acq},
	while \co{bar()} will acquire \co{rcu_gp_lock} on
	\clnref{bar:rrl}.
	\end{fcvref}

\begin{listing}
\begin{fcvlabel}[ln:app:toyrcu:Deadlock in Lock-Based RCU Implementation]
\begin{VerbatimL}[commandchars=\\\[\]]
void foo(void)
{
	spin_lock(&my_lock);		\lnlbl[foo:acq]
	rcu_read_lock();		\lnlbl[foo:rrl]
	do_something();
	rcu_read_unlock();
	do_something_else();
	spin_unlock(&my_lock);
}

void bar(void)
{
	rcu_read_lock();		\lnlbl[bar:rrl]
	spin_lock(&my_lock);		\lnlbl[bar:acq]
	do_some_other_thing();
	spin_unlock(&my_lock);
	do_whatever();
	rcu_read_unlock();
}
\end{VerbatimL}
\end{fcvlabel}
\caption{Deadlock in Lock-Based RCU Implementation}
\label{lst:app:toyrcu:Deadlock in Lock-Based RCU Implementation}
\end{listing}

	\begin{fcvref}[ln:app:toyrcu:Deadlock in Lock-Based RCU Implementation]
	When \co{foo()} advances to \clnref{foo:rrl}, it will attempt to
	acquire \co{rcu_gp_lock}, which is held by \co{bar()}.
	Then when \co{bar()} advances to \clnref{bar:acq}, it will attempt
	to acquire \co{my_lock}, which is held by \co{foo()}.
	\end{fcvref}

	Each function is then waiting for a lock that the other
	holds, a classic deadlock.

	Other RCU implementations neither spin nor block in
	\co{rcu_read_lock()}, hence avoiding deadlocks.
\QuickE{}
\QuickQ{}
	Why not simply use reader-writer locks in the RCU implementation
	in
	\cref{lst:app:toyrcu:Lock-Based RCU Implementation}
	in order to allow RCU readers to proceed in parallel?
\QuickA{}
	One could in fact use reader-writer locks in this manner.
	However, textbook reader-writer locks suffer from memory
	contention, so that the RCU read-side critical sections would
	need to be quite long to actually permit parallel
	execution~\cite{McKenney03a}.

	On the other hand, use of a reader-writer lock that is
	read-acquired in \co{rcu_read_lock()} would avoid the
	deadlock condition noted above.
\QuickE{}
\QuickQ{}
	\begin{fcvref}[ln:defer:rcu_lock_percpu:sync:loop]
	Wouldn't it be cleaner to acquire all the locks, and then
	release them all in the loop from \clnrefrange{b}{e} of
	\cref{lst:app:toyrcu:Per-Thread Lock-Based RCU Implementation}?
	After all, with this change, there would be a point in time
	when there were no readers, simplifying things greatly.
	\end{fcvref}
\QuickA{}
	Making this change would re-introduce the deadlock, so
	no, it would not be cleaner.
\QuickE{}
\QuickQ{}
	Is the implementation shown in
	\cref{lst:app:toyrcu:Per-Thread Lock-Based RCU Implementation}
	free from deadlocks?
	Why or why not?
\QuickA{}
	One deadlock is where a lock is
	held across \co{synchronize_rcu()}, and that same lock is
	acquired within an RCU read-side critical section.
	However, this situation could deadlock any correctly designed
	RCU implementation.
	After all, the \co{synchronize_rcu()} primitive must wait for all
	pre-existing RCU read-side critical sections to complete,
	but if one of those critical sections is spinning on a lock
	held by the thread executing the \co{synchronize_rcu()},
	we have a deadlock inherent in the definition of RCU\@.

	Another deadlock happens when attempting to nest RCU read-side
	critical sections.
	This deadlock is peculiar to this implementation, and might
	be avoided by using recursive locks, or by using reader-writer
	locks that are read-acquired by \co{rcu_read_lock()} and
	write-acquired by \co{synchronize_rcu()}.

	However, if we exclude the above two cases,
	this implementation of RCU does not introduce any deadlock
	situations.
	This is because only time some other thread's lock is acquired is when
	executing \co{synchronize_rcu()}, and in that case, the lock
	is immediately released, prohibiting a deadlock cycle that
	does not involve a lock held across the \co{synchronize_rcu()}
	which is the first case above.
\QuickE{}
\QuickQ{}
	Isn't one advantage of the RCU algorithm shown in
	\cref{lst:app:toyrcu:Per-Thread Lock-Based RCU Implementation}
	that it uses only primitives that are widely available,
	for example, in POSIX pthreads?
\QuickA{}
	This is indeed an advantage, but do not forget that
	\co{rcu_dereference()} and \co{rcu_assign_pointer()}
	are still required, which means \co{volatile} manipulation
	for \co{rcu_dereference()} and memory barriers for
	\co{rcu_assign_pointer()}.
	Of course, many Alpha CPUs require memory barriers for both
	primitives.
\QuickE{}
\QuickQ{}
	But what if you hold a lock across a call to
	\co{synchronize_rcu()}, and then acquire that same lock within
	an RCU read-side critical section?
\QuickA{}
	Indeed, this would deadlock any legal RCU implementation.
	But is \co{rcu_read_lock()} \emph{really} participating in
	the deadlock cycle?
	If you believe that it is, then please
	ask yourself this same question when looking at the
	RCU implementation in
	\cref{sec:app:toyrcu:RCU Based on Quiescent States}.
\QuickE{}
\QuickQ{}
	How can the grace period possibly elapse in 40 nanoseconds when
	\co{synchronize_rcu()} contains a 10-millisecond delay?
\QuickA{}
	The update-side test was run in absence of readers, so the
	\co{poll()} system call was never invoked.
	In addition, the actual code has this \co{poll()}
	system call commented out, the better to evaluate the
	true overhead of the update-side code.
	Any production uses of this code would be better served by
	using the \co{poll()} system call, but then again,
	production uses would be even better served by other implementations
	shown later in this section.
\QuickE{}
\QuickQ{}
	Why not simply make \co{rcu_read_lock()} wait when a concurrent
	\co{synchronize_rcu()} has been waiting too long in
	the RCU implementation in
	\cref{lst:app:toyrcu:RCU Implementation Using Single Global Reference Counter}?
	Wouldn't that prevent \co{synchronize_rcu()} from starving?
\QuickA{}
	Although this would in fact eliminate the starvation, it would
	also mean that \co{rcu_read_lock()} would spin or block waiting
	for the writer, which is in turn waiting on readers.
	If one of these readers is attempting to acquire a lock that
	the spinning/blocking \co{rcu_read_lock()} holds, we again
	have deadlock.

	In short, the cure is worse than the disease.
	See \cref{sec:app:toyrcu:Starvation-Free Counter-Based RCU}
	for a proper cure.
\QuickE{}
\QuickQ{}
	\begin{fcvref}[ln:defer:rcu_rcpg:sync]
	Why the memory barrier on \clnref{mb1} of \co{synchronize_rcu()} in
	\cref{lst:app:toyrcu:RCU Update Using Global Reference-Count Pair}
	given that there is a spin-lock acquisition immediately after?
	\end{fcvref}
\QuickA{}
	The spin-lock acquisition only guarantees that the spin-lock's
	critical section will not ``bleed out'' to precede the
	acquisition.
	It in no way guarantees that code preceding the spin-lock
	acquisition won't be reordered into the critical section.
	Such reordering could cause a removal from an RCU-protected
	list to be reordered to follow the complementing of
	\co{rcu_idx}, which could allow a newly starting RCU
	read-side critical section to see the recently removed
	data element.

	Exercise for the reader:
	Use a tool such as Promela/spin to determine which (if any) of
	the memory barriers in
	\cref{lst:app:toyrcu:RCU Update Using Global Reference-Count Pair}
	are really needed.
	See \cref{chp:Formal Verification}
	for information on using these tools.
	The first correct and complete response will be credited.
\QuickE{}
\QuickQ{}
	Why is the counter flipped twice in
	\cref{lst:app:toyrcu:RCU Update Using Global Reference-Count Pair}?
	Shouldn't a single flip-and-wait cycle be sufficient?
\QuickA{}
	\begin{fcvref}[ln:defer:rcu_rcpg]
	Both flips are absolutely required.
	To see this, consider the following sequence of events:
	\begin{sequence}
	\item	\Clnref{r:lock:cur:b} of \co{rcu_read_lock()} in
		\cref{lst:app:toyrcu:RCU Read-Side Using Global Reference-Count Pair}
		picks up \co{rcu_idx}, finding its value to be zero.
	\item	\Clnref{sync:compl} of \co{synchronize_rcu()} in
		\cref{lst:app:toyrcu:RCU Update Using Global Reference-Count Pair}
		complements the value of \co{rcu_idx}, setting its
		value to one.
	\item	\Clnrefrange{sync:while:b}{sync:while:e}
		of \co{synchronize_rcu()} find that the
		value of \co{rcu_refcnt[0]} is zero, and thus
		returns.
		(Recall that the question is asking what happens if
		\clnrefrange{sync:mb3}{sync:mb5} are omitted.)
	\item	\Clnref{r:lock:set,r:lock:cur:e}
		of \co{rcu_read_lock()} store the
		value zero to this thread's instance of \co{rcu_read_idx}
		and increments \co{rcu_refcnt[0]}, respectively.
		Execution then proceeds into the RCU read-side critical
		section.
		\label{sec:app:toyrcu:rcu_rcgp:RCU Read Side Start}
	\item	Another instance of \co{synchronize_rcu()} again complements
		\co{rcu_idx}, this time setting its value to zero.
		Because \co{rcu_refcnt[1]} is zero, \co{synchronize_rcu()}
		returns immediately.
		(Recall that \co{rcu_read_lock()} incremented
		\co{rcu_refcnt[0]}, not \co{rcu_refcnt[1]}!)
		\label{sec:app:toyrcu:rcu_rcgp:RCU Grace Period Start}
	\item	The grace period that started in
		\cref{sec:app:toyrcu:rcu_rcgp:RCU Grace Period Start}
		has been allowed to end, despite
		the fact that the RCU read-side critical section
		that started beforehand in
		\cref{sec:app:toyrcu:rcu_rcgp:RCU Read Side Start}
		has not completed.
		This violates RCU semantics, and could allow the update
		to free a data element that the RCU read-side critical
		section was still referencing.
	\end{sequence}

	Exercise for the reader:
	What happens if \co{rcu_read_lock()} is preempted for a very long
	time (hours!\@) just after \clnref{r:lock:cur:b}?
	Does this implementation operate correctly in that case?
	Why or why not?
	The first correct and complete response will be credited.
	\end{fcvref}
\QuickE{}
\QuickQ{}
	\begin{fcvref}[ln:defer:rcu_rcpg:r]
	Given that atomic increment and decrement are so expensive,
	why not just use non-atomic increment on \clnref{lock:cur:e} and a
	non-atomic decrement on \clnref{unlock:atmdec} of
	\cref{lst:app:toyrcu:RCU Read-Side Using Global Reference-Count Pair}?
	\end{fcvref}
\QuickA{}
	Using non-atomic operations would cause increments and decrements
	to be lost, in turn causing the implementation to fail.
	See \cref{sec:app:toyrcu:Scalable Counter-Based RCU}
	for a safe way to use non-atomic operations in
	\co{rcu_read_lock()} and \co{rcu_read_unlock()}.
\QuickE{}
\QuickQ{}
	Come off it!
	We can see the \co{atomic_read()} primitive in
	\co{rcu_read_lock()}!!!
	So why are you trying to pretend that \co{rcu_read_lock()}
	contains no atomic operations???
\QuickA{}
	The \co{atomic_read()} primitives does not actually execute
	atomic machine instructions, but rather does a normal load
	from an \co{atomic_t}.
	Its sole purpose is to keep the compiler's type-checking happy.
	If the Linux kernel ran on 8-bit CPUs, it would also need to
	prevent ``store tearing'', which could happen due to the need
	to store a 16-bit pointer with two eight-bit accesses on some
	8-bit systems.
	But thankfully, it seems that no one runs Linux on 8-bit systems.
\QuickE{}
\QuickQ{}
	Great, if we have $N$ threads, we can have $2N$ ten-millisecond
	waits (one set per \co{flip_counter_and_wait()} invocation,
	and even that assumes that we wait only once for each thread).
	Don't we need the grace period to complete \emph{much} more quickly?
\QuickA{}
	Keep in mind that we only wait for a given thread if that thread
	is still in a pre-existing RCU read-side critical section,
	and that waiting for one hold-out thread gives all the other
	threads a chance to complete any pre-existing RCU read-side
	critical sections that they might still be executing.
	So the only way that we would wait for $2N$ intervals
	would be if the last thread still remained in a pre-existing
	RCU read-side critical section despite all the waiting for
	all the prior threads.
	In short, this implementation will not wait unnecessarily.

	However, if you are stress-testing code that uses RCU, you
	might want to comment out the \co{poll()} statement in
	order to better catch bugs that incorrectly retain a reference
	to an RCU-protected data element outside of an RCU
	read-side critical section.
\QuickE{}
\QuickQ{}
	All of these toy RCU implementations have either atomic operations
	in \co{rcu_read_lock()} and \co{rcu_read_unlock()},
	or \co{synchronize_rcu()}
	overhead that increases linearly with the number of threads.
	Under what circumstances could an RCU implementation enjoy
	lightweight implementations for all three of these primitives,
	all having deterministic ($\O{1}$) overheads and latencies?
\QuickA{}
	Special-purpose uniprocessor implementations of RCU can attain
	this ideal~\cite{PaulEMcKenney2009BloatwatchRCU}.
\QuickE{}
\QuickQ{}
	\begin{fcvref}[ln:defer:rcu:read_lock_unlock:unlock]
	If any even value is sufficient to tell \co{synchronize_rcu()}
	to ignore a given task, why don't \clnref{gp1,gp2} of
	\cref{lst:app:toyrcu:Free-Running Counter Using RCU}
	simply assign zero to \co{rcu_reader_gp}?
	\end{fcvref}
\QuickA{}
	Assigning zero (or any other even-numbered constant)
	would in fact work, but assigning the value of
	\co{rcu_gp_ctr} can provide a valuable debugging aid,
	as it gives the developer an idea of when the corresponding
	thread last exited an RCU read-side critical section.
\QuickE{}
\QuickQ{}
	\begin{fcvref}[ln:defer:rcu:synchronize:syn]
	Why are the memory barriers on \clnref{mb1,mb3} of
	\cref{lst:app:toyrcu:Free-Running Counter Using RCU}
	needed?
	Aren't the memory barriers inherent in the locking
	primitives on \clnref{spinlock,spinunlock} sufficient?
	\end{fcvref}
\QuickA{}
	These memory barriers are required because the locking
	primitives are only guaranteed to confine the critical
	section.
	The locking primitives are under absolutely no obligation
	to keep other code from bleeding in to the critical section.
	The pair of memory barriers are therefore requires to prevent
	this sort of code motion, whether performed by the compiler
	or by the CPU\@.
\QuickE{}
\QuickQ{}
	Couldn't the update-side batching optimization described in
	\cref{sec:app:toyrcu:Scalable Counter-Based RCU With Shared Grace Periods}
	be applied to the implementation shown in
	\cref{lst:app:toyrcu:Free-Running Counter Using RCU}?
\QuickA{}
	Indeed it could, with a few modifications.
	This work is left as an exercise for the reader.
\QuickE{}
\QuickQ{}
	\begin{fcvref}[ln:defer:rcu:read_lock_unlock:lock]
	Is the possibility of readers being preempted in
	\clnrefrange{gp1}{gp2} of
	\cref{lst:app:toyrcu:Free-Running Counter Using RCU}
	a real problem, in other words, is there a real sequence
	of events that could lead to failure?
	If not, why not?
	If so, what is the sequence of events, and how can the
	failure be addressed?
	\end{fcvref}
\QuickA{}
	It is a real problem, there is a sequence of events leading to
	failure, and there are a number of possible ways of
	addressing it.
	For more details, see the Quick Quizzes near the end of
	\cref{sec:app:toyrcu:Nestable RCU Based on Free-Running Counter}.
	The reason for locating the discussion there is to (1) give you
	more time to think about it, and (2) because the nesting support
	added in that section greatly reduces the time required to
	overflow the counter.
\QuickE{}
\QuickQ{}
	Why not simply maintain a separate per-thread nesting-level
	variable, as was done in previous section, rather than having
	all this complicated bit manipulation?
\QuickA{}
	The apparent simplicity of the separate per-thread variable
	is a red herring.
	This approach incurs much greater complexity in the guise
	of careful ordering of operations, especially if signal
	handlers are to be permitted to contain RCU read-side
	critical sections.
	But don't take my word for it, code it up and see what you
	end up with!
\QuickE{}
\QuickQ{}
	Given the algorithm shown in
	\cref{lst:app:toyrcu:Nestable RCU Using a Free-Running Counter},
	how could you double the time required to overflow the global
	\co{rcu_gp_ctr}?
\QuickA{}
	\begin{fcvref}[ln:defer:rcu_nest:synchronize:syn]
	One way would be to replace the magnitude comparison on
	\clnref{lt1,lt2} with an inequality check of
	the per-thread \co{rcu_reader_gp} variable against
	\co{rcu_gp_ctr+RCU_GP_CTR_BOTTOM_BIT}.
	\end{fcvref}
\QuickE{}
\QuickQ{}
	Again, given the algorithm shown in
	\cref{lst:app:toyrcu:Nestable RCU Using a Free-Running Counter},
	is counter overflow fatal?
	Why or why not?
	If it is fatal, what can be done to fix it?
\QuickA{}
	It can indeed be fatal.
	To see this, consider the following sequence of events:
	\begin{enumerate}
	\item	Thread~0 enters \co{rcu_read_lock()}, determines
		that it is not nested, and therefore fetches the
		value of the global \co{rcu_gp_ctr}.
		Thread~0 is then preempted for an extremely long time
		(before storing to its per-thread \co{rcu_reader_gp}
		variable).
	\item	Other threads repeatedly invoke \co{synchronize_rcu()},
		so that the new value of the global \co{rcu_gp_ctr}
		is now \co{RCU_GP_CTR_BOTTOM_BIT}
		less than it was when thread~0 fetched it.
	\item	Thread~0 now starts running again, and stores into
		its per-thread \co{rcu_reader_gp} variable.
		The value it stores is
		\co{RCU_GP_CTR_BOTTOM_BIT+1}
		greater than that of the global \co{rcu_gp_ctr}.
	\item	Thread~0 acquires a reference to RCU-protected data
		element~A.
	\item	Thread~1 now removes the data element~A that thread~0
		just acquired a reference to.
	\item	Thread~1 invokes \co{synchronize_rcu()}, which
		increments the global \co{rcu_gp_ctr} by
		\co{RCU_GP_CTR_BOTTOM_BIT}.
		It then checks all of the per-thread \co{rcu_reader_gp}
		variables, but thread~0's value (incorrectly) indicates
		that it started after thread~1's call to
		\co{synchronize_rcu()}, so thread~1 does not wait
		for thread~0 to complete its RCU read-side critical
		section.
	\item	Thread~1 then frees up data element~A, which thread~0
		is still referencing.
	\end{enumerate}

	Note that scenario can also occur in the implementation presented in
	\cref{sec:app:toyrcu:RCU Based on Free-Running Counter}.

	One strategy for fixing this problem is to use 64-bit
	counters so that the time required to overflow them would exceed
	the useful lifetime of the computer system.
	Note that non-antique members of the 32-bit x86 CPU family
	allow atomic manipulation of 64-bit counters via the
	\co{cmpxchg64b} instruction.

	Another strategy is to limit the rate at which grace periods are
	permitted to occur in order to achieve a similar effect.
	For example, \co{synchronize_rcu()} could record the last time
	that it was invoked, and any subsequent invocation would then
	check this time and block as needed to force the desired
	spacing.
	For example, if the low-order four bits of the counter were
	reserved for nesting, and if grace periods were permitted to
	occur at most ten times per second, then it would take more
	than 300 days for the counter to overflow.
	However, this approach is not helpful if there is any possibility
	that the system will be fully loaded with CPU-bound high-priority
	real-time threads for the full 300 days.
	(A remote possibility, perhaps, but best to consider it ahead
	of time.)

	A third approach is to administratively abolish real-time threads
	from the system in question.
	In this case, the preempted process will age up in priority,
	thus getting to run long before the counter had a chance to
	overflow.
	Of course, this approach is less than helpful for real-time
	applications.

	A fourth approach would be for \co{rcu_read_lock()} to recheck
	the value of the global \co{rcu_gp_ctr} after storing to its
	per-thread \co{rcu_reader_gp} counter, retrying if the new
	value of the global \co{rcu_gp_ctr} is inappropriate.
	This works, but introduces non-deterministic execution time
	into \co{rcu_read_lock()}.
	On the other hand, if your application is being preempted long
	enough for the counter to overflow, you have no hope of
	deterministic execution time in any case!

	A fifth approach is for the grace period process to wait for
	all readers to become aware of the new grace period.
	This works nicely in theory, but hangs if a reader blocks
	indefinitely outside of an RCU read-side critical section.

	A final approach is, oddly enough, to use a single-bit
	grace-period counter and for each call to \co{synchronize_rcu()}
	to take two passes through its algorithm.
	This is the approached use by userspace
	RCU~\cite{MathieuDesnoyers2009URCU}, and is described in
	detail in the journal article and supplementary
	materials~\cite[Appendix D]{MathieuDesnoyers2012URCU}.
\QuickE{}
\QuickQ{}
	\begin{fcvref}[ln:defer:rcu_qs:read_lock_unlock:qs]
	Doesn't the additional memory barrier shown on \clnref{mb2} of
	\cref{lst:app:toyrcu:Quiescent-State-Based RCU Read Side}
	greatly increase the overhead of \co{rcu_quiescent_state}?
	\end{fcvref}
\QuickA{}
	\begin{fcvref}[ln:defer:rcu_qs:read_lock_unlock:qs]
	Indeed it does!
	An application using this implementation of RCU should therefore
	invoke \co{rcu_quiescent_state} sparingly, instead using
	\co{rcu_read_lock()} and \co{rcu_read_unlock()} most of the
	time.

	However, this memory barrier is absolutely required so that
	other threads will see the store on
	\clnrefrange{gp1}{gp2} before any
	subsequent RCU read-side critical sections executed by the
	caller.
	\end{fcvref}
\QuickE{}
\QuickQ{}
	\begin{fcvref}[ln:defer:rcu_qs:read_lock_unlock:qs]
	Why are the two memory barriers on \clnref{mb1,mb2} of
	\cref{lst:app:toyrcu:Quiescent-State-Based RCU Read Side}
	needed?
	\end{fcvref}
\QuickA{}
	\begin{fcvref}[ln:defer:rcu_qs:read_lock_unlock:qs]
	The memory barrier on \clnref{mb1} prevents any RCU read-side
	critical sections that might precede the
	call to \co{rcu_thread_offline()} won't be reordered by either
	the compiler or the CPU to follow the assignment on
	\clnrefrange{gp1}{gp2}.
	The memory barrier on \clnref{mb2} is, strictly speaking, unnecessary,
	as it is illegal to have any RCU read-side critical sections
	following the call to \co{rcu_thread_offline()}.
	\end{fcvref}
\QuickE{}
\QuickQ{}
	To be sure, the clock frequencies of \Power{}
	systems in 2008 were quite high, but even a 5\,GHz clock
	frequency is insufficient to allow
	loops to be executed in 50~picoseconds!
	What is going on here?
\QuickA{}
	Since the measurement loop contains a pair of empty functions,
	the compiler optimizes it away.
	The measurement loop takes 1,000 passes between each call to
	\co{rcu_quiescent_state()}, so this measurement is roughly
	one thousandth of the overhead of a single call to
	\co{rcu_quiescent_state()}.
\QuickE{}
\QuickQ{}
	Why would the fact that the code is in a library make
	any difference for how easy it is to use the RCU
	implementation shown in
	\cref{lst:app:toyrcu:Quiescent-State-Based RCU Read Side,%
	lst:app:toyrcu:RCU Update Side Using Quiescent States}?
\QuickA{}
	A library function has absolutely no control over the caller,
	and thus cannot force the caller to invoke \co{rcu_quiescent_state()}
	periodically.
	On the other hand, a library function that made many references
	to a given RCU-protected data structure might be able to invoke
	\co{rcu_thread_online()} upon entry,
	\co{rcu_quiescent_state()} periodically, and
	\co{rcu_thread_offline()} upon exit.
\QuickE{}
\QuickQ{}
	But what if you hold a lock across a call to
	\co{synchronize_rcu()}, and then acquire that same lock within
	an RCU read-side critical section?
	This should be a deadlock, but how can a primitive that
	generates absolutely no code possibly participate in a
	deadlock cycle?
\QuickA{}
	Please note that the RCU read-side critical section is in
	effect extended beyond the enclosing
	\co{rcu_read_lock()} and \co{rcu_read_unlock()}, out to
	the previous and next call to \co{rcu_quiescent_state()}.
	This \co{rcu_quiescent_state} can be thought of as an
	\co{rcu_read_unlock()} immediately followed by an
	\co{rcu_read_lock()}.

	Even so, the actual deadlock itself will involve the lock
	acquisition in the RCU read-side critical section and
	the \co{synchronize_rcu()}, never the \co{rcu_quiescent_state()}.
\QuickE{}
\QuickQ{}
	Given that grace periods are prohibited within RCU read-side
	critical sections, how can an RCU data structure possibly be
	updated while in an RCU read-side critical section?
\QuickA{}
	This situation is one reason for the existence of asynchronous
	grace-period primitives such as \co{call_rcu()}.
	This primitive may be invoked within an RCU read-side critical
	section, and the specified RCU callback will in turn be invoked
	at a later time, after a grace period has elapsed.

	The ability to perform an RCU update while within an RCU read-side
	critical section can be extremely convenient, and is analogous
	to a (mythical) unconditional read-to-write upgrade for
	reader-writer locking.
\QuickE{}
\QuickQAC{chp:app:whymb:Why Memory Barriers?}{Why Memory Barriers?}{qqzwhymb}
\QuickQ{}
	Where does a writeback message originate from and where does
	it go to?
\QuickA{}
	The writeback message originates from a given CPU, or in some
	designs from a given level of a given CPU's cache---or even
	from a cache that might be shared among several CPUs.
	The key point is that a given cache does not have room for
	a given data item, so some other piece of data must be ejected
	from the cache to make room.
	If there is some other piece of data that is duplicated in some
	other cache or in memory, then that piece of data may be simply
	discarded, with no writeback message required.

	On the other hand, if every piece of data that might be ejected
	has been modified so that the only up-to-date copy is in this
	cache, then one of those data items must be copied somewhere
	else.
	This copy operation is undertaken using a ``writeback message''.

	The destination of the writeback message has to be something
	that is able to store the new value.
	This might be main memory, but it also might be some other cache.
	If it is a cache, it is normally a higher-level cache for the
	same CPU, for example, a level-1 cache might write back to a
	level-2 cache.
	However, some hardware designs permit cross-CPU writebacks,
	so that CPU~0's cache might send a writeback message to CPU~1.
	This would normally be done if CPU~1 had somehow indicated
	an interest in the data, for example, by having recently
	issued a read request.

	In short, a writeback message is sent from some part of the
	system that is short of space, and is received by some other
	part of the system that can accommodate the data.
\QuickE{}
\QuickQ{}
	What happens if two CPUs attempt to invalidate the
	same cache line concurrently?
\QuickA{}
	One of the CPUs gains access to the shared bus first,
	and that CPU ``wins''.
	The other CPU must invalidate its copy of the cache line and
	transmit an ``invalidate acknowledge'' message to the other CPU\@.

	Of course, the losing CPU can be expected to immediately issue a
	``read invalidate'' transaction, so the winning CPU's victory will
	be quite ephemeral.
\QuickE{}
\QuickQ{}
	When an ``invalidate'' message appears in a large multiprocessor,
	every CPU must give an ``invalidate acknowledge'' response.
	Wouldn't the resulting ``storm'' of ``invalidate acknowledge''
	responses totally saturate the system bus?
\QuickA{}
	It might, if large-scale multiprocessors were in fact implemented
	that way.
	Larger multiprocessors, particularly NUMA machines,
	tend to use so-called ``directory-based'' cache-coherence
	protocols to avoid this and other problems.
\QuickE{}
\QuickQ{}
	If SMP machines are really using message passing
	anyway, why bother with SMP at all?
\QuickA{}
	There has been quite a bit of controversy on this topic over
	the past few decades.
	One answer is that the cache-coherence
	protocols are quite simple, and therefore can be implemented
	directly in hardware, gaining bandwidths and latencies
	unattainable by software message passing.
	Another answer is that
	the real truth is to be found in economics due to the relative
	prices of large SMP machines and that of clusters of smaller
	SMP machines.
	A third answer is that the SMP programming model is easier to
	use than that of distributed systems, but a rebuttal might note
	the appearance of HPC clusters and MPI\@.
	And so the argument continues.
\QuickE{}
\QuickQ{}
	How does the hardware handle the delayed transitions
	described above?
\QuickA{}
	Usually by adding additional states, though these additional
	states need not be actually stored with the cache line, due to
	the fact that only a few lines at a time will be transitioning.
	The need to delay transitions is but one issue that results in
	real-world cache coherence protocols being much more complex than
	the over-simplified MESI protocol described in this appendix.
	Hennessy and Patterson's classic introduction to computer
	architecture~\cite{Hennessy95a} covers many of these issues.
\QuickE{}
\QuickQ{}
	What sequence of operations would put the CPUs' caches
	all back into the ``invalid'' state?
\QuickA{}
	There is no such sequence, at least in absence of special
	``flush my cache'' instructions in the CPU's instruction set.
	Most CPUs do have such instructions.
\QuickE{}
\QuickQ{}
	But then why do uniprocessors also have store buffers?
\QuickA{}
	Because the purpose of store buffers is not just to hide
	acknowledgement latencies in multiprocessor cache-coherence protocols,
	but to hide memory latencies in general.
	Because memory is much slower than is cache on uniprocessors,
	store buffers on uniprocessors can help to hide write-miss
	memory latencies.
\QuickE{}
\QuickQ{}
	So store-buffer entries are variable length?
	Isn't that difficult to implement in hardware?
\QuickA{}
	Here are two ways for hardware to easily handle variable-length
	stores.

	First, each store-buffer entry could be a single byte wide.
	Then an 64-bit store would consume eight store-buffer entries.
	This approach is simple and flexible, but one disadvantage is
	that each entry would need to replicate much of the address that
	was stored to.

	Second, each store-buffer entry could be double the size of a
	cache line, with half of the bits containing the values stored,
	and the other half indicating which bits had been stored to.
	So, assuming a 32-bit cache line, a single-byte store of 0x5a
	to the low-order byte of a given cache line would result in
	\co{0xXXXXXX5a} for the first half and \co{0x000000ff} for the
	second half, where the values labeled \co{X} are arbitrary
	because they would be ignored.
	This approach allows multiple consecutive stores corresponding to
	a given cache line to be merged into a single store-buffer entry,
	but is space-inefficient for random stores of single bytes.

	Much more complex and efficient schemes are of course used
	by actual hardware designers.
\QuickE{}
\QuickQ{}
	In \cref{seq:app:whymb:Store Buffers and Memory Barriers} above,
	why does CPU~0 need to issue a ``read invalidate''
	rather than a simple ``invalidate''?
	After all, \co{foo()} will overwrite the variable \co{a} in any
	case, so why should it care about the old value of \co{a}?
\QuickA{}
	Because the cache line in question contains more data than just the
	variable \co{a}.
	Issuing ``invalidate'' instead of the needed ``read invalidate''
	would cause that other data to be lost, which would constitute
	a serious bug in the hardware.
\QuickE{}
\QuickQ{}
	In \cref{seq:app:whymb:Store Buffers and Memory Barriers store}
	above, don't systems avoid that store to memory?
\QuickA{}
	Yes, they do.
	But to do so, they add states beyond the MESI quadruple that
	this example is working within.
\QuickE{}
\QuickQ{}
	In \cref{seq:app:whymb:Store Buffers and Memory Barriers victim}
	above, did \co{bar()} read a stale value from \co{a}, or did
	its reads of \co{b} and \co{a} get reordered?
\QuickA{}
	It could be either, depending on the hardware implementation.
	And it really does not matter which.
	After all, the \co{bar()} function's \co{assert()} cannot tell
	the difference!
\QuickE{}
\QuickQ{}
	After \cref{seq:app:whymb:Store buffers: All copies shared}
	in \cref{sec:app:whymb:Store Buffers and Memory Barriers} on
	\cpageref{seq:app:whymb:Store buffers: All copies shared},
	both CPUs might drop the cache line containing the new value of
	``b''.
	Wouldn't that cause this new value to be lost?
\QuickA{}
	It might, and that is why real hardware takes steps to avoid
	this problem.
	A traditional approach, pointed out by Vasilevsky Alexander,
	is to write this cache line back to main memory before marking
	the cache line as ``shared''.
	A more efficient (though more complex) approach is to use
	additional state to indicate whether or not the cache line
	is ``dirty'', allowing the writeback to happen.
	Year-2000 systems went further, using much more state in order to
	avoid redundant writebacks~\cite[Figure 8.42]{DavidECuller1999}.
	It would be reasonable to assume that complexity has not decreased
	in the meantime.
\QuickE{}
\QuickQ{}
	In \cref{seq:app:whymb:Invalidate Queues and Memory Barriers}
	of the first scenario in
	\cref{sec:app:whymb:Invalidate Queues and Memory Barriers},
	why is an ``invalidate'' sent instead of a ''read invalidate''
	message?
	Doesn't CPU~0 need the values of the other variables that share
	this cache line with ``a''?
\QuickA{}
	CPU~0 already has the values of these variables, given that it
	has a read-only copy of the cache line containing ``a''.
	Therefore, all CPU~0 need do is to cause the other CPUs to discard
	their copies of this cache line.
	An ``invalidate'' message therefore suffices.
\QuickE{}
\QuickQ{}
	Say what???
	Why do we need a memory barrier here, given that the CPU cannot
	possibly execute the \co{assert()} until after the
	\co{while} loop completes?
\QuickA{}
	Suppose that memory barrier was omitted.

	Keep in mind that CPUs are free to speculatively execute later
	loads, which can have the effect of executing the assertion
	before the \co{while} loop completes.
	Furthermore, compilers assume that only the currently executing
	thread is updating the variables, and this assumption allows
	the compiler to hoist the load of \co{a} to precede the
	loop.

	In fact, some compilers would transform the loop to a branch
	around an infinite loop as follows:

\begin{VerbatimN}[fontsize=\footnotesize,samepage=true]
void foo(void)
{
	a = 1;
	smp_mb();
	b = 1;
}

void bar(void)
{
	if (b == 0)
		for (;;)
			continue;
	assert(a == 1);
}
\end{VerbatimN}

	Given this optimization, the code would behave in a completely
	different way than the original code.
	If \co{bar()} observed \qco{b == 0}, the assertion could of
	course not be reached at all due to the infinite loop.
	However, if \co{bar()} loaded the value \qco{1} just as
	\qco{foo()} stored it, the CPU might still have the old
	zero value of \qco{a} in its cache, which would cause
	the assertion to fire.
	You should of course use volatile casts (for example, those
	volatile casts implied by the C11 relaxed atomic load operation)
	to prevent the compiler from optimizing your parallel code
	into oblivion.
	But volatile casts would not prevent a weakly ordered CPU
	from loading the old value for \qco{a} from its cache, which
	means that this code also requires the explicit memory barrier
	in \qco{bar()}.

	In short, both compilers and CPUs aggressively apply
	code-reordering optimizations, so you must clearly communicate
	your constraints using the compiler directives and memory barriers
	provided for this purpose.
\QuickE{}
\QuickQ{}
	Instead of all of this marking of invalidation-queue entries
	and stalling of loads, why not simply force an immediate flush
	of the invalidation queue?
\QuickA{}
	An immediate flush of the invalidation queue would do the trick.
	Except that the common-case super-scalar CPU is executing many
	instructions at once, and not necessarily even in the expected
	order.
	So what would ``immediate'' even mean?
	The answer is clearly ``not much''.

	Nevertheless, for simpler CPUs that execute instructions serially,
	flushing the invalidation queue might be a reasonable implementation
	strategy.
\QuickE{}
\QuickQ{}
	But can't full memory barriers impose global ordering?
	After all, isn't that needed to provide the ordering
	shown in \cref{lst:formal:IRIW Litmus Test}?
\QuickA{}
	Sort of.

	Note well that this litmus test has not one but two full
	memory-barrier instructions, namely the two \co{sync} instructions
	executed by \co{P2} and \co{P3}.

	It is the interaction of those two instructions that provides
	the global ordering, not just their individual execution.
	For example, each of those two \co{sync} instructions might stall
	waiting for all CPUs to process their invalidation queues before
	allowing subsequent instructions to execute.\footnote{
		Real-life hardware of course applies many optimizations
		to minimize the resulting stalls.}
\QuickE{}
\QuickQ{}
	Does the guarantee that each CPU sees its own memory accesses
	in order also guarantee that each user-level thread will see
	its own memory accesses in order?
	Why or why not?
\QuickA{}
	No.
	Consider the case where a thread migrates from one CPU to
	another, and where the destination CPU perceives the source
	CPU's recent memory operations out of order.
	To preserve user-mode sanity, kernel hackers must use memory
	barriers in the context-switch path.
	However, the locking already required to safely do a context
	switch should automatically provide the memory barriers needed
	to cause the user-level task to see its own accesses in order.
	That said, if you are designing a super-optimized scheduler,
	either in the kernel or at user level,
	please keep this scenario in mind!
\QuickE{}
\QuickQ{}
	Could this code be fixed by inserting a memory barrier
	between CPU~1's ``while'' and assignment to ``c''?
	Why or why not?
\QuickA{}
	No.
	Such a memory barrier would only force ordering local to CPU~1.
	It would have no effect on the relative ordering of CPU~0's and
	CPU~1's accesses, so the assertion could still fail.
	However, all mainstream computer systems provide one mechanism
	or another to provide ``transitivity'', which provides
	intuitive causal ordering:
	If B saw the effects of A's accesses, and C saw the effects of
	B's accesses, then C must also see the effects of A's accesses.
	In short, hardware designers have taken at least a little pity
	on software developers.
\QuickE{}
\QuickQ{}
	Suppose that lines~3--5 for CPUs~1 and~2 in
	\cref{lst:app:whymb:Memory Barrier Example 3}
	are in an interrupt
	handler, and that the CPU~2's line~9 runs at process level.
	In other words, the code in all three columns of the table
	runs on the same CPU, but the first two columns run in an
	interrupt handler, and the third column runs at process
	level, so that the code in third column can be interrupted
	by the code in the first two columns.
	What changes, if any, are required to enable the code to work
	correctly, in other words, to prevent the assertion from firing?
\QuickA{}
	The assertion must ensure that the load of
	``e'' precedes that of ``a''.
	In the Linux kernel, the \co{barrier()} primitive may be used to
	accomplish this in much the same way that the memory barrier was
	used in the assertions in the previous examples.
	For example, the assertion can be modified as follows:

\begin{VerbatimU}[fontsize=\footnotesize]
r1 = e;
barrier();
assert(r1 == 0 || a == 1);
\end{VerbatimU}

	No changes are needed to the code in the first two columns,
	because interrupt handlers run atomically from the perspective
	of the interrupted code.
\QuickE{}
\QuickQ{}
	If CPU~2 executed an \co{assert(e==0||c==1)} in the example in
	\cref{lst:app:whymb:Memory Barrier Example 3},
	would this assert ever trigger?
\QuickA{}
	The result depends on whether the CPU supports ``transitivity''.
	In other words, CPU~0 stored to ``e'' after seeing CPU~1's
	store to ``c'', with a memory barrier between CPU~0's load
	from ``c'' and store to ``e''.
	If some other CPU sees CPU~0's store to ``e'', is it also
	guaranteed to see CPU~1's store?

	All CPUs I am aware of claim to provide transitivity.
\QuickE{}

	\setlength{\parskip}{0.0pt plus 1.0pt}% return to default

\backmatter

% Glossary
%% Reset footnote count
\setcounter{footnote}{0}
% glossary.tex
% mainfile: perfbook.tex
% SPDX-License-Identifier: CC-BY-SA-3.0

\chapter{Glossary}
\Epigraph{Dictionaries are inherently circular in nature.}
	 {\emph{Self Reference in word definitions},
	        David~Levary~et~al.}

\begin{description}
\item[\IXG{Acquire Load}:]
	A read from memory that has acquire semantics.
	Normal use cases pair an acquire load with a release store,
	in which case if the load returns the value stored, then all
	code executed by the loading CPU after that acquire load will
	see the effects of all memory-reference instructions executed
	by the storing CPU prior to that release store.
	Acquiring a lock provides similar memory-ordering semantics,
	hence the ``acquire'' in ``acquire load''.
	(See also ``memory barrier'' and ``release store''.)
\item[\IXGr{Amdahl's Law}:]
	If sufficient numbers of CPUs are used to run a job that has both
	a sequential portion and a concurrent portion, performance and
	scalability will be limited by the overhead of the sequential
	portion.
\item[\IXGalt{Associativity}{Cache associativity}:]
	The number of cache lines that can be held simultaneously in
	a given cache, when all of these cache lines hash identically
	in that cache.
	A cache that could hold four cache lines for each possible
	hash value would be termed a ``four-way set-associative'' cache,
	while a cache that could hold only one cache line for each
	possible hash value would be termed a ``direct-mapped'' cache.
	A cache whose associativity was equal to its capacity would
	be termed a ``fully associative'' cache.
	Fully associative caches have the advantage of eliminating
	associativity misses, but, due to hardware limitations,
	fully associative caches are normally quite limited in size.
	The associativity of the large caches found on modern microprocessors
	typically range from two-way to eight-way.
\item[\IXGalth{Associativity Miss}{associativity}{cache miss}:]
	A cache miss incurred because the corresponding CPU has recently
	accessed more data hashing to a given set of the cache than will
	fit in that set.
	Fully associative caches are not subject to associativity misses
	(or, equivalently, in fully associative caches, associativity
	and capacity misses are identical).
\item[\IXG{Atomic}:]
	An operation is considered ``atomic'' if it is not possible to
	observe any intermediate state.
	For example, on most CPUs, a store to a properly aligned pointer
	is atomic, because other CPUs will see either the old value or
	the new value, but are guaranteed not to see some mixed value
	containing some pieces of the new and old values.
\item[\IXG{Atomic Read-Modify-Write Operation}:]
	An atomic operation that both reads and writes memory is
	considered an atomic read-modify-write operation, or atomic RMW
	operation for short.
	Although the value written usually depends on the value read,
	\co{atomic_xchg()} is the exception that proves this rule.
\item[\IXGh{Bounded}{Wait Free}:]
	A forward-progress guarantee in which every thread makes
	progress within a specific finite period of time, the specific
	time being the bound.
\item[\IXGh{Bounded Population-Oblivious}{Wait Free}:]
	A forward-progress guarantee in which every thread makes
	progress within a specific finite period of time, the specific
	time being the bound, where this bound is independent of the
	number of threads.
\item[\IXG{Cache}:]
	In modern computer systems, CPUs have caches in which to hold
	frequently used data.
	These caches can be thought of as hardware hash tables with very
	simple hash functions,
	but in which each hash bucket (termed a ``set'' by hardware types)
	can hold only a limited number of data items.
	The number of data items that can be held by each of a cache's hash
	buckets is termed the cache's ``associativity''.
	These data items are normally called ``cache lines'', which
	can be thought of a fixed-length blocks of data that circulate
	among the CPUs and memory.
\item[\IXG{Cache Coherence}:]
	A property of most modern SMP machines where all CPUs will
	observe a sequence of values for a given variable that is
	consistent with at least one global order of values for
	that variable.
	Cache coherence also guarantees that at the end of a group
	of stores to a given variable, all CPUs will agree
	on the final value for that variable.
	Note that cache coherence applies only to the series of values
	taken on by a single variable.
	In contrast, the memory consistency model for a given machine
	describes the order in which loads and stores to groups of
	variables will appear to occur.
	See \cref{sec:memorder:Cache Coherence}
	for more information.
\item[\IXG{Cache-Coherence Protocol}:]
	A communications protocol, normally implemented in hardware,
	that enforces memory consistency and ordering, preventing
	different CPUs from seeing inconsistent views of data held
	in their caches.
\item[\IXG{Cache Geometry}:]
	The size and associativity of a cache is termed its geometry.
	Each cache may be thought of as a two-dimensional array,
	with rows of cache lines (``sets'') that have the same hash
	value, and columns of cache lines (``ways'') in which every
	cache line has a different hash value.
	The associativity of a given cache is its number of
	columns (hence the name ``way''---a two-way set-associative
	cache has two ``ways''), and the size of the cache is its
	number of rows multiplied by its number of columns.
\item[\IXG{Cache Line}:]
	(1) The unit of data that circulates among the CPUs and memory,
	usually a moderate power of two in size.
	Typical cache-line sizes range from 16 to 256 bytes. \\
	(2) A physical location in a CPU cache capable of holding
	one cache-line unit of data. \\
	(3) A physical location in memory capable of holding one
	cache-line unit of data, but that it also aligned
	on a cache-line boundary.
	For example, the address of the first word of a cache line
	in memory will end in 0x00 on systems with 256-byte cache lines.
\item[\IXG{Cache Miss}:]
	A cache miss occurs when data needed by the CPU is not in
	that CPU's cache.
	The data might be missing because of a number of reasons,
	including:
	\begin{enumerate*}[(1)]
	\item This CPU has never accessed the data before
	(``startup'' or ``warmup'' miss),
	\item This CPU has recently accessed more
	data than would fit in its cache, so that some of the older
	data had to be removed (``capacity'' miss),
	\item This CPU
	has recently accessed more data in a given set\footnote{
		In hardware-cache terminology, the word ``set''
		is used in the same way that the word ``bucket''
		is used when discussing software caches.}
	than that set could hold (``associativity'' miss),
	\item Some other CPU has written to the data (or some other
	data in the same cache line) since this CPU has accessed it
	(``communication miss''), or
	\item This CPU attempted to write to a cache line that is
	currently read-only, possibly due to that line being replicated
	in other CPUs' caches.
	\end{enumerate*}
\item[\IXGalth{Capacity Miss}{capacity}{cache miss}:]
	A cache miss incurred because the corresponding CPU has recently
	accessed more data than will fit into the cache.
\item[CAS:]\glsuseriii{cas}
	Compare-and-swap operation, which is an atomic operation that
	takes a pointer, and old value, and a new value.
	If the pointed-to value is equal to the old value, it is atomically
	replaced with the new value.
	There is some variety in CAS API\@.
	One variation returns the actual pointed-to value, so that the
	caller compares the CAS return value to the specified old value,
	with equality indicating a successful CAS operation.
	Another variation returns a boolean success indication, in which
	case a pointer to the old value may be passed in, and if so,
	the old value is updated in the CAS failure case.
\item[\IXG{Clash Free}:]
	A forward-progress guarantee in which, in the absence of
	contention, at least one thread makes progress within a finite
	period of time.
\item[\IXGalth{Code Locking}{code}{locking}:]
	A simple locking design in which a ``global lock'' is used to protect
	a set of critical sections, so that access by a given thread
	to that set is
	granted or denied based only on the set of threads currently
	occupying the set of critical sections, not based on what
	data the thread intends to access.
	The scalability of a code-locked program is limited by the code;
	increasing the size of the data set will normally not increase
	scalability (in fact, will typically \emph{decrease} scalability
	by increasing ``lock contention'').
	Contrast with ``data locking''.
\item[\IXG{Combinatorial Explosion}:]
	Denotes the exponential increase in executions that
	formal-verification tools must analyze as problem size increases.
\item[\IXG{Combinatorial Implosion}:]
	Denotes the exponential decrease in executions that
	formal-verification tools must analyze when a given
	code fragment is partitioned.
\item[\IXGalth{Communication Miss}{communication}{cache miss}:]
	A cache miss incurred because some other CPU has written to
	the cache line since the last time this CPU accessed it.
\item[\IXG{Concurrent}:]
	In this book, a synonym of parallel.
	Please see \cref{sec:app:questions:What is the Difference Between ``Concurrent'' and ``Parallel''?}
	on \cpageref{sec:app:questions:What is the Difference Between ``Concurrent'' and ``Parallel''?}
	for a discussion of the recent distinction between these two
	terms.
\item[\IXG{Critical Section}:]
	A section of code guarded by some synchronization mechanism,
	so that its execution constrained by that primitive.
	For example, if a set of critical sections are guarded by
	the same global lock, then only one of those critical sections
	may be executing at a given time.
	If a thread is executing in one such critical section,
	any other threads must wait until the first thread completes
	before executing any of the critical sections in the set.
\item[\IXGh{Data}{Locking}:]
	A scalable locking design in which each instance of a given
	data structure has its own lock.
	If each thread is using a different instance of the
	data structure, then all of the threads may be executing in
	the set of critical sections simultaneously.
	Data locking has the advantage of automatically scaling to
	increasing numbers of CPUs as the number of instances of
	data grows.
	Contrast with ``code locking''.
\item[\IXG{Data Race}:]
	A race condition in which several CPUs or threads access
	a variable concurrently, and in which at least one of those
	accesses is a store and at least one of those accesses
	is a plain access.
	It is important to note that while the presence of data races
	often indicates the presence of bugs, the absence of data races
	in no way implies the absence of bugs.
	(See ``Plain access'' and ``Race condition''.)
\item[\IXG{Deadlock}:]
	A failure mode in which each of several threads is unable to
	make progress until some other thread makes progress.
	For example, if two threads acquire a pair of locks in opposite
	orders, deadlock can result.
	More information is provided in
	\cref{sec:locking:Deadlock}.
\item[\IXG{Deadlock Free}:]
	A forward-progress guarantee in which, in the absence of
	failures, at least one thread makes progress within a finite
	period of time.
\item[\IXGh{Direct-Mapped}{Cache}:]
	A cache with only one way, so that it may hold only one cache
	line with a given hash value.
\item[\IXG{Efficiency}:]
	A measure of effectiveness normally expressed as a ratio
	of some metric actually achieved to some maximum value.
	The maximum value might be a theoretical maximum, but in
	parallel programming is often based on the corresponding
	measured single-threaded metric.
\item[\IXG{Embarrassingly Parallel}:]
	A problem or algorithm where adding threads does not significantly
	increase the overall cost of the computation, resulting in
	linear speedups as threads are added (assuming sufficient
	CPUs are available).
\item[\IXGalth{Energy Efficiency}{energy}{efficiency}:]
	Shorthand for ``energy-efficient use'' in which the goal is to
	carry out a given computation with reduced energy consumption.
	Sublinear scalability can be an obstacle to energy-efficient use
	of a multicore system.
\item[Epoch-Based Reclamation (EBR):]\glsuseriii{ebr}
	An \acr{rcu} implementation style put forward by
	\ppl{Keir}{Fraser}~\cite{KeirAnthonyFraserPhD,UCAM-CL-TR-579,KeirFraser2007withoutLocks}.
\item[\IXG{Existence Guarantee}:]
	An existence guarantee is provided by a synchronization mechanism
	that prevents a given dynamically allocated object from being
	freed for the duration of that guarantee.
	For example, \acr{rcu} provides existence guarantees for the duration
	of \acr{rcu} read-side critical sections.
	A similar but strictly weaker guarantee is provided by
	type-safe memory.
\item[\IXGh{Exclusive}{Lock}:]
	An exclusive lock is a mutual-exclusion mechanism that
	permits only one thread at a time into the
	set of critical sections guarded by that lock.
\item[\IXG{False Sharing}:]
	If two CPUs each frequently write to one of a pair of data items,
	but the pair of data items are located in the same cache line,
	this cache line will be repeatedly invalidated, ``ping-ponging''
	back and forth between the two CPUs' caches.
	This is a common cause of ``cache thrashing'', also called
	``cacheline bouncing'' (the latter most commonly in the Linux
	community).
	False sharing can dramatically reduce both performance and
	scalability.
\item[\IXG{Forward-Progress Guarantee}:]
	Algorithms or programs that guarantee that execution will
	progress at some rate under specified conditions.
	Academic forward-progress guarantees are grouped into a
	formal hierarchy shown in
	\cref{sec:advsync:Non-Blocking Synchronization}.
	A wide variety of practical forward-progress guarantees are
	provided by real-time systems, as discussed in
	\cref{sec:advsync:Parallel Real-Time Computing}.
\item[\IXG{Fragmentation}:]
	A memory pool that has a large amount of unused memory, but
	not laid out to permit satisfying a relatively small request
	is said to be fragmented.
	External fragmentation occurs when the space is divided up
	into small fragments lying between allocated blocks of memory,
	while internal fragmentation occurs when specific requests or
	types of requests have been allotted more memory than they
	actually requested.
\item[\IXGh{Fully Associative}{Cache}:]
	A fully associative cache contains only
	one set, so that it can hold any subset of
	memory that fits within its capacity.
\item[\IXG{Grace Period}:]
	A grace period is any contiguous time interval such that
	any \acr{rcu} read-side critical section that began before the
	start of that interval has
	completed before the end of that same interval.
	Many \acr{rcu} implementations define a grace period to be a
	time interval during which each thread has passed through at
	least one quiescent state.
	Since \acr{rcu} read-side critical sections by definition cannot
	contain quiescent states, these two definitions are almost
	always interchangeable.
\item[Hardware Transactional Memory (HTM):]\glsuseriii{htm}
	A transactional-memory system based on hardware instructions
	provided for this purpose, as discussed in
	\cref{sec:future:Hardware Transactional Memory}.
	(See ``Transactional memory''.)
\item[\IXG{Hazard Pointer}:]
	A scalable counterpart to a reference counter in which an
	object's reference count is represented implicitly by a count
	of the number of special hazard pointers referencing that object.
\item[\IXG{Heisenbug}:]
	A timing-sensitive bug that disappears from sight when you
	add print statements or tracing in an attempt to track it
	down.
\item[\IXG{Hot Spot}:]
	Data structure that is very heavily used, resulting in high
	levels of contention on the corresponding lock.
	One example of this situation would be a hash table with
	a poorly chosen hash function.
\item[\IXG{Humiliatingly Parallel}:]
	A problem or algorithm where adding threads significantly
	\emph{decreases} the overall cost of the computation, resulting in
	large superlinear speedups as threads are added (assuming sufficient
	CPUs are available).
\item[\IXG{Immutable}:]
	In this book, a synonym for read-mostly.
\item[\IXG{Invalidation}:]
	When a CPU wishes to write to a data item, it must first ensure
	that this data item is not present in any other CPUs' cache.
	If necessary, the item is removed from the other CPUs' caches
	via ``invalidation'' messages from the writing CPUs to any
	CPUs having a copy in their caches.
\item[IPI:]\glsuseriii{ipi}
	Inter-processor interrupt, which is an
	interrupt sent from one CPU to another.
	IPIs are used heavily in the Linux kernel, for example, within
	the scheduler to alert CPUs that a high-priority process is now
	runnable.
\item[IRQ:]\glsuseriii{irq}
	Interrupt request, often used as an abbreviation for ``interrupt''
	within the Linux kernel community, as in ``irq handler''.
\item[\IXG{Latency}:]
	The wall-clock time required for a given operation to complete.
\item[\IXG{Linearizable}:]
	A sequence of operations is ``linearizable'' if there is at
	least one global ordering of the sequence that is consistent
	with the observations of all CPUs and/or threads.
	Linearizability is much prized by many researchers, but less
	useful in practice than one might
	expect~\cite{AndreasHaas2012FIFOisnt}.
\item[\IXG{Livelock}:]
	A failure mode in which each of several threads is able to
	execute, but in which a repeating series of failed operations
	prevents any of the threads from making any useful forward progress.
	For example, incorrect use of conditional locking
	(for example, \co{spin_trylock()} in the Linux kernel)
	can result in livelock.
	More information is provided in
	\cref{sec:locking:Livelock and Starvation}.
\item[\IXG{Lock}:]
	A software abstraction that can be used to guard critical sections,
	as such, an example of a ``mutual exclusion mechanism''.
	An ``exclusive lock'' permits only one thread at a time into the
	set of critical sections guarded by that lock, while a
	``reader-writer lock'' permits any number of reading
	threads, or but one writing thread, into the set of critical
	sections guarded by that lock.
	(Just to be clear, the presence	of a writer thread in any of
	a given reader-writer lock's critical sections will prevent
	any reader from entering any of that lock's critical sections
	and vice versa.)
\item[\IXG{Lock Contention}:]
	A lock is said to be suffering contention when it is being
	used so heavily that there is often a CPU waiting on it.
	Reducing lock contention is often a concern when designing
	parallel algorithms and when implementing parallel programs.
\item[\IXG{Lock Free}:]
	A forward-progress guarantee in which at least one thread makes
	progress within a finite period of time.
\item[\IXG{Marked Access}:]
	A source-code memory access that uses a special function or
	macro, such as \co{READ_ONCE()}, \co{WRITE_ONCE()},
	\co{atomic_inc()}, and so on, in order to protect that access
	from compiler and/or hardware optimizations.
	In contrast, a plain access simply mentions the name of
	the object being accessed, so that in the following, line~2
	is the plain-access equivalent of line~1:
	\begin{VerbatimN}
	WRITE_ONCE(a, READ_ONCE(b) + READ_ONCE(c));
	a = b + c;
	\end{VerbatimN}
\item[\IXG{Memory}:]
	From the viewpoint of memory models, the main memory,
	caches, and store buffers in which values might be stored.
	However, this term is often used to denote the main memory
	itself, excluding caches and store buffers.
\item[\IXG{Memory Barrier}:]
	A compiler directive that might also include a special
	memory-barrier instruction.
	The purpose of a memory barrier is to order memory-reference
	instructions that executed before the memory barrier to precede
	those that will execute following that memory barrier.
	(See also ``read memory barrier'' and ``write memory barrier''.)
\item[\IXGh{Memory}{Consistency}:]
	A set of properties that impose constraints on the order in
	which accesses to groups of variables appear to occur.
	Memory consistency models range from sequential consistency,
	a very constraining model popular in academic circles, through
	process consistency, release consistency, and weak consistency.
\item[\IXGaltr{MESI Protocol}{MESI protocol}:]
	The
	cache-coherence protocol featuring
	modified, exclusive, shared, and invalid (MESI) states,
	so that this protocol is named after the states that the
	cache lines in a given cache can take on.
	A modified line has been recently written to by this CPU,
	and is the sole representative of the current value of
	the corresponding memory location.
	An exclusive cache line has not been written to, but this
	CPU has the right to write to it at any time, as the line
	is guaranteed not to be replicated into any other CPU's cache
	(though the corresponding location in main memory is up to date).
	A shared cache line is (or might be) replicated in some other
	CPUs' cache, meaning that this CPU must interact with those other
	CPUs before writing to this cache line.
	An invalid cache line contains no value, instead representing
	``empty space'' in the cache into which data from memory might
	be loaded.
\item[\IXGaltr{Moore's Law}{Moore's Law}:]
	A 1965 empirical projection by Gordon Moore that
	transistor density increases exponentially over
	time~\cite{GordonMoore1965MooresLaw}.
\item[\IXG{Mutual-Exclusion Mechanism}:]
	A software abstraction that regulates threads' access to
	``critical sections'' and corresponding data.
\item[NMI:]\glsuseriii{nmi}
	Non-maskable interrupt.
	As the name indicates, this is an extremely high-priority
	interrupt that cannot be masked.
	These are used for hardware-specific purposes such as profiling.
	The advantage of using NMIs for profiling is that it allows you
	to profile code that runs with interrupts disabled.
\item[\IXG{Non-Blocking}:]
	A group of academic forward-progress guarantees that includes
	bounded population-oblivious wait free,
	bounded wait free,
	wait free,
	lock free,
	obstruction free,
	clash free,
	starvation free, and
	deadlock free.
	See \cref{sec:advsync:Non-Blocking Synchronization}
	for more information.
\item[Non-Blocking Synchronization (NBS):]\glsuseriii{nbs}
	The use of algorithms, mechanisms, or techniques that provide
	non-blocking forward-progress guarantees.
	NBS is often used in a more restrictive sense of providing one
	of the stronger forward-progress guarantees, usually wait free or
	lock free, but sometimes also obstruction free.
	(See ``Non-blocking''.)
\item[NUCA:]\glsuseriii{nuca}
	Non-uniform cache architecture, where groups of CPUs share
	caches and/or store buffers.
	CPUs in a group can therefore exchange cache lines with each
	other much more quickly than they can with CPUs in other groups.
	Systems comprised of CPUs with hardware threads will generally
	have a NUCA architecture.
\item[NUMA:]\glsuseriii{numa}
	Non-uniform memory architecture, where memory is split into
	banks and each such bank is ``close'' to a group of CPUs,
	the group being termed a ``NUMA node''.
	An example NUMA machine is Sequent's NUMA-Q system, where
	each group of four CPUs had a bank of memory nearby.
	The CPUs in a given group can access their memory much
	more quickly than another group's memory.
\item[\IXGaltr{NUMA Node}{NUMA node}:]
	A group of closely placed CPUs and associated memory within
	a larger NUMA machines.
\item[\IXG{Obstruction Free}:]
	A forward-progress guarantee in which, in the absence of
	contention, every thread makes progress within a finite
	period of time.
\item[\IXG{Overhead}:]
	Operations that must be executed, but which do not contribute
	directly to the work that must be accomplished.
	For example, lock acquisition and release is normally considered
	to be overhead, and specifically to be synchronization overhead.
\item[\IXG{Parallel}:]
	In this book, a synonym of concurrent.
	Please see \cref{sec:app:questions:What is the Difference Between ``Concurrent'' and ``Parallel''?}
	on \cpageref{sec:app:questions:What is the Difference Between ``Concurrent'' and ``Parallel''?}
	for a discussion of the recent distinction between these two
	terms.
\item[\IXG{Performance}:]
	Rate at which work is done, expressed as work per unit time.
	If this work is fully serialized, then the performance will
	be the reciprocal of the mean latency of the work items.
\item[\IXGr{Pipelined CPU}:]
	A CPU with a pipeline, which is
	an internal flow of instructions internal to the CPU that
	is in some way similar to an assembly line, with many of
	the same advantages and disadvantages.
	In the 1960s through the early 1980s, pipelined CPUs were the
	province of supercomputers, but started appearing in microprocessors
	(such as the 80486) in the late 1980s.
\item[\IXG{Plain Access}:]
	A source-code memory access that simply mentions the name of
	the object being accessed.
	(See ``Marked access''.)
\item[\IXGh{Process}{Consistency}:]
	A memory-consistency model in which each CPU's stores appear to
	occur in program order, but in which different CPUs might see
	accesses from more than one CPU as occurring in different orders.
\item[\IXG{Program Order}:]
	The order in which a given thread's instructions
	would be executed by a now-mythical ``in-order'' CPU that
	completely executed each instruction before proceeding to
	the next instruction.
	(The reason such CPUs are now the stuff of ancient myths
	and legends is that they were extremely slow.
	These dinosaurs were one of the many victims of
	Moore's-Law-driven increases in CPU clock frequency.
	Some claim that these beasts will roam the earth once again,
	others vehemently disagree.)
\item[\IXG{Quiescent State}:]
	In \acr{rcu}, a point in the code where there can be no references held
	to \acr{rcu}-protected data structures, which is normally any point
	outside of an \acr{rcu} read-side critical section.
	Any interval of time during which all threads pass through at
	least one quiescent state each is termed a ``grace period''.
\item[Quiescent-State-Based Reclamation (QSBR):]\glsuseriii{qsbr}
	An \acr{rcu} implementation style characterized by explicit quiescent
	states.
	In \acr{qsbr} implementations, read-side markers
	(\co{rcu_read_lock()} and \co{rcu_read_unlock()} in the Linux
	kernel) are no-ops~\cite{McKenney98,Slingwine95}.
	Hooks in other parts of the software (for example, the Linux-kernel
	scheduler) provide the quiescent states.
\item[\IXG{Race Condition}:]
	Any situation where multiple CPUs or threads can interact,
	though this term is often used in cases where such interaction
	is undesirable.
	(See ``Data race''.)
\item[\IXGaltr{RCU-Protected Data}{RCU-protected data}:]
	A block of dynamically allocated memory whose freeing will be
	deferred such that an \acr{rcu} grace period will elapse between the
	time that there were no longer any \acr{rcu}-reader-accessible pointers
	to that block and the time that that block is freed.
	This ensures that no \acr{rcu} readers will have access to that block at
	the time that it is freed.
\item[\IXGaltr{RCU-Protected Pointer}{RCU-protected pointer}:]
	A pointer to \acr{rcu}-protected data.
	Such pointers must be handled carefully, for example, any reader
	that intends to dereference an \acr{rcu}-protected pointer must
	use \co{rcu_dereference()} (or stronger) to load that pointer,
	and any updater must use \co{rcu_assign_pointer()} (or stronger)
	to store to that pointer.
	More information is provided in
	\cref{sec:memorder:Address- and Data-Dependency Difficulties}.
\item[\IXGalthmr{RCU Read-Side Critical Section}{RCU read-side}{critical section}:]
	A section of code protected by \acr{rcu}, for example, beginning with
	\co{rcu_read_lock()} and ending with \co{rcu_read_unlock()}.
	(See ``Read-side critical section''.)
\item[Read-Copy Update (RCU):]\glsuseriii{rcu}
	A synchronization mechanism that can be thought of as a replacement
	for reader-writer locking or reference counting.
	RCU provides extremely low-overhead access for readers, while
	writers incur additional overhead maintaining old versions
	for the benefit of pre-existing readers.
	Readers neither block nor spin, and thus cannot participate in
	deadlocks, however, they also can see stale data and can
	run concurrently with updates.
	RCU is thus best-suited for read-mostly situations where
	stale data can either be tolerated (as in routing tables)
	or avoided (as in the Linux kernel's System V IPC implementation).
\item[\IXGh{Read}{Memory Barrier}:]
	A memory barrier that is only guaranteed to affect the ordering
	of load instructions, that is, reads from memory.
	(See also ``memory barrier'' and ``write memory barrier''.)
\item[\IXG{Read Mostly}:]
	Read-mostly data is (again, as the name implies) rarely updated.
	However, it might be updated at any time.
\item[\IXG{Read Only}:]
	Read-only data is, as the name implies, never updated except
	by beginning-of-time initialization.
	In this book, a synonym for immutable.
\item[\IXGh{Read-Side}{Critical Section}:]
	A section of code guarded by read-acquisition of
	some reader-writer synchronization mechanism.
	For example, if one set of critical sections are guarded by
	read-acquisition of
	a given global reader-writer lock, while a second set of critical
	section are guarded by write-acquisition of that same reader-writer
	lock, then the first set of critical sections will be the
	read-side critical sections for that lock.
	Any number of threads may concurrently execute the read-side
	critical sections, but only if no thread is executing one of
	the write-side critical sections.
	(See also ``RCU read-side critical section''.)
\item[\IXGh{Reader-Writer}{Lock}:]
	A reader-writer lock is a mutual-exclusion mechanism that
	permits any number of reading
	threads, or but one writing thread, into the set of critical
	sections guarded by that lock.
	Threads attempting to write must wait until all pre-existing
	reading threads release the lock, and, similarly, if there
	is a pre-existing writer, any threads attempting to write must
	wait for the writer to release the lock.
	A key concern for reader-writer locks is ``fairness'':
	Can an unending stream of readers starve a writer or vice versa?
\item[\IXG{Real Time}:]
	A situation in which getting the correct result is not sufficient,
	but where this result must also be obtained within a given amount
	of time.
\item[\IXG{Reference Count}:]
	A counter that tracks the number of users of a given object or
	entity.
	Reference counters provide existence guarantees and are sometimes
	used to implement garbage collectors.
\item[\IXG{Release Store}:]
	A write to memory that has release semantics.
	Normal use cases pair an acquire load with a release store,
	in which case if the load returns the value stored, then all
	code executed by the loading CPU after that acquire load will
	see the effects of all memory-reference instructions executed
	by the storing CPU prior to that release store.
	Releasing a lock provides similar memory-ordering semantics,
	hence the ``release'' in ``release store''.
	(See also ``acquire load'' and ``memory barrier''.)
\item[\IXG{Scalability}:]
	A measure of how effectively a given system is able to utilize
	additional resources.
	For parallel computing, the additional resources are usually
	additional CPUs.
\item[\IXGh{Sequence}{Lock}:]
	A reader-writer synchronization mechanism in which readers
	retry their operations if a writer was present.
\item[\IXGh{Sequential}{Consistency}:]
	A memory-consistency model where all memory references appear to occur
	in an order consistent with
	a single global order, and where each CPU's memory references
	appear to all CPUs to occur in program order.
\item[Software Transactional Memory (HTM):]\glsuseriii{stm}
	A transactional-memory system capable running on computer systems
	without special hardware support.
	(See ``Transactional memory''.)
\item[\IXG{Starvation}:]
	A condition where at least one CPU or thread is unable to make
	progress due to an unfortunate series of resource-allocation
	decisions, as discussed in
	\cref{sec:locking:Livelock and Starvation}.
	For example, in a multisocket system, CPUs on one socket having
	privileged access to the data structure implementing a given lock
	could prevent CPUs on other sockets from ever acquiring that lock.
\item[\IXG{Starvation Free}:]
	A forward-progress guarantee in which, in the absence of
	failures, every thread makes progress within a finite
	period of time.
\item[\IXG{Store Buffer}:]
	A small set of internal registers used by a given CPU
	to record pending stores
	while the corresponding cache lines are making their
	way to that CPU\@.
	Also called ``store queue''.
\item[\IXG{Store Forwarding}:]
	An arrangement where a given CPU refers to its store buffer
	as well as its cache so as to ensure that the software sees
	the memory operations performed by this CPU as if they
	were carried out in program order.
\item[\IXGr{Superscalar CPU}:]
	A scalar (non-vector) CPU capable of executing multiple instructions
	concurrently.
	This is a step up from a pipelined CPU that executes multiple
	instructions in an assembly-line fashion---in a superscalar
	CPU, each stage of the pipeline would be capable of handling
	more than one instruction.
	For example, if the conditions were exactly right,
	the Intel Pentium Pro CPU from the mid-1990s could
	execute two (and sometimes three) instructions per clock cycle.
	Thus, a 200\,MHz Pentium Pro CPU could ``retire'', or complete the
	execution of, up to 400 million instructions per second.
\item[\IXG{Synchronization}:]
	Means for avoiding destructive interactions among CPUs or threads.
	Synchronization mechanisms include atomic RMW operations, memory
	barriers, locking, reference counting, hazard pointers, sequence
	locking, RCU, non-blocking synchronization, and transactional
	memory.
\item[\IXG{Teachable}:]
	A topic, concept, method, or mechanism that teachers believe that
	they understand completely and are therefore comfortable teaching.
\item[\IXG{Throughput}:]
	A performance metric featuring work items completed per unit time.
\item[Transactional Lock Elision (TLE):]\glsuseriii{tle}
	The use of transactional memory to emulate locking.
	Synchronization is instead carried out by conflicting accesses
	to the data to be protected by the lock.
	In some cases, this can increase performance because TLE
	avoids contention on the lock
	word~\cite{MartinPohlack2011HTM2TLE,Kleen:2014:SEL:2566590.2576793,PascalFelber2016rwlockElision,SeongJaePark2020HTMRCUlock}.
\item[Transactional Memory (TM):]\glsuseriii{tm}
	A synchronization mechanism that gathers groups of memory
	accesses so as to execute them atomically from the viewpoint
	of transactions on other CPUs or threads, discussed in
	\cref{sec:future:Transactional Memory,sec:future:Hardware Transactional Memory}.
\item[\IXG{Type-Safe Memory}:]
	Type-safe memory~\cite{Cheriton96a} is provided by a
	synchronization mechanism that prevents a given dynamically
	allocated object from changing to an incompatible type.
	Note that the object might well be freed and then reallocated, but
	the reallocated object is guaranteed to be of a compatible type.
	Within the Linux kernel, type-safe memory is provided within
	RCU read-side critical sections for memory allocated from slabs
	marked with the \co{SLAB_TYPESAFE_BY_RCU} flag.
	The strictly stronger existence guarantee also prevents freeing
	of the protected object.
\item[Unbounded Transactional Memory (UTM):]\glsuseriii{utm}
	A transactional-memory system based on hardware instructions
	provided for this purpose, but with special hardware or
	software capabilities that allow a given transaction to
	have a very large memory footprint.
	Such a system would at least partially avoid
	HTM's transaction-size limitations called out in
	\cref{sec:future:Transaction-Size Limitations}.
	(See ``Hardware transactional memory''.)
\item[\IXG{Unfairness}:]
	A condition where the progress of at least one CPU or thread
	is impeded by an unfortunate series of resource-allocation
	decisions, as discussed in
	\cref{sec:locking:Livelock and Starvation}.
	Extreme levels of unfairness are termed ``starvation''.
\item[\IXG{Unteachable}:]
	A topic, concept, method, or mechanism that the teacher does
	not understand well is therefore uncomfortable teaching.
\item[\IXGr{Vector CPU}:]
	A CPU that can apply a single instruction to multiple items of
	data concurrently.
	In the 1960s through the 1980s, only supercomputers had vector
	capabilities, but the advent of MMX in x86 CPUs and VMX in
	PowerPC CPUs brought vector processing to the masses.
\item[\IXG{Wait Free}:]
	A forward-progress guarantee in which every thread makes
	progress within a finite period of time.
\item[\IXGh{Write}{Memory Barrier}:]
	A memory barrier that is only guaranteed to affect the ordering
	of store instructions, that is, writes to memory.
	(See also ``memory barrier'' and ``read memory barrier''.)
\item[\IXGalth{Write Miss}{write}{cache miss}:]
	A cache miss incurred because the corresponding CPU attempted
	to write to a cache line that is read-only, most likely due
	to its being replicated in other CPUs' caches.
\item[\IXG{Write Mostly}:]
	Write-mostly data is (yet again, as the name implies) frequently
	updated.
\item[\IXGh{Write-Side}{Critical Section}:]
	A section of code guarded by write-acquisition of
	some reader-writer synchronization mechanism.
	For example, if one set of critical sections are guarded by
	write-acquisition of
	a given global reader-writer lock, while a second set of critical
	section are guarded by read-acquisition of that same reader-writer
	lock, then the first set of critical sections will be the
	write-side critical sections for that lock.
	Only one thread may execute in the write-side critical section
	at a time, and even then only if there are no threads are
	executing concurrently in any of the corresponding read-side
	critical sections.
\end{description}

% Bibliograpy
\IfTwoColumn{
  \onecolumn\begin{adjustwidth*}{.8in}{.7in}
  \bibliographystyle{alphapf} % Use alpha style customized by urlbst with --inlinelinks option
}{
  \bibliographystyle{alphapf} % Use alpha style customized by urlbst with --inlinelinks option
}
\IfColorLinks{}{
  \hypersetup{pdfborder=0 0 1,urlbordercolor=0.4 1 1}
}
\bibliography{bib/RCU,bib/WFS,bib/hw,bib/os,bib/parallelsys,bib/patterns,bib/perfmeas,bib/refs,bib/syncrefs,bib/search,bib/swtools,bib/realtime,bib/TM,bib/standards,bib/OSS,bib/maze,bib/energy,bib/QC,bib/NP,bib/memorymodel}
\IfColorLinks{}{
  \hypersetup{pdfborder=0 0 0}
}
\IfTwoColumn{
  \end{adjustwidth*}\twocolumn
}{
}

% Credits
\setcounter{secnumdepth}{-1} % surpress section numbering in backmatter
% ack.tex
% mainfile: perfbook.tex
% SPDX-License-Identifier: CC-BY-SA-3.0

\chapter{Credits}
\label{bck:ack:Credits}
\Epigraph{If I have seen further it is by standing on the shoulders of
	  giants.}{Isaac Newton, modernized}

% \section{Authors}

% CCASA 3.0 US wording from John Wiegley.
%  http://www.newartisans.com/blog_assets/git.from.bottom.up.pdf

\section{\LaTeX\ Advisor}

Akira Yokosawa is this book's \LaTeX\ advisor, which perhaps most
notably includes the care and feeding of the style guide laid out
in \cref{chp:app:styleguide:Style Guide}\@.
This work includes table layout, listings, fonts, rendering of math,
acronyms, bibliography formatting, epigraphs, hyperlinks, paper size.
Akira also perfected the cross-referencing of quick quizzes, allowing
easy and exact navigation between quick quizzes and their answers.
He also added build options that permit quick quizzes to be hidden
and to be gathered at the end of each chapter, textbook style.

This role also includes the build system, which Akira has optimized and
made much more user-friendly.
His enhancements have included automating response to bibliography
changes, automatically determining which source files are present,
and automatically generating listings (with automatically generated
hyperlinked line-number references) from the source files.

\section{Reviewers}

\begin{itemize}
\item	Alan Stern (\cref{chp:Advanced Synchronization: Memory Ordering}).
\item	Andy Whitcroft (\cref{sec:defer:RCU Fundamentals},
	\cref{sec:defer:RCU Linux-Kernel API}).
\item	Artem Bityutskiy (\cref{chp:Advanced Synchronization: Memory Ordering},
	\cref{chp:app:whymb:Why Memory Barriers?}).
\item	Dave Keck (\cref{chp:app:whymb:Why Memory Barriers?}).
\item	David S. Horner
	(\cref{sec:formal:Promela Parable: dynticks and Preemptible RCU}).
\item	Gautham Shenoy (\cref{sec:defer:RCU Fundamentals},
	\cref{sec:defer:RCU Linux-Kernel API}).
\item	``jarkao2'', AKA LWN guest \#41960 (\cref{sec:defer:RCU Linux-Kernel API}).
\item	Jonathan Walpole (\cref{sec:defer:RCU Linux-Kernel API}).
\item	Josh Triplett (\cref{chp:Formal Verification}).
\item	Michael Factor (\cref{sec:future:Transactional Memory}).
\item	Mike Fulton (\cref{sec:defer:RCU Fundamentals}).
\item	Peter Zijlstra
	(\cref{sec:defer:RCU Usage}). % Lanin and Shasha citation.
\item	Richard Woodruff (\cref{chp:app:whymb:Why Memory Barriers?}).
\item	Suparna Bhattacharya (\cref{chp:Formal Verification}).
\item	Vara Prasad
	(\cref{sec:formal:Promela Parable: dynticks and Preemptible RCU}).
\end{itemize}

Reviewers whose feedback took the extremely welcome form of a patch
are credited in the git logs.

\section{Machine Owners}

Readers might have noticed some graphs showing scalability data out
to several hundred CPUs, courtesy of my current employer, with special
thanks to Paul Saab, Yashar Bayani, Joe Boyd, and Kyle McMartin.

From back in my time at IBM,
a great debt of thanks goes to Martin Bligh, who originated the
Advanced Build and Test (ABAT) system at IBM's Linux Technology
Center, as well as to Andy Whitcroft, Dustin Kirkland, and many
others who extended this system.
Many thanks go also to a great number of machine owners:
Andrew Theurer,
Andy Whitcroft,
Anton Blanchard,
Chris McDermott,
Cody Schaefer,
Darrick Wong,
David ``Shaggy'' Kleikamp,
Jon M. Tollefson,
Jose R. Santos,
Marvin Heffler,
Nathan Lynch,
Nishanth Aravamudan,
Tim Pepper,
and
Tony Breeds.

\section{Original Publications}

	\begin{enumerate}
	% mainfile: perfbook.tex
\OrigPubItem{Section}{sec:intro:What Makes Parallel Programming Hard?}{What Makes Parallel Programming Hard?}{a Portland State University Technical Report}{PaulEMcKenney2009ProgrammingHard}
\OrigPubItem{Section}{sec:toolsoftrade:Shared-Variable Shenanigans}{Shared-Variable Shenanigans}{Linux Weekly News}{JadeAlglave2019WhoAfraidCompiler}
\OrigPubItem{Section}{sec:SMPdesign:Beyond Partitioning}{Retrofitted Parallelism Considered Grossly Sub-Optimal}{4\textsuperscript{th} USENIX Workshop on Hot Topics on Parallelism}{PaulEMcKenney2012HOTPARsuboptimal}
\OrigPubItem{Section}{sec:defer:RCU Fundamentals}{RCU Fundamentals}{Linux Weekly News}{PaulEMcKenney2007WhatIsRCUFundamentally}
\OrigPubItem{Section}{sec:defer:RCU Linux-Kernel API}{RCU Linux-Kernel API}{Linux Weekly News}{PaulEMcKenney2008WhatIsRCUAPI}
\OrigPubItem{Section}{sec:defer:RCU Usage}{RCU Usage}{Linux Weekly News}{PaulEMcKenney2008WhatIsRCUUsage}
\OrigPubItem{Section}{sec:defer:RCU Related Work}{RCU Related Work}{Linux Weekly News}{PaulEMcKenney2014ReadMostly}
\OrigPubItem{Section}{sec:defer:RCU Related Work}{RCU Related Work}{Linux Weekly News}{PaulEMcKenney2015ReadMostly}
\OrigPubItem{Chapter}{chp:Formal Verification}{Formal Verification}{Linux Weekly News}{PaulEMcKenney2007QRCUspin,PaulEMcKenney2008dynticksRCU,PaulEMcKenney2011ppcmem}
\OrigPubItem{Section}{sec:formal:Axiomatic Approaches}{Axiomatic Approaches}{Linux Weekly News}{PaulEMcKenney2014weakaxiom}
\OrigPubItem{Section}{sec:together:Correlated Fields}{Correlated Fields}{Oregon Graduate Institute}{PaulEdwardMcKenneyPhD}
\OrigPubItem{Chapter}{chp:Advanced Synchronization: Memory Ordering}{Advanced Synchronization: Memory Ordering}{the Linux kernel}{Howells2009membartxt}
\OrigPubItem{Chapter}{chp:Advanced Synchronization: Memory Ordering}{Advanced Synchronization: Memory Ordering}{Linux Weekly News}{JadeAlglave2017LWN-LKMM-1,JadeAlglave2017LWN-LKMM-2}
\OrigPubItem{Chapter}{chp:Advanced Synchronization: Memory Ordering}{Advanced Synchronization: Memory Ordering}{ASPLOS '18}{Alglave:2018:FSC:3173162.3177156}
\OrigPubItem{Section}{sec:memorder:Address- and Data-Dependency Difficulties}{Address- and Data-Dependency Difficulties}{the Linux kernel}{PaulEMcKenney2014rcu-dereference}
\OrigPubItem{Section}{sec:memorder:Hardware Specifics}{Memory-Barrier Instructions For Specific CPUs}{Linux Journal}{PaulMcKenney2005i,PaulMcKenney2005j}

	\end{enumerate}

\section{Figure Credits}

	\begin{enumerate}
	% mainfile: perfbook.tex
\ContribItem{Figure}{fig:cpu:CPU Performance at its Best}{Melissa Broussard}
\ContribItem{Figure}{fig:cpu:CPUs Old and New}{Melissa Broussard}
\ContribItem{Figure}{fig:cpu:CPU Meets a Pipeline Flush}{Melissa Broussard}
\ContribItem{Figure}{fig:cpu:CPU Meets a Memory Reference}{Melissa Broussard}
\ContribItem{Figure}{fig:cpu:CPU Meets an Atomic Operation}{Melissa Broussard}
\ContribItem{Figure}{fig:cpu:CPU Meets a Memory Barrier}{Melissa Broussard}
\ContribItem{Figure}{fig:cpu:Encounters Thermal Throttling}{Melissa Broussard, remixed}
\ContribItem{Figure}{fig:cpu:CPU Meets a Cache Miss}{Melissa Broussard}
\ContribItem{Figure}{fig:cpu:CPU Waits for I/O Completion}{Melissa Broussard}
\ContribItem{Figure}{fig:cpu:Hardware and Software: On Same Side}{Melissa Broussard}
\ContribItem{Figure}{fig:count:Waiting to Count}{Melissa Broussard}
\ContribItem{Figure}{fig:SMPdesign:Dining Philosophers Problem}{Kornilios Kourtis}
\ContribItem{Figure}{fig:cpu:Partial Starvation Is Also Bad}{Melissa Broussard}
\ContribItem{Figure}{fig:SMPdesign:Dining Philosophers Problem; Textbook Solution}{Kornilios Kourtis}
\ContribItem{Figure}{fig:SMPdesign:Dining Philosophers Problem; Partitioned}{Kornilios Kourtis}
\ContribItem{Figure}{fig:SMPdesign:Lock Contention}{Melissa Broussard}
\ContribItem{Figure}{fig:SMPdesign:Data Locking}{Melissa Broussard}
\ContribItem{Figure}{fig:SMPdesign:Data and Skew}{Melissa Broussard}
\ContribItem{Figure}{fig:locking:Locking: Villain or Slob?}{Melissa Broussard}
\ContribItem{Figure}{fig:locking:Locking: Workhorse or Hero?}{Melissa Broussard}
\ContribItem{Figure}{fig:datastruct:Even Veterinarians Disagree}{Melissa Broussard}
\ContribItem{Figure}{fig:datastruct:Partitioning Problems}{Melissa Broussard}
\ContribItem{Figure}{fig:debugging:Validation and the Geneva Convention}{Melissa Broussard}
\ContribItem{Figure}{fig:debugging:Rationalizing Validation}{Melissa Broussard}
\ContribItem{Figure}{fig:cpu:Passed-the-stress-test}{Melissa Broussard}
\ContribItem{Figure}{fig:debugging:Choose Validation Methods Wisely}{Melissa Broussard}
\ContribItem{Figure}{fig:advsync:Hard Real-Time Response; Meet Hammer}{Melissa Broussard}
\ContribItem{Figure}{fig:advsync:Real-Time Response: Hardware Matters}{Melissa Broussard}
\ContribItem{Figure}{fig:advsync:Real-Time Response: Notification Insufficient}{Melissa Broussard}
\ContribItem{Figure}{fig:advsync:Timer Wheel at 1kHz}{Melissa Broussard}
\ContribItem{Figure}{fig:advsync:Timer Wheel at 100kHz}{Melissa Broussard}
\ContribItem{Figure}{fig:advsync:Priority Inversion and User Input}{Melissa Broussard}
\ContribItem{Figure}{fig:advsync:The Dark Side of Real-Time Computing}{Sarah McKenney}
\ContribItem{Figure}{fig:advsync:The Dark Side of Real-Fast Computing}{Sarah McKenney}
\ContribItem{Figure}{fig:memorder:Intel Core 2 Architecture}{Wikipedia user “I, Appaloosa” CC BY-SA 3.0, reformatted}
\ContribItem{Figure}{fig:memorder:Intel Core 2 Architecture Simplified}{Wikipedia user “I, Appaloosa” CC BY-SA 3.0, reformatted}
\ContribItem{Figure}{fig:memorder:CPUs Can Do Things Out of Order}{Melissa Broussard}
\ContribItem{Figure}{fig:memorder:A Variable With More Simultaneous Values}{Akira Yokosawa}
\ContribItem{Figure}{fig:memorder:Half Memory Barrier}{Melissa Brossard}
\ContribItem{Figure}{fig:easy:Shaving the Mandelbrot Set}{Melissa Broussard}
\ContribItem{Figure}{fig:future:Uniprocessor \"Uber Alles}{Melissa Broussard}
\ContribItem{Figure}{fig:future:Multithreaded Mania}{Melissa Broussard}
\ContribItem{Figure}{fig:future:More of the Same}{Melissa Broussard}
\ContribItem{Figure}{fig:future:Crash Dummies Slamming into the Memory Wall}{Melissa Broussard}
\ContribItem{Figure}{fig:future:Astounding Accelerators}{Melissa Broussard, remixed}
\ContribItem{Figure}{fig:future:The STM Vision}{Melissa Broussard}
\ContribItem{Figure}{fig:future:The STM Reality: Conflicts}{Melissa Broussard}
\ContribItem{Figure}{fig:future:The STM Reality: Irrevocable Operations}{Melissa Broussard}
\ContribItem{Figure}{fig:future:The STM Reality: Realtime Response}{Melissa Broussard}
\ContribItem{Figure}{fig:summary:The Most Important Lesson}{Melissa Broussard}
\ContribItem{Figure}{fig:app:questions:What Time Is It?}{Melissa Broussard}
\ContribItem{Figure}{fig:SMPdesign:Dining Philosophers Problem; Fully Partitioned}{Kornilios Kourtis}

	\end{enumerate}

\Cref{fig:defer:RCU Areas of Applicability} was adapted from
\ppl{Fedor}{Pikus}'s ``When to use RCU'' slide~\cite{FedorPikus2017RCUthenWhat}.
The discussion of mechanical reference counters in
\cref{sec:defer:Reference Counting}
stemmed from a private conversation with \ppl{Dave}{Regan}.

\section{Other Support}

We owe thanks to many CPU architects for patiently explaining the
instruction- and memory-reordering features of their CPUs, particularly
Wayne Cardoza, Ed Silha, Anton Blanchard, Tim Slegel, Juergen Probst,
Ingo Adlung, Ravi Arimilli, Cathy May, Derek Williams,
H.~Peter Anvin,
Andy Glew, Leonid Yegoshin,
Richard Grisenthwaite, and Will Deacon.
Wayne deserves special thanks for his patience in explaining Alpha's reordering
of dependent loads, a lesson that Paul resisted quite strenuously!

The bibtex-generation service of the Association for Computing Machinery
has saved us a huge amount of time and effort compiling the bibliography,
for which we are grateful.
Thanks are also due to Stamatis Karnouskos, who convinced me to drag my
antique bibliography database kicking and screaming into the
21\textsuperscript{st} century.
Any technical work of this sort owes thanks to the many individuals and
organizations that keep Internet and the World Wide Web up and running,
and this one is no exception.

Portions of this material are based upon work supported by the National
Science Foundation under Grant No.\@ CNS-0719851.

% Index if enabled
\IfIndexOn{
\IfIndexHier{
\index{Anti-Heisenbug|see{Heisenbug, anti-}}
\index{Associativity|see{Cache associativity}}
\index{Associativity miss|see{Cache miss, associativity}}
\index{Bounded wait free|see{Wait free, bounded}}
\index{Bounded population-oblivious wait free|see{Wait free, bounded population-oblivious}}
\index{Cache-invalidation latency|see{Latency, cache-invalidation}}
\index{Cache-miss latency|see{Latency, cache-miss}}
\index{Capacity miss|see{Cache miss, capacity}}
\index{Code locking|see{Locking, code}}
\index{Communication miss|see{Cache miss, communication}}
\index{Data locking|see{Locking, data}}
\index{Direct-mapped cache|see{Cache, direct-mapped}}
\index{Exclusive lock|see{Lock, exclusive}}
\index{Full memory barrier|see{Memory barrier, full}}
\index{Fully associative cache|see{Cache, fully associative}}
\index{Grace-period latency|see{Latency, grace-period}}
\index{Memory consistency|see{Consistency, memory}}
\index{Memory latency|see{Latency, memory}}
\index{Memory-barrier latency|see{Latency, memory-barrier}}
\index{Memory-barrier overhead|see{Overhead, memory-barrier}}
\index{Message latency|see{Latency, message}}
\index{Process consistency|see{Consistency, process}}
\index{RCU read-side critical section|see{Critical section, RCU read-side}}
\index{Read memory barrier|see{Memory barrier, read}}
\index{Read-side critical section|see{Critical section, read-side}}
\index{Reader-writer lock|see{Lock, reader-writer}}
\index{Scheduling latency|see{Latency, scheduling}}
\index{Sequence lock|see{Lock, sequence}}
\index{Sequential consistency|see{Consistency, sequential}}
\index{Weak consistency|see{Consistency, weak}}
\index{Write memory barrier|see{Memory barrier, write}}
\index{Write miss|see{Cache miss, write}}
\index{Write-side critical section|see{Critical section, write-side}}

}{}
\phantomsection
\IfEbookSize{
\idxlayout{columns=2}
}{
\newgeometry{body={6.5in,8.25in},centering=true,twocolumn,columnsep=0.25in}
}
\printglossary[type=\acronymtype]
\phantomsection
\setindexprenote{%
  \hfill\begin{minipage}{1in}
    \footnotesize%
    \BF{Bold}: Major reference.
    \GL{Underline}: Definition.
    \vspace*{1ex}
  \end{minipage}
}
\printindex
\phantomsection
\setindexprenote{%
  \hfill\begin{minipage}{2.6in}
    \footnotesize (c):~Cxx standard, (g):~GCC extension,
    (k):~Linux kernel, (kh):~Linux kernel historic, (pf):~perfbook CodeSamples,
    (px):~POSIX, (ur):~userspace RCU\@.
    \vspace*{1ex}
  \end{minipage}
}
\printindex[api]
\IfEbookSize{}{
\restoregeometry
}
}{}

% page-layout dimensions
% \cleardoublepage
% ~ \\
% paperheight: \the\paperheight \\
% paperwidth: \the\paperwidth \\
% topmargin: \the\topmargin \\
% headheight: \the\headheight \\
% headsep: \the\headsep \\
% textheight: \the\textheight \\
% textwidth: \the\textwidth \\
% oddsidemargin: \the\oddsidemargin \\
% evensidemargin: \the\evensidemargin \\
% footskip: \the\footskip \\
% topskip: \the\topskip \\

\end{document}